\pdfoutput=1
\RequirePackage[log]{snapshot}
\documentclass[12pt,twoside]{article}

\usepackage[utf8]{inputenc}
\usepackage[T1]{fontenc}
\usepackage{amsmath}
\usepackage{amssymb}
\usepackage[nottoc,notlot,notlof]{tocbibind}
\usepackage{calc}
\usepackage[pdftex]{graphicx} %
\usepackage[figuresright]{rotating}

\usepackage{authblk}
\usepackage{caption}
\usepackage{array}
\usepackage{longtable}
\usepackage{etoolbox}
\usepackage{float}

\usepackage{adjustbox}
\usepackage{tabularx}

\usepackage{multirow}
\usepackage{booktabs}
\usepackage{color, colortbl}
\usepackage{footmisc}
\usepackage{afterpage}
\usepackage{relsize}                                      %
\usepackage{wasysym}
\usepackage[math]{cellspace}
\usepackage{lineno}  %

\definecolor{Gray}{gray}{0.9}
\definecolor{LightGray}{gray}{0.97}

\usepackage{silence}
\usepackage{hyperref}
\usepackage[all]{hypcap} 
\usepackage{cite}    %
\usepackage{mciteplus}

\newcommand{\mysection}[1]{\section{\boldmath #1}}
\newcommand{\mysubsection}[1]{\subsection[#1]{\boldmath #1}}
\newcommand{\mysubsubsection}[1]{\subsubsection[#1]{\boldmath #1}}
\newcommand{\mysubsubsubsection}[1]{\subsubsubsection{\boldmath #1}}

\def\dof{{\rm dof}}

\newcommand\VCKM{{V}}
\newcommand\etacpf{{\eta_f}}
\newcommand\etacp{{\eta}}

\renewcommand\Im{{\rm Im}} 
\renewcommand\Re{{\rm Re}}

\newcommand\Abar{\kern 0.18em\overline{\kern -0.18em A}{}}
\newcommand\Af{A_f}
\newcommand\Abarf{\Abar_f}
\newcommand\Afbar{A_{\bar f}}
\newcommand\Abarfbar{\Abar_{\bar f}}
\newcommand\Acp{{\cal A}}
\newcommand\Adirnoncp{\ensuremath{\langle{\cal A}_{f\bar f}\rangle}\xspace}

\newcommand\mc{\multicolumn}

\newcommand {\cbf}{\ensuremath{{\cal B}}}

\newcommand {\vcb}{\ensuremath{|V_{cb}|}}
\newcommand {\vub}{\ensuremath{|V_{ub}|}}

\def\Bp      {\ensuremath{B^{+}}}
\def\Bm      {\ensuremath{B^{-}}}
\def\Bz      {\ensuremath{B^{0}}}
\def\Bs      {\ensuremath{B_{s}}}

\newcommand{\BzbDplnu}    {\ensuremath{\Bzb \to D^{+}\ell^{-}\nub_\ell}}
\newcommand{\BzbDstarlnu} {\ensuremath{\Bzb \to D^{*+}\ell^{-}\nub_\ell}}

\newcommand {\rhoz} {\ensuremath{\rho^0}\hbox{ }}

\usepackage{relsize}

\def\beq{\begin{equation}}
\def\eeq#1{\label{#1}\end{equation}}
\def\eeqn{\end{equation}}

\def\beqa{\begin{eqnarray}}
\def\eeqa#1{\label{#1}\end{eqnarray}}
\def\eeqan{\end{eqnarray}}

\let\bar=\overbar

\def\ie{{\it i.e.}}
\def\eg{{\it e.g.}}
\def\etc{{\it etc.}}
\def\cf{{\it cf.}}

\def\Dslash{\ensuremath{\not{\hbox{\kern-4pt $D$}}}\xspace}
\def\dslash{\not{\hbox{\kern-2pt $\del$}}}

\def\BR{\mbox{\rm BR}}
\def\ee{e^+e^-}

\def\alphas{\alpha_s}
\def\msb{{\bar{\ssstyle M \kern -1pt S}}}

\RequirePackage{xspace}

\usepackage{relsize}
\def\babar{\mbox{\slshape B\kern-0.1em{\smaller A}\kern-0.1em
    B\kern-0.1em{\smaller A\kern-0.2em R}}\xspace}
\def\belle{\mbox{\normalfont Belle}\xspace}
\def\belletwo   {\mbox{Belle II}\xspace}

\def\dzero{\mbox{\normalfont D0}\xspace} %

\def\epem       {\ensuremath{e^+e^-}\xspace}
\def\ee         {\ensuremath{e^-e^-}\xspace}

\def\mup        {\ensuremath{\mu^+}\xspace}
\def\mun        {\ensuremath{\mu^-}\xspace}    %
\def\mumu       {\ensuremath{\mu^+\mu^-}\xspace}
\def\mtau       {\ensuremath{\tau}\xspace}

\def\nub        {\ensuremath{\overline{\nu}}\xspace}

\def\nub        {\ensuremath{\overline{\nu}}\xspace}

\def\nul        {\ensuremath{\nu_\ell}\xspace}

\def\g     {\ensuremath{\gamma}\xspace}

\def\Z      {\ensuremath{Z^0}\xspace}

\def\ubar  {\ensuremath{\overline u}\xspace}

\def\dbar  {\ensuremath{\overline d}\xspace}
\def\ddbar {\ensuremath{d\overline d}\xspace}

\def\sbar  {\ensuremath{\overline s}\xspace}

\def\b  {\ensuremath{b}\xspace}
\def\bbar  {\ensuremath{\overline b}\xspace}

\def\piz   {\ensuremath{\pi^0}\xspace}

\def\pip   {\ensuremath{\pi^+}\xspace}
\def\pim   {\ensuremath{\pi^-}\xspace}
\def\pipi  {\ensuremath{\pi^+\pi^-}\xspace}
\def\pipm  {\ensuremath{\pi^\pm}\xspace}
\def\pimp  {\ensuremath{\pi^\mp}\xspace}

\def\etapr {\ensuremath{\eta^{\prime}}\xspace}

\def\Kbar  {\kern 0.2em\overline{\kern -0.2em K}{}\xspace}

\def\Kpm   {\ensuremath{K^\pm}\xspace}
\def\Kmp   {\ensuremath{K^\mp}\xspace}
\def\Kp    {\ensuremath{K^+}\xspace}
\def\Km    {\ensuremath{K^-}\xspace}
\def\KS    {\ensuremath{K^0_{\scriptscriptstyle S}}\xspace} 
\def\KL    {\ensuremath{K^0_{\scriptscriptstyle L}}\xspace} 
\def\Kstarz  {\ensuremath{K^{*0}}\xspace}
\def\Kstarzb  {\ensuremath{\Kbar^{*0}}\xspace}
\def\Kstar   {\ensuremath{K^*}\xspace}

\def\Kstarp   {\ensuremath{K^{*+}}\xspace}

\def\Kz   {\ensuremath{K^0}\xspace}
\def\Kzb   {\ensuremath{\Kbar^0}\xspace}
\def\KzKzb {\ensuremath{K^0 \kern -0.16em \Kzb}\xspace}

\def\KorKstar   {\ensuremath{K^{(*)}}\xspace}

\def\KorKstarp  {\ensuremath{K^{(*)+}}\xspace}

\def\Dz    {\ensuremath{D^0}\xspace}
\def\Dbar  {\kern 0.2em\overline{\kern -0.2em D}{}\xspace}

\def\Dzb   {\ensuremath{\Dbar^0}\xspace}
\def\DzDzb {\ensuremath{D^0 {\kern -0.16em \Dzb}}\xspace}
\def\Dp    {\ensuremath{D^+}\xspace}
\def\Dm    {\ensuremath{D^-}\xspace}

\def\Dmp   {\ensuremath{D^\mp}\xspace}

\def\Dstar   {\ensuremath{D^*}\xspace}

\def\Dstarp  {\ensuremath{D^{*+}}}
\def\Dstarm  {\ensuremath{D^{*-}}}
\def\DorDstar   {\ensuremath{D^{(*)}}\xspace}
\def\DorDstarz  {\ensuremath{D^{(*)0}}\xspace}
\def\DorDstarzb {\ensuremath{\Dbar^{(*)0}}\xspace}

\def\Ds    {\ensuremath{D^+_s}\xspace}
\def\Dsp   {\ensuremath{D^+_s}\xspace}
\def\Dsm   {\ensuremath{D^-_s}\xspace}

\def\Bz    {\ensuremath{B^0}\xspace}
\def\B     {\ensuremath{B}\xspace}
\def\Bbar  {\kern 0.18em\overline{\kern -0.18em B}{}\xspace}
\def\Bb    {\ensuremath{\Bbar}\xspace}
\def\Bzb   {\ensuremath{\Bbar^0}\xspace}
\def\Bu    {\ensuremath{B^+}\xspace}

\def\Bpm   {\ensuremath{B^\pm}\xspace}
\def\Bmp   {\ensuremath{B^\mp}\xspace}
\def\Bs    {\ensuremath{B_s}\xspace}
\def\Bsb   {\ensuremath{\Bbar_s^0}\xspace}
\def\BB    {\ensuremath{B\Bbar}\xspace} 
\def\BzBzb {\ensuremath{B^0 {\kern -0.16em \Bzb}}\xspace}

\def\jpsi  {\ensuremath{{J\mskip -3mu/\mskip -2mu\psi\mskip 2mu}}\xspace}

\mathchardef\Upsilon="7107
\def\Y#1S{\ensuremath{\Upsilon{(#1S)}}\xspace}%

\mathchardef\Deltares="7101
\mathchardef\Xi="7104
\mathchardef\Lambda="7103
\mathchardef\Sigma="7106
\mathchardef\Omega="710A
\def\Deltabar   {\kern 0.25em\overline{\kern -0.25em \Deltares}{}\xspace}
\def\Lbar {\kern 0.2em\overline{\kern -0.2em\Lambda\kern 0.05em}\kern-0.05em{}\xspace}
\def\Sigbar{\kern 0.2em\overline{\kern -0.2em \Sigma}{}\xspace}
\def\Xibar{\kern 0.2em\overline{\kern -0.2em \Xi}{}\xspace}
\def\Obar{\kern 0.2em\overline{\kern -0.2em \Omega}{}\xspace}
\def\Nbar{\kern 0.2em\overline{\kern -0.2em N}{}\xspace}
\def\Xb{\kern 0.2em\overline{\kern -0.2em X}{}}

\newcommand{\particle}[1]{\ensuremath{#1}\xspace}
\renewcommand{\ee}{\particle{e^+e^-}}
\newcommand{\Ups}{\particle{\Upsilon(4S)}}
\newcommand{\Upsfive}{\particle{\Upsilon(5S)}}
\renewcommand{\b}{\particle{b}}
\renewcommand{\B}{\particle{B}}
\newcommand{\Bd}{\particle{B^0}}
\renewcommand{\Bs}{\particle{B^0_s}}
\renewcommand{\Bu}{\particle{B^+}}
\newcommand{\Bc}{\particle{B^+_c}}
\newcommand{\Bdbar}{\particle{\bar{B}^0}}
\newcommand{\Bsbar}{\particle{\bar{B}^0_s}}
\newcommand{\Lb}{\particle{\Lambda_b^0}}
\newcommand{\Xib}{\particle{\Xi_b}}
\newcommand{\Xibd}{\particle{\Xi_b^-}}
\newcommand{\Xibu}{\particle{\Xi_b^0}}
\newcommand{\Omegab}{\particle{\Omega_b^-}}
\newcommand{\Lc}{\particle{\Lambda_c^+}}

\def\BR{{\ensuremath{\cal B}}}

\def\Btopilnu   {\ensuremath{B \to \pi\ell\nu}}

\newcommand{\tev}{\ensuremath{\mathrm{Te\kern -0.1em V}}\xspace}
\newcommand{\gev}{\ensuremath{\mathrm{Ge\kern -0.1em V}}\xspace}
\newcommand{\mev}{\ensuremath{\mathrm{Me\kern -0.1em V}}\xspace}
\newcommand{\kev}{\ensuremath{\mathrm{ke\kern -0.1em V}}\xspace}
\newcommand{\ev}{\ensuremath{\mathrm{e\kern -0.1em V}}\xspace}
\newcommand{\gevc}{\ensuremath{{\mathrm{Ge\kern -0.1em V\!/}c}}\xspace}
\newcommand{\mevc}{\ensuremath{{\mathrm{Me\kern -0.1em V\!/}c}}\xspace}
\newcommand{\gevcc}{\ensuremath{{\mathrm{Ge\kern -0.1em V\!/}c^2}}\xspace}
\newcommand{\gevgevcccc}{\ensuremath{{\mathrm{Ge\kern -0.1em V^2\!/}c^4}}\xspace}
\newcommand{\mevcc}{\ensuremath{{\mathrm{Me\kern -0.1em V\!/}c^2}}\xspace}

\def\pb {\ensuremath{\rm \,pb}\xspace}

\def\fb   {\ensuremath{\mbox{\,fb}}\xspace}
\def\invfb   {\ensuremath{\mbox{\,fb}^{-1}}\xspace}
\def\mus  {\ensuremath{\rm \,\mus}\xspace}

\def\ps   {\ensuremath{\rm \,ps}\xspace}

\def\mus        {\ensuremath{\,\mu{\rm s}}\xspace}    %
\def\ps         {\ensuremath{{\rm \,ps}}\xspace}  %

\def\degrees{\ensuremath{^{\circ}}\xspace}

\def\gsim{{~\raise.15em\hbox{$>$}\kern-.85em
          \lower.35em\hbox{$\sim$}~}\xspace}
\def\lsim{{~\raise.15em\hbox{$<$}\kern-.85em
          \lower.35em\hbox{$\sim$}~}\xspace}

\def\CP                 {\ensuremath{C\!P}\xspace}
\def\CPT                {\ensuremath{C\!PT}\xspace}
\def\ra                 {\ensuremath{\to}\xspace}
\def\pep2{PEP-II}

\def\rhobar {\ensuremath{\overline{\rho}}\xspace}
\def\etabar {\ensuremath{\overline{\eta}}\xspace}

\def\Vud  {\ensuremath{|V_{ud}|}\xspace}

\def\Vus  {\ensuremath{|V_{us}|}\xspace}

\def\Vub  {\ensuremath{|V_{ub}|}\xspace}
\def\Vcb  {\ensuremath{|V_{cb}|}\xspace}

\def\stwob{\ensuremath{\sin\! 2 \beta   }\xspace}

\def\deltamd{\ensuremath{{\rm \Delta}m_d}\xspace}

\xspace
\newcommand{\fds}{\ensuremath{f_{D_s}}\xspace}

\def\jetset74   {\mbox{\tt Jetset \hspace{-0.5em}7.\hspace{-0.2em}4}}
\newcommand{\aerr}[4]   {\mbox{${{#1}^{+ #2}_{- #3}\pm #4}$}}
\newcommand{\berr}[4]   {\mbox{${{#1}\pm #2^{+ #3}_{- #4}}$}}
\newcommand{\cerr}[3]   {\mbox{${{#1}^{+ #2}_{- #3}}$}}
\newcommand{\aerrsy}[5] {\mbox{${{#1}^{+ #2 + #4}_{- #3 - #5}}$}}

\newcommand{\cerrsyt}[5] {\mbox{${{#1}^{+ #2}_{- #3}\pm{#4}\pm{#5}}$}}
\newcommand{\derrsyt}[5] {\mbox{${{#1}\pm{#2}\pm{#3}^{+ #4}_{- #5}}$}}
\newcommand{\ferrsyt}[5] {\mbox{${{#1}\pm{#2}\pm{#3}\pm{#4}\pm{#5}}$}}
\newcommand{\gerrsyt}[4] {\mbox{${{#1}\pm{#2}\pm{#3}\pm{#4}}$}}
\newcommand{\err}[3]   {\mbox{${{#1}\pm{#2}\pm{#3}}$}}

\newcommand{\nodata}{$$}
\newcommand{\vs}{\mbox{$vs.$}}

\def\sgline{\noalign{\vskip 0.10truecm\hrule\vskip 0.10truecm}}
\def\sglinespt{\noalign{\vskip 0.05truecm\hrule}}
\def\sglinespb{\noalign{\hrule\vskip 0.05truecm}}
\newcommand{\ks}    {\KS}
\newcommand{\kz}    {\Kz}
\newcommand{\kzb}   {\Kzb}

\newboolean{unpublished}\setboolean{unpublished}{false}

\newboolean{history}\setboolean{history}{false}

\newcommand{\unpublished}[2]{\ifthenelse{\boolean{unpublished}}{#2}{#1}}
\newcommand{\history}[2]{\ifthenelse{\boolean{history}}{#2}{#1}}
\newcommand{\citehistory}[2]{\ifthenelse{\boolean{history}}{\cite{#2}}{\cite{#1}}}

\newcommand{\definemath}[2]{\newcommand{#1}{\ensuremath{#2}\xspace}}

\definemath{\hflavCHIBARLEPval}{0.1259}%
\definemath{\hflavCHIBARLEPerr}{\pm0.0042}%
\definemath{\hflavNSIGMATAULBEXCLSEMI}{3.1}%
\definemath{\hflavTAUBDval}{1.519}%
\definemath{\hflavTAUBDerr}{\pm0.004}%
\definemath{\hflavTAUBUval}{1.638}%
\definemath{\hflavTAUBUerr}{\pm0.004}%
\definemath{\hflavRTAUBUval}{1.076}%
\definemath{\hflavRTAUBUerr}{\pm0.004}%
\definemath{\hflavTAUBSval}{1.510}%
\definemath{\hflavTAUBSerr}{\pm0.004}%
\definemath{\hflavRTAUBSval}{0.994}%
\definemath{\hflavRTAUBSerr}{\pm0.004}%
\definemath{\hflavTAULBval}{1.471}%
\definemath{\hflavTAULBerr}{\pm0.009}%
\definemath{\hflavTAULBSval}{1.247}%
\definemath{\hflavTAULBSerp}{^{+0.071}}%
\definemath{\hflavTAULBSern}{_{-0.069}}%
\definemath{\hflavTAULBEval}{1.471}%
\definemath{\hflavTAULBEerr}{\pm0.009}%
\definemath{\hflavTAUXBDval}{1.572}%
\definemath{\hflavTAUXBDerr}{\pm0.040}%
\definemath{\hflavTAUXBUval}{1.480}%
\definemath{\hflavTAUXBUerr}{\pm0.030}%
\definemath{\hflavTAUOBval}{1.64}%
\definemath{\hflavTAUOBerp}{^{+0.18}}%
\definemath{\hflavTAUOBern}{_{-0.17}}%
\definemath{\hflavTAUBCval}{0.510}%
\definemath{\hflavTAUBCerr}{\pm0.009}%
\definemath{\hflavTAUBSSLval}{1.527}%
\definemath{\hflavTAUBSSLerr}{\pm0.011}%
\definemath{\hflavTAUBSMEANCval}{1.510}%
\definemath{\hflavTAUBSMEANCerr}{\pm0.004}%
\definemath{\hflavTAUBSJFval}{1.480}%
\definemath{\hflavTAUBSJFerr}{\pm0.007}%
\definemath{\hflavRTAUBSSLval}{1.006}%
\definemath{\hflavRTAUBSSLerr}{\pm0.008}%
\definemath{\hflavRTAUBSMEANCval}{0.994}%
\definemath{\hflavRTAUBSMEANCerr}{\pm0.004}%
\definemath{\hflavRTAUBSMEANCsig}{1.6}%
\definemath{\hflavONEMINUSRTAUBSMEANCpercent}{(0.6\pm0.4)\%}%
\definemath{\hflavRTAULBval}{0.969}%
\definemath{\hflavRTAULBerr}{\pm0.006}%
\definemath{\hflavTAUBVTXval}{1.572}%
\definemath{\hflavTAUBVTXerr}{\pm0.009}%
\definemath{\hflavTAUBLEPval}{1.537}%
\definemath{\hflavTAUBLEPerr}{\pm0.020}%
\definemath{\hflavTAUBSMMval}{2.04}%
\definemath{\hflavTAUBSMMerr}{\pm0.44}%
\definemath{\hflavTAUBJPval}{1.533}%
\definemath{\hflavTAUBJPerr}{\pm0.036}%
\definemath{\hflavNSIGMATAULBCDFTWO}{2.4}%
\definemath{\hflavRTAUXBUXBDval}{0.929}%
\definemath{\hflavRTAUXBUXBDerr}{\pm0.028}%
\definemath{\hflavSDGDGDval}{0.001}%
\definemath{\hflavSDGDGDerr}{\pm0.010}%
\definemath{\hflavTAUBSJPSIPIPIval}{1.666}%
\definemath{\hflavTAUBSJPSIPIPIerr}{\pm0.024}%
\definemath{\hflavTAUBSJPSIKSHORTval}{1.75}%
\definemath{\hflavTAUBSJPSIKSHORTerr}{\pm0.14}%
\definemath{\hflavTAUBSLONGval}{1.666}%
\definemath{\hflavTAUBSLONGerr}{\pm0.024}%
\definemath{\hflavTAUBSKKval}{1.408}%
\definemath{\hflavTAUBSKKerr}{\pm0.017}%
\definemath{\hflavTAUBSDSDSval}{1.379}%
\definemath{\hflavTAUBSDSDSerr}{\pm0.031}%
\definemath{\hflavTAUBSJPSIETAval}{1.479}%
\definemath{\hflavTAUBSJPSIETAerr}{\pm0.036}%
\definemath{\hflavTAUBSSHORTval}{1.422}%
\definemath{\hflavTAUBSSHORTerr}{\pm0.023}%
\definemath{\hflavGSval}{0.6640}%
\definemath{\hflavGSerr}{\pm0.0020}%
\definemath{\hflavTAUBSMEANval}{1.506}%
\definemath{\hflavTAUBSMEANerr}{\pm0.005}%
\definemath{\hflavDGSGSval}{+0.128}%
\definemath{\hflavDGSGSerr}{\pm0.009}%
\definemath{\hflavDGSval}{+0.085}%
\definemath{\hflavDGSerr}{\pm0.006}%
\definemath{\hflavRHOGSDGS}{-0.193}%
\definemath{\hflavTAUBSLval}{1.415}%
\definemath{\hflavTAUBSLerr}{\pm0.007}%
\definemath{\hflavTAUBSHval}{1.609}%
\definemath{\hflavTAUBSHerr}{\pm0.010}%
\definemath{\hflavGSCOval}{0.6627}%
\definemath{\hflavGSCOerr}{\pm0.0019}%
\definemath{\hflavTAUBSMEANCOval}{1.509}%
\definemath{\hflavTAUBSMEANCOerr}{\pm0.004}%
\definemath{\hflavDGSGSCOval}{+0.135}%
\definemath{\hflavDGSGSCOerr}{\pm0.008}%
\definemath{\hflavDGSCOval}{+0.089}%
\definemath{\hflavDGSCOerr}{\pm0.006}%
\definemath{\hflavRHOGSDGSCO}{-0.151}%
\definemath{\hflavTAUBSLCOval}{1.414}%
\definemath{\hflavTAUBSLCOerr}{\pm0.006}%
\definemath{\hflavTAUBSHCOval}{1.618}%
\definemath{\hflavTAUBSHCOerr}{\pm0.009}%
\definemath{\hflavGSCONval}{0.6624}%
\definemath{\hflavGSCONerr}{\pm0.0018}%
\definemath{\hflavTAUBSMEANCONval}{1.510}%
\definemath{\hflavTAUBSMEANCONerr}{\pm0.004}%
\definemath{\hflavDGSGSCONval}{+0.135}%
\definemath{\hflavDGSGSCONerr}{\pm0.008}%
\definemath{\hflavDGSCONval}{+0.090}%
\definemath{\hflavDGSCONerr}{\pm0.005}%
\definemath{\hflavRHOGSDGSCON}{-0.080}%
\definemath{\hflavTAUBSLCONval}{1.414}%
\definemath{\hflavTAUBSLCONerr}{\pm0.006}%
\definemath{\hflavTAUBSHCONval}{1.619}%
\definemath{\hflavTAUBSHCONerr}{\pm0.009}%
\definemath{\hflavBETASCOMBval}{+0.011}%
\definemath{\hflavBETASCOMBerr}{\pm0.016}%
\definemath{\hflavPHISCOMBval}{-0.021}%
\definemath{\hflavPHISCOMBerr}{\pm0.031}%
\definemath{\hflavDGSCOMBval}{+0.085}%
\definemath{\hflavDGSCOMBerr}{\pm0.006}%
\definemath{\hflavPHISSMval}{-0.0369}%
\definemath{\hflavPHISSMerp}{^{+0.0007}}%
\definemath{\hflavPHISSMern}{_{-0.0010}}%
\definemath{\hflavPHISTWELVESMval}{0.0046}%
\definemath{\hflavPHISTWELVESMerr}{\pm0.0012}%
\definemath{\hflavPHISTWELVEval}{0.020}%
\definemath{\hflavPHISTWELVEerr}{\pm0.032}%
\definemath{\hflavFCWval}{0.514}%
\definemath{\hflavFCWerr}{\pm0.006}%
\definemath{\hflavFNWval}{0.486}%
\definemath{\hflavFNWerr}{\pm0.006}%
\definemath{\hflavFFWval}{1.058}%
\definemath{\hflavFFWerr}{\pm0.024}%
\definemath{\hflavNSIGMAFFW}{2.4}%
\definemath{\hflavFCNval}{0.513}%
\definemath{\hflavFCNerr}{\pm0.013}%
\definemath{\hflavFNNval}{0.487}%
\definemath{\hflavFNNerr}{\pm0.013}%
\definemath{\hflavFFNval}{1.053}%
\definemath{\hflavFFNerr}{\pm0.054}%
\definemath{\hflavFCval}{0.514}%
\definemath{\hflavFCerr}{\pm0.006}%
\definemath{\hflavFNval}{0.486}%
\definemath{\hflavFNerr}{\pm0.006}%
\definemath{\hflavFFval}{1.059}%
\definemath{\hflavFFerr}{\pm0.027}%
\definemath{\hflavNSIGMAFF}{2.2}%
\definemath{\hflavFPRODval}{0.516}%
\definemath{\hflavFPRODerr}{\pm0.019}%
\definemath{\hflavFSUMval}{1.003}%
\definemath{\hflavFSUMerr}{\pm0.029}%
\definemath{\hflavFSFIVEOSval}{0.205}%
\definemath{\hflavFSFIVEOSsta}{\pm0.010}%
\definemath{\hflavFSFIVEOSsys}{\pm0.024}%
\definemath{\hflavFSFIVEOSerr}{\pm0.027}%
\definemath{\hflavFSFIVERLval}{0.215}%
\definemath{\hflavFSFIVERLerr}{\pm0.031}%
\definemath{\hflavFUDFIVEval}{0.758}%
\definemath{\hflavFUDFIVEerp}{^{+0.027}}%
\definemath{\hflavFUDFIVEern}{_{-0.037}}%
\definemath{\hflavFSFIVEval}{0.198}%
\definemath{\hflavFSFIVEerp}{^{+0.030}}%
\definemath{\hflavFSFIVEern}{_{-0.029}}%
\definemath{\hflavFSFUDFIVEval}{0.261}%
\definemath{\hflavFSFUDFIVEerp}{^{+0.051}}%
\definemath{\hflavFSFUDFIVEern}{_{-0.043}}%
\definemath{\hflavFNBFIVEval}{0.044}%
\definemath{\hflavFNBFIVEerp}{^{+0.044}}%
\definemath{\hflavFNBFIVEern}{_{-0.005}}%
\definemath{\hflavZFSFACTOR}{}%
\definemath{\hflavZFBSNOMIXval}{0.087}%
\definemath{\hflavZFBSNOMIXerr}{\pm0.013}%
\definemath{\hflavZFBBNOMIXval}{0.089}%
\definemath{\hflavZFBBNOMIXerr}{\pm0.012}%
\definemath{\hflavZFBDNOMIXval}{0.412}%
\definemath{\hflavZFBDNOMIXerr}{\pm0.008}%
\definemath{\hflavWFSFACTOR}{1.1}%
\definemath{\hflavWFBSNOMIXval}{0.102}%
\definemath{\hflavWFBSNOMIXerr}{\pm0.005}%
\definemath{\hflavWFBBNOMIXval}{0.090}%
\definemath{\hflavWFBBNOMIXerr}{\pm0.012}%
\definemath{\hflavWFBDNOMIXval}{0.404}%
\definemath{\hflavWFBDNOMIXerr}{\pm0.006}%
\definemath{\hflavTFSFACTOR}{}%
\definemath{\hflavTFBSNOMIXval}{0.101}%
\definemath{\hflavTFBSNOMIXerr}{\pm0.015}%
\definemath{\hflavTFBBNOMIXval}{0.220}%
\definemath{\hflavTFBBNOMIXerr}{\pm0.048}%
\definemath{\hflavTFBDNOMIXval}{0.340}%
\definemath{\hflavTFBDNOMIXerr}{\pm0.021}%
\definemath{\hflavLFSFACTOR}{}%
\definemath{\hflavLFBSNOMIXval}{0.087}%
\definemath{\hflavLFBSNOMIXerr}{\pm0.005}%
\definemath{\hflavLFBBNOMIXval}{0.232}%
\definemath{\hflavLFBBNOMIXerr}{\pm0.020}%
\definemath{\hflavLFBDNOMIXval}{0.341}%
\definemath{\hflavLFBDNOMIXerr}{\pm0.009}%
\definemath{\hflavCHIBARTEVval}{0.147}%
\definemath{\hflavCHIBARTEVerr}{\pm0.011}%
\definemath{\hflavCHIBARSFACTOR}{1.8}%
\definemath{\hflavCHIBARval}{0.1284}%
\definemath{\hflavCHIBARerr}{\pm0.0069}%
\definemath{\hflavWFBSMIXval}{0.117}%
\definemath{\hflavWFBSMIXerr}{\pm0.018}%
\definemath{\hflavTFBSMIXval}{0.165}%
\definemath{\hflavTFBSMIXerr}{\pm0.029}%
\definemath{\hflavZFBSMIXval}{0.111}%
\definemath{\hflavZFBSMIXerr}{\pm0.011}%
\definemath{\hflavCHIDUval}{0.182}%
\definemath{\hflavCHIDUerr}{\pm0.015}%
\definemath{\hflavCHIDWUval}{0.1858}%
\definemath{\hflavCHIDWUerr}{\pm0.0011}%
\definemath{\hflavXDWval}{0.769}%
\definemath{\hflavXDWerr}{\pm0.004}%
\definemath{\hflavXDWUval}{0.769}%
\definemath{\hflavXDWUerr}{\pm0.004}%
\definemath{\hflavDMDWval}{0.5065}%
\definemath{\hflavDMDWsta}{\pm0.0016}%
\definemath{\hflavDMDWsys}{\pm0.0011}%
\definemath{\hflavDMDWerr}{\pm0.0019}%
\definemath{\hflavDMDWUval}{0.5065}%
\definemath{\hflavDMDWUerr}{\pm0.0019}%
\definemath{\hflavDMDLHCbval}{0.5063}%
\definemath{\hflavDMDLHCbsta}{\pm0.0019}%
\definemath{\hflavDMDLHCbsys}{\pm0.0010}%
\definemath{\hflavDMDLHCberr}{\pm0.0022}%
\definemath{\hflavZFBSval}{0.101}%
\definemath{\hflavZFBSerr}{\pm0.008}%
\definemath{\hflavZFBBval}{0.085}%
\definemath{\hflavZFBBerr}{\pm0.011}%
\definemath{\hflavZFBDval}{0.407}%
\definemath{\hflavZFBDerr}{\pm0.007}%
\definemath{\hflavZRHOFBBFBS}{+0.065}%
\definemath{\hflavZRHOFBDFBS}{-0.628}%
\definemath{\hflavZRHOFBDFBB}{-0.817}%
\definemath{\hflavWFBSval}{0.103}%
\definemath{\hflavWFBSerr}{\pm0.005}%
\definemath{\hflavWFBBval}{0.088}%
\definemath{\hflavWFBBerr}{\pm0.012}%
\definemath{\hflavWFBDval}{0.405}%
\definemath{\hflavWFBDerr}{\pm0.006}%
\definemath{\hflavWRHOFBBFBS}{-0.260}%
\definemath{\hflavWRHOFBDFBS}{-0.136}%
\definemath{\hflavWRHOFBDFBB}{-0.922}%
\definemath{\hflavTFBSval}{0.115}%
\definemath{\hflavTFBSerr}{\pm0.013}%
\definemath{\hflavTFBBval}{0.198}%
\definemath{\hflavTFBBerr}{\pm0.046}%
\definemath{\hflavTFBDval}{0.344}%
\definemath{\hflavTFBDerr}{\pm0.021}%
\definemath{\hflavTRHOFBBFBS}{-0.429}%
\definemath{\hflavTRHOFBDFBS}{+0.159}%
\definemath{\hflavTRHOFBDFBB}{-0.960}%
\definemath{\hflavLFBSval}{0.089}%
\definemath{\hflavLFBSerr}{\pm0.005}%
\definemath{\hflavLFBBval}{0.224}%
\definemath{\hflavLFBBerr}{\pm0.020}%
\definemath{\hflavLFBDval}{0.343}%
\definemath{\hflavLFBDerr}{\pm0.009}%
\definemath{\hflavLRHOFBBFBS}{-0.517}%
\definemath{\hflavLRHOFBDFBS}{+0.318}%
\definemath{\hflavLRHOFBDFBB}{-0.976}%
\definemath{\hflavZFBSBDval}{0.249}%
\definemath{\hflavZFBSBDerr}{\pm0.023}%
\definemath{\hflavWFBSBDval}{0.255}%
\definemath{\hflavWFBSBDerr}{\pm0.013}%
\definemath{\hflavTFBSBDval}{0.334}%
\definemath{\hflavTFBSBDerr}{\pm0.040}%
\definemath{\hflavLFBSBDval}{0.260}%
\definemath{\hflavLFBSBDerr}{\pm0.013}%
\definemath{\hflavDMDLval}{0.493}%
\definemath{\hflavDMDLsta}{\pm0.011}%
\definemath{\hflavDMDLsys}{\pm0.009}%
\definemath{\hflavDMDLerr}{\pm0.014}%
\definemath{\hflavDMDTval}{0.509}%
\definemath{\hflavDMDTsta}{\pm0.017}%
\definemath{\hflavDMDTsys}{\pm0.013}%
\definemath{\hflavDMDTerr}{\pm0.022}%
\definemath{\hflavDMDBval}{0.509}%
\definemath{\hflavDMDBsta}{\pm0.003}%
\definemath{\hflavDMDBsys}{\pm0.003}%
\definemath{\hflavDMDBerr}{\pm0.005}%
\definemath{\hflavDMDTWODval}{0.509}%
\definemath{\hflavDMDTWODsta}{\pm0.004}%
\definemath{\hflavDMDTWODsys}{\pm0.004}%
\definemath{\hflavDMDTWODerr}{\pm0.006}%
\definemath{\hflavTAUBDTWODval}{1.527}%
\definemath{\hflavTAUBDTWODsta}{\pm0.006}%
\definemath{\hflavTAUBDTWODsys}{\pm0.008}%
\definemath{\hflavTAUBDTWODerr}{\pm0.010}%
\definemath{\hflavRHOstaDMDTAUBD}{-0.19}%
\definemath{\hflavRHOsysDMDTAUBD}{-0.25}%
\definemath{\hflavRHODMDTAUBD}{-0.23}%
\definemath{\hflavZRHOTAUHTAUL}{-0.398}%
\definemath{\hflavTAUBZCALCval}{1.5662}%
\definemath{\hflavTAUBZCALCerr}{\pm0.0029}%
\definemath{\hflavQPDBval}{1.0009}%
\definemath{\hflavQPDBerr}{\pm0.0013}%
\definemath{\hflavQPDDval}{1.0000}%
\definemath{\hflavQPDDerr}{\pm0.0010}%
\definemath{\hflavQPDWval}{1.0005}%
\definemath{\hflavQPDWerr}{\pm0.0009}%
\definemath{\hflavQPDAval}{1.0005}%
\definemath{\hflavQPDAerr}{\pm0.0009}%
\definemath{\hflavASLDBval}{-0.0019}%
\definemath{\hflavASLDBerr}{\pm0.0027}%
\definemath{\hflavASLDDval}{+0.0001}%
\definemath{\hflavASLDDerr}{\pm0.0020}%
\definemath{\hflavASLDWval}{-0.0010}%
\definemath{\hflavASLDWerr}{\pm0.0018}%
\definemath{\hflavASLDAval}{-0.0010}%
\definemath{\hflavASLDAerr}{\pm0.0018}%
\definemath{\hflavREBDBval}{-0.0005}%
\definemath{\hflavREBDBerr}{\pm0.0007}%
\definemath{\hflavREBDDval}{+0.0000}%
\definemath{\hflavREBDDerr}{\pm0.0005}%
\definemath{\hflavREBDWval}{-0.0002}%
\definemath{\hflavREBDWerr}{\pm0.0004}%
\definemath{\hflavREBDAval}{-0.0002}%
\definemath{\hflavREBDAerr}{\pm0.0004}%
\definemath{\hflavASLLHCBDZERONSIGMA}{2.2}%
\definemath{\hflavASLLHCBDZEROPVALPERCENT}{3.1}%
\definemath{\hflavASLSval}{-0.0006}%
\definemath{\hflavASLSerr}{\pm0.0028}%
\definemath{\hflavQPSval}{1.0003}%
\definemath{\hflavQPSerr}{\pm0.0014}%
\definemath{\hflavASLDval}{-0.0021}%
\definemath{\hflavASLDerr}{\pm0.0017}%
\definemath{\hflavQPDval}{1.0010}%
\definemath{\hflavQPDerr}{\pm0.0008}%
\definemath{\hflavRHOASLSASLD}{-0.054}%
\definemath{\hflavCLPERCENTASLSASLD}{4.5}%
\definemath{\hflavREBDval}{-0.0005}%
\definemath{\hflavREBDerr}{\pm0.0004}%
\definemath{\hflavASLDNOMUval}{+0.0000}%
\definemath{\hflavASLDNOMUerr}{\pm0.0019}%
\definemath{\hflavASLSNOMUval}{+0.0016}%
\definemath{\hflavASLSNOMUerr}{\pm0.0030}%
\definemath{\hflavRHOASLSASLDNOMU}{+0.066}%
\definemath{\hflavASLDASLSNSIGMA}{0.5}%
\definemath{\hflavASLDASLSPVALPERCENT}{61.3}%
\definemath{\hflavTANPHIval}{-0.1}%
\definemath{\hflavTANPHIerr}{\pm0.6}%
\definemath{\hflavDMSval}{17.757}%
\definemath{\hflavDMSsta}{\pm0.020}%
\definemath{\hflavDMSsys}{\pm0.007}%
\definemath{\hflavDMSerr}{\pm0.021}%
\definemath{\hflavXSval}{26.81}%
\definemath{\hflavXSerr}{\pm0.08}%
\definemath{\hflavCHISval}{0.499308}%
\definemath{\hflavCHISerr}{\pm0.000004}%
\definemath{\hflavRATIODGSDMSval}{0.00505}%
\definemath{\hflavRATIODGSDMSerr}{\pm0.00031}%
\definemath{\hflavRATIODMDDMSval}{0.02852}%
\definemath{\hflavRATIODMDDMSerr}{\pm0.00011}%
\definemath{\hflavVTDVTSval}{0.2053}%
\definemath{\hflavVTDVTSexx}{\pm0.0004}%
\definemath{\hflavVTDVTSthe}{\pm0.0029}%
\definemath{\hflavVTDVTSerr}{\pm0.0029}%
\definemath{\hflavXIval}{1.206}%
\definemath{\hflavXIerr}{\pm0.017}%

\newcommand{\unit}[1]{~\ensuremath{\rm #1}\xspace}
\renewcommand{\ps}{\unit{ps}}
\newcommand{\invps}{\unit{ps^{-1}}}
\newcommand{\TeV}{\unit{TeV}}
\newcommand{\MeVcc}{\unit{MeV/\mbox{$c$}^2}}

\definemath{\hflavCHIBARLEP}{\hflavCHIBARLEPval\hflavCHIBARLEPerr}
\definemath{\hflavTAUBD}{\hflavTAUBDval\hflavTAUBDerr\ps}
\definemath{\hflavTAUBDnounit}{\hflavTAUBDval\hflavTAUBDerr}
\definemath{\hflavTAUBU}{\hflavTAUBUval\hflavTAUBUerr\ps}
\definemath{\hflavTAUBUnounit}{\hflavTAUBUval\hflavTAUBUerr}
\definemath{\hflavRTAUBU}{\hflavRTAUBUval\hflavRTAUBUerr}
\definemath{\hflavTAUBS}{\hflavTAUBSval\hflavTAUBSerr\ps}
\definemath{\hflavTAUBSnounit}{\hflavTAUBSval\hflavTAUBSerr}
\definemath{\hflavRTAUBS}{\hflavRTAUBSval\hflavRTAUBSerr}
\definemath{\hflavTAULB}{\hflavTAULBval\hflavTAULBerr\ps}
\definemath{\hflavTAULBnounit}{\hflavTAULBval\hflavTAULBerr}
\definemath{\hflavTAULBSerr}{\hflavTAULBSerp\hflavTAULBSern}
\definemath{\hflavTAULBS}{\hflavTAULBSval\hflavTAULBSerr\ps}
\definemath{\hflavTAULBSnounit}{\hflavTAULBSval\hflavTAULBSerr}
\definemath{\hflavTAULBE}{\hflavTAULBEval\hflavTAULBEerr\ps}
\definemath{\hflavTAULBEnounit}{\hflavTAULBEval\hflavTAULBEerr}
\definemath{\hflavTAUXBD}{\hflavTAUXBDval\hflavTAUXBDerr\ps}
\definemath{\hflavTAUXBDnounit}{\hflavTAUXBDval\hflavTAUXBDerr}
\definemath{\hflavTAUXBU}{\hflavTAUXBUval\hflavTAUXBUerr\ps}
\definemath{\hflavTAUXBUnounit}{\hflavTAUXBUval\hflavTAUXBUerr}
\definemath{\hflavTAUOBerr}{\hflavTAUOBerp\hflavTAUOBern}
\definemath{\hflavTAUOB}{\hflavTAUOBval\hflavTAUOBerr\ps}
\definemath{\hflavTAUOBnounit}{\hflavTAUOBval\hflavTAUOBerr}
\definemath{\hflavTAUBC}{\hflavTAUBCval\hflavTAUBCerr\ps}
\definemath{\hflavTAUBCnounit}{\hflavTAUBCval\hflavTAUBCerr}
\definemath{\hflavTAUBSSL}{\hflavTAUBSSLval\hflavTAUBSSLerr\ps}
\definemath{\hflavTAUBSSLnounit}{\hflavTAUBSSLval\hflavTAUBSSLerr}
\definemath{\hflavTAUBSMEANC}{\hflavTAUBSMEANCval\hflavTAUBSMEANCerr\ps}
\definemath{\hflavTAUBSMEANCnounit}{\hflavTAUBSMEANCval\hflavTAUBSMEANCerr}
\definemath{\hflavTAUBSJF}{\hflavTAUBSJFval\hflavTAUBSJFerr\ps}
\definemath{\hflavTAUBSJFnounit}{\hflavTAUBSJFval\hflavTAUBSJFerr}
\definemath{\hflavRTAUBSSL}{\hflavRTAUBSSLval\hflavRTAUBSSLerr}
\definemath{\hflavRTAUBSMEANC}{\hflavRTAUBSMEANCval\hflavRTAUBSMEANCerr}
\definemath{\hflavRTAULB}{\hflavRTAULBval\hflavRTAULBerr}
\definemath{\hflavTAUBVTX}{\hflavTAUBVTXval\hflavTAUBVTXerr\ps}
\definemath{\hflavTAUBVTXnounit}{\hflavTAUBVTXval\hflavTAUBVTXerr}
\definemath{\hflavTAUBLEP}{\hflavTAUBLEPval\hflavTAUBLEPerr\ps}
\definemath{\hflavTAUBLEPnounit}{\hflavTAUBLEPval\hflavTAUBLEPerr}
\definemath{\hflavTAUBSMM}{\hflavTAUBSMMval\hflavTAUBSMMerr\ps}
\definemath{\hflavTAUBSMMnounit}{\hflavTAUBSMMval\hflavTAUBSMMerr}
\definemath{\hflavTAUBJP}{\hflavTAUBJPval\hflavTAUBJPerr\ps}
\definemath{\hflavTAUBJPnounit}{\hflavTAUBJPval\hflavTAUBJPerr}
\definemath{\hflavRTAUXBUXBD}{\hflavRTAUXBUXBDval\hflavRTAUXBUXBDerr}
\definemath{\hflavSDGDGD}{\hflavSDGDGDval\hflavSDGDGDerr}
\definemath{\hflavTAUBSJPSIPIPI}{\hflavTAUBSJPSIPIPIval\hflavTAUBSJPSIPIPIerr\ps}
\definemath{\hflavTAUBSJPSIPIPInounit}{\hflavTAUBSJPSIPIPIval\hflavTAUBSJPSIPIPIerr}
\definemath{\hflavTAUBSJPSIKSHORT}{\hflavTAUBSJPSIKSHORTval\hflavTAUBSJPSIKSHORTerr\ps}
\definemath{\hflavTAUBSJPSIKSHORTnounit}{\hflavTAUBSJPSIKSHORTval\hflavTAUBSJPSIKSHORTerr}
\definemath{\hflavTAUBSLONG}{\hflavTAUBSLONGval\hflavTAUBSLONGerr\ps}
\definemath{\hflavTAUBSLONGnounit}{\hflavTAUBSLONGval\hflavTAUBSLONGerr}
\definemath{\hflavTAUBSKK}{\hflavTAUBSKKval\hflavTAUBSKKerr\ps}
\definemath{\hflavTAUBSKKnounit}{\hflavTAUBSKKval\hflavTAUBSKKerr}
\definemath{\hflavTAUBSDSDS}{\hflavTAUBSDSDSval\hflavTAUBSDSDSerr\ps}
\definemath{\hflavTAUBSDSDSnounit}{\hflavTAUBSDSDSval\hflavTAUBSDSDSerr}
\definemath{\hflavTAUBSJPSIETA}{\hflavTAUBSJPSIETAval\hflavTAUBSJPSIETAerr\ps}
\definemath{\hflavTAUBSJPSIETAnounit}{\hflavTAUBSJPSIETAval\hflavTAUBSJPSIETAerr}
\definemath{\hflavTAUBSSHORT}{\hflavTAUBSSHORTval\hflavTAUBSSHORTerr\ps}
\definemath{\hflavTAUBSSHORTnounit}{\hflavTAUBSSHORTval\hflavTAUBSSHORTerr}
\definemath{\hflavGS}{\hflavGSval\hflavGSerr\invps}
\definemath{\hflavGSnounit}{\hflavGSval\hflavGSerr}
\definemath{\hflavTAUBSMEAN}{\hflavTAUBSMEANval\hflavTAUBSMEANerr\ps}
\definemath{\hflavTAUBSMEANnounit}{\hflavTAUBSMEANval\hflavTAUBSMEANerr}
\definemath{\hflavDGSGS}{\hflavDGSGSval\hflavDGSGSerr}
\definemath{\hflavDGS}{\hflavDGSval\hflavDGSerr\invps}
\definemath{\hflavDGSnounit}{\hflavDGSval\hflavDGSerr}
\definemath{\hflavTAUBSL}{\hflavTAUBSLval\hflavTAUBSLerr\ps}
\definemath{\hflavTAUBSLnounit}{\hflavTAUBSLval\hflavTAUBSLerr}
\definemath{\hflavTAUBSH}{\hflavTAUBSHval\hflavTAUBSHerr\ps}
\definemath{\hflavTAUBSHnounit}{\hflavTAUBSHval\hflavTAUBSHerr}
\definemath{\hflavGSCO}{\hflavGSCOval\hflavGSCOerr\invps}
\definemath{\hflavGSCOnounit}{\hflavGSCOval\hflavGSCOerr}
\definemath{\hflavTAUBSMEANCO}{\hflavTAUBSMEANCOval\hflavTAUBSMEANCOerr\ps}
\definemath{\hflavTAUBSMEANCOnounit}{\hflavTAUBSMEANCOval\hflavTAUBSMEANCOerr}
\definemath{\hflavDGSGSCO}{\hflavDGSGSCOval\hflavDGSGSCOerr}
\definemath{\hflavDGSCO}{\hflavDGSCOval\hflavDGSCOerr\invps}
\definemath{\hflavDGSCOnounit}{\hflavDGSCOval\hflavDGSCOerr}
\definemath{\hflavTAUBSLCO}{\hflavTAUBSLCOval\hflavTAUBSLCOerr\ps}
\definemath{\hflavTAUBSLCOnounit}{\hflavTAUBSLCOval\hflavTAUBSLCOerr}
\definemath{\hflavTAUBSHCO}{\hflavTAUBSHCOval\hflavTAUBSHCOerr\ps}
\definemath{\hflavTAUBSHCOnounit}{\hflavTAUBSHCOval\hflavTAUBSHCOerr}
\definemath{\hflavGSCON}{\hflavGSCONval\hflavGSCONerr\invps}
\definemath{\hflavGSCONnounit}{\hflavGSCONval\hflavGSCONerr}
\definemath{\hflavTAUBSMEANCON}{\hflavTAUBSMEANCONval\hflavTAUBSMEANCONerr\ps}
\definemath{\hflavTAUBSMEANCONnounit}{\hflavTAUBSMEANCONval\hflavTAUBSMEANCONerr}
\definemath{\hflavDGSGSCON}{\hflavDGSGSCONval\hflavDGSGSCONerr}
\definemath{\hflavDGSCON}{\hflavDGSCONval\hflavDGSCONerr\invps}
\definemath{\hflavDGSCONnounit}{\hflavDGSCONval\hflavDGSCONerr}
\definemath{\hflavTAUBSLCON}{\hflavTAUBSLCONval\hflavTAUBSLCONerr\ps}
\definemath{\hflavTAUBSLCONnounit}{\hflavTAUBSLCONval\hflavTAUBSLCONerr}
\definemath{\hflavTAUBSHCON}{\hflavTAUBSHCONval\hflavTAUBSHCONerr\ps}
\definemath{\hflavTAUBSHCONnounit}{\hflavTAUBSHCONval\hflavTAUBSHCONerr}
\definemath{\hflavBETASCOMB}{\hflavBETASCOMBval\hflavBETASCOMBerr}
\definemath{\hflavPHISCOMB}{\hflavPHISCOMBval\hflavPHISCOMBerr}
\definemath{\hflavDGSCOMB}{\hflavDGSCOMBval\hflavDGSCOMBerr\invps}
\definemath{\hflavDGSCOMBnounit}{\hflavDGSCOMBval\hflavDGSCOMBerr}
\definemath{\hflavPHISSMerr}{\hflavPHISSMerp\hflavPHISSMern}
\definemath{\hflavPHISSM}{\hflavPHISSMval\hflavPHISSMerr}
\definemath{\hflavPHISTWELVESM}{\hflavPHISTWELVESMval\hflavPHISTWELVESMerr}
\definemath{\hflavPHISTWELVE}{\hflavPHISTWELVEval\hflavPHISTWELVEerr}
\definemath{\hflavFCW}{\hflavFCWval\hflavFCWerr}
\definemath{\hflavFNW}{\hflavFNWval\hflavFNWerr}
\definemath{\hflavFFW}{\hflavFFWval\hflavFFWerr}
\definemath{\hflavFCN}{\hflavFCNval\hflavFCNerr}
\definemath{\hflavFNN}{\hflavFNNval\hflavFNNerr}
\definemath{\hflavFFN}{\hflavFFNval\hflavFFNerr}
\definemath{\hflavFC}{\hflavFCval\hflavFCerr}
\definemath{\hflavFN}{\hflavFNval\hflavFNerr}
\definemath{\hflavFF}{\hflavFFval\hflavFFerr}
\definemath{\hflavFPROD}{\hflavFPRODval\hflavFPRODerr}
\definemath{\hflavFSUM}{\hflavFSUMval\hflavFSUMerr}
\definemath{\hflavFSFIVEOS}{\hflavFSFIVEOSval\hflavFSFIVEOSerr}
\definemath{\hflavFSFIVEOSfull}{\hflavFSFIVEOSval\hflavFSFIVEOSsta\hflavFSFIVEOSsys}
\definemath{\hflavFSFIVERL}{\hflavFSFIVERLval\hflavFSFIVERLerr}
\definemath{\hflavFUDFIVEerr}{\hflavFUDFIVEerp\hflavFUDFIVEern}
\definemath{\hflavFUDFIVE}{\hflavFUDFIVEval\hflavFUDFIVEerr}
\definemath{\hflavFSFIVEerr}{\hflavFSFIVEerp\hflavFSFIVEern}
\definemath{\hflavFSFIVE}{\hflavFSFIVEval\hflavFSFIVEerr}
\definemath{\hflavFSFUDFIVEerr}{\hflavFSFUDFIVEerp\hflavFSFUDFIVEern}
\definemath{\hflavFSFUDFIVE}{\hflavFSFUDFIVEval\hflavFSFUDFIVEerr}
\definemath{\hflavFNBFIVEerr}{\hflavFNBFIVEerp\hflavFNBFIVEern}
\definemath{\hflavFNBFIVE}{\hflavFNBFIVEval\hflavFNBFIVEerr}
\definemath{\hflavZFBSNOMIX}{\hflavZFBSNOMIXval\hflavZFBSNOMIXerr}
\definemath{\hflavZFBBNOMIX}{\hflavZFBBNOMIXval\hflavZFBBNOMIXerr}
\definemath{\hflavZFBDNOMIX}{\hflavZFBDNOMIXval\hflavZFBDNOMIXerr}
\definemath{\hflavWFBSNOMIX}{\hflavWFBSNOMIXval\hflavWFBSNOMIXerr}
\definemath{\hflavWFBBNOMIX}{\hflavWFBBNOMIXval\hflavWFBBNOMIXerr}
\definemath{\hflavWFBDNOMIX}{\hflavWFBDNOMIXval\hflavWFBDNOMIXerr}
\definemath{\hflavTFBSNOMIX}{\hflavTFBSNOMIXval\hflavTFBSNOMIXerr}
\definemath{\hflavTFBBNOMIX}{\hflavTFBBNOMIXval\hflavTFBBNOMIXerr}
\definemath{\hflavTFBDNOMIX}{\hflavTFBDNOMIXval\hflavTFBDNOMIXerr}
\definemath{\hflavLFBSNOMIX}{\hflavLFBSNOMIXval\hflavLFBSNOMIXerr}
\definemath{\hflavLFBBNOMIX}{\hflavLFBBNOMIXval\hflavLFBBNOMIXerr}
\definemath{\hflavLFBDNOMIX}{\hflavLFBDNOMIXval\hflavLFBDNOMIXerr}
\definemath{\hflavCHIBARTEV}{\hflavCHIBARTEVval\hflavCHIBARTEVerr}
\definemath{\hflavCHIBAR}{\hflavCHIBARval\hflavCHIBARerr}
\definemath{\hflavWFBSMIX}{\hflavWFBSMIXval\hflavWFBSMIXerr}
\definemath{\hflavTFBSMIX}{\hflavTFBSMIXval\hflavTFBSMIXerr}
\definemath{\hflavZFBSMIX}{\hflavZFBSMIXval\hflavZFBSMIXerr}
\definemath{\hflavCHIDU}{\hflavCHIDUval\hflavCHIDUerr}
\definemath{\hflavCHIDWU}{\hflavCHIDWUval\hflavCHIDWUerr}
\definemath{\hflavXDW}{\hflavXDWval\hflavXDWerr}
\definemath{\hflavXDWU}{\hflavXDWUval\hflavXDWUerr}
\definemath{\hflavDMDW}{\hflavDMDWval\hflavDMDWerr\invps}
\definemath{\hflavDMDWnounit}{\hflavDMDWval\hflavDMDWerr}
\definemath{\hflavDMDWfull}{\hflavDMDWval\hflavDMDWsta\hflavDMDWsys\invps}
\definemath{\hflavDMDWnounitfull}{\hflavDMDWval\hflavDMDWsta\hflavDMDWsys}
\definemath{\hflavDMDWU}{\hflavDMDWUval\hflavDMDWUerr\invps}
\definemath{\hflavDMDWUnounit}{\hflavDMDWUval\hflavDMDWUerr}
\definemath{\hflavDMDLHCb}{\hflavDMDLHCbval\hflavDMDLHCberr\invps}
\definemath{\hflavDMDLHCbnounit}{\hflavDMDLHCbval\hflavDMDLHCberr}
\definemath{\hflavDMDLHCbfull}{\hflavDMDLHCbval\hflavDMDLHCbsta\hflavDMDLHCbsys\invps}
\definemath{\hflavDMDLHCbnounitfull}{\hflavDMDLHCbval\hflavDMDLHCbsta\hflavDMDLHCbsys}
\definemath{\hflavZFBS}{\hflavZFBSval\hflavZFBSerr}
\definemath{\hflavZFBB}{\hflavZFBBval\hflavZFBBerr}
\definemath{\hflavZFBD}{\hflavZFBDval\hflavZFBDerr}
\definemath{\hflavWFBS}{\hflavWFBSval\hflavWFBSerr}
\definemath{\hflavWFBB}{\hflavWFBBval\hflavWFBBerr}
\definemath{\hflavWFBD}{\hflavWFBDval\hflavWFBDerr}
\definemath{\hflavTFBS}{\hflavTFBSval\hflavTFBSerr}
\definemath{\hflavTFBB}{\hflavTFBBval\hflavTFBBerr}
\definemath{\hflavTFBD}{\hflavTFBDval\hflavTFBDerr}
\definemath{\hflavLFBS}{\hflavLFBSval\hflavLFBSerr}
\definemath{\hflavLFBB}{\hflavLFBBval\hflavLFBBerr}
\definemath{\hflavLFBD}{\hflavLFBDval\hflavLFBDerr}
\definemath{\hflavZFBSBD}{\hflavZFBSBDval\hflavZFBSBDerr}
\definemath{\hflavWFBSBD}{\hflavWFBSBDval\hflavWFBSBDerr}
\definemath{\hflavTFBSBD}{\hflavTFBSBDval\hflavTFBSBDerr}
\definemath{\hflavLFBSBD}{\hflavLFBSBDval\hflavLFBSBDerr}
\definemath{\hflavDMDL}{\hflavDMDLval\hflavDMDLerr\invps}
\definemath{\hflavDMDLnounit}{\hflavDMDLval\hflavDMDLerr}
\definemath{\hflavDMDLfull}{\hflavDMDLval\hflavDMDLsta\hflavDMDLsys\invps}
\definemath{\hflavDMDLnounitfull}{\hflavDMDLval\hflavDMDLsta\hflavDMDLsys}
\definemath{\hflavDMDT}{\hflavDMDTval\hflavDMDTerr\invps}
\definemath{\hflavDMDTnounit}{\hflavDMDTval\hflavDMDTerr}
\definemath{\hflavDMDTfull}{\hflavDMDTval\hflavDMDTsta\hflavDMDTsys\invps}
\definemath{\hflavDMDTnounitfull}{\hflavDMDTval\hflavDMDTsta\hflavDMDTsys}
\definemath{\hflavDMDB}{\hflavDMDBval\hflavDMDBerr\invps}
\definemath{\hflavDMDBnounit}{\hflavDMDBval\hflavDMDBerr}
\definemath{\hflavDMDBfull}{\hflavDMDBval\hflavDMDBsta\hflavDMDBsys\invps}
\definemath{\hflavDMDBnounitfull}{\hflavDMDBval\hflavDMDBsta\hflavDMDBsys}
\definemath{\hflavDMDTWOD}{\hflavDMDTWODval\hflavDMDTWODerr\invps}
\definemath{\hflavDMDTWODnounit}{\hflavDMDTWODval\hflavDMDTWODerr}
\definemath{\hflavDMDTWODfull}{\hflavDMDTWODval\hflavDMDTWODsta\hflavDMDTWODsys\invps}
\definemath{\hflavDMDTWODnounitfull}{\hflavDMDTWODval\hflavDMDTWODsta\hflavDMDTWODsys}
\definemath{\hflavTAUBDTWOD}{\hflavTAUBDTWODval\hflavTAUBDTWODerr\ps}
\definemath{\hflavTAUBDTWODnounit}{\hflavTAUBDTWODval\hflavTAUBDTWODerr}
\definemath{\hflavTAUBDTWODfull}{\hflavTAUBDTWODval\hflavTAUBDTWODsta\hflavTAUBDTWODsys\ps}
\definemath{\hflavTAUBDTWODnounitfull}{\hflavTAUBDTWODval\hflavTAUBDTWODsta\hflavTAUBDTWODsys}
\definemath{\hflavTAUBZCALC}{\hflavTAUBZCALCval\hflavTAUBZCALCerr\ps}
\definemath{\hflavTAUBZCALCnounit}{\hflavTAUBZCALCval\hflavTAUBZCALCerr}
\definemath{\hflavQPDB}{\hflavQPDBval\hflavQPDBerr}
\definemath{\hflavQPDD}{\hflavQPDDval\hflavQPDDerr}
\definemath{\hflavQPDW}{\hflavQPDWval\hflavQPDWerr}
\definemath{\hflavQPDA}{\hflavQPDAval\hflavQPDAerr}
\definemath{\hflavASLDB}{\hflavASLDBval\hflavASLDBerr}
\definemath{\hflavASLDD}{\hflavASLDDval\hflavASLDDerr}
\definemath{\hflavASLDW}{\hflavASLDWval\hflavASLDWerr}
\definemath{\hflavASLDA}{\hflavASLDAval\hflavASLDAerr}
\definemath{\hflavREBDB}{\hflavREBDBval\hflavREBDBerr}
\definemath{\hflavREBDD}{\hflavREBDDval\hflavREBDDerr}
\definemath{\hflavREBDW}{\hflavREBDWval\hflavREBDWerr}
\definemath{\hflavREBDA}{\hflavREBDAval\hflavREBDAerr}
\definemath{\hflavASLS}{\hflavASLSval\hflavASLSerr}
\definemath{\hflavQPS}{\hflavQPSval\hflavQPSerr}
\definemath{\hflavASLD}{\hflavASLDval\hflavASLDerr}
\definemath{\hflavQPD}{\hflavQPDval\hflavQPDerr}
\definemath{\hflavREBD}{\hflavREBDval\hflavREBDerr}
\definemath{\hflavASLDNOMU}{\hflavASLDNOMUval\hflavASLDNOMUerr}
\definemath{\hflavASLSNOMU}{\hflavASLSNOMUval\hflavASLSNOMUerr}
\definemath{\hflavTANPHI}{\hflavTANPHIval\hflavTANPHIerr}
\definemath{\hflavDMS}{\hflavDMSval\hflavDMSerr\invps}
\definemath{\hflavDMSnounit}{\hflavDMSval\hflavDMSerr}
\definemath{\hflavDMSfull}{\hflavDMSval\hflavDMSsta\hflavDMSsys\invps}
\definemath{\hflavDMSnounitfull}{\hflavDMSval\hflavDMSsta\hflavDMSsys}
\definemath{\hflavXS}{\hflavXSval\hflavXSerr}
\definemath{\hflavCHIS}{\hflavCHISval\hflavCHISerr}
\definemath{\hflavRATIODGSDMS}{\hflavRATIODGSDMSval\hflavRATIODGSDMSerr}
\definemath{\hflavRATIODMDDMS}{\hflavRATIODMDDMSval\hflavRATIODMDDMSerr}
\definemath{\hflavVTDVTS}{\hflavVTDVTSval\hflavVTDVTSerr}
\definemath{\hflavVTDVTSfull}{\hflavVTDVTSval\hflavVTDVTSexx\hflavVTDVTSthe}
\definemath{\hflavXI}{\hflavXIval\hflavXIerr}

\newcommand{\comment}[1]{}

\newcommand{\fBs}{\ensuremath{f_{\particle{s}}}\xspace}
\newcommand{\fBd}{\ensuremath{f_{\particle{d}}}\xspace}
\newcommand{\fBu}{\ensuremath{f_{\particle{u}}}\xspace}
\newcommand{\fbb}{\ensuremath{f_{\rm baryon}}\xspace}
\newcommand{\fLb}{\ensuremath{f_{\Lb}}\xspace}

\newcommand{\dmd}{\ensuremath{\Delta m_{\particle{d}}}\xspace}
\newcommand{\dms}{\ensuremath{\Delta m_{\particle{s}}}\xspace}
\newcommand{\xd}{\ensuremath{x_{\particle{d}}}\xspace}
\newcommand{\xs}{\ensuremath{x_{\particle{s}}}\xspace}
\newcommand{\yd}{\ensuremath{y_{\particle{d}}}\xspace}
\newcommand{\ys}{\ensuremath{y_{\particle{s}}}\xspace}
\newcommand{\chibar}{\ensuremath{\overline{\chi}}\xspace}
\newcommand{\chid}{\ensuremath{\chi_{\particle{d}}}\xspace}
\newcommand{\chis}{\ensuremath{\chi_{\particle{s}}}\xspace}
\newcommand{\Gd}{\ensuremath{\Gamma_{\particle{d}}}\xspace}
\newcommand{\DGd}{\ensuremath{\Delta\Gd}\xspace}
\newcommand{\DGGd}{\ensuremath{\DGd/\Gd}\xspace}
\newcommand{\Gs}{\ensuremath{\Gamma_{\particle{s}}}\xspace}
\newcommand{\DGs}{\ensuremath{\Delta\Gs}\xspace}

\newcommand{\ASLd}{\ensuremath{{\cal A}_{\rm SL}^\particle{d}}\xspace}
\newcommand{\ASLs}{\ensuremath{{\cal A}_{\rm SL}^\particle{s}}\xspace}
\newcommand{\ASLb}{\ensuremath{{\cal A}_{\rm SL}^\particle{b}}\xspace}

\newcommand{\DG}{\ensuremath{\Delta\Gamma}\xspace}
\newcommand{\phiccbars}{\ensuremath{\phi_s^{c\bar{c}s}}\xspace}

\newcommand{\BRp}[1]{\particle{{\cal B}(#1)}}
\newcommand{\CL}[1]{#1\%~\mbox{CL}}
\newcommand{\Qjet}{\ensuremath{Q_{\rm jet}}\xspace}

\newcommand{\labe}[1]{\label{equ:#1}}
\newcommand{\labs}[1]{\label{sec:#1}}
\newcommand{\labf}[1]{\label{fig:#1}}
\newcommand{\labt}[1]{\label{tab:#1}}
\newcommand{\refe}[1]{\ref{equ:#1}}
\newcommand{\refs}[1]{\ref{sec:#1}}
\newcommand{\reff}[1]{\ref{fig:#1}}
\newcommand{\reft}[1]{\ref{tab:#1}}
\renewcommand{\Ref}[1]{Ref.~\cite{#1}}
\newcommand{\Refs}[1]{Refs.~\cite{#1}}

\newcommand{\Eq}[1]{Eq.~(\refe{#1})}

\newcommand{\Eqss}[2]{Eqs.~(\refe{#1}) and (\refe{#2})}

\newcommand{\Eqsss}[3]{Eqs.~(\refe{#1}), (\refe{#2}), and (\refe{#3})}
\newcommand{\Figure}[1]{Figure~\reff{#1}}

\newcommand{\Fig}[1]{Fig.~\reff{#1}}

\newcommand{\Sec}[1]{Sec.~\refs{#1}}

\newcommand{\Secss}[2]{Secs.~\refs{#1} and \refs{#2}}

\newcommand{\Table}[1]{Table~\reft{#1}}

\newcommand{\Tablesss}[3]{Tables~\reft{#1}, \reft{#2}, and \reft{#3}}

\newcommand{\subsubsubsection}[1]{\vspace{2ex}\par\noindent {\bf\boldmath\em #1} \vspace{2ex}\par}

\input{tau/tau-header} %

\renewcommand{\mysection}[1]{\section[#1]{#1}} %

\newcommand\red[1]{{\color{red}#1}}
\newcommand\blue[1]{{\color{blue}#1}}

\newif\ifref
\newif\ifhtml
\reftrue 

\makeatletter
\patchcmd{\@sect}{#8}{\boldmath #8}{}{}
\let\ori@chapter\@chapter
\def\@chapter[#1]#2{\ori@chapter[\boldmath#1]{\boldmath#2}}
\makeatother

\begin{document}

\setcounter{page}{1}
\thispagestyle{empty}
\renewcommand\Affilfont{\itshape\small}

\title{
  Averages of $b$-hadron, $c$-hadron, and $\tau$-lepton properties
  as of 2018
\vskip0.20in
\large{\it Heavy Flavor Averaging Group (HFLAV):}
\vspace*{-0.20in}}
\author[1]{Y.~Amhis}\affil[1]{Universit\'e Paris-Saclay, CNRS/IN2P3, IJCLab, Orsay, France}
\author[2]{Sw.~Banerjee}\affil[2]{University of Louisville, Louisville, Kentucky, USA}
\author[3]{E.~Ben-Haim}\affil[3]{LPNHE, Sorbonne Universit\'e, Paris Diderot Sorbonne Paris Cit\'e, CNRS/IN2P3, Paris, France}
\author[4]{F.~U.~Bernlochner}\affil[4]{University of Bonn, Bonn, Germany}
\author[5]{M.~Bona}\affil[5]{School of Physics and Astronomy, Queen Mary University of London, London, UK}
\author[6]{A.~Bozek}\affil[6]{H. Niewodniczanski Institute of Nuclear Physics, Krak\'{o}w, Poland}
\author[7]{C.~Bozzi}\affil[7]{INFN Sezione di Ferrara, Ferrara, Italy}
\author[6]{J.~Brodzicka}
\author[6]{M.~Chrzaszcz}
\author[4]{J.~Dingfelder}
\author[4]{S.~Duell}
\author[8]{U.~Egede}\affil[8]{School of Physics and Astronomy, Monash University, Melbourne, Australia}
\author[9]{M.~Gersabeck}\affil[9]{School of Physics and Astronomy, University of Manchester, Manchester, UK}
\author[10]{T.~Gershon}\affil[10]{Department of Physics, University of Warwick, Coventry, UK}
\author[11]{P.~Goldenzweig}\affil[11]{Institut f\"ur Experimentelle Teilchenphysik, Karlsruher Institut f\"ur Technologie, Karlsruhe, Germany}
\author[12]{K.~Hayasaka}\affil[12]{Niigata University, Niigata, Japan}
\author[13]{H.~Hayashii}\affil[13]{Nara Women's University, Nara, Japan}
\author[14]{D.~Johnson}\affil[14]{European Organization for Nuclear Research (CERN), Geneva, Switzerland}
\author[10]{M.~Kenzie}
\author[15]{T.~Kuhr}\affil[15]{Ludwig-Maximilians-University, Munich, Germany}
\author[16]{O.~Leroy}\affil[16]{Aix Marseille Univ, CNRS/IN2P3, CPPM, Marseille, France}
\author[17,18]{H.-B.~Li}\affil[17]{Institute of High Energy Physics, Beijing 100049, People’s Republic of China}\affil[18]{University of Chinese Academy of Sciences, Beijing 100049, People’s Republic of China}
\author[19,20]{A.~Lusiani}\affil[19]{Scuola Normale Superiore, Pisa, Italy}\affil[20]{INFN Sezione di Pisa, Pisa, Italy}
\author[17]{H.-L.~Ma}
\author[13]{K.~Miyabayashi}
\author[21]{P.~Naik}\affil[21]{H.H.~Wills Physics Laboratory, University of Bristol, Bristol, UK}
\author[22]{T.~Nanut}\affil[22]{Institute of Physics, Ecole Polytechnique F\'{e}d\'{e}rale de Lausanne (EPFL), Lausanne, Switzerland}
\author[23]{M.~Patel}\affil[23]{Imperial College London, London, UK}
\author[24,25]{A.~Pompili}\affil[24]{Universit\`a di Bari Aldo Moro, Bari, Italy}\affil[25]{INFN Sezione di Bari, Bari, Italy}
\author[20]{M.~Rama}
\author[26]{M.~Roney}\affil[26]{University of Victoria, Victoria, British Columbia, Canada}
\author[27]{M.~Rotondo}\affil[27]{Laboratori Nazionali dell'INFN di Frascati, Frascati, Italy}
\author[22]{O.~Schneider}
\author[28]{C.~Schwanda}\affil[28]{Institute of High Energy Physics, Vienna, Austria}
\author[29]{A.~J.~Schwartz}\affil[29]{University of Cincinnati, Cincinnati, Ohio, USA}
\author[30,31]{B.~Shwartz}\affil[30]{Budker Institute of Nuclear Physics (SB RAS), Novosibirsk, Russia}\affil[31]{Novosibirsk State University, Novosibirsk, Russia}
\author[16]{J.~Serrano}
\author[32]{A.~Soffer}\affil[32]{Tel Aviv University, Tel Aviv, Israel}
\author[33]{D.~Tonelli}\affil[33]{INFN Sezione di Trieste, Trieste, Italy}
\author[34]{P.~Urquijo}\affil[34]{School of Physics, University of Melbourne, Melbourne, Victoria, Australia}
\author[35]{R.~Van Kooten}\affil[35]{Indiana University, Bloomington, Indiana, USA}
\author[36]{J.~Yelton}\affil[36]{University of Florida, Gainesville, Florida, USA}

\date{\today} %
\maketitle

\begin{abstract}
\noindent
This paper reports world averages of measurements of $b$-hadron, $c$-hadron,
and $\tau$-lepton properties obtained by the Heavy Flavour Averaging Group using results available through September 2018. In rare cases, significant results obtained several months later are also used.
For the averaging, common input parameters used in the various analyses are adjusted (rescaled) to common values, and known correlations are taken into account.
The averages include branching fractions, lifetimes, neutral meson mixing
parameters, \CP~violation parameters, parameters of semileptonic decays, and 
Cabibbo-Kobayashi-Maskawa matrix elements.
\end{abstract}

\newpage
\tableofcontents
\newpage

\section{Executive Summary}
\label{sec:summary}

This paper provides updated world averages of measurements of $b$-hadron, $c$-hadron, and $\tau$-lepton properties using results available by September 2018. In a few cases, later important results are included and clearly labelled as such. While new measurements since the previous version of this paper~\cite{Amhis:2016xyh} have been dominated by the LHCb and the BESIII experiment in, there are new results from many other experiments, and the older results from previous generations of experiments are still very important. The future will provide updated results, with the most important change being that \belletwo has started data taking in 2019.

Since the previous version of the paper, 
the \b-hadron fraction, lifetime and mixing averages have mostly made small
incremental progress in precision, 
with the most significant improvements in several effective \Bs lifetimes. 
In total 14 new results 
(of which 12 from the LHC data and 2 from the KEKB data)
have been incorporated in these averages.

The lifetime hierarchy for the most abundant weakly decaying \b-hadron species
is well established, with impressive precisions of 4~fs
for the most common \Bd, \Bu and \Bs mesons,
and compatible with the expectations from the Heavy Quark Expansion. 
However, small sample sizes still limit the precision for \b baryons heavier 
than \Lb (\Xibd, \Xibu, $\Omega_b$, and all other yet-to-be-discovered 
\b baryons). 
A sizable value of the decay width difference in the $\Bs$--$\Bsb$ system 
is measured with a relative precision of 6\% and is well predicted by the 
Standard Model~(SM). In contrast, 
the experimental results for the decay width difference in the
$\Bd$--$\Bzb$ system are not yet precise enough to distinguish
the small (expected) value from zero.
The mass differences in both systems are known very accurately, to the (few)
per mil level. On the other hand, \CP violation in the mixing of either system 
has not been observed yet, with asymmetries known within a couple per mil but
still consistent both with zero and their SM predictions. 
A similar conclusion holds for the \CP violation induced
by \Bs mixing in the $b\to c\bar{c}s$ transition, although in this case 
the experimental uncertainty on the corresponding weak phase is an order
of magnitude larger, but now smaller than the SM central value. 
Many measurements are still dominated by statistical uncertainties and will improve once new results from the LHC Run~2 become available, and later from LHC Run~3 and \belletwo. 

The measurement of $\sin 2\beta \equiv \sin 2\phi_1$ from $b \to
c\bar{c}s$ transitions such as $\Bz \to \jpsi\KS$ has reached better than $2.5\,\%$
precision: $\sin 2\beta \equiv \sin 2\phi_1 = 0.699 \pm 0.017$.
Measurements of the same parameter using different quark-level processes
provide a consistency test of the SM and allow insight into
possible beyond the Standard Model effects.  
All results among hadronic $b \to s$ penguin dominated decays of \Bz mesons are currently consistent with the SM expectations.  
Measurements of \CP violation parameters in $\Bs \to \phi\phi$ allow a similar comparison to the value of $\phi_s^{c\bar{c}s}$; where results again are consistent with the close to zero SM expectation.
Among measurements related to the Unitarity Triangle angle $\alpha \equiv \phi_2$, results from $B$ decays to $\pi\pi$, $\rho\pi$ and $\rho\rho$ are combined to obtain a world average value of $\left( 84.9\,^{+5.1}_{-4.5} \right)^\circ$.  
Knowledge of the third angle $\gamma \equiv \phi_3$ also continues to improve, with the current world average being $\left( 71.1\,^{+4.6}_{-5.3} \right)^\circ$.
The constraints on the angles of the Unitarity Triangle are summarized in Fig.~\ref{fig:cp_uta:HFLAV_RhoEta}.

In semileptonic $B$~meson decays, the anomalies in the magnitudes of the CKM elements \vcb\ and \vub\ remain at about the same level compared to the previous update: the discrepancy between \vcb\ measured with inclusive and exclusive decays is of the order of $3\sigma$ ($3.3\sigma$ for \vcb\ from $\Bb\to D^*\ell^-\bar\nu_\ell$, $2.0\sigma$ for \vcb\ from $\Bb\to D\ell^-\bar\nu_\ell$). 
The difference between \vub\ measured with
inclusive decays $\Bb\to X_u\ell^-\nub_\ell$ and \vub\ from $\Bb\to\pi\ell^-\nub_\ell$ has fallen to $2.8\sigma$. Some decrease is observed in the the size of the $B\to D^{(*)}\tau\nu_\tau$ decays anomaly: the combined discrepancy of the measured values of ${\cal R}(D^*)$ and ${\cal R}(D)$ to their SM expectations is found to be $3.1\sigma$.

The most important new measurements of rare $b$-hadron decays are coming from the LHC. 
Precision measurements of $\Bs$ decays are particularly noteworthy, including several measurements of the longitudinal polarisation fraction from LHCb. 
ATLAS and LHCb have updated their measurements of the branching fractions of $B^0_{(s)}\to\mumu$ decays, improving the sensitivity. There are more and more measurements of observables related to $b \to s\ell\ell$ transitions.
One of the observables measured by LHCb, $P_5^{\prime}$, differs from the SM prediction by $3.7\sigma$ in one of the squared dimuon mass intervals; results from Belle on this observable are consistent but less precise.
Improved measurements from LHCb and other experiments are keenly anticipated.
A measurement of the ratio of branching fractions of $\Bp \to K^+\mu^+\mu^-$ and $\Bp \to K^+e^+e^-$ decays ($R_K$) has been made by LHCb. 
In the low squared dilepton mass region, it differs from the SM prediction by $2.6\sigma$.
Among the \CP violating observables in rare decays, the ``$K\pi$ puzzle'' persists, and important new results have appeared in three-body decays.
LHCb has produced many other results on a wide variety of decays, including $b$-baryon and \Bc-meson decays. %

About 800 $b$ to charm results from \babar, Belle, CDF, D0, LHCb, CMS, and ATLAS reported in more than 200 papers
are compiled in a list of over 600 averages.
The large samples of $b$ hadrons that are available in contemporary experiments allows measurements of decays to states with open or hidden charm content with unprecedented precision.
In addition to improvements in precision for branching fractions of $\Bzb$ and $B^-$ mesons, many new decay modes have been discovered.
In addition, the set of measurements available for $\Bsb$ and $B_c^-$ mesons as well as for $b$ baryon decays is rapidly increasing.

In the charm sector, the main highlight is the discovery of \CP\ violation.
A global fit to measurements of $D^0\ra K^+K^-/\pi^+\pi^-$ decays gives
$\Delta a^{\rm dir}_{\CP}=(-0.164\pm0.028)\%$, confirming the observation of direct \CP\ violation
with indirect \CP\ violation being compatible with zero.
Measurements of 49 observables from the E791, FOCUS, Belle, \babar, CLEO, BESIII, CDF, 
and LHCb experiments are input into a global fit for 10 underlying parameters, 
and the no-mixing hypothesis is excluded at a confidence level above $11\sigma$. 
The mixing parameters $x$ and $y$ individually differ from zero by 
$3.1\sigma$ and above $11\sigma$, respectively. This is the first time that $x$ has been found to be non-zero at a significance exceeding $3\sigma$.
The world average value for 
the observable $y_{\CP}$ is positive, indicating that the \CP-even 
state is shorter-lived as in the $\Kz$--$\Kzb$ system; however, the \CP-even state also appears to be the heavier one, which differs from the $\Kz$--$\Kzb$ system. 
The \CP\ violation parameters $|q/p|$ and $\phi$ are consistent with the 
\CP\ symmetry hypothesis within~$1\sigma$. Thus there is no evidence
for \CP\ violation arising from mixing ($|q/p|\neq 1$) or 
from a phase difference between the mixing amplitude and 
a direct decay amplitude ($\phi\neq 0$).
The world's most precise measurements of $|V^{}_{cd}|$ and $|V^{}_{cs}|$
are obtained from leptonic $D^+\ra \mu^+\nu$ and $D^+_s\ra\mu^+\nu/\tau^+\nu$ 
decays, respectively. 
These measurements have theoretical uncertainties arising from decay constants.
However, calculations of decay constants within lattice QCD have improved such 
that the theory error is below $15\%$ of the experimental uncertainties of the measurements.

The \mtau branching fraction fit has been updated using 7 new branching
fraction measurements by \babar that were released in 2018.
With respect to the HFLAV Spring 2017 report, there are no
significant changes to the lepton universality tests.
The precision of \Vus from $\BR(\tau\to X_s \nu)$ improved by about 10\%.
There is no significant variation of the significance of the about
$3\,\sigma$ discrepancy
between \Vus from $\BR(\tau\to X_s \nu)$ and \Vus from \Vud and the CKM
matrix unitarity. No additional upper limit on \mtau
lepton-flavour-violating branching fractions has been added.

A small selection of highlights of the results described in 
Sections~\ref{sec:life_mix}--\ref{sec:tau} is given in Table~\ref{tab_summary}.

\renewcommand*{\arraystretch}{1.08}
\begin{longtable}{|l|c|}
\caption{
  Selected world averages.
  Where two uncertainties are given the first is statistical and the second is systematic, except where indicated otherwise.
} %
\label{tab_summary}
\endfirsthead
\multicolumn{2}{c}{Selected world averages -- continued from previous page.}
\endhead
\endfoot
\endlastfoot
\hline
{\bf\boldmath \b-hadron lifetimes} &   \\
~~$\tau(\Bd)$         & \hflavTAUBD \\
~~$\tau(\Bu)$         & \hflavTAUBU \\
~~$\bar{\tau}(\Bs) = 1/\Gs$  & \hflavTAUBSMEANC \\
~~$\tau(B^0_{s\rm L})$ & \hflavTAUBSLCON \\
~~$\tau(B^0_{s\rm H})$  & \hflavTAUBSHCON \\
~~$\tau(\Bc)$         & \hflavTAUBC \\
~~$\tau(\Lb)$         & \hflavTAULB \\
~~$\tau(\Xibd)$       & \hflavTAUXBD  \\
~~$\tau(\Xibu)$       & \hflavTAUXBU  \\
~~$\tau(\Omegab)$     & \hflavTAUOB   \\
\hline
\multicolumn{2}{|l|}{{\bf\boldmath \Bd\ and \Bs\ mixing / \CP violation parameters}}  \\
~~\dmd &  \hflavDMDWU \\
~~\DGGd  & \hflavSDGDGD \\
~~$|q_{\particle{d}}/p_{\particle{d}}|$ & \hflavQPDB  \\
~~\dms  &  \hflavDMS \\
~~\DGs & \hflavDGSCON \\
~~$|q_{\particle{s}}/p_{\particle{s}}|$ & \hflavQPS   \\
~~\phiccbars  & \hflavPHISCOMB rad\\
\hline
\multicolumn{2}{|l|}{{\bf Parameters related to Unitarity Triangle angles}} \\
 ~~ $\stwob \equiv \sin\! 2\phi_1$ & $\phantom{-}0.699 \pm 0.017$ \\
 ~~ $\beta \equiv \phi_1$          & $\phantom{\degrees}\left( 22.2 \pm 0.7 \right)\degrees$ \\
 ~~ $-\etacp S_{\phi \KS}$         & $0.74\,^{+0.11}_{-0.13}$ \\
 ~~ $-\etacp S_{\etapr \Kz}$       & $\phantom{-}0.63 \pm 0.06$ \\
 ~~ $-\etacp S_{\KS \KS \KS}$      & $\phantom{-}0.72 \pm 0.19$ \\
 ~~ $\phi_s(\phi\phi)$             & $-0.06 \pm 0.13 \pm 0.03 \, {\rm rad}$ \\
 ~~ $-\etacp S_{\jpsi \piz}$       & $\phantom{-}0.86 \pm 0.14$ \\
 ~~ $-\etacp S_{\Dp\Dm}$           & $\phantom{-}0.84 \pm 0.12$ \\
 ~~ $-\etacp S_{\jpsi \rhoz}$      & $\phantom{-}0.66 \, ^{+0.13}_{-0.12}\,^{+0.09}_{-0.03}$ \\
 ~~ $S_{K^* \gamma}$               & $-0.16 \pm 0.22$ \\
 ~~ $\left( S_{\pi^+\pi^-}, C_{\pi^+\pi^-} \right)$ & $\left( -0.63 \pm 0.04, -0.32 \pm 0.04 \right)$ \\  
 ~~ $\left( S_{\rho^+\rho^-}, C_{\rho^+\rho^-} \right)$ & $\left( -0.14 \pm 0.13, \phantom{-}0.00 \pm 0.09\right)$ \\
 ~~ $\alpha \equiv \phi_2$         & $\left( 84.9\,^{+5.1}_{-4.5} \right)^\circ$ \\
 ~~ $a(D^{\mp}\pi^{\pm})$, $a(D^{*\mp}\pi^{\pm})$ & $-0.038 \pm 0.013$, $-0.039 \pm 0.010$ \\
 ~~ $A^{}_{\CP}(B\ra D^{}_{\CP+}K)$       & $\phantom{-}0.129 \pm 0.012$ \\
 ~~ $A_{\rm ADS}(B\ra D^{}_{K\pi}K)$     & $-0.415 \pm 0.055$ \\
 ~~ $\gamma \equiv \phi_3$               & $\left( 71.1\,^{+4.6}_{-5.3} \right)^\circ$ \\
\hline
{\bf\boldmath Semileptonic \B decay parameters} & \\
 ~~${\cal B}(\Bzb\to D^{*+}\ell^-\nub_\ell)$ & $(5.06\pm 0.12)\%$\\
 ~~${\cal B}(\B^-\to D^{*0}\ell^-\nub_\ell)$ & $(5.66\pm 0.22)\%$\\
 ~~$\eta_{\rm EW}{\cal F}(1)\vcb$ & $(35.27\pm 0.38)\times 10^{-3}$\\
 ~~$\vcb$ from $\bar B\to D^*\ell^-\bar\nu_\ell$ & $(38.76\pm 0.42_{\rm exp}\pm 0.55_{\rm th})\times 10^{-3}$\\
\hline
 ~~${\cal B}(\Bzb\to D^+\ell^-\nub_\ell)$ & $(2.31\pm 0.10)\%$\\
 ~~${\cal B}(\B^-\to D^0\ell^-\nub_\ell)$ & $(2.35\pm 0.09)\%$\\
 ~~$\eta_{\rm EW}{\cal G}(1)\vcb$ & $(42.00 \pm 1.00)\times 10^{-3}$\\
 ~~$\vcb$ from $\bar B\to D\ell^-\bar\nu_\ell$ & $(39.58 \pm 0.94_{\rm exp}\pm 0.37_{\rm th})\times 10^{-3}$\\
\hline
 ~~${\cal B}(\bar B\to X_c\ell^-\bar\nu_\ell)$ & $(10.65\pm 0.16)\%$\\
 ~~${\cal B}(\bar B\to X\ell^-\bar\nu_\ell)$ & $(10.86\pm 0.16)\%$\\
 ~~$\vcb$ from $\bar B\to X\ell^-\bar\nu_\ell$ & $(42.19\pm 0.78)\times 10^{-3}$\\
\hline
 ~~${\cal B}(\bar B^0\to\pi^+\ell^-\nub_\ell)$ & $(1.50\pm 0.06)\times 10^{-4}$\\
 ~~$\vub$ from $\Bb\to\pi\ell^-\nub_\ell$ & $(3.67\pm 0.15)\times
 10^{-3}$\\
 ~~$\vub$ from $\Bb\to X_u\ell^-\nub_\ell$ & $(4.32\pm 0.12_{\rm exp}\pm 0.13_{\rm th})\times 10^{-3}$\\
 ~~$\vub/\vcb$ from $\Lambda_b^0\to p\mu^-\nub_\mu/\Lambda_b^0\to \Lambda_c^+\mu^-\nub_\mu$ & $0.079\pm 0.004_{\rm exp}\pm 0.004_{\rm th}$\\

\hline
~~${\cal R}(D)={\cal B}(B\to D\tau\nu_\tau)/{\cal B}(B\to
  D\ell\nu_\ell)$ & $0.340\pm 0.030$\\
~~${\cal R}(D^*)={\cal B}(B\to D^*\tau\nu_\tau)/{\cal B}(B\to
  D^*\ell\nu_\ell)$ & $0.295\pm 0.014$\\
\hline
{\bf\boldmath \b-hadron to charmed hadron decays} & \\
 ~~ ${\cal B}(\Bzb \to D^+ \pi^-)$ & $(2.65\pm 0.15) \times 10^{-3}$ \\
 ~~ ${\cal B}(B^- \to D^0 \pi^-)$ & $(4.75 \pm 0.19) \times 10^{-3}$ \\
 ~~ ${\cal B}(\Bsb \to D_s^+ \pi^-)$ & $(3.03 \pm 0.25) \times 10^{-3}$ \\
 ~~ ${\cal B}(\Lambda_b^0 \to \Lambda_c^+ \pi^-)$ & $(4.30^{+0.36}_{-0.35}) \times 10^{-3}$ \\
 ~~ ${\cal B}(\Bzb \to J/\psi \bar{K}^0)$ & $(0.863 \pm 0.035) \times 10^{-3}$ \\
 ~~ ${\cal B}(B^- \to J/\psi K^-)$ & $(1.028 \pm 0.040) \times 10^{-3}$ \\
 ~~ ${\cal B}(\Bsb \to J/\psi \phi)$ & $(1.00 \pm 0.09) \times 10^{-3}$ \\
 ~~ ${\cal B}(\Lambda_b^0 \to J/\psi \Lambda^0)$ & $(0.47 \pm 0.28) \times 10^{-3}$ \\
 ~~ ${\cal B}(B_c^- \to J/\psi D_s^-) / {\cal B}(B_c^- \to J/\psi \pi^-)$ & $3.09 \pm 0.55$ \\
\hline
{\bf\boldmath \b-hadron to charmless final states} &   \\
 ~~ ${\cal B}(\Bs \to \mu^+\mu^-)$ & $\left( 3.1 \pm 0.6  \right) \times 10^{-9}$ \\
 ~~ ${\cal B}(\Bz \to \mu^+\mu^-)$ & $<0.34\times 10^{-9}$ (CL=90\%) \\
  ~~ ${\cal B}(\Bs \to \tau^+\tau^-)$ & $<5.2\times 10^{-3}$ (CL=90\%) \\
 ~~ ${\cal B}(B \to X_s \gamma)$  ($E_{\gamma}>1.6~\gev$) & $(3.32 \pm 0.15) \times 10^{-4}$ \\
 ~~ ${\cal B}(\Bp \to \tau^+ \nu)$ & $(1.06 \pm 0.19) \times 10^{-4}$ \\
 ~~ $R_K = \mathcal{B}(\Bp \to K^+\mu^+\mu^-)/\mathcal{B}(\Bp \to K^+e^+e^-)$ &
 \multirow{2}{*} {$\aerr{0.745}{0.090}{0.074}{0.036}$}\\
 ~~ ~~~~ in $1.0<m^2_{\ell^+\ell^-}<6.0~{\mathrm{Ge\kern -0.1em V}^2/c^4}$ & \\ 
 ~~ $R_{K^\ast} = \mathcal{B}(\Bp \to K^{\ast 0}\mu^+\mu^-)/\mathcal{B}(\Bp \to K^{\ast 0}e^+e^-)$ &
 \multirow{2}{*} {$\cerr{0.69}{0.12}{0.09}$}\\
 ~~ ~~~~ in $1.1<m^2_{\ell^+\ell^-}<6.0~{\mathrm{Ge\kern -0.1em V}^2/c^4}$ & \\ 
 ~~ $A_{\CP}(\Bd\to K^+\pi^-)$, $A_{\CP}(B^+\to K^+\pi^0)$  & $-0.084 \pm 0.004$, $0.040 \pm 0.021$ \\
 ~~ $A_{\CP}(\Bs\to K^-\pi^+)$ & $0.213 \pm 0.017$ \\
 ~~ Longitudinal polarisation of $\Bd \to \phi  \Kstarz$ & $0.497 \pm 0.017$ \\
 ~~ Longitudinal polarisation of $\Bs \to \phi  \phi$ & $0.379 \pm 0.013$ \\
 ~~ Observables in $\Bz \to K^{*0}\mu^+\mu^-$ decays & \multirow{2}{*}{See Sec.~\ref{sec:rare-radll}} \\
 ~~ ~~~~ in bins of $q^2 = m^2(\mu^+\mu^-)$ & \\
\hline
 {\bf\boldmath $D^0$ mixing and \CP violation parameters} &   \\
 ~~$x$ &  $(0.32\,\pm 0.14)\%$  \\
 ~~$y$ &  $(0.69\,^{+0.06}_{-0.07})\%$  \\
~~$\delta^{}_{K\pi}$ &  $(15.2\,^{+7.6}_{-10.0})^\circ$  \\
 ~~$A^{}_D$ &  $(-0.88\,\pm 0.99)\%$  \\
 ~~$|q/p|$ & $0.89\,^{+0.08}_{-0.07}$  \\
 ~~$\phi$ &  $(-12.9\,^{+9.9}_{-8.7})^\circ$  \\
\hline
 ~~$x^{}_{12}$ (no direct \CP violation) &  $(0.41\,^{+0.14}_{-0.15})\%$  \\
 ~~$y^{}_{12}$ (no direct \CP violation) &  $(0.61\,\pm 0.07)\%$  \\
 ~~$\phi^{}_{12}$ (no direct \CP violation) &  $(-0.17\,\pm 1.8)^\circ$  \\
\hline
~~$a^{\rm ind}_{\CP}$ & $(0.030 \pm 0.026)\%$ \\
~~$\Delta a^{\rm dir}_{\CP}$ & $(-0.134 \pm 0.070)\%$ \\
\hline
 {\bf\boldmath Leptonic $D$ decays} &   \\
 ~~$f^{}_D$     & $(203.7\,\pm 4.9)$~MeV  \\
 ~~$f^{}_{D_s}$  & $(257.1\,\pm 4.6)$~MeV  \\
 ~~$|V^{}_{cd}|$ & $0.2164\,\pm 0.0050_{\rm exp} \pm 0.0015_{\rm LQCD}$  \\
 ~~$|V^{}_{cs}|$ & $1.006\,\pm 0.018_{\rm exp} \pm 0.005_{\rm LQCD}$  \\
\hline
 {\bf\boldmath Benchmark charm branching fractions} &   \\
 ~~${\cal B}(D^0\ra K^-\pi^+)$   & 
$(3.962\,\pm 0.017\,\pm 0.038\,\pm 0.027_{\rm FSR})\%$ \\
 ~~${\cal B}(D^0\ra K^+\pi^-)/{\cal B}(D^0\ra K^-\pi^+)$    & 
$(0.349\,^{+0.004}_{-0.003})\%$  \\
 ~~${\cal B}(D^+_s\ra K^+K^-\pi^+)$   &
$(5.44\,\pm 0.09\,\pm 0.11)\%$ \\
\hline
{\bf\boldmath $\tau$ parameters, lepton universality, and $|V_{us}|$} &   \\
~~ $g^{}_{\tau}/g^{}_{\mu}$    & \htuse{gtaubygmu_tau} \\
~~ $g^{}_{\tau}/g^{}_{e}$      & \htuse{gtaubyge_tau} \\
~~ $g^{}_\mu/g^{}_e$           & \htuse{gmubyge_tau} \\
~~ $\BR_e^{\text{uni}}$   & $(\htuse{Be_univ})\%$ \\
~~ $R_{\text{had}}$            & \htuse{R_tau} \\
~~ $|V_{us}|$ from $\BR(\tau^-\to X_s \nu^{}_\tau)$ & \htuse{Vus} \\
~~ $|V_{us}|$ from $\BR(\tau^- \to K^-\nu^{}_\tau)/ \BR(\tau^- \to \pi^-\nu^{}_\tau)$ & \htuse{Vus_tauKpi} \\ 
~~ $|V_{us}|$ from $\BR(\tau^-\to K^-\nu^{}_\tau)$                                    & \htuse{Vus_tauKnu} \\
~~ $|V_{us}|$ \mtau average                                                           & \htuse{Vus_tau} \\
\hline
\end{longtable}

\clearpage

\mysection{Introduction}
\label{sec:intro}

Flavour dynamics plays an important role in elementary particle interactions. 
The accurate knowledge of properties of heavy flavour
hadrons, especially $b$ hadrons, plays an essential role in determination of the elements of the Cabibbo-Kobayashi-Maskawa (CKM)
quark-mixing matrix~\cite{Cabibbo:1963yz,Kobayashi:1973fv}. 
The operation of the \belle\ and \babar\ $e^+e^-$ $B$ factory 
experiments led to a large increase in the size of available 
$B$-meson, $D$-hadron and $\tau$-lepton samples, 
enabling dramatic improvement in the accuracies of related measurements.
The CDF and \dzero\ experiments at the Fermilab Tevatron 
have also provided important results in heavy flavour physics,
most notably in the $B^0_s$ sector.
In the $D$-meson sector, the dedicated $e^+e^-$ charm factory experiments
CLEO-c and BESIII have made significant contributions.
Run~I and Run~II of the CERN Large Hadron Collider delivered high luminosity, 
enabling the collection of even larger samples of $b$ and $c$ hadrons, and
thus a further leap in precision in many areas, at the ATLAS, CMS, and
(especially) LHCb experiments.  
With ongoing analyses of the LHC Run~II data, further improvements are  anticipated.

The Heavy Flavour Averaging Group (HFLAV)\footnote{
  The group was originally known by the acronym ``HFAG.''  
  Following feedback from the community, this was changed to HFLAV in 2017.
} 
was formed in 2002 to 
continue the activities of the LEP Heavy Flavour Steering 
Group~\cite{Abbaneo:2000ej_mod,*Abbaneo:2001bv_mod_cont}, which was responsible for calculating averages of measurements of $b$-flavour related quantities. 
HFLAV has evolved since its inception and currently consists of seven subgroups:
\begin{itemize}
\item the ``$B$ Lifetime and Oscillations'' subgroup provides 
averages for $b$-hadron lifetimes, $b$-hadron fractions in 
$\Upsilon(4S)$ decay and $pp$ or $p\bar{p}$ collisions, and various 
parameters governing $\Bz$--$\Bzb$ and $\Bs$--$\Bsb$ mixing and \CP violation;

\item the ``Unitarity Triangle Angles'' subgroup provides
averages for parameters associated with time-dependent $\CP$ 
asymmetries and $B \to DK$ decays, and resulting determinations 
of the angles of the CKM unitarity triangle;

\item the ``Semileptonic $B$ Decays'' subgroup provides averages
for inclusive and exclusive measurements of $B$-decay branching fractions, and
subsequent determinations of the CKM matrix element magnitudes
$|V_{cb}|$ and $|V_{ub}|$;

\item the ``$B$ to Charm Decays'' subgroup provides averages of 
branching fractions for $b$-hadron decays to final states involving open 
charm or charmonium mesons;

\item the ``Rare Decays'' subgroup provides averages of branching 
fractions and $\CP$ asymmetries for charmless, radiative, 
leptonic, and baryonic $B$-meson and \b-baryon decays;

\item the ``Charm Physics'' subgroup provides averages of numerous 
quantities in the charm sector, including branching fractions, 
properties of excited $D^{**}$ and $D^{}_{sJ}$ mesons, 
properties of charm baryons,
mixing, $\CP$-, and $T$-violation parameters in the $\Dz$--$\Dzb$ system,
and the $D^+$ and $D^+_s$ decay constants $f^{}_{D}$ and~$f^{}_{D_s}$;

\item the ``Tau Physics'' subgroup provides averages for \mtau
  branching fractions using a global fit,  elaborates on the results
  to test lepton universality and to determine the CKM matrix element
  magnitude $|V_{us}|$, and lists and combines branching-fraction upper limits for \mtau lepton-flavour-violating decays.

\end{itemize}
Subgroups consist of representatives from experiments producing 
relevant results in that area, \ie, representatives from
\babar, \belle, \belletwo, BESIII, CLEO(c), CDF, \dzero,  LHCb, ATLAS, and CMS.

This article is an update of the last HFLAV publication, which used results available by summer 2016~\cite{Amhis:2016xyh}. 
Here we report world averages using results available by September 2018.
In some cases, important new results made available later are included, and in others, minor revisions in the September 2018 averages have been made.
All plots carry a timestamp indicating approximately when the results shown were published.
In general, we use all publicly available results, including preliminary results that are supported by
written documentation, such as conference proceedings or publicly available reports from the collaborations.
However, we do not use preliminary results that remain unpublished 
for an extended period of time, or for which no publication is planned. 
Since HFLAV members are also members of the different collaborations, we exploit our close contact with analyzers to ensure that the
results are prepared in a form suitable for combinations.  

Section~\ref{sec:method} describes the methodology used for calculating
averages. In the averaging procedure, common input parameters used in 
the various analyses are adjusted (rescaled) to common values, and, 
where possible, known correlations are taken into account. 
Sections~\ref{sec:life_mix}--\ref{sec:tau} present world 
average values from each of the subgroups listed above. 
A complete listing of the averages and plots,
including updates since this document was prepared,
is available on the HFLAV web site:
\vskip0.15in\hskip0.75in
\vbox{
  \href{https://hflav.web.cern.ch}{\tt https://hflav.web.cern.ch} 
}

\clearpage
\section{Averaging methodology} 
\label{sec:method} 

The main task of HFLAV is to combine independent but possibly
correlated measurements of a parameter to obtain the world's 
best estimate of that parameter's value and uncertainty. These
measurements are typically made by different experiments, or by the
same experiment using different data sets, or by the same
experiment using the same data but using different analysis methods.
In this section, the general approach adopted by HFLAV is outlined.
For some cases, somewhat simplified or more complex algorithms are 
used; these are noted in the corresponding sections.
    
Our methodology focuses on the problem of combining measurements 
obtained with different assumptions about external (or ``nuisance'') 
parameters and with potentially correlated systematic uncertainties.
It is important for any averaging procedure that the quantities
measured by experiments be statistically well-behaved, which in this 
context means having a (one- or multi-dimensional) Gaussian likelihood
function. We let $\boldsymbol{x}$ represent a set of parameters and $\boldsymbol{x}_i$ denotes the $i$th set of measurements of those parameters with the covariance matrix $\boldsymbol{V}_{\!i}$.
In what follows we assume that $\boldsymbol{x}$ does not 
contain redundant information, \ie, if it contains $n$ 
elements then $n$ is the number of parameters being determined.
A $\chi^2$ statistic is constructed as
\begin{equation}
  \chi^2(\boldsymbol{x}) = \sum_i^N 
  \left( \boldsymbol{x}_i - \boldsymbol{x} \right)^{\rm T} 
  \boldsymbol{V}_{\!i}^{-1}  
  \left( \boldsymbol{x}_i - \boldsymbol{x} \right) \, ,
\end{equation}
where the sum is over the $N$ independent determinations of the quantities
$\boldsymbol{x}$, typically coming from different experiments; possible
correlations of the systematic uncertainties are discussed below.
The results of the average are the central values $\boldsymbol{\hat{x}}$, 
which are the values of $\boldsymbol{x}$ at the minimum of
$\chi^2(\boldsymbol{x})$, and their covariance matrix
\begin{equation}
  \boldsymbol{\hat{V}}^{-1} = \sum_i^N \boldsymbol{V}_{\!i}^{-1}
\end{equation}
as a generalisation of the one-dimensional estimate $\sigma^{-2} = \sum_i \sigma_i^{-2}$. We report the covariance matrices or the correlation matrices derived from
the averages whenever possible. 
In some cases where the matrices are large, it is inconvenient to report them
in this document, and they can instead be found on the HFLAV web pages. 

The value of $\chi^2(\boldsymbol{\hat{x}})$ provides a measure of the
consistency of the independent measurements of $\boldsymbol{x}$ after
accounting for the number of degrees of freedom ($\dof$), which is the 
difference between the number of measurements and the number of
fitted parameters: $N\cdot n - n$.
The values of $\chi^2(\boldsymbol{\hat{x}})$ and $\dof$ are typically 
converted to a confidence level (C.L.) and reported together with the 
averages. In cases where $\chi^2/\dof > 1$, 
we do not usually scale the resulting uncertainty, in contrast
to what is done by the Particle Data Group~\cite{PDG_2016}.
Rather, we examine the systematic uncertainties of each measurement 
to better understand them. Unless we find systematic discrepancies 
among the measurements, we do not apply any additional correction 
to the calculated uncertainty. 
If special treatment is necessary in order to calculate an average, or 
if an approximation used in the calculation might not be sufficiently
accurate (\eg, assuming Gaussian uncertainties when the likelihood function 
exhibits non-Gaussian behavior), we point this out. 
Further modifications to the averaging procedures for non-Gaussian
situations are discussed in Sec.~\ref{sec:method:nonGaussian}.

For observables such as branching fractions, experiments typically 
report upper limits when the signal is not significant.  
Sometimes there is insufficient information available to combine 
upper limits on a parameter obtained by different experiments;
in this case we usually report 
only the most restrictive upper limit. 
For branching fractions of lepton-flavour-violating decays of 
tau leptons, we calculate combined upper limits as discussed
in Sec.~\ref{sec:tau:lfv-limits}.

\subsection{Treatment of correlated systematic uncertainties}
\label{sec:method:corrSysts} 

Consider two hypothetical measurements of a parameter $x$, which can
be summarized as
\begin{align*}
 & x_1 \pm \delta x_1 \pm \Delta x_{1,1} \pm \Delta x_{1,2} \ldots \\
 & x_2 \pm \delta x_2 \pm \Delta x_{2,1} \pm \Delta x_{2,2} \ldots \, ,
\end{align*}
where the $\delta x_k$ are statistical uncertainties and
the $\Delta x_{k,i}$ are contributions to the systematic
uncertainty. The simplest approach is to combine statistical 
and systematic uncertainties in quadrature:
\begin{align*}
 & x_1 \pm \left(\delta x_1 \oplus \Delta x_{1,1} \oplus \Delta x_{1,2} \oplus \ldots\right) \\
 & x_2 \pm \left(\delta x_2 \oplus \Delta x_{2,1} \oplus \Delta x_{2,2} \oplus \ldots\right) \, ,
\end{align*}
and then perform a weighted average of $x_1$ and $x_2$ using their
combined uncertainties, treating the measurements as independent. This 
approach suffers from two potential problems that we try to address. 
First, the values $x_k$ may have been obtained using different
assumptions for nuisance parameters; \eg, different values of the \Bz
lifetime may have been used for different measurements of the
oscillation frequency $\deltamd$. The second potential problem 
is that some systematic uncertainties may be correlated
between measurements. For example, different measurements of 
$\deltamd$ may depend on the same branching fraction 
used to model a common background.

The above two problems are related. We can represent the systematic uncertainties as a set of nuisance parameters $y_i$
upon which $x_k$ depends. The uncertainty $\Delta y_i$, which is the uncertainty on $y_i$ coming from external measurements, in this way  gives a contribution $\Delta x_{k,i}$ to the
systematic uncertainty. 
We thus use the values of $y_i$ and
$\Delta y_i$ assumed by each measurement in our averaging (we refer 
to these values as $y_{k,i}$ and $\Delta y_{k,i}$). 
To properly treat correlated systematic uncertainties among measurements,
requires decomposing the overall systematic uncertainties into correlated
and uncorrelated components. Correlated systematic uncertainties are those that depend on a shared nuisance  parameter, \eg a lifetime as mentioned above; Uncorrelated systematic uncertainties do not share a nuisance parameter, \eg the statistical uncertainty resulting from independent limited size simulations of background components.
As different measurements often quote different types of systematic
uncertainties, achieving consistent definitions in order to properly 
treat correlations
requires close coordination between HFLAV and the experiments. 
In some cases, a group of
systematic uncertainties must be combined into a coarser
description in order to obtain an average that is consistent 
among measurements. Systematic uncertainties
that are uncorrelated with any other source of uncertainty are 
combined together with the statistical uncertainty, so that the only
systematic uncertainties treated explicitly are those that are
correlated with at least one other measurement via a consistently-defined
external parameter $y_i$. When asymmetric statistical or systematic
uncertainties are quoted by experiments, we symmetrize them, since our 
combination method implicitly assumes Gaussian likelihoods 
(or parabolic log likelihoods) for each measurement.

\begin{figure}[!tb]
\centering
\includegraphics[width=5.0in]{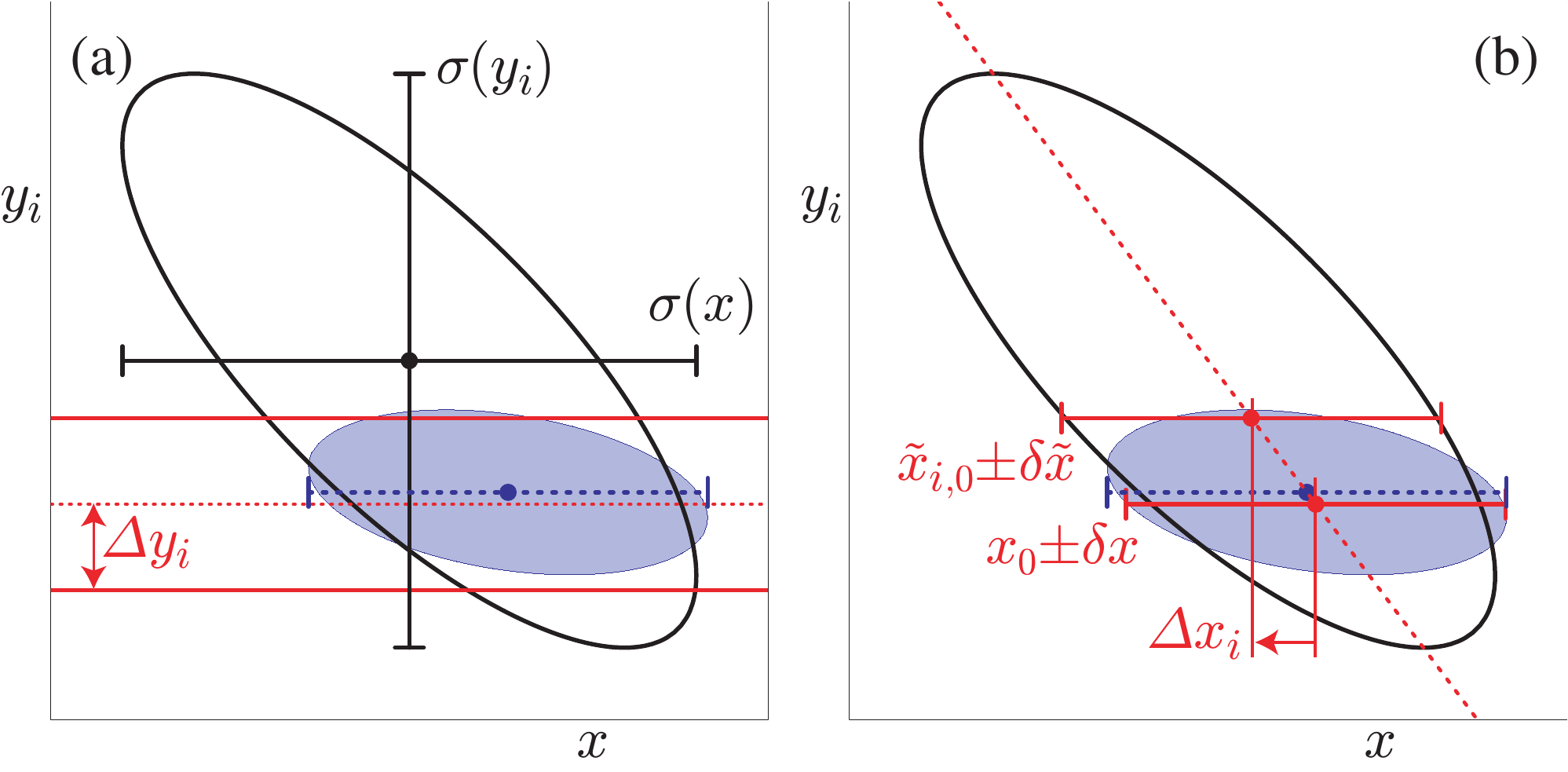}
\caption{
  Illustration of the possible dependence of a measured quantity $x$ on a
  nuisance parameter $y_i$.
  The left-hand plot (a) compares the 68\% confidence level contours of a
  hypothetical measurement's unconstrained (large ellipse) and
  constrained (filled ellipse) likelihoods, using the Gaussian
  constraint on $y_i$ represented by the horizontal band. 
  The solid error bars represent the statistical uncertainties $\sigma(x)$ and
  $\sigma(y_i)$ of the unconstrained likelihood. 
  The dashed error bar shows the statistical uncertainty on $x$ from a
  constrained simultaneous fit to $x$ and $y_i$. 
  The right-hand plot (b) illustrates the method described in the text of
  performing fits to $x$ with $y_i$ fixed at different values. 
  The dashed diagonal line between these fit results has the slope
  $\rho(x,y_i)\sigma(y_i)/\sigma(x)$ in the limit of an unconstrained
  parabolic log likelihood. 
  The result of the constrained simultaneous fit from (a) is shown as a dashed
  error bar on $x$.
}
\label{fig:singlefit}
\end{figure}

The fact that a measurement of $x$ is sensitive to $y_i$
indicates that, in principle, the data used to measure $x$ could
also be used for a simultaneous measurement of $x$ and $y_i$. This
is illustrated by the large contour in Fig.~\ref{fig:singlefit}(a).
However, there often exists an external measurement of $y_i$ with uncertainty $\Delta y_i$ (represented by the horizontal band in
Fig.~\ref{fig:singlefit}(a)) that is more precise than the constraint
$\sigma(y_i)$ from the $x$ data alone. In this case one can perform 
a simultaneous fit to $x$ and $y_i$, including the external 
measurement as a constraint, and obtain the filled $(x,y)$ contour and dashed 
one-dimensional estimate of $x$ shown in Fig.~\ref{fig:singlefit}(a). We call the fit without the external measurement \emph{unconstrained} and when it is included \emph{constrained}. 
For this procedure one usually takes the uncertainty on the external measurement  
$\Delta y_i$ to be Gaussian.

When the external constraints $\Delta y_i$ are significantly more
precise than the sensitivity $\sigma(y_i)$ of the data alone, 
the additional complexity of a constrained fit with extra free 
parameters may not be justified by the resulting increase in
sensitivity. In this case the usual procedure 
is to perform a baseline fit with all $y_i$ fixed
to nominal values $y_{i,0}$, obtaining $x = x_0 \pm \delta x$. 
This baseline fit neglects the uncertainty due to $\Delta y_i$, but
this uncertainty is subsequently recovered by repeating the fit separately 
for each external parameter $y_i$, with its value fixed to 
$y_i = y_{i,0}\pm \Delta y_i$. This gives the result
$x = \tilde{x}_{0,i} \pm \delta\tilde{x}$ as
illustrated in Fig.~\ref{fig:singlefit}(b). The shift
in the central value $\Delta x_i = \tilde{x}_{0,i} - x_0$ is 
usually quoted as the systematic uncertainty due 
to the unknown value of $y_i$. 
If the unconstrained data can be represented by a Gaussian
likelihood function, the shift will equal
\begin{equation}
\Delta x_i = \rho(x,y_i)\frac{\sigma(x)}{\sigma(y_i)}\,\Delta y_i \,,
\end{equation}
where $\sigma(x)$ and $\rho(x,y_i)$ are the statistical uncertainty on
$x$ and the correlation between $x$ and $y_i$ in the unconstrained data,
respectively. 
This procedure gives very similar results to that of the 
constrained fit with extra parameters: 
the central values $x_0$ agree to ${\cal O}(\Delta y_i/\sigma(y_i))^2$, 
and the uncertainties $\delta x \oplus \Delta x_i$ agree to 
${\cal O}(\Delta y_i/\sigma(y_i))^4$.

To combine two or more measurements that share systematic
uncertainty due to the same external parameter(s) $y_i$, we try 
to perform a constrained simultaneous fit of all measurements 
to obtain values of $x$ and $y_i$.
When this is not practical, \eg\ if we do not have sufficient 
information to reconstruct the %
likelihoods corresponding to each measurement, we perform 
the two-step approximate procedure described below.

\begin{figure}[!tb]
\centering
\includegraphics[width=5.0in]{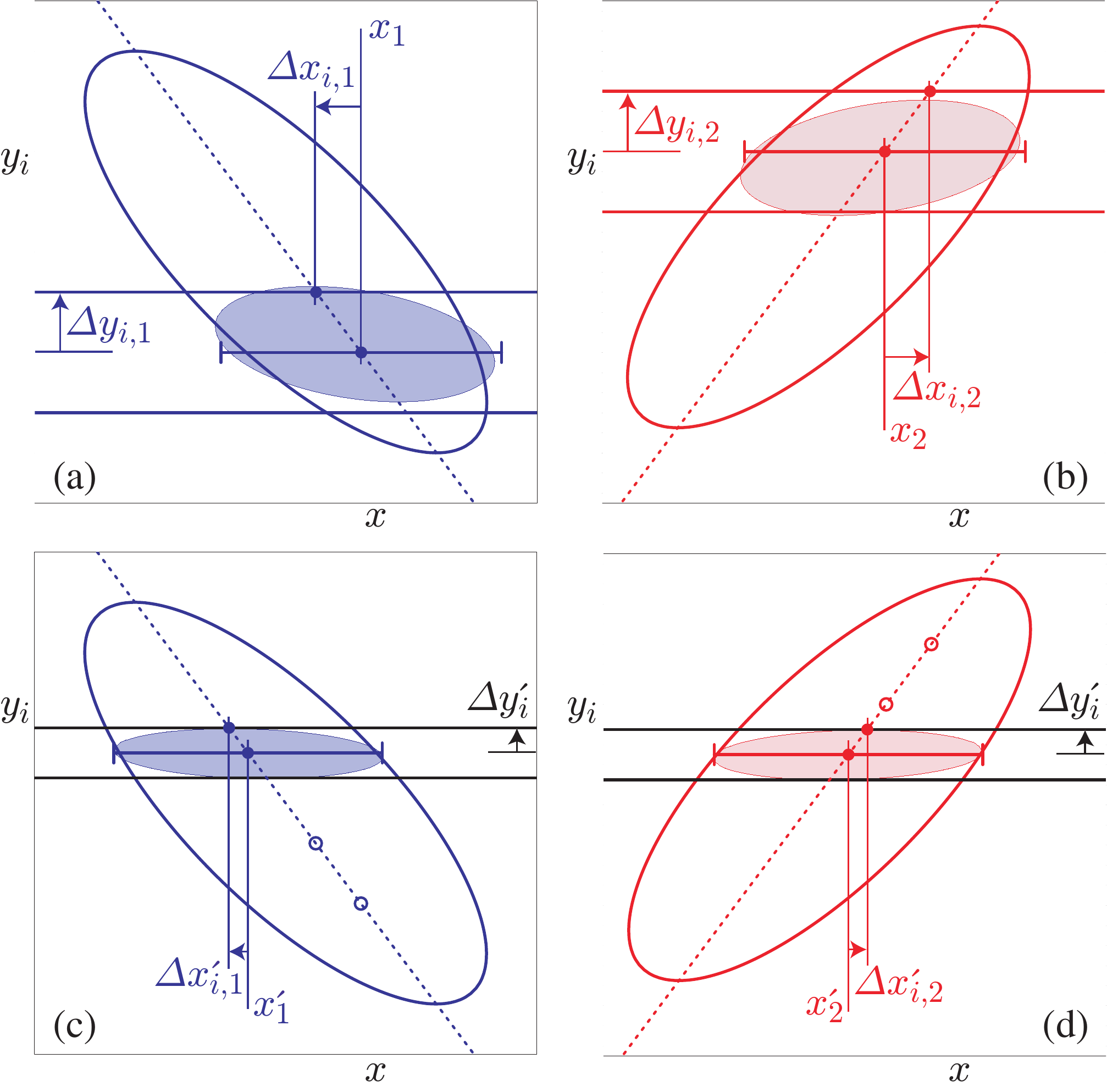}
\caption{
  Illustration of the HFLAV combination procedure for correlated systematic uncertainties.
  Upper plots (a) and (b) show examples of two individual measurements to be
  combined. 
  The large (filled) ellipses represent their unconstrained (constrained)
  likelihoods, while horizontal bands indicate the different assumptions about
  the value and uncertainty of $y_i$ used by each measurement. 
  The error bars show the results of the method described in the text for
  obtaining $x$ by performing fits with $y_i$ fixed to different values. 
  Lower plots (c) and (d) illustrate the adjustments to accommodate updated
  and consistent knowledge of $y_i$. 
  Open circles mark the central values of the unadjusted fits to $x$ with $y$
  fixed; these determine the dashed line used to obtain the adjusted values. 
}
\label{fig:multifit}
\end{figure}

Consider two statistically-independent measurements, 
$x_1 \pm (\delta x_1 \oplus \Delta x_{1,i})$ and 
$x_2\pm(\delta x_2\oplus \Delta x_{2,i})$, of 
the quantity $x$ as shown in Figs.~\ref{fig:multifit}(a,b).
For simplicity we consider only one correlated systematic 
uncertainty for each external parameter $y_i$.
As our knowledge of the $y_i$ improves, 
the measurements of $x$ will shift to different central
values and uncertainties. The first step of our procedure is 
to adjust the values of each measurement to reflect the current 
best knowledge of the external parameters $y_i'$ and their
ranges $\Delta y_i'$, as illustrated in Figs.~\ref{fig:multifit}(c,d). 
We adjust the central values $x_k$ and correlated systematic uncertainties
$\Delta x_{k,i}$ linearly for each measurement (indexed by $k$) and each
external parameter (indexed by $i$):
\begin{align}
x_k' &= x_k + \sum_i\,\frac{\Delta x_{k,i}}{\Delta y_{k,i}}\left(y_i'-y_{k,i}\right)  \label{eq:shiftx} \\
\Delta x_{k,i}'&= \Delta x_{k,i} \frac{\Delta y_i'}{\Delta y_{k,i}} \label{eq:shiftDx} \, .
\end{align}
This procedure is exact in the limit that the unconstrained
likelihood of each measurement is Gaussian and the linear relationships in Eqs.~(\ref{eq:shiftx})and~(\ref{eq:shiftDx}) are valid.

\begin{figure}[!tb]
\centering
\includegraphics[width=3.0in]{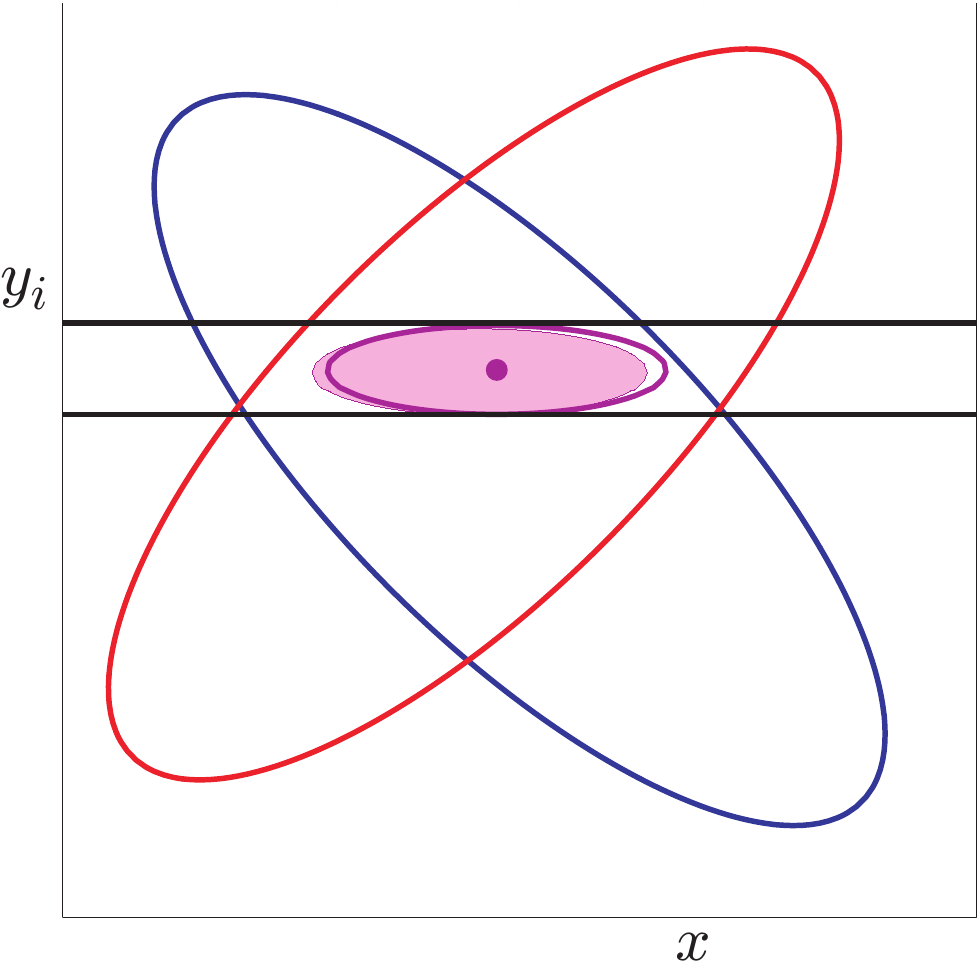}
\caption{
 Illustration of the combination of two hypothetical measurements of $x$
 using the method described in the text. 
 The ellipses represent the unconstrained likelihoods of each measurement,
 and the horizontal band represents the latest knowledge about $y_i$ that is
 used to adjust the individual measurements.
 The filled small ellipse shows the result of the exact method using 
 ${\cal L}_{\text{comb}}$, and the hollow small ellipse and dot show the
 result of the approximate method using $\chi^2_{\text{comb}}$. 
}
\label{fig:fit12}
\end{figure}

The second step is to combine the adjusted
measurements, $x_k'\pm (\delta x_k\oplus \Delta x_{k,1}'\oplus \Delta
x_{k,2}'\oplus\ldots)$ by constructing the goodness-of-fit statistic
\begin{equation}
\chi^2_{\text{comb}}(x,y_1,y_2,\ldots) \equiv \sum_k\,
\frac{1}{\delta x_k^2}\left[
x_k' - \left(x + \sum_i\,(y_i-y_i')\frac{\Delta x_{k,i}'}{\Delta y_i'}\right)
\right]^2 + \sum_i\,
\left(\frac{y_i - y_i'}{\Delta y_i'}\right)^2 \; .
\end{equation}
We minimize this $\chi^2$ to obtain the best values of $x$ and
$y_i$ and their uncertainties, as shown in Fig.~\ref{fig:fit12}. 
Although this method determines new values for
the $y_i$, we typically do not report them.

For comparison, the exact method we perform, if the 
unconstrained likelihoods ${\cal L}_k(x,y_1,y_2,\ldots)$ are
available, is to minimize the simultaneous likelihood
\begin{equation}
{\cal L}_{\text{comb}}(x,y_1,y_2,\ldots) \equiv \prod_k\,{\cal
  L}_k(x,y_1,y_2,\ldots)\,\prod_{i}\,{\cal 
  L}_i(y_i) \; ,
\end{equation}
with an independent Gaussian constraint for each $y_i$:
\begin{equation}
{\cal L}_i(y_i) = \exp\left[-\frac{1}{2}\,\left(\frac{y_i-y_i'}{\Delta
 y_i'}\right)^2\right] \; .
\end{equation}
The results of this exact method 
agree with those of the approximate method when the ${\cal L}_k$ are 
Gaussian, $\Delta y_i' \ll \sigma(y_i)$ and the linear assumption for the approximate method is valid.
If the likelihoods are non-Gaussian,
experiments need to provide ${\cal L}_k$ in order to perform
a combination.
If $\sigma(y_i)\approx \Delta y_i'$, experiments are encouraged 
to perform a simultaneous measurement of $x$ and $y_i$ so that their 
data will improve the world knowledge of~$y_i$. 

For averages where common sources of systematic uncertainty are important,
central values and uncertainties are rescaled to a common set of input 
parameters following the prescription above.
We use the most up-to-date values for common inputs, consistently across
subgroups, taking values from within HFLAV or from the Particle Data Group
when possible.
The parameters and values used are listed in each subgroup section.

\subsection{Treatment of non-Gaussian likelihood functions}
\label{sec:method:nonGaussian} 

For measurements with no correlation between them and with Gaussian uncertainties, the usual estimator for the
average of a set of measurements is obtained by minimizing
\begin{equation}
  \chi^2(x) = \sum_k^N \frac{\left(x_k-x\right)^2}{\sigma^2_k} \, ,
\label{eq:chi2t}
\end{equation}
where $x_k$ is the $k$-th measured value of $x$ and $\sigma_k^2$ is the
variance of the distribution from which $x_k$ was drawn.  
The value $\hat{x}$ at minimum $\chi^2$ is the estimate for the parameter $x$.
The true $\sigma_k$ are unknown but typically the uncertainty as assigned by the
experiment $\sigma_k^{\rm raw}$ is used as an estimator for it.
However, caution is advised when $\sigma_k^{\rm raw}$
depends on the measured value $x_k$. 
Examples of this are multiplicative systematic uncertainties such as those
due to acceptance, or the $\sqrt{N}$
dependence of Poisson statistics for which $x_k \propto N$
and $\sigma_k \propto \sqrt{N}$.
Failing to account for this type of dependence when averaging leads to a
biased average. Such biases can be avoided by minimizing
\begin{equation}
  \chi^2(x) = \sum_k^N \frac{\left(x_k-x\right)^2}{\sigma^2_k(\hat{x})} \,,
\label{eq:chi2that}
\end{equation}
where $\sigma_k(\hat{x})$ is the uncertainty on $x_k$ that includes 
the dependence of the uncertainty on the value measured.  As an example, 
consider the uncertainty due to acceptance for which
$\sigma_k(\hat{x}) = (\hat{x} / x_k)\times\sigma_k^{\rm raw}$.
Inserting this into Eq.~(\ref{eq:chi2that}) leads to
$$ 
\hat{x} = \frac{\sum_k^N x_k^3/(\sigma_k^{\rm raw})^2}
{\sum_k^N x_k^2/(\sigma_k^{\rm raw})^2} \, ,
$$
which is the correct behavior, \ie, weighting by the inverse 
square of the fractional uncertainty $\sigma_k^{\rm raw}/x_k$.
It is sometimes difficult to assess the dependence of $\sigma_k^{\rm raw}$ 
on $\hat{x}$ from the uncertainties quoted by the experiments.

\subsection{Treatment of unknown correlations}
\label{sec:method:unknownCorrelations} 

Another issue that needs careful treatment is that of correlations
among measurements, \eg, due to using the same decay model for 
intermediate states to calculate acceptances.
A common practice is to set the correlation
coefficient to unity to indicate full correlation. However, this is
not necessarily conservative and can result in 
underestimated uncertainty on the average.  
The most conservative choice of correlation coefficient
between two measurements $i$ and $j$
is that which maximizes the uncertainty on $\hat{x}$
due to the pair of measurements,
\begin{equation}
\sigma_{\hat{x}(i,j)}^2 = \frac{\sigma_i^2\,\sigma_j^2\,(1-\rho_{ij}^2)}
   {\sigma_i^2 + \sigma_j^2 - 2\,\rho_{ij}\,\sigma_i\,\sigma_j} \, ,
\label{eq:correlij}
\end{equation}
namely
\begin{equation}
\rho_{ij} =
\mathrm{min}\left(\frac{\sigma_i}{\sigma_j},\frac{\sigma_j}{\sigma_i}\right)
\, .
\label{eq:correlrho}
\end{equation}
This corresponds to setting 
$\sigma_{\hat{x}(i,j)}^2=\mathrm{min}(\sigma_i^2,\sigma_j^2)$.
Setting $\rho_{ij}=1$ when $\sigma_i\ne\sigma_j$ can lead to a significant
underestimate of the uncertainty on $\hat{x}$, as can be seen
from Eq.~(\ref{eq:correlij}). In the absence of better information on the correlation, we always use Eq.~(\ref{eq:correlij}).

\subsection{Splitting uncertainty for an average into components}
\label{sec:method:split} 

We carefully consider the various uncertainties
contributing to the overall uncertainty of an average. The 
covariance matrix describing the uncertainties of different 
measurements and their correlations is constructed, \ie,
$\boldsymbol{V} = 
\boldsymbol{V}_{\rm stat} + \boldsymbol{V}_{\rm sys} + 
\boldsymbol{V}_{\rm theory}$.
If the measurements are from independent data samples, then
$\boldsymbol{V}_{\rm stat}$ is diagonal, but
$\boldsymbol{V}_{\rm sys}$ and $\boldsymbol{V}_{\rm theory}$ 
may contain correlations.
The variance on the average $\hat{x}$ can be written as
\begin{eqnarray}
\sigma^2_{\hat{x}} 
 & = & 
\frac{1}{\sum_{i,j} \boldsymbol{V}^{-1}_{ij}} \ = \ 
\frac{ \sum_{i,j}
  \left(\boldsymbol{V}^{-1}\, 
    \boldsymbol{V} \, \boldsymbol{V}^{-1}\right)_{ij} }
{\left(\sum_{i,j} \boldsymbol{V}^{-1}_{ij}\right)^2} \\
& = & 
\frac{ \sum_{i,j}
  \left(\boldsymbol{V}^{-1}\, 
    \left[ \boldsymbol{V}_{\rm stat}+ \boldsymbol{V}_{\rm sys}+
      \boldsymbol{V}_{\rm theory} \right] \, \boldsymbol{V}^{-1}\right)_{ij} }
{\left(\sum_{i,j} \boldsymbol{V}^{-1}_{ij}\right)^2} 
\ = \ 
\sigma^2_{\text{stat}} + \sigma^2_{\text{sys}} + \sigma^2_{\text{th}} \, ,
\end{eqnarray}
where the last step follows from defining $\sigma^2_{\text{stat}}$, $\sigma^2_{\text{sys}}$ and $\sigma^2_{\text{th}}$ as each of the three parts of the sum in the previous step.
This breakdown of uncertainties is used in certain cases,
but usually only a single, total uncertainty is quoted for an average.

\clearpage

\mysection{Production fractions, lifetimes and mixing parameters of \b hadrons}
\labs{life_mix}

Quantities such as \b-hadron production fractions, \b-hadron lifetimes, 
and neutral \B-meson oscillation frequencies were studied
in the 1990's at LEP and SLC (\ee colliders at $\sqrt{s}=m_{\particle{Z}}$), at DORIS~II and CESR (\ee colliders at $\sqrt{s}=m_{\Ups}$),
as well as at the Tevatron
(\particle{p\bar{p}} collider at $\sqrt{s}=1.8\TeV$). 
This was followed by precise measurements of the \Bd and \Bu mesons
performed at the asymmetric \B factories, KEKB and PEPII
(\ee colliders at $\sqrt{s}=m_{\Ups}$), as well as measurements related 
to the other \b hadrons, in particular \Bs, \Bc and \Lb, 
performed at the upgraded Tevatron ($\sqrt{s}=1.96\TeV$).
Nowadays, the most precise measurements are coming from the 
LHC ($pp$ collider at $\sqrt{s}=7$, $8\TeV$ and $13\TeV$),
in particular the LHCb experiment. 

In most cases, these basic quantities, in addition to being interesting by themselves,
are necessary ingredients for more refined measurements,
such as those of decay-time-dependent \CP-violating asymmetries.
It is therefore important that the best experimental
values of these quantities continue to be kept up-to-date and improved. 

In several cases, the averages presented in this chapter are 
needed and used as input for the results given in the subsequent sections. 
Some averages need the knowledge of other 
averages in a circular way. This coupling, which appears through the 
\b-hadron fractions whenever inclusive or semi-exclusive measurements 
have to be considered, has been reduced drastically in the past several years 
with increasingly precise exclusive measurements becoming available
and dominating practically all averages. 

In addition to \b-hadron fractions, lifetimes and 
oscillation frequencies, this section also deals with \CP violation
in the \Bd and \Bs mixing amplitudes, as well as the
\CP-violating phase 
$\phiccbars$, %
which is the phase 
difference %
between the $\Bs-\Bsbar$ mixing amplitude and the $b\to c\bar{c}s$ decay amplitude.
In the absence of new physics and Penguin contributions this phase is equal to $-2\beta_s = -\arg\left[\left(V_{ts}V^*_{tb}\right)^2/\left(V_{cs}V^*_{cb}\right)^2\right]$.
The angle $\beta$, which is the equivalent of $\beta_s$ for the \Bd 
system, is discussed in Section~\ref{sec:cp_uta}. 

Throughout this section, published results that have been superseded 
by subsequent publications are ignored (\ie, excluded from the averages)
and are only referred to if necessary.

\mysubsection{\b-hadron production fractions}
\labs{fractions}
 
We consider here the relative fractions of the different \b-hadron 
species found in an unbiased sample of weakly decaying \b hadrons 
produced in a specific process. The knowledge of these fractions
is useful to characterize the signal composition in inclusive \b-hadron 
analyses, to predict the background composition in exclusive analyses, 
and to convert (relative) observed event yields into (relative) branching fraction 
measurements. 
We distinguish 
here the following three $b$-hadron production processes: \Ups decays, \Upsfive decays, and 
high-energy collisions (including \Z decays). 

\mysubsubsection{\b-hadron production fractions in \Ups decays}
\labs{fraction_Ups4S}

Only pairs of the two lightest (charged and neutral) \B mesons 
can be produced in \Ups decays. 
Therefore, only the following two branching fractions must be considered: 
\begin{eqnarray}
f^{+-} & = & \frac{\Gamma(\Ups \to \particle{B^+B^-})}{\Gamma_{\rm tot}(\Ups)}  \,, \\
f^{00} & = & \frac{\Gamma(\Ups \to \particle{B^0\bar{B}^0})}{\Gamma_{\rm tot}(\Ups)} \,.
\end{eqnarray}
In practice, most analyses measure their ratio
\begin{equation}
R^{+-/00} = \frac{f^{+-}}{f^{00}} = \frac{\Gamma(\Ups \to \particle{B^+B^-})}{\Gamma(\Ups \to \particle{B^0\bar{B}^0})} \,,
\end{equation}
which is easier to access experimentally.
An inclusive (but separate) reconstruction of 
\Bu and \Bd is difficult. Therefore, $R^{+-/00}$ is measured with exclusive decays ${\Bu} \to f^+$ and ${\Bd} \to f^0$ to specific final states $f^+$ and $f^0$ 
 that are related by isospin symmetry. 
Under the assumption that $\Gamma(\Bu \to f^+) = \Gamma(\Bd \to f^0)$, 
\ie, that isospin invariance holds in relating these \B decays,
the ratio of the number of reconstructed
$\Bu \to f^+$ and $\Bd \to f^0$ mesons, after correcting for efficiency, is
equal to
\begin{equation}
\frac{f^{+-}\, \BRp{\Bu\to f^+}}{f^{00}\, \BRp{\Bd\to f^0}}
= \frac{f^{+-}\, \Gamma({\Bu}\to f^+)\, \tau(\Bu)}%
{f^{00}\, \Gamma({\Bd}\to f^0)\,\tau(\Bd)}
= \frac{f^{+-}}{f^{00}} \, \frac{\tau(\Bu)}{\tau(\Bd)}  \,, 
\end{equation} 
where $\tau(\Bu)$ and $\tau(\Bd)$ are the \Bu and \Bd 
lifetimes, respectively.
Hence the primary quantity measured in these analyses 
is $R^{+-/00} \, \tau(\Bu)/\tau(\Bd)$, 
and the extraction of $R^{+-/00}$ with this method therefore 
requires the knowledge of the $\tau(\Bu)/\tau(\Bd)$ lifetime ratio. 

\begin{table}
\caption{Published measurements of the $\Bu/\Bd$ production ratio
in \Ups decays, together with their average (see text).
Systematic uncertainties due to the imperfect knowledge of 
$\tau(\Bu)/\tau(\Bd)$ are included. 
}
\labt{R_data}
\begin{center}
\begin{tabular}{lccll}
\hline
Experiment, year  & Ref. & Decay modes & Published value of & Assumed value \\
& & or method & $R^{+-/00}=f^{+-}/f^{00}$ & of $\tau(\Bu)/\tau(\Bd)$ \\
\hline
CLEO,   2001 & \cite{Alexander:2000tb}  & \particle{\jpsi K^{(*)}} 
             & $1.04 \pm0.07 \pm0.04$ & $1.066 \pm0.024$ \\
CLEO,   2002 & \cite{Athar:2002mr}  & \particle{D^*\ell\nu}
             & $1.058 \pm0.084 \pm0.136$ & $1.074 \pm0.028$\\
\belle, 2003 & \citehistory{Hastings:2002ff}{Hastings:2002ff,*Abe:2000yh_hist} & Dilepton events 
             & $1.01 \pm0.03 \pm0.09$ & $1.083 \pm0.017$\\
\babar, 2005 & \citehistory{Aubert:2004rz}{Aubert:2004rz,*Aubert:2001xs_hist,*Aubert:2004ur_hist} & \particle{(c\bar{c})K^{(*)}}
             & $1.06 \pm0.02 \pm0.03$ & $1.086 \pm0.017$\\ 
\hline
Average      & & & \hflavFF~(tot) & \hflavRTAUBU \\
\hline
\end{tabular}
\end{center}
\end{table}

The published measurements of $R^{+-/00}$ are listed\footnote{An old and imprecise $R$ measurement from
CLEO~\cite{Barish:1994mu} is
included in neither \Table{R_data} nor the average.} in \Table{R_data}
together with the corresponding values of 
$\tau(\Bu)/\tau(\Bd)$ assumed in each measurement.
All measurements are based on the above-mentioned method, 
except the one from \belle, which is a by-product of the 
\Bd mixing frequency analysis using dilepton events
(but note that it too assumes isospin invariance, 
namely $\Gamma(\Bu \to \ell^+{\rm X}) = \Gamma(\Bd \to \ell^+{\rm X})$).
The latter is therefore treated in a slightly different 
manner in the following procedure used to combine 
these measurements:
\begin{itemize} 
\item each published value of $R^{+-/00}$ from CLEO and \babar
      is first converted back to the original measurement of 
      $R^{+-/00} \, \tau(\Bu)/\tau(\Bd)$, using the value of the 
      lifetime ratio assumed in the corresponding analysis;
\item a simple weighted average of these original
      measurements of $R^{+-/00} \, \tau(\Bu)/\tau(\Bd)$ from 
      CLEO and \babar
      is then computed, assuming no 
      statistical or systematic correlations between them;

\item the weighted average of $R^{+-/00} \, \tau(\Bu)/\tau(\Bd)$ 
      is converted into a value of $R^{+-/00}$, using the latest 
      average of the lifetime ratios, $\tau(\Bu)/\tau(\Bd)=\hflavRTAUBU$ 
      (see \Sec{lifetime_ratio});
\item the \belle measurement of $R^{+-/00}$ is adjusted to the 
      current values of $\tau(\Bd)=\hflavTAUBD$ and 
      $\tau(\Bu)/\tau(\Bd)=\hflavRTAUBU$ (see \Sec{lifetime_ratio}),
      using the procedure described in Sec.~\ref{sec:method:corrSysts}; 
\item the combined value of $R^{+-/00}$ from CLEO and \babar is averaged 
      with the adjusted value of $R^{+-/00}$ from \belle, assuming a 100\% 
      correlation of the systematic uncertainty due to the limited 
      knowledge on $\tau(\Bu)/\tau(\Bd)$; no other correlation is considered. 
\end{itemize} 
The resulting global average, 
\begin{equation}
R^{+-/00} = \frac{f^{+-}}{f^{00}} =  \hflavFF \,,
\labe{Rplusminus}
\end{equation}
is consistent with equal production rate of charged and neutral \B mesons, 
although only at the $\hflavNSIGMAFF\,\sigma$ level.

On the other hand, the \babar collaboration has 
performed a direct measurement of the $f^{00}$ fraction 
using an original method, which neither relies on isospin symmetry nor requires 
the knowledge of $\tau(\Bu)/\tau(\Bd)$. Rather, the method is
based on comparing the number of events where a single 
$B^0 \to D^{*-} \ell^+ \nu$ decay is reconstructed to the number 
of events where two such decays are reconstructed. The result of this measurement is~\cite{Aubert:2005bq}
\begin{equation}
f^{00}= 0.487 \pm 0.010\,\mbox{(stat)} \pm 0.008\,\mbox{(syst)} \,.
\labe{fzerozero}
\end{equation}

The results of \Eqss{Rplusminus}{fzerozero} are obtained with very different methods
and are completely independent of each other. 
Their product yields $f^{+-} = \hflavFPROD$, 
and  combining them into the sum of the  charged and neutral fractions gives $f^{+-} + f^{00}= \hflavFSUM$, 
compatible with unity.

The precision of the fractions can be further improved by setting $f^{+-}+f^{00}= 1$. This approximation is justified by the small branching fractions, of order $10^{-4}$, that have been measured for $\Upsilon(4S)$ decays to several non-$B\bar B$ final states, specifically
$\Upsilon(1S)\pi^+\pi^-$,
$\Upsilon(2S)\pi^+\pi^-$, $\Upsilon(1S)\eta$ and $\Upsilon(1S)\eta'$~\cite{Aubert:2006bm,Sokolov:2006sd,Aubert:2008az,Guido:2018ywg}. These branching fractions correspond to a sum of partial
widths that is several times larger than $\Gamma(\Upsilon(4S)\to \ee)$, yet are much smaller than the uncertainties in the measurements of $f^{+-}$ and $f^{00}$. The approximation is also consistent with CLEO's observation that $\BRp{\Ups\to \BB}>0.96$ at \CL{95}~\cite{Barish:1995cx}.
Assuming $f^{+-}+f^{00}=1$, the results of \Eqss{Rplusminus}{fzerozero}
are averaged (first converting \Eq{Rplusminus} 
into a value of $f^{00}=1/(R^{+-/00}+1)$) 
to yield the following more precise estimates:
\begin{equation}
f^{00} = \hflavFNW  \,,~~~ f^{+-} = 1 -f^{00} =  \hflavFCW \,,~~~
\frac{f^{+-}}{f^{00}} =  \hflavFFW \,.
\end{equation}
The latter ratio differs from unity by $\hflavNSIGMAFFW\,\sigma$.

\mysubsubsection{\b-hadron production fractions at the \Upsfive energy}
\labs{fraction_Ups5S}

\newcommand{\fsfive}{\ensuremath{f^{\Upsfive}_{s}}}
\newcommand{\fudfive}{\ensuremath{f^{\Upsfive}_{u,d}}}
\newcommand{\fnBfive}{\ensuremath{f^{\Upsfive}_{B\!\!\!\!/}}}

Hadronic events produced in $e^+e^-$ collisions at the \Upsfive (also known as
$\Upsilon(10860)$) energy can be classified into three categories: 
light-quark ($u$, $d$, $s$, $c$) continuum events, $b\bar{b}$ continuum events,
and \Upsfive events. The latter two cannot be distinguished and will be called
$b\bar{b}$ events in the following. These $b\bar{b}$ events, including
$b\bar{b}\gamma$ where the photon arises from initial-state radiation, 
can hadronize into different final states.
We define \fudfive\ to be
the fraction of $b\bar{b}$ events with a pair of non-strange 
bottom mesons, namely, $B\bar{B}$, $B\bar{B}^*$, $B^*\bar{B}$, $B^*\bar{B}^*$,
$B\bar{B}\pi$, $B\bar{B}^*\pi$, $B^*\bar{B}\pi$,
$B^*\bar{B}^*\pi$, and $B\bar{B}\pi\pi$, 
where
$B$ denotes a $B^0$ or $B^+$ meson and 
$\bar{B}$ denotes a $\bar{B}^0$ or $B^-$ meson. Similarly, we define \fsfive\ to be
the fraction of $b\bar{b}$ events that hadronize into a pair of strange bottom mesons
($B_s^0\bar{B}_s^0$, $B_s^0\bar{B}_s^{*0}$, $B_s^{*0}\bar{B}_s^0$, and
$B_s^{*0}\bar{B}_s^{*0}$). 
Note that the excited bottom-meson states decay via $B^* \to B \gamma$ and
$B_s^{*0} \to B_s^0 \gamma$.
Lastly, 
\fnBfive\ is defined to be the fraction of $b\bar{b}$ events without 
open-bottom mesons in the final state (which includes \Upsfive decays to light bottomonium).
By construction, these fractions satisfy
\begin{equation}
\fudfive + \fsfive + \fnBfive = 1 \,.
\labe{sum_frac_five}
\end{equation} 

\begin{table}
\caption{Published measurements of \fsfive, obtained 
assuming $\fnBfive=0$. The results are 
quoted as in the original publications, except for the 2010
Belle measurement, which is quoted as 
$1-\fudfive$ with \fudfive\ from \Ref{Drutskoy:2010an}.
The 2012 Belle measurement, reported and used in \Ref{Esen:2012yz}, is an undocumented update of the analysis of \Ref{Drutskoy:2006fg} with the full \Upsfive dataset.
}
\labt{fsFiveS}
\begin{center}
\begin{tabular}{lll}
\hline
Experiment, year, dataset                 & Decay mode or method    & Value of \fsfive\  \\
\hline
CLEO, 2006, 0.42\invfb~\citehistory{Huang:2006em}{Huang:2006em_hist} & $\Upsfive\to D_{s}X$     & $0.168 \pm 0.026\,\,^{+\,0.067}_{-\,0.034}$  \\
             & $\Upsfive \to \phi X$    & $0.246 \pm 0.029\,\,^{+\,0.110}_{-\,0.053}$ \\
             & $\Upsfive \to B\bar{B}X$ & $0.411 \pm 0.100 \pm 0.092$ \\  
             & CLEO average of above 3  & $0.21^{+0.06}_{-0.03}$      \\  \hline
Belle, 2006, 1.86\invfb~\cite{Drutskoy:2006fg} & $\Upsfive \to D_s X$     & $0.179 \pm 0.014 \pm 0.041$ \\
             & $\Upsfive \to D^0 X$     & $0.181 \pm 0.036 \pm 0.075$ \\  
             & Belle average of above 2 & $0.180 \pm 0.013 \pm 0.032$ \\  \hline 
Belle, 2010, 23.6\invfb~\cite{Drutskoy:2010an} & $\Upsfive \to B\bar{B}X$ & $0.263 \pm 0.032 \pm 0.051$ %
\\  \hline 
Belle, 2012, 121.4\invfb~\cite{Esen:2012yz} & $\Upsfive \to D_sX, D^0X$ & $0.172 \pm 0.030$ \\ \hline
\end{tabular}
\end{center}
\end{table}

\begin{table}
\caption{External inputs on which the \fsfive\ averages are based.}
\labt{fsFiveS_external}
\begin{center}
\begin{tabular}{lcl}
\hline
Branching fraction   & Value     & Explanation and reference \\
\hline
${\cal B}(B\to D_s X)\times {\cal B}(D_s \to \phi\pi)$ & 
$0.00374\pm 0.00014$ & Derived from~\cite{PDG_2018} \\
${\cal B}(B^0_s \to D_s X)$ & 
$0.92\pm0.11$ & Model-dependent estimate~\cite{Artuso:2005xw} \\
${\cal B}(D_s \to \phi\pi)$ & 
$0.045\pm0.004$ &\!\!\cite{PDG_2018} \\
${\cal B}(B\to D^0 X)\times {\cal B}(D^0 \to K\pi)$ & 
$0.0242\pm0.0011$ & Derived from~\cite{PDG_2018} \\
${\cal B}(B^0_s \to D^0 X)$ & 
$0.08\pm0.07$ & Model-dependent estimate~\cite{Drutskoy:2006fg,Artuso:2005xw} \\
${\cal B}(D^0 \to K\pi)$ & 
$0.03954\pm0.00031$ & \!\!\cite{PDG_2018} \\
${\cal B}(B \to \phi X)$ & 
$0.0343\pm0.0012$ &\!\!\cite{PDG_2018} \\
${\cal B}(B^0_s \to \phi X)$ &
$0.161\pm0.024$ & Model-dependent estimate~\citehistory{Huang:2006em}{Huang:2006em_hist} \\
\hline
\end{tabular}
\end{center}
\end{table}

The CLEO and Belle collaborations have published %
measurements of several inclusive \Upsfive branching fractions, 
${\cal B}(\Upsfive\to D_s X)$, 
${\cal B}(\Upsfive\to \phi X)$ and 
${\cal B}(\Upsfive\to D^0 X)$, %
from which they extracted the
model-dependent estimates of \fsfive\ reported in \Table{fsFiveS}.
This extraction was performed under the implicit assumption  
$\fnBfive=0$, using the relation 
\begin{equation}
\frac12{\cal B}(\Upsfive\to D_s X)=\fsfive\times{\cal B}(B_s^0\to D_s X) + 
\left(1-\fsfive-\fnBfive\right)\times{\cal B}(B\to D_s X) \,,
\labe{Ds_correct}
\end{equation}
and similar relations for
${\cal B}(\Upsfive\to D^0 X)$ and ${\cal B}(\Upsfive\to \phi X)$.

However, the assumption $\fnBfive=0$ is known to be incorrect, given the observed production in $e^+e^-$ collisions at the \Upsfive\ energy of
the final states
$\Upsilon(1S)\pi^+\pi^-$,
$\Upsilon(2S)\pi^+\pi^-$,
$\Upsilon(3S)\pi^+\pi^-$
and
$\Upsilon(1S)K^+K^-$~\citehistory{Abe:2007tk,Garmash:2014dhx}{Abe:2007tk,Garmash:2014dhx_hist},
$h_b(1P)\pi^+\pi^-$ and 
$h_b(2P)\pi^+\pi^-$~\cite{Adachi:2011ji},
$\Upsilon(1S)\pi^0\pi^0$,
$\Upsilon(2S)\pi^0\pi^0$ 
and
$\Upsilon(3S)\pi^0\pi^0$~\cite{Krokovny:2013mgx},
and more recently 
$\Upsilon_J(1D)\eta$ and $\Upsilon(2S)\eta$~\cite{Tamponi:2018cuf}.
The sum of the visible (i.e., uncorrected for initial-state radiation) cross-sections into these final states, plus those of the unmeasured final states
$\Upsilon(1S)K^0\bar{K}^0$, $h_b(1P)\pi^0\pi^0$ and $h_b(2P)\pi^0\pi^0$, which are obtained by
assuming isospin conservation, amounts to
$$
\sigma^{\rm vis}(e^+e^-\to (\b\bar{\b})X) = 15.0\pm1.4~{\rm pb} \,,
$$
where $(\b\bar{\b})=\Upsilon(1S,2S,3S)$,  $\Upsilon_J(1D)$, $h_b(1P,2P)$, and $X=\pi\pi$, $KK$, $\eta$.
We divide this by the $\b\bar{\b}$ production cross section, 
$\sigma(e^+e^- \to \b\bar{\b} X) = 337 \pm 15$~pb, obtained as the average of the 
CLEO~\cite{Artuso:2005xw} and Belle~\cite{Esen:2012yz}
measurements, to obtain
$$
{\cal B}(\Upsfive\to (\b\bar{\b})X) = 0.045\pm0.005 \,.
$$
This should be taken as a lower bound for \fnBfive. 

To simultaneously extract the fractions under the exact constraints of \Eqss{sum_frac_five}{Ds_correct} and the
one-sided Gaussian constraint $\fnBfive \ge {\cal B}(\Upsfive \to (\b\bar{\b})X)$,
we follow the method described in \Ref{thesis_Louvot}, 
performing a $\chi^2$ fit of the  
measurements of the \Upsfive\ branching fractions of Refs.~%
\citehistory{Huang:2006em,Drutskoy:2006fg,Drutskoy:2010an}{Huang:2006em_hist,Drutskoy:2006fg,Drutskoy:2010an}. The latest Belle measurement of \fsfive~\Ref{Esen:2012yz}
lacks the information needed for the averaging, and is therefore not included.
Taking the inputs of \Table{fsFiveS_external} and all known 
correlations into account, the best fit values are
\begin{eqnarray}
\fudfive &=& \hflavFUDFIVE \,, \labe{fudfive} \\
\fsfive  &=& \hflavFSFIVE  \,, \labe{fsfive} \label{eq:5-Fractions-from-fit} \\
\fnBfive &=& \hflavFNBFIVE \,, \labe{fnBfive}
\end{eqnarray}
where the strongly asymmetric uncertainty on \fnBfive\ is due to the one-sided constraint
from the observed $(\b\bar{\b})X$ decays. These results, together with their correlations, 
imply
\begin{eqnarray}
\fsfive/\fudfive  &=& \hflavFSFUDFIVE  \,. \labe{fsfudfive} 
\end{eqnarray}
This is in fair agreement with \babar
results~\cite{Lees:2011ji}, obtained as a function of centre-of-mass energy and as a by-product of another measurement, and which are not used in our average due to insufficient information.

The production of $B^0_s$ mesons at the \Upsfive
is observed to be dominated by the $B_s^{*0}\bar{B}_s^{*0}$
channel, %
with $\sigma(e^+e^- \to B_s^{*0}\bar{B}_s^{*0})/%
\sigma(e^+e^- \to B_s^{(*)0}\bar{B}_s^{(*)0})
= (87.0\pm 1.7)\%$~\cite{Li:2011pg} %
measured as described in \Ref{Louvot:2008sc}.
The proportions of the various production channels 
for non-strange $B$ mesons have also been measured~\cite{Drutskoy:2010an}.

\mysubsubsection{\b-hadron production fractions at high energy}
\labs{fractions_high_energy}
\labs{chibar}

At high energy, all species of weakly decaying \b hadrons 
may be produced, either directly or in strong and electromagnetic 
decays of excited \b hadrons. Before 2010, it was assumed that the fractions 
of different species in unbiased samples of 
high-$p_{\rm T}$ \b-hadron jets where independent of whether they originated from \particle{Z^0} decays, 
\particle{p\bar{p}} collisions at the Tevatron, or 
\particle{p p} collisions at the LHC.
This hypothesis was plausible under the condition $Q^2 \gg \Lambda_{\rm QCD}^2$, namely, that the square of
the momentum transfer to the produced \b quarks is large compared 
with the square of the hadronization energy scale.
This hypothesis is correct %
in the limit $p_T\to \infty$, in which the production mechanism of a \b hadron is completely described by the fragmentation of the \b quark. For finite $p_T$, however, there are interference effects of the production mechanism of the \b quark and its hadronization. While formally suppressed by inverse powers of $p_T$, these effects may be sizable, especially when the fragmentation probabilities are small as \eg\ in the case of \b baryons.
In fact, the available data show that the fractions depend on the kinematics 
of the produced \b hadron.
Both CDF and LHCb report a $p_{\rm T}$ dependence of the  fractions, with 
the fraction of \Lb baryons
observed at low $p_{\rm T}$ being enhanced with respect to that 
seen at LEP at higher $p_{\rm T}$.

We present here two sets of  averages: one set includes only measurements 
performed at LEP, and the second set includes only measurements performed 
by CDF at the Tevatron.\footnote{The LHC production fractions results 
are still incomplete, lacking measurements of the production of 
weakly-decaying baryons heavier than \Lb. In Ref~\cite{Amhis:2016xyh}, we provided also a third set of averages including measurements performed at LEP, 
Tevatron and LHC, but this was mostly for comparison with previous averages. We have decided to discontinue these ``world averages'', because they mix environments with different fractions.} %
While the first set is well defined and is basically related to branching fractions of inclusive $Z^0$ decays, the other set is somewhat ill-defined, 
since it depends on the kinematic region covered by the experiment and over which the measurements are integrated.

Contrary to what happens in the charm sector, where the fractions of 
\particle{D^+} and \particle{D^0} are different, the relative production rate 
of \Bu and \Bd is not affected by the electromagnetic decays of 
excited $B^{*+}$ and $B^{*0}$ states and strong decays of excited 
$B^{**+}$ and $B^{**0}$ states. Decays of the type 
\particle{B_s^{**0} \to B^{(*)}K} also contribute to the \Bu and \Bd rates, 
but with the same magnitude if mass effects can be neglected.  
We therefore assume equal production of \Bu and \Bd mesons. We also  
neglect the production of weakly decaying states
made of several heavy quarks (such as \Bc or doubly heavy baryons) 
which is much smaller. Hence, for the purpose of determining 
the \b-hadron fractions, we use the constraints
\begin{equation}
\fBu = \fBd ~~~~\mbox{and}~~~ \fBu + \fBd + \fBs + \fbb = 1 \,,
\labe{constraints}
\end{equation}
where \fBu, \fBd, \fBs and \fbb
are the fractions of \Bu, \Bd, \Bs and weakly-decaying \b baryons, respectively.

We note that there are many measurements of the production cross-sections of
different species of \b hadrons.
In principle, these could be included in a global fit to determine the
production fractions.
We do not use these inputs at the current time, and instead average only the
explicit measurements of the production fractions.

The LEP experiments have measured
$\fBs \times \BRp{\Bs\to\particle{D_s^-} \ell^+ \nu_\ell \mbox{$X$}}$~\cite{Abreu:1992rv,Acton:1992zq,Buskulic:1995bd}, 
$\BRp{\b\to\Lb} \times \BRp{\Lb\to\Lc\ell^-\bar{\nu}_\ell \mbox{$X$}}$~\cite{Abreu:1995me,Barate:1997if}
and $\BRp{\b\to\Xib^-} \times \BRp{\Xi_b^- \to \Xi^-\ell^-\overline\nu_\ell 
\mbox{$X$}}$~\citehistory{Buskulic:1996sm,Abdallah:2005cw}{Buskulic:1996sm,Abdallah:2005cw_hist}
using partially reconstructed hadronic final states and a lepton to identify the $b$~hadron. They have also measured \fbb
using protons identified in \b-hadron decays~\cite{Barate:1997ty}, as well as the 
production rate of charged \b hadrons~\cite{Abdallah:2003xp}. 

Ratios of \b-hadron fractions have been measured by CDF using 
lepton+charm final 
states~\cite{Affolder:1999iq,Aaltonen:2008zd,Aaltonen:2008eu}%
\unpublished{ and}{,} double semileptonic decays 
with \particle{K^*\mu\mu} and \particle{\phi\mu\mu}
final states~\cite{Abe:1999ta}\unpublished{.}{,
and fully reconstructed $\Bs\to\jpsi\phi$ decays~\cite{CDFnote10795:2012}.}
In our determination of \fbb at the Tevatron, we include measurements of the production of $\Xi_b$ and $\Omega_b^-$ relative to that of the \Lb~\cite{Abazov:2007am,Abazov:2008qm,Aaltonen:2009ny}
by applying the constraint
\begin{eqnarray}
\fbb & = & f_{\Lb} + f_{\Xi_b^0} + f_{\Xi_b^-} + f_{\Omega_b^-} 
     \nonumber \\
     & = & f_{\Lb}\left(1 + 2\frac{f_{\Xi_b^-}}{f_{\Lb}} 
           + \frac{f_{\Omega_b^-}}{f_{\Lb}}\right),
           \label{eq:fbaryon-tevatron}
\end{eqnarray}
where isospin invariance in the production of $\Xi_b^0$ and 
$\Xi_b^-$ is assumed. Excited \b baryons are expected to decay strongly or 
electromagnetically to the baryons listed in Eq.~(\ref{eq:fbaryon-tevatron}). Both CDF~\cite{Aaltonen:2009ny} and \dzero~\cite{Abazov:2007am,Abazov:2008qm} reconstruct their \b baryons exclusively 
to final states that include a $\jpsi$ and a hyperon, namely, $\Lb\to \jpsi \Lambda$, 
$\Xi_b^- \rightarrow \jpsi \Xi^-$ and 
$\Omega_b^- \rightarrow \jpsi \Omega^-$.  
We assume that the partial decay width of a \b baryon to a $\jpsi$ and the 
corresponding hyperon is equal to the partial width of any other \b baryon to 
a $\jpsi$ and the corresponding hyperon.  
We use the CDF+\dzero average of 
$f_{\Xi_b^-}/f_{\Lb}$ to obtain $f_{\Omega_b^-}/f_{\Lb}$ from the \dzero measurement of $f_{\Omega_b^-}/f_{\Xi_b^-}$, which we  
combine with the CDF measurement of $f_{\Omega_b^-}/f_{\Lb}$ for input  into Eq.~(\ref{eq:fbaryon-tevatron}).

LHCb has also measured
ratios of \b-hadron fractions in charm+lepton final states~\cite{Aaij:2011jp} 
and in the fully reconstructed hadronic two-body decays $\Bd \to D^-\pi^+$, $\Bs \to D_s^- \pi^+$ and 
$\Lb \to \Lc \pi^-$~\citehistory{Aaij:2013qqa,Aaij:2014jyk}{Aaij:2013qqa,*Aaij:2011hi_hist,Aaij:2014jyk}.

Both CDF~\cite{Aaltonen:2008eu} and LHCb~\cite{Aaij:2011jp} observe that the ratio $\fLb/\fBd$ depends on the $p_{\rm T}$
of the charm+lepton system.%
\footnote{
  \label{foot:life_mix:Aaltonen:2008eu}
  CDF compare the $p_{\rm T}$ distribution of fully reconstructed 
  $\Lb \to \Lc \pi^-$ 
  with that of $\Bzb\rightarrow D^+\pi^-$, which 
  gives $\fLb/\fBd$ up to a scale factor. LHCb compares the $p_{\rm T}$ 
  of the charm+lepton system in \Lb, \Bd and \Bu decays, giving
  $R_{\Lb}/2 = \fLb/(\fBu+\fBd) = \fLb/(2\fBd)$.}
In \Ref{Aaltonen:2008eu}, CDF chose to correct an older result\cite{Affolder:1999iq} to account for the $p_{\rm T}$ dependence.
In a second result, CDF binned their data in $p_{\rm T}$ of the charm+electron 
system~\cite{Aaltonen:2008zd}.
In their more recent  measurement using hadronic decays~\cite{Aaij:2014jyk}, LHCb 
obtain the scale for $R_{\Lb} = \fLb/\fBd$ from their previous 
charm + lepton data~\cite{Aaij:2011jp}, bin the data in pseudorapidity ($\eta$) and see a 
linear dependence of $R_{\Lb}$ on $\eta$.  Since $\eta$ is not entirely
independent of $p_{\rm T}$, it is impossible to tell at this time whether 
this dependence is just an artifact of the $p_{\rm T}$ dependence.
\Figure{rlb_comb} shows the ratio $R_{\Lb}$ as a function of 
$p_{\rm T}$ for the \b hadron, as measured by LHCb.\footnote{
  \label{foot:life_mix:Aaltonen:2008zd}
  The CDF results from semileptonic decays~\cite{Aaltonen:2008zd} would require significant corrections to obtain the $p_{\rm T}$ of the \b hadron and be included on the same plot with the LHCb data.
  We do not have these corrections at this time.}  
LHCb fit their
scaled results using hadronic decays to obtain~\cite{Aaij:2014jyk}
\begin{equation}
R_{\Lb} = (0.151\pm 0.030) + 
  \exp{\left\{-(0.57\pm 0.11) - 
  (0.095\pm 0.016)[\gevc]^{-1} \times p_{\rm T}\right\}}.
\end{equation}
Since the two LHCb results for $R_{\Lb}$ are not 
independent, we use only the results with semileptonic final states for the averages.
Note that the $p_{\rm T}$ dependence
of $R_{\Lb}$ combined with the constraint from \Eq{constraints} implies
a compensating $p_{\rm T}$ dependence in one or more of the production fractions, \fBu, \fBd,
or \fBs.

\begin{figure}
 \begin{center}
  \includegraphics[width=\textwidth]{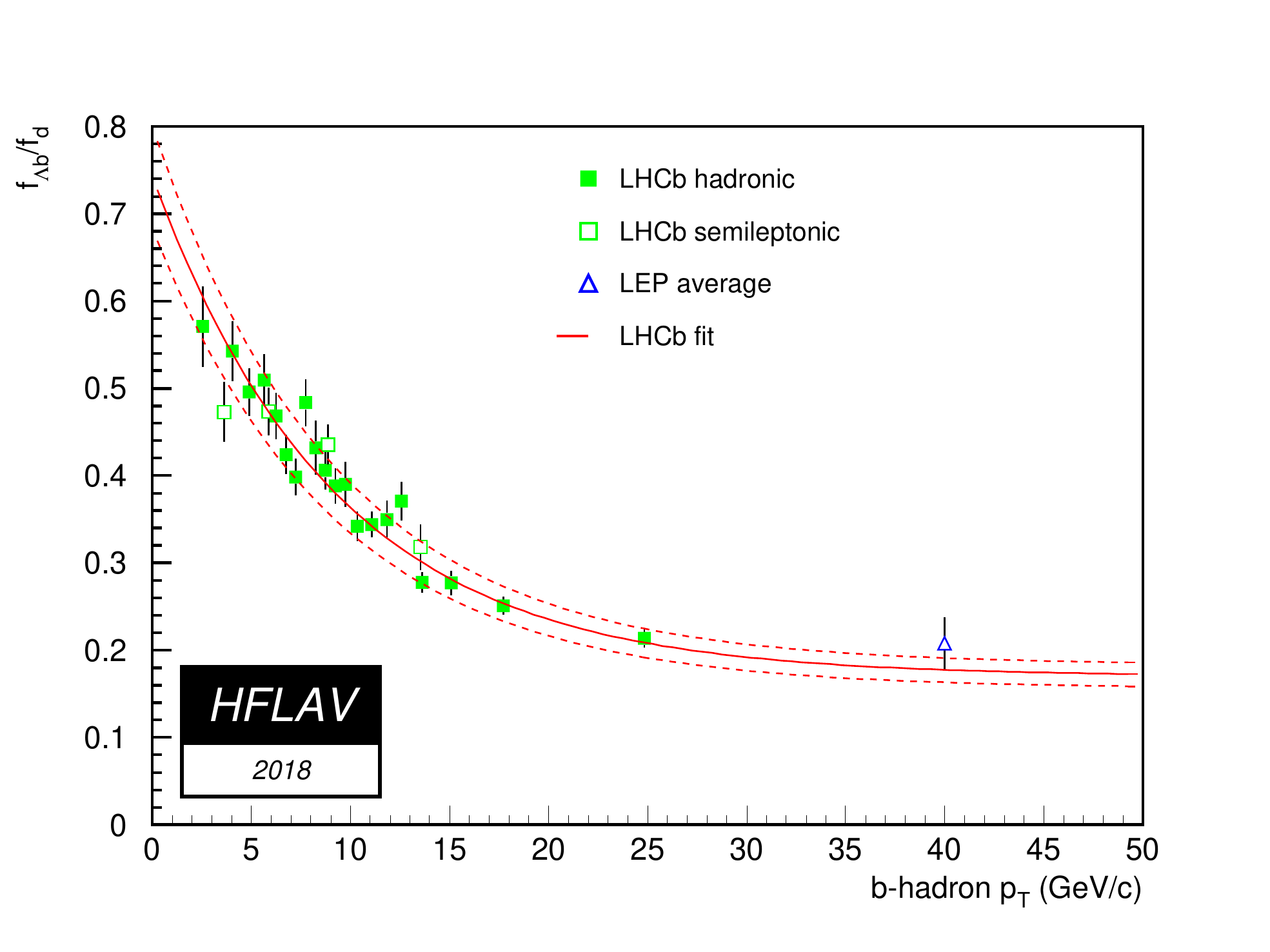}
  \caption{Ratio of production fractions $\fLb/\fBd$ as 
   a function of $p_{\rm T}$ of the \b hadron from
   LHCb data for \b hadrons decaying semileptonically~\cite{Aaij:2011jp}
   and fully reconstructed in hadronic decays~\cite{Aaij:2014jyk}. 
   The curve represents a fit to the LHCb hadronic data~\cite{Aaij:2014jyk}.
   Our LEP average (derived from \Table{fractions}) is displayed at an approximate $p_{\rm T}$ 
   in $Z$ decays, but is not used in the fit.}
  \labf{rlb_comb}
 \end{center}
\end{figure}

\unpublished{}{CDF\footnote{
The analysis of \Ref{CDFnote10795:2012} is unpublished, 
therefore not further discussed here nor included in the averages.},}
LHCb and ATLAS have investigated the $p_{\rm T}$ dependence of the ratio $R_s =\fBs/\fBd$, shown in \Figure{rs_comb}, 
using fully reconstructed $\Bs$ and $\Bd$ decays.
LHCb reported $3\sigma$ evidence that $R_s$ %
decreases with 
$p_{\rm T}$ using 
theoretical predictions 
for branching fractions~\citehistory{Aaij:2013qqa}{Aaij:2013qqa,*Aaij:2011hi_hist}.
The results from the
ATLAS experiment~\cite{Aad:2015cda} %
use theoretical predictions for branching fractions~\cite{Liu:2013nea} and indicate 
that $R_s$ is consistent with no $p_T$ dependence.
From \Figure{rs_comb}, we perform two fits for $R_s$. %
The first fit, using a linear parameterization, yields
$R_s = (0.2701\pm 0.0058) - (0.00139\pm 0.00044)[\gevc]^{-1} \times p_{\rm T}$.  
The second fit, using a simple exponential, yields
\begin{equation}
R_s = \exp\left\{(-1.304\pm 0.024) - (0.0058\pm 0.0019)[\gevc]^{-1} \times p_{\rm T}\right\}.
\end{equation}
The two fits are nearly indistinguishable over the $p_{\rm T}$ range of the results,
but the second fit gives a physical value for all $p_{\rm T}$.  
The $p_{\rm T}$-independent value of $R_s$ published by LHCb~\citehistory{Aaij:2013qqa}{Aaij:2013qqa,*Aaij:2011hi_hist} and our LEP average are also shown in \Figure{rs_comb}.

\begin{figure}
 \begin{center}
  \includegraphics[width=\textwidth]{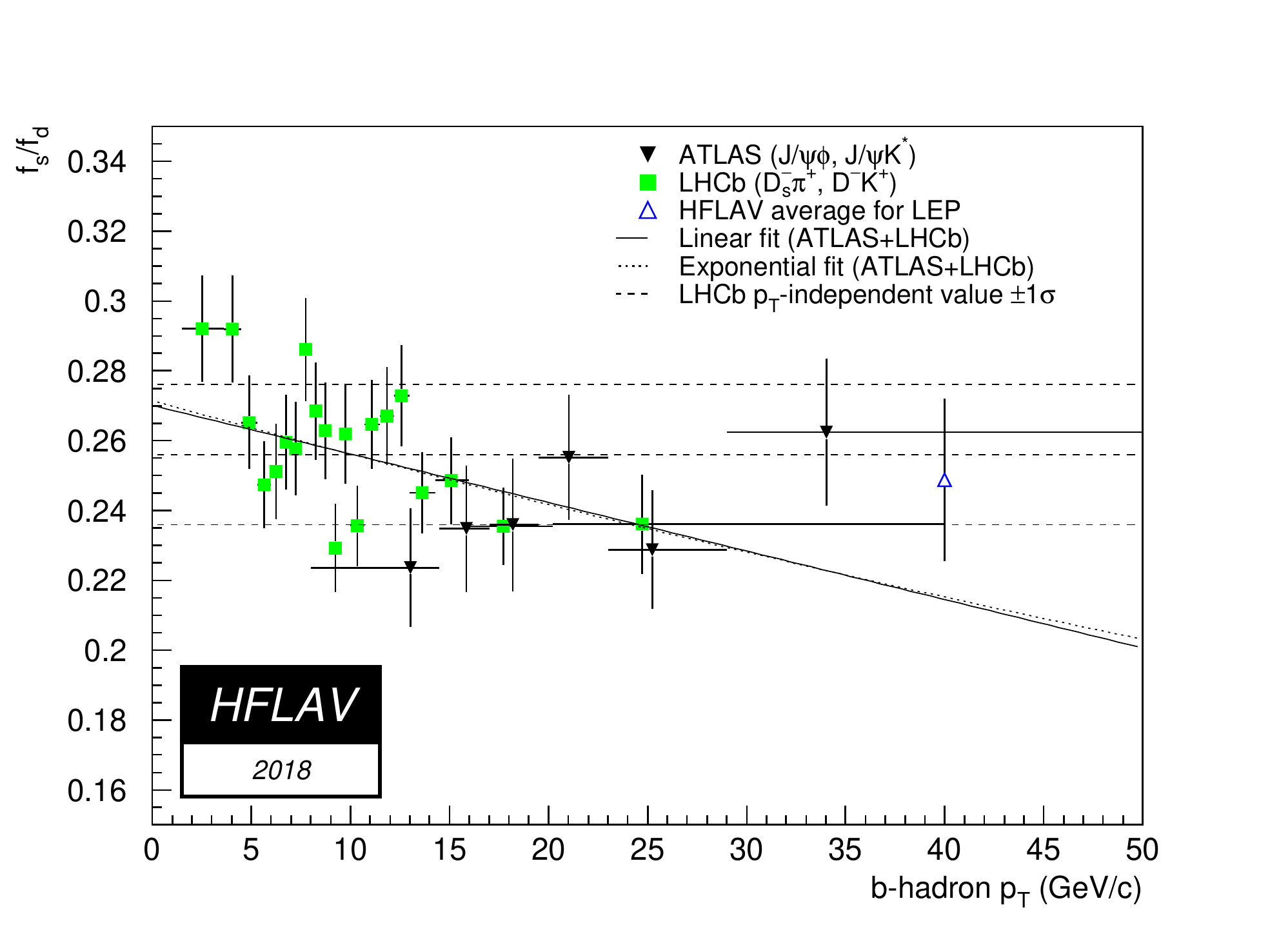}
  \caption{Ratio of production fractions $\fBs/\fBd$ as 
   a function of $p_{\rm T}$ of the reconstructed \b hadrons for the 
   LHCb~\protect\citehistory{Aaij:2013qqa}{Aaij:2013qqa,*Aaij:2011hi_hist}
   (green solid squares)
   and ATLAS~\cite{Aad:2015cda} (black solid triangles)
   data. Note the suppressed zero for the vertical axis.
   The curves represent fits to these data:
   a linear fit (solid curve), and an exponential fit described in the text (dotted curve).
   The $p_{\rm T}$-independent value of $R_s$ published by LHCb~\protect\citehistory{Aaij:2013qqa}{Aaij:2013qqa,*Aaij:2011hi_hist} (dashed lines) 
   and our LEP average of \Table{fractions} (open blue triangle at an 
   approximate $p_{\rm T}$ in $Z$ decays) are shown for comparison, but not used in any fit.}
  \labf{rs_comb}
 \end{center}
\end{figure}

For comparison purposes, a weighted average of the LHCb measurements in bins of $p_{\rm T}$ and $\eta$ is computed, both for $\fBs/(\fBu + \fBd)$ and $\fLb/(\fBu + \fBd)$.\footnote{
  \label{foot:life_mix:Aaij:2011jp}
  In practice, the LHCb data are given in 14 bins in $p_{\rm T}$ and $\eta$ with a full covariance matrix~\cite{Aaij:2011jp}. 
  The weighted average is calculated as
  $D^T C^{-1} M/\sigma$, where $\sigma = D^T C^{-1} D$, $M$ is a vector 
  of measurements, $C^{-1}$ is the inverse covariance matrix and $D^T$ is the 
  transpose of the design matrix (vector of 1's).}
As shown in \Table{LHCBcomp}, the weighted LHCb data and similar averages from CDF appear to be still compatible, at the current level of precision, despite the \b hadrons being produced in different kinematic regimes.

\begin{table}
 \caption{Comparison of average production fraction ratios from 
 CDF~\cite{Aaltonen:2008eu,Aaltonen:2008zd} and LHCb~\cite{Aaij:2011jp}.
 The kinematic regime of the charm+lepton system reconstructed in each
 experiment is also shown.}
 \labt{LHCBcomp}
 \begin{center}
  \begin{tabular}{lccc}
   \hline
   Quantity                         & CDF               & LHCb \\
   \hline
   $\fBs/(\fBu + \fBd)$             & $0.224\pm 0.057$  & $0.134\pm 0.009$   \\
   $\fLb/(\fBu + \fBd)$             & $0.229\pm 0.062$  & $0.240\pm 0.022$    \\
   Average charm+lepton $p_{\rm T}$ & $\sim 13~\gevc$ & $\sim 7~\gevc$ \\
   Pseudorapidity range             & $-1 < \eta < 1$   & $2 < \eta < 5$      \\
   \hline
  \end{tabular}
 \end{center}
\end{table}

Ignoring the $p_{\rm T}$ and $\eta$ dependence, we have adjusted the published results to the latest branching fraction averages~\cite{PDG_2018} and combined them
under the constraints of \Eq{constraints},
following the procedure and 
assumptions described in \Ref{Abbaneo:2000ej_mod,*Abbaneo:2001bv_mod_cont}.
This yield 
$\fBu=\fBd=\hflavZFBDNOMIX$,
$\fBs=\hflavZFBSNOMIX$ and $\fbb=\hflavZFBBNOMIX$ when using LEP data only, and 
$\fBu=\fBd=\hflavTFBDNOMIX$, $\fBs=\hflavTFBSNOMIX$ and
$\fbb = \hflavTFBBNOMIX$ when using Tevatron data only.  
As noted previously,
the LHC data are insufficient to determine a complete set of \b-hadron production
fractions. 
For these combinations other external inputs are used, 
\eg, the branching fractions of \B mesons to final states with a
\particle{D} or \particle{D^*} %
in semileptonic decays, which are needed 
to evaluate the fraction of semileptonic \Bs decays with a \particle{D_s^-} 
in the final state.

Time-integrated mixing analyses performed with lepton pairs 
from \particle{b\bar{b}} 
events produced at high-energy colliders measure the quantity 
\begin{equation}
\chibar = f'_{\particle{d}} \,\chid + f'_{\particle{s}} \,\chis \,,
\labe{chibar}
\end{equation}
where $f'_{\particle{d}}$ and $f'_{\particle{s}}$ are 
the fractions of \Bd and \Bs hadrons 
in a sample of semileptonic \b-hadron decays, and where \chid and \chis 
are the \Bd and \Bs time-integrated mixing probabilities.
Assuming that all \b hadrons have the same semileptonic decay width implies 
$f'_i = f_i R_i$, where $R_i = \tau_i/\tau_{\b}$ is the ratio of the lifetime 
$\tau_i$ of species $i$ to the average \b-hadron lifetime 
$\tau_{\b} = \sum_i f_i \tau_i$.
Hence measurements of the mixing probabilities
\chibar, \chid and \chis can be used to improve our 
knowledge of \fBu, \fBd, \fBs and \fbb.
In practice, the above relations yield another determination of 
\fBs obtained from \fbb and mixing information, 
\begin{equation}
\fBs = \frac{1}{R_{\particle{s}}}
\frac{(1+r)\overline{\chi}-(1-\fbb R_{\rm baryon}) \chid}{(1+r)\chis - \chid} \,,
\labe{fBs-mixing}
\end{equation}
where $r=R_{\particle{u}}/R_{\particle{d}} = \tau(\Bu)/\tau(\Bd)$.

The published measurements of \chibar performed by the LEP
experiments have been combined by the LEP Electroweak Working Group to yield 
$\chibar = \hflavCHIBARLEP$~\cite{ALEPH:2005ab}.%
\footnote{We use the $\bar{\chi}$ average of Eq.~5.39 in \Ref{ALEPH:2005ab}, 
obtained from a 10-parameter global fit of all electroweak data where the
asymmetry measurements have been excluded.}
This can be compared with the our Tevatron average, $\chibar = \hflavCHIBARTEV$,
obtained from \dzero~\cite{Abazov:2006qw} and CDF~\citehistory{Acosta:2003ie}{Acosta:2003ie_hist} measurements.%
\unpublished{}{\footnote{ \label{foot:life_mix:Acosta:2003ie}
The CDF result of Ref.~\citehistory{Acosta:2003ie}{Acosta:2003ie_hist} is from Run~I data.
A preliminary CDF measurement based on Run~II data~\cite{CDFnote10335:2011}
is still unpublished and therefore no longer included in our averages.
}}
The two averages deviate
from each other by $\hflavCHIBARSFACTOR\,\sigma$; 
this could be due to the fact that the production fractions of \b hadrons 
at the \particle{Z} peak or at the Tevatron are not the same. 

\begin{table}
\centering
\caption{Time-integrated mixing probability \chibar (defined in \Eq{chibar}), 
and production fractions of the different \b-hadron species in an unbiased sample of 
weakly decaying \b hadrons, obtained from both direct
and mixing measurements. The correlation coefficients $\rho$ between the fractions are 
also given.
}
\labt{fractions}
\begin{tabular}{@{}l@{}rcccc@{}}
\hline
Quantity            &                      & $Z$ decays      & Tevatron       & 
ATLAS~\cite{Aad:2015cda} & 
LHCb~\citehistory{Aaij:2013qqa}{Aaij:2013qqa,*Aaij:2011hi_hist}
\\ %
\hline
Mixing probability  & $\overline{\chi}$    & \hflavCHIBARLEP  & \hflavCHIBARTEV & & \\ %
\Bu or \Bd fraction & $\fBu = \fBd$        & \hflavZFBD       & \hflavTFBD      & & \\ %
\Bs fraction        & $\fBs$               & \hflavZFBS       & \hflavTFBS      & & \\ %
\b-baryon fraction  & $\fbb$               & \hflavZFBB       & \hflavTFBB      & & \\ %
$\Bs/\Bd$ ratio     & $\fBs/\fBd$          & \hflavZFBSBD     & \hflavTFBSBD    & $0.240 \pm 0.020$ & $0.256 \pm 0.020^u$ \\ %
\multicolumn{2}{@{}r}{$\rho(\fBs,\fBu) = \rho(\fBs,\fBd)$} & \hflavZRHOFBDFBS & \hflavTRHOFBDFBS & & \\ %
\multicolumn{2}{@{}r}{$\rho(\fbb,\fBu) = \rho(\fbb,\fBd)$} & \hflavZRHOFBDFBB & \hflavTRHOFBDFBB & & \\ %
\multicolumn{2}{@{}r}{$\rho(\fbb,\fBs)$}                   & \hflavZRHOFBBFBS & \hflavTRHOFBBFBS & & \\ %
\hline
\multicolumn{6}{l}{$^u$ \footnotesize This value has been updated with new inputs by LHCb to yield $0.259 \pm 0.015$~\cite{LHCb-CONF-2013-011}.} 
\end{tabular}
\end{table}

Using the \chibar average in \Eq{fBs-mixing} together with our world average 
$\chid = \hflavCHIDWU$ (see \Eq{chid} of \Sec{dmd}), the assumption $\chis= 1/2$ 
(justified by \Eq{chis} in \Sec{dms}), the 
best knowledge of the lifetimes (see \Sec{lifetimes}) and the estimate of \fbb given above, 
yields 
$\fBs = \hflavZFBSMIX$ using only LEP data, 
or $\fBs = \hflavTFBSMIX$ using only Tevatron data.%
Taking into account all known correlations (including that introduced by \fbb), 
this result is then combined with the set of fractions obtained from direct measurements 
(given above), to yield the %
improved estimates of \Table{fractions}, 
still under the constraints of \Eq{constraints}.
As can be seen, 
the inclusion of mixing information
reduces the uncertainty on \fBs, quite substantially in the case of LEP data.
\mysubsection{\b-hadron lifetimes}
\labs{lifetimes}

Lifetime calculations are performed in the framework of
the Heavy Quark Expansion
(HQE) \cite{Shifman:1986mx,Chay:1990da,Bigi:1992su}.
In these calculations, 
the total decay rate of a hadron $H_b$ 
is expressed as a series of expectation values of operators of increasing dimension,
\begin{equation}
\Gamma_{H_b} = |{\rm CKM}|^2 \sum_{n,k} \frac{c_{nk}}{m_b^n}
\langle H_b|O_n|H_b\rangle \,,
\labe{hqe}
\end{equation}
where $|{\rm CKM}|^2$ is the relevant combination of CKM matrix elements.
The coefficients $c_{nk}$
are calculated perturbatively~\cite{Wilson:1969zs}, 
\ie\ as a series in $\alpha_s(m_b)$.
The non-perturbative QCD effects are comprised in the matrix elements $\langle H_b| O_{nk} | H_b\rangle \propto \Lambda_{\rm QCD}^n$ of the operators $O_{nk}$. 
One usually encounters several operators of the same dimension $n$,
as indicated by the second index $k$. 
Hence the HQE predicts $\Gamma_{H_b}$ in the
form of an expansion in both $\Lambda_{\rm QCD}/m_b$ and
$\alpha_s(m_b)$.
The leading term in \Eq{hqe} corresponds to the weak
decay of a free \b quark, as in the old spectator model.
At this order all \b-flavoured hadrons have the same lifetime. The concept of the HQE and first calculations of valence quark effects emerged in 1986~\cite{Shifman:1986mx}; in
the early 1990's experiments became sensitive enough to start seeing
lifetime differences among various $H_b$ species. 
The possible existence of exponential contributions to $\Gamma_{H_b}$ is
not captured by the power series encoded in the HQE~\cite{Shifman:2000jv,Bigi:2001ys}. The presence of such terms is dubbed \emph{violation of quark-hadron duality}\ and their size can only be determined
experimentally, by confronting the HQE with data.\footnote{Possible violation of quark-hadron duality has been shown to be severely constrained by experimental results~\cite{Jubb:2016mvq}.}
The matrix elements
can be calculated using lattice QCD or QCD sum rules. In some cases they
can also be related to those appearing in other observables by utilising
symmetries of QCD. One may reasonably expect that powers of
$\Lambda_{\rm QCD}/m_b\sim 0.1$ provide enough suppression that only the
first terms of the sum in \Eq{hqe} matter.  Importantly, starting from
the third power the coefficients are enhanced by a factor of
$16\pi^2$. The dominant contribution to lifetime differences stems from
these terms of order $16\pi^2 (\Lambda_{\rm QCD}/m_b)^3$~\cite{Voloshin:1999pz,*Guberina:1999bw,*Neubert:1996we,*Bigi:1997fj}. 
State-of-the-art calculations of first-order corrections to these predictions exist
in both $\Lambda_{\rm QCD}/m_b$~\cite{Beneke:1996gn,Lenz:2011ti,*Lenz:2006hd} and $\alpha_s(m_b)$~\cite{Beneke:1998sy,Beneke:2002rj,Franco:2002fc,Ciuchini:2003ww,Beneke:2003az}; all subsequent theory
papers use these results.

Theoretical predictions are usually made for the ratios of the lifetimes
(with $\tau(\Bd)$ often chosen as the common denominator) rather than for the
individual lifetimes, since this leads to cancellation of several uncertainties.
The precision of the HQE calculations (see
\Refs{Ciuchini:2001vx,Beneke:2002rj,Franco:2002fc,Tarantino:2003qw,Gabbiani:2003pq,Gabbiani:2004tp}, and \Ref{Lenz:2015dra,Kirk:2017juj} for the latest updates)
is in some instances already surpassed by the measurements,
\eg, in the case of $\tau(\Bu)/\tau(\Bd)$.  
Improvement in the precision of calculations requires progress
along two lines. Firstly, better non-perturbative
matrix elements are needed and one expects precise calculations 
from lattice QCD, where significant advances have been made in the last
decade. Secondly, the coefficients $c_{kn}$ must be calculated in higher
orders of $\alpha_s$, namely the $\alpha_s^2$ and $\alpha_s \Lambda_{\rm
QCD}/m_b $ contributions to the lifetime differences are needed to keep
up with the experimental precision.

The following important conclusions, which are in agreement with experimental observation, can be drawn from the HQE, even in its present state:
\begin{itemize}
\item The larger the mass of the heavy quark, the smaller the
  variation in the lifetimes among different hadrons containing this
  quark, which is to say that, as $m_{\b}\to\infty$, we retrieve the
  spectator picture in which the lifetimes of all $H_b$ states are the same.
   This is well illustrated by the fact that lifetimes are rather
   similar in the \b sector, while they differ by large factors
   in the charm sector ($m_{\particle{c}}<m_{\b}$).
\item 
First corrections to the spectator decay occur at order $\Lambda_{\rm QCD}^2/m_b^2$, leading to lifetime differences around one percent.
\item 
The dominant contribution to the lifetime splittings is of order
$16\pi^2 (\Lambda_{\rm QCD}/m_b)^3$ and typically amounts to several
percent.
\end{itemize}

\mysubsubsection{Overview of  lifetime measurements}

This section gives  an overview of the types of \b-hadron lifetime  measurements, with details  given in subsequent sections. In most cases, the decay time of an $H_b$ state is estimated by measuring its flight
distance and dividing it by  $\beta\gamma c$.  Methods of accessing lifetime
information can roughly be divided into the following five categories:
\begin{enumerate}
\item {\bf\em Inclusive (flavour-blind) measurements}.  Early, low-statistics
  measurements were aimed at extracting the lifetime from a mixture of
  \b-hadron decays, without distinguishing the decaying species.  Often,
  knowledge of the $H_b$ composition was limited, which made the
  measurements experiment-specific.  Also, Monte Carlo simulation was used for estimating the
  $\beta\gamma$ factor, because the decaying hadrons were not fully
  reconstructed.  These were usually the largest-statistics \b-hadron lifetime measurements accessible to a
  given experiment, and could therefore serve as an important
  performance benchmark.
\item {\bf\em Measurements in semileptonic decays of a specific
  {\boldmath $H_b$\unboldmath}}.  The \particle{W} boson from \particle{\b\to Wc}
  produces a $\ell\nu_l$ pair (\particle{\ell=e,\mu}) in about 21\% of the
  cases.  The electron or muon from such decays provides a clean and efficient
  trigger signature.
  The \particle{c} quark and the spectator
  quark(s) combine into a charm hadron $H_c$,
  which is reconstructed in one or more exclusive decay channels.
  Identification of the $H_c$ species allows one to separate, at least
  statistically, different $H_b$ species.  The advantage of these
  measurements is in the sample size,
  which is usually larger than in the case of
  exclusively reconstructed hadronic $H_b$ decays (described  next). The main
  disadvantages are related to the difficulty of estimating the lepton+charm
  sample composition and to the Monte Carlo reliance for
  the momentum (and hence $\beta\gamma$ factor) estimate.
\item {\bf\em Measurements in exclusively reconstructed hadronic decays}.
  These
  have the advantage of complete reconstruction of the decaying $H_b$ state, 
  which allows one to infer the decaying species, as well as to perform precise
  measurement of the $\beta\gamma$ factor.  Both lead to generally
  smaller systematic uncertainties than in the above two categories.
  The downsides are smaller branching fractions and larger combinatorial
  backgrounds in the case of multi-hadron decays, such as $H_b\rightarrow H_c\pi(\pi\pi)$ with
  multi-body $H_c$ decays. This problem is often more serious in a hadron collider environment, which has many hadrons and 
  a non-trivial underlying event.  Decays of the type $H_b\to \jpsi H_s$ are often used, as they are
  relatively clean and easy to trigger on due to the $\jpsi\to \ell^+\ell^-$
  signature.%
\item {\bf\em Measurements at asymmetric B factories}. 
  In the $\Ups\rightarrow B \bar{B}$ decay, the \B mesons (\Bu or \Bd) are
essentially at rest in the \Ups frame.  This makes direct lifetime
measurements impossible in experiments at symmetric-energy colliders, which produce the 
\Ups at rest. 
At asymmetric \B factories the \Ups meson is boosted,
resulting in the \B and \particle{\bar{B}} moving nearly parallel to each 
other with similar boosts. The lifetime is inferred from the distance $\Delta z$        
separating the \B and \particle{\bar{B}} decay vertices along the beam axis 
and from the \Ups boost, which is known from the beam energies. This boost was 
$\beta \gamma \approx 0.55$ (0.43) in the \babar (\belle) experiment,
resulting in an average \B decay length of approximately 250~(190)~$\mu$m. 
While one \Bd or \Bu meson is fully reconstructed in a semileptonic or hadronic decay mode,
the other \B in the event is typically not fully reconstructed, in order to avoid loss of efficiency. Rather, only the position
of its decay vertex is determined from the remaining tracks in the event.
These measurements benefit from large sample sizes, but suffer from poor proper time 
resolution, comparable to the \B lifetime itself. The resolution is dominated by the 
uncertainty on the decay-vertex positions, which is typically 50~(100)~$\mu$m for a
fully (partially) reconstructed \B meson. 
With much larger samples in the future, 
the resolution and purity could be improved (and hence the systematics reduced)
by fully reconstructing both \B mesons in the event. 
 
\item {\bf\em Measurement of lifetime ratios}.  This method, 
  initially applied 
  in the measurement of $\tau(\Bu)/\tau(\Bd)$, is now also used for other 
  \b-hadron species at the LHC. 
  The ratio of the lifetimes is extracted from the proper-time 
  dependence of the ratio of the observed yields of 
  of two different \b-hadron species, 
  both reconstructed in decay modes with similar topologies. 
  The advantage of this method is that subtle efficiency effects and systematic uncertainties
  (partially) cancel in the ratio. 
\end{enumerate}

In some analyses, measurements of two (\eg, $\tau(\Bu)$ and
$\tau(\Bu)/\tau(\Bd)$) or three (\eg\ $\tau(\Bu)$,
$\tau(\Bu)/\tau(\Bd)$, and \dmd) quantities are combined.  This
introduces correlations among measurements.  Another source of
correlations among the measurements is  systematic effects, which
could be common to a number of measurements in the same experiment or to an analysis technique across different
experiments.  When calculating the averages presented below, such known correlations are taken
into account.
\mysubsubsection{Inclusive \b-hadron lifetimes}

The inclusive \b-hadron lifetime is defined as $\tau_{\b} = \sum_i f_i
\tau_i$ where $\tau_i$ are the individual species lifetimes and $f_i$ are
the fractions of the various species present in an unbiased sample of
weakly decaying \b hadrons produced at a high-energy
collider.%
This quantity is experiment-dependent and certainly
less fundamental than the lifetimes of the individual species, which are much more useful for comparison with
the theoretical predictions.  Nonetheless, we perform the averaging of
the inclusive lifetime measurements for completeness and because
they might be of interest as ``technical numbers.''

In practice, an unbiased measurement of the inclusive lifetime is
difficult to achieve, because it would imply that the efficiency is
guaranteed to be identical across $H_b$ species.  As a result, most of the measurements
are biased.  In an attempt to group analyses that are expected to
select the same mixture of \b hadrons, the available results (given in
\Table{lifeincl}) are divided into the following three sets:
\begin{enumerate}
\item measurements at LEP and SLD that include any \b-hadron decay, based 
      on topological reconstruction (secondary vertex or track impact
      parameters);
\item measurements at LEP based on the identification
      of a lepton from a \b decay; and
\item measurements at hadron colliders based on inclusive 
      \particle{H_b\to \jpsi X} reconstruction, where the
      \particle{\jpsi} is fully reconstructed.
\end{enumerate}

\begin{table}[t]
\caption{Measurements of average \b-hadron lifetimes.}
\labt{lifeincl}
\begin{center}
\begin{tabular}{lcccl} \hline
Experiment &Method           &Data set & $\tau_{\b}$ (ps)       &Ref.\\
\hline
ALEPH  &Dipole               & 1991     &$1.511\pm 0.022\pm 0.078$ &\cite{Buskulic:1993gj}\\
DELPHI &All track i.p.\ (2D) &91--92 &$1.542\pm 0.021\pm 0.045$ &\cite{Abreu:1994dr}$^a$\\
DELPHI &Sec.\ vtx            &91--93 &$1.582\pm 0.011\pm 0.027$ &\cite{Abreu:1996hv}$^a$\\
DELPHI &Sec.\ vtx            &94--95 &$1.570\pm 0.005\pm 0.008$ &\cite{Abdallah:2003sb}\\
L3     &Sec.\ vtx + i.p.     &91--94 &$1.556\pm 0.010\pm 0.017$ &\cite{Acciarri:1997tt}$^b$\\
OPAL   &Sec.\ vtx            &91--94 &$1.611\pm 0.010\pm 0.027$ &\cite{Ackerstaff:1996as}\\
SLD    &Sec.\ vtx            & 1993     &$1.564\pm 0.030\pm 0.036$ &\cite{Abe:1995rm}\\ 
\hline
\multicolumn{2}{l}{Average set 1 (\b vertex)} && \hflavTAUBVTXnounit &\\
\hline\hline
ALEPH  &Lepton i.p.\ (3D)    &91--93 &$1.533\pm 0.013\pm 0.022$ &\cite{Buskulic:1995rw}\\
L3     &Lepton i.p.\ (2D)    &91--94 &$1.544\pm 0.016\pm 0.021$ &\cite{Acciarri:1997tt}$^b$\\
OPAL   &Lepton i.p.\ (2D)    &90--91 &$1.523\pm 0.034\pm 0.038$ &\cite{Acton:1993xk}\\ 
\hline
\multicolumn{2}{l}{Average set 2 ($\b\to\ell$)} && \hflavTAUBLEPnounit &\\
\hline\hline
CDF1   &\particle{\jpsi} vtx&92--95 &$1.533\pm 0.015^{+0.035}_{-0.031}$ &\cite{Abe:1997bd} \\ 
\hline
\multicolumn{2}{l}{Average set 3 (\particle{\b\to \jpsi})} && \hflavTAUBJPnounit & \\ 
\hline
\multicolumn{5}{l}{$^a$ \footnotesize The combined DELPHI result quoted in
\cite{Abreu:1996hv} is 1.575 $\pm$ 0.010 $\pm$ 0.026 ps.} \\[-0.5ex]
\multicolumn{5}{l}{$^b$ \footnotesize The combined L3 result quoted in \cite{Acciarri:1997tt} 
is 1.549 $\pm$ 0.009 $\pm$ 0.015 ps.} \\[-0.5ex]
\end{tabular}
\end{center}
\end{table}

The mixtures corresponding to Sets 2 and 3 are better defined than for Set 1, in the limit where the reconstruction
and selection efficiency of a lepton or a \particle{\jpsi} from an
$H_b$ does not depend on the decaying hadron type.  These mixtures are
given by the production fractions and the inclusive branching fractions
for each $H_b$ species to give a lepton or a \particle{\jpsi}.  In
particular, under the assumption that all \b hadrons have the same
semileptonic decay width, the analyses of the second set should measure
$\tau(\b\to\ell) = (\sum_i f_i \tau_i^3) /(\sum_i f_i \tau_i^2)$ which is
necessarily larger than $\tau_{\b}$ if lifetime differences exist.
Given the present knowledge on $\tau_i$ and $f_i$,
$\tau(\b\to\ell)-\tau_{\b}$ is expected to be of the order of 0.003\ps.
On the other hand, the third set measuring $\tau(\b\to\particle{\jpsi})$
is expected to give an average smaller than $\tau_{\b}$ because 
of the \Bc meson, which has a significantly
larger probability to decay to a \particle{\jpsi}
than other \b-hadron species. 

Measurements by SLC and LEP experiments are subject to a number of
common systematic uncertainties, such as those due to (lack of knowledge
of) \b and \particle{c} fragmentation, \b and \particle{c} decay models,
\BRp{B\to\ell}, \BRp{B\to c\to\ell}, \BRp{c\to\ell}, $\tau_{\particle{c}}$,
and $H_b$ decay multiplicity.  In the averaging, these systematic
uncertainties are assumed to be 100\% correlated among the experiments.  The averages for the
sets defined above (also given in \Table{lifeincl})%
\unpublished{}{\footnote{We do not include here an unpublished measurement from ATLAS~\cite{ATLAS-CONF-2011-145}.}}
are
\begin{eqnarray}
\tau(\b~\mbox{vertex}) &=& \hflavTAUBVTX \,, \labe{TAUBVTX} \\
\tau(\b\to\ell) &=& \hflavTAUBLEP  \,, \\
\tau(\b\to\particle{\jpsi}) &=& \hflavTAUBJP\,.
\end{eqnarray}
The differences between these averages are consistent with zero
within less than $2\,\sigma$.

\mysubsubsection{\Bd and \Bu lifetimes and their ratio}
\labs{taubd}
\labs{taubu}
\labs{lifetime_ratio}

After a number of years of dominating these averages, the LEP experiments
yielded the scene to the asymmetric \B~factories and
the Tevatron experiments. The \B~factories have been very successful in
utilizing their potential -- in only a few years of running, \babar and,
to a greater extent, \belle, have struck a balance between the
statistical and the systematic uncertainties, with both being close to
(or even better than) an impressive 1\% level.  In the meanwhile, CDF and
\dzero have emerged as significant contributors to the field as the
Tevatron Run~II data flowed in. In more recent years, the LHCb experiment reached 
a further step in precision, improving by a factor $\sim 2 $ 
over the previous best measurements.

At the present time we are in an interesting position of having three sets
of measurements (from LEP/SLC, \B factories and Tevatron/LHC) that
originate from different environments, are obtained using substantially
different techniques and are precise enough for cross-checking and comparison.

\begin{table}[!t]
\caption{Measurements of the \Bd lifetime.}
\labt{lifebd}
\begin{center}
\begin{tabular}{lcccl} \hline
Experiment &Method                    &Data set &$\tau(\Bd)$ (ps)                  &Ref.\\
\hline
ALEPH  &\particle{D^{(*)} \ell}       &91--95 &$1.518\pm 0.053\pm 0.034$          &\cite{Barate:2000bs}\\
ALEPH  &Exclusive                     &91--94 &$1.25^{+0.15}_{-0.13}\pm 0.05$     &\cite{Buskulic:1996hy}\\
ALEPH  &Partial rec.\ $\pi^+\pi^-$    &91--94 &$1.49^{+0.17+0.08}_{-0.15-0.06}$   &\cite{Buskulic:1996hy}\\
DELPHI &\particle{D^{(*)} \ell}       &91--93 &$1.61^{+0.14}_{-0.13}\pm 0.08$     &\cite{Abreu:1995mc}\\
DELPHI &Charge sec.\ vtx              &91--93 &$1.63 \pm 0.14 \pm 0.13$           &\cite{Adam:1995mb}\\
DELPHI &Inclusive \particle{D^* \ell} &91--93 &$1.532\pm 0.041\pm 0.040$          &\cite{Abreu:1996gb}\\
DELPHI &Charge sec.\ vtx              &94--95 &$1.531 \pm 0.021\pm0.031$          &\cite{Abdallah:2003sb}\\
L3     &Charge sec.\ vtx              &94--95 &$1.52 \pm 0.06 \pm 0.04$           &\cite{Acciarri:1998uv}\\
OPAL   &\particle{D^{(*)} \ell}       &91--93 &$1.53 \pm 0.12 \pm 0.08$           &\cite{Akers:1995pa}\\
OPAL   &Charge sec.\ vtx              &93--95 &$1.523\pm 0.057\pm 0.053$          &\cite{Abbiendi:1998av}\\
OPAL   &Inclusive \particle{D^* \ell} &91--00 &$1.541\pm 0.028\pm 0.023$          &\cite{Abbiendi:2000ec}\\
SLD    &Charge sec.\ vtx $\ell$       &93--95 &$1.56^{+0.14}_{-0.13} \pm 0.10$    &\cite{Abe:1997ys}$^a$\\
SLD    &Charge sec.\ vtx              &93--95 &$1.66 \pm 0.08 \pm 0.08$           &\cite{Abe:1997ys}$^a$\\
CDF1   &\particle{D^{(*)} \ell}       &92--95 &$1.474\pm 0.039^{+0.052}_{-0.051}$ &\cite{Abe:1998wt}\\
CDF1  &Excl.\ \particle{\jpsi K^{*0}}&92--95 &$1.497\pm 0.073\pm 0.032$          &\cite{Acosta:2002nd}\\
CDF2   &Excl.\ \particle {\jpsi K_S^0}, \particle{\jpsi K^{*0}} &02--09 &$1.507\pm 0.010\pm0.008$           &\citehistory{Aaltonen:2010pj}{Aaltonen:2010pj,*Abulencia:2006dr_hist} \\
\dzero &Excl.\ \particle{\jpsi K^{*0}}&03--07 &$1.414\pm0.018\pm0.034$ &\citehistory{Abazov:2008jz}{Abazov:2008jz,*Abazov:2005sa_hist}\\ %
\dzero &Excl.\ \particle {\jpsi K_S^0} &02--11 &$1.508 \pm0.025 \pm0.043$  &\citehistory{Abazov:2012iy}{Abazov:2012iy,*Abazov:2007sf_hist,*Abazov:2004bn_hist} \\
\dzero &Inclusive \particle {D^-\mu^+} &02--11 &$1.534 \pm0.019 \pm0.021$  & \citehistory{Abazov:2014rua}{Abazov:2014rua,*Abazov:2006cb_hist} \\ %
\babar &Exclusive                     &99--00 &$1.546\pm 0.032\pm 0.022$          &\cite{Aubert:2001uw}\\
\babar &Inclusive \particle{D^* \ell} &99--01 &$1.529\pm 0.012\pm 0.029$          &\cite{Aubert:2002gi}\\
\babar &Exclusive \particle{D^* \ell} &99--02 &$1.523^{+0.024}_{-0.023}\pm 0.022$ &\cite{Aubert:2002sh}\\
\babar &Incl.\ \particle{D^*\pi}, \particle{D^*\rho} 
                                      &99--01 &$1.533\pm 0.034 \pm 0.038$         &\cite{Aubert:2002ms}\\
\babar &Inclusive \particle{D^* \ell}
&99--04 &$1.504\pm0.013^{+0.018}_{-0.013}$  &\cite{Aubert:2005kf} \\ 
\belle & Exclusive                     & 00--03 & $1.534\pm 0.008\pm0.010$        &  \citehistory{Abe:2004mz}{Abe:2004mz,*Abe:2002id_hist,*Tomura:2002qs_hist,*Hara:2002mq_hist} \\
ATLAS & Excl.\ \particle {\jpsi K_S^0} & 2011 & $1.509 \pm 0.012 \pm 0.018$ & \cite{Aad:2012bpa} \\
CMS   & Excl.\ \particle{\jpsi K^{*0}} & 2012 & $1.511 \pm0.005 \pm 0.006$ & \cite{Sirunyan:2017nbv}$^b$ \\
CMS   & Excl.\ \particle {\jpsi K_S^0}   & 2012 & $1.527 \pm0.009 \pm 0.009$ & \cite{Sirunyan:2017nbv}$^b$ \\
LHCb  & Excl.\ \particle{\jpsi K^{*0}} & 2011 & $1.524 \pm0.006 \pm 0.004$ & \cite{Aaij:2014owa} \\
LHCb  & Excl.\ \particle {\jpsi K_S^0}   & 2011 & $1.499 \pm0.013 \pm 0.005$ & \cite{Aaij:2014owa} \\
LHCb    & \particle{K^+\pi^-}   & 2011 & $1.524 \pm 0.011 \pm 0.004$ & \citehistory{Aaij:2014fia}{Aaij:2014fia,*Aaij:2012ns_hist} \\
\hline
Average&                               &        & \hflavTAUBDnounit & \\
\hline\hline           
\multicolumn{5}{l}{$^a$ \footnotesize The combined SLD result 
quoted in Ref.~\cite{Abe:1997ys} is 1.64 $\pm$ 0.08 $\pm$ 0.08 ps.}\\[-0.5ex]
\multicolumn{5}{l}{$^b$ \footnotesize The combined CMS result 
quoted in Ref.~\cite{Sirunyan:2017nbv} is 1.515 $\pm$ 0.005 $\pm$ 0.006 ps.} %
\end{tabular}
\end{center}
\end{table}

\begin{table}[p]
\centering
\caption{Measurements of the \Bu lifetime.}
\labt{lifebu}
\begin{tabular}{lcccl} \hline
Experiment &Method                 &Data set &$\tau(\Bu)$ (ps)                 &Ref.\\
\hline
ALEPH  &\particle{D^{(*)} \ell}    &91--95 &$1.648\pm 0.049\pm 0.035$          &\cite{Barate:2000bs}\\
ALEPH  &Exclusive                  &91--94 &$1.58^{+0.21+0.04}_{-0.18-0.03}$   &\cite{Buskulic:1996hy}\\
DELPHI &\particle{D^{(*)} \ell}    &91--93 &$1.61\pm 0.16\pm 0.12$             &\cite{Abreu:1995mc}$^a$\\
DELPHI &Charge sec.\ vtx           &91--93 &$1.72\pm 0.08\pm 0.06$             &\cite{Adam:1995mb}$^a$\\
DELPHI &Charge sec.\ vtx           &94--95 &$1.624\pm 0.014\pm 0.018$          &\cite{Abdallah:2003sb}\\
L3     &Charge sec.\ vtx           &94--95 &$1.66\pm  0.06\pm 0.03$            &\cite{Acciarri:1998uv}\\
OPAL   &\particle{D^{(*)} \ell}    &91--93 &$1.52 \pm 0.14\pm 0.09$            &\cite{Akers:1995pa}\\
OPAL   &Charge sec.\ vtx           &93--95 &$1.643\pm 0.037\pm 0.025$          &\cite{Abbiendi:1998av}\\
SLD    &Charge sec.\ vtx $\ell$    &93--95 &$1.61^{+0.13}_{-0.12}\pm 0.07$     &\cite{Abe:1997ys}$^b$\\
SLD    &Charge sec.\ vtx           &93--95 &$1.67\pm 0.07\pm 0.06$             &\cite{Abe:1997ys}$^b$\\
CDF1   &\particle{D^{(*)} \ell}    &92--95 &$1.637\pm 0.058^{+0.045}_{-0.043}$ &\cite{Abe:1998wt}\\
CDF1   &Excl.\ \particle{\jpsi K} &92--95 &$1.636\pm 0.058\pm 0.025$          &\cite{Acosta:2002nd}\\
CDF2   &Excl.\ \particle{\jpsi K} &02--09 &$1.639\pm 0.009\pm 0.009$          &\citehistory{Aaltonen:2010pj}{Aaltonen:2010pj,*Abulencia:2006dr_hist}\\ 
CDF2   &Excl.\ \particle{D^0 \pi}  &02--06 &$1.663\pm 0.023\pm0.015$           &\cite{Aaltonen:2010ta}\\
\babar &Exclusive                  &99--00 &$1.673\pm 0.032\pm 0.023$          &\cite{Aubert:2001uw}\\
\belle &Exclusive                  &00--03 &$1.635\pm 0.011\pm 0.011$          &\citehistory{Abe:2004mz}{Abe:2004mz,*Abe:2002id_hist,*Tomura:2002qs_hist,*Hara:2002mq_hist} \\
LHCb  & Excl.\ \particle{\jpsi K} & 2011 & $1.637 \pm0.004 \pm 0.003$ & \cite{Aaij:2014owa} \\
\hline
Average&                           &       &\hflavTAUBUnounit &\\
\hline\hline
\multicolumn{5}{l}{$^a$ \footnotesize The combined DELPHI result quoted 
in~\cite{Adam:1995mb} is $1.70 \pm 0.09$ ps.} \\[-0.5ex]
\multicolumn{5}{l}{$^b$ \footnotesize The combined SLD result 
quoted in~\cite{Abe:1997ys} is $1.66 \pm 0.06 \pm 0.05$ ps.}\\[-0.5ex]
\end{tabular}
\end{table}
\begin{table}[t]
\centering
\caption{Measurements of the ratio $\tau(\Bu)/\tau(\Bd)$.}
\labt{liferatioBuBd}
\begin{tabular}{lcccl} 
\hline
Experiment &Method                 &Data set &Ratio $\tau(\Bu)/\tau(\Bd)$      &Ref.\\
\hline
ALEPH  &\particle{D^{(*)} \ell}    &91--95 &$1.085\pm 0.059\pm 0.018$          &\cite{Barate:2000bs}\\
ALEPH  &Exclusive                  &91--94 &$1.27^{+0.23+0.03}_{-0.19-0.02}$   &\cite{Buskulic:1996hy}\\
DELPHI &\particle{D^{(*)} \ell}    &91--93 &$1.00^{+0.17}_{-0.15}\pm 0.10$     &\cite{Abreu:1995mc}\\
DELPHI &Charge sec.\ vtx           &91--93 &$1.06^{+0.13}_{-0.11}\pm 0.10$     &\cite{Adam:1995mb}\\
DELPHI &Charge sec.\ vtx           &94--95 &$1.060\pm 0.021 \pm 0.024$         &\cite{Abdallah:2003sb}\\
L3     &Charge sec.\ vtx           &94--95 &$1.09\pm 0.07  \pm 0.03$           &\cite{Acciarri:1998uv}\\
OPAL   &\particle{D^{(*)} \ell}    &91--93 &$0.99\pm 0.14^{+0.05}_{-0.04}$     &\cite{Akers:1995pa}\\
OPAL   &Charge sec.\ vtx           &93--95 &$1.079\pm 0.064 \pm 0.041$         &\cite{Abbiendi:1998av}\\
SLD    &Charge sec.\ vtx $\ell$    &93--95 &$1.03^{+0.16}_{-0.14} \pm 0.09$    &\cite{Abe:1997ys}$^a$\\
SLD    &Charge sec.\ vtx           &93--95 &$1.01^{+0.09}_{-0.08} \pm0.05$     &\cite{Abe:1997ys}$^a$\\
CDF1   &\particle{D^{(*)} \ell}    &92--95 &$1.110\pm 0.056^{+0.033}_{-0.030}$ &\cite{Abe:1998wt}\\
CDF1   &Excl.\ \particle{\jpsi K} &92--95 &$1.093\pm 0.066 \pm 0.028$         &\cite{Acosta:2002nd}\\
CDF2   &Excl.\ \particle{\jpsi K^{(*)}} &02--09 &$1.088\pm 0.009 \pm 0.004$   &\citehistory{Aaltonen:2010pj}{Aaltonen:2010pj,*Abulencia:2006dr_hist}\\ 
\dzero &\particle{D^{*+} \mu} \particle{D^0 \mu} ratio
	                           &02--04 &$1.080\pm 0.016\pm 0.014$          &\cite{Abazov:2004sa}\\
\babar &Exclusive                  &99--00 &$1.082\pm 0.026\pm 0.012$          &\cite{Aubert:2001uw}\\
\belle &Exclusive                  &00--03 &$1.066\pm 0.008\pm 0.008$          &\citehistory{Abe:2004mz}{Abe:2004mz,*Abe:2002id_hist,*Tomura:2002qs_hist,*Hara:2002mq_hist} \\
LHCb  & Excl.\ \particle{\jpsi K^{(*)}} & 2011 & $1.074 \pm0.005 \pm 0.003$ & \cite{Aaij:2014owa} \\
\hline
Average&                           &       & \hflavRTAUBU & \\   
\hline\hline
\multicolumn{5}{l}{$^a$ \footnotesize The combined SLD result quoted
	   in~\cite{Abe:1997ys} is $1.01 \pm 0.07 \pm 0.06$.}
\end{tabular}
\end{table}

The $\tau(\Bu)$, $\tau(\Bd)$ and $\tau(\Bu)/\tau(\Bd)$
measurements, and their averages, are summarized\unpublished{}{\footnote{%
We do not include the old unpublished measurements of Refs.~\cite{CDFnote7514:2005,CDFnote7386:2005,ATLAS-CONF-2011-092}.}}
in \Tablesss{lifebd}{lifebu}{liferatioBuBd}.
For the average of $\tau(\Bu)/\tau(\Bd)$ we use only direct measurements of this
ratio and not separate measurements of $\tau(\Bu)$ and
$\tau(\Bd)$.
The following sources of correlated (within experiment/machine)
systematic uncertainties have been considered in the averaging:
\begin{itemize}
\item for the SLC and LEP measurements -- \particle{D^{**}} branching fraction uncertainties~\cite{Abbaneo:2000ej_mod,*Abbaneo:2001bv_mod_cont},
estimation of the momentum  of \b mesons produced in \particle{Z^0} decays
(\b-quark fragmentation parameter $\langle X_E \rangle = 0.702 \pm 0.008$~\cite{Abbaneo:2000ej_mod,*Abbaneo:2001bv_mod_cont}),
\Bs and \b-baryon lifetimes (see \Secss{taubs}{taulb}),
and \b-hadron fractions at high energy (see \Table{fractions}); 
\item for the \B-factory measurements -- alignment, $z$ scale, machine boost (separately within each experiment),
sample composition (where applicable);
\item for the Tevatron and LHC measurements -- alignment (separately
within each experiment).
\end{itemize}
The resultant averages are:
\begin{eqnarray}
\tau(\Bd) & = & \hflavTAUBD \,, \\
\tau(\Bu) & = & \hflavTAUBU \,, \\
\tau(\Bu)/\tau(\Bd) & = & \hflavRTAUBU \,.
\end{eqnarray}

\mysubsubsection{\Bs lifetimes}
\labs{taubs}

Like neutral kaons, neutral \B mesons contain
short- and long-lived components, since the
light (L) and heavy (H)
eigenstates %
differ not only
in their masses, but also in their total decay widths. 
While the decay width difference \DGd can be neglected in the \Bd system, 
the \Bs system exhibits a significant value of 
$\DGs = \Gamma_{s\rm L} - \Gamma_{s\rm H}$, where $\Gamma_{s\rm L}$ and $\Gamma_{s\rm H}$
are the total decay widths of the light eigenstate $\B^0_{s\rm L}$ and the heavy eigenstate $\B^0_{s\rm H}$, respectively.
The sign of \DGs is measured to be positive~\cite{Aaij:2012eq}, \ie,
$\B^0_{s\rm H}$ has a longer lifetime than $\B^0_{s\rm L}$. 
Specific measurements of \DGs and 
$\Gs = (\Gamma_{s\rm L} + \Gamma_{s\rm H})/2$ are explained
and averaged in \Sec{DGs}, but the results for
$1/\Gamma_{s\rm L} = 1/(\Gs+\DGs/2)$, $1/\Gamma_{s\rm H}= 1/(\Gs-\DGs/2)$
and the mean \Bs lifetime, defined as $\tau(\Bs) = 1/\Gs$, are also quoted at the end of this section. 
Neglecting \CP violation in $\Bs-\Bsbar$ mixing, 
which is expected to be very
small~\citehistory{Jubb:2016mvq,Artuso:2015swg,Laplace:2002ik,Ciuchini:2003ww,Beneke:2003az}{Jubb:2016mvq,Artuso:2015swg,*Lenz_hist,Laplace:2002ik,Ciuchini:2003ww,Beneke:2003az}
(see also \Sec{qpd}), the mass eigenstates are also \CP eigenstates,
with the short-lived (light) %
state being \CP-even and the long-lived (heavy) %
state being \CP-odd~\cite{Aaij:2012eq}.

Many \Bs lifetime analyses, in particular the early 
ones performed before the non-zero value of \DGs was 
firmly established, ignore \DGs and fit the proper time 
distribution of a sample of \Bs candidates 
reconstructed in a certain final state $f$
with a model assuming a single exponential function 
for the signal.
Such {\em effective lifetime} measurements, which we denote as $\tau_{\rm single}(\Bs\to f)$, are estimates of the expectation value  $\int_0^\infty t\,\Gamma(B_s(t)\to f) dt/\int_0^\infty \Gamma(B_s(t)\to f) dt$ of the total untagged %
time-dependent decay rate $\Gamma(B_s(t)\to f)$~\cite{Hartkorn:1999ga,Dunietz:2000cr,Fleischer:2011cw}; 
this expectation value may lie {\em a priori} anywhere
between $1/\Gamma_{s\rm L}$ %
and $1/\Gamma_{s,\rm H}$, %
depending on the proportion of $B^0_{s\rm L}$ and $B^0_{s\rm H}$
in the final state $f$. 
More recent determinations of effective lifetimes may be interpreted as
measurements of the relative composition of 
$B^0_{s\rm L}$ and $B^0_{s\rm H}$
decaying to the final state $f$. 
\Table{lifebs} summarizes the effective 
lifetime measurements.

Averaging measurements of $\tau_{\rm single}(\Bs\to f)$
over several final states $f$ will yield a result 
corresponding to an ill-defined observable
when the proportions of $B^0_{s\rm L}$ and $B^0_{s\rm H}$
differ. 
Therefore, the effective \Bs lifetime measurements are broken down into
the following categories and averaged separately.

\begin{table}[t]
\centering
\caption{Measurements of the effective \Bs lifetimes obtained from single exponential fits.}
\labt{lifebs}
\begin{tabular}{l@{}c@{}cc@{}rc@{}l} \hline
Experiment & \multicolumn{2}{c}{Final state $f$}           & \multicolumn{2}{c}{Data set} & $\tau_{\rm single}(\Bs\to f)$ (ps) & Ref. \\
\hline \hline
ALEPH  & \particle{D_s h}     & ill-defined & 91--95 & & $1.47\pm 0.14\pm 0.08$           & \cite{Barate:1997ua}          \\
DELPHI & \particle{D_s h}     & ill-defined & 91--95 & & $1.53^{+0.16}_{-0.15}\pm 0.07$   & \citehistory{Abreu:2000ev}{Abreu:2000ev,*Abreu:1996ep_hist} \\
OPAL   & \particle{D_s} incl. & ill-defined & 90--95 & & $1.72^{+0.20+0.18}_{-0.19-0.17}$ & \cite{Ackerstaff:1997ne}          \\ 
\hline
ALEPH  & \particle{D_s^- \ell^+}  & flavour-specific & 91--95 & & $1.54^{+0.14}_{-0.13}\pm 0.04$   & \cite{Buskulic:1996ei}          \\
CDF1   & \particle{D_s^- \ell^+}  & flavour-specific & 92--96 & & $1.36\pm 0.09 ^{+0.06}_{-0.05}$  & \cite{Abe:1998cj}           \\
DELPHI & \particle{D_s^- \ell^+}  & flavour-specific & 92--95 & & $1.42^{+0.14}_{-0.13}\pm 0.03$   & \cite{Abreu:2000sh}          \\
OPAL   & \particle{D_s^- \ell^+}  & flavour-specific & 90--95 & & $1.50^{+0.16}_{-0.15}\pm 0.04$   & \cite{Ackerstaff:1997qi}  \\
\dzero & \particle{D_s^-\mu^+X}   & flavour-specific & 02--11 & 10.4 fb$^{-1}$ & $1.479 \pm 0.010 \pm 0.021$   & \citehistory{Abazov:2014rua}{Abazov:2014rua,*Abazov:2006cb_hist} \\
CDF2   & \particle{D_s^- \pi^+ (X)} 
                              & flavour-specific & 02--06 & 1.3 fb$^{-1}$ & $1.518 \pm 0.041 \pm 0.027     $   & \unpublished{\cite{Aaltonen:2011qsa}}{\citehistory{Aaltonen:2011qsa}{Aaltonen:2011qsa,*Aaltonen:2011qsa_hist}} \\ %
LHCb   &  \particle{D_s^- D^+} & flavour-specific & 11--12 & 3 fb$^{-1}$ & $1.52 \pm 0.15 \pm 0.01$ & \cite{Aaij:2013bvd} \\
LHCb   &  \particle{D_s^- \pi^+} & flavour-specific & 2011 & 1 fb$^{-1}$ & $1.535 \pm 0.015 \pm 0.014$ & \cite{Aaij:2014sua} \\
LHCb    & \particle{\pi^+K^-}   &  flavour-specific & 2011 & 1.0 fb$^{-1}$ & $1.60 \pm 0.06 \pm 0.01$ & \citehistory{Aaij:2014fia}{Aaij:2014fia,*Aaij:2012ns_hist} \\
LHCb    & \particle{D_s^{(*)-}\mu^+\nu_\mu}   &  flavour-specific & 11--12 & 3.0 fb$^{-1}$ & $1.547 \pm 0.013 \pm 0.011$ & \cite{Aaij:2017vqj} \\
\multicolumn{5}{l}{Average of above 10 flavour-specific lifetime measurements} &  \hflavTAUBSSLnounit & \\  
\hline\hline
CDF1     & \particle{\jpsi\phi} & \CP even+odd & 92--95 &  & $1.34^{+0.23}_{-0.19}    \pm 0.05$ & \cite{Abe:1997bd} \\
\dzero   & \particle{\jpsi\phi} & \CP even+odd & 02--04 &  & $1.444^{+0.098}_{-0.090} \pm 0.02$ & \cite{Abazov:2004ce}  \\
LHCb  & \particle{\jpsi\phi} & \CP even+odd & 2011 & 1 fb$^{-1}$ & $1.480 \pm0.011 \pm 0.005$ & \cite{Aaij:2014owa} \\
CMS   & \particle{\jpsi\phi} & \CP even+odd & 2012 & 19.7 fb$^{-1}$ & $1.481 \pm0.007 \pm 0.005$ & \cite{Sirunyan:2017nbv} \\
\multicolumn{5}{l}{Average of above 4 \particle{\jpsi \phi} lifetime measurements} &  \hflavTAUBSJFnounit & \\ 
\hline
LHCb    & \particle{\mu^+\mu^-}   &  \CP even+odd & 11--16 & 4.4 fb$^{-1}$ & $2.04 \pm 0.44 \pm 0.05$ & \cite{Aaij:2017vad} \\
\hline\hline
ALEPH    & \particle{D_s^{(*)+}D_s^{(*)-}} & mostly \CP even & 91--95 & & $1.27 \pm 0.33 \pm 0.08$ & \cite{Barate:2000kd} \\
\hline
LHCb    & \particle{K^+K^-}   &  \CP-even & 2010 & 0.037 fb$^{-1}$ & $1.440 \pm 0.096 \pm 0.009$ & \cite{Aaij:2012kn} \\
LHCb    & \particle{K^+K^-}   &  \CP-even & 2011 & 1.0 fb$^{-1}$ & $1.407 \pm 0.016 \pm 0.007$ & \citehistory{Aaij:2014fia}{Aaij:2014fia,*Aaij:2012ns_hist} \\
\multicolumn{5}{l}{Average of above 2 \particle{K^+K^-} lifetime measurements} &  \hflavTAUBSKKnounit & \\ 
\hline
LHCb   &  \particle{D_s^+ D_s^-} & \CP-even & 11--12 & 3 fb$^{-1}$ & $1.379 \pm 0.026 \pm 0.017$ & \cite{Aaij:2013bvd} \\
LHCb   &  \particle{\jpsi\eta} & \CP-even & 11--12 & 3 fb$^{-1}$ & $1.479 \pm 0.034 \pm 0.011$ &\cite{Aaij:2016dzn} \\
\multicolumn{5}{l}{Average of above 2 measurements of $1/\Gamma_{s\rm L}$} &  \hflavTAUBSSHORTnounit & \\ \hline \hline
LHCb     & \particle{\jpsi K^0_{\rm S}} & \CP-odd & 2011   & 1.0 fb$^{-1}$ & $1.75 \pm 0.12 \pm 0.07$ & \cite{Aaij:2013eia} \\
\hline
CDF2     & \particle{\jpsi f_0(980)} & \CP-odd & 02--08 & 3.8 fb$^{-1}$ & $1.70^{+0.12}_{-0.11} \pm 0.03$ & \cite{Aaltonen:2011nk} \\
\dzero       & \particle{\jpsi f_0(980)} & \CP-odd & 02--11 & 10.4 fb$^{-1}$ & $1.70\pm 0.14 \pm 0.05$ & \cite{Abazov:2016oqi} \\
LHCb     & \particle{\jpsi \pi^+\pi^-} & \CP-odd & 2011   & 1.0 fb$^{-1}$ & $1.652 \pm 0.024 \pm 0.024$ & \citehistory{Aaij:2013oba}{Aaij:2013oba,*LHCb:2011aa_hist,*LHCb:2012ad_hist,*LHCb:2011ab_hist,*Aaij:2012nta_hist} \\
CMS      & \particle{\jpsi \pi^+\pi^-} & \CP-odd & 2012   & 19.7 fb$^{-1}$ & $1.677 \pm 0.034 \pm 0.011$ & \cite{Sirunyan:2017nbv} \\
\multicolumn{5}{l}{Average of above 4 measurements of $1/\Gamma_{s\rm H}$} &  \hflavTAUBSLONGnounit & \\ \hline \hline
\end{tabular}
\end{table}

\afterpage{\clearpage}

\begin{itemize}

\item
{\bf\em \boldmath $\Bs\to D_s^{\mp} X$ decays}
include mostly flavour-specific decays but also decays 
with an unknown mixture of light and heavy components. 
Measurements performed with such inclusive states are
no longer used in averages. 

\item 
{\bf\em Decays to flavour-specific final states}, 
\ie, decays to final states $f$ with decay amplitudes satisfying 
$A(\Bs\to f) \ne 0$, $A(\Bsbar\to \bar{f}) \ne 0$, 
$A(\Bs\to \bar{f}) = 0$ and $A(\Bsbar\to f)=0$, 
have equal 
fractions of $B^0_{s\rm L}$ and $B^0_{s\rm H}$ at time zero.
The corresponding effective lifetime,
called the {\em flavour-specific lifetime}, is equal to~\cite{Hartkorn:1999ga}
\begin{equation}
\tau_{\rm single}(\Bs\to \mbox{flavour specific})
 =  \frac{1/\Gamma_{s\rm L}^2+1/\Gamma_{s\rm H}^2}{1/\Gamma_{s\rm L}+1/\Gamma_{s\rm H}}
 = \frac{1}{\Gs} \,
\frac{{1+\left(\frac{\DGs}{2\Gs}\right)^2}}{{1-\left(\frac{\DGs}{2\Gs}\right)^2}
}\,.
\labe{fslife}
\end{equation}

Because of the fast $\Bs-\Bsbar$ oscillations, 
possible biases of the flavour-specific lifetime due to a
combination of $\Bs/\Bsbar$ production asymmetry,
\CP violation in the decay amplitudes ($|A(\Bs\to f)| \ne |A(\Bsbar\to \bar{f})|$), 
and \CP violation in $\Bs-\Bsbar$ mixing
($|q_{\particle{s}}/p_{\particle{s}}| \ne 1$) 
are strongly suppressed, by a factor $\sim x_s^2$ (given in \Eq{xs}).
The $\Bs/\Bsbar$ production asymmetry at LHCb and the \CP asymmetry due to mixing 
have been measured to be compatible with zero with a precision below 3\%~\cite{Aaij:2014bba} 
and 0.3\% (see \Eq{ASLS}), respectively. The corresponding effects on the flavour-specific lifetime, which therefore have a relative size of the order of $10^{-5}$ or smaller, can be neglected at the current level of experimental precision.
Under the assumption of no production asymmetry 
and no \CP violation in mixing, \Eq{fslife} is exact even for a flavour-specific decay with 
\CP violation in the decay amplitudes. Hence any flavour-specific decay 
mode can be used to measure the flavour-specific lifetime. 

The average of all flavour-specific 
\Bs lifetime measurements%
\unpublished{\citehistory{Buskulic:1996ei,Abe:1998cj,Abreu:2000sh,Ackerstaff:1997qi,Abazov:2014rua,Aaltonen:2011qsa,Aaij:2013bvd,Aaij:2014sua,Aaij:2014fia,Aaij:2017vqj}{Buskulic:1996ei,Abe:1998cj,Abreu:2000sh,Ackerstaff:1997qi,Abazov:2014rua,*Abazov:2006cb_hist,Aaltonen:2011qsa,Aaij:2013bvd,Aaij:2014sua,Aaij:2014fia,*Aaij:2012ns_hist,Aaij:2017vqj}}{\citehistory{Buskulic:1996ei,Abe:1998cj,Abreu:2000sh,Ackerstaff:1997qi,Abazov:2014rua,Aaltonen:2011qsa,Aaij:2013bvd,Aaij:2014sua,Aaij:2014fia,Aaij:2017vqj}{Buskulic:1996ei,Abe:1998cj,Abreu:2000sh,Ackerstaff:1997qi,Abazov:2014rua,*Abazov:2006cb_hist,Aaltonen:2011qsa,*Aaltonen:2011qsa_hist,Aaij:2013bvd,Aaij:2014sua,Aaij:2014fia,*Aaij:2012ns_hist,Aaij:2017vqj}}%
\unpublished{}{\footnote{%
An old unpublished measurement~\cite{CDFnote7757:2005} is not included.}}
is
\begin{equation}
\tau_{\rm single}(\Bs\to \mbox{flavour specific}) = \hflavTAUBSSL \,.
\labe{tau_fs}
\end{equation}

\item
{\bf\em 
{\boldmath $\Bs \to \jpsi\phi$ \unboldmath}decays}
contain a well-measured mixture of \CP-even and \CP-odd states.
The published 
\particle{\Bs\to \jpsi\phi}
effective lifetime measurements~\cite{Abe:1997bd,Abazov:2004ce,Aaij:2014owa,Sirunyan:2017nbv} are combined  
into the average\unpublished{}{\footnote{%
The old unpublished measurements of Refs.~\citehistory{CDFnote8524:2007,ATLAS-CONF-2011-092}{CDFnote8524:2007,*CDFnote8524:2007_hist,ATLAS-CONF-2011-092} are not included.}}
$\tau_{\rm single}(\Bs\to \jpsi \phi) = \hflavTAUBSJF$. %
Analyses that separate the \CP-even and \CP-odd components in
this decay through a full angular study, outlined in \Sec{DGs},
provide directly precise measurements of $1/\Gs$ and $\DGs$ (see \Table{phisDGsGs}).

\item
{\bf\em 
{\boldmath $\Bs \to \mu^+\mu^-$ \unboldmath}decays}
contain an as-yet unknown mixture of \CP-even and \CP-odd states. A first measurement has been 
published by LHCb~\cite{Aaij:2017vad}.

\item
{\bf\em Decays to \boldmath\CP eigenstates} have also 
been measured, in the \CP-even modes 
$\Bs \to D_s^{(*)+}D_s^{(*)-}$ by ALEPH~\cite{Barate:2000kd},
$\Bs \to K^+ K^-$ by LHCb~\citehistory{Aaij:2012kn,Aaij:2014fia}{Aaij:2012kn,Aaij:2014fia,*Aaij:2012ns_hist}%
\unpublished{}{\footnote{An old unpublished measurement of the $\Bs \to K^+ K^-$
effective lifetime by CDF~\cite{Tonelli:2006np} is no longer considered.}},
$\Bs \to D_s^+D_s^-$ by LHCb~\cite{Aaij:2013bvd}
and $\Bs \to J/\psi \eta$ by LHCb~\cite{Aaij:2016dzn}, as well as in the \CP-odd modes 
$\Bs \to \jpsi f_0(980)$ by CDF~\cite{Aaltonen:2011nk}
and \dzero~\cite{Abazov:2016oqi},
$\Bs \to \jpsi \pi^+\pi^-$ by LHCb~\citehistory{Aaij:2013oba}{Aaij:2013oba,*LHCb:2011aa_hist,*LHCb:2012ad_hist,*LHCb:2011ab_hist,*Aaij:2012nta_hist} 
and CMS~\cite{Sirunyan:2017nbv},
and $\Bs \to \jpsi K^0_{\rm S}$ by LHCb~\cite{Aaij:2013eia}.
If these 
decays are dominated by a single weak phase and if \CP violation 
can be neglected, then $\tau_{\rm single}(\Bs \to \mbox{\CP-even}) = 1/\Gamma_{s\rm L}$ 
and  $\tau_{\rm single}(\Bs \to \mbox{\CP-odd}) = 1/\Gamma_{s\rm H}$ 
(see \Eqss{tau_KK_approx}{tau_Jpsif0_approx} for approximate relations in the presence of mixing-induced
\CP violation). 
However, not all these modes can be considered as pure \CP eigenstates:
a small \CP-odd component is most probably present
in $\Bs \to D_s^{(*)+}D_s^{(*)-}$ decays. Furthermore, the decays
$\Bs \to K^+ K^-$ and $\Bs \to \jpsi K^0_{\rm S}$ %
may suffer from direct \CP violation due to interfering tree and loop amplitudes. 
The averages for the effective lifetimes obtained for decays to
pure \CP-even ($D_s^+D_s^-$, $\jpsi\eta$) and \CP-odd ($\jpsi f_0(980)$, $\jpsi \pi^+\pi^-$)
final states where \CP conservation can be assumed are
\begin{eqnarray}
\tau_{\rm single}(\Bs \to \mbox{\CP-even}) & = & \hflavTAUBSSHORT \,,
\labe{tau_KK}
\\
\tau_{\rm single}(\Bs \to \mbox{\CP-odd}) & = & \hflavTAUBSLONG \,.
\labe{tau_Jpsif0}
\end{eqnarray}

\end{itemize}

As described in \Sec{DGs}, 
the effective lifetime averages of \Eqsss{tau_fs}{tau_KK}{tau_Jpsif0}
are used as ingredients to improve the 
determination of $1/\Gs$ and \DGs obtained from the full angular analyses
of $\Bs\to \jpsi\phi$ and $\Bs\to \jpsi K^+K^-$ decays. 
The resulting world averages for the \Bs lifetimes are
\begin{eqnarray}
\tau(B^0_{s\rm L}) = \frac{1}{\Gamma_{s\rm L}}
 = \frac{1}{\Gs+\DGs/2} & = & \hflavTAUBSLCON \,, \\
\tau(B^0_{s\rm H}) = \frac{1}{\Gamma_{s\rm H}}
 = \frac{1}{\Gs-\DGs/2} & = & \hflavTAUBSHCON \,, \\
\tau(\Bs) = \frac{1}{\Gs} = \frac{2}{\Gamma_{s\rm L}+\Gamma_{s\rm H}} & = & \hflavTAUBSMEANCON \,.
\labe{oneoverGs}
\end{eqnarray}

\mysubsubsection{\Bc lifetime}
\labs{taubc}

Early measurements of the \Bc meson lifetime,
from CDF~\unpublished{\cite{Abe:1998wi,Abulencia:2006zu}}{\cite{Abe:1998wi,CDFnote9294:2008,Abulencia:2006zu}} and \dzero~\cite{Abazov:2008rba},
use the semileptonic decay mode \particle{\Bc \to \jpsi \ell^+ \nu} 
and are based on a 
simultaneous fit to the mass and lifetime using the vertex formed
with the leptons from the decay of the \particle{\jpsi} and
the third lepton. Correction factors
to estimate the boost due to the missing neutrino are used.
Correlated systematic uncertainties include the impact
of the uncertainty of the \Bc transverse-momentum spectrum on the correction
factors, the level of feed-down from $\psi(2S)$ decays, 
Monte Carlo modeling of the decay (estimated by varying the decay model from phase space
to the ISGW model), and uncertainties in the \Bc mass.
With more statistics, CDF2 was able to perform the first \Bc lifetime 
based on fully reconstructed
$\Bc \to J/\psi \pi^+$ decays~\cite{Aaltonen:2012yb},
which does not suffer from a missing neutrino.
More recent measurements at the LHC, both with  
\particle{\Bc \to \jpsi \mu^+ \nu} decays from LHCb~\cite{Aaij:2014bva} and 
\particle{\Bc \to \jpsi \pi^+} decays from LHCb~\cite{Aaij:2014gka} and CMS~\cite{Sirunyan:2017nbv},
achieve the highest level of precision. Two of them~\cite{Aaij:2014gka,Sirunyan:2017nbv} are made relative to the \Bu lifetime. Before averaging, they are scaled to our latest \Bu lifetime average, and the induced correlation is taken into account. 

All the measurements\unpublished{}{\footnote{We do not list (nor include in the average) an unpublished result from CDF2~\cite{CDFnote9294:2008}.}}
are summarized in 
\Table{lifebc} and the world average, dominated by the LHCb measurements, is
determined to be
\begin{equation}
\tau(\Bc) = \hflavTAUBC \,.
\end{equation}

\begin{table}[tb]
\centering
\caption{Measurements of the \Bc lifetime.}
\labt{lifebc}
\begin{tabular}{lccrcl} \hline
Experiment & Method                    & \multicolumn{2}{c}{Data set}  & $\tau(\Bc)$ (ps)
      & Ref.\\   \hline
CDF1       & \particle{\jpsi \ell} & 92--95 & 0.11 fb$^{-1}$ & $0.46^{+0.18}_{-0.16} \pm
 0.03$   & \cite{Abe:1998wi}  \\ 
CDF2       & \particle{\jpsi e} & 02--04 & 0.36 fb$^{-1}$ & $0.463^{+0.073}_{-0.065} \pm 0.036$   & \cite{Abulencia:2006zu} \\
 \dzero & \particle{\jpsi \mu} & 02--06 & 1.3 fb$^{-1}$  & $0.448^{+0.038}_{-0.036} \pm 0.032$
   & \cite{Abazov:2008rba}  \\
CDF2       & \particle{\jpsi \pi} & & 6.7 fb$^{-1}$ & $0.452 \pm 0.048 \pm 0.027$  & \cite{Aaltonen:2012yb} \\
LHCb & \particle{\jpsi \mu} & 2012 & 2 fb$^{-1}$  & $0.509 \pm 0.008 \pm 0.012$ & \cite{Aaij:2014bva}  \\
LHCb & \particle{\jpsi \pi} & 11--12 & 3 fb$^{-1}$  & $0.5134 \pm 0.0110 \pm 0.0057$ & \cite{Aaij:2014gka} \\
CMS  & \particle{\jpsi \pi} & 2012 & 19.7 fb$^{-1}$  & $0.541  \pm 0.026  \pm 0.014$ & \cite{Sirunyan:2017nbv} \\
\hline
  \multicolumn{2}{l}{Average} & &  &  \hflavTAUBCnounit
                 &    \\   \hline
\end{tabular}
\end{table}

\mysubsubsection{\Lb and \b-baryon lifetimes}
\labs{taulb}

The first measurements of \b-baryon lifetimes, performed at LEP,
originate from two classes of partially reconstructed decays.
In the first class, decays with a fully
reconstructed \Lc baryon
and a lepton of opposite charge are used. These products are
likely to occur in the decay of \Lb baryons.
In the second class, more inclusive final states with a baryon
(\particle{p}, \particle{\bar{p}}, $\Lambda$, or $\bar{\Lambda}$) 
and a lepton have been used, and these final states can generally
arise from any \b baryon.  With the large \b-hadron samples available
at the Tevatron and the LHC, the most precise measurements of \b baryons now
come from fully reconstructed exclusive decays.

The following sources of correlated systematic uncertainties have 
been accounted for when averaging these measurements:
experimental time resolution within a given experiment, \b-quark
fragmentation distribution into weakly decaying \b baryons,
\Lb polarisation, decay model,
and evaluation of the \b-baryon purity in the selected event samples.
In computing the averages,
the central values of the masses are scaled to 
$M(\Lb) = 5619.60 \pm 0.17\MeVcc$~\cite{PDG_2018}.
For measurements with partially reconstructed decays,
the meaning of the decay model
systematic uncertainties
and the correlation of these uncertainties between measurements
are not always clear.
Uncertainties related to the decay model are dominated by
assumptions on the fraction of $n$-body semileptonic decays.
To be conservative, it is assumed
that these are 100\% correlated whenever given as an uncertainty.
DELPHI varies the fraction of four-body decays from 0.0 to 0.3. 
In computing the average, the DELPHI
result is scaled to a value of $0.2 \pm 0.2$ for this fraction.
Furthermore
the semileptonic decay results from LEP are scaled to a
$\Lb$ polarisation of 
$-0.45^{+0.19}_{-0.17}$~\cite{Abbaneo:2000ej_mod,*Abbaneo:2001bv_mod_cont}
and a $b$ fragmentation parameter
$\langle x_E \rangle_b =0.702\pm 0.008$~\cite{ALEPH:2005ab}.

\begin{table}[!p]
\centering
\caption{Measurements of the \b-baryon lifetimes.
}
\labt{lifelb}
\begin{tabular}{lcccl} 
\hline
Experiment&Method                &Data set& Lifetime (ps) & Ref. \\\hline\hline
ALEPH  &$\Lambda\ell$         & 91--95 &$1.20 \pm 0.08 \pm 0.06$ & \cite{Barate:1997if}\\
DELPHI &$\Lambda\ell\pi$ vtx  & 91--94 &$1.16 \pm 0.20 \pm 0.08$        & \cite{Abreu:1999hu}$^b$\\
DELPHI &$\Lambda\mu$ i.p.     & 91--94 &$1.10^{+0.19}_{-0.17} \pm 0.09$ & \cite{Abreu:1996nt}$^b$ \\
DELPHI &\particle{p\ell}      & 91--94 &$1.19 \pm 0.14 \pm 0.07$        & \cite{Abreu:1999hu}$^b$\\
OPAL   &$\Lambda\ell$ i.p.    & 90--94 &$1.21^{+0.15}_{-0.13} \pm 0.10$ & \cite{Akers:1995ui}$^c$  \\
OPAL   &$\Lambda\ell$ vtx     & 90--94 &$1.15 \pm 0.12 \pm 0.06$        & \cite{Akers:1995ui}$^c$ \\ 
\hline
ALEPH  &$\Lc\ell$             & 91--95 &$1.18^{+0.13}_{-0.12} \pm 0.03$ & \cite{Barate:1997if}$^a$\\
ALEPH  &$\Lambda\ell^-\ell^+$ & 91--95 &$1.30^{+0.26}_{-0.21} \pm 0.04$ & \cite{Barate:1997if}$^a$\\
DELPHI &$\Lc\ell$             & 91--94 &$1.11^{+0.19}_{-0.18} \pm 0.05$ & \cite{Abreu:1999hu}$^b$\\
OPAL   &$\Lc\ell$, $\Lambda\ell^-\ell^+$ 
                                 & 90--95 & $1.29^{+0.24}_{-0.22} \pm 0.06$ & \cite{Ackerstaff:1997qi}\\ 
CDF1   &$\Lc\ell$             & 91--95 &$1.32 \pm 0.15        \pm 0.07$ & \cite{Abe:1996df}\\
\dzero &$\Lc\mu$              & 02--06 &$1.290^{+0.119+0.087}_{-0.110-0.091}$ & \cite{Abazov:2007al} \\
\multicolumn{3}{l}{Average of above 6} & \hflavTAULBSnounit & \\
\hline
CDF2   &$\Lc\pi$              & 02--06 &$1.401 \pm 0.046 \pm 0.035$ & \cite{Aaltonen:2009zn} \\
CDF2   &$\jpsi \Lambda$      & 01--11 &$1.565 \pm 0.035 \pm 0.020$ & \citehistory{Aaltonen:2014wfa}{Aaltonen:2014wfa,*Aaltonen:2014wfa_hist} \\
\dzero &$\jpsi \Lambda$      & 02--11 &$1.303 \pm 0.075 \pm 0.035$ & \citehistory{Abazov:2012iy}{Abazov:2012iy,*Abazov:2007sf_hist,*Abazov:2004bn_hist} \\
ATLAS  &$\jpsi \Lambda$      & 2011   &$1.449 \pm 0.036 \pm 0.017$ & \cite{Aad:2012bpa} \\
CMS    &$\jpsi \Lambda$      & 2011   &$1.503 \pm 0.052 \pm 0.031$ & \cite{Chatrchyan:2013sxa} \\ %
CMS    &$\jpsi \Lambda$      & 2012   &$1.477 \pm 0.027 \pm 0.009$ & \cite{Sirunyan:2017nbv} \\ %
LHCb   &$\jpsi \Lambda$      & 2011   &$1.415 \pm 0.027 \pm 0.006$ & \cite{Aaij:2014owa} \\
LHCb   &$\jpsi pK$ (w.r.t.\ $B^0$)  & 11--12 &$1.479 \pm 0.009 \pm 0.010$ & \citehistory{Aaij:2014zyy}{Aaij:2014zyy,*Aaij:2013oha_hist} \\ %
\multicolumn{3}{l}{Average of above 8: \hfill \Lb lifetime $=$} & \hflavTAULBnounit & \\
\hline\hline
ALEPH  &$\Xi^-\ell^-X$        & 90--95 &$1.35^{+0.37+0.15}_{-0.28-0.17}$ & \cite{Buskulic:1996sm}\\
DELPHI &$\Xi^-\ell^-X$        & 91--93 &$1.5 ^{+0.7}_{-0.4} \pm 0.3$     & \cite{Abreu:1995kt}$^d$ \\
DELPHI &$\Xi^-\ell^-X$        & 92--95 &$1.45 ^{+0.55}_{-0.43} \pm 0.13$     & \cite{Abdallah:2005cw}$^d$ \\
\hline
CDF2   &$\jpsi \Xi^-$        & 01--11 &$1.32 \pm 0.14 \pm 0.02$ & \citehistory{Aaltonen:2014wfa}{Aaltonen:2014wfa,*Aaltonen:2014wfa_hist} \\ %
LHCb   &$\jpsi \Xi^-$         & 11--12 &$1.55 ^{+0.10}_{-0.09} \pm 0.03$ & \cite{Aaij:2014sia} \\ 
LHCb   &$\Xi_c^0\pi^-$ (w.r.t.\ $\Lb$)  & 11--12 &$1.599 \pm 0.041 \pm 0.022$ & \cite{Aaij:2014lxa} \\ 
\multicolumn{3}{l}{Average of above 3: \hfill \Xibd lifetime $=$} & \hflavTAUXBDnounit & \\
\hline\hline
LHCb   &$\Xi_c^+\pi^-$  (w.r.t.\ $\Lb$) & 11--12 &$1.477 \pm 0.026 \pm 0.019$ & \cite{Aaij:2014esa} \\ 
\multicolumn{3}{l}{Average of above 1: \hfill \Xibu lifetime $=$} & \hflavTAUXBUnounit & \\
\hline\hline
CDF2   &$\jpsi \Omega^-$     & 01--11 & $1.66 ^{+0.53}_{-0.40} \pm 0.02$ & \citehistory{Aaltonen:2014wfa}{Aaltonen:2014wfa,*Aaltonen:2014wfa_hist} \\ %
LHCb   &$\jpsi \Omega^-$     & 11--12 &$1.54 ^{+0.26}_{-0.21} \pm 0.05$ & \cite{Aaij:2014sia} \\ 
LHCb   &$\Omega_c^0 \pi^-$ (w.r.t.\ $\Xi_b^-$)  & 11--12 &$1.78 \pm 0.26 \pm 0.05 \pm 0.06$ & \cite{Aaij:2016dls} \\
\multicolumn{3}{l}{Average of above 3: \hfill \Omegab lifetime $=$} & \hflavTAUOBnounit & \\
\hline\hline
\multicolumn{5}{l}{$^a$ \footnotesize The combined ALEPH result quoted 
in \cite{Barate:1997if} is $1.21 \pm 0.11$ ps.} \\[-0.5ex]
\multicolumn{5}{l}{$^b$ \footnotesize The combined DELPHI result quoted 
in \cite{Abreu:1999hu} is $1.14 \pm 0.08 \pm 0.04$ ps.} \\[-0.5ex]
\multicolumn{5}{l}{$^c$ \footnotesize The combined OPAL result quoted 
in \cite{Akers:1995ui} is $1.16 \pm 0.11 \pm 0.06$ ps.} \\[-0.5ex]
\multicolumn{5}{l}{$^d$ \footnotesize The combined DELPHI result quoted 
in \cite{Abdallah:2005cw} is $1.48 ^{+0.40}_{-0.31} \pm 0.12$ ps.}
\end{tabular}
\end{table}

The list of all measurements are given in \Table{lifelb}.
We do not attempt to average measurements performed with $p\ell$ or 
$\Lambda\ell$ combinations, which select unknown mixtures of $b$ baryons. 
Measurements performed with $\Lc\ell$ or $\Lambda\ell^+\ell^-$
combinations can be assumed to correspond to semileptonic \Lb decays. 
Their average (\hflavTAULBS) is significantly different 
from the average using only measurements performed with
exclusively reconstructed hadronic \Lb decays (\hflavTAULBE). 
The latter is much more precise
and less prone to potential biases than the former. 
The discrepancy between the two averages is at the level of
$\hflavNSIGMATAULBEXCLSEMI\sigma$ 
and assumed to be due to a systematic effect in the 
semileptonic measurements, where the \Lb momentum is not determined directly, or to a rare statistical fluctuation.
The best estimate of the \Lb lifetime is therefore taken 
as the average of the exclusive  measurements only. 
The CDF $\Lb \to \jpsi \Lambda$
lifetime result~\citehistory{Aaltonen:2014wfa}{Aaltonen:2014wfa,*Aaltonen:2014wfa_hist} 
is larger than the average of all other exclusive measurements
by $\hflavNSIGMATAULBCDFTWO\sigma$. 
It is nonetheless kept in the average
without adjustment of input uncertainties.
The world average \Lb lifetime is then
\begin{equation}
\tau(\Lb) = \hflavTAULB \,. 
\end{equation}

For the strange \b baryons, we do not include the measurements based on
inclusive $\Xi^{\mp} \ell^{\mp}$
final states, which consist of a mixture of 
$\Xibd$ and $\Xibu$ baryons. Rather, we only average results obtained with 
fully reconstructed $\Xibd$, $\Xibu$ and $\Omegab$ baryons, and obtain
\begin{eqnarray}
\tau(\Xibd) &=& \hflavTAUXBD \,, \\
\tau(\Xibu) &=& \hflavTAUXBU \,, \\
\tau(\Omegab) &=& \hflavTAUOB \,. 
\end{eqnarray}
It should be noted that several $b$-baryon lifetime measurements from LHCb~%
\citehistory{Aaij:2014zyy,Aaij:2014lxa,Aaij:2014esa,Aaij:2016dls}{Aaij:2014zyy,*Aaij:2013oha_hist,Aaij:2014lxa,Aaij:2014esa,Aaij:2016dls}
were made with respect to the lifetime of another $b$ hadron
(\ie, the original measurement is that of a decay width difference).
Before these measurements are included in the averages quoted above, we 
rescale them according to our latest lifetime average
of that reference $b$ hadron. This introduces correlations between 
our averages, in particular between the $\Xibd$ and $\Xibu$ lifetimes. 
Taking this correlation into account leads to 
\begin{equation}
\tau(\Xibu) / \tau(\Xibd) = \hflavRTAUXBUXBD \,.
\end{equation}

\mysubsubsection{Summary and comparison with theoretical predictions}
\labs{lifesummary}

Averages of lifetimes of specific \b-hadron species are collected
in \Table{sumlife}.
\begin{table}[t]
\centering
\caption{Summary of the lifetime averages for the different \b-hadron species.}
\labt{sumlife}
\begin{tabular}{lrc} \hline
\multicolumn{2}{l}{\b-hadron species} & Measured lifetime \\ \hline
\Bu &                       & \hflavTAUBU   \\
\Bd &                       & \hflavTAUBD   \\
\Bs & $1/\Gs~\, =$               & \hflavTAUBSMEANC \\
~~ $B^0_{s\rm L}$ & $1/\Gamma_{s\rm L}=$  & \hflavTAUBSLCON \\
~~ $B^0_{s\rm H}$ & $1/\Gamma_{s\rm H}=$  & \hflavTAUBSHCON \\
\Bc     &                   & \hflavTAUBC   \\ 
\Lb     &                   & \hflavTAULB   \\
\Xibd   &                   & \hflavTAUXBD  \\
\Xibu   &                   & \hflavTAUXBU  \\
\Omegab &                   & \hflavTAUOB   \\
\hline
\end{tabular}
\end{table}
\begin{table}[t]
\centering
\caption{Experimental averages of \b-hadron lifetime ratios and
Heavy-Quark Expansion (HQE) predictions.}
\labt{liferatio}
\begin{tabular}{lcc} \hline
Lifetime ratio & Experimental average & HQE prediction \\ \hline
$\tau(\Bu)/\tau(\Bd)$ & \hflavRTAUBU & $1.082 ^{+0.022}_{-0.026}$~\cite{Kirk:2017juj} \\
$\tau(\Bs)/\tau(\Bd)$ & \hflavRTAUBSMEANC & $0.9994 \pm 0.0025$~\cite{Kirk:2017juj} \\
$\tau(\Lb)/\tau(\Bd)$ & \hflavRTAULB & $0.935 \pm 0.054$~\cite{Lenz:2015dra} \\
$\tau(\Xibu)/\tau(\Xibd)$ & \hflavRTAUXBUXBD & $0.95 \pm 0.06$~\cite{Lenz:2015dra} \\
\hline
\end{tabular}
\end{table}
As described in the introduction to \Sec{lifetimes},
the HQE can be employed to explain the hierarchy of
$\tau(\Bc) \ll \tau(\Lb) < \tau(\Bs) \approx \tau(\Bd) < \tau(\Bu)$,
and used to predict the ratios between lifetimes.
Recent predictions are compared to the measured 
lifetime ratios in \Table{liferatio}.

The predictions of the ratio between the \Bu and \Bd lifetimes,
$1.06 \pm 0.02$~\cite{Beneke:2002rj,Franco:2002fc} %
or
$1.082 ^{+0.022}_{-0.026}$~\cite{Kirk:2017juj},
are in good agreement with experiment. 

The total widths of the \Bs and \Bd mesons
are expected to be very close and differ by at most 
1\%~\cite{Beneke:1996gn,Keum:1998fd,Gabbiani:2004tp,Lenz:2015dra,Kirk:2017juj}.
This prediction is consistent with the
experimental ratio $\tau(\Bs)/\tau(\Bd)=\Gd/\Gs$,
which is smaller than 1 by 
\hflavONEMINUSRTAUBSMEANCpercent. 
The authors of Ref.~\citehistory{Jubb:2016mvq,Artuso:2015swg}{Jubb:2016mvq,Artuso:2015swg,*Lenz_hist} predict
$\tau(\Bs)/\tau(\Bd) = 1.00050 \pm 0.00108 \pm 0.0225\times \delta$,
where $\delta$ quantifies a possible breaking of the quark-hadron duality.
In this context, they interpret the $2.5\sigma$ difference 
between theory and experiment as being due to either new physics
or a sizable duality violation.
The key message is that improved experimental precision
on this ratio is very welcome.

The ratio $\tau(\Lb)/\tau(\Bd)$ has particularly been the source of theoretical
scrutiny since earlier calculations using the HQE~\cite{Shifman:1986mx,Chay:1990da,Bigi:1992su,Voloshin:1999pz,*Guberina:1999bw,*Neubert:1996we,*Bigi:1997fj}
predicted a value larger than 0.90, almost $2\sigma$ 
above the world average at the time. 
Many predictions cluster around a most likely central value
of 0.94~\cite{Uraltsev:1996ta,*Pirjol:1998ur,*Colangelo:1996ta,*DiPierro:1999tb}.
Calculations
of this ratio that include higher-order effects predict a
lower ratio between the
\Lb and \Bd lifetimes~\cite{Franco:2002fc}
and reduce this difference.
Since then, the experimental average has settled at a value 
significantly larger than initially, in agreement with the latest theoretical 
predictions. 
A review~\cite{Lenz:2015dra} concludes that 
the long-standing $\Lb$ lifetime puzzle is resolved, with a
nice agreement between the precise experimental determination
of $\tau(\Lb)/\tau(\Bd)$ and the less precise HQE prediction,
which needs new lattice calculations.
There is also good agreement for the 
$\tau(\Xibu)/\tau(\Xibd)$ ratio, 
for which the prediction is
based on the next-to-leading-order  calculation of \Ref{Beneke:2002rj}.

The lifetimes of the most abundant \b-hadron species are now all known to sub-percent precision. Neglecting the 
contributions of the rarer species (\Bc meson and \b baryons other than the \Lb), one can compute the average 
\b-hadron lifetime from the individual lifetimes and production fractions as 
\begin{equation}
\tau_b = \frac%
{\fBd \tau(\Bd)^2+ \fBu \tau(\Bu)^2+0.5 \fBs \tau(B^0_{s\rm H})^2+0.5 \fBs \tau(B^0_{s\rm L})^2+ \fbb \tau(\Lb)^2}%
{\fBd \tau(\Bd)  + \fBu \tau(\Bu)  +0.5 \fBs \tau(B^0_{s\rm H})  +0.5 \fBs \tau(B^0_{s\rm L})  + \fbb \tau(\Lb)  } \,.
\end{equation}
Using the lifetimes of \Table{sumlife} and the fractions in $Z$ decays of \Table{fractions},
taking into account the correlations between the fractions (\Table{fractions}) as well as the correlation 
between $\tau(B_{s\rm H})$ and $\tau(B_{s\rm L})$ (\hflavZRHOTAUHTAUL), one obtains
\begin{equation}
\tau_b(Z) = \hflavTAUBZCALC \,.
\end{equation}
This is in very good agreement with (and three times more precise than)
the average of \Eq{TAUBVTX} for the inclusive measurements performed at LEP. 
\mysubsection{Neutral \B-meson mixing}
\labs{mixing}

The $\Bd-\Bdbar$ and $\Bs-\Bsbar$ systems
both exhibit the phenomenon of particle-antiparticle mixing. For each of them, 
there are two mass eigenstates which are linear combinations of the two flavour states,
$B^0_q$ and $\bar{B}^0_q$, 
\begin{eqnarray}
| B^0_{q\rm L}\rangle &=& p_q |B^0_q \rangle +  q_q |\bar{B}^0_q \rangle \,, \\
| B^0_{q\rm H}\rangle &=& p_q |B^0_q \rangle -  q_q |\bar{B}^0_q \rangle  \,,
\end{eqnarray}
where the subscript $q=d$ is used for the  $B^0_d$ ($=\Bd$) meson and $q=s$ for the \Bs meson.
The heaviest (lightest) of these mass states is denoted
$B^0_{q\rm H}$ ($B^0_{q\rm L}$),
with mass $m_{q\rm H}$ ($m_{q\rm L}$)
and total decay width $\Gamma_{q\rm H}$ ($\Gamma_{q\rm L}$). We define
\begin{eqnarray}
\Delta m_q = m_{q\rm H} - m_{q\rm L} \,, &~~~~&  x_q = \Delta m_q/\Gamma_q \,, \labe{dm} \\
\Delta \Gamma_q \, = \Gamma_{q\rm L} - \Gamma_{q\rm H} \,, ~ &~~~~&  y_q= \Delta\Gamma_q/(2\Gamma_q) \,, \labe{dg}
\end{eqnarray}
where 
$\Gamma_q = (\Gamma_{q\rm H} + \Gamma_{q\rm L})/2 =1/\bar{\tau}(B^0_q)$ 
is the average decay width.
$\Delta m_q$ is positive by definition, and 
$\Delta \Gamma_q$ is expected to be positive within
the Standard Model.\footnote{
  \label{foot:life_mix:Eqdg}
  For reasons of symmetry in \Eqss{dm}{dg}, 
  $\Delta \Gamma$ is sometimes defined with the opposite sign. 
  The definition adopted in \Eq{dg} is the one used
  by most experimentalists and many phenomenologists in \B physics.}

Four different time-dependent probabilities are needed to describe the 
evolution of a neutral \B meson that is produced as a flavour state and decays without
\CP violation to a flavour-specific final state. 
If \CPT is conserved (which  
will be assumed throughout), they can be written as 
\begin{equation}
\left\{
\begin{array}{rcl}
{\cal P}(B^0_q \to B^0_q) & = &  \frac{1}{2} e^{-\Gamma_q t} 
\left[ \cosh\!\left(\frac{1}{2}\Delta\Gamma_q t\right) + \cos\!\left(\Delta m_q t\right)\right]  \\
{\cal P}(B^0_q \to \bar{B}^0_q) & = &   \frac{1}{2} e^{-\Gamma_q t} 
\left[ \cosh\!\left(\frac{1}{2}\Delta\Gamma_q t\right) - \cos\!\left(\Delta m_q t\right)\right] 
\left|q_q/p_q\right|^2 \\
{\cal P}(\bar{B}^0_q \to B^0_q) & = &  \frac{1}{2} e^{-\Gamma_q t} 
\left[ \cosh\!\left(\frac{1}{2}\Delta\Gamma_q t\right) - \cos\!\left(\Delta m_q t\right)\right] 
\left|p_q/q_q\right|^2 \\
{\cal P}(\bar{B}^0_q \to\bar{B}^0_q) & = &  \frac{1}{2} e^{-\Gamma_q t}  
\left[ \cosh\!\left(\frac{1}{2}\Delta\Gamma_q t\right) + \cos\!\left(\Delta m_q t\right)\right] 
\end{array} \right. \,,
\labe{oscillations}
\end{equation}
where $t$ is the proper time of the system (\ie, the time interval between the production 
and the decay in the rest frame of the \B meson). 
At the \B factories, only the proper-time difference $\Delta t$ between the decays
of the two neutral \B mesons from the \Ups can be determined. However,
since the two \B mesons evolve coherently (keeping opposite flavours as long
as neither of them has decayed), the 
above formulae remain valid 
if $t$ is replaced with $\Delta t$ and the production flavour is replaced by the flavour 
at the time of the decay of the accompanying \B meson into a flavour-specific state.
As can be seen in the above expressions,
the mixing probabilities 
depend on three mixing observables:
$\Delta m_q$, $\Delta\Gamma_q$,
and $|q_q/p_q|^2$. In particular, \CP violation in mixing exists if $|q_q/p_q|^2 \ne 1$.
Another (non independent) observable often used to characterize \CP violation in the mixing 
is the so-called semileptonic asymmetry, defined as
\begin{equation} 
{\cal A}_{\rm SL}^q = 
\frac{|p_{\particle{q}}/q_{\particle{q}}|^2 - |q_{\particle{q}}/p_{\particle{q}}|^2}%
{|p_{\particle{q}}/q_{\particle{q}}|^2 + |q_{\particle{q}}/p_{\particle{q}}|^2} \,.
\labe{ASLq}
\end{equation} 
All  mixing observables depend on two complex numbers, $M^q_{12}$ and $\Gamma^q_{12}$, which are the off-diagonal elements of the $2\times 2$ mass and decay matrices describing the evolution of the $B^0_q-\bar{B}^0_q$ system. In the Standard Model the quantity $|\Gamma^q_{12}/M^q_{12}|$ is small, of the order of $(m_b/m_t)^2$, where $m_b$ and $m_t$ are the bottom and top quark masses. The following relations hold to first order in $|\Gamma^q_{12}/M^q_{12}|$:
\begin{eqnarray}
\Delta m_q & = & 2 |M^q_{12}| \left[1 + {\cal O} \left(|\Gamma^q_{12}/M^q_{12}|^2 \right) \right] \,, \\
\Delta\Gamma_q & = & 2 |\Gamma^q_{12}| \cos\phi^q_{12} \left[1 + {\cal O} \left(|\Gamma^q_{12}/M^q_{12}|^2 \right) \right]   \,, \\
{\cal A}_{\rm SL}^q & = &  \Im \left(\Gamma^q_{12}/M^q_{12} \right) +
{\cal O} \left(|\Gamma^q_{12}/M^q_{12}|^2 \right) =
\frac{\Delta\Gamma_q}{\Delta m_q}\tan\phi^q_{12} +
{\cal O} \left(|\Gamma^q_{12}/M^q_{12}|^2 \right)  \,,
\labe{ALSq_tanphi2}
\end{eqnarray}
where 
\begin{equation}
\phi^q_{12} = \arg \left( -{M^q_{12}}/{\Gamma^q_{12}} \right)
\labe{phi12}
\end{equation}
is the observable phase difference between $-M^q_{12}$ and $\Gamma^q_{12}$ (often called the mixing phase). 
It should be noted that the theoretical predictions for $\Gamma^q_{12}$ are based on the same HQE as the lifetime predictions. 

In the next sections we review in turn the experimental knowledge
on the \Bd decay-width and mass differences, 
the \Bs decay-width and mass differences,  
\CP violation in \Bd and \Bs mixing, and mixing-induced \CP violation in \Bs decays. 

\mysubsubsection{\Bd mixing parameters \DGd and \dmd}
\labs{DGd} \labs{dmd}

\begin{table}
\centering
\caption{Time-dependent measurements included in the \dmd average.
The results obtained from multi-dimensional fits involving also 
the \Bd (and \Bu) lifetime(s)
as free parameter(s)~\protect\citehistory{Aubert:2002sh,Aubert:2005kf,Abe:2004mz}{Aubert:2002sh,Aubert:2005kf,Abe:2004mz,*Abe:2002id_hist,*Tomura:2002qs_hist,*Hara:2002mq_hist} 
have been converted into one-dimensional measurements of \dmd.
All measurements have then been adjusted to a common set of physics
parameters before being combined.}
\labt{dmd}
\begin{tabular}{@{}rc@{}cc@{}c@{}cc@{}c@{}c@{}}
\hline
Experiment & \multicolumn{2}{c}{Method} & \multicolumn{3}{l}{\dmd in\invps}   
                                        & \multicolumn{3}{l}{\dmd in\invps}     \\[-0.8ex]
and Ref.   &  rec. & tag                & \multicolumn{3}{l}{before adjustment} 
                                        & \multicolumn{3}{l}{after adjustment} \\
\hline
 ALEPH~\cite{Buskulic:1996qt}  & \particle{ \ell  } & \particle{ \Qjet  } & $  0.404 $ & $ \pm  0.045 $ & $ \pm  0.027 $ & & & \\
 ALEPH~\cite{Buskulic:1996qt}  & \particle{ \ell  } & \particle{ \ell  } & $  0.452 $ & $ \pm  0.039 $ & $ \pm  0.044 $ & & & \\
 ALEPH~\cite{Buskulic:1996qt}  & \multicolumn{2}{c}{above two combined} & $  0.422 $ & $ \pm  0.032 $ & $ \pm  0.026 $ & $  0.440 $ & $ \pm  0.032 $ & $ ^{+  0.020 }_{-  0.019 } $ \\
 ALEPH~\cite{Buskulic:1996qt}  & \particle{ D^*  } & \particle{ \ell,\Qjet  } & $  0.482 $ & $ \pm  0.044 $ & $ \pm  0.024 $ & $  0.482 $ & $ \pm  0.044 $ & $ \pm  0.024 $ \\
 DELPHI~\cite{Abreu:1997xq}  & \particle{ \ell  } & \particle{ \Qjet  } & $  0.493 $ & $ \pm  0.042 $ & $ \pm  0.027 $ & $  0.500 $ & $ \pm  0.042 $ & $ \pm  0.024 $ \\
 DELPHI~\cite{Abreu:1997xq}  & \particle{ \pi^*\ell  } & \particle{ \Qjet  } & $  0.499 $ & $ \pm  0.053 $ & $ \pm  0.015 $ & $  0.500 $ & $ \pm  0.053 $ & $ \pm  0.015 $ \\
 DELPHI~\cite{Abreu:1997xq}  & \particle{ \ell  } & \particle{ \ell  } & $  0.480 $ & $ \pm  0.040 $ & $ \pm  0.051 $ & $  0.495 $ & $ \pm  0.040 $ & $ ^{+  0.042 }_{-  0.040 } $ \\
 DELPHI~\cite{Abreu:1997xq}  & \particle{ D^*  } & \particle{ \Qjet  } & $  0.523 $ & $ \pm  0.072 $ & $ \pm  0.043 $ & $  0.518 $ & $ \pm  0.072 $ & $ \pm  0.043 $ \\
 DELPHI~\cite{Abdallah:2002mr}  & \particle{ \mbox{vtx}  } & \particle{ \mbox{comb}  } & $  0.531 $ & $ \pm  0.025 $ & $ \pm  0.007 $ & $  0.525 $ & $ \pm  0.025 $ & $ \pm  0.006 $ \\
 L3~\citehistory{Acciarri:1998pq}{Acciarri:1998pq,*Acciarri:1996ia_hist}  & \particle{ \ell  } & \particle{ \ell  } & $  0.458 $ & $ \pm  0.046 $ & $ \pm  0.032 $ & $  0.467 $ & $ \pm  0.046 $ & $ \pm  0.028 $ \\
 L3~\citehistory{Acciarri:1998pq}{Acciarri:1998pq,*Acciarri:1996ia_hist}  & \particle{ \ell  } & \particle{ \Qjet  } & $  0.427 $ & $ \pm  0.044 $ & $ \pm  0.044 $ & $  0.439 $ & $ \pm  0.044 $ & $ \pm  0.042 $ \\
 L3~\citehistory{Acciarri:1998pq}{Acciarri:1998pq,*Acciarri:1996ia_hist}  & \particle{ \ell  } & \particle{ \ell\mbox{(IP)}  } & $  0.462 $ & $ \pm  0.063 $ & $ \pm  0.053 $ & $  0.471 $ & $ \pm  0.063 $ & $ \pm  0.044 $ \\
 OPAL~\cite{Ackerstaff:1997iw}  & \particle{ \ell  } & \particle{ \ell  } & $  0.430 $ & $ \pm  0.043 $ & $ ^{+  0.028 }_{-  0.030 } $ & $  0.467 $ & $ \pm  0.043 $ & $ ^{+  0.017 }_{-  0.016 } $ \\
 OPAL~\cite{Ackerstaff:1997vd}  & \particle{ \ell  } & \particle{ \Qjet  } & $  0.444 $ & $ \pm  0.029 $ & $ ^{+  0.020 }_{-  0.017 } $ & $  0.481 $ & $ \pm  0.029 $ & $ \pm  0.013 $ \\
 OPAL~\cite{Alexander:1996id}  & \particle{ D^*\ell  } & \particle{ \Qjet  } & $  0.539 $ & $ \pm  0.060 $ & $ \pm  0.024 $ & $  0.544 $ & $ \pm  0.060 $ & $ \pm  0.023 $ \\
 OPAL~\cite{Alexander:1996id}  & \particle{ D^*  } & \particle{ \ell  } & $  0.567 $ & $ \pm  0.089 $ & $ ^{+  0.029 }_{-  0.023 } $ & $  0.572 $ & $ \pm  0.089 $ & $ ^{+  0.028 }_{-  0.022 } $ \\
 OPAL~\cite{Abbiendi:2000ec}  & \particle{ \pi^*\ell  } & \particle{ \Qjet  } & $  0.497 $ & $ \pm  0.024 $ & $ \pm  0.025 $ & $  0.496 $ & $ \pm  0.024 $ & $ \pm  0.025 $ \\
 CDF1~\cite{Abe:1997qf,*Abe:1998sq}  & \particle{ D\ell  } & \particle{ \mbox{SST}  } & $  0.471 $ & $ ^{+  0.078 }_{-  0.068 } $ & $ ^{+  0.033 }_{-  0.034 } $ & $  0.470 $ & $ ^{+  0.078 }_{-  0.068 } $ & $ ^{+  0.033 }_{-  0.034 } $ \\
 CDF1~\cite{Abe:1999pv}  & \particle{ \mu  } & \particle{ \mu  } & $  0.503 $ & $ \pm  0.064 $ & $ \pm  0.071 $ & $  0.514 $ & $ \pm  0.064 $ & $ ^{+  0.070 }_{-  0.069 } $ \\
 CDF1~\cite{Abe:1999ds}  & \particle{ \ell  } & \particle{ \ell,\Qjet  } & $  0.500 $ & $ \pm  0.052 $ & $ \pm  0.043 $ & $  0.546 $ & $ \pm  0.052 $ & $ \pm  0.036 $ \\
 CDF1~\cite{Affolder:1999cn}  & \particle{ D^*\ell  } & \particle{ \ell  } & $  0.516 $ & $ \pm  0.099 $ & $ ^{+  0.029 }_{-  0.035 } $ & $  0.523 $ & $ \pm  0.099 $ & $ ^{+  0.028 }_{-  0.035 } $ \\
 \dzero~\cite{Abazov:2006qp}  & \particle{ D^{(*)}\mu  } & \particle{ \mbox{OST}  } & $  0.506 $ & $ \pm  0.020 $ & $ \pm  0.016 $ & $  0.506 $ & $ \pm  0.020 $ & $ \pm  0.016 $ \\
 \babar~\cite{Aubert:2001te,*Aubert:2002rg}  & \particle{ \Bd  } & \particle{ \ell,K,\mbox{NN}  } & $  0.516 $ & $ \pm  0.016 $ & $ \pm  0.010 $ & $  0.521 $ & $ \pm  0.016 $ & $ \pm  0.008 $ \\
 \babar~\cite{Aubert:2001tf}  & \particle{ \ell  } & \particle{ \ell  } & $  0.493 $ & $ \pm  0.012 $ & $ \pm  0.009 $ & $  0.487 $ & $ \pm  0.012 $ & $ \pm  0.006 $ \\
 \babar~\cite{Aubert:2002sh}  & \particle{ D^*\ell\nu  } & \particle{ \ell,K,\mbox{NN}  } & $  0.492 $ & $ \pm  0.018 $ & $ \pm  0.014 $ & $  0.493 $ & $ \pm  0.018 $ & $ \pm  0.013 $ \\
 \babar~\cite{Aubert:2005kf}  & \particle{ D^*\ell\nu\mbox{(part)}  } & \particle{ \ell  } & $  0.511 $ & $ \pm  0.007 $ & $ \pm  0.007 $ & $  0.513 $ & $ \pm  0.007 $ & $ \pm  0.007 $ \\
 \belle~\citehistory{Abe:2004mz}{Abe:2004mz,*Abe:2002id_hist,*Tomura:2002qs_hist,*Hara:2002mq_hist}  & \particle{ \Bd,D^*\ell\nu  } & \particle{ \mbox{comb}  } & $  0.511 $ & $ \pm  0.005 $ & $ \pm  0.006 $ & $  0.514 $ & $ \pm  0.005 $ & $ \pm  0.006 $ \\
 \belle~\cite{Zheng:2002jv}  & \particle{ D^*\pi\mbox{(part)}  } & \particle{ \ell  } & $  0.509 $ & $ \pm  0.017 $ & $ \pm  0.020 $ & $  0.514 $ & $ \pm  0.017 $ & $ \pm  0.019 $ \\
 \belle~\citehistory{Hastings:2002ff}{Hastings:2002ff,*Abe:2000yh_hist}  & \particle{ \ell  } & \particle{ \ell  } & $  0.503 $ & $ \pm  0.008 $ & $ \pm  0.010 $ & $  0.506 $ & $ \pm  0.008 $ & $ \pm  0.008 $ \\
 LHCb~\cite{Aaij:2011qx}  & \particle{ \Bd  } & \particle{ \mbox{OST}  } & $  0.499 $ & $ \pm  0.032 $ & $ \pm  0.003 $ & $  0.499 $ & $ \pm  0.032 $ & $ \pm  0.003 $ \\
 LHCb~\cite{Aaij:2012nt}  & \particle{ \Bd  } & \particle{ \mbox{OST,SST}  } & $  0.5156 $ & $ \pm  0.0051 $ & $ \pm  0.0033 $ & $  0.5156 $ & $ \pm  0.0051 $ & $ \pm  0.0033 $ \\
 LHCb~\cite{Aaij:2013gja}  & \particle{ D\mu  } & \particle{ \mbox{OST,SST}  } & $  0.503 $ & $ \pm  0.011 $ & $ \pm  0.013 $ & $  0.503 $ & $ \pm  0.011 $ & $ \pm  0.013 $ \\
 LHCb~\cite{Aaij:2016fdk}  & \particle{ D^{(*)}\mu  } & \particle{ \mbox{OST}  } & $  0.5050 $ & $ \pm  0.0021 $ & $ \pm  0.0010 $ & $  0.5050 $ & $ \pm  0.0021 $ & $ \pm  0.0010 $ \\
 \hline \\[-2.0ex]
 \multicolumn{6}{l}{World average (all above measurements included):} & $  0.5065 $ & $ \pm  0.0016 $ & $ \pm  0.0011 $ \\

\\[-3.0ex]
\multicolumn{6}{l}{~~~ -- ALEPH, DELPHI, L3 and OPAL only:}
     & \hflavDMDLval & \hflavDMDLsta & \hflavDMDLsys \\[-0.8ex]
\multicolumn{6}{l}{~~~ -- CDF and \dzero only:}
     & \hflavDMDTval & \hflavDMDTsta & \hflavDMDTsys \\[-0.8ex]
\multicolumn{6}{l}{~~~ -- \babar and \belle only:}
     & \hflavDMDBval & \hflavDMDBsta & \hflavDMDBsys \\[-0.8ex]
\multicolumn{6}{l}{~~~ -- LHCb only:} & \hflavDMDLHCbval & \hflavDMDLHCbsta & \hflavDMDLHCbsys \\
\hline
\end{tabular}
\end{table}

A large number of time-dependent \Bd--\Bdbar oscillation analyses
have been performed in the past 20 years by the 
ALEPH, DELPHI, L3, OPAL, CDF, \dzero, \babar, \belle and  LHCb collaborations. 
The corresponding measurements of \dmd are summarized in 
\Table{dmd}\history{.}{,
where only the most recent results
are listed (\ie\ measurements superseded by more recent ones are omitted\unpublished{}{\footnote{
  \label{foot:life_mix:CDFnote8235:2006}
  Two old unpublished CDF2 measurements~\cite{CDFnote8235:2006,CDFnote7920:2005}
  are also omitted from our averages, \Table{dmd} and \Fig{dmd}.}}).}
It is notable that the systematic uncertainties are comparable to the statistical uncertainties;
they are often dominated by sample composition, mistag probability,
or \b-hadron lifetime contributions.
Before being combined, the measurements are adjusted on the basis of a 
common set of input values, including the averages of the 
\b-hadron fractions and lifetimes given in this report 
(see \Secss{fractions}{lifetimes}).
Some measurements are statistically correlated. 
Systematic correlations arise both from common physics sources 
(fractions, lifetimes, branching fractions of \b hadrons), and from purely 
experimental or algorithmic effects (efficiency, resolution, flavour tagging, 
background description). Combining all published measurements
listed in \Table{dmd}
and accounting for all identified correlations
as described in \Ref{Abbaneo:2000ej_mod,*Abbaneo:2001bv_mod_cont} yields $\dmd = \hflavDMDWfull$.

On the other hand, ARGUS and CLEO have published 
measurements of the time-integrated mixing probability 
\chid~\cite{Albrecht:1992yd,*Albrecht:1993gr,Bartelt:1993cf,Behrens:2000qu}, 
which average to $\chid =\hflavCHIDU$.
Following \Ref{Behrens:2000qu}, 
the decay width difference \DGd could 
in principle be extracted from the
measured value of $\Gd=1/\tau(\Bd)$ and the above averages for 
\dmd and \chid 
(provided that \DGd has a negligible impact on 
the \dmd and $\tau(\Bd)$ analyses that have assumed $\DGd=0$), 
using the relation
\begin{equation}
\chid = \frac{\xd^2+\yd^2}{2(\xd^2+1)}
\,,
\labe{chid_definition}
\end{equation}
but $\DGd/\Gd$ is too small to provide any useful sensitivity. 
Direct time-dependent studies provide much stronger constraints: 
$|\DGd|/\Gd < 18\%$ at \CL{95} from DELPHI~\cite{Abdallah:2002mr},
$-6.8\% < {\rm sign}({\rm Re} \lambda_{\CP}) \DGGd < 8.4\%$
at \CL{90} from \babar~\cite{Aubert:2003hd,*Aubert:2004xga},
and ${\rm sign}({\rm Re} \lambda_{\CP})\DGGd = (1.7 \pm 1.8 \pm 1.1)\%$~\cite{Higuchi:2012kx}
from Belle, 
where $\lambda_{\CP} = (q_{\particle{d}}/p_{\particle{d}}) (\bar{A}_{\CP}/A_{\CP})$
is defined for a \CP-even final state. 
The sensitivity to the overall sign of 
${\rm sign}({\rm Re} \lambda_{\CP}) \DGGd$ comes
from the use of \Bd decays to \CP eigenstates.
In addition,
LHCb has obtained $\DGGd=(-4.4 \pm 2.5 \pm 1.1)\%$~\cite{Aaij:2014owa}
by comparing measurements of the lifetime for $\Bd \to \jpsi K^{*0}$ and $\Bd \to \jpsi K^0_{\rm S}$
decays, following the method of Ref.~\cite{Gershon:2010wx}.
Using a similar method, ATLAS and CMS have measured %
$\DGGd=(-0.1 \pm 1.1 \pm 0.9)\%$~\cite{Aaboud:2016bro} and
$\DGGd=(+3.4 \pm 2.3 \pm 2.4)\%$~\cite{Sirunyan:2017nbv}, respectively.
Assuming ${\rm Re} \lambda_{\CP} > 0$, as expected from the global fits
of the Unitarity Triangle within the Standard Model~\cite{Charles:2011va_mod,*Bona:2006ah_mod},
a combination of these six results (after adjusting the DELPHI and \babar results to  
$1/\Gd=\tau(\Bd)=\hflavTAUBD$) yields
\begin{equation}
\DGGd  = \hflavSDGDGD \,,
\end{equation}
an average consistent with zero and with the latest Standard Model prediction of $(3.97\pm0.90)\times 10^{-3}$~\cite{Artuso:2015swg}. 
An independent result, 
$\DGGd=(0.50 \pm 1.38)\%$\citehistory{Abazov:2013uma}{Abazov:2013uma,*Abazov:2011yk_hist,*Abazov:2010hv_hist,*Abazov:2010hj_hist,*Abazov:dimuon_hist},
was obtained by the \dzero collaboration 
from their measurements of the single muon and same-sign dimuon charge asymmetries,
under the interpretation that 
the observed asymmetries are due to \CP violation in neutral $B$-meson mixing and interference.
This indirect determination was called into question~\cite{Nierste_CKM2014}
and is therefore not included in the above average, 
as explained in \Sec{qpd}.%

Assuming $\DGd=0$ 
and using $1/\Gd=\tau(\Bd)=\hflavTAUBD$,
the \dmd and \chid results are combined through \Eq{chid_definition} 
to yield the 
world average
\begin{equation} 
\dmd = \hflavDMDWU \,,
\labe{dmd}
\end{equation} 
or, equivalently,
\begin{equation} 
\xd= \hflavXDWU ~~~ \mbox{and} ~~~ \chid=\hflavCHIDWU \,.  
\labe{chid}
\end{equation}
\Figure{dmd} compares the \dmd values obtained by the different experiments.

\begin{figure}
\begin{center}
\includegraphics[width=\textwidth]{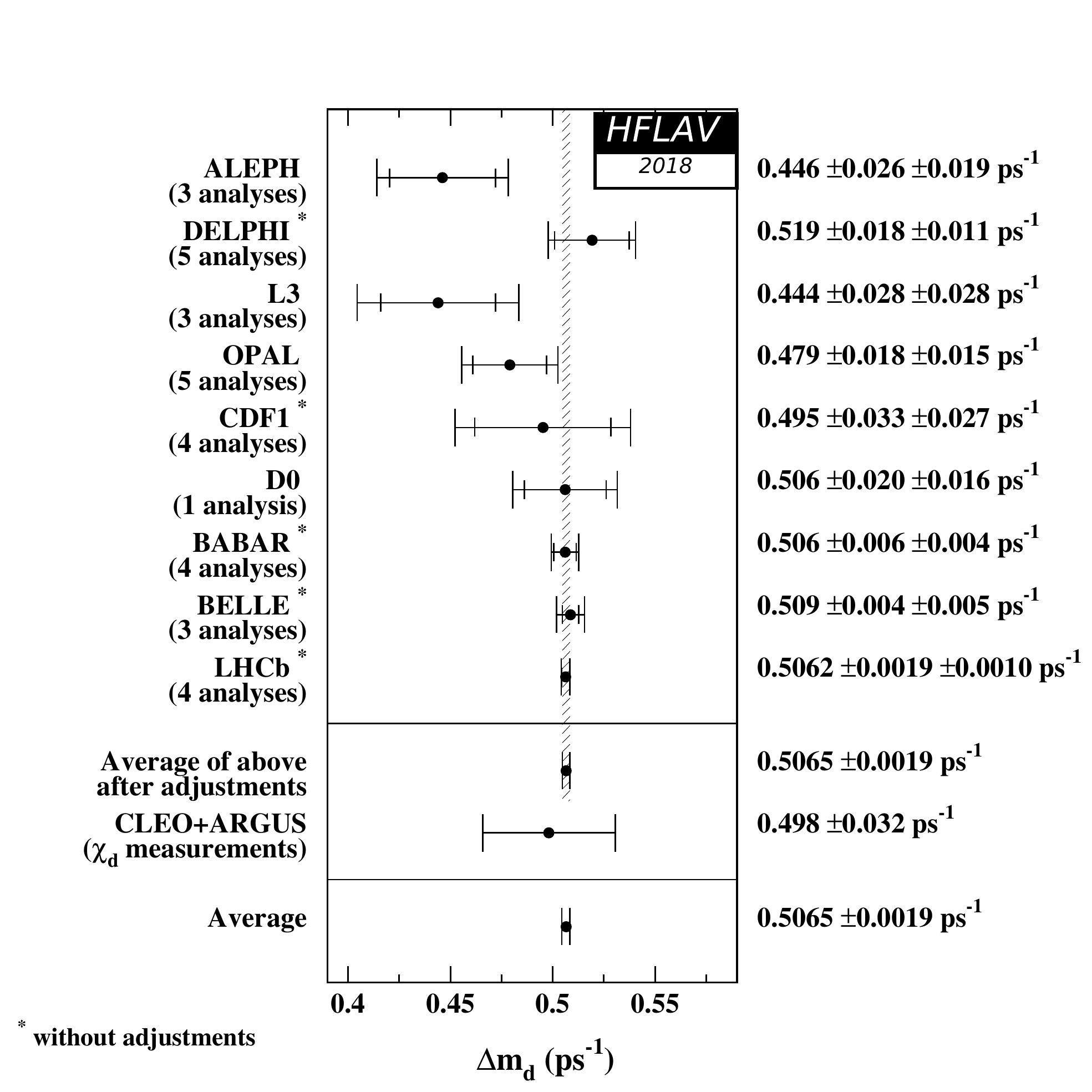}
\caption{The \Bd--\Bdbar oscillation frequency \dmd as measured by the different experiments. 
The averages quoted for ALEPH, L3 and OPAL are taken from the original publications, while the 
ones for DELPHI, CDF, \babar, \belle and LHCb are computed from the individual results 
listed in \Table{dmd} without performing any adjustments. The time-integrated measurements 
of \chid from the symmetric \B factory experiments ARGUS and CLEO are converted 
to a \dmd value using $\tau(\Bd)=\hflavTAUBD$. The two global averages are obtained 
after adjustments of all the individual \dmd results of \Table{dmd} (see text).}
\labf{dmd}
\end{center}
\end{figure}

The \Bd mixing averages given in \Eqss{dmd}{chid}
and the \b-hadron fractions of \Table{fractions} have been obtained in a fully 
consistent way, taking into account the fact that the fractions are computed using 
the \chid value of \Eq{chid} and that many individual measurements of \dmd
at high energy depend on the assumed values for the \b-hadron fractions.
Furthermore, this set of averages is consistent with the lifetime averages 
of \Sec{lifetimes}.

\mysubsubsection{\Bs mixing parameters \DGs and \dms}
\labs{DGs} \labs{dms}

The best sensitivity to \DGs is currently achieved 
by the recent time-dependent measurements
of the $\Bs\to\jpsi\phi$ (or more generally $\Bs\to (c\bar{c}) K^+K^-$) decay rates performed at
CDF~\citehistory{Aaltonen:2012ie}{Aaltonen:2012ie,*CDF:2011af_hist,*Aaltonen:2007he_hist,*Aaltonen:2007gf_hist},
\dzero~\citehistory{Abazov:2011ry}{Abazov:2011ry,*Abazov:2008af_hist,*Abazov:2007tx_hist},
ATLAS~\citehistory{Aad:2014cqa,Aad:2016tdj}{Aad:2014cqa,*Aad:2012kba_hist,Aad:2016tdj}
CMS~\unpublished{\cite{Khachatryan:2015nza}}{\cite{CMS-PAS-BPH-11-006,Khachatryan:2015nza}}
and LHCb~\citehistory{Aaij:2014zsa,Aaij:2017zgz,Aaij:2016ohx}{Aaij:2014zsa,*Aaij:2013oba_supersede2,Aaij:2017zgz,*Aaij:2014zsa_partial_supersede,Aaij:2016ohx},
where the \CP-even and \CP-odd
amplitudes are statistically separated through a full angular analysis.
\unpublished{These}{With the exception of the first CMS analysis~\cite{CMS-PAS-BPH-11-006}%
\footnote{The CMS result of \Ref{CMS-PAS-BPH-11-006}
is statistically independent of that of
Ref.~\cite{Khachatryan:2015nza} but, since it has not
been published, it is not included in \Table{GsDGs} nor in our averages.},
these}
studies use both untagged and tagged \Bs\ candidates and 
are optimized for the measurement of the \CP-violating 
phase \phiccbars, defined later in \Sec{phasebs}.
The LHCb collaboration analyzed the $\Bs \to \jpsi K^+K^-$
decay, considering that the $K^+K^-$ system can be in a $P$-wave or $S$-wave state, 
and measured the dependence of the strong phase difference between the 
$P$-wave and $S$-wave amplitudes as a function of the $K^+K^-$ invariant
mass~\cite{Aaij:2012eq}. 
This allowed, for the first time, the unambiguous determination of the sign of 
$\DGs$, which was found to be positive at the $4.7\,\sigma$ level. 
The following averages present only the $\DGs > 0$ solutions.

The published results~\citehistory%
{Aaltonen:2012ie,Abazov:2011ry,Aad:2014cqa,Aad:2016tdj,Khachatryan:2015nza,Aaij:2014zsa,Aaij:2017zgz,Aaij:2016ohx}%
{Aaltonen:2012ie,*CDF:2011af_hist,*Aaltonen:2007he_hist,*Aaltonen:2007gf_hist,Abazov:2011ry,*Abazov:2008af_hist,*Abazov:2007tx_hist,Aad:2014cqa,*Aad:2012kba_hist,Aad:2016tdj,Khachatryan:2015nza,Aaij:2014zsa,*Aaij:2013oba_supersede2,Aaij:2017zgz,*Aaij:2014zsa_partial_supersede,Aaij:2016ohx}
are shown in \Table{GsDGs}. They are combined taking into account, in each analysis, the correlation between \DGs and \Gs.
The results, displayed as the red contours labelled ``$\Bs \to (c\bar{c}) KK$ measurements'' in the
plots of \Fig{DGs}, are given in the first column of numbers of \Table{tabtauLH}.

\begin{table}
\caption{Measurements of \DGs and \Gs using
$\Bs\to\jpsi\phi$, $\Bs\to\jpsi K^+K^-$ and $\Bs\to\psi(2S)\phi$ decays.
Only the solution with $\DGs > 0$ is shown, since the two-fold ambiguity has been
resolved in \Ref{Aaij:2012eq}. The first error is due to 
statistics, the second one to systematics. The last line gives our average.}
\labt{GsDGs}
\begin{center}
\begin{tabular}{ll@{\,}rll@{\,}l@{\,}} 
\hline
Exp.\ & Mode & Dataset
      & \multicolumn{1}{c}{\DGs (\!\!\invps)}
      & \multicolumn{1}{c}{\Gs  (\!\!\invps)}
      & Ref.\ \\
\hline
CDF    & $\jpsi\phi$ & $9.6\invfb$
       & $+0.068\pm0.026\pm0.009$
       & $0.654\pm0.008\pm0.004$ %
       & \citehistory{Aaltonen:2012ie}{Aaltonen:2012ie,*CDF:2011af_hist,*Aaltonen:2007he_hist,*Aaltonen:2007gf_hist} \\
\dzero & $\jpsi\phi$ & $8.0\invfb$
       & $+0.163^{+0.065}_{-0.064}$ 
       & $0.693^{+0.018}_{-0.017}$
       & \citehistory{Abazov:2011ry}{Abazov:2011ry,*Abazov:2008af_hist,*Abazov:2007tx_hist} \\
ATLAS  & $\jpsi\phi$ & $4.9\invfb$
       & $+0.053 \pm0.021 \pm0.010$
       & $0.677 \pm0.007 \pm0.004$
       & \citehistory{Aad:2014cqa}{Aad:2014cqa,*Aad:2012kba_hist}  \\
ATLAS  & $\jpsi\phi$ & $14.3\invfb$
       & $+0.101 \pm0.013 \pm0.007$
       & $0.676 \pm0.004 \pm0.004$
       & \cite{Aad:2016tdj} \\
ATLAS  & \multicolumn{2}{r}{above 2 combined}
       & $+0.085 \pm0.011 \pm0.007$
       & $0.675 \pm0.003 \pm0.003$
       & \cite{Aad:2016tdj} \\
CMS    & $\jpsi\phi$ & $19.7\invfb$ 
       & $+0.095\pm0.013\pm0.007$
       & $0.6704 \pm0.0043 \pm0.0055$
       & \cite{Khachatryan:2015nza} \\
LHCb   & $\jpsi K^+K^-$ & $3.0\invfb$
       & $+0.0805\pm0.0091\pm0.0032$
       & $0.6603\pm0.0027\pm0.0015$
       & \citehistory{Aaij:2014zsa}{Aaij:2014zsa,*Aaij:2013oba_supersede2} \\
LHCb   & $\jpsi K^+K^-{}^a$ & $3.0\invfb$
       & $+0.066 \pm0.018 \pm0.010$
       & $0.650 \pm0.006 \pm0.004$
       & \citehistory{Aaij:2017zgz}{Aaij:2017zgz,*Aaij:2014zsa_partial_supersede} \\
LHCb   & \multicolumn{2}{r}{above 2 combined}
       & $+0.0813\pm0.0073\pm0.0036$
       & $0.6588\pm0.0022\pm0.0015$
       & \citehistory{Aaij:2017zgz}{Aaij:2017zgz,*Aaij:2014zsa_partial_supersede} \\
LHCb   & $\psi(2S) \phi$ & $3.0\invfb$
       & $+0.066 ^{+0.041}_{-0.044} \pm 0.007$
       & $0.668 \pm0.011 \pm0.006$
       & \cite{Aaij:2016ohx} \\
\hline
\multicolumn{3}{l}{All combined} & \hflavDGSnounit & \hflavGSnounit & \\ 
\hline
\multicolumn{6}{l}{$^a$ {\footnotesize $m(K^+K^-) > 1.05~\gevcc$.}}
\end{tabular}
\end{center}
\end{table}

\begin{figure}
\begin{center}
\includegraphics[width=0.49\textwidth]{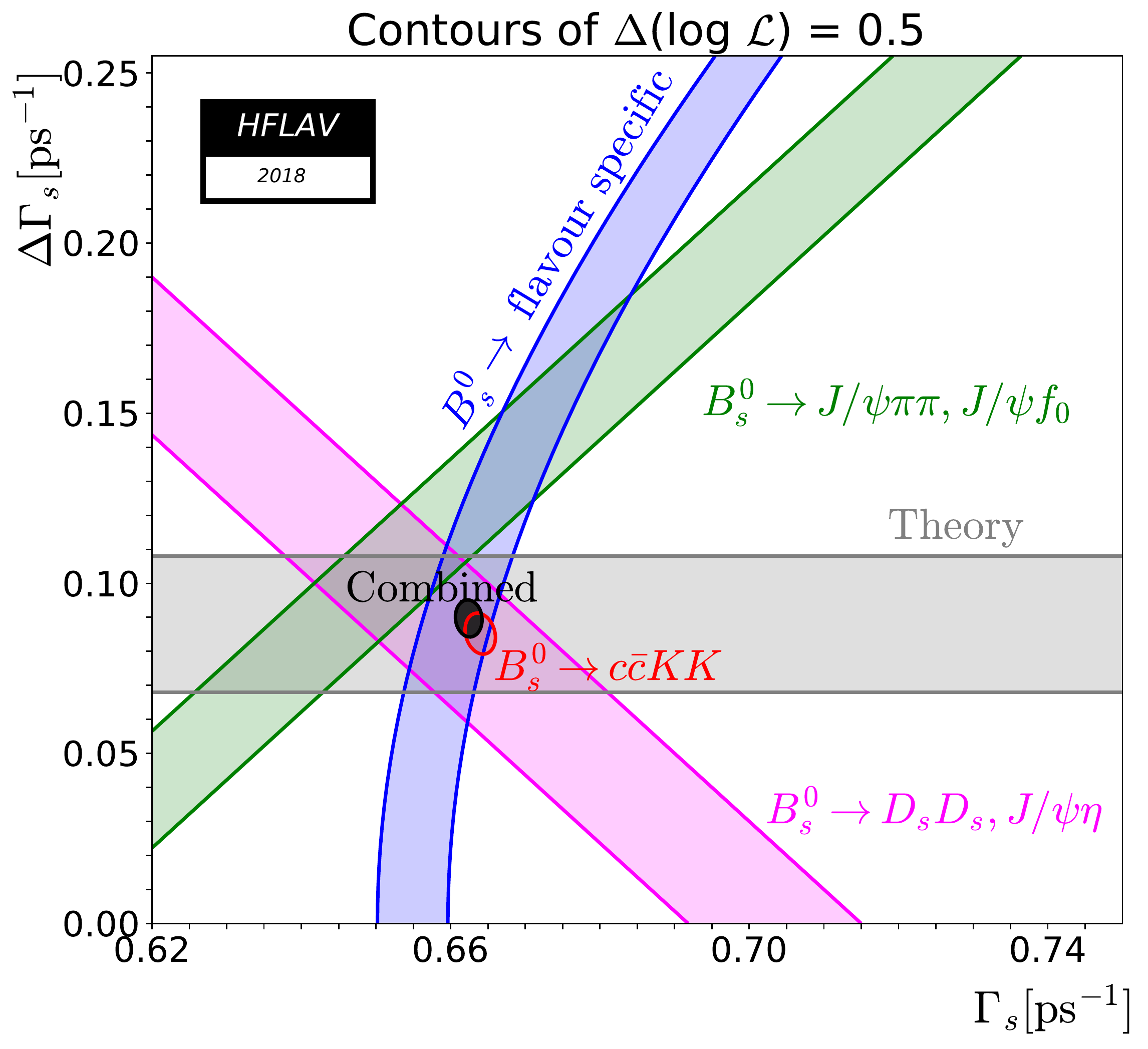}
\hfill
\includegraphics[width=0.49\textwidth]{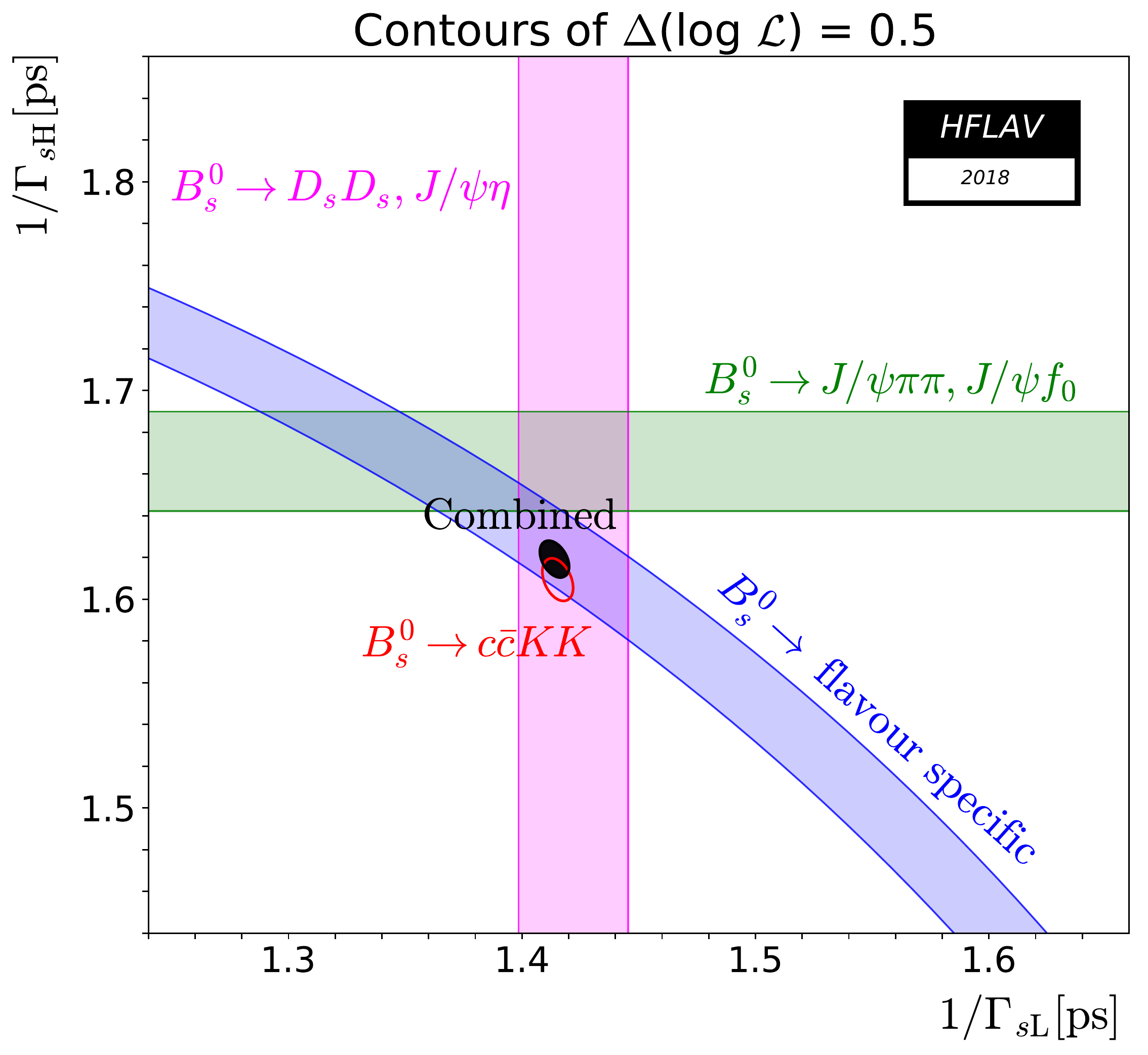}
\caption{Contours of $\Delta \ln L = 0.5$ (39\% CL for the enclosed 2D regions, 68\% CL for the bands)
shown in the $(\Gs,\,\DGs)$ plane on the left
and in the $(1/\Gamma_{s\rm L},\,1/\Gamma_{s\rm H})$ plane on the right. 
The average of all the $\Bs \to \jpsi\phi$, $\Bs\to \jpsi K^+K^-$ and
$\Bs \to \psi(2S)\phi$
results is shown as the red contour,
and the constraints given by the effective lifetime measurements of
\Bs\ to flavour-specific, pure \CP-odd and pure \CP-even final states
are shown as the blue, green and purple bands, 
respectively. The average taking all constraints into account is shown as the dark-filled contour.
The light-grey band is a theory prediction
$\DGs = 0.088 \pm 0.020~\hbox{ps}^{-1}$~\protect\citehistory{Lenz:2011ti,*Lenz:2006hd,Jubb:2016mvq,Artuso:2015swg}{Lenz:2011ti,*Lenz:2006hd,Jubb:2016mvq,Artuso:2015swg,*Lenz_hist}
that assumes no new physics in \Bs\ mixing.}
\labf{DGs}
\end{center}
\end{figure}

\begin{table}
\caption{Averages of \DGs, $\Gs$ and related quantities, obtained from
$\Bs\to\jpsi\phi$, $\Bs\to\jpsi K^+K^-$ and $\Bs\to\psi(2S)\phi$ alone (first column),
adding the constraints from the effective lifetimes measured in pure \CP modes
$\Bs\to D_s^+D_s^-,J/\psi\eta$ and $\Bs \to \jpsi f_0(980), \jpsi \pi^+\pi^-$ (second column),
and adding the constraint from the effective lifetime measured in flavour-specific modes
$\Bs\to D_s^-\ell^+\nu X, \, D_s^-\pi^+, \, D_s^-D^+$ (third column, recommended world averages).}
\labt{tabtauLH}
\begin{center}
\begin{tabular}{c|c|c|c}
\hline
& $\Bs\to (c\bar{c}) K^+K^-$ modes & $\Bs\to (c\bar{c}) K^+K^-$ modes & $\Bs\to (c\bar{c}) K^+K^-$ modes \\
& only (see \Table{GsDGs}) & + pure \CP modes & + pure \CP modes \\
&                          &                  & + flavour-specific modes \\
\hline
\Gs                & \hflavGS        &  \hflavGSCO        &  \hflavGSCON        \\
$1/\Gs$            & \hflavTAUBSMEAN &  \hflavTAUBSMEANCO &  \hflavTAUBSMEANCON \\
$1/\Gamma_{s\rm L}$ & \hflavTAUBSL    &  \hflavTAUBSLCO    &  \hflavTAUBSLCON    \\
$1/\Gamma_{s\rm H}$ & \hflavTAUBSH    &  \hflavTAUBSHCO    &  \hflavTAUBSHCON    \\
\DGs               & \hflavDGS       &  \hflavDGSCO       &  \hflavDGSCON       \\
\DGs/\Gs           & \hflavDGSGS     &  \hflavDGSGSCO     &  \hflavDGSGSCON     \\
$\rho(\Gs,\DGs)$   & \hflavRHOGSDGS  &  \hflavRHOGSDGSCO  &  \hflavRHOGSDGSCON  \\
\hline
\end{tabular}
\end{center}
\end{table}

An alternative approach, which is directly sensitive to first order in 
$\DGs/\Gs$, 
is to determine the effective lifetime of untagged \Bs\ candidates
decaying to %
pure \CP eigenstates; we use here measurements with
$\Bs \to D_s^+D_s^-$~\cite{Aaij:2013bvd}, 
$\Bs \to J/\psi \eta$~\cite{Aaij:2016dzn}, 
$\Bs \to \jpsi f_0(980)$~\cite{Aaltonen:2011nk,Abazov:2016oqi}
and $\Bs\to \jpsi \pi^+\pi^-$~\citehistory{Aaij:2013oba}{Aaij:2013oba,*LHCb:2011aa_hist,*LHCb:2012ad_hist,*LHCb:2011ab_hist,*Aaij:2012nta_hist} decays.
The precise extraction of $1/\Gs$ and $\DGs$
from such measurements, discussed in detail in \Ref{Hartkorn:1999ga,Dunietz:2000cr,Fleischer:2011cw}, 
requires additional information 
in the form of theoretical assumptions or
external inputs on weak phases and hadronic parameters. 
If $f$ denotes a final state into which both \Bs and \Bsbar can decay,
the ratio of the effective $\Bs \to f$
lifetime $\tau_{\rm single}$, found by
fitting the decay-time distribution to a single exponential, relative to the mean
\Bs lifetime is~\cite{Fleischer:2011cw}%
\footnote{%
\label{foot:life_mix:ADG-def}
The definition of $A_f^{\DG}$ given in \Eq{ADG} has the sign opposite to that given in \Ref{Fleischer:2011cw}.}
\begin{equation}
  \frac{\tau_{\rm single}(\Bs \to f)}{\tau(\Bs)} = \frac{1}{1-y_s^2} \left[ \frac{1 - 2A_f^{\DG} y_s + y_s^2}{1 - A_f^{\DG} y_s}\right ] \,,
\labe{tauf_fleisch}
\end{equation}
where
\begin{equation}
A_f^{\DG} = -\frac{2 \Re(\lambda_f)} {1+|\lambda_f|^2} \,.
\labe{ADG}
\end{equation}
To include the measurements of the effective
$\Bs \to D_s^+D_s^-$ (\CP-even), $\Bs \to \jpsi f_0(980)$ (\CP-odd) and
$\Bs \to \jpsi\pi^+\pi^-$ (\CP-odd) 
lifetimes as constraints in the \DGs fit,\footnote{%
The effective lifetimes measured in $\Bs\to K^+ K^-$ (mostly \CP-even) and  $\Bs \to \jpsi K_{\rm S}^0$ (mostly \CP-odd) are not used because we can not quantify the penguin contributions in those modes.}
we neglect sub-leading penguin contributions and possible direct \CP violation. 
Explicitly, in \Eq{ADG}, we set
$A_{\mbox{\scriptsize \CP-even}}^{\DG} = \cos \phiccbars$
and $A_{\mbox{\scriptsize \CP-odd}}^{\DG} = -\cos \phiccbars$.
Given the small value of $\phiccbars$, we have, to first order in $y_s$:
\begin{eqnarray}
\tau_{\rm single}(\Bs \to \mbox{\CP-even})
& \approx & \frac{1}{\Gamma_{s\rm L}} \left(1 + \frac{(\phiccbars)^2 y_s}{2} \right) \,,
\labe{tau_KK_approx}
\\
\tau_{\rm single}(\Bs \to \mbox{\CP-odd})
& \approx & \frac{1}{\Gamma_{s\rm H}} \left(1 - \frac{(\phiccbars)^2 y_s}{2} \right) \,.
\labe{tau_Jpsif0_approx}
\end{eqnarray}
The numerical inputs are taken from \Eqss{tau_KK}{tau_Jpsif0},
and the resulting averages, combined with the $\Bs\to\jpsi K^+K^-$ information,
are indicated in the second column of numbers of \Table{tabtauLH}. 
These averages assume $\phiccbars = 0$, which is compatible with
the \phiccbars average presented in \Sec{phasebs}.

Information on \DGs can also be obtained from the study of the
proper time distribution of untagged samples
of flavour-specific \Bs decays~\cite{Hartkorn:1999ga}, where
the flavour (\ie, \Bs or \Bsbar) at the time of decay can be determined by
the decay products. In such decays,
\eg\ semileptonic \Bs decays, there is
an equal mix of the heavy and light mass eigenstates at time zero.
The proper time distribution is then a superposition 
of two exponential functions with decay constants
$\Gamma_{s\rm L}$ and $\Gamma_{s\rm H}$. %
This provides sensitivity to both $1/\Gs$ and 
$(\DGs/\Gs)^2$. Ignoring \DGs and fitting for 
a single exponential leads to an estimate of \Gs with a 
relative bias proportional to $(\DGs/\Gs)^2$, as shown in \Eq{fslife}. 
Including the constraint from the world-average flavour-specific \Bs 
lifetime, given in \Eq{tau_fs}, leads to the results shown in the last column 
of \Table{tabtauLH}.
These world averages are displayed as the dark-filled contours labelled ``Combined'' in the
plots of \Fig{DGs}. 
They correspond to the lifetime averages
$1/\Gs=\hflavTAUBSMEANCON$,
$1/\Gamma_{s\rm L}=\hflavTAUBSLCON$,
$1/\Gamma_{s\rm H}=\hflavTAUBSHCON$,
and to the decay-width difference
\begin{equation}
\DGs = \hflavDGSCON ~~~~\mbox{and} ~~~~~ \DGs/\Gs = \hflavDGSGSCON \,.
\labe{DGs_DGsGs}
\end{equation}
The good agreement with the Standard Model prediction 
$\DGs = 0.088 \pm 0.020~\hbox{ps}^{-1}$~\citehistory{Lenz:2011ti,*Lenz:2006hd,Jubb:2016mvq,Artuso:2015swg}{Lenz:2011ti,*Lenz:2006hd,Jubb:2016mvq,Artuso:2015swg,*Lenz_hist}
excludes significant quark-hadron duality violation in the HQE~\cite{Lenz:2012mb}. 
Estimates of $\DGs/\Gs$ obtained from measurements of the 
$\Bs \to D_s^{(*)+} D_s^{(*)-}$ branching fraction~\citehistory{Barate:2000kd,Esen:2010jq,Abazov:2008ig,Abulencia:2007zz}{Barate:2000kd,Esen:2010jq,Abazov:2008ig,*Abazov:2007rb_hist,Abulencia:2007zz}
are not used in the average,
since they are based on the questionable~\cite{Lenz:2011ti,*Lenz:2006hd}
assumption that these decays account for all \CP-even final states.
The results of early lifetime analyses that attempted
to measure $\DGs/\Gs$~\citehistory{Acciarri:1998uv,Abreu:2000sh,Abreu:2000ev,Abe:1997bd}{Acciarri:1998uv,Abreu:2000sh,Abreu:2000ev,*Abreu:1996ep_hist,Abe:1997bd}
are not used either.

The strength of \Bs mixing has been known to be large for more than 20 years. 
Indeed the time-integrated measurements of \chibar (see \Sec{chibar}),
when compared to our knowledge
of \chid and the \b-hadron fractions, indicated that 
\chis should be close to its maximal possible value of $1/2$.
Many searches of the time dependence of this mixing 
have been performed by ALEPH~\cite{Heister:2002gk}, %
DELPHI~\citehistory{Abreu:2000sh,Abreu:2000ev,Abdallah:2002mr,Abdallah:2003qga}{Abreu:2000sh,Abreu:2000ev,*Abreu:1996ep_hist,Abdallah:2002mr,Abdallah:2003qga}, %
OPAL~\cite{Abbiendi:1999gm,Abbiendi:2000bh},
SLD~\unpublished{\cite{Abe:2002ua,Abe:2002wfa}}{\cite{Abe:2002ua,Abe:2002wfa,Abe:2000gp}},
CDF (Run~I)~\cite{Abe:1998qj} and
\dzero~\cite{Abazov:2006dm}
but did not have enough statistical power
and proper time resolution to resolve 
the small period of the \Bs\ oscillations.

\Bs oscillations were observed for the first time in 2006
by the CDF collaboration~\citehistory{Abulencia:2006ze}{Abulencia:2006ze,*Abulencia:2006mq_hist},
based on samples of flavour-tagged hadronic and semileptonic \Bs decays
(in flavour-specific final states), partially or fully reconstructed in 
$1\invfb$ of data collected during Tevatron's Run~II. 
\unpublished{}{
This was shortly followed by independent unpublished evidence obtained by the \dzero collaboration
with $2.4\invfb$ of
data~\cite{D0note5618:2008,*D0note5474:2007,*D0note5254:2006}.}
More recently, the LHCb collaboration obtained the most precise results using fully reconstructed 
$\Bs \to D_s^-\pi^+$ and $\Bs \to D_s^-\pi^+\pi^-\pi^+$ decays~\cite{Aaij:2011qx,Aaij:2013mpa}.
LHCb has also observed \Bs oscillations with 
$\Bs\to\jpsi K^+K^-$ decays~\citehistory{Aaij:2014zsa}{Aaij:2014zsa,*Aaij:2013oba_supersede2}
and with semileptonic $\Bs \to D_s^-\mu^+ X$ decays~\cite{Aaij:2013gja}.
The measurements of \dms are summarized in \Table{dms}. 

\begin{table}[t]
\caption{Measurements of \dms.}
\labt{dms}
\begin{center}
\begin{tabular}{ll@{}crl@{\,}l@{\,}ll} \hline
Experiment & Method           & \multicolumn{2}{c}{Data set} & \multicolumn{3}{c}{\dms (\!\!\invps)} & Ref. \\
\hline
CDF2   & \multicolumn{2}{l}{\particle{D_s^{(*)-} \ell^+ \nu}, \particle{D_s^{(*)-} \pi^+}, \particle{D_s^{-} \rho^+}}
       & 1 \invfb & $17.77$ & $\pm 0.10$ & $\pm 0.07~$
       & \citehistory{Abulencia:2006ze}{Abulencia:2006ze,*Abulencia:2006mq_hist} \\
\unpublished{}{%
\dzero & \particle{D_s^- \ell^+ X}, \particle{D_s^- \pi^+ X}
       &  & 2.4 \invfb & $18.53$ & $\pm 0.93$ & $\pm 0.30~$ 
       & \cite{D0note5618:2008,*D0note5474:2007,*D0note5254:2006}$^u$ \\
}%
LHCb   & \particle{D_s^- \pi^+}, \particle{D_s^- \pi^+\pi^-\pi^+}
       & 2010 & 0.034 \invfb & $17.63$ & $\pm 0.11$ & $\pm 0.02~$   
       & \cite{Aaij:2011qx} \\
LHCb   & \particle{D_s^- \mu^+ X}
       & 2011 & 1.0 \invfb & $17.93$ & $\pm 0.22$ & $\pm 0.15$ 
       & \cite{Aaij:2013gja}  \\
LHCb   & \particle{D_s^- \pi^+}
       & 2011 & 1.0 \invfb & $17.768$ & $\pm 0.023$ & $\pm 0.006$ 
       & \cite{Aaij:2013mpa}  \\
LHCb   & \particle{\jpsi K^+K^-}
       & 2011--2012 & 3.0 \invfb & $17.711$ & $^{+0.055}_{-0.057}$ & $\pm 0.011$ 
       & \citehistory{Aaij:2014zsa}{Aaij:2014zsa,*Aaij:2013oba_supersede2}  \\
\hline
\multicolumn{4}{l}{Average \unpublished{}{of CDF and LHCb measurements}} & $\hflavDMSval$ & $\hflavDMSsta$ & $\hflavDMSsys$ & \\  
\hline
\unpublished{}{$^u$ \footnotesize Unpublished. }
\end{tabular}
\end{center}
\end{table}

\begin{figure}
\begin{center}
\includegraphics[width=0.8\textwidth]{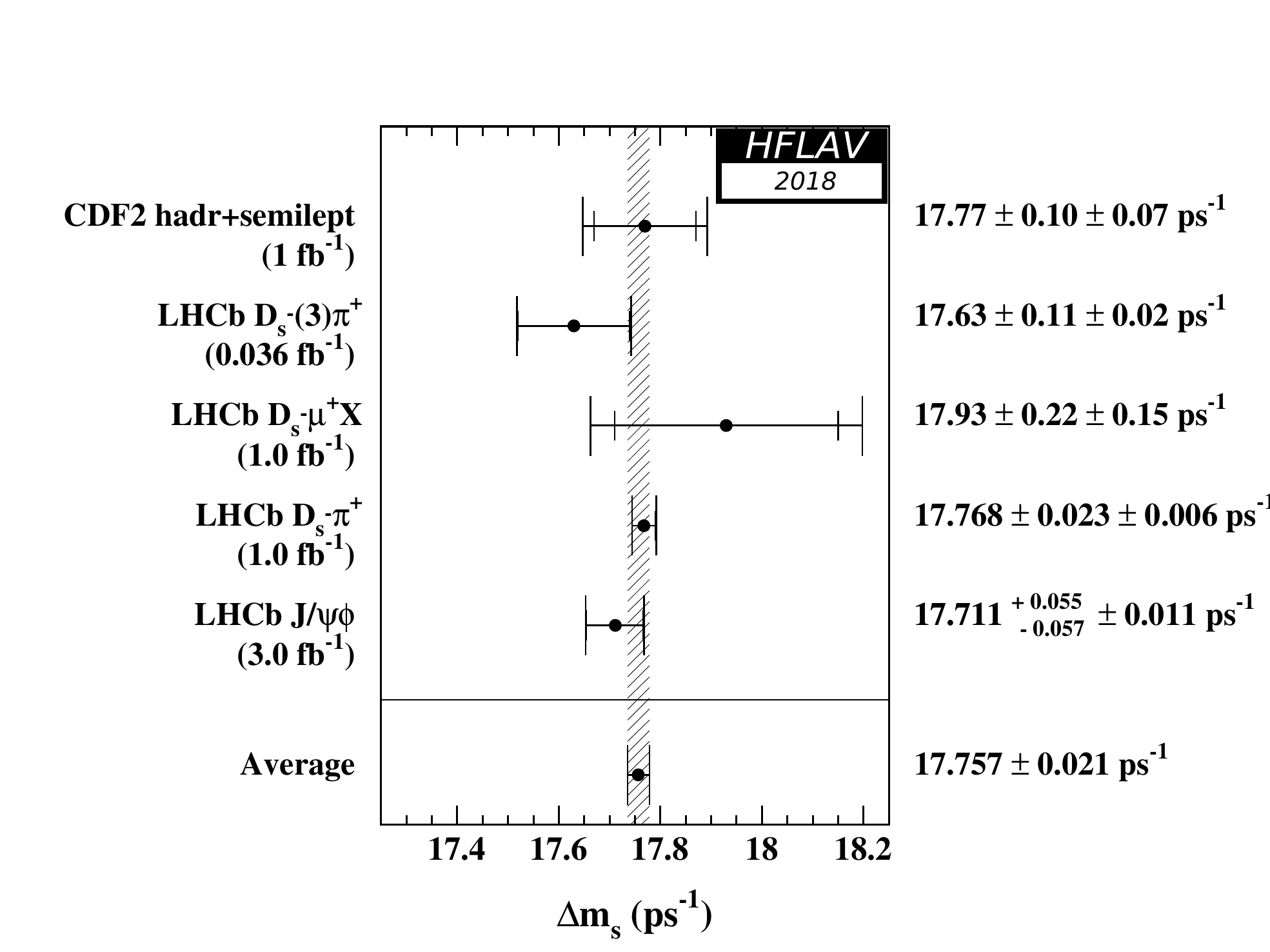}
\caption{Published %
measurements of \dms, together with their average.} 
\labf{dms}
\end{center}
\end{figure}

A simple average of the CDF and LHCb results\unpublished{}{\footnote{
  \label{foot:life_mix:D0note5618:2008}
  We do not include the unpublished
  \dzero~\cite{D0note5618:2008,*D0note5474:2007,*D0note5254:2006} result in the average.}},
taking into account the correlated systematic uncertainties between the three 
LHCb measurements, yields 
\begin{equation}
\dms = \hflavDMSfull = \hflavDMS \labe{dms}
\end{equation}
and is illustrated in \Figure{dms}.
The Standard Model prediction 
$\dms = 18.3 \pm 2.7~\hbox{ps}^{-1}$~\citehistory{Jubb:2016mvq,Artuso:2015swg}{Jubb:2016mvq,Artuso:2015swg,*Lenz_hist} is consistent with the experimental value, but has a much larger uncertainty dominated by the uncertainty on the hadronic matrix elements.
The ratio $\DGs/\dms$ can be predicted more accurately to be $0.0048 \pm 0.0008$~\citehistory{Lenz:2011ti,*Lenz:2006hd,Jubb:2016mvq,Artuso:2015swg}{Lenz:2011ti,*Lenz:2006hd,Jubb:2016mvq,Artuso:2015swg,*Lenz_hist},
in good agreement with the experimental determination of 
\begin{equation}
\DGs/\dms= \hflavRATIODGSDMS \,.
\end{equation}

Multiplying the \dms result of \Eq{dms} by the 
mean \Bs lifetime of \Eq{oneoverGs}, $1/\Gs=\hflavTAUBSMEANCON$,
yields
\begin{equation}
\xs %
= \hflavXS \,. \labe{xs}
\end{equation}
With $2\ys %
=\hflavDGSGSCON$ 
(see \Eq{DGs_DGsGs})
and under the assumption of no \CP violation in \Bs mixing,
this corresponds to
\begin{equation}
\chis = \frac{\xs^2+\ys^2}{2(\xs^2+1)} = \hflavCHIS \,. \labe{chis}
\end{equation}
The ratio 
\begin{equation}
\frac{\dmd}{\dms} = \hflavRATIODMDDMS \,, \labe{dmd_over_dms}
\end{equation}
of the \Bd and \Bs oscillation frequencies, 
obtained from \Eqss{dmd}{dms}, 
can be used to extract the following magnitude of the ratio of CKM matrix elements, 
\begin{equation}
\left|\frac{V_{td}}{V_{ts}}\right| =
\xi \sqrt{\frac{\dmd}{\dms}\frac{m(\Bs)}{m(\Bd)}} = 
\hflavVTDVTSfull \,, \labe{Vtd_over_Vts}
\end{equation}
where the first uncertainty is from experimental uncertainties 
(with the masses $m(\Bs)$ and $m(\Bd)$ taken from \Ref{PDG_2018}),
and the second uncertainty arises from theoretical uncertainties 
in the estimation of the SU(3) flavour-symmetry breaking factor
$\xi %
= \hflavXI$~\cite{Aoki:2019cca},
an average of three-flavour lattice QCD calculations dominated by the results of Ref.~\cite{Bazavov:2016nty}.
Note that \Eq{Vtd_over_Vts} assumes that \dms and \dmd only receive 
Standard Model contributions.
\mysubsubsection{\CP violation in \Bd and \Bs mixing}
\labs{qpd} \labs{qps}

Evidence for \CP violation in \Bd mixing
has been searched for,
both with flavour-specific and inclusive \Bd decays, 
in samples where the initial 
flavour state is tagged. In the case of semileptonic 
(or other flavour-specific) decays, 
where the final state tag is 
also available, the asymmetry
\begin{equation} 
\ASLd = \frac{
N(\hbox{\Bdbar}(t) \to \ell^+      \nu_{\ell} X) -
N(\hbox{\Bd}(t)    \to \ell^- \bar{\nu}_{\ell} X) }{
N(\hbox{\Bdbar}(t) \to \ell^+      \nu_{\ell} X) +
N(\hbox{\Bd}(t)    \to \ell^- \bar{\nu}_{\ell} X) } 
\labe{ASLd}
\end{equation} 
has been measured, either in decay-time-integrated analyses at 
CLEO~\citehistory{Behrens:2000qu,Jaffe:2001hz}{Behrens:2000qu,Jaffe:2001hz,*Jaffe:2001hz_hist},
\babar~\citehistory{Lees:2014kep}{Lees:2014kep,*Lees:2014kep_hist},
CDF~\unpublished{\cite{Abe:1996zt}}{\cite{Abe:1996zt,CDFnote9015:2007}}
and \dzero~\citehistory{Abazov:2013uma}{Abazov:2013uma,*Abazov:2011yk_hist,*Abazov:2010hv_hist,*Abazov:2010hj_hist,*Abazov:dimuon_hist},
or in decay-time-dependent analyses at 
OPAL~\cite{Ackerstaff:1997vd}, ALEPH~\cite{Barate:2000uk}, 
\babar~\unpublished{%
\citehistory%
{Aubert:2003hd,*Aubert:2004xga,Lees:2013sua,Aubert:2006nf}%
{Aubert:2003hd,*Aubert:2004xga,Lees:2013sua,Aubert:2006nf,*Aubert:2002mn_hist}}{%
\citehistory%
{Aubert:2003hd,*Aubert:2004xga,Lees:2013sua,Aubert:2006nf}%
{Aubert:2003hd,*Aubert:2004xga,Lees:2013sua,*Margoni:2013qx_hist,*Aubert:2006sa_hist,Aubert:2006nf,*Aubert:2002mn_hist}}
and \belle~\cite{Nakano:2005jb}.
Note that the asymmetry of time-dependent decay rates in \Eq{ASLd} is 
related to  $|q_d/p_d|$ through \Eq{ASLq} and is therefore time-independent.
In the inclusive case, also investigated and published
by ALEPH~\cite{Barate:2000uk} and OPAL~\cite{Abbiendi:1998av},
no final state tag is used, and the asymmetry~\cite{Beneke:1996hv,*Dunietz:1998av}
\begin{equation} 
\frac{
N(\hbox{\Bd}(t) \to {\rm all}) -
N(\hbox{\Bdbar}(t) \to {\rm all}) }{
N(\hbox{\Bd}(t) \to {\rm all}) +
N(\hbox{\Bdbar}(t) \to {\rm all}) } 
\simeq
\ASLd \left[ \frac{\dmd}{2\Gd} \sin(\dmd \,t) - 
\sin^2\left(\frac{\dmd \,t}{2}\right)\right] 
\labe{ASLincl}
\end{equation} 
must be measured as a function of the proper time to extract information 
on \CP violation.

On the other hand, \dzero~\cite{Abazov:2012hha} and
LHCb~\cite{Aaij:2014nxa} have studied the time-dependence of the 
charge asymmetry of $B^0 \to D^{(*)-}\mu^+\nu_{\mu}X$ decays
without tagging the initial state,
which would be equal to 
\begin{equation} 
\frac{N(D^{(*)-}\mu^+\nu_{\mu}X)-N(D^{(*)+}\mu^-\bar{\nu}_{\mu}X)}%
{N(D^{(*)-}\mu^+\nu_{\mu}X)+N(D^{(*)+}\mu^-\bar{\nu}_{\mu}X)} =
\ASLd \frac{1- \cos(\dmd \,t)}{2}
\label{eq:untagged_ASL}
\end{equation}
in absence of detection and production asymmetries.

\begin{table}[t]
\caption{Measurements\footref{foot:life_mix:Abe:1996zt}
of \CP violation in \Bd mixing and their average
in terms of both \ASLd and $|q_{\particle{d}}/p_{\particle{d}}|$.
The individual results are listed as quoted in the original publications, 
or converted\footref{foot:life_mix:epsilon_B}
to an \ASLd value.
When two errors are quoted, the first one is statistical and the 
second one systematic. The ALEPH and OPAL %
results assume no \CP violation in \Bs mixing.}
\labt{qoverp}
\begin{center}
\resizebox{\textwidth}{!}{
\begin{tabular}{@{}rcl@{$\,\pm$}l@{$\pm$}ll@{$\,\pm$}l@{$\pm$}l@{}}
\hline
Exp.\ \& Ref. & Method & \multicolumn{3}{c}{Measured \ASLd} 
                       & \multicolumn{3}{c}{Measured $|q_{\particle{d}}/p_{\particle{d}}|$} \\
\hline
CLEO   \cite{Behrens:2000qu} & Partial hadronic rec. 
                             & $+0.017$ & 0.070 & 0.014 
                             & \multicolumn{3}{c}{} \\
CLEO   \citehistory{Jaffe:2001hz}{Jaffe:2001hz,*Jaffe:2001hz_hist}   & Dileptons 
                             & $+0.013$ & 0.050 & 0.005 
                             & \multicolumn{3}{c}{} \\
CLEO   \citehistory{Jaffe:2001hz}{Jaffe:2001hz,*Jaffe:2001hz_hist}   & Average of above two 
                             & $+0.014$ & 0.041 & 0.006 
                             & \multicolumn{3}{c}{} \\
\babar \cite{Aubert:2003hd,*Aubert:2004xga}   & Full hadronic rec. 
                             & \multicolumn{3}{c}{}  
                             & 1.029 & 0.013 & 0.011  \\
\babar \unpublished{\cite{Lees:2013sua}}{\citehistory{Lees:2013sua}{Lees:2013sua,*Margoni:2013qx_hist,*Aubert:2006sa_hist}} & Part.\ rec.\ $D^{*}X\ell\nu$ 
                             & $+0.0006$ & \multicolumn{2}{@{}l}{$0.0017 ^{+0.0038}_{-0.0032}$} 
                             & $0.99971$ & $0.00084$ & $0.00175$ \\ 
\babar \citehistory{Lees:2014kep}{Lees:2014kep,*Lees:2014kep_hist}  & Dileptons
                             & $-0.0039$ & 0.0035 & 0.0019 
                             & \multicolumn{3}{c}{} \\
\belle \cite{Nakano:2005jb}  & Dileptons 
                             & $-0.0011$ & 0.0079 & 0.0085 
                             & 1.0005 & 0.0040 & 0.0043 \\
\multicolumn{2}{l}{Average of above 6 \B-factory results} & \multicolumn{3}{l}{\hflavASLDB\ (tot)} 
                             & \multicolumn{3}{l}{\hflavQPDB\  (tot)} \\ 
\hline
\dzero \cite{Abazov:2012hha} & $B^0 \to D^{(*)-}\mu^+\nu X$
                            & $+0.0068$ & 0.0045 & 0.0014 & \multicolumn{3}{c}{} \\
LHCb \cite{Aaij:2014nxa} & $B^0 \to D^{(*)-}\mu^+\nu X$
                            & $-0.0002$ & 0.0019 & 0.0030 & \multicolumn{3}{c}{} \\
\multicolumn{2}{l}{Average of above 8 pure $B^0$ results} & \multicolumn{3}{l}{\hflavASLDD\ (tot)}
                             & \multicolumn{3}{l}{\hflavQPDD\  (tot)} \\
\hline
\dzero  \citehistory{Abazov:2013uma}{Abazov:2013uma,*Abazov:2011yk_hist,*Abazov:2010hv_hist,*Abazov:2010hj_hist,*Abazov:dimuon_hist}  & Muons  \& dimuons
                             & $-0.0062$ & \multicolumn{2}{@{\hspace{0.26em}}l}{0.0043 (tot)}
                             & \multicolumn{3}{c}{} \\
\multicolumn{2}{l}{Average of above 9 direct measurements} & \multicolumn{3}{l}{\hflavASLDW\ (tot)} 
                             & \multicolumn{3}{l}{\hflavQPDW\  (tot)} \\ 
\hline
OPAL   \cite{Ackerstaff:1997vd}   & Leptons     
                             & $+0.008$ & 0.028 & 0.012 
                             & \multicolumn{3}{c}{} \\
OPAL   \cite{Abbiendi:1998av}   & Inclusive (\Eq{ASLincl}) 
                             & $+0.005$ & 0.055 & 0.013 
                             & \multicolumn{3}{c}{} \\
ALEPH  \cite{Barate:2000uk}       & Leptons 
                             & $-0.037$ & 0.032 & 0.007 
                             & \multicolumn{3}{c}{} \\
ALEPH  \cite{Barate:2000uk}       & Inclusive (\Eq{ASLincl}) 
                             & $+0.016$ & 0.034 & 0.009 
                             & \multicolumn{3}{c}{} \\
ALEPH  \cite{Barate:2000uk}       & Average of above two 
                             & $-0.013$ & \multicolumn{2}{@{\hspace{0.26em}}l}{0.026 (tot)} 
                             & \multicolumn{3}{c}{} \\
\multicolumn{2}{l}{Average of above 13 results} & \multicolumn{3}{l}{\hflavASLDA\ (tot)} 
                             & \multicolumn{3}{l}{\hflavQPDA\  (tot)} \\ 
\hline
\multicolumn{5}{l}{Best fit value from 2D combination of} \\
\multicolumn{2}{l}{\ASLd and \ASLs results (see \Eq{ASLD})} & \multicolumn{3}{l}{\hflavASLD\ (tot)} 
                             & \multicolumn{3}{l}{\hflavQPD\  (tot)} \\ 
\hline
\end{tabular}
}
\end{center}
\end{table}

\Table{qoverp} summarizes the different measurements%
\footnote{
\label{foot:life_mix:Abe:1996zt}
A low-statistics result published by CDF using the Run~I data~\cite{Abe:1996zt}
\unpublished{is}{and an unpublished result by CDF using Run~II data~\cite{CDFnote9015:2007} are}
not included in our averages, nor in \Table{qoverp}.
}
of \ASLd and $|q_{\particle{d}}/p_{\particle{d}}|$. 
In all cases asymmetries compatible with zero have been found,  
with precision limited by the available statistics.

A simple average of all measurements performed at the
\B factories~\unpublished%
{\citehistory%
{Behrens:2000qu,Jaffe:2001hz,Aubert:2003hd,*Aubert:2004xga,Lees:2013sua,Lees:2014kep,Nakano:2005jb}%
{Behrens:2000qu,Jaffe:2001hz,*Jaffe:2001hz_hist,Aubert:2003hd,*Aubert:2004xga,Lees:2013sua,Lees:2014kep,*Lees:2014kep_hist,Nakano:2005jb}}%
{\citehistory%
{Behrens:2000qu,Jaffe:2001hz,Aubert:2003hd,*Aubert:2004xga,Lees:2013sua,Lees:2014kep,Nakano:2005jb}%
{Behrens:2000qu,Jaffe:2001hz,*Jaffe:2001hz_hist,Aubert:2003hd,*Aubert:2004xga,Lees:2013sua,*Margoni:2013qx_hist,*Aubert:2006sa_hist,Lees:2014kep,*Lees:2014kep_hist,Nakano:2005jb}}
yields
$\ASLd = \hflavASLDB$.
Adding also the \dzero~\cite{Abazov:2012hha}
and LHCb~\cite{Aaij:2014nxa} measurements obtained with reconstructed 
semileptonic \Bd decays yields $\ASLd = \hflavASLDD$.
As discussed in more detail later in this section, 
the \dzero analysis with single muons and like-sign dimuons~\citehistory{Abazov:2013uma}{Abazov:2013uma,*Abazov:2011yk_hist,*Abazov:2010hv_hist,*Abazov:2010hj_hist,*Abazov:dimuon_hist}
separates the \Bd and \Bs contributions by exploiting the dependence on the muon impact parameter cut; including the 
\ASLd result quoted by \dzero in the average yields
$\ASLd = \hflavASLDW$. %
All the other \Bd analyses performed at high energy, either at LEP or at the Tevatron,
did not separate the contributions from the \Bd and \Bs mesons.
Under the assumption of no \CP violation in \Bs mixing ($\ASLs =0$),
a number of these early analyses~\cite{Abazov:2006qw,Ackerstaff:1997vd,Barate:2000uk,Abbiendi:1998av}
quote a measurement of $\ASLd$ or $|q_{\particle{d}}/p_{\particle{d}}|$ for the \Bd meson.
However, these imprecise determinations no longer improve the world average of 
\ASLd.
Furthermore, the assumption makes sense within the Standard Model, 
since \ASLs is predicted to be much smaller than \ASLd~\citehistory{Jubb:2016mvq,Artuso:2015swg}{Jubb:2016mvq,Artuso:2015swg,*Lenz_hist},
but may not be suitable in the presence of new physics. 

The Tevatron experiments
have measured linear combinations of 
\ASLd and \ASLs
using inclusive semileptonic decays of \b hadrons. CDF (Run~I) finds 
$\ASLb = +0.0015 \pm 0.0038 \mbox{(stat)} \pm 0.0020 \mbox{(syst)}$~\cite{Abe:1996zt},
and D0 obtains $\ASLb = -0.00496 \pm 0.00153 \mbox{(stat)} \pm 0.00072 \mbox{(syst)}$~\citehistory{Abazov:2013uma}{Abazov:2013uma,*Abazov:2011yk_hist,*Abazov:2010hv_hist,*Abazov:2010hj_hist,*Abazov:dimuon_hist}.
While the imprecise CDF result is compatible with no \CP violation%
\unpublished{}{\footnote{
  \label{foot:life_mix:CDFnote9015:2007}
  An unpublished measurement from CDF2, 
  $\ASLb = +0.0080 \pm 0.0090 \mbox{(stat)} \pm 0.0068 \mbox{(syst)}$~\cite{CDFnote9015:2007},
  more precise than the \dzero measurement,
  is also compatible with no \CP violation.
}},
the \dzero result, obtained by measuring
the single muon and like-sign dimuon charge asymmetries,
differs by 2.8 standard
deviations from the Standard Model expectation of
${\cal A}_{\rm SL}^{b,\rm SM} = (-2.3\pm 0.4) \times 10^{-4}$%
~\citehistory{Abazov:2013uma,Lenz:2011ti,*Lenz:2006hd}{Abazov:2013uma,*Abazov:2011yk_hist,*Abazov:2010hv_hist,*Abazov:2010hj_hist,*Abazov:dimuon_hist,Lenz:2011ti,*Lenz:2006hd}
With a more sophisticated analysis in bins of the
muon impact parameters, \dzero conclude that the overall deviation of 
their measurements from the SM is at the level of $3.6\,\sigma$.
Interpreting the observed asymmetries
in bins of the muon impact parameters
in terms of \CP violation in $B$-meson mixing and in interference, 
and using 
the mixing parameters and the world \b-hadron fractions 
of \Ref{Amhis:2012bh}, the \dzero collaboration
extracts~\citehistory{Abazov:2013uma}{Abazov:2013uma,*Abazov:2011yk_hist,*Abazov:2010hv_hist,*Abazov:2010hj_hist,*Abazov:dimuon_hist}
values for \ASLd and \ASLs and their correlation
coefficient\footnote{
\label{foot:life_mix:Abazov:2013uma}
In each impact parameter bin $i$ the measured same-sign dimuon asymmetry is interpreted as  
$A_i = K^s_i \ASLs + K^d_i \ASLd + \lambda K^{\rm int}_i \DGd/\Gd$, where the factors  $K^s_i$, $K^d_i$ and $K^{\rm int_i}$ are obtained by \dzero from Monte Carlo simulation. The \dzero publication~\citehistory{Abazov:2013uma}{Abazov:2013uma,*Abazov:2011yk_hist,*Abazov:2010hv_hist,*Abazov:2010hj_hist,*Abazov:dimuon_hist} assumes $\lambda=1$, but it has been demonstrated 
subsequently that $\lambda \le 0.49$~\cite{Nierste_CKM2014}. This particular point invalidates the $\DGd/\Gd$ result published by \dzero, but not the \ASLd and \ASLs results. As stated by \dzero, their \ASLd and \ASLs results assume the above expression for $A_i$, \ie\ that the observed asymmetries are due to \CP violation in $B$ mixing. As long as this assumption is not shown to be wrong (or withdrawn by \dzero), we include the \ASLd and \ASLs results in our world average.},
as shown in \Table{ASLs_ASLd}.
However, the various 
contributions to the total quoted uncertainties from this analysis and from the
external inputs are not given, so the adjustment of these results to different
or more recent values of the external inputs cannot (easily) be done. 

Finally, direct determinations of \ASLs,
also shown in \Table{ASLs_ASLd},
have been obtained by \dzero~\citehistory{Abazov:2012zz}{Abazov:2012zz,*Abazov:2009wg_hist,*Abazov:2007nw_hist}
and LHCb~\citehistory{Aaij:2016yze}{Aaij:2016yze,*Aaij:2013gta_hist}
from the time-integrated charge asymmetry of
untagged $\Bs \to D_s^- \mu^+\nu X$ decays.

Using a two-dimensional fit, all measurements of \ASLs and \ASLd obtained by 
\dzero and LHCb are combined with the 
\B-factory average of \Table{qoverp}. Correlations are taken into account as 
shown in \Table{ASLs_ASLd}.
The results, displayed graphically in \Fig{ASLs_ASLd}, are

\begin{table}[p]
\caption{Measurements of \CP violation in \Bs and \Bd mixing, together 
with their correlations $\rho(\ASLs,\ASLd)$
and their two-dimensional average. Only total errors are quoted.}
\labt{ASLs_ASLd}
\begin{center}
\begin{tabular}{rcccl}
\hline
Exp.\ \& Ref.\ & Method & Measured \ASLs & Measured \ASLd & $\rho(\ASLs,\ASLd)$ \\
\hline
\multicolumn{2}{l}{\B-factory average  of \Table{qoverp}}
       & & \hflavASLDB & \\ 
\dzero  \citehistory{Abazov:2012zz,Abazov:2012hha}{Abazov:2012zz,*Abazov:2009wg_hist,*Abazov:2007nw_hist,Abazov:2012hha} & $B_{(s)}^0 \to D_{(s)}^{(*)-} \mu^+\nu X$
       & $-0.0112 \pm 0.0076$ %
       & $+0.0068 \pm 0.0047$ %
       & ~~~$+0.$ \\
LHCb \citehistory{Aaij:2016yze,Aaij:2014nxa}{Aaij:2016yze,*Aaij:2013gta_hist,Aaij:2014nxa} & $B_{(s)}^0 \to D_{(s)}^{(*)-} \mu^+\nu X$
       & $+0.0039 \pm 0.0033$ %
       & $-0.0002 \pm 0.0036$ %
       & ~~~$+0.13$ \\
\hline
\multicolumn{2}{l}{Average of above}
       & \hflavASLSNOMU & \hflavASLDNOMU & ~~~\hflavRHOASLSASLDNOMU \\ 
\dzero  \citehistory{Abazov:2013uma}{Abazov:2013uma,*Abazov:2011yk_hist,*Abazov:2010hv_hist,*Abazov:2010hj_hist,*Abazov:dimuon_hist}  & muons \& dimuons
       & $-0.0082 \pm 0.0099$ %
       & $-0.0062 \pm 0.0043$ %
       & ~~~$-0.61$ \\          %
\hline
\multicolumn{2}{l}{Average of all above}
       & \hflavASLS & \hflavASLD & ~~~\hflavRHOASLSASLD \\ 
\hline
\end{tabular}
\end{center}
\end{table}

\begin{figure}[p]
\begin{center}
\includegraphics[width=0.6\textwidth]{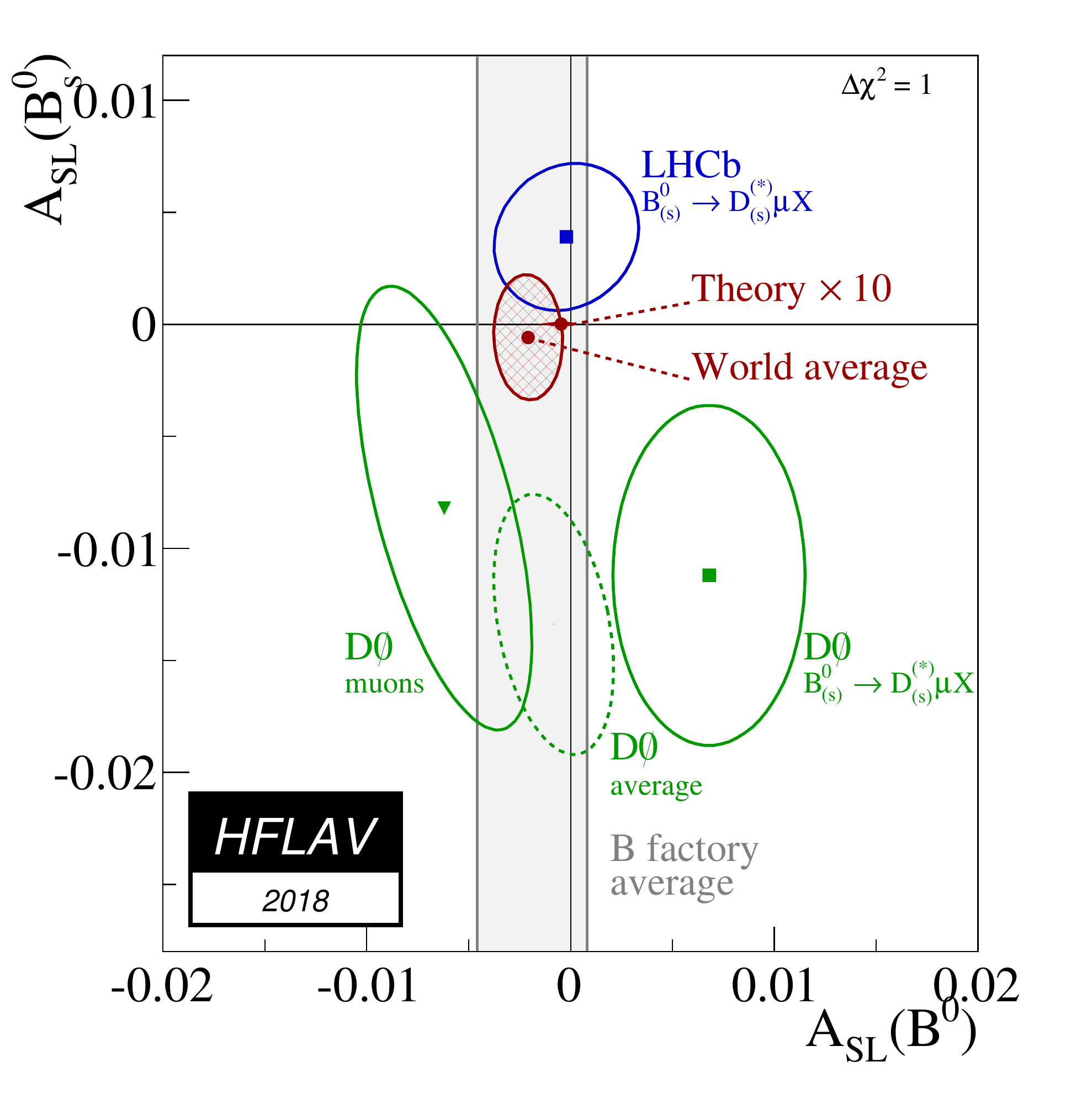}
\end{center}
\vspace{-5mm}
\caption{
Measurements of \ASLs and \ASLd listed in \Table{ASLs_ASLd}
(\B-factory average as the grey band, \dzero measurements as the green ellipses, LHCb measurements as the blue ellipse) 
together with their two-dimensional average (red hatched ellipse).
The red point close to $(0,0)$ is the Standard Model prediction
of Ref.~\protect\citehistory{Jubb:2016mvq,Artuso:2015swg}{Jubb:2016mvq,Artuso:2015swg,*Lenz_hist} with error bars multiplied by 10.
The prediction and the experimental world average deviate from each other by $\hflavASLDASLSNSIGMA\,\sigma$.}
\labf{ASLs_ASLd}
\end{figure}
\begin{eqnarray}
\ASLd & = & \hflavASLD ~~~ \Longleftrightarrow ~~~ |q_{\particle{d}}/p_{\particle{d}}| = \hflavQPD \,,
\labe{ASLD}
\\
\ASLs & = & \hflavASLS ~~~ \Longleftrightarrow ~~~ |q_{\particle{s}}/p_{\particle{s}}| = \hflavQPS \,,
\labe{ASLS}
\\
\rho(\ASLd , \ASLs) & = & \hflavRHOASLSASLD \,,
\labe{rhoASLDASLS}
\end{eqnarray}
where the relation between ${\cal A}_{\rm SL}^q$ and $|q_{\particle{q}}/p_{\particle{q}}|$ is given in \Eq{ASLq}.%
\footnote{
  \label{foot:life_mix:epsilon_B}
  Early analyses and %
  the PDG use the complex
  parameter $\epsilon_{\B} = (p_q-q_q)/(p_q+q_q)$ for the \Bd; if \CP violation in the mixing is small,
  $\ASLd \cong 4 {\rm Re}(\epsilon_{\B})/(1+|\epsilon_{\B}|^2)$ and the 
  average of \Eq{ASLD} corresponds to
  ${\rm Re}(\epsilon_{\B})/(1+|\epsilon_{\B}|^2)= \hflavREBD$.
}
However, the fit $\chi^2$ probability %
is only $\hflavCLPERCENTASLSASLD\%$.
This is mostly due to an overall discrepancy between the \dzero and 
LHCb averages at the level of $\hflavASLLHCBDZERONSIGMA\,\sigma$.
Since the assumptions underlying the inclusion of the \dzero muon results 
in the average\footref{foot:life_mix:Abazov:2013uma}
are somewhat controversial~\cite{Lenz_private_communication}, 
we also provide in  \Table{ASLs_ASLd} an average excluding these results.

The above averages show no evidence of \CP violation in \Bd or \Bs mixing.
They deviate by $\hflavASLDASLSNSIGMA\,\sigma$ from the very small predictions of the Standard Model (SM), 
${\cal A}_{\rm SL}^{d,\rm SM} = -(4.7\pm 0.6)\times 10^{-4}$ and 
${\cal A}_{\rm SL}^{s,\rm SM} = +(2.22\pm 0.27)\times 10^{-5}$~\citehistory{Jubb:2016mvq,Artuso:2015swg}{Jubb:2016mvq,Artuso:2015swg,*Lenz_hist}.
Given the current experimental uncertainties,
there is still significant room for a possible new physics contribution, in particular in the \Bs system. 
In this respect, the deviation of the \dzero dimuon
asymmetry~\citehistory{Abazov:2013uma}{Abazov:2013uma,*Abazov:2011yk_hist,*Abazov:2010hv_hist,*Abazov:2010hj_hist,*Abazov:dimuon_hist}
from expectation has generated significant interest.
However, the recent \ASLs and \ASLd results from LHCb 
are not precise enough yet to settle the issue.
It was pointed out~\cite{DescotesGenon:2012kr}
that the \dzero dimuon result can be reconciled with the SM expectations
of \ASLs and \ASLd if there were non-SM sources of \CP violation
in the semileptonic decays of the $b$ and $c$ quarks. 
A Run~1 ATLAS study~\cite{Aaboud:2016bmk} of charge asymmetries
in muon+jets $t\bar{t}$ events, %
in which a \b-hadron decays semileptonically to a soft muon,
yields results with limited statistical precision, 
compatible both with the D0 dimuon asymmetry and with the SM predictions. 

At the more fundamental level, \CP violation in \Bs
mixing is caused by the weak phase difference $\phi^s_{12}$ defined in \Eq{phi12}.
The SM prediction for this phase is tiny~\citehistory{Jubb:2016mvq,Artuso:2015swg}{Jubb:2016mvq,Artuso:2015swg,*Lenz_hist},
\begin{equation}
\phi_{12}^{s,\rm SM} = \hflavPHISTWELVESM \,.
\labe{phis12SM}
\end{equation}
However, new physics in \Bs mixing could change the observed phase to
\begin{equation}
\phi^s_{12} = \phi_{12}^{s,\rm SM} + \phi_{12}^{s,\rm NP} \,.
\labe{phi12NP}
\end{equation}
Using \Eq{ALSq_tanphi2}, the current knowledge of \ASLs, \DGs and \dms, 
given in \Eqsss{ASLS}{DGs_DGsGs}{dms} respectively, yields an
experimental determination of $ \phi^s_{12}$,
\begin{equation}
\tan\phi^s_{12} = \ASLs \frac{\dms}{\DGs} = \hflavTANPHI \,,
\labe{tanphi12}
\comment{ %
from math import *
asls = -0.010460 ; easls = 0.006400
dms = 17.719032  ; edms = 0.042701
dgs = 0.0951919  ; edgs = 0.01362480381803716 
tanphi12 = asls*dms/dgs
etanphi12 = tanphi12*sqrt((easls/asls)**2+(edms/dms)**2+(edgs/dgs)**2)
print tanphi12, etanphi12
} %
\end{equation}
which represents only a very weak constraint at present.

\mysubsubsection{Mixing-induced \CP violation in \Bs decays}
\labs{phasebs}

\CP violation 
arising in the interference between $\Bs-\Bsbar$ mixing and decay is a very active field and large experimental progress has been achieved in the last decade.
The main observable is the 
\CP-violating phase \phiccbars, defined as 
the weak phase difference between
the $\Bs-\Bsbar$ mixing amplitude $M^s_{12}$
and the $b \to c\bar{c}s$ decay amplitude.

The golden mode for such studies is 
$\Bs \to \jpsi\phi$, followed by $\jpsi \to \mu^+\mu^-$ and 
$\phi\to K^+K^-$, for which a full angular 
analysis of the decay products is performed to 
statistically separate the \CP-even and \CP-odd
contributions in the final state. As already mentioned in 
\Sec{DGs},
CDF~\citehistory{Aaltonen:2012ie}{Aaltonen:2012ie,*CDF:2011af_hist,*Aaltonen:2007he_hist,*Aaltonen:2007gf_hist},
\dzero~\citehistory{Abazov:2011ry}{Abazov:2011ry,*Abazov:2008af_hist,*Abazov:2007tx_hist},
ATLAS~\citehistory{Aad:2014cqa,Aad:2016tdj}{Aad:2014cqa,*Aad:2012kba_hist,Aad:2016tdj},
CMS~\cite{Khachatryan:2015nza}
and LHCb~\citehistory{Aaij:2014zsa,Aaij:2017zgz,Aaij:2016ohx}{Aaij:2014zsa,*Aaij:2013oba_supersede2,Aaij:2017zgz,*Aaij:2014zsa_partial_supersede,Aaij:2016ohx}
have used both untagged and tagged $\Bs \to \jpsi\phi$ (and more generally $\Bs \to (c\bar{c}) K^+K^-$) decays 
for the measurement of \phiccbars.
LHCb~\citehistory{Aaij:2014dka}{Aaij:2014dka,*Aaij:2013oba_supersede}
has used $\Bs \to \jpsi \pi^+\pi^-$ events, 
analyzed with a full amplitude model
including several $\pi^+\pi^-$ resonances (\eg, $f_0(980)$),
although the
$\jpsi \pi^+\pi^-$ final state had already been shown
to be almost \CP pure with a \CP-odd fraction
larger than 0.977 at 95\% CL~\cite{LHCb:2012ae}. 
In addition, LHCb has used the $\Bs \to \Dsp\Dsm$ channel~\cite{Aaij:2014ywt} to measure \phiccbars.

All CDF, \dzero, ATLAS and CMS analyses provide 
two mirror solutions related by the transformation 
$(\DGs, \phiccbars) \to (-\DGs, \pi-\phiccbars)$. However, the
LHCb analysis of $\Bs \to \jpsi K^+K^-$ resolves this ambiguity and 
rules out the solution with negative \DGs~\cite{Aaij:2012eq},
a result in agreement with the Standard Model expectation.
Therefore, in what follows, we only consider the solution with $\DGs > 0$.

\begin{table}
\caption{Direct experimental measurements of \phiccbars, \DGs and \Gs using
$\Bs\to\jpsi\phi$, $\jpsi K^+K^-$, $\psi(2S)\phi$, $\jpsi\pi^+\pi^-$ and $D_s^+D_s^-$ decays.
Only the solution with $\DGs > 0$ is shown, since the two-fold ambiguity has been
resolved in \Ref{Aaij:2012eq}. The first error is due to 
statistics, and the second one is due to systematics. The last line gives our average.}
\labt{phisDGsGs}
\begin{center}
\begin{tabular}{ll@{\,}rll@{\,}r} 
\hline
Exp.\ & Mode & Dataset & \multicolumn{1}{c}{\phiccbars}
                     & \multicolumn{1}{c}{\DGs (\!\!\invps)} & Ref.\ \\
\hline
CDF    & $\jpsi\phi$ & $9.6\invfb$
       & $[-0.60,\, +0.12]$, 68\% CL & $+0.068\pm0.026\pm0.009$
       & \protect\citehistory{Aaltonen:2012ie}{Aaltonen:2012ie,*CDF:2011af_hist,*Aaltonen:2007he_hist,*Aaltonen:2007gf_hist} \\
\dzero & $\jpsi\phi$ & $8.0\invfb$
       & $-0.55^{+0.38}_{-0.36}$ & $+0.163^{+0.065}_{-0.064}$ %
       & \citehistory{Abazov:2011ry}{Abazov:2011ry,*Abazov:2008af_hist,*Abazov:2007tx_hist} \\
ATLAS  & $\jpsi\phi$ & $4.9\invfb$
       & $+0.12 \pm 0.25 \pm 0.05$ & $+0.053 \pm0.021 \pm0.010$ %
       & \citehistory{Aad:2014cqa}{Aad:2014cqa,*Aad:2012kba_hist} \\
ATLAS  & $\jpsi\phi$ & $14.3\invfb$
       & $-0.110 \pm 0.082 \pm 0.042$ & $+0.101 \pm0.013 \pm0.007$ %
       & \cite{Aad:2016tdj} \\
ATLAS  & \multicolumn{2}{r}{above 2 combined}
       & $-0.090 \pm 0.078 \pm 0.041$ & $+0.085 \pm0.011 \pm0.007$ %
       & \cite{Aad:2016tdj} \\
CMS    & $\jpsi\phi$ & $19.7\invfb$ 
       & $-0.075 \pm 0.097 \pm 0.031$ & $+0.095\pm0.013\pm0.007$ %
       & \cite{Khachatryan:2015nza} \\ 
LHCb   & $\jpsi K^+K^-$ & $3.0\invfb$
       & $-0.058\pm0.049\pm0.006$ & $+0.0805\pm0.0091\pm0.0032$ %
       & \citehistory{Aaij:2014zsa}{Aaij:2014zsa,*Aaij:2013oba_supersede2} \\
LHCb   & $\jpsi\pi^+\pi^-$ & $3.0\invfb$
       & $+0.070 \pm0.068 \pm 0.008$ & --- %
       & \citehistory{Aaij:2014dka}{Aaij:2014dka,*Aaij:2013oba_supersede} \\
LHCb   & $\jpsi K^+K^-{}^a$ & $3.0\invfb$
       & $+0.119\pm0.107\pm0.034$ & $+0.066 \pm0.018 \pm0.010$ %
       & \citehistory{Aaij:2017zgz}{Aaij:2017zgz,*Aaij:2014zsa_partial_supersede} \\
LHCb   & \multicolumn{2}{r}{above 3 combined}
       & $+0.001\pm0.037(\rm tot)$ & $+0.0813\pm0.0073\pm0.0036$ %
       & \citehistory{Aaij:2017zgz}{Aaij:2017zgz,*Aaij:2014zsa_partial_supersede} \\
LHCb   & $\psi(2S)\phi$ & $3.0\invfb$
       & $+0.23 ^{+0.29}_{-0.28} \pm0.02$ & $+0.066 ^{+0.41}_{-0.44} \pm0.007$ %
       & \cite{Aaij:2016ohx} \\
LHCb   & $D_s^+D_s^-$ & $3.0\invfb$
       & $+0.02 \pm0.17 \pm 0.02$ & --- %
       & \cite{Aaij:2014ywt} \\
\hline
\multicolumn{3}{l}{All combined} & \hflavPHISCOMB & \hflavDGSCOMBnounit & \\ %
\hline
\multicolumn{6}{l}{$^a$ {\footnotesize $m(K^+K^-) > 1.05~\gevcc$.}}
\end{tabular}
\end{center}
\end{table}

\begin{figure}
\begin{center}
\includegraphics[width=0.65\textwidth]{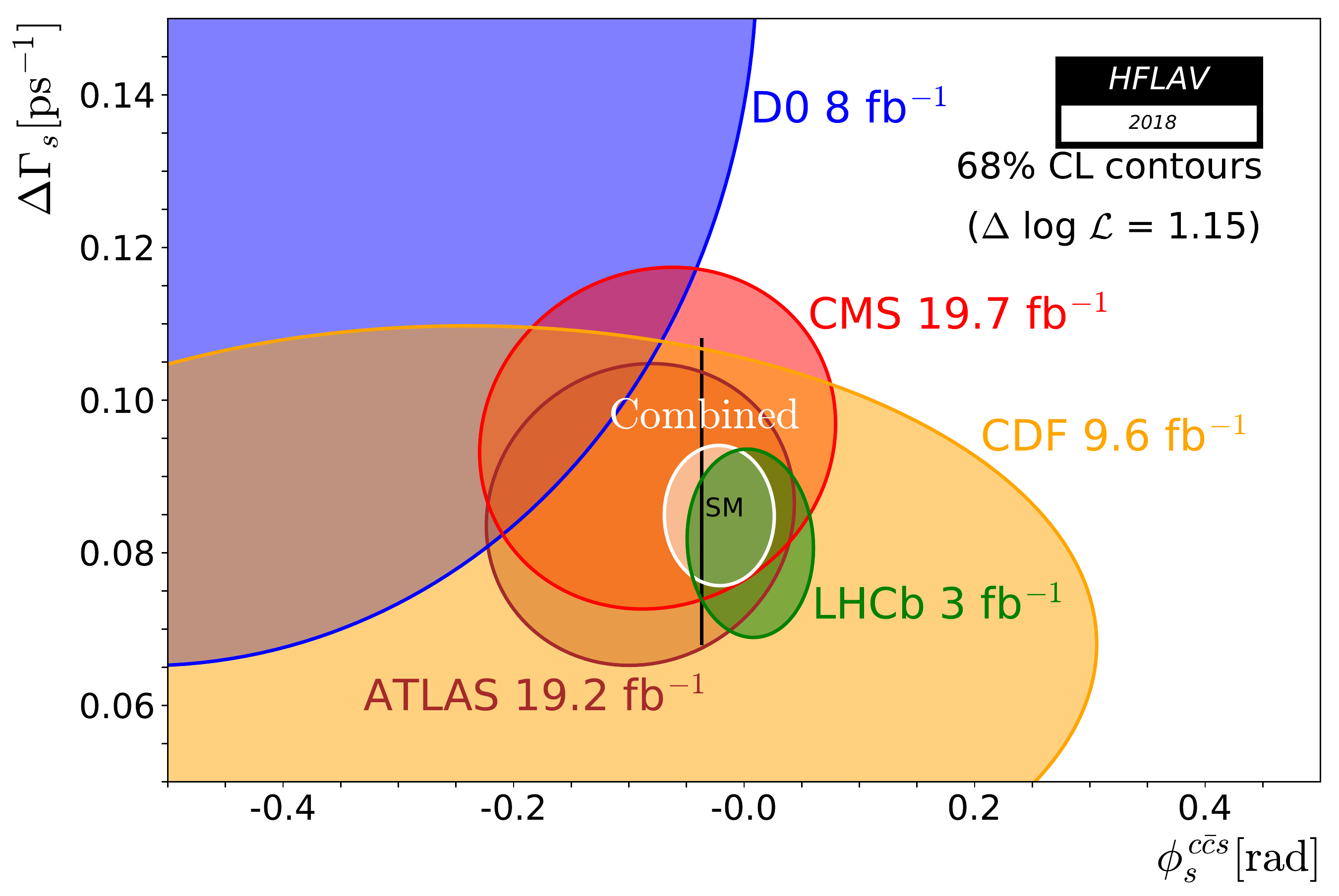}
\caption{
68\% CL regions in \Bs width difference \DGs and weak phase \phiccbars
obtained from individual and combined CDF~\protect\citehistory{Aaltonen:2012ie}{Aaltonen:2012ie,*CDF:2011af_hist,*Aaltonen:2007he_hist,*Aaltonen:2007gf_hist},
\dzero~\protect\citehistory{Abazov:2011ry}{Abazov:2011ry,*Abazov:2008af_hist,*Abazov:2007tx_hist},
ATLAS~\protect\citehistory{Aad:2014cqa,Aad:2016tdj}{Aad:2014cqa,*Aad:2012kba_hist,Aad:2016tdj}, 
CMS~\cite{Khachatryan:2015nza}
and LHCb~\protect\citehistory%
{Aaij:2014zsa,Aaij:2017zgz,Aaij:2016ohx,Aaij:2014dka,Aaij:2014ywt}%
{Aaij:2014zsa,*Aaij:2013oba_supersede2,Aaij:2017zgz,*Aaij:2014zsa_partial_supersede,Aaij:2016ohx,Aaij:2014dka,*Aaij:2013oba_supersede,Aaij:2014ywt}
likelihoods of 
$\Bs\to \jpsi\phi$, $\Bs\to \jpsi K^+K^-$, $\Bs\to \psi(2S) \phi$, $\Bs\to\jpsi\pi^+\pi^-$ and 
$\Bs\to D_s^+D_s^-$ samples. 
The expectation within the Standard Model~\protect\citehistory{Charles:2011va_mod,Lenz:2011ti,*Lenz:2006hd,Jubb:2016mvq,Artuso:2015swg}{Charles:2011va_mod,Lenz:2011ti,*Lenz:2006hd,Jubb:2016mvq,Artuso:2015swg,*Lenz_hist}
is shown as the black rectangle.
}
\labf{DGs_phase}
\end{center}
\end{figure}

We perform a combination of the CDF~\citehistory{Aaltonen:2012ie}{Aaltonen:2012ie,*CDF:2011af_hist,*Aaltonen:2007he_hist,*Aaltonen:2007gf_hist},
\dzero~\citehistory{Abazov:2011ry}{Abazov:2011ry,*Abazov:2008af_hist,*Abazov:2007tx_hist},
ATLAS~\citehistory{Aad:2014cqa,Aad:2016tdj}{Aad:2014cqa,*Aad:2012kba_hist,Aad:2016tdj},
CMS~\cite{Khachatryan:2015nza}
and LHCb~\citehistory%
{Aaij:2014zsa,Aaij:2017zgz,Aaij:2016ohx,Aaij:2014dka}%
{Aaij:2014zsa,*Aaij:2013oba_supersede2,Aaij:2017zgz,*Aaij:2014zsa_partial_supersede,Aaij:2016ohx,Aaij:2014dka,*Aaij:2013oba_supersede}
results summarized in \Table{phisDGsGs}.
This is done by adding the two-dimensional log profile-likelihood scans of
\DGs and \phiccbars from all $\Bs\to\ (c\bar{c})  K^+K^-$ analyses and 
a one-dimensional log profile-likelihood of \phiccbars
from the $\Bs\to\jpsi\pi^+\pi^-$ and $\Bs \to D_s^+ D_s^-$ analyses; 
the combined likelihood is then maximized with respect to \DGs and \phiccbars.

In the $\Bs\to\jpsi\phi$ and $\Bs\to\jpsi K^+K^-$ analyses, \phiccbars and \DGs 
come from a simultaneous fit that determines also the \Bs lifetime,
the polarisation amplitudes, and the strong phases.
While the correlation between \phiccbars and all other parameters is small,
the correlations between \DGs and the polarisation amplitudes are sizable.
However, since the various experiments use different conventions
for the amplitudes and phases, a full combination including all
correlations is not performed. Instead, our average only takes
into account the correlation between \phiccbars and \DGs.

In the LHCb $\Bs \to \jpsi K^+K^-$ analysis~\citehistory{Aaij:2014zsa}{Aaij:2014zsa,*Aaij:2013oba_supersede2}, \phiccbars is measured for the first time separately for each polarisation of the final state. Since the measured values for the different polarisations are compatible, we use the average value of \phiccbars from Ref.~\citehistory{Aaij:2014zsa}{Aaij:2014zsa,*Aaij:2013oba_supersede2} for our world average. 
In the same analysis, the statistical correlation coefficient between \phiccbars and $|\lambda|$
(which signals \CP violation in the decay if $|\lambda|\ne 1$) 
is measured to be very small ($-0.02$). We neglect this correlation in our average. 
Furthermore, the statistical correlation coefficient between \phiccbars and \DGs, measured to be $-0.08$, is also neglected when averaging the 
$\Bs \to \jpsi K^+K^-$,  $\Bs \to \jpsi \pi^+\pi^-$ and $\Bs \to D_s^+ D_s^-$ results of LHCb. 
Given the increasing experimental precision of the LHC results, we have stopped using the two-dimensional $\DGs-\phiccbars$ histograms provided by the CDF and \dzero collaborations, and are now approximating them with two-dimensional Gaussian likelihoods. 

We obtain the individual and combined contours shown in \Fig{DGs_phase}. %
Maximizing the likelihood, we find, as summarized in \Table{phisDGsGs}:  
\begin{eqnarray}
\DGs &=& \hflavDGSCOMB \,, \\    
\phiccbars &=& \hflavPHISCOMB \,.
\labe{phis}
\end{eqnarray}
This \DGs average is consistent but highly correlated with the average
of \Eq{DGs_DGsGs}. Our
final recommended average for \DGs is the one of \Eq{DGs_DGsGs}, which 
includes all available information on this quantity.

In the Standard Model and ignoring sub-leading penguin contributions, 
\phiccbars is expected to be equal to $-2\beta_s$, 
where
$\beta_s = \arg\left[-\left(V_{ts}V^*_{tb}\right)/\left(V_{cs}V^*_{cb}\right)\right]$ %
is a phase analogous to the angle $\beta$ of the usual CKM
unitarity triangle (aside from a sign change). %
An indirect determination via global fits to experimental data
gives~\cite{Charles:2011va_mod}
\begin{equation}
(\phiccbars)^{\rm SM} = -2\beta_s = \hflavPHISSM \,.
\labe{phisSM}
\end{equation}
The average value of \phiccbars from \Eq{phis} is consistent with this
Standard Model expectation.
Penguin contributions to \phiccbars from $\Bs \to \jpsi \phi$ are calculated to be smaller than $0.021$ in magnitude~\cite{Frings:2015eva} but may become relevant if future measurements reduce the error in \Eq{phis}.
There are no reliable estimates of the  penguin contribution to $\Bs \to \jpsi f_0$.

From its measurements of time-dependent \CP violation in $\Bs \to K^+K^-$ decays, the LHCb collaboration has determined the 
\Bs mixing phase to be $-2\beta_s = -0.12^{+0.14}_{-0.12}$~\cite{Aaij:2014xba},
assuming a U-spin relation (with up to 50\% breaking effects) between the decay amplitudes of $\Bs \to K^+K^-$ 
and $\Bd \to \pi^+\pi^-$, and a value of the CKM angle $\gamma$  
of $(70.1 \pm7.1)^{\circ}$. This determination is compatible with, 
and less precise than, the world average of \phiccbars from \Eq{phis}.

New physics could contribute to \phiccbars. Assuming that new physics only 
enters in $M^s_{12}$ (rather than in $\Gamma^s_{12}$),
one can write~\cite{Lenz:2011ti,*Lenz:2006hd}
\begin{equation}
\phiccbars = -  2\beta_s + \phi_{12}^{s,\rm NP} \,,
\end{equation}
where the new physics phase $\phi_{12}^{s,\rm NP}$ is the same as that appearing in \Eq{phi12NP}.
In this case
\begin{equation}
\phi^s_{12} = %
\phi_{12}^{s,\rm SM} +2\beta_s + \phiccbars = \hflavPHISTWELVE \,,
\end{equation}
where the numerical estimation was performed with the values of \Eqsss{phis12SM}{phisSM}{phis}.
Keeping in mind the approximation and assumption mentioned above,
this can serve as a reference value to which the measurement of \Eq{tanphi12} can be compared.

\clearpage
\mysection{Measurements related to Unitarity Triangle angles
}
\label{sec:cp_uta}

We provide averages of measurements obtained from analyses of decay-time-dependent asymmetries and other quantities that are related to the angles of the Unitarity Triangle (UT).
Straightforward interpretations of the averages are given, where possible.
However, no attempt to extract the angles is made in cases where considerable theoretical input is required to do so.

In Sec.~\ref{sec:cp_uta:introduction}
a brief introduction to the relevant phenomenology is given.
In Sec.~\ref{sec:cp_uta:notations}
an attempt is made to clarify the various different notations in use.
In Sec.~\ref{sec:cp_uta:common_inputs}
the common inputs to which experimental results are rescaled in the
averaging procedure are listed.
We also briefly introduce the treatment of experimental uncertainties.
In the remainder of this section,
the experimental results and their averages are given,
divided into subsections based on the underlying quark-level decays.
All the measurements reported are quantities determined from decay-time-dependent analyses, with the exception of several in Sec.~\ref{sec:cp_uta:cus}, which are related to the UT angle $\gamma$ and are obtained from decay-time-integrated analyses.
In the compilations of measurements, indications of the sizes of the data samples used by each experiment are given.
For the $\epem$ $B$ factory experiments, this is quoted in terms of the number of $B\bar{B}$ pairs in the data sample, while the integrated luminosity is given for experiments at hadron colliders.

\mysubsection{Introduction
}
\label{sec:cp_uta:introduction}

In the Standard Model, the Cabibbo-Kobayashi-Maskawa (CKM) quark-mixing matrix is a unitary matrix, conventionally written as the product
of three (complex) rotation matrices~\cite{Chau:1984fp}.
The rotations are parametrised by the Euler mixing angles between the generations, $\theta_{12}$, $\theta_{13}$ and $\theta_{23}$, and one overall phase $\delta$,
\begin{equation}
\label{eq:ckmPdg}
\VCKM =
        \left(
          \begin{array}{ccc}
            V_{ud} & V_{us} & V_{ub} \\
            V_{cd} & V_{cs} & V_{cb} \\
            V_{td} & V_{ts} & V_{tb} \\
          \end{array}
        \right)
        =
        \left(
        \begin{array}{ccc}
        c_{12}c_{13}
                &    s_{12}c_{13}
                        &   s_{13}e^{-i\delta}  \\
        -s_{12}c_{23}-c_{12}s_{23}s_{13}e^{i\delta}
                &  c_{12}c_{23}-s_{12}s_{23}s_{13}e^{i\delta}
                        & s_{23}c_{13} \\
        s_{12}s_{23}-c_{12}c_{23}s_{13}e^{i\delta}
                &  -c_{12}s_{23}-s_{12}c_{23}s_{13}e^{i\delta}
                        & c_{23}c_{13}
        \end{array}
        \right) \, ,
\end{equation}
where $c_{ij}=\cos\theta_{ij}$, $s_{ij}=\sin\theta_{ij}$ for
$i<j=1,2,3$.

The often used Wolfenstein parametrisation~\cite{Wolfenstein:1983yz}
involves the replacements~\cite{Buras:1994ec}
\begin{eqnarray}
  \label{eq:burasdef}
  s_{12}             &\equiv& \lambda\,,\nonumber \\
  s_{23}             &\equiv& A\lambda^2\,, \\
  s_{13}e^{-i\delta} &\equiv& A\lambda^3(\rho -i\eta)\,.\nonumber
\end{eqnarray}
The observed hierarchy among the CKM matrix elements is captured by the small value of $\lambda$, in which a Taylor expansion of $\VCKM$ leads to the familiar approximation
\begin{equation}
  \label{eq:cp_uta:ckm}
  \VCKM
  =
  \left(
    \begin{array}{ccc}
      1 - \lambda^2/2 & \lambda & A \lambda^3 ( \rho - i \eta ) \\
      - \lambda & 1 - \lambda^2/2 & A \lambda^2 \\
      A \lambda^3 ( 1 - \rho - i \eta ) & - A \lambda^2 & 1 \\
    \end{array}
  \right) + {\cal O}\left( \lambda^4 \right) \, .
\end{equation}
At order $\lambda^{5}$, the CKM matrix in this parametrisation is
{\small
  \begin{equation}
    \label{eq:cp_uta:ckm_lambda5}
    \VCKM
    =
    \left(
      \begin{array}{ccc}
        1 - \frac{1}{2}\lambda^{2} - \frac{1}{8}\lambda^4 &
        \lambda &
        A \lambda^{3} (\rho - i \eta) \\
        - \lambda + \frac{1}{2} A^2 \lambda^5 \left[ 1 - 2 (\rho + i \eta) \right] &
        1 - \frac{1}{2}\lambda^{2} - \frac{1}{8}\lambda^4 (1+4A^2) &
        A \lambda^{2} \\
        A \lambda^{3} \left[ 1 - (1-\frac{1}{2}\lambda^2)(\rho + i \eta) \right] &
        -A \lambda^{2} + \frac{1}{2}A\lambda^4 \left[ 1 - 2(\rho + i \eta) \right] &
        1 - \frac{1}{2}A^2 \lambda^4
      \end{array}
    \right) + {\cal O}\left( \lambda^{6} \right)\,.
  \end{equation}
}

\vspace{-5mm}
\noindent
A non-zero value of $\eta$ implies that the CKM matrix is not purely real, and is the source of $\CP$ violation in the Standard Model.
This is encapsulated in a parametrisation-invariant way through the Jarlskog parameter $J = \Im\left(V_{us}V_{cb}V^*_{ub}V^*_{cs}\right)$~\cite{Jarlskog:1985ht},
which is non-zero if and only if $\CP$ violation exists.

The unitarity relation $\VCKM^\dagger\VCKM = {\mathit 1}$
results in a total of nine equations, which can be written as
$\sum_{i=u,c,t} V_{ij}^*V_{ik} = \delta_{jk}$,
where $\delta_{jk}$ is the Kronecker symbol.
Of the off-diagonal expressions ($j \neq k$),
three can be transformed into the other three
(under $j \leftrightarrow k$, corresponding to complex conjugation).
This leaves three relations in which three complex numbers sum to zero,
which therefore can be expressed as triangles in the complex plane.
The diagonal terms yield three relations, in which the squares of the elements in each column of the CKM matrix sum to unity.
Similar relations are obtained for the rows of the matrix from $\VCKM\VCKM^\dagger = {\mathit 1}$.
Thus, there are in total six triangle relations and six sums to unity.
More details about unitarity triangles can be found in Refs.~\cite{Jarlskog:2005uq,Harrison:2009bz,Frampton:2010ii,Frampton:2010uq}.

\begin{figure}[t]
  \begin{center}
    \resizebox{0.55\textwidth}{!}{\includegraphics{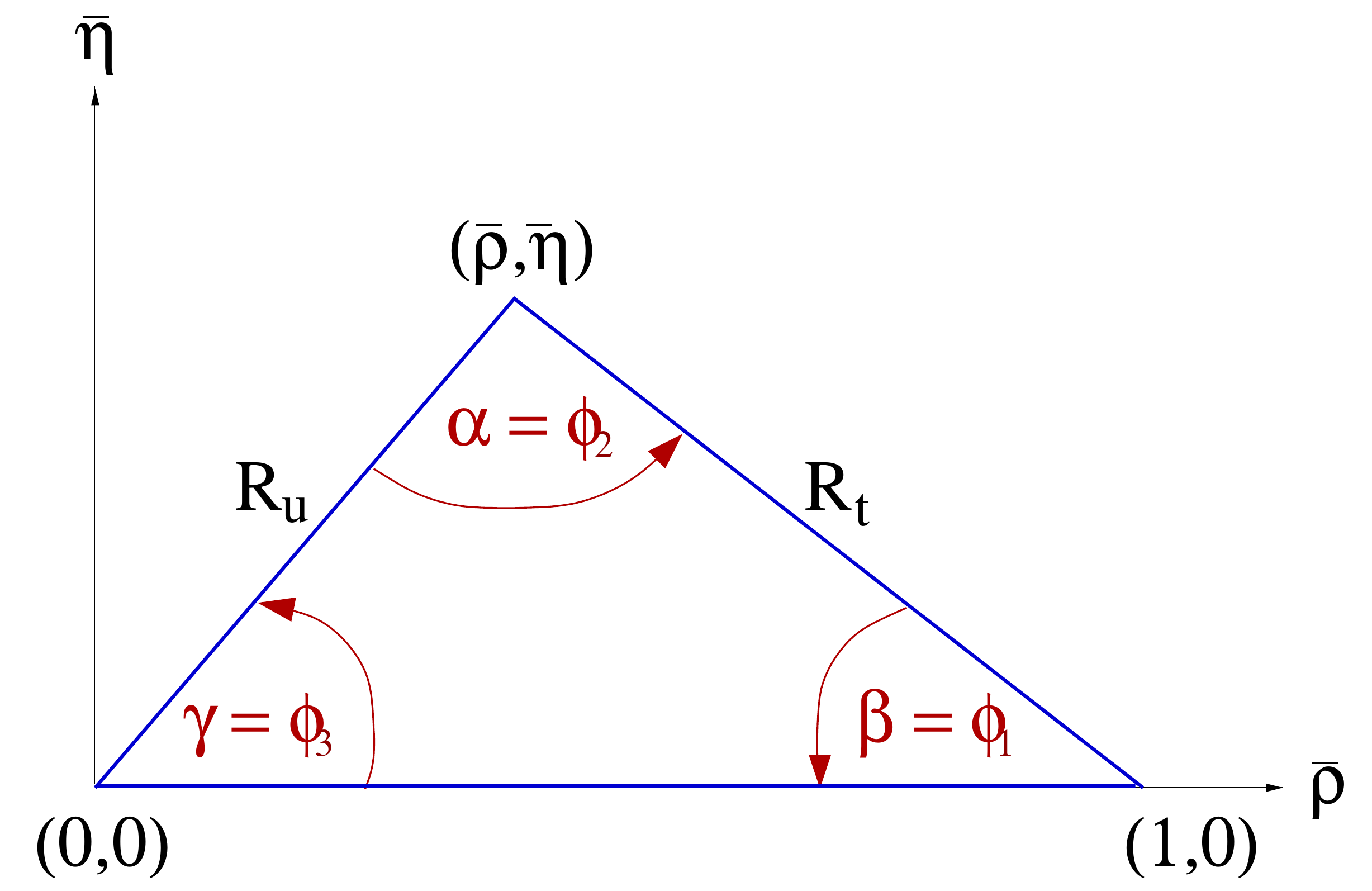}}
    \caption{The Unitarity Triangle.}
    \label{fig:cp_uta:ut}
  \end{center}
\end{figure}

One of the triangle relations,
\begin{equation}
  \label{eq:cp_uta:ut}
  V_{ud}V_{ub}^* + V_{cd}V_{cb}^* + V_{td}V_{tb}^* = 0\,,
\end{equation}
is of particular importance to the $\B$ system,
being specifically related to flavour-changing neutral-current $b \to d$ transitions, and since the three terms in Eq.~(\ref{eq:cp_uta:ut}) are of the same order,
${\cal O}\left( \lambda^3 \right)$.
This relation is commonly known as the Unitarity Triangle (UT).
For presentational purposes,
it is convenient to rescale the triangle by $(V_{cd}V_{cb}^*)^{-1}$,
so that one of its sides becomes $1$, as shown in Fig.~\ref{fig:cp_uta:ut}.

Two popular naming conventions for the UT angles exist in the literature,
\begin{equation}
  \label{eq:cp_uta:abc}
  \alpha  \equiv  \phi_2  =
  \arg\left[ - \frac{V_{td}V_{tb}^*}{V_{ud}V_{ub}^*} \right]\,,
  \hspace{0.5cm}
  \beta   \equiv   \phi_1 =
  \arg\left[ - \frac{V_{cd}V_{cb}^*}{V_{td}V_{tb}^*} \right]\,,
  \hspace{0.5cm}
  \gamma  \equiv   \phi_3  =
  \arg\left[ - \frac{V_{ud}V_{ub}^*}{V_{cd}V_{cb}^*} \right]\,.
\end{equation}
In this document the $\left( \alpha, \beta, \gamma \right)$ set is used.
The sides $R_u$ and $R_t$ of the UT (see Fig.~\ref{fig:cp_uta:ut}) are given by
\begin{equation}
  \label{eq:ru_rt}
  R_u =
  \left|\frac{V_{ud}V_{ub}^*}{V_{cd}V_{cb}^*} \right|
  = \sqrt{\rhobar^2+\etabar^2} \,,
  \hspace{0.5cm}
  R_t =
  \left|\frac{V_{td}V_{tb}^*}{V_{cd}V_{cb}^*}\right|
  = \sqrt{(1-\rhobar)^2+\etabar^2} \,.
\end{equation}
Determinations of $R_u$ rely on measurements of semileptonic \B decays and are discussed in Sec.~\ref{sec:slbdecays}, while $R_t$ is constrained by measurements of \B meson oscillation frequencies (Sec.~\ref{sec:life_mix}) and of rare decays (Sec.~\ref{sec:rare}).
The parameters $\rhobar$ and $\etabar$ define the apex of the UT, and are given by~\cite{Buras:1994ec}
\begin{equation}
  \label{eq:rhoetabar}
  \rhobar + i\etabar
  \equiv -\frac{V_{ud}V_{ub}^*}{V_{cd}V_{cb}^*}
  \equiv 1 + \frac{V_{td}V_{tb}^*}{V_{cd}V_{cb}^*}
  = \frac{\sqrt{1-\lambda^2}\,(\rho + i \eta)}{\sqrt{1-A^2\lambda^4}+\sqrt{1-\lambda^2}A^2\lambda^4(\rho+i\eta)} \, .
\end{equation}
The inverse relation between $\left( \rho, \eta \right)$ and
$\left( \rhobar, \etabar \right)$ is
\begin{equation}
  \label{eq:rhoetabarinv}
  \rho + i\eta \;=\;
  \frac{
    \sqrt{ 1-A^2\lambda^4 }(\rhobar+i\etabar)
  }{
    \sqrt{ 1-\lambda^2 } \left[ 1-A^2\lambda^4(\rhobar+i\etabar) \right]
  } \, .
\end{equation}
By expanding in powers of $\lambda$, several useful approximate expressions
can be obtained, including
\begin{equation}
  \label{eq:rhoeta_approx}
  \rhobar = \rho \left(1 - \frac{1}{2}\lambda^{2}\right) + {\cal O}(\lambda^4) \, ,
  \hspace{0.5cm}
  \etabar = \eta \left(1 - \frac{1}{2}\lambda^{2}\right) + {\cal O}(\lambda^4) \, ,
  \hspace{0.5cm}
  V_{td} = A \lambda^{3} (1-\rhobar -i\etabar) + {\cal O}(\lambda^6) \, .
\end{equation}
Recent world-average values for the Wolfenstein parameters, evaluated using many of the measurements reported in this document, are~\cite{Charles:2004jd}

\begin{equation}
  A = 0.8403 \,^{+0.0056}_{-0.0201}\, , \quad
  \lambda = 0.224747 \,^{+0.000254}_{-0.000059} \, , \quad
  \rhobar = 0.1577 \,^{+0.0096}_{-0.0074} \, , \quad
  \etabar = 0.3493 \,^{+0.0095}_{-0.0071} \, .
\end{equation}

The relevant unitarity triangle for the $b \to s$ transition is obtained
by replacing $d \leftrightarrow s$ in Eq.~(\ref{eq:cp_uta:ut}).
Definitions of the set of angles $( \alpha_s, \beta_s, \gamma_s )$
can be obtained using equivalent relations to those of Eq.~(\ref{eq:cp_uta:abc}).
However, this gives a value of $\beta_s$ that is negative in the Standard Model, so that the sign is usually flipped in the literature; this convention, \ie\ $\beta_s = \arg\left[ - (V_{ts}V_{tb}^*)/(V_{cs}V_{cb}^*) \right]$, is also followed here and in Sec.~\ref{sec:life_mix}.
Since the sides of the $b \to s$ unitarity triangle are not all of the same order in $\lambda$, the triangle is squashed, and $\beta_s \sim \lambda^2\eta$.

\mysubsection{Notations
}
\label{sec:cp_uta:notations}

Several different notations for $\CP$ violation parameters
are commonly used.
This section reviews those found in the experimental literature,
in the hope of reducing the potential for confusion,
and to define the frame that is used for the averages.

In some cases, when $\B$ mesons decay into
multibody final states via broad resonances ($\rho$, $\Kstar$, \etc),
the experimental analyses ignore the effects of interference
between the overlapping structures.
This is referred to as the quasi-two-body (Q2B) approximation in the following.

\mysubsubsection{$\CP$ asymmetries
}
\label{sec:cp_uta:notations:pra}

The $\CP$ asymmetry is defined as the difference between the rate of a decay
involving a $b$ quark and that involving a $\bar b$ quark, divided
by the sum. For example, the partial rate asymmetry for a charged $\B$ decay
would be given as
\begin{equation}
  \label{eq:cp_uta:pra}
  \Acp_{f} \;\equiv\;
  \frac{\Gamma(\Bm \to f)-\Gamma(\Bp \to \bar{f})}{\Gamma(\Bm \to f)+\Gamma(\Bp \to \bar{f})},
\end{equation}
where $f$ and $\bar f$ are $\CP$-conjugate final states.

\mysubsubsection{Time-dependent \CP asymmetries in decays to $\CP$ eigenstates
}
\label{sec:cp_uta:notations:cp_eigenstate}

In the case of decays to a final state $f$, which is a $\CP$ eigenstate with eigenvalue $\etacpf$, the $\Bz$ and $\Bzb$ decay amplitudes can be written as $\Af$ and $\Abarf$, respectively.
The time-dependent decay rates for neutral $\B$ mesons, with known (\ie\ ``tagged'') flavour at time $\Delta t =0$, are then given by
\begin{eqnarray}
  \label{eq:cp_uta:td_cp_asp1}
  \Gamma_{\Bzb \to f} (\Delta t) & = &
  \frac{e^{-| \Delta t | / \tau(\Bz)}}{4\tau(\Bz)}
  \left[
    1 +
    \frac{2\, \Im(\lambda_f)}{1 + |\lambda_f|^2} \sin(\Delta m \Delta t) -
    \frac{1 - |\lambda_f|^2}{1 + |\lambda_f|^2} \cos(\Delta m \Delta t)
  \right], \\
  \label{eq:cp_uta:td_cp_asp2}
  \Gamma_{\Bz \to f} (\Delta t) & = &
  \frac{e^{-| \Delta t | / \tau(\Bz)}}{4\tau(\Bz)}
  \left[
    1 -
    \frac{2\, \Im(\lambda_f)}{1 + |\lambda_f|^2} \sin(\Delta m \Delta t) +
    \frac{1 - |\lambda_f|^2}{1 + |\lambda_f|^2} \cos(\Delta m \Delta t)
  \right].
\end{eqnarray}
This formulation assumes $\CPT$ invariance and neglects a possible lifetime difference between the two physical states.
The case where non-zero lifetime differences are taken into account, which must be considered for \Bs\ decays, is discussed in Sec.~\ref{sec:cp_uta:notations:Bs}.

The notation and normalisation used here are relevant for the $e^+e^-$ $B$ factory experiments.
In this case, neutral $B$ mesons are produced via the $e^+e^- \to \Upsilon(4S) \to \B\Bbar$ process, and the wavefunction of the produced $\B\Bbar$ pair evolves coherently until one meson decays.
When one of the pair decays into a final state that tags its flavour, the flavour of the other at that instant is known.
The evolution of the other neutral $B$ meson is therefore described in terms of $\Delta t$, the difference between the decay times of the two mesons in the pair.
At hadron collider experiments, $t$ is usually used in place of $\Delta t$, since the flavour tagging is done at production ($t = 0$);
due to the nature of the production in hadron colliders (incoherent $b\bar{b}$ quark pair production with many additional associated particles), very different methods are used for tagging compared to those in $e^+e^-$ experiments.
Moreover, since negative values of $t$ are not possible, the normalisation is such that
$\int_0^{+\infty} \left(
\Gamma_{\Bzb \to f} (t) + \Gamma_{\Bz \to f} (t) \right) dt = 1$,
rather than
the $\int_{-\infty}^{+\infty} \left(
\Gamma_{\Bzb \to f} (\Delta t) + \Gamma_{\Bz \to f} (\Delta t) \right) d(\Delta t) = 1$
normalization in Eqs.~(\ref{eq:cp_uta:td_cp_asp1}) and~(\ref{eq:cp_uta:td_cp_asp2}).

The term
\begin{equation}
  \label{eq:cp_uta:lambda_f}
  \lambda_f = \frac{q}{p} \frac{\Abarf}{\Af}
\end{equation}
contains factors related to the decay amplitudes and to $\Bz$--$\Bzb$ mixing, which originates from the fact that the Hamiltonian eigenstates with physical masses and lifetimes
are $\left| B_\pm \right> = p \left| \Bz \right> \pm q \left| \Bzb \right>$
(see Sec.~\ref{sec:mixing}, where the mass difference $\Delta m$ is also defined).
The definition of $\lambda_f$ in Eq.~(\ref{eq:cp_uta:lambda_f}) allows three different categories of \CP\ violation to be distinguished, both in the \Bz\ and \Bs\ systems.
\begin{itemize}\setlength{\itemsep}{0.5ex}
\item \CP\ violation in mixing, where $\left| \frac{q}{p} \right| \neq 1$.
  The strongest constraints on the associated parameters are obtained using semileptonic decays, and are discussed in Sec.~\ref{sec:life_mix}.
  There is currently no evidence for \CP\ violation mixing in either of the $\Bz$--$\Bzb$ or $\Bs$--$\Bsb$ systems; therefore $\left| \frac{q}{p} \right| = 1$ is assumed throughout the discussion in this Section.
\item \CP\ violation in decay, where $\left| \frac{\Abarf}{\Af} \right| \neq 1$.
  This is the only possible category of \CP\ violation for charged \B mesons and $b$ baryons (see, for example, results reported in Sec.~\ref{sec:rare}).
  Several parameters measured in time-dependent analyses are also sensitive to \CP\ violation in decay, and are discussed in this Section.
\item \CP\ violation in the interference between mixing and decay, where $\Im\left(\lambda_f \right) \neq 0$.
  Results related to this category, also referred to as mixing-induced \CP violation, are reported in this Section.
\end{itemize}

The time-dependent $\CP$ asymmetry,
again defined as the normalized difference between the decay rate
involving a $b$ quark and that involving a $\bar b$ quark,
is then given by
\begin{equation}
  \label{eq:cp_uta:td_cp_asp}
  \Acp_{f} \left(\Delta t\right) \; \equiv \;
  \frac{
    \Gamma_{\Bzb \to f} (\Delta t) - \Gamma_{\Bz \to f} (\Delta t)
  }{
    \Gamma_{\Bzb \to f} (\Delta t) + \Gamma_{\Bz \to f} (\Delta t)
  } \; = \;
  \frac{2\, \Im(\lambda_f)}{1 + |\lambda_f|^2} \sin(\Delta m \Delta t) -
  \frac{1 - |\lambda_f|^2}{1 + |\lambda_f|^2} \cos(\Delta m \Delta t).
\end{equation}
While the coefficient of the $\sin(\Delta m \Delta t)$ term in
Eq.~(\ref{eq:cp_uta:td_cp_asp}) is customarily\footnote
{
  Occasionally one also finds Eq.~(\ref{eq:cp_uta:td_cp_asp}) written as
  $\Acp_{f} \left(\Delta t\right) =
  {\cal A}^{\rm mix}_f \sin(\Delta m \Delta t) + {\cal A}^{\rm dir}_f \cos(\Delta m \Delta t)$,
  or similar.
} denoted $S_f$:
\begin{equation}
  \label{eq:cp_uta:s_def}
  S_f \;\equiv\; \frac{2\, \Im(\lambda_f)}{1 + \left|\lambda_f\right|^2}\,,
\end{equation}
different notations are in use for the
coefficient of the $\cos(\Delta m \Delta t)$ term:
\begin{equation}
  \label{eq:cp_uta:c_def}
  C_f \;\equiv\; - A_f \;\equiv\; \frac{1 - \left|\lambda_f\right|^2}{1 + \left|\lambda_f\right|^2}\,.
\end{equation}
The $C$ notation has been used by the \babar\ collaboration (see \eg\ Ref.~\cite{Aubert:2001sp}),
and subsequently by the LHCb collaboration (see \eg\ Ref.~\cite{Aaij:2012ke}),
and is also adopted in this document.
The $A$ notation has been used by the \belle\ collaboration
(see \eg\ Ref.~\cite{Abe:2001xe}).
For the case when the final state is a \CP eigenstate, as is being considered here, the notation $S_{\CP}$ and $C_{\CP}$ is widely used, including in this document, instead of specifying the final state $f$.
In addition, the $S$, $C$ notation with a subscript indicating the transition is used, particularly when grouping together measurements with different final states mediated by the same quark-level transition.

Neglecting effects due to $\CP$ violation in mixing,
if the decay amplitude contains terms with a single weak (\ie\ $\CP$-violating) phase then $\left|\lambda_f\right| = 1$,
and one finds
$S_f = -\etacpf \sin(\phi_{\rm mix} + \phi_{\rm dec})$, $C_f = 0$,
where $\phi_{\rm mix}=\arg(q/p)$ and $\phi_{\rm dec}=\arg(\Abarf/\Af)$.
The $\Bz$--$\Bzb$ mixing phase $\phi_{\rm mix}$ is approximately equal to $2\beta$ in the Standard Model (in the usual phase convention)~\cite{Carter:1980tk,Bigi:1981qs}.

If amplitudes with different weak phases contribute to the decay,
no clean interpretation of $S_f$ in terms of UT angles is possible without further input.
In this document, only the theoretically cleanest channels are interpreted as measurements of the weak phase (\eg\ $b \to c\bar{c}s$ transitions for $\sin(2\beta)$), although even in these cases some care is necessary.
In channels in which a second amplitude with a different weak phase to the leading amplitude contributes but is expected to be suppressed, the concept of an effective weak phase difference is sometimes used, \eg\ $\sin(2\beta^{\rm eff})$ in $b \to q\bar{q}s$ transitions. %

If, in addition to having a weak phase difference, two contributing decay amplitudes have different
strong (\ie\ $\CP$-conserving)
phases, then $\left| \lambda_f \right| \neq 1$.
Additional input is required for interpretation of the results.
The coefficient of the cosine term becomes non-zero, indicating $\CP$ violation in decay.

Due to the fact that $\sin(\Delta m \Delta t)$ and $\cos(\Delta m \Delta t)$
are, respectively, odd and even functions of $\Delta t$, only small correlations
(that can be induced by backgrounds, for example) between $S_f$ and $C_f$ are
expected at an $e^+e^-$ $B$ factory experiment, where the range of $\Delta t$
is $-\infty < \Delta t < +\infty$.
The situation is different for measurements at hadron collider experiments, where the range of the time variable is $0 < t < +\infty$, so that more sizable correlations can be expected.
We include the correlations in the averages where available.

Frequently, we are interested in combining measurements
governed by similar or identical short-distance physics,
but with different final states
(\eg, $\Bz \to \jpsi \KS$ and $\Bz \to \jpsi \KL$).
In this case, we remove the dependence on the $\CP$ eigenvalue
of the final state by quoting $-\etacp S_f$.
In cases where the final state is not a $\CP$ eigenstate but has
an effective $\CP$ content (see Sec.~\ref{sec:cp_uta:notations:vv}),
the reported $-\etacp S$ is corrected by the effective $\CP$.

\mysubsubsection{Time-dependent distributions with non-zero decay width difference}
\label{sec:cp_uta:notations:Bs}

A complete analysis of the time-dependent decay rates of neutral $B$ mesons must also take into account the difference between the widths of the Hamiltonian eigenstates, denoted $\Delta \Gamma$.
This is particularly important in the $\Bs$ system, where a non-negligible value of $\Delta \Gamma_s$ has been established (see Sec.~\ref{sec:mixing}).
The formalism given here is appropriate for measurements of $\Bs$ decays to a $\CP$ eigenstate $f$ as studied at hadron colliders, but appropriate modifications for $\Bz$ mesons or for the $e^+e^-$ environment are straightforward to make.

Neglecting $\CP$ violation in mixing, the relevant replacements for
Eqs.~(\ref{eq:cp_uta:td_cp_asp1})~and~(\ref{eq:cp_uta:td_cp_asp2})
are~\cite{Dunietz:2000cr}
\begin{equation}
  \label{eq:cp_uta:td_cp_bs_asp1}
  \begin{array}{l@{\hspace{50mm}}cr}
    \mc{2}{l}{
      \Gamma_{\Bsbar \to f} (t) =
      {\cal N} %
      \frac{e^{-t/\tau(\Bs)}}{2\tau(\Bs)}
      \Big[
      \cosh(\frac{\Delta \Gamma_s t}{2}) + {}
    } & \hspace{40mm} \\
    \hspace{40mm} &
    \mc{2}{r}{
      S_f \sin(\Delta m_s t) - C_f \cos(\Delta m_s t) +
      A^{\Delta \Gamma}_f \sinh(\frac{\Delta \Gamma_s t}{2})
      \Big]\,
    } \\
  \end{array}
\end{equation}
and
\begin{equation}
  \label{eq:cp_uta:td_cp_bs_asp2}
  \begin{array}{l@{\hspace{50mm}}cr}
    \mc{2}{l}{
      \Gamma_{\Bs \to f} (t) =
      {\cal N} %
      \frac{e^{-t/\tau(\Bs)}}{2\tau(\Bs)}
      \Big[
      \cosh(\frac{\Delta \Gamma_s t}{2}) - {}
    } & \hspace{40mm} \\
    \hspace{40mm} &
    \mc{2}{r}{
      S_f \sin(\Delta m_s t) + C_f \cos(\Delta m_s t) +
      A^{\Delta \Gamma}_f \sinh(\frac{\Delta \Gamma_s t}{2})
      \Big]\,,
    } \\
  \end{array}
\end{equation}
where $S_f$ and $C_f$ are as defined in Eqs.~(\ref{eq:cp_uta:s_def}) and~(\ref{eq:cp_uta:c_def}), respectively, $\tau(\Bs) = 1/\Gamma_s$ is defined in Sec.~\ref{sec:taubs}, and the coefficient of the $\sinh$ term is\footnote{
  As ever, alternative and conflicting notations appear in the literature.
  One popular alternative notation for this parameter is
  ${\cal A}_{\Delta \Gamma}$.
  Particular care must be taken regarding the signs.
}
\begin{equation}
  A^{\Delta \Gamma}_f = - \frac{2\, \Re(\lambda_f)}{1 + |\lambda_f|^2} \, .
\end{equation}
With the requirement
$\int_{0}^{+\infty} \left[ \Gamma_{\Bsbar \to f} (t) + \Gamma_{\Bs \to f} (t) \right] dt = 1$,
the normalisation factor is fixed to ${\cal N} = \left(1 - (\frac{\Delta \Gamma_s}{2\Gamma_s})^2\right)/\left(1 +\frac{A^{\Delta \Gamma}_f \Delta \Gamma_s}{2\Gamma_s}\right)$.\footnote{
  The prefactor of ${\cal N}/2\tau(\Bs)$ in Eqs.~(\ref{eq:cp_uta:s_def}) and~(\ref{eq:cp_uta:c_def}) has been chosen so that ${\cal N} = 1$ in the limit $\Delta \Gamma_s = 0$.
  In the $e^+e^-$ environment, where the range is $-\infty < \Delta t < \infty$, the prefactor should be ${\cal N}/4\tau(\Bs)$ and ${\cal N} = 1 - (\frac{\Delta \Gamma_s}{2\Gamma_s})^2$.
}

A time-dependent analysis of \CP asymmetries in flavour-tagged
$\Bs$ decays to a \CP eigenstate $f$ can thus determine the parameters
$S_f$, $C_f$ and $A^{\Delta \Gamma}_f$.
Note that, by definition,
\begin{equation}
  \left( S_f \right)^2 + \left( C_f \right)^2 + \left( A^{\Delta \Gamma}_f \right)^2 = 1 \, ,
\end{equation}
and this constraint may or may not be imposed in the fits.
Since these parameters have sensitivity to both
$\Im(\lambda_f)$ and $\Re(\lambda_f)$,
alternative choices of parametrisation,
including those directly involving \CP violating phases (such as $\beta_s$),
are possible.
These can also be adopted for vector-vector final states (see Sec.~\ref{sec:cp_uta:notations:vv}).

The {\it untagged} time-dependent decay rate is given by
\begin{equation}
  \Gamma_{\Bsbar \to f} (t) + \Gamma_{\Bs \to f} (t)
  =
  {\cal N} %
  \frac{e^{-t/\tau(\Bs)}}{\tau(\Bs)}
  \Big[
  \cosh\left(\frac{\Delta \Gamma_s t}{2}\right)
  + A^{\Delta \Gamma}_f \sinh\left(\frac{\Delta \Gamma_s t}{2}\right)
  \Big] \, .
\end{equation}
Thus, an untagged time-dependent analysis can probe $\lambda_f$, through the dependence of $A^{\Delta \Gamma}_f $ on $\Re(\lambda_f)$, given that $\Delta \Gamma_s \neq 0$.
This is equivalent to determining the ``{\it effective lifetime}''~\cite{Fleischer:2011cw}, as discussed in Sec.~\ref{sec:taubs}.
The analysis of flavour-tagged \Bs\ mesons is, of course, more sensitive.

The discussion in this and the previous section is relevant for decays to \CP\ eigenstates.
In the remainder of Sec.~\ref{sec:cp_uta:notations}, various cases of time-dependent \CP\ asymmetries in decays to non-\CP eigenstates are considered.
For brevity, equations will usually be given assuming that the decay width difference $\Delta \Gamma$ is negligible.
Modifications similar to those described here can be made to take into account a non-zero decay width difference.

\mysubsubsection{Time-dependent \CP asymmetries in decays to vector-vector final states
}
\label{sec:cp_uta:notations:vv}

Consider \B decays to states consisting of two spin-1 particles,
such as $\jpsi K^{*0}(\to\KS\piz)$, $\jpsi\phi$, $D^{*+}D^{*-}$ and $\rho^+\rho^-$,
which are eigenstates of charge conjugation but not of parity.\footnote{
  \noindent
  This is not true for all vector-vector final states,
  \eg, $D^{*\pm}\rho^{\mp}$ is clearly not an eigenstate of
  charge conjugation.
}
For such a system, there are three possible final states.
In the helicity basis, these are denoted $h_{-1}, h_0, h_{+1}$.
The $h_0$ state is an eigenstate of parity, and hence of $\CP$.
By contrast, $\CP$ transforms $h_{+1} \leftrightarrow h_{-1}$ (up to an unobservable phase).
These states are transformed into the transversity basis states
$h_\parallel =  (h_{+1} + h_{-1})/2$ and $h_\perp = (h_{+1} - h_{-1})/2$.
In this basis all three states are $\CP$ eigenstates, and $h_\perp$ has the opposite $\CP$ to the others.

The amplitude for decays to the transversity basis states are usually given by $A_{0,\perp,\parallel}$, with normalisation such that $| A_0 |^2 + | A_\perp |^2 + | A_\parallel |^2 = 1$.
Given the relation between the $\CP$ eigenvalues of the states, the effective $\CP$ content of the vector-vector state is known if $| A_\perp |^2$ is measured.
An alternative strategy is to measure just the longitudinally polarised component,  $| A_0 |^2$ (sometimes denoted by $f_{\rm long}$), which allows a limit to be set on the effective $\CP$ content,
since $| A_\perp |^2 \leq | A_\perp |^2 + | A_\parallel |^2 = 1 - | A_0 |^2$.
The value of the effective $\CP$ content can be used to treat the decay with the same formalism as for $\CP$ eigenstates.
The most complete treatment for neutral $\B$ decays to vector-vector final states is, however, time-dependent angular analysis (also known as time-dependent transversity analysis).
In such an analysis, interference between $\CP$-even and $\CP$-odd states provides additional sensitivity to the weak and strong phases involved.

In most analyses of time-dependent \CP asymmetries in decays to vector-vector final states carried out to date, an assumption has been made that each helicity (or transversity) amplitude has the same weak phase.
This is a good approximation for decays that are dominated by
amplitudes with a single weak phase, such $\Bz \to \jpsi K^{*0}$,
and is a reasonable approximation in any mode for which only small sample sizes are available.
However, for modes that have contributions from amplitudes with different
weak phases, the relative size of these contributions can be different
for each helicity (or transversity) amplitude,
and therefore the time-dependent \CP asymmetry parameters can also differ.
The most generic analysis, suitable for analyses with sufficiently large samples,
allows for this effect; such an analysis has been carried out by LHCb for the $\Bz \to \jpsi \rhoz$ decay~\cite{Aaij:2014vda}.
An intermediate analysis can allow different parameters for the $\CP$-even and $\CP$-odd components; such an analysis has been carried out by \babar\ for the decay $\Bz \to D^{*+}D^{*-}$~\cite{Aubert:2008ah}.
The independent treatment of each helicity (or transversity) amplitude, as in the study of $\Bs \to \jpsi\phi$~\cite{Aaij:2014zsa} (discussed in Sec.~\ref{sec:life_mix}), becomes increasingly important for high precision measurements.

\mysubsubsection{Time-dependent asymmetries: self-conjugate multiparticle final states
}
\label{sec:cp_uta:notations:dalitz}

Amplitudes for neutral \B decays into
self-conjugate multiparticle final states
such as $\pi^+\pi^-\pi^0$, $K^+K^-\KS$, $\pi^+\pi^-\KS$,
$\jpsi \pi^+\pi^-$ or $D\pi^0$ with $D \to \KS\pi^+\pi^-$
may be written in terms of \CP-even and \CP-odd amplitudes.
As above, the interference between these terms
provides additional sensitivity to the weak and strong phases
involved in the decay,
and the time-dependence depends on both the sine and cosine
of the weak phase difference.
In order to perform unbinned maximum likelihood fits,
and thereby extract as much information as possible from the distributions,
it is necessary to choose a model for the multiparticle decay,
and therefore the results acquire some model dependence.
In certain cases, model-independent methods are also possible, but the resulting need to bin the Dalitz plot leads to some loss of statistical precision.
The number of observables depends on the final state (and on the model used);
the key feature is that as long as there are kinematic regions where both
\CP-even and \CP-odd amplitudes contribute,
the interference terms will be sensitive to the cosine
of the weak phase difference.
Therefore, these measurements allow distinction between multiple solutions
for, \eg, the two values of $2\beta$ from the measurement of $\sin(2\beta)$.

In model-dependent analysis of multibody decays, the decay amplitude is typically described as a coherent sum of contributions that proceed via different intermediate resonances and through nonresonant interactions.
It is therefore of interest to present results in terms of the \CP\ violation parameters associated with each resonant amplitude, \eg\ $\rhoz\KS$ in the case of the $\pi^+\pi^-\KS$ final state.
These are referred to as Q2B parameters, since in the limit that there was no other contribution to the multibody decay, the amplitude analysis and the Q2B analysis would give the same results.

We now consider the various notations that have been used in experimental studies of
time-dependent asymmetries in decays to self-conjugate multiparticle final states.

\mysubsubsubsection{$\Bz \to D^{(*)}h^0$ with $D \to \KS\pi^+\pi^-$
}
\label{sec:cp_uta:notations:dalitz:dh0}

The states $D\pi^0$, $D^*\pi^0$, $D\eta$, $D^*\eta$, $D\omega$
are collectively denoted $D^{(*)}h^0$.
When the $D$ decay model is fixed,
fits to the time-dependent decay distributions can be performed
to extract the weak phase difference.
However, it is experimentally advantageous to use the sine and cosine of
this phase as fit parameters, since these behave as essentially
independent parameters, with low correlations and (potentially)
rather different uncertainties.
A parameter representing $\CP$ violation in the $B$ decay
can be simultaneously determined.
For consistency with other analyses, this could be chosen to be $C_f$,
but could equally well be $\left| \lambda_f \right|$, or other possibilities.

\belle\ performed an analysis of these channels
with $\sin(2\beta)$ and $\cos(2\beta)$ as free parameters~\cite{Krokovny:2006sv}.
\babar\ has performed an analysis in which $\left| \lambda_f \right|$ was also determined~\cite{Aubert:2007rp}.
A joint analysis of the final \babar\ and \belle\ data samples supersedes these earlier measurements, and uses $\sin(2\beta)$ and $\cos(2\beta)$ as free parameters~\cite{Adachi:2018itz,Adachi:2018jqe}.
\belle\ has in addition performed a model-independent analysis~\cite{Vorobyev:2016npn} using as input information about the average strong phase difference between symmetric bins of the Dalitz plot determined by CLEO-c~\cite{Libby:2010nu}.\footnote{
  The external input needed for this analysis is the same as in the model-independent analysis of $\Bp \to D\Kp$ with $D \to \KS\pip\pim$, discussed in Sec.~\ref{sec:cp_uta:cus:dalitz:modInd}.
}
The results of this analysis are measurements of $\sin(2\phi_1)$ and $\cos(2\phi_1)$.

\mysubsubsubsection{$\Bz \to D^{*+}D^{*-}\KS$
}
\label{sec:cp_uta:notations:dalitz:dstardstarks}

The hadronic structure of the $\Bz \to D^{*+}D^{*-}\KS$ decay
is not sufficiently well understood to perform a full
time-dependent Dalitz-plot analysis.
Instead, following Ref.~\cite{Browder:1999ng},
\babar~\cite{Aubert:2006fh} and \belle~\cite{Dalseno:2007hx} divide the Dalitz plane into two regions:
$m(D^{*+}\KS)^2 > m(D^{*-}\KS)^2$ (labelled $\eta_y = +1$) and
$m(D^{*+}\KS)^2 < m(D^{*-}\KS)^2$ $(\eta_y = -1)$;
and then fit to a decay-time distribution with asymmetry given by
\begin{equation}
  \Acp_{f} \left(\Delta t\right) =
  \eta_y \frac{J_c}{J_0} \cos(\Delta m \Delta t) -
  \left[
    \frac{2J_{s1}}{J_0} \sin(2\beta) + \eta_y \frac{2J_{s2}}{J_0} \cos(2\beta)
  \right] \sin(\Delta m \Delta t) \, .
\end{equation}
The fitted observables are $\frac{J_c}{J_0}$, $\frac{2J_{s1}}{J_0} \sin(2\beta)$ and $\frac{2J_{s2}}{J_0} \cos(2\beta)$,
where the parameters $J_0$, $J_c$, $J_{s1}$ and $J_{s2}$ are the integrals
over the half Dalitz plane $m(D^{*+}\KS)^2 < m(D^{*-}\KS)^2$
of the functions $|a|^2 + |\bar{a}|^2$, $|a|^2 - |\bar{a}|^2$,
$\Re(\bar{a}a^*)$ and $\Im(\bar{a}a^*)$, respectively,
where $a$ and $\bar{a}$ are the decay amplitudes of
$\Bz \to D^{*+}D^{*-}\KS$ and $\Bzb \to D^{*+}D^{*-}\KS$, respectively.
The parameter $J_{s2}$ (and hence $J_{s2}/J_0$) is predicted to be positive~\cite{Browder:1999ng};
assuming this prediction to be correct, it is possible to determine the sign of $\cos(2\beta)$.

\mysubsubsubsection{$\Bz \to \jpsi \pip\pim$
}
\label{sec:cp_uta:notations:dalitz:jpsipipi}

Amplitude analyses of $\Bz \to \jpsi \pip\pim$ decays~\cite{Aaij:2014siy,Aaij:2014vda} show large contributions from the $\rho(770)^0$ and $f_0(500)$ states, together with smaller contributions from higher resonances.
Since modelling the $f_0(500)$ structure is challenging~\cite{Pelaez:2015qba}, it is difficult to determine reliably its associated \CP violation parameters.
Corresponding parameters for the $\jpsi\rhoz$ decay can, however, be determined.
In the LHCb analysis~\cite{Aaij:2014vda}, $2\beta^{\rm eff}$ is determined from the fit; results are then converted into values for $S_{\CP}$ and $C_{\CP}$ to allow comparison with other modes.
Here, the notation $S_{\CP}$ and $C_{\CP}$ denotes parameters obtained for the $\jpsi\rhoz$ final state accounting for the composition of \CP-even and \CP-odd amplitudes (while assuming that all amplitudes involve the same phases), so that no dilution occurs.
Possible \CP violation effects in the other amplitudes contributing to the Dalitz plot are treated as a source of systematic uncertainty.

Amplitude analyses have also been done for the $\Bs \to \jpsi\pip\pim$ decay, where the final state is dominated by scalar resonances, including the $f_0(980)$~\cite{LHCb:2012ae,Aaij:2014dka}. %
Time-dependent analyses of this \Bs decay allow a determination of $2\beta_s$, as discussed in Sec.~\ref{sec:life_mix}.

\mysubsubsubsection{$\Bz \to K^+K^-\Kz$
}
\label{sec:cp_uta:notations:dalitz:kkk0}

Studies of $\Bz \to K^+K^-\Kz$~\cite{Aubert:2007sd,Nakahama:2010nj,Lees:2012kxa}
and of the related decay
$\Bp \to K^+K^-K^+$~\cite{Garmash:2004wa,Aubert:2006nu,Lees:2012kxa},
show that the decay is dominated by a large nonresonant contribution
with significant components from the
intermediate $K^+K^-$ resonances $\phi(1020)$, $f_0(980)$,
and other higher resonances,
as well as a contribution from $\chi_{c0}$.

The full time-dependent Dalitz plot analysis allows
the complex amplitudes of each contributing term to be determined from data,
including $\CP$ violation effects
(\ie\ allowing the complex amplitude for the $\Bz$ decay to be independent
from that for $\Bzb$ decay), although one amplitude must be fixed
to serve as a reference.
There are several choices for parametrisation of the complex amplitudes
(\eg\ real and imaginary part, or magnitude and phase).
Similarly, there are various approaches to the inclusion of $\CP$ violation effects.
Note that the use of positive definite parameters such as magnitudes are
disfavoured in certain circumstances
(it inevitably leads to biases for small values).
In order to compare results between analyses,
it is useful for each experiment to present results in terms of the
parameters that can be measured in a Q2B analysis
(such as $\Acp_{f}$, $S_f$, $C_f$,
$\sin(2\beta^{\rm eff})$, $\cos(2\beta^{\rm eff})$, \etc)

In the \babar\ analysis of the $\Bz \to K^+K^-\Kz$ decay~\cite{Lees:2012kxa},
the complex amplitude for each resonant contribution was written as
\begin{equation}
  A_f = c_f ( 1 + b_f ) e^{i ( \phi_f + \delta_f )}
  \ , \ \ \ \
  \bar{A}_f = c_f ( 1 - b_f ) e^{i ( \phi_f - \delta_f )} \, ,
\end{equation}
where $b_f$ and $\delta_f$ parametrize $\CP$ violation in the magnitude
and phase, respectively.
Belle~\cite{Nakahama:2010nj} used the same parametrisation but with a different notation for the parameters.\footnote{
  $(c, b, \phi, \delta) \leftrightarrow (a, c, b, d)$.
  See Eq.~(\ref{eq:cp_uta:BelleDPCPparam}).
}
The Q2B parameter of $\CP$ violation in decay is directly related to $b_f$,
\begin{equation}
  \Acp_{f} = \frac{-2b_f}{1+b_f^2} \approx C_f \, ,
\end{equation}
and the mixing-induced $\CP$ violation parameter can be used to obtain
$\sin(2\beta^{\rm eff})$,
\begin{equation}
  \label{eq:cp_uta:sin2betaeff-def}
  -\eta_f S_f \approx \frac{1-b_f^2}{1+b_f^2}\sin(2\beta^{\rm eff}_f) \, ,
\end{equation}
where the approximations are exact in the case that $\left| q/p \right| = 1$.

Both \babar~\cite{Lees:2012kxa} and \belle~\cite{Nakahama:2010nj} present results for $c_f$ and $\phi_f$,
for each resonant contribution,
and in addition present results for $\Acp_{f}$ and $\beta^{\rm eff}_{f}$ for $\phi(1020) \Kz$, $f_0(980) \Kz$ and for the remainder of the contributions to the $K^+K^-\Kz$ Dalitz plot combined.
\babar also presents results for the Q2B parameter $S_{f}$ for these channels.
The models used to describe the resonant structure of the Dalitz plot differ, however.
Both analyses suffer from symmetries in the likelihood that lead to multiple solutions, from which we select only one for averaging.

\mysubsubsubsection{$\Bz \to \pi^+\pi^-\KS$
}
\label{sec:cp_uta:notations:dalitz:pipik0}

Studies of $\Bz \to \pi^+\pi^-\KS$~\cite{Aubert:2009me,Dalseno:2008wwa}
and of the related decay
$\Bp \to \pi^+\pi^-K^+$~\cite{Garmash:2004wa,Garmash:2005rv,Aubert:2005ce,Aubert:2008bj}
show that the decay is dominated by components from intermediate resonances
in the $K\pi$ ($K^*(892)$, $K^*_0(1430)$)
and $\pi\pi$ ($\rho(770)$, $f_0(980)$, $f_2(1270)$) spectra,
together with a poorly understood scalar structure that peaks near
$m(\pi\pi) \sim 1300 \ {\rm MeV}/c^2$ and is denoted $f_X$,\footnote{
  The $f_X$ component may originate from either the $f_0(1370)$ or $f_0(1500)$ resonances, or from interference between those or other states and nonresonant amplitudes in this region.
}
as well as a large nonresonant component.
There is also a contribution from the $\chi_{c0}$ state.

The full time-dependent Dalitz plot analysis allows
the complex amplitudes of each contributing term to be determined from data,
including $\CP$ violation effects.
In the \babar\ analysis~\cite{Aubert:2009me},
the magnitude and phase of each component (for both $\Bz$ and $\Bzb$ decays)
are measured relative to $\Bz \to f_0(980)\KS$, using the following
parametrisation:
\begin{equation}
  A_f = \left| A_f \right| e^{i\,{\rm arg}(A_f)}
  \ , \ \ \ \
  \bar{A}_f = \left| \bar{A}_f \right| e^{i\,{\rm arg}(\bar{A}_f)} \, .
\end{equation}
In the \belle\ analysis~\cite{Dalseno:2008wwa}, the $\Bz \to K^{*+}\pi^-$ amplitude
is chosen as the reference, and the amplitudes are parametrised as
\begin{equation}
  \label{eq:cp_uta:BelleDPCPparam}
  A_f = a_f ( 1 + c_f ) e^{i ( b_f + d_f )}
  \ , \ \ \ \
  \bar{A}_f = a_f ( 1 - c_f ) e^{i ( b_f - d_f )} \, .
\end{equation}
In both cases, the results are translated into Q2B parameters
such as $2\beta^{\rm eff}_f$, $S_f$, $C_f$ for each \CP\ eigenstate $f$,
and parameters of \CP\ violation in decay for each flavour-specific state.
Relative phase differences between resonant terms are also extracted.

\mysubsubsubsection{$\Bz \to \pi^+\pi^-\pi^0$
}
\label{sec:cp_uta:notations:dalitz:pipipi0}

The $\Bz \to \pi^+\pi^-\pi^0$ decay is dominated by
intermediate $\rho$ resonances.
Although it is possible, as above,
to directly determine the complex amplitudes for each component,
an alternative approach~\cite{Snyder:1993mx,Quinn:2000by}
has been used by both \babar~\cite{Aubert:2007jn,Lees:2013nwa}
and \belle~\cite{Kusaka:2007dv,Kusaka:2007mj}.
The amplitudes for $\Bz$ and $\Bzb$ decays to $\pi^+\pi^-\pi^0$ are written as
\begin{equation}
  A_{3\pi} = f_+ A_+ + f_- A_- + f_0 A_0
  \, , \ \ \
  \bar{A}_{3\pi} = f_+ \bar{A}_+ + f_- \bar{A}_- + f_0 \bar{A}_0 \, ,
\end{equation}
respectively.
The symbols $A_+$, $A_-$ and $A_0$
represent the complex decay amplitudes for
$\Bz \to \rho^+\pi^-$, $\Bz \to \rho^-\pi^+$ and $\Bz \to \rho^0\pi^0$
while
$\bar{A}_+$, $\bar{A}_-$ and $\bar{A}_0$
represent those for
$\Bzb \to \rho^+\pi^-$, $\Bzb \to \rho^-\pi^+$ and $\Bzb \to \rho^0\pi^0$,
respectively.
The terms $f_+$, $f_-$ and $f_0$ incorporate kinematic and dynamical factors
and depend on the Dalitz plot coordinates.
The full decay-time-dependent distribution can then be written
in terms of 27 free parameters,
one for each coefficient of the form factor bilinears,
as listed in Table~\ref{tab:cp_uta:pipipi0:uandi}.
These parameters are sometimes referred to as ``the $U$s and $I$s'',
and can be expressed in terms of
$A_+$, $A_-$, $A_0$, $\bar{A}_+$, $\bar{A}_-$ and $\bar{A}_0$.
If the full set of parameters is determined,
together with their correlations,
other parameters, such as weak and strong phases,
parameters of $\CP$ violation in decay, \etc,
can be subsequently extracted.
Note that one of the parameters (typically $U_+^+$, the coefficient of $|f_+|^2$) is often fixed to unity to provide a reference; this does not affect the analysis.

\begin{table}[htbp]
  \begin{center}
    \caption{
      Definitions of the $U$ and $I$ coefficients.
      Modified from Ref.~\cite{Aubert:2007jn}.
    }
    \label{tab:cp_uta:pipipi0:uandi}
    \setlength{\tabcolsep}{0.3pc}
    \begin{tabular}{l@{\extracolsep{5mm}}l}
      \hline
      Parameter   & Description \\
      \hline
      $U_+^+$          & Coefficient of $|f_+|^2$ \\
      $U_0^+$          & Coefficient of $|f_0|^2$ \\
      $U_-^+$          & Coefficient of $|f_-|^2$ \\
      [0.15cm]
      $U_0^-$          & Coefficient of $|f_0|^2\cos(\Delta m\Delta t)$ \\
      $U_-^-$          & Coefficient of $|f_-|^2\cos(\Delta m\Delta t)$ \\
      $U_+^-$          & Coefficient of $|f_+|^2\cos(\Delta m\Delta t)$ \\
      [0.15cm]
      $I_0$            & Coefficient of $|f_0|^2\sin(\Delta m\Delta t)$ \\
      $I_-$            & Coefficient of $|f_-|^2\sin(\Delta m\Delta t)$ \\
      $I_+$            & Coefficient of $|f_+|^2\sin(\Delta m\Delta t)$ \\
      [0.15cm]
      $U_{+-}^{+,\Im}$ & Coefficient of $\Im[f_+f_-^*]$ \\
      $U_{+-}^{+,\Re}$ & Coefficient of $\Re[f_+f_-^*]$ \\
      $U_{+-}^{-,\Im}$ & Coefficient of $\Im[f_+f_-^*]\cos(\Delta m\Delta t)$ \\
      $U_{+-}^{-,\Re}$ & Coefficient of $\Re[f_+f_-^*]\cos(\Delta m\Delta t)$ \\
      $I_{+-}^{\Im}$   & Coefficient of $\Im[f_+f_-^*]\sin(\Delta m\Delta t)$ \\
      $I_{+-}^{\Re}$   & Coefficient of $\Re[f_+f_-^*]\sin(\Delta m\Delta t)$ \\
      [0.15cm]
      $U_{+0}^{+,\Im}$ & Coefficient of $\Im[f_+f_0^*]$ \\
      $U_{+0}^{+,\Re}$ & Coefficient of $\Re[f_+f_0^*]$ \\
      $U_{+0}^{-,\Im}$ & Coefficient of $\Im[f_+f_0^*]\cos(\Delta m\Delta t)$ \\
      $U_{+0}^{-,\Re}$ & Coefficient of $\Re[f_+f_0^*]\cos(\Delta m\Delta t)$ \\
      $I_{+0}^{\Im}$   & Coefficient of $\Im[f_+f_0^*]\sin(\Delta m\Delta t)$ \\
      $I_{+0}^{\Re}$   & Coefficient of $\Re[f_+f_0^*]\sin(\Delta m\Delta t)$ \\
      [0.15cm]
      $U_{-0}^{+,\Im}$ & Coefficient of $\Im[f_-f_0^*]$ \\
      $U_{-0}^{+,\Re}$ & Coefficient of $\Re[f_-f_0^*]$ \\
      $U_{-0}^{-,\Im}$ & Coefficient of $\Im[f_-f_0^*]\cos(\Delta m\Delta t)$ \\
      $U_{-0}^{-,\Re}$ & Coefficient of $\Re[f_-f_0^*]\cos(\Delta m\Delta t)$ \\
      $I_{-0}^{\Im}$   & Coefficient of $\Im[f_-f_0^*]\sin(\Delta m\Delta t)$ \\
      $I_{-0}^{\Re}$   & Coefficient of $\Re[f_-f_0^*]\sin(\Delta m\Delta t)$ \\
      \hline
    \end{tabular}
  \end{center}
\end{table}

\mysubsubsection{Time-dependent \CP asymmetries in decays to non-$\CP$ eigenstates
}
\label{sec:cp_uta:notations:non_cp}

Consider a non-$\CP$ eigenstate $f$, and its conjugate $\bar{f}$.
For neutral $\B$ decays to these final states,
there are four amplitudes to consider:
those for $\Bz$ to decay to $f$ and $\bar{f}$
($\Af$ and $\Afbar$, respectively),
and the equivalents for $\Bzb$
($\Abarf$ and $\Abarfbar$).
If $\CP$ is conserved in the decay, then
$\Af = \Abarfbar$ and $\Afbar = \Abarf$.

The decay-time-dependent distributions can be written in many different ways.
Here, we follow Sec.~\ref{sec:cp_uta:notations:cp_eigenstate}
and define $\lambda_f = \frac{q}{p}\frac{\Abarf}{\Af}$ and
$\lambda_{\bar f} = \frac{q}{p}\frac{\Abarfbar}{\Afbar}$.
The time-dependent \CP asymmetries that are sensitive to mixing-induced
$\CP$ violation effects then follow Eq.~(\ref{eq:cp_uta:td_cp_asp}):
\begin{eqnarray}
\label{eq:cp_uta:non-cp-obs}
  {\cal A}_f (\Delta t) \; \equiv \;
  \frac{
    \Gamma_{\Bzb \to f} (\Delta t) - \Gamma_{\Bz \to f} (\Delta t)
  }{
    \Gamma_{\Bzb \to f} (\Delta t) + \Gamma_{\Bz \to f} (\Delta t)
  } & = & S_f \sin(\Delta m \Delta t) - C_f \cos(\Delta m \Delta t), \\
  {\cal A}_{\bar{f}} (\Delta t) \; \equiv \;
  \frac{
    \Gamma_{\Bzb \to \bar{f}} (\Delta t) - \Gamma_{\Bz \to \bar{f}} (\Delta t)
  }{
    \Gamma_{\Bzb \to \bar{f}} (\Delta t) + \Gamma_{\Bz \to \bar{f}} (\Delta t)
  } & = & S_{\bar{f}} \sin(\Delta m \Delta t) - C_{\bar{f}} \cos(\Delta m \Delta t),
\end{eqnarray}
with the definitions of the parameters
$C_f$, $S_f$, $C_{\bar{f}}$ and $S_{\bar{f}}$,
following Eqs.~(\ref{eq:cp_uta:s_def}) and~(\ref{eq:cp_uta:c_def}).

The time-dependent decay rates are given by
\begin{eqnarray}
  \label{eq:cp_uta:non-CP-TD1}
  \Gamma_{\Bzb \to f} (\Delta t) & = &
  \frac{e^{-\left| \Delta t \right| / \tau(\Bz)}}{8\tau(\Bz)}
  ( 1 + \Adirnoncp )
  \left[
    1 + S_f \sin(\Delta m \Delta t) - C_f \cos(\Delta m \Delta t)
  \right],
  \\
  \label{eq:cp_uta:non-CP-TD2}
  \Gamma_{\Bz \to f} (\Delta t) & = &
  \frac{e^{-\left| \Delta t \right| / \tau(\Bz)}}{8\tau(\Bz)}
  ( 1 + \Adirnoncp )
  \left[
    1 - S_f \sin(\Delta m \Delta t) + C_f \cos(\Delta m \Delta t)
  \right],
  \\
  \label{eq:cp_uta:non-CP-TD3}
  \Gamma_{\Bzb \to \bar{f}} (\Delta t) & = &
  \frac{e^{-\left| \Delta t \right| / \tau(\Bz)}}{8\tau(\Bz)}
  ( 1 - \Adirnoncp )
  \left[
    1 + S_{\bar{f}} \sin(\Delta m \Delta t) - C_{\bar{f}} \cos(\Delta m \Delta t)
  \right],
  \\
  \label{eq:cp_uta:non-CP-TD4}
  \Gamma_{\Bz \to \bar{f}} (\Delta t) & = &
    \frac{e^{-\left| \Delta t \right| / \tau(\Bz)}}{8\tau(\Bz)}
  ( 1 - \Adirnoncp )
  \left[
    1 - S_{\bar{f}} \sin(\Delta m \Delta t) + C_{\bar{f}} \cos(\Delta m \Delta t)
  \right],
\end{eqnarray}
where the time-independent parameter \Adirnoncp
represents an overall asymmetry in the production of the
$f$ and $\bar{f}$ final states,\footnote{
  This parameter is often denoted ${\cal A}_f$ (or ${\cal A}_{\CP}$),
  but here we avoid this notation to prevent confusion with the
  time-dependent $\CP$ asymmetry.
}
\begin{equation}
  \Adirnoncp =
  \frac{
    \left(
      \left| \Af \right|^2 + \left| \Abarf \right|^2
    \right) -
    \left(
      \left| \Afbar \right|^2 + \left| \Abarfbar \right|^2
    \right)
  }{
    \left(
      \left| \Af \right|^2 + \left| \Abarf \right|^2
    \right) +
    \left(
      \left| \Afbar \right|^2 + \left| \Abarfbar \right|^2
    \right)
  }.
\end{equation}
Assuming $|q/p| = 1$, \ie\ absence of \CP violation in mixing,
the parameters $C_f$ and $C_{\bar{f}}$
can also be written in terms of the decay amplitudes as
\begin{equation}
  C_f =
  \frac{
    \left| \Af \right|^2 - \left| \Abarf \right|^2
  }{
    \left| \Af \right|^2 + \left| \Abarf \right|^2
  }
  \hspace{5mm}
  {\rm and}
  \hspace{5mm}
  C_{\bar{f}} =
  \frac{
    \left| \Afbar \right|^2 - \left| \Abarfbar \right|^2
  }{
    \left| \Afbar \right|^2 + \left| \Abarfbar \right|^2
  },
\end{equation}
giving rise to asymmetries in the decay amplitudes for the final states $f$ and $\bar{f}$.
In this notation, the conditions for absence of $\CP$ violation in decay are
$\Adirnoncp = 0$ and $C_f = - C_{\bar{f}}$.
Note that $C_f$ and $C_{\bar{f}}$ are typically non-zero;
\eg, for a
flavour-specific final state where $\Abarf = \Afbar = 0$,
they take the values $C_f = - C_{\bar{f}} = 1$.

The coefficients of the sine terms contain information about the weak phase.
In the case that each decay amplitude contains only a single weak phase
(\ie, no $\CP$ violation in decay as well as none in mixing),
these terms can be written as
\begin{equation}
  S_f =
  \frac{
    - 2 \left| \Af \right| \left| \Abarf \right|
    \sin( \phi_{\rm mix} + \phi_{\rm dec} - \delta_f )
  }{
    \left| \Af \right|^2 + \left| \Abarf \right|^2
  }
  \hspace{5mm}
  {\rm and}
  \hspace{5mm}
  S_{\bar{f}} =
  \frac{
    - 2 \left| \Afbar \right| \left| \Abarfbar \right|
    \sin( \phi_{\rm mix} + \phi_{\rm dec} + \delta_f )
  }{
    \left| \Afbar \right|^2 + \left| \Abarfbar \right|^2
  },
\end{equation}
where $\delta_f$ is the strong phase difference between the decay amplitudes.
If there is no $\CP$ violation, the condition $S_f = - S_{\bar{f}}$ holds.
If decay amplitudes with different weak and strong phases contribute,
no straightforward interpretation of $S_f$ and $S_{\bar{f}}$ is possible.

The conditions for $\CP$ invariance $C_f = - C_{\bar{f}}$ and $S_f = - S_{\bar{f}}$ motivate a rotation of the parameters:
\begin{equation}
\label{eq:cp_uta:non-cp-s_and_deltas}
  S_{f\bar{f}} = \frac{S_{f} + S_{\bar{f}}}{2},
  \hspace{4mm}
  \Delta S_{f\bar{f}} = \frac{S_{f} - S_{\bar{f}}}{2},
  \hspace{4mm}
  C_{f\bar{f}} = \frac{C_{f} + C_{\bar{f}}}{2},
  \hspace{4mm}
  \Delta C_{f\bar{f}} = \frac{C_{f} - C_{\bar{f}}}{2}.
\end{equation}
With these parameters, the $\CP$ invariance conditions become
$S_{f\bar{f}} = 0$ and $C_{f\bar{f}} = 0$.
The parameter $\Delta C_{f\bar{f}}$ gives a measure of the ``flavour-specificity''
of the decay:
$\Delta C_{f\bar{f}}=\pm1$ corresponds to a completely flavour-specific decay,
in which no interference between decays with and without mixing can occur,
while $\Delta C_{f\bar{f}} = 0$ results in
maximum sensitivity to mixing-induced $\CP$ violation.
The parameter $\Delta S_{f\bar{f}}$ is related to the strong phase difference
between the decay amplitudes of the $\Bz$ meson to the $f$ and to $\bar f$ final states.
We note that the observables of Eq.~(\ref{eq:cp_uta:non-cp-s_and_deltas})
exhibit experimental correlations
(typically of $\sim 20\%$, depending on the tagging purity, and other effects)
between $S_{f\bar{f}}$ and  $\Delta S_{f\bar{f}}$,
and between $C_{f\bar{f}}$ and $\Delta C_{f\bar{f}}$.
On the other hand,
the final-state-specific observables of Eqs.~(\ref{eq:cp_uta:non-CP-TD1})--(\ref{eq:cp_uta:non-CP-TD4}) tend to have low correlations.

Alternatively, if we recall that the $\CP$ invariance
conditions at the decay amplitude level are
$\Af = \Abarfbar$ and $\Afbar = \Abarf$,
we are led to consider the parameters~\cite{Charles:2004jd}
\begin{equation}
  \label{eq:cp_uta:non-cp-directcp}
  {\cal A}_{f\bar{f}} =
  \frac{
    \left| \Abarfbar \right|^2 - \left| \Af \right|^2
  }{
    \left| \Abarfbar \right|^2 + \left| \Af \right|^2
  }
  \hspace{5mm}
  {\rm and}
  \hspace{5mm}
  {\cal A}_{\bar{f}f} =
  \frac{
    \left| \Abarf \right|^2 - \left| \Afbar \right|^2
  }{
    \left| \Abarf \right|^2 + \left| \Afbar \right|^2
  }.
\end{equation}
These are sometimes considered more physically intuitive parameters,
since they characterise $\CP$ violation in decay
in decays with particular topologies.
For example, in the case of $\Bz \to \rho^\pm\pi^\mp$
(choosing $f =  \rho^+\pi^-$ and $\bar{f} = \rho^-\pi^+$),
${\cal A}_{f\bar{f}}$ (also denoted ${\cal A}^{+-}_{\rho\pi}$)
parametrises $\CP$ violation
in decays in which the produced $\rho$ meson does not contain the
spectator quark,
while ${\cal A}_{\bar{f}f}$ (also denoted ${\cal A}^{-+}_{\rho\pi}$)
parametrises $\CP$ violation in decays in which it does.
Note that we have again followed the sign convention that the asymmetry
is the difference between the rate involving a $b$ quark and that
involving a $\bar{b}$ quark, \cf\ Eq.~(\ref{eq:cp_uta:pra}).
Of course, these parameters are not independent of the
other sets of parameters given above, and can be written as
\begin{equation}
  {\cal A}_{f\bar{f}} =
  - \frac{
    \Adirnoncp + C_{f\bar{f}} + \Adirnoncp \Delta C_{f\bar{f}}
  }{
    1 + \Delta C_{f\bar{f}} + \Adirnoncp C_{f\bar{f}}
  }
  \hspace{5mm}
  {\rm and}
  \hspace{5mm}
  {\cal A}_{\bar{f}f} =
  \frac{
    - \Adirnoncp + C_{f\bar{f}} + \Adirnoncp \Delta C_{f\bar{f}}
  }{
    - 1 + \Delta C_{f\bar{f}} + \Adirnoncp C_{f\bar{f}}
  }.
\end{equation}
They usually exhibit strong correlations.

We now consider the various notations used in experimental studies of
time-dependent $\CP$ asymmetries in decays to non-$\CP$ eigenstates.

\mysubsubsubsection{$\Bz \to D^{*\pm}D^\mp$
}
\label{sec:cp_uta:notations:non_cp:dstard}

The ($\Adirnoncp$, $C_f$, $S_f$, $C_{\bar{f}}$, $S_{\bar{f}}$)
set of parameters was used in early publications by both \babar~\cite{Aubert:2007pa} and \belle~\cite{Aushev:2004uc} (albeit with slightly different notations), with $f = D^{*+}D^-$, $\bar{f} = D^{*-}D^+$.
In a more recent paper on this topic, \belle~\cite{Rohrken:2012ta} instead uses the parametrisation ($A_{D^*D}$, $S_{D^*D}$, $\Delta S_{D^*D}$, $C_{D^*D}$, $\Delta C_{D^*D}$), while \babar~\cite{Aubert:2008ah} gives results in both sets of parameters.
We therefore use the ($A_{D^*D}$, $S_{D^*D}$, $\Delta S_{D^*D}$, $C_{D^*D}$, $\Delta C_{D^*D}$) set.

\mysubsubsubsection{$\Bz \to \rho^{\pm}\pi^\mp$
}
\label{sec:cp_uta:notations:non_cp:rhopi}

In the $\rho^\pm\pi^\mp$ system, the
($\Adirnoncp$, $C_{f\bar{f}}$, $S_{f\bar{f}}$, $\Delta C_{f\bar{f}}$,
$\Delta S_{f\bar{f}}$)
set of parameters was originally used by \babar~\cite{Aubert:2003wr} and \belle~\cite{Wang:2004va} in the Q2B approximation;
the exact names\footnote{
  \babar\ has used the notations
  $A_{\CP}^{\rho\pi}$~\cite{Aubert:2003wr} and
  ${\cal A}_{\rho\pi}$~\cite{Aubert:2007jn}
  in place of ${\cal A}_{\CP}^{\rho\pi}$.
}
used in this case were
$\left(
  {\cal A}_{\CP}^{\rho\pi}, C_{\rho\pi}, S_{\rho\pi}, \Delta C_{\rho\pi}, \Delta S_{\rho\pi}
\right)$,
and these names are also used in this document.

Since $\rho^\pm\pi^\mp$ is reconstructed in the final state $\pi^+\pi^-\pi^0$,
the interference between the $\rho$ resonances
can provide additional information about the phases
(see Sec.~\ref{sec:cp_uta:notations:dalitz}).
Both \babar~\cite{Aubert:2007jn}
and \belle~\cite{Kusaka:2007dv,Kusaka:2007mj}
have performed time-dependent Dalitz-plot analyses,
from which the weak phase $\alpha$ is directly extracted.
In such an analysis, the measured Q2B parameters are
also naturally corrected for interference effects.

\mysubsubsubsection{$\Bz \to D^{\mp}\pi^{\pm}, D^{*\mp}\pi^{\pm}, D^{\mp}\rho^{\pm}$
}
\label{sec:cp_uta:notations:non_cp:dstarpi}

Time-dependent $\CP$ analyses have also been performed for the
final states $D^{\mp}\pi^{\pm}$, $D^{*\mp}\pi^{\pm}$ and $D^{\mp}\rho^{\pm}$.
In these theoretically clean cases, no penguin contributions are possible,
so there is no $\CP$ violation in decay.
Furthermore, due to the smallness of the ratio of the magnitudes of the
suppressed ($b \to u$) and favoured ($b \to c$) amplitudes (denoted $R_f$),
to a very good approximation, $C_f = - C_{\bar{f}} = 1$
(using $f = D^{(*)-}h^+$, $\bar{f} = D^{(*)+}h^-$, $h = \pi,\rho$),
and the coefficients of the sine terms are given by
\begin{equation}
  S_f = - 2 R_f \sin( \phi_{\rm mix} + \phi_{\rm dec} - \delta_f )
  \quad
  \text{and}
  \quad
  S_{\bar{f}} = - 2 R_f \sin( \phi_{\rm mix} + \phi_{\rm dec} + \delta_f ).
\end{equation}
Thus, weak phase information can be obtained from measurements of $S_f$ and $S_{\bar{f}}$,
although external information on at least one of $R_f$ or $\delta_f$ is necessary,
constituting a source of theoretical uncertainty.
Note that $\phi_{\rm mix} + \phi_{\rm dec} = 2\beta + \gamma \equiv 2\phi_1 + \phi_3$ for all the decay modes in question, while $R_f$ and $\delta_f$ depend on the decay mode.

Again, different notations have been used in the literature.
\babar~\cite{Aubert:2006tw,Aubert:2005yf}
defines the time-dependent probability function by
\begin{equation}
  f^\pm (\eta, \Delta t) = \frac{e^{-|\Delta t|/\tau}}{4\tau}
  \left[
    1 \mp S_\zeta \sin (\Delta m \Delta t) \mp \eta C_\zeta \cos(\Delta m \Delta t)
  \right],
\end{equation}
where the upper (lower) sign corresponds to the tagging meson being a $\Bz$ ($\Bzb$).
The parameter $\eta$ takes the value $+1$ ($-1$) and $\zeta$ denotes $+$ ($-$)
when the final state is, \eg, $D^-\pi^+$ ($D^+\pi^-$).
However, in the fit, the substitutions $C_\zeta = 1$ and
$S_\zeta = a \mp \eta b_i - \eta c_i$ are made, where the subscript $i$ denotes the flavour tagging category.
These are motivated by the possibility of $\CP$ violation on the tag side~\cite{Long:2003wq}.
The parameter $a$ is not affected by tag-side $\CP$ violation.
The parameter $b$ only depends on tag-side $\CP$ violation parameters
and is not directly useful for determining UT angles.
A clean interpretation of the $c$ parameter is only possible for lepton-tagged events, which are not affected by tag-side $\CP$ violation effects,
so the \babar\ measurements report $c$ measured with those events only.
Neglecting $b$ terms,
\begin{equation}
  \label{eq:cp_uta:aandc}
    S_+ = a - c \quad \text{and} \quad S_- = a + c \ \Leftrightarrow \
    a = (S_+ + S_-)/2 \quad \text{and} \quad c = (S_- - S_+)/2 \, ,
\end{equation}
in analogy to the parameters of Eq.~(\ref{eq:cp_uta:non-cp-s_and_deltas}).

The parameters used by \belle\ in the analysis using
partially reconstructed $\B$ decays~\cite{Bahinipati:2011yq},
are similar to the $S_\zeta$ parameters defined above.
However, in the \belle\ convention,
a tagging $\Bz$ corresponds to a $+$ sign in front of the sine coefficient;
furthermore the correspondence between the super/subscript
and the final state is opposite, so that $S_\pm$ (\babar) = $- S^\mp$ (\belle).
In this analysis, only lepton tags are used,
so there is no effect from tag-side $\CP$ violation.
In the \belle\ analysis that used
fully reconstructed $\B$ decays~\cite{Ronga:2006hv},
this effect is measured and taken into account using $\Dstar \ell \nu$ decays;
in neither \belle\ analysis are the $a$, $b$ and $c$ parameters used.
The parameters measured by \belle\ are
$2 R_{D^{(*)}\pi} \sin( 2\phi_1 + \phi_3 \pm \delta_{D^{(*)}\pi} )$;
the definition is such that
$S^\pm (\text{\belle}) = - 2 R_{\Dstar \pi} \sin( 2\phi_1 + \phi_3 \pm \delta_{\Dstar \pi} )$.
This definition includes an angular momentum factor $(-1)^L$~\cite{Fleischer:2003yb},
and so for the results in the $D\pi$ system,
there is an additional factor of $-1$ in the conversion.

LHCb has also measured the parameters of $\Bz \to \Dmp\pipm$ decays~\cite{Aaij:2018kpq}.
The convention used is essentially the same as Belle, but with the notation $(S_f, S_{\bar{f}}) = (S_-, S_+)$.
For the averages in this document, we use the $a$ and $c$ parameters.
Correlations are taken into account in the LHCb case, where significant correlations are reported.
Explicitly, the conversion reads: $a = -(S_+ + S_-)/2$, $c = -(S_+ - S_-)/2$.

\mysubsubsubsection{$\Bs \to D_s^{\mp}K^\pm$}
\label{sec:cp_uta:notations:non_cp:dsk}

The phenomenology of $\Bs \to D_s^{\mp}K^\pm$ decays is similar to that of $\Bz \to D^{\mp}\pi^{\pm}$, with some important caveats.
The two amplitudes for $b \to u$ and $b \to c$ transitions have the same level of Cabibbo-suppression (\ie\ are of the same order in $\lambda$) though the former is suppressed by $\sqrt{\rho^2+\eta^2}$.
The large value of the ratio $R$ of their magnitudes allows it to be determined from data, as the deviation of $|C_f|$ and $|C_{\bar{f}}|$ from unity can be observed.
Moreover, the non-zero value of $\Delta \Gamma_s$ allows the determination of additional terms, $A^{\Delta\Gamma}_f$ and $A^{\Delta\Gamma}_{\bar{f}}$ (see Sec.~\ref{sec:cp_uta:notations:Bs}), that break ambiguities in the solutions for $\phi_{\rm mix} + \phi_{\rm dec}$, which for $\Bs \to D_s^{\mp}K^\pm$ decays is equal to $\gamma-2\beta_s$.

LHCb~\cite{Aaij:2014fba,Aaij:2017lff} has performed such an analysis with $\Bs \to D_s^{\mp}K^\pm$ decays.
The absence of \CP violation in decay was assumed, and the parameters determined from the fit were labelled $C$, $A^{\Delta\Gamma}$, $\bar{A}{}^{\Delta\Gamma}$, $S$, $\bar{S}$.
These are trivially related to the definitions used in this section.

\mysubsubsubsection{Time-dependent asymmetries in radiative $\B$ decays
}
\label{sec:cp_uta:notations:non_cp:radiative}

As a special case of decays to non-$\CP$ eigenstates,
let us consider radiative $\B$ decays.
Here, the emitted photon has a distinct helicity,
which is in principle observable, but in practice is not usually measured.
Thus, the measured time-dependent decay rates for
neutral $B$ meson decays are given by sums of the expressions of Eqs.~(\ref{eq:cp_uta:non-CP-TD1})--(\ref{eq:cp_uta:non-CP-TD4}) for the final states with left-handed ($\gamma_L$) and right-handed ($\gamma_R$) photon helicity ~\cite{Atwood:1997zr,Atwood:2004jj}
\begin{eqnarray}
  \label{eq:cp_uta:non-cp-radiative1}
  \Gamma_{\Bzb \to X \gamma} (\Delta t) & = &
  \Gamma_{\Bzb \to X \gamma_L} (\Delta t) + \Gamma_{\Bzb \to X \gamma_R} (\Delta t) \\ \nonumber
  & = &
  \frac{e^{-\left| \Delta t \right| / \tau(\Bz)}}{4\tau(\Bz)}
  \left[
    1 +
    \left( S_L + S_R \right) \sin(\Delta m \Delta t) -
    \left( C_L + C_R \right) \cos(\Delta m \Delta t)
  \right]\,,
  \\
  \label{eq:cp_uta:non-cp-radiative2}
  \Gamma_{\Bz \to X \gamma} (\Delta t) & = &
  \Gamma_{\Bz \to X \gamma_L} (\Delta t) + \Gamma_{\Bz \to X \gamma_R} (\Delta t) \\ \nonumber
  & = &
  \frac{e^{-\left| \Delta t \right| / \tau(\Bz)}}{4\tau(\Bz)}
  \left[
    1 -
    \left( S_L + S_R \right) \sin(\Delta m \Delta t) +
    \left( C_L + C_R \right) \cos(\Delta m \Delta t)
  \right]\,.
\end{eqnarray}
Here, in place of the subscripts $f$ and $\bar{f}$, we have used $L$ and $R$
to indicate the photon helicity.
In order for interference between decays with and without $\Bz$-$\Bzb$ mixing
to occur, the $X$ system must not be flavour-specific,
\eg, in the case of $\Bz \to K^{*0}\gamma$, the final state must be $\KS \pi^0 \gamma$.
The sign of the sine term depends on the $C$ eigenvalue of the $X$ system.
At leading order, the photons from
$b \to q \gamma$ ($\bar{b} \to \bar{q} \gamma$) are predominantly
left (right) polarised, with corrections of order of $m_q/m_b$,
and thus interference effects are suppressed.
Higher-order effects can lead to corrections of order
$\Lambda_{\rm QCD}/m_b$~\cite{Grinstein:2004uu,Grinstein:2005nu},
although explicit calculations indicate that such corrections may be small
for exclusive final states~\cite{Matsumori:2005ax,Ball:2006cva}.
The predicted smallness of the $S$ terms in the Standard Model
results in sensitivity to new physics contributions.

The formalism discussed above is valid for any radiative decay to a final state where the hadronic system is an eigenstate of $C$.
In addition to $\KS\piz\gamma$, experiments have presented results using $\Bz$ decays to $\KS\eta\gamma$, $\KS\rho^0\gamma$ and $\KS\phi\gamma$.
For the case of the $\KS\rho^0\gamma$ final state, particular care is needed, as due to the non-negligible width of the $\rho^0$ meson, decays selected as $\Bz \to \KS\rho^0\gamma$ can include a significant contribution from $K^{*\pm}\pimp\gamma$ decays, which are flavour-specific and do not have the same oscillation phenomenology.
It is therefore necessary to correct the fitted asymmetry parameter for a ``dilution factor''.

In the case of radiative \Bs\ decays, the time-dependent decay rates of Eqs.~(\ref{eq:cp_uta:non-cp-radiative1}) and~(\ref{eq:cp_uta:non-cp-radiative2}) must be modified, in a similar way to that discussed in Sec.~\ref{sec:cp_uta:notations:Bs}, to account for the non-zero value of \DGs.
Thus, for decays such as $\Bs\to\phi\gamma$, there is an additional observable, $A^{\Delta \Gamma}_{\phi\gamma}$, which can be determined from an untagged effective lifetime measurement~\cite{Muheim:2008vu}.

\mysubsubsection{Asymmetries in $\B \to \DorDstar K^{(*)}$ decays
}
\label{sec:cp_uta:notations:cus}

$\CP$ asymmetries in $\B \to \DorDstar K^{(*)}$ decays are sensitive to $\gamma$.
The neutral $D^{(*)}$ meson produced
is an admixture of $\DorDstarz$ (produced by a $b \to c$ transition) and
$\DorDstarzb$ (produced by a colour-suppressed $b \to u$ transition) states.
If the final state is chosen so that both $\DorDstarz$ and $\DorDstarzb$
can contribute, the two amplitudes interfere,
and the resulting observables are sensitive to $\gamma$,
the relative weak phase between
the two $\B$ decay amplitudes~\cite{Bigi:1988ym}.
Various methods have been proposed to exploit this interference,
including those where the neutral $D$ meson is reconstructed
as a $\CP$ eigenstate (GLW)~\cite{Gronau:1990ra,Gronau:1991dp},
in a suppressed final state (ADS)~\cite{Atwood:1996ci,Atwood:2000ck},
or in a self-conjugate three-body final state,
such as $\KS \pi^+\pi^-$ (BPGGSZ or Dalitz)~\cite{Giri:2003ty,Poluektov:2004mf}.
While each method differs in the choice of $D$ decay,
they are all sensitive to the same parameters of the $B$ decay,
and can be considered as variations of the same technique.

Consider the case of $\Bmp \to D \Kmp$, with $D$ decaying to a final state $f$, which is accessible from both $\Dz$ and $\Dzb$.
We can write the decay rates $\Gamma_\mp$ for $\Bm$ and $\Bp$,
the charge averaged rate $\Gamma = (\Gamma_- + \Gamma_+)/2$,
and the charge asymmetry
$A = (\Gamma_- - \Gamma_+)/(\Gamma_- + \Gamma_+)$ (see Eq.~(\ref{eq:cp_uta:pra})) as
\begin{eqnarray}
  \label{eq:cp_uta:dk:rate_def}
  \Gamma_\mp  & \propto &
  r_B^2 + r_D^2 + 2 r_B r_D \cos \left( \delta_B + \delta_D \mp \gamma \right), \\
  \label{eq:cp_uta:dk:av_rate_def}
  \Gamma & \propto &
  r_B^2 + r_D^2 + 2 r_B r_D \cos \left( \delta_B + \delta_D \right) \cos \left( \gamma \right), \\
  \label{eq:cp_uta:dk:acp_def}
  A & = &
  \frac{
    2 r_B r_D \sin \left( \delta_B + \delta_D \right) \sin \left( \gamma \right)
  }{
    r_B^2 + r_D^2 + 2 r_B r_D \cos \left( \delta_B + \delta_D \right) \cos \left( \gamma \right)
  },
\end{eqnarray}
where the ratio of $\B$ decay amplitudes\footnote{
  Note that here we use the notation $r_B$ to denote the ratio
  of $\B$ decay amplitudes,
  whereas in Sec.~\ref{sec:cp_uta:notations:non_cp:dstarpi}
  we used, \eg, $R_{D\pi}$, for a rather similar quantity.
  The reason is that here we need to be concerned also with
  $D$ decay amplitudes,
  and so it is convenient to use the subscript to denote the decaying particle.
  Hopefully, using $r$ in place of $R$ will reduce the potential for confusion.
}
\begin{equation}
  \label{eq:cp_uta:dk:rb_def}
  r_B =
  \left|
    \frac{A\left( \Bm \to \Dzb K^- \right)}{A\left( \Bm \to \Dz  K^- \right)}
  \right| =
  \left|
    \frac{A\left( \Bp \to \Dz K^+ \right)}{A\left( \Bp \to \Dzb  K^+ \right)}
  \right| ,
\end{equation}

is usually defined to be less than one,
and the ratio of $D$ decay amplitudes is correspondingly defined by
\begin{equation}
  \label{eq:cp_uta:dk:rd_def}
  r_D =
  \left|
    \frac{A\left( \Dz  \to f \right)}{A\left( \Dzb \to f \right)}
  \right| .
\end{equation}
The relation between $\Bm$ and $\Bp$ amplitudes given in Eq.~(\ref{eq:cp_uta:dk:rb_def}) is a result of there being only one weak phase contributing to each amplitude in the Standard Model, which is the source of the theoretical cleanliness of this approach for measuring $\gamma$~\cite{Brod:2013sga}.
The strong phase differences between the $\B$ and $D$ decay amplitudes
are denoted by $\delta_B$ and $\delta_D$, respectively.
The values of $r_D$ and $\delta_D$ depend on the final state $f$:
for the GLW analysis, $r_D = 1$ and $\delta_D$ is trivial (either zero or $\pi$);
for other modes, values of $r_D$ and $\delta_D$ are not trivial, and for multibody final states they vary across the phase space.
This can be quantified either by an explicit $D$ decay amplitude model or
by model-independent information.
In the case that the multibody final state is treated inclusively, the formalism is modified by the inclusion of a coherence factor, usually denoted $\kappa$, while $r_D$ and $\delta_D$ become effective parameters corresponding to amplitude-weighted averages across the phase space.

Note that, for given values of $r_B$ and $r_D$,
the maximum size of $A$ (at $\sin \left( \delta_B + \delta_D \right) = 1$)
is $2 r_B r_D \sin \left( \gamma \right) / \left( r_B^2 + r_D^2 \right)$.
Thus, even for $D$ decay modes with small $r_D$,
large asymmetries, and hence sensitivity to $\gamma$,
may occur for $B$ decay modes with similar
values of $r_B$.
For this reason, the ADS analysis of the decay $B^\mp \to D \pi^\mp$ is also of interest.

The expressions of Eq.~(\ref{eq:cp_uta:dk:rate_def})--(\ref{eq:cp_uta:dk:rd_def}) are for a specific point in phase space, and therefore are relevant where both $B$ and $D$ decays are to two-body final states.
Additional coherence factors enter the expressions when the $B$ decay is to a multibody final state (further discussion of multibody $D$ decays can be found below).
In particular, experiments have studied $B^+ \to DK^*(892)^+$, $B^0 \to DK^*(892)^0$ and $B^+ \to DK^+\pi^+\pi^-$ decays.
Considering, for concreteness, the $B \to DK^*(892)$ case, the non-negligible width of the $K^*(892)$ resonance implies that contributions from other $B \to DK\pi$ decays can pass the selection requirements.
Their effect on the Q2B analysis can be accounted for with a coherence factor~\cite{Gronau:2002mu}, usually denoted $\kappa$, which tends to unity in the limit that the $K^*(892)$ resonance is the only signal amplitude contributing in the selected region of phase space.
In this case, the hadronic parameters $r_B$ and $\delta_B$ become effectively weighted averages across the selected phase space of the magnitude ratio and relative strong phase between the CKM-suppressed and -favoured amplitudes; these effective parameters are denoted $\bar{r}_B$ and $\bar{\delta}_B$ (the notations $r_s$, $\delta_s$ and $r_S$, $\delta_S$ are also found in the literature).
An alternative, and in certain cases more advantageous, approach is Dalitz plot analysis of the full $B \to DK\pi$ phase space~\cite{Aleksan:2002mh,Gershon:2008pe,Gershon:2009qc}.

We now consider the various notations used in experimental studies of
$\CP$ asymmetries in $\B \to \DorDstar K^{(*)}$ decays.
To simplify the notation the $\Bp \to D\Kp$ decay is considered; the extension to other modes mediated by the same quark-level transitions is straightforward.

\mysubsubsubsection{$\B \to \DorDstar K^{(*)}$ with $D \to$ \CP\ eigenstate decays
}
\label{sec:cp_uta:notations:cus:glw}

In the GLW analysis, the measured quantities are the
partial rate asymmetry
\begin{equation}
  \label{eq:cp_uta:dk:glw-adef}
  A_{\CP} = \frac{
    \Gamma\left(\Bm\to D_{\CP}\Km\right) - \Gamma\left(\Bp\to D_{\CP}\Kp\right)
  }{
    \Gamma\left(\Bm\to D_{\CP}\Km\right) + \Gamma\left(\Bp\to D_{\CP}\Kp\right)
  } \,
\end{equation}
and the charge-averaged rate
\begin{equation}
  \label{eq:cp_uta:dk:glw-rdef}
  R_{\CP} =
  \frac{2 \, \Gamma \left( \Bp \to D_{\CP} \Kp  \right)}
  {\Gamma\left( \Bp \to \Dzb \Kp \right)} \, ,
\end{equation}
which are measured for $D$ decays to both $\CP$-even and $\CP$-odd final states.
It is often experimentally convenient to measure $R_{\CP}$ using a double ratio,
\begin{equation}
  \label{eq:cp_uta:dk:double_ratio}
  R_{\CP} =
  \frac{
    \Gamma\left( \Bp \to D_{\CP} \Kp  \right) \, / \, \Gamma\left( \Bp \to \Dzb \Kp \right)
  }{
    \Gamma\left( \Bp \to D_{\CP} \pip \right) \, / \, \Gamma\left( \Bp \to \Dzb \pip \right)
  }
\end{equation}
that is normalised both to the rate for the favoured $\Dzb \to \Kp\pim$ decay,
and to the equivalent quantities for $\Bp \to D\pip$ decays
(charge conjugate processes are implicitly included in
Eqs.~(\ref{eq:cp_uta:dk:glw-rdef}) and~(\ref{eq:cp_uta:dk:double_ratio})).
In this way the constant of proportionality drops out of
Eq.~(\ref{eq:cp_uta:dk:av_rate_def}).
Eq.~(\ref{eq:cp_uta:dk:double_ratio}) is exact in the limit that the
contribution of the $b \to u$ decay amplitude to $\Bp \to D \pip$ vanishes and
when the flavour-specific rates $\Gamma\left( \Bp \to \Dzb h^+ \right)$ ($h =
\pi,K$) are determined using appropriately flavour-specific $D$ decays.
In reality, the Cabibbo-favoured $D \to K\pi$ decay
is used, leading to a small source of systematic uncertainty.

\mysubsubsubsection{$\B \to \DorDstar K^{(*)}$ with $D \to$ non-\CP\ eigenstate two-body decays
}
\label{sec:cp_uta:notations:cus:ads}

For the ADS analysis, which is based on a suppressed $D \to f$ decay,
the measured quantities are again the partial rate asymmetry
and the charge-averaged rate.
In this case it is sufficient to measure the rate in a single ratio
(normalised to the favoured $D \to \bar{f}$ decay)
since potential systematic uncertainties related to detection cancel naturally;
the observed charge-averaged rate is then
\begin{equation}
  \label{eq:cp_uta:dk:r_ads}
  R_{\rm ADS} =
  \frac{
    \Gamma \left( \Bm \to \left[\,f\,\right]_D \Km \right) +
    \Gamma \left( \Bp \to \left[\,\bar{f}\,\right]_D \Kp \right)
  }{
    \Gamma \left( \Bm \to \left[\,\bar{f}\,\right]_D \Km \right) +
    \Gamma \left( \Bp \to \left[\,f\,\right]_D \Kp \right)
  } \, ,
\end{equation}
where the inclusion of charge-conjugate modes has been made explicit.
The \CP\ asymmetry is defined as
\begin{equation}
  \label{eq:cp_uta:dk:a_ads}
  A_{\rm ADS} =
  \frac{
    \Gamma\left(\Bm\to\left[\,f\,\right]_D\Km\right)-
    \Gamma\left(\Bp\to\left[\,f\,\right]_D\Kp\right)
  }{
    \Gamma\left(\Bm\to\left[\,f\,\right]_D\Km\right)+
    \Gamma\left(\Bp\to\left[\,f\,\right]_D\Kp\right)
  } \, .
\end{equation}
Since the uncertainty of $A_{\rm ADS}$ depends on the central value of $R_{\rm ADS}$, for some statistical treatments it is preferable to use an alternative pair of parameters~\cite{Bondar:2004bi}
\begin{equation}
  R_- = \frac{
    \Gamma \left( \Bm \to \left[\,f\,\right]_D \Km \right)
  }{
    \Gamma \left( \Bm \to \left[\,\bar{f}\,\right]_D \Km \right)
  } \,
  \hspace{5mm}
  R_+ = \frac{
    \Gamma \left( \Bp \to \left[\,\bar{f}\,\right]_D \Kp \right)
  }{
    \Gamma \left( \Bp \to \left[\,f\,\right]_D \Kp \right)
  } \, ,
\end{equation}
where there is no implied inclusion of charge-conjugate processes.
These parameters are statistically uncorrelated but may be affected by common sources of systematic uncertainty.
We use the $(R_{\rm ADS}, A_{\rm ADS})$ set in our compilation where available.

In the ADS analysis, there are two additional unknowns ($r_D$ and $\delta_D$) compared to the GLW case.
Additional constraints are therefore required in order to obtain sensitivity to $\gamma$.
Generally, one needs access to two different linear admixtures of $\Dz$ and $\Dzb$ states in order to determine the relative phase: one such sample can be flavour tagged $D$ mesons, which are available in abundant quantities in many experiments; the other can be \CP-tagged $D$ mesons from $\psi(3770)$ decays,
or a superposition of $\Dz$ and $\Dzb$ from $D^0$--$\Dzb$ mixing or from production in $B \to DK$ decays.
In fact, the most precise information on both $r_D$ and $\delta_D$ for $D \to K\pi$ currently comes from global fits to charm mixing data, as discussed in Sec.~\ref{sec:charm:mixcpv}.

The relation of $A_{\rm ADS}$ to the underlying parameters given in Eq.~(\ref{eq:cp_uta:dk:acp_def}) and Table~\ref{tab:cp_uta:notations:dk} is exact for a two-body $D$ decay.
For multibody decays, a similar formalism can be used with the introduction of a coherence factor~\cite{Atwood:2003mj}.
This is most appropriate for doubly-Cabibbo-suppressed decays to non-self-conjugate final states, but can also be modified for use with singly-Cabibbo-suppressed decays~\cite{Grossman:2002aq}.
For multibody self-conjugate final states, such as $\KS\pi^+\pi^-$, a Dalitz plot analysis (discussed below) is often more appropriate.
However, in certain cases where the final state can be approximated as a \CP\ eigenstate, a modified version of the GLW formalism can be used~\cite{Nayak:2014tea}.
In such cases the observables are denoted $A_{\rm qGLW}$ and $R_{\rm qGLW}$ to indicate that the final state is not a pure \CP eigenstate.

\mysubsubsubsection{$\B \to \DorDstar K^{(*)}$ with $D \to$ multibody final state decays
}
\label{sec:cp_uta:notations:cus:ggsz}

In the model-dependent Dalitz-plot (or BPGGSZ) analysis of $D$ decays to multibody self-conjugate final states, the values of $r_D$ and $\delta_D$ across the Dalitz plot are given by an amplitude model (with parameters typically obtained from data).
A simultaneous fit to the $B^+$ and $B^-$ samples can then be used to obtain $\gamma$, $r_B$ and $\delta_B$ directly.
The uncertainties on the phases depend approximately inversely on $r_B$, which is positive definite and therefore tends to be overestimated leading to an underestimation of the uncertainty on $\gamma$ that must be corrected statistically (unless $\sigma(r_B) \ll r_B$).
An alternative approach is to fit for the ``Cartesian'' variables
\begin{equation}
  \label{eq:cp_uta:cartesian}
  \left( x_\pm, y_\pm \right) =
  \left( \Re(r_B e^{i(\delta_B\pm\gamma)}), \Im(r_B e^{i(\delta_B\pm\gamma)}) \right) =
  \left( r_B \cos(\delta_B\pm\gamma), r_B \sin(\delta_B\pm\gamma) \right).
\end{equation}
These variables tend to be statistically well-behaved, and are therefore appropriate for combination of results obtained from independent $B^\pm$ data samples.

The assumption of a model for the $D$ decay leads to a non-negligible, and hard to quantify, source of uncertainty.
To obviate this, it is possible to use instead a model-independent approach, in which the Dalitz plot (or, more generally, the phase space) is binned~\cite{Giri:2003ty,Bondar:2005ki,Bondar:2008hh}.
In this case, hadronic parameters describing the average strong phase difference in each bin between the interfering decay amplitudes enter the equations.
These parameters can
be determined from interference effects in decays of quantum-correlated $D\bar{D}$ pairs produced at the $\psi(3770)$ resonance.
Measurements of such parameters have been made for several hadronic $D$ decays by CLEO-c and BESIII.

When a multibody $D$ decay is dominated by one $\CP$ state, additional sensitivity to $\gamma$ is obtained from the relative widths of the $\Bp \to D\Kp$ and $\Bm \to D\Km$ decays.
This can be taken into account in various ways.
One possibility is to perform a GLW-like analysis, as mentioned above.
An alternative approach proceeds by defining
\begin{equation}
z_\pm = x_\pm + i y_\pm \, , \ \ \
x_0 = - \int \Re \left[ f(s_1,s_2)f^*(s_2,s_1) \right] ds_1ds_2\, ,
\end{equation}
where $s_1, s_2$ are the coordinates of invariant mass squared that
define the Dalitz plot and $f$ is the complex amplitude for $D$ decay
as a function of the Dalitz plot coordinates.\footnote{
  The $x_0$ parameter gives a model-dependent measure of the net \CP content of the final state~\cite{Nayak:2014tea,Gershon:2015xra}.
  It is closely related to the $c_i$ parameters of the model dependent Dalitz plot analysis~\cite{Giri:2003ty,Bondar:2005ki,Bondar:2008hh},
  and the coherence factor of inclusive ADS-type analyses~\cite{Atwood:2003mj}, integrated over the entire Dalitz plot.
}
The fitted parameters ($\rho^\pm, \theta^\pm$) are then defined by
\begin{equation}
  \rho^\pm e^{i \theta^\pm} = z_\pm - x_0 \, .
\end{equation}
Note that the yields of $B^\pm$ decays are proportional
to $1 + (\rho^\pm)^2 - (x_0)^2$.
This choice of variables has been used by \babar\ in the analysis of
$\Bp \to D\Kp$ with $D \to \pi^+\pi^-\pi^0$~\cite{Aubert:2007ii};
for this $D$ decay, and with the assumed amplitude model, a value of $x_0 = 0.850$ is obtained.

The relations between the measured quantities and the
underlying parameters are summarised in Table~\ref{tab:cp_uta:notations:dk}.
It must be emphasised that the hadronic factors $r_B$ and $\delta_B$
are different, in general, for each $\B$ decay mode.

\begin{table}[htbp]
  \begin{center}
    \caption{
      Summary of relations between measured and physical parameters
      in GLW, ADS and Dalitz analyses of $\B \to \DorDstar K^{(*)}$ decays.
    }
    \vspace{0.2cm}
    \setlength{\tabcolsep}{1.0pc}
    \begin{tabular}{cc} \hline
      \mc{2}{l}{GLW analysis} \\
      $R_{\CP\pm}$ & $1 + r_B^2 \pm 2 r_B \cos \left( \delta_B \right) \cos \left( \gamma \right)$ \\
      $A_{\CP\pm}$ & $\pm 2 r_B \sin \left( \delta_B \right) \sin \left( \gamma \right) / R_{\CP\pm}$ \\
      \hline
      \mc{2}{l}{ADS analysis} \\
      $R_{\rm ADS}$ & $r_B^2 + r_D^2 + 2 r_B r_D \cos \left( \delta_B + \delta_D \right) \cos \left( \gamma \right)$ \\
      $A_{\rm ADS}$ & $2 r_B r_D \sin \left( \delta_B + \delta_D \right) \sin \left( \gamma \right) / R_{\rm ADS}$ \\
      \hline
      \mc{2}{l}{BPGGSZ Dalitz analysis ($D \to \KS \pi^+\pi^-$)} \\
      $x_\pm$ & $r_B \cos(\delta_B\pm\gamma)$ \\
      $y_\pm$ & $r_B \sin(\delta_B\pm\gamma)$ \\
      \hline
      \mc{2}{l}{Dalitz analysis ($D \to \pi^+\pi^-\pi^0$)} \\
      $\rho^\pm$ & $|z_\pm - x_0|$ \\
      $\theta^\pm$ & $\tan^{-1}(\Im(z_\pm)/(\Re(z_\pm) - x_0))$ \\
      \hline
    \end{tabular}
    \label{tab:cp_uta:notations:dk}
  \end{center}
\end{table}

\mysubsection{Common inputs and uncertainty treatment
}
\label{sec:cp_uta:common_inputs}

As described in Sec.~\ref{sec:method}, where measurements combined in an average depend on external parameters, it can be important to rescale to the latest values of those parameters in order to obtain the most precise and accurate results.
In practice, this is only necessary for modes with reasonably small statistical errors, so that the systematic uncertainty associated with the knowledge of the external parameter is not negligible.
Among the averages in this section, rescaling to common inputs is only done for $b \to c\bar{c}s$ transitions of \Bz\ mesons.
Correlated sources of systematic uncertainty are also taken into account in these averages.
For most other modes, the effects of common inputs and sources of systematic uncertainty are currently negligible, however similar considerations are applied when combining results to obtain constraints on $\alpha \equiv \phi_2$ and $\gamma \equiv \phi_3$ as discussed in Sec.~\ref{sec:cp_uta:uud:alpha} and~\ref{sec:cp_uta:cus:gamma}, respectively.

The common inputs used for calculating the averages
are listed in Table~\ref{tab:cp_uta:common_inputs}.
The average values for the $\Bz$ lifetime ($\tau(\Bz)$), mixing parameter ($\Delta m_d$) and relative width difference ($\Delta\Gamma_d / \Gamma_d$)
averages are discussed in Sec.~\ref{sec:life_mix}.
The fraction of the perpendicularly polarised component
($\left| A_{\perp} \right|^2$) in $\B \to \jpsi \Kstar(892)$ decays,
which determines the $\CP$ composition in these decays,
is averaged from results by
\babar~\cite{Aubert:2007hz}, \belle~\cite{Itoh:2005ks}, CDF~\cite{Acosta:2004gt}, D0~\cite{Abazov:2008jz} and LHCb~\cite{Aaij:2013cma}
(see also Sec.~\ref{sec:b2c}).

\begin{table}[htbp]
  \begin{center}
    \caption{
      Common inputs used in calculating the averages.
    }
    \vspace{0.2cm}
    \setlength{\tabcolsep}{1.0pc}
    \begin{tabular}{cr@{$\,\pm\,$}l} \hline
      $\tau(\Bz)$  & $1.520$ & $0.004\,{\rm ps}$ \\
      $\Delta m_d$ & $0.5065$ & $0.0019\,{\rm ps}^{-1}$ \\
      $\Delta\Gamma_d / \Gamma_d$ & $-0.002$ & $0.010$ \\
      $\left| A_{\perp} \right|^2 (\jpsi \Kstar)$ & $0.209$ & $0.006$ \\
      \hline
    \end{tabular}
    \label{tab:cp_uta:common_inputs}
  \end{center}
\end{table}

As explained in Sec.~\ref{sec:intro},
we do not apply a rescaling factor on the uncertainty of an average
that has $\chi^2/\dof > 1$
(unlike the procedure currently used by the PDG~\cite{PDG_2018}).
We provide a confidence level of the fit so that
one can know the consistency of the measurements included in the average,
and attach comments in case some care needs to be taken in the interpretation.
Note that, in general, results obtained from small data samples
will exhibit some non-Gaussian behaviour.
We average measurements with asymmetric uncertainties
using the PDG~\cite{PDG_2018} prescription.
In cases where several measurements are correlated
(\eg\ $S_f$ and $C_f$ in measurements of time-dependent $\CP$ violation
in $B$ decays to a particular $\CP$ eigenstate)
we take these into account in the averaging procedure
if the uncertainties are sufficiently Gaussian.
For measurements where one uncertainty is given,
it represents the total error,
where statistical and systematic uncertainties have been added in quadrature.
If two uncertainties are given, the first is statistical and the second systematic.
If more than two errors are given,
the origin of the additional uncertainty will be explained in the text.

\mysubsection{Time-dependent asymmetries in $b \to c\bar{c}s$ transitions
}
\label{sec:cp_uta:ccs}

\mysubsubsection{Time-dependent $\CP$ asymmetries in $b \to c\bar{c}s$ decays to $\CP$ eigenstates
}
\label{sec:cp_uta:ccs:cp_eigen}

\begin{figure}[htbp]
  \begin{center}
    \resizebox{0.51\textwidth}{!}{
      \includegraphics{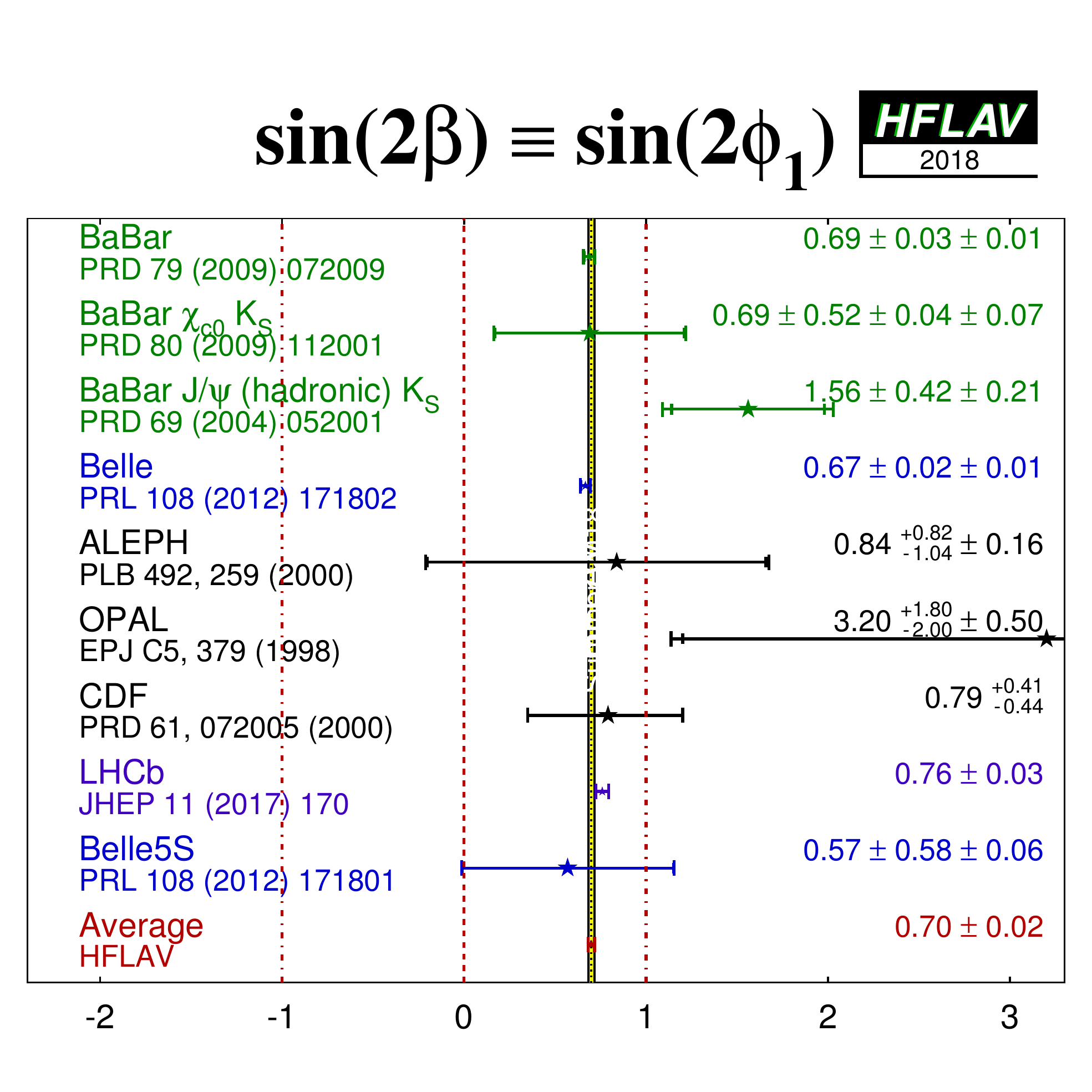}
    }
    \hfill
    \resizebox{0.48\textwidth}{!}{
      \includegraphics{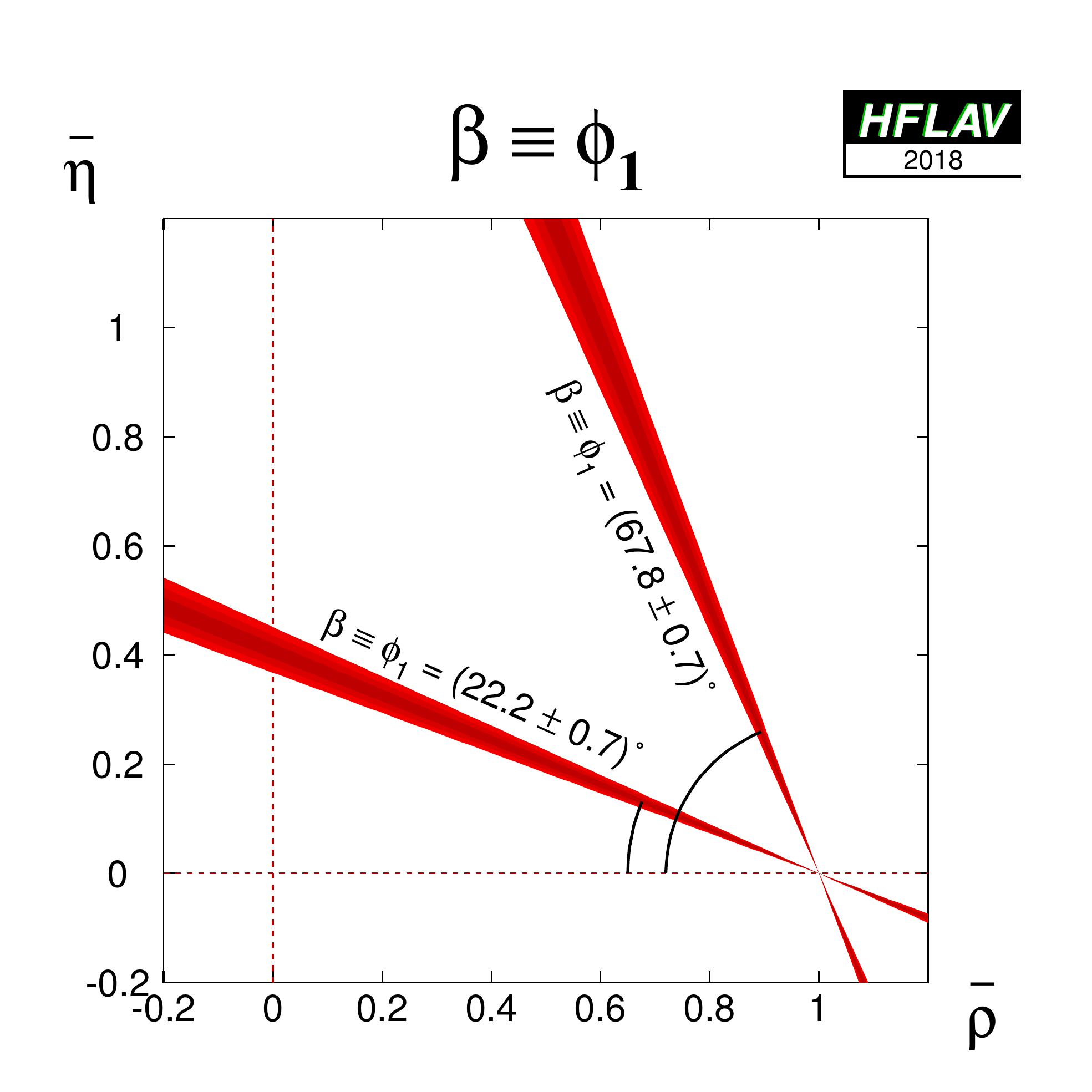}
    }
  \end{center}
  \vspace{-0.5cm}
  \caption{
    (Left) Average of measurements of $S_{b \to c\bar c s}$, interpreted as $\sin(2\beta)$.
    (Right) Constraints on the $(\rhobar,\etabar)$ plane,
    obtained from the average of $-\etacp S_{b \to c\bar c s}$
    and Eq.~(\ref{eq:cp_uta:sin2beta}).
    Note that the solution with the smaller (larger) value of $\beta$ has $\cos(2\beta)>0$ ($<0$).
  }
  \label{fig:cp_uta:ccs}
\end{figure}

In the Standard Model, the time-dependent parameters for $\Bz$ decays governed by $b \to c\bar c s$ transitions are predicted to be
$S_{b \to c\bar c s} = - \etacp \sin(2\beta)$ and $C_{b \to c\bar c s} = 0$ to very good accuracy.
Deviations from this relation are currently limited to the level of $\lsim 1^\circ$ on $2\beta$~\cite{Jung:2012mp,DeBruyn:2014oga,Frings:2015eva}.
The averages for $-\etacp S_{b \to c\bar c s}$ and $C_{b \to c\bar c s}$
are provided in Table~\ref{tab:cp_uta:ccs}.
The averages for $-\etacp S_{b \to c\bar c s}$
are shown in Fig.~\ref{fig:cp_uta:ccs}.

Both \babar\  and \belle\ have used the $\etacp = -1$ modes
$\jpsi \KS$, $\psi(2S) \KS$, $\chi_{c1} \KS$ and $\eta_c \KS$,
as well as $\jpsi \KL$, which has $\etacp = +1$
and $\jpsi K^{*0}(892)$, which is found to have $\etacp$ close to $+1$
based on the measurement of $\left| A_\perp \right|$
(see Sec.~\ref{sec:cp_uta:common_inputs}).
The most recent \belle\ result does not use $\eta_c \KS$ or $\jpsi K^{*0}(892)$ decays.\footnote{
  Previous analyses from \belle\ did include these channels~\cite{Abe:2004mz},
  but it is not possible to obtain separate results for those modes from the
  published information.
}
LHCb has used $\jpsi \KS$ (data with $\jpsi \to \mumu$ and $\epem$ are reported in different publications) and $\psi(2S) \KS$ decays.
ALEPH, OPAL, and CDF have used only the $\jpsi \KS$ final state.
\babar\ has also determined the \CP violation parameters of the
$\Bz\to\chi_{c0} \KS$ decay from the time-dependent Dalitz-plot analysis of
the $\Bz \to \pi^+\pi^-\KS$ mode (see Sec.~\ref{sec:cp_uta:qqs:dp}).
In addition, \belle\ has performed a measurement with data accumulated at the $\Upsilon(5S)$ resonance, using the $\jpsi\KS$ final state -- this involves a different flavour tagging method compared to the measurements performed with data accumulated at the $\Upsilon(4S)$ resonance.
A breakdown of results in each charmonium-kaon final state is given in
Table~\ref{tab:cp_uta:ccs-BF}.

\begin{table}[htb]
	\begin{center}
		\caption{
                        Results and averages for $S_{b \to c\bar c s}$ and $C_{b \to c\bar c s}$.
                        The averages are given from a combination of the most precise results only, and also including less precise measurements.
                }
		\vspace{0.2cm}
    \resizebox{\textwidth}{!}{
\renewcommand{\arraystretch}{1.2}
		\begin{tabular}{@{\extracolsep{2mm}}lrccc} \hline
      \mc{2}{l}{Experiment} & Sample size & $- \etacp S_{b \to c\bar c s}$ & $C_{b \to c\bar c s}$ \\
      \hline
	\babar\ $b \to c\bar c s$ & \cite{:2009yr} & $N(B\bar{B})$ = 465M & $0.687 \pm 0.028 \pm 0.012$ & $0.024 \pm 0.020 \pm 0.016$ \\
	\belle\ $b \to c\bar c s$ & \cite{Adachi:2012et} & $N(B\bar{B})$ = 772M & $0.667 \pm 0.023 \pm 0.012$ & $-0.006 \pm 0.016 \pm 0.012$ \\
        LHCb $J/\psi \KS$ & \hspace{-4ex}\cite{Aaij:2015vza,Aaij:2017yld} & $\int {\cal L} \, dt = 3\ {\rm fb}^{-1}$ & $0.75 \pm 0.04$ & $-0.014 \pm 0.030$ \\
        LHCb $\psi(2S) \KS$ & \cite{Aaij:2017yld} & $\int {\cal L} \, dt = 3\ {\rm fb}^{-1}$ & $0.84 \pm 0.10 \pm 0.01$ & $-0.05 \pm 0.10 \pm 0.01$ \\
	\mc{3}{l}{\bf \boldmath Average} & $0.698 \pm 0.017$ & $-0.005 \pm 0.015$ \\
	\mc{3}{l}{\small Confidence level} & {\small $0.09~(1.7\sigma)$} & {\small $0.54~(0.6\sigma)$} \\
        \hline
	\babar\ $\chi_{c0} \KS$ & \cite{Aubert:2009me} & $N(B\bar{B})$ = 383M & $0.69 \pm 0.52 \pm 0.04 \pm 0.07$ & $-0.29 \,^{+0.53}_{-0.44} \pm 0.03 \pm 0.05$ \\
	\babar\ $J/\psi \KS$ ($^{*}$) & \cite{Aubert:2003xn} & $N(B\bar{B})$ = 88M & $1.56 \pm 0.42 \pm 0.21$ &  \textendash{} \\
        ALEPH & \cite{Barate:2000tf} &  $N(Z \to \text{hadrons})$ = 4M & $0.84 \, ^{+0.82}_{-1.04} \pm 0.16$ &  \textendash{} \\
        OPAL  & \cite{Ackerstaff:1998xz} & $N(Z \to \text{hadrons})$ = 4.4M & $3.2 \, ^{+1.8}_{-2.0} \pm 0.5$ &  \textendash{} \\
        CDF   & \cite{Affolder:1999gg} & $\int {\cal L} \, dt = 110\ {\rm pb}^{-1}$ & $0.79 \, ^{+0.41}_{-0.44}$ &  \textendash{} \\
	Belle $\Upsilon(5S)$ & \cite{Sato:2012hu} & $\int {\cal L} \, dt = 121\ {\rm fb}^{-1}$ & $0.57 \pm 0.58 \pm 0.06$ &  \textendash{} \\
        \mc{3}{l}{\bf Average} & $0.699 \pm 0.017$ & $-0.005 \pm 0.015$ \\
		\hline \\ [-1.8ex]
                  \mc{5}{l}{\begin{minipage}{1.2\textwidth} %
{\footnotesize 
  ($^{*}$) This result uses ``{\it hadronic and previously unused muonic decays of the $J/\psi$}''.
  We neglect a small possible correlation of this result with the main \babar\ result~\cite{:2009yr} that could be caused by reprocessing of the data.
}
\end{minipage}}
		\end{tabular}
}
                \label{tab:cp_uta:ccs}
        \end{center}
\end{table}

\begin{table}[htb]
	\begin{center}
		\caption{
                        Breakdown of results on $S_{b \to c\bar c s}$ and $C_{b \to c\bar c s}$.
                }
		\vspace{0.2cm}
    \resizebox{\textwidth}{!}{
\renewcommand{\arraystretch}{1.2}
		\begin{tabular}{@{\extracolsep{2mm}}lrccc} \hline
        \mc{2}{l}{Mode} & Sample size & $- \etacp S_{b \to c\bar c s}$ & $C_{b \to c\bar c s}$ \\
        \hline
        \mc{5}{c}{\babar} \\
        $J/\psi \KS$ & \cite{:2009yr} & $N(B\bar{B}) =$ 465M & $0.657 \pm 0.036 \pm 0.012$ & $\phantom{-}0.026 \pm 0.025 \pm 0.016$ \\
        $J/\psi \KL$ & \cite{:2009yr} & $N(B\bar{B}) =$ 465M & $0.694 \pm 0.061 \pm 0.031$ & $-0.033 \pm 0.050 \pm 0.027$ \\
        {\bf \boldmath $J/\psi K^0$} & \cite{:2009yr} & $N(B\bar{B}) =$ 465M & $0.666 \pm 0.031 \pm 0.013$ & $\phantom{-}0.016 \pm 0.023 \pm 0.018$ \\
        $\psi(2S) \KS$ & \cite{:2009yr} & $N(B\bar{B}) =$ 465M & $0.897 \pm 0.100 \pm 0.036$ & $\phantom{-}0.089 \pm 0.076 \pm 0.020$ \\
        $\chi_{c1} \KS$ & \cite{:2009yr} & $N(B\bar{B}) =$ 465M & $0.614 \pm 0.160 \pm 0.040$ & $\phantom{-}0.129 \pm 0.109 \pm 0.025$ \\
        $\eta_c \KS$ & \cite{:2009yr} & $N(B\bar{B}) =$ 465M & $0.925 \pm 0.160 \pm 0.057$ & $\phantom{-}0.080 \pm 0.124 \pm 0.029$ \\
        $\jpsi K^{*0}(892)$ & \cite{:2009yr} & $N(B\bar{B}) =$ 465M & $0.601 \pm 0.239 \pm 0.087$ & $\phantom{-}0.025 \pm 0.083 \pm 0.054$ \\
        {\bf All} & \cite{:2009yr} & $N(B\bar{B}) =$ 465M & $0.687 \pm 0.028 \pm 0.012$ & $\phantom{-}0.024 \pm 0.020 \pm 0.016$ \\
	\hline
	\mc{5}{c}{\bf \belle} \\
        $J/\psi \KS$ & \cite{Adachi:2012et} & $N(B\bar{B}) =$ 772M & $0.670 \pm 0.029 \pm 0.013$ & $\phantom{-}0.015 \pm 0.021 \,^{+0.023}_{-0.045}$ \\
        $J/\psi \KL$ & \cite{Adachi:2012et} & $N(B\bar{B}) =$ 772M & $0.642 \pm 0.047 \pm 0.021$ & $-0.019 \pm 0.026 \,^{+0.041}_{-0.017}$ \\
	$\psi(2S) \KS$ & \cite{Adachi:2012et} & $N(B\bar{B}) =$ 772M & $0.738 \pm 0.079 \pm 0.036$ & $-0.104 \pm 0.055 \,^{+0.027}_{-0.047}$ \\
	$\chi_{c1} \KS$ & \cite{Adachi:2012et} & $N(B\bar{B}) =$ 772M & $0.640 \pm 0.117 \pm 0.040$ & $\phantom{-}0.017 \pm 0.083 \,^{+0.026}_{-0.046}$ \\
        {\bf All} & \cite{Adachi:2012et} & $N(B\bar{B}) =$ 772M & $0.667 \pm 0.023 \pm 0.012$ & $-0.006 \pm 0.016 \pm 0.012$ \\
	\hline
	\mc{5}{c}{\bf LHCb} \\
        $J/\psi(\to \mumu) \KS$ & \cite{Aaij:2015vza} & $\int {\cal L} \, dt = 3\ {\rm fb}^{-1}$ & $0.731 \pm 0.035 \pm 0.020$ & $-0.038 \pm 0.032 \pm 0.005$ \\
        $J/\psi(\to \epem) \KS$ & \cite{Aaij:2017yld} & $\int {\cal L} \, dt = 3\ {\rm fb}^{-1}$ & $0.83 \pm 0.08 \pm 0.01$ & $0.12 \pm 0.07 \pm 0.02$ \\
        $\psi(2S) \KS$ & \cite{Aaij:2017yld} & $\int {\cal L} \, dt = 3\ {\rm fb}^{-1}$ & $0.84 \pm 0.10 \pm 0.01$ & $-0.05 \pm 0.10 \pm 0.01$ \\
	\hline
	\mc{5}{c}{\bf Averages} \\
        \mc{3}{l}{$J/\psi \KS$} & $0.695 \pm 0.019$ & $\phantom{-}0.000 \pm 0.020$ \\
        \mc{3}{l}{$J/\psi \KL$} & $0.663 \pm 0.041$ & $-0.023 \pm 0.030$ \\
        \mc{3}{l}{$\psi(2S) \KS$} & $0.817 \pm 0.056$ & $-0.019 \pm 0.048$ \\
        \mc{3}{l}{$\chi_{c1} \KS$} & $0.632 \pm 0.099$ & $\phantom{-}0.066 \pm 0.074$ \\
		\hline
		\end{tabular}
}
                \label{tab:cp_uta:ccs-BF}
        \end{center}
\end{table}

While the uncertainty in the average for $-\etacp S_{b \to c\bar c s}$ is limited by statistical error,
the precision for $C_{b \to c\bar c s}$ is close to being dominated by the systematic uncertainty, particularly for measurements from the $\epem$ $B$ factory experiments.
This occurs due to the possible effect of tag-side interference~\cite{Long:2003wq} on the $C_{b \to c\bar c s}$ measurement, an effect which is correlated between different $e^+e^- \to \Upsilon(4S) \to B\bar{B}$ experiments.
Understanding of this effect may continue to improve in future, allowing the uncertainty to reduce.

From the average for $-\etacp S_{b \to c\bar c s}$ above,
we obtain the following solutions for $\beta$
(in $\left[ 0, \pi \right]$):
\begin{equation}
  \beta = \left( 22.2 \pm 0.7 \right)^\circ
  \hspace{5mm}
  {\rm or}
  \hspace{5mm}
  \beta = \left( 67.8 \pm 0.7 \right)^\circ \, .
  \label{eq:cp_uta:sin2beta}
\end{equation}

This result gives a precise constraint on the $(\rhobar,\etabar)$ plane,
as shown in Fig.~\ref{fig:cp_uta:ccs}.
The measurement is in remarkable agreement with other constraints from
$\CP$-conserving quantities,
and with $\CP$ violation in the kaon system, in the form of the parameter $\epsilon_K$.
Such comparisons have been performed by various phenomenological groups,
such as CKMfitter~\cite{Charles:2004jd} and UTFit~\cite{Bona:2005vz} (see also Refs.~\cite{Lunghi:2008aa,Eigen:2013cv}).

\mysubsubsection{Time-dependent transversity analysis of $\Bz \to J/\psi K^{*0}$ decays
}
\label{sec:cp_uta:ccs:vv}

\begin{figure}[htbp]
  \begin{center}
    \resizebox{0.46\textwidth}{!}{
      \includegraphics{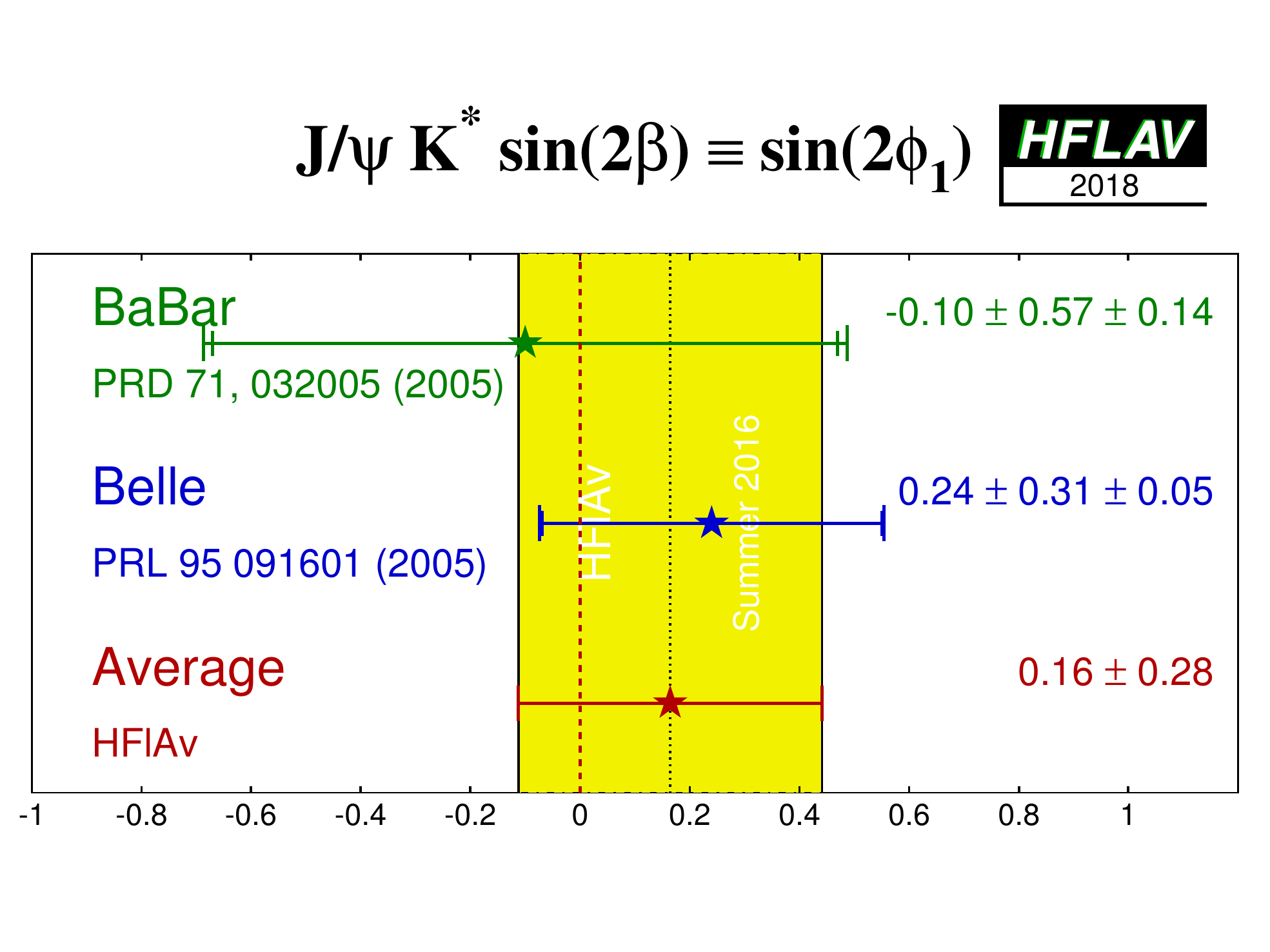}
    }
    \hfill
    \resizebox{0.46\textwidth}{!}{
      \includegraphics{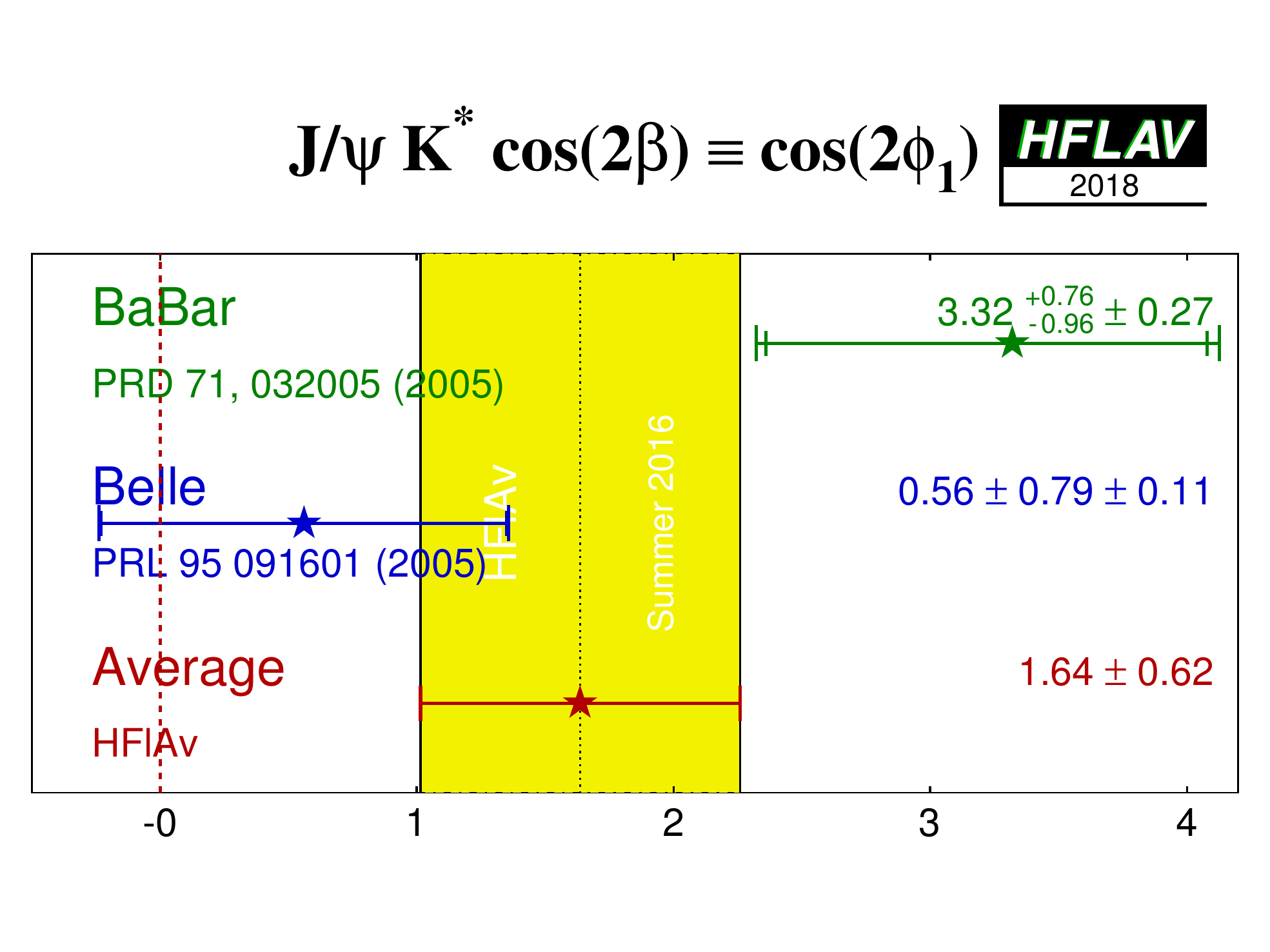}
    }
  \end{center}
  \vspace{-0.5cm}
  \caption{
    Averages of
    (left) $\sin(2\beta) \equiv \sin(2\phi_1)$ and
    (right) $\cos(2\beta) \equiv \cos(2\phi_1)$
    from time-dependent analyses of $\Bz \to \jpsi K^{*0}$ decays.
  }
  \label{fig:cp_uta:JpsiKstar}
\end{figure}

$\B$ meson decays to the vector-vector final state $J/\psi K^{*0}$
are also mediated by the $b \to c \bar c s$ transition.
When a final state that is not flavour-specific ($K^{*0} \to \KS \pi^0$) is used,
a time-dependent transversity analysis can be performed,
yielding sensitivity to both
$\sin(2\beta)$ and $\cos(2\beta)$~\cite{Dunietz:1990cj}.
Such analyses have been performed by both $\B$ factory experiments.
In principle, the strong phases between the transversity amplitudes
are not uniquely determined by such an analysis,
leading to a discrete ambiguity in the sign of $\cos(2\beta)$.
The \babar\ collaboration resolves
this ambiguity using the known
variation~\cite{Aston:1987ir}
of the P-wave phase (fast) relative to that of the S-wave phase (slow)
with
the invariant mass of the $K\pi$ system
in the vicinity of the $K^*(892)$ resonance.
The result is in agreement with the prediction from
$s$-quark helicity conservation,
and corresponds to Solution II defined by Suzuki~\cite{Suzuki:2001za}.
We include only the solutions consistent with this phase variation in
Table~\ref{tab:cp_uta:ccs:psi_kstar} and Fig.~\ref{fig:cp_uta:JpsiKstar}.

\begin{table}[htb]
	\begin{center}
		\caption{
			Averages from $\Bz \to J/\psi K^{*0}$ transversity analyses.
		}
		\vspace{0.2cm}
		\setlength{\tabcolsep}{0.0pc}
\renewcommand{\arraystretch}{1.1}
		\begin{tabular*}{\textwidth}{@{\extracolsep{\fill}}lrcccc} \hline
		\mc{2}{l}{Experiment} & $N(B\bar{B})$ & $\sin 2\beta$ & $\cos 2\beta$ & Correlation \\
		\hline
	\babar & \cite{Aubert:2004cp} & 88M & $-0.10 \pm 0.57 \pm 0.14$ & $3.32 ^{+0.76}_{-0.96} \pm 0.27$ & $-0.37$ \\
	\belle & \cite{Itoh:2005ks} & 275M & $0.24 \pm 0.31 \pm 0.05$ & $0.56 \pm 0.79 \pm 0.11$ & $0.22$ \\
	\mc{3}{l}{\bf Average} & $0.16 \pm 0.28$ & $1.64 \pm 0.62$ &  \hspace{-8mm} {\small uncorrelated averages}  \\
        \mc{3}{l}{\small Confidence level} & {\small $0.61~(0.5\sigma)$} & {\small $0.03~(2.2\sigma)$} & \\
		\hline
		\end{tabular*}
		\label{tab:cp_uta:ccs:psi_kstar}
	\end{center}
\end{table}

At present, the results are dominated by large and non-Gaussian statistical uncertainties, and exhibit significant correlations.
We perform uncorrelated averages,
which necessitates care in the interpretation of these averages.
Nonetheless, it is clear that $\cos(2\beta)>0$ is preferred
by the experimental data in $J/\psi \Kstarz$
(for example, \babar~\cite{Aubert:2004cp} finds a confidence level for $\cos(2\beta)>0$ of $89\%$).

\mysubsubsection{Time-dependent $\CP$ asymmetries in $\Bz \to \Dstarp \Dstarm \KS$ decays
}
\label{sec:cp_uta:ccs:DstarDstarKs}

\begin{figure}[htbp]
  \begin{center}
    \begin{tabular}{c@{\hspace{-1mm}}c@{\hspace{-1mm}}c}
      \resizebox{0.32\textwidth}{!}{
        \includegraphics{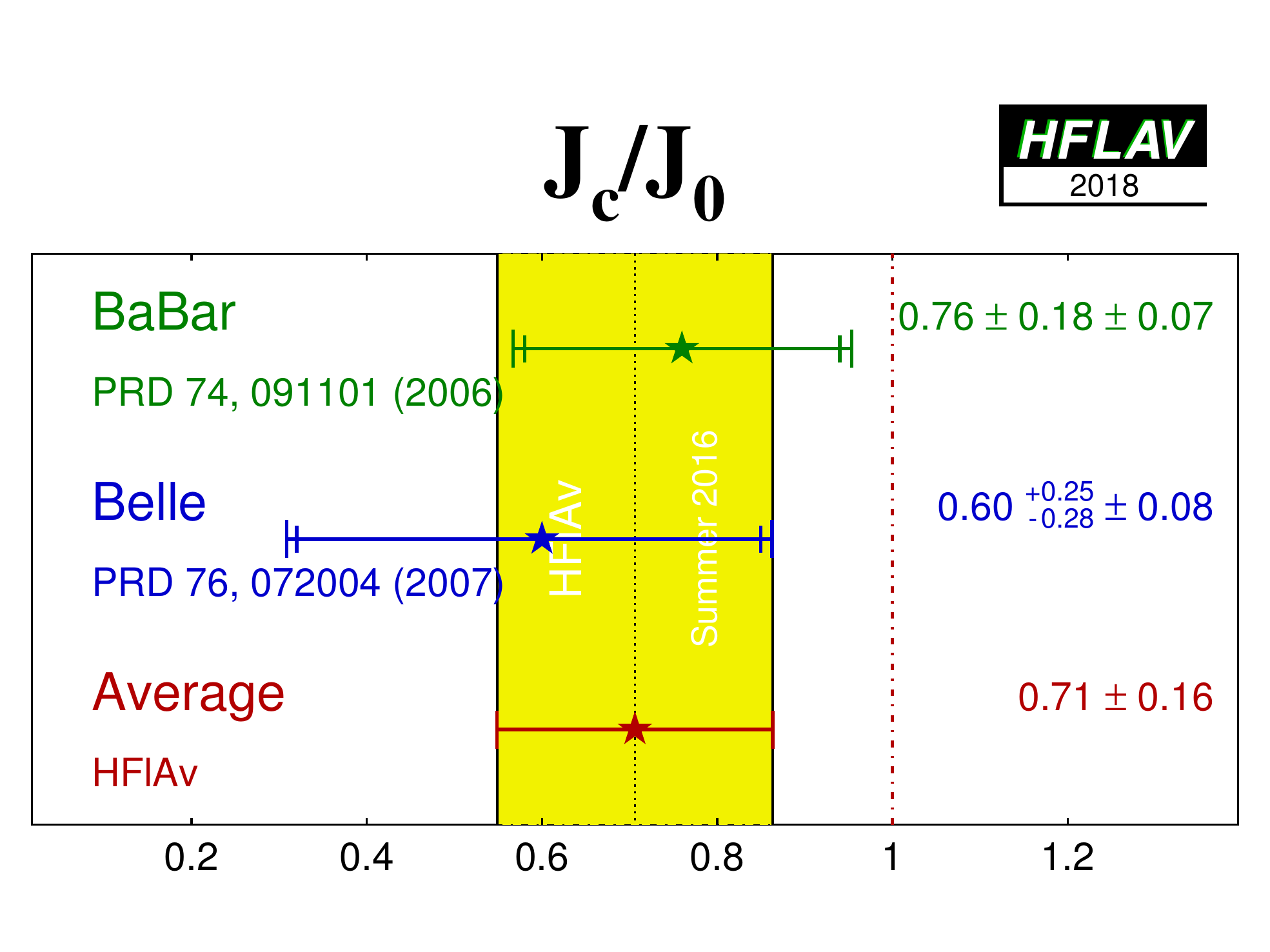}
      }
      &
      \resizebox{0.32\textwidth}{!}{
        \includegraphics{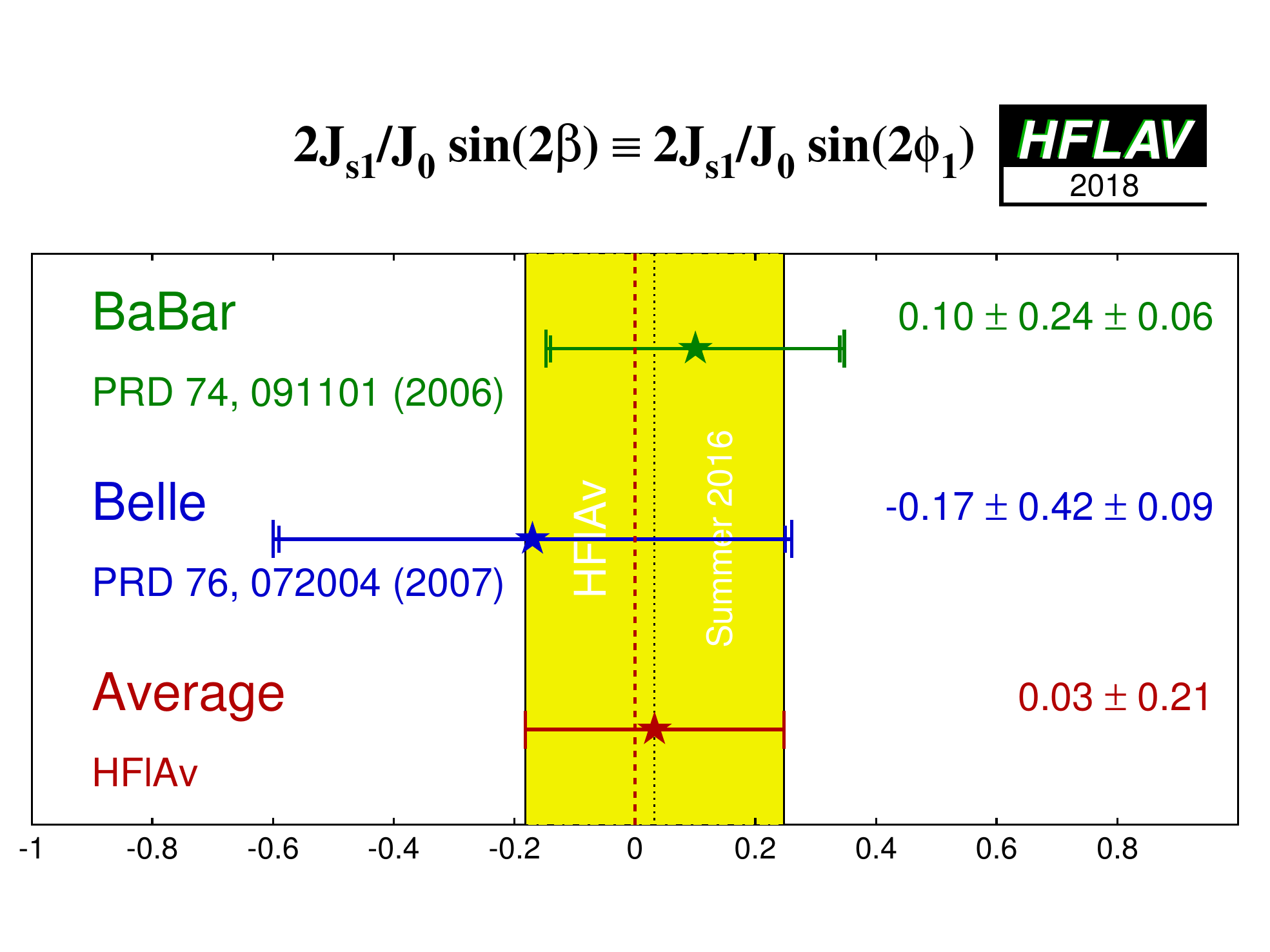}
      }
      &
      \resizebox{0.32\textwidth}{!}{
        \includegraphics{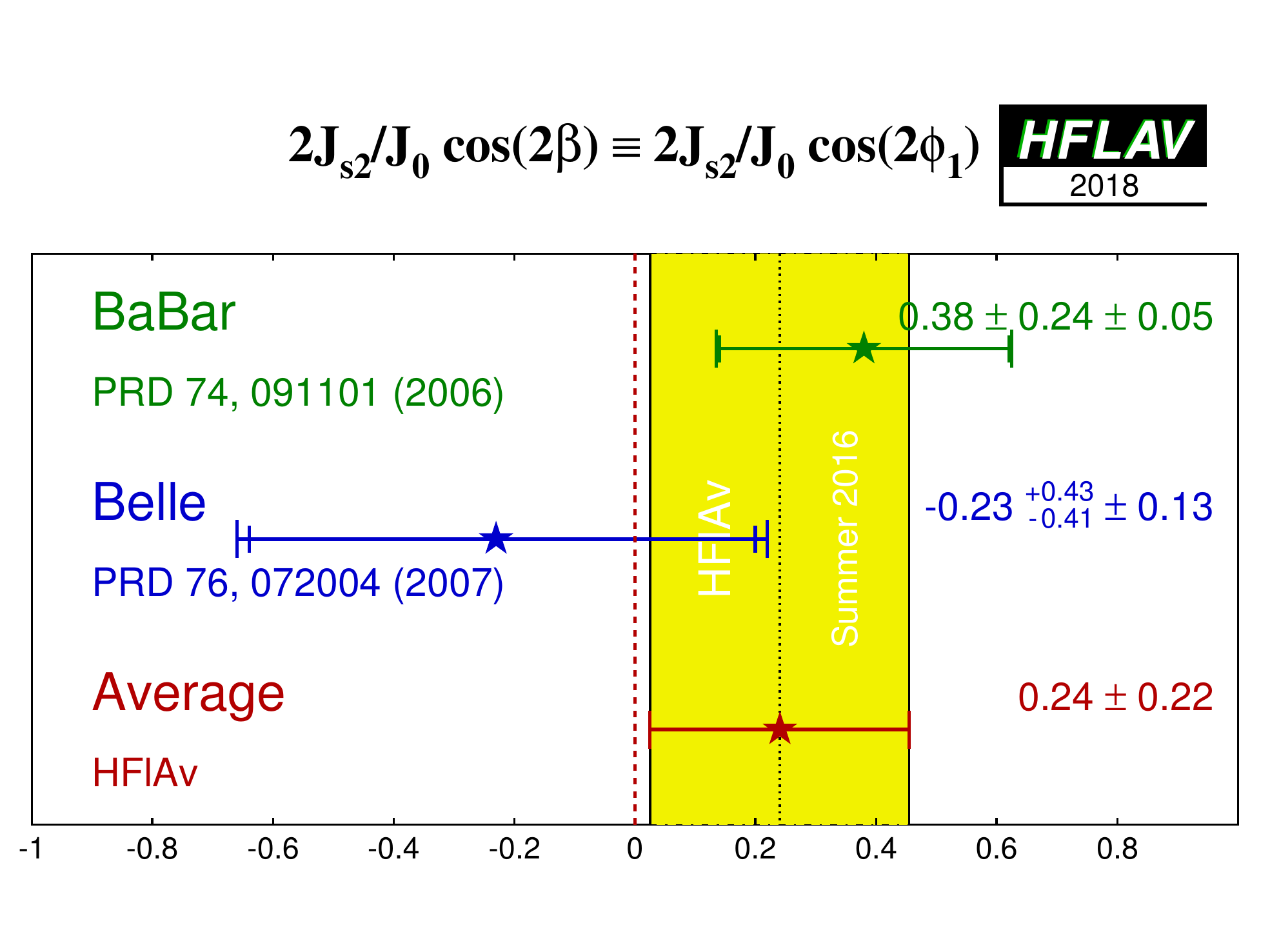}
      }
    \end{tabular}
  \end{center}
  \vspace{-0.8cm}
  \caption{
    Averages of
    (left) $(J_c/J_0)$, (middle) $(2J_{s1}/J_0) \sin(2\beta)$ and
    (right) $(2J_{s2}/J_0) \cos(2\beta)$
    from time-dependent analyses of $\Bz \to \Dstarp \Dstarm \KS$ decays.
  }
  \label{fig:cp_uta:ccs:dstardstarks}
\end{figure}

Both \babar~\cite{Aubert:2006fh} and \belle~\cite{Dalseno:2007hx} have performed time-dependent analyses of the $\Bz \to \Dstarp \Dstarm \KS$ decay, to obtain information on the sign of $\cos(2\beta)$.
More information can be found in Sec.~\ref{sec:cp_uta:notations:dalitz:dstardstarks}.
The results are given in Table~\ref{tab:cp_uta:ccs:dstardstarks}, and shown in Fig.~\ref{fig:cp_uta:ccs:dstardstarks}.
From the above result and the assumption that $J_{s2}>0$, \babar\ infers that $\cos(2\beta)>0$ at the $94\%$ confidence level~\cite{Aubert:2006fh}.

\begin{table}[htb]
	\begin{center}
		\caption{
                        Results from time-dependent analysis of $\Bz \to \Dstarp \Dstarm \KS$.
		}
		\vspace{0.2cm}
		\setlength{\tabcolsep}{0.0pc}
\renewcommand{\arraystretch}{1.2}
		\begin{tabular*}{\textwidth}{@{\extracolsep{\fill}}lrcccc} \hline
                \mc{2}{l}{Experiment} & $N(B\bar{B})$ & $\frac{J_c}{J_0}$ & $\frac{2J_{s1}}{J_0} \sin(2\beta)$ &  $\frac{2J_{s2}}{J_0} \cos(2\beta)$ \\
		\hline
	\babar & \cite{Aubert:2006fh} & 230M & $0.76 \pm 0.18 \pm 0.07$ & $0.10 \pm 0.24 \pm 0.06$ & $0.38 \pm 0.24 \pm 0.05$ \\
	\belle & \cite{Dalseno:2007hx} & 449M & $0.60 \,^{+0.25}_{-0.28} \pm 0.08$ & $-0.17 \pm 0.42 \pm 0.09$ & $-0.23 \,^{+0.43}_{-0.41} \pm 0.13$ \\
	\mc{3}{l}{\bf Average} & $0.71 \pm 0.16$ & $0.03 \pm 0.21$ & $0.24 \pm 0.22$ \\
	\mc{3}{l}{\small Confidence level} & {\small $0.63~(0.5\sigma)$} & {\small $0.59~(0.5\sigma)$} & {\small $0.23~(1.2\sigma)$} \\
		\hline
		\end{tabular*}
		\label{tab:cp_uta:ccs:dstardstarks}
	\end{center}
\end{table}

\mysubsubsection{Time-dependent analysis of $\Bs$ decays through the $b \to c\bar{c}s$ transition
}
\label{sec:cp_uta:ccs:jpsiphi}

As described in Sec.~\ref{sec:cp_uta:notations:Bs},
time-dependent analysis of decays such as $\Bs \to J/\psi \phi$ probes the
$\CP$ violating phase of $\Bs$--$\Bsbar$ oscillations, $\phi_s$.\footnote{
  We use $\phi_s$ here to denote the same quantity labelled $\phi_s^{c\bar{c}s}$ in Sec.~\ref{sec:life_mix}.
  It should not be confused with the parameter $\phi_{12} \equiv \arg\left[ -M_{12}/\Gamma_{12} \right]$, which historically was also often referred to as $\phi_s$.
}
The combination of results on $\Bs \to \jpsi \phi$ decays, including also results from channels such a $\Bs \to \jpsi \pi^+\pi^-$ and $\Bs \to D_s^+D_s^-$ decays, is discussed in Sec.~\ref{sec:life_mix}.

\mysubsection{Time-dependent $\CP$ asymmetries in colour-suppressed $b \to c\bar{u}d$ transitions
}
\label{sec:cp_uta:cud_beta}

\mysubsubsection{Time-dependent $\CP$ asymmetries: $b \to c\bar{u}d$ decays to
  \CP eigenstates
}
\label{sec:cp_uta:cud_beta:cp}

Decays of $\B$ mesons to final states such as $D\pi^0$ are
governed by $b \to c\bar{u}d$ transitions.
If the final state is a $\CP$ eigenstate, \eg\ $D_{\CP}\pi^0$,
the usual time-dependent formulae are recovered,
with the sine coefficient sensitive to $\sin(2\beta)$.
Since there is no penguin contribution to these decays,
there is even less associated theoretical uncertainty
than for $b \to c\bar{c}s$ decays such as $\B \to \jpsi \KS$.
Such measurements therefore allow to test the Standard Model prediction
that the $\CP$ violation parameters in $b \to c\bar{u}d$ transitions
are the same as those in $b \to c\bar{c}s$~\cite{Grossman:1996ke}.
Although there is an additional contribution from CKM suppressed $b \to u \bar{c} d$ amplitudes, which have a different weak phase compared to the leading $b \to c\bar{u}d$ transition, the effect is small and can be taken into account in the analysis~\cite{Fleischer:2003ai,Fleischer:2003aj}.

Results are available from a joint analysis of \babar\ and \belle\ data~\cite{Abdesselam:2015gha}.
The following \CP-even final states are included: $D\piz$ and $D\eta$ with $D \to \KS\piz$ and $D \to \KS\omega$; $D\omega$ with $D \to \KS\piz$; $\Dstar\piz$ and $\Dstar\eta$ with $\Dstar \to D\piz$ and $D \to \Kp\Km$.
The following \CP-odd final states are included: $D\piz$, $D\eta$ and $D\omega$ with $D \to \Kp\Km$, $\Dstar\piz$ and $\Dstar\eta$ with $\Dstar \to D\piz$ and $D \to \KS\piz$.
All $\Bd \to \DorDstar h^0$ decays are analysed together, taking into account the different $\CP$ factors (denoted $\DorDstar_{\CP} h^0$).
The results are summarised in Table~\ref{tab:cp_uta:cud_cp_beta}.

\begin{table}[htb]
	\begin{center}
		\caption{
			Results from analyses of $\Bz \to D^{(*)}h^0$, $D \to \CP$ eigenstates decays.
		}
		\vspace{0.2cm}
		\setlength{\tabcolsep}{0.0pc}
\renewcommand{\arraystretch}{1.1}
		\begin{tabular*}{\textwidth}{@{\extracolsep{\fill}}lrcccc} \hline
	\mc{2}{l}{Experiment} & $N(B\bar{B})$ & $S_{\CP}$ & $C_{\CP}$ & Correlation \\
	\hline
	\babar\ \& \belle & \cite{Abdesselam:2015gha} & 1243M & $0.66 \pm 0.10 \pm 0.06$ & $-0.02 \pm 0.07 \pm 0.03$ & $-0.05$ \\
	\hline
		\end{tabular*}
		\label{tab:cp_uta:cud_cp_beta}
	\end{center}
\end{table}

\mysubsubsection{Time-dependent Dalitz-plot analyses of $b \to c\bar{u}d$ decays
}
\label{sec:cp_uta:cud_beta:dalitz}

When multibody $D$ decays, such as $D \to \KS\pi^+\pi^-$ are used,
a time-dependent analysis of the Dalitz plot of the neutral $D$ decay
allows for a direct determination of the weak phase $2\beta$
or, equivalently, of both $\sin(2\beta)$ and $\cos(2\beta)$.
This information can be used to resolve the ambiguity in the
measurement of $2\beta$ from $\sin(2\beta)$~\cite{Bondar:2005gk}.

Results are available from a joint analysis of \babar\ and \belle\ data~\cite{Adachi:2018itz,Adachi:2018jqe}.
The decays $\B \to D\pi^0$, $\B \to D\eta$, $\B \to D\omega$,
$\B \to D^*\pi^0$ and $\B \to D^*\eta$ are used.
(This collection of states is denoted by $D^{(*)}h^0$.)
The daughter decays are $D^* \to D\pi^0$ and $D \to \KS\pi^+\pi^-$.
These results supersede those from previous analyses done separately by
\belle~\cite{Krokovny:2006sv} and \babar~\cite{Aubert:2007rp} and are given in Table~\ref{tab:cp_uta:cud_beta}.
Treating $\beta$ as a free parameter in the fit, the result $\beta = (22.5 \pm 4.4 \pm 1.2 \pm 0.6)^\circ$ is obtained.
This corresponds to an observation of \CP\ violation ($\beta \neq 0$) at $5.1\sigma$ significance, and evidence for $\cos(2\beta) > 0$ at $3.7\sigma$.
The ambiguous solution with $\cos(2\beta) < 0$, corresponding to the solution for $\sin(2\beta)$ from $b \to c\bar c s$ transitions, is ruled out at $7.3\sigma$.

\begin{figure}[htbp]
  \begin{center}
    \resizebox{0.46\textwidth}{!}{
      \includegraphics{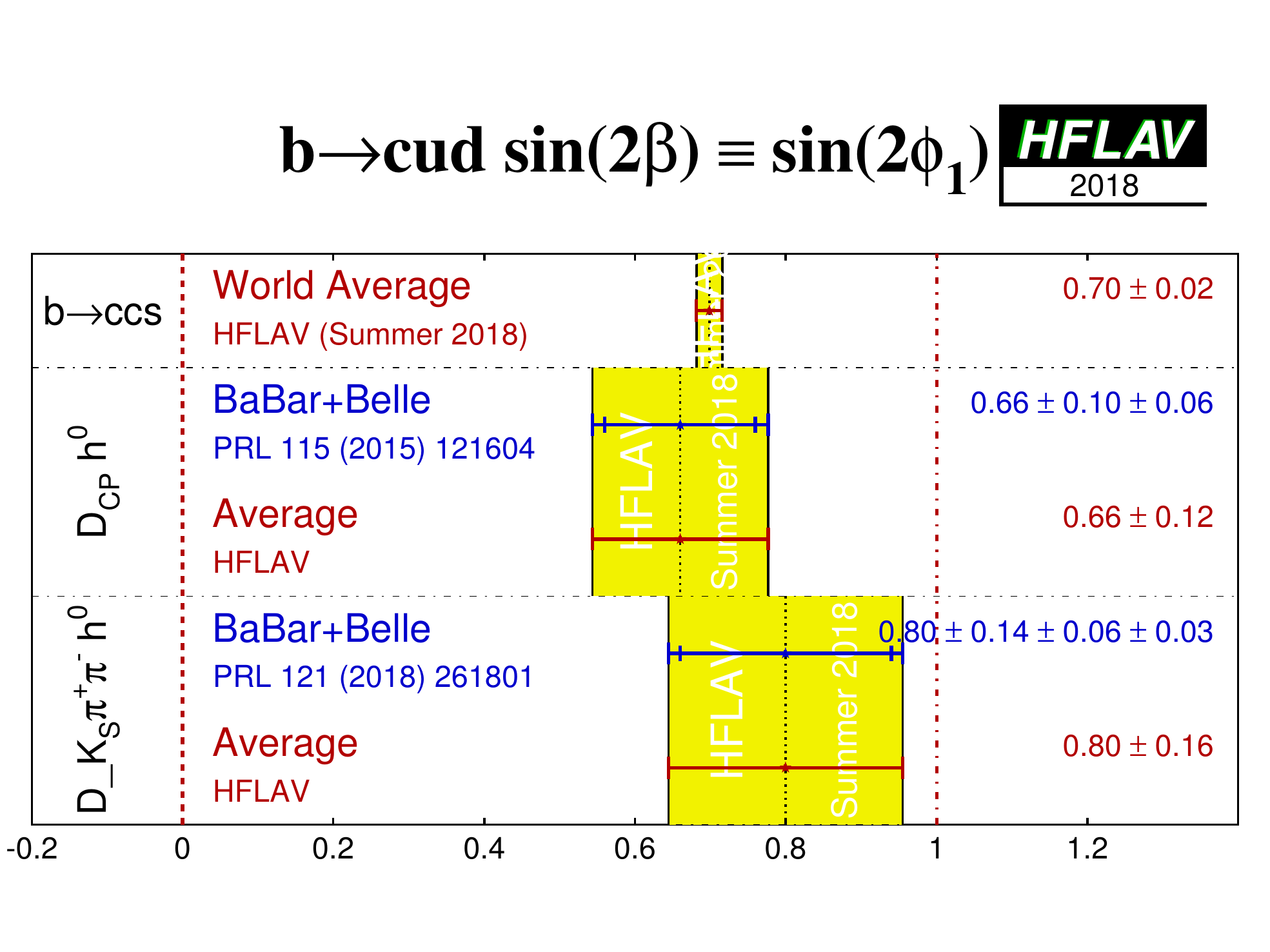} %
    }
  \end{center}
  \vspace{-0.8cm}
  \caption{
    Averages of $\sin(2\beta)$ measured in colour-suppressed $b \to c\bar{u}d$ transitions.
  }
  \label{fig:cp_uta:cud_beta}
\end{figure}

A comparison of the results for $\sin(2\beta)$ from  $\Bz \to \DorDstar h^0$ decays, with $D$ decays to \CP\ eigenstates or to $D \to \KS\pi^+\pi^-$, is shown in Fig.~\ref{fig:cp_uta:cud_beta}.
Averaging these results gives $\sin(2\beta) = 0.71 \pm 0.09$, which is consistent with, but not as precise as, the value from $b \to c\bar c s$ transitions.

\begin{table}[htb]
	\begin{center}
		\caption{
			Averages from $\Bz \to D^{(*)}h^0$, $D \to \KS\pi^+\pi^-$ analyses.
		}
		\vspace{0.2cm}
		\setlength{\tabcolsep}{0.0pc}
    \resizebox{\textwidth}{!}{
\renewcommand{\arraystretch}{1.1}
      		\begin{tabular*}{\textwidth}{@{\extracolsep{\fill}}lrccc} \hline
	\mc{2}{l}{Experiment} & $N(B\bar{B})$ & $\sin 2\beta$ & $\cos 2\beta$ \\
		\hline
                \mc{5}{c}{Model-dependent} \\
        \babar\ \& \belle & \cite{Adachi:2018itz,Adachi:2018jqe} & 1240M & $0.80 \pm 0.14 \pm 0.06 \pm 0.03$ & $0.91 \pm 0.22 \pm 0.09 \pm 0.07$ \\
		\hline
                \mc{5}{c}{Model-independent} \\
	\belle & \cite{Vorobyev:2016npn} & 772M & $0.43 \pm 0.27 \pm 0.08$ & $1.06 \pm 0.33 \,^{+0.21}_{-0.15}$ \\
		\hline        
		\end{tabular*}
    }
		\label{tab:cp_uta:cud_beta}
	\end{center}
\end{table}

A model-independent time-dependent analysis of $\Bd \to \DorDstar h^0$ decays, with $D \to \KS\pip\pim$, has been performed by \belle~\cite{Vorobyev:2016npn}.
The decays $\Bd \to D\piz$, $\Bd \to D\eta$, $\Bd \to D\etapr$, $\Bd \to D\omega$, $\Bd \to \Dstar\piz$ and $\Bd \to \Dstar\eta$ are used.
The results are also included in Table~\ref{tab:cp_uta:cud_beta}.
From these results, \belle\ disfavours the $\cos(2\phi_1)<0$ solution that corresponds to the $\sin(2\phi_1)$ results from $b \to c\bar{c}s$ transitions at $5.1\,\sigma$ significance.
The solution with $\cos(2\phi_1)>0$ is consistent with the data at the level of $1.3\,\sigma$.
Note that due to the strong statistical and systematic correlations, model-dependent results and model-independent results from the same experiment cannot be combined.

\mysubsection{Time-dependent $\CP$ asymmetries in $b \to c\bar{c}d$ transitions
}
\label{sec:cp_uta:ccd}

The transition $b \to c\bar c d$ can occur via either a $b \to c$ tree
or a $b \to d$ penguin amplitude.
The flavour changing neutral current $b \to d$ penguin can be mediated by any up-type quark in the loop, and hence the amplitude can be written as
\begin{equation}
  \label{eq:cp_uta:b_to_d}
  \begin{array}{ccccc}
    A_{b \to d} & = &
    \mc{3}{l}{F_u V_{ub}V^*_{ud} + F_c V_{cb}V^*_{cd} + F_t V_{tb}V^*_{td}} \\
    & = & (F_u - F_c) V_{ub}V^*_{ud} & + & (F_t - F_c) V_{tb}V^*_{td} \, , \\
  \end{array}
\end{equation}
where $F_{u,c,t}$ describe all factors, except CKM suppression, in each quark loop diagram.
In the last line, both terms are ${\cal O}(\lambda^3)$, exposing that the $b \to d$ penguin amplitude contains terms with different weak phases at the same order of CKM suppression.

In Eq.~(\ref{eq:cp_uta:b_to_d}), we have chosen to eliminate the $F_c$ term using unitarity.
However, we could equally well write
\begin{equation}
  \label{eq:cp_uta:b_to_d_alt}
  \begin{array}{ccccc}
    A_{b \to d}
    & = & (F_u - F_t) V_{ub}V^*_{ud} & + & (F_c - F_t) V_{cb}V^*_{cd} \\
    & = & (F_c - F_u) V_{cb}V^*_{cd} & + & (F_t - F_u) V_{tb}V^*_{td} \, . \\
  \end{array}
\end{equation}
Since the $b \to c\bar{c}d$ tree amplitude has the weak phase of $V_{cb}V^*_{cd}$,
either of the above expressions allows the penguin amplitude to be decomposed into a part with
the same weak phase as the tree amplitude and a part with another weak phase, which can be chosen to be either $\beta$ or $\gamma$.
The choice of parametrisation cannot, of course, affect the physics~\cite{Botella:2005ks}.
In any case, if the tree amplitude dominates, there is little sensitivity to any phase other than that from $\Bz$--$\Bzb$ mixing.

\begin{figure}[htbp]
  \begin{center}
    \begin{tabular}{cc}
      \resizebox{0.46\textwidth}{!}{
        \includegraphics{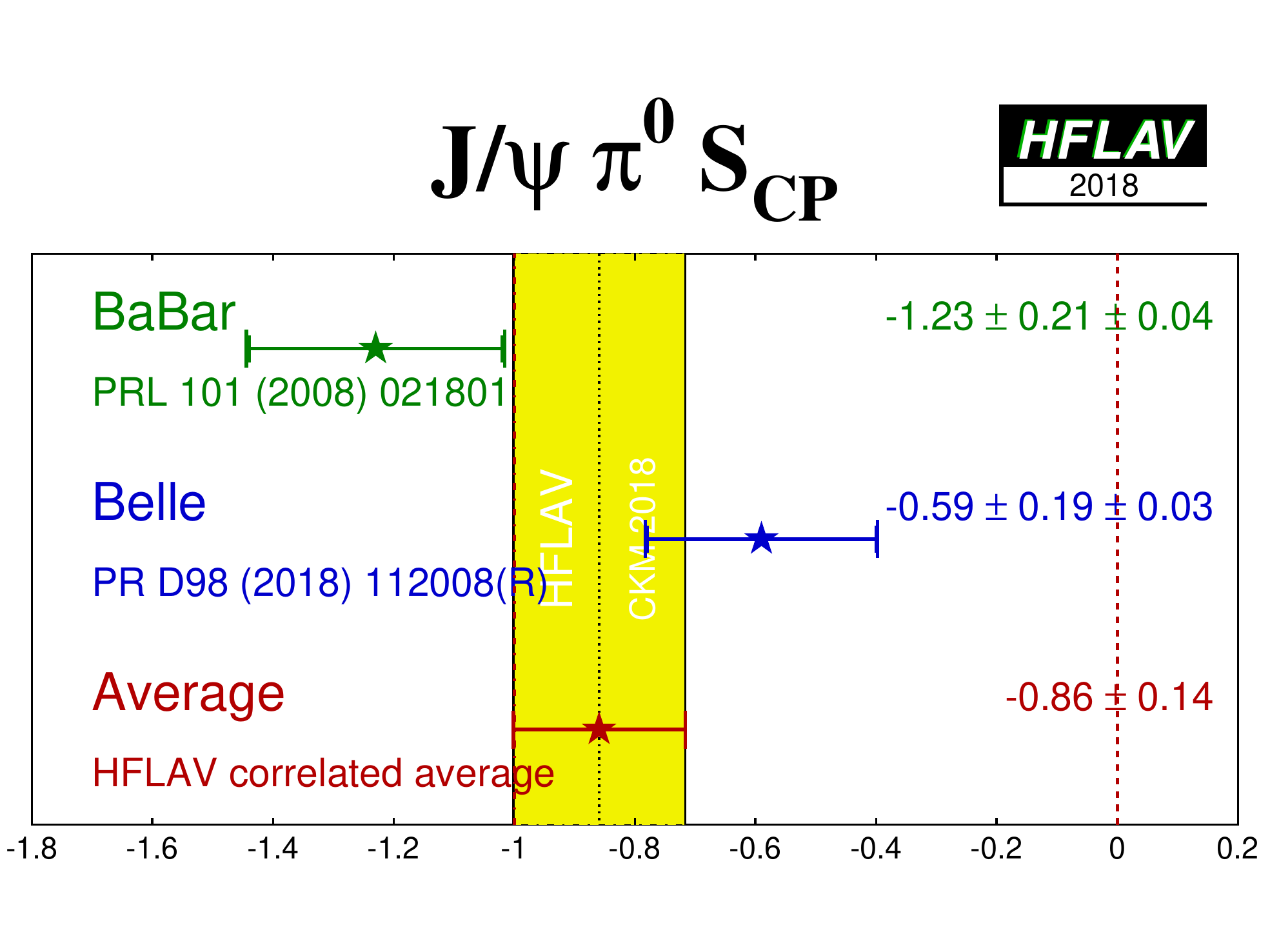}
      }
      &
      \resizebox{0.46\textwidth}{!}{
        \includegraphics{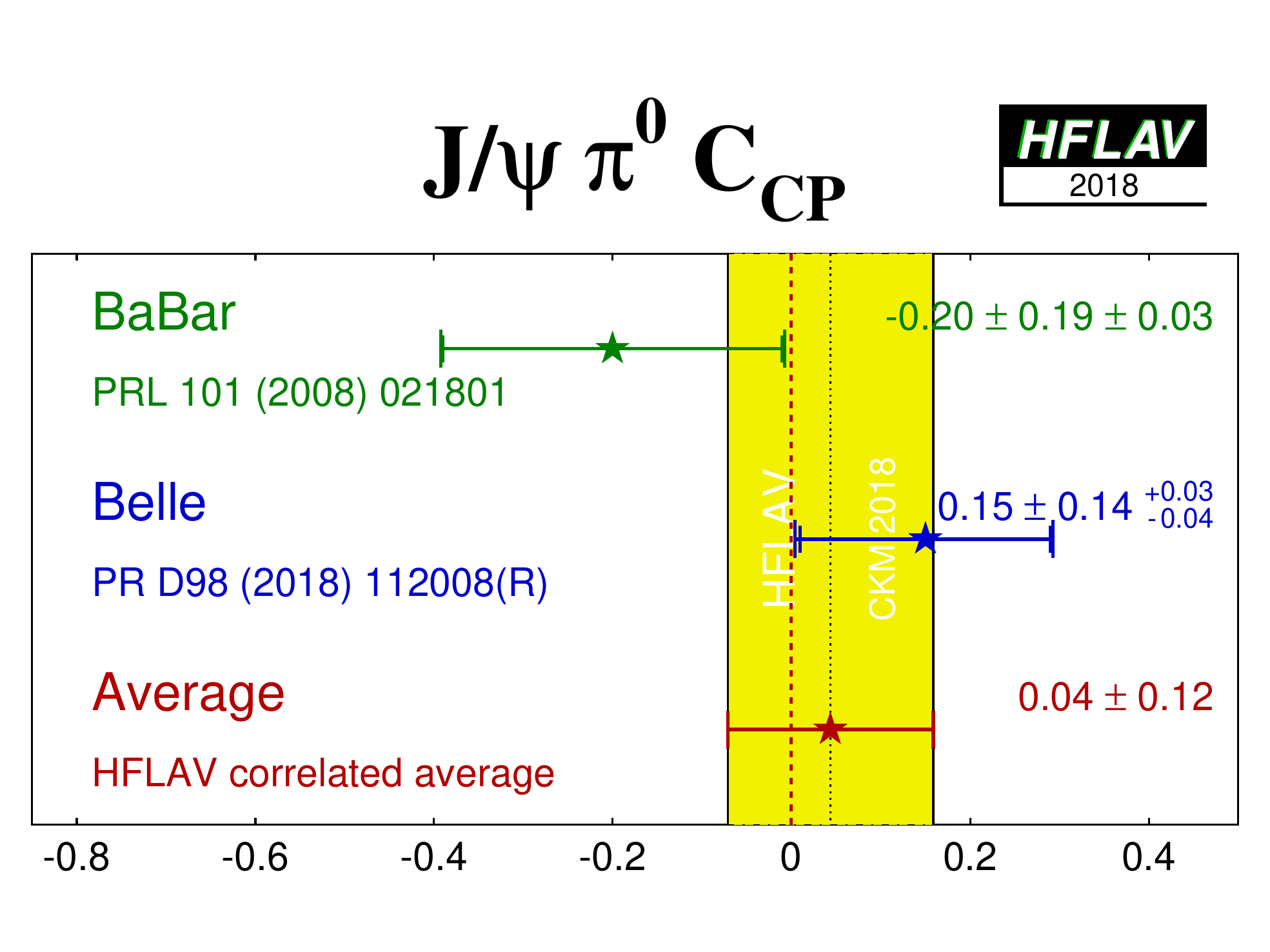}
      }
    \end{tabular}
  \end{center}
  \vspace{-0.8cm}
  \caption{
    Averages of
    (left) $S_{b \to c\bar c d}$ and (right) $C_{b \to c\bar c d}$
    for the mode $\Bz \to J/ \psi \pi^0$.
  }
  \label{fig:cp_uta:ccd:psipi0}
\end{figure}

\begin{figure}[htbp]
  \begin{center}
    \begin{tabular}{cc}
      \resizebox{0.46\textwidth}{!}{
        \includegraphics{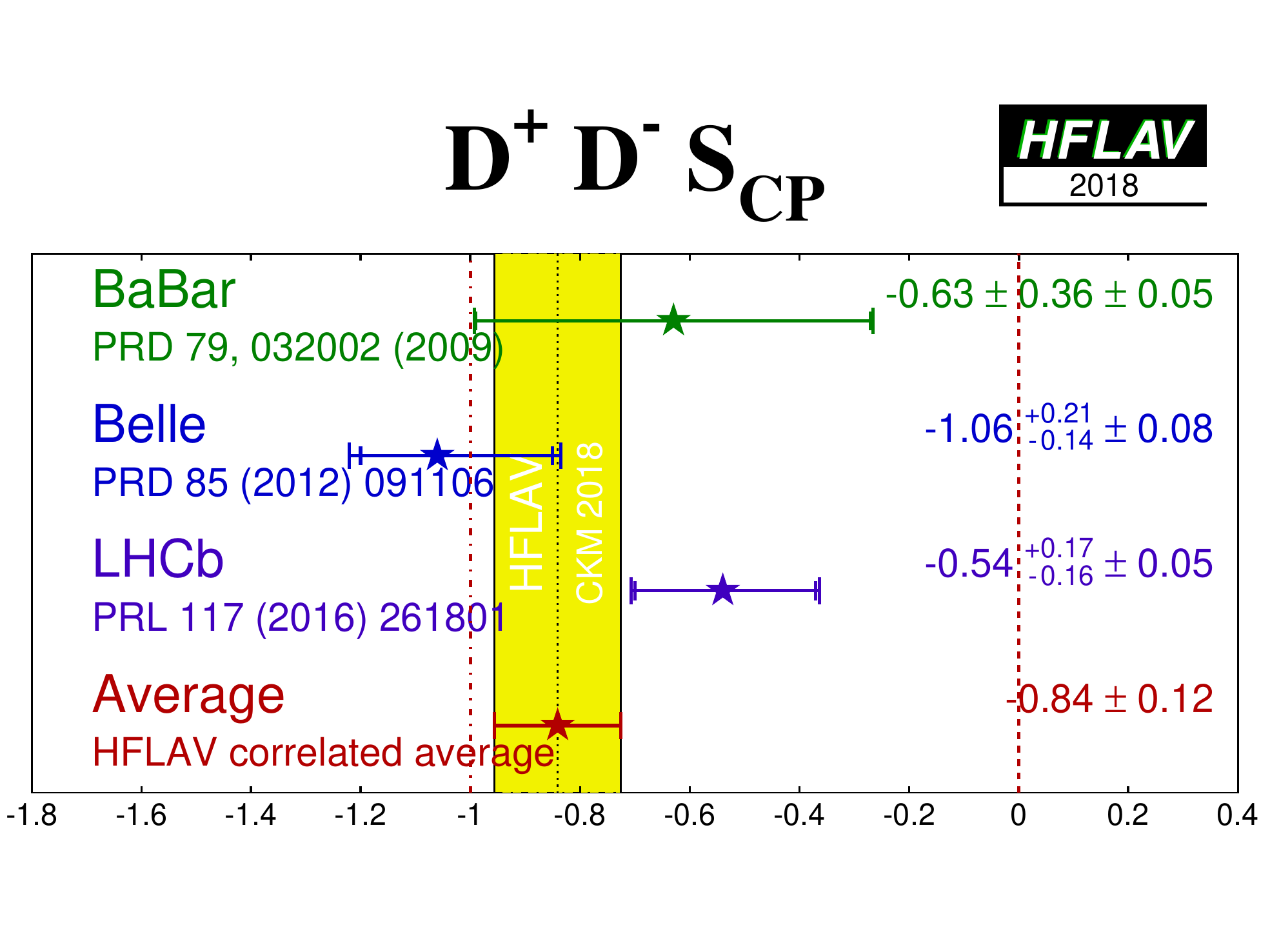}
      }
      &
      \resizebox{0.46\textwidth}{!}{
        \includegraphics{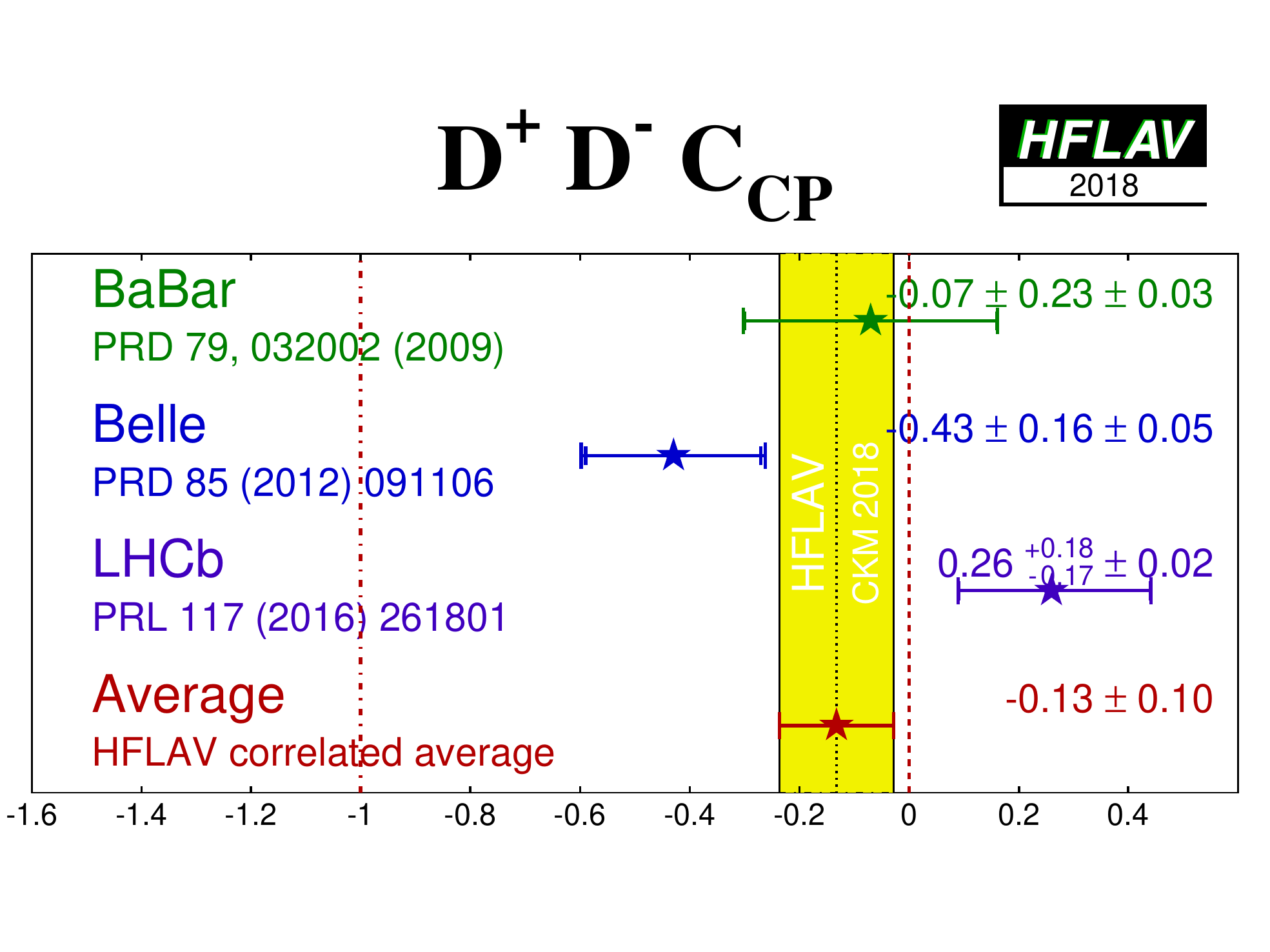}
      }
    \end{tabular}
  \end{center}
  \vspace{-0.8cm}
  \caption{
    Averages of
    (left) $S_{b \to c\bar c d}$ and (right) $C_{b \to c\bar c d}$
    for the mode $\Bz \to D^+D^-$.
  }
  \label{fig:cp_uta:ccd:dd}
\end{figure}

\begin{figure}[htbp]
  \begin{center}
    \begin{tabular}{cc}
      \resizebox{0.46\textwidth}{!}{
        \includegraphics{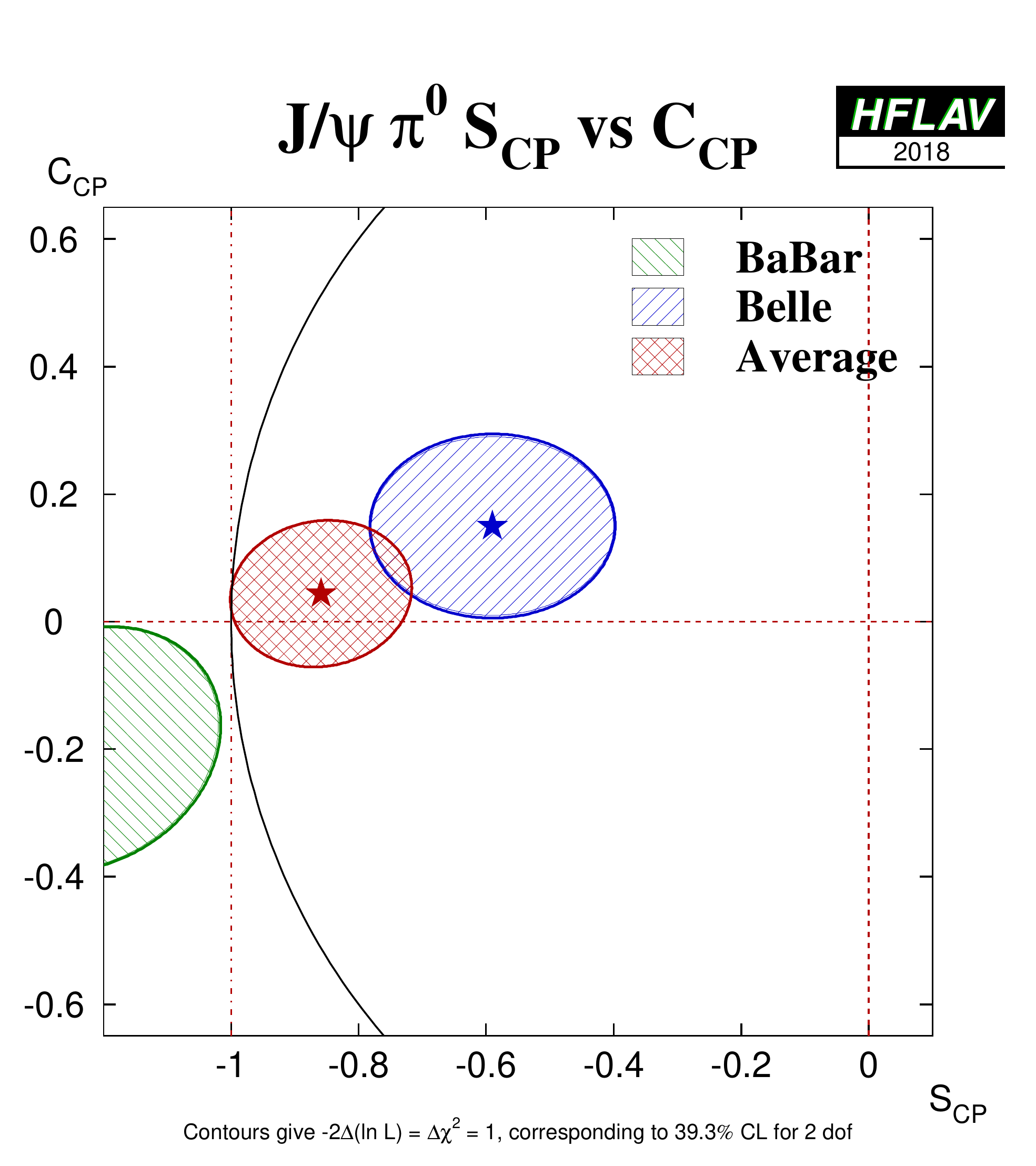}
      }
      &
      \resizebox{0.46\textwidth}{!}{
        \includegraphics{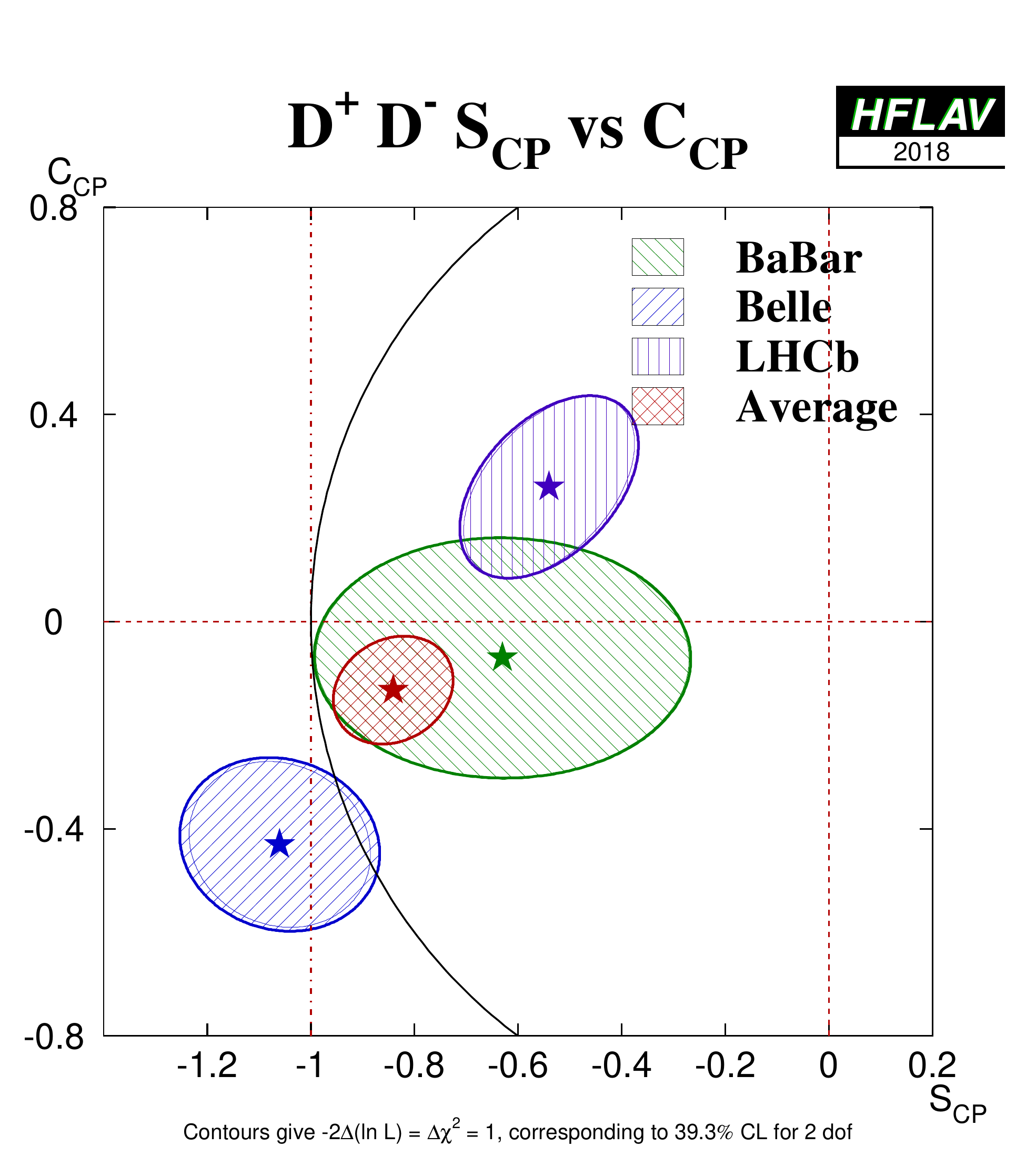}
      }
    \end{tabular}
  \end{center}
  \vspace{-0.8cm}
  \caption{
    Averages of two $b \to c\bar c d$ dominated channels,
    for which correlated averages are performed,
    in the $S_{\CP}$ \vs\ $C_{\CP}$ plane.
    (Left) $\Bz \to J/ \psi \pi^0$ and (right) $\Bz \to D^+D^-$.
  }
  \label{fig:cp_uta:ccd_SvsC}
\end{figure}

The $b \to c\bar{c}d$ transitions can be investigated with studies
of various final states.
Results are available from both \babar\ and \belle\
using the final states $\jpsi \pi^0$, $D^+D^-$, $D^{*+}D^{*-}$ and $D^{*\pm}D^{\mp}$,
and from LHCb using the final states $\jpsi \rho^0$ and $D^+D^-$;
the averages of these results are given in Tables~\ref{tab:cp_uta:ccd1} and~\ref{tab:cp_uta:ccd2}.
The results using the
$\CP$-even modes $\jpsi \pi^0$ and $D^+D^-$
are shown in Fig.~\ref{fig:cp_uta:ccd:psipi0} and
Fig.~\ref{fig:cp_uta:ccd:dd} respectively,
with two-dimensional constraints shown in Fig.~\ref{fig:cp_uta:ccd_SvsC}.

\begin{table}[htb]
	\begin{center}
		\caption{
     Averages for the $b \to c\bar{c}d$ modes,
     $\Bz \to J/\psi \pi^{0}$ and $D^+D^-$.
		}
		\vspace{0.2cm}
		\setlength{\tabcolsep}{0.0pc}
\renewcommand{\arraystretch}{1.1}
		\begin{tabular*}{\textwidth}{@{\extracolsep{\fill}}lrcccc} \hline
	\mc{2}{l}{Experiment} & Sample size & $S_{\CP}$ & $C_{\CP}$ & Correlation \\
	\hline
        \mc{6}{c}{$J/\psi \pi^{0}$} \\
	\babar & \cite{Aubert:2008bs} & $N(B\bar{B})$ = 466M & $-1.23 \pm 0.21 \pm 0.04$ & $-0.20 \pm 0.19 \pm 0.03$ & $0.20$ \\
	\belle & \cite{Pal:2018olx} & $N(B\bar{B})$ = 772M & $-0.59 \pm 0.19 \pm 0.03$ & $0.15 \pm 0.14 \,^{+0.03}_{-0.04}$ & $0.01$ \\
	\mc{3}{l}{\bf Average} & $-0.86 \pm 0.14$ & $0.04 \pm 0.12$ & $0.08$ \\
	\mc{3}{l}{\small Confidence level} & \mc{2}{c}{\small $0.04~(2.0\sigma)$} & \\
		\hline
        \mc{6}{c}{$D^{+} D^{-}$} \\
	\babar & \cite{Aubert:2008ah} & $N(B\bar{B})$ = 467M & $-0.65 \pm 0.36 \pm 0.05$ & $-0.07 \pm 0.23 \pm 0.03$ & $-0.01$ \\
	\belle & \cite{Rohrken:2012ta} & $N(B\bar{B})$ = 772M & $-1.06 \,^{+0.21}_{-0.14} \pm 0.08$ & $-0.43 \pm 0.16 \pm 0.05$ & $-0.12$ \\
	LHCb & \cite{Aaij:2016yip} & $\int {\cal L}\,dt = 3 \, {\rm fb}^{-1}$ & $-0.54 \,^{+0.17}_{-0.16} \pm 0.05$ & $0.26 \,^{+0.18}_{-0.17} \pm 0.02$ & $0.48$ \\
	\mc{3}{l}{\bf Average} & $-0.84 \pm 0.12$ & $-0.13 \pm 0.10$ & $0.18$ \\
	\mc{3}{l}{\small Confidence level} & \mc{2}{c}{\small $0.027~(2.2\sigma)$} & \\
		\hline
 		\end{tabular*}
 		\label{tab:cp_uta:ccd1}
 	\end{center}
 \end{table}

\begin{sidewaystable}
 	\begin{center}
 		\caption{
      Averages for the $b \to c\bar{c}d$ modes,
      $\jpsi\rho^0$, $D^{*+} D^{*-}$ and $D^{*\pm}D^\mp$.
 		}

\renewcommand{\arraystretch}{1.1}
 		\begin{tabular*}{\textwidth}{@{\extracolsep{\fill}}lrcccc} \hline
 		\mc{2}{l}{Experiment} & $N(B\bar{B})$ & $S_{\CP}$ & $C_{\CP}$ & $R_\perp$ \\
	\hline
        \mc{6}{c}{$\jpsi\rho^0$} \\
	LHCb & \cite{Aaij:2014vda} & 3 ${\rm fb}^{-1}$ & $-0.66 \,^{+0.13}_{-0.12} \,^{+0.09}_{-0.03}$ & $-0.06 \pm 0.06 \,^{+0.02}_{-0.01}$ & $0.198 \pm 0.017$ \\
		\hline
        \mc{6}{c}{$D^{*+} D^{*-}$} \\
	\babar & \cite{Aubert:2008ah} & 467M & $-0.70 \pm 0.16 \pm 0.03$ & $0.05 \pm 0.09 \pm 0.02$ & $0.17 \pm 0.03$ \\
	\babar part. rec. & \cite{Lees:2012px} & 471M & $-0.49 \pm 0.18 \pm 0.07 \pm 0.04$ & $0.15 \pm 0.09 \pm 0.04$ & \textemdash{} \\
	\belle & \cite{Kronenbitter:2012ha} & 772M & $-0.79 \pm 0.13 \pm 0.03$ & $-0.15 \pm 0.08 \pm 0.02$ & $0.14 \pm 0.02 \pm 0.01$ \\
	\mc{3}{l}{\bf Average} & $-0.71 \pm 0.09$ & $-0.01 \pm 0.05$ & $0.15 \pm 0.02$ \\
	\mc{3}{l}{\small Confidence level} & \mc{3}{c}{\small $0.72~(0.4\sigma)$} \\
		\hline
		\end{tabular*}

                \vspace{2ex}

    \resizebox{\textwidth}{!}{
		\begin{tabular}{@{\extracolsep{2mm}}lrcccccc} \hline
	\mc{2}{l}{Experiment} & $N(B\bar{B})$ & $S_{\CP+}$ & $C_{\CP+}$ & $S_{\CP-}$ & $C_{\CP-}$ & $R_\perp$ \\
	\hline
        \mc{8}{c}{$D^{*+} D^{*-}$} \\
	\babar & \cite{Aubert:2008ah} & 467M & $-0.76 \pm 0.16 \pm 0.04$ & $0.02 \pm 0.12 \pm 0.02$ & $-1.81 \pm 0.71 \pm 0.16$ & $0.41 \pm 0.50 \pm 0.08$ & $0.15 \pm 0.03$ \\
		\hline
		\end{tabular}
    }

                \vspace{2ex}

    \resizebox{\textwidth}{!}{
		\begin{tabular}{@{\extracolsep{2mm}}lrcccccc} \hline
	\mc{2}{l}{Experiment} & $N(B\bar{B})$ & $S$ & $C$ & $\Delta S$ & $\Delta C$ & ${\cal A}$ \\
        \hline
        \mc{8}{c}{$D^{*\pm} D^{\mp}$} \\
	\babar & \cite{Aubert:2008ah} & 467M & $-0.68 \pm 0.15 \pm 0.04$ & $0.04 \pm 0.12 \pm 0.03$ & $0.05 \pm 0.15 \pm 0.02$ & $0.04 \pm 0.12 \pm 0.03$ & $0.01 \pm 0.05 \pm 0.01$ \\
	\belle & \cite{Rohrken:2012ta} & 772M & $-0.78 \pm 0.15 \pm 0.05$ & $-0.01 \pm 0.11 \pm 0.04$ & $-0.13 \pm 0.15 \pm 0.04$ & $0.12 \pm 0.11 \pm 0.03$ & $0.06 \pm 0.05 \pm 0.02$ \\
	\mc{3}{l}{\bf Average} & $-0.73 \pm 0.11$ & $0.01 \pm 0.09$ & $-0.04 \pm 0.11$ & $0.08 \pm 0.08$ & $0.03 \pm 0.04$ \\
	\mc{3}{l}{\small Confidence level} & {\small $0.65~(0.5\sigma)$} & {\small $0.77~(0.3\sigma)$} & {\small $0.41~(0.8\sigma)$} & {\small $0.63~(0.5\sigma)$} & {\small $0.48~(0.7\sigma)$} \\
        \hline
                \end{tabular}
    }
		\label{tab:cp_uta:ccd2}
	\end{center}
\end{sidewaystable}

Results for the vector-vector mode $\jpsi \rho^0$ are obtained from a full time-dependent amplitude analysis of $\Bz \to \jpsi \pip\pim$ decays.
LHCb~\cite{Aaij:2014vda} finds a $\jpsi \rhoz$ fit fraction of $65.6 \pm 1.9\%$ and a longitudinal polarisation fraction of $56.7 \pm 1.8\%$ (uncertainties are statistical only; both results are consistent with those from a time-integrated amplitude analysis~\cite{Aaij:2014siy} where systematic uncertainties were also evaluated).
Fits are performed to obtain $2\beta^{\rm eff}$ in the cases that all transversity amplitudes are assumed to have the same \CP violation parameter.
A separate fit is performed allowing different parameters.
The results in the former case are presented in terms of $S_{\CP}$ and $C_{\CP}$ in Table~\ref{tab:cp_uta:ccd2}.

\begin{figure}[htbp]
  \begin{center}
    \begin{tabular}{cc}
      \resizebox{0.46\textwidth}{!}{
        \includegraphics{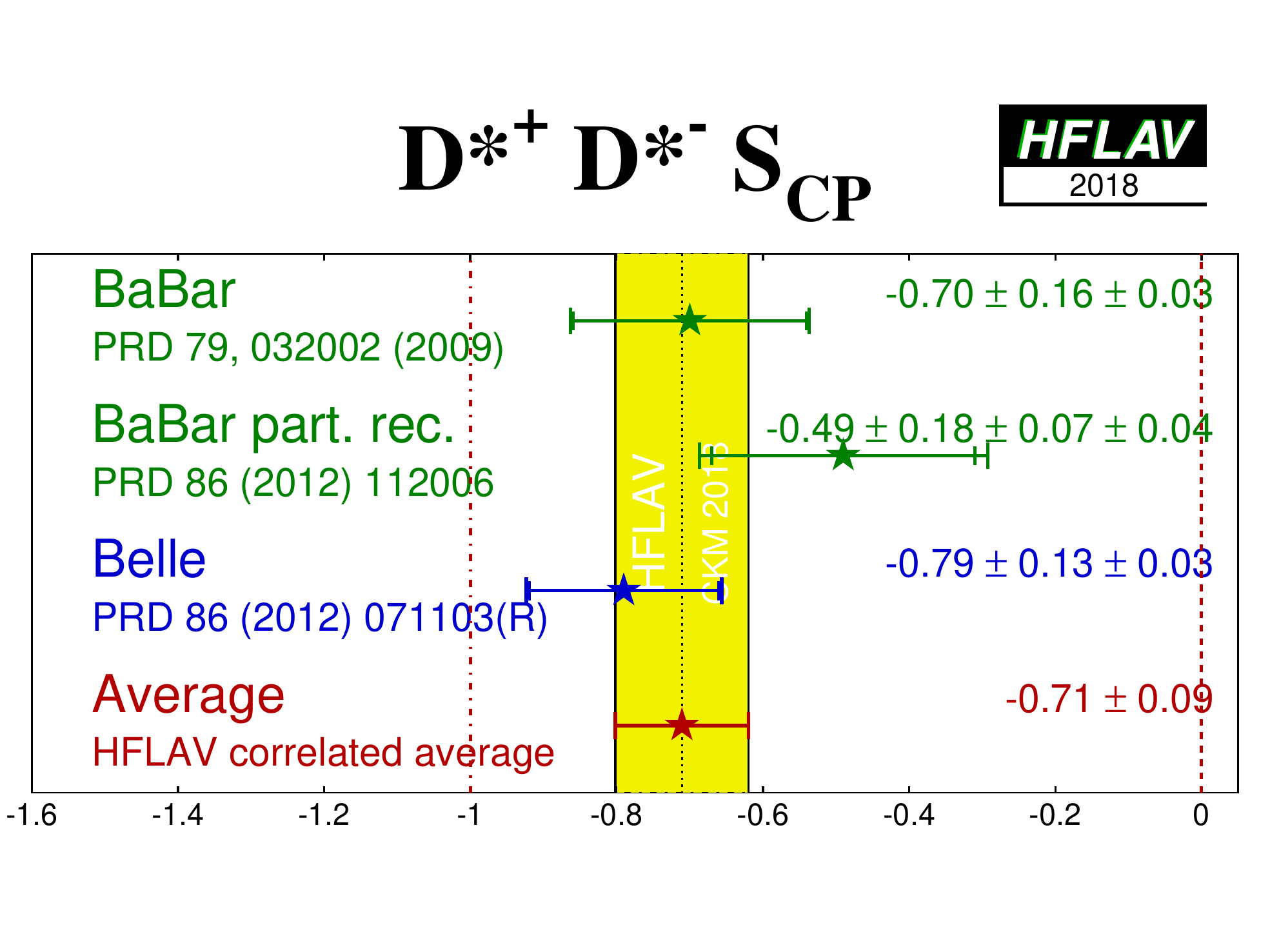}
      }
      &
      \resizebox{0.46\textwidth}{!}{
        \includegraphics{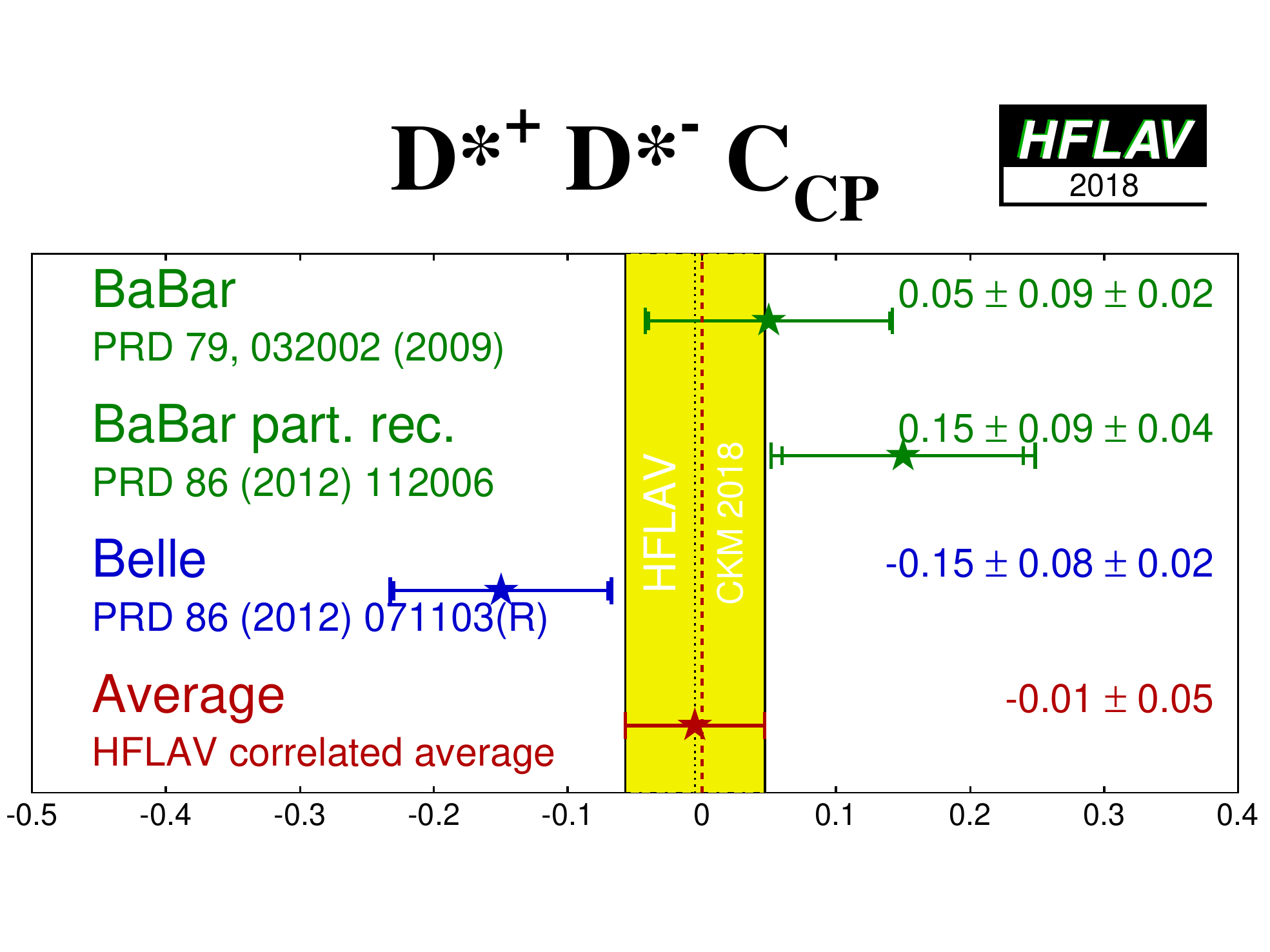}
      }
    \end{tabular}
  \end{center}
  \vspace{-0.8cm}
  \caption{
    Averages of
    (left) $S_{b \to c\bar c d}$ and (right) $C_{b \to c\bar c d}$
    for the mode $\Bz \to D^{*+}D^{*-}$.
  }
  \label{fig:cp_uta:ccd:dstardstar}
\end{figure}

The vector-vector mode $D^{*+}D^{*-}$
is found to be dominated by the $\CP$-even, longitudinally polarised component.
\babar\ measures a $\CP$-odd fraction of
$0.158 \pm 0.028 \pm 0.006$~\cite{Aubert:2008ah}, and
\belle\ measures
$0.138 \pm 0.024 \pm 0.006$~\cite{Kronenbitter:2012ha}.
These values are listed as $R_\perp$ in Table~\ref{tab:cp_uta:ccd2}, and are included in the averages so that correlations are taken into account.\footnote{
  Note that the \babar\ value given in Table~\ref{tab:cp_uta:ccd2} differs from
  the value quoted here, since that in the table is not corrected for efficiency.
}
\babar\ has also performed an additional fit in which the
$\CP$-even and $\CP$-odd components have independent pairs of
$\CP$ violation parameters $S$ and $C$.
These results are included in Table~\ref{tab:cp_uta:ccd2}.
Results using $D^{*+}D^{*-}$ are shown in Fig.~\ref{fig:cp_uta:ccd:dstardstar}.

As discussed in Sec.~\ref{sec:cp_uta:notations:non_cp}, the most recent papers on the non-$\CP$ eigenstate mode $D^{*\pm}D^{\mp}$ use the ($A$, $S$, $\Delta S$, $C$, $\Delta C$) set of parameters.
Therefore, we perform the averages with this choice, with results presented in Table~\ref{tab:cp_uta:ccd2}.

In the absence of the penguin contribution (so-called tree dominance),
the time-dependent parameters are given by
$S_{b \to c\bar c d} = - \etacp \sin(2\beta)$,
$C_{b \to c\bar c d} = 0$,
$S_{+-} = \sin(2\beta + \delta)$,
$S_{-+} = \sin(2\beta - \delta)$,
$C_{+-} = - C_{-+}$ and
${\cal A} = 0$,
where $\delta$ is the strong phase difference between the
$D^{*+}D^-$ and $D^{*-}D^+$ decay amplitudes.
In the presence of the penguin contribution,
there is no straightforward interpretation in terms of CKM parameters;
however,
$\CP$ violation in decay may be observed through any of
$C_{b \to c\bar c d} \neq 0$, $C_{+-} \neq - C_{-+}$ or $A_{+-} \neq 0$.

\begin{figure}[htbp]
  \begin{center}
    \begin{tabular}{cc}
      \resizebox{0.46\textwidth}{!}{
        \includegraphics{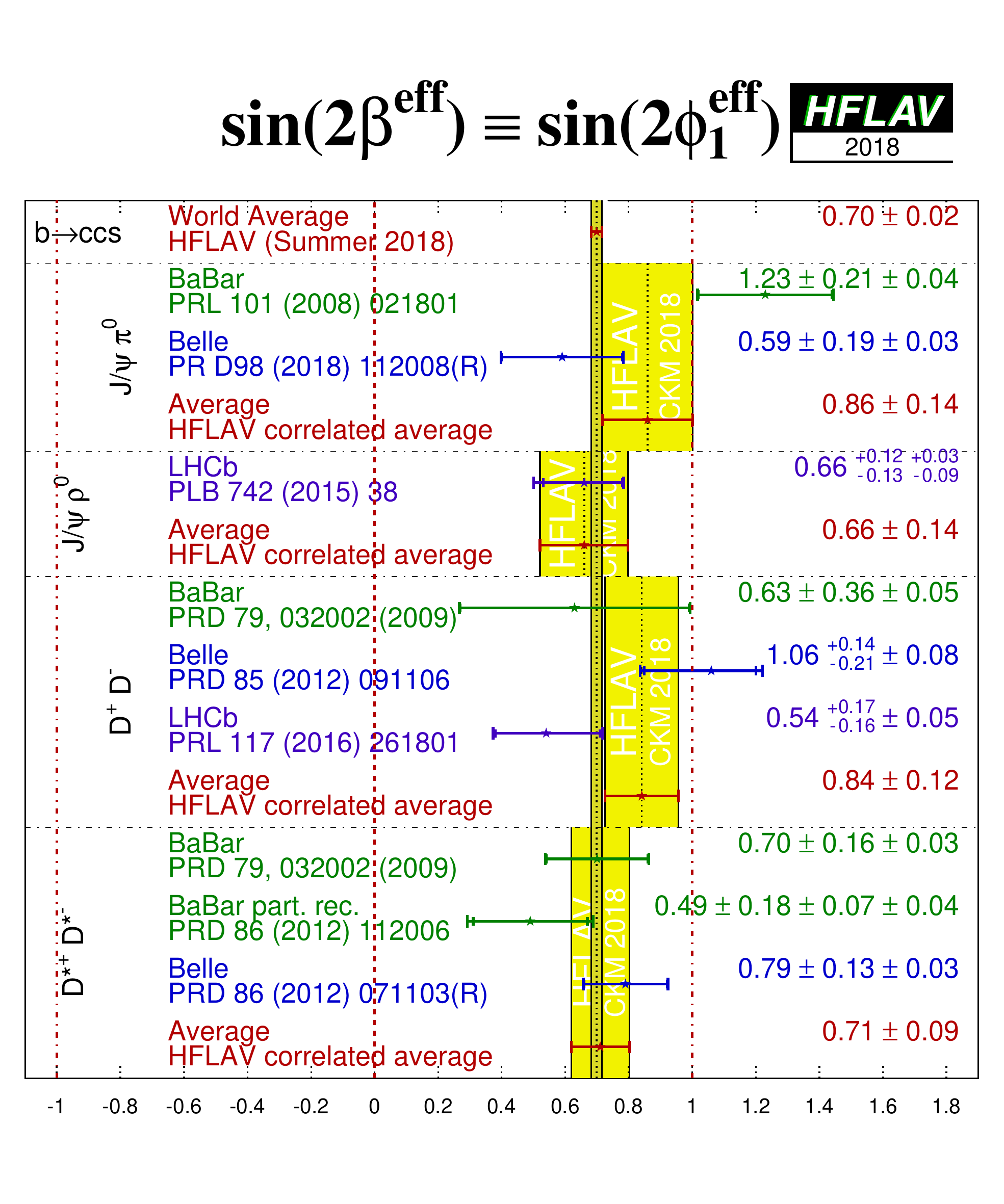}
      }
      &
      \resizebox{0.46\textwidth}{!}{
        \includegraphics{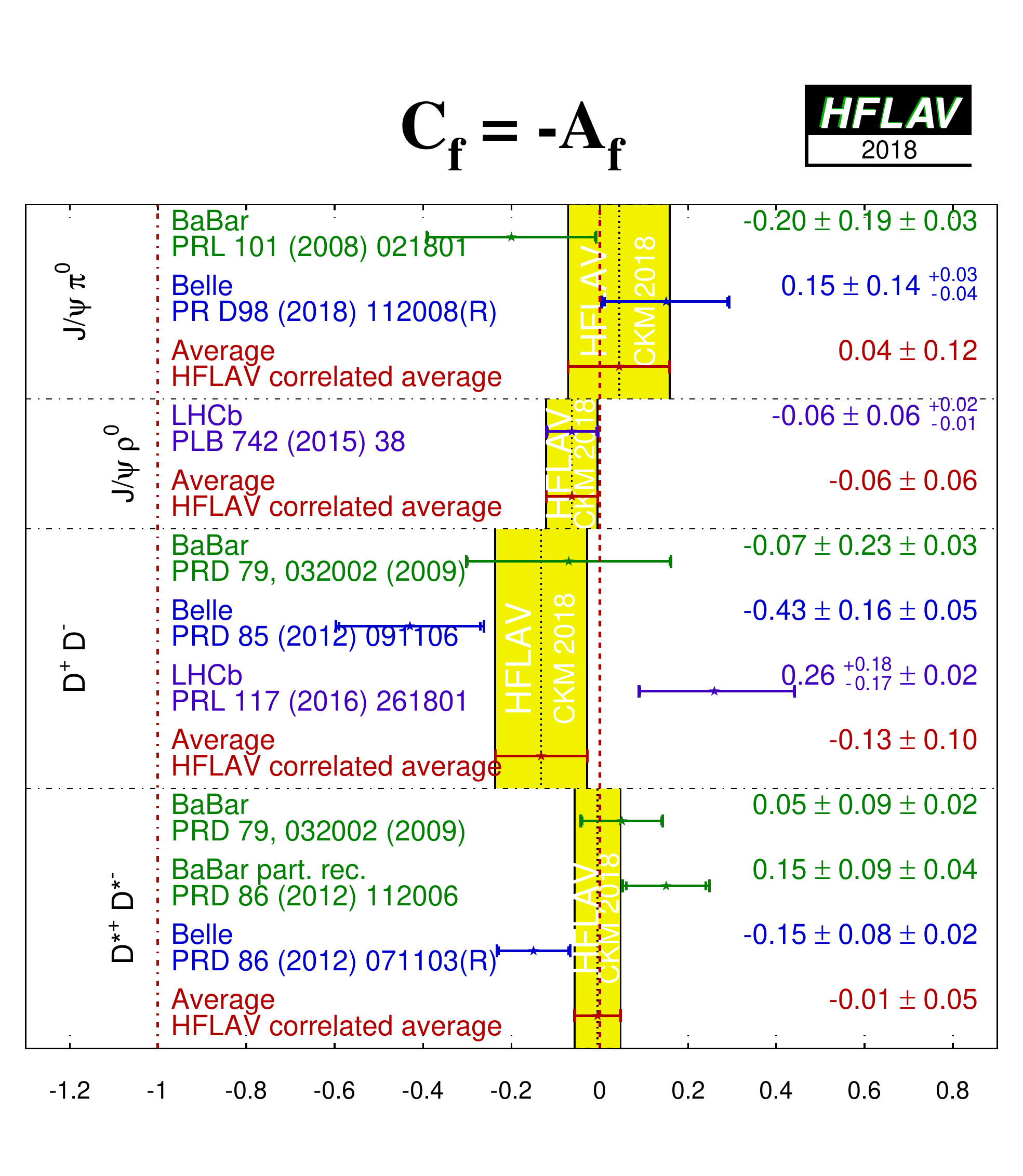}
      }
    \end{tabular}
  \end{center}
  \vspace{-0.8cm}
  \caption{
    Averages of
    (left) $-\etacp S_{b \to c\bar c d}$ interpreted as $\sin(2\beta^{\rm eff})$ and (right) $C_{b \to c\bar c d}$.
    The $-\etacp S_{b \to c\bar c d}$ figure compares the results to
    the world average
    for $-\etacp S_{b \to c\bar c s}$ (see Sec.~\ref{sec:cp_uta:ccs:cp_eigen}).
  }
  \label{fig:cp_uta:ccd}
\end{figure}

\begin{figure}[htbp]
  \begin{center}
    \resizebox{0.66\textwidth}{!}{
      \includegraphics{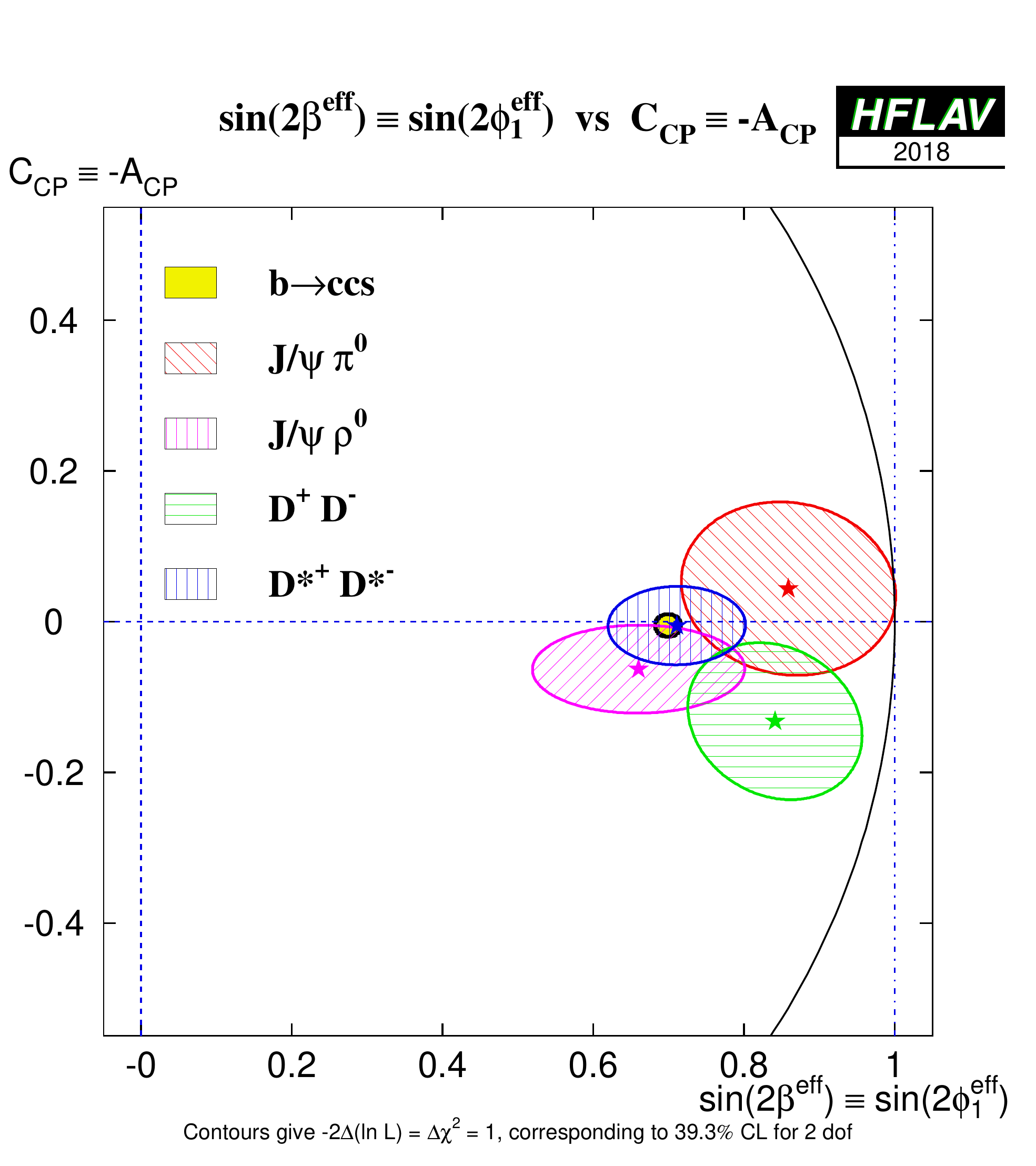}
    }
  \end{center}
  \vspace{-0.8cm}
  \caption{
    Compilation of constraints in the
    $-\etacp S_{b \to c\bar c d}$, interpreted as $\sin(2\beta^{\rm eff})$, \vs\ $C_{b \to c\bar c d}$ plane.
  }
  \label{fig:cp_uta:ccd_SvsC-all}
\end{figure}

The averages for the $b \to c\bar c d$ modes
are shown in Figs.~\ref{fig:cp_uta:ccd} and~\ref{fig:cp_uta:ccd_SvsC-all}.
Results are consistent with tree dominance and with the Standard Model,
although the \belle\ results in $\Bz \to D^+D^-$~\cite{Fratina:2007zk}
show an indication of $\CP$ violation in decay,
and hence a non-zero penguin contribution.
The average of $S_{b \to c\bar c d}$ in each of the $J/\psi \pi^{0}$, $\Dp\Dm$ and $D^{*+}D^{*-}$ final states is more than $5\sigma$ away from zero, corresponding to observations of \CP violation in these decay channels.
Possible non-Gaussian effects due to some of the input measurements being outside the physical region ($S_{\CP}^2 + C_{\CP}^2 \leq 1$) should, however, be borne in mind.

\mysubsubsection{Time-dependent $\CP$ asymmetries in $\Bs$ decays mediated by $b \to c\bar{c}d$ transitions
}
\label{sec:cp_uta:ccd:Bs}

Time-dependent \CP asymmetries in \Bs\ decays mediated by $b \to c\bar{c}d$ transitions provide a determination of $2\beta_s^{\rm eff}$, where possible effects from penguin amplitudes may cause a shift from the value of $2\beta_s$ seen in $b \to c\bar{c}s$ transitions.
Results in the $b \to c\bar{c}d$ case, with larger penguin effects, can be used together with flavour symmetries to derive limits on the possible size of penguin effects in the $b \to c\bar{c}s$ transitions~\cite{Fleischer:1999nz,DeBruyn:2010hh}.

The parameters have been measured in $\Bs \to \jpsi\KS$ decays by LHCb, as summarised in Table~\ref{tab:cp_uta:ccd:Bs}.
The results supersede an earlier measurement of the effective lifetime, which is directly related to $A^{\Delta\Gamma}$, in the same mode~\cite{Aaij:2013eia}. %

\begin{table}[!htb]
	\begin{center}
		\caption{
     Measurements of \CP violation parameters from $\Bs \to \jpsi\KS$.
		}
		\vspace{0.2cm}
		\setlength{\tabcolsep}{0.0pc}
\renewcommand{\arraystretch}{1.1}
		\begin{tabular*}{\textwidth}{@{\extracolsep{\fill}}lrcccc} \hline
	\mc{2}{l}{Experiment} & $\int {\cal L}\,dt$ & $S_{\CP}$ & $C_{\CP}$ & $A^{\Delta\Gamma}$ \\
	\hline
	LHCb & \cite{Aaij:2015tza} & $3 \ {\rm fb}^{-1}$ & $0.49 \,^{+0.77}_{-0.65} \pm 0.06$ & $-0.28 \pm 0.41 \pm 0.08$ & $-0.08 \pm 0.40 \pm 0.08$ \\
		\hline
		\end{tabular*}
		\label{tab:cp_uta:ccd:Bs}
	\end{center}
\end{table}

\mysubsection{Time-dependent $\CP$ asymmetries in charmless $b \to q\bar{q}s$ transitions
}
\label{sec:cp_uta:qqs}

Similarly to Eq.~(\ref{eq:cp_uta:b_to_d}), the $b \to s$ penguin amplitude can be written as
\begin{equation}
  \label{eq:cp_uta:b_to_s}
  \begin{array}{ccccc}
    A_{b \to s} & = &
    \mc{3}{l}{F_u V_{ub}V^*_{us} + F_c V_{cb}V^*_{cs} + F_t V_{tb}V^*_{ts}} \\
    & = & (F_u - F_c) V_{ub}V^*_{us} & + & (F_t - F_c) V_{tb}V^*_{ts} \, , \\
  \end{array}
\end{equation}
using the unitarity of the CKM matrix to eliminate the $F_c$ term.
In this case, the first term in the last line is ${\cal O}(\lambda^4)$ while the second is ${\cal O}(\lambda^2)$.
Therefore, in the Standard Model, this amplitude is dominated by $V_{tb}V^*_{ts}$, and to within a few degrees ($\left| \delta\beta^{\rm eff} \right| \equiv \left| \beta^{\rm eff}-\beta \right| \lesssim 2^\circ$ for $\beta \approx 20^\circ$) the time-dependent parameters can be written as\footnote{
  The presence of a small (${\cal O}(\lambda^2)$) weak phase in the dominant amplitude of the $s$ penguin decays introduces a phase shift given by
  $S_{b \to q\bar q s} = -\eta\sin(2\beta)(1 + \Delta)$.
  Using the CKMfitter results for the Wolfenstein parameters~\cite{Charles:2004jd}, one finds $\Delta \simeq 0.033$, which corresponds to a shift of $2\beta$ of $+2.1^\circ$.
  Nonperturbative contributions can alter this result.
}
$S_{b \to q\bar q s} \approx - \etacp \sin(2\beta)$,
$C_{b \to q\bar q s} \approx 0$,
assuming $b \to s$ penguin contributions only ($q = u,d,s$).

Due to the suppression of the Standard Model amplitude, contributions of additional diagrams from physics beyond the Standard Model,
with heavy virtual particles in the penguin loops, may have observable effects.
In general, these contributions will affect the values of
$S_{b \to q\bar q s}$ and $C_{b \to q\bar q s}$.
A discrepancy between the values of
$S_{b \to c\bar c s}$ and $S_{b \to q\bar q s}$
can therefore provide a solid indication of non-Standard Model  physics~\cite{Grossman:1996ke,Fleischer:1996bv,London:1997zk,Ciuchini:1997zp}.

However, there is an additional consideration to take into account.
The above argument assumes that only the $b \to s$ penguin contributes
to the $b \to q\bar q s$ transition.
For $q = s$ this is a good assumption, which neglects only rescattering effects.
However, for $q = u$ there is a colour-suppressed $b \to u$ tree diagram
(of order ${\cal O}(\lambda^4)$),
which has a different weak (and possibly strong) phase.
In the case $q = d$, any light neutral meson that is formed from $d \bar{d}$
also has a $u \bar{u}$ component, and so again there is ``tree pollution''.
The \Bz decays to $\piz\KS$, $\rho^0\KS$ and $\omega\KS$ belong to this category.
The mesons $\phi$, $f_0$ and $\etapr$ are expected to have predominant
$s\bar{s}$ composition, which reduces the relative size of the possible tree
pollution.
If the inclusive decay $\Bz\to\Kp\Km\Kz$ (excluding $\phi\Kz$) is dominated by
a nonresonant three-body transition,
an Okubo-Zweig-Iizuka-suppressed~\cite{Okubo:1963fa,Zweig:1964jf,Iizuka:1966fk} tree-level diagram can occur through insertion of an $s\sbar$ pair.
The corresponding penguin-type transition
proceeds via insertion of a $u\ubar$ pair, which is expected
to be favoured over the $s\sbar$ insertion by fragmentation models.
Neglecting rescattering, the final state $\Kz\Kzb\Kz$
(reconstructed as $\KS\KS\KS$) has no tree pollution~\cite{Gershon:2004tk}.
Various estimates, using different theoretical approaches,
of the values of $\Delta S = S_{b \to q\bar q s} - S_{b \to c\bar c s}$
exist in the literature~\cite{Grossman:2003qp,Gronau:2003ep,Gronau:2003kx,Gronau:2004hp,Cheng:2005bg,Gronau:2005gz,Buchalla:2005us,Beneke:2005pu,Engelhard:2005hu,Cheng:2005ug,Engelhard:2005ky,Gronau:2006qh,Silvestrini:2007yf,Dutta:2008xw}.
In general, there is agreement that the modes
$\phi\Kz$, $\etapr\Kz$ and $\Kz\Kzb\Kz$ are the cleanest,
with values of $\left| \Delta S \right|$ at or below the few percent level,
with $\Delta S$ usually predicted to be positive.
Nonetheless, the uncertainty is sufficient that interpretation is given here in terms of $\sin(2\beta^{\rm eff})$.

\mysubsubsection{Time-dependent $\CP$ asymmetries: $b \to q\bar{q}s$ decays to $\CP$ eigenstates
}
\label{sec:cp_uta:qqs:cp_eigen}

\begin{figure}[htbp]
  \begin{center}
    \resizebox{0.45\textwidth}{!}{
      \includegraphics{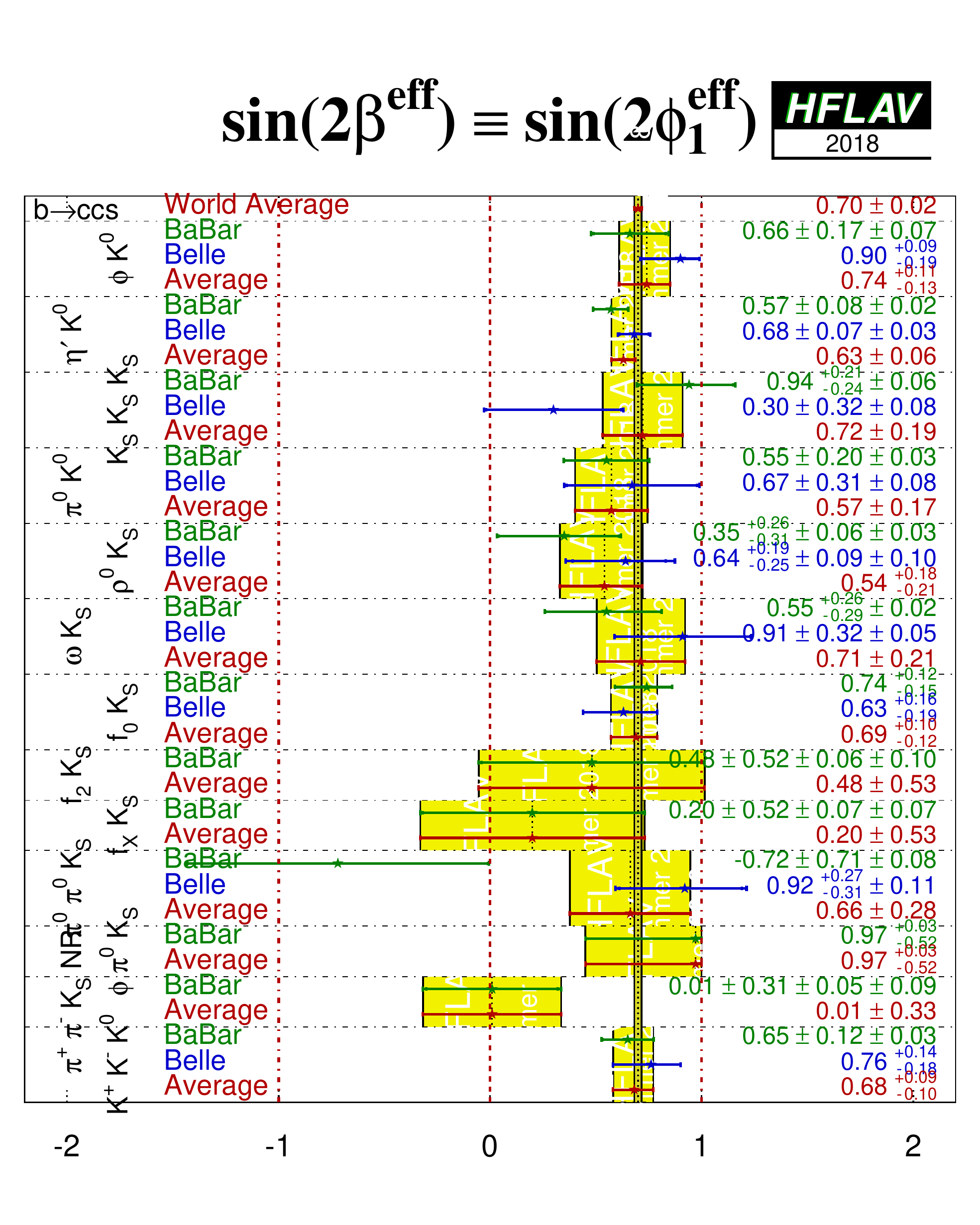}
    }
    \hfill
    \resizebox{0.45\textwidth}{!}{
      \includegraphics{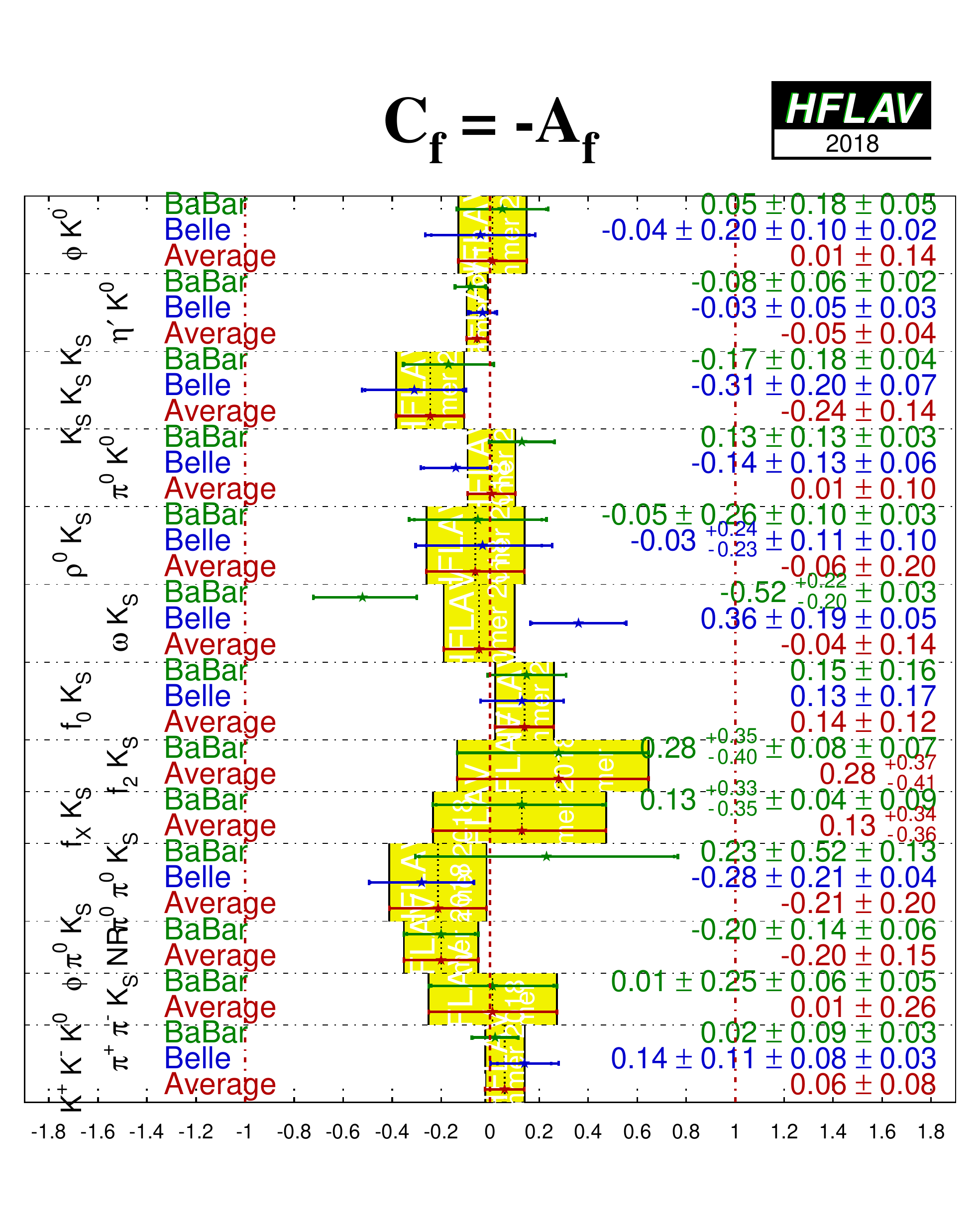}
    }
    \\
    \resizebox{0.45\textwidth}{!}{
      \includegraphics{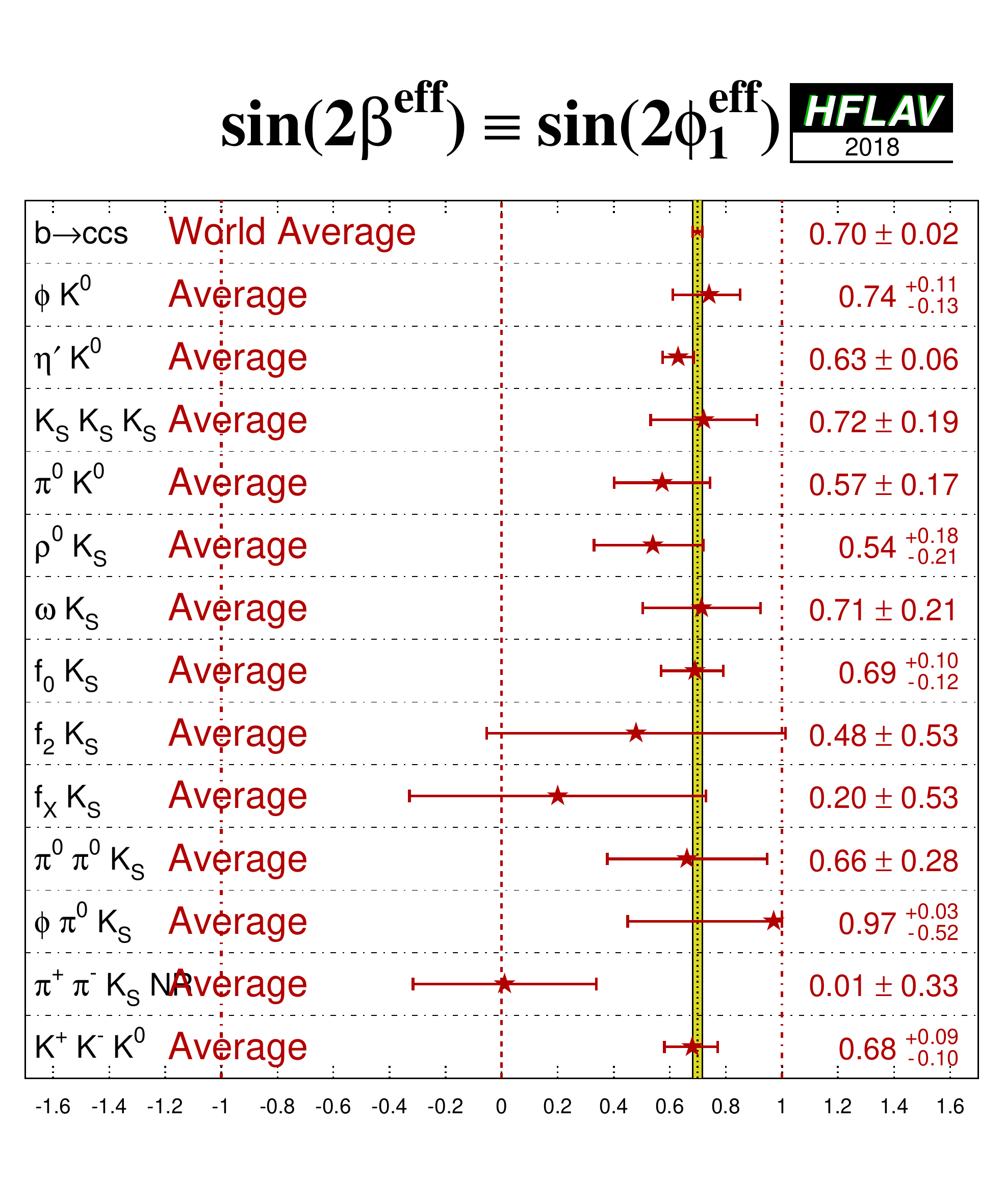}
    }
    \hfill
    \resizebox{0.45\textwidth}{!}{
      \includegraphics{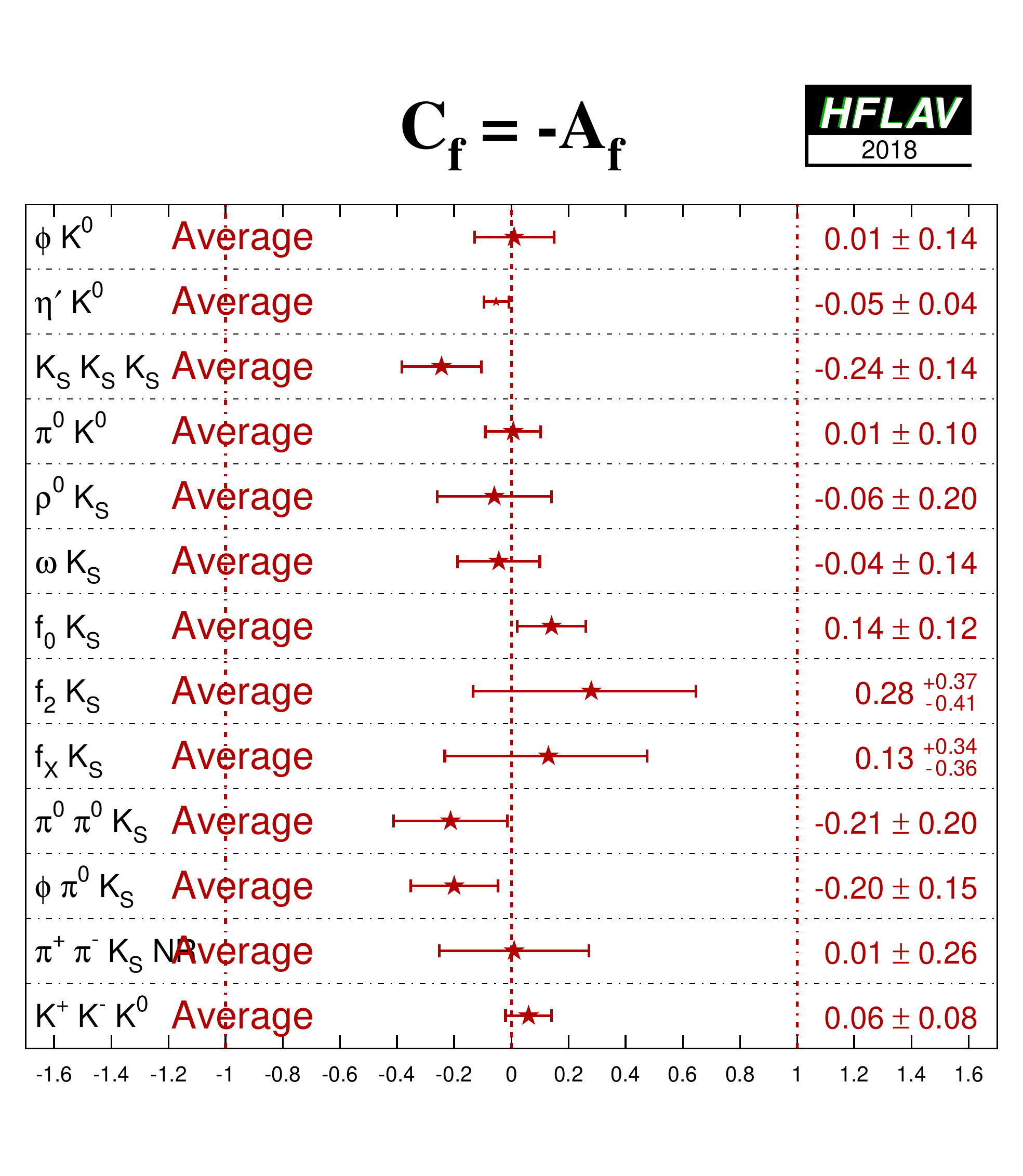}
    }
  \end{center}
  \vspace{-0.8cm}
  \caption{
    (Top)
    Averages of
    (left) $-\etacp S_{b \to q\bar q s}$, interpreted as $\sin(2\beta^{\rm eff}$ and (right) $C_{b \to q\bar q s}$.
    The $-\etacp S_{b \to q\bar q s}$ figure compares the results to
    the world average
    for $-\etacp S_{b \to c\bar c s}$ (see Sec.~\ref{sec:cp_uta:ccs:cp_eigen}).
    (Bottom) Same, but only averages for each mode are shown.
    More figures are available from the HFLAV web pages.
  }
  \label{fig:cp_uta:qqs}
\end{figure}

\begin{figure}[htbp]
  \begin{center}
    \resizebox{0.38\textwidth}{!}{
      \includegraphics{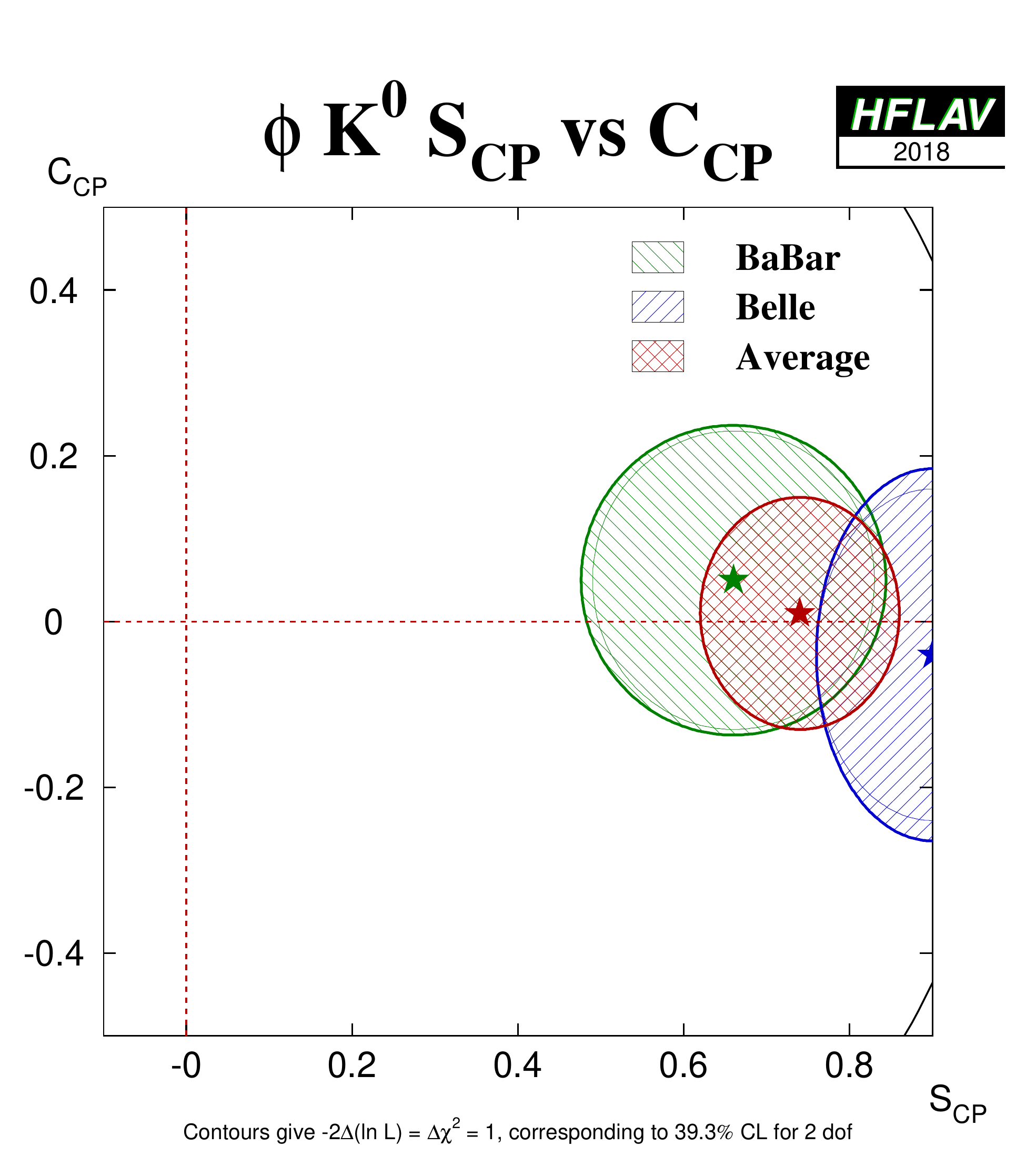}
    }
    \hspace{0.04\textwidth}
    \resizebox{0.38\textwidth}{!}{
      \includegraphics{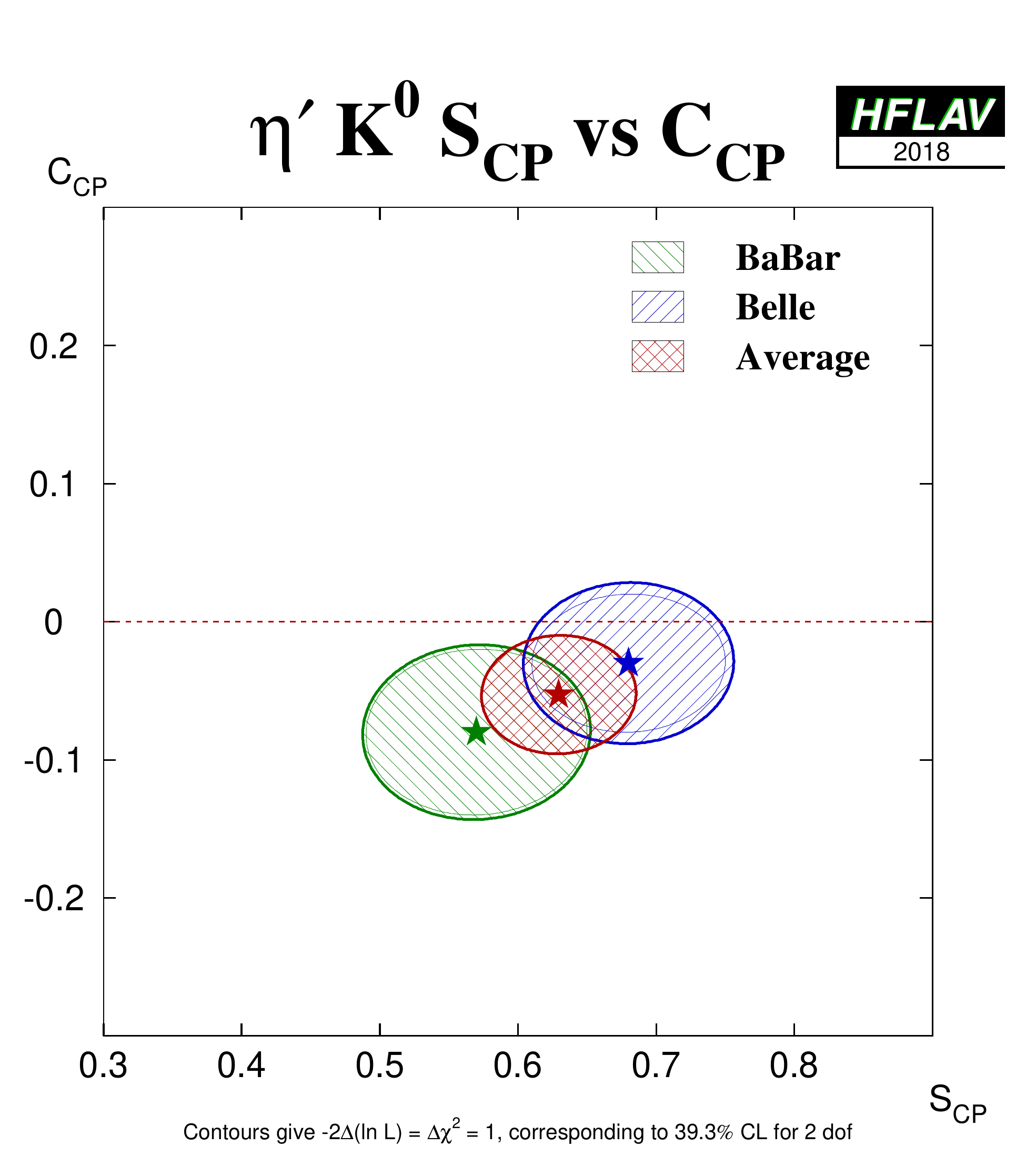}
    }
    \\
    \resizebox{0.38\textwidth}{!}{
      \includegraphics{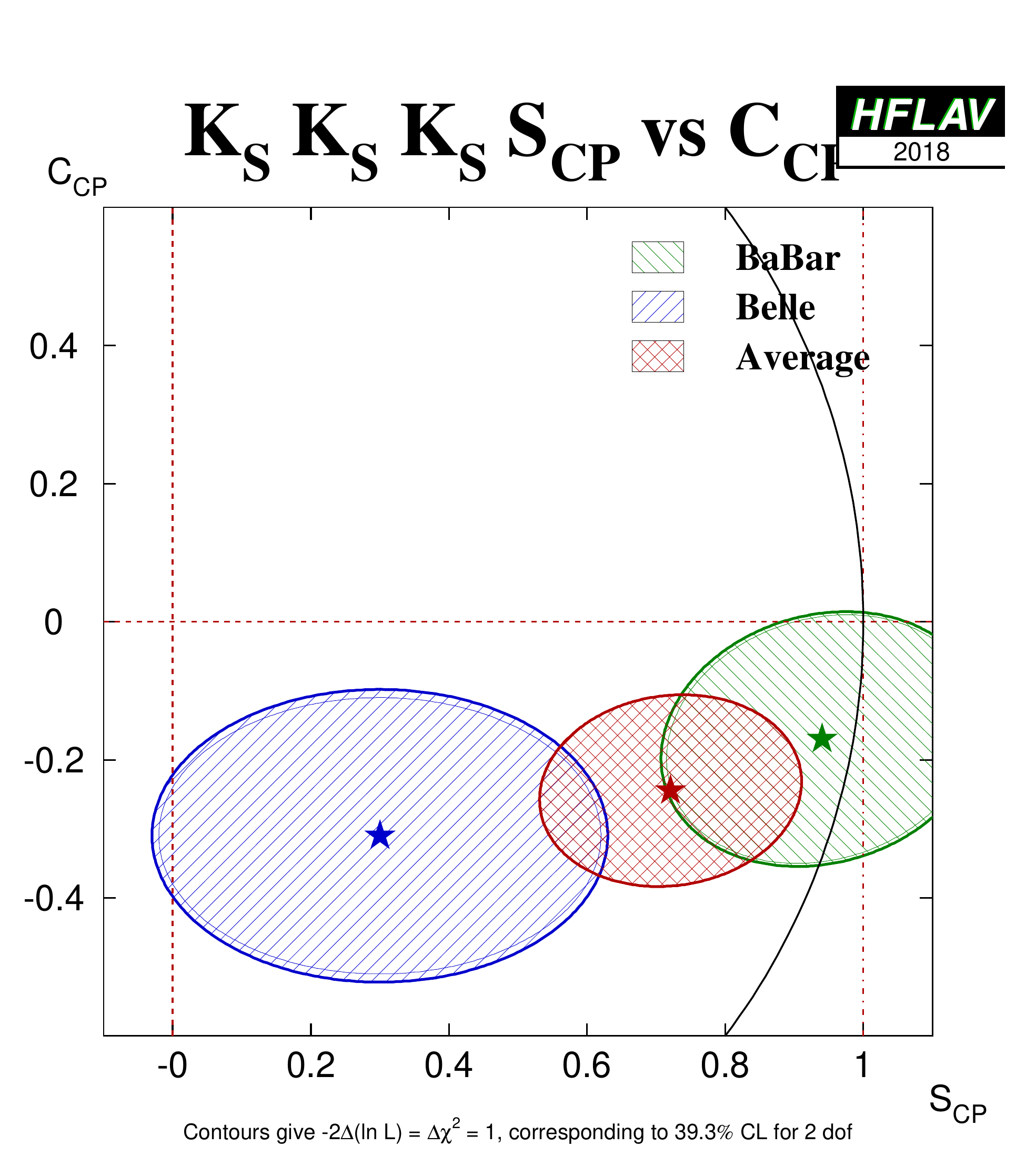}
    }
    \hspace{0.04\textwidth}
    \resizebox{0.38\textwidth}{!}{
      \includegraphics{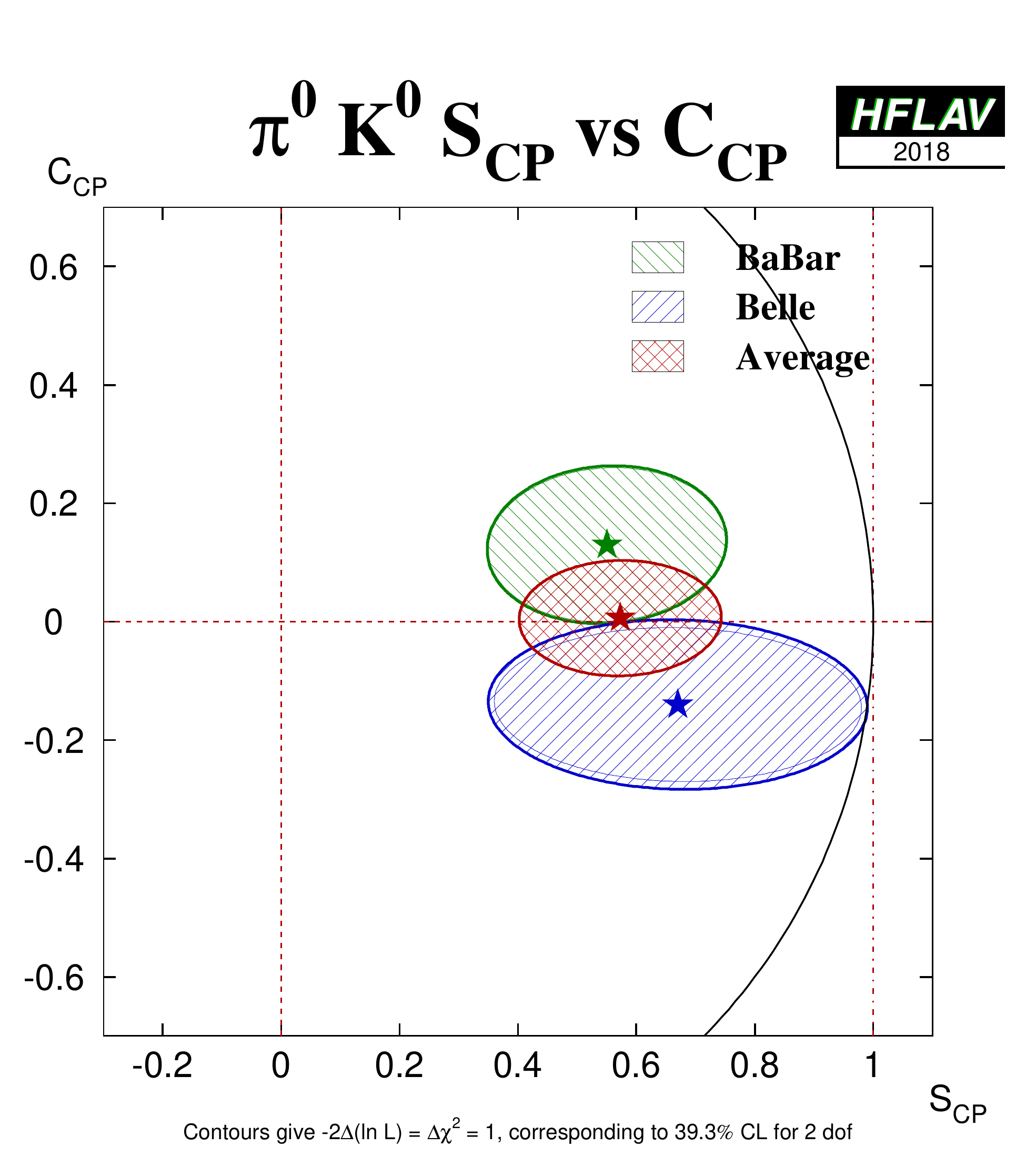}
    }
  \end{center}
  \vspace{-0.5cm}
  \caption{
    Averages of four $b \to q\bar q s$ dominated channels,
    for which correlated averages are performed,
    in the $S_{\CP}$ \vs\ $C_{\CP}$ plane,
    where $S_{\CP}$ has been corrected by the $\CP$ eigenvalue to give
    $\sin(2\beta^{\rm eff})$.
    (Top left) $\Bz \to \phi\Kz$,
    (top right) $\Bz \to \eta^\prime\Kz$,
    (bottom left) $\Bz \to \KS\KS\KS$,
    (bottom right) $\Bz \to \pi^0\KS$.
    More figures are available from the HFLAV web pages.
  }
  \label{fig:cp_uta:qqs_SvsC}
\end{figure}

\begin{figure}[htbp]
  \begin{center}
    \resizebox{0.66\textwidth}{!}{
      \includegraphics{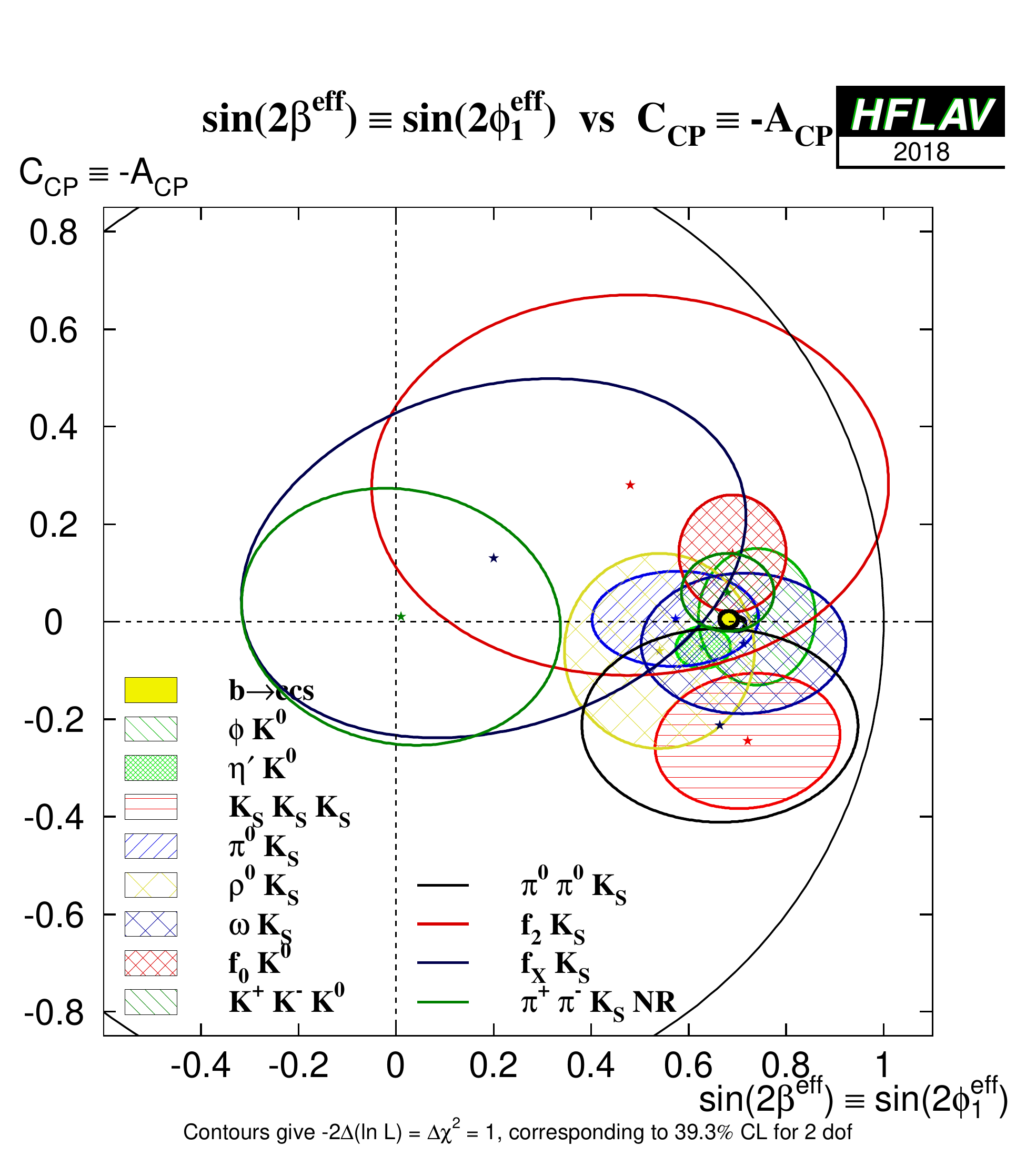}
    }
  \end{center}
  \vspace{-0.8cm}
  \caption{
    Compilation of constraints in the $-\etacp S_{b \to q\bar q s}$, interpreted as $\sin(2\beta^{\rm eff})$, \vs\ $C_{b \to q\bar q s}$ plane.
  }
  \label{fig:cp_uta:qqs_SvsC-all}
\end{figure}

The averages for $-\etacp S_{b \to q\bar q s}$ and $C_{b \to q\bar q s}$
can be found in Tables~\ref{tab:cp_uta:qqs} and~\ref{tab:cp_uta:qqs2},
and are shown in Figs.~\ref{fig:cp_uta:qqs},~\ref{fig:cp_uta:qqs_SvsC}
and~\ref{fig:cp_uta:qqs_SvsC-all}.
Results from both \babar\  and \belle\ are averaged for the modes
$\etapr\Kz$ ($\Kz$ indicates that both $\KS$ and $\KL$ are used)
$\KS\KS\KS$, $\pi^0 \KS$ and $\omega\KS$.\footnote{
  \belle~\cite{Fujikawa:2008pk} includes the $\pi^0\KL$ final state together with $\pi^0 \KS$ in order to improve the constraint on the parameter of \CP\ violation in decay; these events cannot be used for time-dependent analysis.
}
Results on $\phi\KS$ and $\Kp\Km\KS$ (implicitly excluding $\phi\KS$ and $f_0\KS$) are taken from time-dependent Dalitz plot analyses of $\Kp\Km\KS$;
results on $\rho^0\KS$, $f_2\KS$, $f_X\KS$ and $\pip\pim\KS$ nonresonant are taken from time-dependent Dalitz-plot analyses of $\pip\pim\KS$ (see Sec.~\ref{sec:cp_uta:qqs:dp}).\footnote{
  Throughout this section, $f_0 \equiv f_0(980)$ and $f_2 \equiv f_2(1270)$. Details of the assumed lineshapes of these states, and of the $f_X$ (which is taken to have even spin), can be found in the relevant experimental papers~\cite{Lees:2012kxa,Aubert:2009me,Nakahama:2010nj,Dalseno:2008wwa}.}
The results on $f_0\KS$ are from combinations of both Dalitz plot analyses.
\babar\ has also presented results with the final states
$\pi^0\pi^0\KS$ and $\phi \KS \pi^0$.

Of these final states,
$\phi\KS$, $\etapr\KS$, $\pi^0 \KS$, $\rho^0\KS$, $\omega\KS$ and $f_0\KL$
have $\CP$ eigenvalue $\etacp = -1$,
while $\phi\KL$, $\etapr\KL$, $\KS\KS\KS$, $f_0\KS$, $f_2\KS$, $f_X\KS$, $\pi^0\pi^0\KS$ and $\pi^+ \pi^- \KS$ nonresonant have $\etacp = +1$.
The final state $K^+K^-\KS$ (with $\phi\KS$ and $f_0\KS$ implicitly excluded)
is not a $\CP$ eigenstate, but the \CP~content can be absorbed in the amplitude analysis to allow the determination of a single effective $S$ parameter.
(In earlier analyses of the $K^+K^-\Kz$ final state,
its $\CP$ composition was determined using an isospin argument~\cite{Abe:2006gy}
and a moments analysis~\cite{Aubert:2005ja}.)

\begin{table}[!htb]
	\begin{center}
		\caption{
      Averages of $-\etacp S_{b \to q\bar q s}$ and $C_{b \to q\bar q s}$.
      Where a third source of uncertainty is given, it is due to model
      uncertainties arising in Dalitz plot analyses.
		}
		\vspace{0.2cm}
    \resizebox{\textwidth}{!}{
\renewcommand{\arraystretch}{1.1}
		\begin{tabular}{@{\extracolsep{2mm}}lrccc@{\hspace{-3pt}}c} \hline
        \mc{2}{l}{Experiment} & $N(B\bar{B})$ & $- \etacp S_{b \to q\bar q s}$ & $C_{b \to q\bar q s}$ & Correlation \\
	\hline
      \mc{6}{c}{$\phi \Kz$} \\
	\babar & \cite{Lees:2012kxa} & 470M & $0.66 \pm 0.17 \pm 0.07$ & $0.05 \pm 0.18 \pm 0.05$ & \textendash{} \\
	\belle & \cite{Nakahama:2010nj} & 657M & $0.90 \,^{+0.09}_{-0.19}$ & $-0.04 \pm 0.20 \pm 0.10 \pm 0.02$ & \textendash{} \\
	\mc{3}{l}{\bf Average} & $0.74 \,^{+0.11}_{-0.13}$ & $0.01 \pm 0.14$ & {\small uncorrelated averages} \\
		\hline
      \mc{6}{c}{$\etapr \Kz$} \\
	\babar & \cite{:2008se} & 467M & $0.57 \pm 0.08 \pm 0.02$ & $-0.08 \pm 0.06 \pm 0.02$ & $0.03$ \\
	\belle & \cite{Santelj:2014sja} & 772M & $0.68 \pm 0.07 \pm 0.03$ & $-0.03 \pm 0.05 \pm 0.03$ & $0.03$ \\
	\mc{3}{l}{\bf Average} & $0.63 \pm 0.06$ & $-0.05 \pm 0.04$ & $0.02$ \\
	\mc{3}{l}{\small Confidence level} & \mc{2}{c}{\small $0.53~(0.6\sigma)$} & \\
		\hline
      \mc{6}{c}{$\KS\KS\KS$} \\
	\babar & \cite{Lees:2011nf} & 468M & $0.94 \,^{+0.21}_{-0.24} \pm 0.06$ & $-0.17 \pm 0.18 \pm 0.04$ & $0.16$ \\
	\belle & \cite{Chen:2006nk} & 535M & $0.30 \pm 0.32 \pm 0.08$ & $-0.31 \pm 0.20 \pm 0.07$ & \textendash{} \\
	\mc{3}{l}{\bf Average} & $0.72 \pm 0.19$ & $-0.24 \pm 0.14$ & $0.09$ \\
	\mc{3}{l}{\small Confidence level} & \mc{2}{c}{\small $0.26~(1.1\sigma)$} & \\
		\hline
      \mc{6}{c}{$\pi^0 K^0$} \\
	\babar & \cite{:2008se} & 467M & $0.55 \pm 0.20 \pm 0.03$ & $0.13 \pm 0.13 \pm 0.03$ & $0.06$ \\
	\belle & \cite{Fujikawa:2008pk} & 657M & $0.67 \pm 0.31 \pm 0.08$ & $-0.14 \pm 0.13 \pm 0.06$ & $-0.04$ \\
	\mc{3}{l}{\bf Average} & $0.57 \pm 0.17$ & $0.01 \pm 0.10$ & $0.02$ \\
	\mc{3}{l}{\small Confidence level} & \mc{2}{c}{\small $0.37~(0.9\sigma)$} & \\
		\hline
		\hline
      \mc{6}{c}{$\rho^0 \KS$} \\
	\babar & \cite{Aubert:2009me} & 383M & $0.35 \,^{+0.26}_{-0.31} \pm 0.06 \pm 0.03$ & $-0.05 \pm 0.26 \pm 0.10 \pm 0.03$ & \textendash{} \\
	\belle & \cite{Dalseno:2008wwa} & 657M & $0.64 \,^{+0.19}_{-0.25} \pm 0.09 \pm 0.10$ & $-0.03 \,^{+0.24}_{-0.23} \pm 0.11 \pm 0.10$ & \textendash{} \\
	\mc{3}{l}{\bf Average} & $0.54 \,^{+0.18}_{-0.21}$ & $-0.06 \pm 0.20$ & {\small uncorrelated averages} \\
		\hline
      \mc{6}{c}{$\omega \KS$} \\
	\babar & \cite{:2008se} & 467M & $0.55 \,^{+0.26}_{-0.29} \pm 0.02$ & $-0.52 \,^{+0.22}_{-0.20} \pm 0.03$ & $0.03$ \\
	\belle & \cite{Chobanova:2013ddr} & 772M & $0.91 \pm 0.32 \pm 0.05$ & $0.36 \pm 0.19 \pm 0.05$ & $-0.00$ \\
	\mc{3}{l}{\bf Average} & $0.71 \pm 0.21$ & $-0.04 \pm 0.14$ & $0.01$ \\
	\mc{3}{l}{\small Confidence level} & \mc{2}{c}{\small $0.007~(2.7\sigma)$} & \\
		\hline
      \mc{6}{c}{$f_0 \Kz$} \\
	\babar & \cite{Lees:2012kxa,Aubert:2009me} & \textendash{} & $0.74 \,^{+0.12}_{-0.15}$ & $0.15 \pm 0.16$ & \textendash{} \\
	\belle & \cite{Nakahama:2010nj,Dalseno:2008wwa} & \textendash{} & $0.63 \,^{+0.16}_{-0.19}$ & $0.13 \pm 0.17$ & \textendash{} \\
	\mc{3}{l}{\bf Average} & $0.69 \,^{+0.10}_{-0.12}$ & $0.14 \pm 0.12$ & {\small uncorrelated averages} \\
		\hline
      \mc{6}{c}{$f_2 \KS$} \\
	\babar & \cite{Aubert:2009me} & 383M & $0.48 \pm 0.52 \pm 0.06 \pm 0.10$ & $0.28 \,^{+0.35}_{-0.40} \pm 0.08 \pm 0.07$ & \textendash{} \\
		\hline
      \mc{6}{c}{$f_{X} \KS$} \\
	\babar & \cite{Aubert:2009me} & 383M & $0.20 \pm 0.52 \pm 0.07 \pm 0.07$ & $0.13 \,^{+0.33}_{-0.35} \pm 0.04 \pm 0.09$ & \textendash{} \\
		\hline
 		\end{tabular}
}
		\label{tab:cp_uta:qqs}
	\end{center}
\end{table}

\begin{table}[!htb]
	\begin{center}
		\caption{
      Averages of $-\etacp S_{b \to q\bar q s}$ and $C_{b \to q\bar q s}$ (continued).
      Where a third source of uncertainty is given, it is due to model
      uncertainties arising in Dalitz plot analyses.
		}
		\vspace{0.2cm}
		\setlength{\tabcolsep}{0.0pc}
\renewcommand{\arraystretch}{1.1}
		\begin{tabular*}{\textwidth}{@{\extracolsep{\fill}}lrccc@{\hspace{-3pt}}c} \hline
        \mc{2}{l}{Experiment} & $N(B\bar{B})$ & $- \etacp S_{b \to q\bar q s}$ & $C_{b \to q\bar q s}$ & Correlation \\
	\hline
      \mc{6}{c}{$\pi^0 \pi^0 \KS$} \\
	\babar & \cite{Aubert:2007ub} & 227M & $-0.72 \pm 0.71 \pm 0.08$ & $0.23 \pm 0.52 \pm 0.13$ & $-0.02$ \\
	\belle & \cite{Yusa:2018hmz} & 772M & $0.92 \,^{+0.27}_{-0.31} \pm 0.11$ & $-0.28 \pm 0.21 \pm 0.04$ & $0.00$ \\
	\mc{3}{l}{\bf Average} & $0.66 \pm 0.28$ & $-0.21 \pm 0.20$ & $0.00$ \\
	\mc{3}{l}{\small Confidence level} & \mc{2}{c}{\small $0.08~(1.8\sigma)$} & \\
		\hline
      \mc{6}{c}{$\phi \KS \pi^0$} \\
	\babar & \cite{Aubert:2008zza} & 465M & $0.97 \,^{+0.03}_{-0.52}$ & $-0.20 \pm 0.14 \pm 0.06$ & \textendash{} \\
 		\hline
      \mc{6}{c}{$\pi^+ \pi^- \KS$ nonresonant} \\
	\babar & \cite{Aubert:2009me} & 383M & $0.01 \pm 0.31 \pm 0.05 \pm 0.09$ & $0.01 \pm 0.25 \pm 0.06 \pm 0.05$ & \textendash{} \\
 		\hline
      \mc{6}{c}{$K^+K^- \Kz$} \\
	\babar & \cite{Lees:2012kxa} & 470M & $0.65 \pm 0.12 \pm 0.03$ & $0.02 \pm 0.09 \pm 0.03$ & \textendash{} \\
	\belle & \cite{Nakahama:2010nj} & 657M & $0.76 \,^{+0.14}_{-0.18}$ & $0.14 \pm 0.11 \pm 0.08 \pm 0.03$ & \textendash{} \\
	\mc{3}{l}{\bf Average} & $0.68 \,^{+0.09}_{-0.10}$ & $0.06 \pm 0.08$ & {\small uncorrelated averages} \\
		\hline

		\hline
		\end{tabular*}
		\label{tab:cp_uta:qqs2}
	\end{center}
\end{table}

The final state $\phi \KS \pi^0$ is also not a \CP eigenstate but its
\CP-composition can be determined from an angular analysis.
Since the parameters are common to the $\Bz\to\phi \KS \pi^0$ and
$\Bz\to \phi \Kp\pim$ decays (because only $K\pi$ resonances contribute),
\babar\ performed a simultaneous analysis of the two final
states~\cite{Aubert:2008zza} (see Sec.~\ref{sec:cp_uta:qqs:vv}).

It must be noted that Q2B parameters extracted from Dalitz-plot analyses
are constrained to lie within the physical boundary ($S_{\CP}^2 + C_{\CP}^2 < 1$).
Consequently, the obtained uncertainties are highly non-Gaussian when
the central value is close to the boundary.
This is particularly evident in the \babar\ results for
$\Bz \to f_0\Kz$ with $f_0 \to \pi^+\pi^-$~\cite{Aubert:2009me}.
These results must be treated with caution.

As explained above, each of the modes listed in Tables~\ref{tab:cp_uta:qqs} and~\ref{tab:cp_uta:qqs2} has potentially different subleading contributions within the Standard Model,
and thus each may have a different value of $-\etacp S_{b \to q\bar q s}$.
Therefore, there is no strong motivation to make a combined average
over the different modes.
We refer to such an average as a ``na\"\i ve $s$-penguin average.''
It is na\"\i ve not only because the theoretical uncertainties are neglected,
but also since possible correlations of systematic effects
between different modes are not included.
In spite of these caveats, there remains interest in the value of this quantity
and therefore it is given here:
$\langle -\etacp S_{b \to q\bar q s} \rangle = 0.648 \pm 0.038$,
with confidence level $0.63~(0.5\sigma)$.
This value is in agreement with the average
$-\etacp S_{b \to c\bar c s}$ given in Sec.~\ref{sec:cp_uta:ccs:cp_eigen}.
The average for $C_{b \to q\bar q s}$ is $\langle C_{b \to q\bar q s} \rangle = -0.003 \pm 0.029$ with a confidence level of $0.43~(0.8\sigma)$.

From Table~\ref{tab:cp_uta:qqs} it may be noted
that the averages for $-\etacp S_{b \to q\bar q s}$ in
$\phi\KS$, $\etapr \Kz$, $f_0\KS$ and $\Kp\Km\KS$
are all now more than $5\sigma$ away from zero,
so that $\CP$ violation in these modes can be considered well established.
There is no evidence (above $2\sigma$) for $\CP$ violation in decay in any of these $b \to q \bar q s$ transitions.

\mysubsubsection{Time-dependent Dalitz plot analyses: $\Bz \to K^+K^-\Kz$ and $\Bz \to \pi^+\pi^-\KS$}
\label{sec:cp_uta:qqs:dp}

\begin{figure}[htbp]
  \begin{center}
    \resizebox{0.45\textwidth}{!}{
      \includegraphics{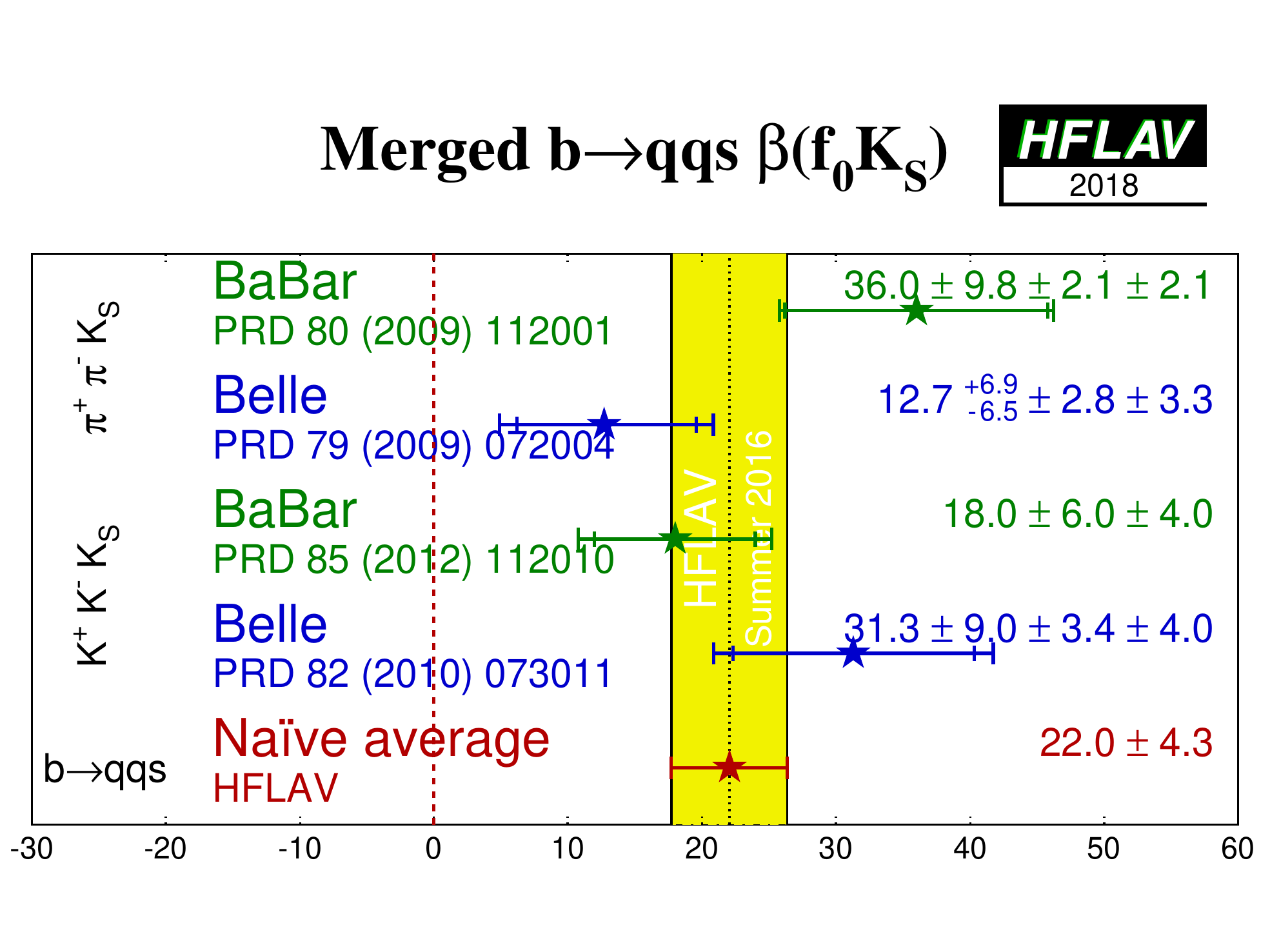}
    }
    \hfill
    \resizebox{0.45\textwidth}{!}{
      \includegraphics{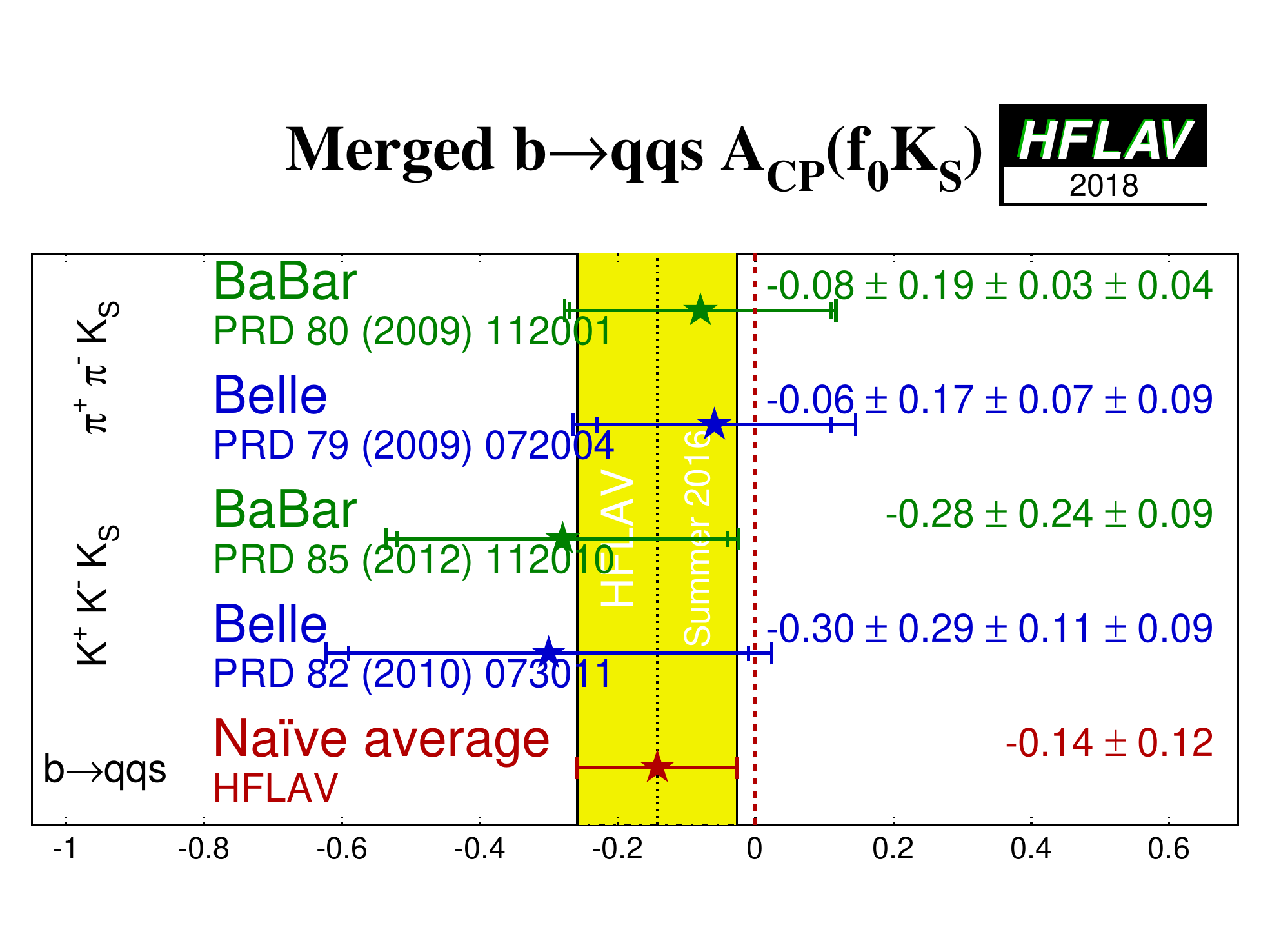}
    }
  \end{center}
  \vspace{-0.8cm}
  \caption{
    Averages of
    (left) $\beta^{\rm eff} \equiv \phi_1^{\rm eff}$ and (right) $A_{\CP}$
    for the $\Bz\to f_0\KS$ decay including measurements from Dalitz plot analyses of both $\Bz\to K^+K^-\KS$ and $\Bz\to \pi^+\pi^-\KS$.
  }
  \label{fig:cp_uta:qqs:f0KS}
\end{figure}

As mentioned in Sec.~\ref{sec:cp_uta:notations:dalitz:kkk0} and above,
both \babar\ and \belle\ have performed time-dependent Dalitz plot analyses of
$\Bz \to K^+K^-\Kz$ and $\Bz \to \pi^+\pi^-\KS$ decays.
The results are summarised in Tables~\ref{tab:cp_uta:kkk0_tddp}
and~\ref{tab:cp_uta:pipik0_tddp}.
Averages for the $\Bz\to f_0 \KS$ decay, which contributes to both Dalitz
plots, are shown in Fig.~\ref{fig:cp_uta:qqs:f0KS}.
Results are presented in terms of the effective weak phase (from mixing and
decay) difference $\beta^{\rm eff}$ and the parameter of $\CP$ violation in decay
$\Acp$ ($\Acp = -C$) for each of the resonant contributions.
Note that Dalitz-plot analyses, including all those included in these
averages, often suffer from ambiguous solutions -- we quote the results
corresponding to those presented as "solution 1" in all cases.
Results on flavour-specific amplitudes that may contribute to these Dalitz
plots (such as $K^{*+}\pi^-$) are given in Sec.~\ref{sec:rare}.

For the $\Bz \to K^+K^-\Kz$ decay, both \babar\ and \belle\ measure the \CP
violation parameters for the $\phi\Kz$, $f_0\Kz$ and ``other $\Kp\Km\Kz$''
amplitudes, where the latter includes all remaining resonant and nonresonant
contributions to the charmless three-body decay.
For the $\Bz \to \pi^+\pi^-\KS$ decay, \babar\ reports \CP violation parameters for all of the \CP eigenstate components in the Dalitz plot model ($\rhoz\KS$, $f_0\KS$, $f_2\KS$, $f_X\KS$ and nonresonant decays; see Sec.~\ref{sec:cp_uta:notations:dalitz:pipik0}),
while \belle\ reports the \CP violation parameters for only the $\rhoz\KS$ and $f_0\KS$ amplitudes, although the Dalitz-plot models used by the two collaborations are rather similar.

\begin{sidewaystable}
  \begin{center}
    \caption{
      Results from time-dependent Dalitz plot analyses of 
      the $\Bz \to K^+K^-\Kz$ decay.
      Correlations (not shown) are taken into account in the average.
    }
    \vspace{0.2cm}
    \setlength{\tabcolsep}{0.0pc}
    \resizebox{\textwidth}{!}{
\renewcommand{\arraystretch}{1.2}
      \begin{tabular}{l@{\hspace{2mm}}r@{\hspace{2mm}}c@{\hspace{2mm}}|@{\hspace{2mm}}c@{\hspace{2mm}}c@{\hspace{2mm}}|@{\hspace{2mm}}c@{\hspace{2mm}}c@{\hspace{2mm}}|@{\hspace{2mm}}c@{\hspace{2mm}}c} 
        \hline 
        \mc{2}{l}{Experiment} & $N(B\bar{B})$ &
        \mc{2}{c}{$\phi\KS$} & \mc{2}{c}{$f_0\KS$} & \mc{2}{c}{$K^+K^-\KS$} \\
        & & & $\beta^{\rm eff}\,(^\circ)$ & $\Acp$ & $\beta^{\rm eff}\,(^\circ)$ & $\Acp$ & $\beta^{\rm eff}\,(^\circ)$ & $\Acp$ \\
	\babar & \cite{Lees:2012kxa} & 470M & $21 \pm 6 \pm 2$ & $-0.05 \pm 0.18 \pm 0.05$ & $18 \pm 6 \pm 4$ & $-0.28 \pm 0.24 \pm 0.09$ & $20.3 \pm 4.3 \pm 1.2$ & $-0.02 \pm 0.09 \pm 0.03$ \\
	\belle & \cite{Nakahama:2010nj} & 657M & $32.2 \pm 9.0 \pm 2.6 \pm 1.4$ & $0.04 \pm 0.20 \pm 0.10 \pm 0.02$ & $31.3 \pm 9.0 \pm 3.4 \pm 4.0$ & $-0.30 \pm 0.29 \pm 0.11 \pm 0.09$ & $24.9 \pm 6.4 \pm 2.1 \pm 2.5$ & $-0.14 \pm 0.11 \pm 0.08 \pm 0.03$ \\
	\mc{2}{l}{\bf Average} & & $24 \pm 5$ & $-0.01 \pm 0.14$ & $22 \pm 6$ & $-0.29 \pm 0.20$ & $21.6 \pm 3.7$ & $-0.06 \pm 0.08$ \\
	\mc{3}{l}{\small Confidence level} & \mc{6}{c}{\small $0.93~(0.1\sigma)$} \\
        \hline
      \end{tabular}
    }
    
    \label{tab:cp_uta:kkk0_tddp}
  \end{center}
\end{sidewaystable}
\begin{sidewaystable}
  \begin{center}
    \caption{
      Results from time-dependent Dalitz plot analysis of 
      the $\Bz \to \pi^+\pi^-\KS$ decay.
      Correlations (not shown) are taken into account in the average.
    }
    \vspace{0.2cm}
    \setlength{\tabcolsep}{0.0pc}
    \resizebox{\textwidth}{!}{
\renewcommand{\arraystretch}{1.2}
      \begin{tabular}{l@{\hspace{2mm}}r@{\hspace{2mm}}c@{\hspace{2mm}}|@{\hspace{2mm}}c@{\hspace{2mm}}c@{\hspace{2mm}}|@{\hspace{2mm}}c@{\hspace{2mm}}c} 
        \hline 
        \mc{2}{l}{Experiment} & $N(B\bar{B})$ & 
        \mc{2}{c}{$\rho^0\KS$} & \mc{2}{c}{$f_0\KS$} \\
        & & & $\beta^{\rm eff}$ & $\Acp$ & $\beta^{\rm eff}$ & $\Acp$ \\
        \hline
        \babar & \cite{Aubert:2009me} & 383M & $(10.2 \pm 8.9 \pm 3.0 \pm 1.9)^\circ$ & $0.05 \pm 0.26 \pm 0.10 \pm 0.03$ & $(36.0 \pm 9.8 \pm 2.1 \pm 2.1)^\circ$ & $-0.08 \pm 0.19 \pm 0.03 \pm 0.04$ \\
        \belle & \cite{Dalseno:2008wwa} & 657M & $(20.0 \,^{+8.6}_{-8.5} \pm 3.2 \pm 3.5)^\circ$ & $0.03 \,^{+0.23}_{-0.24} \pm 0.11 \pm 0.10$ & $(12.7 \,^{+6.9}_{-6.5} \pm 2.8 \pm 3.3)^\circ$ & $-0.06 \pm 0.17 \pm 0.07 \pm 0.09$ \\
        \hline
        \mc{2}{l}{\bf Average} & & $16.4 \pm 6.8$ & $0.06 \pm 0.20$ & $20.6 \pm 6.2$ & $-0.07 \pm 0.14$  \\
        \mc{3}{l}{\small Confidence level} & \mc{4}{c}{\small $0.39~(0.9\sigma)$} \\
        \hline
      \end{tabular}
    }

    \vspace{2ex}

    \setlength{\tabcolsep}{0.0pc}
    \resizebox{\textwidth}{!}{
\renewcommand{\arraystretch}{1.2}
      \begin{tabular}{l@{\hspace{2mm}}r@{\hspace{2mm}}c@{\hspace{2mm}}|@{\hspace{2mm}}c@{\hspace{2mm}}c@{\hspace{2mm}}|@{\hspace{2mm}}c@{\hspace{2mm}}c} 
        \hline 
        \mc{2}{l}{Experiment} & $N(B\bar{B})$ & 
        \mc{2}{c}{$f_2\KS$} & \mc{2}{c}{$f_{\rm X}\KS$} \\
        & & & $\beta^{\rm eff}$ & $\Acp$ & $\beta^{\rm eff}$ & $\Acp$ \\
        \babar & \cite{Aubert:2009me} & 383M & $(14.9 \pm 17.9 \pm 3.1 \pm 5.2)^\circ$ & $-0.28 \,^{+0.40}_{-0.35} \pm 0.08 \pm 0.07$ & $(5.8 \pm 15.2 \pm 2.2 \pm 2.3)^\circ$ & $-0.13 \,^{+0.35}_{-0.33} \pm 0.04 \pm 0.09$ \\
        \hline
      \end{tabular}
    }

    \vspace{2ex}

    \setlength{\tabcolsep}{0.0pc}
    \resizebox{\textwidth}{!}{
\renewcommand{\arraystretch}{1.1}
      \begin{tabular}{l@{\hspace{2mm}}r@{\hspace{2mm}}c@{\hspace{2mm}}|@{\hspace{2mm}}c@{\hspace{2mm}}c@{\hspace{2mm}}|@{\hspace{2mm}}c@{\hspace{2mm}}c} 
        \hline 
        \mc{2}{l}{Experiment} & $N(B\bar{B})$ & 
        \mc{2}{c}{$\Bz \to \pi^+\pi^-\KS$ nonresonant} & \mc{2}{c}{$\chi_{c0}\KS$} \\
        & & & $\beta^{\rm eff}$ & $\Acp$ & $\beta^{\rm eff}$ & $\Acp$ \\
        \babar & \cite{Aubert:2009me} & 383M & $(0.4 \pm 8.8 \pm 1.9 \pm 3.8)^\circ$ & $-0.01 \pm 0.25 \pm 0.06 \pm 0.05$ & $(23.2 \pm 22.4 \pm 2.3 \pm 4.2)^\circ$ & $0.29 \,^{+0.44}_{-0.53} \pm 0.03 \pm 0.05$ \\
        \hline
      \end{tabular}
    }

    \label{tab:cp_uta:pipik0_tddp}
  \end{center}
\end{sidewaystable}

\mysubsubsection{Time-dependent analyses of $\Bz \to \phi \KS \pi^0$}
\label{sec:cp_uta:qqs:vv}

The final state in the decay $\Bz \to \phi \KS \pi^0$ is a mixture of \CP-even
and \CP-odd amplitudes. However, since only $\phi K^{*0}$ resonant states
contribute (in particular, $\phi K^{*0}(892)$, $\phi K^{*0}_0(1430)$ and $\phi
K^{*0}_2(1430)$ are seen), the composition can be determined from the analysis
of $B \to \phi K^+ \pi^-$ decays, assuming only that the ratio of branching fractions ${\cal B}(K^{*0} \to \KS \pi^0)/{\cal B}(K^{*0} \to K^+ \pi^-)$ is the same
for each excited kaon state.

\babar~\cite{Aubert:2008zza} has performed a simultaneous analysis of
$\Bz \to \phi \KS \pi^0$ and $\Bz \to \phi K^+ \pi^-$ decays that is time-dependent for the former mode and time-integrated for the latter.
Such an analysis allows, in principle, all parameters of the $\Bz \to \phi K^{*0}$ system to be determined, including mixing-induced \CP violation effects.
The latter is determined to be $\Delta\phi_{00} = 0.28 \pm 0.42 \pm 0.04$, where $\Delta\phi_{00}$ is half the weak phase difference between $\Bz$ and $\Bzb$ decays to the $\phi K^{*0}_0(1430)$ final state.
As discussed above, this can also be presented in terms of the Q2B parameter $\sin(2\beta^{\rm eff}_{00}) = \sin(2\beta+2\Delta\phi_{00}) = 0.97 \,^{+0.03}_{-0.52}$.
The highly asymmetric uncertainty arises due to the conversion from the phase to the sine of the phase, and the proximity of the physical boundary.

Similar $\sin(2\beta^{\rm eff})$ parameters can be defined for each of the helicity amplitudes for both $\phi K^{*0}(892)$ and $\phi K^{*0}_2(1430)$.
However, the relative phases between these decays are constrained due to the nature of the simultaneous analysis of $\Bz \to \phi \KS \pi^0$ and $\Bz \to \phi K^+ \pi^-$, decays and therefore these measurements are highly correlated.
Instead of quoting all these results, \babar provides an illustration of the measurements with the following differences:
\begin{eqnarray}
  \sin(2\beta - 2\Delta\delta_{01}) - \sin(2\beta) & = & -0.42\,^{+0.26}_{-0.34} \, , \\
  \sin(2\beta - 2\Delta\phi_{\parallel1}) - \sin(2\beta) & = & -0.32\,^{+0.22}_{-0.30} \, , \\
  \sin(2\beta - 2\Delta\phi_{\perp1}) - \sin(2\beta) & = & -0.30\,^{+0.23}_{-0.32} \, , \\
  \sin(2\beta - 2\Delta\phi_{\perp1}) - \sin(2\beta - 2\Delta\phi_{\parallel1}) & = & 0.02 \pm 0.23 \, , \\
  \sin(2\beta - 2\Delta\delta_{02}) - \sin(2\beta) & = & -0.10\,^{+0.18}_{-0.29} \, ,
\end{eqnarray}
where the first subscript indicates the helicity amplitude and the second
indicates the spin of the kaon resonance.
For the complete definitions of the
$\Delta\delta$ and $\Delta\phi$ parameters, refer to the \babar\ paper~\cite{Aubert:2008zza}.

Parameters of \CP violation in decay for each of the contributing helicity amplitudes can also be measured.
Again, these are determined from a simultaneous fit of $\Bz \to \phi \KS \pi^0$ and $\Bz \to \phi K^+ \pi^-$ decays, with the precision being dominated by the statistics of the latter mode.
Measurements of \CP violation in decay, obtained from decay-time-integrated analyses, are tabulated in Sec.~\ref{sec:rare}.

\mysubsubsection{Time-dependent \CP asymmetries in $\Bs \to \Kp\Km$}
\label{sec:cp_uta:qqs:BstoKK}

The decay $\Bs \to \Kp\Km$ involves a $b \to u\bar{u}s$ transition, and hence has both penguin and tree contributions.
Both mixing-induced and \CP violation in decay effects may arise, and additional input is needed to disentangle the contributions and determine $\gamma$ and $\beta_s^{\rm eff}$.
For example, the observables in $\Bd \to \pip\pim$ can be related using U-spin, as proposed in Refs.~\cite{Dunietz:1993rm,Fleischer:1999pa}.

The observables are $A_{\rm mix} = S_{\CP}$, $A_{\rm dir} = -C_{\CP}$, and $A_{\Delta\Gamma}$.
They are related by $A_{\rm mix}^2 + A_{\rm dir}^2 + A_{\Delta\Gamma}^2 = 1$, but are usually treated as independent (albeit correlated) free parameters in experimental analyses, since this approach yields results with better statistical behavior.
Note that the untagged decay distribution, from which an ``effective lifetime'' can be measured, retains sensitivity to $A_{\Delta\Gamma}$; measurements of the $\Bs \to \Kp\Km$ effective lifetime have been made by LHCb~\cite{Aaij:2012kn,Aaij:2014fia}.
Compilations and averages of effective lifetimes are performed by the HFLAV Lifetimes and Oscillations subgroup, see Sec.~\ref{sec:life_mix}.

The observables in $\Bs \to \Kp\Km$ have been measured by LHCb~\cite{Aaij:2018tfw}. %
The results are shown in Table~\ref{tab:cp_uta:BstoKK}, and correspond to evidence for \CP violation both in the interference between mixing and decay, and in the $\Bs \to \Kp\Km$ decay.

\begin{table}[!htb]
	\begin{center}
		\caption{
      Results from time-dependent analysis of the $\Bs \to K^{+} K^{-}$ decay.
		}
		\vspace{0.2cm}
		\setlength{\tabcolsep}{0.0pc}
\renewcommand{\arraystretch}{1.1}
		\begin{tabular*}{\textwidth}{@{\extracolsep{\fill}}lrcccc} \hline
	\mc{2}{l}{Experiment} & Sample size & $S_{\CP}$ & $C_{\CP}$ & $A^{\Delta\Gamma}$ \\
	\hline 	 	
	LHCb & \cite{Aaij:2018tfw} & $\int {\cal L} \, dt = 3.0 \ {\rm fb}^{-1}$ & $0.18 \pm 0.06 \pm 0.02$ & $0.20 \pm 0.06 \pm 0.02$ & $-0.79 \pm 0.07 \pm 0.10$ \\
	\hline
		\end{tabular*}
		\label{tab:cp_uta:BstoKK}
	\end{center}
\end{table}

Interpretations of an earlier set of results~\cite{Aaij:2013tna}, in terms of constraints on $\gamma$ and $2\beta_s$, have been separately published by LHCb~\cite{Aaij:2014xba}.

\mysubsubsection{Time-dependent \CP asymmetries in $\Bs \to \phi\phi$}
\label{sec:cp_uta:qqs:Bstophiphi}

 The decay $\Bs \to \phi\phi$ involves a $b \to s\bar{s}s$ transition, and hence is a ``pure penguin'' mode (in the limit that the $\phi$ meson is considered a pure $s\bar{s}$ state).
Since the mixing phase and the decay phase are expected to cancel in the Standard Model, the phase from the interference of mixing and decay is predicted to be $\phi_s(\phi\phi) = 0$ with low uncertainty~\cite{Raidal:2002ph}.
Due to the vector-vector nature of the final state, angular analysis is needed to separate the \CP-even and \CP-odd contributions. Such an analysis also makes it possible to fit directly for $\phi_s(\phi\phi)$.

A constraint on $\phi_s(\phi\phi)$ has been obtained by LHCb using $5 \,{\rm fb}^{-1}$~\cite{Aaij:2019uld}. %
The result is $\phi_s(\phi\phi) = -0.06 \pm 0.13 \pm 0.03 \, {\rm rad}$, where the first uncertainty is statistical and the second is systematic.

\mysubsection{Time-dependent $\CP$ asymmetries in $b \to q\bar{q}d$ transitions
}
\label{sec:cp_uta:qqd}

Decays such as $\Bz\to\KS\KS$ are pure $b \to q\bar{q}d$ penguin transitions.
As shown in Eq.~(\ref{eq:cp_uta:b_to_d}),
this diagram has different contributing weak phases,
and therefore the observables are sensitive to their difference
(which can be chosen to be either $\beta$ or $\gamma$).
Note that if the contribution with the top quark in the loop dominates,
the weak phase from the decay amplitudes should cancel that from mixing,
so that no $\CP$ violation (neither mixing-induced nor in decay) occurs.
Non-zero contributions from loops with intermediate up and charm quarks
can result in both types of effect
(as usual, a strong phase difference is required for $\CP$ violation in decay
to occur).

\begin{figure}[htbp]
  \begin{center}
    \begin{tabular}{cc}
      \resizebox{0.46\textwidth}{!}{
        \includegraphics{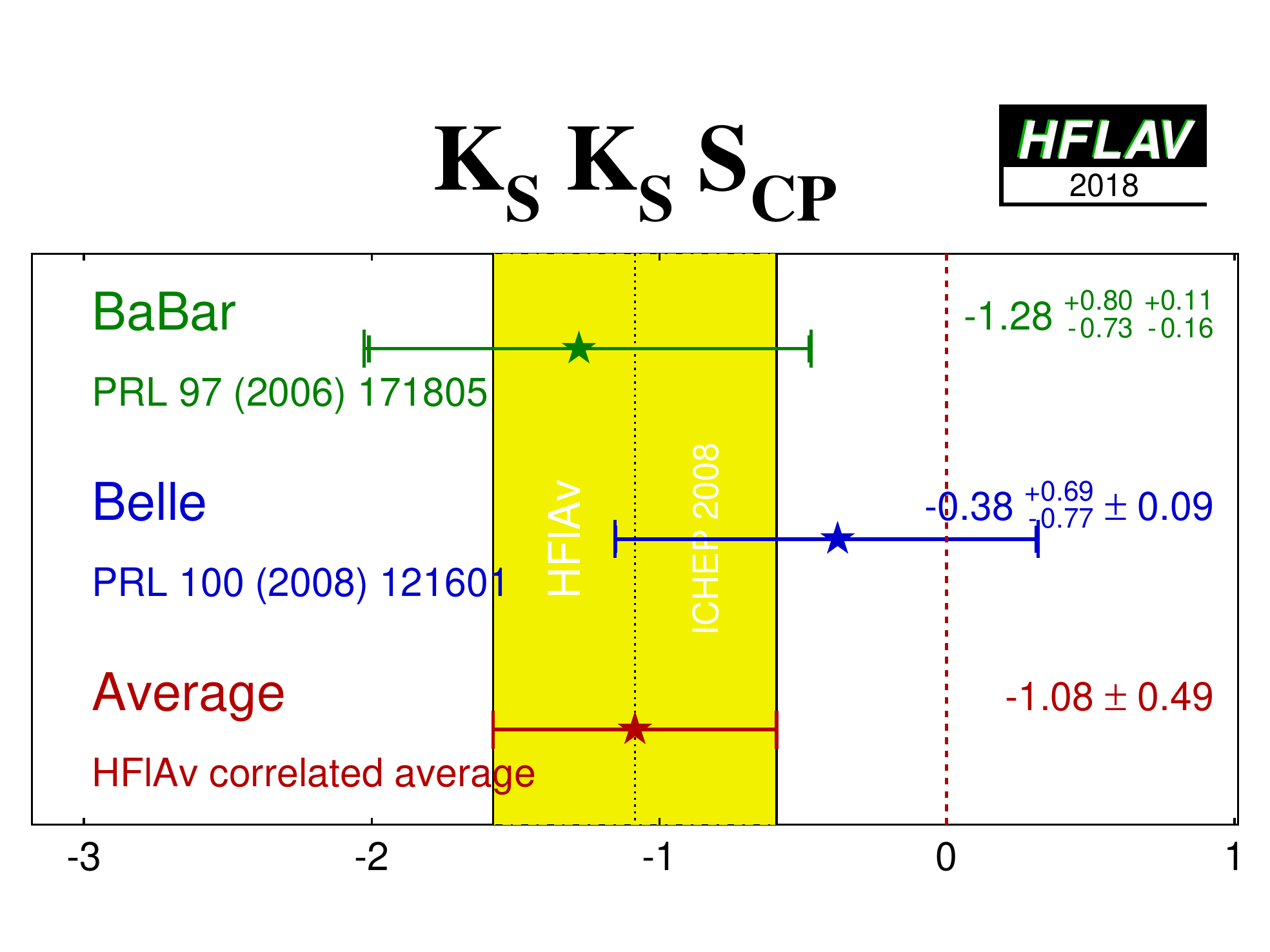}
      }
      &
      \resizebox{0.46\textwidth}{!}{
        \includegraphics{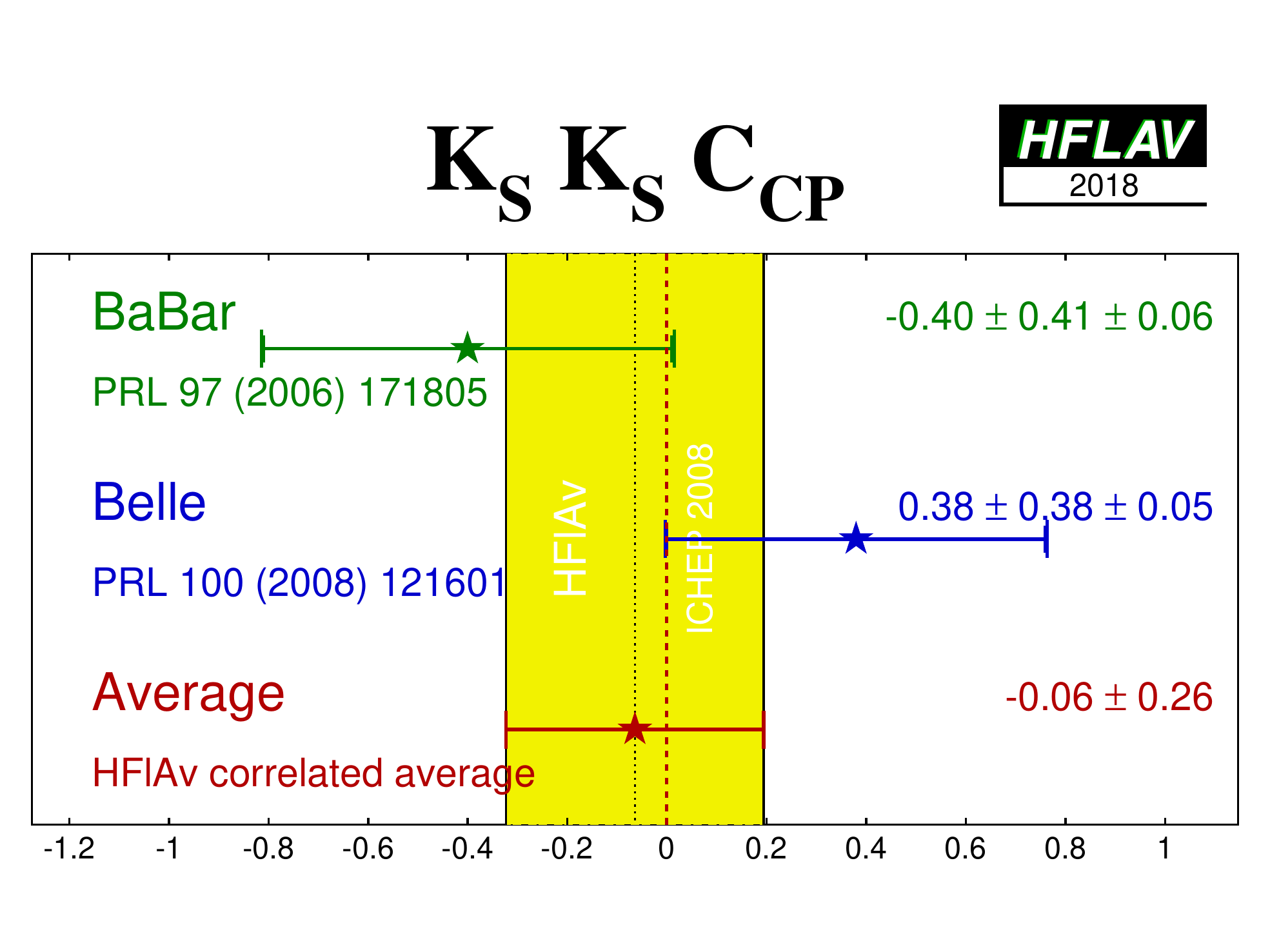}
      }
    \end{tabular}
  \end{center}
  \vspace{-0.8cm}
  \caption{
    Averages of (left) $S_{\CP}$ and (right) $C_{\CP}$ for the mode $\Bz \to \KS\KS$.
  }
  \label{fig:cp_uta:qqd:ksks}
\end{figure}

Both \babar~\cite{Aubert:2006gm} and \belle~\cite{Nakahama:2007dg}
have performed time-dependent analyses of $\Bz\to\KS\KS$ decays.
The results are given in Table~\ref{tab:cp_uta:qqd}
and shown in Fig.~\ref{fig:cp_uta:qqd:ksks}.

\begin{table}[htb]
	\begin{center}
		\caption{
			Results for $\Bz \to \KS\KS$.
		}
		\vspace{0.2cm}
		\setlength{\tabcolsep}{0.0pc}
\renewcommand{\arraystretch}{1.1}
		\begin{tabular*}{\textwidth}{@{\extracolsep{\fill}}lrcccc} \hline
	\mc{2}{l}{Experiment} & $N(B\bar{B})$ & $S_{\CP}$ & $C_{\CP}$ & Correlation \\
	\hline
	\babar & \cite{Aubert:2006gm} & 350M & $-1.28 \,^{+0.80}_{-0.73} \,^{+0.11}_{-0.16}$ & $-0.40 \pm 0.41 \pm 0.06$ & $-0.32$ \\
	\belle & \cite{Nakahama:2007dg} & 657M & $-0.38 \,^{+0.69}_{-0.77} \pm 0.09$ & $0.38 \pm 0.38 \pm 0.05$ & $0.48$ \\
	\hline
	\mc{3}{l}{\bf Average} & $-1.08 \pm 0.49$ & $-0.06 \pm 0.26$ & $0.14$ \\
	\mc{3}{l}{\small Confidence level} & \mc{2}{c}{\small $0.29~(1.1\sigma)$} & \\
		\hline
		\end{tabular*}
		\label{tab:cp_uta:qqd}
	\end{center}
\end{table}

\mysubsection{Time-dependent asymmetries in $b \to s\gamma$ transitions
}
\label{sec:cp_uta:bsg}

The radiative decays $b \to s\gamma$ produce photons that
are highly polarised in the Standard Model.
The decays $\Bz \to F \gamma$ and $\Bzb \to F \gamma$,
where $F$ is a strange hadronic system,
produce photons with opposite helicities,
and since the polarisation is, in principle, observable,
these final states cannot interfere.
The finite mass of the $s$ quark introduces small corrections
to the limit of maximum polarisation,
but any large mixing-induced $\CP$ violation would be a signal for new physics.
Since a single weak phase dominates the $b \to s \gamma$ transition in the
Standard Model, the cosine term is also expected to be small.

Atwood {\it et al.}~\cite{Atwood:2004jj} have shown that
an inclusive analysis of $\KS\pi^0\gamma$ can be performed,
since the properties of the decay amplitudes
are independent of the angular momentum of the $\KS\pi^0$ system.
However, if non-dipole operators contribute significantly to the amplitudes,
then the Standard Model mixing-induced $\CP$ violation could be larger
than the na\"\i ve expectation
$S \simeq -2 (m_s/m_b) \sin \left(2\beta\right)$~\cite{Grinstein:2004uu,Grinstein:2005nu}.
In this case,
the $\CP$ parameters may vary over the $\KS\pi^0\gamma$ Dalitz plot,
for example, as a function of the $\KS\pi^0$ invariant mass.

With the above in mind, we quote two averages: one for the final state $K^*(892)\gamma$ only, and one for the inclusive $\KS\pi^0\gamma$ final state (including  $K^*(892)\gamma$).
If the Standard Model dipole operator is dominant,
both should give the same $\CP$-violation parameters
(the latter, naturally, with smaller statistical uncertainties).
If not, care needs to be taken in interpretation of the inclusive parameters,
while the results on the $K^*(892)$ resonance remain relatively clean.
Results from \babar\ and \belle\ are
used for both averages; both experiments use the invariant-mass range
$0.60 < M_{\KS\pi^0} < 1.80~\gevcc$
in the inclusive analysis.

In addition to the $\KS\pi^0\gamma$ decay, both \babar\ and \belle\ have presented results using the $\KS\rho\gamma$ mode, while \babar\ (\belle) has in addition presented results using the $\KS\eta\gamma$ ($\KS\phi\gamma$) channel.
For the $\KS\rho\gamma$ case, due to the non-negligible width of the $\rho^0$ meson, decays selected as $\Bz \to \KS\rho^0\gamma$ can include a significant contribution from $K^{*\pm}\pi^\mp\gamma$ decays, which are flavour-specific and do not have the same oscillation phenomenology.
Both \babar\ and \belle\ measure $S_{\rm eff}$ for all \B decay candidates with the $\rho^0$ selection being $0.6 < m(\pip\pim) < 0.9~\gevcc$, obtaining $0.14 \pm 0.25 \,^{+0.04}_{-0.03}$ (\babar) and $0.09 \pm 0.27 \,^{+0.04}_{-0.07}$ (\belle).
These values are then corrected for a ``dilution factor''~\cite{Akar:2018zhv}, that is evaluated with different methods in the two experiments: \babar~\cite{Akar:2013ima,Sanchez:2015pxu} obtains a dilution factor of $-0.78 \,^{+0.19}_{-0.17}$, while \belle~\cite{Li:2008qma} obtains $+0.83 \,^{+0.19}_{-0.03}$.
Until the discrepancy between these values is understood, the average of the results should be treated with caution.

\begin{table}[htb]
	\begin{center}
		\caption{
      Averages for $b \to s \gamma$ modes.
		}
		\vspace{0.2cm}
		\setlength{\tabcolsep}{0.0pc}
\renewcommand{\arraystretch}{1.1}
		\begin{tabular*}{\textwidth}{@{\extracolsep{\fill}}lrcccc} \hline
	\mc{2}{l}{Experiment} & $N(B\bar{B})$ & $S_{\CP} (b \to s \gamma)$ & $C_{\CP} (b \to s \gamma)$ & Correlation \\
        \hline
        \mc{6}{c}{$\Kstar(892)\gamma$} \\
	\babar & \cite{Aubert:2008gy} & 467M & $-0.03 \pm 0.29 \pm 0.03$ & $-0.14 \pm 0.16 \pm 0.03$ & $0.05$ \\
	\belle & \cite{Ushiroda:2006fi} & 535M & $-0.32 \,^{+0.36}_{-0.33} \pm 0.05$ & $0.20 \pm 0.24 \pm 0.05$ & $0.08$ \\
	\mc{3}{l}{\bf Average} & $-0.16 \pm 0.22$ & $-0.04 \pm 0.14$ & $0.06$ \\
	\mc{3}{l}{\small Confidence level} & \mc{2}{c}{\small $0.40~(0.9\sigma)$} & \\
		\hline
        \mc{6}{c}{$\KS \pi^0 \gamma$ (including $\Kstar(892)\gamma$)} \\
	\babar & \cite{Aubert:2008gy} & 467M & $-0.17 \pm 0.26 \pm 0.03$ & $-0.19 \pm 0.14 \pm 0.03$ & $0.04$ \\
	\belle & \cite{Ushiroda:2006fi} & 535M & $-0.10 \pm 0.31 \pm 0.07$ & $0.20 \pm 0.20 \pm 0.06$ & $0.08$ \\
	\mc{3}{l}{\bf Average} & $-0.15 \pm 0.20$ & $-0.07 \pm 0.12$ & $0.05$ \\
        \mc{3}{l}{\small Confidence level} & \mc{2}{c}{\small $0.30~(1.0\sigma)$} & \\

		\hline
        \mc{6}{c}{$\KS \eta \gamma$} \\
	\babar & \cite{Aubert:2008js} & 465M & $-0.18 \,^{+0.49}_{-0.46} \pm 0.12$ & $-0.32 \,^{+0.40}_{-0.39} \pm 0.07$ & $-0.17$ \\
	\belle & \cite{Nakano:2018lqo} & 772M & $-1.32 \pm 0.77 \pm 0.36$ & $0.48 \pm 0.41 \pm 0.07$ & $-0.15$ \\
	\mc{3}{l}{\bf Average} & $-0.49 \pm 0.42$ & $0.06 \pm 0.29$ & $-0.15$ \\
	\mc{3}{l}{\small Confidence level} & \mc{2}{c}{\small $0.24~(1.2\sigma)$} & \\
 		\hline
        \mc{6}{c}{$\KS \rho^0 \gamma$} \\
	\babar & \cite{Sanchez:2015pxu} & 471M & $-0.18 \pm 0.32 \,^{+0.06}_{-0.05}$ & $-0.39 \pm 0.20 \,^{+0.03}_{-0.02}$ & $-0.09$ \\
	\belle & \cite{Li:2008qma} & 657M & $0.11 \pm 0.33 \,^{+0.05}_{-0.09}$ & $-0.05 \pm 0.18 \pm 0.06$ & $\phantom{-}0.04$ \\
	\mc{3}{l}{\bf Average} & $-0.06 \pm 0.23$ & $-0.22 \pm 0.14$ & $-0.02$ \\
	\mc{3}{l}{\small Confidence level} & \mc{2}{c}{\small $0.38~(0.9\sigma)$} & \\
 		\hline
        \mc{6}{c}{$\KS \phi \gamma$} \\
	\belle & \cite{Sahoo:2011zd} & 772M & $0.74 \,^{+0.72}_{-1.05} \,^{+0.10}_{-0.24}$ & $-0.35 \pm 0.58 \,^{+0.10}_{-0.23}$ & \textendash{} \\
	\hline
		\end{tabular*}
		\label{tab:cp_uta:bsg}
	\end{center}
\end{table}

\begin{figure}[htbp]
  \begin{center}
    \begin{tabular}{cc}
      \resizebox{0.46\textwidth}{!}{
        \includegraphics{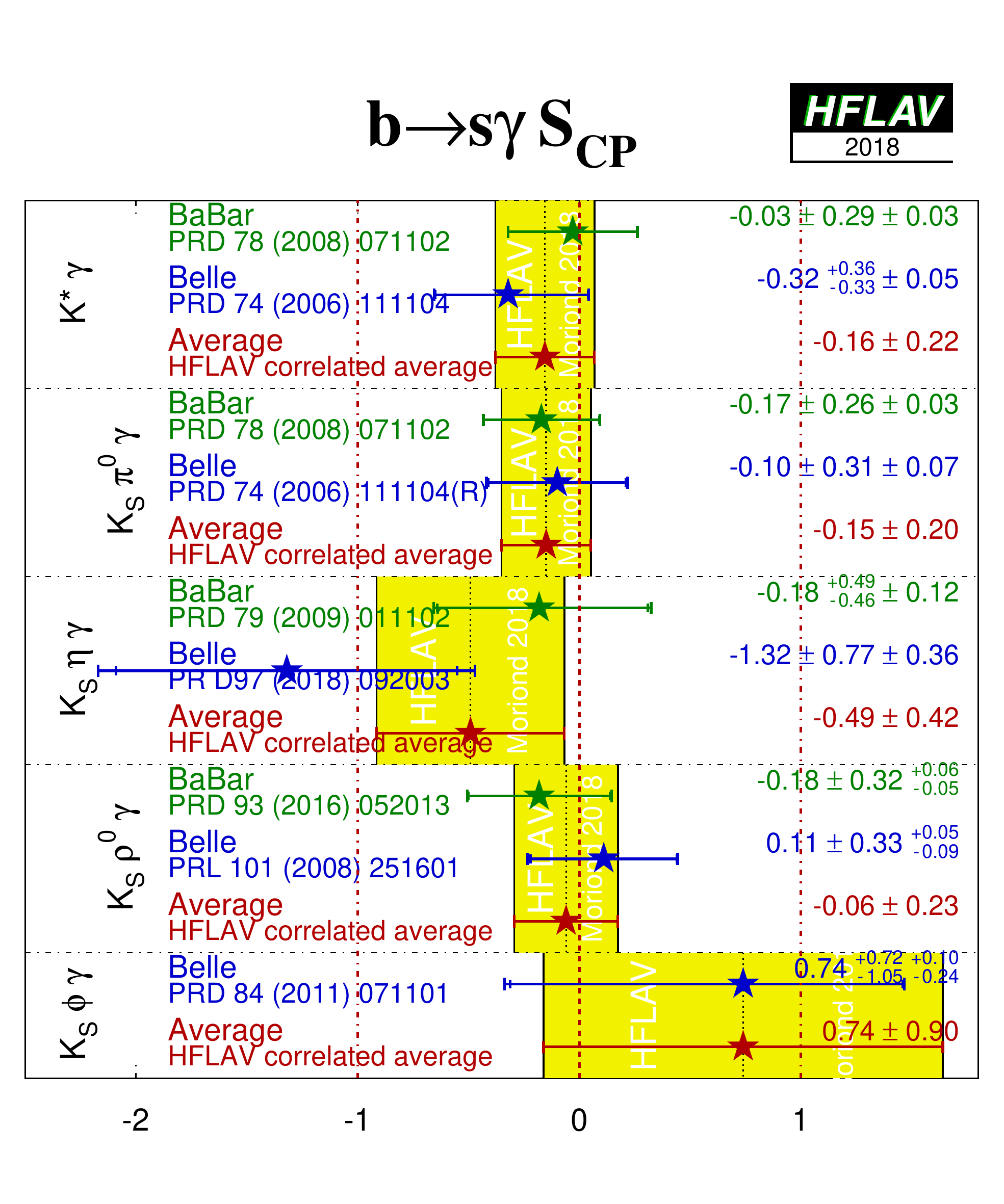}
      }
      &
      \resizebox{0.46\textwidth}{!}{
        \includegraphics{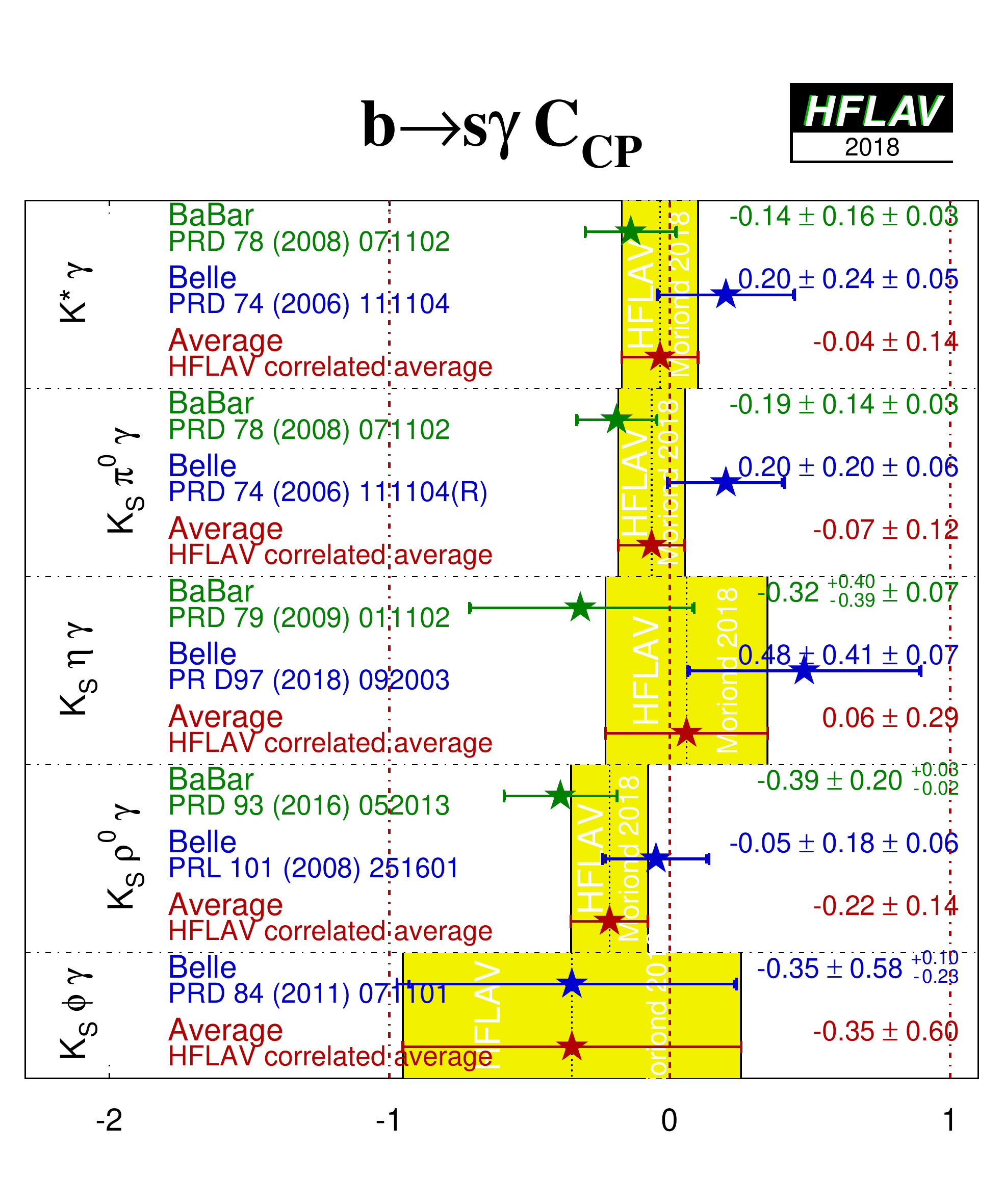}
      }
    \end{tabular}
  \end{center}
  \vspace{-0.8cm}
  \caption{
    Averages of (left) $S_{b \to s \gamma}$ and (right) $C_{b \to s \gamma}$.
    Recall that the data for $K^*\gamma$ is a subset of that for $\KS\pi^0\gamma$.
  }
  \label{fig:cp_uta:bsg}
\end{figure}

\begin{figure}[htbp]
  \centering
    \resizebox{0.38\textwidth}{!}{
      \includegraphics{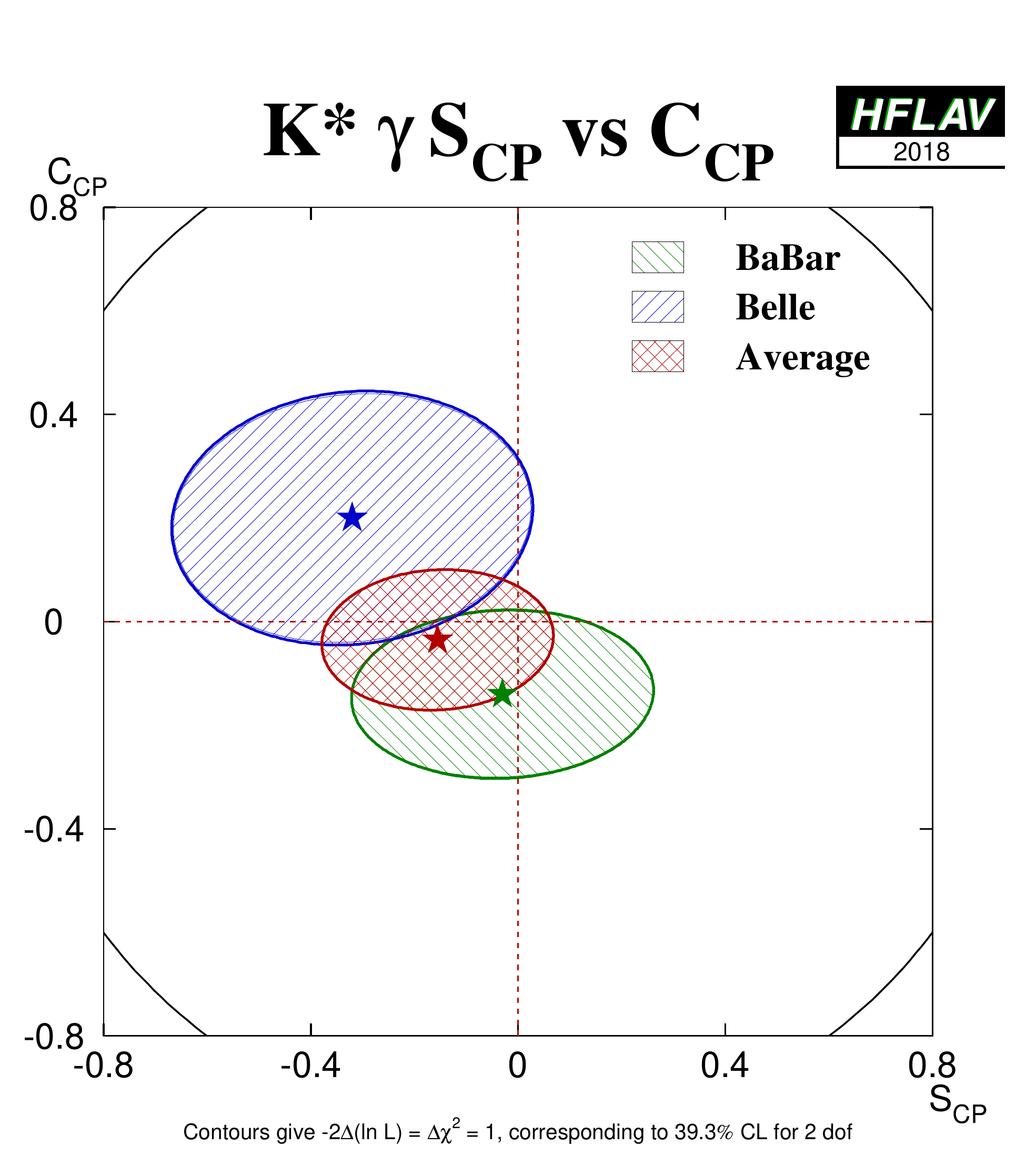}
    }
    \resizebox{0.38\textwidth}{!}{
      \includegraphics{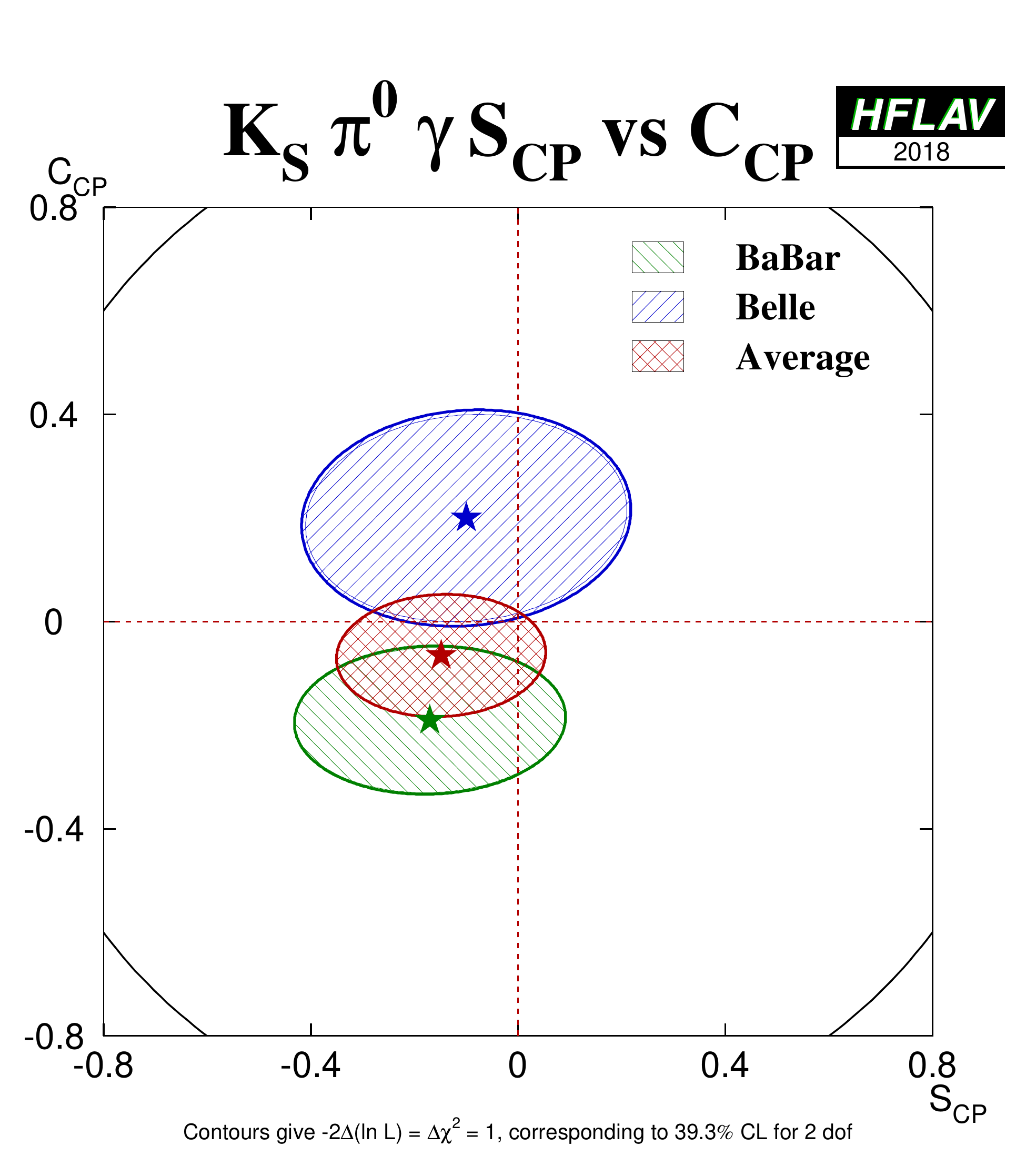}
    }\\
    \resizebox{0.38\textwidth}{!}{
      \includegraphics{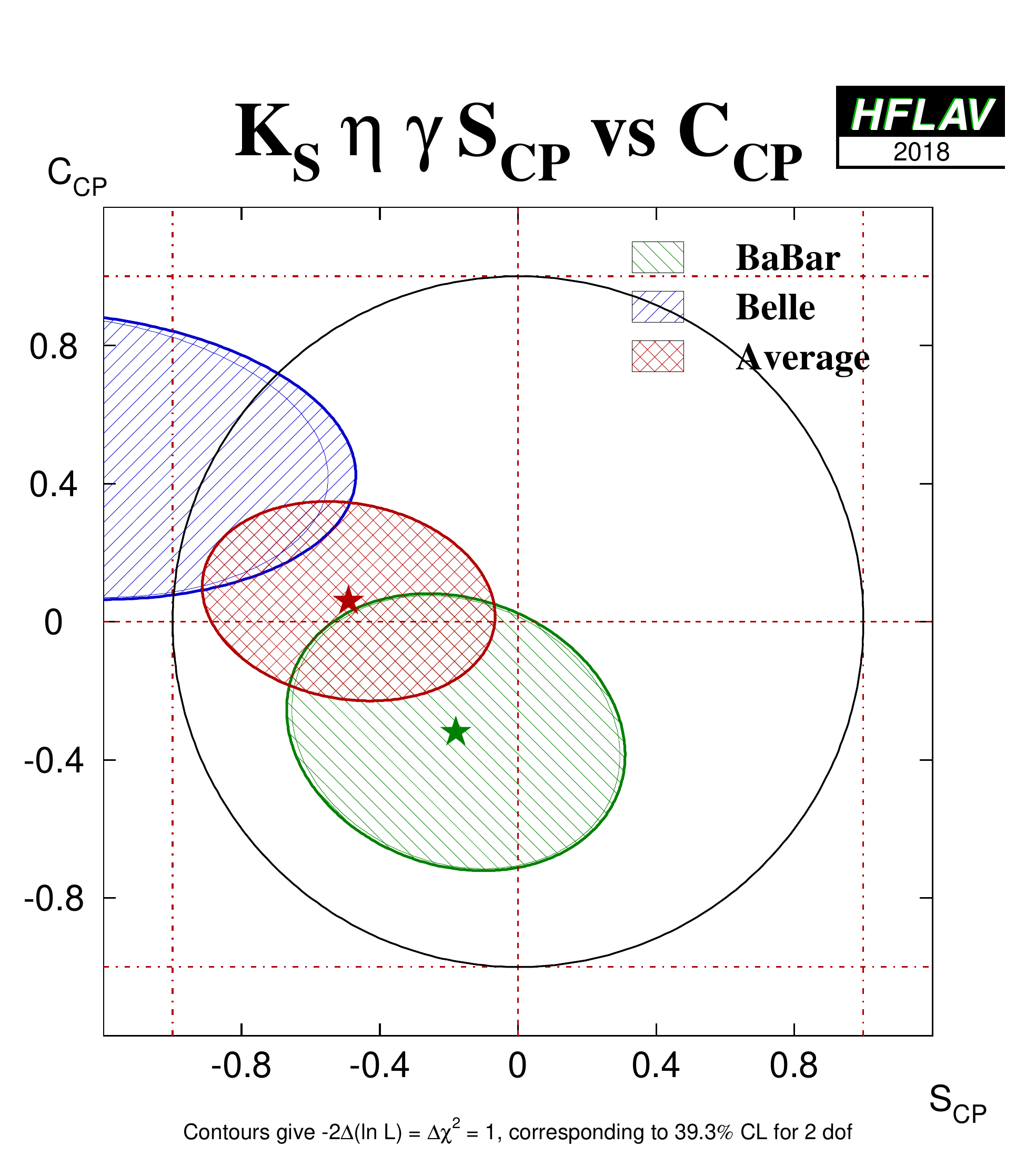}
    }
    \resizebox{0.38\textwidth}{!}{
      \includegraphics{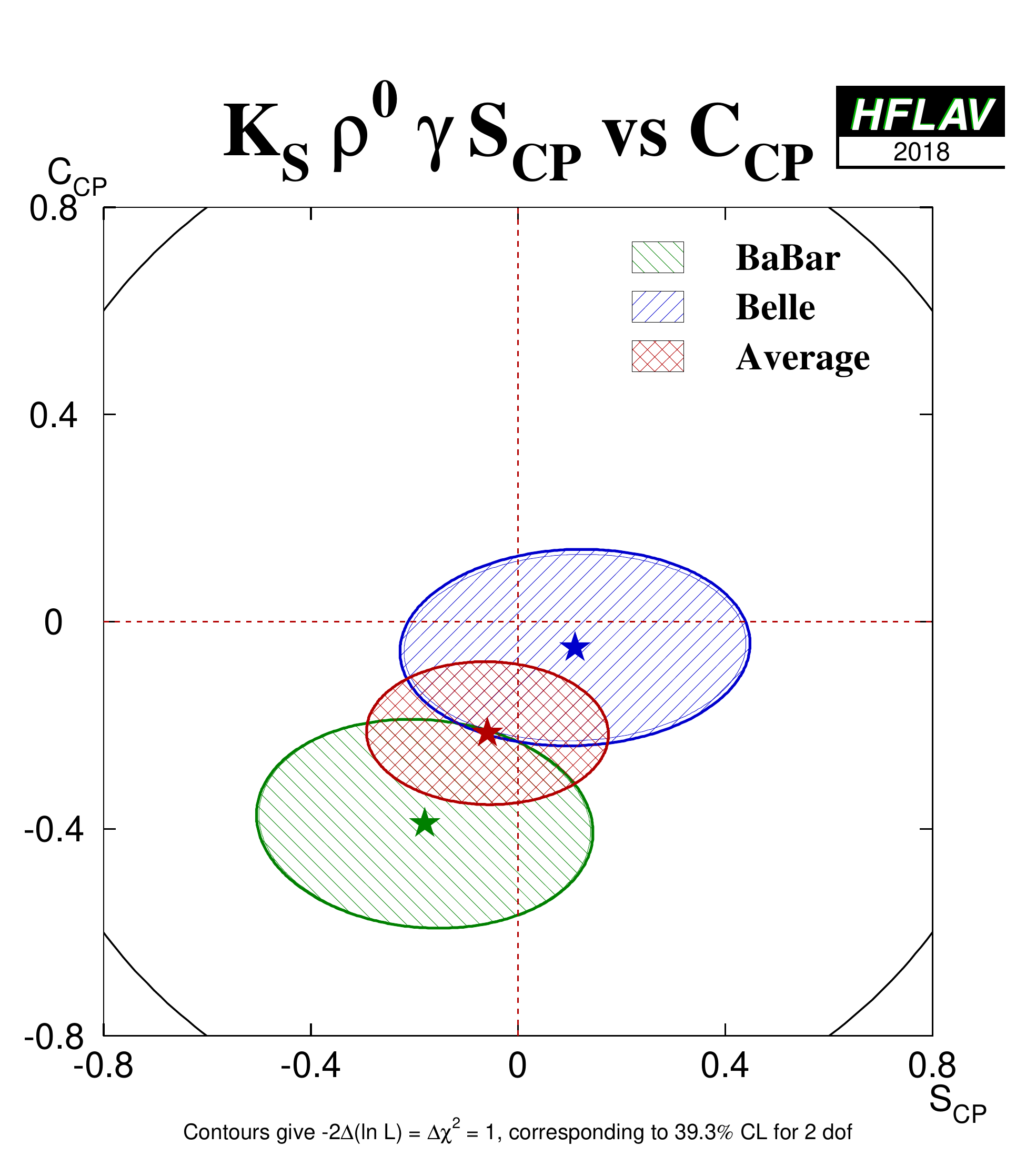}
    }
  \caption{
    Averages of four $b \to s\gamma$ dominated channels
    in the $S_{\CP}$ \vs\ $C_{\CP}$ plane.
    (Top left) $\Bz \to K^*\gamma$, (top right) $\Bz \to \KS\pi^0\gamma$ (including $K^*\gamma$), (bottom left) $\Bz \to \KS\eta\gamma$, (bottom right) $\Bz \to \KS\rhoz\gamma$.
  }
  \label{fig:cp_uta:bsg_SvsC}
\end{figure}

The results are given in Table~\ref{tab:cp_uta:bsg},
and shown in Figs.~\ref{fig:cp_uta:bsg} and~~\ref{fig:cp_uta:bsg_SvsC}.
No significant $\CP$ violation is seen;
the results are consistent with the Standard Model
and with other measurements in the $b \to s\gamma$ system (see Sec.~\ref{sec:rare}).

A similar analysis can be performed for radiative \Bs decays to, for example, the $\phi\gamma$ final state.
As for other observables determined with self-conjugate final states produced in \Bs decays, the effective lifetime also provides sensitivity to the underlying amplitudes,
and can be determined without tagging the initial flavour of the decaying meson.
The LHCb collaboration has determined the associated parameter $A_{\Delta\Gamma}(\phi\gamma) = -0.98 \,^{+0.46}_{-0.52}\,^{+0.23}_{-0.20}$~\cite{Aaij:2016ofv}.

\mysubsection{Time-dependent asymmetries in $b \to d\gamma$ transitions
}
\label{sec:cp_uta:bdg}

The formalism for the radiative decays $b \to d\gamma$ is much the same
as that for $b \to s\gamma$ discussed above.
Assuming dominance of the top quark in the loop,
the weak phase in decay should cancel with that from mixing,
so that the mixing-induced \CP\ violation parameter $S_{\CP}$
should be very small.
Corrections due to the finite light-quark mass are smaller compared to $b \to s\gamma$, since $m_d < m_s$, but QCD corrections of ${\cal O}\left(\Lambda_{\rm QCD}/m_b\right)$ may be sizable~\cite{Grinstein:2004uu}.
Large \CP\ violation effects could be seen through a non-zero value of $C_{b \to d \gamma}$, since the top loop is not the only contribution.

Results using the mode $\Bz \to \rho^0\gamma$ are available from
\belle\ and are given in Table~\ref{tab:cp_uta:bdg}.

\begin{table}[htb]
	\begin{center}
		\caption{
			Averages for $\Bz \to \rho^{0} \gamma$.
		}
		\vspace{0.2cm}
		\setlength{\tabcolsep}{0.0pc}
\renewcommand{\arraystretch}{1.1}
		\begin{tabular*}{\textwidth}{@{\extracolsep{\fill}}lrcccc} \hline
	\mc{2}{l}{Experiment} & $N(B\bar{B})$ & $S_{\CP}$ & $C_{\CP}$ & Correlation \\
	\hline
	\belle & \cite{Ushiroda:2007jf} & 657M & $-0.83 \pm 0.65 \pm 0.18$ & $0.44 \pm 0.49 \pm 0.14$ & $-0.08$ \\
		\hline
		\end{tabular*}
		\label{tab:cp_uta:bdg}
	\end{center}
\end{table}

\mysubsection{Time-dependent $\CP$ asymmetries in $b \to u\bar{u}d$ transitions
}
\label{sec:cp_uta:uud}

The $b \to u \bar u d$ transition can be mediated by either
a $b \to u$ tree amplitude or a $b \to d$ penguin amplitude.
These transitions can be investigated using
the time dependence of $\Bz$ decays to final states containing light mesons.
Results are available from both \babar\ and \belle\ for the
$\CP$ eigenstate ($\etacp = +1$) $\pi^+\pi^-$ final state
and for the vector-vector final state $\rho^+\rho^-$,
which is found to be dominated by the $\CP$-even
longitudinally polarised component
(\babar\ measures $f_{\rm long} =
0.992 \pm 0.024 \, ^{+0.026}_{-0.013}$~\cite{Aubert:2007nua},
and \belle\ measures $f_{\rm long} =
0.988 \pm 0.012 \pm 0.023$~\cite{Vanhoefer:2015ijw}).
\babar\ has also performed a time-dependent analysis of the
vector-vector final state $\rho^0\rho^0$~\cite{Aubert:2008au},
in which $f_{\rm long} = 0.70 \pm 0.14 \pm 0.05$ is determined;
\belle\ measures a smaller branching fraction than \babar\ for
$\Bz\to\rho^0\rho^0$~\cite{Adachi:2012cz} with corresponding signal yields too small to perform a time-dependent analysis, and finds $f_{\rm long} = 0.21 \,^{+0.18}_{-0.22} \pm 0.13$ for the longitudinal polarisation.
LHCb has measured the branching fraction and longitudinal polarisation for $\Bz\to\rho^0\rho^0$, and for the latter finds $f_{\rm long} = 0.745 \,^{+0.048}_{-0.058} \pm 0.034$~\cite{Aaij:2015ria}, but has not yet performed a time-dependent analysis of this decay.
The \belle\ measurement for $f_{\rm long}$ is thus in some tension with the other results.
Both \babar\ and \belle\ have furthermore performed time-dependent analyses of the $\Bz \to a_1^\pm \pi^\mp$ decay~\cite{Aubert:2006gb,Dalseno:2012hp};
\babar\ in addition has reported further experimental input for the extraction of $\alpha$ from this channel in a later publication~\cite{Aubert:2009ab}.

\begin{figure}[htbp]
  \begin{center}
    \begin{tabular}{cc}
      \resizebox{0.46\textwidth}{!}{
        \includegraphics{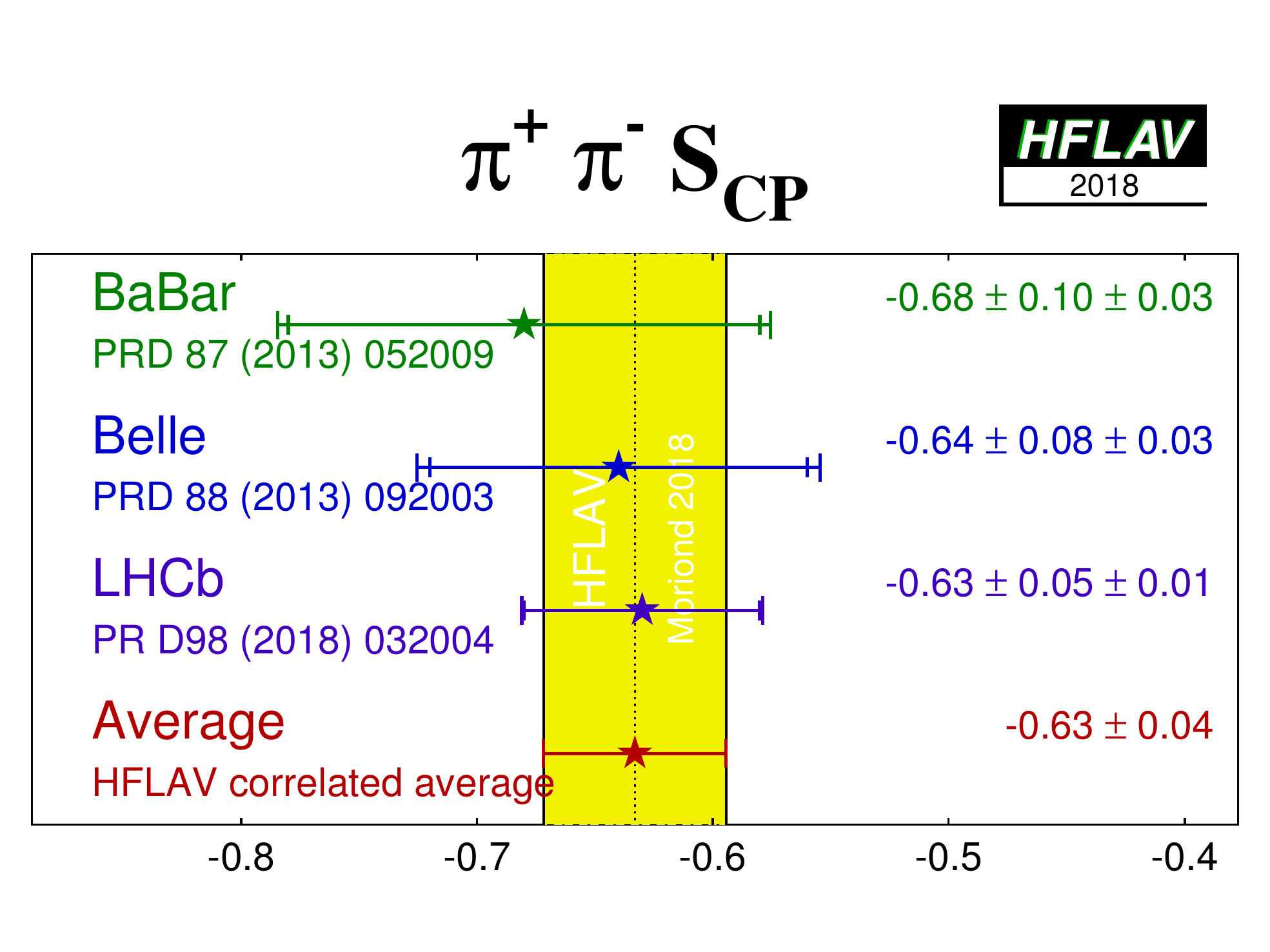}
      }
      &
      \resizebox{0.46\textwidth}{!}{
        \includegraphics{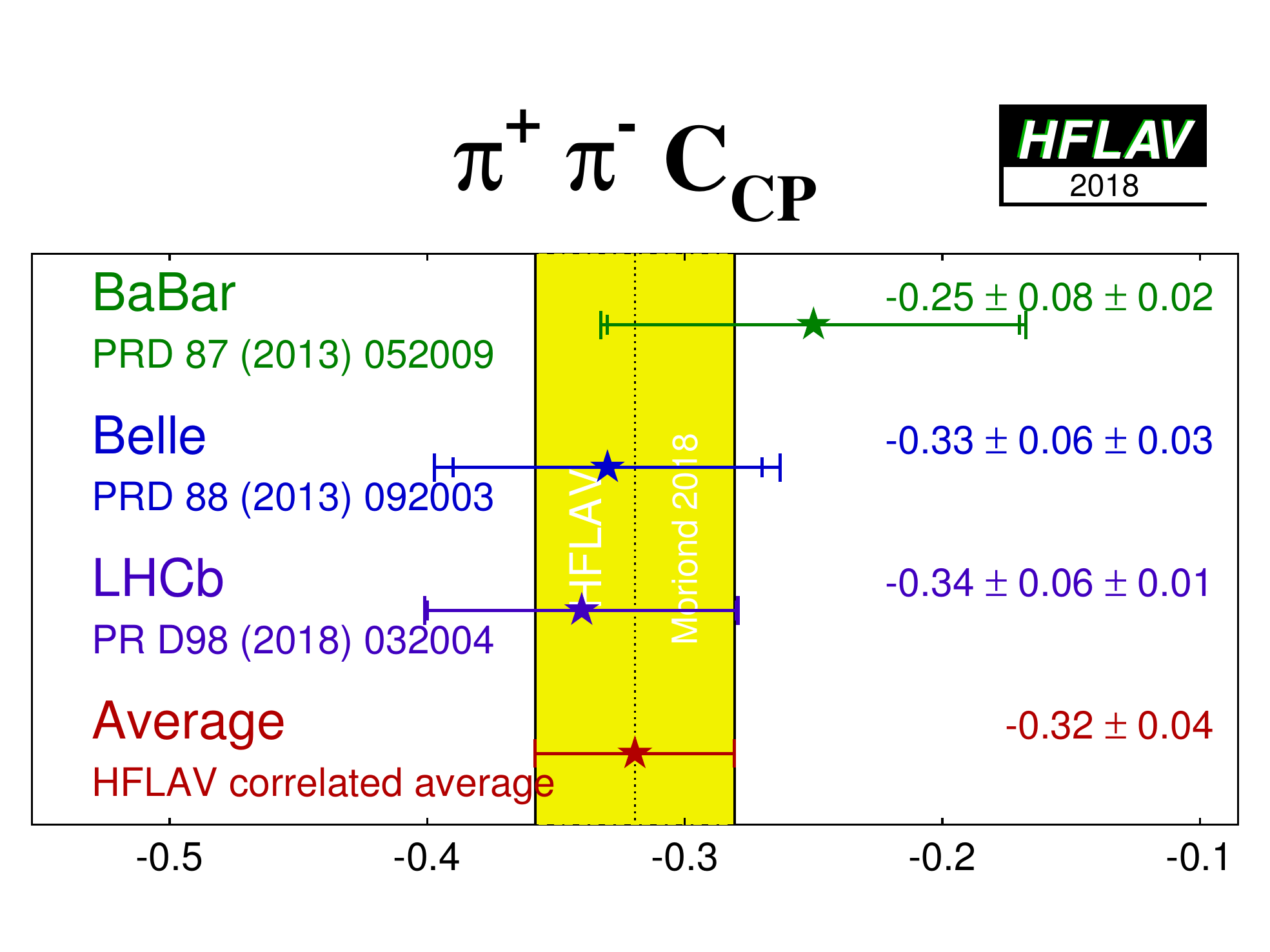}
      }
    \end{tabular}
  \end{center}
  \vspace{-0.8cm}
  \caption{
    Averages of (left) $S_{\CP}$ and (right) $C_{\CP}$ for the mode $\Bz \to \pi^+\pi^-$.
  }
  \label{fig:cp_uta:uud:pipi}
\end{figure}

\begin{figure}[htbp]
  \begin{center}
    \begin{tabular}{cc}
      \resizebox{0.46\textwidth}{!}{
        \includegraphics{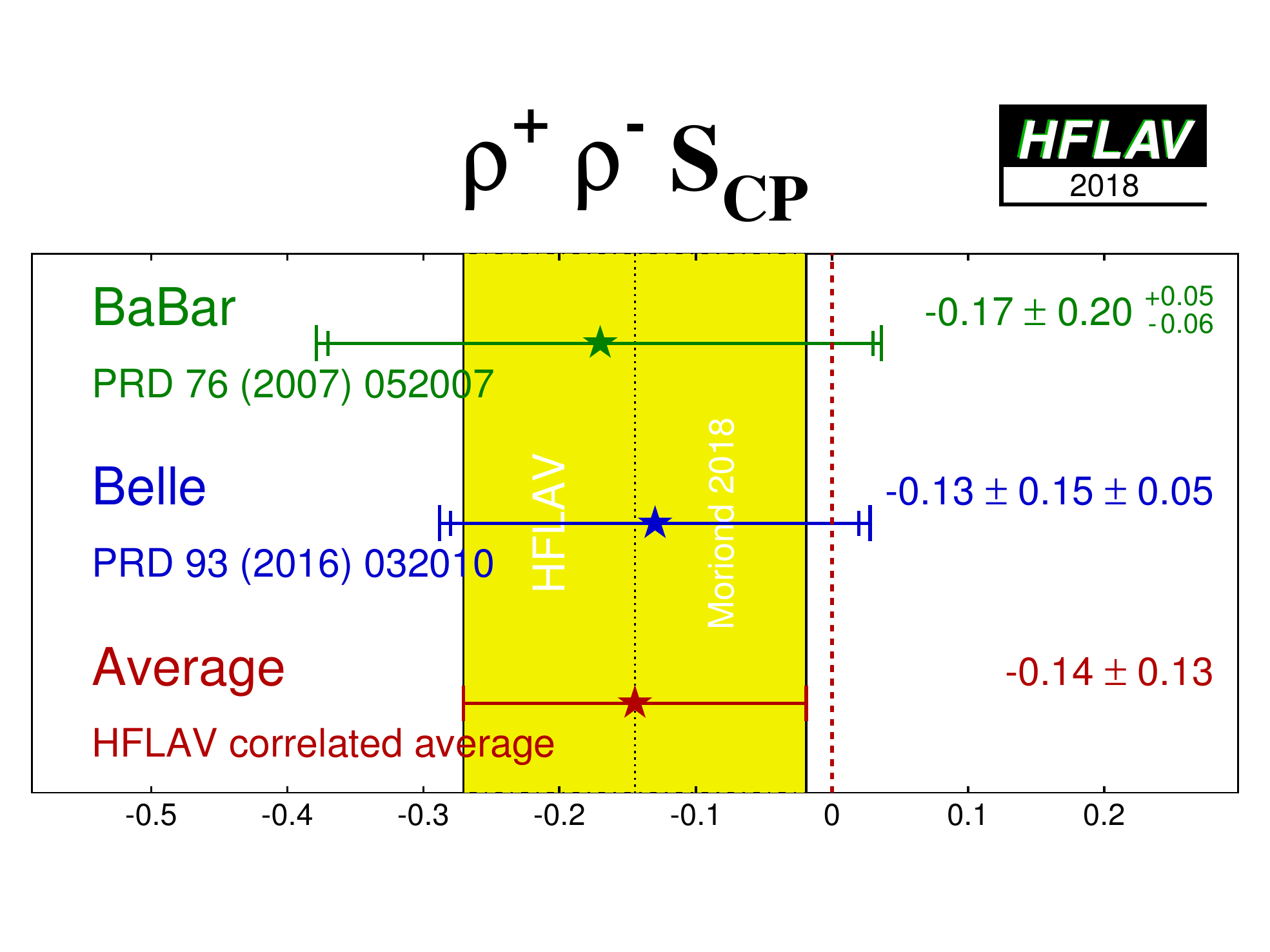}
      }
      &
      \resizebox{0.46\textwidth}{!}{
        \includegraphics{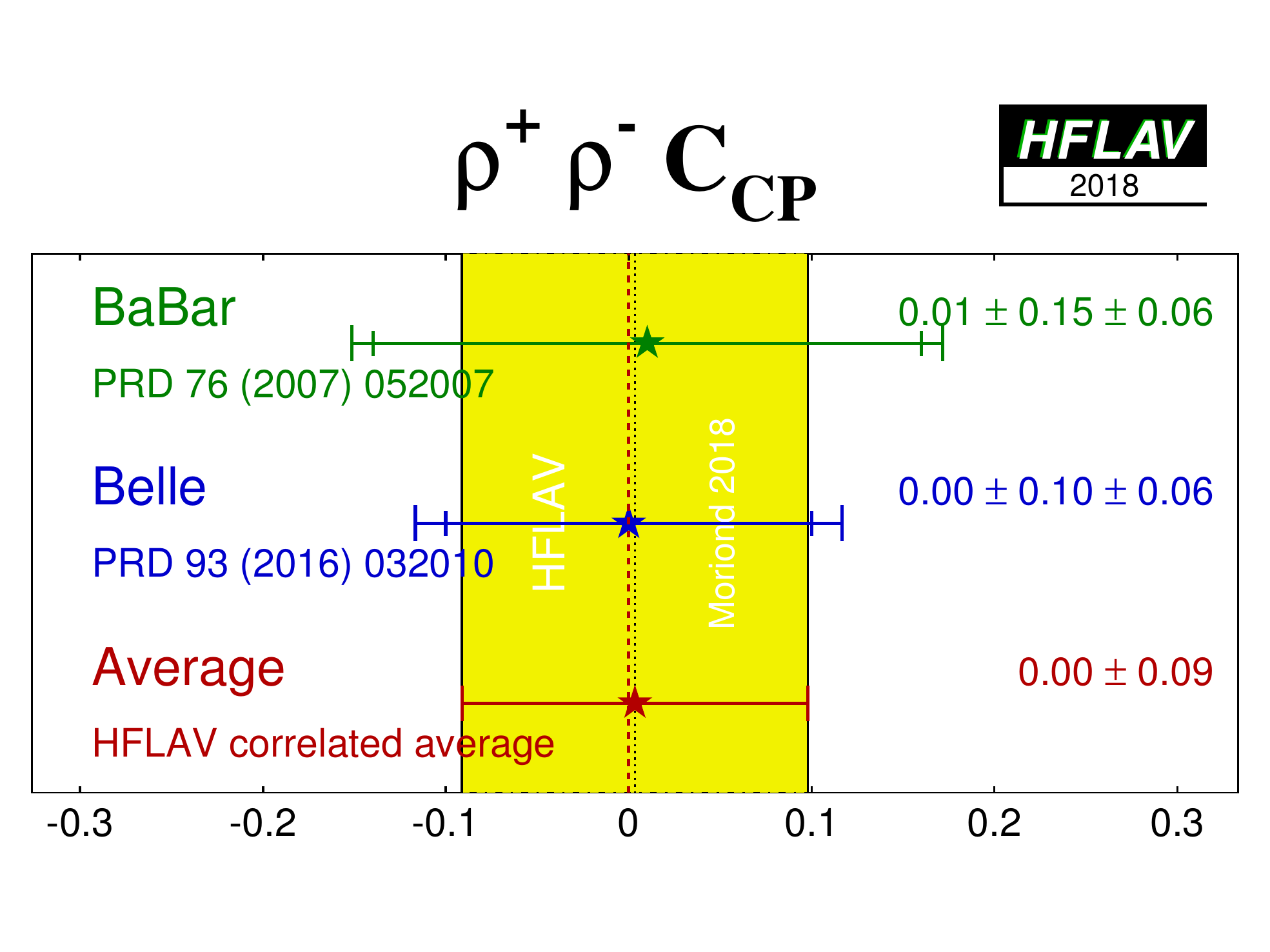}
      }
    \end{tabular}
  \end{center}
  \vspace{-0.8cm}
  \caption{
    Averages of (left) $S_{\CP}$ and (right) $C_{\CP}$ for the mode $\Bz \to \rho^+\rho^-$.
  }
  \label{fig:cp_uta:uud:rhorho}
\end{figure}

\begin{figure}[htbp]
  \begin{center}
    \begin{tabular}{cc}
      \resizebox{0.46\textwidth}{!}{
        \includegraphics{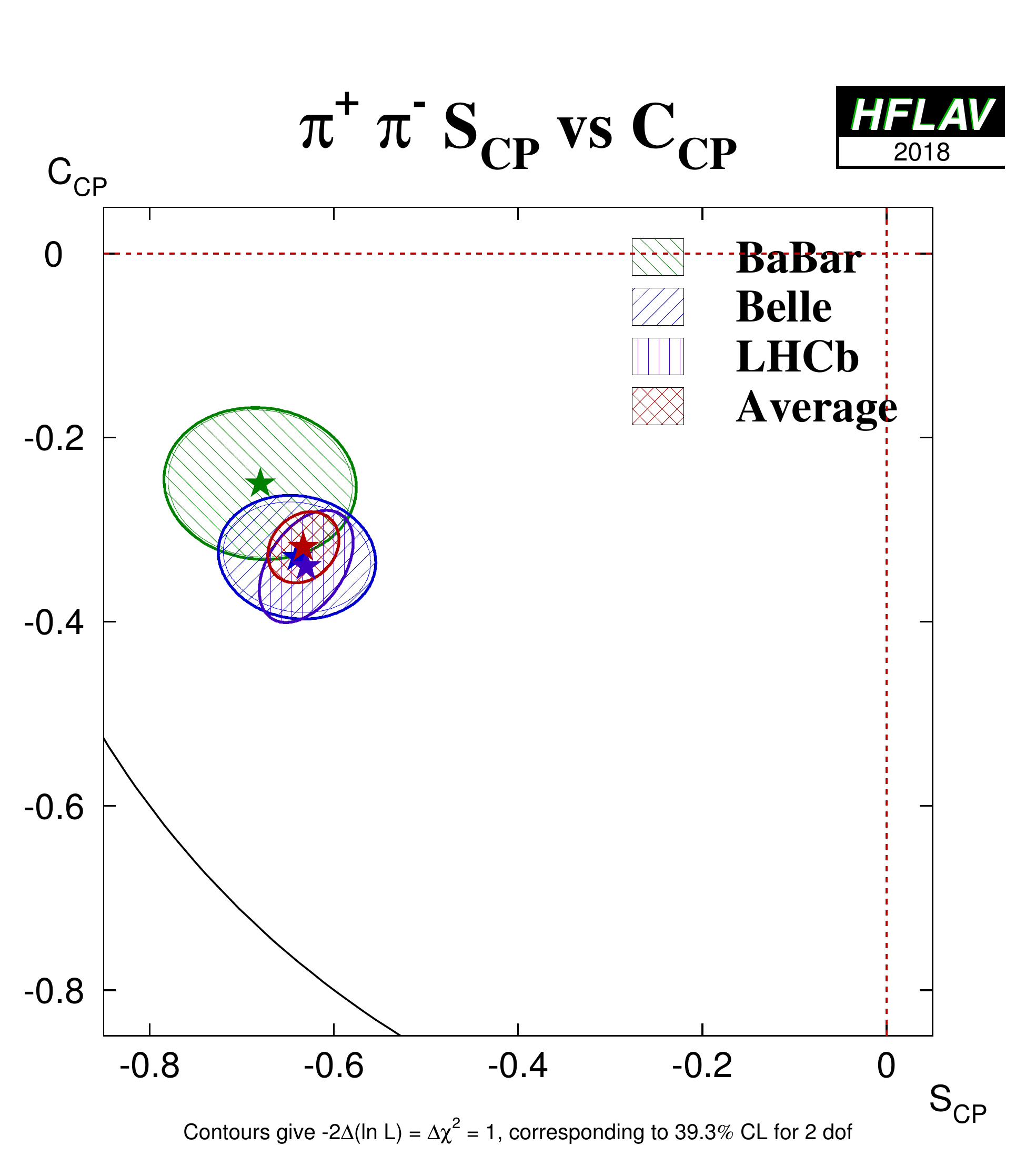}
      }
      &
      \resizebox{0.46\textwidth}{!}{
        \includegraphics{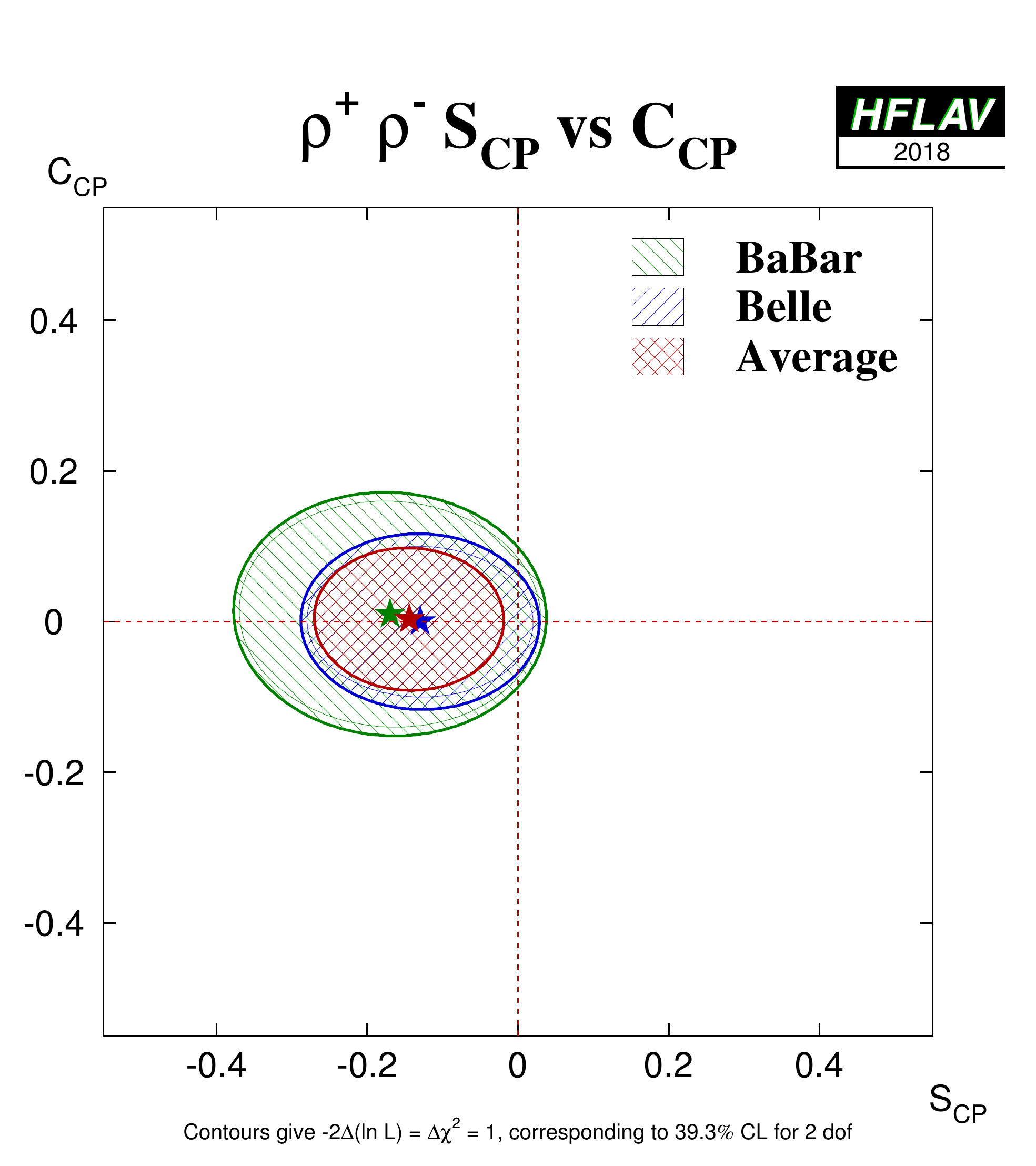}
      }
    \end{tabular}
  \end{center}
  \vspace{-0.8cm}
  \caption{
    Averages of $b \to u\bar u d$ dominated channels,
    for which correlated averages are performed,
    in the $S_{\CP}$ \vs\ $C_{\CP}$ plane.
    (Left) $\Bz \to \pi^+\pi^-$ and (right) $\Bz \to \rho^+\rho^-$.
  }
  \label{fig:cp_uta:uud_SvsC}
\end{figure}

\begin{figure}[htbp]
  \begin{center}
    \resizebox{0.46\textwidth}{!}{
      \includegraphics{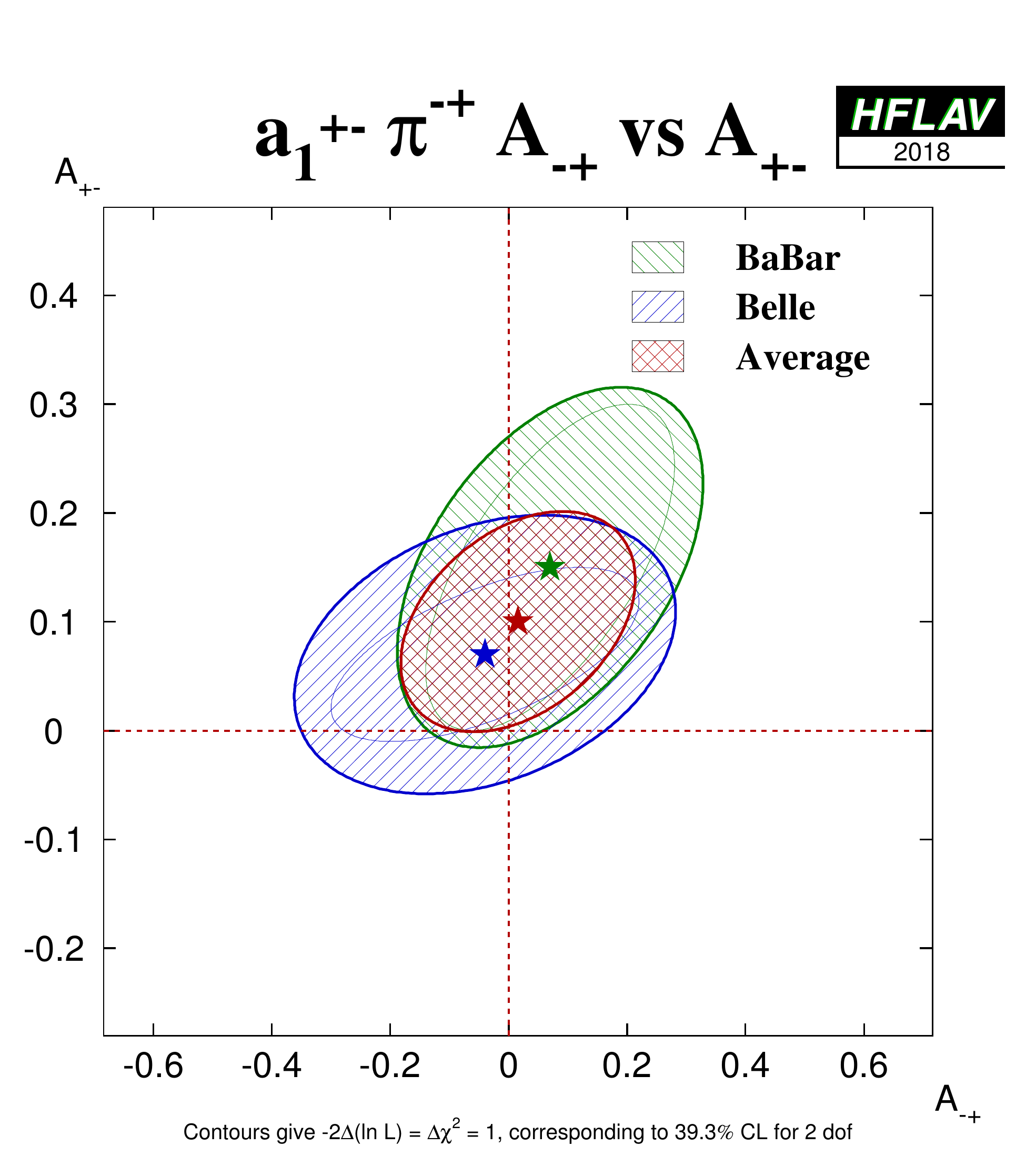}
    }
    \vspace{-0.3cm}
    \caption{
      Averages of \CP violation parameters in $\Bz \to a_1^\pm\pi^\mp$ in
      ${\cal A}^{-+}_{a_1\pi}$ \vs\ ${\cal A}^{+-}_{a_1\pi}$ space.
    }
    \label{fig:cp_uta:a1pi}
  \end{center}
\end{figure}

Results and averages of time-dependent \CP violation parameters in
$b \to u \bar u d$ transitions are listed in Table~\ref{tab:cp_uta:uud}.
The averages for $\pi^+\pi^-$ are shown in Fig.~\ref{fig:cp_uta:uud:pipi},
and those for $\rho^+\rho^-$ are shown in Fig.~\ref{fig:cp_uta:uud:rhorho},
with the averages in the $S_{\CP}$ \vs\ $C_{\CP}$ plane
shown in Fig.~\ref{fig:cp_uta:uud_SvsC}, and
averages of \CP violation parameters in $\Bz \to a_1^\pm \pi^\mp$ decay shown in Fig.~\ref{fig:cp_uta:a1pi}.

\begin{sidewaystable}
	\begin{center}
		\caption{
      Averages for $b \to u \bar u d$ modes.
		}
		\vspace{0.2cm}
		\setlength{\tabcolsep}{0.0pc}
\renewcommand{\arraystretch}{1.1}
		\begin{tabular*}{\textwidth}{@{\extracolsep{\fill}}lrcccc} \hline
	\mc{2}{l}{Experiment} & Sample size & $S_{\CP}$ & $C_{\CP}$ & Correlation \\
	\hline
      \mc{6}{c}{$\pi^{+} \pi^{-}$} \\
	\babar & \cite{Lees:2012mma} & $N(B\bar{B})$ = 467M & $-0.68 \pm 0.10 \pm 0.03$ & $-0.25 \pm 0.08 \pm 0.02$ & $-0.06$ \\
	\belle & \cite{Adachi:2013mae} & $N(B\bar{B})$ = 772M & $-0.64 \pm 0.08 \pm 0.03$ & $-0.33 \pm 0.06 \pm 0.03$ & $-0.10$ \\
	LHCb & \cite{Aaij:2018tfw} & $\int {\cal L}\,dt = 3.0 \, {\rm fb}^{-1}$ & $-0.63 \pm 0.05 \pm 0.01$ & $-0.34 \pm 0.06 \pm 0.01$ & $0.45$ \\
	\mc{3}{l}{\bf Average} & $-0.63 \pm 0.04$ & $-0.32 \pm 0.04$ & $0.21$ \\
	\mc{3}{l}{\small Confidence level} & \mc{2}{c}{\small $0.90~(0.1\sigma)$} & \\
		\hline
      \mc{6}{c}{$\rho^{+} \rho^{-}$} \\
	\babar & \cite{Aubert:2007nua} & $N(B\bar{B})$ = 387M & $-0.17 \pm 0.20 \,^{+0.05}_{-0.06}$ & $0.01 \pm 0.15 \pm 0.06$ & $-0.04$ \\
	\belle & \cite{Vanhoefer:2015ijw} & $N(B\bar{B})$ = 772M & $-0.13 \pm 0.15 \pm 0.05$ & $0.00 \pm 0.10 \pm 0.06$ & $-0.02$ \\
	\mc{3}{l}{\bf Average} & $-0.14 \pm 0.13$ & $0.00 \pm 0.09$ & $-0.02$ \\
	\mc{3}{l}{\small Confidence level} & \mc{2}{c}{\small $0.99~(0.02\sigma)$} & \\
		\hline
      \mc{6}{c}{$\rho^{0} \rho^{0}$} \\
	\babar & \cite{Aubert:2008au} & $N(B\bar{B}) =$ 465M & $0.3 \pm 0.7 \pm 0.2$ & $0.2 \pm 0.8 \pm 0.3$ & $-0.04$ \\
 		\hline
 		\end{tabular*}

                \vspace{2ex}

    \resizebox{\textwidth}{!}{
 		\begin{tabular}{@{\extracolsep{2mm}}lrcccccc} \hline
 		\mc{2}{l}{Experiment} & $N(B\bar{B})$ & $A_{\CP}^{a_1\pi}$ & $C_{a_1\pi}$ & $S_{a_1\pi}$ & $\Delta C_{a_1\pi}$ & $\Delta S_{a_1\pi}$ \\
 		\hline
      \mc{8}{c}{$a_1^{\pm} \pi^{\mp}$} \\
	\babar & \cite{Aubert:2006gb} & 384M & $-0.07 \pm 0.07 \pm 0.02$ & $-0.10 \pm 0.15 \pm 0.09$ & $0.37 \pm 0.21 \pm 0.07$ & $0.26 \pm 0.15 \pm 0.07$ & $-0.14 \pm 0.21 \pm 0.06$ \\
	\belle & \cite{Dalseno:2012hp} & 772M & $-0.06 \pm 0.05 \pm 0.07$ & $-0.01 \pm 0.11 \pm 0.09$ & $-0.51 \pm 0.14 \pm 0.08$ & $0.54 \pm 0.11 \pm 0.07$ & $-0.09 \pm 0.14 \pm 0.06$ \\ 
 	\hline
	\mc{3}{l}{\bf Average} & $-0.06 \pm 0.06$ & $-0.05 \pm 0.11$ & $-0.20 \pm 0.13$ & $0.43 \pm 0.10$ & $-0.10 \pm 0.12$ \\
	\mc{3}{l}{\small Confidence level} & \mc{5}{c}{\small $0.03~(2.1\sigma)$} \\
        \hline
		\end{tabular}
              }

                \vspace{2ex}

		\begin{tabular*}{\textwidth}{@{\extracolsep{\fill}}lrcccc} \hline
		\mc{2}{l}{Experiment} & $N(B\bar{B})$ & ${\cal A}^{-+}_{a_1\pi}$ & ${\cal A}^{+-}_{a_1\pi}$ & Correlation \\
		\hline
	\babar & \cite{Aubert:2006gb} & 384M & $0.07 \pm 0.21 \pm 0.15$ & $0.15 \pm 0.15 \pm 0.07$ & 0.63 \\
	\belle & \cite{Dalseno:2012hp} & 772M & $-0.04 \pm 0.26 \pm 0.19$ & $0.07 \pm 0.08 \pm 0.10$ & 0.61 \\
	\mc{3}{l}{\bf Average} & $0.02 \pm 0.20$ & $0.10 \pm 0.10$ & 0.38 \\
        \mc{3}{l}{\small Confidence level} & \mc{2}{c}{\small $0.92~(0.1\sigma)$} \\
		\hline
		\end{tabular*}

		\label{tab:cp_uta:uud}
	\end{center}
\end{sidewaystable}

If the penguin contribution is negligible,
the time-dependent parameters for $\Bz \to \pi^+\pi^-$
and $\Bz \to \rho^+\rho^-$ are given by
$S_{b \to u\bar u d} = \etacp \sin(2\alpha)$ and
$C_{b \to u\bar u d} = 0$.
In the presence of the penguin contribution,
$\CP$ violation in decay may arise,
and there is no straightforward interpretation
of $S_{b \to u\bar u d}$ and $C_{b \to u\bar u d}$.
An isospin analysis~\cite{Gronau:1990ka}
can be used to disentangle the contributions and extract $\alpha$, as discussed further in Sec.~\ref{sec:cp_uta:uud:alpha}.

\begin{figure}[htbp]
  \begin{center}
    \begin{tabular}{cc}
      \resizebox{0.46\textwidth}{!}{
        \includegraphics{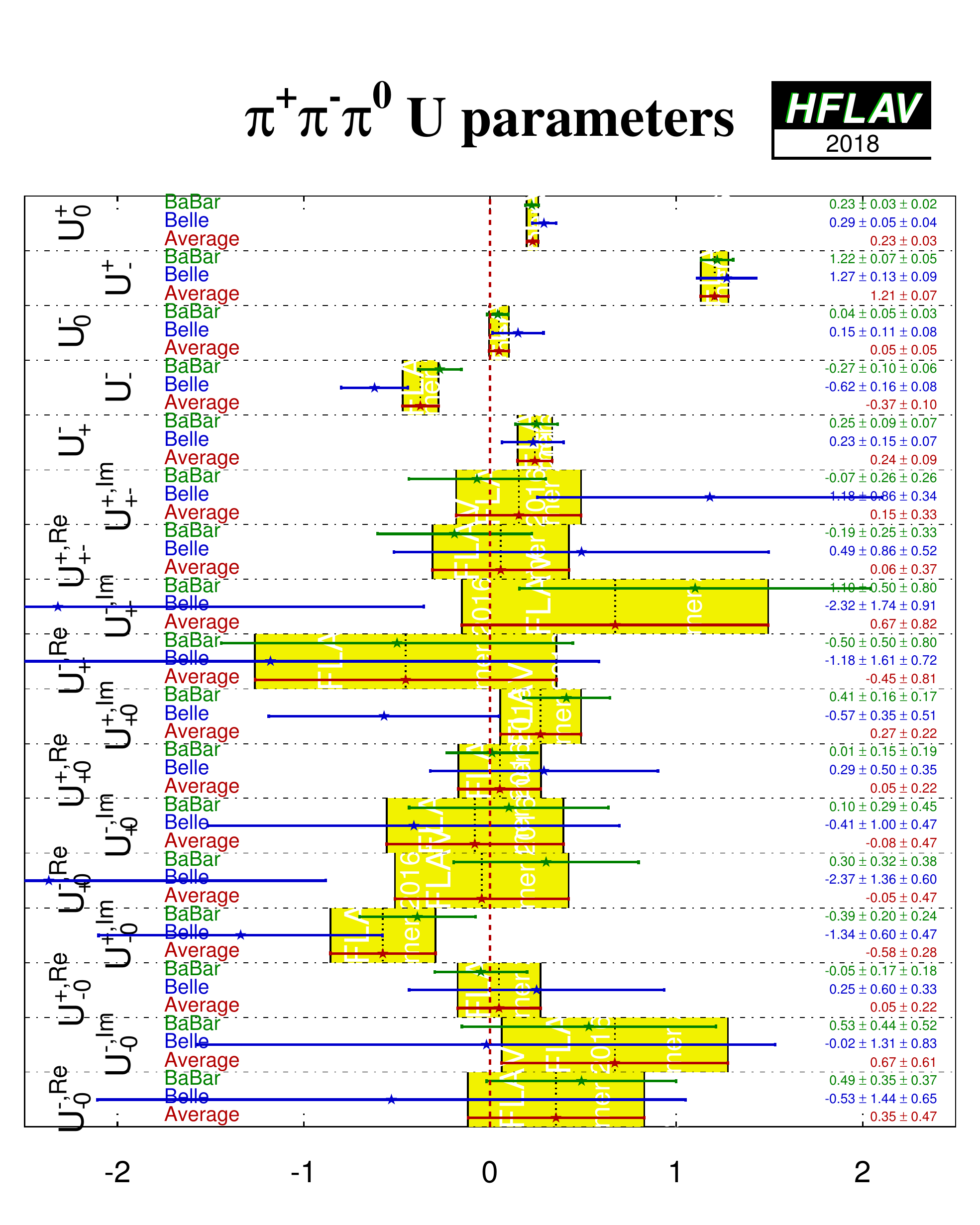}
      }
      &
      \resizebox{0.46\textwidth}{!}{
        \includegraphics{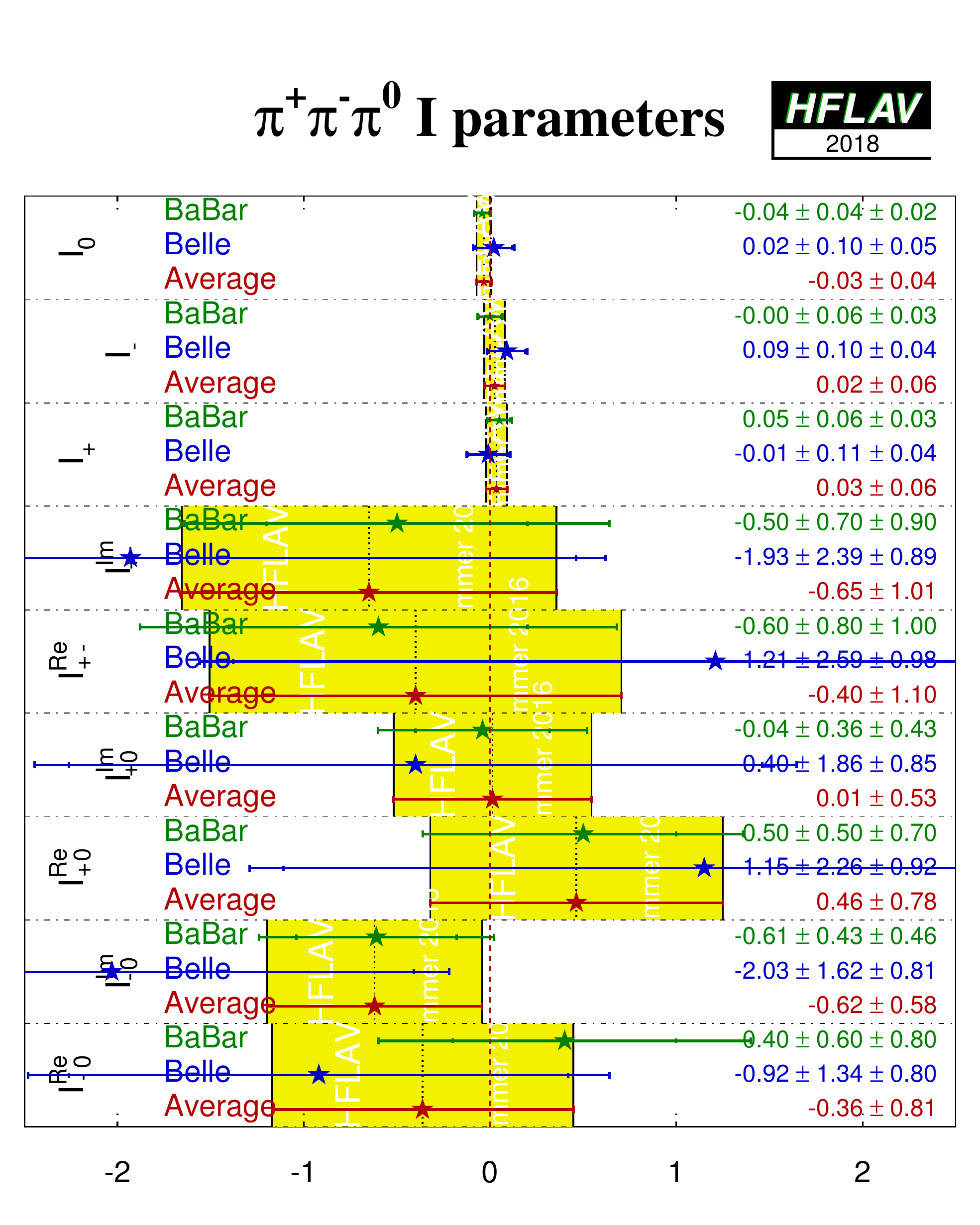}
      }
    \end{tabular}
  \end{center}
  \vspace{-0.8cm}
  \caption{
    Summary of the $U$ and $I$ parameters measured in the
    time-dependent $\Bz \to \pi^+\pi^-\pi^0$ Dalitz plot analysis.
  }
  \label{fig:cp_uta:uud:uandi}
\end{figure}

For the non-$\CP$ eigenstate $\rho^{\pm}\pi^{\mp}$,
both \babar~\cite{Aubert:2007jn}
and \belle~\cite{Kusaka:2007dv,Kusaka:2007mj} have performed
time-dependent Dalitz-plot analyses
of the $\pi^+\pi^-\pi^0$ final state~\cite{Snyder:1993mx};
such analyses allow direct measurements of the phases.
Both experiments have measured the $U$ and $I$ parameters discussed in
Sec.~\ref{sec:cp_uta:notations:dalitz:pipipi0} and defined in
Table~\ref{tab:cp_uta:pipipi0:uandi}.
We have performed a full correlated average of these parameters,
the results of which are summarised in Fig.~\ref{fig:cp_uta:uud:uandi}.

\begin{figure}[htbp]
  \begin{center}
    \begin{tabular}{cc}
      \resizebox{0.46\textwidth}{!}{
        \includegraphics{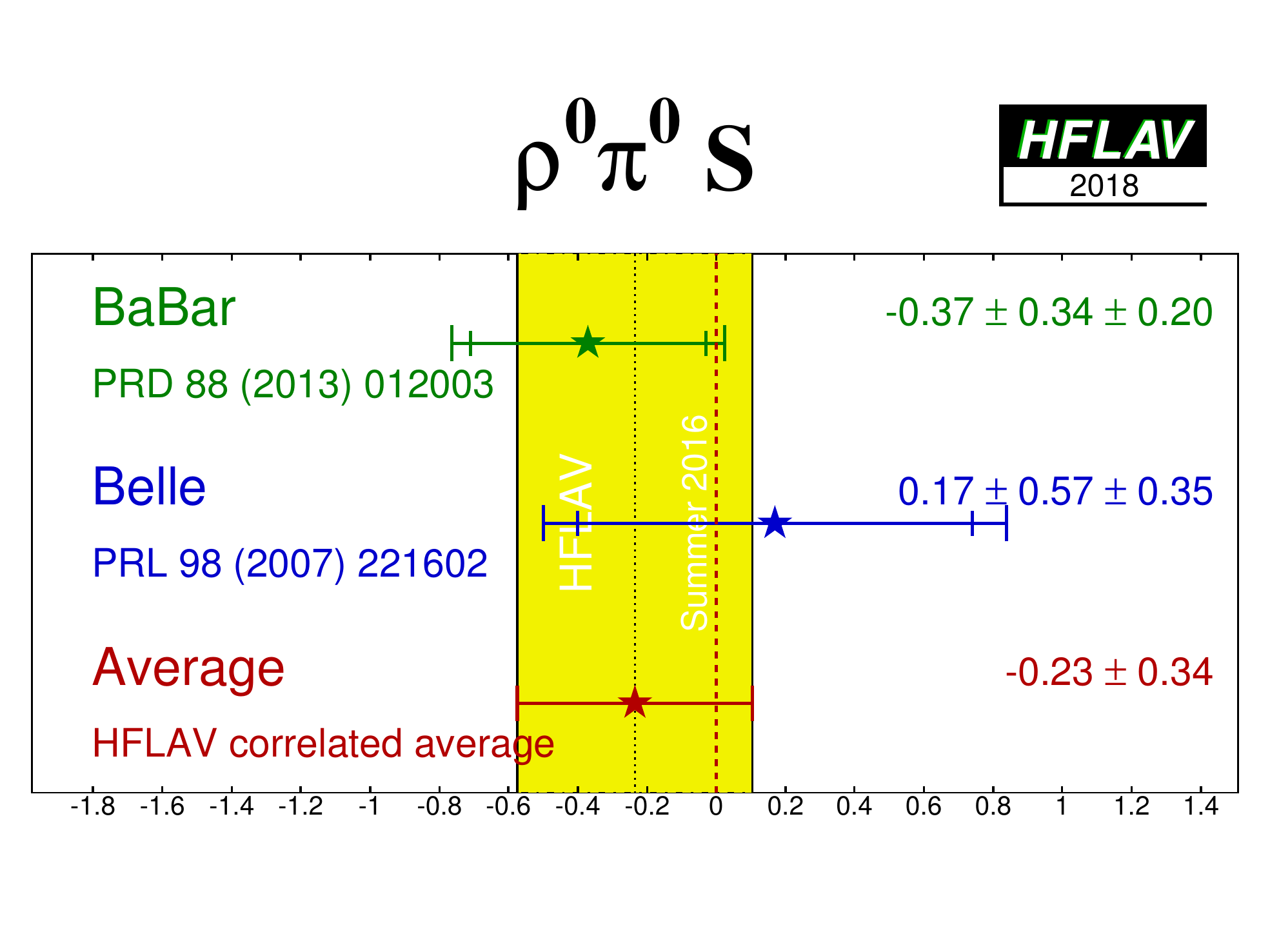}
      }
      &
      \resizebox{0.46\textwidth}{!}{
        \includegraphics{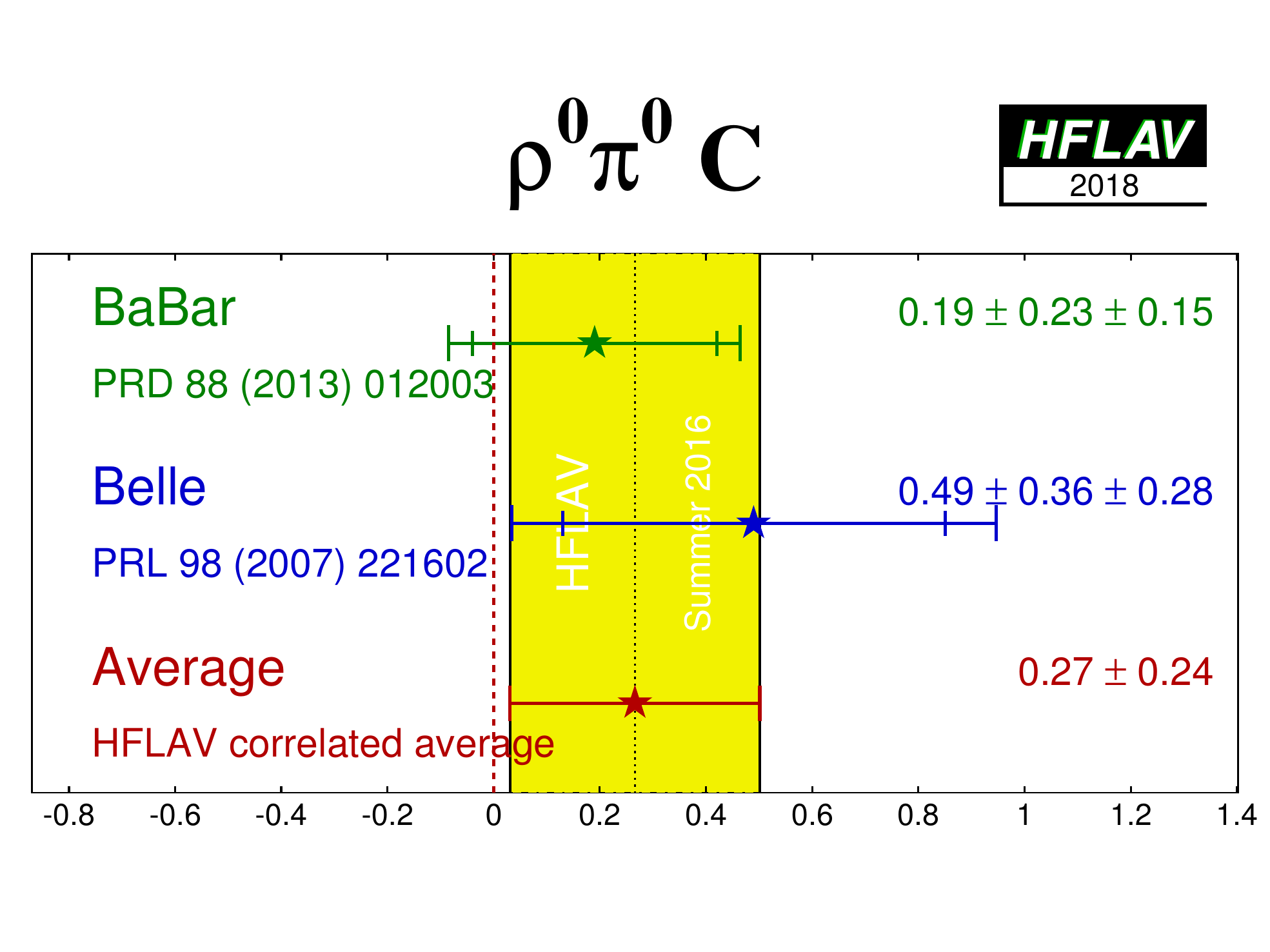}
      }
    \end{tabular}
  \end{center}
  \vspace{-0.8cm}
  \caption{
    Averages of (left) $S_{b \to u\bar u d}$ and (right) $C_{b \to u\bar u d}$
    for the mode $\Bz \to \rho^0\pi^0$.
  }
  \label{fig:cp_uta:uud:rho0pi0}
\end{figure}

\begin{figure}[htbp]
  \begin{center}
    \resizebox{0.46\textwidth}{!}{
      \includegraphics{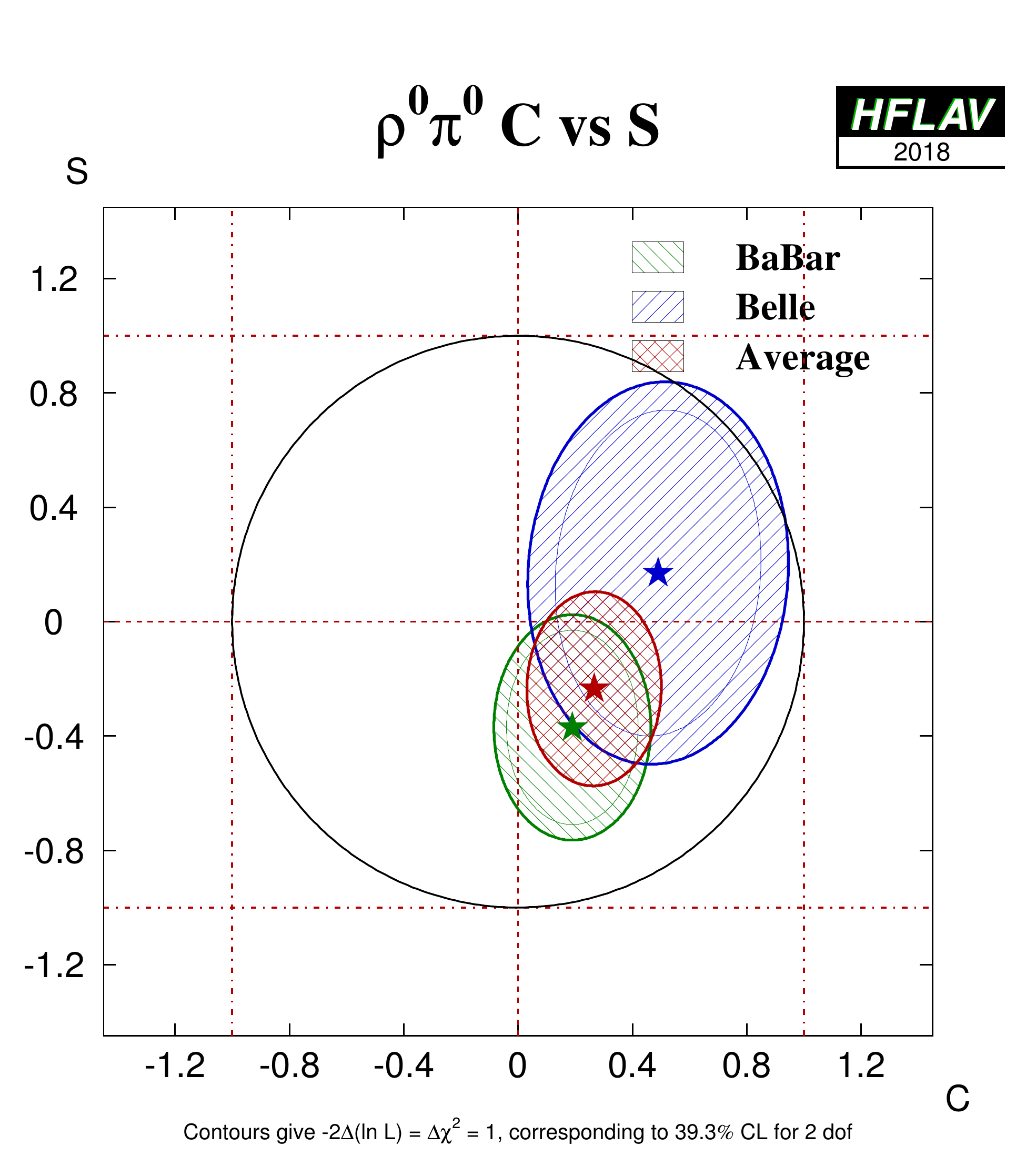}
    }
  \end{center}
  \vspace{-0.8cm}
  \caption{
    Averages of $b \to u\bar u d$ dominated channels,
    for the mode $\Bz \to \rho^0\pi^0$
    in the $S_{\CP}$ \vs\ $C_{\CP}$ plane.
  }
  \label{fig:cp_uta:uud:rho0pi0_SvsC}
\end{figure}

Both experiments have also extracted the Q2B parameters for the $\rho\pi$ channels.
We have performed a full correlated average of these parameters,
which is equivalent to determining the values from the
averaged $U$ and $I$ parameters.
The results are given in Table~\ref{tab:cp_uta:uud:rhopi_q2b}.\footnote{
  The $\Bz \to \rho^\pm \pi^\mp$ Q2B parameters are comparable to the
  parameters used for $\Bz \to a_1^\pm \pi^\mp$ decays, reported in
  Table~\ref{tab:cp_uta:uud}.
  For the $\Bz \to a_1^\pm \pi^\mp$ case there has not yet been a full
  amplitude analysis of $\Bz \to \pi^+\pi^-\pi^+\pi^-$ and therefore only the
  Q2B parameters are available.
}
Averages of the $\Bz \to \rho^0\pi^0$ Q2B parameters are shown in
Figs.~\ref{fig:cp_uta:uud:rho0pi0} and~\ref{fig:cp_uta:uud:rho0pi0_SvsC}.

\begin{sidewaystable}
	\begin{center}
		\caption{
                  Averages of quasi-two-body parameters extracted
                  from time-dependent Dalitz plot analysis of 
                  $\Bz \to \pi^+\pi^-\pi^0$.
		}
		\vspace{0.2cm}
		\setlength{\tabcolsep}{0.0pc}
    \resizebox{\textwidth}{!}{
\renewcommand{\arraystretch}{1.1}
		\begin{tabular}{@{\extracolsep{2mm}}lrcccccc} \hline
		\mc{2}{l}{Experiment} & $N(B\bar{B})$ & ${\cal A}_{\CP}^{\rho\pi}$ & $C_{\rho\pi}$ & $S_{\rho\pi}$ & $\Delta C_{\rho\pi}$ & $\Delta S_{\rho\pi}$ \\
	\hline
	\babar & \cite{Lees:2013nwa} & 471M & $-0.10 \pm 0.03 \pm 0.02$ & $0.02 \pm 0.06 \pm 0.04$ & $0.05 \pm 0.08 \pm 0.03$ & $0.23 \pm 0.06 \pm 0.05$ & $0.05 \pm 0.08 \pm 0.04$ \\
	\belle & \cite{Kusaka:2007dv,Kusaka:2007mj} & 449M & $-0.12 \pm 0.05 \pm 0.04$ & $-0.13 \pm 0.09 \pm 0.05$ & $0.06 \pm 0.13 \pm 0.05$ & $0.36 \pm 0.10 \pm 0.05$ & $-0.08 \pm 0.13 \pm 0.05$ \\
	\mc{3}{l}{\bf Average} & $-0.11 \pm 0.03$ & $-0.03 \pm 0.06$ & $0.06 \pm 0.07$ & $0.27 \pm 0.06$ & $0.01 \pm 0.08$ \\
	\mc{3}{l}{\small Confidence level} & \mc{5}{c}{\small $0.63~(0.5\sigma)$} \\
        \hline
		\end{tabular}
              }

                \vspace{2ex}

		\begin{tabular*}{\textwidth}{@{\extracolsep{\fill}}lrcccc} \hline
		\mc{2}{l}{Experiment} & $N(B\bar{B})$ & ${\cal A}^{-+}_{\rho\pi}$ & ${\cal A}^{+-}_{\rho\pi}$ & Correlation \\
		\hline
	\babar & \cite{Lees:2013nwa} & 471M & $-0.12 \pm 0.08 \,^{+0.04}_{-0.05}$ & $0.09 \,^{+0.05}_{-0.06} \pm 0.04$ & $0.55$ \\
	\belle & \cite{Kusaka:2007dv,Kusaka:2007mj} & 449M & $0.08 \pm 0.16 \pm 0.11$ & $0.21 \pm 0.08 \pm 0.04$ & $0.47$ \\
	\mc{3}{l}{\bf Average} & $-0.08 \pm 0.08$ & $0.13 \pm 0.05$ & $0.37$ \\
        \mc{3}{l}{\small Confidence level} & \mc{2}{c}{\small $0.47~(0.7\sigma)$} \\
		\hline
		\end{tabular*}

                \vspace{2ex}

		\begin{tabular*}{\textwidth}{@{\extracolsep{\fill}}lrcccc} \hline
		\mc{2}{l}{Experiment} & $N(B\bar{B})$ & $C_{\rho^0\pi^0}$ & $S_{\rho^0\pi^0}$ & Correlation \\
		\hline
	\babar & \cite{Lees:2013nwa} & 471M & $0.19 \pm 0.23 \pm 0.15$ & $-0.37 \pm 0.34 \pm 0.20$ & $0.00$ \\
	\belle & \cite{Kusaka:2007dv,Kusaka:2007mj} & 449M & $0.49 \pm 0.36 \pm 0.28$ & $0.17 \pm 0.57 \pm 0.35$ & $0.08$ \\
	\mc{3}{l}{\bf Average} & $0.27 \pm 0.24$ & $-0.23 \pm 0.34$ & $0.02$ \\
	\mc{3}{l}{\small Confidence level} & \mc{2}{c}{\small $0.68~(0.4\sigma)$} \\
		\hline
		\end{tabular*}
		\label{tab:cp_uta:uud:rhopi_q2b}
	\end{center}
\end{sidewaystable}

\begin{figure}[htbp]
  \begin{center}
    \resizebox{0.46\textwidth}{!}{
      \includegraphics{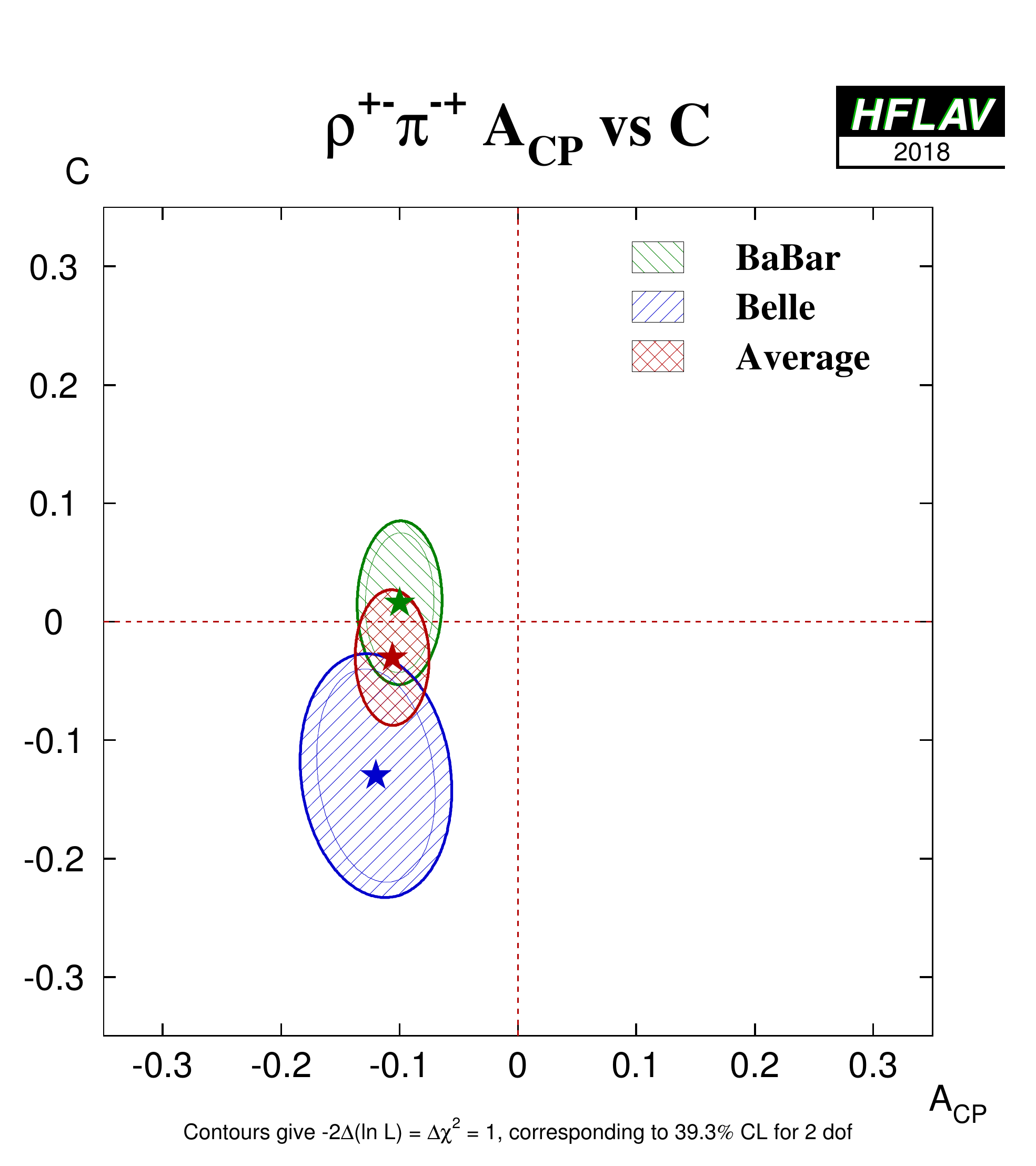}
    }
    \hfill
    \resizebox{0.46\textwidth}{!}{
      \includegraphics{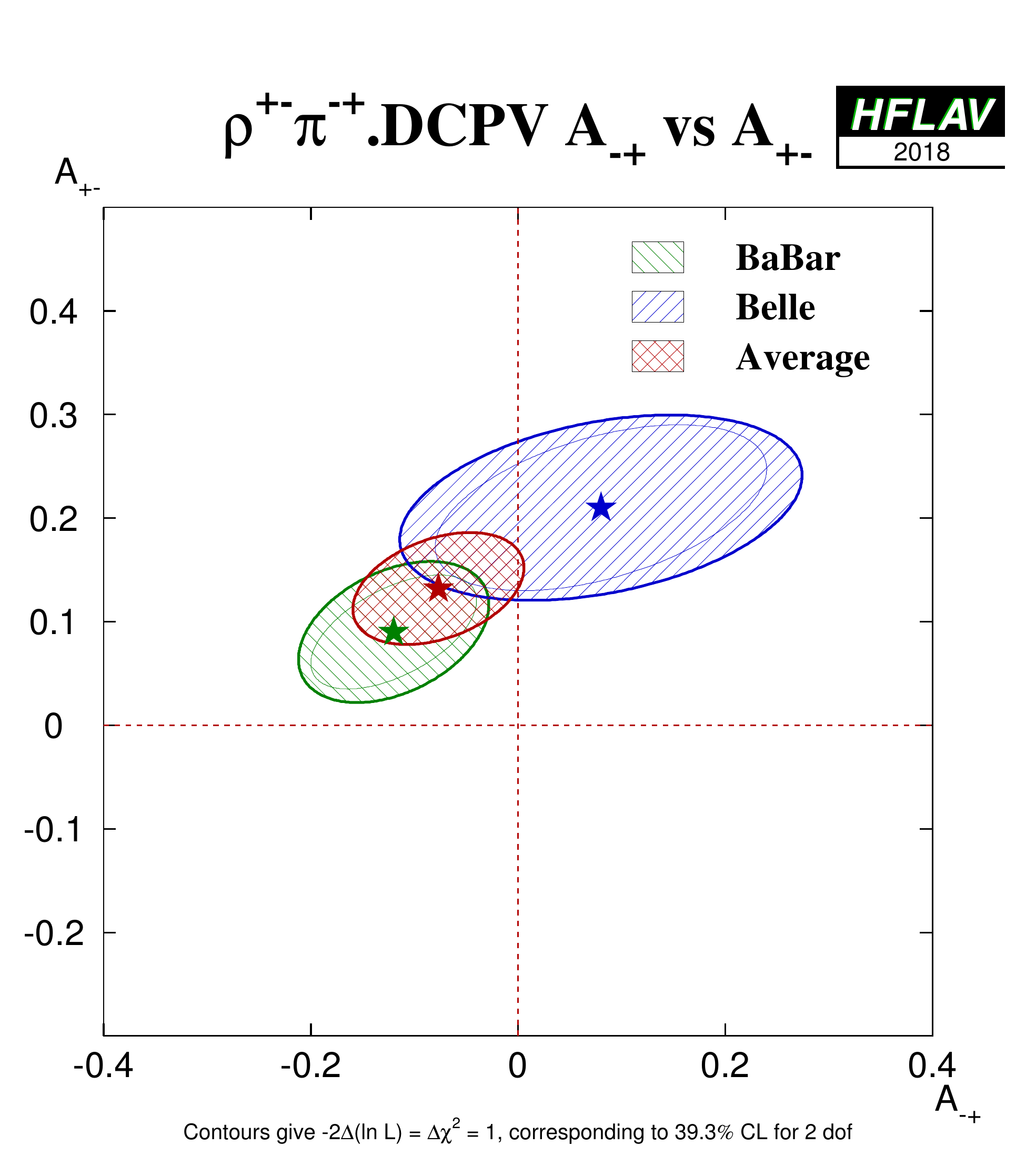}
    }
  \end{center}
  \vspace{-0.8cm}
  \caption{
    $\CP$ violation in $\Bz\to\rho^\pm\pi^\mp$ decays.
    (Left) ${\cal A}^{\rho\pi}_{\CP}$ \vs\ $C_{\rho\pi}$ space,
    (right) ${\cal A}^{-+}_{\rho\pi}$ \vs\ ${\cal A}^{+-}_{\rho\pi}$ space.
  }
  \label{fig:cp_uta:uud:rhopi-dircp}
\end{figure}

With the notation described in Sec.~\ref{sec:cp_uta:notations}
(Eq.~(\ref{eq:cp_uta:non-cp-s_and_deltas})),
the time-dependent parameters for the Q2B $\Bz \to \rho^\pm\pi^\mp$ analysis are,
in the limit of negligible penguin contributions, given by
\begin{equation}
  S_{\rho\pi} =
  \sqrt{1 - \left(\frac{\Delta C}{2}\right)^2}\sin(2\alpha)\cos(\delta)
  \ , \ \ \
  \Delta S_{\rho\pi} =
  \sqrt{1 - \left(\frac{\Delta C}{2}\right)^2}\cos(2\alpha)\sin(\delta)
\end{equation}
and $C_{\rho\pi} = {\cal A}_{\CP}^{\rho\pi} = 0$,
where $\delta=\arg(A_{-+}A^*_{+-})$ is the strong phase difference
between the $\rho^-\pi^+$ and $\rho^+\pi^-$ decay amplitudes.
In the presence of penguin contributions, there is no straightforward
interpretation of the Q2B observables in the $\Bz \to \rho^\pm\pi^\mp$ system
in terms of CKM parameters.
However, $\CP$ violation in decay may arise,
resulting in either or both of $C_{\rho\pi} \neq 0$ and ${\cal A}_{\CP}^{\rho\pi} \neq 0$.
Equivalently, $\CP$ violation in decay may be detected via deviation from zero of either of the decay-type-specific observables ${\cal A}^{+-}_{\rho\pi}$ and ${\cal A}^{-+}_{\rho\pi}$, defined in Eq.~(\ref{eq:cp_uta:non-cp-directcp}).
Results and averages for these parameters
are also given in Table~\ref{tab:cp_uta:uud:rhopi_q2b}.
Averages of $\CP$ violation parameters in $\Bz \to \rho^\pm\pi^\mp$ decays
are shown in Fig.~\ref{fig:cp_uta:uud:rhopi-dircp},
both in
${\cal A}^{\rho\pi}_{\CP}$ \vs\ $C_{\rho\pi}$ space and in
${\cal A}^{-+}_{\rho\pi}$ \vs\ ${\cal A}^{+-}_{\rho\pi}$ space.

The averages for $S_{b \to u\bar u d}$ and $C_{b \to u\bar u d}$
in $\Bz \to \pi^+\pi^-$ decays are both more than $5\sigma$ away from zero,
suggesting that both mixing-induced and $\CP$ violation in decay
are well-established in this channel.
The discrepancy between results from \babar\ and Belle that used to exist in
this channel (see, for example, Ref.~\cite{Asner:2010qj}) is no longer
apparent, and the results from LHCb are also fully consistent with other
measurements.
Some difference is, however, seen between the \babar\ and \belle\ measurements
in the $a_1^\pm\pi^\mp$ system.
The confidence level of the five-dimensional average is $0.03$,
which corresponds to a $2.1\sigma$ discrepancy.
As seen in Table~\ref{tab:cp_uta:uud}, this discrepancy is primarily in the
values of $S_{a_1\pi}$, and is not evident in the ${\cal A}^{-+}_{a_1\pi}$
\vs\ ${\cal A}^{+-}_{a_1\pi}$ projection shown in Fig.~\ref{fig:cp_uta:a1pi}.
Since there is no evidence of underestimation of uncertainties in either analysis, we do not rescale the uncertainties of the averages.

In $\Bz \to \rho^\pm\pi^\mp$ decays,
both experiments see an indication of $\CP$ violation in the
${\cal A}^{\rho\pi}_{\CP}$ parameter
(as seen in Fig.~\ref{fig:cp_uta:uud:rhopi-dircp}).
The average is more than $3\sigma$ from zero,
providing evidence of $\CP$ violation in decay in this channel.
In $\Bz \to \rho^+\rho^-$ decays there is no evidence for $\CP$ violation,
either mixing-induced or in decay.
The absence of evidence of penguin contributions in this mode leads to
strong constraints on $\alpha \equiv \phi_2$.

\mysubsubsection{Constraints on $\alpha \equiv \phi_2$}
\label{sec:cp_uta:uud:alpha}

The precision of the measured $\CP$ violation parameters in
$b \to u\bar{u}d$ transitions allows
constraints to be set on the UT angle $\alpha \equiv \phi_2$.
Constraints have been obtained with various methods:
\begin{itemize}\setlength{\itemsep}{0.5ex}
\item
  Both \babar~\cite{Lees:2012mma}
  and  \belle~\cite{Adachi:2013mae} have performed
  isospin analyses in the $\pi\pi$ system.
  \belle\ excludes $23.8^\circ < \phi_2 < 66.8^\circ$ at 68\% CL while
  \babar\ gives a confidence level interpretation for $\alpha$, and constrain
  $\alpha \in \left[ 71^\circ, 109^\circ \right]$ at 68\% CL.
  Values in the range $\left[ 23^\circ, 67^\circ \right]$ are excluded at 90\% CL.
  In both cases, only solutions in $0^\circ$--$180^\circ$ are quoted.

\item
  Both experiments have also performed isospin analyses in the $\rho\rho$
  system.
  The most recent result from \babar\ is given in an update of the
  measurements of the $B^+\to\rho^+\rho^0$ decay~\cite{Aubert:2009it}, and
  sets the constraint $\alpha = \left( 92.4 \,^{+6.0}_{-6.5}\right)^\circ$.
  The most recent result from \belle\ is given in their paper on time-dependent \CP violation parameters in $\Bz \to \rho^+\rho^-$ decays, and sets the constraint
  $\phi_2 = \left( 93.7 \pm 10.6 \right)^\circ$~\cite{Vanhoefer:2015ijw}.

\item
  The time-dependent Dalitz-plot analysis of the $\Bz \to \pi^+\pi^-\pi^0$
  decay allows a determination of $\alpha$ without input from any other
  channels.
  \babar~\cite{Lees:2013nwa} presents a scan, but not an interval, for $\alpha$, since
  their studies indicate that the scan is not statistically robust and cannot
  be interpreted in terms of 1$-$CL.
  \belle~\cite{Kusaka:2007dv,Kusaka:2007mj}
  has obtained a constraint on $\alpha$ using additional information from SU(2) relations between $B \to \rho\pi$ decay amplitudes, which can be used to constrain $\alpha$ via an isospin pentagon relation~\cite{Lipkin:1991st}.
  With this analysis,
  \belle\ obtains the constraint $\phi_2 = (83 \, ^{+12}_{-23})^\circ$.

\item
  The results from \babar\ on $\Bz \to a_1^\pm \pi^\mp$~\cite{Aubert:2006gb} can be
  combined with results from modes related by flavour symmetries ($a_1K$ and $K_1\pi$)~\cite{Gronau:2005kw}.
  This has been done by \babar~\cite{Aubert:2009ab}, resulting in the constraint
  $\alpha = \left( 79 \pm 7 \pm 11 \right)^\circ$, where the first uncertainty is from the analysis of $\Bz \to a_1^\pm \pi^\mp$ that obtains $\alpha^{\rm eff}$, and the second is due to the constraint on $\left| \alpha^{\rm eff} - \alpha \right|$.
  This approach gives a result with several ambiguous solutions;
  only the one that is consistent with other determinations of $\alpha$ and with global fits to the CKM matrix parameters is quoted here.

\item
  The CKMfitter~\cite{Charles:2004jd} and
  UTFit~\cite{Bona:2005vz} groups use the measurements
  from \belle\ and \babar\ given above
  with other branching fractions and \CP asymmetries in
  $\B\to\pi\pi$, $\pi\pi\piz$ and $\rho\rho$ modes
  to perform isospin analyses for each system,
  and to obtain combined constraints on $\alpha$.

\item
  The \babar\ and \belle\ collaborations have combined their results on $B \to \pi\pi$, $\pi\pi\piz$ and $\rho\rho$ decays to obtain~\cite{Bevan:2014iga}
  \begin{equation}
    \alpha \equiv \phi_2 = (88 \pm 5)^\circ \, .
  \end{equation}
  The above solution is
  that consistent with the Standard Model
  (there exists an ambiguous solution, shifted by $180^\circ$).
  The strongest constraint currently comes from the $B \to \rho\rho$ system. The inclusion of results from $\Bz \to a_1^\pm \pi^\mp$ does not significantly affect the average.

\item
  All results for $\alpha \equiv \phi_2$ based on isospin symmetry have a theoretical uncertainty due to possible isospin-breaking effects.
  This is expected to be small, $\lesssim 1^\circ$~\cite{Gronau:2005pq,Gardner:2005pq,Charles:2017evz}, but is hard to quantify reliably and is usually not included in the quoted uncertainty.

\end{itemize}

Note that methods based on isospin symmetry make extensive use of
measurements of branching fractions and $\CP$ asymmetries,
for which averages are reported in Sec.~\ref{sec:rare}.
Note also that each method suffers from discrete ambiguities in the solutions.
The model assumption in the $\Bz \to \pi^+\pi^-\pi^0$ analysis
helps resolve some of the multiple solutions,
and results in a single preferred value for $\alpha$ in $\left[ 0, \pi \right]$.
All the above measurements correspond to the choice that is in agreement with the global CKM fit.

Independently from the constraints on $\alpha \equiv \phi_2$ obtained by the
experiments, the results summarised in Sec.~\ref{sec:cp_uta:uud} are statistically combined to produce world average constraints on $\alpha \equiv \phi_2$.
The combination is performed with the \textsc{GammaCombo} framework~\cite{gammacombo} and follows a frequentist procedure, similar to that used by \babar\ and \belle~\cite{Bevan:2014iga}, and described in detail in Ref.~\cite{Charles:2017evz}.

The input measurements used in the combination are those listed above and are summarised in Table~\ref{tab:cp_uta:alpha:inputs}.
Additional inputs, summarised in Table~\ref{tab:cp_uta:alpha:inputs_aux}, for the branching fractions and (for $\rho\rho$) polarisation fractions, for
the relevant modes and their isospin partners are taken from Sec.~\ref{sec:rare}, whilst the ratio of $\Bp$ to $\Bz$ lifetimes is taken from Sec.~\ref{sec:life_mix}.
Individual measurements are used as inputs, rather than the HFLAV averages, in order to facilitate cross-checks and to ensure the most appropriate treatment of correlations.
A combination based on HFLAV averages gives consistent results.
Results on $\Bz \to a_1^\pm \pi^\mp$ decays are not included, as to do so requires additional theoretical assumptions, but as shown in Ref.~\cite{Bevan:2014iga} this does not significantly affect the average.

\newcommand{\rhop}{\ensuremath{\rho^{+}}\xspace}
\newcommand{\rhom}{\ensuremath{\rho^{-}}\xspace}
\newcommand{\rhopm}{\ensuremath{\rho^{\pm}}\xspace}
\newcommand{\Bdpippim}{\ensuremath{\Bd\to\pip\pim}\xspace}
\newcommand{\Bdpizpiz}{\ensuremath{\Bd\to\piz\piz}\xspace}
\newcommand{\Bpmpipmpiz}{\ensuremath{\Bpm\to\pipm\piz}\xspace}
\newcommand{\Bdpippimpiz}{\ensuremath{\Bd\to\pip\pim\piz}\xspace}
\newcommand{\Bdrhoprhom}{\ensuremath{\Bd\to\rhop\rhom}\xspace}
\newcommand{\Bdrhozrhoz}{\ensuremath{\Bd\to\rhoz\rhoz}\xspace}
\newcommand{\Bpmrhopmrhoz}{\ensuremath{\Bpm\to\rhopm\rhoz}\xspace}
\newcommand{\fL}{\ensuremath{f_{L}}\xspace}

\begin{table}[b]
  \caption{
    List of measurements used in the $\alpha$ combination.
    Results are obtained from either time-dependent (TD) \CP asymmetries of decays to \CP eigenstates or vector-vector final states, or time-integrated \CP asymmetry measurements (CP).
    Results from time-dependent asymmetries in decays to self-conjugate three-body final states (TD-Dalitz) are also used in the form of the $U$ and $I$ parameters defined in Tab.~\ref{tab:cp_uta:pipipi0:uandi}.
  }
  \label{tab:cp_uta:alpha:inputs}
  \centering
    \begin{tabular}{l l l l l}
       \hline
        $B$ decay & Method & Parameters & Experiment & Ref. \\
        \hline
        \multirow{3}{*}{\Bdpippim}      & \multirow{3}{*}{TD}      & \multirow{3}{*}{$S_{\CP}$, $C_{\CP}$} & \babar   & \cite{Lees:2012mma} \\
              & & & \belle   & \cite{Adachi:2013mae} \\
              & & & LHCb     & \cite{Aaij:2018tfw} \\
        \hline
        \multirow{2}{*}{\Bdpizpiz}      & \multirow{2}{*}{CP}      & \multirow{2}{*}{$C_{\CP}$} & \babar   & \cite{Lees:2012mma} \\
        & & & \belle   & \cite{Julius:2017jso} \\
        \hline
        \multirow{2}{*}{\Bdrhoprhom}    & \multirow{2}{*}{TD}       & \multirow{2}{*}{$S_{\CP}$, $C_{\CP}$} & \babar   & \cite{Aubert:2007nua} \\
         &  &  & \belle   & \cite{Vanhoefer:2015ijw} \\
        \hline
        \Bdrhozrhoz    & TD       & $S_{\CP}$, $C_{\CP}$ & \babar   & \cite{Aubert:2008au} \\
        \hline
        \multirow{2}{*}{\Bdpippimpiz} & \multirow{2}{*}{TD-Dalitz} & \multirow{2}{*}{$\left\{ U, I \right\}$} & \babar    & \cite{Lees:2013nwa} \\
        & & & \belle    & \cite{Kusaka:2007dv} \\
        \hline
  \end{tabular}
\end{table}

\begin{table}[b]
  \caption{List of the auxiliary inputs used in the $\alpha$ combination.}
  \label{tab:cp_uta:alpha:inputs_aux}
  \centering
  \renewcommand{\arraystretch}{1.1}
      \begin{tabular}{l l l l }
        \hline
        Particle / Decay    & Parameters                           & Source & Ref. \\
        \hline \\[-2.5ex]
        \Bp/\Bd             & $\tau(\Bp)/\tau(\Bd)$                & HFLAV      & Sec.~\ref{sec:life_mix} \\
        \hline
        \Bdpippim           & BR                                   & HFLAV      & Sec.~\ref{sec:rare} \\
        \Bdpizpiz           & BR                                   & HFLAV      & Sec.~\ref{sec:rare} \\
        \Bpmpipmpiz         & BR                                   & HFLAV      & Sec.~\ref{sec:rare} \\
        \hline
        \Bdrhoprhom         & BR, \fL                              & HFLAV      & Sec.~\ref{sec:rare} \\
        \Bdrhozrhoz         & BR, \fL                              & HFLAV      & Sec.~\ref{sec:rare} \\
        \Bpmrhopmrhoz       & BR, \fL                              & HFLAV      & Sec.~\ref{sec:rare} \\
        \hline
      \end{tabular}
\end{table}

The fit has a $\chi^2$ of 16.4 with 51 observables and 24 parameters.
Using the $\chi^2$ distribution, this corresponds to a p-value of $94.4\%$ (or $0.1\sigma$).
A coverage check with pseudoexperiments gives a p-value of $(92.9 \pm 0.3)\%$.

\begin{figure}
    \centering
    \includegraphics[width=0.6\textwidth]{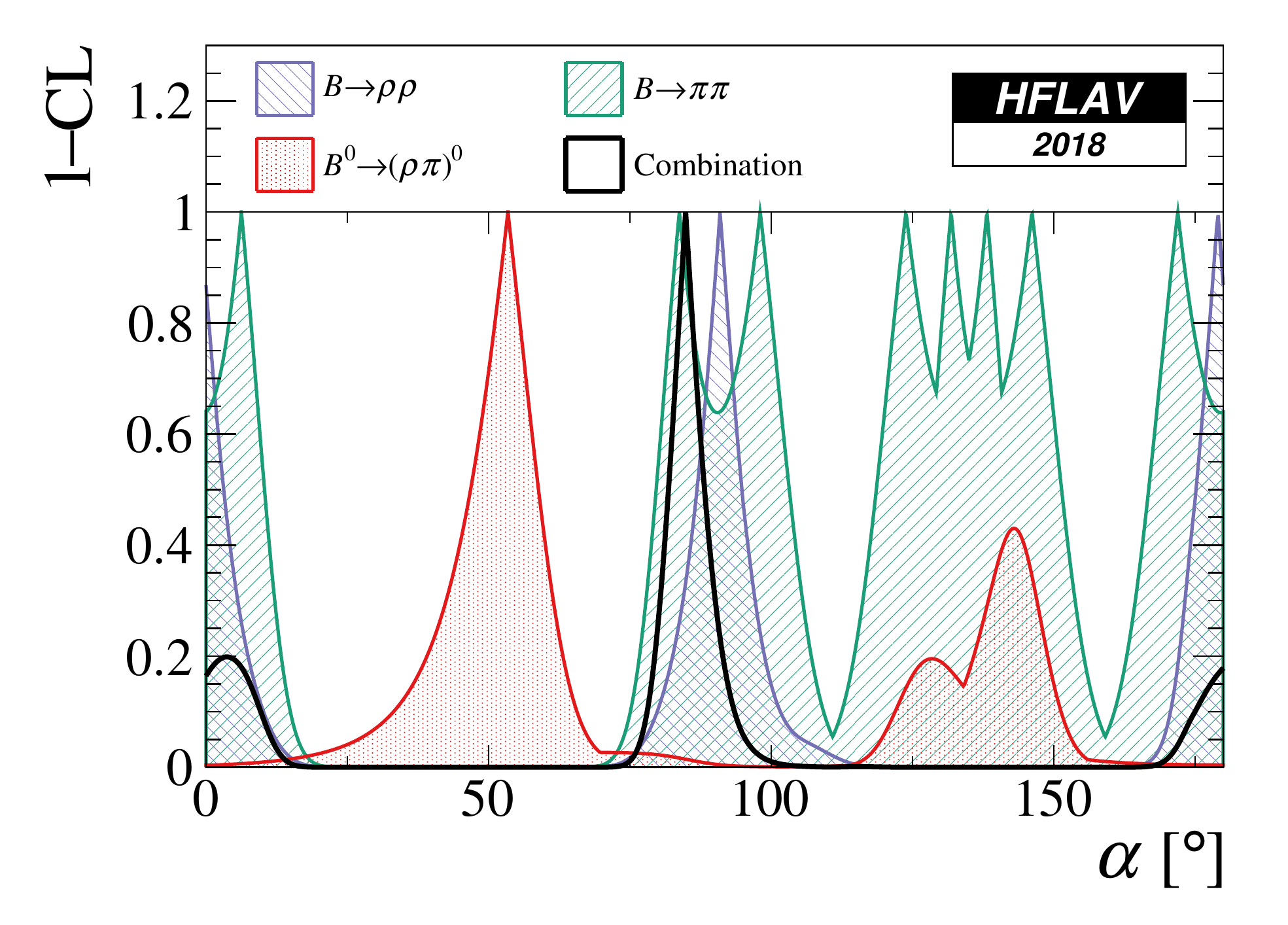}
    \caption{World average of $\alpha\equiv\phi_{2}$, in terms of 1$-$CL, split by decay mode.}
    \label{fig:cp_uta:alpha_comb}
\end{figure}

The obtained world average for the Unitarity Triangle angle $\alpha \equiv \phi_2$ is
\begin{equation}
  \alpha \equiv \phi_2 = \left( 84.9 \,^{+5.1}_{-4.5} \right)^\circ \, .
\end{equation}
An ambiguous solution also exists at $\alpha \equiv \phi_2 \Leftrightarrow \alpha + \pi \equiv \phi_2 +\pi$.
The quoted uncertainty does not include effects due to isospin-breaking.
A secondary minimum close to zero is disfavoured, as discussed in Ref.~\cite{Charles:2017evz}.
Results split by decay mode are shown in Table~\ref{tab:cp_uta:alpha_comb} and Fig.~\ref{fig:cp_uta:alpha_comb}.

 \begin{table}[t]
   \caption{Averages of $\alpha \equiv \phi_2$ split by \B meson decay mode. Only solutions consistent with the obtained world average are shown.}
   \label{tab:cp_uta:alpha_comb}
   \centering
   \renewcommand{\arraystretch}{1.1}
   \begin{tabular}{l c}
     \hline
     Decay Mode & Value \\
     \hline
     $B\to\pi\pi$        &  $(84 \,^{+21}_{-6})^\circ$ \\
                         &  $(98 \,^{+7}_{-20})^\circ$ \\
     $B\to\rho\rho$      &  $(91 \pm 6)^\circ$       \\
     $\Bd\to(\rho\pi)^0$ &  $(53 \,^{+8}_{-10})^\circ$ \\
     \hline
   \end{tabular}
 \end{table}

\afterpage{\clearpage}
\mysubsection{Time-dependent $\CP$ asymmetries in $b \to c\bar{u}d / u\bar{c}d$ transitions
}
\label{sec:cp_uta:cud}

Non-$\CP$ eigenstates such as $D^\mp\pi^\pm$, $D^{*\mp}\pi^\pm$ and $D^\mp\rho^\pm$ can be produced
in decays of $\Bz$ mesons either via Cabibbo-favoured ($b \to c$) or
doubly-Cabibbo-suppressed ($b \to u$) tree amplitudes.
Since no penguin contribution is possible,
these modes are theoretically clean.
The ratio of the magnitudes of the suppressed and favoured amplitudes, $R$,
is sufficiently small (predicted to be about $0.02$), that ${\cal O}(R^2)$ terms can be neglected,
and the sine terms give sensitivity to the combination of UT angles $2\beta+\gamma$.

\begin{figure}[htbp]
  \begin{center}
    \begin{tabular}{cc}
      \resizebox{0.46\textwidth}{!}{
        \includegraphics{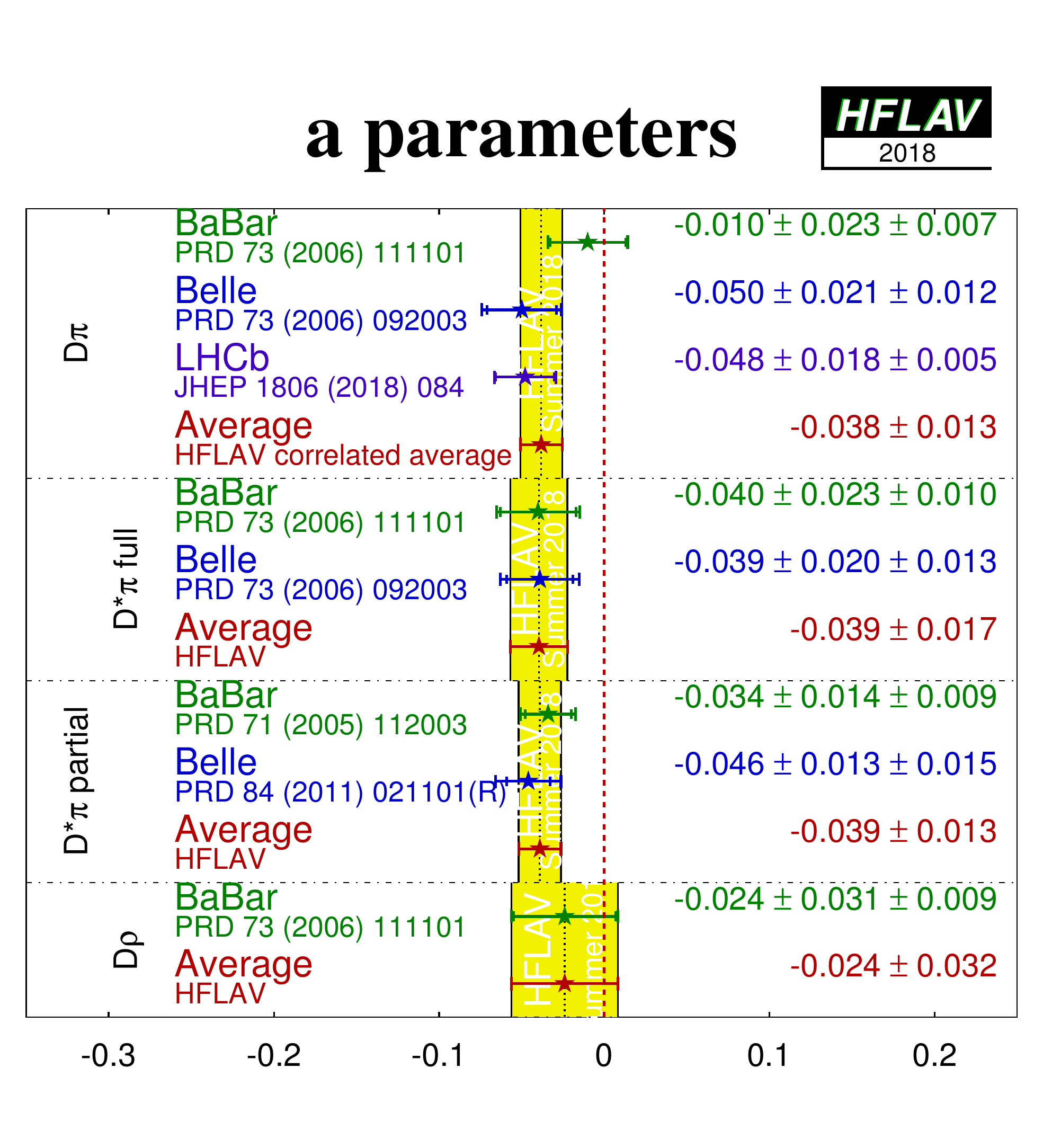}
      }
      &
      \resizebox{0.46\textwidth}{!}{
        \includegraphics{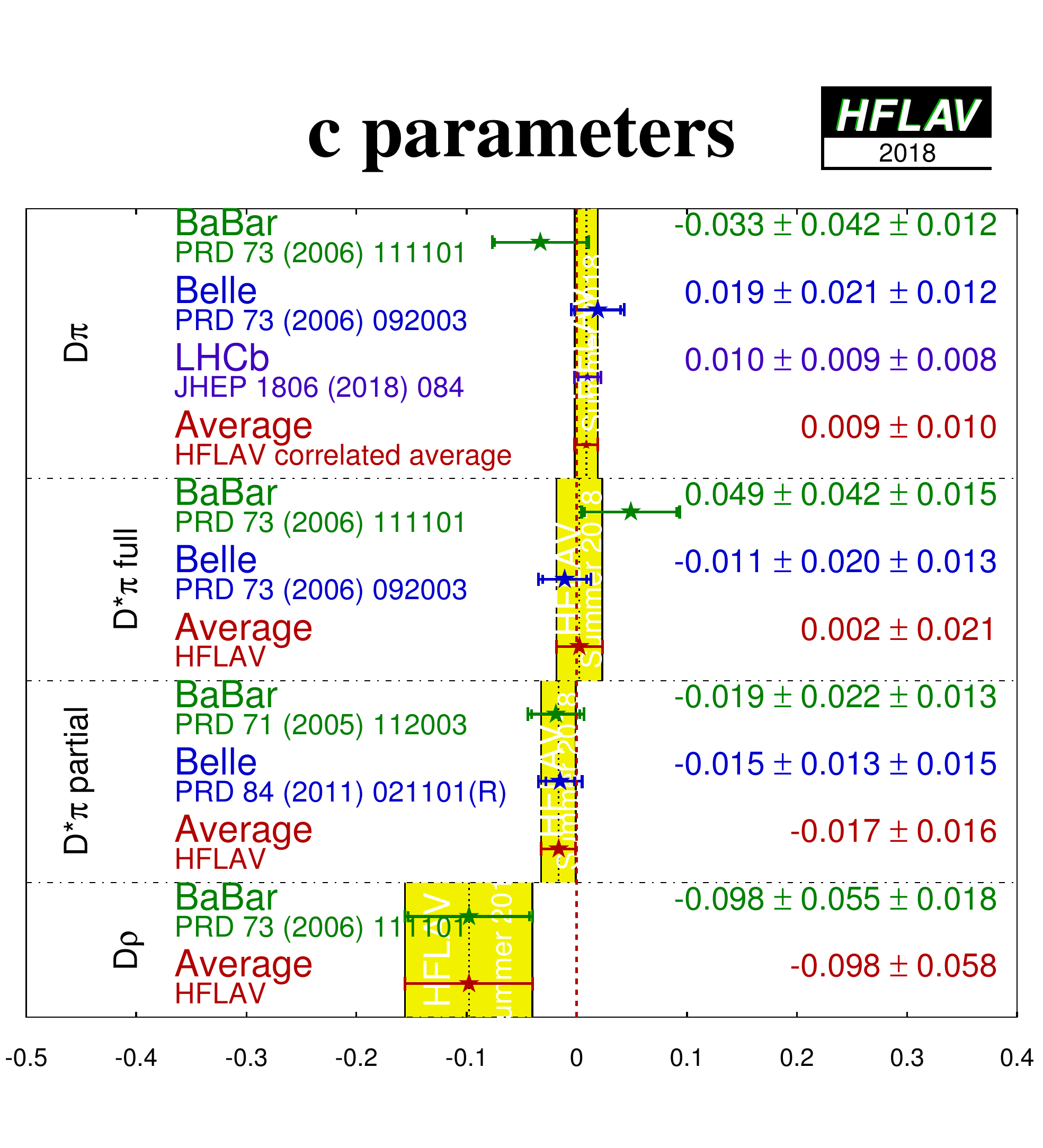}
      }
    \end{tabular}
  \end{center}
  \vspace{-0.8cm}
  \caption{
    Averages for $b \to c\bar{u}d / u\bar{c}d$ modes.
  }
  \label{fig:cp_uta:cud}
\end{figure}

As described in Sec.~\ref{sec:cp_uta:notations:non_cp:dstarpi},
the averages are given in terms of the parameters $a$ and $c$ of Eq.~(\ref{eq:cp_uta:aandc}).
$\CP$ violation would appear as $a \neq 0$.
Results for the $D^\mp\pi^\pm$ mode are available from \babar, \belle\ and LHCb,
while for $D^{*\mp}\pi^\pm$ \babar\ and \belle\ have results with both full and partial reconstruction techniques.
Results are also available from \babar\ using $D^\mp\rho^\pm$.
These results, and their averages, are listed in Table~\ref{tab:cp_uta:cud}
and shown in Fig.~\ref{fig:cp_uta:cud}.
It is notable that the average value of $a$ from $D^*\pi$ is more than
$3\sigma$ from zero, providing evidence of $\CP$ violation in this channel.

\begin{table}[htb]
	\begin{center}
		\caption{
      Averages for $b \to c\bar{u}d / u\bar{c}d$ modes.
                }
                \vspace{0.2cm}
    \resizebox{\textwidth}{!}{
                \setlength{\tabcolsep}{0.0pc}
\renewcommand{\arraystretch}{1.2}
                \begin{tabular}{@{\extracolsep{2mm}}lrcccc} \hline 
	\mc{2}{l}{Experiment} & Sample size & $a$ & $c$ & Correlation \\
	\hline
      \mc{6}{c}{$D^{\mp}\pi^{\pm}$} \\
	\babar (full rec.) & \cite{Aubert:2006tw} & $N(B\bar{B})$ = 232M & $-0.010 \pm 0.023 \pm 0.007$ & $-0.033 \pm 0.042 \pm 0.012$ & --- \\
	\belle (full rec.) & \cite{Ronga:2006hv} & $N(B\bar{B})$ = 386M & $-0.050 \pm 0.021 \pm 0.012$ & $\phantom{-}0.019 \pm 0.021 \pm 0.012$ & ---\\
        LHCb & \cite{Aaij:2018kpq} & $\int {\cal L}\,dt = 3.0 \, {\rm fb}^{-1}$ & $-0.048 \pm 0.018 \pm 0.005$ & $\phantom{-}0.010 \pm 0.009 \pm 0.008$ & $-0.46$ {\small (syst)} \\
        \mc{3}{l}{\bf Average} & $-0.038 \pm 0.013$ & $0.009 \pm 0.010$ & $-0.05$ \\
        \mc{3}{l}{\small Confidence level} & \mc{2}{c}{\small $0.56~(0.6\sigma)$} & \\
        \hline
      \mc{6}{c}{$D^{*\mp}\pi^{\pm}$} \\
      \babar (full rec.) & \cite{Aubert:2006tw} & $N(B\bar{B})$ = 232M & $-0.040 \pm 0.023 \pm 0.010$ & $\phantom{-}0.049 \pm 0.042 \pm 0.015$ \\
      \babar (partial rec.)  & \hspace{-5mm}\cite{Aubert:2005yf} & $N(B\bar{B})$ = 232M & $-0.034 \pm 0.014 \pm 0.009$ & $-0.019 \pm 0.022 \pm 0.013$ \\
      \belle (full rec.) & \cite{Ronga:2006hv} & $N(B\bar{B})$ = 386M & $-0.039 \pm 0.020 \pm 0.013$ & $-0.011 \pm 0.020 \pm 0.013$ \\
      \belle (partial rec.) & \cite{Bahinipati:2011yq} & $N(B\bar{B})$ = 657M & $-0.046 \pm 0.013 \pm 0.015$ & $-0.015 \pm 0.013 \pm 0.015$ \\
	\mc{3}{l}{\bf Average} & $-0.039 \pm 0.010$ & $-0.010 \pm 0.013$ \\
      \mc{3}{l}{\small Confidence level} & {\small $0.97~(0.03\sigma)$} & {\small $0.59~(0.6\sigma)$} \\
      \hline
      \mc{6}{c}{$D^{\mp}\rho^{\pm}$} \\
      \babar (full rec.) & \cite{Aubert:2006tw} & $N(B\bar{B})$ = 232M & $-0.024 \pm 0.031 \pm 0.009$ & $-0.098 \pm 0.055 \pm 0.018$ \\
      \hline 
    \end{tabular}
}
    \label{tab:cp_uta:cud}
  \end{center}
\end{table}

For each mode, $D\pi$, $D^*\pi$ and $D\rho$,
there are two measurements ($a$ and $c$, or $S^+$ and $S^-$)
that depend on three unknowns ($R$, $\delta$ and $2\beta+\gamma$),
of which two are different for each decay mode.
Therefore, there is not enough information to solve directly for $2\beta+\gamma$.
Constraints can be obtained if one is willing
to use theoretical input on the values of $R$ and/or $\delta$.
One popular choice is the use of SU(3) symmetry to obtain
$R$ by relating the suppressed decay mode to $\B$ decays
involving $D_s$ mesons.
More details can be found in Refs.~\cite{Dunietz:1997in,Fleischer:2003yb,Baak:2007gp,DeBruyn:2012jp,Kenzie:2016yee}.

\mysubsection{Time-dependent $\CP$ asymmetries in $b \to c\bar{u}s / u\bar{c}s$ transitions
}
\label{sec:cp_uta:cus-td}

\mysubsubsection{Time-dependent $\CP$ asymmetries in $\Bz \to D^\mp \KS \pi^\pm$}
\label{sec:cp_uta:cus-td-DKSpi}

Time-dependent analyses of transitions such as $\Bz \to D^\mp \KS \pi^\pm$ can
be used to probe $\sin(2\beta+\gamma)$ in a similar way to that discussed
above (Sec.~\ref{sec:cp_uta:cud}). Since the final state contains three
particles, a Dalitz-plot analysis is necessary to maximise the sensitivity.
\babar~\cite{Aubert:2007qe} has carried out such an analysis, finding $2\beta+\gamma = \left( 83 \pm 53 \pm 20 \right)^\circ$
(with an ambiguity $2\beta+\gamma \leftrightarrow 2\beta+\gamma+\pi$) assuming
the ratio of the $b \to u$ and $b \to c$ amplitude to be constant across the
Dalitz plot at 0.3.

\mysubsubsection{Time-dependent $\CP$ asymmetries in $\Bs \to D_s^\mp K^\pm$}
\label{sec:cp_uta:cus-td-DsK}

Time-dependent analysis of $\Bs \to D_s^\mp K^\pm$ decays can be used to determine $\gamma-2\beta_s$~\cite{Dunietz:1987bv,Aleksan:1991nh}.
Compared to the situation for $\Bz \to D^{(*)\mp} \pi^\pm$ decays discussed in Sec.~\ref{sec:cp_uta:cud}, the larger value of the ratio $R$ of the magnitudes of the suppressed and favoured amplitudes allows it to be determined from the data.
Moreover, the non-zero value of $\Delta \Gamma_s$ allows the determination of additional terms, labelled $A^{\Delta\Gamma}$ and $\bar{A}{}^{\Delta\Gamma}$, that break ambiguities in the solutions for $\gamma-2\beta_s$.

LHCb~\cite{Aaij:2017lff} has measured the time-dependent \CP violation parameters in $\Bs \to D_s^\mp K^\pm$ decays, using $3.0 \ {\rm fb}^{-1}$ of data.
The results are given in Table~\ref{tab:cp_uta:DsK}, and correspond to $3.8\,\sigma$ evidence for \CP violation in the interference between mixing and $\Bs \to D_s^\mp K^\pm$ decays.
From these results, and the world average constraint on $2\beta_s$~\cite{Amhis:2016xyh}, LHCb determine $\gamma = (128 \,^{+17}_{-22})^\circ$, $\delta_{D_sK} = (358 \,^{+13}_{-14})^\circ$ and $R_{D_sK} = 0.37 \,^{+0.10}_{-0.09}$.

\begin{table}[!htb]
	\begin{center}
		\caption{
			Results for $\Bs \to D_s^\mp K^\pm$.
		}
    \resizebox{\textwidth}{!}{
\renewcommand{\arraystretch}{1.2}
		\begin{tabular}{@{\extracolsep{2mm}}lrcccccc} \hline
	\mc{2}{l}{Experiment} & $\int {\cal L}\,dt$ & $C$ & $A^{\Delta\Gamma}$ & $\bar{A}{}^{\Delta\Gamma}$ & $S$ & $\bar{S}$ \\
	\hline
	LHCb & \cite{Aaij:2017lff} & 3 ${\rm fb}^{-1}$ & $0.73 \pm 0.14 \pm 0.05$ & $0.39 \pm 0.28 \pm 0.15$ & $0.31 \pm 0.28 \pm 0.15$ & $-0.52 \pm 0.20 \pm 0.07$ & $-0.49 \pm 0.20 \pm 0.07$ \\
		\hline
		\end{tabular}
    }
		\label{tab:cp_uta:DsK}
	\end{center}
\end{table}

\mysubsection{Rates and asymmetries in $\B \to \DorDstar K^{(*)}$ decays
}
\label{sec:cp_uta:cus}

As explained in Sec.~\ref{sec:cp_uta:notations:cus},
rates and asymmetries in $\Bp \to \DorDstar K^{(*)+}$ decays
are sensitive to $\gamma$, and have negligible theoretical uncertainty~\cite{Brod:2013sga}.
Various methods using different $\DorDstar$ final states have been used.

\mysubsubsection{$D$ decays to $\CP$ eigenstates}
\label{sec:cp_uta:cus:glw}

\begin{figure}[htbp]
  \begin{center}
    \begin{tabular}{cc}
      \resizebox{0.46\textwidth}{!}{
        \includegraphics{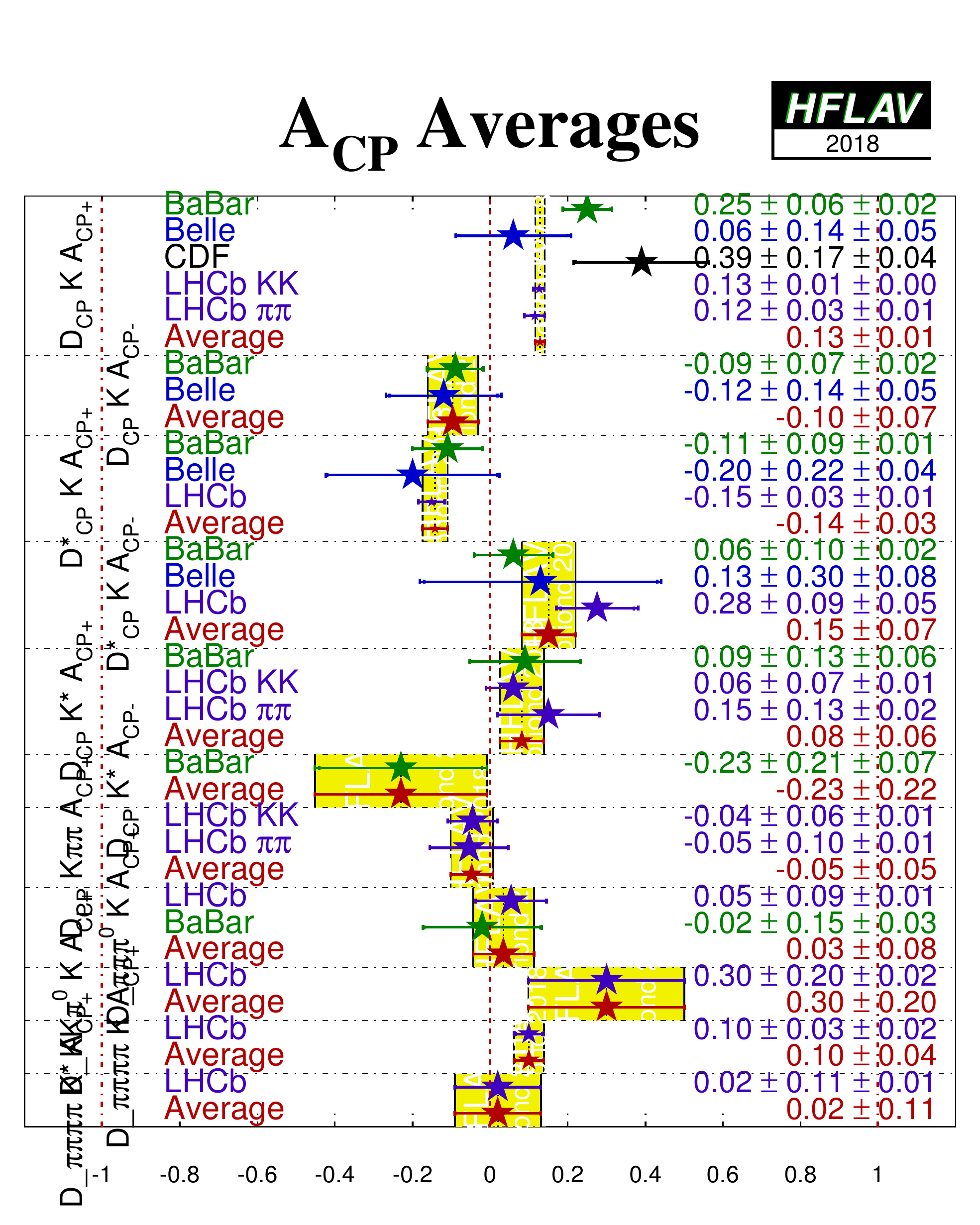}
      }
      &
      \resizebox{0.46\textwidth}{!}{
        \includegraphics{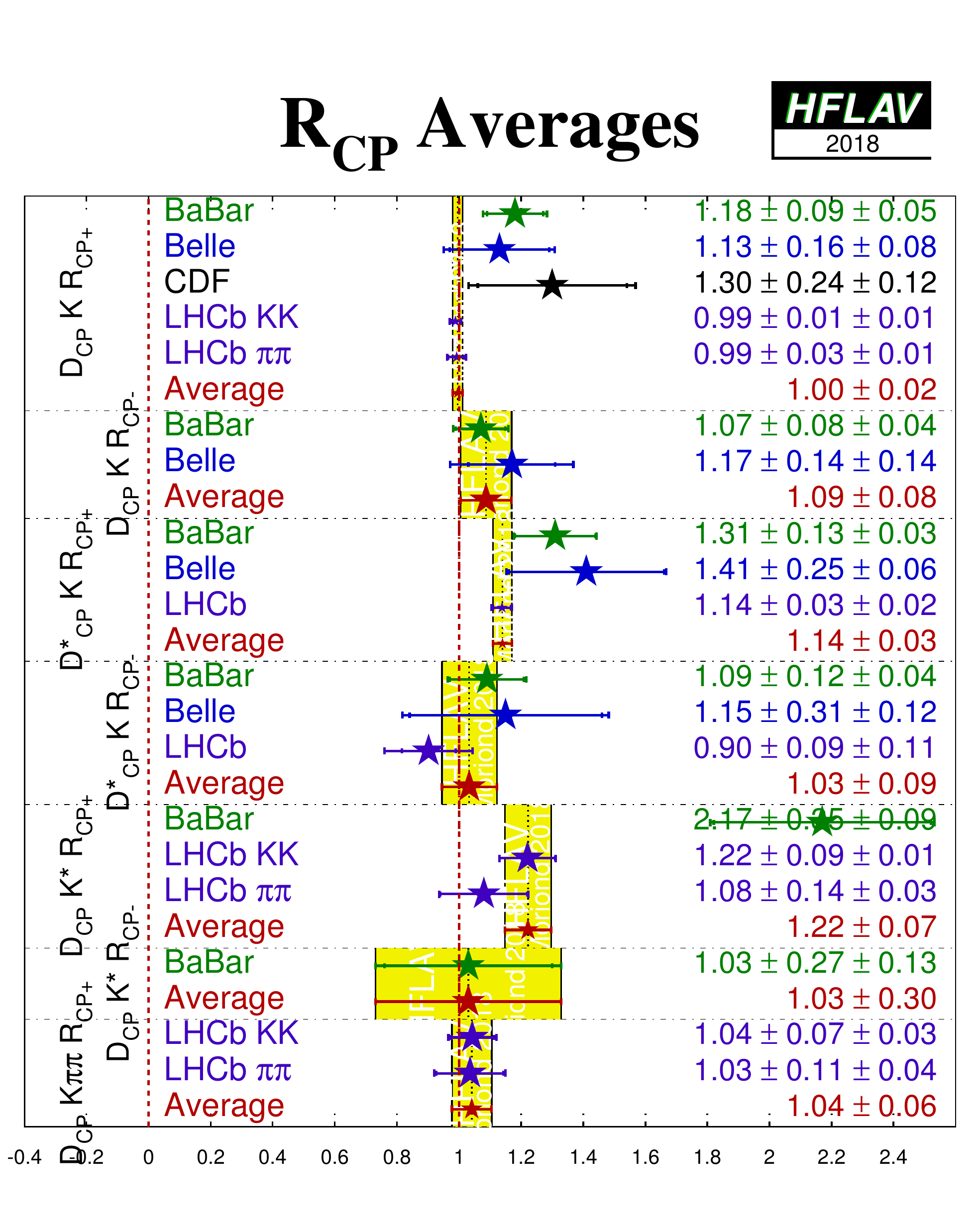}
      }
    \end{tabular}
 \end{center}
  \vspace{-0.8cm}
  \caption{
    Averages of $A_{\CP}$ and $R_{\CP}$ from GLW analyses.
  }
  \label{fig:cp_uta:cus:glw}
\end{figure}

Results are available from \babar, \belle, CDF and LHCb on GLW analyses in the
decay mode $\Bp \to D\Kp$.
All experiments use the $\CP$-even $D$ decay final states $K^+K^-$ and
$\pi^+\pi^-$; \babar\ and \belle\ in addition use the \CP-odd
decay modes $\KS\pi^0$, $\KS\omega$ and $\KS\phi$, though care is taken to
avoid statistical overlap with the $\KS K^+K^-$ sample used for Dalitz plot
analyses (see Sec.~\ref{sec:cp_uta:cus:dalitz}).
\babar\ and \belle\ also have results in the decay mode $\Bp \to \Dstar\Kp$,
using both the $\Dstar \to D\pi^0$ decay, for which $\CP(\Dstar) = \CP(D)$,
and the $\Dstar \to D\gamma$ decay, for which $\CP(\Dstar) = -\CP(D)$.
LHCb also has results in the $\Bp \to \Dstar\Kp$ decay mode, exploiting a partial reconstruction technique in which the $\piz$ or $\gamma$ produced in the $\Dstar$ decay is not explicitly reconstructed.
Results obtained with this technique have significant correlations, and therefore a correlated average is performed for the $\Bp \to \Dstar\Kp$ observables.
In addition, \babar\ and LHCb have results in the decay mode $\Bp \to D\Kstarp$,
and LHCb has results in the decay mode $\Bp \to D\Kp\pi^+\pi^-$.
In many cases LHCb presents results separately for the cases of $D$ decay to $K^+K^-$ and $\pipi$ to allow for possible effects related to $\Dz$--$\Dzb$ mixing and \CP violation in charm decays~\cite{Rama:2013voa}, which, however, are known to be small and are neglected in our averages.
These separate results are presented together with their combination, as provided in the LHCb publications, where possible.
The results and averages are given in Table~\ref{tab:cp_uta:cus:glw}
and shown in Fig.~\ref{fig:cp_uta:cus:glw}.
LHCb has performed a GLW analysis using the $B^0 \to DK^{*0}$ decay with the \CP-even $D \to K^+K^-$ and $D \to \pi^+\pi^-$ channels, which are also included in Table~\ref{tab:cp_uta:cus:glw}.

\begin{table}[htb]
	\begin{center}
		\caption{
                        Averages from GLW analyses of $b \to c\bar{u}s / u\bar{c}s$ modes.
                        The sample size is given in terms of number of $B\bar{B}$ pairs, $N(B\bar{B})$, for the $\epem$ $B$ factory experiments \babar\ and \belle, and in terms of integrated luminosity, $\int {\cal L}\,dt$, for the hadron collider experiments CDF and LHCb.
                }
                \vspace{0.2cm}
    \resizebox{\textwidth}{!}{
 \renewcommand{\arraystretch}{1.2}
     \setlength{\tabcolsep}{0.0pc}
      \begin{tabular}{@{\extracolsep{2mm}}lrccccc} \hline 
        \mc{2}{l}{Experiment} & Sample size & $A_{\CP+}$ & $A_{\CP-}$ & $R_{\CP+}$ & $R_{\CP-}$ \\
        & & \hspace{-5mm}$N(B\bar{B})$ or $\int {\cal L}\,dt$\hspace{-5mm} \\
        \hline
        \mc{7}{c}{$B^+ \to D_{\CP} K^+$} \\
	\babar & \cite{delAmoSanchez:2010ji} & 467M & $0.25 \pm 0.06 \pm 0.02$ & $-0.09 \pm 0.07 \pm 0.02$ & $1.18 \pm 0.09 \pm 0.05$ & $1.07 \pm 0.08 \pm 0.04$ \\
	\belle & \cite{Abe:2006hc} & 275M & $0.06 \pm 0.14 \pm 0.05$ & $-0.12 \pm 0.14 \pm 0.05$ & $1.13 \pm 0.16 \pm 0.08$ & $1.17 \pm 0.14 \pm 0.14$ \\
	CDF & \cite{Aaltonen:2009hz} & $1 \, {\rm fb}^{-1}$ & $0.39 \pm 0.17 \pm 0.04$ & \textendash{} & $1.30 \pm 0.24 \pm 0.12$ &  \textendash{} \\
	LHCb $KK$ & \cite{Aaij:2017ryw} & $5 \, {\rm fb}^{-1}$ & $0.126 \pm 0.014 \pm 0.002$ &  \textendash{} & $0.988 \pm 0.015 \pm 0.011$ &  \textendash{} \\
	LHCb $\pi\pi$ & \cite{Aaij:2017ryw} & $5 \, {\rm fb}^{-1}$ & $0.115 \pm 0.025 \pm 0.007$ &  \textendash{} & $0.992 \pm 0.027 \pm 0.015$ &  \textendash{} \\
	LHCb average & \cite{Aaij:2017ryw} & $5 \, {\rm fb}^{-1}$ & $0.124 \pm 0.012 \pm 0.002$ & \textendash{} & $0.989 \pm 0.013 \pm 0.010$ & \textendash{} \\
	\mc{3}{l}{\bf Average} & $0.129 \pm 0.012$ & $-0.10 \pm 0.07$ & $0.996 \pm 0.016$ & $1.09 \pm 0.08$ \\
	\mc{3}{l}{\small Confidence level} & {\small $0.17~(1.4\sigma)$} & {\small $0.86~(0.2\sigma)$} & {\small $0.26~(1.1\sigma)$} & {\small $0.65~(0.5\sigma)$} \\
		\hline
        \mc{7}{c}{$B^+ \to \Dstar_{\CP} K^+$} \\
	\babar & \cite{:2008jd} & 383M & $-0.11 \pm 0.09 \pm 0.01$ & $0.06 \pm 0.10 \pm 0.02$ & $1.31 \pm 0.13 \pm 0.03$ & $1.09 \pm 0.12 \pm 0.04$ \\
	\belle & \cite{Abe:2006hc} & 275M & $-0.20 \pm 0.22 \pm 0.04$ & $0.13 \pm 0.30 \pm 0.08$ & $1.41 \pm 0.25 \pm 0.06$ & $1.15 \pm 0.31 \pm 0.12$ \\
	LHCb & \cite{Aaij:2017ryw} & $5 \, {\rm fb}^{-1}$ & $-0.151 \pm 0.033 \pm 0.011$ & $0.276 \pm 0.094 \pm 0.047$ & $1.138 \pm 0.029 \pm 0.016$ & $0.902 \pm 0.087 \pm 0.112$ \\ 
	\mc{3}{l}{\bf Average} & $-0.142 \pm 0.032$ & $0.15 \pm 0.07$ & $1.140 \pm 0.031$ & $1.03 \pm 0.09$ \\ 
	\mc{3}{l}{\small Confidence level} & \mc{4}{c}{\small $0.67~(0.4\sigma)$} \\
		\hline
        \mc{7}{c}{$B^+ \to D_{\CP} K^{*+}$} \\
	\babar & \cite{Aubert:2009yw} & 379M & $0.09 \pm 0.13 \pm 0.06$ & $-0.23 \pm 0.21 \pm 0.07$ & $2.17 \pm 0.35 \pm 0.09$ & $1.03 \pm 0.27 \pm 0.13$ \\
        LHCb $KK$ & \cite{Aaij:2017glf} & $4.8 \, {\rm fb}^{-1}$ & $0.06 \pm 0.07 \pm 0.01$ & \textendash{} & $1.22 \pm 0.09 \pm 0.01$ & \textendash{} \\
	LHCb $\pi\pi$ & \cite{Aaij:2017glf} & $4.8 \, {\rm fb}^{-1}$ & $0.15 \pm 0.13 \pm 0.02$ & \textendash{} & $1.08 \pm 0.14 \pm 0.03$ & \textendash{} \\
        LHCb average & \cite{Aaij:2017glf} & $4.8 \, {\rm fb}^{-1}$ & $0.08 \pm 0.06 \pm 0.01$ & \textendash{} & $1.18 \pm 0.08 \pm 0.02$ & \textendash{} \\
	\mc{3}{l}{\bf Average} & $0.08 \pm 0.06$ & $-0.23 \pm 0.22$ & $1.22 \pm 0.07$ & $1.03 \pm 0.30$ \\
	\mc{3}{l}{\small Confidence level} & {\small $0.83~(0.2\sigma)$} & & {\small $0.02~(2.3\sigma)$} \\
		\hline
        \mc{7}{c}{$B^+ \to D_{\CP} K^+\pi^+\pi^-$} \\
	LHCb $KK$ & \cite{Aaij:2015ina} & $3 \, {\rm fb}^{-1}$ & $-0.045 \pm 0.064 \pm 0.011$ & \textendash{} & $1.043 \pm 0.069 \pm 0.034$ & \textendash{} \\
	LHCb $\pi\pi$ & \cite{Aaij:2015ina} & $3 \, {\rm fb}^{-1}$ & $-0.054 \pm 0.101 \pm 0.011$ & \textendash{} & $1.035 \pm 0.108 \pm 0.038$ & \textendash{} \\
        LHCb average & \cite{Aaij:2015ina} & $3 \, {\rm fb}^{-1}$ & $-0.048 \pm 0.055$ & \textendash{} & $1.040 \pm 0.064$ & \textendash{} \\
	\hline
        \mc{7}{c}{$B^0 \to D_{\CP} K^{*0}$} \\
        LHCb $KK$ & \cite{Aaij:2014eha} & $3 \, {\rm fb}^{-1}$ & $-0.20 \pm 0.15 \pm 0.02$ & \textendash{} & $1.05 \,^{+0.17}_{-0.15} \pm 0.04$ & \textendash{} \\
        LHCb $\pi\pi$ & \cite{Aaij:2014eha} & $3 \, {\rm fb}^{-1}$ & $-0.09 \pm 0.22 \pm 0.02$ & \textendash{} & $1.21 \,^{+0.28}_{-0.25} \pm 0.05$ & \textendash{} \\
        \mc{3}{l}{\bf Average} & $-0.16 \pm 0.12$ & \textendash{} & $1.10 \pm 0.14$ & \textendash{} \\
                \hline

      \end{tabular}
    }
    \label{tab:cp_uta:cus:glw}
	\end{center}
\end{table}

As pointed out in Refs.~\cite{Gershon:2008pe,Gershon:2009qc}, a Dalitz plot analysis of $B^0 \to DK^+\pi^-$ decays provides more sensitivity to $\gamma \equiv \phi_3$ than the Q2B $DK^{*0}$ approach.
The analysis provides direct sensitivity to the hadronic parameters $r_B$ and $\delta_B$ associated with the $B^0 \to DK^{*0}$ decay amplitudes, rather than effective hadronic parameters averaged over the $K^{*0}$ selection window as in the Q2B case.

Such an analysis has been performed by LHCb.
A simultaneous fit is performed to the $B^0 \to DK^+\pi^-$ Dalitz plots with the neutral $D$ meson reconstructed in the $K^+\pi^-$, $K^+K^-$ and $\pi^+\pi^-$ final states.
The reported results in Table~\ref{tab:cp_uta:cus:DKpiDalitz} are for the Cartesian parameters, defined in Eq.~(\ref{eq:cp_uta:cartesian}) associated with the $B^0 \to DK^*(892)^0$ decay.
Note that, since the measurements use overlapping data samples, these results cannot be combined with the LHCb results for GLW observables in $B^0 \to DK^*(892)^0$ decays reported in Table~\ref{tab:cp_uta:cus:glw}.

\begin{table}[!htb]
	\begin{center}
		\caption{
			Results from Dalitz plot analysis of $\B^0 \to DK^+\pi^-$ decays with $D \to \Kp\Km$ and $\pip\pim$.
		}
		\vspace{0.2cm}
		\setlength{\tabcolsep}{0.0pc}
    \resizebox{\textwidth}{!}{
\renewcommand{\arraystretch}{1.2}
		\begin{tabular}{@{\extracolsep{2mm}}lrccccc} \hline
	\mc{2}{l}{Experiment} & $\int {\cal L}\,dt$ & $x_+$ & $y_+$ & $x_-$ & $y_-$ \\
	\hline
	LHCb & \cite{Aaij:2016bqv} & $3 \, {\rm fb}^{-1}$ & $0.04 \pm 0.16 \pm 0.11$ & $-0.47 \pm 0.28 \pm 0.22$ & $-0.02 \pm 0.13 \pm 0.14$ & $-0.35 \pm 0.26 \pm 0.41$ \\
		\hline
		\end{tabular}
}
		\label{tab:cp_uta:cus:DKpiDalitz}
	\end{center}
\end{table}

LHCb uses these results to obtain confidence levels for $\gamma$, $r_B(DK^{*0})$ and $\delta_B(DK^{*0})$.
In addition, results are reported for the hadronic parameters needed to relate these results to Q2B measurements of $B^0 \to DK^*(892)^0$ decays, where a selection window of $m(K^+\pi^-)$ within $50~\mevcc$ of the pole mass and helicity angle satisfying $\left|\cos(\theta_{K^{*0}})\right|>0.4$ is assumed.
These parameters are the coherence factor $\kappa$, the ratio of Q2B and amplitude level $r_B$ values, $\bar{R}_B = \bar{r}_B/r_B$, and the difference between Q2B and amplitude level $\delta_B$ values, $\Delta \bar{\delta}_B = \bar{\delta}_B-\delta_B$.
LHCb~\cite{Aaij:2016bqv} obtains
\begin{equation}
  \kappa = 0.958 \,^{+0.005}_{-0.010}\,^{+0.002}_{-0.045}\,,\quad
  \bar{R}_B = 1.02 \,^{+0.03}_{-0.01} \pm 0.06\,,\quad
  \Delta \bar{\delta}_B = 0.02 \,^{+0.03}_{-0.02} \pm 0.11\,.
\end{equation}

\mysubsubsection{$D$ decays to quasi-$\CP$ eigenstates}
\label{sec:cp_uta:cus:quasi-glw}

As discussed in Sec.~\ref{sec:cp_uta:notations:cus}, if a multibody neutral $D$ meson decay can be shown to be dominated by one $\CP$ eigenstate, it can be used in a ``GLW-like'' (sometimes called ``quasi-GLW'') analysis~\cite{Nayak:2014tea}.
The same observables $R_{\CP}$, $A_{\CP}$ as for the GLW case are measured, but an additional factor of $(2F_+-1)$, where $F_+$ is the fractional $\CP$-even content, enters the expressions relating these observables to $\gamma \equiv \phi_3$.
The $F_+$ factors have been measured using CLEO-c data to be $F_+(\pi^+\pi^-\pi^0) = 0.973 \pm 0.017$, $F_+(K^+K^-\pi^0) = 0.732 \pm 0.055$, $F_+(\pi^+\pi^-\pi^+\pi^-) = 0.737 \pm 0.028$~\cite{Malde:2015mha}.

The GLW-like observables for $\Bp \to D\Kp$ with $D\to\pi^+\pi^-\pi^0$, $K^+K^-\pi^0$ and $D\to\pi^+\pi^-\pi^+\pi^-$ have been measured by LHCb.
The $A_{\rm qGLW}$ observable for $\Bp \to D\Kp$ with $D\to\pi^+\pi^-\pi^0$ was measured in an earlier analysis by \babar, from which additional observables, discussed in Sec.~\ref{sec:cp_uta:notations:cus} and reported in Table~\ref{tab:cp_uta:cus:dalitz} below, were reported.
The observables for $\Bp \to D\Kstarp$ with $D\to\pi^+\pi^-\pi^+\pi^-$ have also been measured by LHCb.
The results are given in Table~\ref{tab:cp_uta:cus:glwLike}.

\begin{table}[!htb]
	\begin{center}
		\caption{
                        Averages from GLW-like analyses of $b \to c\bar{u}s / u\bar{c}s$ modes.
		}
		\vspace{0.2cm}
		\setlength{\tabcolsep}{0.0pc}
\renewcommand{\arraystretch}{1.1}
		\begin{tabular*}{\textwidth}{@{\extracolsep{\fill}}lrccc} \hline
	\mc{2}{l}{Experiment} & Sample size & $A_{\rm qGLW}$ & $R_{\rm qGLW}$ \\
	\hline
        \mc{5}{c}{$D_{\pi^+\pi^-\pi^0} K^+$} \\
	LHCb & \cite{Aaij:2015jna} & $\int {\cal L}\,dt = 3 \, {\rm fb}^{-1}$ & $0.05 \pm 0.09 \pm 0.01$ & $0.98 \pm 0.11 \pm 0.05$ \\
	\babar & \cite{Aubert:2007ii} & $N(B\bar{B}) =$ 324M & $-0.02 \pm 0.15 \pm 0.03$ &  \textendash{} \\
	\mc{3}{l}{\bf Average} & $0.03 \pm 0.08$ & $0.98 \pm 0.12$ \\
	\mc{3}{l}{\small Confidence level} & {\small $0.68~(0.4\sigma)$} & \textendash{} \\
		\hline
        \mc{5}{c}{$D_{K^+K^-\pi^0} K^+$} \\
	LHCb & \cite{Aaij:2015jna} & $\int {\cal L}\,dt = 3 \, {\rm fb}^{-1}$ & $0.30 \pm 0.20 \pm 0.02$ & $0.95 \pm 0.22 \pm 0.04$ \\
		\hline
        \mc{5}{c}{$D_{\pi^+\pi^-\pi^+\pi^-} K^+$} \\
	LHCb & \cite{Aaij:2016oso} & $\int {\cal L}\,dt = 3 \, {\rm fb}^{-1}$ & $0.10 \pm 0.03 \pm 0.02$ & $0.97 \pm 0.04 \pm 0.02$ \\
		\hline
        \mc{5}{c}{$D_{\pi^+\pi^-\pi^+\pi^-} K^{*+}$} \\
	LHCb & \cite{Aaij:2017glf} & $\int {\cal L}\,dt = 4.8 \, {\rm fb}^{-1}$ & $0.02 \pm 0.11 \pm 0.01$ & $1.08 \pm 0.13 \pm 0.03$ \\
                \hline
		\end{tabular*}
		\label{tab:cp_uta:cus:glwLike}
	\end{center}
\end{table}

\mysubsubsection{$D$ decays to suppressed final states}
\label{sec:cp_uta:cus:ads}

\begin{figure}[htbp]
  \begin{center}
    \begin{tabular}{cc}
      \resizebox{0.46\textwidth}{!}{
        \includegraphics{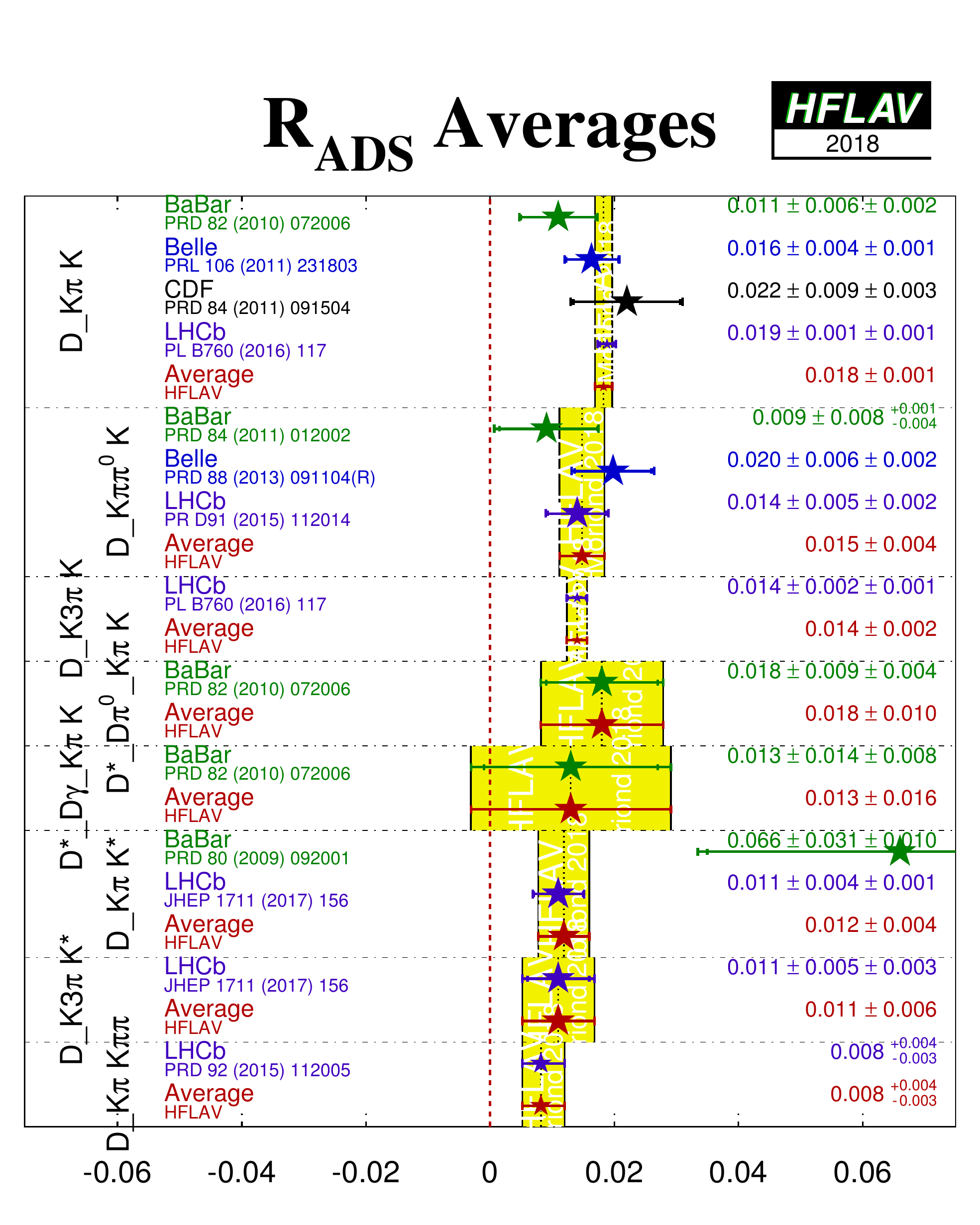}
      }
      &
      \resizebox{0.46\textwidth}{!}{
        \includegraphics{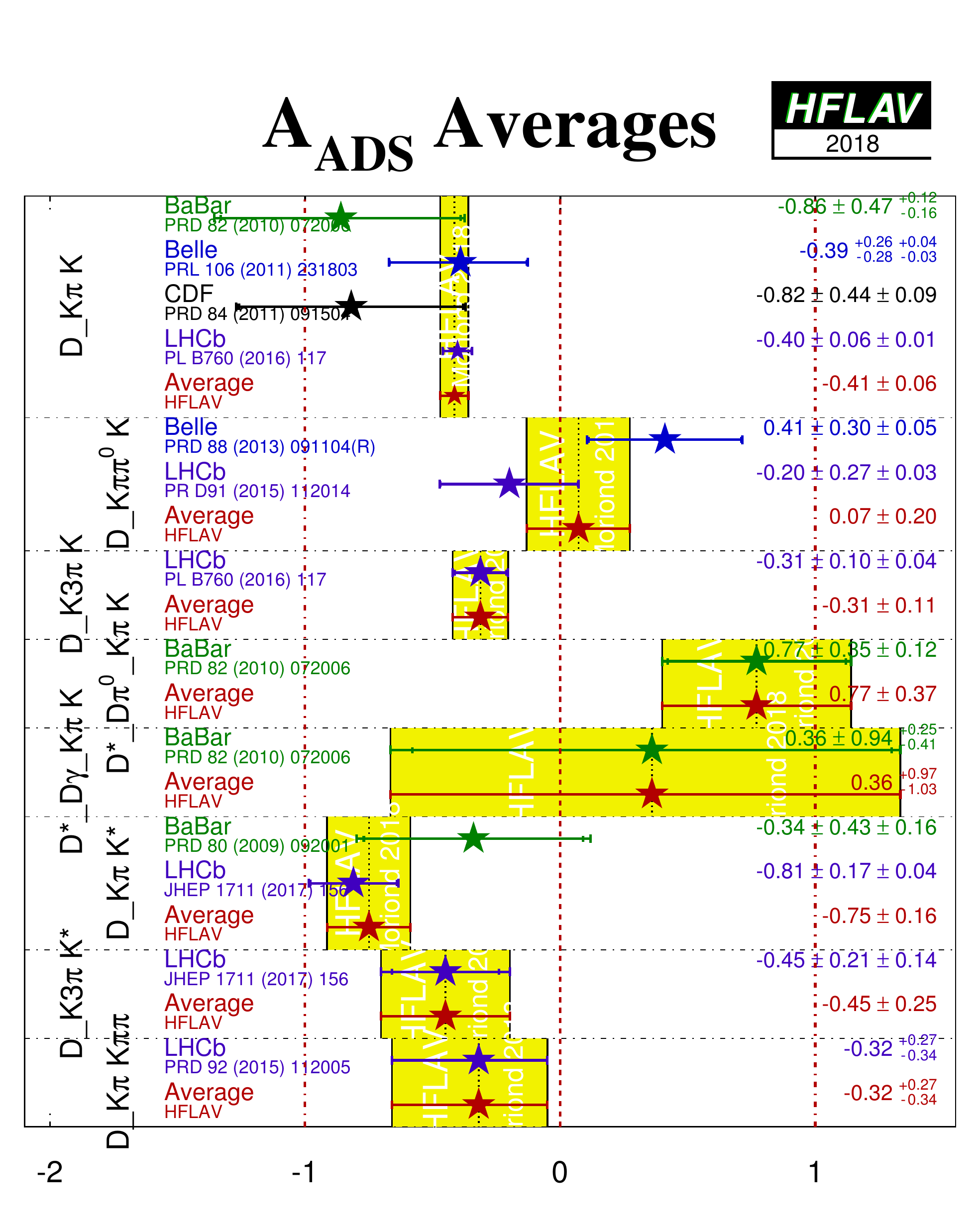}
      }
    \end{tabular}
  \end{center}
  \vspace{-0.8cm}
  \caption{
    Averages of $R_{\rm ADS}$ and $A_{\rm ADS}$ for $B \to D^{(*)}K^{(*)}$ decays.
  }
  \label{fig:cp_uta:cus:ads}
\end{figure}

For ADS analyses, all of \babar, \belle, CDF and LHCb have studied the modes
$\Bp \to D\Kp$ and $\Bp \to D\pip$.
\babar\ has also analysed the $\Bp \to \Dstar\Kp$ mode.
There is an effective shift of $\pi$ in the strong phase difference between
the cases that the $\Dstar$ is reconstructed as $D\pi^0$ and
$D\gamma$~\cite{Bondar:2004bi}, therefore these modes are studied separately.
In addition, \babar\ has studied the $\Bp \to D\Kstarp$ mode,
where $\Kstarp$ is reconstructed as $\KS\pip$,
and LHCb has studied the $\Bp \to D\Kp\pip\pim$ mode.
In all the above cases the suppressed decay $D \to K^-\pi^+$ has been used.
\babar, \belle\ and LHCb also have results using $\Bp \to D\Kp$ with $D \to K^-\pi^+\pi^0$,
while LHCb has results using $\Bp \to D\Kp$ with $D \to K^-\pi^+\pi^+\pi^-$.
The results and averages are given in Table~\ref{tab:cp_uta:cus:ads}
and shown in Fig.~\ref{fig:cp_uta:cus:ads}.

\begin{figure}[htbp]
  \begin{center}
    \begin{tabular}{cc}
      \resizebox{0.46\textwidth}{!}{
        \includegraphics{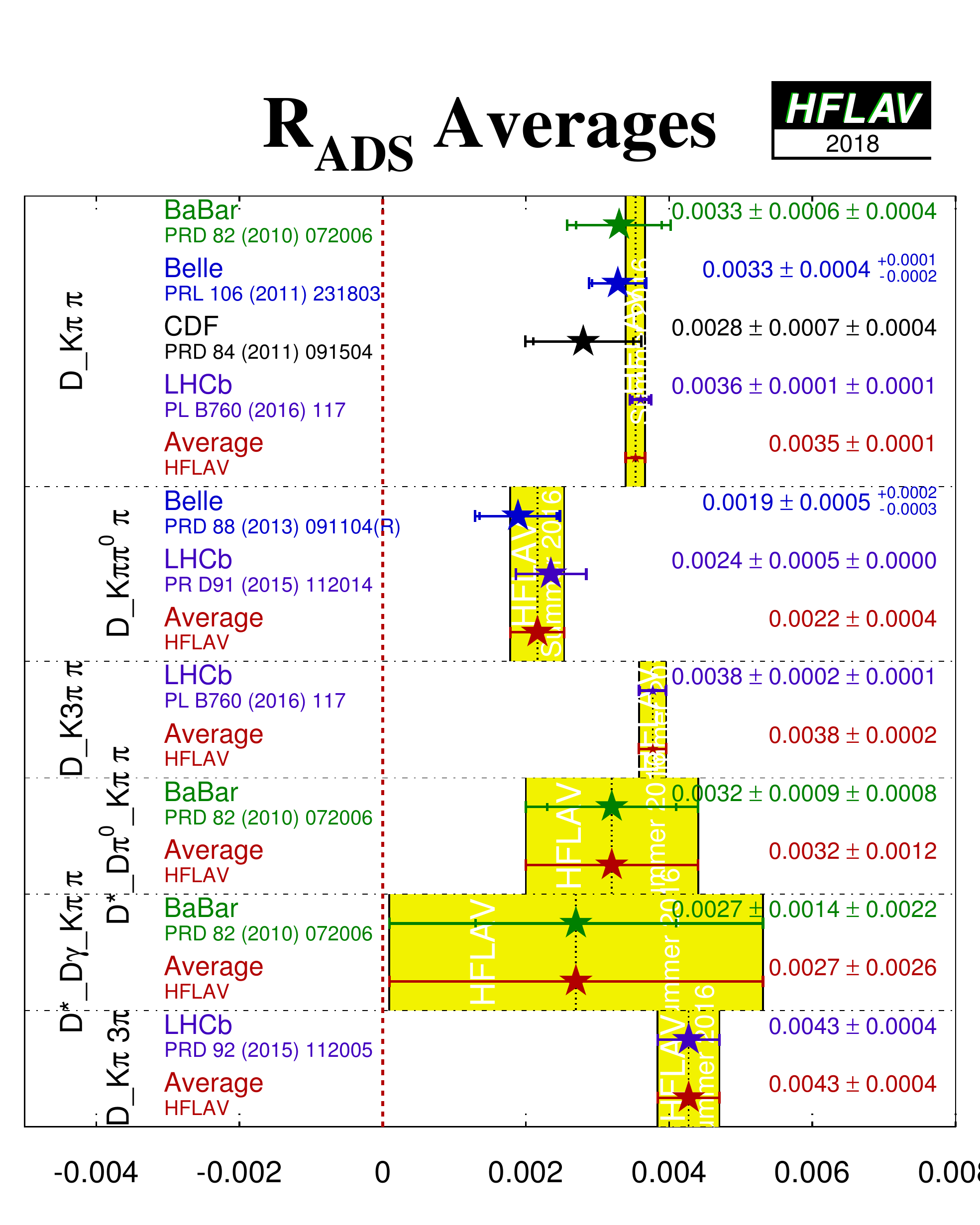}
      }
      &
      \resizebox{0.46\textwidth}{!}{
        \includegraphics{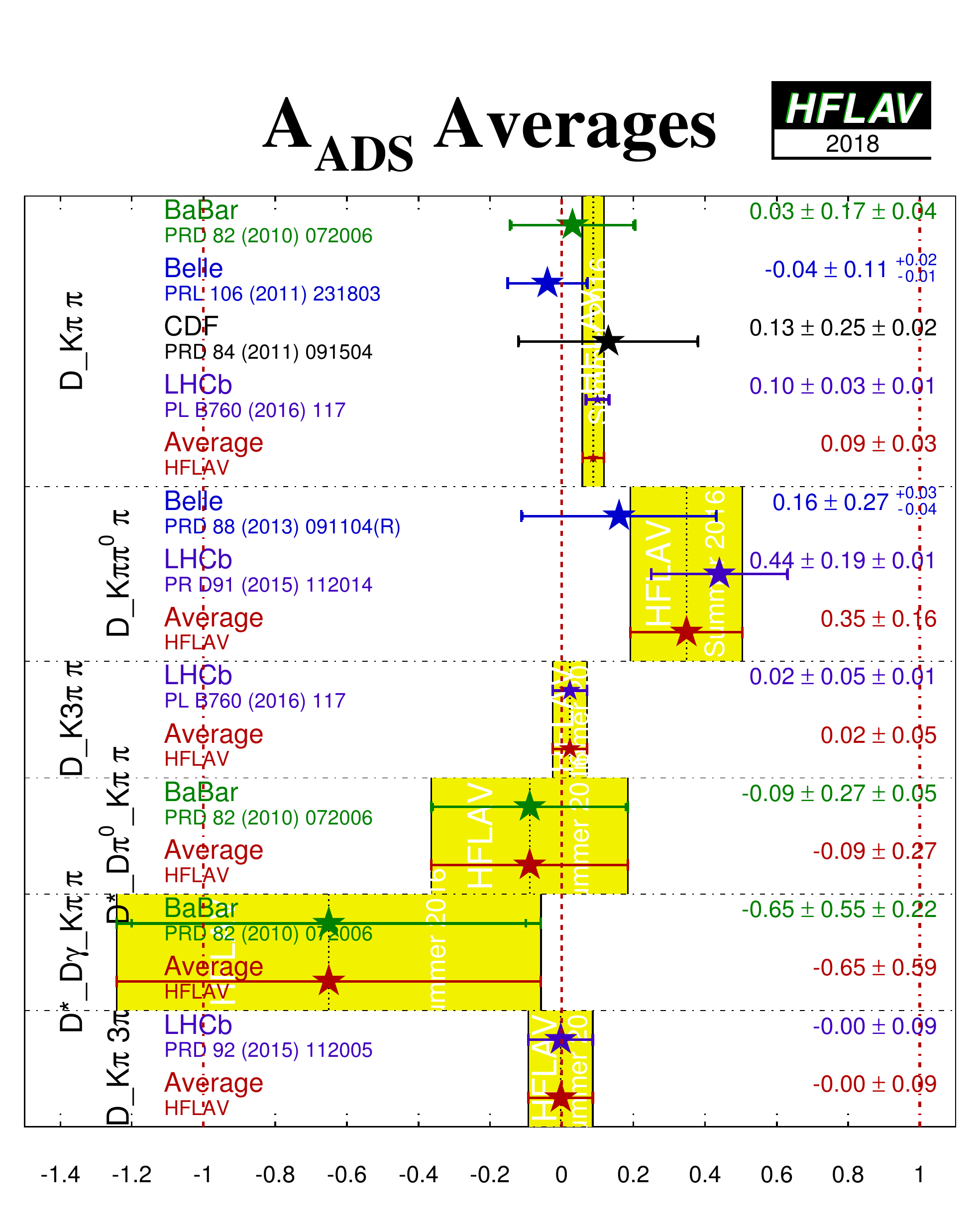}
      }
    \end{tabular}
  \end{center}
  \vspace{-0.8cm}
  \caption{
    Averages of $R_{\rm ADS}$ and $A_{\rm ADS}$ for $B \to D^{(*)}\pi$ decays.
  }
  \label{fig:cp_uta:cus:ads-Dpi}
\end{figure}

Similar phenomenology as for $B \to DK$ decays holds for $B \to D\pi$ decays, although in this case the interference is between $b \to c\bar{u}d$ and $b \to u\bar{c}d$ transitions, and the ratio of suppressed to favoured amplitudes is expected to be much smaller, ${\cal O}(1\%)$.
For most $D$ meson final states this implies that the interference effect is too small to be of interest, but in the case of the ADS analysis it is possible that effects due to $\gamma$ may be observable.
Accordingly, the experiments now measure the corresponding observables in the $D\pi$ final states.
The results and averages are given in Table~\ref{tab:cp_uta:cus:ads2}
and shown in Fig.~\ref{fig:cp_uta:cus:ads-Dpi}.

\begin{table}[htb]
	\begin{center}
		\caption{
      Averages from ADS analyses of $b \to c\bar{u}s / u\bar{c}s$ modes.
                }
                \vspace{0.2cm}
                \setlength{\tabcolsep}{0.0pc}
\renewcommand{\arraystretch}{1.1}
                \begin{tabular*}{\textwidth}{@{\extracolsep{\fill}}lrccc} \hline 
        \mc{2}{l}{Experiment} & Sample size & $A_{\rm ADS}$ & $R_{\rm ADS}$ \\
        & & \hspace{-5mm}$N(B\bar{B})$ or $\int {\cal L}\,dt$\hspace{-5mm} \\
	\hline
        \mc{5}{c}{$D K^+$, $D \to K^-\pi^+$} \\
	\babar & \cite{delAmoSanchez:2010dz} & 467M & $-0.86 \pm 0.47 \,^{+0.12}_{-0.16}$ & $0.011 \pm 0.006 \pm 0.002$ \\
	\belle & \cite{Belle:2011ac} & 772M & $-0.39 \,^{+0.26}_{-0.28} \,^{+0.04}_{-0.03}$ & $0.0163 \,^{+0.0044}_{-0.0041} \,^{+0.0007}_{-0.0013}$ \\
	CDF & \cite{Aaltonen:2011uu} & $7 \, {\rm fb}^{-1}$ & $-0.82 \pm 0.44 \pm 0.09$ & $0.0220 \pm 0.0086 \pm 0.0026$ \\
	LHCb & \cite{Aaij:2016oso} & $3 \, {\rm fb}^{-1}$ & $-0.403 \pm 0.056 \pm 0.011$ & $0.0188 \pm 0.0011 \pm 0.0010$ \\
	\mc{3}{l}{\bf Average} & $-0.415 \pm 0.055$ & $0.0183 \pm 0.0014$ \\
	\mc{3}{l}{\small Confidence level} & {\small $0.64~(0.5\sigma)$} & {\small $0.61~(0.5\sigma)$} \\
		\hline
        \mc{5}{c}{$D K^+$, $D \to K^-\pi^+\pi^0$} \\
	\babar & \cite{Lees:2011up} & 474M & \textendash{} & $0.0091 \,^{+0.0082}_{-0.0076} \,^{+0.0014}_{-0.0037}$ \\
	\belle & \cite{Nayak:2013tgg} & 772M & $0.41 \pm 0.30 \pm 0.05$ & $0.0198 \pm 0.0062 \pm 0.0024$ \\
	LHCb & \cite{Aaij:2015jna} & $3 \, {\rm fb}^{-1}$ & $-0.20 \pm 0.27 \pm 0.03$ & $0.0140 \pm 0.0047 \pm 0.0019$ \\
	\mc{3}{l}{\bf Average} & $0.07 \pm 0.20$ & $0.0148 \pm 0.0036$ \\
 	\mc{3}{l}{\small Confidence level} & {\small $0.13~(1.5\sigma)$} & {\small $0.59~(0.5\sigma)$} \\
 	\hline
        \mc{5}{c}{$D K^+$, $D \to K^-\pi^+\pi^+\pi^-$} \\
	LHCb & \cite{Aaij:2016oso} & $3 \, {\rm fb}^{-1}$ & $-0.313 \pm 0.102 \pm 0.038$ & $0.0140 \pm 0.0015 \pm 0.0006$ \\
        \hline
        \mc{5}{c}{$\Dstar K^+$, $\Dstar \to D\pi^0$, $D \to K^-\pi^+$} \\
	\babar & \cite{delAmoSanchez:2010dz} & 467M & $0.77 \pm 0.35 \pm 0.12$ & $0.018 \pm 0.009 \pm 0.004$ \\
 	\hline
        \mc{5}{c}{$\Dstar K^+$, $\Dstar \to D\gamma$, $D \to K^-\pi^+$} \\
	\babar & \cite{delAmoSanchez:2010dz} & 467M & $0.36 \pm 0.94 \,^{+0.25}_{-0.41}$ & $0.013 \pm 0.014 \pm 0.008$ \\
		\hline
        \mc{5}{c}{$D K^{*+}$, $D \to K^-\pi^+$, $K^{*+} \to \KS \pi^+$} \\
	\babar & \cite{Aubert:2009yw} & 379M & $-0.34 \pm 0.43 \pm 0.16$ & $0.066 \pm 0.031 \pm 0.010$ \\
        LHCb   & \cite{Aaij:2017glf} & $4.8 \, {\rm fb}^{-1}$ & $-0.81 \pm 0.17 \pm 0.04$ & $0.011 \pm 0.004 \pm 0.001$ \\
 	\mc{3}{l}{\bf Average} & $-0.75 \pm 0.16$ & $0.012 \pm 0.004$ \\
 	\mc{3}{l}{\small Confidence level} & {\small $0.34~(1.0\sigma)$} & {\small $0.09~(1.7\sigma)$} \\
 		\hline
        \mc{5}{c}{$D K^{*+}$, $D \to K^-\pi^+\pi^+\pi^-$, $K^{*+} \to \KS \pi^+$} \\
        LHCb   & \cite{Aaij:2017glf} & $4.8 \, {\rm fb}^{-1}$ & $-0.45 \pm 0.21 \pm 0.14$ & $0.011 \pm 0.005 \pm 0.003$ \\
        \hline
        \mc{5}{c}{$D K^+\pi^+\pi^-$, $D \to K^-\pi^+$} \\
	LHCb & \cite{Aaij:2015ina} & $3 \, {\rm fb}^{-1}$ & $-0.32 \,^{+0.27}_{-0.34}$ & $0.0082 \,^{+0.0038}_{-0.0030}$ \\
        \hline
 		\end{tabular*}
                \label{tab:cp_uta:cus:ads}
 	\end{center}
\end{table}

\afterpage{\clearpage}

\begin{table}[htb]
	\begin{center}
		\caption{
      Averages from ADS analyses of $b \to c\bar{u}d / u\bar{c}d$ modes.
                }
                \vspace{0.2cm}
                \setlength{\tabcolsep}{0.0pc}
\renewcommand{\arraystretch}{1.1}
                \begin{tabular*}{\textwidth}{@{\extracolsep{\fill}}lrccc} \hline 
        \mc{2}{l}{Experiment} & Sample size & $A_{\rm ADS}$ & $R_{\rm ADS}$ \\
        & & \hspace{-5mm}$N(B\bar{B})$ or $\int {\cal L}\,dt$\hspace{-5mm} \\
        \hline
       \mc{5}{c}{$D \pi^+$, $D \to K^-\pi^+$} \\
	\babar & \cite{delAmoSanchez:2010dz} & 467M & $0.03 \pm 0.17 \pm 0.04$ & $0.0033 \pm 0.0006 \pm 0.0004$ \\
	\belle & \cite{Belle:2011ac} & 772M & $-0.04 \pm 0.11 \,^{+0.02}_{-0.01}$ & $0.00328 \,^{+0.00038}_{-0.00036} \,^{+0.00012}_{-0.00018}$ \\
	CDF & \cite{Aaltonen:2011uu} & $7 \, {\rm fb}^{-1}$ & $0.13 \pm 0.25 \pm 0.02$ & $0.0028 \pm 0.0007 \pm 0.0004$ \\
	LHCb & \cite{Aaij:2016oso} & $3 \, {\rm fb}^{-1}$ & $0.100 \pm 0.031 \pm 0.009$ & $0.00360 \pm 0.00012 \pm 0.00009$ \\
	\mc{3}{l}{\bf Average} & $0.088 \pm 0.030$ & $0.00353 \pm 0.00014$ \\
	\mc{3}{l}{\small Confidence level} & {\small $0.66~(0.4\sigma)$} & {\small $0.68~(0.4\sigma)$} \\
        \hline 
        \mc{5}{c}{$D \pi^+$, $D \to K^-\pi^+\pi^0$} \\
	\belle & \cite{Nayak:2013tgg} & 772M & $0.16 \pm 0.27 \,^{+0.03}_{-0.04}$ & $0.00189 \pm 0.00054 \,^{+0.00022}_{-0.00025}$ \\
	LHCb & \cite{Aaij:2015jna} & $3 \, {\rm fb}^{-1}$ & $0.44 \pm 0.19 \pm 0.01$ & $0.00235 \pm 0.00049 \pm 0.00004$ \\
	\mc{3}{l}{\bf Average} & $0.35 \pm 0.16$ & $0.00216 \pm 0.00038$ \\
	\mc{3}{l}{\small Confidence level} & {\small $0.40~(0.8\sigma)$} & {\small $0.55~(0.6\sigma)$} \\
        \hline 
        \mc{5}{c}{$D \pi^+$, $D \to K^-\pi^+\pi^+\pi^-$} \\
 	LHCb & \cite{Aaij:2016oso} & $3 \, {\rm fb}^{-1}$ & $0.023 \pm 0.048 \pm 0.005$ & $0.00377 \pm 0.00018 \pm 0.00006$ \\
        \hline
       \mc{5}{c}{$\Dstar \pi^+$, $\Dstar \to D\pi^0$, $D \to K^-\pi^+$} \\
	\babar & \cite{delAmoSanchez:2010dz} & 467M & $-0.09 \pm 0.27 \pm 0.05$ & $0.0032 \pm 0.0009 \pm 0.0008$ \\
        \hline 
        \mc{5}{c}{$\Dstar \pi^+$, $\Dstar \to D\gamma$, $D \to K^-\pi^+$} \\
	\babar & \cite{delAmoSanchez:2010dz} & 467M & $-0.65 \pm 0.55 \pm 0.22$ & $0.0027 \pm 0.0014 \pm 0.0022$ \\
        \hline
        \mc{5}{c}{$D \pi^+\pi^+\pi^-$, $D \to K^-\pi^+$} \\
        LHCb & \cite{Aaij:2015ina} & $3 \, {\rm fb}^{-1}$ & $-0.003 \pm 0.090$ & $0.00427 \pm 0.00043$ \\
        \hline
 		\end{tabular*}
                \label{tab:cp_uta:cus:ads2}
	\end{center}
\end{table}

\babar, \belle\ and LHCb have also presented results from a similar analysis method with self-tagging neutral $B$ decays:
$\Bz \to DK^{*0}$ with $D \to K^-\pi^+$ (all), $D \to K^-\pi^+\pi^0$ and $D \to K^-\pi^+\pi^+\pi^-$ (\babar\ only).
All these results are obtained with the $K^{*0} \to K^+\pi^-$ decay.
Effects due to the natural width of the $K^{*0}$ are
handled using the parametrisation suggested by Gronau~\cite{Gronau:2002mu}.

The following 95\% CL limits are set by \babar~\cite{:2009au}:
\begin{equation}
  R_{\rm ADS}(K\pi) < 0.244 \hspace{5mm}
  R_{\rm ADS}(K\pi\pi^0) < 0.181 \hspace{5mm}
  R_{\rm ADS}(K\pi\pi\pi) < 0.391 \, ,
\end{equation}
while \belle~\cite{Negishi:2012uxa} obtains
\begin{equation}
  R_{\rm ADS}(K\pi) < 0.16 \, .
\end{equation}
The results from LHCb, which are presented in terms of the parameters $R_+$ and $R_-$ instead of $R_{\rm ADS}$ and $A_{\rm ADS}$, are given in Table~\ref{tab:cp_uta:ads-DKstar}.

\begin{table}[!htb]
        \begin{center}
                \caption{
      Results from ADS analysis of $\Bz \to D\Kstarz$, $D \to K^-\pi^+$.
                }
                \vspace{0.2cm}
                \setlength{\tabcolsep}{0.0pc}
\renewcommand{\arraystretch}{1.1}
                \begin{tabular*}{\textwidth}{@{\extracolsep{\fill}}lrccc} \hline
        \mc{2}{l}{Experiment} & Sample size & $R_{+}$ & $R_{-}$ \\
        \hline
        LHCb & \cite{Aaij:2014eha} & $\int {\cal L}\,dt = 3 {\rm fb}^{-1}$ & $0.06 \pm 0.03 \pm 0.01$ & $0.06 \pm 0.03 \pm 0.01$ \\
                \hline
                \end{tabular*}
                \label{tab:cp_uta:ads-DKstar}
        \end{center}
\end{table}

Combining the results and using additional input from
CLEO-c~\cite{Asner:2008ft,Lowery:2009id} a limit on the ratio between the
$b \to u$ and $b \to c$ amplitudes of $\bar{r}_B(DK^{*0}) \in \left[ 0.07,0.41 \right]$
at 95\% CL limit is set by \babar.
Belle sets a limit of $\bar{r}_B < 0.4$ at 95\% CL.
LHCb takes input from Sec.~\ref{sec:charm_physics} and obtain $\bar{r}_B = 0.240 \,^{+0.055}_{-0.048}$ (different from zero with $2.7\sigma$ significance).

\mysubsubsection{$D$ decays to multiparticle self-conjugate final states (model-dependent analysis)}
\label{sec:cp_uta:cus:dalitz}

\begin{figure}[htbp]
  \begin{center}
    \resizebox{0.30\textwidth}{!}{
      \includegraphics{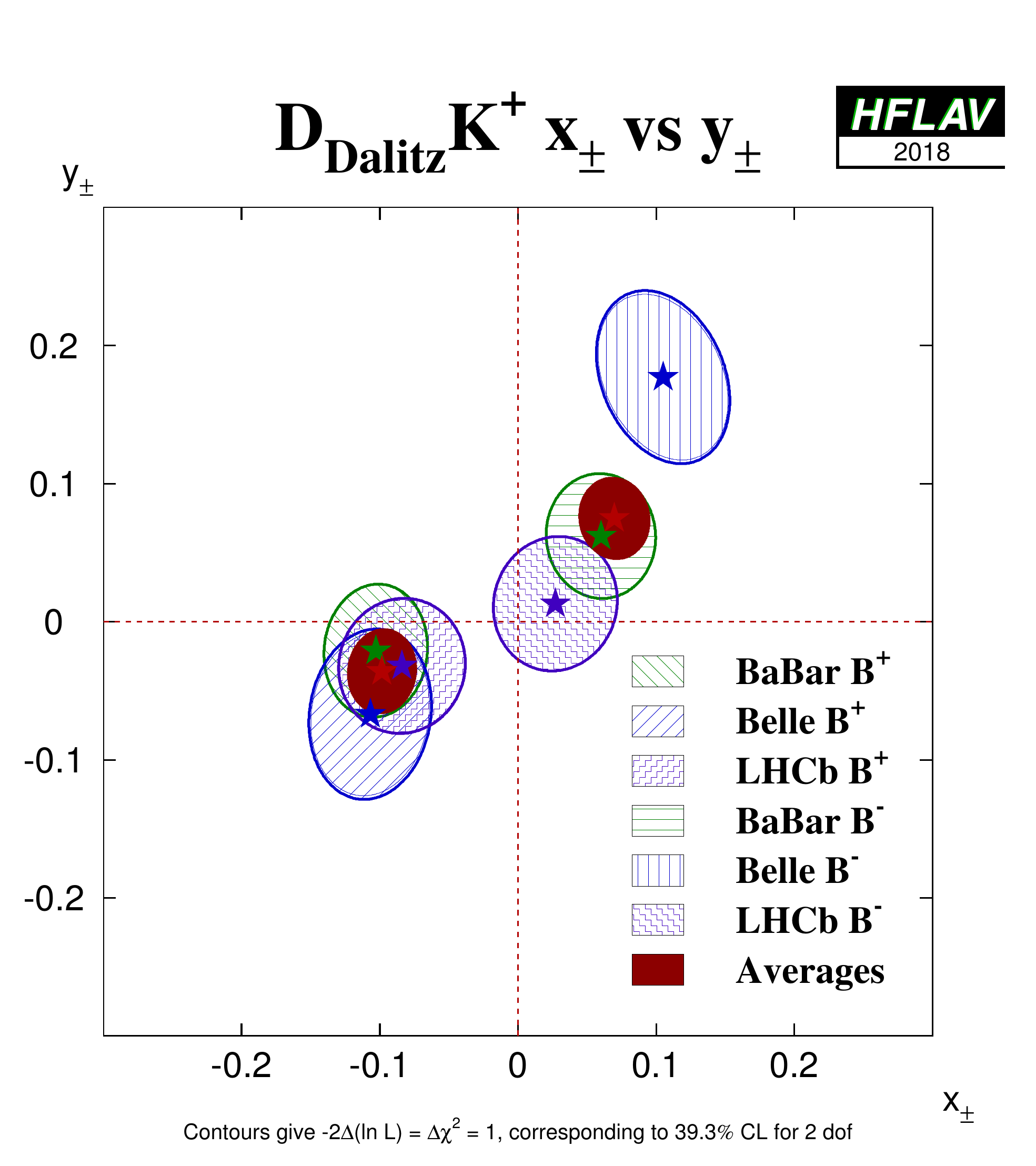}
    }
    \hfill
    \resizebox{0.30\textwidth}{!}{
      \includegraphics{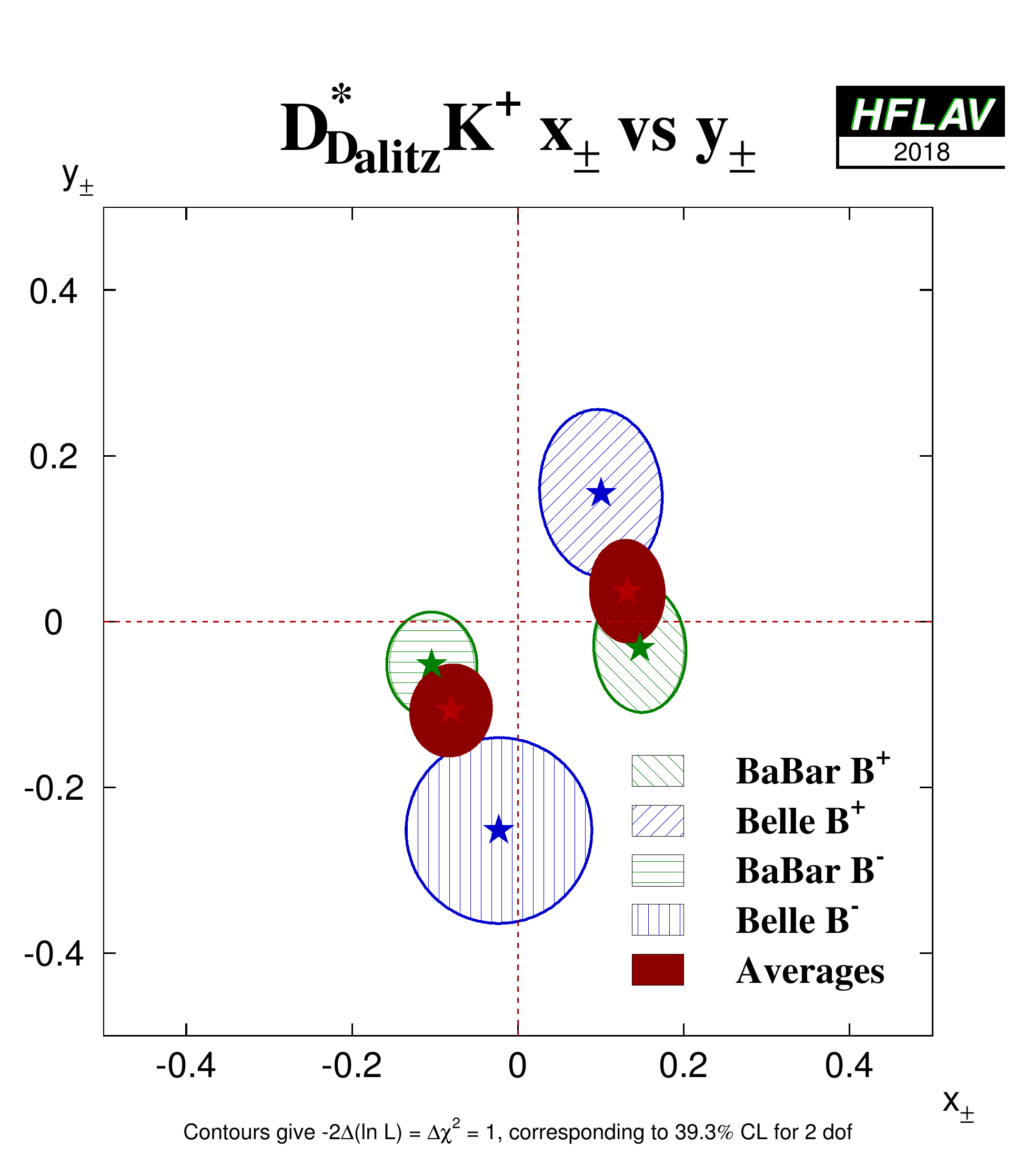}
    }
    \hfill
    \resizebox{0.30\textwidth}{!}{
      \includegraphics{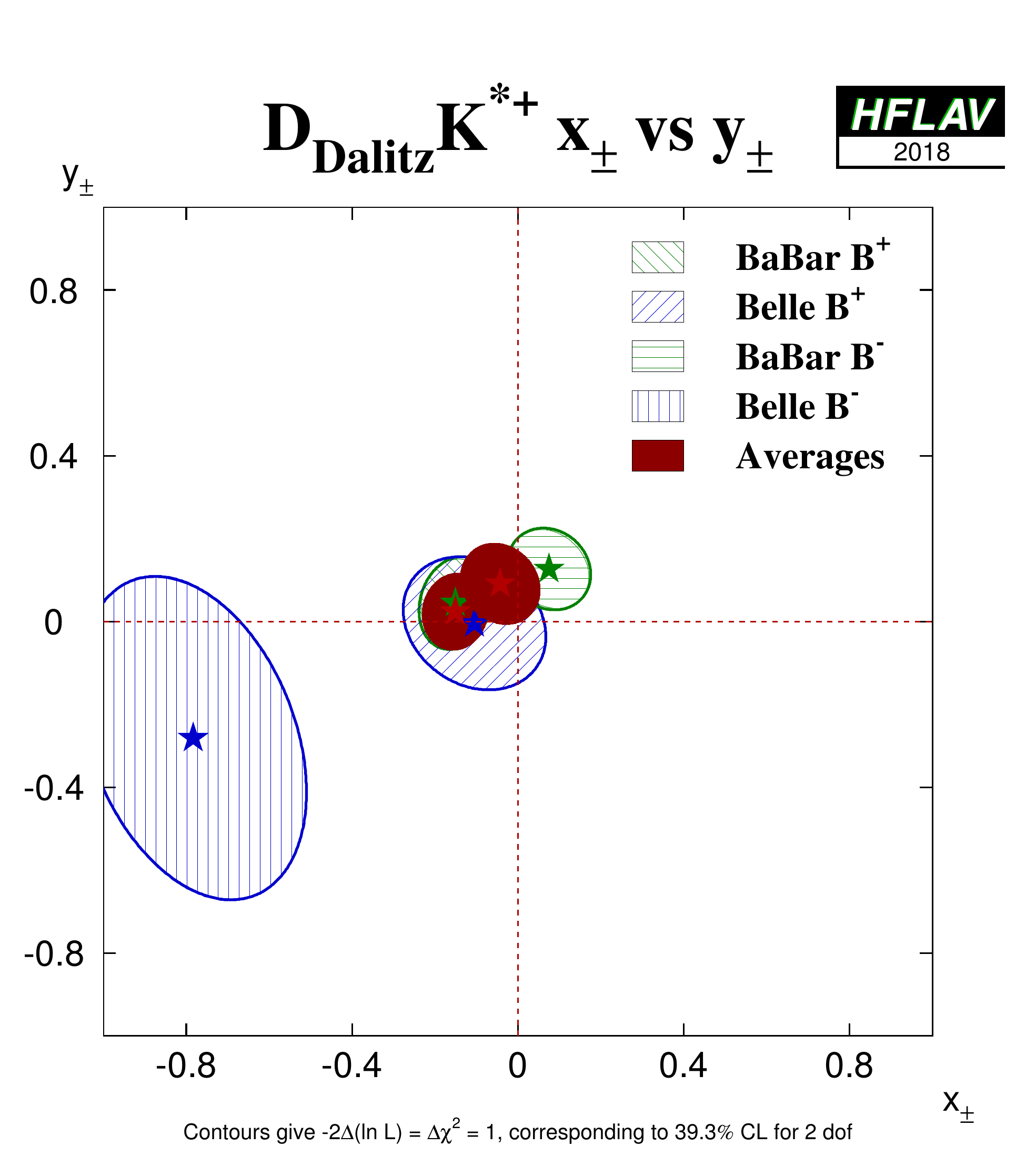}
    }
  \end{center}
  \vspace{-0.5cm}
  \caption{
    Contours in the $(x_\pm, y_\pm)$ from model-dependent analysis of $\Bp \to D^{(*)}K^{(*)+}$, $D \to \KS h^+ h^-$ ($h = \pi,K$).
    (Left) $\Bp \to D\Kp$,
    (middle) $\Bp \to \Dstar\Kp$,
    (right) $\Bp \to D\Kstarp$.
    Note that the uncertainties assigned to the averages given in these plots
    do not include model uncertainties.
  }
  \label{fig:cp_uta:cus:dalitz_2d}
\end{figure}

\begin{figure}[htbp]
  \begin{center}
    \resizebox{0.40\textwidth}{!}{
      \includegraphics{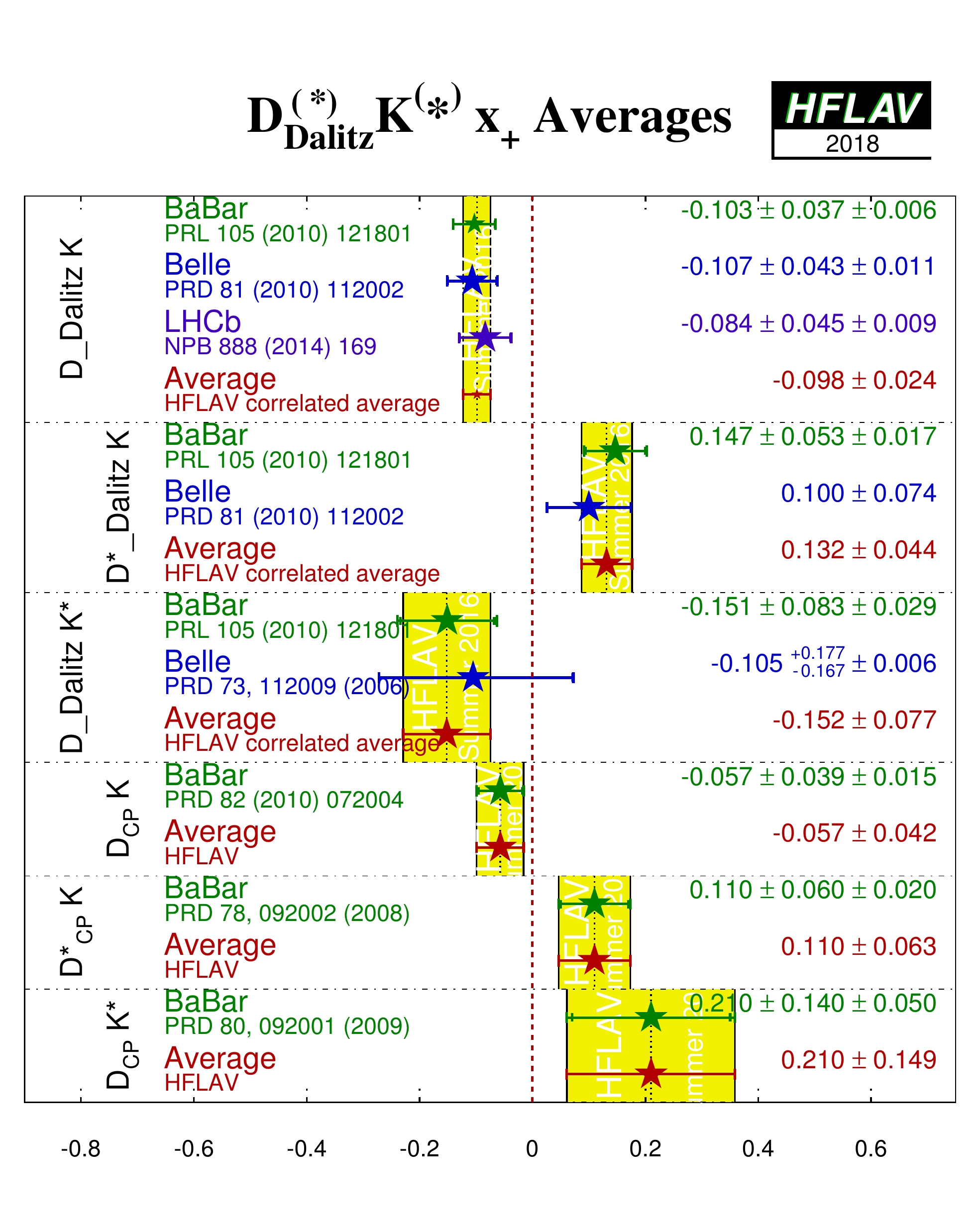}
    }
    \hspace{0.1\textwidth}
    \resizebox{0.40\textwidth}{!}{
      \includegraphics{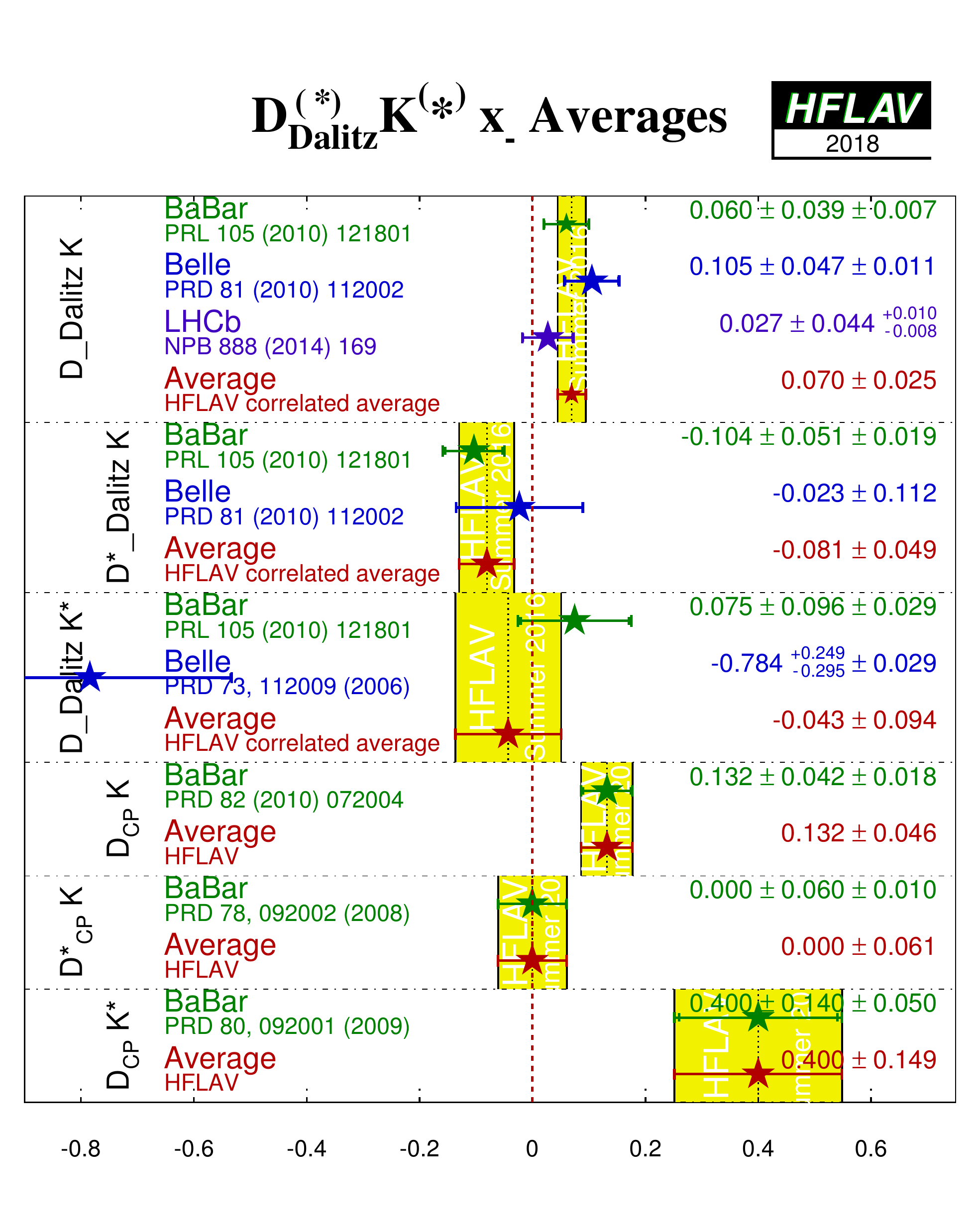}
    }
    \\
    \resizebox{0.40\textwidth}{!}{
      \includegraphics{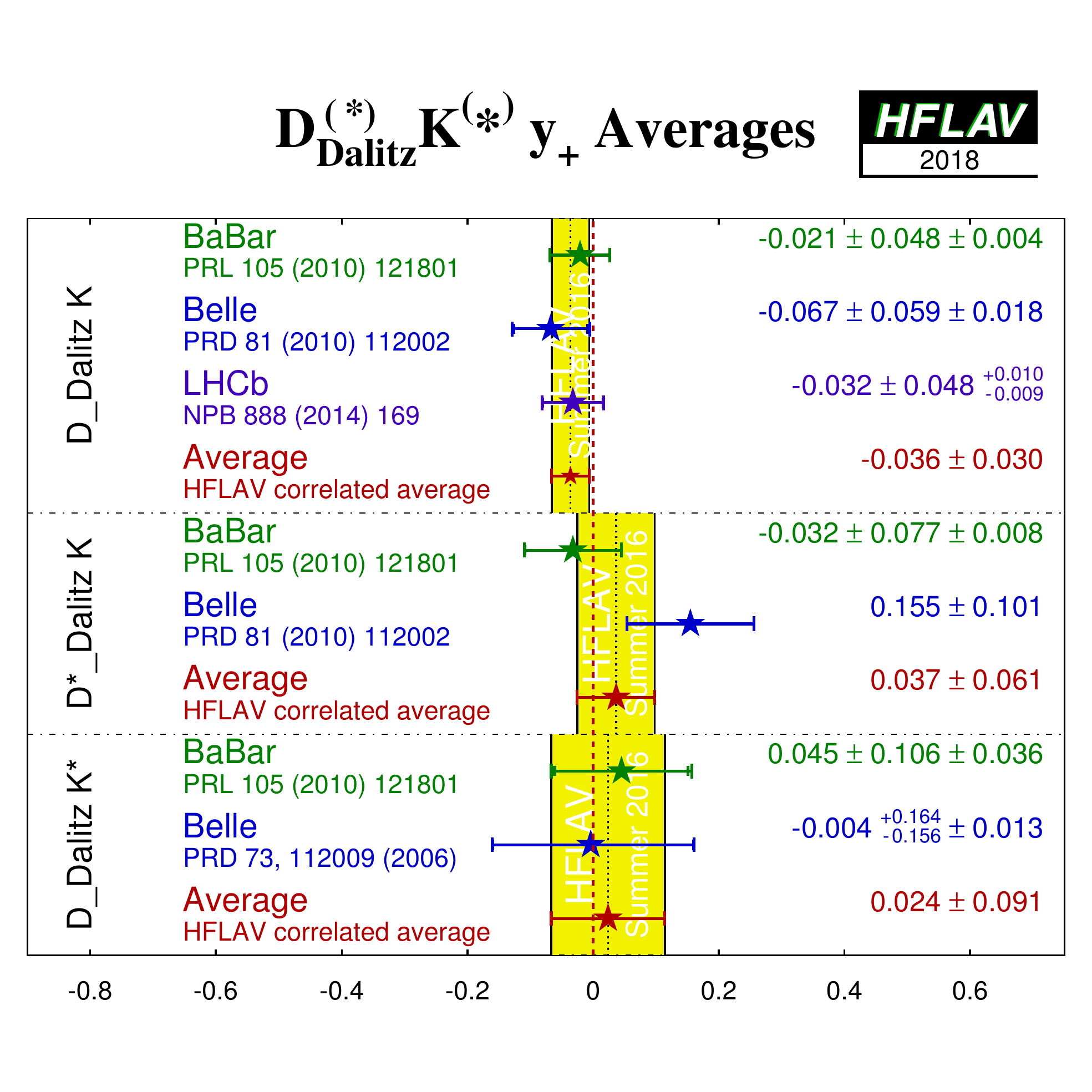}
    }
    \hspace{0.1\textwidth}
    \resizebox{0.40\textwidth}{!}{
      \includegraphics{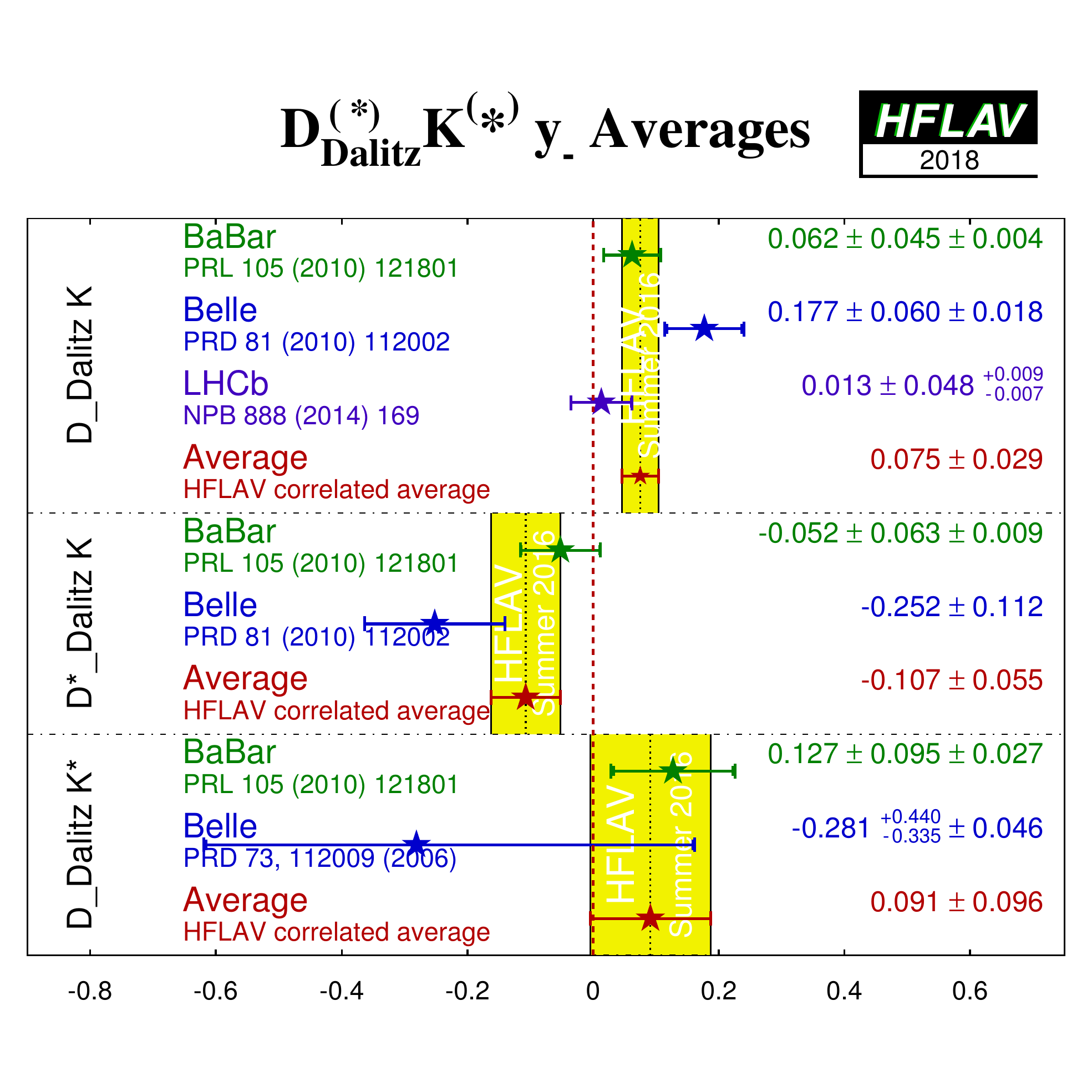}
    }
  \end{center}
  \vspace{-0.8cm}
  \caption{
    Averages of $(x_\pm, y_\pm)$ from model-dependent analyses of $\Bp \to D^{(*)}K^{(*)+}$ with $D \to \KS h^+h^-$ ($h=\pi,K$).
    (Top left) $x_+$, (top right) $x_-$,
    (bottom left) $y_+$, (bottom right) $y_-$.
    The top plots include constraints on $x_{\pm}$ obtained from GLW analyses (see Sec.~\ref{sec:cp_uta:cus:glw}).
    Note that the uncertainties assigned to the averages given in these plots
    do not include model uncertainties.
  }
  \label{fig:cp_uta:cus:dalitz_1d}
\end{figure}

For the model-dependent Dalitz plot analysis, both \babar\ and \belle\ have studied the modes
$\Bp \to D\Kp$, $\Bp \to \Dstar\Kp$ and $\Bp \to D\Kstarp$.
For $\Bp \to \Dstar\Kp$,
both experiments have used both $\Dstar$ decay modes, $\Dstar \to D\pi^0$ and
$\Dstar \to D\gamma$, taking the effective shift in the strong phase
difference into account.\footnote{
  \belle~\cite{Poluektov:2010wz} quotes separate results for $\Bp \to \Dstar\Kp$ with  $\Dstar \to D\pi^0$ and $\Dstar \to D\gamma$.
  The results presented in Table~\ref{tab:cp_uta:cus:dalitz} are from our average, performed using the statistical correlations provided, and neglecting all systematic correlations; model uncertainties are not included.
  The first uncertainty on the given results is combined statistical and systematic, the second is the model error (taken from the Belle results on $\Bp \to \Dstar\Kp$ with  $\Dstar \to D\pi^0$).
}
In all cases the decay $D \to \KS\pi^+\pi^-$ has been used.
\babar\ also used the decay $D \to \KS K^+K^-$.
LHCb has also studied $\Bp \to D\Kp$ decays with $D \to \KS\pi^+\pi^-$.
\babar\ has also performed an analysis of $\Bp \to D\Kp$ with $D \to \pi^+\pi^-\pi^0$.
Results and averages are given in Table~\ref{tab:cp_uta:cus:dalitz},
and shown in Figs.~\ref{fig:cp_uta:cus:dalitz_2d} and~\ref{fig:cp_uta:cus:dalitz_1d}.
The third error on each measurement is due to $D$ decay model uncertainty.

The parameters measured in the analyses are explained in
Sec.~\ref{sec:cp_uta:notations:cus}.
All experiments measure the Cartesian variables, defined in Eq.~(\ref{eq:cp_uta:cartesian}), and perform frequentist statistical procedures, to convert these into measurements of $\gamma$, $r_B$ and $\delta_B$.
In the $\Bp \to D\Kp$ with $D \to \pi^+\pi^-\pi^0$ analysis,
the parameters $(\rho^{\pm}, \theta^\pm)$ are used instead.

In the $\Bp \to D\Kstarp$ analysis both \babar\ and \belle\ experiments reconstruct $\Kstarp$ as $\KS\pip$,
but the treatment of possible nonresonant $\KS\pip$ differs:
\belle\ assigns an additional model uncertainty,
while \babar\ uses a parametrisation suggested by Gronau~\cite{Gronau:2002mu}
in which the parameters $r_B$ and $\delta_B$ are replaced with
effective parameters $\kappa \bar{r}_B$ and $\bar{\delta}_B$.
In this case no attempt is made to extract the true hadronic parameters
of the $\Bp \to D\Kstarp$ decay.

We perform averages using the following procedure, which is based on a set of
reasonable, though imperfect, assumptions.

\begin{itemize}\setlength{\itemsep}{0.5ex}
\item
  It is assumed that effects due to differences in the $D$ decay models
  used by the two experiments are negligible.
  Therefore, we do not rescale the results to a common model.
\item
  It is further assumed that the $D$ decay model uncertainty is $100\%$ correlated between experiments.
  (This approximation is compromised by the fact that the \babar\ results
  include $D \to \KS K^+K^-$ decays in addition to $D \to \KS\pi^+\pi^-$.)
  Other than the $D$ decay model, we do not consider common sources of systematic uncertainty.
\item
  We include in the average the effect of correlations
  within each experiment's set of measurements.
\item
  At present it is unclear how to assign a model uncertainty to the average.
  We have not attempted to do so.
  An unknown amount of model uncertainty should be added to the final error.
\item
  We follow the suggestion of Gronau~\cite{Gronau:2002mu}
  in making the $DK^*$ averages.
  Explicitly, we assume that the selection of $K^{*+} \to \KS\pip$
  is the same across experiments
  (so that $\kappa$, $\bar{r}_B$ and $\bar{\delta}_B$ are the same),
  and drop the additional source of model uncertainty
  assigned by Belle due to possible nonresonant decays.
\end{itemize}

\begin{sidewaystable}
	\begin{center}
		\caption{
      Averages from model-dependent Dalitz plot analyses of $b \to c\bar{u}s / u\bar{c}s$ modes.
      Note that the uncertainities assigned to the averages do not include model errors.	
		}
		\vspace{0.2cm}
		\setlength{\tabcolsep}{0.0pc}
    \resizebox{\textwidth}{!}{
\renewcommand{\arraystretch}{1.2}
		\begin{tabular}{@{\extracolsep{2mm}}lrccccc} \hline
	\mc{2}{l}{Experiment} & Sample size & $x_+$ & $y_+$ & $x_-$ & $y_-$ \\
	\hline
        \mc{7}{c}{$D K^+$, $D \to \KS \pi^+\pi^-$} \\
	\babar & \cite{delAmoSanchez:2010rq} & $N(B\bar{B}) =$ 468M & $-0.103 \pm 0.037 \pm 0.006 \pm 0.007$ & $-0.021 \pm 0.048 \pm 0.004 \pm 0.009$ & $0.060 \pm 0.039 \pm 0.007 \pm 0.006$ & $0.062 \pm 0.045 \pm 0.004 \pm 0.006$ \\
	\belle & \cite{Poluektov:2010wz} & $N(B\bar{B}) =$ 657M & $-0.107 \pm 0.043 \pm 0.011 \pm 0.055$ & $-0.067 \pm 0.059 \pm 0.018 \pm 0.063$ & $0.105 \pm 0.047 \pm 0.011 \pm 0.064$ & $0.177 \pm 0.060 \pm 0.018 \pm 0.054$ \\
	LHCb & \cite{Aaij:2014iba} & $\int {\cal L}\,dt = 1 {\rm fb}^{-1}$ & $-0.084 \pm 0.045 \pm 0.009 \pm 0.005$ & $-0.032 \pm 0.048 \,^{+0.010}_{-0.009} \pm 0.008$ & $0.027 \pm 0.044 \,^{+0.010}_{-0.008} \pm 0.001$ & $0.013 \pm 0.048 \,^{+0.009}_{-0.007} \pm 0.003$ \\
	\mc{3}{l}{\bf Average} & $-0.098 \pm 0.024$ & $-0.036 \pm 0.030$ & $0.070 \pm 0.025$ & $0.075 \pm 0.029$ \\
        \mc{3}{l}{\small Confidence level} &  \mc{4}{c}{\small $0.52~(0.7\sigma)$} \\
 		\hline
                \mc{7}{c}{$\Dstar K^+$, $\Dstar \to D\pi^0$ or $D\gamma$, $D \to \KS \pi^+\pi^-$} \\
	\babar & \cite{delAmoSanchez:2010rq} & $N(B\bar{B}) =$ 468M & $0.147 \pm 0.053 \pm 0.017 \pm 0.003$ & $-0.032 \pm 0.077 \pm 0.008 \pm 0.006$ & $-0.104 \pm 0.051 \pm 0.019 \pm 0.002$ & $-0.052 \pm 0.063 \pm 0.009 \pm 0.007$ \\
	\belle & \cite{Poluektov:2010wz} & $N(B\bar{B}) =$ 657M & $0.100 \pm 0.074 \pm 0.081$ & $0.155 \pm 0.101 \pm 0.063$ & $-0.023 \pm 0.112 \pm 0.090$ & $-0.252 \pm 0.112 \pm 0.049$ \\
	\mc{3}{l}{\bf Average} & $0.132 \pm 0.044$ & $0.037 \pm 0.061$ & $-0.081 \pm 0.049$ & $-0.107 \pm 0.055$ \\
        \mc{3}{l}{\small Confidence level} & \mc{4}{c}{\small $0.22~(1.2\sigma)$} \\
 		\hline
                \mc{7}{c}{$D K^{*+}$, $D \to \KS \pi^+\pi^-$} \\
	\babar & \cite{delAmoSanchez:2010rq} & $N(B\bar{B}) =$ 468M & $-0.151 \pm 0.083 \pm 0.029 \pm 0.006$ & $0.045 \pm 0.106 \pm 0.036 \pm 0.008$ & $0.075 \pm 0.096 \pm 0.029 \pm 0.007$ & $0.127 \pm 0.095 \pm 0.027 \pm 0.006$ \\
 	\belle & \cite{Poluektov:2006ia} & $N(B\bar{B}) =$ 386M & $-0.105 \,^{+0.177}_{-0.167} \pm 0.006 \pm 0.088$ & $-0.004 \,^{+0.164}_{-0.156} \pm 0.013 \pm 0.095$ & $-0.784 \,^{+0.249}_{-0.295} \pm 0.029 \pm 0.097$ & $-0.281 \,^{+0.440}_{-0.335} \pm 0.046 \pm 0.086$ \\
	\mc{3}{l}{\bf Average} & $-0.152 \pm 0.077$ & $0.024 \pm 0.091$ & $-0.043 \pm 0.094$ & $0.091 \pm 0.096$ \\
        \mc{3}{l}{\small Confidence level} & \mc{4}{c}{\small $0.011~(2.5\sigma)$} \\
 		\hline
                \mc{7}{c}{$D K^{*0}$, $D \to \KS \pi^+\pi^-$, $K^{*0} \to \Kp\pim$} \\
	LHCb & \cite{Aaij:2016zlt} & $\int {\cal L}\,dt = 3 \, {\rm fb}^{-1}$ & $0.05 \pm 0.24 \pm 0.04 \pm 0.01$ & $-0.65 \,^{+0.24}_{-0.23} \pm 0.08 \pm 0.01$ & $-0.15 \pm 0.14 \pm 0.03 \pm 0.01$ & $0.25 \pm 0.15 \pm 0.06 \pm 0.01$ \\
		\hline

                \vspace{1ex} \\

	\hline
	\mc{2}{l}{Experiment} & $N(B\bar{B})$ & $\rho^{+}$ & $\theta^+$ & $\rho^{-}$ & $\theta^-$ \\
	\hline
        \mc{7}{c}{$D K^+$, $D \to \pi^+\pi^-\pi^0$} \\
	\babar & \cite{Aubert:2007ii} & 324M & $0.75 \pm 0.11 \pm 0.04$ & $147 \pm 23 \pm 1$ & $0.72 \pm 0.11 \pm 0.04$ & $173 \pm 42 \pm 2$ \\
	\hline
		\end{tabular}
              }
		\label{tab:cp_uta:cus:dalitz}
	\end{center}
\end{sidewaystable}

\vspace{3ex}

\noindent
\underline{Constraints on $\gamma \equiv \phi_3$}

The measurements of $(x_\pm, y_\pm)$ can be used to obtain constraints on
$\gamma \equiv \phi_3$, as well as the hadronic parameters $r_B$ and $\delta_B$.
\babar~\cite{delAmoSanchez:2010rq},
\belle~\cite{Poluektov:2010wz,Poluektov:2006ia} and
LHCb~\cite{Aaij:2014iba} have all done so using a frequentist procedure,
with some differences in the details of the techniques used.

\begin{itemize}\setlength{\itemsep}{0.5ex}

\item
  \babar\ obtains $\gamma = (68 \,^{+15}_{-14} \pm 4 \pm 3)^\circ$
  from $D\Kp$, $\Dstar\Kp$ and $D\Kstarp$.

\item
  \belle\ obtains $\phi_3 = (78 \,^{+11}_{-12} \pm 4 \pm 9)^\circ$
  from $D\Kp$ and $\Dstar\Kp$.

\item
  LHCb obtains $\gamma = (84 \,^{+49}_{-42})^\circ$
  from $D\Kp$ using 1 ${\rm fb}^{-1}$ of data (a more precise result using 3 ${\rm fb}^{-1}$ and the model-independent method is reported below).

\item
  The experiments also obtain values for the hadronic parameters as detailed
  in Table~\ref{tab:cp_uta:rBdeltaB_summary}.

\item
  In the \babar\ analysis of $\Bp \to D\Kp$ with
  $D \to \pi^+\pi^-\pi^0$ decays~\cite{Aubert:2007ii},
  a constraint of $-30^\circ < \gamma < 76^\circ$ is obtained
  at the 68\% confidence level.

\item
  The results discussed here are included in the HFLAV combination to obtain a world average value for $\gamma \equiv \phi_3$, as discussed in Sec.~\ref{sec:cp_uta:cus:gamma}.

\end{itemize}

\begin{table}
  \begin{center}
  \caption{
    Summary of constraints on hadronic parameters from model-dependent analyses of $\Bp \to \DorDstar\KorKstarp$ and $\Bz \to D\Kstarz$ decays.
    Note the alternative parametrisation of the hadronic parameters used by \babar\ in the $D\Kstarp$ mode.
  }
  \label{tab:cp_uta:rBdeltaB_summary}
  \renewcommand{\arraystretch}{1.1}
  \begin{tabular}{@{\extracolsep{\fill}}lrccc}
    \hline
    \mc{2}{l}{Experiment} & Sample size & $r_B$ & $\delta_B$ \\
    \hline
    & & \multicolumn{3}{c}{In $D\Kp$} \\
    \babar & \cite{delAmoSanchez:2010rq} & $N(B\bar{B}) =$ 468M & $0.096 \pm 0.029 \pm 0.005 \pm 0.004$ & $(119 \,^{+19}_{-20} \pm 3 \pm 3)^\circ$ \\
    \belle & \cite{Poluektov:2010wz} & $N(B\bar{B}) =$ 657M & $0.160 \,^{+0.040}_{-0.038} \pm 0.011 \,^{+0.05}_{-0.010}$ & $(138 \,^{+13}_{-16} \pm 4 \pm 23)^\circ$ \\
    LHCb & \cite{Aaij:2014iba} & $\int {\cal L}\,dt = 1 {\rm fb}^{-1}$ & $0.06 \pm 0.04$ & $(115 \,^{+41}_{-51})^\circ$ \\
    \hline
    & & \multicolumn{3}{c}{In $\Dstar\Kp$} \\
    \babar & \cite{delAmoSanchez:2010rq} & $N(B\bar{B}) =$ 468M & $0.133 \,^{+0.042}_{-0.039} \pm 0.014 \pm 0.003$ & $(-82 \pm 21 \pm 5 \pm 3)^\circ$ \\
    \belle & \cite{Poluektov:2010wz} & $N(B\bar{B}) =$ 657M & $0.196 \,^{+0.072}_{-0.069} \pm 0.012 \,^{+0.062}_{-0.012}$ & $(342 \,^{+19}_{-21} \pm 3 \pm 23)^\circ$ \\
    \hline \\ [-2.4ex]
    & & & $\bar{r}_B$ & $\bar{\delta}_B$ \\
    \hline
    & & \multicolumn{3}{c}{In $D\Kstarp$} \\
    \babar & \cite{delAmoSanchez:2010rq} & $N(B\bar{B}) =$ 468M & $\kappa \bar{r}_B = 0.149 \,^{+0.066}_{-0.062} \pm 0.026 \pm 0.006$ & $(111 \pm 32 \pm 11 \pm 3)^\circ$ \\
    \belle & \cite{Poluektov:2006ia} & $N(B\bar{B}) =$ 386M & $0.56 \,^{+0.22}_{-0.16} \pm 0.04 \pm 0.08$ &
    $(243 \,^{+20}_{-23} \pm 3 \pm 50)^\circ$ \\
    \hline
    & & \multicolumn{3}{c}{In $D\Kstarz$} \\
    \babar & \cite{Aubert:2008yn} & $N(B\bar{B}) =$ 371M & $< 0.55$ at 95\% probability & $(62 \pm 57)^\circ$ \\
    LHCb & \cite{Aaij:2016zlt} & $\int {\cal L}\,dt = 3 \, {\rm fb}^{-1}$ & $0.39 \pm 0.13$ & $(197 \,^{+24}_{-20})^\circ$ \\
    \hline
  \end{tabular}
  \end{center}
\end{table}

\babar\ and LHCb have performed a similar analysis using the self-tagging neutral $B$ decay $\Bz \to DK^{*0}$ (with $K^{*0} \to K^+\pi^-$).
Effects due to the natural width of the $K^{*0}$ are handled using the parametrisation suggested by Gronau~\cite{Gronau:2002mu}.
LHCb~\cite{Aaij:2016zlt} gives results in terms of the Cartesian parameters, as shown in Table~\ref{tab:cp_uta:cus:dalitz}.
\babar~\cite{Aubert:2008yn} presents results only in terms of $\gamma$ and the hadronic parameters.
The obtained constraints are:
\begin{itemize}\setlength{\itemsep}{0.5ex}
\item
  \babar\ obtains $\gamma = (162 \pm 56)^\circ$;
\item
  LHCb obtains $\gamma = (80 \,^{+21}_{-22})^\circ$;
\item
  Values for the hadronic parameters are given in Table~\ref{tab:cp_uta:rBdeltaB_summary}.
\end{itemize}

\mysubsubsection{$D$ decays to multiparticle self-conjugate final states (model-independent analysis)}
\label{sec:cp_uta:cus:dalitz:modInd}

A model-independent approach to the analysis of $\Bp \to \DorDstar \Kp$ with multibody $D$ decays was proposed by Giri, Grossman, Soffer and Zupan~\cite{Giri:2003ty}, and further developed by Bondar and Poluektov~\cite{Bondar:2005ki,Bondar:2008hh}.
The method relies on information on the average strong phase difference between $\Dz$ and $\Dzb$ decays in bins of Dalitz plot position that can be obtained from quantum-correlated $\psi(3770) \to \Dz\Dzb$ events.
This information is measured in the form of parameters $c_i$ and $s_i$ that are the weighted averages of the cosine and sine of the strong phase difference in a Dalitz plot bin labelled by $i$, respectively.
These quantities have been obtained for $D \to \KS \pi^+\pi^-$ (and $D \to \KS K^+K^-$) decays by CLEO-c~\cite{Briere:2009aa,Libby:2010nu}.

\belle~\cite{Aihara:2012aw} and LHCb~\cite{Aaij:2014uva,Aaij:2018uns} have used the model-independent Dalitz-plot analysis approach to study the mode $\Bp \to D\Kp$.
LHCb has presented results separately for two subsamples of their data, with the averaged result also given.
Both \belle~\cite{Negishi:2015vqa} and LHCb~\cite{Aaij:2016nao} have also used this approach to study $\Bz \to DK^*(892)^0$ decays.
In both cases, the experiments use $D \to \KS\pi^+\pi^-$ decays, and LHCb has also included the $D \to \KS K^+K^-$ decay.
The Cartesian variables $(x_\pm, y_\pm)$, defined in Eq.~(\ref{eq:cp_uta:cartesian}), were determined from the data.
Note that due to the strong statistical and systematic correlations with the model-dependent results given in Sec.~\ref{sec:cp_uta:cus:dalitz}, these sets of results cannot be combined.

\begin{figure}[htbp]
  \begin{center}
    \resizebox{0.45\textwidth}{!}{
      \includegraphics{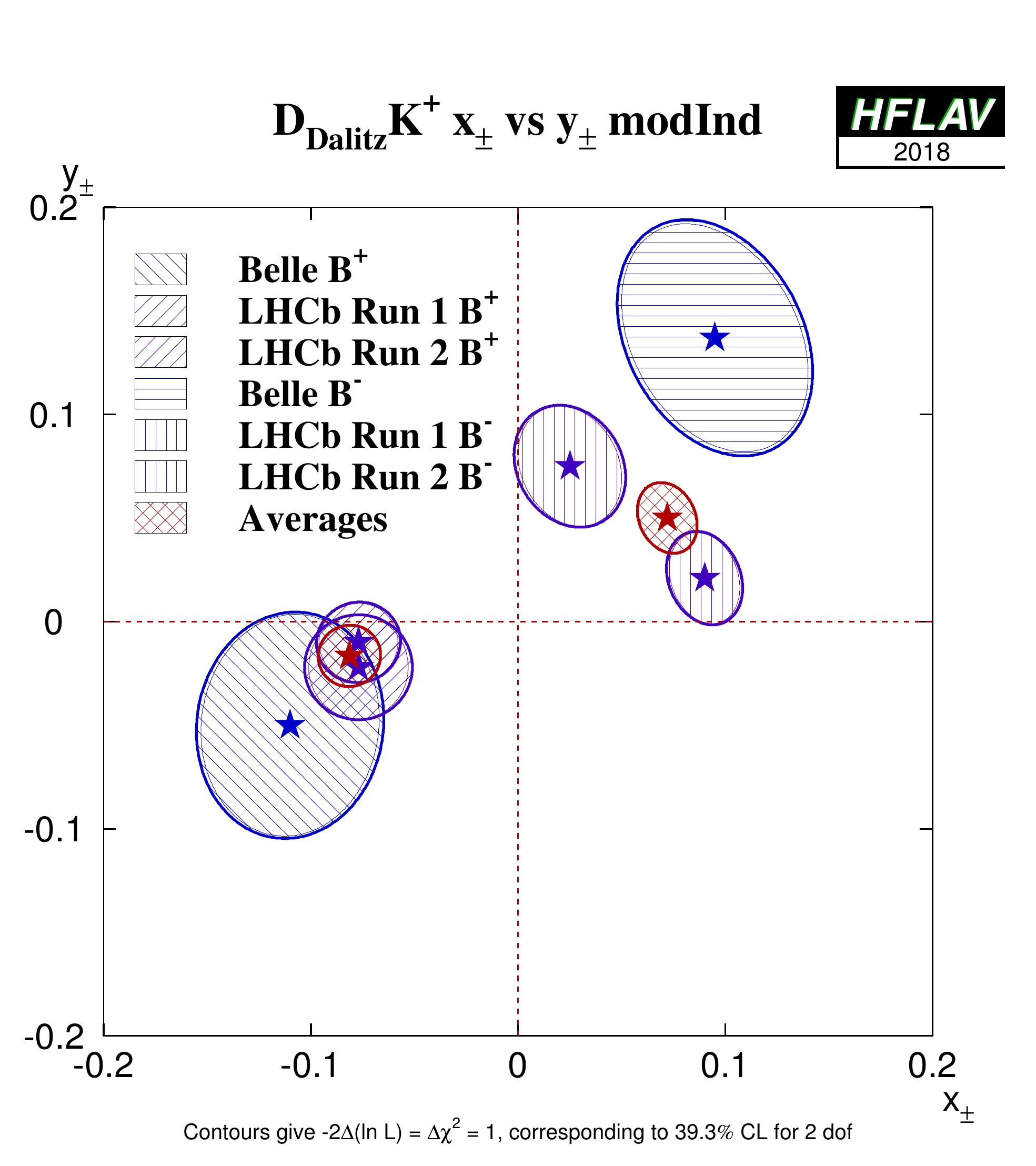}
    }
  \end{center}
  \vspace{-0.5cm}
  \caption{
    Contours in the $(x_\pm, y_\pm)$ plane from model-independent analysis of
    $\Bp \to D\Kp$ with $D \to \KS h^+ h^-$ ($h = \pi,K$).
  }
  \label{fig:cp_uta:cus:dalitz-modInd_2d}
\end{figure}

The results and averages are given in Table~\ref{tab:cp_uta:cus:dalitz-modInd}, and shown in Figs.~\ref{fig:cp_uta:cus:dalitz-modInd_2d}.
Most results have three sets of errors, which are, respectively, statistical, systematic, and the uncertainty coming from the knowledge of $c_i$ and $s_i$.
To perform the average, we first remove the last uncertainty, which should be 100\% correlated between the measurements.
Since the size of the uncertainty from $c_i$ and $s_i$ is found to depend on the size of the $B \to DK$ data sample, we assign the LHCb uncertainties (which are mostly the smaller of the Belle and LHCb values) to the averaged result.
This procedure should be conservative.
In the LHCb $\Bz \to DK^*(892)^0$ results~\cite{Aaij:2016nao}, the values of $c_i$ and $s_i$ are constrained to their measured values within uncertainties in the fit to data, and hence the systematic uncertainties associated with the knowledge of these parameters is absorbed in their statistical uncertainties.
The $\Bz \to DK^*(892)^0$ average is performed neglecting the model uncertainties on the Belle results.

\begin{sidewaystable}
	\begin{center}
		\caption{
      Averages from model-independent Dalitz plot analyses of $b \to c\bar{u}s / u\bar{c}s$ modes.
		}
		\vspace{0.2cm}
		\setlength{\tabcolsep}{0.0pc}
    \resizebox{\textwidth}{!}{
\renewcommand{\arraystretch}{1.2}
		\begin{tabular}{@{\extracolsep{2mm}}lrccccc} \hline
	\mc{2}{l}{Experiment} & Sample size & $x_+$ & $y_+$ & $x_-$ & $y_-$ \\
	\hline
        \mc{7}{c}{$D K^+$, $D \to \KS \pi^+\pi^-$} \\
	\belle & \cite{Aihara:2012aw} & $N(B\bar{B}) =$ 772M & $-0.110 \pm 0.043 \pm 0.014 \pm 0.007$ & $-0.050 \,^{+0.052}_{-0.055} \pm 0.011 \pm 0.007$ & $0.095 \pm 0.045 \pm 0.014 \pm 0.010$ & $0.137 \,^{+0.053}_{-0.057} \pm 0.015 \pm 0.023$ \\
	LHCb Run 1 & \cite{Aaij:2014uva} & $\int {\cal L}\,dt = 3 {\rm fb}^{-1}$ & $-0.077 \pm 0.024 \pm 0.010 \pm 0.004$ & $-0.022 \pm 0.025 \pm 0.004 \pm 0.010$ & $0.025 \pm 0.025 \pm 0.010 \pm 0.005$ & $0.075 \pm 0.029 \pm 0.005 \pm 0.014$ \\
        LHCb Run 2 & \cite{Aaij:2018uns} & $\int {\cal L}\,dt = 2 {\rm fb}^{-1}$ & $-0.077 \pm 0.019 \pm 0.007 \pm 0.004$ & $-0.010 \pm 0.019 \pm 0.004 \pm 0.009$ & $0.090 \pm 0.017 \pm 0.007 \pm 0.004$ & $0.021 \pm 0.022 \pm 0.005 \pm 0.011$ \\
        LHCb combined & \cite{Aaij:2018uns} & $\int {\cal L}\,dt = 5 {\rm fb}^{-1}$ & $-0.078 \pm 0.017$ & $-0.014 \pm 0.017$ & $0.070 \pm 0.017$ & $0.041 \pm 0.020$ \\
	\mc{3}{l}{\bf Average} & $-0.081 \pm 0.015 \pm 0.004$ & $-0.017 \pm 0.015 \pm 0.009$ & $0.072 \pm 0.014 \pm 0.004$ & $0.050 \pm 0.017 \pm 0.011$ \\ 
        \mc{3}{l}{\small Confidence level} &  \mc{4}{c}{\small $0.28~(1.1\sigma)$} \\
 		\hline
        \mc{7}{c}{$D K^{*0}$, $D \to \KS \pi^+\pi^-$} \\
	\belle & \cite{Negishi:2015vqa} & $N(B\bar{B}) =$ 772M & $0.1 \,^{+0.7}_{-0.4} \,^{+0.0}_{-0.1} \pm 0.1$ & $0.3 \,^{+0.5}_{-0.8} \,^{+0.0}_{-0.1} \pm 0.1$ & $0.4 \,^{+1.0}_{-0.6} \,^{+0.0}_{-0.1} \pm 0.0$ & $-0.6 \,^{+0.8}_{-1.0} \,^{+0.1}_{-0.0} \pm 0.1$ \\
	LHCb & \cite{Aaij:2016nao} & $\int {\cal L}\,dt = 3 {\rm fb}^{-1}$ & $0.05 \pm 0.35 \pm 0.02$ & $-0.81 \pm 0.28 \pm 0.06$ & $-0.31 \pm 0.20 \pm 0.04$ & $0.31 \pm 0.21 \pm 0.05$ \\
	\mc{3}{l}{\bf Average} & $0.10 \pm 0.30$ & $-0.63 \pm 0.26$ & $-0.27 \pm 0.20$ & $0.27 \pm 0.21$ \\
	\mc{3}{l}{\small Confidence level} & \mc{4}{c}{\small $0.38~(0.9\sigma)$} \\
 		\hline
		\end{tabular}
              }
		\label{tab:cp_uta:cus:dalitz-modInd}
	\end{center}
\end{sidewaystable}

\begin{sidewaystable}
	\begin{center}
		\caption{
      Results from model-independent Dalitz plot analysis of $B^+ \to DK^+$, $D \to \KS\Kpm\pimp$.
		}
		\setlength{\tabcolsep}{0.0pc}
    \resizebox{\textwidth}{!}{
\renewcommand{\arraystretch}{1.2}
		\begin{tabular}{@{\extracolsep{4mm}}lrccccccc} \hline
	\mc{2}{l}{Experiment} & $\int {\cal L}\,dt$ & $R_{\rm SS}$ & $R_{\rm OS}$ & $A_{{\rm SS},DK}$ & $A_{{\rm OS},DK}$ & $A_{{\rm SS},D\pi}$ & $A_{{\rm OS},D\pi}$ \\
	\hline
        \mc{9}{c}{$D \to \KS\Kpm\pimp$ (whole Dalitz plot)} \\
	LHCb & \cite{Aaij:2014dia} & $3 \ {\rm fb}^{-1}$ & $0.092 \pm 0.009 \pm 0.004$ & $0.066 \pm 0.009 \pm 0.002$ & $0.040 \pm 0.091 \pm 0.018$ & $0.233 \pm 0.129 \pm 0.024$ & $-0.025 \pm 0.024 \pm 0.010$ & $-0.052 \pm 0.029 \pm 0.017$ \\
	\hline
        \mc{9}{c}{$D \to K^*(892)^\pm\Kmp$} \\
	LHCb & \cite{Aaij:2014dia} & $3 \ {\rm fb}^{-1}$ & $0.084 \pm 0.011 \pm 0.003$ & $0.056 \pm 0.013 \pm 0.002$ & $0.026 \pm 0.109 \pm 0.029$ & $0.336 \pm 0.208 \pm 0.026$ & $-0.012 \pm 0.028 \pm 0.010$ & $-0.054 \pm 0.043 \pm 0.017$ \\
        \hline
                \end{tabular}
    }
		\label{tab:cp_uta:cus:KSKpi-modInd}
	\end{center}
\end{sidewaystable}

\vspace{3ex}

\noindent
\underline{Constraints on $\gamma \equiv \phi_3$}

The measurements of $(x_\pm, y_\pm)$ can be used to obtain constraints on
$\gamma$, as well as the hadronic parameters $r_B$ and $\delta_B$.
The experiments have done so using frequentist procedures, with some differences in the details of the techniques used.

\begin{itemize}\setlength{\itemsep}{0.5ex}

\item
  From $\Bp \to D\Kp$, \belle~\cite{Aihara:2012aw} obtains
  $\phi_3 = (77.3 \,^{+15.1}_{-14.9} \pm 4.1 \pm 4.3)^\circ$.

\item
  From $\Bp \to D\Kp$, LHCb~\cite{Aaij:2018uns} obtains
  $\gamma = (80\,^{+10}_{-9})^\circ$.

\item
  From $\Bz \to DK^*(892)^0$, LHCb~\cite{Aaij:2016nao} obtains
  $\gamma = (71 \pm 20)^\circ$.

\item
  The experiments also obtain values for the hadronic parameters as detailed
  in Table~\ref{tab:cp_uta:rBdeltaB_summary-modInd}.

\item
  The results discussed here are included in the HFLAV combination to obtain a world average value for $\gamma \equiv \phi_3$, as discussed in Sec.~\ref{sec:cp_uta:cus:gamma}.

\end{itemize}

\begin{table}
  \begin{center}
  \caption{
    Summary of constraints on hadronic parameters from model-independent analyses of $\Bp \to D\Kp$ and $\Bz \to D\Kstarz$, $D \to \KS h^+h^-$ ($h=\pi,K$) decays.
  }
  \label{tab:cp_uta:rBdeltaB_summary-modInd}
  \renewcommand{\arraystretch}{1.1}
  \begin{tabular}{@{\extracolsep{\fill}}lrccc}
    \hline
    \mc{2}{l}{Experiment} & Sample size & $r_B(D\Kp)$ & $\delta_B(D\Kp)$ \\
    \hline
    \belle & \cite{Aihara:2012aw} & $N(B\bar{B}) =$ 772M & $0.145 \pm 0.030 \pm 0.010 \pm 0.011$ & $(129.9 \pm 15.0 \pm 3.8 \pm 4.7)^\circ$ \\
    LHCb & \cite{Aaij:2018uns} & $\int {\cal L}\,dt = 5 {\rm fb}^{-1}$ & $0.080 \pm 0.011$ & $(110 \pm 10)^\circ$ \\
    \hline
    & & & $\bar{r}_B(DK^{*0})$ & $\bar{\delta}_B(DK^{*0})$ \\
    \belle & \cite{Negishi:2015vqa} & $N(B\bar{B}) =$ 772M & $< 0.87 \ \text{at 68\% confidence level}$ & \\
    LHCb & \cite{Aaij:2016nao} & $\int {\cal L}\,dt = 3 {\rm fb}^{-1}$ & $0.56 \pm 0.17$ & $(204 \,^{+21}_{-20})^\circ$ \\
    \hline
  \end{tabular}
  \end{center}
\end{table}

\mysubsubsection{$D$ decays to multiparticle non-self-conjugate final states (model-independent analysis)}
\label{sec:cp_uta:cus:dalitz:KsKpi}

Following the original suggestion of Grossman, Ligeti and Soffer~\cite{Grossman:2002aq}, decays of $D$ mesons to $\KS\Kpm\pimp$ can be used in a similar approach to that discussed above to determine $\gamma \equiv \phi_3$.
Since these decays are less abundant, the event samples available to date have not been sufficient for a fine binning of the Dalitz plots, but the analysis can be performed using only an overall coherence factor and related strong phase difference for the decay.
These quantities have been determined by CLEO-c~\cite{Insler:2012pm} both for the full Dalitz plots and in a restricted region $\pm 100 \ {\rm MeV}/c^2$ around the peak of the $K^*(892)^\pm$ resonance.

LHCb~\cite{Aaij:2014dia} has reported results of an analysis of $B^+\to D K^+$ and $B^+ \to D \pi^+$ decays with $D \to \KS\Kpm\pimp$.
The decays with different final states of the $D$ meson are distinguished by the charge of the kaon from the decay of the $D$ meson relative to the charge of the $B$ meson, and are labelled ``same sign'' (SS) and ``opposite sign'' (OS).
Six observables potentially sensitive to $\gamma \equiv \phi_3$ are measured: two ratios of rates for $DK$ and $D\pi$ decays (one each for SS and OS) and four asymmetries (for $DK$ and $D\pi$, SS and OS).
This is done both for the full Dalitz plot of the $D$ decay and for the $K^*(892)^\pm$-dominated region (with the same boundaries as used by CLEO-c).
Note that there is a significant overlap of events between the two samples.
The results, shown in Table~\ref{tab:cp_uta:cus:KSKpi-modInd}, do not yet have sufficient precision to set significant constraints on $\gamma \equiv \phi_3$.

\mysubsubsection{Combinations of results on rates and asymmetries in $B \to \DorDstar K^{(*)}$ decays to obtain constraints on $\gamma \equiv \phi_3$}
\label{sec:cp_uta:cus:gamma}

\babar\ and LHCb have both produced constraints on $\gamma \equiv \phi_3$ from combinations of their results on $B^+ \to DK^+$ and related processes.
The experiments use a frequentist procedure,
with some differences in the details of the techniques used.

\begin{itemize}\setlength{\itemsep}{0.5ex}

\item
  \babar~\cite{Lees:2013nha} uses results from $DK$, $D^*K$ and $DK^*$ modes with GLW, ADS and BPGGSZ analyses, to obtain $\gamma = (69 \,^{+17}_{-16})^\circ$.

\item
  LHCb~\cite{Aaij:2016kjh,LHCb-CONF-2018-002} uses results from the $D\Kp$ mode with GLW, GLW-like, ADS, BPGGSZ ($\KS h^+h^-$) and GLS ($\KS\Kpm\pimp$) analyses, as well as $DK^{*0}$ with GLW, ADS and BPGGSZ analyses, $D\Kp\pim$ GLW Dalitz plot analysis, $D\Kp\pim\pip$ with GLW and ADS analyses and $\Bs \to D_s^\mp\Kpm$ decays.
  The LHCb combination takes into account subleading effects due to charm mixing and \CP violation~\cite{Rama:2013voa}.
  The result is $\gamma = (74.0 \,^{+5.0}_{-5.8})^\circ$.

\item
  All the combinations use inputs determined from $\psi(3770)\to \Dz\Dzb$ data samples (and/or from the HFLAV global fits on charm mixing parameters; see Sec.~\ref{sec:charm:mixcpv}) to constrain the hadronic parameters in the charm system.

\item
  Constraints are also obtained on the hadronic parameters involved in the decays.
  A summary of these is given in Table~\ref{tab:cp_uta:rBdeltaB_combination}.

\item
  The CKMfitter~\cite{Charles:2004jd} and
  UTFit~\cite{Bona:2005vz} groups perform similar combinations of all available results to obtain combined constraints on $\gamma \equiv \phi_3$.

\end{itemize}

\begin{table}
  \begin{center}
  \caption{
    Summary of constraints on hadronic parameters obtained from global combinations of results in $\Bp \to \DorDstar\KorKstarp$ and $\Bz \to D\Kstarz$ decays.
    Results for parameters associated with the other decay modes discussed in this section are less precise and are not included in this summary.
  }
  \label{tab:cp_uta:rBdeltaB_combination}
  \renewcommand{\arraystretch}{1.1}
  \begin{tabular}{@{\extracolsep{5mm}}lrcccc}
    \hline
    \mc{2}{l}{Experiment} & $r_B(D\Kp)$ & $\delta_B(D\Kp)$ & $r_B(\Dstar\Kp)$ & $\delta_B(\Dstar\Kp)$ \\
    \hline
    \babar & \cite{Lees:2013nha} & $0.092 \,^{+0.013}_{-0.012}$ & $(105 \,^{+16}_{-17})^\circ$ & $0.106 \,^{+0.019}_{-0.036} $ & $(294 \,^{+21}_{-31})^\circ$ \\
    LHCb & \cite{LHCb-CONF-2018-002} & $0.1019 \pm 0.0056$ & $(142.6 \,^{+5.7}_{-6.6})^\circ$ & $0.191 \,^{+0.045}_{-0.038}$ & $(332 \,^{+8}_{-10})^\circ$ \\
    \hline
  \end{tabular}
  \end{center}
\end{table}

\newcommand{\Dsmp}{\ensuremath{D_{s}^{\mp}}\xspace}
\newcommand{\phis}{\ensuremath{\phi_{s}}\xspace}

\newcommand{\rb}{\ensuremath{r_{B}}\xspace}
\newcommand{\db}{\ensuremath{\delta_{B}}\xspace}
\newcommand{\rbdk}{\ensuremath{r_{B}(D\Kp)}\xspace}
\newcommand{\dbdk}{\ensuremath{\delta_{B}(D\Kp)}\xspace}
\newcommand{\rbdstk}{\ensuremath{r_{B}(\Dstar\Kp)}\xspace}
\newcommand{\dbdstk}{\ensuremath{\delta_{B}(\Dstar\Kp)}\xspace}
\newcommand{\kbdstk}{\ensuremath{\kappa_{B}(\Dstar\Kp)}\xspace}
\newcommand{\rbdkst}{\ensuremath{r_{B}(D\Kstarp)}\xspace}
\newcommand{\dbdkst}{\ensuremath{\delta_{B}(D\Kstarp)}\xspace}
\newcommand{\kbdkst}{\ensuremath{\kappa_{B}(D\Kstarp)}\xspace}
\newcommand{\rbdkstz}{\ensuremath{r_{B}(D\Kstarz)}\xspace}
\newcommand{\dbdkstz}{\ensuremath{\delta_{B}(D\Kstarz)}\xspace}
\newcommand{\kbdkstz}{\ensuremath{\kappa_{B}^{D\Kstarz}}\xspace}
\newcommand{\RbDKstz}{\ensuremath{\bar{R}_{B}^{DK^{*0}}}\xspace}
\newcommand{\DbDKstz}{\ensuremath{\bar{\Delta}_{B}^{DK^{*0}}}\xspace}

\newcommand{\rD}{\ensuremath{r_{D}}\xspace}
\newcommand{\dD}{\ensuremath{\delta_{D}}\xspace}
\newcommand{\kD}{\ensuremath{\kappa_{D}}\xspace}
\newcommand{\rdKpi}  {\ensuremath{r_D^{K\pi}}\xspace}
\newcommand{\ddKpi}  {\ensuremath{\delta_D^{K\pi}}\xspace}
\newcommand{\Fp}{\texorpdfstring{\ensuremath{F_{+}}}{F+}\xspace}
\newcommand{\rDKpi}{\texorpdfstring{\ensuremath{r_{D}^{K\pi}}}{rDKPi}\xspace}
\newcommand{\dDKpi}{\texorpdfstring{\ensuremath{\delta_{D}^{K\pi}}}{dDKPi}\xspace}
\newcommand{\rdKpp}{\ensuremath{r_{D}^{K2\pi}}\xspace}
\newcommand{\rdKppsq}{\ensuremath{(r_{D}^{K2\pi})^{2}}\xspace}
\newcommand{\ddKpp}{\ensuremath{\delta_{D}^{K2\pi}}\xspace}
\newcommand{\kdKpp}{\ensuremath{\kappa_{D}^{K2\pi}}\xspace}
\newcommand{\Fppp}{\ensuremath{F_+(\pip\pim\piz)}\xspace}
\newcommand{\FKKp}{\ensuremath{F_+(\Kp\Km\piz)}\xspace}
\newcommand{\kdppp}{\ensuremath{\kappa_{\pi\pi\piz}}\xspace}
\newcommand{\kdkkp}{\ensuremath{\kappa_{KK\piz}}\xspace}
\newcommand{\Aprod}{\ensuremath{A_{B}^{\rm{prod}}}\xspace}
\newcommand{\rdKskpi}{\ensuremath{r_D^{K_SK\pi}}\xspace}
\newcommand{\rdKskpisq}{\ensuremath{(r_D^{K_SK\pi})^2}\xspace}
\newcommand{\ddKskpi}{\ensuremath{\delta_D^{K_SK\pi}}\xspace}
\newcommand{\kdKskpi}{\ensuremath{\kappa_D^{K_SK\pi}}\xspace}
\newcommand{\RdKskpi}{\ensuremath{R_D^{K_SK\pi}}\xspace}
\newcommand{\rdKppp}{\texorpdfstring{\ensuremath{r_D^{K3\pi}}}{rD(K3pi)}\xspace}
\newcommand{\rdKpppsq}{\ensuremath{(r_D^{K3\pi})^2}\xspace}
\newcommand{\ddKppp}{\ensuremath{\delta_D^{K3\pi}}\xspace}
\newcommand{\kdKppp}{\ensuremath{\kappa_D^{K3\pi}}\xspace}
\newcommand{\Fpppp}{\ensuremath{F_+(\pip\pim\pip\pim)}\xspace}
\newcommand{\kdpppp}{\ensuremath{\kappa_{\pi\pi\pi\pi}}\xspace}

\newcommand{\BuDK}    {\ensuremath{\Bp\to D \Kp}\xspace}
\newcommand{\BuDstK}  {\ensuremath{\Bp\to \Dstar \Kp}\xspace}
\newcommand{\BuDKst}  {\ensuremath{\Bp\to D \Kstarp}\xspace}
\newcommand{\BuDKpipi}{\ensuremath{\Bp\to D \Kp\pip\pim}\xspace}
\newcommand{\BuDhhpizK}    {\ensuremath{\Bp\to D_{hh\piz} \Kp}\xspace}
\newcommand{\BuDppppK}    {\ensuremath{\Bp\to D_{\pi\pi\pi\pi} \Kp}\xspace}
\newcommand{\BdDKstz} {\ensuremath{\Bd\to D \Kstarz}\xspace}
\newcommand{\BdDKpi}  {\ensuremath{\Bd\to D \Kp\pim} }
\newcommand{\BsDsK}	  {\ensuremath{\Bs\to \Dsmp \Kpm}\xspace}
\newcommand{\DKpi}     {\ensuremath{D\to K^{\pm}\pi^{\mp}}\xspace}
\newcommand{\Dhh}      {\ensuremath{D\to h^+h^-}\xspace}
\newcommand{\Dhhh}     {\ensuremath{D\to hhhh}\xspace}
\newcommand{\DKpipipi} {\texorpdfstring{\ensuremath{D\to K^{\pm}\pi^{\mp}\pip\pim}}{D -> K3pi}\xspace}
\newcommand{\DKpipiz}  {\texorpdfstring{\ensuremath{D\to K^{\pm}\pi^{\mp}\piz}}{D -> K2pi}\xspace}
\newcommand{\Dhpipipi} {\ensuremath{D\to h^+\pim\pip\pim}\xspace}
\newcommand{\Dpipipipi}{\ensuremath{D\to \pip\pim\pip\pim}\xspace}
\newcommand{\Dhhpiz}   {\ensuremath{D\to h^+h^-\piz}\xspace}
\newcommand{\Dzhh}     {\ensuremath{\Dz\to h^+h^-}\xspace}
\newcommand{\DzKK}     {\ensuremath{\Dz\to \Kp\Km}\xspace}
\newcommand{\Dzpipi}   {\ensuremath{\Dz\to\pip\pim}\xspace}
\newcommand{\DzKShh}   {\ensuremath{\Dz\to\KS h^+h^-}\xspace}
\newcommand{\DKSpipi}  {\ensuremath{D\to\KS\pip\pim}\xspace}
\newcommand{\DKSKK}    {\ensuremath{D\to\KS \Kp\Km}\xspace}
\newcommand{\DKShh}    {\ensuremath{D\to\KS h^+h^-}\xspace}
\newcommand{\DKSKpi}   {\ensuremath{D\to \KS \Kpm\pimp}\xspace}
\newcommand{\Dzpipipiz}{\ensuremath{\Dz\to\pip\pim\piz}\xspace}
\newcommand{\DzKKpiz}  {\ensuremath{\Dz\to \Kp \Km\piz}\xspace}
\newcommand{\Dpipipiz} {\ensuremath{D\to\pip\pim\piz}\xspace}
\newcommand{\DKKpiz}   {\ensuremath{D\to \Kp \Km\piz}\xspace}
\newcommand{\Dshhh}    {\ensuremath{\Ds\to h^+h^-\pip}}
\newcommand{\DstD}     {\ensuremath{\Dstar\to D\piz(\g)}\xspace}
\newcommand{\DstDg}    {\ensuremath{\Dstar\to D\g}\xspace}
\newcommand{\DstDpiz}  {\ensuremath{\Dstar\to D\piz}\xspace}
\newcommand{\DKK}      {\ensuremath{D\to \Kp\Km}\xspace}
\newcommand{\Dpipi}      {\ensuremath{D\to \pip\pim}\xspace}
\newcommand{\DKSpi}      {\ensuremath{D\to \KS\piz}\xspace}
\newcommand{\DKSw}      {\ensuremath{D\to \KS\omega}\xspace}
\newcommand{\DKSphi}      {\ensuremath{D\to \KS\phi}\xspace}

Independently from the constraints on $\gamma \equiv \phi_3$ obtained by the
experiments, the results summarised in Sec.~\ref{sec:cp_uta:cus} are statistically combined to produce world average constraints on $\gamma \equiv \phi_3$ and the hadronic parameters involved.
The combination is performed with the \textsc{GammaCombo} framework~\cite{gammacombo} and follows a frequentist procedure, identical to that used in Ref.~\cite{Aaij:2013zfa}.

The input measurements used in the combination are listed in Table~\ref{tab:cp_uta:gamma:inputs}.
Individual measurements are used as inputs, rather than the averages presented in Sec.~\ref{sec:cp_uta:cus}, in order to facilitate cross-checks and to ensure the most appropriate treatment of correlations.
A combination based on our averages for each of the quantities measured by experiments gives consistent results.

All results from GLW and GLW-like analyses of $\Bp\to \DorDstar \KorKstarp$ modes, as listed in Tables~\ref{tab:cp_uta:cus:glw} and~\ref{tab:cp_uta:cus:glwLike}, are used.
All results from ADS analyses of $\Bp\to \DorDstar \KorKstarp$ as listed in Table~\ref{tab:cp_uta:cus:ads} are also used.
Regarding $\Bd\to D\Kstarz$ decays, the results of the $\Bd\to D\Kp\pim$ GLW-Dalitz analysis (Table~\ref{tab:cp_uta:cus:DKpiDalitz}) are included, as are the LHCb results of the ADS analysis of $\Bd\to D\Kstarz$ (Table~\ref{tab:cp_uta:ads-DKstar}).
Concerning results of BPGGSZ analyses of $\Bp\to \DorDstar \KorKstarp$ with $D\to\KS h^{+}h^{-}$, the model-dependent results, as listed in Table~\ref{tab:cp_uta:cus:dalitz}, are used for the \babar\ and \belle\ experiments, whilst the model-independent results, as listed in Table~\ref{tab:cp_uta:cus:dalitz-modInd}, are used for LHCb.
This choice is made in order to maintain consistency of the approach across experiments whilst maximising the size of the samples used to obtain inputs for the combination.
For BPGGSZ analyses of $\Bd\to D\Kstarz$ with $D\to\KS h^{+}h^{-}$, the model-independent result from LHCb (given in Table~\ref{tab:cp_uta:cus:dalitz-modInd}) is used for consistency with the treatment of the LHCb $\Bp \to D\Kp$ BPGGSZ result; the model-independent result by Belle is also included.
Finally, results from the time-dependent analysis of $\Bs\to\Dsmp\Kpm$ from LHCb (Table~\ref{tab:cp_uta:DsK}) are used.

Several results with sensitivity to $\gamma$ are not included in the combination.
Results from time-dependent analyses of $\Bz \to D^{(*)\mp}\pi^\pm$ and $D^\mp\rho^\pm$ (Table~\ref{tab:cp_uta:cud}) are not used, as there are insufficient constraints on the associated hadronic parameters.
Similarly, results from $\Bz \to \Dmp\KS\pipm$ (Sec.~\ref{sec:cp_uta:cus-td-DKSpi}) are not used.
Results from the LHCb $\Bz \to D\Kstarz$ GLW analysis (Table~\ref{tab:cp_uta:cus:glw}) are not used because of the statistical overlap with the GLW-Dalitz analysis, which is used instead.
Limits on ADS parameters reported in Sec.~\ref{sec:cp_uta:cus:ads} are not used.
Results on $\Bp \to D\pip$ decays, given in Table~\ref{tab:cp_uta:cus:ads2}, are not used, since the small value of $r_B(D\pip)$ means that these channels have less sensitivity to $\gamma$ and are more vulnerable to biases from subleading effects~\cite{Aaij:2016kjh}.
Results from the \babar\ Dalitz plot analysis of $\Bp \to D\Kp$ with $D \to \pip\pim\piz$ (given in Table~\ref{tab:cp_uta:cus:dalitz}) are not included due to their limited sensitivity.
Results from the $\Bp \to D\Kp$, $D \to \KS \pip\pim$ BPGGSZ model-dependent analysis by LHCb (given in Table~\ref{tab:cp_uta:cus:dalitz}), and of the model-independent analysis of the same decay by Belle (given in Table~\ref{tab:cp_uta:cus:dalitz-modInd}) are not included due to the statistical overlap with results from model-(in)dependent analyses of the same data.

\begin{longtable}{l l l l l}
  \caption{List of measurements used in the $\gamma$ combination.}
  \label{tab:cp_uta:gamma:inputs}
\endfirsthead
\multicolumn{5}{c}{List of measurements used in the $\gamma$ combination -- continued from previous page.}
\endhead
\endfoot
\endlastfoot
       \hline
        $B$ decay & $D$ decay & Method & Experiment & Ref. \\
        \hline
        \BuDK     & \DKK, \Dpipi,                        & GLW         & \babar & \cite{delAmoSanchez:2010ji} \\
                  & \DKSpi, \DKSw, \DKSphi               &             &       &         \\
        \BuDK     & \DKK, \Dpipi,                        & GLW         & \belle & \cite{Abe:2006hc} \\
                  & \DKSpi, \DKSw, \DKSphi               &             &       &         \\
        \BuDK     & \DKK, \Dpipi & GLW         & CDF   & \cite{Aaltonen:2009hz} \\
        \BuDK     & \DKK, \Dpipi &GLW         & LHCb  & \cite{Aaij:2017ryw} \\
        \hline
        \BuDstK   & \DKK, \Dpipi,                        & GLW         & \babar & \cite{:2008jd} \\
        $\;\;$\DstDg(\piz)  & \DKSpi, \DKSw, \DKSphi               &             &       &         \\
        \BuDstK   & \DKK, \Dpipi,                        & GLW         & \belle & \cite{Abe:2006hc} \\
        $\;\;$\DstDg(\piz)  & \DKSpi, \DKSw, \DKSphi               &             &       &         \\
        \BuDstK   & \DKK, \Dpipi                        & GLW         & LHCb & \cite{Aaij:2017ryw} \\
        $\;\;$\DstDg(\piz)  &               &             &       &         \\
        \hline
        \BuDKst   & \DKK, \Dpipi,                        & GLW         & \babar & \cite{Aubert:2009yw} \\
                  & \DKSpi, \DKSw, \DKSphi               &             &       &         \\
        \BuDKst   & \DKK, \Dpipi                        & GLW         & LHCb & \cite{Aaij:2017glf} \\
        \hline
        \BuDKpipi & \DKK, \Dpipi & GLW         & LHCb  & \cite{Aaij:2015ina} \\
        \hline
        \BuDK     & \Dpipipiz             & GLW-like & \babar & \cite{Aubert:2007ii} \\
        \BuDK     & \DKKpiz, \Dpipipiz    & GLW-like & LHCb & \cite{Aaij:2015jna} \\
        \BuDK     & \Dpipipipi            & GLW-like & LHCb & \cite{Aaij:2016oso} \\
        \hline
        \BuDKst   & \Dpipipipi            & GLW-like & LHCb & \cite{Aaij:2017glf} \\
        \hline
        \BdDKpi  & \DKK, \Dpipi & GLW-Dalitz & LHCb & \cite{Aaij:2016bqv} \\
        \hline
        \BuDK  & \DKpi & ADS & \babar & \cite{delAmoSanchez:2010dz} \\
        \BuDK  & \DKpi & ADS & \belle & \cite{Belle:2011ac} \\
        \BuDK  & \DKpi & ADS & CDF & \cite{Aaltonen:2011uu} \\
        \BuDK  & \DKpi & ADS & LHCb & \cite{Aaij:2016oso} \\
        \hline
        \BuDK  & \DKpipiz & ADS & \babar & \cite{Lees:2011up} \\
        \BuDK  & \DKpipiz & ADS & \belle & \cite{Nayak:2013tgg} \\
        \BuDK  & \DKpipiz & ADS & LHCb & \cite{Aaij:2015jna} \\
        \hline
        \BuDK  & \DKpipipi & ADS & LHCb & \cite{Aaij:2016oso} \\
        \hline
        \BuDstK  & \DKpi & ADS & \babar & \cite{delAmoSanchez:2010dz} \\
        $\;\;$\DstDg(\piz)  & & & &  \\
        \hline
        \BuDKst  & \DKpi  & ADS & \babar & \cite{Aubert:2009yw} \\
        \BuDKst  & \DKpi  & ADS & LHCb & \cite{Aaij:2017glf} \\
        \hline
        \BuDKst  & \DKpipipi & ADS & LHCb & \cite{Aaij:2017glf} \\
        \hline
        \BdDKstz  & \DKpi & ADS & LHCb & \cite{Aaij:2014eha} \\
        \hline
        \BuDKpipi  & \DKpi & ADS & LHCb & \cite{Aaij:2015ina} \\
        \hline
        \BuDK  & \DKSpipi & GGSZ MD & \babar & \cite{delAmoSanchez:2010rq} \\
        \BuDK  & \DKSpipi & GGSZ MD & \belle & \cite{Poluektov:2010wz} \\
        \BuDK  & \DKSpipi, \DKSKK & GGSZ MI & LHCb & \cite{Aaij:2014uva,Aaij:2018uns} \\
        \hline
        \BuDstK  & \DKSpipi & GGSZ MD & \babar & \cite{delAmoSanchez:2010rq} \\
        $\;\;$\DstDg(\piz)& &         &       &         \\
        \BuDstK  & \DKSpipi & GGSZ MD & \belle & \cite{Poluektov:2010wz} \\
        $\;\;$\DstDg(\piz)& &         &       &         \\
        \hline
        \BuDKst  & \DKSpipi & GGSZ MD & \babar & \cite{delAmoSanchez:2010rq} \\
        \BuDKst  & \DKSpipi & GGSZ MD & \belle & \cite{Poluektov:2006ia} \\
        \hline
        \BdDKstz  & \DKSpipi & GGSZ MI & Belle & \cite{Negishi:2015vqa} \\
        \BdDKstz  & \DKSpipi, \DKSKK & GGSZ MI & LHCb & \cite{Aaij:2016nao} \\
        \hline
        \BsDsK  & \Dshhh & TD & LHCb & \cite{Aaij:2017lff} \\
        \hline
\end{longtable}

\begin{table}[b]
  \caption{List of the auxiliary inputs used in the combinations.}
  \label{tab:cp_uta:gamma:inputs_aux}
  \centering
  \renewcommand{\arraystretch}{1.1}
      \begin{tabular}{l l l l }
        \hline
        Decay      & Parameters                  & Source & Ref. \\
        \hline \\[-2.5ex]
         \DKpi              & \rdKpi, \ddKpi                       & HFLAV       &  Sec.~\ref{sec:charm_physics}       \\
         \DKpipipi          & \ddKppp, \kdKppp, \rdKppp            & CLEO+LHCb  &  \cite{Evans:2016tlp}       \\
         \Dpipipipi         & \Fpppp                               & CLEO       &  \cite{Malde:2015mha}       \\
         \DKpipiz           & \ddKpp, \kdKpp, \rdKpp               & CLEO+LHCb  &  \cite{Evans:2016tlp}       \\
         \Dhhpiz            & \Fppp, \FKKp                         & CLEO       &  \cite{Malde:2015mha}       \\
         \multirow{2}{*}{\DKSKpi}            & \ddKskpi, \kdKskpi, \rdKskpi         & CLEO       &  \cite{Insler:2012pm}       \\
                     & \rdKskpi                             & LHCb       &  \cite{Aaij:2015lsa} \\
         \BdDKstz           & \kbdkstz, \RbDKstz, \DbDKstz         & LHCb       &  \cite{Aaij:2016bqv} \\
         \BsDsK             & \phis                                & HFLAV       &  Sec.~\ref{sec:life_mix} \\
        \hline
      \end{tabular}
\end{table}

Auxiliary inputs are used in the combination in order to constrain the $D$ system parameters and subsequently improve the determination of $\gamma \equiv \phi_3$.
These include the ratio of suppressed to favoured decay amplitudes and the strong phase difference for $D\to\Kpm\pimp$ decays, taken from the charm global fits (see Sec.~\ref{sec:charm_physics}).
The amplitude ratios, strong phase differences and coherence factors of $D\to\Kpm\pimp\piz$, $D\to\Kpm\pimp\pip\pim$ and $D\to\KS\Kpm\pipm$ decays are taken from CLEO-c and LHCb measurements~\cite{Evans:2016tlp,Insler:2012pm,Aaij:2015lsa}.
The fraction of \CP-even content for the GLW-like $D\to\pip\pim\pip\pim$, $D\to\Kp\Km\piz$ and $D\to\pip\pim\piz$ decays are taken from CLEO-c measurements~\cite{Malde:2015mha}.
Constraints required to relate the hadronic parameters of the $\Bd\to D\Kstarz$ GLW-Dalitz analysis to the effective hadronic parameters of the Q2B approaches are taken from LHCb measurements~\cite{Aaij:2016bqv}.
Finally, the value of $-2\beta_{s}$ is taken from the HFLAV averages (see Sec.~\ref{sec:life_mix}); this is required to obtain sensitivity to $\gamma \equiv \phi_3$ from the time-dependent analysis of $\Bs\to\Dsmp\Kpm$ decays.
A summary of the auxiliary constraints is given in Table~\ref{tab:cp_uta:gamma:inputs_aux}.

The following reasonable, although imperfect, assumptions are made when performing the averages.
\begin{itemize}
  \item{\CP violation in \DKK and \Dpipi decays is assumed to be zero. The results of Sec.~\ref{sec:charm_physics} anyhow suggest such effects to be negligible.}
  \item{The combination is potentially sensitive to subleading effects from $\Dz$--$\Dzb$ mixing~\cite{Silva:1999bd,Grossman:2005rp,Rama:2013voa}, but these are expected to have little impact and are not accounted for.}
  \item{All \BuDKst modes are treated as two-body decays. In other words any dilution caused by non-\Kstarp\ contributions in the selected regions of the $D\KS\pip$ or $D\Kp\piz$ Dalitz plots is assumed to be negligible.  As a check of this assumption, it was found that including a coherence factor for \BuDKst modes, $\kbdkst = 0.9$, had negligible impact on the results.}
  \item{Each individual set of input measurements listed in Table 38 is assumed to be completely uncorrelated, though correlations between observables in a set are used if provided by the experiment. Whilst this assumption is true for the statistical uncertainties, it is not necessarily the case for systematic uncertainties. In particular, the model uncertainties for different model-dependent BPGGSZ analyses are fully correlated (when the same model is used). Similarly, the model-independent BPGGSZ analyses have correlated systematic uncertainties originating from the measurement of the strong phase variation across the Dalitz plot. The effect of including these correlations is estimated to be $<1^\circ$.}
\end{itemize}

\begin{figure}
    \centering
    \includegraphics[width=0.6\textwidth]{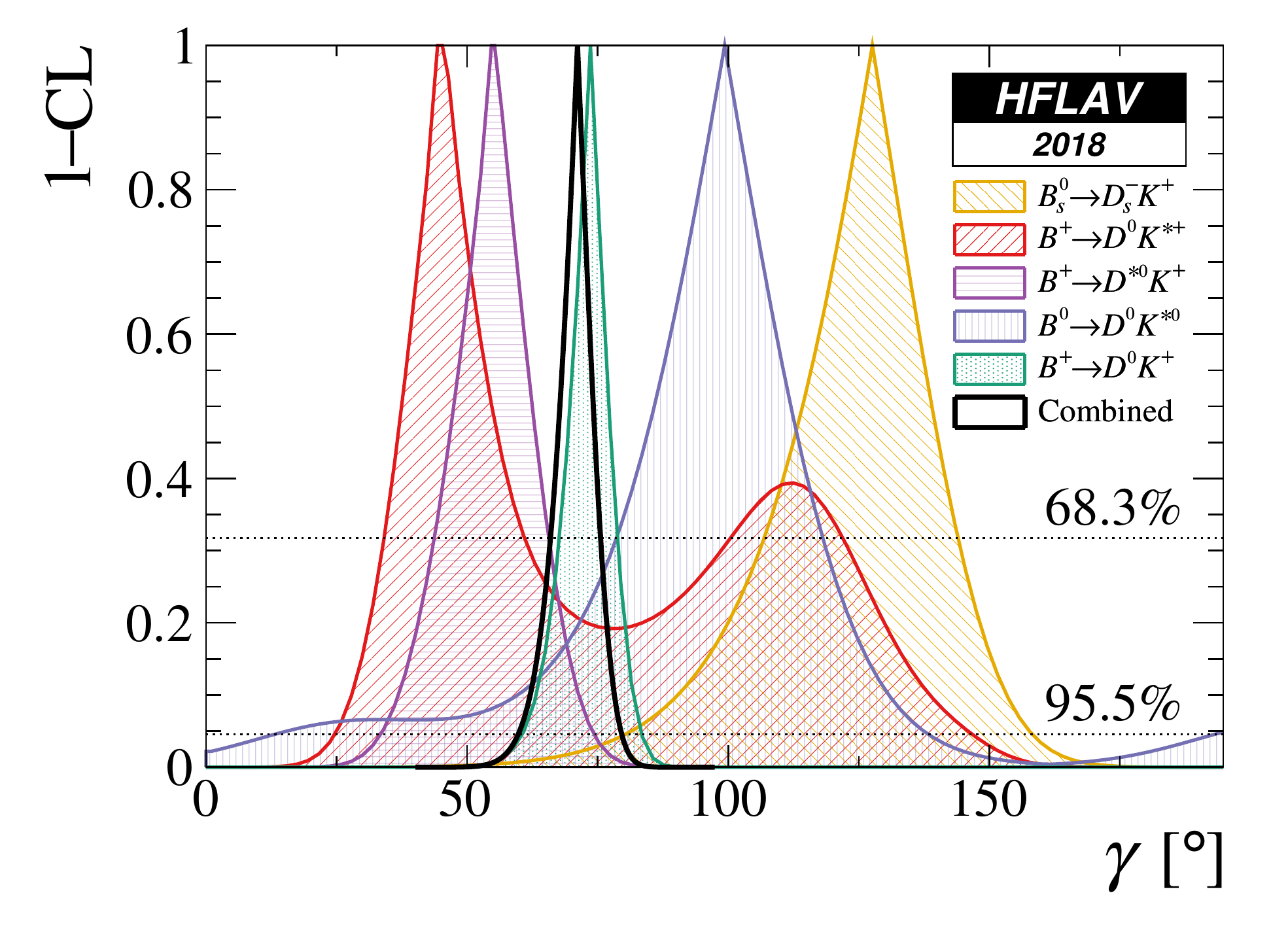}
    \caption{World average of $\gamma\equiv\phi_{3}$, in terms of 1$-$CL, split by decay mode.}
    \label{fig:cp_uta:gamma:results_mode}
\end{figure}

\begin{figure}
    \centering
    \includegraphics[width=0.6\textwidth]{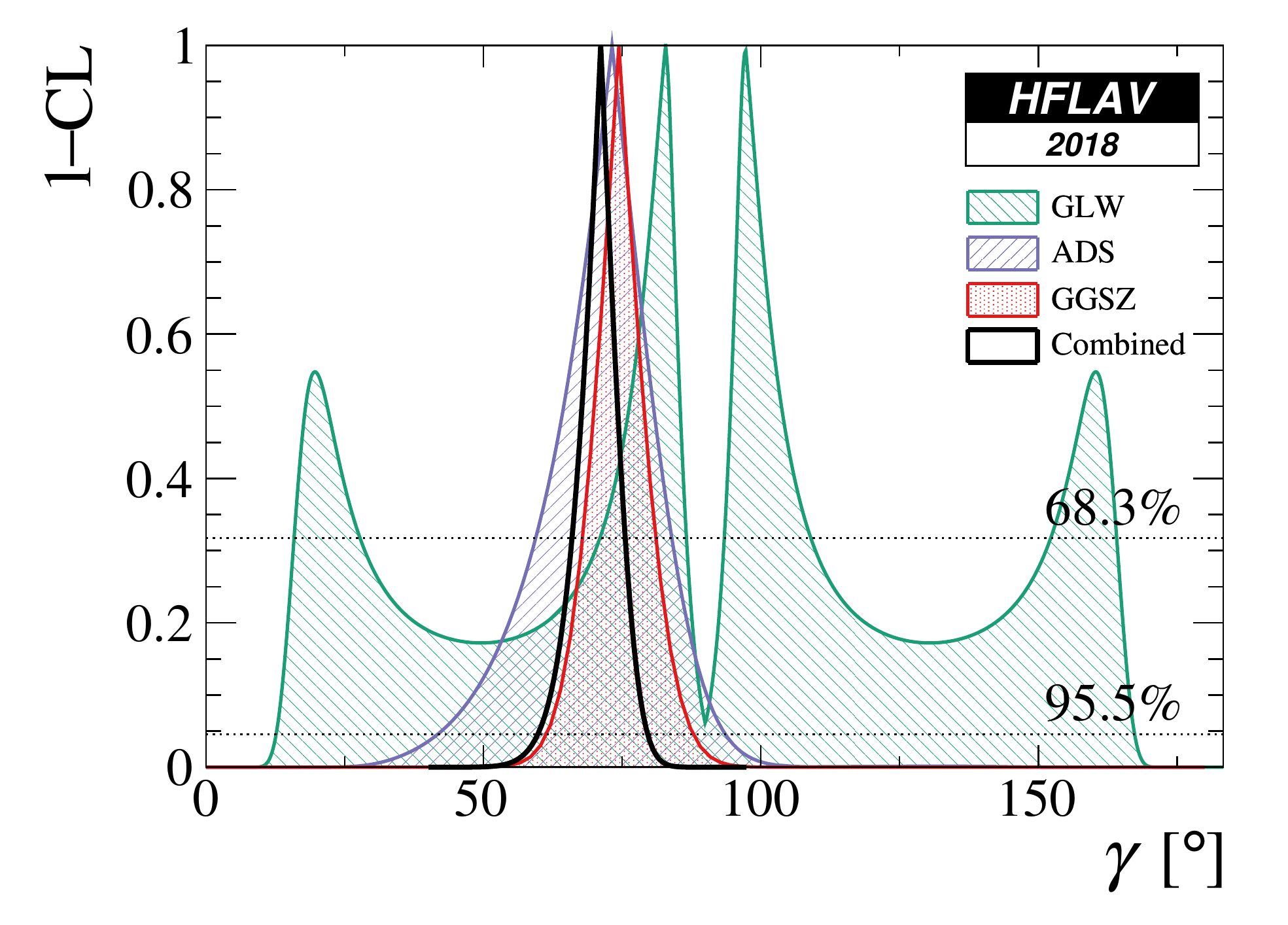}
    \caption{World average of $\gamma\equiv\phi_{3}$, in terms of 1$-$CL, split by analysis method.}
    \label{fig:cp_uta:gamma:results_method}
\end{figure}

\begin{figure}
    \centering
    \includegraphics[width=0.6\textwidth]{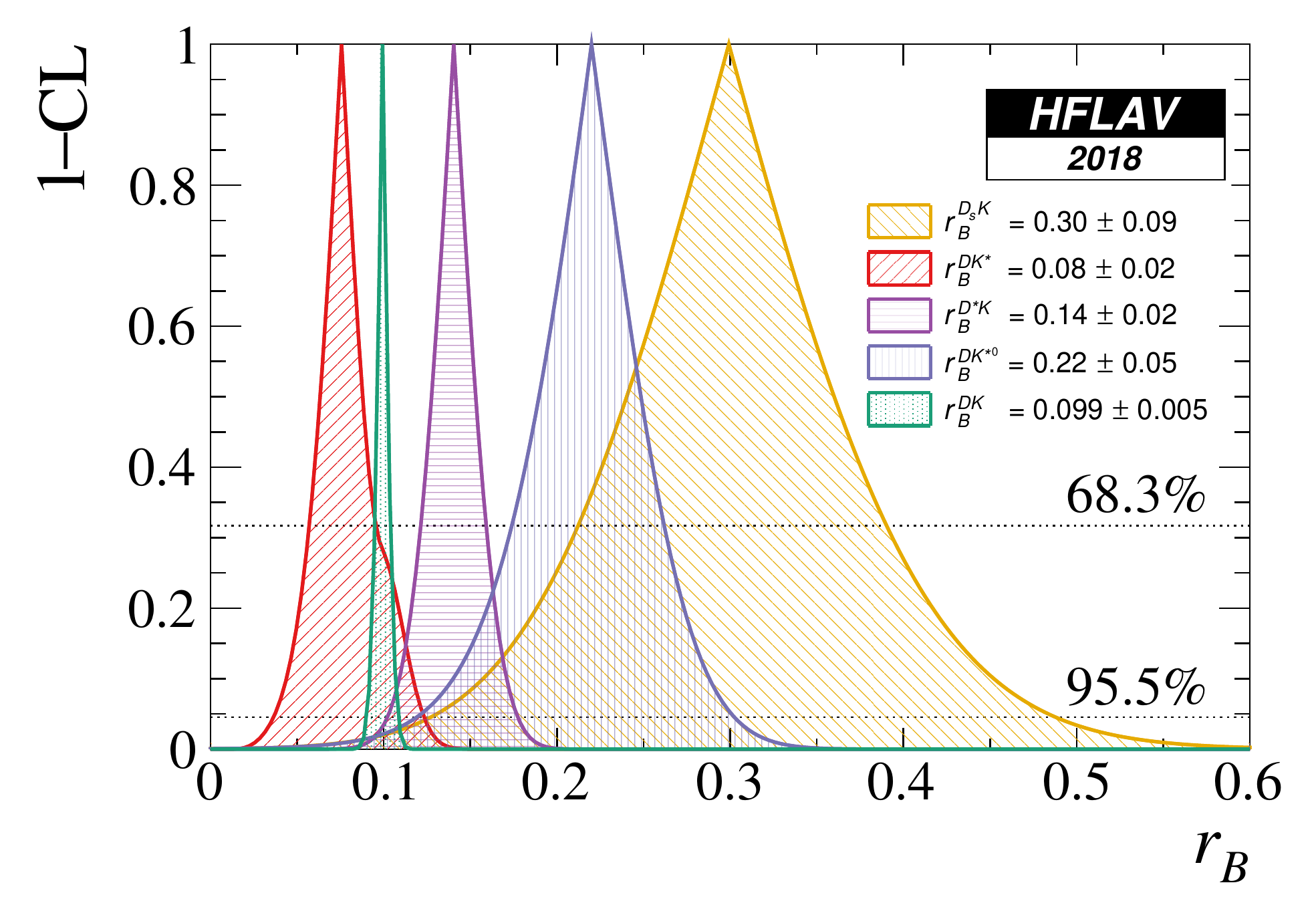}
    \caption{World averages for the hadronic parameters \rb\ in the different decay modes, in terms of 1$-$CL.}
    \label{fig:cp_uta:gamma:results_rb}
\end{figure}

\begin{figure}
    \centering
    \includegraphics[width=0.48\textwidth]{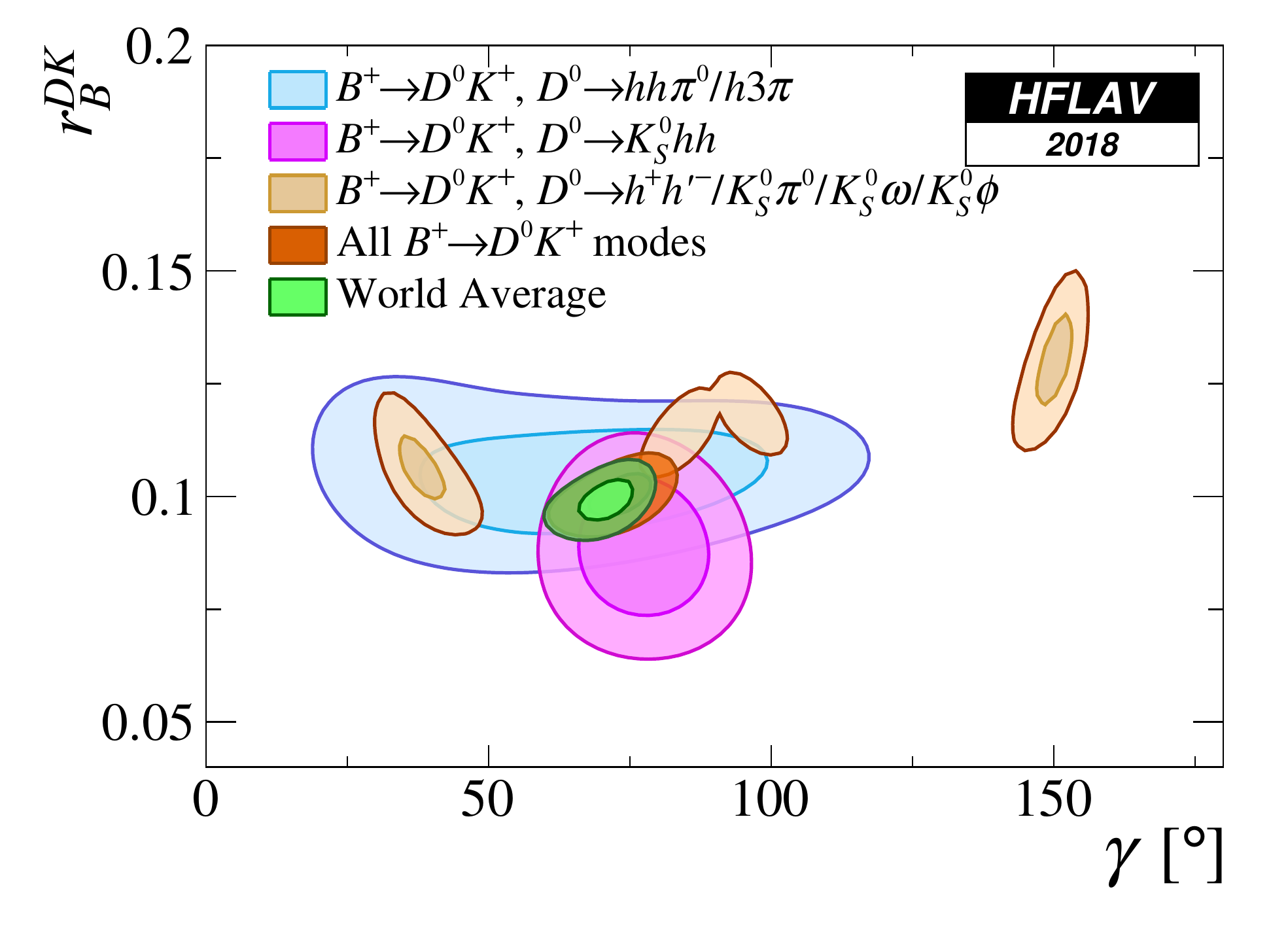}
    \includegraphics[width=0.48\textwidth]{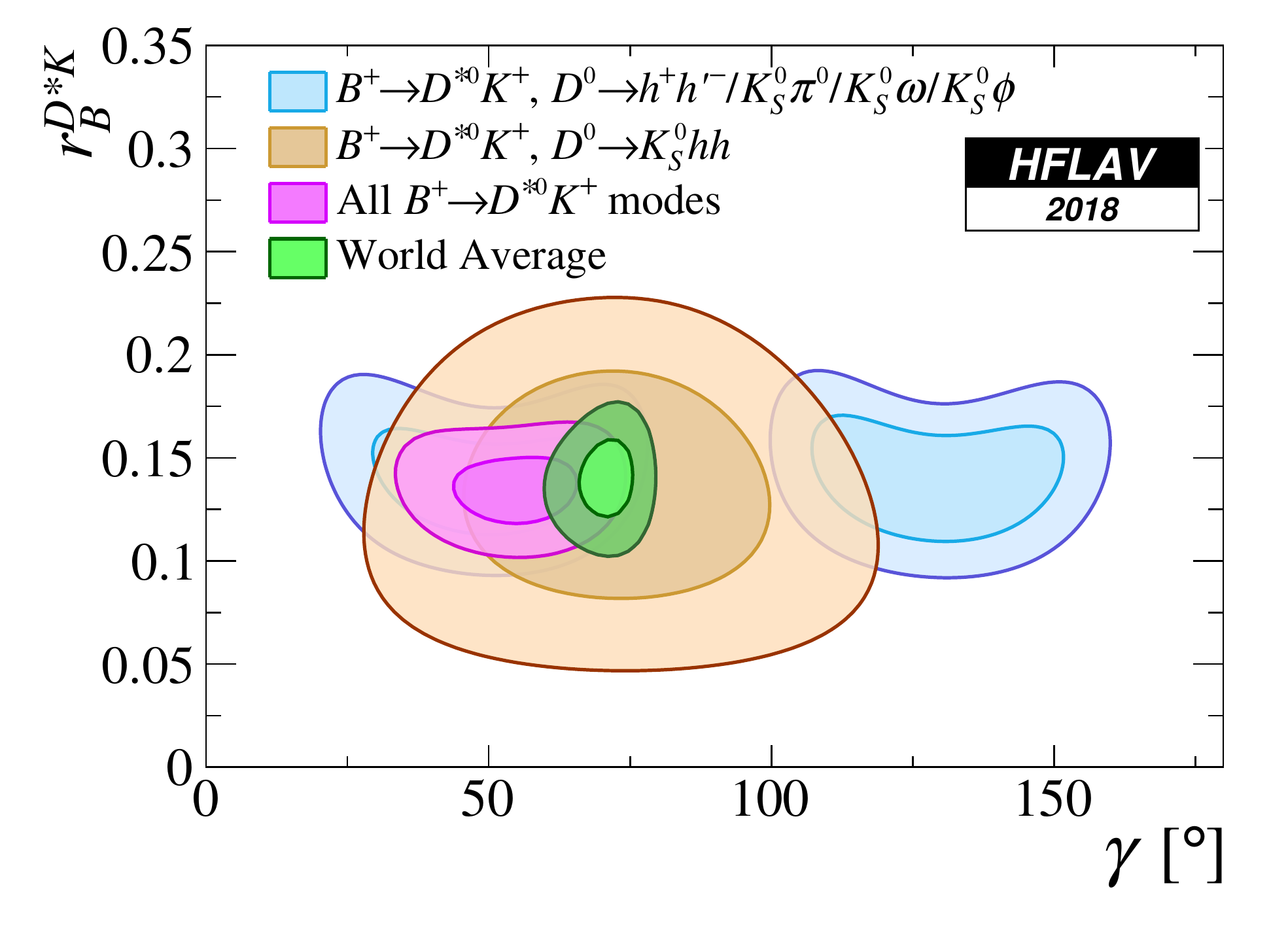} \\
    \includegraphics[width=0.48\textwidth]{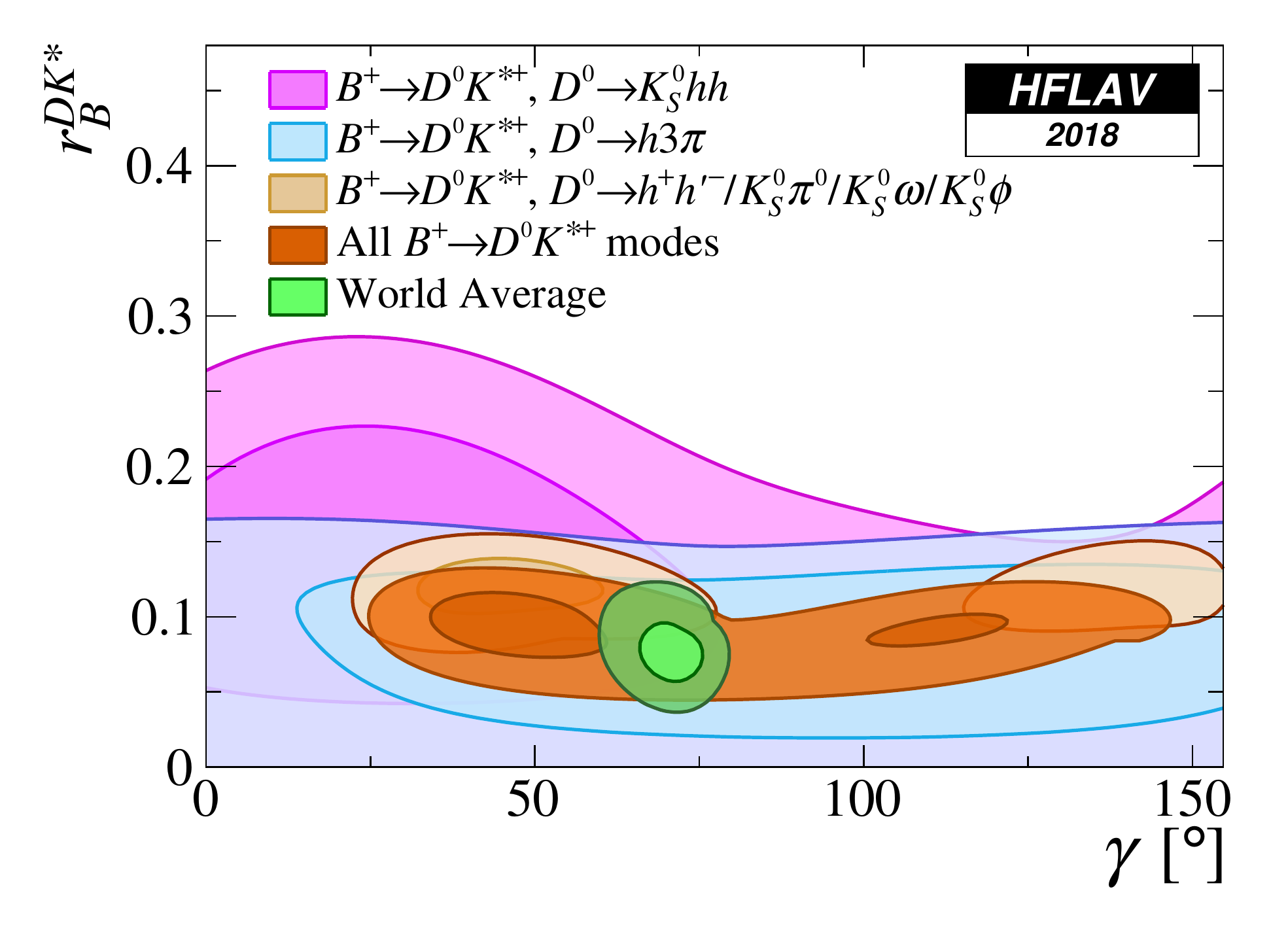}
    \includegraphics[width=0.48\textwidth]{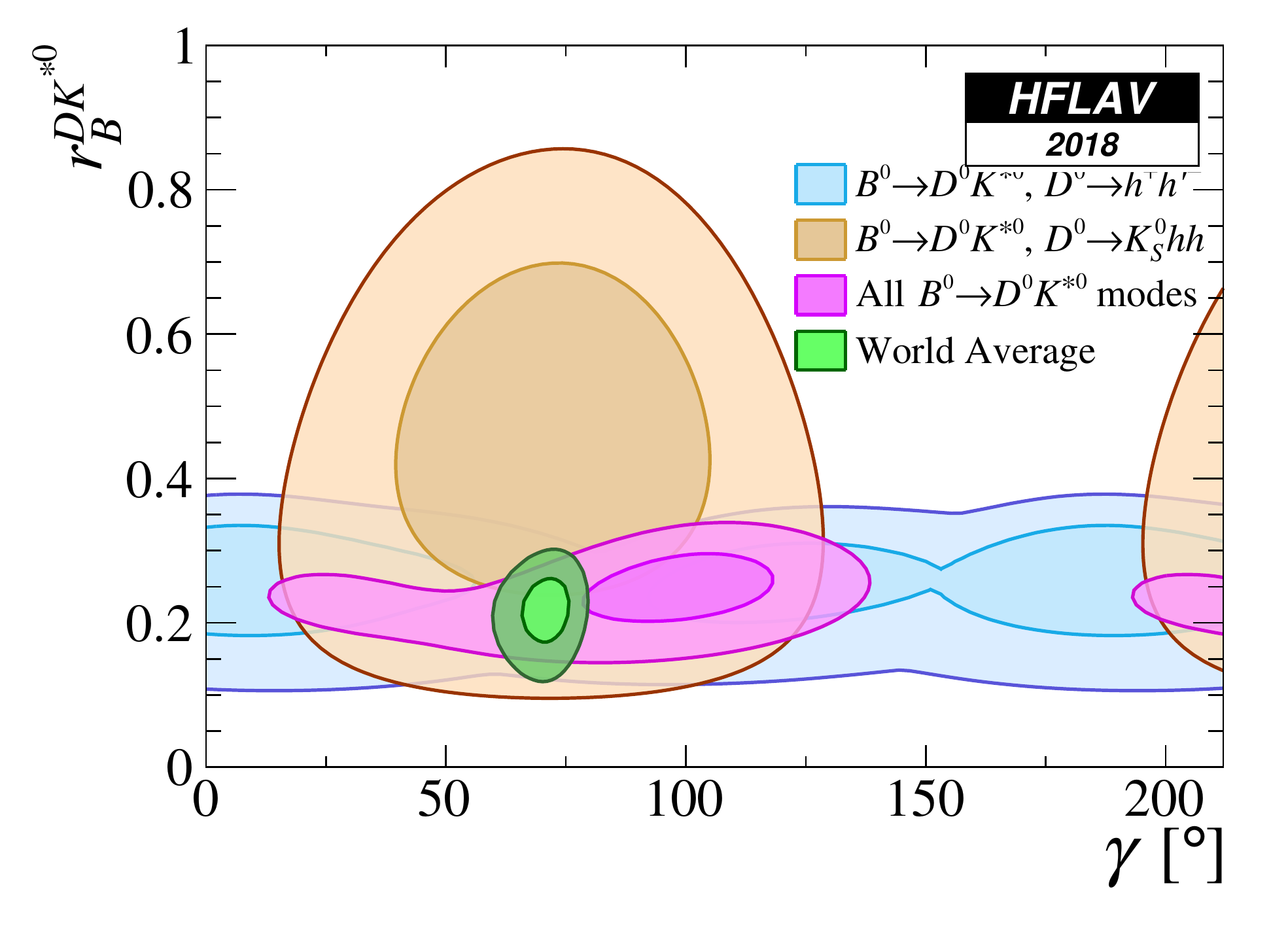}
  \caption{Contributions to the combination from different input measurements, shown in the plane of the relevant \rb parameter \vs\ $\gamma \equiv \phi_3$.
    From left to right, top to bottom: \BuDK, \BuDstK, \BuDKst\ and \BdDKstz.
    Contours show the two-dimensional $68\,\%$ and $95\,\%$ CL regions.
}
  \label{fig:cp_uta:gamma:results_contribs}
\end{figure}

In total, there are 136 observables and 29 free parameters.
The combination has a $\chi^2$ value of 123.4, which corresponds to a global p-value of 0.133.
A coverage check with pseudoexperiments gives a p-value of $(11.4 \pm 0.3)\%$.
The obtained world average for the Unitarity Triangle angle $\gamma \equiv \phi_3$ is
\begin{equation}
  \gamma \equiv \phi_3 = \left( 71.1\,^{+4.6}_{-5.3} \right)^\circ \, .
\end{equation}
An ambiguous solution at $\gamma \equiv \phi_3 \longrightarrow \gamma \equiv \phi_3+\pi$ also exists.
The results for the hadronic parameters are listed in Table~\ref{tab:cp_uta:gamma:results}.
Results for input analyses split by \B meson decay mode are shown in Table~\ref{tab:cp_uta:gamma:results_mode} and Fig.~\ref{fig:cp_uta:gamma:results_mode}.
Results for input analyses split by the method are shown in Table~\ref{tab:cp_uta:gamma:results_method} and Fig.~\ref{fig:cp_uta:gamma:results_method}.
Results for the hadronic ratios, \rb, are shown in Fig.~\ref{fig:cp_uta:gamma:results_rb}.
A demonstration of how the various analyses contribute to the combination is shown in Fig.~\ref{fig:cp_uta:gamma:results_contribs}.

\begin{table}
  \caption{Averages values obtained for the hadronic parameters in $\B \to \DorDstar\KorKstar$ decays.}
  \label{tab:cp_uta:gamma:results}
  \centering
  \renewcommand{\arraystretch}{1.1}
  \begin{tabular}{l c}
    \hline
    Parameter & Value \\
    \hline
    \rbdk     &  $0.0993 \pm 0.0046$ \\
    \rbdstk   &  $0.140 \pm 0.019$ \\
    \rbdkst   &  $0.076 \pm 0.020$ \\
    \rbdkstz  &  $0.220 \,^{+0.041}_{-0.047}$ \\
    \dbdk     &  $(129.6 \,^{+5.0}_{-6.0})^\circ$ \\
    \dbdstk   &  $(319\,^{+8}_{-9})^\circ$ \\
    \dbdkst   &  $(98 \,^{+18}_{-37})^\circ$ \\
    \dbdkstz  &  $(194\,^{+30}_{-22})^\circ$ \\
    \hline
  \end{tabular}
\end{table}

 \begin{table}
   \caption{Averages of $\gamma \equiv \phi_3$ split by \B meson decay mode.}
   \label{tab:cp_uta:gamma:results_mode}
   \centering
   \renewcommand{\arraystretch}{1.1}
   \begin{tabular}{l c}
     \hline
     Decay Mode & Value \\
     \hline
     \BsDsK   &  $(128 \,^{+18}_{-22})^\circ$ \\
     \BuDKst  &  $(45 \,^{+16}_{-12})^\circ$ \\
     \BuDstK  &  $(55 \,^{+11}_{-12})^\circ$ \\
     \BdDKstz &  $(99 \,^{+19}_{-21})^\circ$ \\
     \BuDK    &  $(73.6 \,^{+5.4}_{-6.2})^\circ$ \\
     \hline
   \end{tabular}
 \end{table}

\begin{table}
  \caption{Averages of $\gamma \equiv \phi_3$ split by method. For GLW method only the solution nearest the combined average is shown.}
  \label{tab:cp_uta:gamma:results_method}
  \centering
  \renewcommand{\arraystretch}{1.1}
  \begin{tabular}{l c}
    \hline
    Method & Value \\
    \hline
    GLW  &  $(82.8\,^{+4.9}_{-12.3})^\circ$ \\
    ADS  &  $(73\,^{+12}_{-13})^\circ$ \\
    BPGGSZ &  $(74.2\,^{+6.9}_{-6.8})^\circ$ \\
    \hline
  \end{tabular}
\end{table}

\mysubsection{Summary of the constraints on the angles of the Unitarity Triangle}
\label{sec:cp_uta:HFLAV_RhoEta}

\begin{figure}[htbp]
  \centering
  \includegraphics[width=0.6\textwidth]{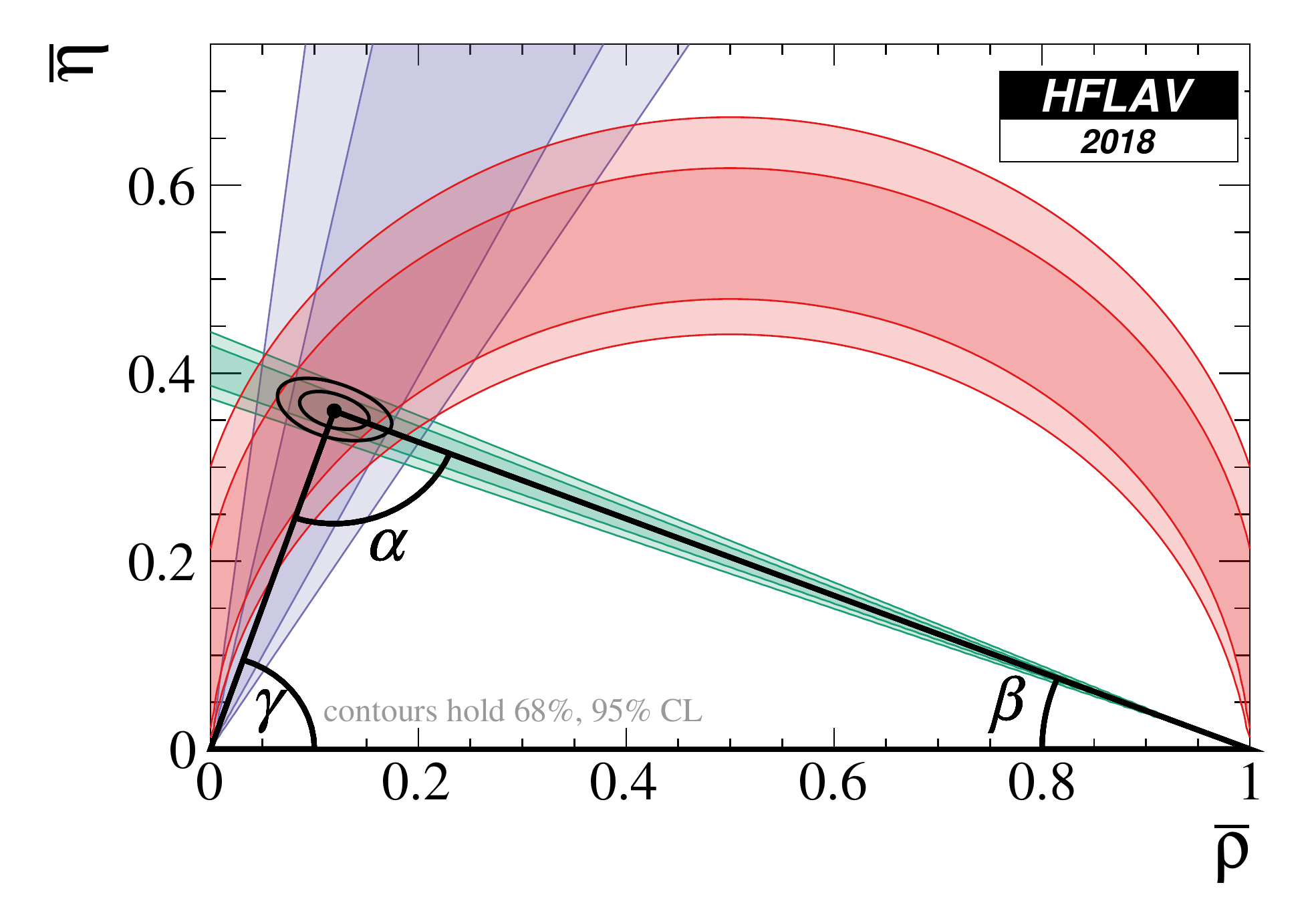}
  \caption{Summary of the constraints on the angles of the Unitarity Triangle.}
  \label{fig:cp_uta:HFLAV_RhoEta}
\end{figure}

World averages for the angles of the Unitarity Triangle $\beta \equiv \phi_1$, $\alpha \equiv \phi_2$ and $\gamma \equiv \phi_3$ are given in Sec.~\ref{sec:cp_uta:ccs:cp_eigen}, Sec.~\ref{sec:cp_uta:uud:alpha} and Sec.~\ref{sec:cp_uta:cus:gamma}, respectively.
These constraints are summarised in Fig.~\ref{fig:cp_uta:HFLAV_RhoEta} in terms of the CKM parameters $\bar{\rho}$ and $\bar{\eta}$ defined in Eq.~(\ref{eq:rhoetabar}) using the relations,
$\tan\g = \bar{\eta}/\bar{\rho}$, $\tan\beta = \bar{\eta}/(1-\bar{\rho})$, $\alpha = \tan^{-1}( \bar{\rho}/\bar{\eta}) + \tan^{-1}( (1-\bar{\rho}) / \bar{\eta})$.
The overlap of the constraints demonstrates agreement with the unitarity of the CKM matrix as predicted in the Standard Model.
The obtained values of $\bar{\rho}$ and $\bar{\eta}$ from this angles only combination are
\begin{equation}
 \bar{\rho} = 0.119 \pm 0.022\,, \qquad \bar{\eta} = 0.360\pm0.013\,,
\end{equation}
with a correlation of $-0.42$.

\clearpage

\section{Semileptonic $B$ decays}
\label{sec:slbdecays}

This section contains averages for semileptonic $B$~meson decays,
\ie\ decays of the type $B\to X\ell\nu_\ell$, where $X$ refers to one
or more hadrons, $\ell$ to a charged lepton and $\nu_\ell$ to its associated
neutrino. Unless otherwise stated, $\ell$ stands for an electron
or a muon, lepton universality is assumed, and both charge
conjugate states are combined. Some averages assume isospin symmetry,
explicitly mentioned at every instance.

Averages are presented separately for CKM favored $b\to c$ quark transitions
and CKM suppressed $b\to u$ transitions. We further
distinguish \emph{exclusive} decays involving a specific meson ($X=D, D^*, \pi,
\rho,\dots$) from \emph{inclusive} decay modes, \ie\ the sum over all possible
hadronic states. Semileptonic decays proceed
via first order weak interactions and are well described in the framework of
the SM. Their decay rates are sensitive to the magnitude
squared of the CKM elements $V_{cb}$ and $V_{ub}$, the determination of which is
one of the primary goals for the study of these decays. Semileptonic decays
involving the $\tau$~lepton might be more sensitive to beyond-SM processes, because
the high $\tau$~mass might result in enhanced couplings 
to a hypothetical charged Higgs boson or leptoquarks.

The technique for obtaining the averages follows the general HFLAV
procedure (Sec.~\ref{sec:method}) unless otherwise stated. More
information on the averages, in particular on the common input parameters,
is available on the HFLAV semileptonic webpage~\cite{HFLAV_sl:webpage}. In general, averages in this
section use experimental results available through September 2018. Some
averages include more recent results and the corresponding figures are labeled \emph{Spring 2019} for easier identification.
\subsection{Exclusive CKM-favoured decays}
\label{slbdecays_b2cexcl}

\mysubsubsection{$\bar B\to D^*\ell^-\bar\nu_\ell$}
\label{slbdecays_dstarlnu}

$\bar B\to D^*\ell^-\bar\nu_\ell$
decays are described in terms of the recoil variable  $w=v_B\cdot v_{D^{(*)}}$, the product of the four-velocities of the initial and final state
mesons. The differential decay rate for massless
fermions as a function of $w$ is given by (see, \eg,~\cite{Neubert:1993mb})
\begin{equation}
  \frac{d\Gamma(\bar B\to D^*\ell^-\bar\nu_\ell)}{dw} = \frac{G^2_\mathrm{F} m^3_{D^*}}{48\pi^3}(m_B-m_{D^*})^2\chi(w)\eta_\mathrm{EW}^2\mathcal{F}^2(w)\vcb^2~,
  \label{eq:gamma_dslnu}
\end{equation}
where $G_\mathrm{F}$ is Fermi's constant, $m_B$ and $m_{D^*}$ are the $B$ and
$D^*$ meson masses, $\chi(w)$ is a known phase-space factor, and
$\eta_\mathrm{EW}$ is a small electroweak correction~\cite{Sirlin:1981ie}.
Some authors also include a long-distance EM radiation effect (Coulomb
correction) in this factor.
The form factor $\mathcal{F}(w)$ for the $\bar B\to D^*\ell^-\bar\nu_\ell$
decay contains three independent functions, $h_{A_1}(w)$, $R_1(w)$ and $R_2(w)$,
\begin{eqnarray}
  &&\chi(w)\mathcal{F}^2(w)=\\
  && \phantom{\mathcal{F}^2(w)} h_{A_1}^2(w)\sqrt{w^2-1}(w+1)^2 \left\{2\left[\frac{1-2wr+r^2}{(1-r)^2}\right]\left[1+R^2_1(w)\frac{w-1}{w+1}\right]+\right. \nonumber\\
  && \phantom{\mathcal{F}^2(w)} \left.\left[1+(1-R_2(w))\frac{w-1}{1-r}\right]^2\right\}~, \nonumber
\end{eqnarray}
where $r=m_{D^*}/m_B$.

\mysubsubsubsection{Branching fraction}

\begin{figure}[!ht]
  \begin{center}
  \unitlength1.0cm %
  \begin{picture}(14.,9.0)  %
    \put( -1.5, 0.0){\includegraphics[width=9.0cm]{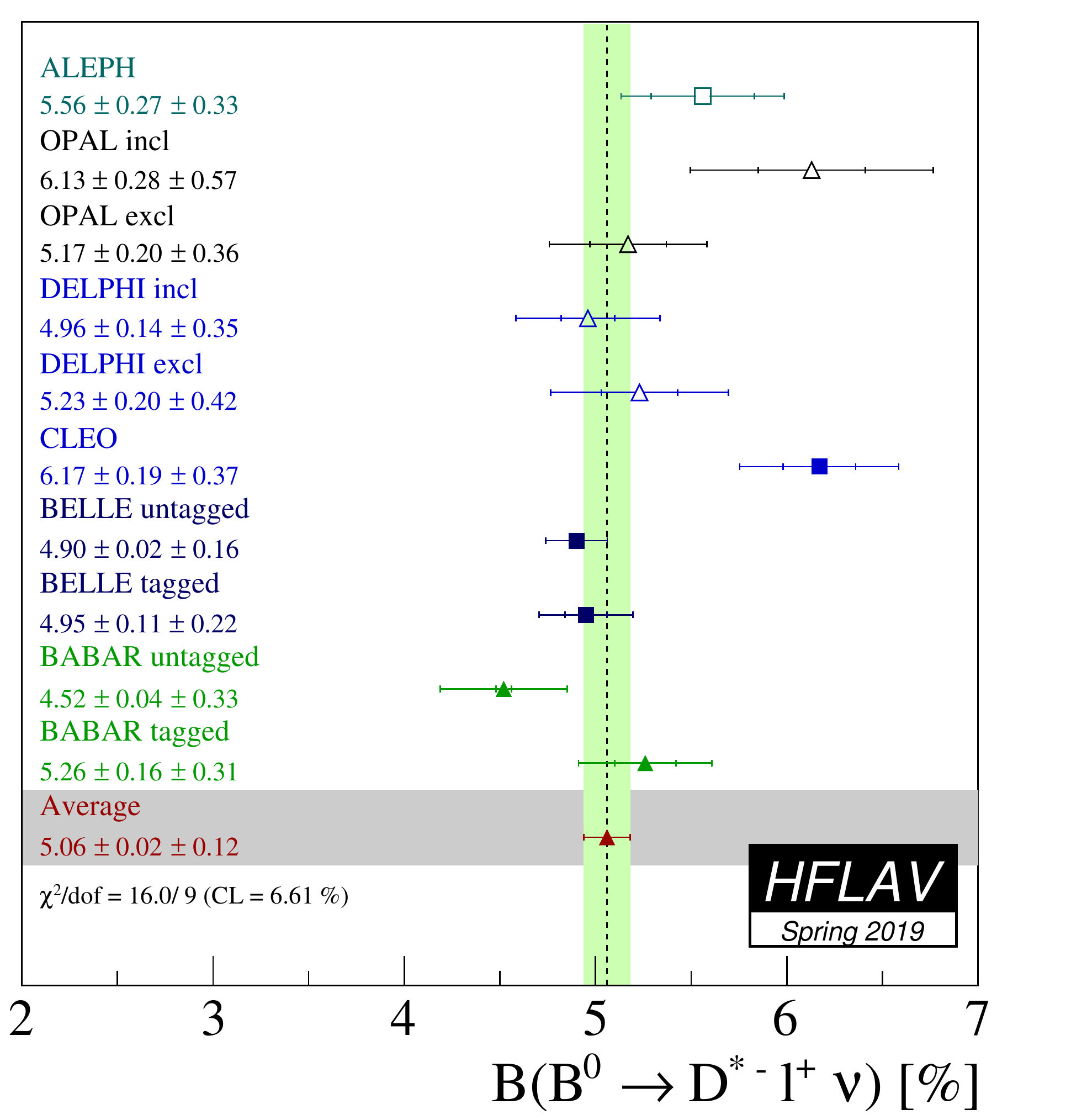}
    }
    \put(  7.5, 0.0){\includegraphics[width=9.0cm]{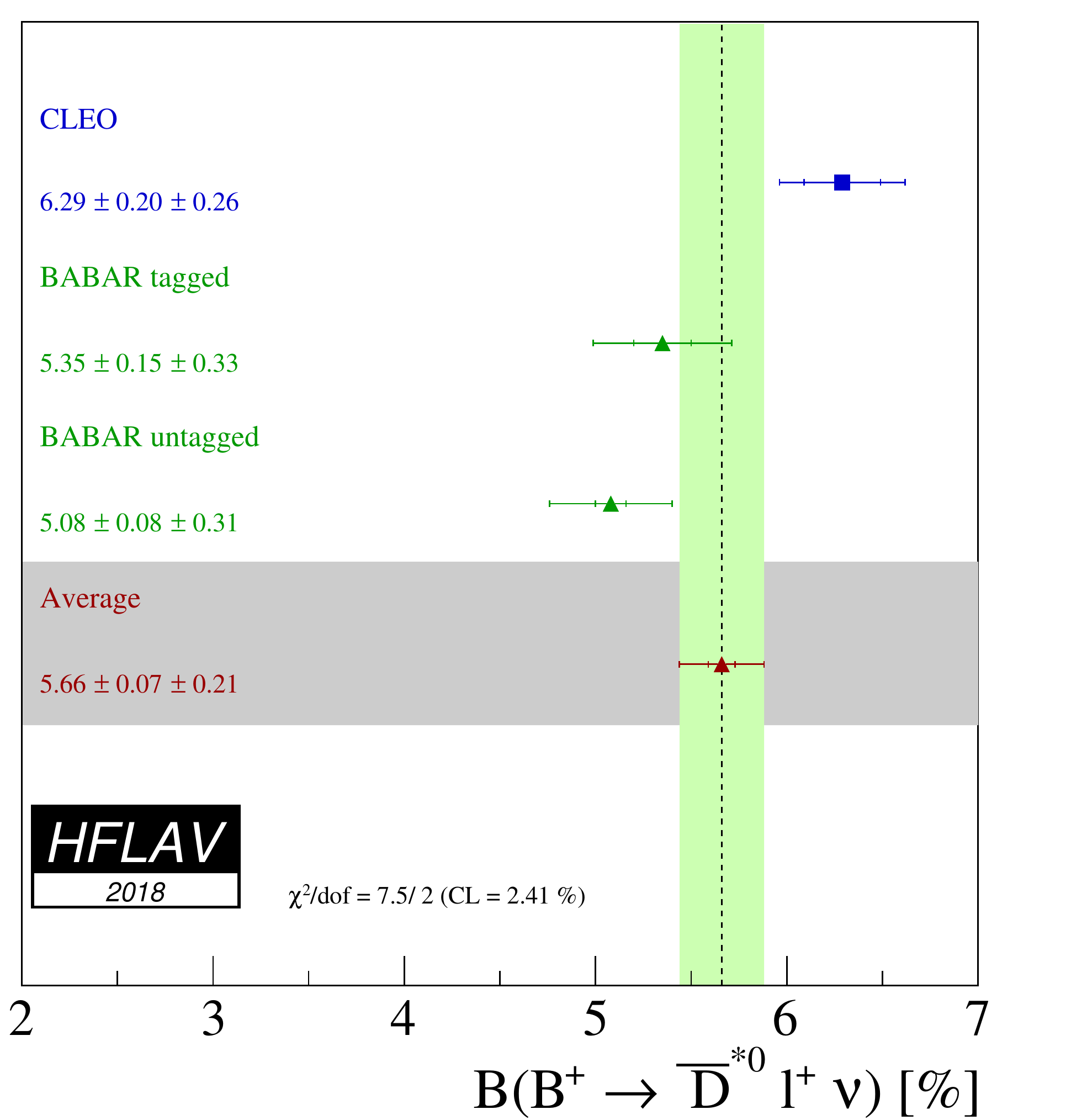}
    }
    \put(  5.8, 8.5){{\large\bf a)}}
    \put( 14.8, 8.5){{\large\bf b)}}
  \end{picture}
  \caption{Branching fractions of exclusive semileptonic $B$ decays:
    (a) $\BzbDstarlnu$ (Table~\ref{tab:dstarlnu}) and (b) $B^-\to
    D^{*0}\ell^-\bar\nu_\ell$ (Table~\ref{tab:dstar0lnu}).} \label{fig:brdsl}
  \end{center}
\end{figure}

First, we perform separate one-dimensional averages of the
$\BzbDstarlnu$ and $B^-\to D^{*0}\ell^-\bar\nu_\ell$
branching fractions. The measurements listed in Tables~\ref{tab:dstarlnu}
and \ref{tab:dstar0lnu} are rescaled to the latest values of the input
parameters (mainly branching fractions of charmed
mesons)~\cite{HFLAV_sl:inputparams} and the following results are obtained
\begin{eqnarray}
  \cbf(\BzbDstarlnu) & = & (5.06\pm 0.02\pm 0.12)\%~, \label{eq:br_dstarlnu} \\
  \cbf(B^-\to D^{*0}\ell^-\bar\nu_\ell) & = & (5.66\pm 0.07\pm 0.21)\%~,
   \label{eq:br_dstar0lnu}
\end{eqnarray}
where the first uncertainty is statistical and the second one is systematic.
The results of these two fits are also shown in Fig.~\ref{fig:brdsl}.
\begin{table}[!htb]
  \caption{Average of the $\BzbDstarlnu$ branching fraction measurements.}
  \begin{center} 
\resizebox{0.99\textwidth}{!}{
\begin{tabular}{|l|c|c|}\hline
  Experiment & $\cbf(\BzbDstarlnu)$ [\%] (calculated) &
  $\cbf(\BzbDstarlnu)$ [\%] (published)\\
  \hline\hline
  ALEPH~\cite{Buskulic:1996yq}
  & $5.56\pm 0.27_{\rm stat} \pm 0.33_{\rm syst}$
  & $5.53\pm 0.26_{\rm stat} \pm 0.52_{\rm syst}$\\
  OPAL incl~\cite{Abbiendi:2000hk}
  & $6.13\pm 0.28_{\rm stat} \pm 0.57_{\rm syst}$
  & $5.92\pm 0.27_{\rm stat} \pm 0.68_{\rm syst}$\\
  OPAL excl~\cite{Abbiendi:2000hk}
  & $5.17\pm 0.20_{\rm stat} \pm 0.36_{\rm syst}$
  & $5.11\pm 0.19_{\rm stat} \pm 0.49_{\rm syst}$\\
  DELPHI incl~\cite{Abreu:2001ic}
  & $4.96\pm 0.14_{\rm stat} \pm 0.35_{\rm syst}$
  & $4.70\pm 0.13_{\rm stat} \ {}^{+0.36}_{-0.31}\ {}_{\rm syst}$\\
  DELPHI excl~\cite{Abdallah:2004rz}
  & $5.23\pm 0.20_{\rm stat} \pm 0.42_{\rm syst}$
  & $5.90\pm 0.22_{\rm stat} \pm 0.50_{\rm syst}$\\
  CLEO~\cite{Adam:2002uw}
  & $6.17\pm 0.19_{\rm stat} \pm 0.37_{\rm syst}$
  & $6.09\pm 0.19_{\rm stat} \pm 0.40_{\rm syst}$\\
  Belle untagged~\cite{Waheed:2018djm}
  & $4.90\pm 0.02_{\rm stat} \pm 0.16_{\rm syst}$
  & $4.90\pm 0.02_{\rm stat} \pm 0.16_{\rm syst}$\\
  Belle tagged~\cite{Abdesselam:2017kjf}
  & $4.95\pm 0.11_{\rm stat} \pm 0.22_{\rm syst}$
  & $4.95\pm 0.11_{\rm stat} \pm 0.22_{\rm syst}$\\
  \babar\ untagged~\cite{Aubert:2006mb}
  & $4.52\pm 0.04_{\rm stat}\pm 0.33_{\rm syst}$
  & $4.69\pm 0.04_{\rm stat} \pm 0.34_{\rm syst}$\\
  \babar\ tagged~\cite{Aubert:vcbExcl}
  & $5.26\pm 0.16_{\rm stat}\pm 0.31_{\rm syst}$
  & $5.49\pm 0.16_{\rm stat} \pm 0.25_{\rm syst}$\\
  \hline 
  {\bf Average} & \mathversion{bold}$5.06\pm 0.02_{\rm stat}\pm
  0.12_{\rm syst}$ & \mathversion{bold}$\chi^2/\dof = 16.0/9$ (CL=$6.61\%$)\\
  \hline 
\end{tabular}
}
\end{center}
\label{tab:dstarlnu}
\end{table}
\begin{table}[!htb]
\caption{Average of the $B^-\to D^{*0}\ell^-\bar\nu_\ell$ branching
  fraction measurements.}
\begin{center}
\begin{tabular}{|l|c|c|}
  \hline
  Experiment & $\cbf(B^-\to D^{*0}\ell^-\bar\nu_\ell)$ [\%] (rescaled) &
  $\cbf(B^-\to D^{*0}\ell^-\bar\nu_\ell)$ [\%] (published)\\
  \hline \hline
  CLEO~\cite{Adam:2002uw}
  & $6.29\pm 0.20_{\rm stat}\pm 0.26_{\rm syst}$
  & $6.50\pm 0.20_{\rm stat}\pm 0.43_{\rm syst}$\\
  \babar tagged~\cite{Aubert:vcbExcl}
  & $5.35\pm 0.15_{\rm stat}\pm 0.33_{\rm syst}$
  & $5.83\pm 0.15_{\rm stat}\pm 0.30_{\rm syst}$\\
  \babar~untagged\cite{Aubert:2009_3}
  & $5.08\pm 0.08_{\rm stat}\pm 0.31_{\rm syst}$
  & $5.56\pm 0.08_{\rm stat}\pm 0.41_{\rm syst}$\\
  \hline
  {\bf Average} & \mathversion{bold}$5.66\pm 0.07_{\rm stat}\pm
  0.21_{\rm syst}$ & \mathversion{bold}$\chi^2/\dof = 7.45/2$ (CL=$2.41\%$)\\
  \hline 
\end{tabular}
\end{center}
\label{tab:dstar0lnu}
\end{table}

\mysubsubsubsection{Extraction of $\vcb$ based on the CLN form factor}

To extract $\vcb$, we consider the parametrizations of the form factor
functions $h_{A_1}(w)$, $R_1(w)$ and $R_2(w)$ by Caprini, Lellouch and
Neubert (CLN)~\cite{CLN},
\begin{eqnarray}
  h_{A_1}(w) & = &
  h_{A_1}(1)\big[1-8\rho^2z+(53\rho^2-15)z^2-(231\rho^2-91)z^3\big]~, \\
  R_1(w) & = & R_1(1)-0.12(w-1)+0.05(w-1)^2~, \\ 
  R_2(w) & = & R_2(1)+0.11(w-1)-0.06(w-1)^{2}~,
\end{eqnarray}
where $z=(\sqrt{w+1}-\sqrt{2})/(\sqrt{w+1}+\sqrt{2})$. The form factor
${\cal F}(w)$ in Eq.~\ref{eq:gamma_dslnu} is thus described by the slope
$\rho^2$ and the ratios $R_1(1)$ and $R_2(1)$.

We use the measurements of these form factor parameters, shown in
Table~\ref{tab:vcbf1}, and rescale them as described above. Most of the
measurements in Table~\ref{tab:vcbf1} are based on the decay $\bar B^0\to
D^{*+}\ell^-\bar\nu_\ell$. Some measurements~\cite{Adam:2002uw,Aubert:2009_1}
are sensitive also to  $B^-\to D^{*0}\ell^-\bar\nu_\ell$, and one
measurement~\cite{Aubert:2009_3} is based on the decay
$B^-\to D^{*0}\ell^-\bar\nu_\ell$. Isospin symmetry is assumed in this average.
We note that the earlier results from the LEP experiments and CLEO required significant
rescaling and have significantly larger uncertainties than the recent
measurements by Belle and \babar.
\begin{table}[!htb]
\caption{Measurements of the Caprini, Lellouch and Neubert
  (CLN)~\cite{CLN} form factor parameters in $\bar B\to
  D^*\ell^-\bar\nu_\ell$ before and after rescaling. Most analyses
  (except \cite{Aubert:2006mb}) measure only
  $\eta_\mathrm{EW}{\cal F}(1)\vcb$, and $\rho^2$, so only these two
  parameters are shown here.}
\begin{center}
\resizebox{0.99\textwidth}{!}{
\begin{tabular}{|l|c|c|}
  \hline
  Experiment
  & $\eta_\mathrm{EW}{\cal F}(1)\vcb [10^{-3}]$ (rescaled)
  & $\rho^2$ (rescaled)\\
  & $\eta_\mathrm{EW}{\cal F}(1)\vcb [10^{-3}]$ (published)
  & $\rho^2$ (published)\\
  \hline\hline
  ALEPH~\cite{Buskulic:1996yq}
  & $31.78\pm 1.83_{\rm stat}\pm 1.21_{\rm syst}$
  & $0.489\pm 0.226_{\rm stat}\pm 0.145_{\rm syst}$\\
  & $31.9\pm 1.8_{\rm stat}\pm 1.9_{\rm syst}$
  & $0.37\pm 0.26_{\rm stat}\pm 0.14_{\rm syst}$\\
  \hline
  CLEO~\cite{Adam:2002uw}
  & $40.47\pm 1.25_{\rm stat}\pm 1.55_{\rm syst}$
  & $1.363\pm 0.084_{\rm stat}\pm 0.087_{\rm syst}$\\
  & $43.1\pm 1.3_{\rm stat}\pm 1.8_{\rm syst}$
  & $1.61\pm 0.09_{\rm stat}\pm 0.21_{\rm syst}$\\
  \hline
  OPAL excl~\cite{Abbiendi:2000hk}
  & $36.50\pm 1.60_{\rm stat}\pm 1.46_{\rm syst}$
  & $1.212\pm 0.209_{\rm stat}\pm 0.148_{\rm syst}$\\
  & $36.8\pm 1.6_{\rm stat}\pm 2.0_{\rm syst}$
  & $1.31\pm 0.21_{\rm stat}\pm 0.16_{\rm syst}$\\
  \hline
  OPAL partial reco~\cite{Abbiendi:2000hk}
  & $37.44\pm 1.20_{\rm stat}\pm 2.32_{\rm syst}$
  & $1.091\pm 0.138_{\rm stat}\pm 0.297_{\rm syst}$\\
  & $37.5\pm 1.2_{\rm stat}\pm 2.5_{\rm syst}$
  & $1.12\pm 0.14_{\rm stat}\pm 0.29_{\rm syst}$\\
  \hline
  DELPHI partial reco~\cite{Abreu:2001ic}
  & $35.64\pm 1.41_{\rm stat}\pm 2.29_{\rm syst}$
  & $1.144\pm 0.123_{\rm stat} \pm 0.381_{\rm syst}$\\
  & $35.5\pm 1.4_{\rm stat}\ {}^{+2.3}_{-2.4}{}_{\rm syst}$
  & $1.34\pm 0.14_{\rm stat}\ {}^{+0.24}_{-0.22}{}_{\rm syst}$\\
  \hline
  DELPHI excl~\cite{Abdallah:2004rz}
  & $36.29\pm 1.71_{\rm stat}\pm 1.94_{\rm syst}$
  & $1.079\pm 0.142_{\rm stat} \pm 0.152_{\rm syst}$\\
  & $39.2\pm 1.8_{\rm stat}\pm 2.3_{\rm syst}$
  & $1.32\pm 0.15_{\rm stat}\pm 0.33_{\rm syst}$\\
  \hline
  \belle~\cite{Waheed:2018djm}
  & $35.07\pm 0.15_{\rm stat}\pm 0.56_{\rm syst}$
  & $1.106\pm 0.031_{\rm stat}\pm 0.008_{\rm syst}$\\
  & $35.06\pm 0.15_{\rm stat}\pm 0.56_{\rm syst}$
  & $1.106\pm 0.031_{\rm stat} \pm 0.007_{\rm syst}$\\
  \hline
  \babar\ excl~\cite{Aubert:2006mb}
  & $33.77\pm 0.29_{\rm stat}\pm 0.98_{\rm syst}$
  & $1.184\pm 0.048_{\rm stat}\pm 0.029_{\rm syst}$\\
  & $34.7\pm 0.3_{\rm stat}\pm 1.1_{\rm syst}$
  & $1.18\pm 0.05_{\rm stat}\pm 0.03_{\rm syst}$\\
  \hline
  \babar\ $D^{*0}$~\cite{Aubert:2009_3}
  & $34.81\pm 0.58_{\rm stat}\pm 1.06_{\rm syst}$
  & $1.125\pm 0.058_{\rm stat}\pm 0.053_{\rm syst}$\\
  & $35.9\pm 0.6_{\rm stat}\pm 1.4_{\rm syst}$
  & $1.16\pm 0.06_{\rm stat}\pm 0.08_{\rm syst}$\\
  \hline
  \babar\ global fit~\cite{Aubert:2009_1}
  & $35.75\pm 0.20_{\rm stat}\pm 1.09_{\rm syst}$
  & $1.180\pm 0.020_{\rm stat}\pm 0.061_{\rm syst}$\\
  & $35.7\pm 0.2_{\rm stat}\pm 1.2_{\rm syst}$
  & $1.21\pm 0.02_{\rm stat}\pm 0.07_{\rm syst}$\\
  \hline
  {\bf Average}
  & \mathversion{bold} $35.27\pm 0.11_{\rm stat}\pm 0.36_{\rm syst}$ &
  \mathversion{bold} $1.122\pm 0.015_{\rm stat}\pm 0.019_{\rm syst}$\\
  \hline 
\end{tabular}
}
\end{center}
\label{tab:vcbf1}
\end{table}

\begin{figure}[!ht]
  \begin{center}
  \unitlength 1.0cm %
  \begin{picture}(14.,7.0)
    \put( 10.9,-0.2){\includegraphics[width=6.2cm]{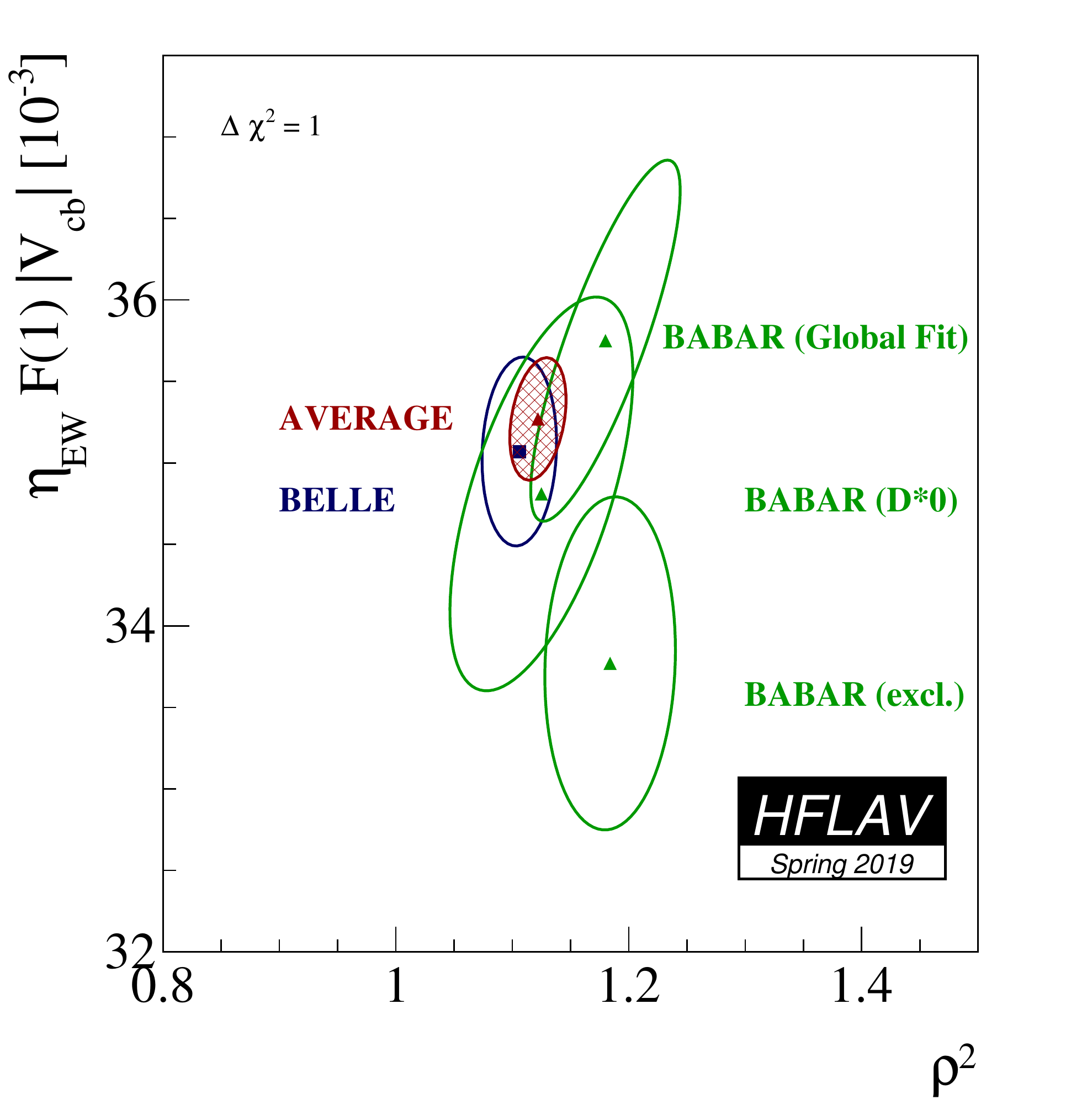}
    }
    \put(  4.7,-0.2){\includegraphics[width=6.2cm]{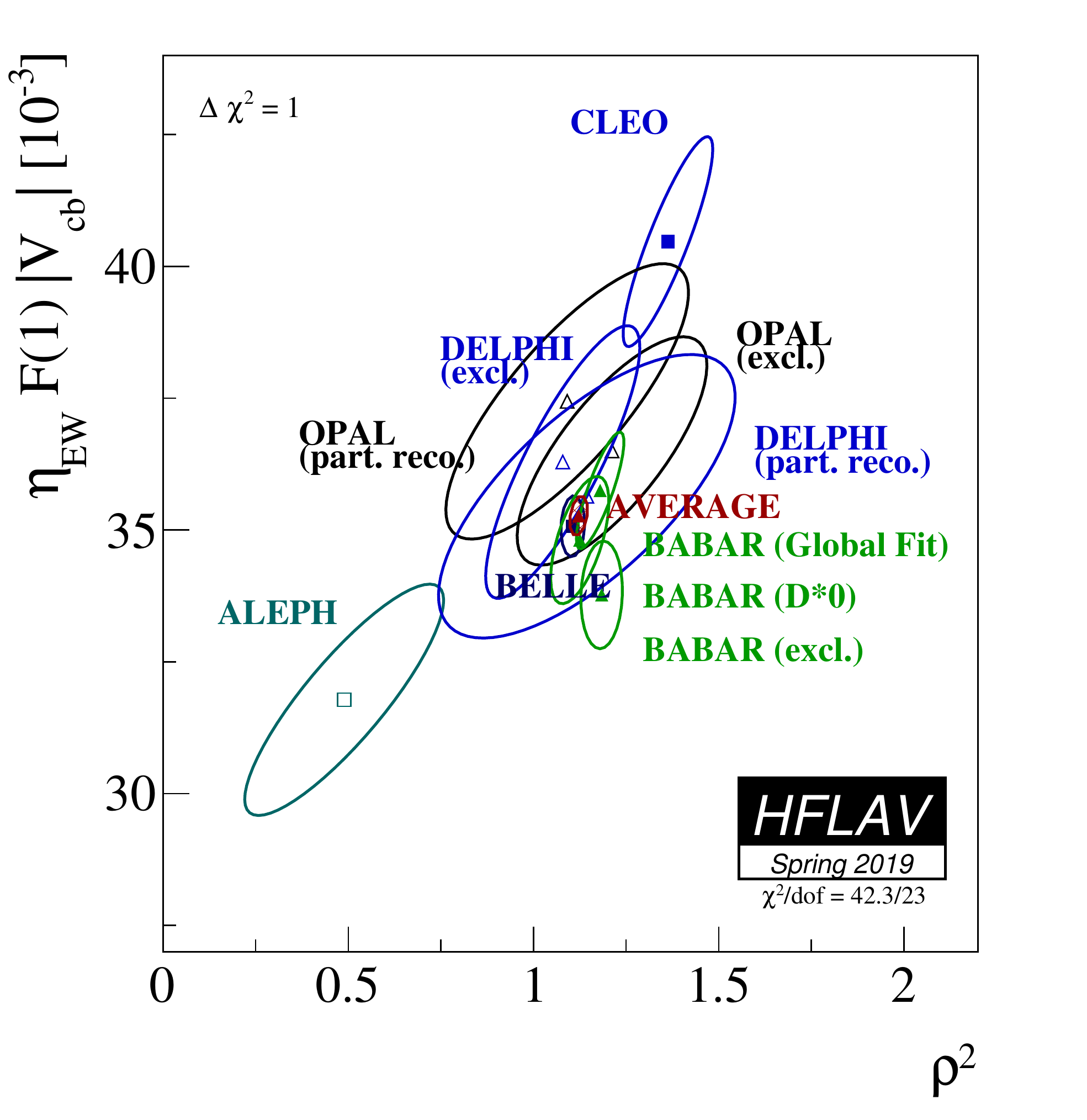}
    }
    \put( -1.5, 0.0){\includegraphics[width=5.9cm]{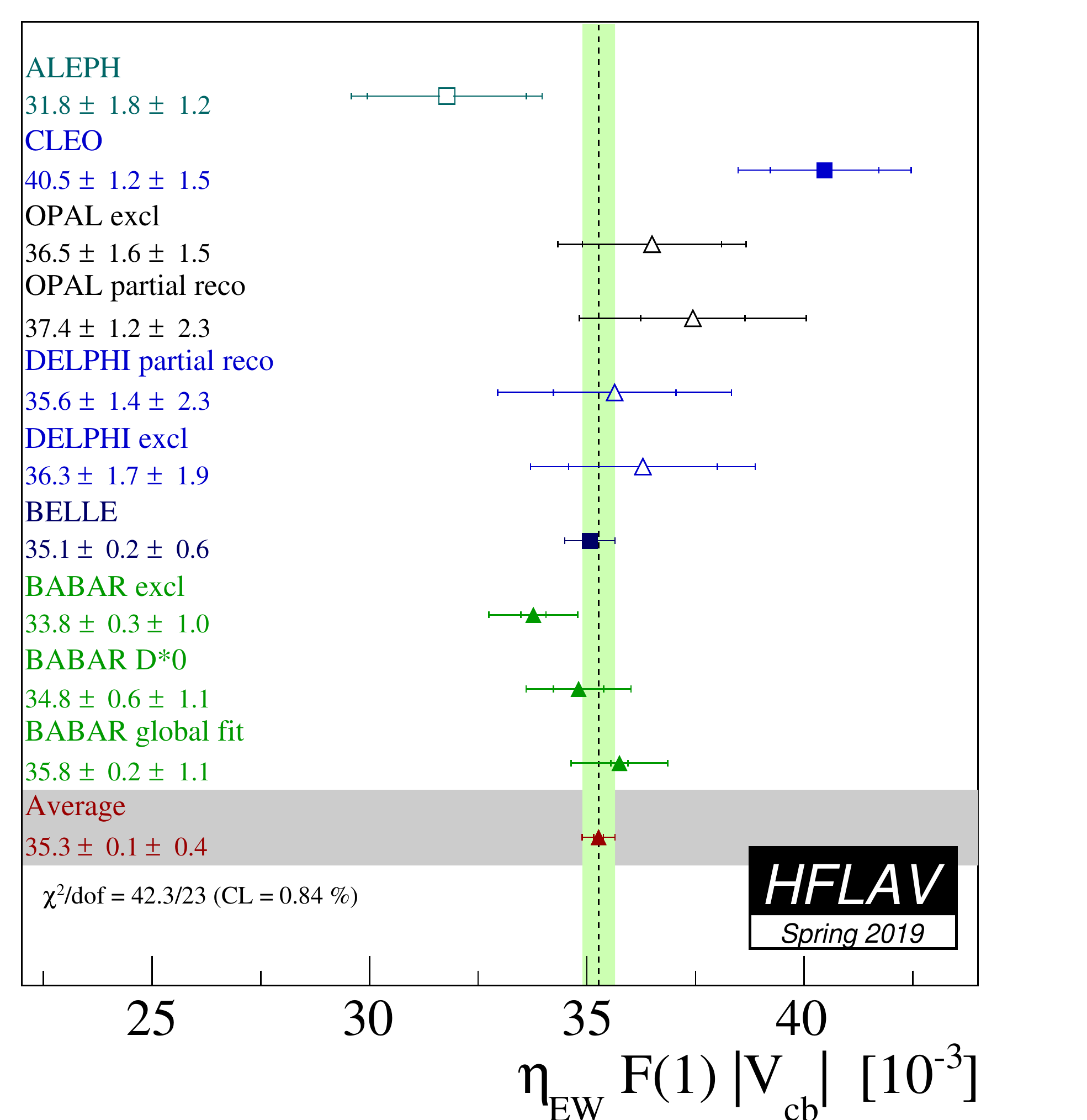}
    }
    \put(  3.2,  5.4){{\large\bf a)}}  
    \put(  9.6,  5.4){{\large\bf b)}}
    \put( 15.9,  5.4){{\large\bf c)}}
  \end{picture}
  \caption{Illustration of (a) the average and (b) the dependence of
    $\eta_\mathrm{EW}{\cal F}(1)\vcb$ on $\rho^2$. The error ellipses
    correspond to $\Delta\chi^2 = 1$ (CL=39\%). Figure (c) is a zoomed in view of the Belle and BaBar measurements.} \label{fig:vcbf1}
  \end{center}
\end{figure}

In the next step, we perform a four-parameter fit of
$\eta_\mathrm{EW}{\cal F}(1)\vcb$, $\rho^2$, $R_1(1)$ and $R_2(1)$
to the rescaled measurements, taking into account correlated
statistical and systematic uncertainties. Only two measurements
constrain all four parameters~\cite{Aubert:2006mb,Waheed:2018djm}, and the remaining
measurements determine only the normalization $\eta_\mathrm{EW}{\cal
  F}(1)\vcb$ and the slope $\rho^2$. The result of the fit is
\begin{eqnarray}
  \eta_\mathrm{EW}{\cal F}(1)\vcb & = & (35.27\pm 0.38)\times
  10^{-3}~, \label{eq:vcbf1} \\
  \rho^2 & = & 1.122\pm 0.024~,\\
  R_1(1) & = & 1.270\pm 0.026~, \label{eq:r1} \\
  R_2(1) & = & 0.852\pm 0.018~, \label{eq:r2}
\end{eqnarray}
and the correlation coefficients are
\begin{eqnarray}
  \rho_{\eta_\mathrm{EW}{\cal F}(1)\vcb,\rho^2} & = & 0.313~,\\
  \rho_{\eta_\mathrm{EW}{\cal F}(1)\vcb,R_1(1)} & = & -0.097~,\\
  \rho_{\eta_\mathrm{EW}{\cal F}(1)\vcb,R_2(1)} & = & -0.076~,\\
  \rho_{\rho^2,R_1(1)} & = & 0.566~,\\
  \rho_{\rho^2,R_2(1)} & = & -0.824~,\\
  \rho_{R_1(1),R_2(1)} & = & -0.715~.
\end{eqnarray}
The uncertainties and correlations quoted here include both
statistical and systematic contributions. The $\chi^2$ of the fit is
42.3 for 23 degrees of freedom, which corresponds to a confidence
level of 0.84\%. The largest contribution to the $\chi^2$ of the average is due to the ALEPH and CLEO measurements~\cite{Buskulic:1996yq,Adam:2002uw}. An illustration of this fit result is given in
Fig.~\ref{fig:vcbf1}.

To convert this result into $\vcb$, theory input for the form factor
normalization is required. We use the result of the FLAG 2019 average~\cite{Aoki:2019cca}, which LQCD results from Refs.~\cite{Bailey:2014tva,Harrison:2017fmw},
\begin{equation}
  \eta_\mathrm{EW}{\cal F}(1) = 0.910\pm 0.013~,
\end{equation}
where $\eta_\mathrm{EW}=1.0066\pm 0.0050$ has been used. The central value of
the latter corresponds to the electroweak correction only. The uncertainty
has been increased to accommodate the Coulomb effect~\cite{Bailey:2014tva,Lattice:2015rga}. With
Eq.~(\ref{eq:vcbf1}), this gives
\begin{equation}
  \vcb = (38.76\pm 0.42_{\rm exp}\pm 0.55_{\rm th})\times
  10^{-3}~, \label{eq:vcbdstar}
\end{equation}
where the first uncertainty is experimental and the second is
theoretical (lattice QCD calculation and electro-weak correction).

\mysubsubsubsection{Extraction of $\vcb$ based on the BGL form factor}

A more general parameterization of the $\bar B\to D^*\ell^-\bar\nu_\ell$ form factor is provided by BGL~\cite{Boyd:1997kz,Grinstein:2017nlq,Bigi:2017njr}. Both Belle~\cite{Waheed:2018djm} and BaBar~\cite{Dey:2019bgc} have recently published analyses of $\bar B\to D^*\ell^-\bar\nu_\ell$ using the BGL form factor: While Belle performs an extraction of $\vcb$ using BGL, the BaBar analysis only fits the BGL form factor parameters but not the normalization. Due to the limited set of input measurements we do not perform a combination of the BGL form factor parameters or $\vcb$ obtained with the BGL form factor at this point. We simply note that $\vcb$ obtained in Refs.~\cite{Waheed:2018djm,Dey:2019bgc} using BGL is consistent with Eq.~\ref{eq:vcbdstar}.

\mysubsubsection{$\bar B\to D\ell^-\bar\nu_\ell$}
\label{slbdecays_dlnu}

The differential decay rate for massless fermions as a function of $w$
(introduced in the previous section) is given by (see, \eg,~\cite{Neubert:1993mb})
\begin{equation}
  \frac{\bar B\to D\ell^-\bar\nu_\ell}{dw} = \frac{G^2_\mathrm{F} m^3_D}{48\pi^3}(m_B+m_D)^2(w^2-1)^{3/2}\eta_\mathrm{EW}^2\mathcal{G}^2(w)|V_{cb}|^2~,
\end{equation}
where $G_\mathrm{F}$ is Fermi's constant, and $m_B$ and $m_D$ are the $B$ and $D$
meson masses. Again, $\eta_\mathrm{EW}$ is the electroweak correction
introduced in the previous section. In contrast to
$\bar B\to D^*\ell^-\bar\nu_\ell$, $\mathcal{G}(w)$ contains a single
form-factor function $f_+(w)$,
\begin{equation}
  \mathcal{G}^2(w) = \frac{4r}{(1+r)^2} f^2_+(w)~,
\end{equation}
where $r=m_D/m_B$.

\mysubsubsubsection{Branching fraction}

\begin{figure}[!ht]
  \begin{center}
  \unitlength1.0cm %
  \begin{picture}(14.,9.0)  %
    \put( -1.5, 0.0){\includegraphics[width=9.0cm]{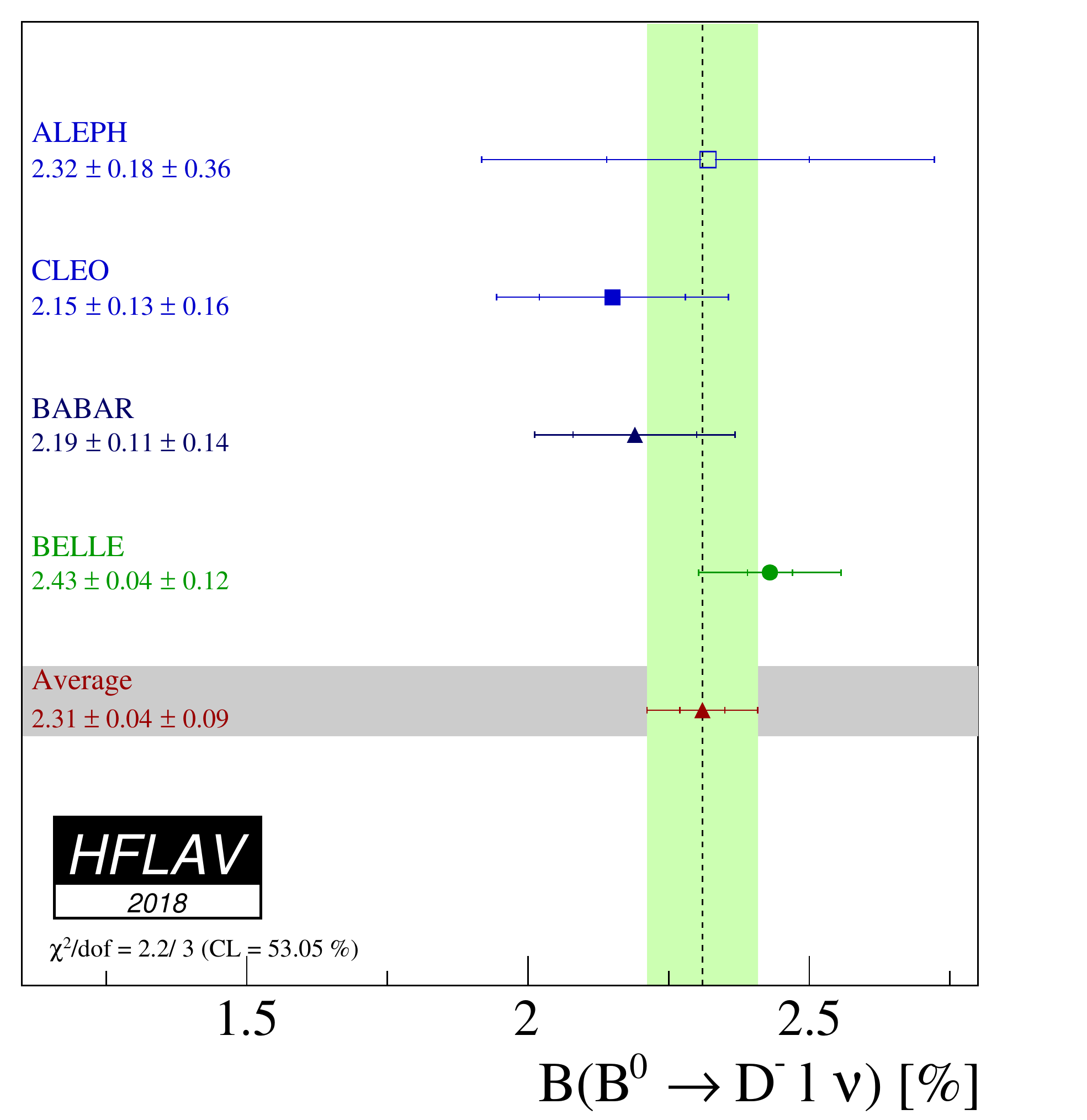}
    }
    \put(  7.5, 0.0){\includegraphics[width=9.0cm]{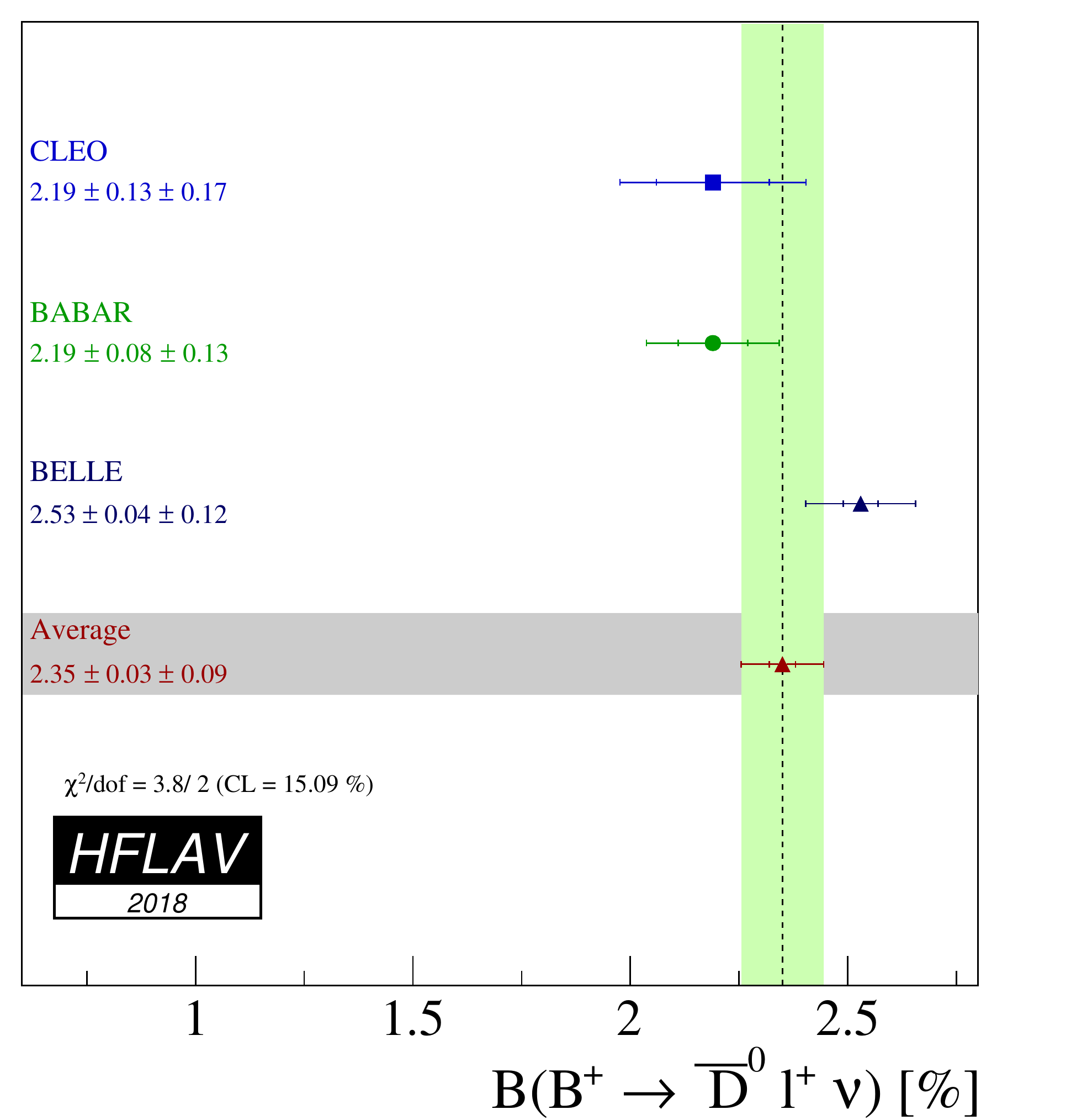}
    }
    \put(  5.8, 8.5){{\large\bf a)}}
    \put( 14.8, 8.5){{\large\bf b)}}
  \end{picture}
  \caption{Branching fractions of exclusive semileptonic $B$ decays:
    (a) $\BzbDplnu$ (Table~\ref{tab:dlnu}) and (b) $B^-\to
    D^0\ell^-\bar\nu_\ell$ (Table~\ref{tab:d0lnu}).} \label{fig:brdl}
  \end{center}
\end{figure}

Separate one-dimensional averages of the
$\BzbDplnu$ and $B^-\to D^0\ell^-\bar\nu_\ell$ branching fractions are shown
in Tables~\ref{tab:dlnu} and \ref{tab:d0lnu}. We obtain
\begin{eqnarray}
  \cbf(\BzbDplnu) & = & (2.31\pm 0.04\pm 0.09)\%~, \label{eq:br_dlnu} \\
  \cbf(B^-\to D^0\ell^-\bar\nu_\ell) & = & (2.35\pm 0.03\pm 0.09)\%~,
   \label{eq:br_d0lnu}
\end{eqnarray}
where the first uncertainty is statistical and the second one is systematic.
These fits are also shown in Fig.~\ref{fig:brdl}.
\begin{table}[!htb]
\caption{Average of $\BzbDplnu$ branching fraction
  measurements.}
\begin{center}
\begin{tabular}{|l|c|c|}
  \hline
  Experiment
  & $\cbf(\BzbDplnu)$ [\%] (rescaled)
  & $\cbf(\BzbDplnu)$ [\%] (published)\\
  \hline \hline
  ALEPH~\cite{Buskulic:1996yq}
  & $2.32\pm 0.18_{\rm stat}\pm 0.36_{\rm syst}$
  & $2.35\pm 0.20_{\rm stat}\pm 0.44_{\rm syst}$\\
  CLEO~\cite{Bartelt:1998dq}
  & $2.15\pm 0.13_{\rm stat}\pm 0.16_{\rm syst}$
  & $2.20\pm 0.16_{\rm stat}\pm 0.19_{\rm syst}$\\
  \babar~\cite{Aubert:2009_2}
  & $2.19\pm 0.11_{\rm stat}\pm 0.14_{\rm syst}$
  & $2.23\pm 0.11_{\rm stat}\pm 0.11_{\rm syst}$\\
  \belle~\cite{Glattauer:2015teq}
  & $2.43\pm 0.04_{\rm stat}\pm 0.12_{\rm syst}$
  & $2.39\pm 0.04_{\rm stat}\pm 0.11_{\rm syst}$\\
  \hline 
  {\bf Average}
  & \mathversion{bold}$2.31\pm 0.04_{\rm stat}\pm 0.09_{\rm syst}$
  & \mathversion{bold}$\chi^2/\dof = 2.20/3$ (CL=$53.1\%$)\\
  \hline 
\end{tabular}
\end{center}
\label{tab:dlnu}
\end{table}
\begin{table}[!htb]
\caption{Average of $B^-\to D^0\ell^-\bar\nu_\ell$ branching fraction
  measurements.}
\begin{center}
\begin{tabular}{|l|c|c|}
  \hline
  Experiment
  & $\cbf(B^-\to D^0\ell^-\bar\nu_\ell)$ [\%] (rescaled)
  & $\cbf(B^-\to D^0\ell^-\bar\nu_\ell)$ [\%] (published)\\
  \hline \hline
  CLEO~\cite{Bartelt:1998dq}
  & $2.19\pm 0.13_{\rm stat}\pm 0.17_{\rm syst}$
  & $2.32\pm 0.17_{\rm stat}\pm 0.20_{\rm syst}$\\
  \babar~\cite{Aubert:2009_2}
  & $2.19\pm 0.08_{\rm stat}\pm 0.13_{\rm syst}$
  & $2.31\pm 0.08_{\rm stat}\pm 0.09_{\rm syst}$\\
  \belle~\cite{Glattauer:2015teq}
  & $2.53\pm 0.04_{\rm stat}\pm 0.12_{\rm syst}$
  & $2.54\pm 0.04_{\rm stat}\pm 0.13_{\rm syst}$\\
  \hline
  {\bf Average}
  & \mathversion{bold}$2.35\pm 0.03_{\rm stat}\pm 0.09_{\rm syst}$
  & \mathversion{bold}$\chi^2/\dof = 3.78/2$ (CL=$15.1\%$)\\
  \hline
\end{tabular}
\end{center}
\label{tab:d0lnu}
\end{table}

\mysubsubsubsection{Extraction of $\vcb$ based on the CLN form factor}

As for $\bar B\to D^*\ell^-\bar\nu_\ell$ decays, we adopt the prescription by
Caprini, Lellouch and Neubert~\cite{CLN}, which describes the shape and
normalization of the measured decay distributions in terms of two parameters:
the normalization ${\cal G}(1)$ and the slope $\rho^2$,
\begin{equation}
  \mathcal{G}(w)= \mathcal{G}(1)\big[1 - 8 \rho^2 z + (51 \rho^2 - 10 )
  z^2 - (252 \rho^2 - 84 ) z^3\big]~,
\end{equation}
where $z=(\sqrt{w+1}-\sqrt{2})/(\sqrt{w+1}+\sqrt{2})$.

Table~\ref{tab:vcbg1} shows experimental measurements of the two CLN
parameters, which are corrected to match the latest values of the input
parameters~\cite{HFLAV_sl:inputparams}. Both measurements of $\BzbDplnu$ and
$B^-\to D^0\ell^-\bar\nu_\ell$ are used and isospin symmetry is assumed in the
analysis.
\begin{table}[!htb]
\caption{Measurements of the Caprini, Lellouch and Neubert
  (CLN)~\cite{CLN} form factor parameters in $\bar B\to
  D\ell^-\bar\nu_\ell$ before and after rescaling.}
\begin{center}
\begin{tabular}{|l|c|c|}
  \hline
  Experiment
  & $\eta_\mathrm{EW}{\cal G}(1)\vcb$ [10$^{-3}$] (rescaled)
  & $\rho^2$ (rescaled)\\
  & $\eta_\mathrm{EW}{\cal G}(1)\vcb$ [10$^{-3}$] (published)
  & $\rho^2$ (published)\\
  \hline \hline
  ALEPH~\cite{Buskulic:1996yq}
  & $38.75\pm 9.51_{\rm stat}\pm 6.93_{\rm syst}$
  & $0.955\pm 0.834_{\rm stat}\pm 0.425_{\rm syst}$\\
  & $31.1\pm 9.9_{\rm stat}\pm 8.6_{\rm syst}$
  & $0.70\pm 0.98_{\rm stat}\pm 0.50_{\rm syst}$\\
  \hline
  CLEO~\cite{Bartelt:1998dq}
  & $44.97\pm 5.70_{\rm stat}\pm 3.47_{\rm syst}$
  & $1.270\pm 0.215_{\rm stat}\pm 0.121_{\rm syst}$\\
  & $44.8\pm 6.1_{\rm stat}\pm 3.7_{\rm syst}$
  & $1.30\pm 0.27_{\rm stat}\pm 0.14_{\rm syst}$\\
  \hline
  \belle~\cite{Glattauer:2015teq}
  & $42.22\pm 0.60_{\rm stat}\pm 1.21_{\rm syst}$
  & $1.090\pm 0.036_{\rm stat}\pm 0.019_{\rm syst}$\\
  & $42.29\pm 1.37$ & $1.09\pm 0.05$\\
  \hline
  \babar global fit~\cite{Aubert:2009_1}
  & $43.84\pm 0.76_{\rm stat}\pm 2.19_{\rm syst}$
  & $1.215\pm 0.035_{\rm stat}\pm 0.062_{\rm syst}$\\
  & $43.1\pm 0.8_{\rm stat}\pm 2.3_{\rm syst}$
  & $1.20\pm 0.04_{\rm stat}\pm 0.07_{\rm syst}$\\
  \hline
  \babar tagged~\cite{Aubert:2009_2}
  & $42.76\pm 1.71_{\rm stat}\pm 1.26_{\rm syst}$
  & $1.200\pm 0.088_{\rm stat}\pm 0.043_{\rm syst}$\\
  & $42.3\pm 1.9_{\rm stat}\pm 1.0_{\rm syst}$
  & $1.20\pm 0.09_{\rm stat}\pm 0.04_{\rm syst}$\\
  \hline 
  {\bf Average }
  & \mathversion{bold}$42.00\pm 0.45_{\rm stat}\pm 0.89_{\rm syst}$
  & \mathversion{bold}$1.131\pm 0.024_{\rm stat}\pm 0.023_{\rm syst}$\\
  \hline 
\end{tabular}
\end{center}
\label{tab:vcbg1}
\end{table}

\begin{figure}[!ht]
  \begin{center}
  \unitlength1.0cm %
  \begin{picture}(14.,7.) %
    \put( 10.9, -0.2){\includegraphics[width=6.2cm]{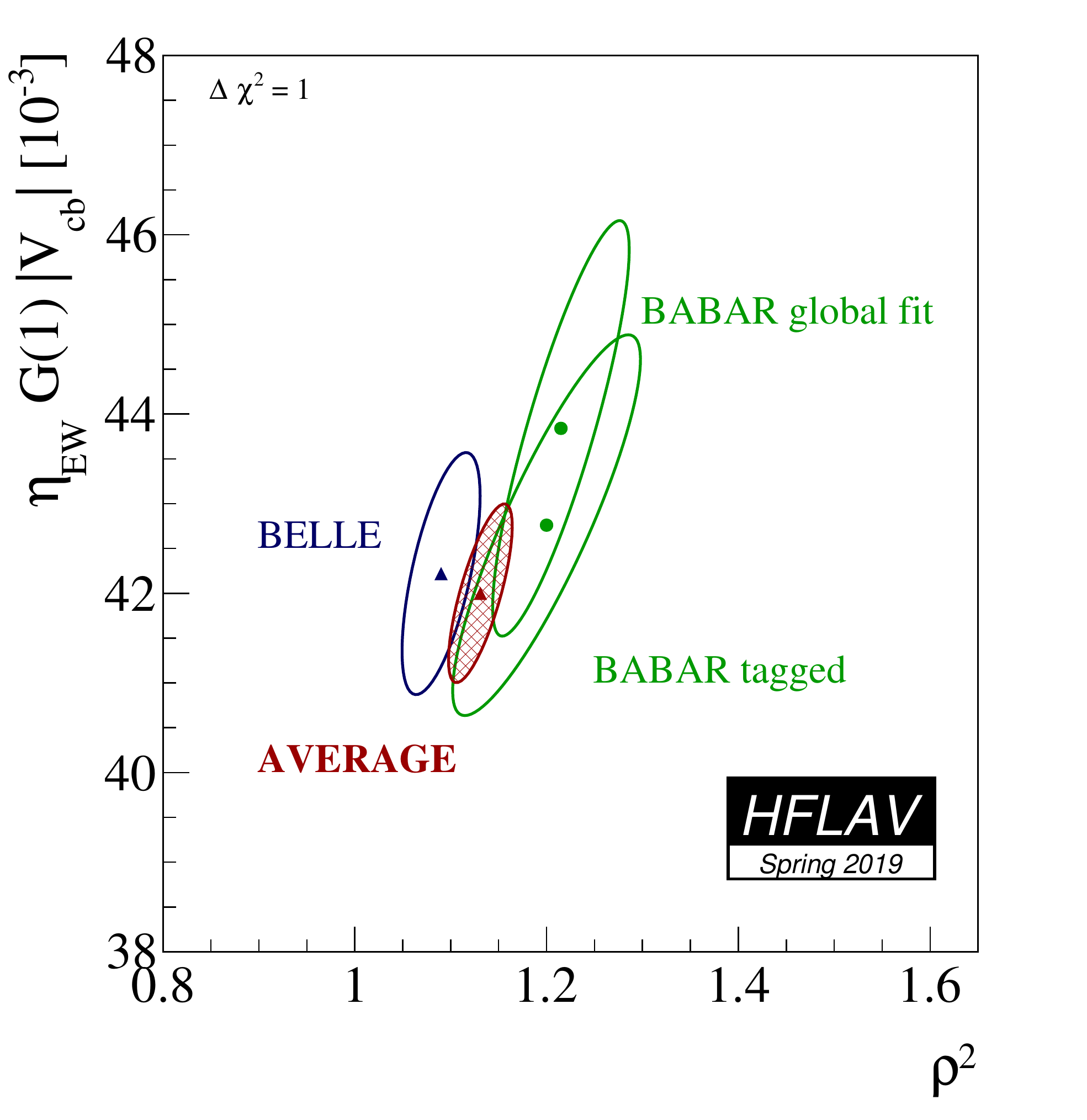}
    }
    \put(  4.7, -0.2){\includegraphics[width=6.2cm]{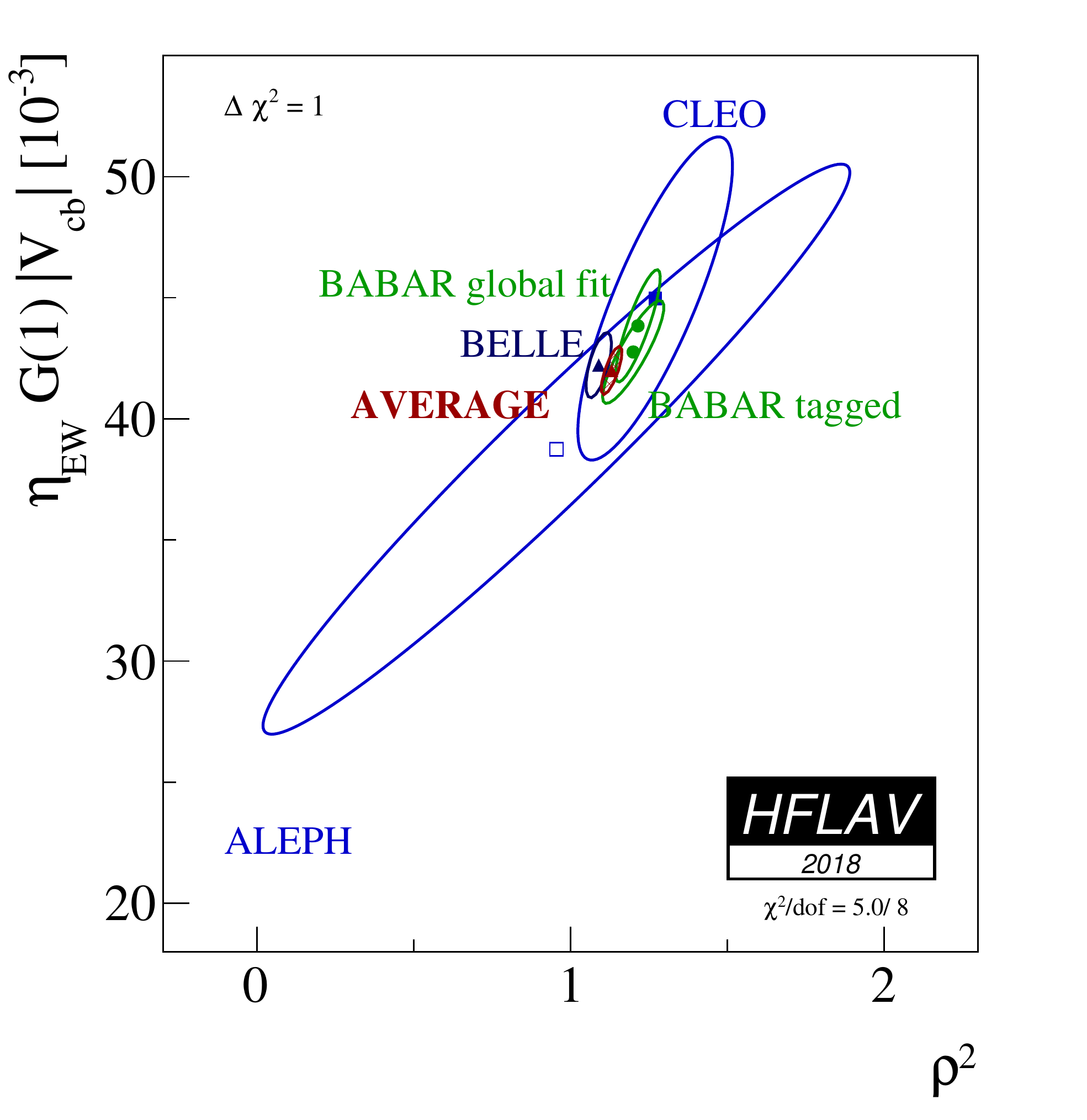}
    }
    \put( -1.5, 0.0){\includegraphics[width=5.9cm]{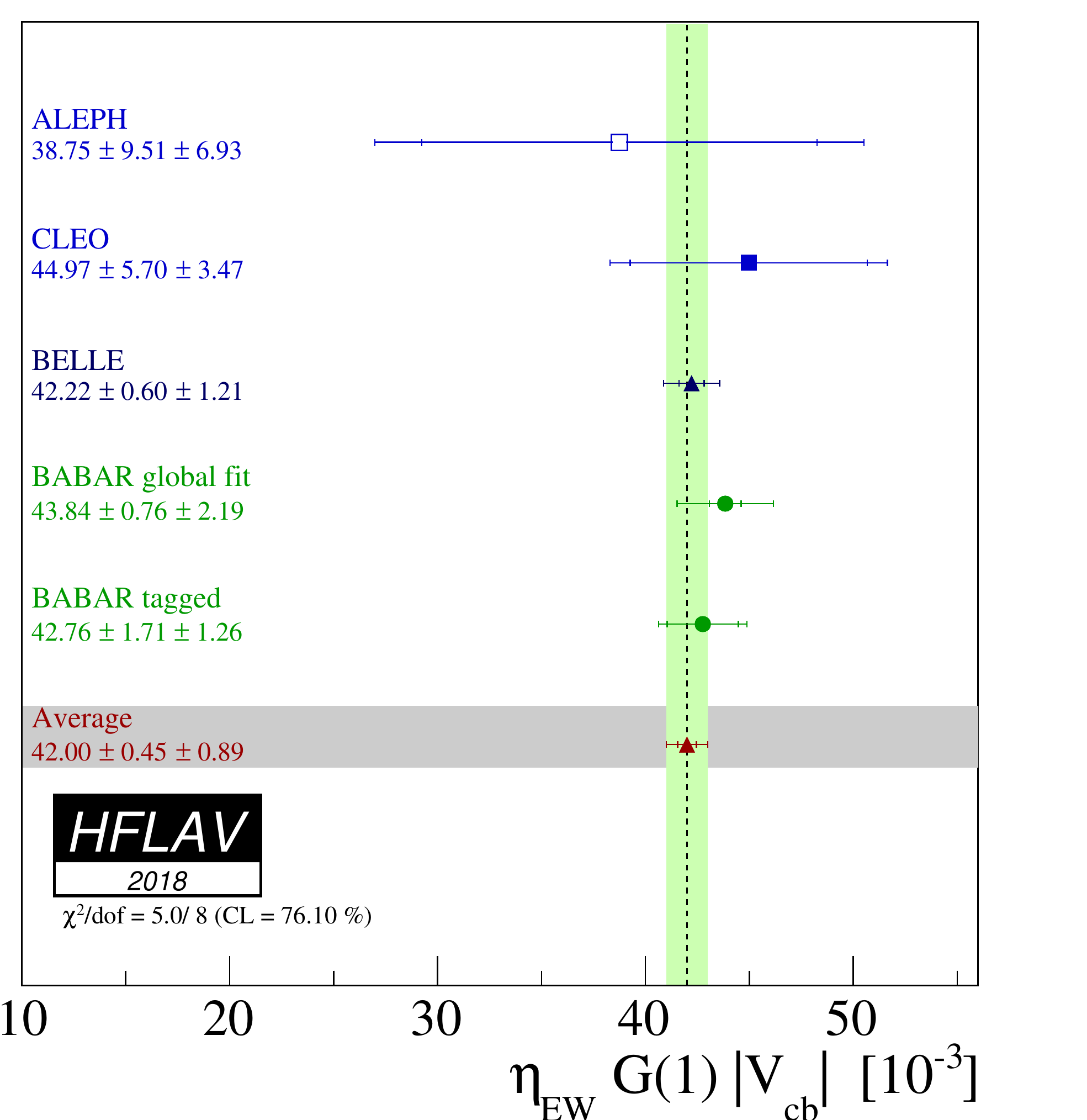}
    }
    \put(  3.2, 5.4){{\large\bf a)}}
    \put(  9.6, 5.4){{\large\bf b)}}
    \put( 15.9, 5.4){{\large\bf c)}}
  \end{picture}
  \caption{Illustration of (a) the average and (b) dependence of
    $\eta_\mathrm{EW}{\cal G}(w)\vcb$ on $\rho^2$. The error ellipses correspond
    to $\Delta\chi^2 = 1$ (CL=39\%). Figure (c) is a zoomed in view of the Belle and BaBar measurements.}
  \label{fig:vcbg1}
  \end{center}
\end{figure}

The form factor parameters are extracted by a two-parameter fit to
the rescaled measurements of $\eta_\mathrm{EW}{\cal G}(1)\vcb$ and
$\rho^2$ taking into account correlated statistical and systematic
uncertainties. The result of the fit is
\begin{eqnarray}
  \eta_\mathrm{EW}{\cal G}(1)\vcb & = & (42.00\pm 1.00)\times
  10^{-3}~, \label{eq:vcbg1} \\
  \rho^2 & = & 1.131\pm 0.033~,
\end{eqnarray}
with a correlation of
\begin{equation}
  \rho_{\eta_\mathrm{EW}{\cal G}(1)\vcb,\rho^2} = 0.751~.
\end{equation}
The uncertainties and the correlation coefficient include both
statistical and systematic contributions. The $\chi^2$ of the fit is
5.0 for 8 degrees of freedom, which corresponds to a probability of 
76.1\%. An illustration of this fit result is given in
Fig.~\ref{fig:vcbg1}.

The most recent lattice QCD result obtained for the form factor
normalization is~\cite{Lattice:2015rga}
\begin{equation}
  {\cal G}(1) = 1.0541\pm 0.0083~.
\end{equation}
Using again $\eta_\mathrm{EW}=1.0066\pm 0.0050$, we determine $\vcb$ from
Eq.~(\ref{eq:vcbg1}),
\begin{equation}
  \vcb = (39.58\pm 0.94_{\rm exp}\pm 0.37_{\rm th})\times 10^{-3}~,
\end{equation}
where the first error is experimental and the second theoretical. This
number is in excellent agreement with $\vcb$ obtained from 
$\bar B\to D^*\ell^-\bar\nu_\ell$ decays given in Eq.~(\ref{eq:vcbdstar}).

\mysubsubsubsection{Extraction of $\vcb$ based on the BGL form factor}

A more general expression for the $\bar B\to D\ell^-\bar\nu_\ell$ form factor is again BGL. If experimental data on the $w$~spectrum is available, a BGL fit allows to include available lattice QCD data at non-zero recoil $w>1$~\cite{Lattice:2015rga,Na:2015kha} to improve the extrapolation to the zero recoil point~$w=1$. A $w$~spectrum of $\bar B\to D\ell^-\bar\nu_\ell$ has been published by BaBar~\cite{Aubert:2009ac} and Belle~\cite{Glattauer:2015teq}. As the BaBar result does not include the full error matrix of the $w$~spectrum, we refrain from performing a combined BGL fit at this point. Instead we refer to \cite{Glattauer:2015teq} for the impact of the non-zero recoil lattice data on the value of $\vcb$ from $\bar B\to D\ell^-\bar\nu_\ell$.

\mysubsubsection{$\bar{B} \to D^{(*)}\pi \ell^-\bar{\nu}_{\ell}$}
\label{slbdecays_dpilnu}

\begin{figure}[!ht]
 \begin{center}
  \unitlength1.0cm %
  \begin{picture}(14.,9.5)  %
   \put( -1.5,  0.0){\includegraphics[width=8.7cm]{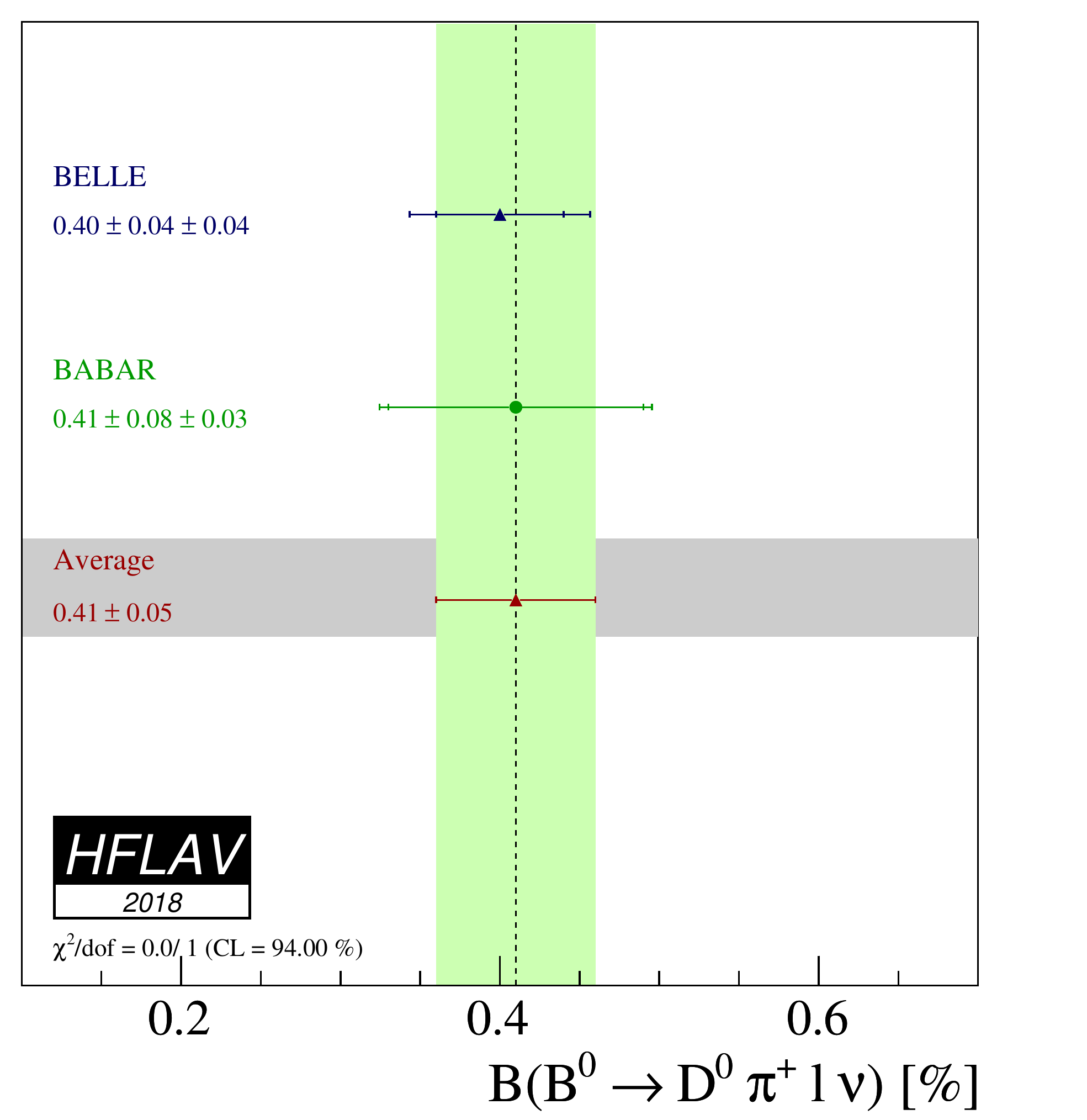}
   }
   \put(  7.5,  0.0){\includegraphics[width=8.7cm]{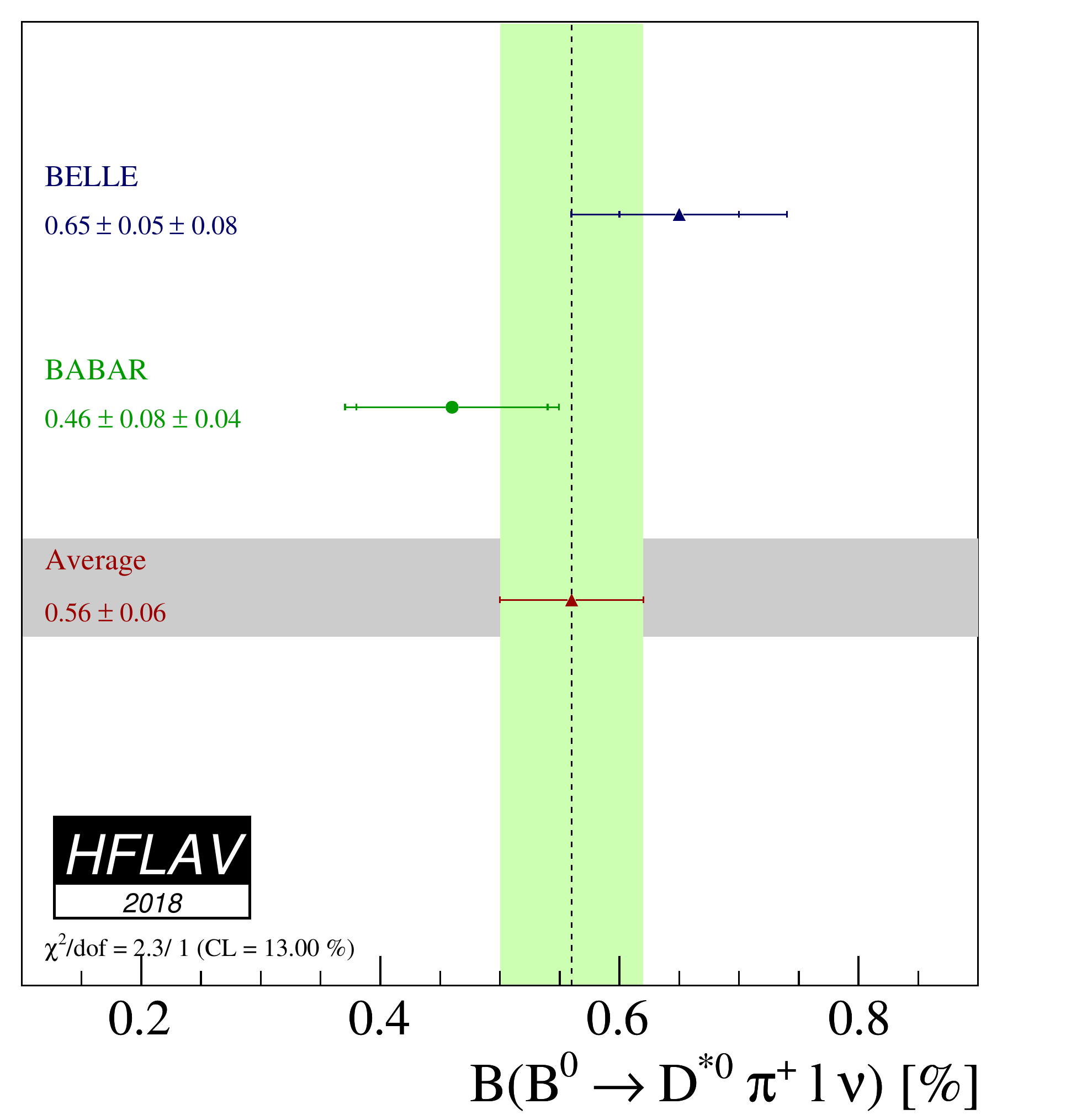}
   }
   \put(  5.5,  8.2){{\large\bf a)}}
   \put( 14.5,  8.2){{\large\bf b)}}
  \end{picture}
  \begin{picture}(14.,9.5)  %
   \put( -1.5,  0.0){\includegraphics[width=8.7cm]{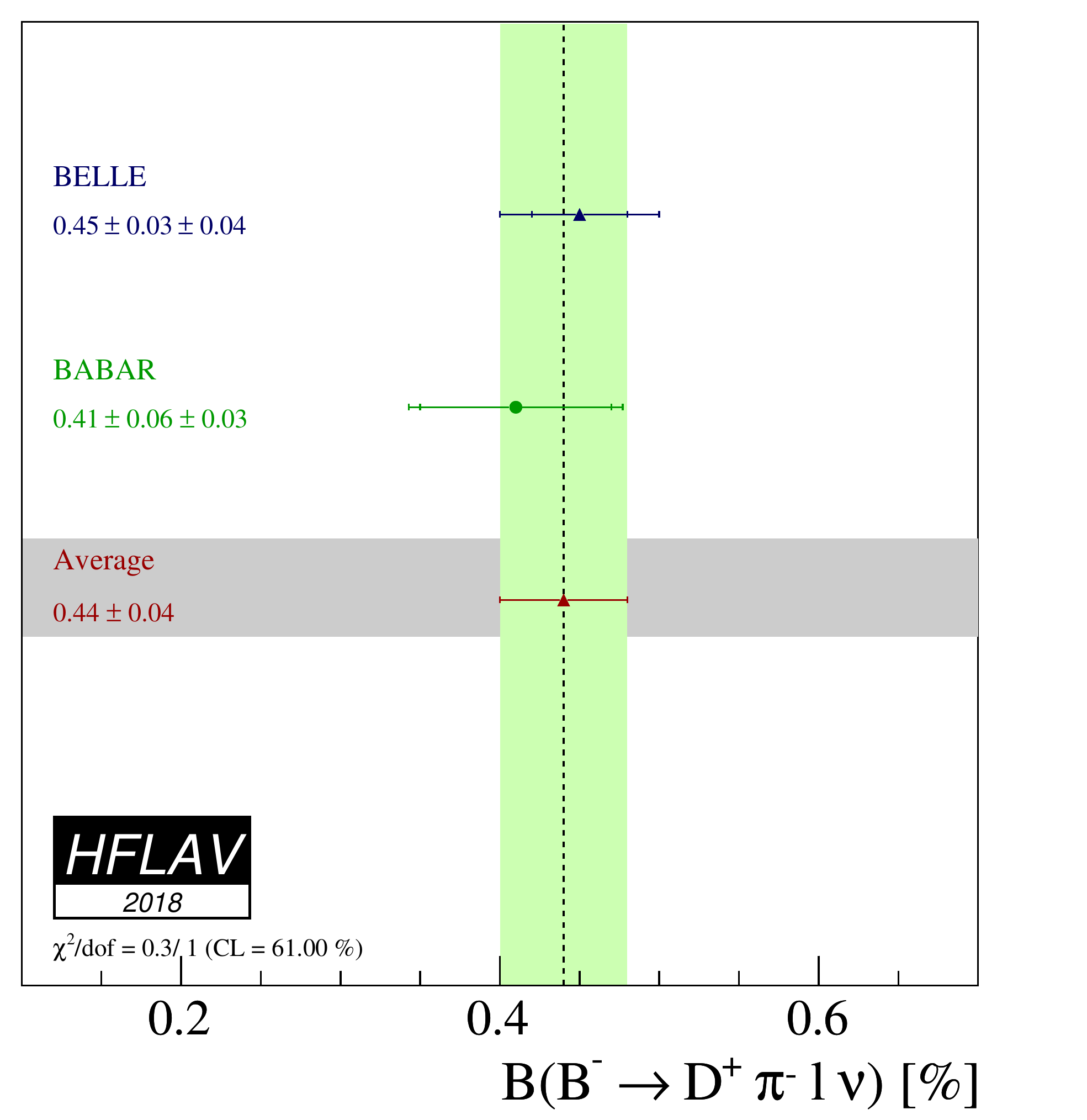}
   }
   \put(  7.5,  0.0){\includegraphics[width=8.7cm]{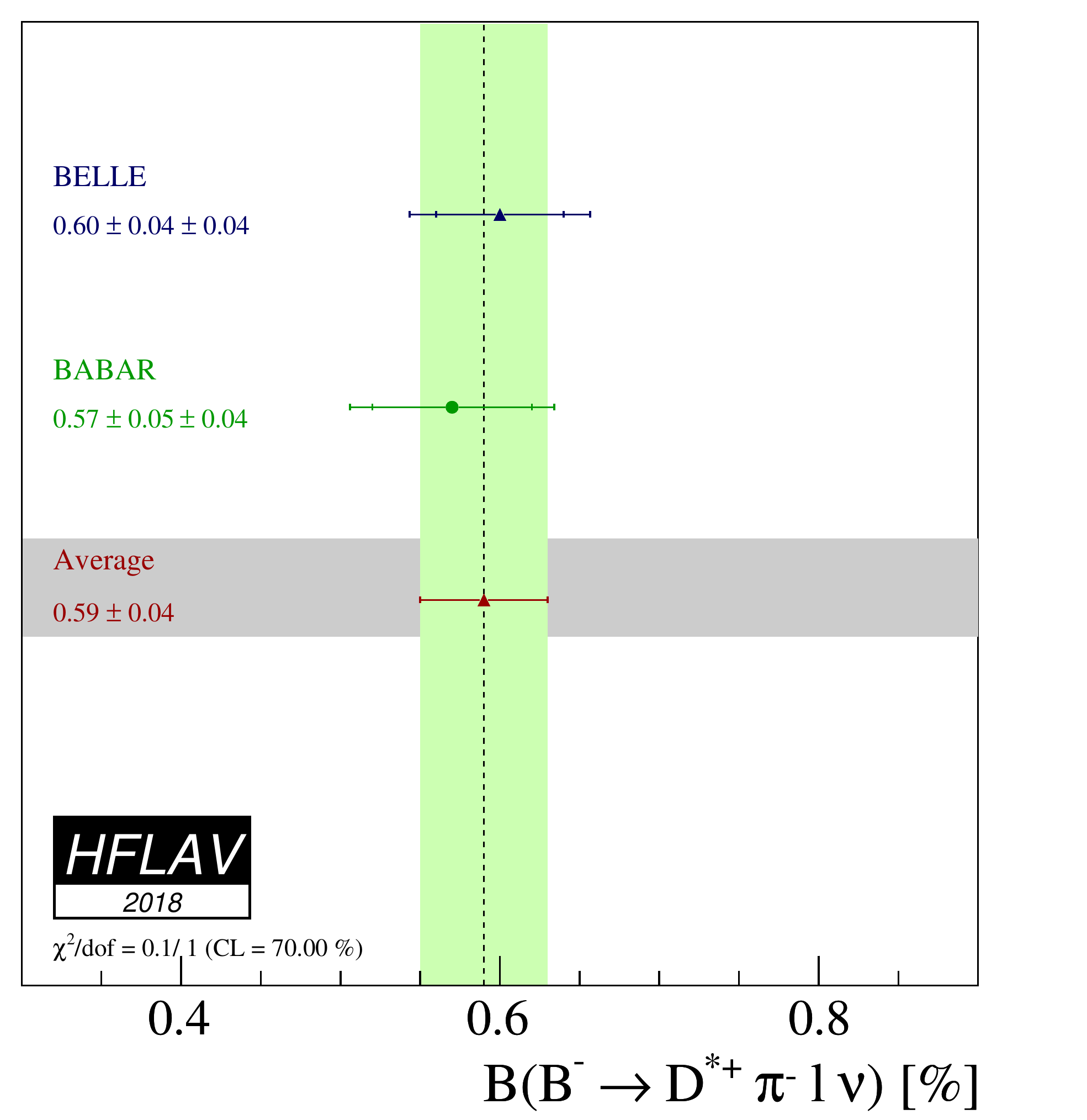}
   }
   \put(  5.5,  8.2){{\large\bf c)}}
   \put( 14.5,  8.2){{\large\bf d)}}
  \end{picture}
  \caption{Average branching fraction  of exclusive semileptonic $B$ decays
(a) $\bar{B}^0 \to D^0 \pi^+ \ell^-\bar{\nu}_{\ell}$, (b) $\bar{B}^0 \to D^{*0} \pi^+
\ell^-\bar{\nu}_{\ell}$, (c) $B^- \to D^+ \pi^-
\ell^-\bar{\nu}_{\ell}$, and (d) $B^- \to D^{*+} \pi^- \ell^-\bar{\nu}_{\ell}$.
The corresponding individual
  results are also shown.}
  \label{fig:brdpil}
 \end{center}
\end{figure}

The average inclusive branching fractions for $\bar{B} \to D^{(*)}\pi\ell^-\bar{\nu}_{\ell}$ 
decays, where no constraint is applied to the mass of the $D^{(*)}\pi$ system, are determined by the
combination of the results provided in Table~\ref{tab:dpilnu} for 
$\bar{B}^0 \to D^0 \pi^+ \ell^-\bar{\nu}_{\ell}$, $\bar{B}^0 \to D^{*0} \pi^+\ell^-\bar{\nu}_{\ell}$,  $B^- \to D^+ \pi^- \ell^-\bar{\nu}_{\ell}$, and $B^- \to D^{*+} \pi^- \ell^-\bar{\nu}_{\ell}$ decays. For the $\bar{B}^0 \to D^0 \pi^+ \ell^-\bar{\nu}_{\ell}$ decays a veto to reject the $D^{*+}\to D^0 \pi^+$ decays is applied.
The measurements included in the average 
are scaled to a consistent set of input
parameters and their uncertainties~\cite{HFLAV_sl:inputparams}.
For both the \babar\ and Belle results, the $B$ semileptonic signal yields are
 extracted from a fit to the missing mass squared distribution for a sample of fully
 reconstructed \BB\ events.  
Figure~\ref{fig:brdpil} shows the measurements and the resulting average for the 
four decay modes.

\begin{table}[!htb]
\caption{Averages of the $B \to D^{(*)} \pi^- \ell^-\bar{\nu}_{\ell}$  branching fractions and individual results.}
\begin{center}
\begin{tabular}{|l|c c|}\hline
Experiment                                 &$\cbf(B^- \to D^+ \pi^- \ell^-\bar{\nu}_{\ell}) [\%]$ (rescaled) & $\cbf(B^- \to D^+ \pi^- \ell^-\bar{\nu}_{\ell}) [\%]$ (published)\\
\hline
\belle  ~\cite{Vossen:2018zeg}             &$0.455 \pm0.027_{\rm stat} \pm0.039_{\rm syst}$  & $0.455 \pm0.027_{\rm stat} \pm0.039_{\rm syst}$\\
\babar  ~\cite{Aubert:vcbExcl}       &$0.415 \pm0.060_{\rm stat} \pm0.031_{\rm syst}$ & $0.42 \pm0.06_{\rm stat} \pm0.03_{\rm syst}$  \\
\hline 
{\bf Average}                              &\mathversion{bold}$0.443 \pm0.037$ &\mathversion{bold}$\chi^2/\dof = 0.25$ (CL=$61.4\%$) \\
\hline\hline

Experiment                                 &$\cbf(B^- \to D^{*+} \pi^- \ell^-\bar{\nu}_{\ell}) [\%]$ (rescaled) & $\cbf(B^- \to D^{*+} \pi^- \ell^-\bar{\nu}_{\ell}) [\%]$ (published) \\
\hline 
\belle  ~\cite{Vossen:2018zeg}           &$0.603 \pm0.043_{\rm stat} \pm0.038_{\rm syst}$   & $0.604 \pm0.043_{\rm stat} \pm0.038_{\rm syst}$  \\
\babar  ~\cite{Aubert:vcbExcl}       &$0.569 \pm0.050_{\rm stat} \pm0.045_{\rm syst}$   & $0.59 \pm0.05_{\rm stat} \pm0.04_{\rm syst}$ \\
\hline 
{\bf Average}                              &\mathversion{bold}$0.589 \pm0.044$   & \mathversion{bold}$\chi^2/\dof = 0.145$ (CL=$70.3\%$) \\
\hline \hline

Experiment                               &$\cbf(\bar{B}^0 \to D^0 \pi^+ \ell^-\bar{\nu}_{\ell}) [\%]$ (rescaled) & $\cbf(\bar{B}^0 \to D^0 \pi^+ \ell^-\bar{\nu}_{\ell}) [\%]$ (published)\\
\hline 
\belle  ~\cite{Vossen:2018zeg}           &$0.405 \pm0.036_{\rm stat} \pm0.041_{\rm syst}$ & $0.405 \pm0.036_{\rm stat} \pm0.041_{\rm syst}$ \\
\babar  ~\cite{Aubert:vcbExcl}     &$0.410 \pm0.080_{\rm stat} \pm0.035_{\rm syst}$ & $0.43 \pm0.08_{\rm stat} \pm0.03_{\rm syst}$ \\
\hline 
{\bf Average}                              &\mathversion{bold}$0.406 \pm0.047$  &\mathversion{bold}$\chi^2/\dof = 0.002$ (CL=$96.4\%$) \\
\hline\hline

Experiment                                 &$\cbf(\bar{B}^0 \to D^{*0} \pi^+\ell^-\bar{\nu}_{\ell}) [\%]$ (rescaled) & $\cbf(\bar{B}^0 \to D^{*0} \pi^+\ell^-\bar{\nu}_{\ell}) [\%]$ (published) \\
\hline 
\belle  ~\cite{Vossen:2018zeg}           &$0.646 \pm0.053_{\rm stat} \pm0.052_{\rm syst}$  & $0.646 \pm0.053_{\rm stat} \pm0.052_{\rm syst}$ \\
\babar  ~\cite{Aubert:vcbExcl}       &$0.462 \pm0.080_{\rm stat} \pm0.044_{\rm syst}$ &$0.48 \pm0.08_{\rm stat} \pm0.04_{\rm syst}$ \\ 
\hline 
{\bf Average}                              &\mathversion{bold}$0.565 \pm0.061$ &\mathversion{bold}$\chi^2/\dof = 2.25$ (CL=$13.3\%$) \\
\hline

\end{tabular}
\end{center}
\label{tab:dpilnu}
\end{table}

\mysubsubsection{$\bar{B} \to D^{**} \ell^-\bar{\nu}_{\ell}$}
\label{slbdecays_dsslnu}

$D^{**}$ mesons contain one charm quark and one light anti-quark with relative angular momentum $L=1$. 
According to Heavy Quark Symmetry (HQS)~\cite{Isgur:1991wq}, they form one doublet of states with 
angular momentum $j \equiv s_q + L= 3/2$  $\left[D_1(2420), D_2^*(2460)\right]$ and another doublet 
with $j=1/2$ $\left[D^*_0(2400), D_1'(2430)\right]$, where $s_q$ is the light quark spin. 
Parity and angular momentum conservation constrain the decays allowed for each state. The $D_1$ and $D_2^*$ 
states decay via D-wave to $D^*\pi$ and $D^{(*)}\pi$, respectively, and have small decay widths, 
while the $D_0^*$ and $D_1'$  states decay via S-wave to $D\pi$ and $D^*\pi$ and are very broad.
For the narrow states, the averages are determined by the
combination of the results provided in Table~\ref{tab:dss1lnu} and \ref{tab:dss2lnu} for 
$\cbf(B^- \to D_1^0\ell^-\bar{\nu}_{\ell})
\times \cbf(D_1^0 \to D^{*+}\pi^-)$ and $\cbf(B^- \to D_2^0\ell^-\bar{\nu}_{\ell})
\times \cbf(D_2^0 \to D^{*+}\pi^-)$. 
For the broad states, the averages are determined by the
combination of the results provided in Table~\ref{tab:dss1plnu} and \ref{tab:dss0lnu} for 
$\cbf(B^- \to D_1'^0\ell^-\bar{\nu}_{\ell})
\times \cbf(D_1'^0 \to D^{*+}\pi^-)$ and $\cbf(B^- \to D_0^{*0}\ell^-\bar{\nu}_{\ell})
\times \cbf(D_0^{*0} \to D^{+}\pi^-)$. 
The measurements are scaled to a consistent set of input
parameters and their uncertainties~\cite{HFLAV_sl:inputparams}.  The results are reported for $B^-$, and when measurements for both $B^0$ and $B^-$ are available, the combination assumes the validity of the isospin. 
It is worth to notice that, while the results for the narrow resonances and the $D_0^{*}$ are consistent between the various experiments, the available measurements for $B^- \to D_1'^0\ell^-\bar{\nu}_{\ell}$ obtained by \babar \cite{Aubert:2009_4}, Belle \cite{Live:Dss} and DELPHI \cite{Abdallah:2005cx}, are not compatible. In particular Belle did not observed a significant $B^- \to D_1'^0\ell^-\bar{\nu}_{\ell}$ contribution and put an upper limit on the presence of the $D_1'^0$ state. 

\begin{figure}[!ht]
 \begin{center}
  \unitlength1.0cm %
  \begin{picture}(14.,9.0)  %
   \put( -1.5,  0.0){\includegraphics[width=8.7cm]{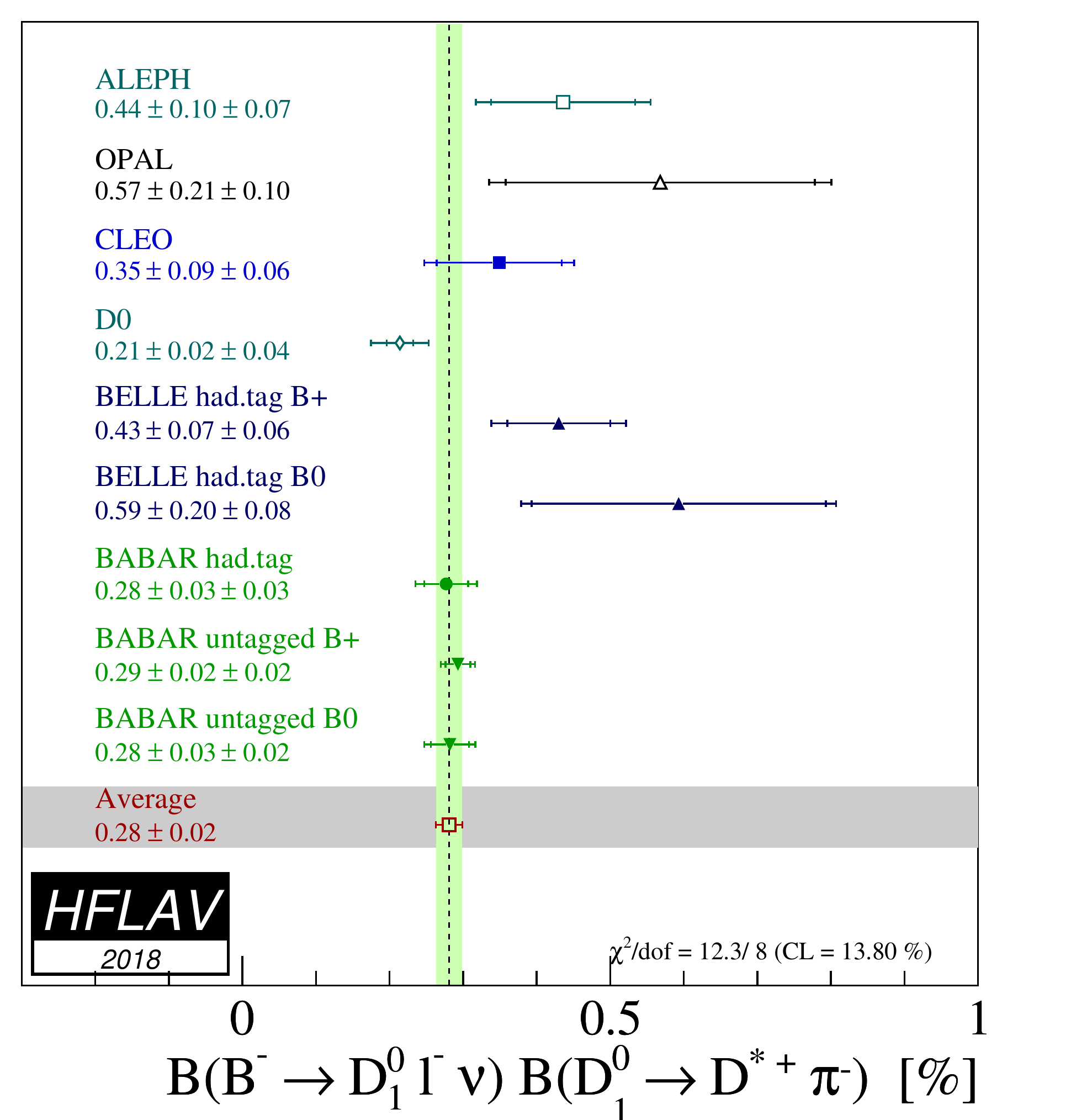}
   }
   \put(  7.5,  0.0){\includegraphics[width=8.7cm]{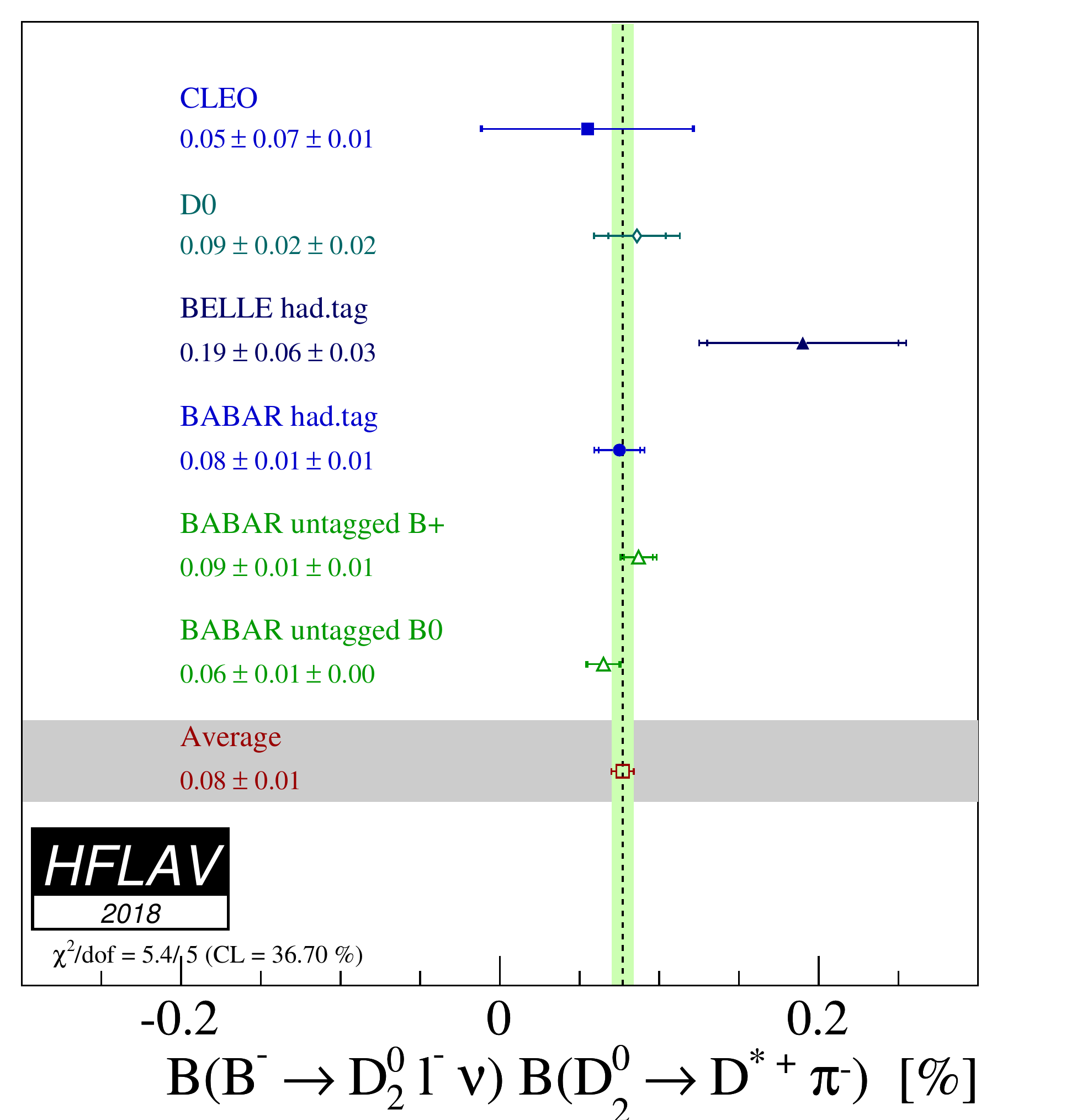}
   }
   \put(  5.6,  8.0){{\large\bf a)}}
   \put( 14.2,  8.0){{\large\bf b)}}
  \end{picture}
  \caption{Rescaled individual measurements and their averages for (a) 
  $\cbf(B^- \to D_1^0\ell^-\bar{\nu}_{\ell})
\times \cbf(D_1^0 \to D^{*+}\pi^-)$ and (b) $\cbf(B^- \to D_2^0\ell^-\bar{\nu}_{\ell})
\times \cbf(D_2^0 \to D^{*+}\pi^-)$.}
  \label{fig:brdssl}
 \end{center}
\end{figure}

\begin{figure}[!ht]
 \begin{center}
  \unitlength1.0cm %
  \begin{picture}(14.,9.0)  %
   \put( -1.5,  0.0){\includegraphics[width=8.7cm]{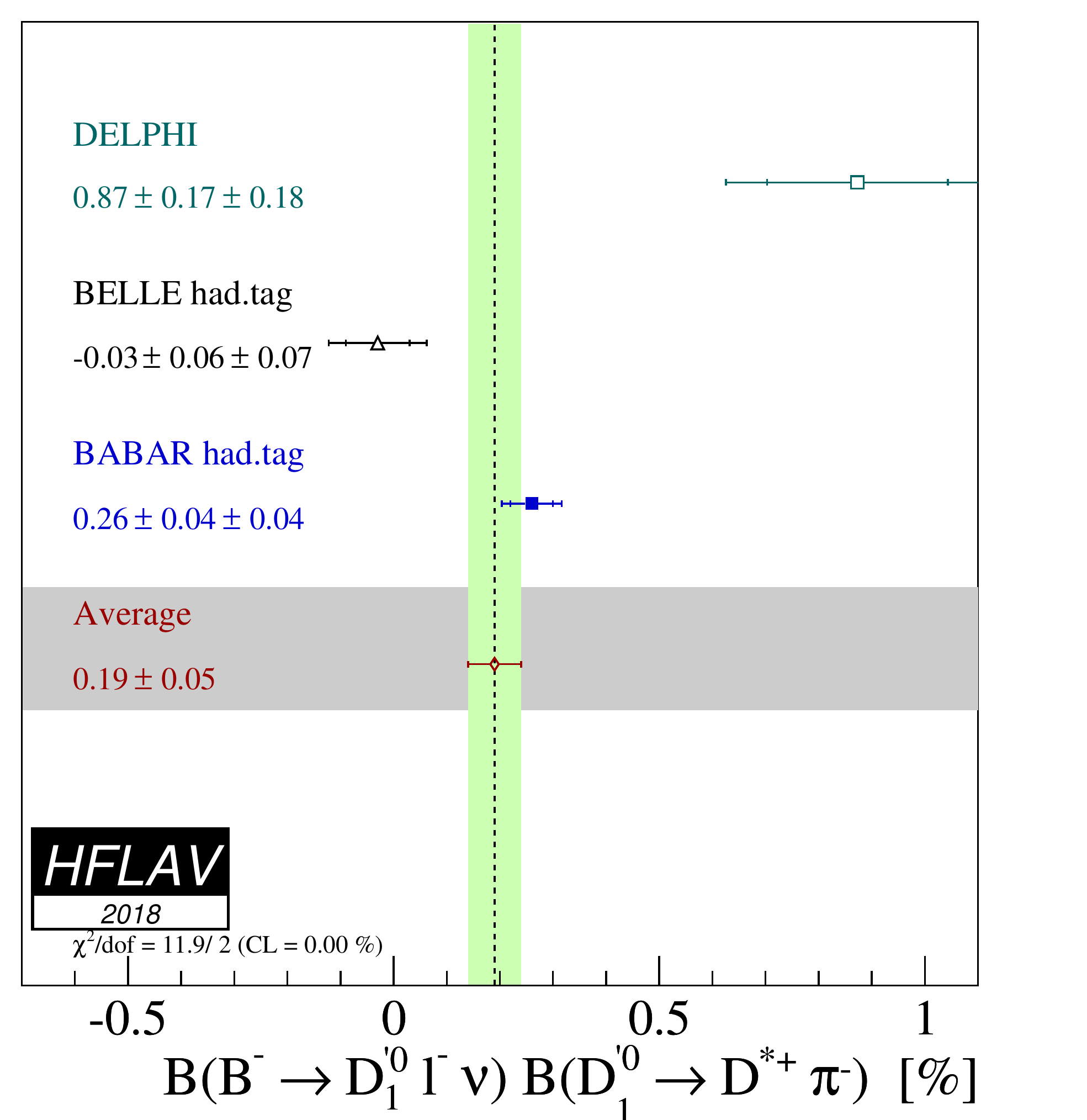}
   }
   \put(  7.5,  0.0){\includegraphics[width=8.7cm]{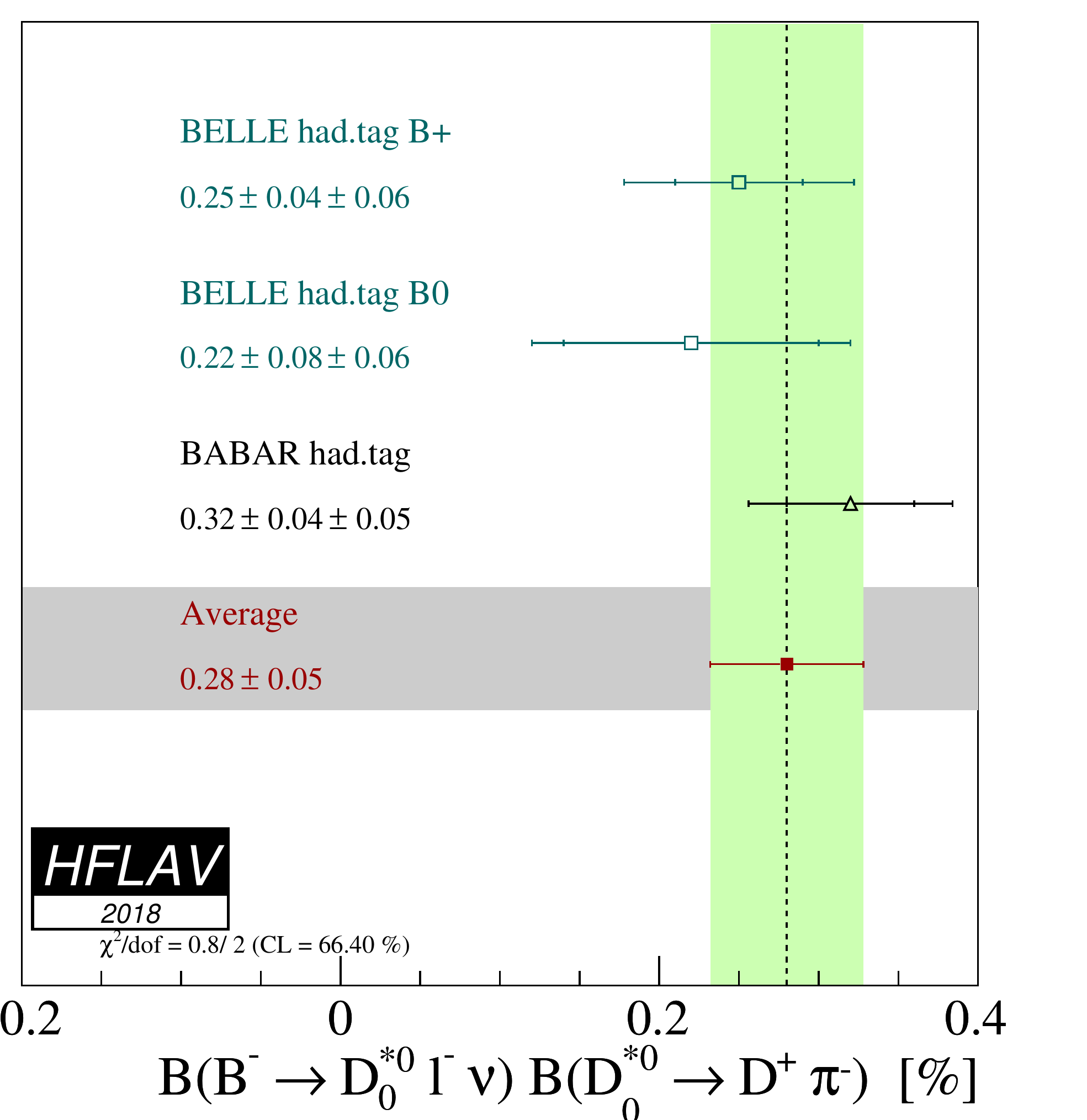}
   }
   \put(  5.8,  8.3){{\large\bf a)}}
   \put( 14.7,  8.3){{\large\bf b)}}
  \end{picture}
  \caption{Rescaled individual measurements and their averages for (a) 
  $\cbf(B^- \to D_1'^0\ell^-\bar{\nu}_{\ell})
\times \cbf(D_1'^0 \to D^{*+}\pi^-)$ and (b) $\cbf(B^- \to D_0^{*0}\ell^-\bar{\nu}_{\ell})
\times \cbf(D_0^{*0} \to D^{+}\pi^-)$.}
  \label{fig:brdssl2}
 \end{center}
\end{figure}

For both the B-factory and the LEP and Tevatron results, the $B$ semileptonic 
signal yields are extracted from a fit to the invariant mass distribution of the $D^{(*)+}\pi^-$ system.
 The LEP and Tevatron measurements 
 are for the inclusive decays $\bar{B} \to D^{**}(D^*\pi^-)X \ell^- \bar{\nu}_{\ell}$. In the average with the results from the B-Factories, we use these measurements assuming that no particles are left in the $X$ system.
 The \babar tagged analysis of $\bar{B} \to D_2^* \ell^- \bar{\nu}_{\ell}$ was performed selecting $D_2^*\to D\pi$ decays.
 The \babar result reported in Table~\ref{tab:dss2lnu} is translated in a branching fraction for the $D_2^*\to D^*\pi$ decay mode assuming
 ${\cal B}(D_2^*\to D\pi)/{\cal B}(D_2^*\to D^*\pi)=1.54\pm 0.15$~\cite{PDG_2018}. 
Figure~\ref{fig:brdssl} and ~\ref{fig:brdssl2} show the measurements and the resulting averages.

\begin{table}[!htb]
\caption{Published and rescaled individual measurements and their averages for
of the branching fraction $\cbf(B^- \to D_1^0\ell^-\bar{\nu}_{\ell})\times \cbf(D_1^0 \to D^{*+}\pi^-)$. 
}
\begin{center}
\resizebox{0.99\textwidth}{!}{
\begin{tabular}{|l|c|c|}\hline
Experiment                                 &$\cbf(B^- \to D_1^0(D^{*+}\pi^-)\ell^-\bar{\nu}_{\ell})
 [\%]$  &$\cbf(B^- \to D_1^0(D^{*+}\pi^-)\ell^-\bar{\nu}_{\ell})
 [\%]$  \\
                                                & (rescaled) & (published) \\

\hline\hline 
ALEPH ~\cite{Aleph:Dss}        &$0.436 \pm0.085_{\rm stat} \pm0.056_{\rm syst}$ 
 &$0.47 \pm0.10_{\rm stat} \pm0.07_{\rm syst}$ \\
OPAL  ~\cite{opal:Dss}         &$0.568 \pm0.210_{\rm stat} \pm0.100_{\rm syst}$  
&$0.70 \pm0.21_{\rm stat} \pm0.10_{\rm syst}$ \\
CLEO  ~\cite{cleo:Dss}         &$0.349 \pm0.085_{\rm stat} \pm0.056_{\rm syst}$ 
 &$0.373 \pm0.085_{\rm stat} \pm0.057_{\rm syst}$ \\
D0  ~\cite{D0:Dss}         &$0.214 \pm0.018_{\rm stat} \pm0.035_{\rm syst}$  
&$0.219 \pm0.018_{\rm stat} \pm0.035_{\rm syst}$ \\
\belle Tagged $B^-$ ~\cite{Live:Dss}           &$0.430 \pm0.070_{\rm stat} \pm0.059_{\rm syst}$  
&$0.42 \pm0.07_{\rm stat} \pm0.07_{\rm syst}$ \\
\belle Tagged $B^0$ ~\cite{Live:Dss}           &$0.593 \pm0.200_{\rm stat} \pm0.076_{\rm syst}$  
&$0.42 \pm0.07_{\rm stat} \pm0.07_{\rm syst}$ \\ 
\babar Tagged ~\cite{Aubert:2009_4}           &$0.277 \pm0.030_{\rm stat} \pm0.029_{\rm syst}$
&$0.29 \pm0.03_{\rm stat} \pm0.03_{\rm syst}$ \\
\babar Untagged $B^-$ ~\cite{Aubert:2008zc}           &$0.293 \pm0.017_{\rm stat} \pm0.016_{\rm syst}$
&$0.30 \pm0.02_{\rm stat} \pm0.02_{\rm syst}$ \\
\babar Untagged $B^0$ ~\cite{Aubert:2008zc}           &$0.282 \pm0.026_{\rm stat} \pm0.023_{\rm syst}$
&$0.30 \pm0.02_{\rm stat} \pm0.02_{\rm syst}$ \\
\hline
{\bf Average}                              &\mathversion{bold}$0.281 \pm0.010 \pm 0.015$ 
    &\mathversion{bold}$\chi^2/\dof = 12.3/8$ (CL=$13.8\%$)  \\
\hline 
\end{tabular}
}
\end{center}
\label{tab:dss1lnu}
\end{table}
\begin{table}[!htb]
\caption{Published and rescaled individual measurements and their averages for 
$\cbf(B^- \to D_2^0\ell^-\bar{\nu}_{\ell})\times \cbf(D_2^0 \to D^{*+}\pi^-)$. 
}
\begin{center}
\resizebox{0.99\textwidth}{!}{
\begin{tabular}{|l|c|c|}\hline
Experiment                                 &$\cbf(B^- \to D_2^0(D^{*+}\pi^-)\ell^-\bar{\nu}_{\ell})
 [\%]$  &$\cbf(B^- \to D_2^0(D^{*+}\pi^-)\ell^-\bar{\nu}_{\ell})
 [\%]$  \\
                                                & (rescaled) & (published) \\
\hline\hline 
CLEO  ~\cite{cleo:Dss}         &$0.055 \pm0.066_{\rm stat} \pm0.011_{\rm syst}$ 
 &$0.059 \pm0.066_{\rm stat} \pm0.011_{\rm syst}$ \\
D0  ~\cite{D0:Dss}         &$0.086 \pm0.018_{\rm stat} \pm0.020_{\rm syst}$  
&$0.088 \pm0.018_{\rm stat} \pm0.020_{\rm syst}$ \\
\belle tagged  ~\cite{Live:Dss}           &$0.190 \pm0.060_{\rm stat} \pm0.025_{\rm syst}$  
&$0.18 \pm0.06_{\rm stat} \pm0.03_{\rm syst}$ \\
\babar tagged ~\cite{Aubert:2009_4}           &$0.075 \pm0.013_{\rm stat} \pm0.009_{\rm syst}$
&$0.078 \pm0.013_{\rm stat} \pm0.010_{\rm syst}$ \\
\babar untagged $B^-$ ~\cite{Aubert:2008zc}           &$0.087 \pm0.009_{\rm stat} \pm0.007_{\rm syst}$
&$0.087 \pm0.013_{\rm stat} \pm0.007_{\rm syst}$ \\
\babar untagged $B^0$ ~\cite{Aubert:2008zc}           &$0.065 \pm0.010_{\rm stat} \pm0.004_{\rm syst}$
&$0.087 \pm0.013_{\rm stat} \pm0.007_{\rm syst}$ \\
\hline
{\bf Average}                              &\mathversion{bold}$0.077 \pm0.006 \pm 0.004$ 
    &\mathversion{bold}$\chi^2/\dof = 5.4/5$ (CL=$36.7\%$)  \\
\hline 
\end{tabular}
}
\end{center}
\label{tab:dss2lnu}
\end{table}
\begin{table}[!htb]
\caption{
Published and rescaled individual measurements and their averages for 
$\cbf(B^- \to D_1^{'0}\ell^-\bar{\nu}_{\ell})\times \cbf(D_1^{'0} \to D^{*+}\pi^-)$. 
}
\begin{center}
\begin{tabular}{|l|c|c|}\hline
Experiment                                 &$\cbf(B^- \to D_1^{'0}(D^{*+}\pi^-)\ell^-\bar{\nu}_{\ell})
 [\%]$  &$\cbf(B^- \to D_1^{'0}(D^{*+}\pi^-)\ell^-\bar{\nu}_{\ell})
 [\%]$  \\
                                                & (rescaled) & (published) \\
\hline\hline 
DELPHI ~\cite{Abdallah:2005cx}        &$0.73 \pm0.17_{\rm stat} \pm0.18_{\rm syst}$ 
 &$0.83 \pm0.17_{\rm stat} \pm0.18_{\rm syst}$ \\
\belle  ~\cite{Live:Dss}           &$-0.03 \pm0.06_{\rm stat} \pm0.07_{\rm syst}$  
&$-0.03 \pm0.06_{\rm stat} \pm0.07_{\rm syst}$ \\
\babar  ~\cite{Aubert:2009_4}           &$0.26 \pm0.04_{\rm stat} \pm0.04_{\rm syst}$
&$0.27 \pm0.04_{\rm stat} \pm0.05_{\rm syst}$ \\
\hline
{\bf Average}                              &\mathversion{bold}$0.19 \pm 0.03 \pm0.04$ 
    &\mathversion{bold}$\chi^2/\dof = 11.9/2$ (CL=$0.003\%$)  \\
\hline 
\end{tabular}
\end{center}
\label{tab:dss1plnu}
\end{table}
\begin{table}[!htb]
\caption{Published and rescaled individual measurements and their averages for  
$\cbf(B^- \to D_0^{*0}\ell^-\bar{\nu}_{\ell})\times \cbf(D_0^{*0} \to D^{+}\pi^-)$. }
\begin{center}
\begin{tabular}{|l|c|c|}\hline
Experiment                                 &$\cbf(B^- \to D_0^{*0}(D^{+}\pi^-)\ell^-\bar{\nu}_{\ell})
 [\%]$  &$\cbf(B^- \to D_0^{*0}(D^{+}\pi^-)\ell^-\bar{\nu}_{\ell})
 [\%]$ \\
						& (rescaled) & (published) \\
\hline\hline 
\belle Tagged $B^-$ ~\hfill\cite{Live:Dss}           &$0.25 \pm0.04_{\rm stat} \pm0.06_{\rm syst}$  
&$0.24 \pm0.04_{\rm stat} \pm0.06_{\rm syst}$ \\
\belle Tagged $B^0$ ~\hfill\cite{Live:Dss}           &$0.22 \pm0.08_{\rm stat} \pm0.06_{\rm syst}$  
&$0.24 \pm0.04_{\rm stat} \pm0.06_{\rm syst}$ \\
\babar Tagged ~\hfill\cite{Aubert:2009_4}            &$0.32 \pm0.04_{\rm stat} \pm0.05_{\rm syst}$
&$0.26 \pm0.05_{\rm stat} \pm0.04_{\rm syst}$ \\
\hline
{\bf Average}                              &\mathversion{bold}$0.28 \pm 0.03 \pm0.04$ 
    &\mathversion{bold}$\chi^2/\dof = 0.82/2$ (CL=$66.4\%$)  \\
\hline 
\end{tabular}
\end{center}
\label{tab:dss0lnu}
\end{table}

\subsection{Inclusive CKM-favored decays}
\label{slbdecays_b2cincl}

\subsubsection{Global analysis of $\bar B\to X_c\ell^-\bar\nu_\ell$}

The semileptonic decay width $\Gamma(\bar B\to X_c\ell^-\bar\nu_\ell)$ has
been calculated in the framework of the operator production expansion
(OPE)~\cite{Shifman:1986mx,Chay:1990da,Bigi:1992su}.
The result is a double-expansion in $\Lambda_{\rm QCD}/m_b$ and
$\alpha_s$, which depends on a number of non-perturbative
parameters. These parameters describe the dynamics of the
$b$-quark inside the $B$~hadron and can be measured using
observables in $\bar B\to X_c\ell^-\bar\nu_\ell$ decays, such as the
moments of the lepton energy and the hadronic mass spectrum.

Two renormalization schemes are commonly used to define the $b$-quark mass
and other theoretical quantities: the
kinetic~\cite{Benson:2003kp,Gambino:2004qm,Gambino:2011cq,Alberti:2014yda}
and the 1S~\cite{Bauer:2004ve} schemes. An independent set of theoretical
expressions is available for each, with several non-perturbative parameters.
The non-perturbative parameters in the kinetic scheme
are: the quark masses $m_b$ and $m_c$, $\mu^2_\pi$ and
$\mu^2_G$ at $O(1/m^2_b)$, and $\rho^3_D$ and $\rho^3_{LS}$ at
$O(1/m^3_b)$. In the 1S scheme, the parameters are: $m_b$, $\lambda_1$
at $O(1/m^2_b)$, and $\rho_1$, $\tau_1$, $\tau_2$ and $\tau_3$ at
$O(1/m^3_b)$.
Note that the numerical values of the kinetic and 1S $b$-quark masses cannot
be compared without converting one or the other, or both, to the same
renormalization scheme.

We use two sets of inclusive observables in $\bar B\to X_c\ell^-\bar\nu_\ell$ decays to constrain OPE parameters: the moments of the hadronic system effective mass $\langle M^n_X\rangle$ of order $n=2,4,6$, and the moments of the charged lepton
momentum $\langle E^n_\ell\rangle$ of order $n=0,1,2,3$.
Moments are determined for different values of $E_\mathrm{cut}$, the lower
limit on the lepton momentum.  Moments derived from the same
spectrum with different value of $E_\mathrm{cut}$ are highly correlated.
The list of measurements used in our analysis is given in Table~\ref{tab:gf_input}. The only external input is the average lifetime~$\tau_B$ of neutral and charged $B$~mesons,
taken to be $(1.579\pm 0.004)$~ps (Sec.~\ref{sec:life_mix}).
\begin{table}[!htb]
\caption{Experimental inputs used in the global analysis of $\bar B\to
  X_c\ell^-\bar\nu_\ell$. $n$ is the order of the moment, $c$ is the
  threshold value of the lepton momentum in GeV. In total, there are
  23 measurements from \babar, 15 measurements from Belle and 12 from
  other experiments.} \label{tab:gf_input}
\begin{center}
\begin{tabular}{|l|l|l|}
  \hline
  Experiment
  & Hadron moments $\langle M^n_X\rangle$
  & Lepton moments $\langle E^n_\ell\rangle$\\
  \hline \hline
  \babar & $n=2$, $c=0.9,1.1,1.3,1.5$ & $n=0$, $c=0.6,1.2,1.5$\\
  & $n=4$, $c=0.8,1.0,1.2,1.4$ & $n=1$, $c=0.6,0.8,1.0,1.2,1.5$\\
  & $n=6$, $c=0.9,1.3$~\cite{Aubert:2009qda} & $n=2$, $c=0.6,1.0,1.5$\\
  & & $n=3$, $c=0.8,1.2$~\cite{Aubert:2009qda,Aubert:2004td}\\
  \hline
  Belle & $n=2$, $c=0.7,1.1,1.3,1.5$ & $n=0$, $c=0.6,1.4$\\
  & $n=4$, $c=0.7,0.9,1.3$~\cite{Schwanda:2006nf} & $n=1$,
  $c=1.0,1.4$\\
  & & $n=2$, $c=0.6,1.4$\\
  & & $n=3$, $c=0.8,1.2$~\cite{Urquijo:2006wd}\\
  \hline
  CDF & $n=2$, $c=0.7$ & \\
  & $n=4$, $c=0.7$~\cite{Acosta:2005qh} & \\
  \hline
  CLEO & $n=2$, $c=1.0,1.5$ & \\
  & $n=4$, $c=1.0,1.5$~\cite{Csorna:2004kp} & \\
  \hline
  DELPHI & $n=2$, $c=0.0$ & $n=1$, $c=0.0$ \\
  & $n=4$, $c=0.0$ & $n=2$, $c=0.0$ \\
  & $n=6$, $c=0.0$~\cite{Abdallah:2005cx} & $n=3$,
  $c=0.0$~\cite{Abdallah:2005cx}\\
  \hline
\end{tabular}
\end{center}
\end{table}

In the kinetic and 1S schemes, the moments in $\bar B\to
X_c\ell^-\bar\nu_\ell$ are not sufficient to determine the $b$-quark
mass precisely. In the kinetic scheme analysis only a combination of $m_b$ and $m_c$ is well determined and we constrain the $c$-quark
mass (defined in the $\overline{\rm MS}$ scheme) to the value of
Ref.~\cite{Chetyrkin:2009fv} to pinpoint $m_b$,
\begin{equation}
  m_c^{\overline{\rm MS}}(3~{\rm GeV})=0.986\pm 0.013~{\rm GeV}~.
\end{equation}
In the 1S~scheme analysis, the $b$-quark mass is constrained by
measurements of the photon energy moments in $B\to
X_s\gamma$~\cite{Aubert:2005cua,Aubert:2006gg,Limosani:2009qg,Chen:2001fja}.

\subsubsection{Analysis in the kinetic scheme}
\label{globalfitsKinetic}

\begin{figure}
\begin{center}
  \includegraphics[width=8.2cm]{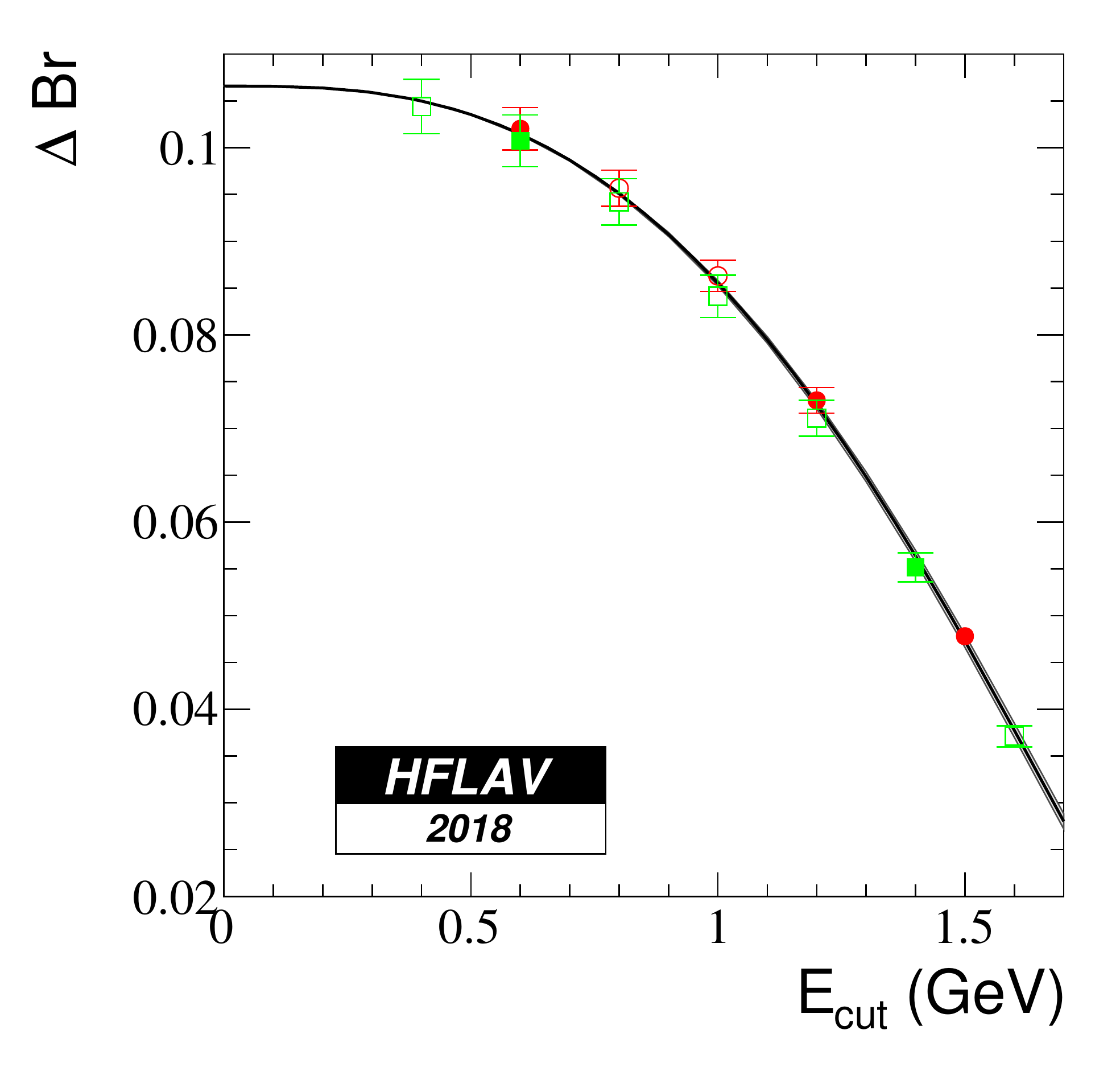}
  \includegraphics[width=8.2cm]{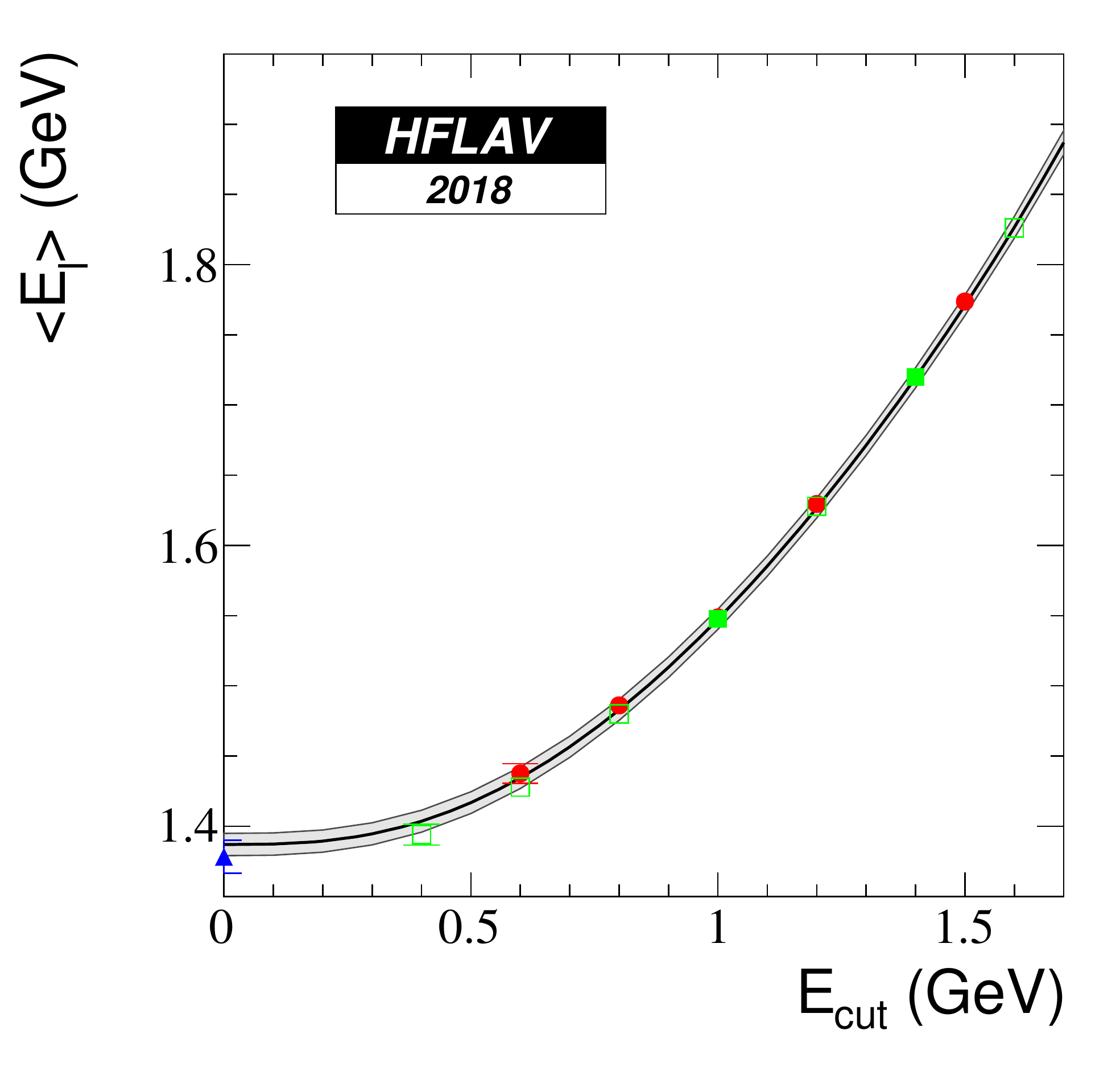}\\
  \includegraphics[width=8.2cm]{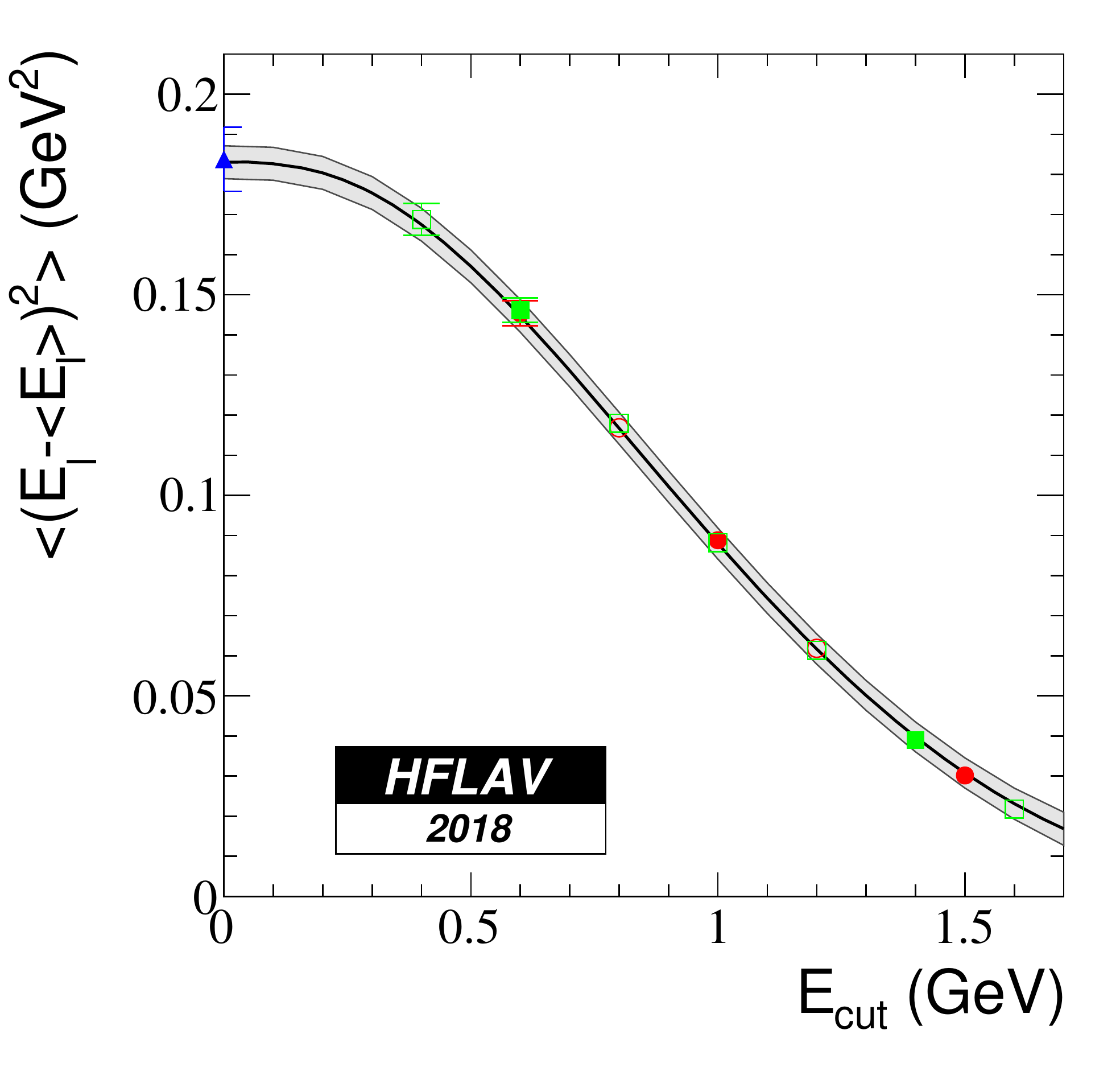}
  \includegraphics[width=8.2cm]{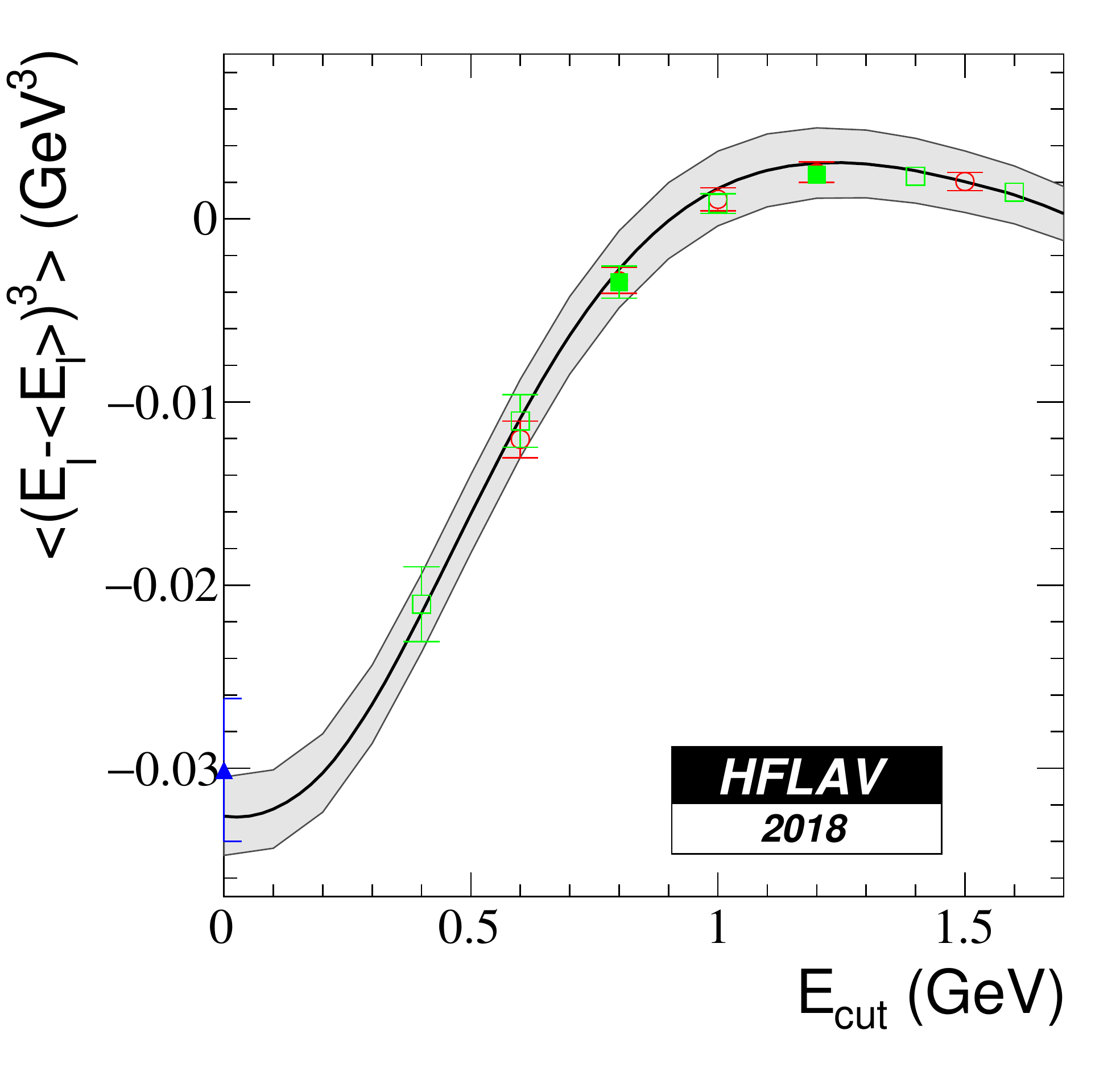}
\end{center}
\caption{Fit to the inclusive partial semileptonic branching fractions and to
  the lepton energy moments in the kinetic mass scheme. In all plots, the
  grey band is the theory prediction with total theory error. \babar
  data are shown by circles, Belle by squares and other experiments
  (DELPHI, CDF, CLEO) by triangles. Filled symbols mean that the point
  was used in the fit. Open symbols are measurements that were not
  used in the fit.} \label{fig:gf_res_kin_el}
\end{figure}

\begin{figure}
\begin{center}
  \includegraphics[width=8.2cm]{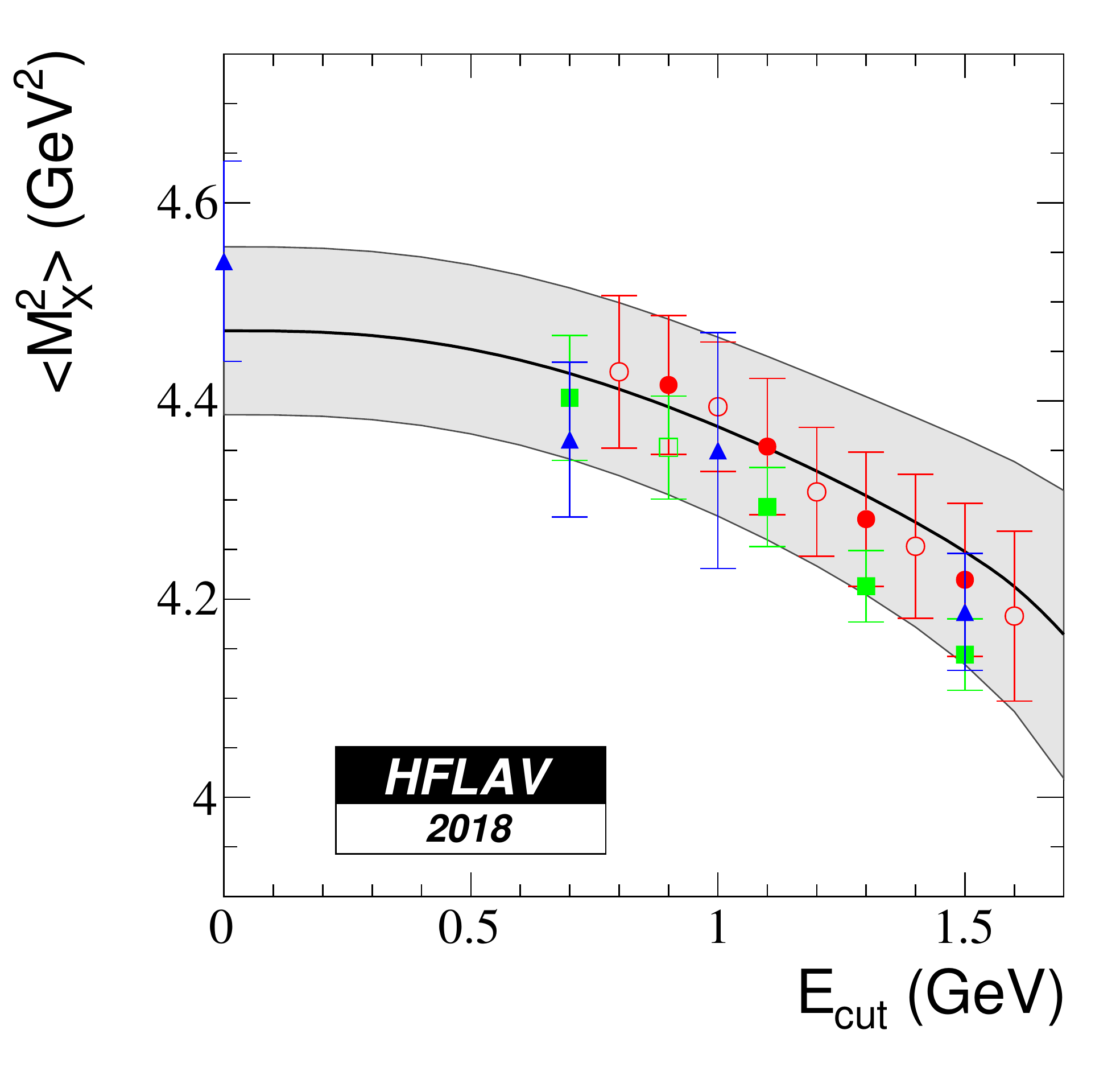}
  \includegraphics[width=8.2cm]{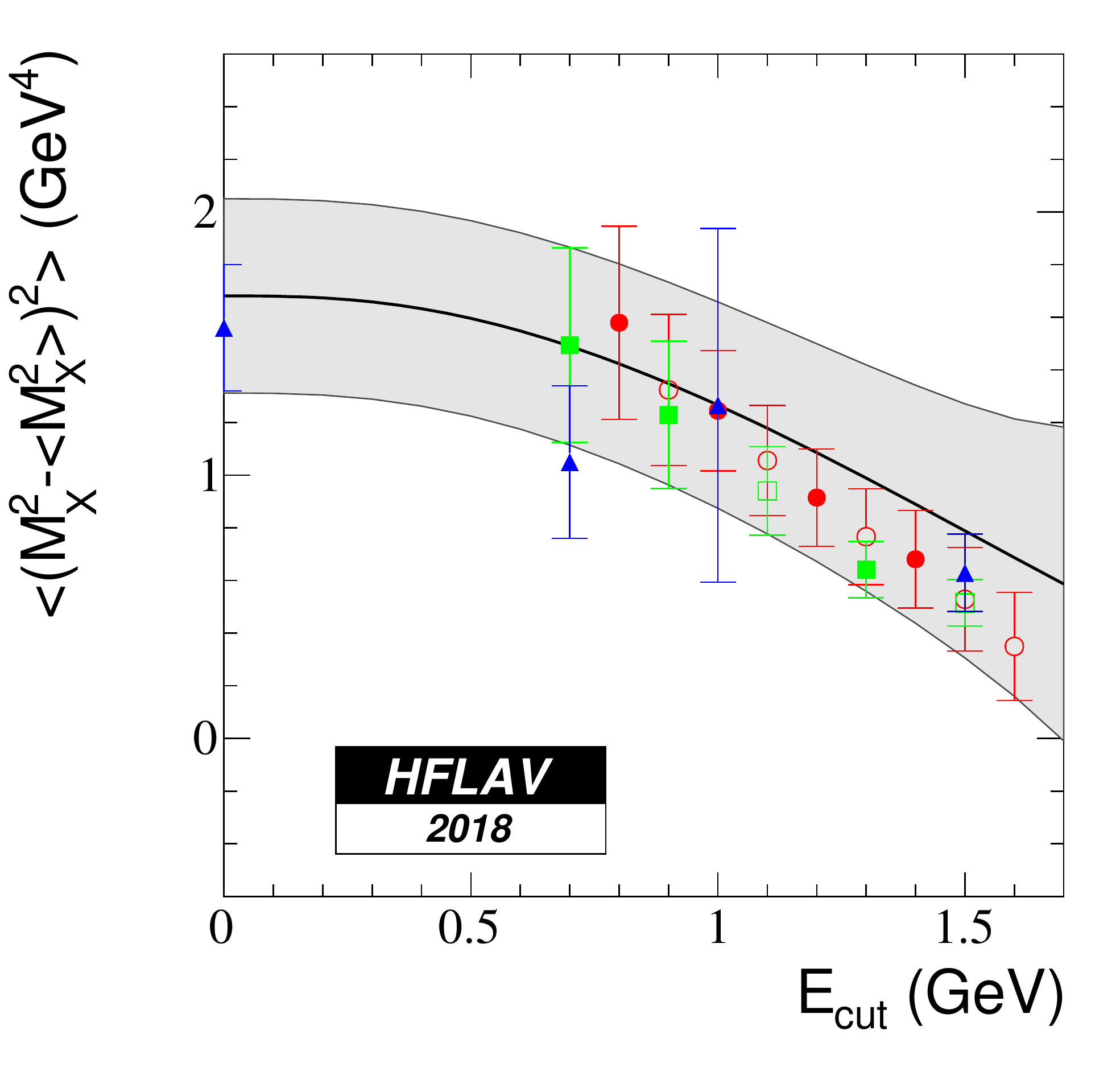}\\
  \includegraphics[width=8.2cm]{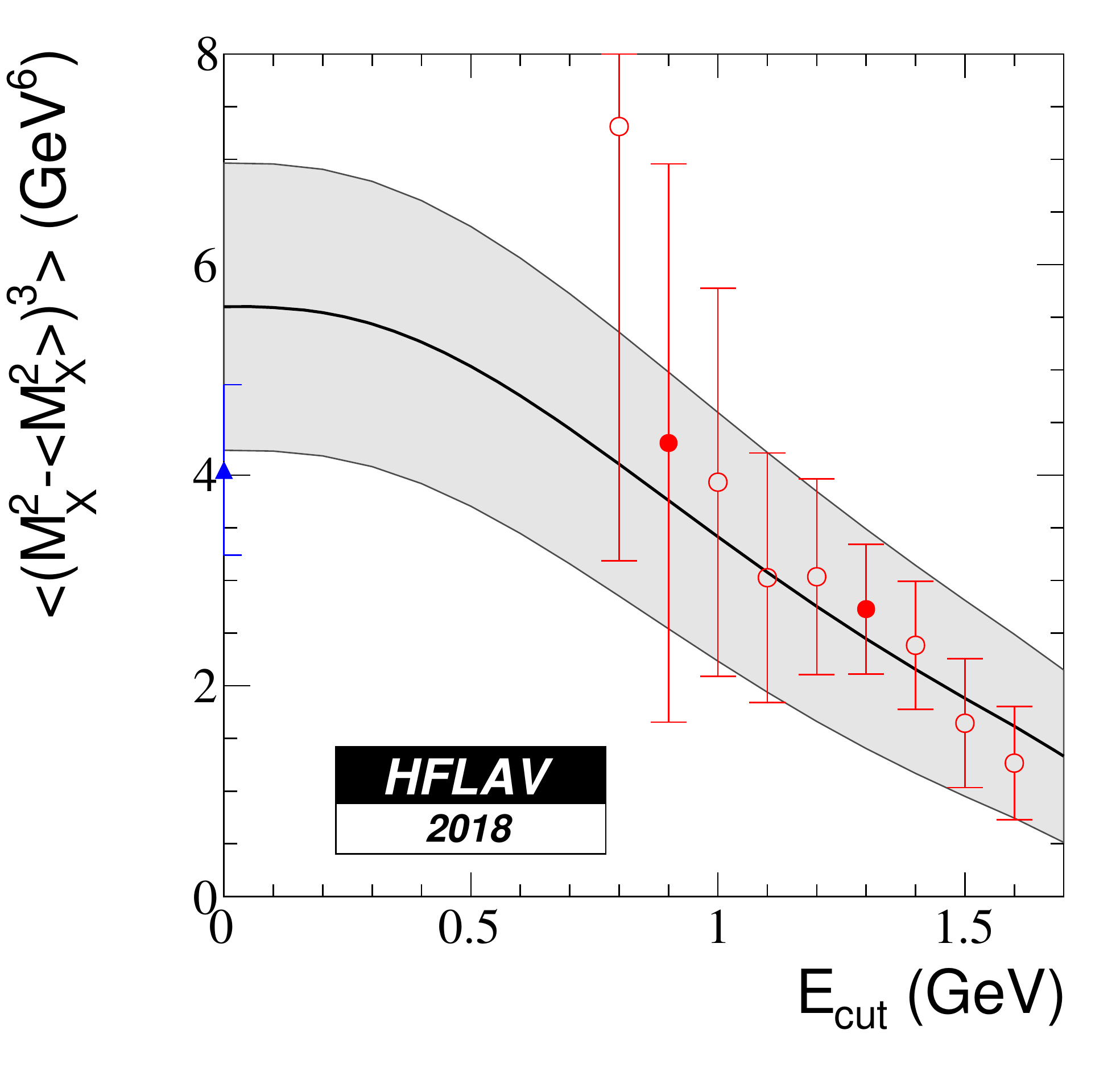}
\end{center}
\caption{Same as Fig.~\ref{fig:gf_res_kin_el} for the fit to the
  hadronic mass moments in the kinetic mass
  scheme.} \label{fig:gf_res_kin_mx}
\end{figure}

We obtain $\vcb$ and the six non-perturbative parameters mentioned above with a  fit that follows closely the procedure described in Ref.~\cite{Gambino:2013rza} and relies on the calculations of the lepton energy and hadronic mass
moments in $\bar B\to X_c\ell^-\bar\nu_\ell$~decays described in
Ref.~\cite{Gambino:2011cq,Alberti:2014yda}. 
The detailed fit result and the matrix of the correlation coefficients is
given in Table~\ref{tab:gf_res_mc_kin}. Projections of the fit onto the lepton energy and
hadronic mass moments are shown in Figs.~\ref{fig:gf_res_kin_el} and
\ref{fig:gf_res_kin_mx}, respectively. The result in terms of the main
parameters is
\begin{eqnarray}
  \vcb & = & (42.19\pm 0.78)\times 10^{-3}~, \\
  m_b^{\rm kin} & = & 4.554\pm 0.018~{\rm GeV}~, \\
  \mu^2_\pi & = & 0.464\pm 0.076~{\rm GeV^2}~,
\end{eqnarray}
with a $\chi^2$ of 15.6 for $43$ degrees of freedom. The scale $\mu$ of the
quantities in the kinetic scheme is 1~GeV.
\begin{table}[!htb]
\caption{Fit result in the kinetic scheme, using a precise $c$-quark
  mass constraint. The error matrix of the fit contains
  experimental and theoretical contributions. In the lower part of the
  table, the correlation matrix of the parameters is
  given. The scale $\mu$ of the quantities in the kinematic scheme is 1~GeV.}
\label{tab:gf_res_mc_kin}
\begin{center}
\resizebox{0.99\textwidth}{!}{
\begin{tabular}{|l|ccccccc|}
  \hline
  & \vcb\ [10$^{-3}$] & $m_b^{\rm kin}$ [GeV] &
  $m_c^{\overline{\rm MS}}$ [GeV] & $\mu^2_\pi$ [GeV$^2$]
  & $\rho^3_D$ [GeV$^3$] & $\mu^2_G$ [GeV$^2$] & $\rho^3_{LS}$ [GeV$^3$]\\
  \hline \hline
  value & 42.19 & \phantom{$-$}4.554 & \phantom{$-$}0.987 &
  \phantom{$-$}0.464 & \phantom{$-$}0.169 & \phantom{$-$}0.333 &
  $-$0.153\\
  error & 0.78 & \phantom{$-$}0.018 &
  \phantom{$-$}0.015 & \phantom{$-$}0.076 & \phantom{$-$}0.043 &
  \phantom{$-$}0.053 & \phantom{$-$}0.096\\
  \hline
  $|V_{cb}|$ & 1.000 & $-$0.257 & $-$0.078 &
  \phantom{$-$}0.354 & \phantom{$-$}0.289 & $-$0.080 &
  $-$0.051\\
  $m_b^{\rm kin}$ & & \phantom{$-$}1.000 & \phantom{$-$}0.769 &
  $-$0.054 & \phantom{$-$}0.097 & \phantom{$-$}0.360 & $-$0.087\\
  $m_c^{\overline{\rm MS}}$ & & & \phantom{$-$}1.000
  & $-$0.021 & \phantom{$-$}0.027 & \phantom{$-$}0.059 & $-$0.013\\
  $\mu^2_\pi$ & & & & \phantom{$-$}1.000 & \phantom{$-$}0.732 &
  \phantom{$-$}0.012 & \phantom{$-$}0.020\\
  $\rho^3_D$ & & & & & \phantom{$-$}1.000 & $-$0.173 & $-$0.123\\
  $\mu^2_G$ & & & & & & \phantom{$-$}1.000 & \phantom{$-$}0.066\\
  $\rho^3_{LS}$ & & & & & & & \phantom{$-$}1.000\\
  \hline
\end{tabular}
}
\end{center}
\end{table}

The inclusive $\bar B\to X_c\ell^-\bar\nu_\ell$ branching fraction
determined by this analysis is
\begin{equation}
  \cbf(\bar B\to X_c\ell^-\bar\nu_\ell)=(10.65\pm 0.16)\%~.
\end{equation}
Including the branching fraction of charmless semileptonic decays
(Sec.~\ref{slbdecays_b2uincl}), $\cbf(\bar B\to
X_u\ell^-\bar\nu_\ell)=(2.13\pm 0.30)\times 10^{-3}$, we obtain the
semileptonic branching fraction,
\begin{equation}
  \cbf(\bar B\to X\ell^-\bar\nu_\ell)=(10.86\pm 0.16)\%~.
\end{equation}

\subsubsection{Analysis in the 1S scheme}
\label{globalfits1S}

The fit relies on the same set of moment measurements and the calculations of
the spectral moments described in
Ref.~\cite{Bauer:2004ve}. The theoretical uncertainties are estimated
as explained in Ref.~\cite{Schwanda:2008kw}. No theory error correlations between different moments are assumed (except between identical moments, {\it i.e.}, moments with same values of $n$ and $c$).
The detailed result of the fit using the $B\to X_s\gamma$ constraint is given
in Table~\ref{tab:gf_res_xsgamma_1s}. The result in terms of the main
parameters is
\begin{eqnarray}
  \vcb & = & (41.98\pm 0.45)\times 10^{-3}~, \\
  m_b^{1S} & = & 4.691\pm 0.037~{\rm GeV}~, \\
  \lambda_1 & = & -0.362\pm 0.067~{\rm GeV^2}~,
\end{eqnarray}
with a $\chi^2$ of 23.0 for $59$ degrees of freedom. We find a good agreement
in the central values of \vcb\ between the kinetic and 1S scheme analyses. No
conclusion should, however, been drawn regarding the uncertainties in \vcb, as
the two approaches are not equivalent in the number of higher-order corrections
that are included.

\begin{table}[!htb]
\caption{Fit result in the 1S scheme, using $B\to X_s\gamma$~moments
  as a constraint. In the lower part of the table, the correlation
  matrix of the parameters is given.} \label{tab:gf_res_xsgamma_1s}
\begin{center}
\begin{tabular}{|l|ccccccc|}
  \hline
  & $m_b^{1S}$ [GeV] & $\lambda_1$ [GeV$^2$] & $\rho_1$ [GeV$^3$] &
  $\tau_1$ [GeV$^3$] & $\tau_2$ [GeV$^3$] & $\tau_3$ [GeV$^3$] &
  $\vcb$ [10$^{-3}$]\\
  \hline \hline
  value & 4.691 & $-0.362$ & \phantom{$-$}0.043 &
  \phantom{$-$}0.161 & $-0.017$ & \phantom{$-$}0.213 &
  \phantom{$-$}41.98\\
  error & 0.037 & \phantom{$-$}0.067 & \phantom{$-$}0.048 &
  \phantom{$-$}0.122 & \phantom{$-$}0.062 & \phantom{$-$}0.102 &
  \phantom{$-$}0.45\\
  \hline
  $m_b^{1S}$ & 1.000 & \phantom{$-$}0.434 & \phantom{$-$}0.213 &
  $-0.058$ & $-0.629$ & $-0.019$ & $-0.215$\\
  $\lambda_1$ & & \phantom{$-$}1.000 & $-0.467$ & $-0.602$ & $-0.239$
  & $-0.547$ & $-0.403$\\
  $\rho_1$ & & & \phantom{$-$}1.000 & \phantom{$-$}0.129 & $-0.624$ &
  \phantom{$-$}0.494 & \phantom{$-$}0.286\\
  $\tau_1$ & & & & \phantom{$-$}1.000 & \phantom{$-$}0.062 & $-0.148$ &
  \phantom{$-$}0.194\\
  $\tau_2$ & & & & & \phantom{$-$}1.000 & $-0.009$ & $-0.145$\\
  $\tau_3$ & & & & & & \phantom{$-$}1.000 & \phantom{$-$}0.376\\
  $\vcb$ & & & & & & & \phantom{$-$}1.000\\
  \hline
\end{tabular}
\end{center}
\end{table}

\subsection{Exclusive CKM-suppressed decays}
\label{slbdecays_b2uexcl}
In this section, we give results on exclusive charmless semileptonic branching fractions
and the determination of $\Vub$ based on \Btopilnu\ decays.
The measurements are based on two different event selections: tagged
events, in which the second $B$ meson in the event is fully (or partially)
reconstructed, and untagged events, for which the momentum
of the undetected neutrino is inferred from measurements of the total 
momentum sum of the detected particles and the knowledge of the initial state.
The LHCb experiment has reported a direct measurement of 
$|V_{ub}|/|V_{cb}|$ \cite{Aaij:2015bfa}, reconstructing the 
$\Lb\to p\mu\nu$ decays and normalizing the branching fraction to the 
$\Lb\to\Lc(\to pK\pi)\mu\nu$ decays. 
We show a combination of $\Vub$ and $\Vcb$ using the LHCb constraint on $|V_{ub}|/|V_{cb}|$, 
the exclusive determination of $\Vub$ from \Btopilnu, and $\Vcb$ from both $B\to D^*\ell\nu$ 
and $B\to D\ell\nu$. 
We also present branching fraction averages for 
$\Bz\to\rho\ell^+\nu$, $\Bp\to\omega\ell^+\nu$, $\Bp\to\eta\ell^+\nu$ and $\Bp\to\etapr\ell^+\nu$.

\subsubsection{\Btopilnu\ branching fraction and $q^2$ spectrum}

\begin{figure} 
 \centering
 \includegraphics[width=0.8\textwidth,page=1]{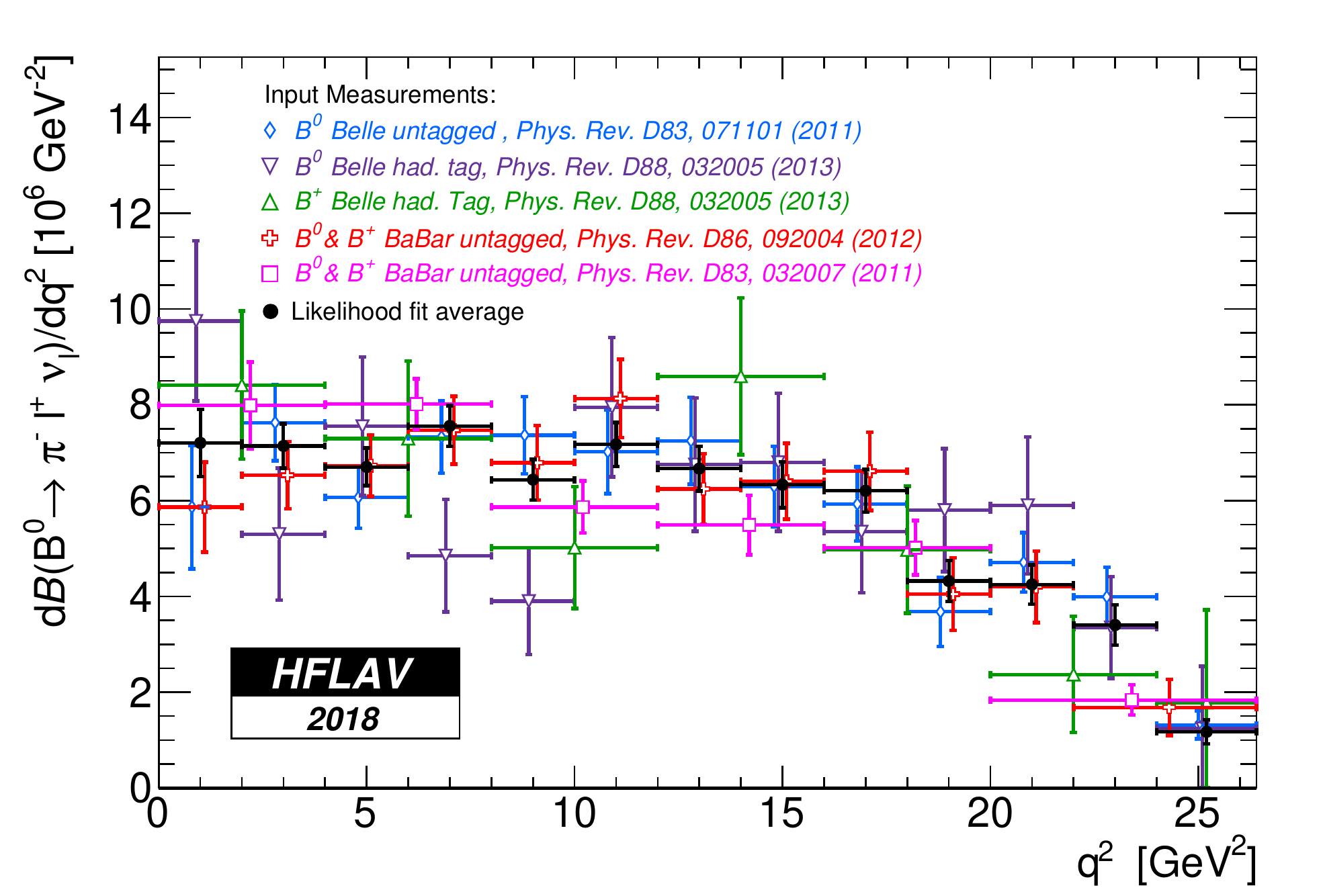}
\caption{The \Btopilnu\ $q^2$ spectrum measurements and the average spectrum obtained 
from the likelihood combination (shown in black). \label{fig:avg}}
\end{figure}

We use the four most precise measurements of the differential \Btopilnu\ decay rate as a function of the four-momentum transfer squared, $q^2$,
from \babar and Belle~\cite{Ha:2010rf,Sibidanov:2013rkk,delAmoSanchez:2010af,Lees:2012vv}
to obtain an average $q^2$ spectrum and an average for the total branching fraction. 
The measurements are presented in Fig.~\ref{fig:avg}.
From the two untagged \babar\ analyses~\cite{delAmoSanchez:2010af,Lees:2012vv},
the combined results for $B^0 \to \pi^- \ell^+ \nu$ and $B^+ \to \pi^0 \ell^+ \nu$ decays based on isospin symmetry are used.
The hadronic-tag analysis by Belle~\cite{Sibidanov:2013rkk} provides results for $B^0 \to \pi^- \ell^+ \nu$ and
$B^+ \to \pi^0 \ell^+ \nu$ separately, but not for the combination of both channels.
In the untagged analysis by Belle~\cite{Ha:2010rf}, only $B^0 \to \pi^- \ell^+ \nu$ decays were measured.
The experimental measurements use different binnings in $q^2$, but have matching bin edges, which allows
them to be easily combined.

To arrive at an average $q^2$ spectrum, a binned maximum-likelihood fit to determine the average partial branching fraction
in each $q^2$ interval is performed, differentiating between common and individual uncertainties and correlations for the
various measurements.
Shared sources of systematic uncertainty of all measurements are included in the likelihood
as nuisance parameters constrained using normal distributions.
The most important shared sources of uncertainty are due to 
continuum subtraction, 
branching fractions, 
the number of $B$-meson pairs (only correlated among measurement by the same experiment), 
tracking efficiency (only correlated among measurements by the same experiment), 
uncertainties from modelling the $b \to u \, \ell \, \bar\nu_\ell$ contamination,
modelling of final state radiation, 
and contamination from $b \to c \, \ell \bar \nu_\ell$ decays. 

The averaged $q^2$ spectrum is shown in Fig.~\ref{fig:avg}. 
The probability of the average is computed as the $\chi^2$ probability quantifying the agreement between
the input spectra and the averaged spectrum and amounts to $6\%$.
The partial branching fractions and the full covariance matrix obtained from the likelihood fit 
are given in Tables \ref{tab:average} and ~\ref{tab:Cov}.
The average for the total $B^0 \to \pi^- \ell^+ \nu_\ell$ branching fraction is obtained by summing up the
partial branching fractions:
\begin{align}
{\cal B}(B^0 \to \pi^- \ell^+ \nu_\ell) = (1.50 \pm 0.02_{\rm stat} \pm 0.06_{\rm syst}) \times 10^{-4}.
\end{align}

\begin{table}
\centering
\small
\caption{Partial $B^0 \to \pi^- \ell^+ \nu_\ell$ branching fractions per GeV$^2$ for the input measurements and the average
obtained from the likelihood fit. The uncertainties are the combined statistical and systematic uncertainties.\label{tab:average}}
\begin{tabular}{|c | c | c | c | c | c || c|}
\hline
\multirow{2}{2cm}{$\Delta q^2$ [GeV${}^2$] }&\multicolumn{6}{ c |}{ $\Delta \mathcal{B}(B^0 \to \pi^- \ell^+ \nu_\ell) / \Delta q^2 $\quad$ [10^{-7}]$}\\\cline{2-7}
& Belle & Belle & Belle & \babar & \babar & Average\\
& untagged & tagged & tagged & untagged & untagged & \\
& ($B^0$) & ($B^0$) & ($B^+$) & ($B^{0,+}$, 12 bins) & ($B^{0,+}$, 6 bins) & \\\hline
$ 0 - 2 $ & $ 58.7 \pm 12.9 $ & $ 97.5 \pm 16.7 $ &\multirow{2}{2cm}{ $ 84.1 \pm 15.5 $ } & $ 58.7 \pm 9.4 $ &\multirow{2}{2cm}{ $ 79.9 \pm 9.1 $} & $ 72.0 \pm 7.0 $\\
$ 2 - 4 $ & $ 76.3 \pm 8.0 $ & $ 53.0 \pm 13.8 $ & & $ 65.3 \pm 7.1 $ & & $ 71.4 \pm 4.6 $\\
$ 4 - 6 $ & $ 60.6 \pm 6.4 $ & $ 75.5 \pm 14.5 $ &\multirow{2}{2cm}{ $ 73.0 \pm 16.2 $} & $ 67.3 \pm 6.4 $ &\multirow{2}{2cm}{ $ 80.1 \pm 5.3 $ }& $ 67.0 \pm 3.9 $\\
$ 6 - 8 $ & $ 73.3 \pm 7.6 $ & $ 48.5 \pm 11.8 $ & & $ 74.7 \pm 7.1 $ & & $ 75.6 \pm 4.3 $\\
$ 8 - 10 $ & $ 73.7 \pm 8.1 $ & $ 39.0 \pm 11.2 $ &\multirow{2}{2cm}{ $ 50.2 \pm 12.8 $ }& $ 67.9 \pm 7.8 $ &\multirow{2}{2cm}{ $ 58.7 \pm 5.5 $} & $ 64.4 \pm 4.3 $\\
$ 10 - 12 $ & $ 70.2 \pm 8.8 $ & $ 79.5 \pm 14.6 $ & & $ 81.3 \pm 8.2 $ & & $ 71.7 \pm 4.6 $\\
$ 12 - 14 $ & $ 72.5 \pm 9.1 $ & $ 67.5 \pm 13.9 $ &\multirow{2}{2cm}{ $ 86.0 \pm 16.4 $ }& $ 62.4 \pm 7.4 $ &\multirow{2}{2cm}{ $ 54.9 \pm 6.2 $} & $ 66.7 \pm 4.7 $\\
$ 14 - 16 $ & $ 63.0 \pm 8.4 $ & $ 68.0 \pm 14.4 $ & & $ 64.0 \pm 7.9 $ & & $ 63.3 \pm 4.8 $\\
$ 16 - 18 $ & $ 59.3 \pm 7.8 $ & $ 53.5 \pm 12.8 $ &\multirow{2}{2cm}{ $ 49.7 \pm 13.3 $} & $ 66.1 \pm 8.2 $ &\multirow{2}{2cm}{ $ 50.2 \pm 5.7 $ }& $ 62.0 \pm 4.4 $\\
$ 18 - 20 $ & $ 36.8 \pm 7.2 $ & $ 58.0 \pm 12.8 $ & & $ 40.5 \pm 7.6 $ & & $ 43.2 \pm 4.3 $\\
$ 20 - 22 $ & $ 47.1 \pm 6.2 $ & $ 59.0 \pm 14.3 $ &\multirow{2}{2cm}{ $ 23.7 \pm 12.1 $ }& $ 42.0 \pm 7.5 $ &\multirow{3}{2cm}{ $ 18.4 \pm 3.2 $ }& $ 42.5 \pm 4.1 $\\
$ 22 - 24 $ & $ 39.9 \pm 6.2 $ & $ 33.5 \pm 10.6 $ & & \multirow{2}{2cm}{ $ 16.8 \pm 5.9 $ }& & $ 34.0 \pm 4.2 $\\
$ 24 - 26.4 $ & $ 13.2 \pm 2.9 $ & $ 12.4 \pm 13.0 $ & $ 17.8 \pm 19.4 $ & & & $ 11.7 \pm 2.6 $\\\hline
\end{tabular}
\end{table}

\begin{table}
\caption{Covariance matrix of the averaged partial branching fractions per GeV$^2$ in units of $10^{-14}$.\label{tab:Cov}}
\tiny
\begin{tabular}{c | c c c c c c c c c c c c c c }
$\Delta q^2$ [GeV${}^{2}$] &$0-2$ &$ 2-4$ &$4-6$ &$6-8 $&$8-10$ &$10-12$ &$12-14$ &$14-16$ &$16-18 $&$18-20$ &$20-22$ &$22-24$ &$24-26.4$ \\\hline
$0-2$&$49.091$&$1.164$&$8.461$&$7.996$&$7.755$&$9.484$&$7.604$&$9.680$&$8.868$&$7.677$&$7.374$&$7.717$&$2.877$\\
$2-4$&        &$21.487$&$-0.0971$&$7.155$&$4.411$&$5.413$&$4.531$&$4.768$&$4.410$&$3.442$&$3.597$&$3.388$&$1.430$\\
$4-6$&        &        &$15.489$&$-0.563$&$5.818$&$4.449$&$4.392$&$4.157$&$4.024$&$3.185$&$3.169$&$3.013$&$1.343$\\
$6-8$&        &        &        &$18.2$&$2.377$&$7.889$&$6.014$&$5.938$&$5.429$&$4.096$&$3.781$&$3.863$&$1.428$\\
$8-10$&       &        &        &      &$18.124$&$1.540$&$7.496$&$5.224$&$5.441$&$4.197$&$3.848$&$4.094$&$1.673$\\
$10-12$&      &        &        &      &        &$21.340$&$4.213$&$7.696$&$6.493$&$5.170$&$4.686$&$4.888$&$1.950$\\
$12-14$&      &        &        &      &        &        &$21.875$&$0.719$&$6.144$&$3.846$&$3.939$&$3.922$&$1.500$\\
$14-16$&      &        &        &      &        &        &        &$23.040$&$5.219$&$6.123$&$4.045$&$4.681$&$1.807$\\
$16-18$&      &        &        &      &        &        &        &        &$19.798$&$1.662$&$4.362$&$4.140$&$1.690$\\
$18-20$&      &        &        &      &        &        &        &        &        &$18.0629$&$2.621$&$3.957$&$1.438$\\
$20-22$&      &        &        &      &        &        &        &        &        &         &$16.990$&$1.670$&$1.127$\\
$22-24$&      &        &        &      &        &        &        &        &        &         &        &$17.774$&$-0.293$\\
$24-26.4$&    &        &        &      &        &        &        &        &        &         &        &        &$6.516$\\
\end{tabular}
\end{table}

\subsubsection{\Vub from \Btopilnu}

The \Vub average can be determined from the averaged $q^2$ spectrum in combination with a prediction for
the normalization of the $\B \to \pi$ form factor.
The differential decay rate for light leptons ($e$, $\mu$) is given by
\begin{align}
 \Delta \Gamma =  \Delta \Gamma(q^2_{\rm low}, q^2_{\rm high})  = \int_{q^2_{\rm low}}^{q^2_{\rm high}} \text{d} q^2 \bigg[ \frac{8 \left| \vec p_\pi \right|  }{3} \frac{ G_F^2 \, \left| V_{ub} \right|^2 q^2 }{256 \, \pi^3 \, m_B^2}  H_0^2(q^2) \bigg]  \, ,
\end{align}
where $G_F$ is Fermi's constant, $\left| \vec p_\pi \right|$ is the magnitude of the three-momentum of the 
final state $\pi$ (a function of $q^2$), $m_B$ the $B^0$-meson mass, 
and $H_0(q^2)$ the only non-zero helicity amplitude. 
The helicity amplitude is a function of the form factor $f_+$, 
\begin{align}
 H_0 = \frac{2 m_B \, \left| \vec p_\pi \right| }{\sqrt{q^2}}\, f_+(q^2) .
\end{align} 
The form factor $f_{+}$ can be calculated with non-perturbative methods, but its general form can be constrained by the differential \Btopilnu\ spectrum. 
Here, we parametrize the form factor using the BCL parametrization~\cite{Bourrely:2008za}.

The decay rate is proportional to $\Vub^2 |f_+(q^2)|^2$. Thus to extract \Vub one needs to determine $f_+(q^2)$
(at least at one value of $q^2$). In order to enhance the precision, a binned $\chi^2$ fit is performed
using a $\chi^2$ function of the form
\begin{align} \label{eq:chi2}
 \chi^2 & = \left( \vec{\cal B} - \Delta \vec{\Gamma} \, \tau \right)^T 
            C^{-1} 
            \left( \vec{\cal B} - \Delta \vec{\Gamma} \, \tau \right) + \chi^2_{\rm LQCD} + \chi^2_{\rm LCSR}
\end{align}
where $C$ denotes the covariance matrix given in Table~\ref{tab:Cov}, $\vec{\cal B}$ is the vector of 
averaged partial branching fractions, and $\Delta \vec{\Gamma} \, \tau$ is the product of the vector of 
theoretical predictions of the partial decay rates and the $B^0$-meson lifetime. 
The form factor normalization is included in the fit by the two extra terms in Eq.~(\ref{eq:chi2}): 
$\chi_{\rm LQCD}$ uses the latest FLAG lattice average~\cite{Aoki:2016frl} from 
two state-of-the-art unquenched lattice QCD calculations~\cite{Lattice:2015tia, Flynn:2015mha}. 
The resulting constraints are quoted directly in terms of the coefficients $b_j$ of the BCL parameterization 
and enter Eq.~(\ref{eq:chi2}) as
\begin{align}
 \chi^2_{\rm LQCD} & = \left( {\vec{b}} - {\vec{b}_{\rm LQCD}} \right)^T \, C_{\rm LQCD}^{-1} \,  \left( {\vec{b}} - {\vec{b}_{\rm LQCD}} \right) \, ,
\end{align}
with ${\vec{b}}$ the vector containing the free parameters of the $\chi^2$ fit 
constraining the form factor, 
${\vec{b}_{\rm LQCD}}$ the averaged values from Ref.~\cite{Aoki:2016frl}, 
and $C_{\rm LQCD}$ their covariance matrix. 
Additional information about the form factor can be obtained from light-cone sum rule calculations. 
The state-of-the-art calculation includes up to two-loop contributions~\cite{Bharucha:2012wy}. 
It is included in Eq.~(\ref{eq:chi2}) via
\begin{align}
\chi^2_{\rm LQCR} & = \left( f_+^{\rm LCSR} - f_+(q^2 = 0; {\vec{b}}) \right)^2 / \sigma_{f_+^{\rm LCSR} }^2 \, .
\end{align}

\begin{figure} 
\centering
 \includegraphics[width=0.8\textwidth]{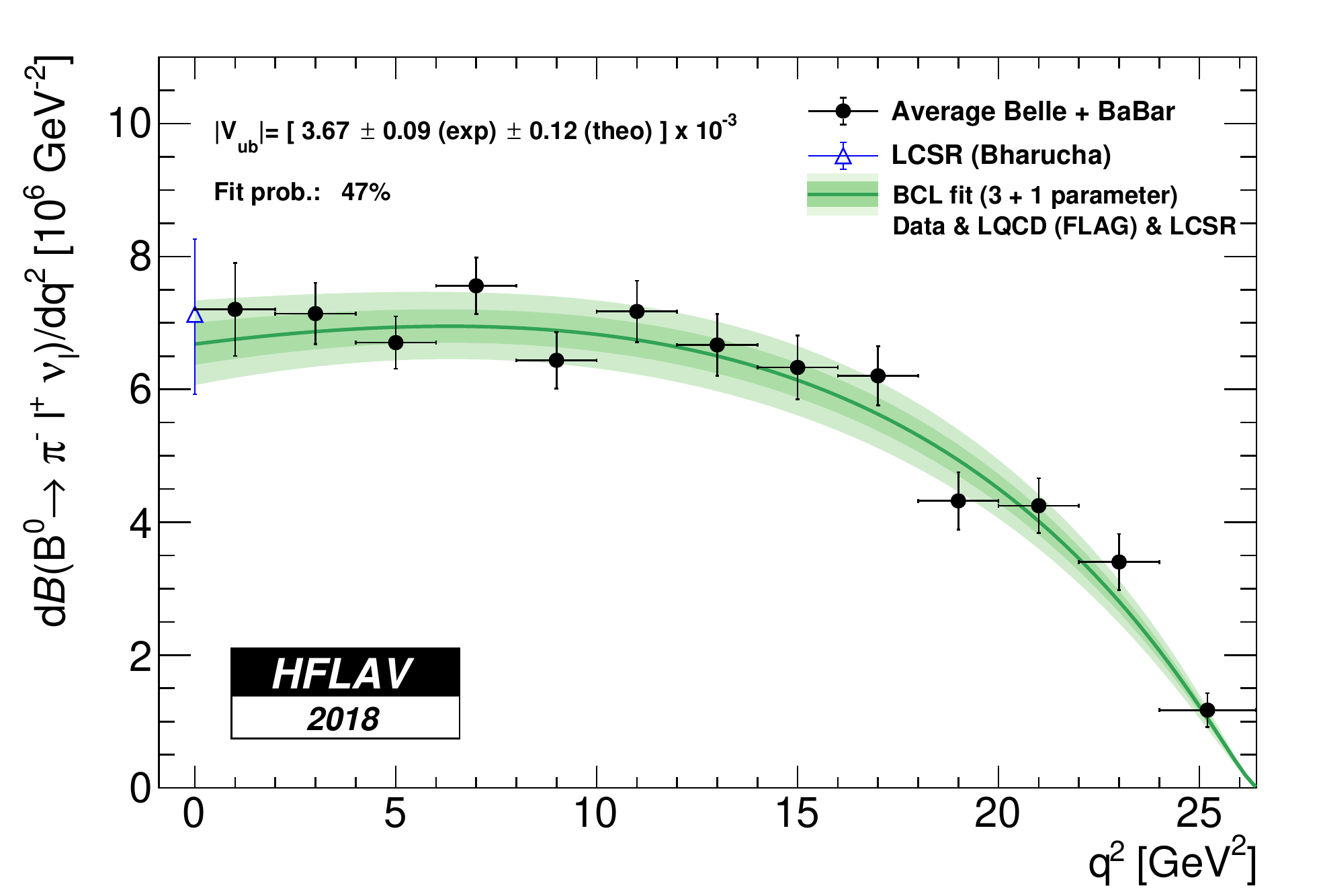}
\caption{Fit of the BCL parametrization to the averaged $q^2$ spectrum from \babar and Belle and the LQCD and LCSR calculations. 
The error bands represent the $1~\sigma$ (dark green) and $2~\sigma$ (light green) uncertainties 
of the fitted spectrum. \label{fig:vub}}
\end{figure}

The \Vub average is obtained for two versions: the first combines the data with the LQCD constraints  
and the second additionally includes the information from the LCSR calculation. 
The resulting values for $\Vub$ are
\begin{align}
 \Vub & = \left( 3.70 \pm 0.10 \, _\text{exp} \pm 0.12 \, _\text{theo} \right) \times 10^{-3} \, \ \rm (data+LQCD), \\
 \Vub & = \left( 3.67 \pm 0.09 \, _\text{exp} \pm 0.12 \, _\text{theo} \right) \times 10^{-3} \, \ \rm (data+LQCD+LCSR),
\end{align}
for the first and second fit version, respectively. 
The result of the fit including both LQCD and LCSR is shown in Figure~\ref{fig:vub}. 
The $\chi^2$ probability of the fit is $47\%$.
We quote the result of the fit including both LQCD and LCSR calculations as our average for \Vub. 
The best fit values for \Vub and the BCL parameters and their covariance matrix 
are given in Tables~\ref{tab:fitres2} and ~\ref{tab:fitcov2}. 

\begin{table}
\centering
\caption{Best fit values and uncertainties for the combined fit to data, LQCD and LCSR results. \label{tab:fitres2}}
\begin{tabular}{c | c }
Parameter & Value\\\hline
$\Vub  $&$(3.67 \pm 0.15)\times 10^{-3}$\\
$b_0$   & $0.418 \pm 0.012$\\
$b_1$   & $-0.399 \pm 0.033$\\
$b_2$   &$-0.578 \pm 0.130$\\
\end{tabular}
\end{table}

\begin{table}
\centering
\caption{Covariance matrix for the combined fit to data, LQCD and LCSR results. \label{tab:fitcov2}}

\begin{tabular}{c | c c c c}
Parameter & $\Vub$ & $b_0$ & $b_1$ & $b_2$\\\hline
$\Vub$    &$ 2.064\times 10^{-8} $&$ -1.321\times 10^{-6}$&$ -1.881\times 10^{-6}$&$ 7.454\times 10^{-6}$ \\
$b_0$   &                       &$ 1.390\times 10^{-4} $&$ 8.074\times 10^{-5}$ &$ -8.953\times 10^{-4} $\\
$b_1$   &                       &                       &$ 1.053\times 10^{-3}$ &$ -2.879\times 10^{-3}$ \\
$b_2$   &                       &                       &                       &$ 1.673\times 10^{-2}$\\
\end{tabular}

\end{table}

\subsubsection{Combined extraction of $\Vub$ and $\Vcb$}

The LHCb experiment reported the first observation of the CKM suppressed decay $\Lb\to p\mu\nu$
\cite{Aaij:2015bfa} and the measurement of the ratio of partial branching fractions at high $q^2$
for $\Lb\to p\mu\nu$ and $\Lb\to \Lc(\to pK\pi)\mu\nu$ decays,
\begin{align}
R = \dfrac{{\cal B}(\Lb\to p\mu\nu)_{q^2>15~GeV^2} }{{\cal B}(\Lb\to \Lc\mu\nu)_{q^2>7~GeV^2} }=(1.00\pm 0.04\pm 0.08)\times 10^{-2}.
\end{align}

\begin{figure} 
\centering
\includegraphics[width=0.8\textwidth]{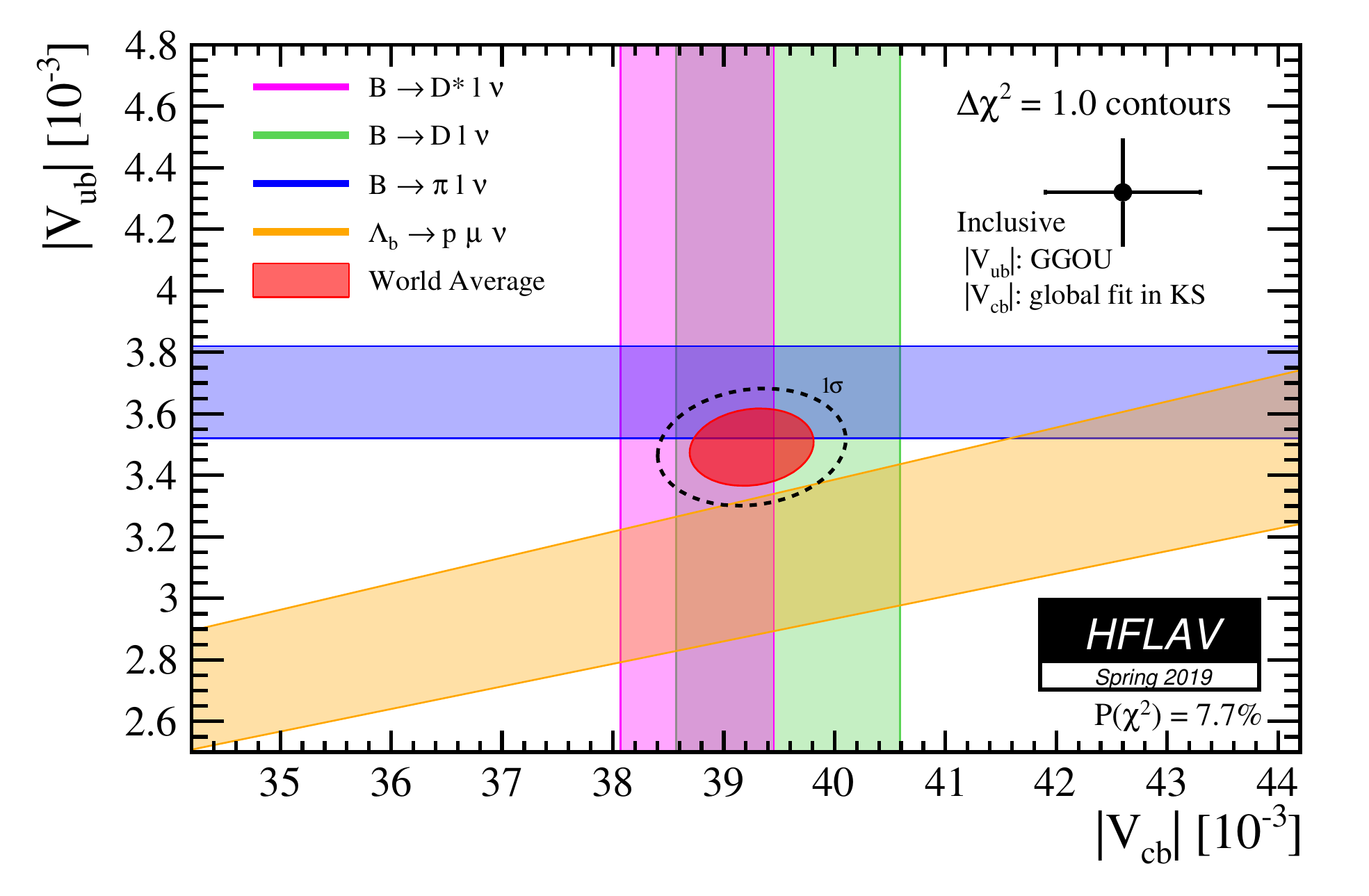}
 \caption{Combined average on \Vub and \Vcb including the LHCb measurement of $\Vub/\Vcb$, the exclusive $\Vub$ measurement from \Btopilnu, and \Vcb measurements from
both  $B\to D^*\ell\nu$ and $B\to D\ell\nu$. The dashed ellipse corresponds to a 1$\sigma$ two-dimensional contour (68\% of CL). The point with the error bars corresponds to the inclusive \Vcb from the kinetic scheme (Sec. \ref{globalfitsKinetic}), and the 
 inclusive \Vub from GGOU calculation (Sec. \ref{subsec:ggou}). \label{fig:vubvc}}
\end{figure}

\noindent The ratio $R$ is proportional to $(|V_{ub}|/|V_{cb}|)^2$ and sensitive to the form factors 
of $\Lb\to p$ and $\Lb\to \Lc$ transitions that have to be computed with non-perturbative
methods, such as lattice QCD.
The uncertainty on ${\cal B}(\Lc\to p K \pi)$ is the largest source of systematic uncertainties
on $R$.
Using the recent average of ${\cal B}(\Lc\to p K \pi)=(6.28\pm 0.32)\%$ \cite{PDG_2018}, the rescaled value for $R$ is  
\begin{align}
R = (0.92\pm 0.04\pm 0.07)\times 10^{-2}.
\end{align}
\noindent Using the precise lattice QCD prediction \cite{Detmold:2015aaa} of the form factors in the 
experimentally interesting $q^2$ region considered, we obtain
\begin{align}
\dfrac{|V_{ub}|}{|V_{cb}|} = 0.079\pm \, 0.004_\text{exp} \pm \, 0.004_\text{FF}
\end{align}

\noindent where the first uncertainty is the total experimental uncertainty, and the second one is due to the 
knowledge of the form factors. A combined fit for \Vub and \Vcb that includes the constraint from LHCb, 
and the determination of \Vub and \Vcb from exclusive $B$ meson decays, results in 
\begin{align}
 \Vub & = \left( 3.49 \pm 0.13 \right) \times 10^{-3}\,  \\
 \Vcb  & = \left( 39.25 \pm 0.56 \right) \times 10^{-3} \, \\
\rho(\Vub,\Vcb) & =0.14\,,
\end{align}
\noindent where the uncertainties in the inputs are considered uncorrelated. 
The $\chi^2$ of the fit is $5.1$ for $2$ d.o.f., corresponding 
to a $P(\chi^2)$ of 7.7\%. The fit result is shown in Fig.~\ref{fig:vubvc}, where both 
the $\Delta\chi^2$ and the two-dimensional $68\%$ C.L. contours are indicated. 
The $\Vub/\Vcb$ value extracted from $R$ is more compatible with the exclusive determinations
of $\Vub$. Another calculation, by Faustov and Galkin \cite{Faustov:2016pal}, based on a relativistic quark model,
gives a value of $\Vub/\Vcb$  closer to the inclusive determinations. 

\subsubsection{Other exclusive charmless semileptonic \B decays}

\begin{figure}[!ht]
 \begin{center}
  \unitlength1.0cm %
  \begin{picture}(14.,10.0)  %
   \put( -1.5,  0.0){\includegraphics[width=9.0cm]{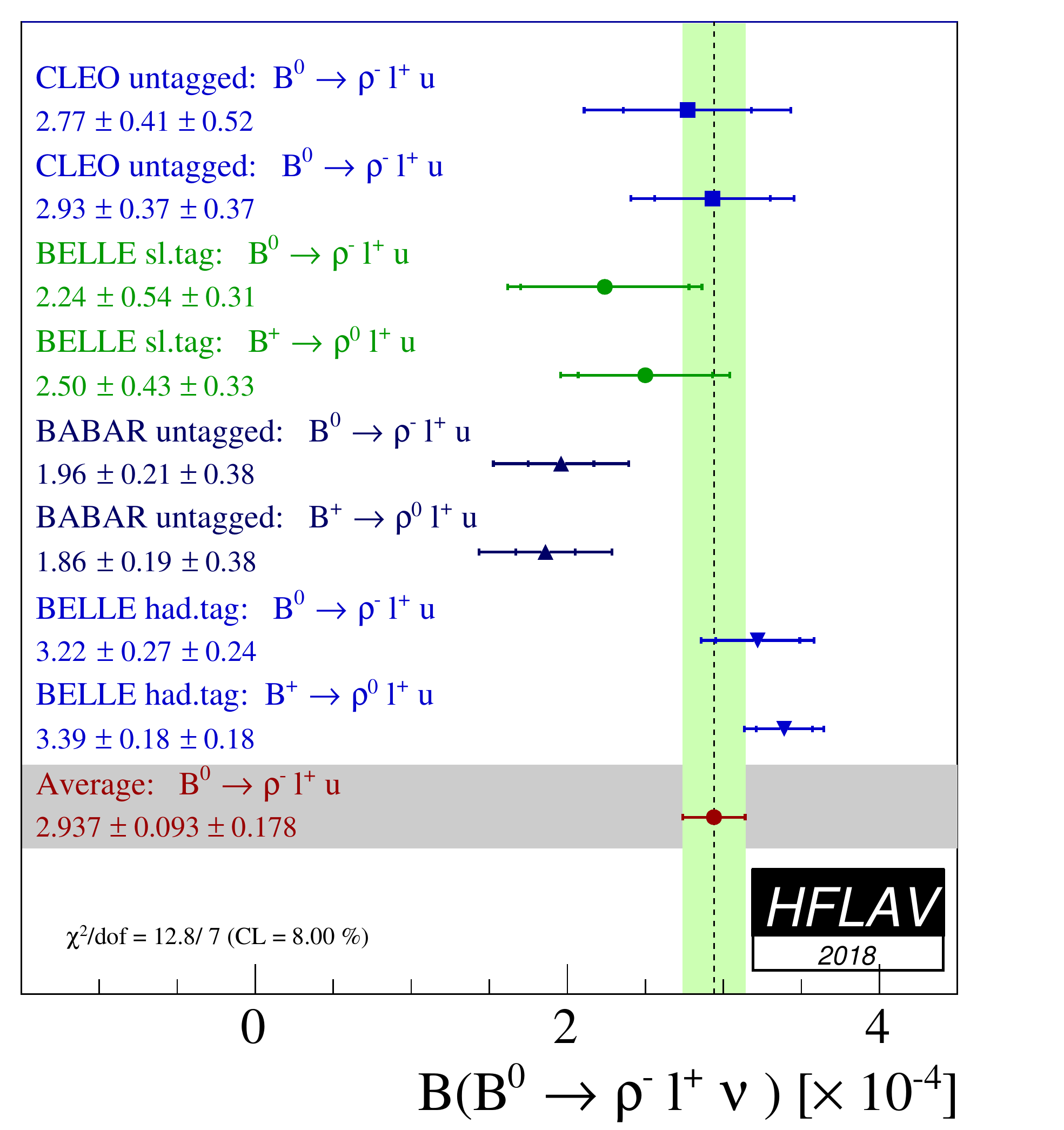}}
   \put( 7.5,  0.0){\includegraphics[width=9.0cm]{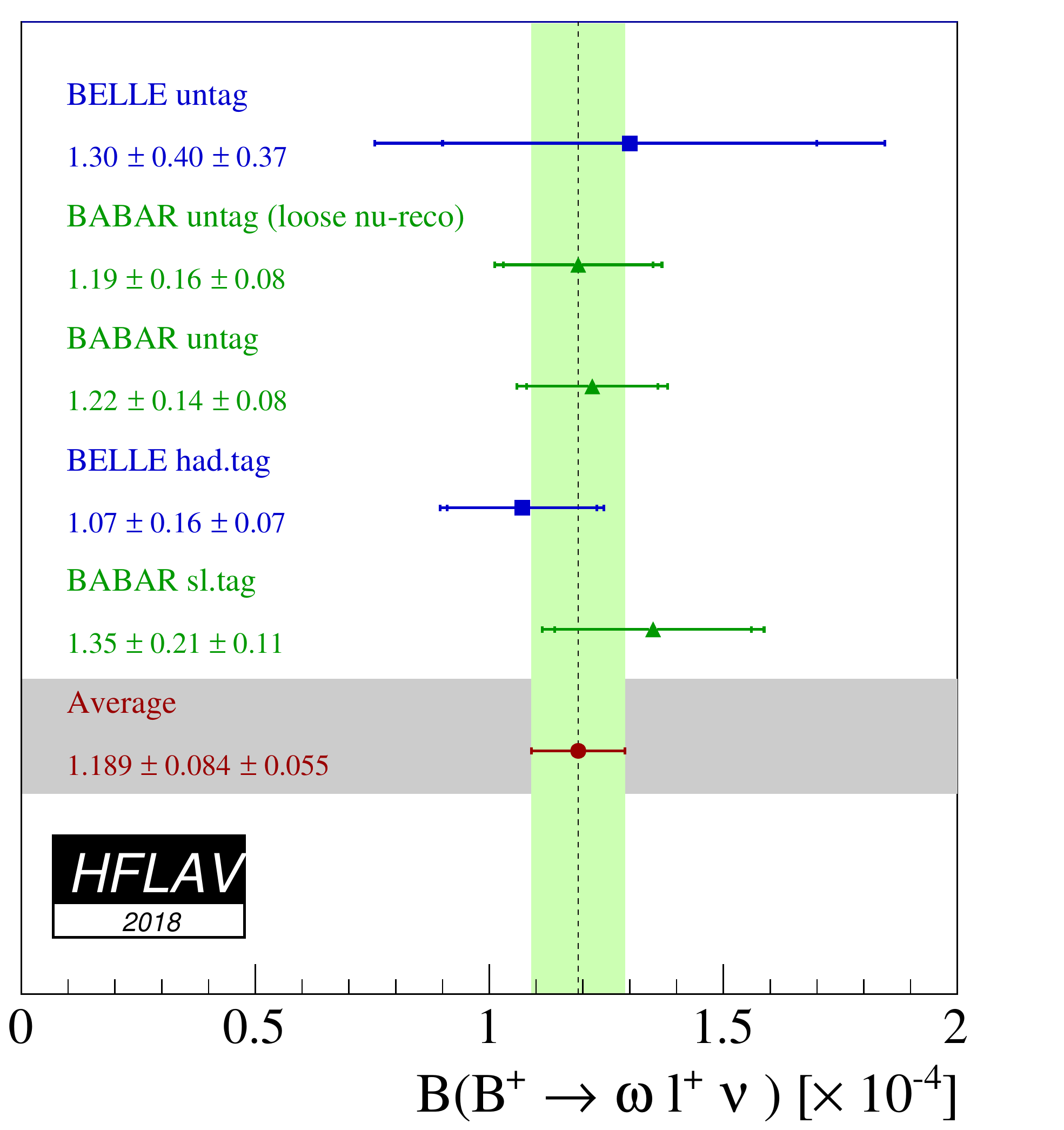}} 
   \put(  5.8,  8.5){{\large\bf a)}}  
   \put( 14.6,  8.5){{\large\bf b)}}
   \end{picture} \caption{
 (a) Summary of exclusive determinations of $\cbf(\Bz\to\rho\ell^+\nu)$ and their average. Measurements
 of $B^+ \to \rho^0\ell^+\nu$ branching fractions have been multiplied by $2\tau_{B^0}/\tau_{B^+}$ 
 in accordance with isospin symmetry.    
(b) Summary of exclusive determinations of $\Bp\to\omega\ell^+\nu$ and their average.
}
\label{fig:xulnu1}
\end{center}
\end{figure}

\begin{figure}[!ht]
 \begin{center}
  \unitlength1.0cm %
  \begin{picture}(14.,10.0)  %
   \put( -1.5,  0.0){\includegraphics[width=9.0cm]{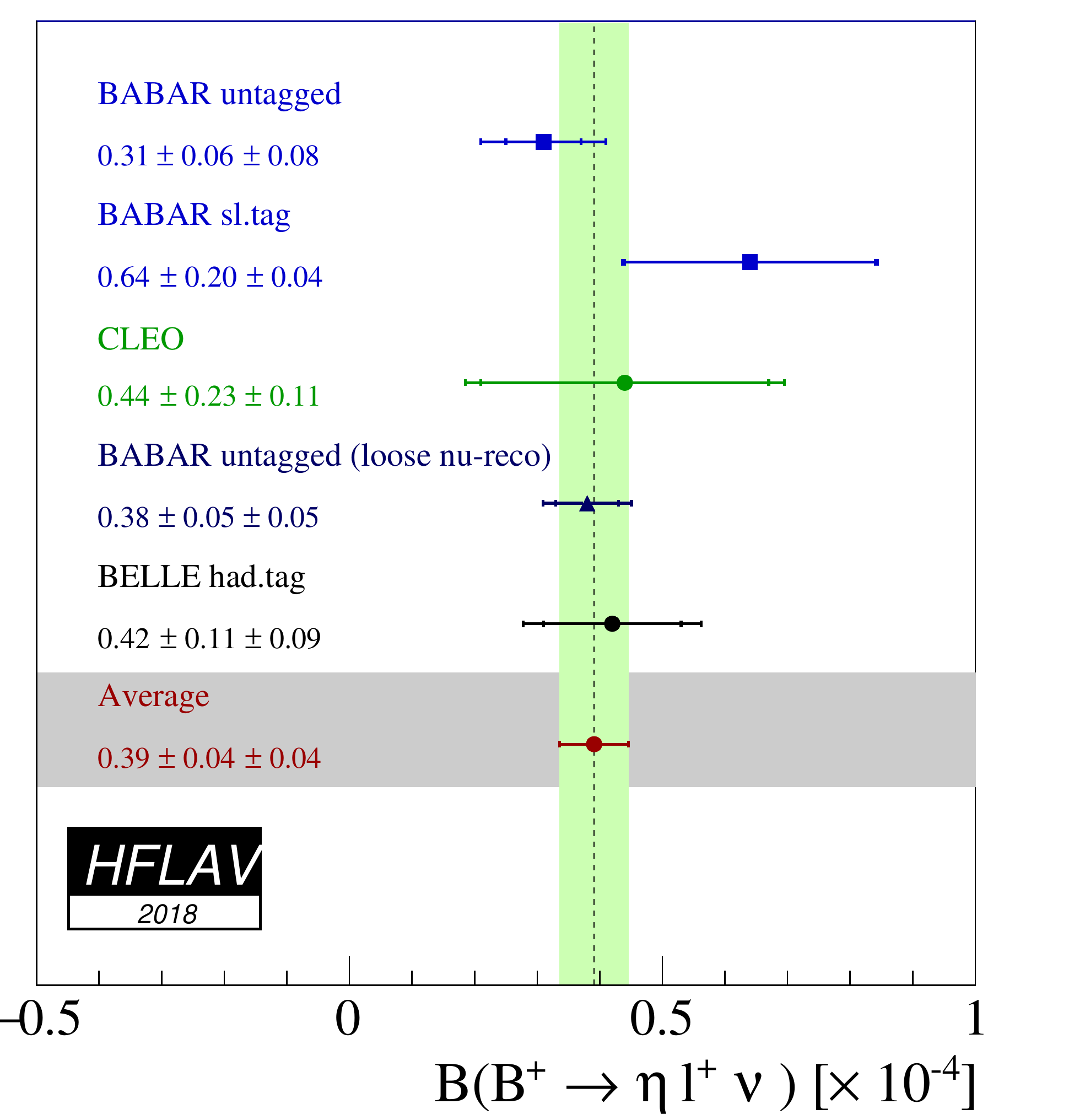}}
   \put( 7.5,  0.0){\includegraphics[width=9.0cm]{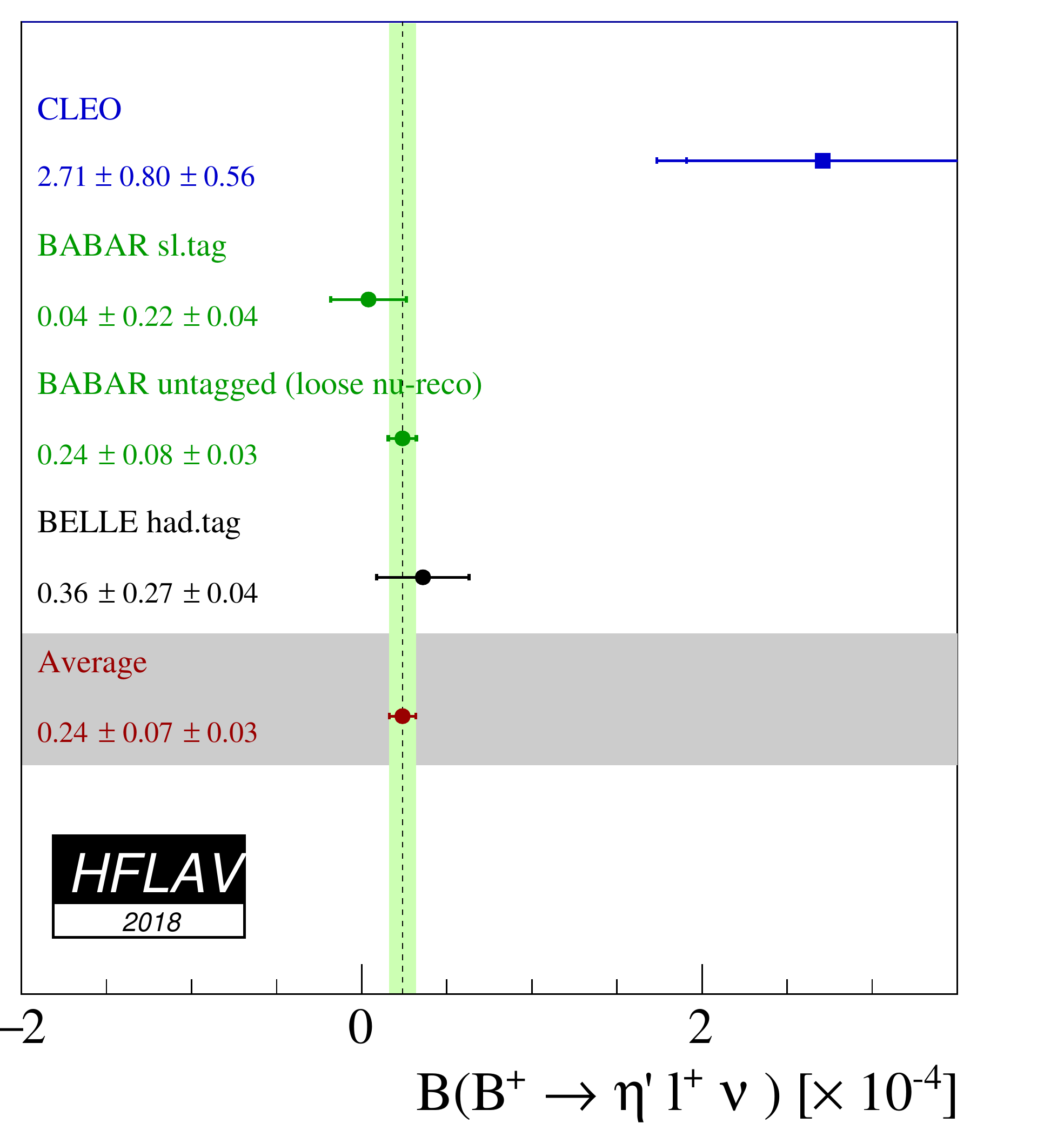}} 
   \put(  5.8,  8.5){{\large\bf a)}}     
   \put( 14.6,  8.5){{\large\bf b)}}
   
   \end{picture} \caption{
(a) Summary of exclusive determinations of $\cbf(\Bp\to\eta\ell^+\nu)$ and their average.
(b) Summary of exclusive determinations of $\cbf(\Bp\to\etapr\ell^+\nu)$ and their average.
}
\label{fig:xulnu2}
\end{center}
\end{figure}

We report the branching fraction averages for $\Bz\to\rho\ell^+\nu$, $\Bp\to\omega\ell^+\nu$, $\Bp\to\eta\ell^+\nu$ 
and $\Bp\to\etapr\ell^+\nu$ decays. The measurements and their averages are listed in 
Tables~\ref{tab:rholnu},~\ref{tab:omegalnu},~\ref{tab:etalnu},~\ref{tab:etaprimelnu}, 
and presented in Figures ~\ref{fig:xulnu1} and ~\ref{fig:xulnu2}. 
In the $\Bz\to\rho^-\ell^+\nu$ average, both the $\Bz\to\rho^-\ell^+\nu$ and $\Bp\to\rho^0\ell^+\nu$ decays are used, 
where the $\Bp\to\rho^0\ell^+\nu$ are rescaled by $2\tau_{B^0}/\tau_{B^+}$ assuming the isospin symmetry.
For  $\Bp\to\omega\ell^+\nu$ and $\Bp\to\eta\ell^+\nu$ decays, the agreement between the different measurements 
is good. $\Bp\to\etapr\ell^+\nu$ shows a discrepancy between the old CLEO measurement and the \babar untagged 
analysis, but the statistical uncertainties of the CLEO measurement are large.
The $\Bz\to\rho\ell^+\nu$ results, instead, show significant differences, in particular the \babar untagged analysis 
gives a branching fraction significantly lower (by about 2$\sigma$) than the Belle measurement based on the hadronic-tag. 
A possible reason for such discrepancy could be the broad nature of the $\rho$ resonance that makes the 
control of the background under the  $\rho$ mass peak more difficult in the untagged analysis than in the hadronic-tag
analysis.

We do not report \vub~ for these exclusive charmless decays, because the form factor calculations have not yet reached 
the precision achieved for $B\to\pi\ell\nu$ decays. 
Unquenched lattice QCD calculations of the form factors are not available for these decays, 
but LCSR calculations exist for all these decay modes. The most recent of these calculations for the  $B\to\rho\ell\nu$ 
and  $B\to\omega\ell\nu$ decays are reported in Ref.\cite{Ball:2004ye} and \cite{Straub:2015ica}.

\begin{table}[!htb]
\begin{center}
\caption{Summary of exclusive determinations of $\Bz\to\rho\ell^+\nu$. The errors quoted
correspond to statistical and systematic uncertainties, respectively.}
\label{tab:rholnu}
\begin{small}
\begin{tabular}{lc}
\hline
& $\cbf [10^{-4}]$
\\
\hline\hline
CLEO (Untagged) $\rho^+$~\cite{Behrens:1999vv}
& $2.77\pm 0.41\pm 0.52\ $ 
\\ 
CLEO (Untagged) $\rho^+$~\cite{Adam:2007pv}
& $2.93\pm 0.37\pm 0.37\ $ 
\\ 
Belle (Hadronic Tag) $\rho^+$~\cite{Sibidanov:2013rkk}
& $3.22\pm 0.27\pm 0.24\ $
\\
Belle (Hadronic Tag) $\rho^0$~\cite{Sibidanov:2013rkk}
& $3.39\pm 0.18\pm 0.18\ $
\\
Belle (Semileptonic Tag) $\rho^+$~\cite{Hokuue:2006nr}
& $2.24\pm 0.54\pm 0.31\ $
\\
Belle (Semileptonic Tag) $\rho^0$~\cite{Hokuue:2006nr}
& $2.50\pm 0.43\pm 0.33\ $
\\
\babar (Untagged) $\rho^+$~\cite{delAmoSanchez:2010af}
& $1.96\pm 0.21\pm 0.38\ $
\\
\babar (Untagged) $\rho^0$~\cite{delAmoSanchez:2010af}
& $1.86\pm 0.19\pm 0.32\ $

\\  \hline
{\bf Average}
& \mathversion{bold}$2.937 \pm 0.093\pm 0.178 $
\\ 
\hline
\end{tabular}\\
\end{small}
\end{center}
\end{table}

\begin{table}[!htb]
\begin{center}
\caption{Summary of exclusive determinations of $\Bp\to\omega\ell^+\nu$. The errors quoted
correspond to statistical and systematic uncertainties, respectively.}
\label{tab:omegalnu}
\begin{small}
\begin{tabular}{lc}
\hline
& $\cbf [10^{-4}]$
\\
\hline\hline
Belle (Untagged) ~\cite{Schwanda:2004fa}
& $1.30\pm 0.40\pm 0.36\ $
\\
\babar (Loose $\nu$ reco.) ~\cite{Lees:2012vv}
& $1.19\pm 0.16\pm 0.09\ $
\\  
\babar (Untagged) ~\cite{Lees:2012mq}
& $1.21\pm 0.14\pm 0.08\ $
\\  
Belle (Hadronic Tag) ~\cite{Sibidanov:2013rkk}
& $1.07\pm 0.16\pm 0.07 $
\\
\babar (Semileptonic Tag) ~\cite{Lees:2013gja}
& $1.35\pm 0.21\pm 0.11\ $
\\  

\hline

{\bf Average}
& \mathversion{bold}$1.189 \pm 0.084 \pm 0.055\ $
\\ 
\hline
\end{tabular}\\
\end{small}
\end{center}
\end{table}

\begin{table}[!htb]
\begin{center}
\caption{Summary of exclusive determinations of $\Bp\to\eta\ell^+\nu$. 
The errors quoted correspond to statistical and systematic uncertainties, respectively.}
\label{tab:etalnu}
\begin{small}
\begin{tabular}{lc}
\hline
& $\cbf [10^{-4}]$
\\
\hline\hline
CLEO ~\cite{Gray:2007pw}
& $0.45\pm 0.23\pm 0.11\ $
\\
\babar\ (Untagged) ~\cite{Aubert:2008ct}
& $0.31\pm 0.06\pm 0.08\ $
\\ 
\babar\ (Semileptonic Tag) ~\cite{Aubert:2008bf}
& $0.64\pm 0.20\pm 0.04\ $
\\
\babar\ (Loose $\nu$-reco.) ~\cite{Lees:2012vv}
& $0.38\pm 0.05\pm 0.05\ $
\\  
\belle\ (Hadronic Tag) ~\cite{Beleno:2017cao}
& $0.42\pm 0.11\pm 0.09\ $
\\  
 \hline
{\bf Average}
& \mathversion{bold}$0.39 \pm 0.04 \pm 0.04 $
\\ 
\hline
\end{tabular}\\
\end{small}
\end{center}
\end{table}

\begin{table}[!htb]
\begin{center}
\caption{Summary of exclusive determinations of  $\Bp\to\eta'\ell^+\nu$. The errors quoted
correspond to statistical and systematic uncertainties, respectively.}
\label{tab:etaprimelnu}
\begin{small}
\begin{tabular}{lc}
\hline
& $\cbf [10^{-4}]$
\\
\hline\hline
CLEO ~\cite{Gray:2007pw} & $2.71\pm 0.80\pm 0.56\ $
\\
\babar\ (Semileptonic Tag) ~\cite{Aubert:2008bf}
& $0.04\pm 0.22\pm 0.04$, $(<0.47 ~~@~ 90\% C.L.)$
\\ 
\babar\ (Untagged) ~\cite{Lees:2012vv} & $0.24\pm 0.08\pm 0.03$
\\  
\belle\ (Hadronic Tag) ~\cite{Beleno:2017cao}
& $0.36\pm 0.27\pm 0.04\ $
\\  
 \hline
{\bf Average}
& \mathversion{bold}$0.24 \pm 0.07 \pm 0.03 $
\\ 
\hline
\end{tabular}\\
\end{small}
\end{center}
\end{table}

\clearpage

\subsection{Inclusive CKM-suppressed decays}
\label{slbdecays_b2uincl}
Measurements of $B \to X_u \ell^+ \nu$  decays are very challenging because of background from the Cabibbo-favoured 
$B \to X_c \ell^+ \nu$ decays, whose branching fraction is about 50 times larger than that of the signal.  Cuts designed to suppress this dominant background severely complicate the perturbative 
QCD calculations required to extract $\vub$.  Tight cuts necessitate parameterization of the so-called 
shape functions in order to describe the unmeasured regions of  phase space.  
We use several theoretical calculations to extract \vub~ and  
do not advocate the use of one method over another.
The authors of the different calculations have provided 
codes to compute the partial rates in limited regions of phase space covered by the measurements. 
Belle~\cite{ref:belle-multivariate} and \babar~\cite{Lees:2011fv} produced measurements that 
explore large portions of phase space, with consequent reduction of the theoretical 
uncertainties. 

In the averages, the systematic uncertainties associated with the
modeling of $\B\to X_c\ell^+\nul$ and $\B\to X_u\ell^+\nul$ decays and the theoretical
uncertainties are taken as fully correlated among all measurements.
Reconstruction-related uncertainties are taken as fully correlated within a given experiment.
Measurements of partial branching fractions for $\B\to X_u\ell^+\nul$
transitions from $\Upsilon(4S)$ decays, together with the corresponding selected region, 
are given in Table~\ref{tab:BFbulnu}.  
The signal yields for all the measurements shown in Table~\ref{tab:BFbulnu}
are not rescaled to common input values of the $B$ meson lifetime (see
Sec.~\ref{sec:life_mix}) and the semileptonic width~\cite{PDG_2018}.
We use all results published by \babar\ in Ref.~\cite{Lees:2011fv}, since the 
statistical correlations are given. 
To make use of the theoretical calculations of Ref.~\cite{ref:BLL}, we restrict the
kinematic range of the invariant mass of the hadronic system, $M_X$, 
and the square of the invariant mass of the lepton pair, $q^2$. This reduces the size of the data
sample significantly, but also the theoretical uncertainty, as stated by the
authors~\cite{ref:BLL}.
The dependence of the quoted error on the measured value for each source of uncertainty 
is taken into account in the calculation of the averages.

It was first suggested by Neubert~\cite{Neubert:1993um} and later detailed by Leibovich, 
Low, and Rothstein (LLR)~\cite{Leibovich:1999xf} and Lange, Neubert and Paz (LNP)~\cite{Lange:2005qn}, 
that the uncertainty of
the leading shape functions can be eliminated by comparing inclusive rates for
$\B\to X_u\ell^+\nul$ decays with the inclusive photon spectrum in $\B\to X_s\gamma$,
based on the assumption that the shape functions for transitions to light
quarks, $u$ or $s$, are the same at first order.
However, shape function uncertainties are only eliminated at the leading order
and they still enter via the signal models used for the determination of efficiency.

In the following, the different theoretical methods and the resulting averages are described.

In a recent paper by \babar~\cite{TheBABAR:2016lja}, detailed studies are performed to assess the impact of four QCD-based theoretical predictions, used also below, on the measurements of the electron spectrum, the branching fraction, and the extraction of $\vub$, where the lower limit on the electron momentum is varied from 0.8\gevc to the kinematic endpoint. An important difference of this paper with respect to the other ones is that the dependency on the theoretical models enters primarily through the partial branching fractions, as the fit is sensitive to signal decays only in regions with good signal-to-noise such as the endpoint region. All other measurements instead determine a partial branching fraction by using a single model, and this partial branching fraction is then converted into a \vub\ measurement by taking the corresponding partial rate predicted by the theory calculations. 
Due to this difference, the $\vub$ results obtained in this paper, with a lower limit of 0.8\gevc on the electron momentum, are directly used as input to the BLNP, DGE and GGOU averages. These determinations supersede the previous \babar endpoint measurement~\cite{ref:babar-endpoint}.  
The partial branching ratio quoted in Table~\ref{tab:BFbulnu} for Ref.~\cite{TheBABAR:2016lja} is taken as that obtained with the GGOU calculation. 
\begin{table}[!htb]
\caption{\label{tab:BFbulnu}
Summary of measurements  of partial branching
fractions for $B\rightarrow X_u \ell^+ \nu_{\ell}$ decays.
The errors quoted on $\Delta\cbf$ correspond to
statistical and systematic uncertainties.
$E_e$ is the electron
energy in the $B$~rest frame, $p^*$ the lepton momentum in the
$B$~frame and $m_X$ is the invariant mass of the hadronic system. The
light-cone momentum $P_+$ is defined in the $B$ rest frame as
$P_+=E_X-|\vec p_X|$.
The $s_\mathrm{h}^{\mathrm{max}}$ variable is described in Refs.~\cite{ref:shmax,ref:babar-elq2}. }
\begin{center}
\begin{small}
\begin{tabular}{|llcl|}
\hline
Measurement & Accepted region &  $\Delta\cbf [10^{-4}]$ & Notes\\
\hline\hline
CLEO~\cite{ref:cleo-endpoint}
& $E_e>2.1\,\gev$ & $3.3\pm 0.2\pm 0.7$ &  \\ 
\babar~\cite{ref:babar-elq2}
& $E_e>2.0~\gev$, $s_\mathrm{h}^{\mathrm{max}}<3.5\,\mathrm{GeV}^2$ & $4.4\pm 0.4\pm 0.4$ & \\
\babar~\cite{TheBABAR:2016lja}
& $E_e>0.8\,\gev$  & $1.55\pm 0.08\pm 0.09$ &  Using the GGOU model\\
Belle~\cite{ref:belle-endpoint}
& $E_e>1.9\,\gev$  & $8.5\pm 0.4\pm 1.5$ & \\
\babar~\cite{Lees:2011fv}
& $M_X<1.7\,\gevcc, q^2>8\,\gevgevcccc$ & $6.9\pm 0.6\pm 0.4$ & 
\\
Belle~\cite{ref:belle-mxq2Anneal}
& $M_X<1.7\,\gevcc, q^2>8\,\gevgevcccc$ & $7.4\pm 0.9\pm 1.3$ & \\
Belle~\cite{ref:belle-mx}
& $M_X<1.7\,\gevcc, q^2>8\,\gevgevcccc$ & $8.5\pm 0.9\pm 1.0$ & Used only in BLL average\\
\babar~\cite{Lees:2011fv}
& $P_+<0.66\,\gev$  & $9.9\pm 0.9\pm 0.8 $ & 
\\
\babar~\cite{Lees:2011fv}
& $M_X<1.7\,\gevcc$ & $11.6\pm 1.0\pm 0.8 $ &
\\ 
\babar~\cite{Lees:2011fv}
& $M_X<1.55\,\gevcc$ & $10.9\pm 0.8\pm 0.6 $ & 
\\ 
Belle~\cite{ref:belle-multivariate}
& ($M_X, q^2$) fit, $p^*_{\ell} > 1~\gev/c$ & $19.6\pm 1.7\pm 1.6$ & \\
\babar~\cite{Lees:2011fv}
& ($M_X, q^2$) fit, $p^*_{\ell} > 1~\gev/c$  & $18.2\pm 1.3\pm 1.5$ & 
\\ 
\babar~\cite{Lees:2011fv}
& $p^*_{\ell} > 1.3~\gev/c$  & $15.5\pm 1.3\pm 1.4$ & 
\\ \hline
\end{tabular}\\
\end{small}
\end{center}
\end{table}

\subsubsection{BLNP}
\begin{figure}[!ht]
 \begin{center}
  \unitlength1.0cm %
  \begin{picture}(14.,10.0)  %
   \put( -1.5,  0.0){\includegraphics[width=9.4cm]{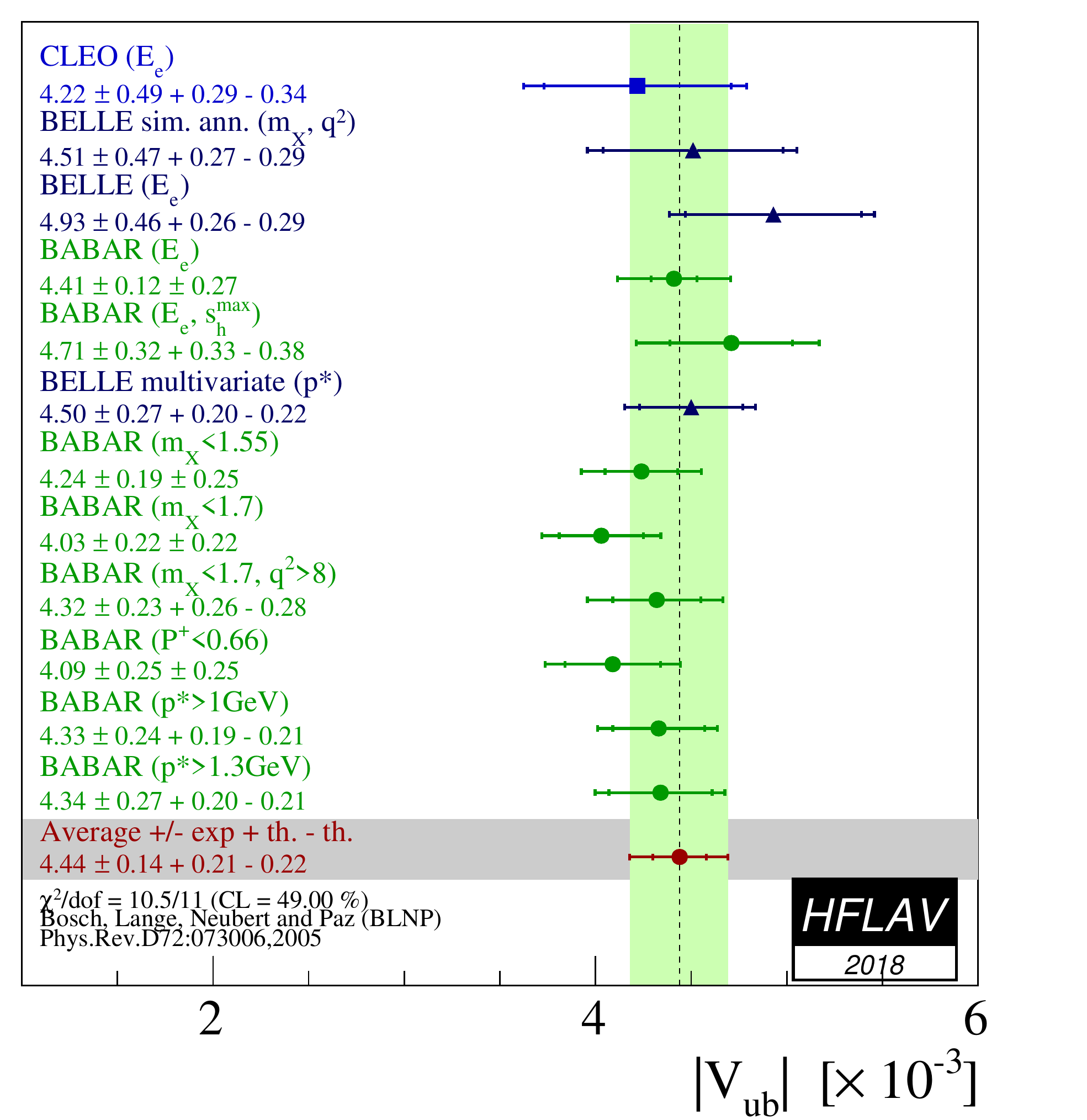}
   }
   \put(  7.4,  0.0){\includegraphics[width=9.4cm]{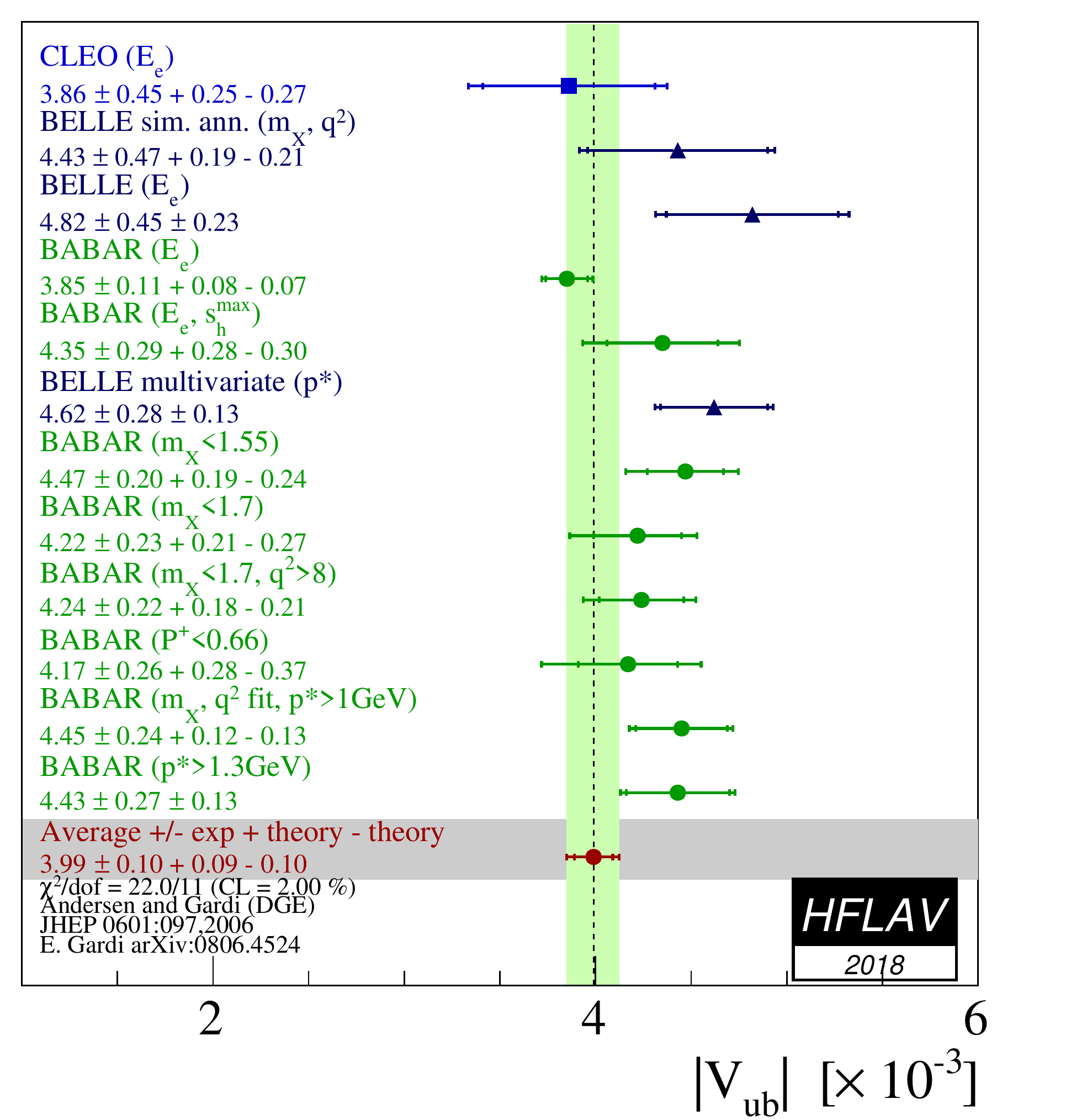}
   }
   \put(  6.0,  8.7){{\large\bf a)}}
   \put( 14.8,  8.7){{\large\bf b)}}
  \end{picture}
  \caption{Measurements of $\vub$ from inclusive semileptonic decays 
and their average based on the BLNP (a) and DGE (b) prescription. The
labels indicate the variables and selections used to define the
signal regions in the different analyses.
} \label{fig:BLNP_DGE}
 \end{center}
\end{figure}

Bosch, Lange, Neubert and Paz (BLNP)~\cite{ref:BLNP,
  ref:Neubert-new-1,ref:Neubert-new-2,ref:Neubert-new-3}
provide theoretical expressions for the triple
differential decay rate for $B\to X_u \ell^+ \nul$ events, incorporating all known
contributions, whilst smoothly interpolating between the 
``shape-function region'' of large hadronic
energy and small invariant mass, and the ``OPE region'' in which all
hadronic kinematical variables scale with the $b$-quark mass. BLNP assign
uncertainties to the $b$-quark mass, which enters through the leading shape function, 
to sub-leading shape function forms, to possible weak annihilation
contribution, and to matching scales. 
The BLNP calculation uses the shape function renormalization scheme; the heavy quark parameters determined  
from the global fit in the kinetic scheme, described in \ref{globalfitsKinetic}, were therefore 
translated into the shape function scheme by using a prescription by Neubert 
\cite{Neubert:2004sp,Neubert:2005nt}. The resulting parameters are 
$m_b({\rm SF})=(4.582 \pm 0.023 \pm 0.018)~\gev$, 
$\mu_\pi^2({\rm SF})=(0.202 \pm 0.089 ^{+0.020}_{-0.040})~\gevcc$, 
where the second uncertainty is due to the scheme translation. 
The extracted values of \vub\, for each measurement along with their average are given in
Table~\ref{tab:bulnu} and illustrated in Fig.~\ref{fig:BLNP_DGE}(a). 
The total uncertainty is $^{+5.6}_{-5.7}\%$ and is due to:
statistics ($^{+1.8}_{-1.9}\%$),
detector effects ($^{+1.7}_{-1.7}\%$),
$B\to X_c \ell^+ \nul$ model ($^{+0.9}_{-1.0}\%$),
$B\to X_u \ell^+ \nul$ model ($^{+1.5}_{-1.5}\%$),
heavy quark parameters ($^{+2.7}_{-2.8}\%$),
SF functional form ($^{+0.1}_{-0.3}\%$),
sub-leading shape functions ($^{+0.8}_{-0.8}\%$),
BLNP theory: matching scales $\mu,\mu_i,\mu_h$ ($^{+3.8}_{-3.8}\%$), and
weak annihilation ($^{+0.0}_{-0.7}\%$).
The error assigned to the matching scales 
is the source of the largest uncertainty, while the
uncertainty due to HQE parameters ($b$-quark mass and $\mu_\pi^2)$ is second. The uncertainty due to 
weak annihilation is assumed to be asymmetric, \ie\ it only tends to decrease \vub.

\begin{table}[!htb]
\caption{\label{tab:bulnu}
Summary of input parameters used by the different theory calculations,
corresponding inclusive determinations of $\vub$ and their average.
The errors quoted on \vub\ correspond to
experimental and theoretical uncertainties, respectively.}
\begin{center}
\resizebox{0.99\textwidth}{!}{
\begin{tabular}{|lccccc|}
\hline
 & BLNP &DGE & GGOU & ADFR &BLL \\
\hline\hline
\multicolumn{6}{|c|}{Input parameters}\\ \hline
scheme & SF           & $\overline{MS}$ & kinetic &  $\overline{MS}$ & $1S$ \\ 
Ref.       & \cite{Neubert:2004sp,Neubert:2005nt} & Ref.~\cite{ref:DGE} & 
see Sec.~\ref{globalfitsKinetic}  & Ref.~\cite{Aglietti:2006yb} & Ref.~\cite{ref:BLL} \\
$m_b$ (GeV)           & 4.582 $\pm$ 0.026 & 4.188 $\pm 0.043$ & 4.554 $\pm 0.018$ & 4.188 $\pm 0.043$ & 4.704 $\pm 0.029$ \\
$\mu_\pi^2$ (GeV$^2$) & 0.145 $^{+0.091}_{-0.097}$ & -                 & 0.414 $\pm 0.078$ & - &  - \\
\hline\hline
Ref. & \multicolumn{5}{c|}{$|V_{ub}|$ values $[10^{-3}]$}\\ 
\hline
CLEO $E_e$~\cite{ref:cleo-endpoint} &
$4.22\pm 0.49 ^{+0.29}_{-0.34}$ &
$3.86\pm 0.45 ^{+0.25}_{-0.27}$ &
$4.23\pm 0.49 ^{+0.22}_{-0.31}$ &
$3.42\pm 0.40 ^{+0.17}_{-0.17}$ &
- \\

Belle $M_X, q^2$~\cite{ref:belle-mxq2Anneal}&
$4.51\pm 0.47 ^{+0.27}_{-0.29}$ &
$4.43\pm 0.47 ^{+0.19}_{-0.21}$ &
$4.52\pm 0.48 ^{+0.25}_{-0.28}$ &
$3.93\pm 0.41 ^{+0.18}_{-0.17}$ &
$4.68\pm 0.49 ^{+0.30}_{-0.30}$ \\

Belle $E_e$~\cite{ref:belle-endpoint}&
$4.93\pm 0.46 ^{+0.26}_{-0.29}$ &
$4.82\pm 0.45 ^{+0.23}_{-0.23}$ &
$4.95\pm 0.46 ^{+0.16}_{-0.21}$ &
$4.48\pm 0.42 ^{+0.20}_{-0.20}$ &
-\\

\babar $E_e$~\cite{TheBABAR:2016lja}&
$4.41\pm 0.12 ^{+0.27}_{-0.27}$ &
$3.85\pm 0.11 ^{+0.08}_{-0.07}$ &
$3.96\pm 0.10 ^{+0.17}_{-0.17}$ &
- &
-\\

\babar $E_e,s_\mathrm{h}^{\mathrm{max}}$~\cite{ref:babar-elq2}&
$4.71\pm 0.32 ^{+0.33}_{-0.38}$ &
$4.35\pm 0.29 ^{+0.28}_{-0.30}$ &
- &
$3.81\pm 0.19 ^{+0.19}_{-0.18}$ &
 \\ 

Belle $p^*_{\ell}$, $(M_X,q^2)$ fit~\cite{ref:belle-multivariate}&
$4.50\pm 0.27 ^{+0.20}_{-0.22}$ &
$4.62\pm 0.28 ^{+0.13}_{-0.13}$ &
$4.62\pm 0.28 ^{+0.09}_{-0.10}$ &
$4.50\pm 0.30 ^{+0.20}_{-0.20}$ &
- \\

\babar $M_X$~\cite{Lees:2011fv}&
$4.24\pm 0.19 ^{+0.25}_{-0.25}$ &
$4.47\pm 0.20 ^{+0.19}_{-0.24}$ &
$4.30\pm 0.20 ^{+0.20}_{-0.21}$ &
$3.83\pm 0.18 ^{+0.20}_{-0.19}$ &
- \\
\babar $M_X$~\cite{Lees:2011fv}&
$4.03\pm 0.22 ^{+0.22}_{-0.22}$ &
$4.22\pm 0.23 ^{+0.21}_{-0.27}$ &
$4.10\pm 0.23 ^{+0.16}_{-0.17}$ &
$3.75\pm 0.21 ^{+0.18}_{-0.18}$ &
- \\

\babar $M_X,q^2$~\cite{Lees:2011fv}&
$4.32\pm 0.23 ^{+0.26}_{-0.28}$  &
$4.24\pm 0.22 ^{+0.18}_{-0.21}$  &
$4.33\pm 0.23 ^{+0.24}_{-0.27}$  &
$3.75\pm 0.20 ^{+0.17}_{-0.17}$  &
$4.50\pm 0.24 ^{+0.29}_{-0.29}$ \\

\babar $P_+$~\cite{Lees:2011fv}&
$4.09\pm 0.25 ^{+0.25}_{-0.25}$  &
$4.17\pm 0.25 ^{+0.28}_{-0.37}$  &
$4.25\pm 0.26 ^{+0.26}_{-0.27}$  &
$3.57\pm 0.22 ^{+0.19}_{-0.18}$  &
- \\

\babar $p^*_{\ell}$, $(M_X,q^2)$ fit~\cite{Lees:2011fv}&
$4.33\pm 0.24 ^{+0.19}_{-0.21}$  &
$4.45\pm 0.24 ^{+0.12}_{-0.13}$  &
$4.44\pm 0.24 ^{+0.09}_{-0.10}$  &
$4.33\pm 0.24 ^{+0.19}_{-0.19}$  &
- \\

\babar $p^*_{\ell}$~\cite{Lees:2011fv}&
$4.34\pm 0.27 ^{+0.20}_{-0.21}$  &
$4.43\pm 0.27 ^{+0.13}_{-0.13}$  &
$4.43\pm 0.27 ^{+0.09}_{-0.11}$  &
$4.28\pm 0.27 ^{+0.19}_{-0.19}$  &
- \\

Belle $M_X,q^2$~\cite{ref:belle-mx}&
- &
- &
- &
- &
$5.01\pm 0.39 ^{+0.32}_{-0.32}$ \\
\hline
Average &
$4.44 {^{+0.13}_{-0.14}} {^{+0.21}_{-0.22}}$ &
$3.99\pm 0.10 ^{+0.09}_{-0.10}$ &
$4.32\pm 0.12 ^{+0.12}_{-0.13}$ &
$3.99\pm 0.13 ^{+0.18}_{-0.12}$ &
$4.62\pm 0.20 ^{+0.29}_{-0.29}$ \\
\hline
\end{tabular}
}
\end{center}
\end{table}

\begin{figure}[!ht]
 \begin{center}
  \unitlength1.0cm %
  \begin{picture}(14.,10.0)  %
   \put( -1.5,  0.0){\includegraphics[width=9.4cm]{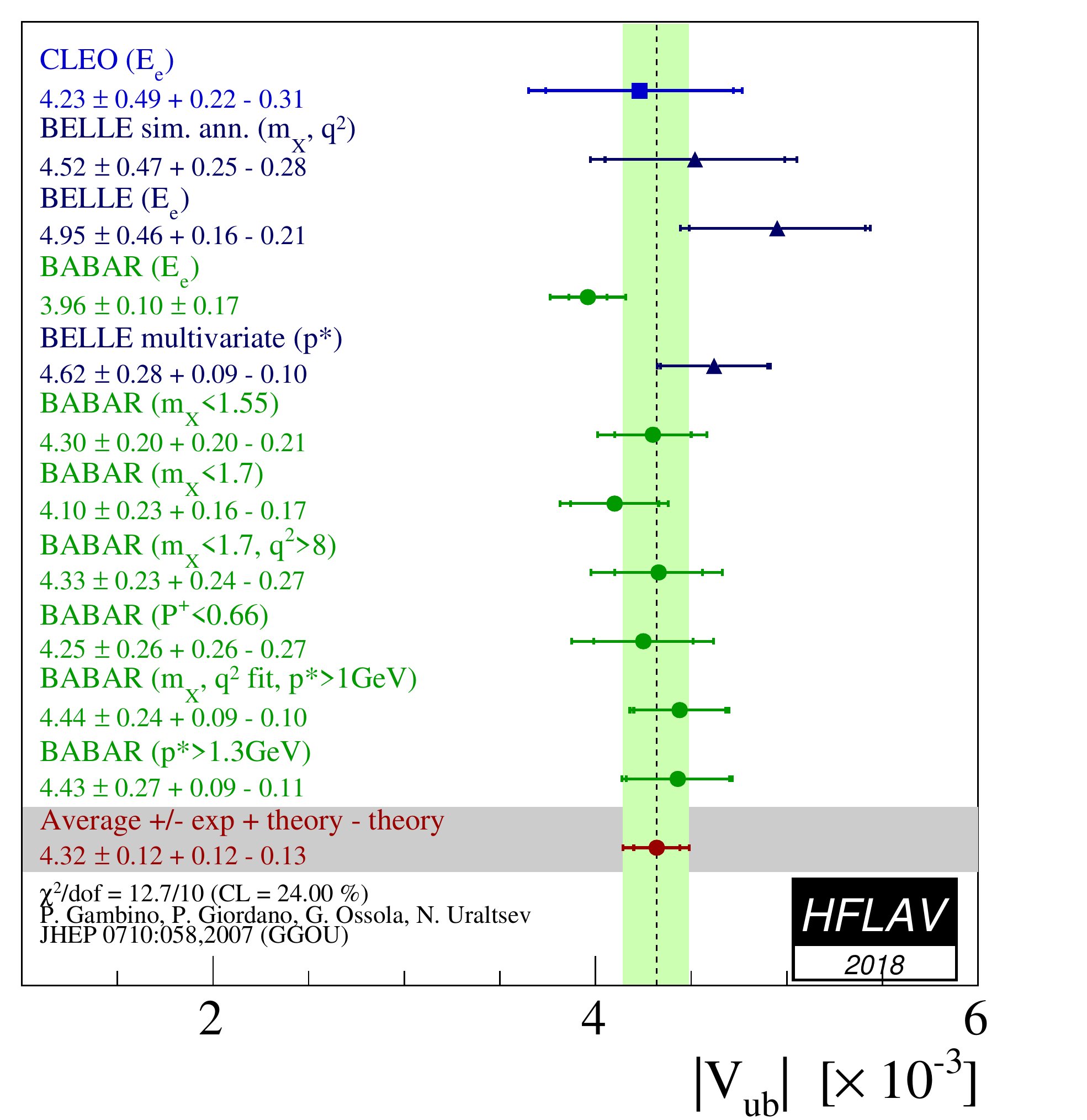}
   }
   \put(  7.4,  0.0){\includegraphics[width=9.4cm]{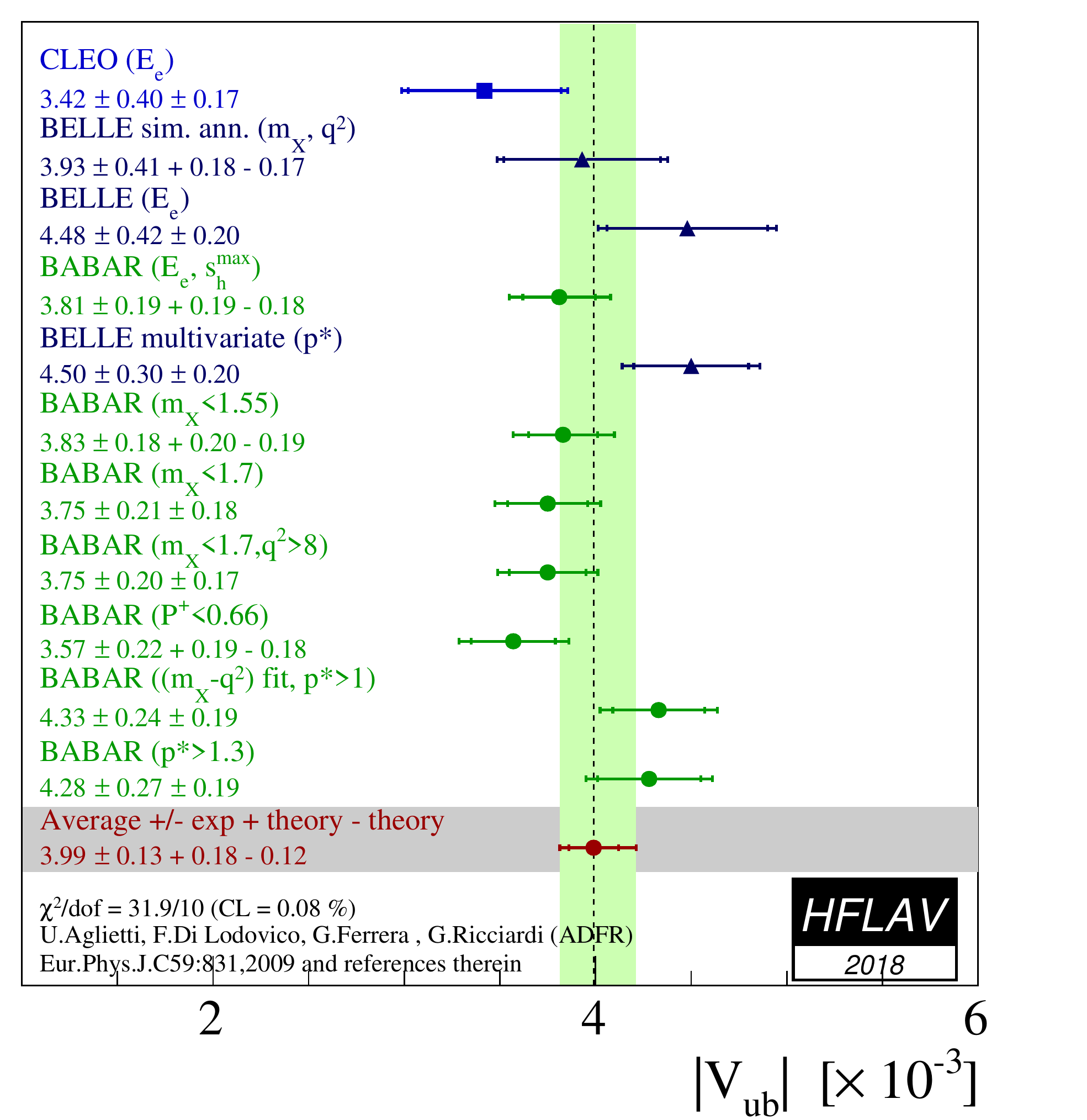}
   }
   \put(  6.0,  8.7){{\large\bf a)}}
   \put( 14.8,  8.7){{\large\bf b)}}
  \end{picture}
  \caption{Measurements of $\vub$ from inclusive semileptonic decays 
and their average based on the GGOU (a) and ADFR (b) prescription. The
labels indicate the variables and selections used to define the
signal regions in the different analyses
.} \label{fig:GGOU_ADFR}
 \end{center}
\end{figure}

\subsubsection{DGE}
Andersen and Gardi (Dressed Gluon Exponentiation, DGE)~\cite{ref:DGE} provide
a framework where the on-shell $b$-quark calculation, converted into hadronic variables, is
directly used as an approximation to the meson decay spectrum without
the use of a leading-power non-perturbative function (or, in other words,
a shape function). The on-shell mass of the $b$-quark within the $B$-meson ($m_b$) is
required as input. 
The DGE calculation uses the $\overline{MS}$ renormalization scheme. The heavy quark parameters determined  
from the global fit in the kinetic scheme, described in \ref{globalfitsKinetic}, were therefore 
translated into the $\overline{MS}$ scheme by using 
code provided by Einan Gardi (based on Refs.\cite{Blokland:2005uk,Gambino:2008fj}),
giving $m_b({\overline{MS}})=(4.188 \pm 0.043)~\gev$.
The extracted values
of \vub\, for each measurement along with their average are given in
Table~\ref{tab:bulnu} and illustrated in Fig.~\ref{fig:BLNP_DGE}(b).
The total error is $^{+3.3}_{-3.4}\%$, whose breakdown is:
statistics ($^{+1.8}_{-1.8}\%$),
detector effects ($^{+1.7}_{-1.7}\%$),
$B\to X_c \ell^+ \nul$ model ($^{+1.3}_{-1.3}\%$),
$B\to X_u \ell^+ \nul$ model ($^{+2.1}_{-1.7}\%$),
strong coupling $\alpha_s$ ($^{+0.5}_{-0.6}\%$),
$m_b$ ($^{+3.2}_{-2.9}\%$),
weak annihilation ($^{+0.0}_{-1.1}\%$),
matching scales in DGE ($^{+0.5}_{-0.4}\%$).
The largest contribution to the total error is due to the effect of the uncertainty 
on $m_b$. 
The uncertainty due to 
weak annihilation has been assumed to be asymmetric, \ie\ it only tends to decrease \vub.

\subsubsection{GGOU}
\label{subsec:ggou}
Gambino, Giordano, Ossola and Uraltsev (GGOU)~\cite{Gambino:2007rp} 
compute the triple differential decay rates of $B \to X_u \ell^+ \nul$, 
including all perturbative and non--perturbative effects through $O(\alphas^2 \beta_0)$ 
and $O(1/m_b^3)$. 
The Fermi motion is parameterized in terms of a single light--cone function 
for each structure function and for any value of $q^2$, accounting for all subleading effects. 
The calculations are performed in the kinetic scheme, a framework characterized by a Wilsonian 
treatment with a hard cutoff $\mu \sim 1~\gev$.
GGOU have not included calculations for the ``($E_e,s^{max}_h$)'' analysis~\cite{ref:babar-elq2}. 
The heavy quark parameters determined  
from the global fit in the kinetic scheme, described in Section~\ref{globalfitsKinetic}, are used as inputs: 
$m_b^{kin}=(4.554 \pm 0.018)~\gev$, 
$\mu_\pi^2=(0.464 \pm 0.076)~\gevcc$. 
The extracted values
of \vub\, for each measurement along with their average are given in
Table~\ref{tab:bulnu} and illustrated in Fig.~\ref{fig:GGOU_ADFR}(a).
The total error is $^{+4.0}_{-4.0}\%$ whose breakdown is:
statistics ($^{+1.6}_{-1.6}\%$),
detector effects ($^{+1.6}_{-1.6}\%$),
$B\to X_c \ell^+ \nul$ model ($^{+0.9}_{-0.9}\%$),
$B\to X_u \ell^+ \nul$ model ($^{+1.5}_{-1.5}\%$),
$\alpha_s$, $m_b$ and other non--perturbative parameters ($^{+1.9}_{-1.9}\%$), 
higher order perturbative and non--perturbative corrections ($^{+1.5}_{-1.5}\%$), 
modelling of the $q^2$ tail
($^{+1.3}_{-1.3}\%$), 
weak annihilations matrix element ($^{+0.0}_{-1.1}\%$), 
functional form of the distribution functions ($^{+0.1}_{-0.1}\%$).  
The leading uncertainties
on  \vub\ are both from theory, and are due to perturbative and non--perturbative
parameters and the modelling of the $q^2$ tail.
The uncertainty due to 
weak annihilation has been assumed to be asymmetric, \ie\ it only tends to decrease \vub.

\subsubsection{ADFR}
Aglietti, Di Lodovico, Ferrera and Ricciardi (ADFR)~\cite{Aglietti:2007ik}
use an approach to extract \vub, that makes use of the ratio
of the  $B \to X_c \ell^+ \nul$ and $B \to X_u \ell^+ \nul$ widths. 
The normalized triple differential decay rate for 
$B \to X_u \ell^+ \nul$~\cite{Aglietti:2006yb,Aglietti:2005mb, Aglietti:2005bm, Aglietti:2005eq}
is calculated with a model based on (i) soft--gluon resummation 
to next--to--next--leading order and (ii) an effective QCD coupling without a
Landau pole. This coupling is constructed by means of an extrapolation to low
energy of the high--energy behaviour of the standard coupling. More technically,
an analyticity principle is used.
The lower cut on the electron energy for the endpoint analyses is 2.3~GeV~\cite{Aglietti:2006yb}.
The ADFR calculation uses the $\overline{MS}$ renormalization scheme; the heavy quark parameters determined  
from the global fit in the kinetic scheme, described in \ref{globalfitsKinetic}, were therefore 
translated into the $\overline{MS}$ scheme 
by using code provided by Einan Gardi (based on Refs.\cite{Blokland:2005uk,Gambino:2008fj}), giving $m_b({\overline{MS}})=(4.188 \pm 0.043)~\gev$.
The extracted values
of \vub\, for each measurement along with their average are given in
Table~\ref{tab:bulnu} and illustrated in Fig.~\ref{fig:GGOU_ADFR}(b).
The total error is $^{+5.6}_{-5.6}\%$ whose breakdown is:
statistics ($^{+1.9}_{-1.9}\%$),
detector effects ($^{+1.7}_{-1.7}\%$),
$B\to X_c \ell^+ \nul$ model ($^{+1.4}_{-1.4}\%$),
$B\to X_u \ell^+ \nul$ model ($^{+1.5}_{-1.4}\%$),
$\alpha_s$ ($^{+1.1}_{-1.0}\%$), 
$|V_{cb}|$ ($^{+1.9}_{-1.9}\%$), 
$m_b$ ($^{+0.7}_{-0.7}\%$), 
$m_c$ ($^{+1.3}_{-1.3}\%$), 
semileptonic branching fraction ($^{+0.8}_{-0.7}\%$), 
theory model ($^{+3.6}_{-3.6}\%$).
The leading uncertainty is due to the theory model.

\subsubsection{BLL}
\begin{figure}
\begin{center}
\includegraphics[width=0.52\textwidth]{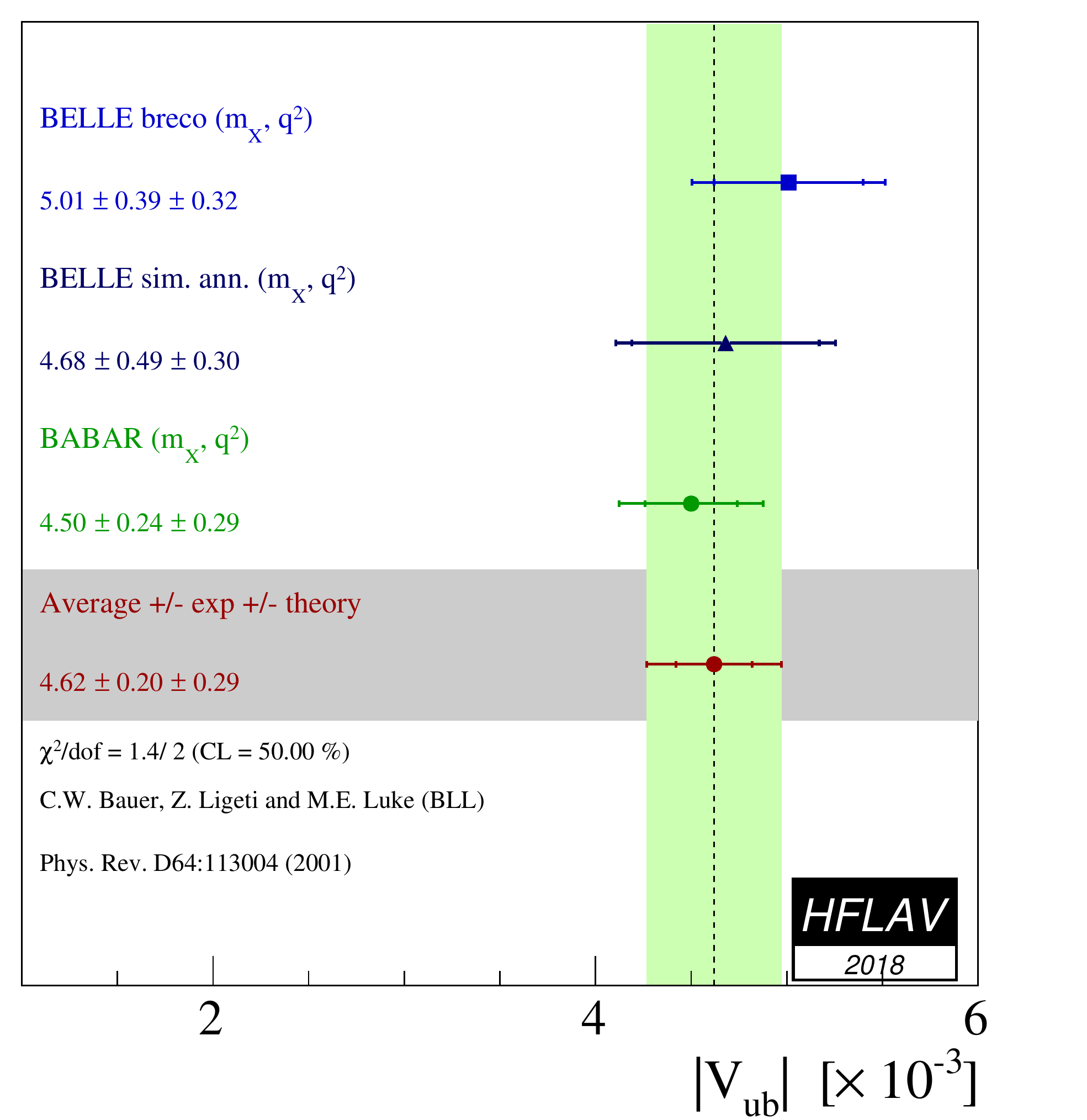}
\end{center}
\caption{Measurements of $\vub$ from inclusive semileptonic decays 
and their average in the BLL prescription.
}
\label{fig:BLL}
\end{figure}

Bauer, Ligeti, and Luke (BLL)~\cite{ref:BLL} give a
HQET-based prescription that advocates combined cuts on the dilepton invariant mass, $q^2$,
and hadronic mass, $m_X$, to minimise the overall uncertainty on \vub.
In their reckoning a cut on $m_X$ only, although most efficient at
preserving phase space ($\sim$80\%), makes the calculation of the partial
rate untenable due to uncalculable corrections
to the $b$-quark distribution function or shape function. These corrections are
suppressed if events in the low $q^2$ region are removed. The cut combination used
in measurements is $M_x<1.7~\gevcc$ and $q^2 > 8~\gevgevcccc$.  
The extracted values
of \vub\, for each measurement along with their average are given in
Table~\ref{tab:bulnu} and illustrated in Fig.~\ref{fig:BLL}.
The total error is $^{+7.7}_{-7.7}\%$ whose breakdown is:
statistics ($^{+3.3}_{-3.3}\%$),
detector effects ($^{+3.0}_{-3.0}\%$),
$B\to X_c \ell^+ \nul$ model ($^{+1.6}_{-1.6}\%$),
$B\to X_u \ell^+ \nul$ model ($^{+1.1}_{-1.1}\%$),
spectral fraction ($m_b$) ($^{+3.0}_{-3.0}\%$),
perturbative approach: strong coupling $\alpha_s$ ($^{+3.0}_{-3.0}\%$),
residual shape function ($^{+2.5}_{-2.5}\%$),
third order terms in the OPE ($^{+4.0}_{-4.0}\%$). 
The leading
uncertainties, both from theory, are due to residual shape function
effects and third order terms in the OPE expansion. The leading
experimental uncertainty is due to statistics. 

\subsubsection{Summary}
The averages presented in several different
frameworks %
are presented in 
Table~\ref{tab:vubcomparison}.
In summary, we recognize that the experimental and theoretical uncertainties play out
differently between the schemes and the theoretical assumptions for the
theory calculations are different. Therefore, it is difficult to perform an average 
between the various determinations of \vub. 
Since the methodology is similar to that used to determine the 
inclusive \vcb\ average, we choose to quote as reference value the average determined 
by the GGOU calculation, which gives \vub $= (4.32 \pm 0.12 ^{+0.12}_{-0.13}) \times 10^{-3}$. 

\begin{table}[!htb]
\caption{\label{tab:vubcomparison}
Summary of inclusive determinations of $\vub$.
The errors quoted on \vub\ correspond to experimental and theoretical uncertainties.
}
\begin{center}
\begin{small}
\begin{tabular}{|lc|}
\hline
Framework
&  $\Vub [10^{-3}]$\\
\hline\hline
BLNP
& $4.44 {^{+0.13}_{-0.14}} {^{+0.21}_{-0.22}}$ \\ 
DGE
& $4.52 \pm 0.10 ^{+0.09}_{-0.10}$ \\
GGOU
& $4.32 \pm 0.12 ^{+0.12}_{-0.13}$ \\
ADFR
& $3.99 \pm 0.13 ^{+0.18}_{-0.12}$ \\
BLL ($m_X/q^2$ only)
& $4.62 \pm 0.20 \pm 0.29$ \\ 
\hline
\end{tabular}\\
\end{small}
\end{center}
\end{table}

\subsection{$B\to D^{(*)}\tau \nu_\tau$ decays}
\label{slbdecays_b2dtaunu}

In the SM the semileptonic decay are tree level processes which proceed via coupling to the $W^{\pm}$ boson.
These couplings are assumed to be universal for all leptons and are well understood theoretically, (see Section 5.1 and 5.2.).
This universality has been tested in purely leptonic and semileptonic $B$ meson decays involving a $\tau$ lepton, which might 
be sensitive to a hypothetical charged Higgs boson or other non-SM processes.

Compared to $B^+\to\tau\nu_\tau$, the $B\to D^{(*)}\tau \nu_\tau$ decay has advantages: the branching fraction is 
relatively high, because it is not Cabibbo-suppressed, and it is a three-body decay allowing access to many 
observables besides the branching fraction, such as $D^{(*)}$ momentum, $q^2$ distributions, and measurements of the 
$D^*$ and $\tau$ polarisations (see Ref.~\cite{Duraisamy:2014sna} and references therein for recent calculations).

Experiments have measured two ratios of branching fractions defined as 
\begin{eqnarray}
{\cal R}(D)&=&\dfrac{ {\cal B}(B\to D\tau\nu_\tau) }{ {\cal B}(B\to D\ell\nu_\ell) },\\
{\cal R}(D^*)&=&\dfrac{ {\cal B}(B\to D^*\tau\nu_\tau) }{ {\cal B}(B\to D^*\ell\nu_\ell) } %
\end{eqnarray}
where $\ell$ refers either to electron or $\mu$. These ratios are independent of  $|V_{cb}|$ and to a large extent, also of 
the $B\to D^{(*)}$ form factors. As a consequence, the SM predictions for these ratios are quite precise:

\begin{itemize}
\item ${\cal R}(D)=0.299\pm 0.003$: which is  
an average of the predictions from Refs.\cite{Bigi:2016mdz, Bernlochner:2017jka, Jaiswal:2017rve}. These predictions use as input the latest results on the $B\to D\ell\nu$ form factors from \babar and Belle, and the most recent lattice calculations ~\cite{Lattice:2015rga,Na:2015kha}.

\item ${\cal R}(D^*)=0.258\pm 0.005$: where the central value and the uncertainty are obtained from an arithmetic average of the predictions from Refs. \cite{Bernlochner:2017jka, Bigi:2017jbd, Jaiswal:2017rve}. 
These calculations are in good agreement between each other, and consistent with the old prediction \cite{Fajfer:2012vx} extensively used in the past, but more robust. There are differences in the evaluation of the theoretical uncertainty associated mainly with assumptions on the pseudoscalar form factor. 
\end{itemize}

\noindent In Ref.~\cite{Gambino:2019sif}, Gambino, Jung and Schacht re-analysed the recent Belle results of $\B\to D^*\ell\nu$ form factors \cite{Waheed:2018djm}, obtaining ${\cal R}(D^*)=0.254^{+0.007}_{-0.006}$, compatible with the predictions mentioned before.
Another calculation, based on the full angular analysis of $\B\to D^*\ell\nu$ decay by \babar \cite{Dey:2019bgc}, gives an independent prediction of ${\cal R}(D^*)=0.253\pm 0.005$. 
Recently, the authors of Ref.\cite{Bordone:2019vic} obtained predictions with and without using experimental inputs. The results for $R(D)$ are consistent with the other predictions, while $R(D^*)$ are slightly shifted toward lower value, resulting in $R(D^*)=0.250\pm 0.003$ and $R(D^*)=0.247\pm 0.006$ using and not using the experimental results, respectively. 

On the experimental side, in the case of the leptonic $\tau$ decay, the ratios  ${\cal R}(D^{(*)})$ can be 
directly measured, and many systematic uncertainties cancel in the measurement. 
The $B^0\to D^{*+}\tau\nu_\tau$ decay was first observed by Belle~\cite{Matyja:2007kt} performing 
an "inclusive" reconstruction, which is based on the reconstruction of the $B_{tag}$ from all the particles
of the events, other than the $D^{(*)}$ and the lepton candidate, without looking for any specific $B_{tag}$ decay chain.  
Since then, both \babar and Belle have published improved measurements 
and have observed the $B\to D\tau\nu_\tau$ decays~\cite{Aubert:2007dsa,Bozek:2010xy}. %

The most powerful way to study these decays at the B-Factories exploits the hadronic or semileptonic $B_{tag}$.
Using the full dataset and an improved hadronic $B_{tag}$ selection, \babar measured~\cite{Lees:2012xj}:  
\begin{equation}
{\cal R}(D)=0.440\pm 0.058\pm 0.042, ~ {\cal R}(D^*)=0.332\pm 0.024\pm 0.018 %
\end{equation}
where decays to both $e^\pm$ and $\mu^\pm$ were summed, and results for $B^0$ and $B^-$ decays were combined in an isospin-constrained fit. The fact that the \babar result exceeded SM predictions by $3.4\sigma$ raised considerable interest.

Belle, exploiting the full dataset, published measurements using both the hadronic \cite{Huschle:2015rga} and the semileptonic tag \cite{Belle:2019rba}. Belle also performed a combined measurement of ${\cal R}(D^*)$ and $\tau$ polarization by reconstructing the $\tau$ in the hadronic $\tau\to\pi\nu$ and $\tau\to\rho\nu$ decay modes \cite{Hirose:2016wfn}. 
LHCb measurements of $R(D^*)$ use both the muonic $\tau$ decay \cite{Aaij:2015yra}, and the three-prong  hadronic $\tau\to 3\pi(\pi^0)\nu$ decays \cite{Aaij:2017uff}.
The latter is a direct measurement of the ratio
${\cal B}(B^0\to D^{*-}\tau^+\nu_\tau)/{\cal B}(B^0\to D^{*-}\pi^+\pi^-\pi^+)$, and is translated into a measurement of $R(D^*)$ using the independently measured branching fractions ${\cal B}(B^0\to D^{*-}\pi^+\pi^-\pi^+)$ and ${\cal B}(B^0\to D^{*-}\mu^+\nu_\mu)$.

The most important source of systematic uncertainties that are correlated among the different measurement  
is the $B\to D^{**}$ background components, which are difficult to disentangle from the signal. 
In our average, the systematic uncertainties 
due to the $B\to D^{**}$ composition and kinematics are considered fully correlated among the measurements. 

\begin{figure}
\centering
\includegraphics[width=0.9\textwidth]{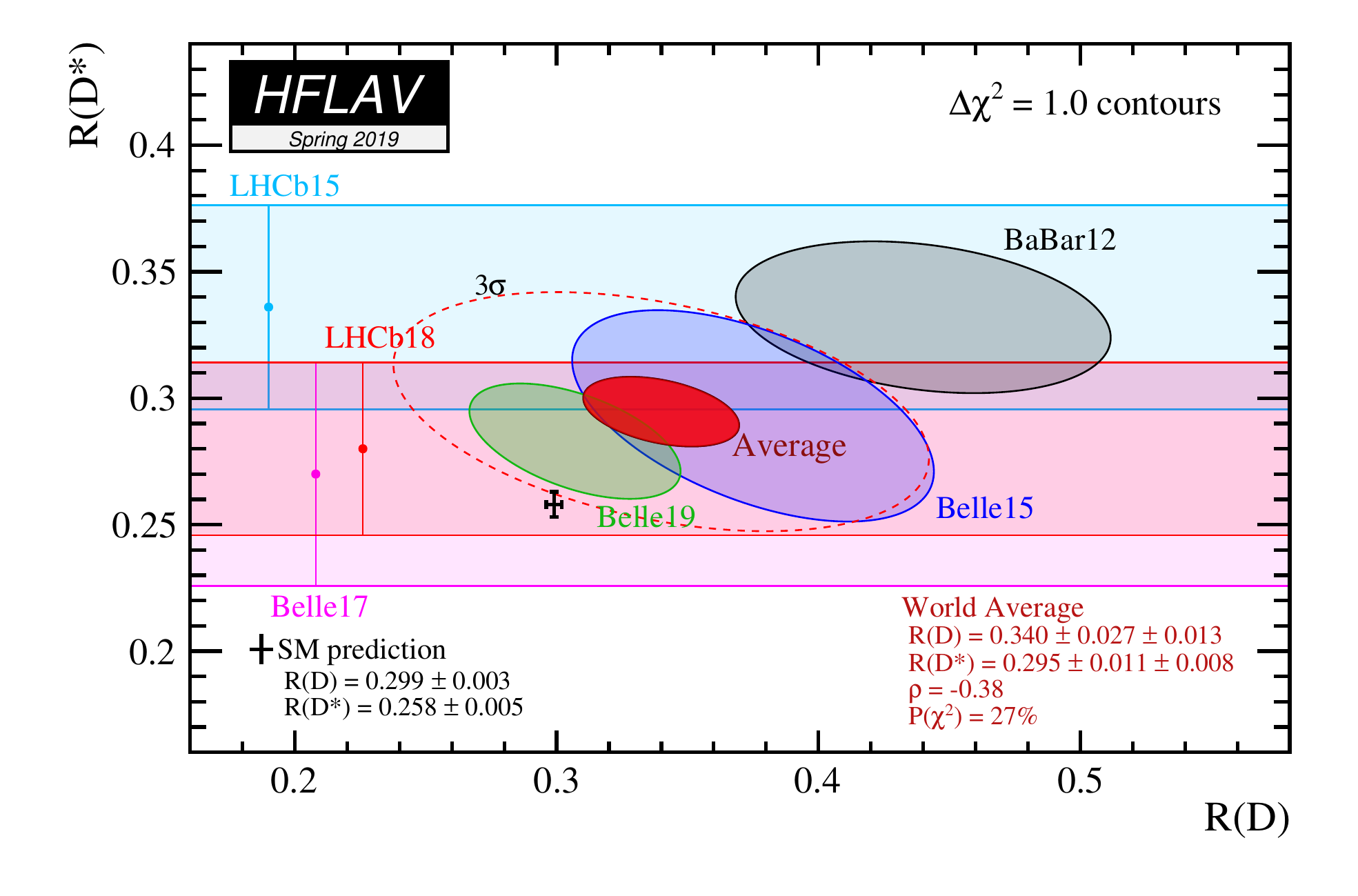}
\caption{Measurements of ${\cal R}(D)$ and ${\cal R}(D^*)$ listed in Table \ref{tab:dtaunu} and their two-dimensional average. Contours correspond to $\Delta \chi^2 = 1$, {\it i.e.}, 68\% CL for the bands and 39\% CL  for the ellipses. The black point with errors is the SM prediction for ${\cal R}(D^*)$ and ${\cal R}(D)$. The SM prediction is based on results from Refs.\cite{Bigi:2016mdz, Bernlochner:2017jka, Jaiswal:2017rve}, as explained in the text. 
The prediction and the experimental average deviate from each other by 3.08$\sigma$. 
The dashed ellipse correspond to a $3\sigma$ contour (99.73\% CL).
\label{fig:rds}}
\end{figure}

\begin{figure}[!ht]
  \begin{center}
  \unitlength 1.0cm %
  \begin{picture}(14.,11.0)
    \put(  -1.5, 0.0){\includegraphics[width=9.2cm]{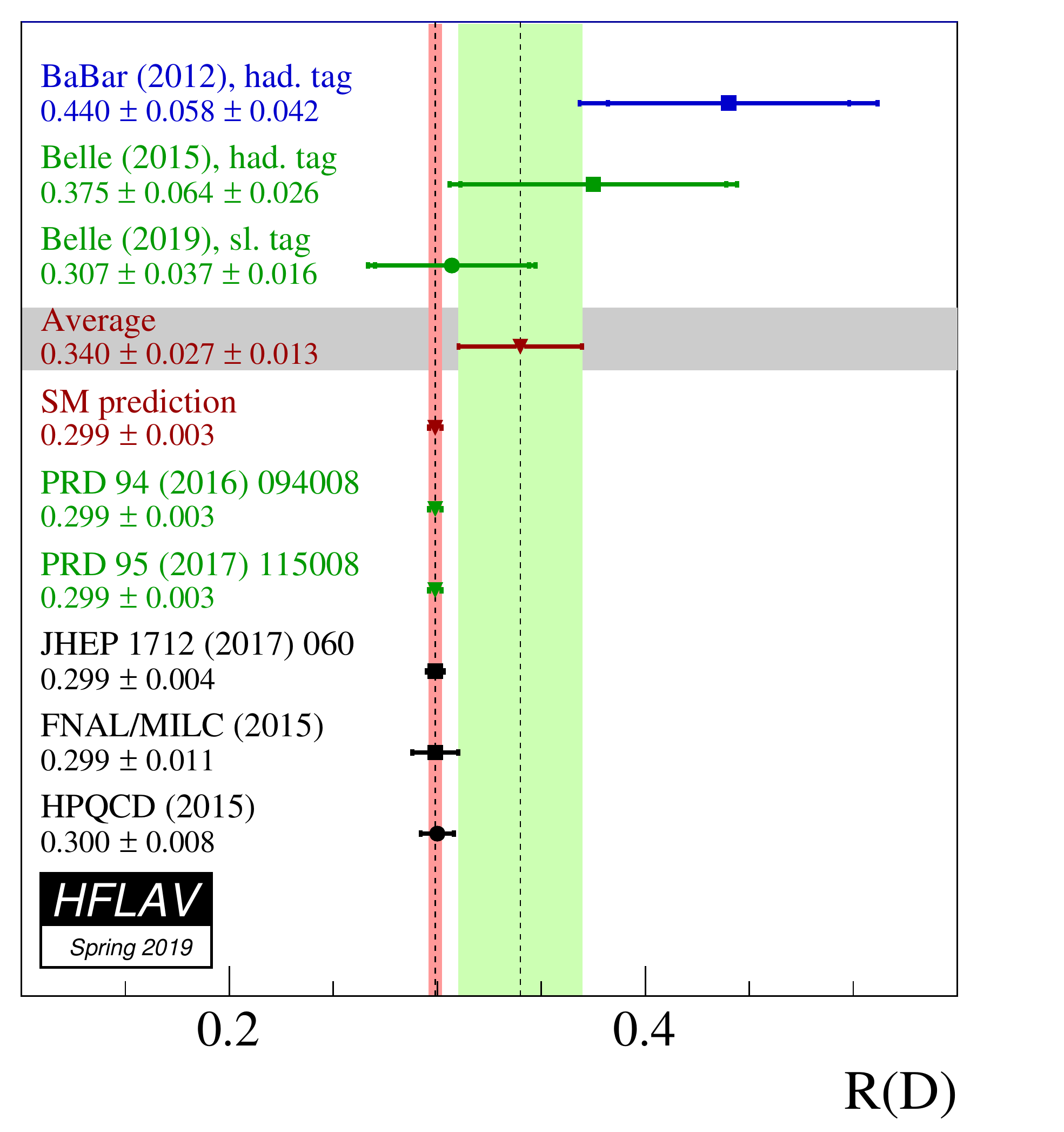}
    }
    \put(  7.5, 0.0){\includegraphics[width=8.8cm]{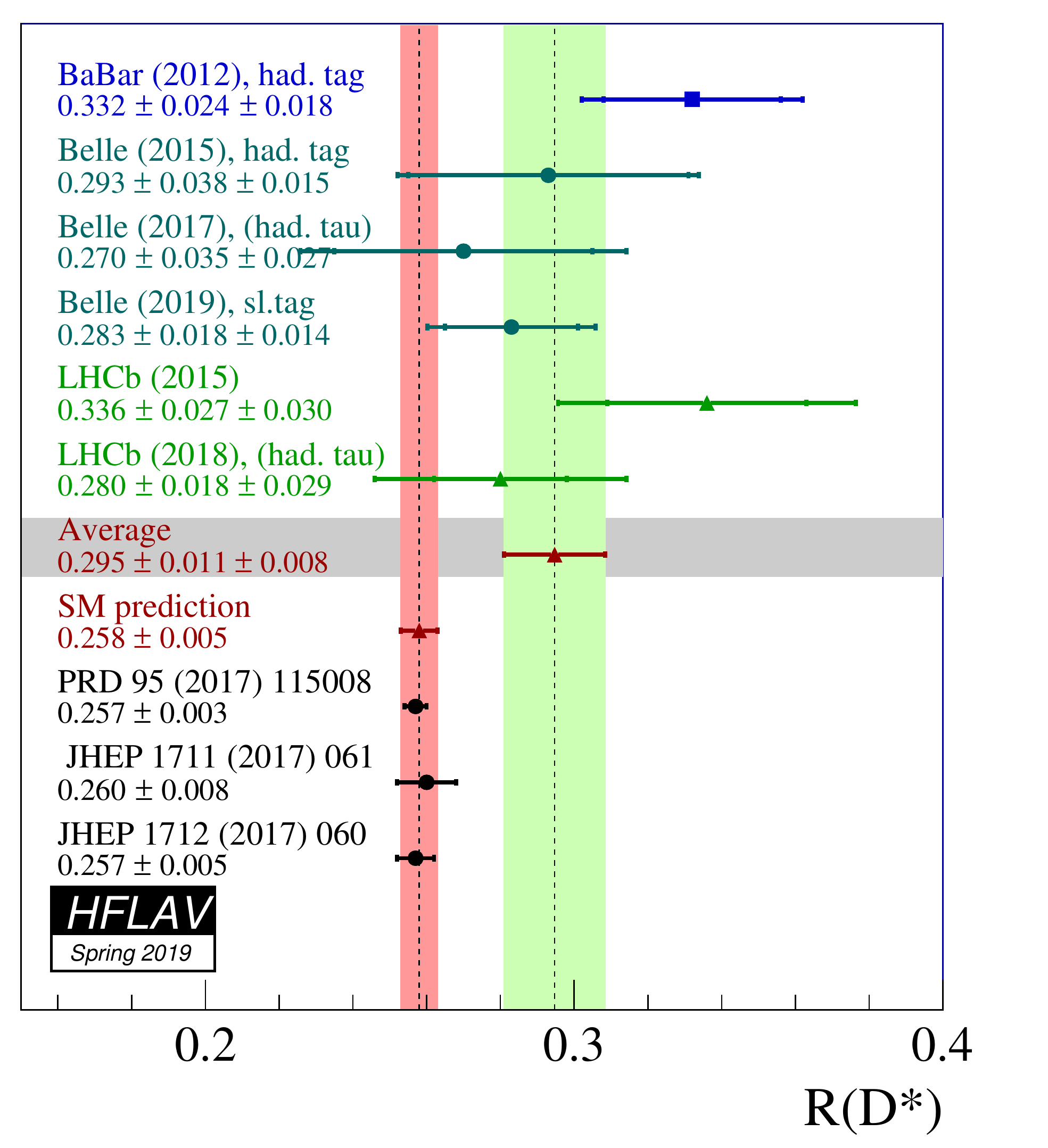}
    }
    \put(  5.8,  8.4){{\large\bf a)}}  
    \put( 14.7,  8.4){{\large\bf b)}}
  \end{picture}
  \caption{(a) Measurements of ${\cal R}(D)$ and (b) ${\cal R}(D^*)$. The green bands are the averages
  obtained from the combined fit. The red bands are the averages of the theoretical predictions obtained 
  as explained in the text.} 
 \label{fig:rd}
 \end{center}
\end{figure}

The results of the individual measurements, their averages and correlations are presented in Table \ref{tab:dtaunu} and Fig.\ref{fig:rds}. 
The combined results, projected separately on ${\cal R}(D)$ 
and ${\cal R}(D^*)$, are reported in Fig.\ref{fig:rd}(a) and  Fig.\ref{fig:rd}(b) respectively. 

The averaged ${\cal R}(D)$ and ${\cal R}(D^*)$ exceed the SM predictions by 1.4$\sigma$ and 2.5$\sigma$ respectively. 
Considering the ${\cal R}(D)$ and ${\cal R}(D^*)$ total correlation of $-0.38$, the difference with respect to the 
SM is about 3.08~$\sigma$, and the combined $\chi^2=12.33$ for 2 degrees of freedom corresponds to a $p$-value 
of $2.07\times 10^{-3}$, assuming Gaussian error distributions.

\begin{table}[!htb]
\caption{Measurements of ${\cal R}(D^*)$ and ${\cal R}(D)$, their correlations and the combined average.}
\begin{center}
\resizebox{0.99\textwidth}{!}{
\begin{tabular}{l|c|c|c}\hline
Experiment  &${\cal R}(D^*)$ & ${\cal R}(D)$ & $\rho$ \\

\hline\hline 
\babar ~\cite{Lees:2012xj,Lees:2013uzd} &$0.332 \pm0.024_{\rm stat} \pm0.018_{\rm syst}$  &$0.440 \pm0.058_{\rm stat} \pm0.042_{\rm syst}$ & $-0.27$\\
Belle  ~\cite{Huschle:2015rga}         &$0.293 \pm0.038_{\rm stat} \pm0.015_{\rm syst}$  &$0.375 \pm0.064_{\rm stat} \pm0.026_{\rm syst}$ & $-0.49$ \\
LHCb   ~\cite{Aaij:2015yra}            &$0.336 \pm0.027_{\rm stat} \pm0.030_{\rm syst}$  &   &\\
Belle  ~\cite{Hirose:2016wfn}          &$0.270 \pm0.035_{\rm stat} \, {^{+0.028} _{-0.025}}_{\rm syst}$  & & \\
LHCb   ~\cite{Aaij:2017uff,Aaij:2017deq}            &$0.280 \pm0.018_{\rm stat} \pm0.029_{\rm syst}$  &   &\\
Belle  ~\cite{Belle:2019rba}            &$0.283 \pm0.018_{\rm stat} \pm0.014_{\rm syst}$  &   $0.307 \pm0.037_{\rm stat} \pm0.016_{\rm syst}$  & -0.51 \\
\hline
{\bf Average} &\mathversion{bold}$0.295 \pm0.011 \pm 0.008$ & \mathversion{bold}$0.340 \pm0.027 \pm 0.013$ & $-0.38$  \\
\hline 
\end{tabular}
}
\end{center}
\label{tab:dtaunu}
\end{table}

\clearpage
\section{Decays of $b$-hadrons into open or hidden charm hadrons}
\label{sec:b2c}
Ground state $B$ mesons and $b$ baryons dominantly decay to particles containing a charm quark via the $b \rightarrow c$ quark transition.
In this section, measurements of such decays to hadronic final states are summarized. %
The use of such decays for studying fundamental properties of the bottom hadrons and for obtaining parameters of the CKM matrix is discussed in Sections~\ref{sec:life_mix} and~\ref{sec:cp_uta}, respectively. 
The properties of certain $b$ hadron decays to open or hidden charm hadrons, such as small $Q$ values and similar topologies for different modes, allow the minimization of systematic uncertainties in measurements of particle and decay properties.

The fact that decays to final states containing open or hidden charm hadrons dominate the $b$-hadron widths makes them a very important part of the experimental programme in heavy flavour physics.
Understanding the rate of charm production in $b$-hadron decays is crucial for
validation of the heavy-quark expansion (HQE) that underpins much of the theoretical framework for $b$
physics (see, for example, Ref.~\cite{Lenz:2014nka} for a review).
Moreover, such decays are often used as normalization modes for
measurements of rarer decays. 
In addition, they are the dominant background in many analyses.
To model accurately such backgrounds with simulated data, it is essential to have precise knowledge of the contributing decay modes.
In particular, with the expected increase in the data samples at LHCb and
Belle~II, the enhanced statistical sensitivity has to be matched by low
systematic uncertainties due to knowledge of the dominant $b$-hadron decay
modes. 
For multibody decays, knowledge of the distribution of decays across the
phase-space (\eg,\ the Dalitz plot density for three-body decays or the
polarization amplitudes for vector-vector final states) is required in
addition to the total branching fraction.

The large yields of $b \to c$ decays to multibody final states 
make them ideal for studying the spectroscopy of both open and hidden charm hadrons.
In particular, they have been used to both discover, and measure the properties
of exotic particles, such as the $X(3872)$~\cite{Choi:2003ue,Aaij:2015eva},
$Z(4430)^+$~\cite{Choi:2007wga,Aaij:2014jqa} and $P_c(4450)^+$~\cite{Aaij:2015tga} states.
Similarly, $b \to c$ transitions are very useful for studying baryon-antibaryon pair production in $B$-meson decays.

In addition to the dominant $b \to c$ decays, there are several decays in this category that are expected
to be highly suppressed in the Standard Model.
These are of interest for probing particular decay topologies (\eg,\ the annihilation diagram,
which dominates the $B^- \to \Dsm \phi$ decay),
which thereby constrain effects in other hadronic decays, or for searching for new physics.
There are also open charm production modes that involve $b \to u$ transitions, such as $\Bzb \to \Dsm \pip$, which are
mediated by the $W$ emission involving the $|V_{ub}|$ CKM matrix element.
Finally, $b \to c$ decays involving lepton flavour or number violation are
extremely suppressed in the Standard Model, and therefore provide highly
sensitive tests of new physics.

In this section, we give an exhaustive list of measured branching ratios of decay modes to
hadrons containing charm quarks.
The averaging procedure follows the methodology described in Section~\ref{sec:method}.
Where available, correlations between measurements are taken into account.
If an insignificant measurement and a limit for the same parameter are provided,
the former is quoted, so that it can be included in averages.
In case of asymmetric uncertainties, a variable width Gaussian likelihood
with linear variance in the range [$\sigma_-, \sigma_+$] around the central value
is assumed, following a suggestion in \cite{Barlow:2004wg}.
The confidence level of an average is quoted if it is below 1\%.
We provide averages of the polarization amplitudes of $B$ meson decays to
vector-vector states, but we do not currently provide detailed averages of
quantities obtained from Dalitz plot analyses, due to the complications
arising from the dependence on the model used.

The results are presented in subsections organized according to the type of decaying bottom
hadron: $\Bzb$ (Sec.~\ref{sec:b2c:Bd}), $B^-$ (Sec.~\ref{sec:b2c:Bu}), $\Bzb/B^-$ admixture (Sec.~\ref{sec:b2c:B}), $\Bsb$ (Sec.~\ref{sec:b2c:Bs}), $B_c^-$ (Sec.~\ref{sec:b2c:Bc}), $b$ baryons (Sec.~\ref{sec:b2c:Bbaryon}).
For each subsection, the measurements are arranged according to the final state
into the following groups: a single charmed meson, two charmed mesons, a
charmonium state, a charm baryon, or other states, such as, e.g. the $X(3872)$.
The individual measurements and averages are shown as numerical values in tables followed by a graphical representation of the averages.
The symbol $\mathcal{B}$ is used for branching ratios, and $f$ for production fractions (see Section~\ref{sec:life_mix}).
The decay amplitudes for longitudinal, parallel, and perpendicular transverse polarization in pseudoscalar to vector-vector decays are denoted ${\cal{A}}_0$, ${\cal{A}}_\parallel$, and ${\cal{A}}_\perp$, respectively, and the definitions $\delta_\parallel = \arg({\cal{A}}_\parallel/{\cal{A}}_0)$ and $\delta_\perp = \arg({\cal{A}}_\perp/{\cal{A}}_0)$ are used for their relative phases.
For normalized P-wave amplitudes we use the notation $f_i = |{\cal{A}}_i|^2 / (|{\cal{A}}_0|^2 + |{\cal{A}}_\parallel|^2+ |{\cal{A}}_\perp|^2)$.
Broad orbitally excited states are denoted by a trailing $(H)$.
The inclusion of charge conjugate modes is always implied.

Following the approach used by the PDG~\cite{PDG_2018}, for decays that involve
neutral kaons we mainly quote results in terms of final states including
either a $\Kz$ or $\Kzb$ meson (instead of a \KS or \KL),
although the flavour of the neutral kaon is never determined experimentally.
The specification as \Kz or \Kzb simply follows the quark model
expectation for the dominant decay
and the inclusion of the conjugate final state neutral kaon is implied.
An exception occurs for some \Bs decays, specifically those to \CP
eigenstates, where the width difference between the mass eigenstates (see
Sec.~\ref{sec:life_mix}) means that the measured branching fraction,
integrated over decay time, is specific to the studied final
state~\cite{DeBruyn:2012wj}. 
In such cases it is appropriate to quote the branching fraction for, \eg, $\Bsb \to
\jpsi \KS$ instead of $\Bsb \to \jpsi \Kzb$.

Several measurements assume $\Gamma(\Upsilon(4S) \to B^+B^-) = \Gamma(\Upsilon(4S) \to B^0\bar{B}^0)$.
While there is no evidence for isospin violation in $\Upsilon(4S)$ decays,
deviations from this assumption can be of the order of a few percent,
see Section~\ref{sec:fraction_Ups4S} and Ref.~\cite{Jung:2015yma}.
As the effect is negligible for many averages, we take the quoted values without applying a correction 
or additional systematic uncertainty. However, we note that this can be relevant
for averages with  percent-level uncertainty.

\newenvironment{btocharmtab}[2]{\begin{table}[H]\begin{center}\caption{#2.}\label{tab:b2c:#1}\begin{tabular}{|Sl l l |}}{\end{tabular}\end{center}\end{table}}
\newenvironment{widebtocharmtab}[2]{\begin{table}[H]\begin{center}\caption{#2.}\label{tab:b2c:#1}\begin{adjustbox}{width=\textwidth,center}\begin{tabular}{|Sl l l |}}{\end{tabular}\end{adjustbox}\end{center}\end{table}}
\newcommand{\btocharmfig}[1]{}
\newcommand{\btocharm}[1]{\input{b2charm/#1.tex}}
\newcommand{\btocharmparam}[3]{#1#3}
\newcommand{\btocharmparaml}[3]{\multicolumn{3}{|Sl|}{#2#3}\\}

\subsection{Decays of $\bar{B}^0$ mesons}
\label{sec:b2c:Bd}
Measurements of $\bar{B}^0$ decays to charmed hadrons are summarized in Sections~\ref{sec:b2c:Bd_D} to~\ref{sec:b2c:Bd_other}.

\subsubsection{Decays to a single open charm meson}
\label{sec:b2c:Bd_D}
Averages of $\bar{B}^0$ decays to a single open charm meson are shown in Tables~\ref{tab:b2c:Bd_D_1}--\ref{tab:b2c:Bd_D_14}.
In this section $D^{**}$ refers to the sum of all the non-strange charm meson
states with masses in the range $2.2-2.8~\gevcc$.
\begin{btocharmtab}{Bd_D_1}{Branching fractions to a $D^{(*)}$ meson and one or more pions, I}
\hline
\textbf{Parameter} & 
\begin{tabular}{l}\textbf{Measurements}\end{tabular} $[10^{-3}]$ & \textbf{Average} $[10^{-3}]$ \\
\hline
\hline
\btocharmparam{${\cal{B}} ( \bar{B}^{0} \to D^{+} \pi^{-} )$}{${\cal{B}} ( \bar{B}^{0} \to D^{+} \pi^{-} )$}{} & \begin{tabular}{l} \babar \cite{Aubert:2006cd}: $2.55 \pm 0.05 \pm 0.16$ \\ \babar \cite{Aubert:2006jc}: $3.03 \pm 0.23 \pm 0.23$ \\ \end{tabular} & $2.65 \pm 0.15$ \\
\hline
\btocharmparam{${\cal{B}} ( \bar{B}^{0} \to D^{*}(2010)^{+} \pi^{-} )$}{${\cal{B}} ( \bar{B}^{0} \to D^{*}(2010)^{+} \pi^{-} )$}{} & \begin{tabular}{l} Belle \cite{Kuzmin:2006mw}: $2.22 \pm 0.04 \pm 0.19$ \\ \babar \cite{Aubert:2006cd}: $2.79 \pm 0.08 \pm 0.17$ \\ \babar \cite{Aubert:2006jc}: $2.99 \pm 0.23 \pm 0.24$ \\ \end{tabular} & $2.58 \pm 0.13$ \\
\hline
\btocharmparaml{${\cal{B}} ( \bar{B}^{0} \to D^{*}(2010)^{+} \pi^{-} \pi^{+} \pi^{-} )$}{${\cal{B}} ( \bar{B}^{0} \to D^{*}(2010)^{+} \pi^{-} \pi^{+} \pi^{-} )$}{} & \begin{tabular}{l} Belle \cite{Majumder:2004su}: $6.81 \pm 0.23 \pm 0.72$ \\ \babar \cite{TheBABAR:2016vzj}: $7.26 \pm 0.11 \pm 0.31$ \\ \end{tabular} & $7.19 \pm 0.30$ \\
\hline
\btocharmparaml{${\cal{B}} ( \bar{B}^{0} \to D^{*}(2007)^{0} \pi^{-} \pi^{+} \pi^{-} \pi^{+} )$}{${\cal{B}} ( \bar{B}^{0} \to D^{*}(2007)^{0} \pi^{-} \pi^{+} \pi^{-} \pi^{+} )$}{} & \begin{tabular}{l} Belle \cite{Majumder:2004su}: $2.60 \pm 0.47 \pm 0.37$ \\ \end{tabular} & $2.60 \pm 0.60$ \\
\hline
\btocharmparaml{${\cal{B}} ( \bar{B}^{0} \to D^{*}(2010)^{+} \pi^{-} \pi^{+} \pi^{-} \pi^{+} \pi^{-} )$}{${\cal{B}} ( \bar{B}^{0} \to D^{*}(2010)^{+} \pi^{-} \pi^{+} \pi^{-} \pi^{+} \pi^{-} )$}{} & \begin{tabular}{l} Belle \cite{Majumder:2004su}: $4.72 \pm 0.59 \pm 0.71$ \\ \end{tabular} & $4.72 \pm 0.92$ \\
\hline
\btocharmparaml{${\cal{B}} ( \bar{B}^{0} \to D^{*}(2010)^{+} \omega(782) \pi^{-} )$}{${\cal{B}} ( \bar{B}^{0} \to D^{*}(2010)^{+} \omega(782) \pi^{-} )$}{} & \begin{tabular}{l} Belle \cite{Matvienko:2015gqa}: $2.31 \pm 0.11 \pm 0.14$ \\ \babar \cite{Aubert:2006zb}: $2.88 \pm 0.21 \pm 0.31$ \\ \end{tabular} & $2.41 \pm 0.16$ \\
\hline
\end{btocharmtab}
\btocharmfig{Bd_D_1}

\begin{btocharmtab}{Bd_D_2}{Branching fractions to a $D^{(*)}$ meson and one or more pions, II}
\hline
\textbf{Parameter} & \begin{tabular}{l}\textbf{Measurements}\end{tabular} $[10^{-4}]$ & \textbf{Average} $[10^{-4}]$ \\
\hline
\hline
\btocharmparam{${\cal{B}} ( \bar{B}^{0} \to D^{0} \pi^{0} )$}{${\cal{B}} ( \bar{B}^{0} \to D^{0} \pi^{0} )$}{} & \begin{tabular}{l} Belle \cite{Blyth:2006at}: $2.25 \pm 0.14 \pm 0.35$ \\ \babar \cite{Lees:2011gw}: $2.69 \pm 0.09 \pm 0.13$ \\ \end{tabular} & $2.62 \pm 0.15$ \\
\hline
\btocharmparam{${\cal{B}} ( \bar{B}^{0} \to D^{*}(2007)^{0} \pi^{0} )$}{${\cal{B}} ( \bar{B}^{0} \to D^{*}(2007)^{0} \pi^{0} )$}{} & \begin{tabular}{l} Belle \cite{Blyth:2006at}: $1.39 \pm 0.18 \pm 0.26$ \\ \babar \cite{Lees:2011gw}: $3.05 \pm 0.14 \pm 0.28$ \\ \end{tabular} & \begin{tabular}{@{}l@{}} $2.23 \pm 0.22$ \\[-5pt] {\tiny CL=0.2\permil} \end{tabular} \\
\hline
\btocharmparam{${\cal{B}} ( \bar{B}^{0} \to D^{0} \pi^{+} \pi^{-} )$}{${\cal{B}} ( \bar{B}^{0} \to D^{0} \pi^{+} \pi^{-} )$}{} & \begin{tabular}{l} LHCb \cite{Aaij:2015sqa}: $8.46 \pm 0.14 \pm 0.49$ \\ Belle \cite{Kuzmin:2006mw}: $8.4 \pm 0.4 \pm 0.8$ \\ \end{tabular} & $8.45 \pm 0.45$ \\
\hline
\btocharmparam{${\cal{B}} ( \bar{B}^{0} \to D^{*}(2007)^{0} \pi^{+} \pi^{-} )$}{${\cal{B}} ( \bar{B}^{0} \to D^{*}(2007)^{0} \pi^{+} \pi^{-} )$}{} & \begin{tabular}{l} Belle \cite{Satpathy:2002js}: $6.2 \pm 1.2 \pm 1.8$ \\ \end{tabular} & $6.2 \pm 2.2$ \\
\hline
\end{btocharmtab}
\btocharmfig{Bd_D_2}

\begin{btocharmtab}{Bd_D_3}{Branching fractions to a $D^{(*)0}$ meson and a light meson}
\hline
\textbf{Parameter} & \begin{tabular}{l}\textbf{Measurements}\end{tabular} $[10^{-4}]$ & \textbf{Average} $[10^{-4}]$ \\
\hline
\hline
\btocharmparam{${\cal{B}} ( \bar{B}^{0} \to D^{0} \rho(770)^{0} )$}{${\cal{B}} ( \bar{B}^{0} \to D^{0} \rho(770)^{0} )$}{} & \begin{tabular}{l} Belle \cite{Kuzmin:2006mw}: $3.19 \pm 0.20 \pm 0.45$ \\ \end{tabular} & $3.19 \pm 0.49$ \\
\hline
\btocharmparam{${\cal{B}} ( \bar{B}^{0} \to D^{*}(2007)^{0} \rho(770)^{0} )$}{${\cal{B}} ( \bar{B}^{0} \to D^{*}(2007)^{0} \rho(770)^{0} )$}{} & \begin{tabular}{l} Belle \cite{Satpathy:2002js}: $< 5.1$ \\ \end{tabular} & $< 5.1$ \\
\hline
\btocharmparam{${\cal{B}} ( \bar{B}^{0} \to D^{0} \eta )$}{${\cal{B}} ( \bar{B}^{0} \to D^{0} \eta )$}{} & \begin{tabular}{l} Belle \cite{Blyth:2006at}: $1.77 \pm 0.16 \pm 0.21$ \\ \babar \cite{Lees:2011gw}: $2.53 \pm 0.09 \pm 0.11$ \\ \end{tabular} & $2.36 \pm 0.13$ \\
\hline
\btocharmparam{${\cal{B}} ( \bar{B}^{0} \to D^{*}(2007)^{0} \eta )$}{${\cal{B}} ( \bar{B}^{0} \to D^{*}(2007)^{0} \eta )$}{} & \begin{tabular}{l} Belle \cite{Blyth:2006at}: $1.40 \pm 0.28 \pm 0.26$ \\ \babar \cite{Lees:2011gw}: $2.69 \pm 0.14 \pm 0.23$ \\ \end{tabular} & \begin{tabular}{@{}l@{}} $2.26 \pm 0.22$ \\[-5pt] {\tiny CL=5.8\permil} \end{tabular} \\
\hline
\btocharmparam{${\cal{B}} ( \bar{B}^{0} \to D^{0} \eta^\prime(958) )$}{${\cal{B}} ( \bar{B}^{0} \to D^{0} \eta^\prime(958) )$}{} & \begin{tabular}{l} Belle \cite{Schumann:2005ej}: $1.14 \pm 0.20 \,^{+0.10}_{-0.13}$ \\ \babar \cite{Lees:2011gw}: $1.48 \pm 0.13 \pm 0.07$ \\ \end{tabular} & $1.38 \pm 0.12$ \\
\hline
\btocharmparam{${\cal{B}} ( \bar{B}^{0} \to D^{*}(2007)^{0} \eta^\prime(958) )$}{${\cal{B}} ( \bar{B}^{0} \to D^{*}(2007)^{0} \eta^\prime(958) )$}{} & \begin{tabular}{l} Belle \cite{Schumann:2005ej}: $1.21 \pm 0.34 \pm 0.22$ \\ \babar \cite{Lees:2011gw}: $1.48 \pm 0.22 \pm 0.13$ \\ \end{tabular} & $1.40 \pm 0.22$ \\
\hline
\btocharmparam{${\cal{B}} ( \bar{B}^{0} \to D^{0} \omega(782) )$}{${\cal{B}} ( \bar{B}^{0} \to D^{0} \omega(782) )$}{} & \begin{tabular}{l} LHCb \cite{Aaij:2015sqa}: $2.81 \pm 0.72 \,^{+0.30}_{-0.33}$ \\ Belle \cite{Blyth:2006at}: $2.37 \pm 0.23 \pm 0.28$ \\ \babar \cite{Lees:2011gw}: $2.57 \pm 0.11 \pm 0.14$ \\ \end{tabular} & $2.54 \pm 0.16$ \\
\hline
\btocharmparam{${\cal{B}} ( \bar{B}^{0} \to D^{*}(2007)^{0} \omega(782) )$}{${\cal{B}} ( \bar{B}^{0} \to D^{*}(2007)^{0} \omega(782) )$}{} & \begin{tabular}{l} Belle \cite{Blyth:2006at}: $2.29 \pm 0.39 \pm 0.40$ \\ \babar \cite{Lees:2011gw}: $4.55 \pm 0.24 \pm 0.39$ \\ \end{tabular} & \begin{tabular}{@{}l@{}} $3.64 \pm 0.35$ \\[-5pt] {\tiny CL=1.8\permil} \end{tabular} \\
\hline
\btocharmparam{${\cal{B}} ( \bar{B}^{0} \to D^{0} f_2(1270) )$}{${\cal{B}} ( \bar{B}^{0} \to D^{0} f_2(1270) )$}{} & \begin{tabular}{l} LHCb \cite{Aaij:2015sqa}: $1.61 \pm 0.11 \,^{+0.19}_{-0.18}$ \\ Belle \cite{Kuzmin:2006mw}: $1.20 \pm 0.18 \pm 0.38$ \\ \end{tabular} & $1.57 \pm 0.21$ \\
\hline
\end{btocharmtab}
\btocharmfig{Bd_D_3}

\begin{btocharmtab}{Bd_D_4}{Branching fractions to a $D^{(*)+}$ meson and one or more kaons}
\hline
\textbf{Parameter} & \begin{tabular}{l}\textbf{Measurements}\end{tabular} $[10^{-3}]$ & \textbf{Average} $[10^{-3}]$ \\
\hline
\hline
\btocharmparam{${\cal{B}} ( \bar{B}^{0} \to D^{+} K^{-} )$}{${\cal{B}} ( \bar{B}^{0} \to D^{+} K^{-} )$}{} & \begin{tabular}{l} LHCb \cite{Aaij:2013qqa}: $0.220 \pm 0.003 \pm 0.013$ \\ Belle \cite{Abe:2001waa}: $0.204 \pm 0.045 \pm 0.034$ \\ \end{tabular} & $0.219 \pm 0.013$ \\
\hline
\btocharmparaml{${\cal{B}} ( \bar{B}^{0} \to D^{*}(2010)^{+} K^{-} )$}{${\cal{B}} ( \bar{B}^{0} \to D^{*}(2010)^{+} K^{-} )$}{} & \begin{tabular}{l} Belle \cite{Abe:2001waa}: $0.204 \pm 0.041 \pm 0.023$ \\ \end{tabular} & $0.204 \pm 0.047$ \\
\hline
\btocharmparam{${\cal{B}} ( \bar{B}^{0} \to D^{+} K^{*}(892)^{-} )$}{${\cal{B}} ( \bar{B}^{0} \to D^{+} K^{*}(892)^{-} )$}{} & \begin{tabular}{l} \babar \cite{Aubert:2004at}: $0.46 \pm 0.06 \pm 0.05$ \\ \end{tabular} & $0.46 \pm 0.08$ \\
\hline
\btocharmparam{${\cal{B}} ( \bar{B}^{0} \to D^{*}(2010)^{+} K^{*}(892)^{-} )$}{${\cal{B}} ( \bar{B}^{0} \to D^{*}(2010)^{+} K^{*}(892)^{-} )$}{} & \begin{tabular}{l} \babar \cite{Aubert:2004at}: $0.32 \pm 0.06 \pm 0.03$ \\ \end{tabular} & $0.32 \pm 0.07$ \\
\hline
\btocharmparam{${\cal{B}} ( \bar{B}^{0} \to D^{+} K^{0} \pi^{-} )$}{${\cal{B}} ( \bar{B}^{0} \to D^{+} K^{0} \pi^{-} )$}{} & \begin{tabular}{l} \babar \cite{Aubert:2004at}: $0.49 \pm 0.07 \pm 0.05$ \\ \end{tabular} & $0.49 \pm 0.09$ \\
\hline
\btocharmparam{${\cal{B}} ( \bar{B}^{0} \to D^{*}(2010)^{+} K^{0} \pi^{-} )$}{${\cal{B}} ( \bar{B}^{0} \to D^{*}(2010)^{+} K^{0} \pi^{-} )$}{} & \begin{tabular}{l} \babar \cite{Aubert:2004at}: $0.30 \pm 0.07 \pm 0.03$ \\ \end{tabular} & $0.30 \pm 0.08$ \\
\hline
\btocharmparam{${\cal{B}} ( \bar{B}^{0} \to D^{+} K^{-} K^{0} )$}{${\cal{B}} ( \bar{B}^{0} \to D^{+} K^{-} K^{0} )$}{} & \begin{tabular}{l} Belle \cite{Drutskoy:2002ib}: $< 0.31$ \\ \end{tabular} & $< 0.31$ \\
\hline
\btocharmparam{${\cal{B}} ( \bar{B}^{0} \to D^{*}(2010)^{+} K^{-} K^{0} )$}{${\cal{B}} ( \bar{B}^{0} \to D^{*}(2010)^{+} K^{-} K^{0} )$}{} & \begin{tabular}{l} Belle \cite{Drutskoy:2002ib}: $< 0.47$ \\ \end{tabular} & $< 0.47$ \\
\hline
\btocharmparam{${\cal{B}} ( \bar{B}^{0} \to D^{+} K^{-} K^{*}(892)^{0} )$}{${\cal{B}} ( \bar{B}^{0} \to D^{+} K^{-} K^{*}(892)^{0} )$}{} & \begin{tabular}{l} Belle \cite{Drutskoy:2002ib}: $0.88 \pm 0.11 \pm 0.15$ \\ \end{tabular} & $0.88 \pm 0.19$ \\
\hline
\btocharmparaml{${\cal{B}} ( \bar{B}^{0} \to D^{*}(2010)^{+} K^{-} K^{*}(892)^{0} )$}{${\cal{B}} ( \bar{B}^{0} \to D^{*}(2010)^{+} K^{-} K^{*}(892)^{0} )$}{} & \begin{tabular}{l} Belle \cite{Drutskoy:2002ib}: $1.29 \pm 0.22 \pm 0.25$ \\ \end{tabular} & $1.29 \pm 0.33$ \\
\hline
\end{btocharmtab}
\btocharmfig{Bd_D_4}

\begin{btocharmtab}{Bd_D_5}{Branching fractions to a $D^{(*)0}$ meson and a kaon}
\hline
\textbf{Parameter} & \begin{tabular}{l}\textbf{Measurements}\end{tabular} $[10^{-4}]$ & \textbf{Average} $[10^{-4}]$ \\
\hline
\hline
\btocharmparam{${\cal{B}} ( \bar{B}^{0} \to D^{0} \bar{K}^{0} )$}{${\cal{B}} ( \bar{B}^{0} \to D^{0} \bar{K}^{0} )$}{} & \begin{tabular}{l} Belle \cite{Krokovny:2002ua}: $0.50 \,^{+0.13}_{-0.12} \pm 0.06$ \\ \babar \cite{Aubert:2006qn}: $0.53 \pm 0.07 \pm 0.03$ \\ \end{tabular} & $0.52 \pm 0.07$ \\
\hline
\btocharmparam{${\cal{B}} ( \bar{B}^{0} \to D^{*}(2007)^{0} \bar{K}^{0} )$}{${\cal{B}} ( \bar{B}^{0} \to D^{*}(2007)^{0} \bar{K}^{0} )$}{} & \begin{tabular}{l} Belle \cite{Krokovny:2002ua}: $< 0.66$ \\ \babar \cite{Aubert:2006qn}: $0.36 \pm 0.12 \pm 0.03$ \\ \end{tabular} & $0.36 \pm 0.12$ \\
\hline
\btocharmparam{${\cal{B}} ( \bar{B}^{0} \to D^{0} K^{-} \pi^{+} )$}{${\cal{B}} ( \bar{B}^{0} \to D^{0} K^{-} \pi^{+} )$}{} & \begin{tabular}{l} \babar \cite{Aubert:2005yt}: $0.88 \pm 0.15 \pm 0.09$ \\ \end{tabular} & $0.88 \pm 0.17$ \\
\hline
\btocharmparam{${\cal{B}} ( \bar{B}^{0} \to D^{0} \bar{K}^{*}(892)^{0} )$}{${\cal{B}} ( \bar{B}^{0} \to D^{0} \bar{K}^{*}(892)^{0} )$}{} & \begin{tabular}{l} Belle \cite{Krokovny:2002ua}: $0.48 \,^{+0.11}_{-0.10} \pm 0.05$ \\ \babar \cite{Aubert:2006qn}: $0.40 \pm 0.07 \pm 0.03$ \\ \end{tabular} & $0.42 \pm 0.06$ \\
\hline
\btocharmparaml{${{\cal{B}} ( \bar{B}^{0} \to D^{0} \bar{K}^{*}(892)^{0} )\times {\cal{B}} ( \bar{K}^{*}(892)^{0} \to K^{-} \pi^{+} )}$}{${{\cal{B}} ( \bar{B}^{0} \to D^{0} \bar{K}^{*}(892)^{0} )\times {\cal{B}} ( \bar{K}^{*}(892)^{0} \to K^{-} \pi^{+} )}$}{} & \begin{tabular}{l} \babar \cite{Aubert:2005yt}: $0.38 \pm 0.06 \pm 0.04$ \\ \end{tabular} & $0.38 \pm 0.07$ \\
\hline
\btocharmparam{${\cal{B}} ( \bar{B}^{0} \to D^{*}(2007)^{0} \bar{K}^{*}(892)^{0} )$}{${\cal{B}} ( \bar{B}^{0} \to D^{*}(2007)^{0} \bar{K}^{*}(892)^{0} )$}{} & \begin{tabular}{l} Belle \cite{Krokovny:2002ua}: $< 0.69$ \\ \end{tabular} & $< 0.69$ \\
\hline
\btocharmparam{${\cal{B}} ( \bar{B}^{0} \to \bar{D}^{*}(2007)^{0} \bar{K}^{*}(892)^{0} )$}{${\cal{B}} ( \bar{B}^{0} \to \bar{D}^{*}(2007)^{0} \bar{K}^{*}(892)^{0} )$}{} & \begin{tabular}{l} Belle \cite{Krokovny:2002ua}: $< 0.40$ \\ \end{tabular} & $< 0.40$ \\
\hline
\btocharmparam{${\cal{B}} ( \bar{B}^{0} \to \bar{D}^{0} K^{-} \pi^{+} )$}{${\cal{B}} ( \bar{B}^{0} \to \bar{D}^{0} K^{-} \pi^{+} )$}{} & \begin{tabular}{l} \babar \cite{Aubert:2005yt}: $< 0.19$ \\ \end{tabular} & $< 0.19$ \\
\hline
\btocharmparam{${\cal{B}} ( \bar{B}^{0} \to \bar{D}^{0} \bar{K}^{*}(892)^{0} )$}{${\cal{B}} ( \bar{B}^{0} \to \bar{D}^{0} \bar{K}^{*}(892)^{0} )$}{} & \begin{tabular}{l} Belle \cite{Krokovny:2002ua}: $< 0.18$ \\ \babar \cite{Aubert:2006qn}: $0.00 \pm 0.05 \pm 0.03$ \\ \end{tabular} & $0.00 \pm 0.06$ \\
\hline
\end{btocharmtab}
\btocharmfig{Bd_D_5}

\begin{btocharmtab}{Bd_D_6}{Branching fractions to a $D^{(*)0}$ meson and more than one kaon or a $\phi$}
\hline
\textbf{Parameter} & \begin{tabular}{l}\textbf{Measurements}\end{tabular} $[10^{-5}]$ & \textbf{Average} $[10^{-5}]$ \\
\hline
\hline
\btocharmparam{${\cal{B}} ( \bar{B}^{0} \to \bar{D}^{0} K^{-} K^{+} )$}{${\cal{B}} ( \bar{B}^{0} \to \bar{D}^{0} K^{-} K^{+} )$}{} & \begin{tabular}{l} LHCb \cite{Aaij:2018rol}: $6.1 \pm 0.4 \pm 0.4$ \\ \end{tabular} & $6.1 \pm 0.6$ \\
\hline
\btocharmparam{${\cal{B}} ( \bar{B}^{0} \to D^{0} \phi(1020) )$}{${\cal{B}} ( \bar{B}^{0} \to D^{0} \phi(1020) )$}{} & \begin{tabular}{l} LHCb \cite{Aaij:2018jqv}: $< 0.20$ \\ \end{tabular} & $< 0.20$ \\
\hline
\end{btocharmtab}
\btocharmfig{Bd_D_6}

\begin{btocharmtab}{Bd_D_7}{Branching fractions to a $D_s^{(*)}$ meson}
\hline
\textbf{Parameter} & \begin{tabular}{l}\textbf{Measurements}\end{tabular} $[10^{-4}]$ & \textbf{Average} $[10^{-4}]$ \\
\hline
\hline
\btocharmparam{${\cal{B}} ( \bar{B}^{0} \to D_s^{-} \pi^{+} )$}{${\cal{B}} ( \bar{B}^{0} \to D_s^{-} \pi^{+} )$}{} & \begin{tabular}{l} Belle \cite{Das:2010be}: $0.199 \pm 0.026 \pm 0.018$ \\ \babar \cite{Aubert:2008zi}: $0.25 \pm 0.04 \pm 0.02$ \\ \end{tabular} & $0.216 \pm 0.026$ \\
\hline
\btocharmparam{${\cal{B}} ( \bar{B}^{0} \to D_s^{*-} \pi^{+} )$}{${\cal{B}} ( \bar{B}^{0} \to D_s^{*-} \pi^{+} )$}{} & \begin{tabular}{l} Belle \cite{Joshi:2009yv}: $0.175 \pm 0.034 \pm 0.020$ \\ \babar \cite{Aubert:2008zi}: $0.26 \,^{+0.05}_{-0.04} \pm 0.02$ \\ \end{tabular} & $0.214 \pm 0.031$ \\
\hline
\btocharmparam{${\cal{B}} ( \bar{B}^{0} \to D_s^{-} \rho(770)^{+} )$}{${\cal{B}} ( \bar{B}^{0} \to D_s^{-} \rho(770)^{+} )$}{} & \begin{tabular}{l} \babar \cite{Aubert:2008zi}: $0.11 \,^{+0.09}_{-0.08} \pm 0.03$ \\ \end{tabular} & $0.11 \,^{+0.09}_{-0.09}$ \\
\hline
\btocharmparam{${\cal{B}} ( \bar{B}^{0} \to D_s^{*-} \rho(770)^{+} )$}{${\cal{B}} ( \bar{B}^{0} \to D_s^{*-} \rho(770)^{+} )$}{} & \begin{tabular}{l} \babar \cite{Aubert:2008zi}: $0.41 \,^{+0.13}_{-0.12} \pm 0.04$ \\ \end{tabular} & $0.41 \,^{+0.14}_{-0.13}$ \\
\hline
\btocharmparam{${\cal{B}} ( \bar{B}^{0} \to D_s^{-} a_0(980)^{+} )$}{${\cal{B}} ( \bar{B}^{0} \to D_s^{-} a_0(980)^{+} )$}{} & \begin{tabular}{l} \babar \cite{Aubert:2005qt}: $0.06 \,^{+0.14}_{-0.11} \pm 0.01$ \\ \end{tabular} & $0.06 \,^{+0.14}_{-0.11}$ \\
\hline
\btocharmparam{${\cal{B}} ( \bar{B}^{0} \to D_s^{*-} a_0(980)^{+} )$}{${\cal{B}} ( \bar{B}^{0} \to D_s^{*-} a_0(980)^{+} )$}{} & \begin{tabular}{l} \babar \cite{Aubert:2005qt}: $0.14 \,^{+0.21}_{-0.16} \pm 0.03$ \\ \end{tabular} & $0.14 \,^{+0.21}_{-0.16}$ \\
\hline
\btocharmparam{${\cal{B}} ( \bar{B}^{0} \to D_s^{-} a_2(1320)^{+} )$}{${\cal{B}} ( \bar{B}^{0} \to D_s^{-} a_2(1320)^{+} )$}{} & \begin{tabular}{l} \babar \cite{Aubert:2005qt}: $0.64 \,^{+1.04}_{-0.57} \pm 0.15$ \\ \end{tabular} & $0.64 \,^{+1.05}_{-0.59}$ \\
\hline
\btocharmparam{${\cal{B}} ( \bar{B}^{0} \to D_s^{*-} a_2(1320)^{+} )$}{${\cal{B}} ( \bar{B}^{0} \to D_s^{*-} a_2(1320)^{+} )$}{} & \begin{tabular}{l} \babar \cite{Aubert:2005qt}: $< 2.0$ \\ \end{tabular} & $< 2.0$ \\
\hline
\btocharmparam{${\cal{B}} ( \bar{B}^{0} \to D_s^{+} K^{-} )$}{${\cal{B}} ( \bar{B}^{0} \to D_s^{+} K^{-} )$}{} & \begin{tabular}{l} Belle \cite{Das:2010be}: $0.191 \pm 0.024 \pm 0.017$ \\ \babar \cite{Aubert:2008zi}: $0.29 \pm 0.04 \pm 0.02$ \\ \end{tabular} & $0.221 \pm 0.025$ \\
\hline
\btocharmparam{${\cal{B}} ( \bar{B}^{0} \to D_s^{*+} K^{-} )$}{${\cal{B}} ( \bar{B}^{0} \to D_s^{*+} K^{-} )$}{} & \begin{tabular}{l} Belle \cite{Joshi:2009yv}: $0.202 \pm 0.033 \pm 0.022$ \\ \babar \cite{Aubert:2008zi}: $0.24 \pm 0.04 \pm 0.02$ \\ \end{tabular} & $0.219 \pm 0.031$ \\
\hline
\btocharmparam{${\cal{B}} ( \bar{B}^{0} \to D_s^{+} K^{*}(892)^{-} )$}{${\cal{B}} ( \bar{B}^{0} \to D_s^{+} K^{*}(892)^{-} )$}{} & \begin{tabular}{l} \babar \cite{Aubert:2008zi}: $0.35 \,^{+0.10}_{-0.09} \pm 0.04$ \\ \end{tabular} & $0.35 \,^{+0.11}_{-0.10}$ \\
\hline
\btocharmparam{${\cal{B}} ( \bar{B}^{0} \to D_s^{*+} K^{*}(892)^{-} )$}{${\cal{B}} ( \bar{B}^{0} \to D_s^{*+} K^{*}(892)^{-} )$}{} & \begin{tabular}{l} \babar \cite{Aubert:2008zi}: $0.32 \,^{+0.14}_{-0.12} \pm 0.04$ \\ \end{tabular} & $0.32 \,^{+0.15}_{-0.13}$ \\
\hline
\btocharmparam{${\cal{B}} ( \bar{B}^{0} \to D_s^{+} K_S^{0} \pi^{-} )$}{${\cal{B}} ( \bar{B}^{0} \to D_s^{+} K_S^{0} \pi^{-} )$}{} & \begin{tabular}{l} \babar \cite{Aubert:2007xma}: $0.55 \pm 0.13 \pm 0.10$ \\ \end{tabular} & $0.55 \pm 0.17$ \\
\hline
\btocharmparam{${\cal{B}} ( \bar{B}^{0} \to D_s^{*+} K^{0} \pi^{-} )$}{${\cal{B}} ( \bar{B}^{0} \to D_s^{*+} K^{0} \pi^{-} )$}{} & \begin{tabular}{l} \babar \cite{Aubert:2007xma}: $< 0.55$ \\ \end{tabular} & $< 0.55$ \\
\hline
\end{btocharmtab}
\btocharmfig{Bd_D_7}

\begin{btocharmtab}{Bd_D_8}{Branching fraction ratios, I}
\hline
\textbf{Parameter} & \begin{tabular}{l}\textbf{Measurements}\end{tabular} & \textbf{Average} \\
\hline
\hline
\btocharmparam{$\frac{{\cal{B}} ( \bar{B}^{0} \to D^{0} \rho(770)^{0} )}{{\cal{B}} ( \bar{B}^{0} \to D^{0} \omega(782) )}$}{${{\cal{B}} ( \bar{B}^{0} \to D^{0} \rho(770)^{0} )}/{{\cal{B}} ( \bar{B}^{0} \to D^{0} \omega(782) )}$}{} & \begin{tabular}{l} Belle \cite{Satpathy:2002js}: $1.6 \pm 0.8$ \\ \end{tabular} & $1.6 \pm 0.8$ \\
\hline
\btocharmparam{$\frac{{\cal{B}} ( \bar{B}^{0} \to D^{+} \pi^{+} \pi^{-} \pi^{-} ) }{ {\cal{B}} ( \bar{B}^{0} \to D^{+} \pi^{-} )}$}{${{\cal{B}} ( \bar{B}^{0} \to D^{+} \pi^{+} \pi^{-} \pi^{-} ) }/{ {\cal{B}} ( \bar{B}^{0} \to D^{+} \pi^{-} )}$}{} & \begin{tabular}{l} LHCb \cite{Aaij:2011rj}: $2.38 \pm 0.11 \pm 0.21$ \\ \end{tabular} & $2.38 \pm 0.24$ \\
\hline
\btocharmparam{$\frac{{\cal{B}} ( \bar{B}^{0} \to D^{*}(2010)^{+} \pi^{-} )}{{\cal{B}} ( \bar{B}^{0} \to D^{+} \pi^{-} )}$}{${{\cal{B}} ( \bar{B}^{0} \to D^{*}(2010)^{+} \pi^{-} )}/{{\cal{B}} ( \bar{B}^{0} \to D^{+} \pi^{-} )}$}{} & \begin{tabular}{l} \babar \cite{Aubert:2006jc}: $0.99 \pm 0.11 \pm 0.08$ \\ \end{tabular} & $0.99 \pm 0.14$ \\
\hline
\btocharmparam{$\frac{{\cal{B}} ( \bar{B}^{0} \to D^{**+} \pi^{-} )}{{\cal{B}} ( \bar{B}^{0} \to D^{+} \pi^{-} )}$}{${{\cal{B}} ( \bar{B}^{0} \to D^{**+} \pi^{-} )}/{{\cal{B}} ( \bar{B}^{0} \to D^{+} \pi^{-} )}$}{} & \begin{tabular}{l} \babar \cite{Aubert:2006jc}: $0.77 \pm 0.22 \pm 0.29$ \\ \end{tabular} & $0.77 \pm 0.36$ \\
\hline
\btocharmparam{$\frac{{\cal{B}} ( \bar{B}^{0} \to D_s^{+} K^{-}  \pi^{+}  \pi^{-} )}{{\cal{B}} ( \bar{B}_s^{0} \to D_s^{+} K^{-}  \pi^{+}  \pi^{-} )}$}{${{\cal{B}} ( \bar{B}^{0} \to D_s^{+} K^{-}  \pi^{+}  \pi^{-} )}/{{\cal{B}} ( \bar{B}_s^{0} \to D_s^{+} K^{-}  \pi^{+}  \pi^{-} )}$}{} & \begin{tabular}{l} LHCb \cite{Aaij:2012mra}: $0.54 \pm 0.07 \pm 0.07$ \\ \end{tabular} & $0.54 \pm 0.10$ \\
\hline
\btocharmparam{$\frac{{\cal{B}} ( \bar{B}^{0} \to D^{*}(2010)^{+} \pi^{+} \pi^{-} \pi^{-} ) }{ {\cal{B}} ( \bar{B}^{0} \to D^{*}(2010)^{+} \pi^{-} )}$}{${{\cal{B}} ( \bar{B}^{0} \to D^{*}(2010)^{+} \pi^{+} \pi^{-} \pi^{-} ) }/{ {\cal{B}} ( \bar{B}^{0} \to D^{*}(2010)^{+} \pi^{-} )}$}{} & \begin{tabular}{l} LHCb \cite{Aaij:2013xca}: $2.64 \pm 0.04 \pm 0.13$ \\ \end{tabular} & $2.64 \pm 0.14$ \\
\hline
\end{btocharmtab}
\btocharmfig{Bd_D_8}

\begin{btocharmtab}{Bd_D_9}{Branching fraction ratios, II}
\hline
\textbf{Parameter} & \begin{tabular}{l}\textbf{Measurements}\end{tabular} & \textbf{Average} \\
\hline
\hline
\btocharmparam{$\frac{ {\cal{B}} ( \bar{B}^{0} \to D^{0} K^{-} \pi^{+} )}  {{\cal{B}} ( \bar{B}^{0} \to D^{0} \pi^{-} \pi^{+} )}$}{${ {\cal{B}} ( \bar{B}^{0} \to D^{0} K^{-} \pi^{+} )}/  {{\cal{B}} ( \bar{B}^{0} \to D^{0} \pi^{-} \pi^{+} )}$}{} & \begin{tabular}{l} LHCb \cite{Aaij:2013pua}: $0.106 \pm 0.007 \pm 0.008$ \\ \end{tabular} & $0.106 \pm 0.011$ \\
\hline
\btocharmparam{$\frac{ {\cal{B}} ( \bar{B}^{0} \to \bar{D}^{0} K^{-} K^{+} )}  {{\cal{B}} ( \bar{B}^{0} \to \bar{D}^{0} \pi^{-} \pi^{+} )}$}{${ {\cal{B}} ( \bar{B}^{0} \to \bar{D}^{0} K^{-} K^{+} )}/  {{\cal{B}} ( \bar{B}^{0} \to \bar{D}^{0} \pi^{-} \pi^{+} )}$}{} & \begin{tabular}{l} LHCb \cite{Aaij:2018rol}: $0.069 \pm 0.004 \pm 0.003$ \\ \end{tabular} & $0.069 \pm 0.005$ \\
\hline
\btocharmparam{$\frac{ {\cal{B}} ( \bar{B}^{0} \to D^{0} \phi(1020) ) }{  {\cal{B}} ( \bar{B}^{0} \to D^{0} \pi^+ \pi^- )}$}{${ {\cal{B}} ( \bar{B}^{0} \to D^{0} \phi(1020) ) }/{  {\cal{B}} ( \bar{B}^{0} \to D^{0} \pi^+ \pi^- )}$}{} & \begin{tabular}{l} LHCb \cite{Aaij:2018jqv}: $0.0012 \pm 0.0007 \pm 0.0003$ \\ \end{tabular} & $0.0012 \pm 0.0008$ \\
\hline
\btocharmparam{$\frac{{\cal{B}} ( \bar{B}^{0} \to D^{+} K^{-} )}{{\cal{B}} ( \bar{B}^{0} \to D^{+} \pi^{-} )}$}{${{\cal{B}} ( \bar{B}^{0} \to D^{+} K^{-} )}/{{\cal{B}} ( \bar{B}^{0} \to D^{+} \pi^{-} )}$}{} & \begin{tabular}{l} LHCb \cite{Aaij:2013qqa}: $0.0822 \pm 0.0011 \pm 0.0025$ \\ Belle \cite{Abe:2001waa}: $0.068 \pm 0.015 \pm 0.007$ \\ \end{tabular} & $0.0818 \pm 0.0027$ \\
\hline
\btocharmparam{$\frac{{\cal{B}} ( \bar{B}^{0} \to D^{+} K^{-} \pi^{+} \pi^{-} ) }{ {\cal{B}} ( \bar{B}^{0} \to D^{+} \pi^{+} \pi^{-} \pi^{-} )}$}{${{\cal{B}} ( \bar{B}^{0} \to D^{+} K^{-} \pi^{+} \pi^{-} ) }/{ {\cal{B}} ( \bar{B}^{0} \to D^{+} \pi^{+} \pi^{-} \pi^{-} )}$}{} & \begin{tabular}{l} LHCb \cite{Aaij:2012bw}: $0.059 \pm 0.011 \pm 0.005$ \\ \end{tabular} & $0.059 \pm 0.012$ \\
\hline
\btocharmparam{$\frac{{\cal{B}} ( \bar{B}^{0} \to D^{*}(2010)^{+} K^{-} )}{{\cal{B}} ( \bar{B}^{0} \to D^{*}(2010)^{+} \pi^{-} )}$}{${{\cal{B}} ( \bar{B}^{0} \to D^{*}(2010)^{+} K^{-} )}/{{\cal{B}} ( \bar{B}^{0} \to D^{*}(2010)^{+} \pi^{-} )}$}{} & \begin{tabular}{l} Belle \cite{Abe:2001waa}: $0.074 \pm 0.015 \pm 0.006$ \\ \babar \cite{Aubert:2005yt}: $0.0776 \pm 0.0034 \pm 0.0029$ \\ \end{tabular} & $0.0773 \pm 0.0043$ \\
\hline
\btocharmparam{$\frac{{\cal{B}} ( \bar{B}^{0} \to D_s^{+} K^{-} )}{{\cal{B}} ( \bar{B}^{0} \to D_s^{+} \pi^{-} )}$}{${{\cal{B}} ( \bar{B}^{0} \to D_s^{+} K^{-} )}/{{\cal{B}} ( \bar{B}^{0} \to D_s^{+} \pi^{-} )}$}{} & \begin{tabular}{l} LHCb \cite{Aaij:2014jpa}: $0.0129 \pm 0.0005 \pm 0.0008$ \\ \end{tabular} & $0.0129 \pm 0.0009$ \\
\hline
\btocharmparaml{$\frac{{\cal{B}} ( \bar{B}^{0} \to D^{*}(2010)^{+} \pi^{+} K^{-} \pi^{-} ) }{ {\cal{B}} ( \bar{B}^{0} \to D^{*}(2010)^{+} \pi^{-} \pi^{-} \pi^{+})}$}{${{\cal{B}} ( \bar{B}^{0} \to D^{*}(2010)^{+} \pi^{+} K^{-} \pi^{-} ) }/{ {\cal{B}} ( \bar{B}^{0} \to D^{*}(2010)^{+} \pi^{-} \pi^{-} \pi^{+})}$}{} & \begin{tabular}{l} LHCb \cite{Aaij:2013xca}: $0.0647 \pm 0.0037 \pm 0.0035$ \\ \end{tabular} & $0.0647 \pm 0.0051$ \\
\hline
\btocharmparaml{$\frac{{\cal{B}} ( \bar{B}^{0} \to D_1(2420)^0 \pi^{+} \pi^{-} ) \times {\cal{B}} (D_1(2420)^0\to D^{*}(2010)^{+} \pi^{-})  }{ {\cal{B}} ( \bar{B}^{0} \to D^{*}(2010)^{+} \pi^{-} \pi^{+} \pi^{-})}$}{$[{{\cal{B}} ( \bar{B}^{0} \to D_1(2420)^0 \pi^{+} \pi^{-} ) \times {\cal{B}} (D_1(2420)^0\to D^{*}(2010)^{+} \pi^{-})  }]/{ {\cal{B}} ( \bar{B}^{0} \to D^{*}(2010)^{+} \pi^{-} \pi^{+} \pi^{-})}$}{} & \begin{tabular}{l} LHCb \cite{Aaij:2013xca}: $0.0204 \pm 0.0042 \pm 0.0022$ \\ \end{tabular} & $0.0204 \pm 0.0047$ \\
\hline
\end{btocharmtab}
\btocharmfig{Bd_D_9}

\begin{btocharmtab}{Bd_D_10}{Product branching fractions to excited $D$ mesons, I}
\hline
\textbf{Parameter} & \begin{tabular}{l}\textbf{Measurements}\end{tabular} $[10^{-4}]$ & \textbf{Average} $[10^{-4}]$ \\
\hline
\hline
\btocharmparaml{${{\cal{B}} ( \bar{B}^{0} \to D_1(2420)^{+} \pi^{-} )\times {\cal{B}} ( D_1(2420)^{+} \to D^{+} \pi^{-} \pi^{+} )}$}{${{\cal{B}} ( \bar{B}^{0} \to D_1(2420)^{+} \pi^{-} )\times {\cal{B}} ( D_1(2420)^{+} \to D^{+} \pi^{-} \pi^{+} )}$}{} & \begin{tabular}{l} Belle \cite{Abe:2004sm}: $0.89 \pm 0.15 \,^{+0.17}_{-0.31}$ \\ \end{tabular} & $0.89 \,^{+0.23}_{-0.34}$ \\
\hline
\btocharmparaml{${{\cal{B}} ( \bar{B}^{0} \to D_1^{0}(H) \omega(782) )\times {\cal{B}} ( D_1^{0}(H) \to D^{*}(2010)^{+} \pi^{-} )}$}{${{\cal{B}} ( \bar{B}^{0} \to D_1^{0}(H) \omega(782) )\times {\cal{B}} ( D_1^{0}(H) \to D^{*}(2010)^{+} \pi^{-} )}$}{} & \begin{tabular}{l} \babar \cite{Aubert:2006zb}: $4.1 \pm 1.2 \pm 1.1$ \\ \end{tabular} & $4.1 \pm 1.6$ \\
\hline
\btocharmparaml{${{\cal{B}} ( \bar{B}^{0} \to D_0^{*}(2400)^{+} \pi^{-} )\times {\cal{B}} ( D_0^{*}(2400)^{+} \to D^{0} \pi^{+} )}$}{${{\cal{B}} ( \bar{B}^{0} \to D_0^{*}(2400)^{+} \pi^{-} )\times {\cal{B}} ( D_0^{*}(2400)^{+} \to D^{0} \pi^{+} )}$}{} & \begin{tabular}{l} Belle \cite{Kuzmin:2006mw}: $0.60 \pm 0.13 \pm 0.27$ \\ \end{tabular} & $0.60 \pm 0.30$ \\
\hline
\btocharmparaml{${{\cal{B}} ( \bar{B}^{0} \to D_2^{*}(2460)^{+} \pi^{-} )\times {\cal{B}} ( D_2^{*}(2460)^{+} \to D^{0} \pi^{+} )}$}{${{\cal{B}} ( \bar{B}^{0} \to D_2^{*}(2460)^{+} \pi^{-} )\times {\cal{B}} ( D_2^{*}(2460)^{+} \to D^{0} \pi^{+} )}$}{} & \begin{tabular}{l} Belle \cite{Kuzmin:2006mw}: $2.15 \pm 0.17 \pm 0.31$ \\ \end{tabular} & $2.15 \pm 0.36$ \\
\hline
\end{btocharmtab}
\btocharmfig{Bd_D_10}

\begin{btocharmtab}{Bd_D_11}{Product branching fractions to excited $D$ mesons, II}
\hline
\textbf{Parameter} & \begin{tabular}{l}\textbf{Measurements}\end{tabular} $[10^{-5}]$ & \textbf{Average} $[10^{-5}]$ \\
\hline
\hline
\btocharmparaml{${{\cal{B}} ( \bar{B}^{0} \to D_1(2420)^{+} \pi^{-} )\times {\cal{B}} ( D_1(2420)^{+} \to D^{*}(2010)^{+} \pi^{-} \pi^{+} )}$}{${{\cal{B}} ( \bar{B}^{0} \to D_1(2420)^{+} \pi^{-} )\times {\cal{B}} ( D_1(2420)^{+} \to D^{*}(2010)^{+} \pi^{-} \pi^{+} )}$}{} & \begin{tabular}{l} Belle \cite{Abe:2004sm}: $< 3.3$ \\ \end{tabular} & $< 3.3$ \\
\hline
\btocharmparaml{${{\cal{B}} ( \bar{B}^{0} \to D_2^{*}(2460)^{+} \pi^{-} )\times {\cal{B}} ( D_2^{*}(2460)^{+} \to D^{*}(2010)^{+} \pi^{-} \pi^{+} )}$}{${{\cal{B}} ( \bar{B}^{0} \to D_2^{*}(2460)^{+} \pi^{-} )\times {\cal{B}} ( D_2^{*}(2460)^{+} \to D^{*}(2010)^{+} \pi^{-} \pi^{+} )}$}{} & \begin{tabular}{l} Belle \cite{Abe:2004sm}: $< 2.4$ \\ \end{tabular} & $< 2.4$ \\
\hline
\btocharmparaml{${{\cal{B}} ( \bar{B}^{0} \to D_2^{*}(2460)^{+} K^{-} )\times {\cal{B}} ( D_2^{*}(2460)^{+} \to D^{0} \pi^{+} )}$}{${{\cal{B}} ( \bar{B}^{0} \to D_2^{*}(2460)^{+} K^{-} )\times {\cal{B}} ( D_2^{*}(2460)^{+} \to D^{0} \pi^{+} )}$}{} & \begin{tabular}{l} \babar \cite{Aubert:2005yt}: $1.83 \pm 0.40 \pm 0.31$ \\ \end{tabular} & $1.83 \pm 0.51$ \\
\hline
\btocharmparaml{${{\cal{B}} ( \bar{B}^{0} \to D_{sJ}(2460)^{-} \pi^{+} )\times {\cal{B}} ( D_{sJ}(2460)^{-} \to D_s^{-} \gamma )}$}{${{\cal{B}} ( \bar{B}^{0} \to D_{sJ}(2460)^{-} \pi^{+} )\times {\cal{B}} ( D_{sJ}(2460)^{-} \to D_s^{-} \gamma )}$}{} & \begin{tabular}{l} Belle \cite{Abe:2004wz}: $< 0.40$ \\ \end{tabular} & $< 0.40$ \\
\hline
\btocharmparaml{${{\cal{B}} ( \bar{B}^{0} \to D_{sJ}^{+}(2460) K^{-} )\times {\cal{B}} ( D_{sJ}^{+}(2460) \to D_s^{+} \gamma )}$}{${{\cal{B}} ( \bar{B}^{0} \to D_{sJ}^{+}(2460) K^{-} )\times {\cal{B}} ( D_{sJ}^{+}(2460) \to D_s^{+} \gamma )}$}{} & \begin{tabular}{l} Belle \cite{Abe:2004wz}: $< 0.94$ \\ \end{tabular} & $< 0.94$ \\
\hline
\btocharmparaml{${{\cal{B}} ( \bar{B}^{0} \to D_{sJ}^{*}(2317)^{-} \pi^{+} )\times {\cal{B}} ( D_{sJ}^{*}(2317)^{-} \to D_s^{-} \pi^{0} )}$}{${{\cal{B}} ( \bar{B}^{0} \to D_{sJ}^{*}(2317)^{-} \pi^{+} )\times {\cal{B}} ( D_{sJ}^{*}(2317)^{-} \to D_s^{-} \pi^{0} )}$}{} & \begin{tabular}{l} Belle \cite{Abe:2004wz}: $< 2.5$ \\ \end{tabular} & $< 2.5$ \\
\hline
\btocharmparaml{${{\cal{B}} ( \bar{B}^{0} \to D_{sJ}^{*}(2317)^{+} K^{-} )\times {\cal{B}} ( D_{sJ}^{*}(2317)^{+} \to D_s^{+} \pi^{0} )}$}{${{\cal{B}} ( \bar{B}^{0} \to D_{sJ}^{*}(2317)^{+} K^{-} )\times {\cal{B}} ( D_{sJ}^{*}(2317)^{+} \to D_s^{+} \pi^{0} )}$}{} & \begin{tabular}{l} Belle \cite{Abe:2004wz}: $5.3 \,^{+1.5}_{-1.3} \pm 1.6$ \\ \end{tabular} & $5.3 \,^{+2.2}_{-2.0}$ \\
\hline
\end{btocharmtab}
\btocharmfig{Bd_D_11}

\begin{btocharmtab}{Bd_D_12}{Branching fractions and ratios to excited $D$ mesons}
\hline
\textbf{Parameter} & \begin{tabular}{l}\textbf{Measurements}\end{tabular} $[10^{-2}]$ & \textbf{Average} $[10^{-2}]$ \\
\hline
\hline
\btocharmparam{${\cal{B}} ( \bar{B}^{0} \to D^{**+} \pi^{-} )$}{${\cal{B}} ( \bar{B}^{0} \to D^{**+} \pi^{-} )$}{} & \begin{tabular}{l} \babar \cite{Aubert:2006jc}: $0.234 \pm 0.065 \pm 0.088$ \\ \end{tabular} & $0.234 \pm 0.109$ \\
\hline
\btocharmparam{$\frac{{\cal{B}} ( \bar{B}^{0} \to D_1(2420)^{+} \pi^{-} ) \times {\cal{B}} ( D_1(2420)^{+} \to D^{+} \pi^{+} \pi^{-} ) }{ {\cal{B}} ( \bar{B}^{0} \to D^{+} \pi^{+} \pi^{-} \pi^{-} )}$}{$[{{\cal{B}} ( \bar{B}^{0} \to D_1(2420)^{+} \pi^{-} ) \times {\cal{B}} ( D_1(2420)^{+} \to D^{+} \pi^{+} \pi^{-} ) }]/{ {\cal{B}} ( \bar{B}^{0} \to D^{+} \pi^{+} \pi^{-} \pi^{-} )}$}{} & \begin{tabular}{l} LHCb \cite{Aaij:2011rj}: $2.1 \pm 0.5 \,^{+0.3}_{-0.5}$ \\ \end{tabular} & $2.1 \,^{+0.6}_{-0.7}$ \\
\hline
\end{btocharmtab}
\btocharmfig{Bd_D_12}

\begin{btocharmtab}{Bd_D_13}{Branching fractions to baryonic decays, I}
\hline
\textbf{Parameter} & \begin{tabular}{l}\textbf{Measurements}\end{tabular} $[10^{-4}]$ & \textbf{Average} $[10^{-4}]$ \\
\hline
\hline
\btocharmparam{${\cal{B}} ( \bar{B}^{0} \to D^{+} p \bar{p} \pi^{-} )$}{${\cal{B}} ( \bar{B}^{0} \to D^{+} p \bar{p} \pi^{-} )$}{} & \begin{tabular}{l} \babar \cite{delAmoSanchez:2011gi}: $3.32 \pm 0.10 \pm 0.29$ \\ \end{tabular} & $3.32 \pm 0.31$ \\
\hline
\btocharmparam{${\cal{B}} ( \bar{B}^{0} \to D^{*}(2010)^{+} p \bar{p} \pi^{-} )$}{${\cal{B}} ( \bar{B}^{0} \to D^{*}(2010)^{+} p \bar{p} \pi^{-} )$}{} & \begin{tabular}{l} \babar \cite{delAmoSanchez:2011gi}: $4.55 \pm 0.16 \pm 0.39$ \\ \end{tabular} & $4.55 \pm 0.42$ \\
\hline
\btocharmparam{${\cal{B}} ( \bar{B}^{0} \to D^{0} p \bar{p} \pi^{-} \pi^{+} )$}{${\cal{B}} ( \bar{B}^{0} \to D^{0} p \bar{p} \pi^{-} \pi^{+} )$}{} & \begin{tabular}{l} \babar \cite{delAmoSanchez:2011gi}: $2.99 \pm 0.21 \pm 0.45$ \\ \end{tabular} & $2.99 \pm 0.50$ \\
\hline
\btocharmparam{${\cal{B}} ( \bar{B}^{0} \to D^{*}(2007)^{0} p \bar{p} \pi^{-} \pi^{+} )$}{${\cal{B}} ( \bar{B}^{0} \to D^{*}(2007)^{0} p \bar{p} \pi^{-} \pi^{+} )$}{} & \begin{tabular}{l} \babar \cite{delAmoSanchez:2011gi}: $1.91 \pm 0.36 \pm 0.29$ \\ \end{tabular} & $1.91 \pm 0.46$ \\
\hline
\end{btocharmtab}
\btocharmfig{Bd_D_13}

\begin{btocharmtab}{Bd_D_14}{Branching fractions to baryonic decays, II}
\hline
\textbf{Parameter} & \begin{tabular}{l}\textbf{Measurements}\end{tabular} $[10^{-5}]$ & \textbf{Average} $[10^{-5}]$ \\
\hline
\hline
\btocharmparam{${\cal{B}} ( \bar{B}^{0} \to D^{0} p \bar{p} )$}{${\cal{B}} ( \bar{B}^{0} \to D^{0} p \bar{p} )$}{} & \begin{tabular}{l} Belle \cite{Abe:2002tw}: $11.8 \pm 1.5 \pm 1.6$ \\ \babar \cite{delAmoSanchez:2011gi}: $10.2 \pm 0.4 \pm 0.6$ \\ \end{tabular} & $10.4 \pm 0.7$ \\
\hline
\btocharmparam{${\cal{B}} ( \bar{B}^{0} \to D^{*}(2007)^{0} p \bar{p} )$}{${\cal{B}} ( \bar{B}^{0} \to D^{*}(2007)^{0} p \bar{p} )$}{} & \begin{tabular}{l} Belle \cite{Abe:2002tw}: $12.0 \,^{+3.3}_{-2.9} \pm 2.1$ \\ \babar \cite{delAmoSanchez:2011gi}: $9.7 \pm 0.7 \pm 0.9$ \\ \end{tabular} & $9.9 \pm 1.1$ \\
\hline
\btocharmparam{${\cal{B}} ( \bar{B}^{0} \to D_s^{+} \Lambda \bar{p} )$}{${\cal{B}} ( \bar{B}^{0} \to D_s^{+} \Lambda \bar{p} )$}{} & \begin{tabular}{l} Belle \cite{Medvedeva:2007af}: $2.9 \pm 0.7 \pm 0.6$ \\ \end{tabular} & $2.9 \pm 0.9$ \\
\hline
\btocharmparam{${\cal{B}} ( \bar{B}^{0} \to D^{0} \Lambda^{0} \bar{\Lambda}^{0} )$}{${\cal{B}} ( \bar{B}^{0} \to D^{0} \Lambda^{0} \bar{\Lambda}^{0} )$}{} & \begin{tabular}{l} Belle \cite{Chang:2008yw}: $1.05 \,^{+0.57}_{-0.44} \pm 0.14$ \\ \babar \cite{Lees:2014mka}: $0.98 \,^{+0.29}_{-0.26} \pm 0.19$ \\ \end{tabular} & $1.00 \pm 0.28$ \\
\hline
\btocharmparam{${\cal{B}}(\bar{B}^{0}\to D^{0}\Sigma^{0}\bar\Lambda+\bar{B}^{0}\to D^{0}\Lambda\bar\Sigma^{0})$}{${\cal{B}}(\bar{B}^{0}\to D^{0}\Sigma^{0}\bar\Lambda+\bar{B}^{0}\to D^{0}\Lambda\bar\Sigma^{0})$}{} & \begin{tabular}{l} \babar \cite{Lees:2014mka}: $1.5 \,^{+0.9}_{-0.8} \pm 0.3$ \\ \end{tabular} & $1.5 \,^{+0.9}_{-0.9}$ \\
\hline
\btocharmparam{${\cal{B}} ( \bar{B}^{0} \to D^{+} \Lambda \bar{p} )$}{${\cal{B}} ( \bar{B}^{0} \to D^{+} \Lambda \bar{p} )$}{} & \begin{tabular}{l} Belle \cite{Chang:2015fja}: $3.36 \pm 0.63 \pm 0.44$ \\ \end{tabular} & $3.36 \pm 0.77$ \\
\hline
\btocharmparam{${\cal{B}} ( \bar{B}^{0} \to D^{*+} \Lambda \bar{p} )$}{${\cal{B}} ( \bar{B}^{0} \to D^{*+} \Lambda \bar{p} )$}{} & \begin{tabular}{l} Belle \cite{Chang:2015fja}: $2.51 \pm 0.26 \pm 0.35$ \\ \end{tabular} & $2.51 \pm 0.44$ \\
\hline
\end{btocharmtab}
\btocharmfig{Bd_D_14}

\subsubsection{Decays to two open charm mesons}
\label{sec:b2c:Bd_DD}
Averages of $\bar{B}^0$ decays to two open charm mesons are shown in Tables~\ref{tab:b2c:Bd_DD_1}--\ref{tab:b2c:Bd_DD_9}.
\begin{btocharmtab}{Bd_DD_1}{Branching fractions to $D^{(*)+}D^{(*)-}$}
\hline
\textbf{Parameter} & \begin{tabular}{l}\textbf{Measurements}\end{tabular} $[10^{-3}]$ & \textbf{Average} $[10^{-3}]$ \\
\hline
\hline
\btocharmparam{${\cal{B}} ( \bar{B}^{0} \to D^{+} D^{-} )$}{${\cal{B}} ( \bar{B}^{0} \to D^{+} D^{-} )$}{} & \begin{tabular}{l} Belle \cite{Rohrken:2012ta}: $0.212 \pm 0.016 \pm 0.018$ \\ \babar \cite{Aubert:2006ia}: $0.28 \pm 0.04 \pm 0.05$ \\ \end{tabular} & $0.220 \pm 0.023$ \\
\hline
\btocharmparaml{${\cal{B}} ( \bar{B}^{0} \to D^{*}(2010)^{-} D^{+} )$}{${\cal{B}} ( \bar{B}^{0} \to D^{*}(2010)^{-} D^{+} )$}{} & \begin{tabular}{l} Belle \cite{Rohrken:2012ta}: $0.614 \pm 0.029 \pm 0.050$ \\ \babar \cite{Aubert:2006ia}: $0.57 \pm 0.07 \pm 0.07$ \\ \end{tabular} & $0.603 \pm 0.050$ \\
\hline
\btocharmparaml{${\cal{B}} ( \bar{B}^{0} \to D^{*}(2010)^{-} D^{*}(2010)^{+} )$}{${\cal{B}} ( \bar{B}^{0} \to D^{*}(2010)^{-} D^{*}(2010)^{+} )$}{} & \begin{tabular}{l} Belle \cite{Kronenbitter:2012ha}: $0.782 \pm 0.038 \pm 0.060$ \\ \babar \cite{Aubert:2006ia}: $0.81 \pm 0.06 \pm 0.10$ \\ \end{tabular} & $0.790 \pm 0.061$ \\
\hline
\btocharmparam{${\cal{B}} ( \bar{B}^{0} \to D^{0} \bar{D}^{0} )$}{${\cal{B}} ( \bar{B}^{0} \to D^{0} \bar{D}^{0} )$}{} & \begin{tabular}{l} Belle \cite{Adachi:2008cj}: $< 0.043$ \\ \babar \cite{Aubert:2006ia}: $< 0.06$ \\ \end{tabular} & $< 0.043$ \\
\hline
\btocharmparam{${\cal{B}} ( \bar{B}^{0} \to D^{0} \bar{D}^{*}(2007)^{0} )$}{${\cal{B}} ( \bar{B}^{0} \to D^{0} \bar{D}^{*}(2007)^{0} )$}{} & \begin{tabular}{l} \babar \cite{Aubert:2006ia}: $< 0.29$ \\ \end{tabular} & $< 0.29$ \\
\hline
\btocharmparam{${\cal{B}} ( \bar{B}^{0} \to D^{*}(2007)^{0} \bar{D}^{*}(2007)^{0} )$}{${\cal{B}} ( \bar{B}^{0} \to D^{*}(2007)^{0} \bar{D}^{*}(2007)^{0} )$}{} & \begin{tabular}{l} \babar \cite{Aubert:2006ia}: $< 0.09$ \\ \end{tabular} & $< 0.09$ \\
\hline
\end{btocharmtab}
\btocharmfig{Bd_DD_1}

\begin{btocharmtab}{Bd_DD_2}{Branching fractions to two $D$ mesons and a kaon}
\hline
\textbf{Parameter} & \begin{tabular}{l}\textbf{Measurements}\end{tabular} $[10^{-3}]$ & \textbf{Average} $[10^{-3}]$ \\
\hline
\hline
\btocharmparam{${\cal{B}} ( \bar{B}^{0} \to D^{*}(2010)^{+} D^{-} \bar{K}^{0} )$}{${\cal{B}} ( \bar{B}^{0} \to D^{*}(2010)^{+} D^{-} \bar{K}^{0} )$}{} & \begin{tabular}{l} \babar \cite{delAmoSanchez:2010pg}: $6.41 \pm 0.36 \pm 0.39$ \\ \end{tabular} & $6.41 \pm 0.53$ \\
\hline
\btocharmparam{${\cal{B}} ( \bar{B}^{0} \to D^{*}(2010)^{-} D^{*}(2010)^{+} \bar{K}^{0} )$}{${\cal{B}} ( \bar{B}^{0} \to D^{*}(2010)^{-} D^{*}(2010)^{+} \bar{K}^{0} )$}{} & \begin{tabular}{l} \babar \cite{delAmoSanchez:2010pg}: $8.26 \pm 0.43 \pm 0.67$ \\ \end{tabular} & $8.26 \pm 0.80$ \\
\hline
\btocharmparam{${\cal{B}} ( \bar{B}^{0} \to D^{*}(2010)^{+} D^{*}(2010)^{-} K_S^{0} )$}{${\cal{B}} ( \bar{B}^{0} \to D^{*}(2010)^{+} D^{*}(2010)^{-} K_S^{0} )$}{} & \begin{tabular}{l} Belle \cite{Dalseno:2007hx}: $3.4 \pm 0.4 \pm 0.7$ \\ \babar \cite{Aubert:2006fh}: $4.4 \pm 0.4 \pm 0.7$ \\ \end{tabular} & $3.9 \pm 0.6$ \\
\hline
\btocharmparam{${\cal{B}} ( \bar{B}^{0} \to D^{*}(2010)^{+} \bar{D}^{0} K^{-} )$}{${\cal{B}} ( \bar{B}^{0} \to D^{*}(2010)^{+} \bar{D}^{0} K^{-} )$}{} & \begin{tabular}{l} \babar \cite{delAmoSanchez:2010pg}: $2.47 \pm 0.10 \pm 0.18$ \\ \end{tabular} & $2.47 \pm 0.21$ \\
\hline
\btocharmparam{${\cal{B}} ( \bar{B}^{0} \to D^{+} \bar{D}^{*}(2007)^{0} K^{-} )$}{${\cal{B}} ( \bar{B}^{0} \to D^{+} \bar{D}^{*}(2007)^{0} K^{-} )$}{} & \begin{tabular}{l} \babar \cite{delAmoSanchez:2010pg}: $3.46 \pm 0.18 \pm 0.37$ \\ \end{tabular} & $3.46 \pm 0.41$ \\
\hline
\btocharmparam{${\cal{B}} ( \bar{B}^{0} \to D^{*}(2010)^{+} \bar{D}^{*}(2007)^{0} K^{-} )$}{${\cal{B}} ( \bar{B}^{0} \to D^{*}(2010)^{+} \bar{D}^{*}(2007)^{0} K^{-} )$}{} & \begin{tabular}{l} \babar \cite{delAmoSanchez:2010pg}: $10.6 \pm 0.3 \pm 0.9$ \\ \end{tabular} & $10.6 \pm 0.9$ \\
\hline
\btocharmparam{${\cal{B}} ( \bar{B}^{0} \to D^{0} \bar{D}^{*}(2007)^{0} \bar{K}^{0} )$}{${\cal{B}} ( \bar{B}^{0} \to D^{0} \bar{D}^{*}(2007)^{0} \bar{K}^{0} )$}{} & \begin{tabular}{l} \babar \cite{delAmoSanchez:2010pg}: $1.08 \pm 0.32 \pm 0.36$ \\ \end{tabular} & $1.08 \pm 0.48$ \\
\hline
\btocharmparam{${\cal{B}} ( \bar{B}^{0} \to D^{*}(2007)^{0} \bar{D}^{*}(2007)^{0} \bar{K}^{0} )$}{${\cal{B}} ( \bar{B}^{0} \to D^{*}(2007)^{0} \bar{D}^{*}(2007)^{0} \bar{K}^{0} )$}{} & \begin{tabular}{l} \babar \cite{delAmoSanchez:2010pg}: $2.40 \pm 0.55 \pm 0.67$ \\ \end{tabular} & $2.40 \pm 0.87$ \\
\hline
\btocharmparam{${\cal{B}} ( \bar{B}^{0} \to D^{+} D^{-} \bar{K}^{0} )$}{${\cal{B}} ( \bar{B}^{0} \to D^{+} D^{-} \bar{K}^{0} )$}{} & \begin{tabular}{l} \babar \cite{delAmoSanchez:2010pg}: $0.75 \pm 0.12 \pm 0.12$ \\ \end{tabular} & $0.75 \pm 0.17$ \\
\hline
\btocharmparam{${\cal{B}} ( \bar{B}^{0} \to D^{+} \bar{D}^{0} K^{-} )$}{${\cal{B}} ( \bar{B}^{0} \to D^{+} \bar{D}^{0} K^{-} )$}{} & \begin{tabular}{l} \babar \cite{delAmoSanchez:2010pg}: $1.07 \pm 0.07 \pm 0.09$ \\ \end{tabular} & $1.07 \pm 0.11$ \\
\hline
\btocharmparam{${\cal{B}} ( \bar{B}^{0} \to D^{0} \bar{D}^{0} \bar{K}^{0} )$}{${\cal{B}} ( \bar{B}^{0} \to D^{0} \bar{D}^{0} \bar{K}^{0} )$}{} & \begin{tabular}{l} \babar \cite{delAmoSanchez:2010pg}: $0.27 \pm 0.10 \pm 0.05$ \\ \end{tabular} & $0.27 \pm 0.11$ \\
\hline
\btocharmparam{${\cal{B}} ( \bar{B}^{0} \to D^{0} \bar{D}^{0} \pi^{0} \bar{K}^{0} )$}{${\cal{B}} ( \bar{B}^{0} \to D^{0} \bar{D}^{0} \pi^{0} \bar{K}^{0} )$}{} & \begin{tabular}{l} Belle \cite{Gokhroo:2006bt}: $0.173 \pm 0.070 \,^{+0.031}_{-0.053}$ \\ \end{tabular} & $0.173 \,^{+0.077}_{-0.088}$ \\
\hline
\end{btocharmtab}
\btocharmfig{Bd_DD_2}

\begin{btocharmtab}{Bd_DD_3}{Branching fractions to $D_s^{(*)-}D^{(*)+}$, I}
\hline
\textbf{Parameter} & \begin{tabular}{l}\textbf{Measurements}\end{tabular} $[10^{-3}]$ & \textbf{Average} $[10^{-3}]$ \\
\hline
\hline
\btocharmparam{${\cal{B}} ( \bar{B}^{0} \to D_s^{-} D^{+} )$}{${\cal{B}} ( \bar{B}^{0} \to D_s^{-} D^{+} )$}{} & \begin{tabular}{l} Belle \cite{Zupanc:2007pu}: $7.5 \pm 0.2 \pm 1.1$ \\ \babar \cite{Aubert:2006nm}: $9.0 \pm 1.8 \pm 1.4$ \\ \end{tabular} & $7.8 \pm 1.0$ \\
\hline
\btocharmparam{${\cal{B}} ( \bar{B}^{0} \to D_s^{-} D^{*}(2010)^{+} )$}{${\cal{B}} ( \bar{B}^{0} \to D_s^{-} D^{*}(2010)^{+} )$}{} & \begin{tabular}{l} \babar \cite{Aubert:2006nm}: $5.7 \pm 1.6 \pm 0.9$ \\ \babar \cite{Aubert:2003jj}: $10.3 \pm 1.4 \pm 2.9$ \\ \end{tabular} & $6.8 \pm 1.6$ \\
\hline
\btocharmparam{${\cal{B}} ( \bar{B}^{0} \to D_s^{*-} D^{*}(2010)^{+} )$}{${\cal{B}} ( \bar{B}^{0} \to D_s^{*-} D^{*}(2010)^{+} )$}{} & \begin{tabular}{l} \babar \cite{Aubert:2006nm}: $16.5 \pm 2.3 \pm 1.9$ \\ \babar \cite{Aubert:2005xu}: $18.8 \pm 0.9 \pm 1.7$ \\ \babar \cite{Aubert:2003jj}: $19.7 \pm 1.5 \pm 5.7$ \\ \end{tabular} & $18.2 \pm 1.6$ \\
\hline
\btocharmparam{${\cal{B}} ( \bar{B}^{0} \to D_s^{*-} D^{+} )$}{${\cal{B}} ( \bar{B}^{0} \to D_s^{*-} D^{+} )$}{} & \begin{tabular}{l} \babar \cite{Aubert:2006nm}: $6.7 \pm 2.0 \pm 1.1$ \\ \end{tabular} & $6.7 \pm 2.3$ \\
\hline
\end{btocharmtab}
\btocharmfig{Bd_DD_3}

\begin{btocharmtab}{Bd_DD_4}{Branching fractions to $D_s^{(*)-}D^{(*)+}$, II}
\hline
\textbf{Parameter} & \begin{tabular}{l}\textbf{Measurements}\end{tabular} $[10^{-4}]$ & \textbf{Average} $[10^{-4}]$ \\
\hline
\hline
\btocharmparaml{${{\cal{B}} ( \bar{B}^{0} \to D_s^{-} D^{+} )\times {\cal{B}} ( D_s^{-} \to \phi(1020) \pi^{-} )}$}{${{\cal{B}} ( \bar{B}^{0} \to D_s^{-} D^{+} )\times {\cal{B}} ( D_s^{-} \to \phi(1020) \pi^{-} )}$}{} & \begin{tabular}{l} \babar \cite{Aubert:2006nm}: $2.67 \pm 0.61 \pm 0.47$ \\ \end{tabular} & $2.67 \pm 0.77$ \\
\hline
\btocharmparaml{${{\cal{B}} ( \bar{B}^{0} \to D_s^{-} D^{*}(2010)^{+} )\times {\cal{B}} ( D_s^{-} \to \phi(1020) \pi^{-} )}$}{${{\cal{B}} ( \bar{B}^{0} \to D_s^{-} D^{*}(2010)^{+} )\times {\cal{B}} ( D_s^{-} \to \phi(1020) \pi^{-} )}$}{} & \begin{tabular}{l} \babar \cite{Aubert:2006nm}: $5.11 \pm 0.94 \pm 0.72$ \\ \end{tabular} & $5.11 \pm 1.18$ \\
\hline
\btocharmparaml{${{\cal{B}} ( \bar{B}^{0} \to D_s^{*-} D^{+} )\times {\cal{B}} ( D_s^{-} \to \phi(1020) \pi^{-} )}$}{${{\cal{B}} ( \bar{B}^{0} \to D_s^{*-} D^{+} )\times {\cal{B}} ( D_s^{-} \to \phi(1020) \pi^{-} )}$}{} & \begin{tabular}{l} \babar \cite{Aubert:2006nm}: $4.14 \pm 1.19 \pm 0.94$ \\ \end{tabular} & $4.14 \pm 1.52$ \\
\hline
\btocharmparaml{${{\cal{B}} ( \bar{B}^{0} \to D_s^{*-} D^{*}(2010)^{+} )\times {\cal{B}} ( D_s^{-} \to \phi(1020) \pi^{-} )}$}{${{\cal{B}} ( \bar{B}^{0} \to D_s^{*-} D^{*}(2010)^{+} )\times {\cal{B}} ( D_s^{-} \to \phi(1020) \pi^{-} )}$}{} & \begin{tabular}{l} \babar \cite{Aubert:2006nm}: $12.2 \pm 2.2 \pm 2.2$ \\ \end{tabular} & $12.2 \pm 3.1$ \\
\hline
\end{btocharmtab}
\btocharmfig{Bd_DD_4}

\begin{btocharmtab}{Bd_DD_5}{Branching fractions to $D_s^{(*)+}D_s^{(*)-}$}
\hline
\textbf{Parameter} & \begin{tabular}{l}\textbf{Measurements}\end{tabular} $[10^{-3}]$ & \textbf{Average} $[10^{-3}]$ \\
\hline
\hline
\btocharmparam{${\cal{B}} ( \bar{B}^{0} \to D_s^{-} D_s^{+} )$}{${\cal{B}} ( \bar{B}^{0} \to D_s^{-} D_s^{+} )$}{} & \begin{tabular}{l} Belle \cite{Zupanc:2007pu}: $< 0.036$ \\ \babar \cite{Aubert:2005jv}: $< 0.10$ \\ \end{tabular} & $< 0.036$ \\
\hline
\btocharmparam{${\cal{B}} ( \bar{B}^{0} \to D_s^{-} D_s^{*+} )$}{${\cal{B}} ( \bar{B}^{0} \to D_s^{-} D_s^{*+} )$}{} & \begin{tabular}{l} \babar \cite{Aubert:2005jv}: $< 0.13$ \\ \end{tabular} & $< 0.13$ \\
\hline
\btocharmparam{${\cal{B}} ( \bar{B}^{0} \to D_s^{*+} D_s^{*-} )$}{${\cal{B}} ( \bar{B}^{0} \to D_s^{*+} D_s^{*-} )$}{} & \begin{tabular}{l} \babar \cite{Aubert:2005jv}: $< 0.24$ \\ \end{tabular} & $< 0.24$ \\
\hline
\end{btocharmtab}
\btocharmfig{Bd_DD_5}

\begin{btocharmtab}{Bd_DD_6}{Branching fraction ratios}
\hline
\textbf{Parameter} & \begin{tabular}{l}\textbf{Measurements}\end{tabular} $[10^{-3}]$ & \textbf{Average} $[10^{-3}]$ \\
\hline
\hline
\btocharmparam{$\frac{{\cal{B}} ( \bar{B}^{0} \to D^{0} \bar{D}^{0} )}{ {\cal{B}} ( \bar{B}^{-} \to D^{0} \bar{D}_s^{-} )}$}{${{\cal{B}} ( \bar{B}^{0} \to D^{0} \bar{D}^{0} )}/{ {\cal{B}} ( \bar{B}^{-} \to D^{0} \bar{D}_s^{-} )}$}{} & \begin{tabular}{l} LHCb \cite{Aaij:2013fha}: $1.4 \pm 0.6 \pm 0.2$ \\ \end{tabular} & $1.4 \pm 0.6$ \\
\hline
\end{btocharmtab}

\begin{btocharmtab}{Bd_DD_7}{Branching fractions to excited $D_s$ mesons}
\hline
\textbf{Parameter} & \begin{tabular}{l}\textbf{Measurements}\end{tabular} $[10^{-3}]$ & \textbf{Average} $[10^{-3}]$ \\
\hline
\hline
\btocharmparam{${\cal{B}} ( \bar{B}^{0} \to D_{sJ}(2460)^{-} D^{+} )$}{${\cal{B}} ( \bar{B}^{0} \to D_{sJ}(2460)^{-} D^{+} )$}{} & \begin{tabular}{l} \babar \cite{Aubert:2006nm}: $2.6 \pm 1.5 \pm 0.7$ \\ \end{tabular} & $2.6 \pm 1.7$ \\
\hline
\btocharmparam{${\cal{B}} ( \bar{B}^{0} \to D_{sJ}(2460)^{-} D^{*}(2010)^{+} )$}{${\cal{B}} ( \bar{B}^{0} \to D_{sJ}(2460)^{-} D^{*}(2010)^{+} )$}{} & \begin{tabular}{l} \babar \cite{Aubert:2006nm}: $8.8 \pm 2.0 \pm 1.4$ \\ \end{tabular} & $8.8 \pm 2.4$ \\
\hline
\btocharmparam{${\cal{B}} ( \bar{B}^{0} \to D_{s1}(2536)^{-} D^{*}(2010)^{+} )$}{${\cal{B}} ( \bar{B}^{0} \to D_{s1}(2536)^{-} D^{*}(2010)^{+} )$}{} & \begin{tabular}{l} \babar \cite{Aubert:2006fh}: $92 \pm 24 \pm 1$ \\ \end{tabular} & $92 \pm 24$ \\
\hline
\end{btocharmtab}
\btocharmfig{Bd_DD_7}

\begin{btocharmtab}{Bd_DD_8}{Product branching fractions to excited $D_s$ mesons, I}
\hline
\textbf{Parameter} & \begin{tabular}{l}\textbf{Measurements}\end{tabular} $[10^{-3}]$ & \textbf{Average} $[10^{-3}]$ \\
\hline
\hline
\btocharmparaml{${{\cal{B}} ( \bar{B}^{0} \to D^{+} D_{sJ}(2460)^{-} )\times {\cal{B}} ( D_{sJ}(2460)^{-} \to D_s^{-} \gamma )}$}{${{\cal{B}} ( \bar{B}^{0} \to D^{+} D_{sJ}(2460)^{-} )\times {\cal{B}} ( D_{sJ}(2460)^{-} \to D_s^{-} \gamma )}$}{} & \begin{tabular}{l} Belle \cite{Krokovny:2003zq}: $0.82 \,^{+0.22}_{-0.19} \pm 0.25$ \\ \babar \cite{Aubert:2004pw}: $0.8 \pm 0.2 \,^{+0.3}_{-0.2}$ \\ \end{tabular} & $0.81 \pm 0.23$ \\
\hline
\btocharmparaml{${{\cal{B}} ( \bar{B}^{0} \to D^{+} D_{sJ}(2460)^{-} )\times {\cal{B}} ( D_{sJ}(2460)^{-} \to D_s^{*-} \pi^{0} )}$}{${{\cal{B}} ( \bar{B}^{0} \to D^{+} D_{sJ}(2460)^{-} )\times {\cal{B}} ( D_{sJ}(2460)^{-} \to D_s^{*-} \pi^{0} )}$}{} & \begin{tabular}{l} Belle \cite{Krokovny:2003zq}: $2.27 \,^{+0.73}_{-0.62} \pm 0.68$ \\ \babar \cite{Aubert:2004pw}: $2.8 \pm 0.8 \,^{+1.1}_{-0.8}$ \\ \end{tabular} & $2.48 \pm 0.75$ \\
\hline
\btocharmparaml{${{\cal{B}} ( \bar{B}^{0} \to D_{sJ}(2460)^{-} D^{*}(2010)^{+} )\times {\cal{B}} ( D_{sJ}(2460)^{-} \to D_s^{*-} \pi^{0} )}$}{${{\cal{B}} ( \bar{B}^{0} \to D_{sJ}(2460)^{-} D^{*}(2010)^{+} )\times {\cal{B}} ( D_{sJ}(2460)^{-} \to D_s^{*-} \pi^{0} )}$}{} & \begin{tabular}{l} \babar \cite{Aubert:2004pw}: $5.5 \pm 1.2 \,^{+2.1}_{-1.6}$ \\ \end{tabular} & $5.5 \,^{+2.5}_{-2.0}$ \\
\hline
\btocharmparaml{${{\cal{B}} ( \bar{B}^{0} \to D_{sJ}(2460)^{-} D^{*}(2010)^{+} )\times {\cal{B}} ( D_{sJ}(2460)^{-} \to D_s^{-} \gamma )}$}{${{\cal{B}} ( \bar{B}^{0} \to D_{sJ}(2460)^{-} D^{*}(2010)^{+} )\times {\cal{B}} ( D_{sJ}(2460)^{-} \to D_s^{-} \gamma )}$}{} & \begin{tabular}{l} \babar \cite{Aubert:2004pw}: $2.3 \pm 0.3 \,^{+0.9}_{-0.6}$ \\ \end{tabular} & $2.3 \,^{+0.9}_{-0.7}$ \\
\hline
\btocharmparaml{${{\cal{B}} ( \bar{B}^{0} \to D_{sJ}^{*}(2317)^{-} D^{*}(2010)^{+} )\times {\cal{B}} ( D_{sJ}^{*}(2317)^{-} \to D_s^{-} \pi^{0} )}$}{${{\cal{B}} ( \bar{B}^{0} \to D_{sJ}^{*}(2317)^{-} D^{*}(2010)^{+} )\times {\cal{B}} ( D_{sJ}^{*}(2317)^{-} \to D_s^{-} \pi^{0} )}$}{} & \begin{tabular}{l} \babar \cite{Aubert:2004pw}: $1.5 \pm 0.4 \,^{+0.5}_{-0.4}$ \\ \end{tabular} & $1.5 \,^{+0.7}_{-0.5}$ \\
\hline
\btocharmparaml{${{\cal{B}} ( \bar{B}^{0} \to D^{+} D_{sJ}^{*}(2317)^{-} )\times {\cal{B}} ( D_{sJ}^{*}(2317)^{-} \to D_s^{-} \pi^{0} )}$}{${{\cal{B}} ( \bar{B}^{0} \to D^{+} D_{sJ}^{*}(2317)^{-} )\times {\cal{B}} ( D_{sJ}^{*}(2317)^{-} \to D_s^{-} \pi^{0} )}$}{} & \begin{tabular}{l} Belle \cite{Choi:2015lpc}: $1.02 \,^{+0.13}_{-0.12} \pm 0.11$ \\ \babar \cite{Aubert:2004pw}: $1.8 \pm 0.4 \,^{+0.7}_{-0.5}$ \\ \end{tabular} & $1.05 \pm 0.16$ \\
\hline
\btocharmparaml{${{\cal{B}} ( \bar{B}^{0} \to D^{+} D_{sJ}^{*}(2317)^{-} )\times {\cal{B}} ( D_{sJ}^{*}(2317)^{-} \to D_s^{*-} \gamma )}$}{${{\cal{B}} ( \bar{B}^{0} \to D^{+} D_{sJ}^{*}(2317)^{-} )\times {\cal{B}} ( D_{sJ}^{*}(2317)^{-} \to D_s^{*-} \gamma )}$}{} & \begin{tabular}{l} Belle \cite{Krokovny:2003zq}: $< 0.95$ \\ \end{tabular} & $< 0.95$ \\
\hline
\end{btocharmtab}
\btocharmfig{Bd_DD_8}

\begin{btocharmtab}{Bd_DD_9}{Product branching fractions to excited $D_s$ mesons, II}
\hline
\textbf{Parameter} & \begin{tabular}{l}\textbf{Measurements}\end{tabular} $[10^{-4}]$ & \textbf{Average} $[10^{-4}]$ \\
\hline
\hline
\btocharmparaml{${{\cal{B}} ( \bar{B}^{0} \to D_{s1}(2536)^{-} D^{+} )\times {\cal{B}} ( D_{s1}(2536)^{-} \to D^{*}(2010)^{-} \bar{K}^{0} )}$}{${{\cal{B}} ( \bar{B}^{0} \to D_{s1}(2536)^{-} D^{+} )\times {\cal{B}} ( D_{s1}(2536)^{-} \to D^{*}(2010)^{-} \bar{K}^{0} )}$}{} & \begin{tabular}{l} \babar \cite{Aubert:2007rva}: $2.61 \pm 1.03 \pm 0.31$ \\ \end{tabular} & $2.61 \pm 1.08$ \\
\hline
\btocharmparaml{${{\cal{B}} ( \bar{B}^{0} \to D_{s1}(2536)^{-} D^{+} )\times {\cal{B}} ( D_{s1}(2536)^{-} \to K^{-} \bar{D}^{*}(2007)^{0} )}$}{${{\cal{B}} ( \bar{B}^{0} \to D_{s1}(2536)^{-} D^{+} )\times {\cal{B}} ( D_{s1}(2536)^{-} \to K^{-} \bar{D}^{*}(2007)^{0} )}$}{} & \begin{tabular}{l} \babar \cite{Aubert:2007rva}: $1.71 \pm 0.48 \pm 0.32$ \\ \end{tabular} & $1.71 \pm 0.58$ \\
\hline
\btocharmparaml{${{\cal{B}} ( \bar{B}^{0} \to D_{s1}(2536)^{-} D^{*}(2010)^{+} )\times {\cal{B}} ( D_{s1}(2536)^{-} \to D^{*}(2010)^{-} \bar{K}^{0} )}$}{${{\cal{B}} ( \bar{B}^{0} \to D_{s1}(2536)^{-} D^{*}(2010)^{+} )\times {\cal{B}} ( D_{s1}(2536)^{-} \to D^{*}(2010)^{-} \bar{K}^{0} )}$}{} & \begin{tabular}{l} \babar \cite{Aubert:2007rva}: $5.00 \pm 1.51 \pm 0.67$ \\ \end{tabular} & $5.00 \pm 1.65$ \\
\hline
\btocharmparaml{${{\cal{B}} ( \bar{B}^{0} \to D_{s1}(2536)^{-} D^{*}(2010)^{+} )\times {\cal{B}} ( D_{s1}(2536)^{-} \to \bar{D}^{*}(2007)^{0} K^{+} )}$}{${{\cal{B}} ( \bar{B}^{0} \to D_{s1}(2536)^{-} D^{*}(2010)^{+} )\times {\cal{B}} ( D_{s1}(2536)^{-} \to \bar{D}^{*}(2007)^{0} K^{+} )}$}{} & \begin{tabular}{l} \babar \cite{Aubert:2007rva}: $3.32 \pm 0.88 \pm 0.66$ \\ \end{tabular} & $3.32 \pm 1.10$ \\
\hline
\btocharmparaml{${{\cal{B}} ( \bar{B}^{0} \to D_{s1}(2536)^{+} D^{-} )\times {\cal{B}} ( D_{s1}(2536)^{+} \to D^{*}(2007)^{0} K^{+} + D^{*}(2010)^{+} K^{0} )}$}{${{\cal{B}} ( \bar{B}^{0} \to D_{s1}(2536)^{+} D^{-} )\times {\cal{B}} ( D_{s1}(2536)^{+} \to D^{*}(2007)^{0} K^{+} + D^{*}(2010)^{+} K^{0} )}$}{} & \begin{tabular}{l} Belle \cite{Belle:2011ad}: $2.75 \pm 0.62 \pm 0.36$ \\ \end{tabular} & $2.75 \pm 0.72$ \\
\hline
\btocharmparaml{${{\cal{B}} ( \bar{B}^{0} \to D_{s1}(2536)^{+} D^{*}(2010)^{-} )\times {\cal{B}} ( D_{s1}(2536)^{+} \to D^{*}(2007)^{0} K^{+} + D^{*}(2010)^{+} K^{0} )}$}{${{\cal{B}} ( \bar{B}^{0} \to D_{s1}(2536)^{+} D^{*}(2010)^{-} )\times {\cal{B}} ( D_{s1}(2536)^{+} \to D^{*}(2007)^{0} K^{+} + D^{*}(2010)^{+} K^{0} )}$}{} & \begin{tabular}{l} Belle \cite{Belle:2011ad}: $5.01 \pm 1.21 \pm 0.70$ \\ \end{tabular} & $5.01 \pm 1.40$ \\
\hline
\btocharmparaml{${{\cal{B}} ( \bar{B}^{0} \to D_{s1}(2536)^{+} D^{*}(2010)^{-} )\times {\cal{B}} ( D_{s1}(2536)^{+} \to D^{*}(2010)^{+} K_S^{0} )}$}{${{\cal{B}} ( \bar{B}^{0} \to D_{s1}(2536)^{+} D^{*}(2010)^{-} )\times {\cal{B}} ( D_{s1}(2536)^{+} \to D^{*}(2010)^{+} K_S^{0} )}$}{} & \begin{tabular}{l} Belle \cite{Dalseno:2007hx}: $< 6.0$ \\ \end{tabular} & $< 6.0$ \\
\hline
\btocharmparaml{${{\cal{B}} ( \bar{B}^{0} \to D^{+} D_{sJ}(2460)^{-} )\times {\cal{B}} ( D_{sJ}(2460)^{-} \to D_s^{-} \pi^{+} \pi^{-} )}$}{${{\cal{B}} ( \bar{B}^{0} \to D^{+} D_{sJ}(2460)^{-} )\times {\cal{B}} ( D_{sJ}(2460)^{-} \to D_s^{-} \pi^{+} \pi^{-} )}$}{} & \begin{tabular}{l} Belle \cite{Krokovny:2003zq}: $< 2.0$ \\ \end{tabular} & $< 2.0$ \\
\hline
\btocharmparaml{${{\cal{B}} ( \bar{B}^{0} \to D^{+} D_{sJ}(2460)^{-} )\times {\cal{B}} ( D_{sJ}(2460)^{-} \to D_s^{-} \pi^{0} )}$}{${{\cal{B}} ( \bar{B}^{0} \to D^{+} D_{sJ}(2460)^{-} )\times {\cal{B}} ( D_{sJ}(2460)^{-} \to D_s^{-} \pi^{0} )}$}{} & \begin{tabular}{l} Belle \cite{Krokovny:2003zq}: $< 3.6$ \\ \end{tabular} & $< 3.6$ \\
\hline
\btocharmparaml{${{\cal{B}} ( \bar{B}^{0} \to D^{+} D_{sJ}(2460)^{-} )\times {\cal{B}} ( D_{sJ}(2460)^{-} \to D_s^{*-} \gamma )}$}{${{\cal{B}} ( \bar{B}^{0} \to D^{+} D_{sJ}(2460)^{-} )\times {\cal{B}} ( D_{sJ}(2460)^{-} \to D_s^{*-} \gamma )}$}{} & \begin{tabular}{l} Belle \cite{Krokovny:2003zq}: $< 6.0$ \\ \end{tabular} & $< 6.0$ \\
\hline
\end{btocharmtab}
\btocharmfig{Bd_DD_9}

\subsubsection{Decays to charmonium states}
\label{sec:b2c:Bd_cc}
Averages of $\bar{B}^0$ decays to charmonium states are shown in Tables~\ref{tab:b2c:Bd_cc_1}--\ref{tab:b2c:Bd_cc_8}.
\begin{btocharmtab}{Bd_cc_1}{Branching fractions to $J/\psi$ and one kaon}
\hline
\textbf{Parameter} & \begin{tabular}{l}\textbf{Measurements}\end{tabular} $[10^{-3}]$ & \textbf{Average} $[10^{-3}]$ \\
\hline
\hline
\btocharmparam{${\cal{B}} ( \bar{B}^{0} \to J/\psi \bar{K}^{0} )$}{${\cal{B}} ( \bar{B}^{0} \to J/\psi \bar{K}^{0} )$}{} & \begin{tabular}{l} CDF \cite{Abe:1995aw}: $1.15 \pm 0.23 \pm 0.17$ \\ Belle \cite{Abe:2002rc}: $0.79 \pm 0.04 \pm 0.09$ \\ \babar \cite{Aubert:2004rz}: $0.869 \pm 0.022 \pm 0.030$ \\ \end{tabular} & $0.863 \pm 0.035$ \\
\hline
\btocharmparam{${\cal{B}} ( \bar{B}^{0} \to J/\psi K^{-} \pi^{+}  )$}{${\cal{B}} ( \bar{B}^{0} \to J/\psi K^{-} \pi^{+}  )$}{} & \begin{tabular}{l} Belle \cite{Chilikin:2014bkk}: $1.15 \pm 0.01 \pm 0.05$ \\ \end{tabular} & $1.15 \pm 0.05$ \\
\hline
\btocharmparam{${\cal{B}} ( \bar{B}^{0} \to J/\psi \bar{K}^{*}(892)^{0} )$}{${\cal{B}} ( \bar{B}^{0} \to J/\psi \bar{K}^{*}(892)^{0} )$}{} & \begin{tabular}{l} CDF \cite{Abe:1998yu}: $1.74 \pm 0.20 \pm 0.18$ \\ Belle \cite{Chilikin:2014bkk}: $1.19 \pm 0.01 \pm 0.08$ \\ \babar \cite{Aubert:2004rz}: $1.309 \pm 0.026 \pm 0.077$ \\ \end{tabular} & $1.270 \pm 0.056$ \\
\hline
\btocharmparaml{${\cal{B}} ( \bar{B}^{0} \to J/\psi \bar{K}^{0} \pi^{+} \pi^{-} )$}{${\cal{B}} ( \bar{B}^{0} \to J/\psi \bar{K}^{0} \pi^{+} \pi^{-} )$}{} & \begin{tabular}{l} LHCb \cite{Aaij:2014naa}: $0.430 \pm 0.030 \pm 0.037$ \\ CDF \cite{Affolder:2001qi}: $1.03 \pm 0.33 \pm 0.15$ \\ \end{tabular} & $0.440 \pm 0.047$ \\
\hline
\btocharmparam{${\cal{B}} ( \bar{B}^{0} \to J/\psi \bar{K}^{0} \rho(770)^{0} )$}{${\cal{B}} ( \bar{B}^{0} \to J/\psi \bar{K}^{0} \rho(770)^{0} )$}{} & \begin{tabular}{l} CDF \cite{Affolder:2001qi}: $0.54 \pm 0.29 \pm 0.09$ \\ \end{tabular} & $0.54 \pm 0.30$ \\
\hline
\btocharmparam{${\cal{B}} ( \bar{B}^{0} \to J/\psi K^{*}(892)^{-} \pi^{+} )$}{${\cal{B}} ( \bar{B}^{0} \to J/\psi K^{*}(892)^{-} \pi^{+} )$}{} & \begin{tabular}{l} CDF \cite{Affolder:2001qi}: $0.77 \pm 0.41 \pm 0.13$ \\ \end{tabular} & $0.77 \pm 0.43$ \\
\hline
\btocharmparam{${\cal{B}} ( \bar{B}^{0} \to J/\psi \omega(782) \bar{K}^{0} )$}{${\cal{B}} ( \bar{B}^{0} \to J/\psi \omega(782) \bar{K}^{0} )$}{} & \begin{tabular}{l} \babar \cite{delAmoSanchez:2010jr}: $0.23 \pm 0.03 \pm 0.03$ \\ \end{tabular} & $0.23 \pm 0.04$ \\
\hline
\btocharmparam{${\cal{B}} ( \bar{B}^{0} \to J/\psi \phi(1020) \bar{K}^{0} )$}{${\cal{B}} ( \bar{B}^{0} \to J/\psi \phi(1020) \bar{K}^{0} )$}{} & \begin{tabular}{l} \babar \cite{Aubert:2003ii}: $0.102 \pm 0.038 \pm 0.010$ \\ \end{tabular}%
 & $0.66 \pm 0.22$ \\
\hline
\end{btocharmtab}
\btocharmfig{Bd_cc_1}

\begin{btocharmtab}{Bd_cc_2}{Branching fractions to charmonium other than $J/\psi$ and one kaon, I}
\hline
\textbf{Parameter} & \begin{tabular}{l}\textbf{Measurements}\end{tabular}%
 & $0.573 \pm 0.071$ \\
\hline
\end{btocharmtab}
\btocharmfig{Bd_cc_2}

\begin{btocharmtab}{Bd_cc_3}{Branching fractions to charmonium other than $J/\psi$ and one kaon, II}
\hline
\textbf{Parameter} & \begin{tabular}{l}\textbf{Measurements}\end{tabular}%
 & $0.72 \pm 0.10$ \\
\hline
\end{btocharmtab}
\btocharmfig{Bd_cc_3}

\begin{btocharmtab}{Bd_cc_4}{Branching fractions to charmonium and light mesons}
\hline
\textbf{Parameter} & \begin{tabular}{l}\textbf{Measurements}\end{tabular}%
 & $< 0.019$ \\
\hline
\end{btocharmtab}
\btocharmfig{Bd_cc_4}

\begin{btocharmtab}{Bd_cc_5}{Branching fractions to  $J/\psi$ and photons, baryons, or heavy mesons}
\hline
\textbf{Parameter} & %
 & $< 1.3$ \\
\hline
\end{btocharmtab}
\btocharmfig{Bd_cc_5}

\begin{btocharmtab}{Bd_cc_6}{Branching fraction ratios, I}
\hline
\textbf{Parameter} & %
 & $0.357 \pm 0.017$ \\
\hline
\end{btocharmtab}
\btocharmfig{Bd_cc_6}

\begin{btocharmtab}{Bd_cc_7}{Branching fraction ratios, II}
\hline
\textbf{Parameter} & %
 & $< 0.26$ \\
\hline
\end{btocharmtab}
\btocharmfig{Bd_cc_8}

\subsubsection{Decays to charm baryons}
\label{sec:b2c:Bd_baryon}
Averages of $\bar{B}^0$ decays to charm baryons are shown in Tables~\ref{tab:b2c:Bd_baryon_1}--\ref{tab:b2c:Bd_baryon_4}.
\begin{btocharmtab}{Bd_baryon_1}{Branching fractions, I}
\hline
\textbf{Parameter} & %
 & $7.9 \pm 2.1$ \\
\hline
\end{btocharmtab}
\btocharmfig{Bd_baryon_1}

\begin{btocharmtab}{Bd_baryon_2}{Branching fractions, II}
\hline
\textbf{Parameter} & %
 & $4.33 \pm 1.43$ \\
\hline
\end{btocharmtab}
\btocharmfig{Bd_baryon_2}

\begin{btocharmtab}{Bd_baryon_3}{Branching fractions, III}
\hline
\textbf{Parameter} & %
 & $< 0.14$ \\
\hline
\end{btocharmtab}
\btocharmfig{Bd_baryon_3}

\begin{btocharmtab}{Bd_baryon_4}{Branching fraction ratios}
\hline
\textbf{Parameter} & %
 & $< 2$ \\
\hline
\end{btocharmtab}

\subsubsection{Decays to $XYZP$ states}
\label{sec:b2c:Bd_other}
New charmonium-like states that are not clearly identified as charmonium with specific quantum numbers are often labeled by $X$, $Y$, or $Z$.
Averages of $\bar{B}^0$ decays to %
such 
states are shown in Tables~\ref{tab:b2c:Bd_other_1}--\ref{tab:b2c:Bd_other_4}.
\begin{btocharmtab}{Bd_other_1}{Branching fractions to $X(3872)$}
\hline
\textbf{Parameter} & %
 & $< 43.7$ \\
\hline
\end{btocharmtab}
\btocharmfig{Bd_other_1}

\begin{btocharmtab}{Bd_other_2}{Branching fractions to neutral states other than $X(3872)$}
\hline
\textbf{Parameter} & %
 & $0.40 \,^{+1.98}_{-0.10}$ \\
\hline
\end{btocharmtab}
\btocharmfig{Bd_other_2}

\begin{btocharmtab}{Bd_other_3}{Branching fractions to charged states}
\hline
\textbf{Parameter} & %
 & $2.2 \,^{+1.3}_{-0.8}$ \\
\hline
\end{btocharmtab}
\btocharmfig{Bd_other_3}

\begin{btocharmtab}{Bd_other_4}{Branching fraction ratios}
\hline
\textbf{Parameter} & %
 & $0.7 \,^{+0.4}_{-0.3}$ \\
\hline
\end{btocharmtab}
\btocharmfig{Bd_other_4}

\subsection{Decays of $B^-$ mesons}
\label{sec:b2c:Bu}
Measurements of $B^-$ decays to charmed hadrons are summarized in Sections~\ref{sec:b2c:Bu_D} to~\ref{sec:b2c:Bu_other}.

\subsubsection{Decays to a single open charm meson}
\label{sec:b2c:Bu_D}
Averages of $B^-$ decays to a single open charm meson are shown in Tables~\ref{tab:b2c:Bu_D_1}--\ref{tab:b2c:Bu_D_15}.
In this section $D^{**}$ refers to the sum of all the non-strange charm meson
states with masses in the range $2.2-2.8~\gevcc$.
\begin{btocharmtab}{Bu_D_1}{Branching fractions to a $D^{(*)}$ meson and one or more pions}
\hline
\textbf{Parameter} & 
 & $5.67 \pm 1.25$ \\
\hline
\end{btocharmtab}
\btocharmfig{Bu_D_1}

\begin{btocharmtab}{Bu_D_2}{Branching fractions to a $D^{(*)0}$ meson and one or more kaons}
\hline
\textbf{Parameter} & %
 & $0.0731 \pm 0.0049$ \\
\hline
\end{btocharmtab}
\btocharmfig{Bu_D_2}

\begin{btocharmtab}{Bu_D_3}{Branching fractions to a $D^{(*)-}$ meson and a neutral kaon or a kaon and a pion}
\hline
\textbf{Parameter} & %
 & $< 9$ \\
\hline
\end{btocharmtab}
\btocharmfig{Bu_D_3}

\begin{btocharmtab}{Bu_D_4}{Branching fraction ratios to $D^{0}$ mesons, I}
\hline
\textbf{Parameter} & %
 & $1.27 \pm 0.13$ \\
\hline
\end{btocharmtab}
\btocharmfig{Bu_D_4}

\begin{btocharmtab}{Bu_D_5}{Branching fraction ratios to $D^{0}$ mesons, II}
\hline
\textbf{Parameter} & %
 & $9.4 \pm 1.6$ \\
\hline
\end{btocharmtab}
\btocharmfig{Bu_D_5}

\begin{btocharmtab}{Bu_D_6}{Branching fractions to excited $D$ mesons}
\hline
\textbf{Parameter} & %
 & $5.50 \pm 1.16$ \\
\hline
\end{btocharmtab}

\begin{btocharmtab}{Bu_D_7}{Product branching fractions to excited $D$ mesons, I}
\hline
\textbf{Parameter} & %
 & $0.35 \pm 0.05$ \\
\hline
\end{btocharmtab}
\btocharmfig{Bu_D_7}

\begin{btocharmtab}{Bu_D_8}{Product branching fractions to excited $D$ mesons, II}
\hline
\textbf{Parameter} & %
 & $< 2.2$ \\
\hline
\end{btocharmtab}
\btocharmfig{Bu_D_8}

\begin{btocharmtab}{Bu_D_9}{Branching fraction ratios to excited $D$ mesons}
\hline
\textbf{Parameter} & %
 & $0.0639 \pm 0.0055$ \\
\hline
\end{btocharmtab}
\btocharmfig{Bu_D_9}

\begin{btocharmtab}{Bu_D_10}{Relative product branching fractions to excited $D$ mesons}
\hline
\textbf{Parameter} & %
 & $0.014 \pm 0.006$ \\
\hline
\end{btocharmtab}
\btocharmfig{Bu_D_10}

\begin{btocharmtab}{Bu_D_11}{Branching fractions to $D_s^{(*)}$ mesons}
\hline
\textbf{Parameter} & %
 & $< 1.2$ \\
\hline
\end{btocharmtab}
\btocharmfig{Bu_D_11}

\begin{btocharmtab}{Bu_D_12}{Branching fraction ratios to $D_s^{(*)}$ mesons}
\hline
\textbf{Parameter} & %
 & $0.197 \pm 0.023$ \\
\hline
\end{btocharmtab}

\begin{btocharmtab}{Bu_D_13}{Branching fractions to baryonic decays, I}
\hline
\textbf{Parameter} & %
 & $1.86 \pm 0.25$ \\
\hline
\end{btocharmtab}
\btocharmfig{Bu_D_13}

\begin{btocharmtab}{Bu_D_14}{Branching fractions to baryonic decays, II}
\hline
\textbf{Parameter} & %
 & $< 1.5$ \\
\hline
\end{btocharmtab}
\btocharmfig{Bu_D_14}

\begin{btocharmtab}{Bu_D_15}{Branching fractions to lepton number violating decays}
\hline
\textbf{Parameter} & %
 & $< 1.0$ \\
\hline
\end{btocharmtab}
\btocharmfig{Bu_D_15}

\subsubsection{Decays to two open charm mesons}
\label{sec:b2c:Bu_DD}
Averages of $B^-$ decays to two open charm mesons are shown in Tables~\ref{tab:b2c:Bu_DD_1}--\ref{tab:b2c:Bu_DD_8}.
\begin{btocharmtab}{Bu_DD_1}{Branching fractions to $D^{(*)-}D^{(*)0}$}
\hline
\textbf{Parameter} & %
 & $0.81 \pm 0.17$ \\
\hline
\end{btocharmtab}
\btocharmfig{Bu_DD_1}

\begin{btocharmtab}{Bu_DD_2}{Branching fractions to two $D$ mesons and a kaon}
\hline
\textbf{Parameter} & %
 & $1.55 \pm 0.21$ \\
\hline
\end{btocharmtab}
\btocharmfig{Bu_DD_2}

\begin{btocharmtab}{Bu_DD_3}{Branching fractions to $D_s^{(*)-}D^{(*)+}$}
\hline
\textbf{Parameter} & %
 & $1.70 \pm 0.35$ \\
\hline
\end{btocharmtab}
\btocharmfig{Bu_DD_3}

\begin{btocharmtab}{Bu_DD_4}{Product branching fractions to $D_s^{(*)-}D^{(*)+}$}
\hline
\textbf{Parameter} & %
 & $8.57 \pm 1.86$ \\
\hline
\end{btocharmtab}
\btocharmfig{Bu_DD_4}

\begin{btocharmtab}{Bu_DD_5}{Branching fraction ratios}
\hline
\textbf{Parameter} & %
 & $1.22 \pm 0.07$ \\
\hline
\end{btocharmtab}

\begin{btocharmtab}{Bu_DD_6}{Branching fractions to excited $D_s$ mesons}
\hline
\textbf{Parameter} & %
 & $11.2 \pm 3.3$ \\
\hline
\end{btocharmtab}
\btocharmfig{Bu_DD_6}

\begin{btocharmtab}{Bu_DD_7}{Product branching fractions to excited $D_s$ mesons, I}
\hline
\textbf{Parameter} & %
 & $7.6 \,^{+3.6}_{-2.9}$ \\
\hline
\end{btocharmtab}
\btocharmfig{Bu_DD_7}

\begin{btocharmtab}{Bu_DD_8}{Product branching fractions to excited $D_s$ mesons, II}
\hline
\textbf{Parameter} & %
 & $< 1.069$ \\
\hline
\end{btocharmtab}
\btocharmfig{Bu_DD_8}

\subsubsection{Decays to charmonium states}
\label{sec:b2c:Bu_cc}
Averages of $B^-$ decays to charmonium states are shown in Tables~\ref{tab:b2c:Bu_cc_1}--\ref{tab:b2c:Bu_cc_9}.
\begin{btocharmtab}{Bu_cc_1}{Branching fractions to $J/\psi$ and one kaon}
\hline
\textbf{Parameter} & %
 & $0.44 \pm 0.15$ \\
\hline
\end{btocharmtab}
\btocharmfig{Bu_cc_1}

\begin{btocharmtab}{Bu_cc_2}{Branching fractions to charmonium other than $J/\psi$ and one kaon, I}
\hline
\textbf{Parameter} & %
\end{center}
\btocharmfig{Bu_cc_2}

\begin{btocharmtab}{Bu_cc_3}{Branching fractions to charmonium other than $J/\psi$ and one kaon, II}
\hline
\textbf{Parameter} & %
 & $< 3.4$ \\
\hline
\end{btocharmtab}
\btocharmfig{Bu_cc_3}

\begin{btocharmtab}{Bu_cc_4}{Branching fractions to charmonium and light mesons}
\hline
\textbf{Parameter} & %
 & $2.2 \pm 0.5$ \\
\hline
\end{btocharmtab}
\btocharmfig{Bu_cc_4}

\begin{btocharmtab}{Bu_cc_5}{Branching fractions to $J/\psi$ and a heavy mesons}
\hline
\textbf{Parameter} & %
 & $< 0.25$ \\
\hline
\end{btocharmtab}
\btocharmfig{Bu_cc_5}

\begin{btocharmtab}{Bu_cc_6}{Branching fractions to $J/\psi$ and baryons}
\hline
\textbf{Parameter} & %
 & $2.21 \pm 0.13$ \\
\hline
\end{btocharmtab}
\btocharmfig{Bu_cc_6}

\begin{btocharmtab}{Bu_cc_7}{Branching fraction ratios, I}
\hline
\textbf{Parameter} & %
 & $0.578 \pm 0.044$ \\
\hline
\end{btocharmtab}
\btocharmfig{Bu_cc_7}

\begin{btocharmtab}{Bu_cc_8}{Branching fraction ratios, II}
\hline
\textbf{Parameter} & %
 & $< 0.052$ \\
\hline
\end{btocharmtab}
\btocharmfig{Bu_cc_8}

\begin{btocharmtab}{Bu_cc_9}{Direct CP violation parameters}
\hline
\textbf{Parameter} & %
 & $-0.042 \pm 0.045$ \\
\hline
\end{btocharmtab}
\btocharmfig{Bu_cc_9}

\subsubsection{Decays to charm baryons}
\label{sec:b2c:Bu_baryon}
Averages of $B^-$ decays to charm baryons are shown in Tables~\ref{tab:b2c:Bu_baryon_1}--\ref{tab:b2c:Bu_baryon_2}.
\begin{btocharmtab}{Bu_baryon_1}{(Product) branching fractions}
\hline
\textbf{Parameter} & %
 & $2.98 \pm 0.80$ \\
\hline
\end{btocharmtab}
\btocharmfig{Bu_baryon_1}

\begin{btocharmtab}{Bu_baryon_2}{Branching fraction ratios}
\hline
\textbf{Parameter} & %
 & $0.117 \pm 0.033$ \\
\hline
\end{btocharmtab}
\btocharmfig{Bu_baryon_2}

\subsubsection{Decays to $XYZP$ states}
\label{sec:b2c:Bu_other}
Averages of $B^-$ decays to $XYZP$ states are shown in Tables~\ref{tab:b2c:Bu_other_1}--\ref{tab:b2c:Bu_other_6}.
\begin{btocharmtab}{Bu_other_1}{Branching fractions}
\hline
\textbf{Parameter} & %
 & $0.4 \pm 1.6$ \\
\hline
\end{btocharmtab}
\btocharmfig{Bu_other_1}

\begin{btocharmtab}{Bu_other_2}{Product branching fractions to $X(3872)$, I}
\hline
\textbf{Parameter} & %
 & $< 0.4$ \\
\hline
\end{btocharmtab}
\btocharmfig{Bu_other_2}

\begin{btocharmtab}{Bu_other_3}{Product branching fractions to $X(3872)$, II}
\hline
\textbf{Parameter} & %
 & $< 0.67$ \\
\hline
\end{btocharmtab}
\btocharmfig{Bu_other_3}

\begin{btocharmtab}{Bu_other_4}{Product branching fractions to neutral states other than $X(3872)$}
\hline
\textbf{Parameter} & %
 & $< 20$ \\
\hline
\end{btocharmtab}
\btocharmfig{Bu_other_4}

\begin{btocharmtab}{Bu_other_5}{Relative product branching fractions to states with $s\bar{s}$ component}
\hline
\textbf{Parameter} & %
 & $0.071 \,^{+0.043}_{-0.035}$ \\
\hline
\end{btocharmtab}
\btocharmfig{Bu_other_5}

\begin{btocharmtab}{Bu_other_6}{Product branching fractions to charged states}
\hline
\textbf{Parameter} & %
 & $< 4.7$ \\
\hline
\end{btocharmtab}
\btocharmfig{Bu_other_6}

\subsection{Decays of admixtures of $\bar{B}^0$/$B^-$ mesons}
\label{sec:b2c:B}
Measurements of $\bar{B}^0$/$B^-$ decays to charmed hadrons are summarized in Sections~\ref{sec:b2c:B_DD} to~\ref{sec:b2c:B_other}. These results reflect the $\bar{B}^0$/$B^-$ production admixture in $\Upsilon(4S)$ decays.

\subsubsection{Decays to two open charm mesons}
\label{sec:b2c:B_DD}
Averages of $\bar{B}^0$/$B^-$ decays to two open charm mesons are shown in Table~\ref{tab:b2c:B_DD_1}.
\begin{btocharmtab}{B_DD_1}{Branching fractions to double charm}
\hline
\textbf{Parameter} & \begin{tabular}{l}\textbf{Measurements}\end{tabular} $[10^{-4}]$ & \textbf{Average} $[10^{-4}]$ \\
\hline
\hline
\btocharmparam{${\cal{B}} ( B \to D^{0} \bar{D}^{0} \pi^{0} K )$}{${\cal{B}} ( B \to D^{0} \bar{D}^{0} \pi^{0} K )$}{} & \begin{tabular}{l} Belle \cite{Gokhroo:2006bt}: $1.27 \pm 0.31 \,^{+0.22}_{-0.39}$ \\ \end{tabular} & $1.27 \,^{+0.38}_{-0.50}$ \\
\hline
\end{btocharmtab}

\subsubsection{Decays to charmonium states}
\label{sec:b2c:B_cc}
Averages of $\bar{B}^0$/$B^-$ decays to charmonium states are shown in Tables~\ref{tab:b2c:B_cc_1}--\ref{tab:b2c:B_cc_5}.
The Belle and LHCb results quoted in this section are only for $\bar{B}^0$ decays.
Assuming isospin symmetry we combine them with the BaBar measurements for the admixture.
\begin{btocharmtab}{B_cc_1}{Decay amplitudes for parallel transverse polarization}
\hline
\textbf{Parameter} & \begin{tabular}{l}\textbf{Measurements}\end{tabular} & \textbf{Average} \\
\hline
\hline
\btocharmparam{$\vert{\cal{A}}_{\parallel}\vert^{2} ( B \to J/\psi K^{*} )$}{$\vert{\cal{A}}_{\parallel}\vert^{2} ( B \to J/\psi K^{*} )$}{} & \begin{tabular}{l} LHCb \cite{Aaij:2013cma}: $0.227 \pm 0.004 \pm 0.011$ \\ Belle \cite{Itoh:2005ks}: $0.231 \pm 0.012 \pm 0.008$ \\ \babar \cite{Aubert:2007hz}: $0.211 \pm 0.010 \pm 0.006$ \\ \end{tabular} & $0.222 \pm 0.007$ \\
\hline
\btocharmparam{$\vert{\cal{A}}_{\parallel}\vert^{2} ( B \to \chi_{c1} K^{*} )$}{$\vert{\cal{A}}_{\parallel}\vert^{2} ( B \to \chi_{c1} K^{*} )$}{} & \begin{tabular}{l} \babar \cite{Aubert:2007hz}: $0.20 \pm 0.07 \pm 0.04$ \\ \end{tabular} & $0.20 \pm 0.08$ \\
\hline
\btocharmparam{$\vert{\cal{A}}_{\parallel}\vert^{2} ( B \to \psi(2S) K^{*} )$}{$\vert{\cal{A}}_{\parallel}\vert^{2} ( B \to \psi(2S) K^{*} )$}{} & \begin{tabular}{l} \babar \cite{Aubert:2007hz}: $0.22 \pm 0.06 \pm 0.02$ \\ \end{tabular} & $0.22 \pm 0.06$ \\
\hline
\end{btocharmtab}
\btocharmfig{B_cc_1}

\begin{btocharmtab}{B_cc_2}{Decay amplitudes for perpendicular transverse polarization}
\hline
\textbf{Parameter} & \begin{tabular}{l}\textbf{Measurements}\end{tabular} & \textbf{Average} \\
\hline
\hline
\btocharmparam{$\vert{\cal{A}}_{\perp}\vert^{2} ( B \to J/\psi K^{*} )$}{$\vert{\cal{A}}_{\perp}\vert^{2} ( B \to J/\psi K^{*} )$}{} & \begin{tabular}{l} LHCb \cite{Aaij:2013cma}: $0.201 \pm 0.004 \pm 0.008$ \\ Belle \cite{Itoh:2005ks}: $0.195 \pm 0.012 \pm 0.008$ \\ \babar \cite{Aubert:2007hz}: $0.233 \pm 0.010 \pm 0.005$ \\ \end{tabular} & $0.210 \pm 0.006$ \\
\hline
\btocharmparam{$\vert{\cal{A}}_{\perp}\vert^{2} ( B \to \chi_{c1} K^{*} )$}{$\vert{\cal{A}}_{\perp}\vert^{2} ( B \to \chi_{c1} K^{*} )$}{} & \begin{tabular}{l} \babar \cite{Aubert:2007hz}: $0.03 \pm 0.04 \pm 0.02$ \\ \end{tabular} & $0.03 \pm 0.04$ \\
\hline
\btocharmparam{$\vert{\cal{A}}_{\perp}\vert^{2} ( B \to \psi(2S) K^{*} )$}{$\vert{\cal{A}}_{\perp}\vert^{2} ( B \to \psi(2S) K^{*} )$}{} & \begin{tabular}{l} \babar \cite{Aubert:2007hz}: $0.30 \pm 0.06 \pm 0.02$ \\ \end{tabular} & $0.30 \pm 0.06$ \\
\hline
\end{btocharmtab}
\btocharmfig{B_cc_2}

\begin{btocharmtab}{B_cc_3}{Decay amplitudes for longitudinal polarization}
\hline
\textbf{Parameter} & \begin{tabular}{l}\textbf{Measurements}\end{tabular} & \textbf{Average} \\
\hline
\hline
\btocharmparam{$\vert{\cal{A}}_{0}\vert^{2} ( B \to J/\psi K^{*} )$}{$\vert{\cal{A}}_{0}\vert^{2} ( B \to J/\psi K^{*} )$}{} & \begin{tabular}{l} Belle \cite{Itoh:2005ks}: $0.574 \pm 0.012 \pm 0.009$ \\ \babar \cite{Aubert:2007hz}: $0.556 \pm 0.009 \pm 0.010$ \\ \end{tabular} & $0.564 \pm 0.010$ \\
\hline
\btocharmparam{$\vert{\cal{A}}_{0}\vert^{2} ( B \to \chi_{c1} K^{*} )$}{$\vert{\cal{A}}_{0}\vert^{2} ( B \to \chi_{c1} K^{*} )$}{} & \begin{tabular}{l} \babar \cite{Aubert:2007hz}: $0.77 \pm 0.07 \pm 0.04$ \\ \end{tabular} & $0.77 \pm 0.08$ \\
\hline
\btocharmparam{$\vert{\cal{A}}_{0}\vert^{2} ( B \to \psi(2S) K^{*} )$}{$\vert{\cal{A}}_{0}\vert^{2} ( B \to \psi(2S) K^{*} )$}{} & \begin{tabular}{l} \babar \cite{Aubert:2007hz}: $0.48 \pm 0.05 \pm 0.02$ \\ \end{tabular} & $0.48 \pm 0.05$ \\
\hline
\end{btocharmtab}
\btocharmfig{B_cc_3}

\begin{btocharmtab}{B_cc_4}{Relative phases of parallel transverse polarization decay amplitudes}
\hline
\textbf{Parameter} & \begin{tabular}{l}\textbf{Measurements}\end{tabular} & \textbf{Average} \\
\hline
\hline
\btocharmparam{$\delta_{\parallel} ( B \to J/\psi K^{*} )$}{$\delta_{\parallel} ( B \to J/\psi K^{*} )$}{} & \begin{tabular}{l} LHCb \cite{Aaij:2013cma}: $-2.94 \pm 0.02 \pm 0.03$ \\ Belle \cite{Itoh:2005ks}: $-2.887 \pm 0.090 \pm 0.008$ \\ \babar \cite{Aubert:2007hz}: $-2.93 \pm 0.08 \pm 0.04$ \\ \end{tabular} & $-2.932 \pm 0.031$ \\
\hline
\btocharmparam{$\delta_{\parallel} ( B \to \chi_{c1} K^{*} )$}{$\delta_{\parallel} ( B \to \chi_{c1} K^{*} )$}{} & \begin{tabular}{l} \babar \cite{Aubert:2007hz}: $0.0 \pm 0.3 \pm 0.1$ \\ \end{tabular} & $0.0 \pm 0.3$ \\
\hline
\btocharmparam{$\delta_{\parallel} ( B \to \psi(2S) K^{*} )$}{$\delta_{\parallel} ( B \to \psi(2S) K^{*} )$}{} & \begin{tabular}{l} \babar \cite{Aubert:2007hz}: $-2.8 \pm 0.4 \pm 0.1$ \\ \end{tabular} & $-2.8 \pm 0.4$ \\
\hline
\end{btocharmtab}
\btocharmfig{B_cc_4}

\begin{btocharmtab}{B_cc_5}{Relative phases of perpendicular transverse polarization decay amplitudes}
\hline
\textbf{Parameter} & \begin{tabular}{l}\textbf{Measurements}\end{tabular} & \textbf{Average} \\
\hline
\hline
\btocharmparam{$\delta_{\perp} ( B \to J/\psi K^{*} )$}{$\delta_{\perp} ( B \to J/\psi K^{*} )$}{} & \begin{tabular}{l} LHCb \cite{Aaij:2013cma}: $2.94 \pm 0.02 \pm 0.02$ \\ Belle \cite{Itoh:2005ks}: $2.938 \pm 0.064 \pm 0.010$ \\ \babar \cite{Aubert:2007hz}: $2.91 \pm 0.05 \pm 0.03$ \\ \end{tabular} & $2.935 \pm 0.024$ \\
\hline
\btocharmparam{$\delta_{\perp} ( B \to \psi(2S) K^{*} )$}{$\delta_{\perp} ( B \to \psi(2S) K^{*} )$}{} & \begin{tabular}{l} \babar \cite{Aubert:2007hz}: $2.8 \pm 0.3 \pm 0.1$ \\ \end{tabular} & $2.8 \pm 0.3$ \\
\hline
\end{btocharmtab}
\btocharmfig{B_cc_5}

\subsubsection{Decays to $XYZP$ states}
\label{sec:b2c:B_other}
Averages of $\bar{B}^0$/$B^-$ decays to $XYZP$ states are shown in Table~\ref{tab:b2c:B_other_1}.
\begin{btocharmtab}{B_other_1}{Branching fractions to $X$/$Y$ states}
\hline
\textbf{Parameter} & \begin{tabular}{l}\textbf{Measurements}\end{tabular} $[10^{-4}]$ & \textbf{Average} $[10^{-4}]$ \\
\hline
\hline
\btocharmparaml{${{\cal{B}} ( B \to X(3872) K )\times {\cal{B}} ( X(3872) \to D^{*}(2007)^{0} \bar{D}^{0} )}$}{${{\cal{B}} ( B \to X(3872) K )\times {\cal{B}} ( X(3872) \to D^{*}(2007)^{0} \bar{D}^{0} )}$}{} & \begin{tabular}{l} Belle \cite{Adachi:2008sua}: $0.80 \pm 0.20 \pm 0.10$ \\ \end{tabular} & $0.80 \pm 0.22$ \\
\hline
\btocharmparaml{${{\cal{B}} ( B \to Y(3940) K )\times {\cal{B}} ( Y(3940) \to D^{*}(2007)^{0} \bar{D}^{0} )}$}{${{\cal{B}} ( B \to Y(3940) K )\times {\cal{B}} ( Y(3940) \to D^{*}(2007)^{0} \bar{D}^{0} )}$}{} & \begin{tabular}{l} Belle \cite{Adachi:2008sua}: $< 0.67$ \\ \end{tabular} & $< 0.67$ \\
\hline
\btocharmparaml{${{\cal{B}} ( B \to K Y(3940) )\times {\cal{B}} ( Y(3940) \to J/\psi \omega(782) )}$}{${{\cal{B}} ( B \to K Y(3940) )\times {\cal{B}} ( Y(3940) \to J/\psi \omega(782) )}$}{} & \begin{tabular}{l} Belle \cite{Abe:2004zs}: $0.71 \pm 0.13 \pm 0.31$ \\ \end{tabular} & $0.71 \pm 0.34$ \\
\hline
\end{btocharmtab}
\btocharmfig{B_other_1}

\subsection{Decays of $\bar{B}_s^0$ mesons}
\label{sec:b2c:Bs}
Measurements of $\bar{B}_s^0$ decays to charmed hadrons are summarized in Sections~\ref{sec:b2c:Bs_D} to~\ref{sec:b2c:Bs_baryon}.
These measurements require knowledge of the production rates of $\bar{B}_s^0$ mesons, usually measured relative to those of $\bar{B}^0$ and $B^-$ mesons, in the same experimental environment. 
Since these production fractions are reasonably well known, see Sec.~\ref{sec:fractions}, they can be corrected for allowing the results to be presented in terms of a $\bar{B}_s^0$ branching fraction.
This is usually done in the publications; we do not attempt to rescale results according to more recent determinations of the relative production fractions. 
Ratios of branching fractions of two decays of the same hadron do not require any such correction.

\subsubsection{Decays to a single open charm meson}
\label{sec:b2c:Bs_D}
Averages of $\bar{B}_s^0$ decays to a single open charm meson are shown in Tables~\ref{tab:b2c:Bs_D_1}--\ref{tab:b2c:Bs_D_5}.
\begin{btocharmtab}{Bs_D_1}{Branching fractions to a $D_s^{(*)}$ and a light meson}
\hline
\textbf{Parameter} & 
 & $0.163 \,^{+0.050}_{-0.050}$ \\
\hline
\end{btocharmtab}
\btocharmfig{Bs_D_1}

\begin{btocharmtab}{Bs_D_2}{Branching fractions to a $D^{(*)}$ and light mesons}
\hline
\textbf{Parameter} & %
 & $0.57 \pm 0.08$ \\
\hline
\end{btocharmtab}
\btocharmfig{Bs_D_2}

\begin{btocharmtab}{Bs_D_3}{Branching fraction ratios, I}
\hline
\textbf{Parameter} & %
 & $0.930 \pm 0.113$ \\
\hline
\end{btocharmtab}
\btocharmfig{Bs_D_3}

\begin{btocharmtab}{Bs_D_4}{Branching fraction ratios, II}
\hline
\textbf{Parameter} & %
 & $4.2 \pm 0.6$ \\
\hline
\end{btocharmtab}
\btocharmfig{Bs_D_4}

\begin{btocharmtab}{Bs_D_5}{Longitudinal polarisation fraction}
\hline
\textbf{Parameter} & %
 & $0.73 \pm 0.15$ \\
\hline
\end{btocharmtab}

\subsubsection{Decays to two open charm mesons}
\label{sec:b2c:Bs_DD}
Averages of $\bar{B}_s^0$ decays to two open charm mesons are shown in Tables~\ref{tab:b2c:Bs_DD_1}--\ref{tab:b2c:Bs_DD_3}.
\begin{btocharmtab}{Bs_DD_1}{Branching fractions}
\hline
\textbf{Parameter} & %
 & $3.19 \pm 0.37$ \\
\hline
\end{btocharmtab}
\btocharmfig{Bs_DD_1}

\begin{btocharmtab}{Bs_DD_2}{Branching fraction ratios, I}
\hline
\textbf{Parameter} & %
 & $0.56 \pm 0.05$ \\
\hline
\end{btocharmtab}
\btocharmfig{Bs_DD_2}

\begin{btocharmtab}{Bs_DD_3}{Branching fraction ratios, II}
\hline
\textbf{Parameter} & %
 & $1.9 \pm 0.4$ \\
\hline
\end{btocharmtab}
\btocharmfig{Bs_DD_3}

\subsubsection{Decays to charmonium states}
\label{sec:b2c:Bs_cc}
Averages of $\bar{B}_s^0$ decays to charmonium states are shown in Tables~\ref{tab:b2c:Bs_cc_1}--\ref{tab:b2c:Bs_cc_6}.
\begin{btocharmtab}{Bs_cc_1}{Branching fractions, I}
\hline
\textbf{Parameter} & %
 & $1.16 \,^{+0.43}_{-0.32}$ \\
\hline
\end{btocharmtab}
\btocharmfig{Bs_cc_1}

\begin{btocharmtab}{Bs_cc_2}{Branching fractions, II}
\hline
\textbf{Parameter} & %
 & $< 0.73$ \\
\hline
\end{btocharmtab}
\btocharmfig{Bs_cc_2}

\begin{btocharmtab}{Bs_cc_3}{Branching fraction ratios, I}
\hline
\textbf{Parameter} & %
 & $2.12 \pm 0.23$ \\
\hline
\end{btocharmtab}
\btocharmfig{Bs_cc_3}

\begin{btocharmtab}{Bs_cc_4}{Branching fraction ratios, II}
\hline
\textbf{Parameter} & %
 & $0.171 \pm 0.033$ \\
\hline
\end{btocharmtab}
\btocharmfig{Bs_cc_4}

\begin{btocharmtab}{Bs_cc_5}{Branching fraction ratios, III}
\hline
\textbf{Parameter} & %
 & $< 3.4$ \\
\hline
\end{btocharmtab}
\btocharmfig{Bs_cc_5}

\begin{btocharmtab}{Bs_cc_6}{Decay amplitudes and relative phases}
\hline
\textbf{Parameter} & %
 & $0.01 \,^{+0.16}_{-0.17}$ \\
\hline
\end{btocharmtab}
\btocharmfig{Bs_cc_6}

\subsubsection{Decays to charm baryons}
\label{sec:b2c:Bs_baryon}
Averages of $\bar{B}_s^0$ decays to charm baryons are shown in Tables~\ref{tab:b2c:Bs_baryon_1}--\ref{tab:b2c:Bs_baryon_2}.
\begin{btocharmtab}{Bs_baryon_1}{Branching fractions to one charm baryon}
\hline
\textbf{Parameter} & %
 & $3.6 \,^{+1.6}_{-1.7}$ \\
\hline
\end{btocharmtab}

\begin{btocharmtab}{Bs_baryon_2}{Branching fractions to two charm baryons}
\hline
\textbf{Parameter} & %
 & $< 0.30$ \\
\hline
\end{btocharmtab}

\subsection{Decays of $B_c^-$ mesons}
\label{sec:b2c:Bc}
Measurements of $B_c^-$ decays to charmed hadrons are summarized in Sections~\ref{sec:b2c:Bc_D} to~\ref{sec:b2c:Bc_B}.
Since the absolute cross-section for $B_c^-$ meson production in any production environment is currently not known, it is not possible to determine absolute branching fractions. 
Instead, results are presented either as ratios of branching fractions of different $B_c^-$ decays, or are normalised to the branching fraction of the decay of a lighter $B$ meson (usually $B^-$).
In the latter case the measured quantity is the absolute or relative $B_c^-$ branching fraction multiplied by the ratio of cross-sections (or, equivalently, production fractions) of the $B_c^-$ and the lighter $B$ meson.

It should be noted that the ratio of cross-sections for different $b$ hadron species can depend on production environment, and on the fiducial region accessed by each experiment.
While this has been studied for certain $b$ hadron species (see Sec.~\ref{sec:fractions}), there is currently little published data that would allow to investigate the effect for $B_c^-$ mesons.
Therefore, we do not attempt to apply any correction for this effect.

\subsubsection{Decays to a single open charm meson}
\label{sec:b2c:Bc_D}
Averages of $B_c^-$ decays to a single open charm meson are shown in Table~\ref{tab:b2c:Bc_D_1}.
\begin{btocharmtab}{Bc_D_1}{Branching fractions to $D^{(*)0}$ meson and one or more kaons}
\hline
\textbf{Parameter} & \begin{tabular}{l}\textbf{Measurements}\end{tabular} $[10^{-7}]$ & \textbf{Average} $[10^{-7}]$ \\
\hline
\hline
\btocharmparam{$\frac{ f_c \times {\cal B} ( B_c^{-} \to  \bar{D}^{0} K^{-} ) }{ f_u }$}{$[{ f_c \times {\cal B} ( B_c^{-} \to  \bar{D}^{0} K^{-} ) }]/{ f_u }$}{} & \begin{tabular}{l} LHCb \cite{Aaij:2017kea}: $9.3 \,^{+2.8}_{-2.5} \pm 0.6$ \\ \end{tabular} & $9.3 \,^{+2.9}_{-2.6}$ \\
\hline
\end{btocharmtab}

\subsubsection{Decays to two open charm mesons}
\label{sec:b2c:Bc_DD}
Averages of $B_c^-$ decays to two open charm mesons are shown in Tables~\ref{tab:b2c:Bc_DD_1}--\ref{tab:b2c:Bc_DD_2}.
\begin{btocharmtab}{Bc_DD_1}{Branching fraction ratios, I}
\hline
\textbf{Parameter} & \begin{tabular}{l}\textbf{Measurements}\end{tabular} $[10^{-3}]$ & \textbf{Average} $[10^{-3}]$ \\
\hline
\hline
\btocharmparam{$\frac{ f_c \times {\cal{B}} ( B_c^- \to D_s^- D^0 ) } { f_u \times {\cal{B}} ( B^- \to D_s^- D^0 ) }$}{$[{ f_c \times {\cal{B}} ( B_c^- \to D_s^- D^0 ) }]/[ { f_u \times {\cal{B}} ( B^- \to D_s^- D^0 ) }]$}{} & \begin{tabular}{l} LHCb \cite{Aaij:2017gon}: $0.30 \pm 0.37$ \\ \end{tabular} & $0.30 \pm 0.37$ \\
\hline
\btocharmparam{$\frac{ f_c \times {\cal{B}} ( B_c^- \to D_s^- \bar{D}^0 ) } { f_u \times {\cal{B}} ( B^- \to D_s^- D^0 ) }$}{$[{ f_c \times {\cal{B}} ( B_c^- \to D_s^- \bar{D}^0 ) }]/[ { f_u \times {\cal{B}} ( B^- \to D_s^- D^0 ) }]$}{} & \begin{tabular}{l} LHCb \cite{Aaij:2017gon}: $-0.38 \pm 0.26$ \\ \end{tabular} & $-0.38 \pm 0.26$ \\
\hline
\btocharmparam{$\frac{ f_c \times {\cal{B}} ( B_c^- \to D^- D^0 ) } { f_u \times {\cal{B}} ( B^- \to D^- D^0 ) }$}{$[{ f_c \times {\cal{B}} ( B_c^- \to D^- D^0 ) }]/[ { f_u \times {\cal{B}} ( B^- \to D^- D^0 ) }]$}{} & \begin{tabular}{l} LHCb \cite{Aaij:2017gon}: $8.0 \pm 7.5$ \\ \end{tabular} & $8.0 \pm 7.5$ \\
\hline
\btocharmparam{$\frac{ f_c \times {\cal{B}} ( B_c^- \to D^- \bar{D}^0 ) } { f_u \times {\cal{B}} ( B^- \to D^- D^0 ) }$}{$[{ f_c \times {\cal{B}} ( B_c^- \to D^- \bar{D}^0 ) }]/[ { f_u \times {\cal{B}} ( B^- \to D^- D^0 ) }]$}{} & \begin{tabular}{l} LHCb \cite{Aaij:2017gon}: $2.9 \pm 5.3$ \\ \end{tabular} & $2.9 \pm 5.3$ \\
\hline
\btocharmparam{$\frac{ f_c \times ( {\cal{B}} ( B_c^- \to D_s^{*-} D^0 ) + {\cal{B}} ( B_c^+ \to D_s^- D^{*0} ) ) } { f_u \times {\cal{B}} ( B^- \to D_s^- D^0 ) }$}{$[{ f_c \times ( {\cal{B}} ( B_c^- \to D_s^{*-} D^0 ) + {\cal{B}} ( B_c^+ \to D_s^- D^{*0} ) ) }]/[ { f_u \times {\cal{B}} ( B^- \to D_s^- D^0 ) }]$}{} & \begin{tabular}{l} LHCb \cite{Aaij:2017gon}: $-0.1 \pm 1.5$ \\ \end{tabular} & $-0.1 \pm 1.5$ \\
\hline
\btocharmparam{$\frac{ f_c \times ( {\cal{B}} ( B_c^- \to D_s^{*-} \bar{D}^0 ) + {\cal{B}} ( B_c^+ \to D_s^- \bar{D}^{*0} ) ) } { f_u \times {\cal{B}} ( B^- \to D_s^- D^0 ) }$}{$[{ f_c \times ( {\cal{B}} ( B_c^- \to D_s^{*-} \bar{D}^0 ) + {\cal{B}} ( B_c^+ \to D_s^- \bar{D}^{*0} ) ) }]/[ { f_u \times {\cal{B}} ( B^- \to D_s^- D^0 ) }]$}{} & \begin{tabular}{l} LHCb \cite{Aaij:2017gon}: $-0.3 \pm 1.9$ \\ \end{tabular} & $-0.3 \pm 1.9$ \\
\hline
\btocharmparam{$\frac{ f_c \times  {\cal{B}} ( B_c^- \to D_s^{*-} D^{*0} ) } { f_u \times {\cal{B}} ( B^- \to D_s^- D^0 ) }$}{$[{ f_c \times  {\cal{B}} ( B_c^- \to D_s^{*-} D^{*0} ) }]/[ { f_u \times {\cal{B}} ( B^- \to D_s^- D^0 ) }]$}{} & \begin{tabular}{l} LHCb \cite{Aaij:2017gon}: $3.2 \pm 4.3$ \\ \end{tabular} & $3.2 \pm 4.3$ \\
\hline
\btocharmparam{$\frac{ f_c \times  {\cal{B}} ( B_c^- \to D_s^{*-} \bar{D}^{*0} ) } { f_u \times {\cal{B}} ( B^- \to D_s^- D^0 ) }$}{$[{ f_c \times  {\cal{B}} ( B_c^- \to D_s^{*-} \bar{D}^{*0} ) }]/[ { f_u \times {\cal{B}} ( B^- \to D_s^- D^0 ) }]$}{} & \begin{tabular}{l} LHCb \cite{Aaij:2017gon}: $7.0 \pm 9.2$ \\ \end{tabular} & $7.0 \pm 9.2$ \\
\hline
\end{btocharmtab}
\btocharmfig{Bc_DD_1}

\begin{btocharmtab}{Bc_DD_2}{Branching fraction ratios, II}
\hline
\textbf{Parameter} & \begin{tabular}{l}\textbf{Measurements}\end{tabular} $[10^{-1}]$ & \textbf{Average} $[10^{-1}]$ \\
\hline
\hline
\btocharmparaml{$\frac{ f_c \times  ( {\cal{B}} ( B_c^- \to D^{*-} D^0 ) \times {\cal{B}} (D^{*-}\to D^- (\pi^0,\gamma)) + {\cal{B}} (B_c^{-}\to D^- D^{*0}) } { f_u \times {\cal{B}} ( B^- \to D^- D^0 ) }$}{$[{ f_c \times  ( {\cal{B}} ( B_c^- \to D^{*-} D^0 ) \times {\cal{B}} (D^{*-}\to D^- (\pi^0,\gamma)) + {\cal{B}} (B_c^{-}\to D^- D^{*0}) }]/[ { f_u \times {\cal{B}} ( B^- \to D^- D^0 ) }]$}{} & \begin{tabular}{l} LHCb \cite{Aaij:2017gon}: $0.02 \pm 0.32$ \\ \end{tabular} & $0.02 \pm 0.32$ \\
\hline
\btocharmparaml{$\frac{ f_c \times  ( {\cal{B}} ( B_c^- \to D^{*-} \bar{D}^0 ) \times {\cal{B}} (D^{*-}\to D^- (\pi^0,\gamma)) + {\cal{B}} (B_c^{-}\to D^- \bar{D}^{*0}) } { f_u \times {\cal{B}} ( B^- \to D^- D^0 ) }$}{$[{ f_c \times  ( {\cal{B}} ( B_c^- \to D^{*-} \bar{D}^0 ) \times {\cal{B}} (D^{*-}\to D^- (\pi^0,\gamma)) + {\cal{B}} (B_c^{-}\to D^- \bar{D}^{*0}) }]/[ { f_u \times {\cal{B}} ( B^- \to D^- D^0 ) }]$}{} & \begin{tabular}{l} LHCb \cite{Aaij:2017gon}: $-0.15 \pm 0.17$ \\ \end{tabular} & $-0.15 \pm 0.17$ \\
\hline
\btocharmparam{$\frac{ f_c \times  {\cal{B}} ( B_c^- \to D_s^{*-} \bar{D}^{*0} ) } { f_u \times {\cal{B}} ( B^- \to D^- D^0 ) }$}{$[{ f_c \times  {\cal{B}} ( B_c^- \to D_s^{*-} \bar{D}^{*0} ) }]/[ { f_u \times {\cal{B}} ( B^- \to D^- D^0 ) }]$}{} & \begin{tabular}{l} LHCb \cite{Aaij:2017gon}: $-0.41 \pm 0.91$ \\ \end{tabular} & $-0.41 \pm 0.91$ \\
\hline
\btocharmparam{$\frac{ f_c \times  {\cal{B}} ( B_c^- \to D_s^{*-} D^{*0} ) } { f_u \times {\cal{B}} ( B^- \to D^- D^0 ) }$}{$[{ f_c \times  {\cal{B}} ( B_c^- \to D_s^{*-} D^{*0} ) }]/[ { f_u \times {\cal{B}} ( B^- \to D^- D^0 ) }]$}{} & \begin{tabular}{l} LHCb \cite{Aaij:2017gon}: $3.4 \pm 2.3$ \\ \end{tabular} & $3.4 \pm 2.3$ \\
\hline
\end{btocharmtab}
\btocharmfig{Bc_DD_2}

\subsubsection{Decays to charmonium states}
\label{sec:b2c:Bc_cc}
Averages of $B_c^-$ decays to charmonium states are shown in Tables~\ref{tab:b2c:Bc_cc_1}--\ref{tab:b2c:Bc_cc_3}.
\begin{btocharmtab}{Bc_cc_1}{Branching fraction ratios}
\hline
\textbf{Parameter} & \begin{tabular}{l}\textbf{Measurements}\end{tabular} & \textbf{Average} \\
\hline
\hline
\btocharmparam{$\frac {{\cal{B}} (B_c^{-} \to J/\psi D_s^{-})} {{\cal{B}} (B_c^{-} \to J/\psi \pi^{-})}$}{${{\cal{B}} (B_c^{-} \to J/\psi D_s^{-})}/ {{\cal{B}} (B_c^{-} \to J/\psi \pi^{-})}$}{} & \begin{tabular}{l} LHCb \cite{Aaij:2013gia}: $2.90 \pm 0.57 \pm 0.24$ \\ ATLAS \cite{Aad:2015eza}: $3.8 \pm 1.1 \pm 0.4$ \\ \end{tabular} & $3.09 \pm 0.55$ \\
\hline
\btocharmparam{$\frac {{\cal{B}} (B_c^{-} \to J/\psi D_s^{*-})} {{\cal{B}} (B_c^{-} \to J/\psi D_s^{-})}$}{${{\cal{B}} (B_c^{-} \to J/\psi D_s^{*-})}/ {{\cal{B}} (B_c^{-} \to J/\psi D_s^{-})}$}{} & \begin{tabular}{l} ATLAS \cite{Aad:2015eza}: $2.8 \,^{+1.2}_{-0.8} \pm 0.3$ \\ \end{tabular} & $2.8 \,^{+1.2}_{-0.9}$ \\
\hline
\btocharmparam{$\frac {{\cal{B}} (B_c^{-} \to J/\psi D_s^{*-})} {{\cal{B}} (B_c^{-} \to J/\psi \pi^{-})}$}{${{\cal{B}} (B_c^{-} \to J/\psi D_s^{*-})}/ {{\cal{B}} (B_c^{-} \to J/\psi \pi^{-})}$}{} & \begin{tabular}{l} ATLAS \cite{Aad:2015eza}: $10.4 \pm 3.1 \pm 1.6$ \\ \end{tabular} & $10.4 \pm 3.5$ \\
\hline
\btocharmparam{$\frac{{\cal{B}} ( B_c^{-} \to J/\psi \pi^{+} \pi^{-} \pi^{-} )} {{\cal{B}} ( B_c^{-} \to J/\psi \pi^{-} )}$}{${{\cal{B}} ( B_c^{-} \to J/\psi \pi^{+} \pi^{-} \pi^{-} )}/ {{\cal{B}} ( B_c^{-} \to J/\psi \pi^{-} )}$}{} & \begin{tabular}{l} LHCb \cite{LHCb:2012ag}: $2.41 \pm 0.30 \pm 0.33$ \\ CMS \cite{Khachatryan:2014nfa}: $2.55 \pm 0.80 \,^{+0.33}_{-0.33}$ \\ \end{tabular} & $2.44 \pm 0.40$ \\
\hline
\btocharmparam{$\frac {{\cal{B}} (B_c^{-} \to J/\psi \bar{D}^{*0} K^- ) } {{\cal{B}} (B_c^{-} \to J/\psi \bar{D}^0 K^-)}$}{${{\cal{B}} (B_c^{-} \to J/\psi \bar{D}^{*0} K^- ) }/ {{\cal{B}} (B_c^{-} \to J/\psi \bar{D}^0 K^-)}$}{} & \begin{tabular}{l} LHCb \cite{Aaij:2016qlz}: $5.1 \pm 1.8 \pm 0.4$ \\ \end{tabular} & $5.1 \pm 1.8$ \\
\hline
\btocharmparam{$\frac {{\cal{B}} (B_c^{-} \to J/\psi D^{*-} \bar{K}^{*0} ) } {{\cal{B}} (B_c^{-} \to J/\psi \bar{D}^0 K^-)}$}{${{\cal{B}} (B_c^{-} \to J/\psi D^{*-} \bar{K}^{*0} ) }/ {{\cal{B}} (B_c^{-} \to J/\psi \bar{D}^0 K^-)}$}{} & \begin{tabular}{l} LHCb \cite{Aaij:2016qlz}: $2.10 \pm 1.08 \pm 0.34$ \\ \end{tabular} & $2.10 \pm 1.13$ \\
\hline
\btocharmparam{$\frac{{\cal{B}}( B_c^{-} \to J/\psi K^{-})} { {\cal{B}}( B_c^{-} \to J/\psi \pi^{-})}$}{${{\cal{B}}( B_c^{-} \to J/\psi K^{-})}/ { {\cal{B}}( B_c^{-} \to J/\psi \pi^{-})}$}{} & \begin{tabular}{l} LHCb \cite{Aaij:2013vcx}: $0.069 \pm 0.019 \pm 0.005$ \\ \end{tabular} & $0.069 \pm 0.020$ \\
\hline
\btocharmparam{$\frac{{\cal{B}} ( B_c^{-} \to J/\psi K^{-} K^{+} \pi^{-} )} { {\cal{B}} ( B_c^{-} \to J/\psi \pi^{-} )}$}{${{\cal{B}} ( B_c^{-} \to J/\psi K^{-} K^{+} \pi^{-} )}/ { {\cal{B}} ( B_c^{-} \to J/\psi \pi^{-} )}$}{} & \begin{tabular}{l} LHCb \cite{Aaij:2013gxa}: $0.53 \pm 0.10 \pm 0.05$ \\ \end{tabular} & $0.53 \pm 0.11$ \\
\hline
\btocharmparam{$\frac{{\cal{B}} ( B_c^{-} \to \psi(2S) \pi^{-} ) }  {{\cal{B}} ( B_c^{-} \to J/\psi \pi^{-} ) }$}{${{\cal{B}} ( B_c^{-} \to \psi(2S) \pi^{-} ) }/  {{\cal{B}} ( B_c^{-} \to J/\psi \pi^{-} ) }$}{} & \begin{tabular}{l} LHCb \cite{Aaij:2015xga}: $0.268 \pm 0.032 \pm 0.009$ \\ \end{tabular} & $0.268 \pm 0.033$ \\
\hline
\btocharmparam{$\frac {{\cal{B}} (B_c^{-} \to J/\psi \bar{D}^0 K^- )} {{\cal{B}} (B_c^{-} \to J/\psi \pi^-)}$}{${{\cal{B}} (B_c^{-} \to J/\psi \bar{D}^0 K^- )}/ {{\cal{B}} (B_c^{-} \to J/\psi \pi^-)}$}{} & \begin{tabular}{l} LHCb \cite{Aaij:2016qlz}: $0.432 \pm 0.136 \pm 0.028$ \\ \end{tabular} & $0.432 \pm 0.139$ \\
\hline
\btocharmparam{$\frac {{\cal{B}} (B_c^{-} \to J/\psi D^{-} \bar{K}^{*0}) } {{\cal{B}} (B_c^{-} \to J/\psi \bar{D}^0 K^-)}$}{${{\cal{B}} (B_c^{-} \to J/\psi D^{-} \bar{K}^{*0}) }/ {{\cal{B}} (B_c^{-} \to J/\psi \bar{D}^0 K^-)}$}{} & \begin{tabular}{l} LHCb \cite{Aaij:2016qlz}: $0.63 \pm 0.39 \pm 0.08$ \\ \end{tabular} & $0.63 \pm 0.40$ \\
\hline
\end{btocharmtab}
\btocharmfig{Bc_cc_1}

\begin{btocharmtab}{Bc_cc_2}{Production times branching fraction ratios}
\hline
\textbf{Parameter} & \begin{tabular}{l}\textbf{Measurements}\end{tabular} $[10^{-3}]$ & \textbf{Average} $[10^{-3}]$ \\
\hline
\hline
\btocharmparam{$\frac{{f_c \times \cal{B}} ( B_c^{-} \to J/\psi \pi^{-} )}  {f_u \times {\cal{B}} ( B^{-} \to J/\psi K^{-} )}$}{$[{{f_c \times \cal{B}} ( B_c^{-} \to J/\psi \pi^{-} )}]/[  {f_u \times {\cal{B}} ( B^{-} \to J/\psi K^{-} )}]$}{} & \begin{tabular}{l} LHCb \cite{Aaij:2014ija}: $6.83 \pm 0.18 \pm 0.09$ \\ LHCb \cite{Aaij:2012dd}: $6.8 \pm 1.0 \pm 0.6$ \\ CMS \cite{Khachatryan:2014nfa}: $4.8 \pm 0.5 \pm 0.6$ \\ \end{tabular} & $6.72 \pm 0.19$ \\
\hline
\end{btocharmtab}

\begin{btocharmtab}{Bc_cc_3}{Branching fractions times production ratios}
\hline
\textbf{Parameter} & \begin{tabular}{l}\textbf{Measurements}\end{tabular} $[10^{-6}]$ & \textbf{Average} $[10^{-6}]$ \\
\hline
\hline
\btocharmparam{${\frac{f_c}{f_u}} \times {\cal{B}} ( B_c^{-} \to \chi_{c0} \pi^{-} )$}{${\frac{f_c}{f_u}} \times {\cal{B}} ( B_c^{-} \to \chi_{c0} \pi^{-} )$}{} & \begin{tabular}{l} LHCb \cite{Aaij:2016xas}: $9.8 \,^{+3.4}_{-3.0} \pm 0.8$ \\ \end{tabular} & $9.8 \,^{+3.5}_{-3.1}$ \\
\hline
\end{btocharmtab}

\subsubsection{Decays to a $B$ meson}
\label{sec:b2c:Bc_B}
Averages of $B_c^-$ decays to a $B$ meson are shown in Table~\ref{tab:b2c:Bc_B_1}.
\begin{btocharmtab}{Bc_B_1}{Branching fractions to $B_s^{0}$ meson}
\hline
\textbf{Parameter} & \begin{tabular}{l}\textbf{Measurements}\end{tabular} $[10^{-3}]$ & \textbf{Average} $[10^{-3}]$ \\
\hline
\hline
\btocharmparam{$\frac{f_c}{f_s}  \times {\cal B} ( B_c^{+} \to  B_s^{0}\pi^{+} )$}{$[{f_c}/{f_s}]  \times {\cal B} ( B_c^{+} \to  B_s^{0}\pi^{+} )$}{} & \begin{tabular}{l} LHCb \cite{Aaij:2013cda}: $2.37 \pm 0.31 \,^{+0.20}_{-0.17}$ \\ \end{tabular} & $2.37 \,^{+0.37}_{-0.35}$ \\
\hline
\end{btocharmtab}

\subsection{Decays of $b$ baryons}
\label{sec:b2c:Bbaryon}
Measurements of $b$ baryons decays to charmed hadrons are summarized in Sections~\ref{sec:b2c:Bbaryon_D} to~\ref{sec:b2c:Bbaryon_other}.
Comments regarding the production rates of $\bar{B}_s^0$ and $B_c^-$ mesons relative to lighter $B$ mesons, in Sec.~\ref{sec:b2c:Bs} and Sec.~\ref{sec:b2c:Bc} respectively, are also appropriate here.
Specifically, since the cross-section for production of $\Lb$ baryons is reasonably well-known, it is possible to determine absolute or relative branching fractions for its decays (although some older measurements are presented as products involving the cross-section).
The cross-sections for production of heavier $b$ baryons are not known, and therefore measured quantities are presented as absolute or relative branching fraction multiplied by a ratio of cross-sections (or, equivalently, production fractions).

\subsubsection{Decays to a single open charm meson}
\label{sec:b2c:Bbaryon_D}
Averages of $b$ baryons decays to a single open charm meson are shown in Table~\ref{tab:b2c:Bbaryon_D_1}.
\begin{btocharmtab}{Bbaryon_D_1}{Branching fraction ratios to $D^{0}$ mesons}
\hline
\textbf{Parameter} & \begin{tabular}{l}\textbf{Measurements}\end{tabular} & \textbf{Average} \\
\hline
\hline
\btocharmparam{$\frac{{\cal{B}} ( \Lambda_b^{0} \to D^{0} p K^{-} ) }{ {\cal{B}} ( \Lambda_b^{0} \to D^{0} p \pi^{-} )}$}{${{\cal{B}} ( \Lambda_b^{0} \to D^{0} p K^{-} ) }/{ {\cal{B}} ( \Lambda_b^{0} \to D^{0} p \pi^{-} )}$}{} & \begin{tabular}{l} LHCb \cite{Aaij:2013pka}: $0.073 \pm 0.008 \,^{+0.005}_{-0.006}$ \\ \end{tabular} & $0.073 \,^{+0.009}_{-0.010}$ \\
\hline
\btocharmparaml{$\frac{ {\cal{B}} ( \Lambda_b^{0} \to D^{0} p \pi^{-})  \times {\cal{B}} ( D^{0} \to K^{+} \pi^{-} )} {  {\cal{B}} ( \Lambda_b^{0} \to \Lambda_c^{+} \pi^{-} )   \times   {\cal{B}}  (\Lambda_c^{+} \to  p K^{-} \pi^{+}) }$}{$[{ {\cal{B}} ( \Lambda_b^{0} \to D^{0} p \pi^{-})  \times {\cal{B}} ( D^{0} \to K^{+} \pi^{-} )}]/[ {  {\cal{B}} ( \Lambda_b^{0} \to \Lambda_c^{+} \pi^{-} )   \times   {\cal{B}}  (\Lambda_c^{+} \to  p K^{-} \pi^{+}) }]$}{} & \begin{tabular}{l} LHCb \cite{Aaij:2013pka}: $0.0806 \pm 0.0023 \pm 0.0035$ \\ \end{tabular} & $0.0806 \pm 0.0042$ \\
\hline
\btocharmparam{$\frac{ f_{\Xi_b^{0}} \times{\cal{B}} ( \Xi_b^{0} \to D^{0} p K^{-} )   } { f_{\Lambda_b^{0}} \times{\cal{B}} ( \Lambda_b^{0} \to D^{0} p K^{-})  }$}{$[{ f_{\Xi_b^{0}} \times{\cal{B}} ( \Xi_b^{0} \to D^{0} p K^{-} )   }]/[ { f_{\Lambda_b^{0}} \times{\cal{B}} ( \Lambda_b^{0} \to D^{0} p K^{-})  }]$}{} & \begin{tabular}{l} LHCb \cite{Aaij:2013pka}: $0.44 \pm 0.09 \pm 0.06$ \\ \end{tabular} & $0.44 \pm 0.11$ \\
\hline
\end{btocharmtab}
\btocharmfig{Bbaryon_D_1}

\subsubsection{Decays to charmonium states}
\label{sec:b2c:Bbaryon_cc}
Averages of $b$ baryons decays to charmonium states are shown in Tables~\ref{tab:b2c:Bbaryon_cc_1}--\ref{tab:b2c:Bbaryon_cc_6}.
\begin{btocharmtab}{Bbaryon_cc_1}{$\Lambda_b^{0}$ branching fractions to charmonium}
\hline
\textbf{Parameter} & \begin{tabular}{l}\textbf{Measurements}\end{tabular} $[10^{-4}]$ & \textbf{Average} $[10^{-4}]$ \\
\hline
\hline
\btocharmparam{${\cal{B}} ( \Lambda_b^{0} \to J/\psi p K^{-} )$}{${\cal{B}} ( \Lambda_b^{0} \to J/\psi p K^{-} )$}{} & \begin{tabular}{l} LHCb \cite{Aaij:2015fea}: $3.17 \pm 0.04 \,^{+0.46}_{-0.29}$ \\ \end{tabular} & $3.17 \,^{+0.46}_{-0.29}$ \\
\hline
\btocharmparam{${\cal{B}} ( \Lambda_b^{0} \to J/\psi \Lambda )$}{${\cal{B}} ( \Lambda_b^{0} \to J/\psi \Lambda )$}{} & \begin{tabular}{l} CDF \cite{Abe:1996tr}: $4.7 \pm 2.1 \pm 1.9$ \\ \end{tabular} & $4.7 \pm 2.8$ \\
\hline
\end{btocharmtab}
\btocharmfig{Bbaryon_cc_1}

\begin{btocharmtab}{Bbaryon_cc_2}{Production times branching fraction to charmonium}
\hline
\textbf{Parameter} & \begin{tabular}{l}\textbf{Measurements}\end{tabular} $[10^{-5}]$ & \textbf{Average} $[10^{-5}]$ \\
\hline
\hline
\btocharmparam{${f_{\Lambda_b} \times \cal{B}} ( \Lambda_b^{0} \to J/\psi \Lambda )$}{${f_{\Lambda_b} \times \cal{B}} ( \Lambda_b^{0} \to J/\psi \Lambda )$}{} & \begin{tabular}{l} \dzero \cite{Abazov:2011wt}: $6.01 \pm 0.60 \pm 0.64$ \\ \end{tabular} & $6.01 \pm 0.88$ \\
\hline
\end{btocharmtab}

\begin{btocharmtab}{Bbaryon_cc_3}{$\Lambda_b^{0}$ branching fraction ratios}
\hline
\textbf{Parameter} & \begin{tabular}{l}\textbf{Measurements}\end{tabular} & \textbf{Average} \\
\hline
\hline
\btocharmparam{$\frac{{\cal{B}} ( \Lambda_b^{0} \to \psi(2S) \Lambda )}{{\cal{B}} ( \Lambda_b^{0} \to J/\psi \Lambda )}$}{${{\cal{B}} ( \Lambda_b^{0} \to \psi(2S) \Lambda )}/{{\cal{B}} ( \Lambda_b^{0} \to J/\psi \Lambda )}$}{} & \begin{tabular}{l} ATLAS \cite{Aad:2015msa}: $0.501 \pm 0.033 \pm 0.019$ \\ \end{tabular} & $0.501 \pm 0.038$ \\
\hline
\btocharmparam{$\frac{{\cal{B}} ( \Lambda_b^{0} \to J/\psi p \pi^{-} )}{{\cal{B}} ( \Lambda_b^{0} \to J/\psi p K^{-} )}$}{${{\cal{B}} ( \Lambda_b^{0} \to J/\psi p \pi^{-} )}/{{\cal{B}} ( \Lambda_b^{0} \to J/\psi p K^{-} )}$}{} & \begin{tabular}{l} LHCb \cite{Aaij:2014zoa}: $0.0824 \pm 0.0025 \pm 0.0042$ \\ \end{tabular} & $0.0824 \pm 0.0049$ \\
\hline
\btocharmparam{$\frac{{\cal{B}} ( \Lambda_b^{0} \to J/\psi \pi^{+} \pi^{-} p K^{-} )}{{\cal{B}} ( \Lambda_b^{0} \to J/\psi p K^{-} )}$}{${{\cal{B}} ( \Lambda_b^{0} \to J/\psi \pi^{+} \pi^{-} p K^{-} )}/{{\cal{B}} ( \Lambda_b^{0} \to J/\psi p K^{-} )}$}{} & \begin{tabular}{l} LHCb \cite{Aaij:2016wxd}: $0.2086 \pm 0.0096 \pm 0.0134$ \\ \end{tabular} & $0.2086 \pm 0.0165$ \\
\hline
\btocharmparam{$\frac{{\cal{B}} ( \Lambda_b^{0} \to \psi(2S)  p K^{-} )}{{\cal{B}} ( \Lambda_b^{0} \to J/\psi p K^{-} )}$}{${{\cal{B}} ( \Lambda_b^{0} \to \psi(2S)  p K^{-} )}/{{\cal{B}} ( \Lambda_b^{0} \to J/\psi p K^{-} )}$}{} & \begin{tabular}{l} LHCb \cite{Aaij:2016wxd}: $0.2070 \pm 0.0076 \pm 0.0059$ \\ \end{tabular} & $0.2070 \pm 0.0096$ \\
\hline
\btocharmparam{$\frac{ {\cal{B}}( \Lambda_b^0 \to \chi_{c1} p K^- ) }{ {\cal{B}}( \Lambda_b^0 \to J/\psi p K^- ) }$}{${ {\cal{B}}( \Lambda_b^0 \to \chi_{c1} p K^- ) }/{ {\cal{B}}( \Lambda_b^0 \to J/\psi p K^- ) }$}{} & \begin{tabular}{l} LHCb \cite{Aaij:2017awb}: $0.242 \pm 0.014 \pm 0.016$ \\ \end{tabular} & $0.242 \pm 0.021$ \\
\hline
\btocharmparam{$\frac{ {\cal{B}}( \Lambda_b^0 \to \chi_{c2} p K^- ) }{ {\cal{B}}( \Lambda_b^0 \to J/\psi p K^- ) }$}{${ {\cal{B}}( \Lambda_b^0 \to \chi_{c2} p K^- ) }/{ {\cal{B}}( \Lambda_b^0 \to J/\psi p K^- ) }$}{} & \begin{tabular}{l} LHCb \cite{Aaij:2017awb}: $0.248 \pm 0.020 \pm 0.017$ \\ \end{tabular} & $0.248 \pm 0.026$ \\
\hline
\btocharmparam{$\frac{ {\cal{B}}( \Lambda_b^0 \to \chi_{c2} p K^- ) }{ {\cal{B}}( \Lambda_b^0 \to \chi_{c1} p K^- ) }$}{${ {\cal{B}}( \Lambda_b^0 \to \chi_{c2} p K^- ) }/{ {\cal{B}}( \Lambda_b^0 \to \chi_{c1} p K^- ) }$}{} & \begin{tabular}{l} LHCb \cite{Aaij:2017awb}: $1.02 \pm 0.10 \pm 0.05$ \\ \end{tabular} & $1.02 \pm 0.11$ \\
\hline
\end{btocharmtab}
\btocharmfig{Bbaryon_cc_3}

\begin{btocharmtab}{Bbaryon_cc_4}{$\Xi_b^{-}$ and $\Omega_b^{-}$ production times branching fraction ratios to charmonium}
\hline
\textbf{Parameter} & \begin{tabular}{l}\textbf{Measurements}\end{tabular} & \textbf{Average} \\
\hline
\hline
\btocharmparam{$\frac{f_{\Xi_b^{-}} \times {\cal{B}} ( \Xi_b^{-} \to J/\psi \Xi^{-} ) }{ f_{\Lambda_b^{0}} \times {\cal{B}} ( \Lambda_b^{0} \to J/\psi \Lambda )}$}{$[{f_{\Xi_b^{-}} \times {\cal{B}} ( \Xi_b^{-} \to J/\psi \Xi^{-} ) }]/[{ f_{\Lambda_b^{0}} \times {\cal{B}} ( \Lambda_b^{0} \to J/\psi \Lambda )}]$}{} & \begin{tabular}{l} CDF \cite{Aaltonen:2009ny}: $0.167 \,^{+0.037}_{-0.025} \pm 0.012$ \\ \end{tabular} & $0.167 \,^{+0.039}_{-0.028}$ \\
\hline
\btocharmparam{$\frac{f_{\Omega_b^{-}} \times {\cal{B}} ( \Omega_b^{-} \to J/\psi \Omega^{-} ) }{ f_{\Lambda_b^{0}}  \times   {\cal{B}} ( \Lambda_b^{0} \to J/\psi \Lambda )}$}{$[{f_{\Omega_b^{-}} \times {\cal{B}} ( \Omega_b^{-} \to J/\psi \Omega^{-} ) }]/[{ f_{\Lambda_b^{0}}  \times   {\cal{B}} ( \Lambda_b^{0} \to J/\psi \Lambda )}]$}{} & \begin{tabular}{l} CDF \cite{Aaltonen:2009ny}: $0.045 \,^{+0.017}_{-0.012} \pm 0.004$ \\ \end{tabular} & $0.045 \,^{+0.017}_{-0.013}$ \\
\hline
\end{btocharmtab}
\btocharmfig{Bbaryon_cc_4}

\begin{btocharmtab}{Bbaryon_cc_5}{Transverse polarization of $\Lambda_b^{0}$ produced in $pp$ collisions}
\hline
\textbf{Parameter} & \begin{tabular}{l}\textbf{Measurements}\end{tabular} & \textbf{Average} \\
\hline
\hline
\btocharmparam{${\cal{P}}_b ( \Lambda_b^{0} \to J/\psi \Lambda )$}{${\cal{P}}_b ( \Lambda_b^{0} \to J/\psi \Lambda )$}{} & \begin{tabular}{l} LHCb \cite{Aaij:2013oxa}: $0.06 \pm 0.07 \pm 0.02$ \\ CMS \cite{Sirunyan:2018bfd}: $0.00 \pm 0.06 \pm 0.06$ \\ \end{tabular} & $0.03 \pm 0.06$ \\
\hline
\end{btocharmtab}

\begin{btocharmtab}{Bbaryon_cc_6}{Parity-violating asymmetry in $\Lambda_b^{0}$ decays to charmonium}
\hline
\textbf{Parameter} & \begin{tabular}{l}\textbf{Measurements}\end{tabular} & \textbf{Average} \\
\hline
\hline
\btocharmparam{$\alpha_b ( \Lambda_b^{0} \to J/\psi \Lambda )$}{$\alpha_b ( \Lambda_b^{0} \to J/\psi \Lambda )$}{} & \begin{tabular}{l} LHCb \cite{Aaij:2013oxa}: $0.05 \pm 0.17 \pm 0.07$ \\ CMS \cite{Sirunyan:2018bfd}: $-0.14 \pm 0.14 \pm 0.10$ \\ ATLAS \cite{Aad:2014iba}: $0.30 \pm 0.16 \pm 0.06$ \\ \end{tabular} & $0.07 \pm 0.10$ \\
\hline
\end{btocharmtab}

\subsubsection{Decays to charm baryons}
\label{sec:b2c:Bbaryon_baryon}
Averages of $b$ baryons decays to charm baryons are shown in Tables~\ref{tab:b2c:Bbaryon_baryon_1}--\ref{tab:b2c:Bbaryon_baryon_5}.
\begin{btocharmtab}{Bbaryon_baryon_1}{$\Lambda_b$ branching fractions}
\hline
\textbf{Parameter} & \begin{tabular}{l}\textbf{Measurements}\end{tabular} $[10^{-2}]$ & \textbf{Average} $[10^{-2}]$ \\
\hline
\hline
\btocharmparam{${\cal{B}} ( \Lambda_b^{0} \to \Lambda_c^{+} \pi^{-} )$}{${\cal{B}} ( \Lambda_b^{0} \to \Lambda_c^{+} \pi^{-} )$}{} & \begin{tabular}{l} LHCb \cite{Aaij:2014jyk}: $0.430 \pm 0.003 \,^{+0.036}_{-0.035}$ \\ \end{tabular} & $0.430 \,^{+0.036}_{-0.035}$ \\
\hline
\btocharmparam{${\cal{B}} ( \Lambda_b^{0} \to \Lambda_c^{+} \pi^{+} \pi^{-} \pi^{-} )$}{${\cal{B}} ( \Lambda_b^{0} \to \Lambda_c^{+} \pi^{+} \pi^{-} \pi^{-} )$}{} & \begin{tabular}{l} CDF \cite{CDF:2011aa}: $2.68 \pm 0.29 \,^{+1.15}_{-1.09}$ \\ \end{tabular} & $2.68 \,^{+1.19}_{-1.12}$ \\
\hline
\end{btocharmtab}
\btocharmfig{Bbaryon_baryon_1}

\begin{btocharmtab}{Bbaryon_baryon_2}{Branching fraction ratios, I}
\hline
\textbf{Parameter} & \begin{tabular}{l}\textbf{Measurements}\end{tabular} & \textbf{Average} \\
\hline
\hline
\btocharmparam{$\frac{{\cal{B}} ( \Lambda_b^{0} \to \Lambda_c^{+} \pi^{-} )}{{\cal{B}} ( \bar{B}^{0} \to D^{+} \pi^{-} )}$}{${{\cal{B}} ( \Lambda_b^{0} \to \Lambda_c^{+} \pi^{-} )}/{{\cal{B}} ( \bar{B}^{0} \to D^{+} \pi^{-} )}$}{} & \begin{tabular}{l} CDF \cite{Abulencia:2006df}: $3.3 \pm 0.3 \pm 1.2$ \\ \end{tabular} & $3.3 \pm 1.2$ \\
\hline
\btocharmparam{$\frac{{\cal{B}} ( \Lambda_b^{0} \to \Lambda_c^{+} \pi^{+} \pi^{-} \pi^{-} ) }{ {\cal{B}} ( \Lambda_b^{0} \to \Lambda_c^{+} \pi^{-} )}$}{${{\cal{B}} ( \Lambda_b^{0} \to \Lambda_c^{+} \pi^{+} \pi^{-} \pi^{-} ) }/{ {\cal{B}} ( \Lambda_b^{0} \to \Lambda_c^{+} \pi^{-} )}$}{} & \begin{tabular}{l} LHCb \cite{Aaij:2011rj}: $1.43 \pm 0.16 \pm 0.13$ \\ CDF \cite{CDF:2011aa}: $3.04 \pm 0.33 \,^{+0.70}_{-0.55}$ \\ \end{tabular} & $1.58 \pm 0.20$ \\
\hline
\btocharmparam{$\frac{ {\cal{B}} ( \Xi_b^{0} \to \Lambda_c^{+} K^{-} )   \times   {\cal{B}}  (\Lambda_c^{+} \to  p K^{-} \pi^{+}) } { {\cal{B}} ( \Xi_b^{0} \to D^{0} p K^{-})  \times {\cal{B}} ( D^{0} \to K^{+} \pi^{-} ) }$}{$[{ {\cal{B}} ( \Xi_b^{0} \to \Lambda_c^{+} K^{-} )   \times   {\cal{B}}  (\Lambda_c^{+} \to  p K^{-} \pi^{+}) }]/[ { {\cal{B}} ( \Xi_b^{0} \to D^{0} p K^{-})  \times {\cal{B}} ( D^{0} \to K^{+} \pi^{-} ) }]$}{} & \begin{tabular}{l} LHCb \cite{Aaij:2013pka}: $0.57 \pm 0.22 \pm 0.21$ \\ \end{tabular} & $0.57 \pm 0.30$ \\
\hline
\btocharmparam{$\frac{{\cal{B}} ( \Lambda_b^{0} \to \Lambda_c(2860)^{+} \pi^{-} ) \times {\cal{B}} ( \Lambda_c(2860)^{+} \to D^0 p ) }{{\cal{B}} ( \Lambda_b^{0} \to \Lambda_c(2880)^{+} \pi^{-} ) \times {\cal{B}} ( \Lambda_c(2880)^{+} \to D^0 p ) }$}{$[{{\cal{B}} ( \Lambda_b^{0} \to \Lambda_c(2860)^{+} \pi^{-} ) \times {\cal{B}} ( \Lambda_c(2860)^{+} \to D^0 p ) }]/[{{\cal{B}} ( \Lambda_b^{0} \to \Lambda_c(2880)^{+} \pi^{-} ) \times {\cal{B}} ( \Lambda_c(2880)^{+} \to D^0 p ) }]$}{} & \begin{tabular}{l} LHCb \cite{Aaij:2017vbw}: $4.51 \,^{+0.51}_{-0.39} \,^{+0.21}_{-0.45}$ \\ \end{tabular} & $4.51 \,^{+0.55}_{-0.59}$ \\
\hline
\btocharmparam{$\frac{{\cal{B}} ( \Lambda_b^{0} \to \Lambda_c(2940)^{+} \pi^{-} ) \times {\cal{B}} ( \Lambda_c(2940)^{+} \to D^0 p ) }{{\cal{B}} ( \Lambda_b^{0} \to \Lambda_c(2880)^{+} \pi^{-} ) \times {\cal{B}} ( \Lambda_c(2880)^{+} \to D^0 p ) }$}{$[{{\cal{B}} ( \Lambda_b^{0} \to \Lambda_c(2940)^{+} \pi^{-} ) \times {\cal{B}} ( \Lambda_c(2940)^{+} \to D^0 p ) }]/[{{\cal{B}} ( \Lambda_b^{0} \to \Lambda_c(2880)^{+} \pi^{-} ) \times {\cal{B}} ( \Lambda_c(2880)^{+} \to D^0 p ) }]$}{} & \begin{tabular}{l} LHCb \cite{Aaij:2017vbw}: $0.83 \,^{+0.31}_{-0.10} \,^{+0.18}_{-0.43}$ \\ \end{tabular} & $0.83 \,^{+0.36}_{-0.45}$ \\
\hline
\end{btocharmtab}
\btocharmfig{Bbaryon_baryon_2}

\begin{btocharmtab}{Bbaryon_baryon_3}{Branching fraction ratios, II}
\hline
\textbf{Parameter} & \begin{tabular}{l}\textbf{Measurements}\end{tabular} $[10^{-2}]$ & \textbf{Average} $[10^{-2}]$ \\
\hline
\hline
\btocharmparam{$\frac{ {\cal{B}} ( \Lambda_b^{0} \to \Lambda_c^{+} K^{-} )  }{ {\cal{B}} ( \Lambda_b^{0} \to \Lambda_c^{+} \pi^{-} )  }$}{${ {\cal{B}} ( \Lambda_b^{0} \to \Lambda_c^{+} K^{-} )  }/{ {\cal{B}} ( \Lambda_b^{0} \to \Lambda_c^{+} \pi^{-} )  }$}{} & \begin{tabular}{l} LHCb \cite{Aaij:2013pka}: $7.31 \pm 0.16 \pm 0.16$ \\ \end{tabular} & $7.31 \pm 0.23$ \\
\hline
\btocharmparam{$\frac{ {\cal{B}} ( \Lambda_b^{0} \to \Lambda_c^{+} D^{-} )  }{ {\cal{B}} ( \Lambda_b^{0} \to \Lambda_c^{+} D_s^{-} )  }$}{${ {\cal{B}} ( \Lambda_b^{0} \to \Lambda_c^{+} D^{-} )  }/{ {\cal{B}} ( \Lambda_b^{0} \to \Lambda_c^{+} D_s^{-} )  }$}{} & \begin{tabular}{l} LHCb \cite{Aaij:2014pha}: $4.2 \pm 0.3 \pm 0.3$ \\ \end{tabular} & $4.2 \pm 0.4$ \\
\hline
\btocharmparam{$\frac{ {\cal{B}} ( \Lambda_b^{0} \to \Lambda_c^{+} p \bar{p} \pi^{-} )  }{ {\cal{B}} ( \Lambda_b^{0} \to \Lambda_c^{+} \pi^{-} )  }$}{${ {\cal{B}} ( \Lambda_b^{0} \to \Lambda_c^{+} p \bar{p} \pi^{-} )  }/{ {\cal{B}} ( \Lambda_b^{0} \to \Lambda_c^{+} \pi^{-} )  }$}{} & \begin{tabular}{l} LHCb \cite{Aaij:2018bre}: $5.40 \pm 0.23 \pm 0.32$ \\ \end{tabular} & $5.40 \pm 0.39$ \\
\hline
\btocharmparam{$\frac{ {\cal{B}} ( \Lambda_b^{0} \to \Sigma_c(2455)^{0} p \bar{p} ) \times {\cal{B}} ( \Sigma_c(2455)^0 \to \Lambda_c^{+} \pi^{-} )  }{ {\cal{B}} ( \Lambda_b^{0} \to \Lambda_c^{+} p \bar{p} \pi^{-} ) }$}{$[{ {\cal{B}} ( \Lambda_b^{0} \to \Sigma_c(2455)^{0} p \bar{p} ) \times {\cal{B}} ( \Sigma_c(2455)^0 \to \Lambda_c^{+} \pi^{-} )  }]/{ {\cal{B}} ( \Lambda_b^{0} \to \Lambda_c^{+} p \bar{p} \pi^{-} ) }$}{} & \begin{tabular}{l} LHCb \cite{Aaij:2018bre}: $8.9 \pm 1.5 \pm 0.6$ \\ \end{tabular} & $8.9 \pm 1.6$ \\
\hline
\btocharmparaml{$\frac { {\cal{B}} ( \Lambda_b^{0} \to \Sigma_c(2520)^{*0} p \bar{p} ) \times {\cal{B}} ( \Sigma_c(2520)^{*0} \to \Lambda_c^{+} \pi^{-} )  }{ {\cal{B}} ( \Lambda_b^{0} \to \Lambda_c^{+} p \bar{p} \pi^{-} ) }$}{$[{ {\cal{B}} ( \Lambda_b^{0} \to \Sigma_c(2520)^{*0} p \bar{p} ) \times {\cal{B}} ( \Sigma_c(2520)^{*0} \to \Lambda_c^{+} \pi^{-} )  }]/{ {\cal{B}} ( \Lambda_b^{0} \to \Lambda_c^{+} p \bar{p} \pi^{-} ) }$}{} & \begin{tabular}{l} LHCb \cite{Aaij:2018bre}: $11.9 \pm 2.0 \pm 1.4$ \\ \end{tabular} & $11.9 \pm 2.4$ \\
\hline
\end{btocharmtab}
\btocharmfig{Bbaryon_baryon_3}

\begin{btocharmtab}{Bbaryon_baryon_4}{Branching fraction ratios, III}
\hline
\textbf{Parameter} & \begin{tabular}{l}\textbf{Measurements}\end{tabular} & \textbf{Average} \\
\hline
\hline
\btocharmparaml{$\frac{{\cal{B}} ( \Lambda_b^{0} \to \Lambda_c(2595)^{+} \pi^{-} ) \times {\cal{B}} ( \Lambda_c(2595)^{+} \to \Lambda_c^{+} \pi^{+} \pi^{-} ) }{ {\cal{B}} ( \Lambda_b^{0} \to \Lambda_c^{+} \pi^{+} \pi^{-} \pi^{-} )}$}{$[{{\cal{B}} ( \Lambda_b^{0} \to \Lambda_c(2595)^{+} \pi^{-} ) \times {\cal{B}} ( \Lambda_c(2595)^{+} \to \Lambda_c^{+} \pi^{+} \pi^{-} ) }]/{ {\cal{B}} ( \Lambda_b^{0} \to \Lambda_c^{+} \pi^{+} \pi^{-} \pi^{-} )}$}{} & \begin{tabular}{l} LHCb \cite{Aaij:2011rj}: $0.044 \pm 0.017 \,^{+0.006}_{-0.004}$ \\ \end{tabular} & $0.044 \,^{+0.018}_{-0.017}$ \\
\hline
\btocharmparaml{$\frac{{\cal{B}} ( \Lambda_b^{0} \to \Lambda_c(2625)^{+} \pi^{-} ) \times {\cal{B}} ( \Lambda_c(2625)^{+} \to \Lambda_c^{+} \pi^{+} \pi^{-} ) }{ {\cal{B}} ( \Lambda_b^{0} \to \Lambda_c^{+} \pi^{+} \pi^{-} \pi^{-} )}$}{$[{{\cal{B}} ( \Lambda_b^{0} \to \Lambda_c(2625)^{+} \pi^{-} ) \times {\cal{B}} ( \Lambda_c(2625)^{+} \to \Lambda_c^{+} \pi^{+} \pi^{-} ) }]/{ {\cal{B}} ( \Lambda_b^{0} \to \Lambda_c^{+} \pi^{+} \pi^{-} \pi^{-} )}$}{} & \begin{tabular}{l} LHCb \cite{Aaij:2011rj}: $0.043 \pm 0.015 \pm 0.004$ \\ \end{tabular} & $0.043 \pm 0.016$ \\
\hline
\btocharmparaml{$\frac{{\cal{B}} ( \Lambda_b^{0} \to \Sigma_c^{0} \pi^{+} \pi^{-} ) \times {\cal{B}} ( \Sigma_c^{0} \to \Lambda_c^{+} \pi^{-} ) }{ {\cal{B}} ( \Lambda_b^{0} \to \Lambda_c^{+} \pi^{+} \pi^{-} \pi^{-} )}$}{$[{{\cal{B}} ( \Lambda_b^{0} \to \Sigma_c^{0} \pi^{+} \pi^{-} ) \times {\cal{B}} ( \Sigma_c^{0} \to \Lambda_c^{+} \pi^{-} ) }]/{ {\cal{B}} ( \Lambda_b^{0} \to \Lambda_c^{+} \pi^{+} \pi^{-} \pi^{-} )}$}{} & \begin{tabular}{l} LHCb \cite{Aaij:2011rj}: $0.074 \pm 0.024 \pm 0.012$ \\ \end{tabular} & $0.074 \pm 0.027$ \\
\hline
\btocharmparaml{$\frac{{\cal{B}} ( \Lambda_b^{0} \to \Sigma_c^{++} \pi^{-} \pi^{-} ) \times {\cal{B}} ( \Sigma_c^{++} \to \Lambda_c^{+} \pi^{+} ) }{ {\cal{B}} ( \Lambda_b^{0} \to \Lambda_c^{+} \pi^{+} \pi^{-} \pi^{-} )}$}{$[{{\cal{B}} ( \Lambda_b^{0} \to \Sigma_c^{++} \pi^{-} \pi^{-} ) \times {\cal{B}} ( \Sigma_c^{++} \to \Lambda_c^{+} \pi^{+} ) }]/{ {\cal{B}} ( \Lambda_b^{0} \to \Lambda_c^{+} \pi^{+} \pi^{-} \pi^{-} )}$}{} & \begin{tabular}{l} LHCb \cite{Aaij:2011rj}: $0.042 \pm 0.018 \pm 0.007$ \\ \end{tabular} & $0.042 \pm 0.019$ \\
\hline
\end{btocharmtab}
\btocharmfig{Bbaryon_baryon_4}

\begin{btocharmtab}{Bbaryon_baryon_5}{$\Xi_b$ branching fractions}
\hline
\textbf{Parameter} & \begin{tabular}{l}\textbf{Measurements}\end{tabular} $[10^{-4}]$ & \textbf{Average} $[10^{-4}]$ \\
\hline
\hline
\btocharmparam{$\frac{f_{\Xi_b^{-}}}{f_{\Lambda_b^{0}}} \times {\cal{B}} ( \Xi_b^{-} \to \Lambda_b^{0}\pi^{-})$}{$[{f_{\Xi_b^{-}}}/{f_{\Lambda_b^{0}}}] \times {\cal{B}} ( \Xi_b^{-} \to \Lambda_b^{0}\pi^{-})$}{} & \begin{tabular}{l} LHCb \cite{Aaij:2015yoy}: $5.7 \pm 1.8 \,^{+0.8}_{-0.9}$ \\ \end{tabular} & $5.7 \,^{+2.0}_{-2.0}$ \\
\hline
\end{btocharmtab}

\subsubsection{Decays to $XYZP$ states}
\label{sec:b2c:Bbaryon_other}
Averages of $b$ baryons decays to $XYZP$ states are shown in Table~\ref{tab:b2c:Bbaryon_other_1}.
\begin{btocharmtab}{Bbaryon_other_1}{Branching fraction ratios involving pentaquarks}
\hline
\textbf{Parameter} & \begin{tabular}{l}\textbf{Measurements}\end{tabular} & \textbf{Average} \\
\hline
\hline
\btocharmparam{$\frac{ {\cal{B}} ( \Lambda_b^0 \to \pi^- P_c(4380)^+ ) }{ {\cal{B}} ( \Lambda_b^0 \to K^- P_c(4380)^+ ) }$}{${ {\cal{B}} ( \Lambda_b^0 \to \pi^- P_c(4380)^+ ) }/{ {\cal{B}} ( \Lambda_b^0 \to K^- P_c(4380)^+ ) }$}{} & \begin{tabular}{l} LHCb \cite{Aaij:2016ymb}: $0.050 \pm 0.016 \,^{+0.036}_{-0.030}$ \\ \end{tabular} & $0.050 \,^{+0.039}_{-0.034}$ \\
\hline
\btocharmparam{$\frac{ {\cal{B}} ( \Lambda_b^0 \to \pi^- P_c(4450)^+ ) }{ {\cal{B}} ( \Lambda_b^0 \to K^- P_c(4450)^+ ) }$}{${ {\cal{B}} ( \Lambda_b^0 \to \pi^- P_c(4450)^+ ) }/{ {\cal{B}} ( \Lambda_b^0 \to K^- P_c(4450)^+ ) }$}{} & \begin{tabular}{l} LHCb \cite{Aaij:2016ymb}: $0.033 \,^{+0.016}_{-0.014} \,^{+0.014}_{-0.013}$ \\ \end{tabular} & $0.033 \,^{+0.021}_{-0.019}$ \\
\hline
\btocharmparam{$\frac{ {\cal{B}} ( \Lambda_b^0 \to \pi^- P_c(4380)^+ ) }{ {\cal{B}} ( \Lambda_b^0 \to K^- J/\psi p ) }$}{${ {\cal{B}} ( \Lambda_b^0 \to \pi^- P_c(4380)^+ ) }/{ {\cal{B}} ( \Lambda_b^0 \to K^- J/\psi p ) }$}{} & \begin{tabular}{l} LHCb \cite{Aaij:2016ymb}: $0.051 \pm 0.015 \,^{+0.026}_{-0.016}$ \\ \end{tabular} & $0.051 \,^{+0.030}_{-0.022}$ \\
\hline
\btocharmparam{$\frac{ {\cal{B}} ( \Lambda_b^0 \to \pi^- P_c(4450)^+ ) }{ {\cal{B}} ( \Lambda_b^0 \to K^- J/\psi p ) }$}{${ {\cal{B}} ( \Lambda_b^0 \to \pi^- P_c(4450)^+ ) }/{ {\cal{B}} ( \Lambda_b^0 \to K^- J/\psi p ) }$}{} & \begin{tabular}{l} LHCb \cite{Aaij:2016ymb}: $0.016 \,^{+0.008}_{-0.006} \,^{+0.006}_{-0.005}$ \\ \end{tabular} & $0.016 \,^{+0.010}_{-0.008}$ \\
\hline
\btocharmparam{$\frac{ {\cal{B}} ( \Lambda_b^0 \to p Z_c(4200)^- ) }{ {\cal{B}} ( \Lambda_b^0 \to K^- J/\psi p ) }$}{${ {\cal{B}} ( \Lambda_b^0 \to p Z_c(4200)^- ) }/{ {\cal{B}} ( \Lambda_b^0 \to K^- J/\psi p ) }$}{} & \begin{tabular}{l} LHCb \cite{Aaij:2016ymb}: $0.077 \pm 0.028 \,^{+0.034}_{-0.040}$ \\ \end{tabular} & $0.077 \,^{+0.044}_{-0.049}$ \\
\hline
\end{btocharmtab}
\btocharmfig{Bbaryon_other_1}

\clearpage
\mysection{$B$ decays to charmless final states}

\label{sec:rare}

This section provides branching fractions (BF), polarization 
fractions, partial rate asymmetries ($A_{\CP}$) and other observables of 
$B$ decays to final states that do not contain charm hadrons or charmonia mesons.
The order of entries in the tables corresponds to that in
the 2017 Review of Particle Physics
(PDG2017)~\cite{PDG_2016}, and the quoted RPP numbers are the PDG numbers of the corresponding branching fractions.
The \CP asymmetry is defined as
\begin{equation}
	A_{\CP} = \frac{N_b - N_{\bbar}}{N_b + N_{\bbar}},
\end{equation}
where $N_b$ ($N_{\bbar}$) is the number of hadrons containing a $b$ ($\bbar$) quark 
decaying into a specific final state.
This definition is consistent with that of Eq.~(\ref{eq:cp_uta:pra}) in Sec.~\ref{sec:cp_uta:notations:pra}.
Four different $\Bz$ and $\Bp$ decay categories are considered: 
charmless mesonic (\ie, final states containing only mesons), baryonic (only hadrons, but including a baryon-antibaryon pair), radiative (including a photon or a lepton-antilepton pair) and semileptonic/leptonic (including/only leptons).
We also include measurements of $\Bs$, $\Bc$ and $b$-baryon decays.
Results from  $A_{\CP}$ measurements  obtained from time-dependent analyses 
are listed and described in Sec.~\ref{sec:cp_uta}.
Measurements supported with public notes are accepted in  
the averages; public notes include journal papers, 
conference contributed papers, preprints or conference proceedings.
In all the tables of this section, values in red (blue) are new published (preliminary) results since PDG2017 (considering the publication status at the time of the closing of this report, September 2018).

Most of the branching fractions from \babar\ and Belle assume equal production 
of charged and neutral $B$ pairs.  The best measurements to date show that this
is still a reasonable approximation (see Sec.~\ref{sec:life_mix}).
For branching fractions, we provide either averages or the most stringent upper limits.
If one or more experiments have measurements with a significance of more than three standard deviations ($\sigma$) for a decay channel, all available central values for that channel are used in the averaging. The most stringent limit will be used for branching fractions that do not satisfy this criterion.
For $A_{\CP}$ we provide averages in all cases. 
At the end of some of the tables we give a list of results that were not
included. Typical cases are the measurements of distributions, such as differential
branching fractions or longitudinal polarizations, which are measured in different
binning schemes by the different collaborations, and thus cannot be directly
used to obtain averages.

Our averaging is performed by maximizing the likelihood,
   $\displaystyle {\mathcal L} = \prod_i {\mathcal P}_i(x),$  
where ${\mathcal P_i}$ is the probability density function (PDF) of the
$i^{\rm th}$  measurement, and $x$ is, \eg, the branching fraction or $A_{\CP}$.
The PDF is modelled by an asymmetric Gaussian function with the measured
central value as its most probable value and the quadratic sum of the statistical
and systematic errors, eventually asymmetric, as the standard deviations on both sides of the central value. The experimental
uncertainties of results from different experiments are assumed to be uncorrelated with each other when the 
averaging is performed. As mentioned in Sec.~\ref{sec:method}, no error scaling is applied when the fit $\chi^2$ is 
greater than 1,
except for cases of extreme disagreement (at present we have no such cases).

The largest improvement since the last report has come from the inclusion of a
variety of new measurements from the LHC, especially LHCb. The
measurements of $\Bs$ decays are particularly noteworthy.

Sections \ref{sec:rare-charmless} and \ref{sec:rare-bary} provide compilations of branching fractions of $\Bz$ and $\Bp$ to mesonic and baryonic charmless final states, respectively, 
while Secs.~\ref{sec:rare-lb} and~\ref{sec:rare-bs} give branching fractions of $b$-baryon and \Bs-meson charmless decays, respectively.
In Sec.~\ref{sec:rare-radll} various observables of interest are given in addition to branching fractions. These observables are related to
radiative decays and FCNC decays with leptons of \Bz and \Bp\ mesons, including
limits from searches for lepton-flavour/number-violating decays.
Sections~\ref{sec:rare-acp} and \ref{sec:rare-polar} give \CP\ asymmetries and results of polarization measurements, respectively, in various $b$-hadron charmless decays.
Finally, Sec.~\ref{sec:rare-bc} gives branching fractions of $\Bc$ meson decays to charmless final states.

\mysubsection{Mesonic decays of \Bz and \Bp\ mesons}
\label{sec:rare-charmless}

This section provides branching fractions of charmless mesonic decays:
Tables~\ref{tab:charmless_BpFirst} to \ref{tab:charmless_BpLast} for \Bp\
and Tables~\ref{tab:charmless_BdFirst} to
\ref{tab:charmless_BdLast}
for \Bz\ mesons.
The tables are separated according to the presence or absence of strange mesons in the final state. 
Finally, Table~\ref{tab:charmless_Ratio} details several relative branching fractions of \Bz decays.

Figure~\ref{fig:rare-mostprec} gives a graphic representation of a selection of high-precision branching fractions given in this section.
Footnote symbols indicate that the footnote in the corresponding table should be consulted.

\input{rare/charmless}

\begin{figure}[htbp!]
\centering
\includegraphics[width=0.50\textwidth]{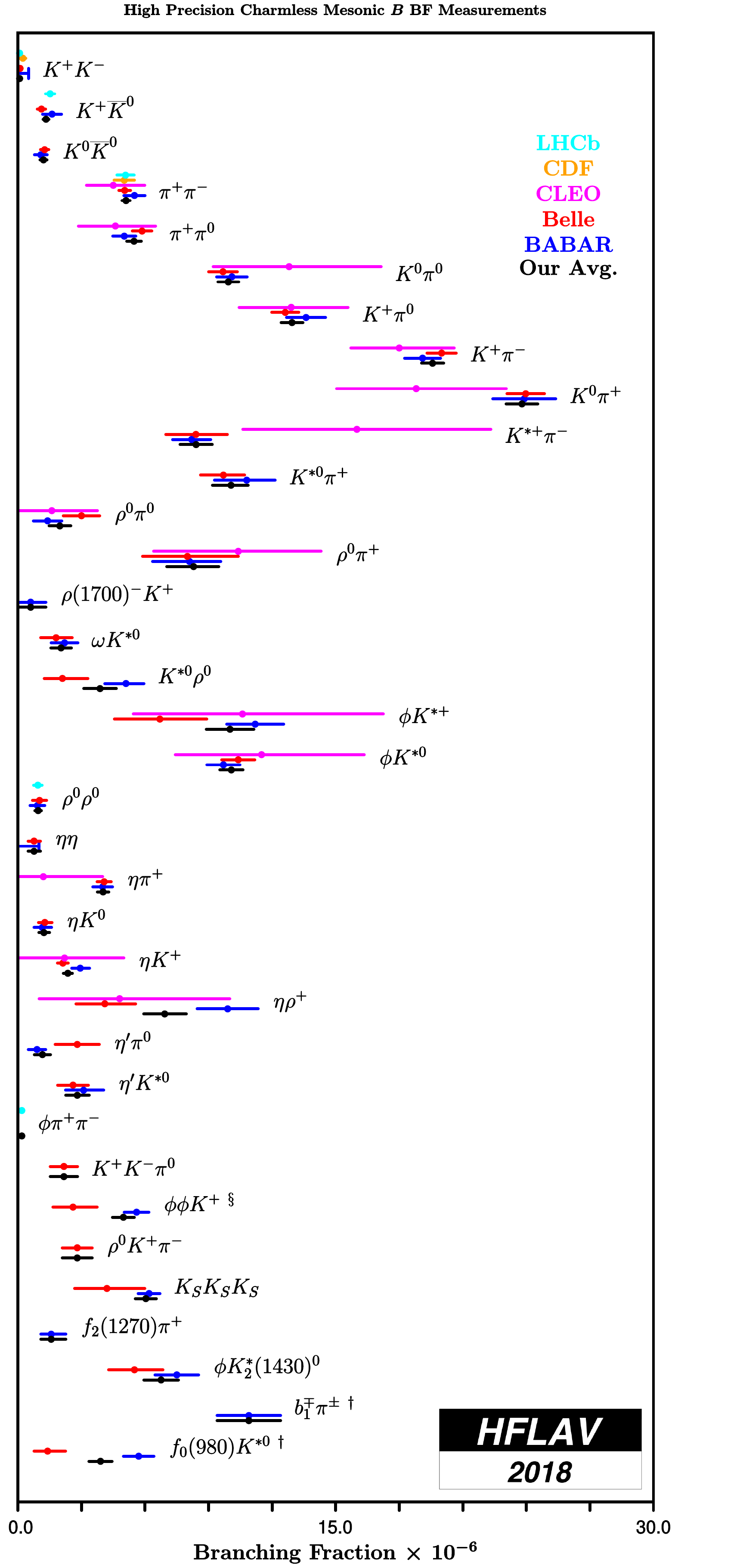}
\caption{Selection of high-precision charmless mesonic $B$ meson branching fraction measurements.}
\label{fig:rare-mostprec}
\end{figure}

\clearpage

\mysubsection{Baryonic decays of \Bp\ and \Bz mesons}
\label{sec:rare-bary}

This section provides branching fractions of charmless baryonic decays of \Bp\ and \Bz mesons in Tables~\ref{tab:bary_Bp} and~\ref{tab:bary_Bz}, respectively. Relative branching fractions are given in Table~\ref{tab:bary_Ratio}.

Figures~\ref{fig:rare-baryns} and~\ref{fig:rare-bary} show graphic representations of a selection of results given in this section.
Footnote symbols indicate that the footnote in the corresponding table should be consulted.

\begin{table}[!htbp]
\begin{center}
\caption{Branching fractions of charmless baryonic  
$\Bp$ decays in units of $\times 10^{-6}$. Upper limits are
at 90\% CL.
Where values are shown in \red{red} (\blue{blue}), this indicates that
they are new \red{published} (\blue{preliminary}) results since PDG2017.}
\label{tab:bary_Bp}
\resizebox{\textwidth}{!}{
\begin{tabular}{|lccc @{}c c @{}c c @{}c c|}
\sgline
RPP\#   & Mode & PDG2017 Avg. & \babar & & Belle & & LHCb & & Our Avg. \\
\sgline
484                                               & %
$p \overline p \pi^+$                             & %
$1.62\pm0.20$                                     & %
{$\err{1.69}{0.29}{0.26}~^\dag$}                  & %
\ifref {\cite{Aubert:2007qea}} \fi \phantom{.}    & %
{$\aerr{1.60}{0.22}{0.19}{0.12}$}                 & %
\ifref {\cite{Wei:2007fg}} \fi \phantom{.}        & %
\nodata                                           & %
\phantom{.}                                       & %
$\cerr{1.62}{0.21}{0.20}$                         \\

484                                               & %
$p \overline p \pi^+~^\S$                         & %
\nodata                                           & %
\nodata                                           & %
\phantom{.}                                       & %
\nodata                                           & %
\phantom{.}                                       & %
$\err{1.07}{0.11}{0.11}$                          & %
\ifref {\cite{Aaij:2014tua}} \fi \phantom{.}      & %
$1.07 \pm 0.16$                                   \\

487                                               & %
$p \overline p K^+$                               & %
$5.9\pm0.5$                                       & %
{$\err{6.7}{0.5}{0.4}~^\dag$}                     & %
\ifref {\cite{Aubert:2005gw}} \fi \phantom{.}     & %
{$\aerr{5.54}{0.27}{0.25}{0.36}$}                 & %
\ifref {\cite{Wei:2007fg}} \fi \phantom{.}        & %
\red{$\err{4.46}{0.21}{0.27}$}~$^\P$              & %
\ifref {\cite{Aaij:2013rha}} \fi \phantom{.}      & %
$5.14 \pm 0.25$                                   \\

488                                               & %
$\Theta^{++} \overline p$ $^1$                    & %
$<0.091$                                          & %
{$<0.09$}                                         & %
\ifref {\cite{Aubert:2005gw}} \fi \phantom{.}     & %
{$<0.091$}                                        & %
\ifref {\cite{Wang:2005fc}} \fi \phantom{.}       & %
\nodata                                           & %
\phantom{.}                                       & %
{$<0.09$}                                         \\

489                                               & %
$f_J(2220) K^+$ $^2$                              & %
$<0.41$                                           & %
\nodata                                           & %
\phantom{.}                                       & %
{$<0.41$}                                         & %
\ifref {\cite{Wang:2005fc}} \fi \phantom{.}       & %
\nodata                                           & %
\phantom{.}                                       & %
{$<0.41$}                                         \\

490                                               & %
$p \overline\Lambda(1520)$                        & %
$0.31\pm0.06$                                     & %
{$< 1.5$}                                         & %
\ifref {\cite{Aubert:2005gw}} \fi \phantom{.}     & %
\nodata                                           & %
\phantom{.}                                       & %
\err{0.315}{0.048}{0.027}                         & %
\ifref {\cite{Aaij:2014tua}} \fi \phantom{.}      & %
$0.315 \pm 0.055$                                 \\

492                                               & %
$p \overline p K^{*+}$                            & %
$\cerr{3.6}{0.8}{0.7}$                            & %
{$\err{5.3}{1.5}{1.3}~^\dag$}                     & %
\ifref {\cite{Aubert:2007qea}} \fi \phantom{.}    & %
{$\aerr{3.38}{0.73}{0.60}{0.39}~^\ddag$}          & %
\ifref {\cite{Chen:2008jy}} \fi \phantom{.}       & %
\nodata                                           & %
\phantom{.}                                       & %
$\cerr{3.64}{0.79}{0.70}$                         \\

493                                               & %
$f_J(2220) K^{*+}$ $^2$                           & %
$<0.77$                                           & %
{$<0.77$}                                         & %
\ifref {\cite{Aubert:2007qea}} \fi \phantom{.}    & %
\nodata                                           & %
\phantom{.}                                       & %
\nodata                                           & %
\phantom{.}                                       & %
{$<0.77$}                                         \\

494                                               & %
$p \overline\Lambda$                              & %
$<0.32$                                           & %
\nodata                                           & %
\phantom{.}                                       & %
{$< 0.32$}                                        & %
\ifref {\cite{Tsai:2007pp}} \fi \phantom{.}       & %
\red{$\aerr{0.24}{0.10}{0.08}{0.03}$}             & %
\ifref {\cite{Aaij:2016xfa}} \fi \phantom{.}      & %
$\cerr{0.24}{0.10}{0.09}$                         \\

496                                               & %
$p \overline\Lambda \pi^0$                        & %
$\cerr{3.00}{0.7}{0.6}$                           & %
\nodata                                           & %
\phantom{.}                                       & %
{$\aerr{3.00}{0.61}{0.53}{0.33}$}                 & %
\ifref {\cite{Wang:2007as}} \fi \phantom{.}       & %
\nodata                                           & %
\phantom{.}                                       & %
$\cerr{3.00}{0.69}{0.62}$                         \\

497                                               & %
$p \overline\Sigma(1385)^0$                       & %
$<0.47$                                           & %
\nodata                                           & %
\phantom{.}                                       & %
$<0.47$                                           & %
\ifref {\cite{Wang:2007as}} \fi \phantom{.}       & %
\nodata                                           & %
\phantom{.}                                       & %
$<0.47$                                           \\

498                                               & %
$\Delta^+\overline \Lambda$                       & %
$<0.82$                                           & %
\nodata                                           & %
\phantom{.}                                       & %
$<0.82$                                           & %
\ifref {\cite{Wang:2007as}} \fi \phantom{.}       & %
\nodata                                           & %
\phantom{.}                                       & %
$<0.82$                                           \\

500                                               & %
$p \overline{\Lambda} \pi^+\pi^-$ (NR)            & %
$5.9\pm1.1$                                       & %
\nodata                                           & %
\phantom{.}                                       & %
$\aerr{5.92}{0.88}{0.84}{0.69}$                   & %
\ifref {\cite{Chen:2009xg}} \fi \phantom{.}       & %
\nodata                                           & %
\phantom{.}                                       & %
$\cerr{5.92}{1.12}{1.09}$                         \\

501                                               & %
$p \overline{\Lambda} \rho^0$                     & %
$4.8\pm0.9$                                       & %
\nodata                                           & %
\phantom{.}                                       & %
$\aerr{4.78}{0.67}{0.64}{0.60}$                   & %
\ifref {\cite{Chen:2009xg}} \fi \phantom{.}       & %
\nodata                                           & %
\phantom{.}                                       & %
$\cerr{4.78}{0.90}{0.88}$                         \\

502                                               & %
$p \overline{\Lambda} f_2(1270)$                  & %
$2.0\pm0.8$                                       & %
\nodata                                           & %
\phantom{.}                                       & %
$\aerr{2.03}{0.77}{0.72}{0.27}$                   & %
\ifref {\cite{Chen:2009xg}} \fi \phantom{.}       & %
\nodata                                           & %
\phantom{.}                                       & %
$\cerr{2.03}{0.82}{0.77}$                         \\

503                                               & %
$\Lambda \overline{\Lambda} \pi^+$                & %
$<0.94$                                           & %
\nodata                                           & %
\phantom{.}                                       & %
$<0.94~\S$                                        & %
\ifref {\cite{Chang:2008yw}} \fi \phantom{.}      & %
\nodata                                           & %
\phantom{.}                                       & %
$<0.94~\S$                                        \\

504                                               & %
$\Lambda \overline{\Lambda} K^+$                  & %
$3.4\pm0.6$                                       & %
\nodata                                           & %
\phantom{.}                                       & %
$\aerr{3.38}{0.41}{0.36}{0.41}~^\ddag$            & %
\ifref {\cite{Chang:2008yw}} \fi \phantom{.}      & %
\nodata                                           & %
\phantom{.}                                       & %
$\cerr{3.38}{0.58}{0.55}$                         \\

505                                               & %
$\Lambda \overline{\Lambda} K^{*+}$               & %
$\cerr{2.2}{1.2}{0.9}$                            & %
\nodata                                           & %
\phantom{.}                                       & %
$\aerr{2.19}{1.13}{0.88}{0.33}~^\S$               & %
\ifref {\cite{Chang:2008yw}} \fi \phantom{.}      & %
\nodata                                           & %
\phantom{.}                                       & %
$\cerr{2.19}{1.18}{0.94}$                         \\

506                                               & %
$\overline{\Delta}^0 p$                           & %
$<1.38$                                           & %
\nodata                                           & %
\phantom{.}                                       & %
{$<1.38$} $^\S$                                   & %
\ifref {\cite{Wei:2007fg}} \fi \phantom{.}        & %
\nodata                                           & %
\phantom{.}                                       & %
{$<1.38$} $^\S$                                   \\

507                                               & %
$\Delta^{++} \overline p$                         & %
$<0.14$                                           & %
\nodata                                           & %
\phantom{.}                                       & %
{$<0.14$} $^\S$                                   & %
\ifref {\cite{Wei:2007fg}} \fi \phantom{.}        & %
\nodata                                           & %
\phantom{.}                                       & %
{$<0.14$} $^\S$                                   \\

\nodata                                           & %
$p \overline{\Lambda} K^+ K^-$ (NR)               & %
\nodata                                           & %
\nodata                                           & %
\phantom{.}                                       & %
$\aerr{4.10}{0.45}{0.43}{0.50}$                   & %
\ifref {\cite{Lu:2018qbw}} \fi \phantom{.}        & %
\nodata                                           & %
\phantom{.}                                       & %
$\cerr{4.10}{0.67}{0.66}$                         \\

\nodata                                           & %
$\bar{p} \Lambda K^+ K^-$ (NR)                    & %
\nodata                                           & %
\nodata                                           & %
\phantom{.}                                       & %
$\aerr{3.70}{0.39}{0.37}{0.44}$                   & %
\ifref {\cite{Lu:2018qbw}} \fi \phantom{.}        & %
\nodata                                           & %
\phantom{.}                                       & %
$\cerr{3.70}{0.59}{0.57}$                         \\

\nodata                                           & %
$p \bar{\Lambda} \phi$                            & %
\nodata                                           & %
\nodata                                           & %
\phantom{.}                                       & %
$\err{0.795}{0.209}{0.077}$                       & %
\ifref {\cite{Lu:2018qbw}} \fi \phantom{.}        & %
\nodata                                           & %
\phantom{.}                                       & %
$0.795 \pm 0.223$                                 \\

\nodata                                           & %
$\Lambda(1520) \bar{\Lambda} K^+$                 & %
\nodata                                           & %
\nodata                                           & %
\phantom{.}                                       & %
$\err{2.23}{0.63}{0.25}$                          & %
\ifref {\cite{Lu:2018qbw}} \fi \phantom{.}        & %
\nodata                                           & %
\phantom{.}                                       & %
$2.23 \pm 0.68$                                   \\

\nodata                                           & %
$\bar{\Lambda}(1520) \Lambda K^+$                 & %
\nodata                                           & %
\nodata                                           & %
\phantom{.}                                       & %
$<2.08$                                           & %
\ifref {\cite{Lu:2018qbw}} \fi \phantom{.}        & %
\nodata                                           & %
\phantom{.}                                       & %
$<2.08$                                           \\

\hline
\end{tabular}
}
\end{center}
\scriptsize
Channels with no RPP\# were not included in PDG Live as of Dec. 31, 2017. \\ %
Results for LHCb are relative BFs converted to absolute BFs.\\    %
$^\dag$ Charmonium decays to $p\bar p$ have been statistically subtracted.\\    %
$^\ddag$ The charmonium mass region has been vetoed.\\    %
$^\S$~Di-baryon mass is less than 2.85~\gevcc.\\    %
$^\P$~Includes contribution where $p \bar{p}$ is produced in charmonia decays.\\    %
$^1$~$\Theta(1540)^{++}\to K^+p$ (pentaquark candidate). \\    %
$^2$~In this product of BFs, all daughter BFs not shown are set to 100\%.     %
\end{table}
\clearpage

\begin{table}[!htbp]
\begin{center}
\caption{Branching fractions of charmless baryonic  
$\Bz$ decays in units of $\times 10^{-6}$. Upper limits are
at 90\% CL. 
Where values are shown in \red{red} (\blue{blue}), this indicates that
they are new \red{published} (\blue{preliminary}) results since PDG2017.}
\label{tab:bary_Bz}
\resizebox{\textwidth}{!}{
\begin{tabular}{|lccc @{}c c @{}c c @{}c c|}
\sgline
RPP\#   & Mode & PDG2017 Avg. & \babar & & Belle & & LHCb & & Our Avg. \\
\sgline

439                                               & %
$p \overline{p}$                                  & %
$\cerr{0.015}{0.007}{0.005}$                      & %
{$<0.27$}                                         & %
\ifref {\cite{Aubert:2004fy}} \fi \phantom{.}     & %
{$<0.11$}                                         & %
\ifref {\cite{Tsai:2007pp}} \fi \phantom{.}       & %
\red{\err{0.0125}{0.0027}{0.0018}}                & %
\ifref {\cite{Aaij:2017gum}} \fi \phantom{.}      & %
$0.0130 \pm 0.0030$                               \\

440                                               & %
$p \overline p \pi^+ \pi^-$                       & %
$<250$                                            & %
\nodata                                           & %
\phantom{.}                                       & %
\nodata                                           & %
\phantom{.}                                       & %
\red{\gerrsyt{2.7}{0.1}{0.1}{0.2}}                & %
\ifref {\cite{Aaij:2017pgn}} \fi \phantom{.}      & %
$2.7 \pm 0.2$                                     \\

441                                               & %
$p \overline{p} K^0$                              & %
$2.66\pm0.32$                                     & %
{$\err{3.0}{0.5}{0.3}~^\dag$}                     & %
\ifref {\cite{Aubert:2007qea}} \fi \phantom{.}    & %
$\aerr{2.51}{0.35}{0.29}{0.21}~^\ddag$            & %
\ifref {\cite{Chen:2008jy}} \fi \phantom{.}       & %
\nodata                                           & %
\phantom{.}                                       & %
$\cerr{2.66}{0.34}{0.32}$                         \\

442                                               & %
$\Theta^+ \overline{p}$~$^\S$                     & %
$<0.05$                                           & %
{$<0.05$}                                         & %
\ifref {\cite{Aubert:2007qea}} \fi \phantom{.}    & %
{$<0.23$}                                         & %
\ifref {\cite{Wang:2005fc}} \fi \phantom{.}       & %
\nodata                                           & %
\phantom{.}                                       & %
{$<0.05$}                                         \\

443                                               & %
$f_J(2220) K^0$~$^\P$                             & %
$<0.45$                                           & %
{$<0.45$}                                         & %
\ifref {\cite{Aubert:2007qea}} \fi \phantom{.}    & %
\nodata                                           & %
\phantom{.}                                       & %
\nodata                                           & %
\phantom{.}                                       & %
{$<0.45$}                                         \\

444                                               & %
$p \overline{p} K^{*0}$                           & %
$\cerr{1.24}{0.28}{0.25}$                         & %
{$\err{1.47}{0.45}{0.40}~^\dag$}                  & %
\ifref {\cite{Aubert:2007qea}} \fi \phantom{.}    & %
$\aerr{1.18}{0.29}{0.25}{0.11}~^\ddag$            & %
\ifref {\cite{Chen:2008jy}} \fi \phantom{.}       & %
\nodata                                           & %
\phantom{.}                                       & %
$\cerr{1.24}{0.28}{0.25}$                         \\

445                                               & %
$f_J(2220) K^{*0}$~$^\P$                          & %
$<0.15$                                           & %
{$<0.15$}                                         & %
\ifref {\cite{Aubert:2007qea}} \fi \phantom{.}    & %
\nodata                                           & %
\phantom{.}                                       & %
\nodata                                           & %
\phantom{.}                                       & %
{$<0.15$}                                         \\

446                                               & %
$p \overline\Lambda \pi^-$                        & %
$3.14\pm0.29$                                     & %
$\err{3.07}{0.31}{0.23}$                          & %
\ifref {\cite{Aubert:2009am}} \fi \phantom{.}     & %
{$\aerr{3.23}{0.33}{0.29}{0.29}$}                 & %
\ifref {\cite{Wang:2007as}} \fi \phantom{.}       & %
\nodata                                           & %
\phantom{.}                                       & %
$\cerr{3.14}{0.29}{0.28}$                         \\

448                                               & %
$p \overline\Sigma(1385)^-$                       & %
$<0.26$                                           & %
\nodata                                           & %
\phantom{.}                                       & %
$<0.26$                                           & %
\ifref {\cite{Wang:2007as}} \fi \phantom{.}       & %
\nodata                                           & %
\phantom{.}                                       & %
$<0.26$                                           \\

449                                               & %
$\Delta^0 \overline\Lambda$                       & %
$<0.93$                                           & %
\nodata                                           & %
\phantom{.}                                       & %
$<0.93$                                           & %
\ifref {\cite{Wang:2007as}} \fi \phantom{.}       & %
\nodata                                           & %
\phantom{.}                                       & %
$<0.93$                                           \\

450                                               & %
$p \overline\Lambda K^-$                          & %
$<0.82$                                           & %
\nodata                                           & %
\phantom{.}                                       & %
$< 0.82$                                          & %
\ifref {\cite{Wang:2003yi}} \fi \phantom{.}       & %
\nodata                                           & %
\phantom{.}                                       & %
$< 0.82$                                          \\

453                                               & %
$p \overline\Sigma^0 \pi^-$                       & %
$<3.8$                                            & %
\nodata                                           & %
\phantom{.}                                       & %
$< 3.8$                                           & %
\ifref {\cite{Wang:2003yi}} \fi \phantom{.}       & %
\nodata                                           & %
\phantom{.}                                       & %
$< 3.8$                                           \\

454                                               & %
$\overline\Lambda \Lambda$                        & %
$<0.32$                                           & %
\nodata                                           & %
\phantom{.}                                       & %
{$<0.32$}                                         & %
\ifref {\cite{Tsai:2007pp}} \fi \phantom{.}       & %
\nodata                                           & %
\phantom{.}                                       & %
{$<0.32$}                                         \\

455                                               & %
$\overline\Lambda \Lambda K^0$                    & %
$\cerr{4.8}{1.0}{0.9}$                            & %
\nodata                                           & %
\phantom{.}                                       & %
$\aerr{4.76}{0.84}{0.68}{0.61}~^\ddag$            & %
\ifref {\cite{Chang:2008yw}} \fi \phantom{.}      & %
\nodata                                           & %
\phantom{.}                                       & %
$\cerr{4.76}{1.04}{0.91}$                         \\

456                                               & %
$\Lambda \overline{\Lambda} K^{*0}$               & %
$\cerr{2.5}{0.9}{0.8}$                            & %
\nodata                                           & %
\phantom{.}                                       & %
$\aerr{2.46}{0.87}{0.72}{0.34}~^\ddag$            & %
\ifref {\cite{Chang:2008yw}} \fi \phantom{.}      & %
\nodata                                           & %
\phantom{.}                                       & %
$\cerr{2.46}{0.93}{0.80}$                         \\

\nodata                                           & %
$p \overline p K^+ K^-$                           & %
\nodata                                           & %
\nodata                                           & %
\phantom{.}                                       & %
\nodata                                           & %
\phantom{.}                                       & %
\gerrsyt{0.113}{0.028}{0.011}{0.008}              & %
\ifref {\cite{Aaij:2017pgn}} \fi \phantom{.}      & %
$0.113 \pm 0.031$                                 \\

\nodata                                           & %
$p \overline p K^+ \pi^-$                         & %
\nodata                                           & %
\nodata                                           & %
\phantom{.}                                       & %
\nodata                                           & %
\phantom{.}                                       & %
\gerrsyt{5.9}{0.3}{0.3}{0.4}                      & %
\ifref {\cite{Aaij:2017pgn}} \fi \phantom{.}      & %
$5.9 \pm 0.6$                                     \\

\nodata                                           & %
$p p \overline p \overline p$                     & %
\nodata                                           & %
$<0.20$                                           & %
\ifref {\cite{BABAR:2018erd}} \fi \phantom{.}     & %
\nodata                                           & %
\phantom{.}                                       & %
\nodata                                           & %
\phantom{.}                                       & %
$<0.20$                                           \\

\hline
\end{tabular}
}
\end{center}
\scriptsize
Channels with no RPP\# were not included in PDG Live as of Dec. 31, 2017. \\ %
Results for LHCb are relative BFs converted to absolute BFs.\\    %
$^\dag$ Charmonium decays to $p\bar p$ have been statistically subtracted.\\    %
$^\ddag$ The charmonium mass region has been vetoed.\\    %
$^\S~\Theta(1540)^+\to p K^0$ (pentaquark candidate).\\    %
$^\P$~In this product of BFs, all daughter BFs not shown are set to 100\%.     %
\end{table}

\begin{table}[!htbp]
\begin{center}
\caption{Relative branching fractions of charmless baryonic  
$\B$ decays.
Where values are shown in \red{red} (\blue{blue}), this indicates that
they are new \red{published} (\blue{preliminary}) results since PDG2017.}
\label{tab:bary_Ratio}
\resizebox{\textwidth}{!}{
\begin{tabular}{|lccc @{}c c|} \hline
RPP\# & Mode & PDG2017 Avg. & LHCb & & Our Avg.  \\ \sglinespb
\nodata                                           & %
$\mathcal{B}(B^+\rightarrow p \overline p \pi^+, m_{p \overline p}<2.85~\gevcc)/\mathcal{B}(B^+\rightarrow J/\psi(\to p \bar{p})\pi^+)$& %
\nodata                                           & %
$\err{12.0}{1.2}{0.3}$                            & %
\ifref {\cite{Aaij:2014tua}} \fi \phantom{.}      & %
$12.0 \pm 1.2$                                    \\

\nodata                                           & %
$\mathcal{B}(B^+\rightarrow p \overline p K^+)/\mathcal{B}(B^+\rightarrow J/\psi(\to p \bar{p})K^+)$& %
\nodata                                           & %
$\err{4.91}{0.19}{0.14}~^\dag$                    & %
\ifref {\cite{Aaij:2013rha}} \fi \phantom{.}      & %
$4.91 \pm 0.24$                                   \\

487                                               & %
$\mathcal{B}(B^+\rightarrow p \overline p K^+)/\mathcal{B}(B^+\rightarrow J/\psi K^+)$& %
$\err{0.0104}{0.0005}{0.0001}$                    & %
$\err{0.0104}{0.0005}{0.0001}$~$^{\dag\ddag}$     & %
\ifref {\cite{Aaij:2013rha}} \fi \phantom{.}      & %
$0.0100 \pm 0.0010$                               \\

\nodata                                           & %
$\mathcal{B}(B^+\rightarrow \overline\Lambda(1520)(\to K^+\bar{p}) p)/\mathcal{B}(B^+\rightarrow J/\psi(\to p \bar{p})\pi^+)$& %
\nodata                                           & %
$\err{0.033}{0.005}{0.007}$                       & %
\ifref {\cite{Aaij:2014tua}} \fi \phantom{.}      & %
$0.033 \pm 0.009$                                 \\

\nodata                                           & %
$\mathcal{B}(B^0\rightarrow p \overline p K^+ K^-)/\mathcal{B}(B^0\rightarrow p \overline p K^+ \pi^-)$& %
\nodata                                           & %
$\err{0.019}{0.005}{0.002}$                       & %
\ifref {\cite{Aaij:2017pgn}} \fi \phantom{.}      & %
$0.019 \pm 0.005$                                 \\

\nodata                                           & %
$\mathcal{B}(B^0\rightarrow p \overline p \pi^+ \pi^-)/\mathcal{B}(B^0\rightarrow p \overline p K^+ \pi^-)$& %
\nodata                                           & %
$\err{0.46}{0.02}{0.02}$                          & %
\ifref {\cite{Aaij:2017pgn}} \fi \phantom{.}      & %
$0.46 \pm 0.03$                                   \\

\sglinespt
\end{tabular}
}
\end{center}
\scriptsize
Channels with no RPP\# were not included in PDG Live as of Dec. 31, 2017. \\ %
$^\dag$~Includes contribution where $p \bar{p}$ is produced in charmonia decays.\\
$^\ddag$~Original experimental relative BF multiplied by the best values (PDG2014) of certain reference BFs. The first error is experimental, and the second is from the reference BFs.\\
\end{table}

\begin{figure}[htbp!]
\centering
\includegraphics[width=0.5\textwidth]{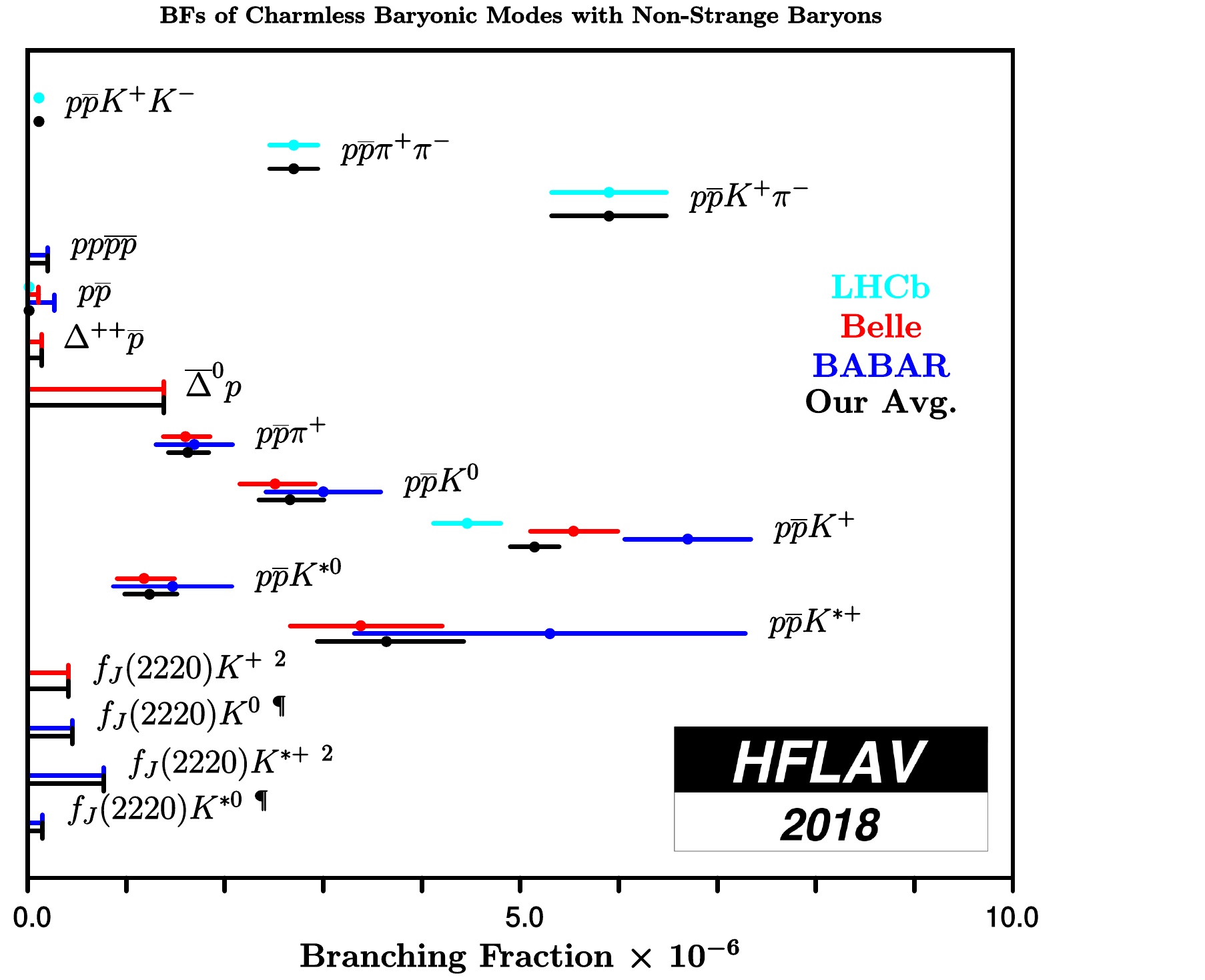}
\caption{Branching fractions of charmless baryonic modes with non-strange baryons.}
\label{fig:rare-baryns}
\end{figure}

\begin{figure}[htbp!]
\centering
\includegraphics[width=0.5\textwidth]{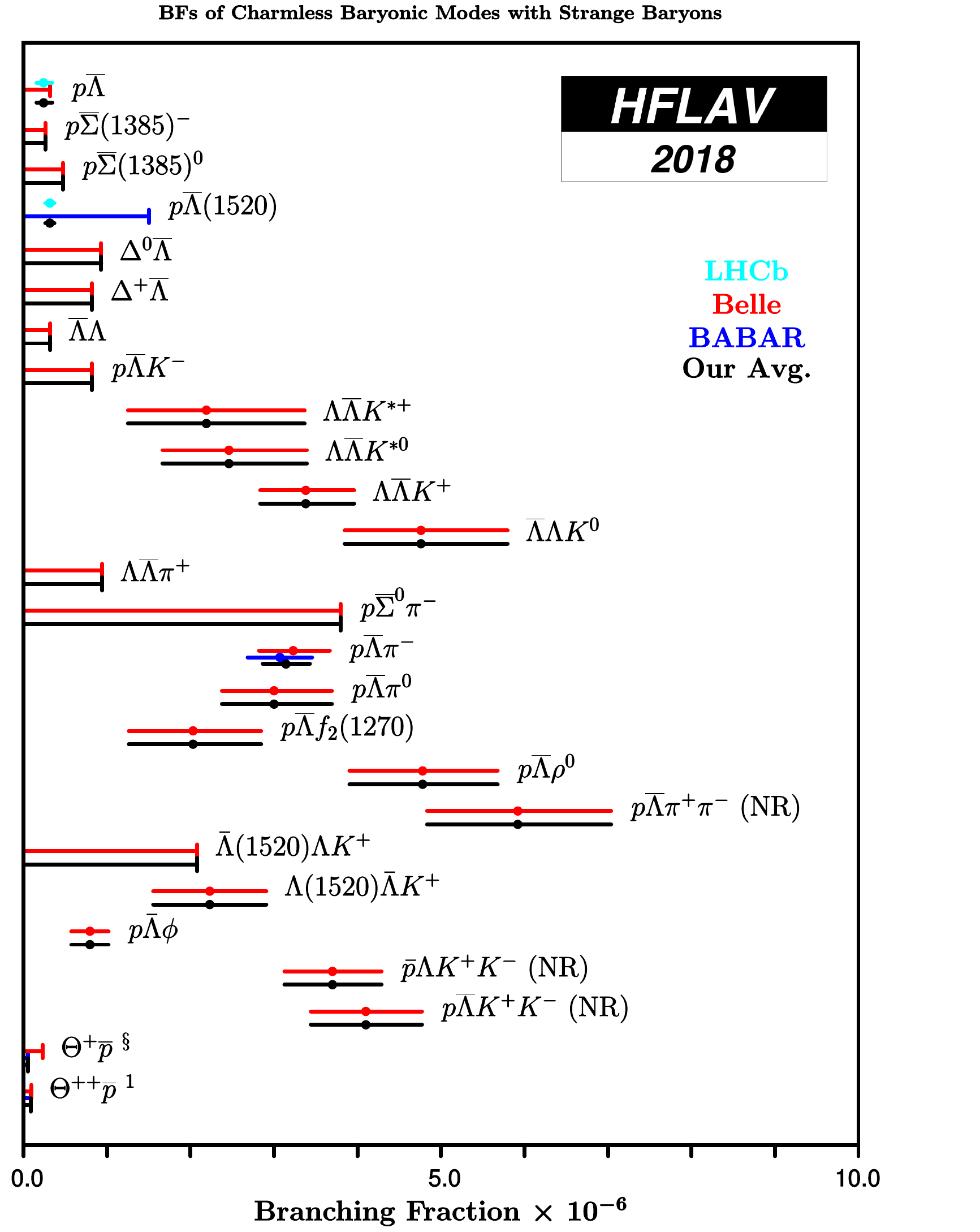}
\caption{Branching fractions of charmless baryonic modes with strange baryons.}
\label{fig:rare-bary}
\end{figure}

\clearpage

\mysubsection{Decays of \b baryons}
\label{sec:rare-lb}

A compilation of branching fractions of $\Lb$ baryon decays is given in Table~\ref{tab:bbaryons_Lb}. Table~\ref{tab:bbaryons_LbPartialBF} provides the partial branching fractions of $\Lb\to\Lambda\mup\mun$ decays in intervals of $q^2=m^2(\mu^+\mu^-)$. Compilations of branching fractions of $\Xi^{0}_{b}$, $\Xi^{-}_{b}$ and $\Omega^{-}_{b}$ baryon decays are given in Tables~\ref{tab:bbaryons_Xhibz}, \ref{tab:bbaryons_Xhibm}, and~\ref{tab:bbaryons_Omegabm}, respectively.

Figure~\ref{fig:rare-lb} shows a graphic representation of branching fractions of $\Lb$ decays.
Footnote symbols indicate that the footnote in the corresponding table should be consulted.

\begin{table}[h!]
\begin{center}
\caption{Branching fractions of charmless $\Lb$ decays in units of
$\times 10^{-6}$.
Upper limits are at 90\% CL.
Where values are shown in \red{red} (\blue{blue}), this indicates that
they are new \red{published} (\blue{preliminary}) results since PDG2017.}
\label{tab:bbaryons_Lb}
\resizebox{\textwidth}{!}{
\begin{tabular}{|lccc @{}c c @{}c c|} \hline
RPP\# &Mode & PDG2017 Avg. & CDF & & LHCb & & Our Avg.  \\ \sglinespb
$~12$                                             & %
$\bar{K}^0 p \pi^{-}$                             & %
$13.0 \pm 4.0$                                    & %
\nodata                                           & %
\phantom{.}                                       & %
$\ferrsyt{12.6}{1.9}{0.9}{3.4}{0.5}$~$^\S$        & %
\ifref {\cite{Aaij:2014lpa}} \fi \phantom{.}      & %
$12.6 \pm 4.0$                                    \\

$~13$                                             & %
$K^0 p K^{-}$                                     & %
$<3.5$                                            & %
\nodata                                           & %
\phantom{.}                                       & %
$<3.5$                                            & %
\ifref {\cite{Aaij:2014lpa}} \fi \phantom{.}      & %
$<3.5$                                            \\

$~33$                                             & %
$p\pi^-$                                          & %
$4.3\pm0.8$                                       & %
$\err{3.5}{0.6}{0.9}$                             & %
\ifref {\cite{Aaltonen:2008hg}} \fi \phantom{.}   & %
\nodata                                           & %
\phantom{.}                                       & %
$3.5 \pm 1.1$                                     \\

$~34$                                             & %
$p K^-$                                           & %
$5.1\pm0.9$                                       & %
$\err{5.6}{0.8}{1.5}$                             & %
\ifref {\cite{Aaltonen:2008hg}} \fi \phantom{.}   & %
\nodata                                           & %
\phantom{.}                                       & %
$5.6 \pm 1.7$                                     \\

$~37$                                             & %
$\Lambda \mu^+\mu^-$                              & %
$1.08\pm0.28$                                     & %
$\err{1.73}{0.42}{0.55}$                          & %
\ifref {\cite{Aaltonen:2011qs}} \fi \phantom{.}   & %
{$\err{0.96}{0.16}{0.25}$}                        & %
\ifref {\cite{Aaij:2013mna}} \fi \phantom{.}      & %
$1.08 \pm 0.27$                                   \\

$~38$                                             & %
$\Lambda \gamma$                                  & %
$<1300$                                           & %
$<1300$                                           & %
\ifref {\cite{Acosta:2002fh}} \fi \phantom{.}     & %
\nodata                                           & %
\phantom{.}                                       & %
$<1300$                                           \\

$~39$                                             & %
$\Lambda \eta$                                    & %
\cerr{9}{7}{5}                                    & %
\nodata                                           & %
\phantom{.}                                       & %
$\cerr{9.3}{7.3}{5.3}$~$^\P$                      & %
\ifref {\cite{Aaij:2015eqa}} \fi \phantom{.}      & %
$\cerr{9.3}{7.3}{5.3}$                            \\

$~40$                                             & %
$\Lambda \etapr$                                  & %
$<3.1$                                            & %
\nodata                                           & %
\phantom{.}                                       & %
$<3.1$                                            & %
\ifref {\cite{Aaij:2015eqa}} \fi \phantom{.}      & %
$<3.1$                                            \\

$~41$                                             & %
$\Lambda \pi^+\pi^-$                              & %
$4.7\pm1.9$                                       & %
\nodata                                           & %
\phantom{.}                                       & %
$\gerrsyt{4.6}{1.2}{1.4}{0.6}$~$^\dag$~$^2$       & %
\ifref {\cite{Aaij:2016nrq}} \fi \phantom{.}      & %
$4.6 \pm 1.9$                                     \\

$~42$                                             & %
$\Lambda K^+\pi^-$                                & %
$5.7\pm1.3$                                       & %
\nodata                                           & %
\phantom{.}                                       & %
$\gerrsyt{5.6}{0.8}{0.8}{0.7}$~$^\dag$~$^2$       & %
\ifref {\cite{Aaij:2016nrq}} \fi \phantom{.}      & %
$5.6 \pm 1.3$                                     \\

$~43$                                             & %
$\Lambda K^+K^-$                                  & %
$16.1\pm2.3$                                      & %
\nodata                                           & %
\phantom{.}                                       & %
$\gerrsyt{15.9}{1.2}{1.2}{2.0}$~$^\dag$~$^2$      & %
\ifref {\cite{Aaij:2016nrq}} \fi \phantom{.}      & %
$15.9 \pm 2.6$                                    \\

$~44$                                             & %
$\Lambda \phi$                                    & %
$2.0 \pm 0.5$                                     & %
\nodata                                           & %
\phantom{.}                                       & %
$\derrsyt{5.18}{1.04}{0.35}{0.67}{0.62}$~$^\ddag$~$^3$& %
\ifref {\cite{Aaij:2016zhm}} \fi \phantom{.}      & %
$\cerr{5.18}{1.29}{1.26}$                         \\

\nodata                                           & %
$p \pi^-\mu^+\mu^-$                               & %
\nodata                                           & %
\nodata                                           & %
\phantom{.}                                       & %
$\derrsyt{0.069}{0.019}{0.011}{0.013}{0.010}$~$^\dag$& %
\ifref {\cite{Aaij:2017ewm}} \fi \phantom{.}      & %
$\cerr{0.069}{0.026}{0.024}$                      \\

\nodata                                           & %
$p \pi^-\pi^+\pi^-$                               & %
\nodata                                           & %
\nodata                                           & %
\phantom{.}                                       & %
$\ferrsyt{19.0}{0.6}{1.0}{1.6}{0.7}$~$^1$         & %
\ifref {\cite{Aaij:2017pgy}} \fi \phantom{.}      & %
$19.0 \pm 2.1$                                    \\

\nodata                                           & %
$p K^-\pi^+\pi^-$                                 & %
\nodata                                           & %
\nodata                                           & %
\phantom{.}                                       & %
$\ferrsyt{45.5}{0.8}{2.0}{3.9}{1.7}$~$^1$         & %
\ifref {\cite{Aaij:2017pgy}} \fi \phantom{.}      & %
$45.5 \pm 4.8$                                    \\

\nodata                                           & %
$p K^- K^+\pi^-$                                  & %
\nodata                                           & %
\nodata                                           & %
\phantom{.}                                       & %
$\ferrsyt{3.7}{0.3}{0.4}{0.3}{0.1}$~$^1$          & %
\ifref {\cite{Aaij:2017pgy}} \fi \phantom{.}      & %
$3.7 \pm 0.6$                                     \\

\nodata                                           & %
$p K^- K^+ K^-$                                   & %
\nodata                                           & %
\nodata                                           & %
\phantom{.}                                       & %
$\ferrsyt{11.4}{0.3}{0.7}{1.0}{0.5}$~$^1$         & %
\ifref {\cite{Aaij:2017pgy}} \fi \phantom{.}      & %
$11.4 \pm 1.4$                                    \\

\nodata                                           & %
$\Psi(2S) p \pi^-$                                & %
\nodata                                           & %
\nodata                                           & %
\phantom{.}                                       & %
$\derrsyt{7.17}{0.82}{0.33}{1.30}{1.03}$~$^4$     & %
\ifref {\cite{Aaij:2018jlf}} \fi \phantom{.}      & %
$\cerr{7.17}{1.57}{1.36}$                         \\

\hline
\end{tabular}
}
\end{center}
\scriptsize
Channels with no RPP\# were not included in PDG Live as of Dec. 31, 2017. \\ %
Results for CDF and LHCb are relative BFs converted to absolute BFs.\\   %
$^\dag$~Last quoted uncertainty is due to the precision with which the normalization channel branching fraction is known. \\     %
$^\ddag$~Third uncertainty is related to external inputs. \\     %
$^\S$~Third uncertainty is from the ratio of fragmentation fractions $f_{\Lambda^{0}_{b}}/f_d$, and the fourth is due to the uncertainty on ${\cal B}(B^{0}\rightarrow K^{0} \pi^+ \pi^-)$. \\     %
$^\P$~Result at 68\% CL. \\     %
$^1$~Third uncertainty is from ${\cal B}(\Lambda_b \to \Lambda_c^+  \pi^-)$, and the fourth is due to the uncertainty on ${\cal B}(\Lambda_c^+ \to p K^- \pi^+ )$. \\     %
$^2$~Normalization taken directly from LHCb paper.\\ %
$^3$~Difference w.r.t. PDG value due to different values for the production rate ratio $f_{\Lambda_b}/f_d$.\\ %
$^4$~Calculated using the value of ${\cal B}(\Lambda_b^0 \to \Psi(2S)p K^-)=(6.29\pm0.23\pm0.14^{+1.14}_{-0.90})\times 10^{-6}$.
\end{table}

\begin{table}[hp!]
\begin{center}
\caption{Partial branching fractions of $\Lb \to \Lambda\mu^+\mu^-$
decays in intervals of $q^2=m^2(\mu^+\mu^-)$ in units of
$\times 10^{-7}$. 
Where values are shown in \red{red} (\blue{blue}), this indicates that
they are new \red{published} (\blue{preliminary}) results since PDG2017.}
\label{tab:bbaryons_LbPartialBF}
\resizebox{\textwidth}{!}{
\begin{tabular}{|cccc @{}c c @{}c c|}
\sgline
Mode & $q^2~[\gevgevcccc]$~$^\dag$~$^\ddag $ & PDG2017 Avg. & CDF & & LHCb & & Our Avg. \\
\sglinespb
$\Lambda\mu^+\mu^-         $                      & %
$< 2.0$                                           & %
$0.71\pm0.27$                                     & %
{$\err{0.15}{2.01}{0.05}$}                        & %
\ifref {\cite{Aaltonen:2011qs}} \fi \phantom{.}   & %
$\aerr{0.72}{0.24}{0.22}{0.14}$                   & %
\ifref {\cite{Aaij:2015xza}} \fi \phantom{.}      & %
$\cerr{0.71}{0.27}{0.26}$                         \\

$\Lambda\mu^+\mu^-$                               & %
$[2.0,4.3]$                                       & %
{$\cerr{0.28}{0.28}{0.21}$}                       & %
{$\err{1.8}{1.7}{0.6}$}                           & %
\phantom{.}                                       & %
$\aerr{0.253}{0.276}{0.207}{0.046}$               & %
\ifref {\cite{Aaij:2015xza}} \fi \phantom{.}      & %
$\cerr{0.281}{0.286}{0.218}$                      \\

$\Lambda\mu^+\mu^-$                               & %
$[4.3,8.68]$                                      & %
$0.5\pm0.7$                                       & %
{$\err{-0.2}{1.6}{0.1}$}                          & %
\phantom{.}                                       & %
$\err{0.66}{0.72}{0.16}$                          & %
\ifref {\cite{Aaij:2013mna}} \fi \phantom{.}      & %
$0.51 \pm 0.67$                                   \\

$\Lambda\mu^+\mu^-$                               & %
$[10.09,12.86]$                                   & %
$2.2\pm0.6$                                       & %
{$\err{3.0}{1.5}{1.0}$}                           & %
\phantom{.}                                       & %
$\aerr{2.08}{0.42}{0.39}{0.42}$                   & %
\ifref {\cite{Aaij:2015xza}} \fi \phantom{.}      & %
$\cerr{2.17}{0.57}{0.55}$                         \\

$\Lambda\mu^+\mu^-$                               & %
$[14.18,16.00]$                                   & %
$1.7\pm0.5$                                       & %
{$\err{1.0}{0.7}{0.3}$}                           & %
\phantom{.}                                       & %
$\aerr{2.04}{0.35}{0.33}{0.42}$                   & %
\ifref {\cite{Aaij:2015xza}} \fi \phantom{.}      & %
$1.70 \pm 0.44$                                   \\

$\Lambda\mu^+\mu^-$                               & %
$>16.00$                                          & %
$7.0\pm2.9$                                       & %
{$\err{7.0}{1.9}{2.2}$}                           & %
\phantom{.}                                       & %
\nodata                                           & %
\phantom{.}                                       & %
$7.0 \pm 2.9$                                     \\

\sglinespt
\end{tabular}
}
\end{center}
\scriptsize
Results for CDF and LHCb are relative BFs converted to absolute BFs.\\    %
$^\dag$ ~See the original paper for the exact $m^2(\mu^+\mu^-)$ selection.\\     %
$^\ddag$~The two LHCb measurements include additional binning not reported here.
\end{table}

\begin{table}[hp!]
\begin{center}
\caption{Branching fractions of charmless $\Xi^{0}_{b}$ decays in units of
$\times 10^{-6}$.
Upper limits are at 90\% CL.
Where values are shown in \red{red} (\blue{blue}), this indicates that
they are new \red{published} (\blue{preliminary}) results since PDG2017.}
\label{tab:bbaryons_Xhibz}
\resizebox{\textwidth}{!}{
\begin{tabular}{|lccc @{}c c|} \hline
RPP\# &Mode & PDG2017 Avg. & LHCb & & Our Avg.  \\ \sglinespb
4                                                 & %
$\it{f}_{\Xi^{0}_{b}}/\it{f}_d\mathcal{B}(\Xi^{0}_{b} \to\bar{K}^0  p \pi^{-})$& %
$<1.6$                                            & %
$<1.6$                                            & %
\ifref {\cite{Aaij:2014lpa}} \fi \phantom{.}      & %
$<1.6$                                            \\

5                                                 & %
$\it{f}_{\Xi^{0}_{b}}/\it{f}_d\mathcal{B}(\Xi^{0}_{b} \to\bar{K}^0  p K^{-})$& %
$<1.1$                                            & %
$<1.1$                                            & %
\ifref {\cite{Aaij:2014lpa}} \fi \phantom{.}      & %
$<1.1$                                            \\

10                                                & %
$\it{f}_{\Xi^{0}_{b}}/\it{f}_{\Lambda^{0}_{b}} \mathcal{B}(\Xi^{0}_{b} \to \Lambda  \pi^+\pi^-)$& %
$<1.7$                                            & %
$<1.7$                                            & %
\ifref {\cite{Aaij:2016nrq}} \fi \phantom{.}      & %
$<1.7$                                            \\

11                                                & %
$\it{f}_{\Xi^{0}_{b}}/\it{f}_{\Lambda^{0}_{b}}\mathcal{B}(\Xi^{0}_{b} \to\Lambda  K^+\pi^-)$& %
$<0.8$                                            & %
$<0.8$                                            & %
\ifref {\cite{Aaij:2016nrq}} \fi \phantom{.}      & %
$<0.8$                                            \\

12                                                & %
$\it{f}_{\Xi^{0}_{b}}/\it{f}_{\Lambda^{0}_{b}}\mathcal{B}(\Xi^{0}_{b} \to\Lambda  K^+K^-)$& %
$<0.3$                                            & %
$<0.3$                                            & %
\ifref {\cite{Aaij:2016nrq}} \fi \phantom{.}      & %
$<0.3$                                            \\

\nodata                                           & %
$\it{f}_{\Xi^{0}_{b}}/\it{f}_{\Lambda^{0}_{b}}\mathcal{B}(\Xi^{0}_{b} \to p  K^- \pi^+ \pi^-)$& %
\nodata                                           & %
\ferrsyt{1.72}{0.21}{0.25}{0.15}{0.07}            & %
\ifref {\cite{Aaij:2017pgy}} \fi \phantom{.}      & %
$1.72 \pm 0.37$                                   \\

\nodata                                           & %
$\it{f}_{\Xi^{0}_{b}}/\it{f}_{\Lambda^{0}_{b}}\mathcal{B}(\Xi^{0}_{b} \to p  K^- \pi^+ K^-)$& %
\nodata                                           & %
\ferrsyt{1.56}{0.16}{0.19}{0.13}{0.06}            & %
\ifref {\cite{Aaij:2017pgy}} \fi \phantom{.}      & %
$1.56 \pm 0.29$                                   \\

\nodata                                           & %
$\it{f}_{\Xi^{0}_{b}}/\it{f}_{\Lambda^{0}_{b}}\mathcal{B}(\Xi^{0}_{b} \to p  K^- K^+ K^-)$& %
\nodata                                           & %
$<0.25$                                           & %
\ifref {\cite{Aaij:2017pgy}} \fi \phantom{.}      & %
$<0.25$                                           \\

\hline
\end{tabular}
}
\end{center}
\scriptsize
Channels with no RPP\# were not included in PDG Live as of Dec. 31, 2017. \\ %
Results for LHCb are relative BFs converted to absolute BFs.\\ \\    %
\end{table}

\begin{table}[hp!]
\begin{center}
\caption{Branching fractions of charmless $\Xi^{-}_{b}$ decays in units of
$\times 10^{-5}$.
Upper limits are at 90\% CL.
Where values are shown in \red{red} (\blue{blue}), this indicates that
they are new \red{published} (\blue{preliminary}) results since PDG2017.}
\label{tab:bbaryons_Xhibm}
\resizebox{\textwidth}{!}{
\begin{tabular}{|lccc @{}c c|} \hline
RPP\# &Mode & PDG2017 Avg. & LHCb & & Our Avg.  \\ \sglinespb

6                                                 & %
$\it{f}_{\Xi^{-}_{b}} \mathcal{B}(\Xi^{-}_{b} \to p  K^- K^-)/(\it{f}_{u}\mathcal{B}(B ^- \to K^+  K^- K^-))$& %
$\dag$                                            & %
{\err{265}{35}{47}}                               & %
\ifref {\cite{Aaij:2016zab}} \fi \phantom{.}      & %
$265 \pm 58$                                      \\

\nodata                                           & %
$\it{f}_{\Xi^{-}_{b}} \mathcal{B}(\Xi^{-}_{b} \to p  K^- \pi^-)/(\it{f}_{u}\mathcal{B}(B ^- \to K^+  K^- K^-))$& %
\nodata                                           & %
{\err{259}{64}{49}}                               & %
\ifref {\cite{Aaij:2016zab}} \fi \phantom{.}      & %
$259 \pm 80$                                      \\

8                                                 & %
$\mathcal{B}(\Xi^{-}_{b} \to p  \pi^- \pi^-)/(\mathcal{B}(\Xi^{-}_{b} \to p  K^- K^-))$& %
$<0.56$                                           & %
{$<0.56$}                                         & %
\ifref {\cite{Aaij:2016zab}} \fi \phantom{.}      & %
{$<0.56$}                                         \\

\nodata                                           & %
$\it{f}_{\Xi^{-}_{b}} \mathcal{B}(\Xi^{-}_{b} \to p  \pi^- \pi^-)/(\it{f}_{u}\mathcal{B}(B ^- \to K^+  K^- K^-))$& %
\nodata                                           & %
{$<147$}                                          & %
\ifref {\cite{Aaij:2016zab}} \fi \phantom{.}      & %
{$<147$}                                          \\

9                                                 & %
$\mathcal{B}(\Xi^{-}_{b} \to p  K^- \pi^-)/(\mathcal{B}(\Xi^{-}_{b} \to p  K^- K^-))$& %
$\err{0.98}{0.27}{0.09}$                          & %
{$\err{0.98}{0.27}{0.09}$}                        & %
\ifref {\cite{Aaij:2016zab}} \fi \phantom{.}      & %
$0.98 \pm 0.28$                                   \\

\hline
\end{tabular}
}
\end{center}
\scriptsize
Channels with no RPP\# were not included in PDG Live as of Dec. 31, 2017. \\ %
$\dag$~PDG reports results multiplied by ${\cal B}(B^+ \to K^+ K^- K^+)$ and ${\cal B}(\overline{b} \to B^+)$.
\end{table}

\begin{table}[ht!]
\begin{center}
\caption{Branching fractions of charmless $\Omega^{-}_{b}$ decays in units of
$\times 10^{-5}$.
Upper limits are at 90\% CL.
Where values are shown in \red{red} (\blue{blue}), this indicates that
they are new \red{published} (\blue{preliminary}) results since PDG2017.}
\label{tab:bbaryons_Omegabm}
\resizebox{\textwidth}{!}{
\begin{tabular}{|lccc @{}c c|} \hline
RPP\# &Mode & PDG2017 Avg. & LHCb & & Our Avg.  \\ \sglinespb

2                                                 & %
$\it{f}_{\Omega^{-}_{b}} \mathcal{B}(\Omega^{-}_{b} \to p  K^- K^-)/(\it{f}_{u}\mathcal{B}(B ^- \to K^+  K^- K^-))$& %
$\dag$                                            & %
$<18$                                             & %
\ifref {\cite{Aaij:2016zab}} \fi \phantom{.}      & %
$<18$                                             \\

3                                                 & %
$\it{f}_{\Omega^{-}_{b}} \mathcal{B}(\Omega^{-}_{b} \to p  K^- \pi^-)/(\it{f}_{u}\mathcal{B}(B ^- \to K^+  K^- K^-))$& %
$\dag$                                            & %
$<51$                                             & %
\ifref {\cite{Aaij:2016zab}} \fi \phantom{.}      & %
$<51$                                             \\

4                                                 & %
$\it{f}_{\Omega^{-}_{b}} \mathcal{B}(\Omega^{-}_{b} \to p  \pi^- \pi^-)/(\it{f}_{u}\mathcal{B}(B ^- \to K^+  K^- K^-))$& %
$\dag$                                            & %
$<109$                                            & %
\ifref {\cite{Aaij:2016zab}} \fi \phantom{.}      & %
$<109$                                            \\

\hline
\end{tabular}
}
\end{center}
\scriptsize
$\dag$~PDG reports results multiplied by ${\cal B}(B^+ \to K^+ K^- K^+)$ and ${\cal B}(\overline{b} \to B^+)$.
\end{table}

\begin{figure}[h!]
\centering
\includegraphics[width=0.45\textwidth]{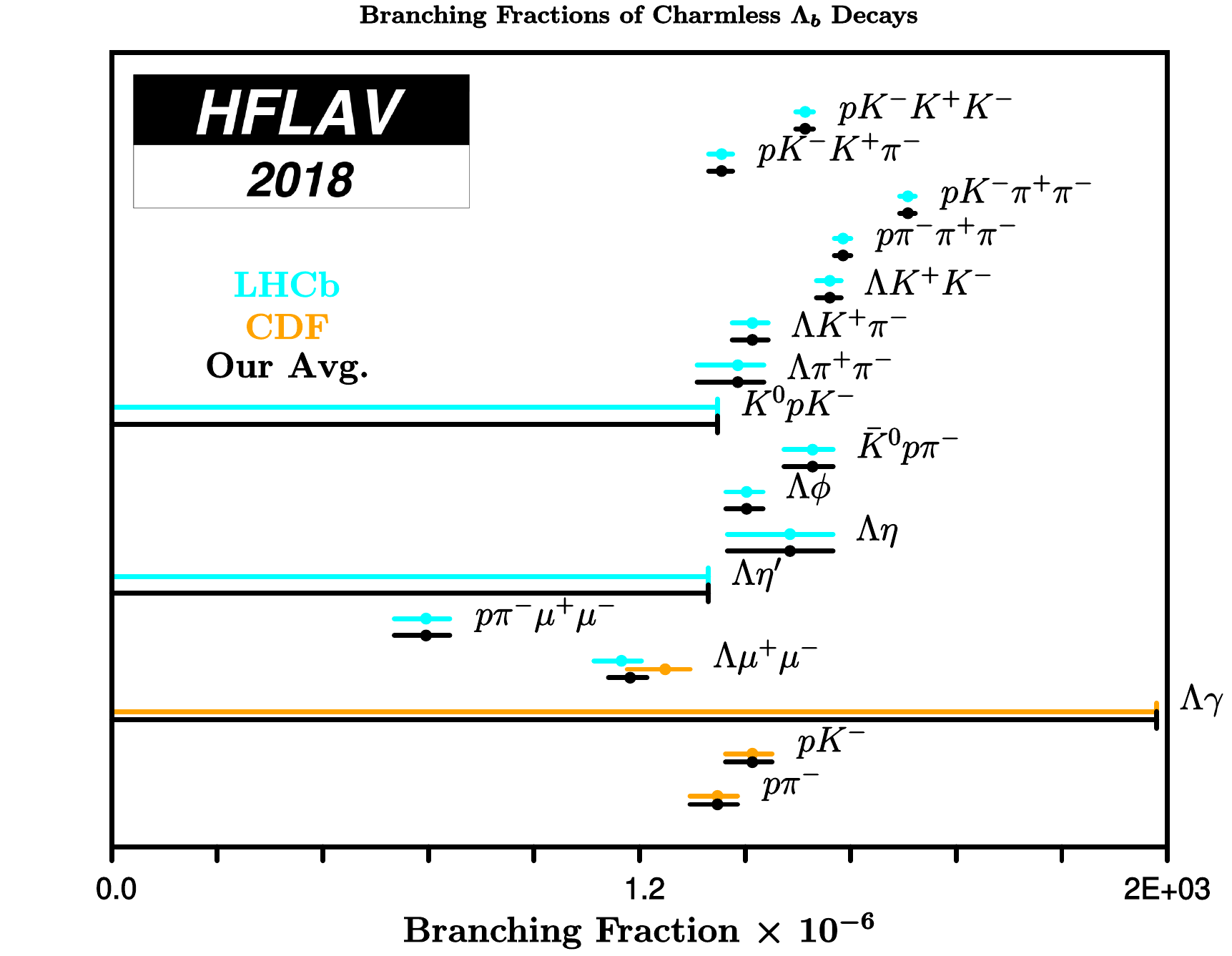}
\caption{Branching fractions of charmless $\Lb$ decays.}
\label{fig:rare-lb}
\end{figure}

\newpage
\noindent List of other measurements that are not included in the tables:
\begin{itemize}
\item In Ref.~\cite{Aaij:2015xza}, LHCb provides a measurement of the differential $\Lb \to \Lambda \mu^+\mu^-$ branching fraction. It is given in bins of $m^2(\mu^+\mu^-)$ that are different from those used in the past by the LHCb and CDF collaborations (see table of differential branching fractions).
\item In Ref.~\cite{Aaij:2018gwm}, LHCb measures angular observables of the decay $\Lb \to \Lambda \mu^+\mu^-$, including the lepton-side, hadron-side and combined forward-backward asymmetries of the decay.

\item In Ref.~\cite{Aaij:2014yka},
LHCb measures the ratios
$$
\frac{\sigma(pp\rightarrow \Xi^{\prime -}_{b}X) {\cal B}(\Xi^{\prime -}_{b}\rightarrow\Xi^{0}_{b}\pi^-)}{\sigma(pp\rightarrow \Xi^{0}_{b}X)},
\frac{\sigma(pp\rightarrow \Xi^{\prime -}_{b}X) {\cal B}(\Xi^{\ast -}_{b}\rightarrow\Xi^{0}_{b}\pi^-)}{\sigma(pp\rightarrow \Xi^{\prime -}_{b}X) {\cal B}(\Xi^{\prime -}_{b}\rightarrow\Xi^{0}_{b}\pi^-)}.
$$
\item In Ref.~\cite{Aaij:2016jnn},
LHCb measures the ratio $$
\frac{\sigma(pp\rightarrow \Xi^{\ast -}_{b}X) {\cal B}(\Xi^{\ast -}_{b}\rightarrow\Xi^{0}_{b}\pi^-)}{\sigma(pp\rightarrow \Xi^{0}_{b}X)}.
$$
\item In Ref.~\cite{Aaij:2017inn}, LHCb performs a search for baryon-number-violating $\Xi^0_b$ oscillations and set an upper limit of $\omega<0.08\ {\rm ps^{-1}}$ on the oscillation rate.
\end{itemize}

\mysubsection{Decays of \Bs mesons}
\label{sec:rare-bs}

Tables~\ref{tab:Bs_BF} and~\ref{tab:Bs_BF_rel} detail branching fractions and relative branching fractions of \Bs meson decays, respectively. 

Figures~\ref{fig:rare-bsleptonic} and~\ref{fig:rare-bs} show graphic representations of a selection of results given in this section.
Footnote symbols indicate that the footnote in the corresponding table should be consulted.
\begin{table}[htbp!]
\begin{center}
\caption{Branching fractions of charmless 
$\Bs$ decays in units of $\times 10^{-6}$. Upper limits are
at 90\% CL.
Where values are shown in \red{red} (\blue{blue}), this indicates that
they are new \red{published} (\blue{preliminary}) results since PDG2017.}
\label{tab:Bs_BF}
\resizebox{\textwidth}{!}{

}
\end{center}
\scriptsize{
Channels with no RPP\# were not included in PDG Live as of Dec. 31, 2017. \\ %
Results for CDF, D0, LHCb, CMS and ATLAS are relative BFs converted to absolute BFs.\\     %
$^\dag$~The first error is experimental, and the second is from the reference BF.\\[0.1cm]     %
$^\ddag$~Last error represents the uncertainty due to the total number of $\Bs \Bsb$ pairs.\\     %
$^\S$~Last error takes into account error the reference BF and $f_d/f_s$.\\     %
\quad $^\P$~Includes two distinct decay processes: ${\cal B}(\Bs \to f)+{\cal B}(\Bs \to \bar{f})$.\\    %
$^1$ UL at 95\% CL.\\     %
$^2$ Muon pairs do not originate from resonances and $0.5<m(\pi^+\pi^-)<1.3~\gevcc$.\\     %
$^3$ In the mass range $400<m(\pi^+\pi^-)<1600~\gevcc$.\\     %
$^4$ The third error is due to the reference BF and the fourth to  $f_d/f_s$. \\     %
}
\end{table}

\begin{table}
\begin{center}
\caption{Relative branching fractions of charmless 
$\Bs$ decays. Upper limits are
at 90\% CL.
Where values are shown in \red{red} (\blue{blue}), this indicates that
they are new \red{published} (\blue{preliminary}) results since PDG2017.}
\label{tab:Bs_BF_rel}
\resizebox{\textwidth}{!}{
\begin{tabular}{|lccc @{}c c @{}c c|} \hline
RPP\# & Mode & PDG2017 Avg. & CDF & & LHCb & & Our Avg.  \\ \sglinespb
$85/257$                                          & %
$\it{f}_s\mathcal{B}(B^0_s\rightarrow\pi^+\pi^-)/\it{f}_d\mathcal{B}(B^0\rightarrow K^+\pi^-)$& %
\nodata                                           & %
\err{0.008}{0.002}{0.001}                         & %
\ifref    {\cite{Aaltonen:2011jv}} \fi \phantom{.}& %
\err{0.00915}{0.00071}{0.00083}                   & %
\ifref    {\cite{Aaij:2016elb}} \fi \phantom{.}   & %
$0.00880 \pm 0.00090$                             \\

$85/387$                                          & %
$\it{f}_s\mathcal{B}(B^0_s\rightarrow\pi^+\pi^-)/\it{f}_d\mathcal{B}(B^0\rightarrow \pi^+\pi^-)$& %
\nodata                                           & %
\nodata                                           & %
\phantom{.}                                       & %
$\aerr{0.050}{0.011}{0.009}{0.004}$               & %
\ifref     {\cite{Aaij:2012as}} \fi \phantom{.}   & %
$\cerr{0.050}{0.012}{0.010}$                      \\

$95/46$                                           & %
$\mathcal{B}(B^0_s\rightarrow\phi\phi)/\mathcal{B}(B^0_s\rightarrow J/\psi\phi)$& %
\nodata                                           & %
$\err{0.0178}{0.0014}{0.0020}$                    & %
\ifref        {\cite{Aaltonen:2011rs}} \fi \phantom{.}& %
\nodata                                           & %
\phantom{.}                                       & %
$0.0180 \pm 0.0020$                               \\

$95/343$                                          & %
$\mathcal{B}(B^0_s\rightarrow\phi\phi)/\mathcal{B}(B^0\rightarrow \phi K^*)$& %
\nodata                                           & %
\nodata                                           & %
\phantom{.}                                       & %
$1.84 \pm 0.05 \pm 0.13$                          & %
\ifref {\cite{Aaij:2014lba}} \fi \phantom{.}      & %
$1.84 \pm 0.14$                                   \\

$96/257$                                          & %
$\it{f}_s\mathcal{B}(B^0_s\rightarrow K^+\pi^-)/\it{f}_d\mathcal{B}(B^0_d\rightarrow K^+\pi^-)$& %
\nodata                                           & %
$0.071\pm0.010\pm0.007 $                          & %
\ifref {\cite{Aaltonen:2008hg}} \fi \phantom{.}   & %
$0.074\pm0.006\pm0.006$                           & %
\ifref {\cite{Aaij:2012as}} \fi \phantom{.}       & %
$0.073 \pm 0.007$                                 \\

$97/257$                                          & %
$\it{f}_s\mathcal{B}(B^0_s\rightarrow K^+K^-)/\it{f}_d\mathcal{B}(B^0_d\rightarrow K^+\pi^-)$& %
\nodata                                           & %
$0.347\pm0.020\pm0.021 $                          & %
\ifref {\cite{Aaltonen:2011qt}} \fi \phantom{.}   & %
\err{0.316}{0.009}{0.019}                         & %
\ifref {\cite{Aaij:2012as}} \fi \phantom{.}       & %
$0.327 \pm 0.017$                                 \\

$99/291$                                          & %
$\mathcal{B}(B^0_s\to K^0\pi^+\pi^-)/\mathcal{B}(B^0 \to K^0\pi^+\pi^-)$& %
\nodata                                           & %
\nodata                                           & %
\phantom{.}                                       & %
$\gerrsyt{0.191}{0.027}{0.031}{0.011}$            & %
\ifref {\cite{Aaij:2017zpx}} \fi \phantom{.}      & %
$0.191 \pm 0.043$                                 \\

$100/322$                                         & %
$\mathcal{B}(B^0_s\to K^0 K^- \pi^+)/\it\mathcal{B}(B^0 \to K^0 K^- \pi^+)$~$^\dag$& %
\nodata                                           & %
\nodata                                           & %
\phantom{.}                                       & %
$\gerrsyt{1.70}{0.07}{0.11}{0.10}$                & %
\ifref {\cite{Aaij:2017zpx}} \fi \phantom{.}      & %
$1.70 \pm 0.16$                                   \\

$101/294$                                         & %
${\cal B}(B^0_s\to K^{*-}\pi^+)/{\cal B}(B^0 \to K^{*+}\pi^-)$& %
\nodata                                           & %
\nodata                                           & %
\phantom{.}                                       & %
$\err{0.39}{0.13}{0.05}$                          & %
\ifref {\cite{Aaij:2014aaa}} \fi \phantom{.}      & %
$0.39 \pm 0.14$                                   \\

$102/294$                                         & %
${\cal B}(B^0_s\to K^{*-}K^+)/{\cal B}(B^0 \to K^{*+}\pi^-)$& %
\nodata                                           & %
\nodata                                           & %
\phantom{.}                                       & %
$\err{1.49}{0.22}{0.18}$                          & %
\ifref {\cite{Aaij:2014aaa}} \fi \phantom{.}      & %
$1.49 \pm 0.28$                                   \\

$103/291$                                         & %
$\mathcal{B}(B^0_s\to K^0_S K^{*0})/\mathcal{B}(B^0\to K^0_S \pi^+ \pi^-)$~$^\dag$& %
\nodata                                           & %
\nodata                                           & %
\phantom{.}                                       & %
$\err{0.33}{0.07}{0.04}$                          & %
\ifref {\cite{Aaij:2015asa}} \fi \phantom{.}      & %
$0.33 \pm 0.08$                                   \\

$104/329$                                         & %
$\mathcal{B}(B^0_s\to K^0 K^+ K^-)/\mathcal{B}(B^0 \to K^0 K^+ K^-)$& %
\nodata                                           & %
\nodata                                           & %
\phantom{.}                                       & %
$<0.051$                                          & %
\ifref {\cite{Aaij:2017zpx}} \fi \phantom{.}      & %
$<0.051$                                          \\

$106/294$                                         & %
${\cal B}(B^0_s\to K^{*0}\overline K^{*0})/{\cal B}(B^0 \to K^{*+}\pi^-)$& %
\nodata                                           & %
\nodata                                           & %
\phantom{.}                                       & %
$\err{1.11}{0.22}{0.13}$                          & %
\ifref {\cite{Aaij:2015kba}} \fi \phantom{.}      & %
$1.11 \pm 0.26$                                   \\

$107/343$                                         & %
${\cal B}(B^0_s\to \phi \overline{K}^{*0})/{\cal B}(B^0 \to \phi K^{*0})$& %
\nodata                                           & %
\nodata                                           & %
\phantom{.}                                       & %
$\err{0.113}{0.024}{0.016}$                       & %
\ifref {\cite{Aaij:2013gga}} \fi \phantom{.}      & %
$0.113 \pm 0.029$                                 \\

$112/371$                                         & %
${\cal B}(B^0_s\to \phi \gamma)/{\cal B}(B^0 \to K^{*0}\gamma)$& %
\nodata                                           & %
\nodata                                           & %
\phantom{.}                                       & %
\err{0.81}{0.04}{0.07}                            & %
\ifref {\cite{Aaij:2012ita}} \fi \phantom{.}      & %
$0.81 \pm 0.08$                                   \\

$117/46$                                          & %
$\mathcal{B}(B^0_s\to \phi\mu^+\mu^-)/\mathcal{B}(B^0_s\to J/\psi\phi)\times10^3$& %
$0.76 \pm 0.09$~$^\diamond$                       & %
$\cerr{1.13}{0.19}{0.07}$                         & %
\ifref {\cite{Aaltonen:2011qs}} \fi \phantom{.}   & %
$\aerr{0.741}{0.042}{0.040}{0.029}$               & %
\ifref {\cite{Aaij:2015esa}} \fi \phantom{.}      & %
$0.876 \pm 0.041$                                 \\

\nodata                                           & %
$\mathcal{B}(B^0_s\to p \overline{p} K^+\pi^-)/\mathcal{B}(B^0\to p \overline{p} K^+\pi^-)$& %
\nodata                                           & %
\nodata                                           & %
\phantom{.}                                       & %
$\gerrsyt{0.22}{0.04}{0.02}{0.01}$                & %
\ifref {\cite{Aaij:2017pgn}} \fi \phantom{.}      & %
$0.22 \pm 0.05$                                   \\

\nodata                                           & %
$\mathcal{B}(B^0_s\to p \overline{p} K^+\pi^-)/\mathcal{B}(B^0_s\to p \overline{p} K^+ K^-)$& %
\nodata                                           & %
\nodata                                           & %
\phantom{.}                                       & %
$\err{0.31}{0.05}{0.02}$                          & %
\ifref {\cite{Aaij:2017pgn}} \fi \phantom{.}      & %
$0.31 \pm 0.05$                                   \\

\nodata                                           & %
$\mathcal{B}(B^0_s\to\overline K^{*0} \mu^+\mu^-)/\mathcal{B}(B^0_s\to J/\psi\overline K^{*0})$~ $^\P$& %
\nodata                                           & %
\nodata                                           & %
\phantom{.}                                       & %
\gerrsyt{0.014}{0.004}{0.001}{0.001}~$^\ddag$     & %
\ifref {\cite{Aaij:2018jhg}} \fi \phantom{.}      & %
$0.014 \pm 0.004$                                 \\

\nodata                                           & %
$\mathcal{B}(B^0_s\to\overline K^{*0} \mu^+\mu^-)/(\mathcal{B}(\overline B^0\to \overline K^{*0}  \mu^+\mu^-)$& %
\nodata                                           & %
\nodata                                           & %
\phantom{.}                                       & %
\gerrsyt{0.033}{0.011}{0.003}{0.002}~$^\S$        & %
\ifref {\cite{Aaij:2018jhg}} \fi \phantom{.}      & %
$0.033 \pm 0.012$                                 \\

\sglinespt
\end{tabular}
}
\end{center}
\scriptsize{
Channels with no RPP\# were not included in PDG Live as of Dec. 31, 2017. \\ %
\quad $^\dag$~Numerator includes two distinct decay processes: ${\cal B}(\Bs \to f)+{\cal B}(\Bs \to \bar{f})$.\\    %
$^\P$~The denominator is multiplied by $\mathcal{B}(J/\psi\to\mu^+\mu^-)$.\\    %
$^\ddag$~Last error is from the S-wave fraction in $B^0_s\to\overline K^{*0} \mu^+\mu^-$ and $B^0_s\to J/\psi\overline K^{*0}$.\\     %
$^\S$~Last error is from the S-wave fraction in $B^0_s\to\overline K^{*0} \mu^+\mu^-$ and $\overline B^0\to \overline K^{*0}  \mu^+\mu^-$, and $f_d/f_s$.  \\   %
$^\diamond$~PDG also uses the denominator as input when computng the average. \\   %
}
\end{table}

\newpage
\noindent List of other measurements that are not included in the tables:
\begin{itemize}
\item  $\Bs \to \phi\mu^+\mu^-$ : LHCb measures the differential BF in bins of $m^2(\mu^+\mu^-)$. It also performs an angular analysis and measures $F_L$, $S_3$, $S_4$, $S_7$, $A_5$, $A_6$, $A_8$ and $A_9$ in bins of $m^2(\mu^+\mu^-)$ \cite{Aaij:2015esa}.
\item  $\Bs \to \phi \gamma$ : LHCb has measured the photon polarization \cite{Aaij:2016ofv}.
\item  $\Bs \to \mu^+\mu^-$ : LHCb also measures the effective lifetime \cite{Aaij:2017vad}.
\end{itemize}

\begin{figure}[htbp!]
\centering
\includegraphics[width=0.5\textwidth]{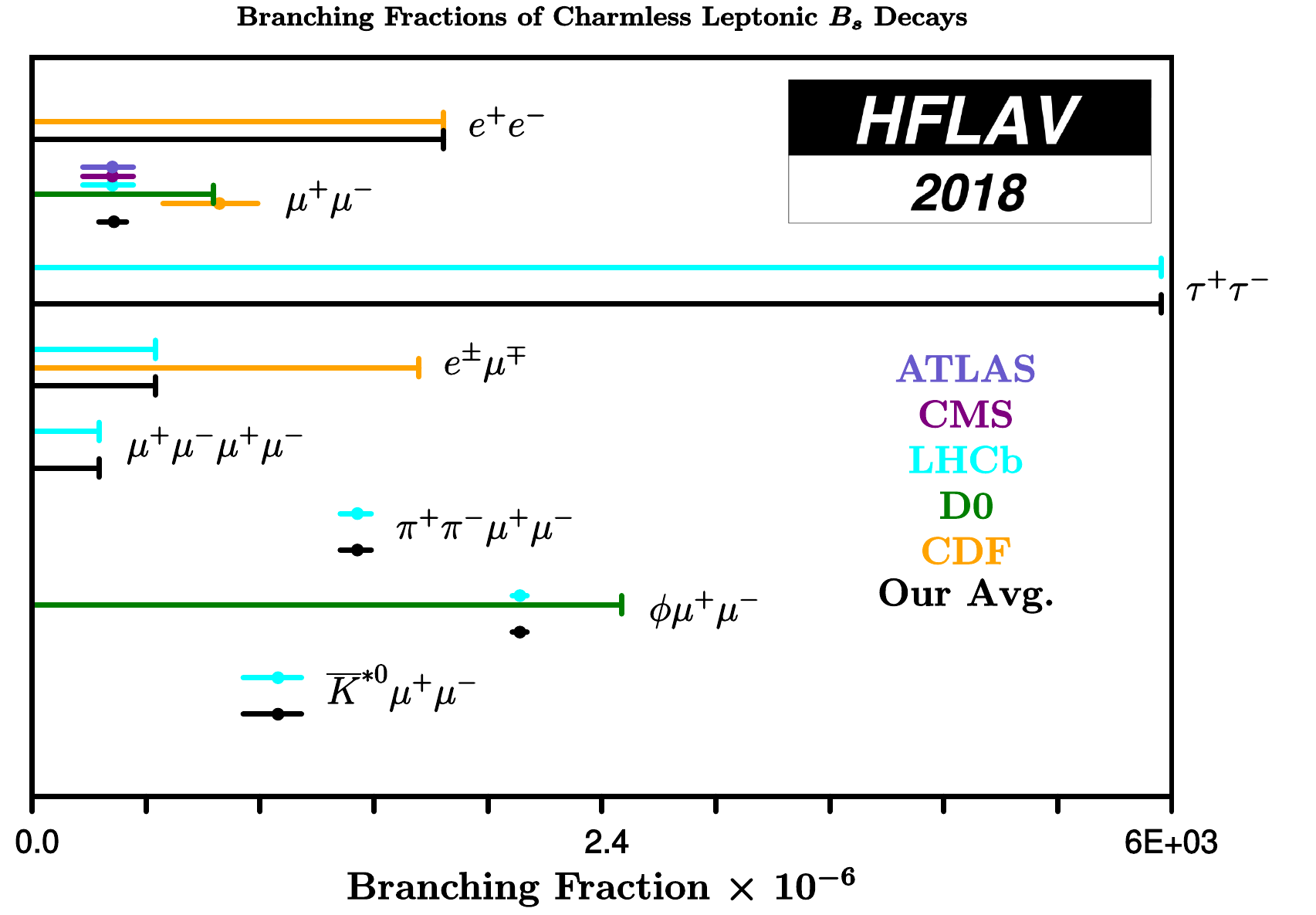}
\caption{Branching fractions of charmless leptonic $\Bs$ decays.}
\label{fig:rare-bsleptonic}
\end{figure}

\begin{figure}[ht!]
\centering
\includegraphics[width=0.5\textwidth]{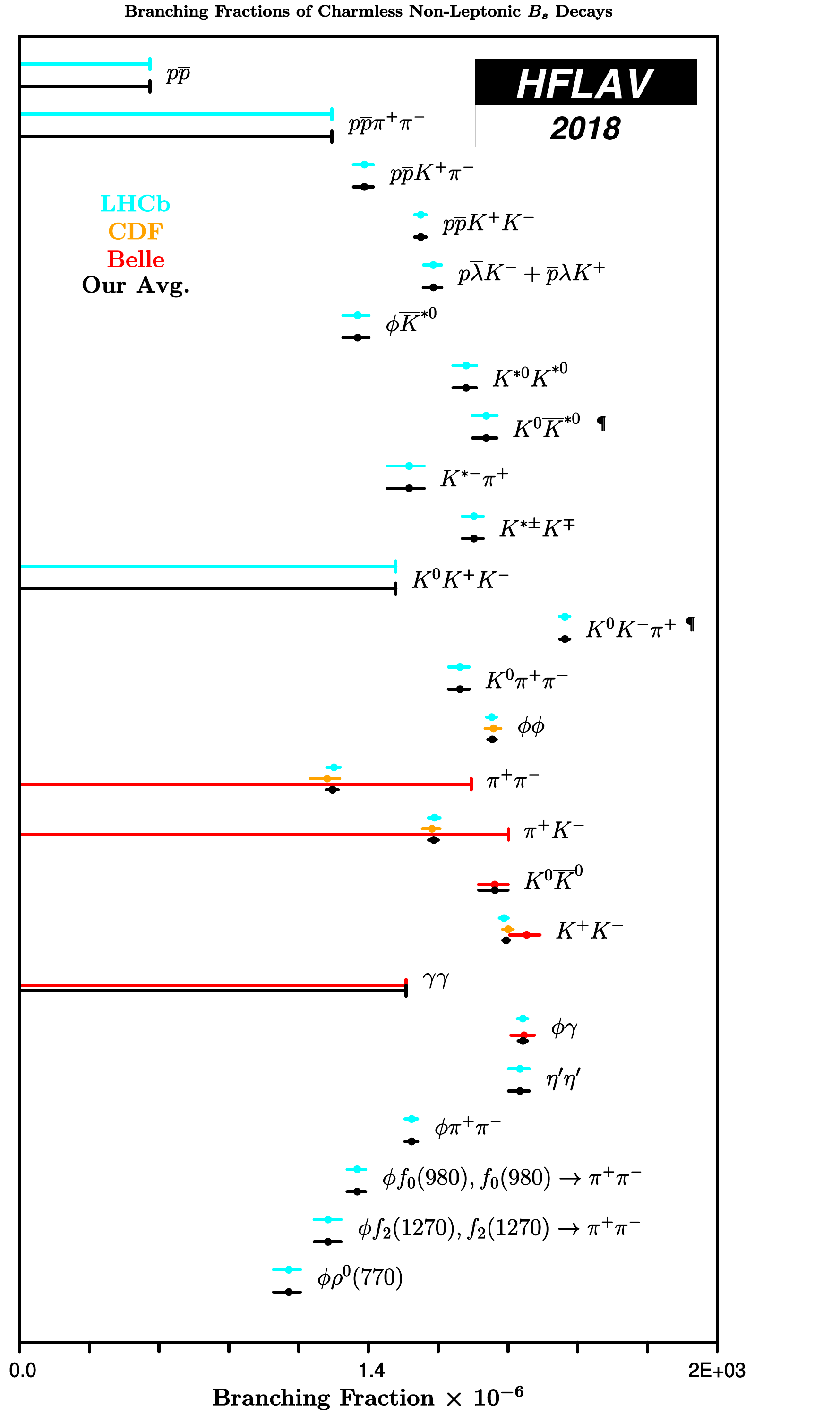}
\caption{Branching fractions of charmless non-leptonic $\Bs$ decays.}
\label{fig:rare-bs}
\end{figure}

\newpage 

\mysubsection{Rare decays of \Bz and \Bp\ mesons with photons and/or leptons}
\label{sec:rare-radll}

This section reports different observables for radiative decays, lepton-flavour/number-violating (LFV/LNV) decays and FCNC decays with leptons of \Bz and \Bp\ mesons. 
Tables~\ref{tab:radll_Bp}, \ref{tab:radll_Bz} and~\ref{tab:radll_B} provide compilations of branching fractions of radiative and FCNC decays with leptons of \Bp\ mesons, \Bz\ mesons and their admixture, respectively.
Table~\ref{tab:radll_B} also includes LFV/LNV decays.
Table~\ref{tab:radll_lep}, contains branching fractions of leptonic
and radiative-leptonic $\Bp$ and $\Bz$ decays. It is followed by Tables~\ref{tab:radll_rel} and~\ref{tab:radll_gluon}, which give relative branching fractions of $\Bp$ and $\Bz$ decays and a compilation of inclusive decays, respectively. In the modes listed in the last table, which is an exception in this section, the radiated particle is a gluon. Table~\ref{tab:radll_AI} contains isospin asymmetry measurements. Finally, Tables~\ref{tab:radll_Bp_lnv} and~\ref{tab:radll_Bz_lnv} provide compilations of branching fractions of \Bp\ and \Bz\ mesons to lepton-flavour/number-violating final states, respectively. 

Figures \ref{fig:rare-btosll} to~\ref{fig:rare-neutrino} show graphic representations of a selection of results given in this section.
Footnote symbols indicate that the footnote in the corresponding table should be consulted.

\begin{table}[!htbp]
\begin{center}
\caption{Branching fractions of charmless radiative and FCNC decays with leptons of
$B^+$ mesons in units of $\times 10^{-6}$. Upper limits are
at 90\% CL. 
Where values are shown in \red{red} (\blue{blue}), this indicates that
they are new \red{published} (\blue{preliminary}) results since PDG2017.}
\label{tab:radll_Bp}
\resizebox{\textwidth}{!}{

}
\end{center}
\scriptsize{
Channels with no RPP\# were not included in PDG Live as of Dec. 31, 2017. \\ %
Results for LHCb are relative BFs converted to absolute BFs.\\     %
CLEO upper limits that have been greatly superseded are not shown. \\     %
$^\dag$ $M_{K\pi\pi} < 1.8$~\gevcc.\\     %
$^\ddag$~$1.0 < M_{K\pi\pi} < 2.0$~\gevcc.\\     %
$^\S$~$M_{K\pi\pi} < 2.4$~\gevcc.\\     %
$^\P$ Average of \babar\ results from~\cite{Sanchez:2015pxu} %
and~\cite{Aubert:2005xk}.\\ %
$^\diamond$ Average of \babar\ results from~\cite{Sanchez:2015pxu} %
and~\cite{Aubert:2003zs}.\\ %
$^1$~Differential BF in bins of $m(\mu^+\mu^-)$ is also available.\\     %
$^2$ Result from ARGUS. Cited in the \babar\ column to avoid adding a column to the table. \\    %
}
\end{table}
\clearpage

\begin{table}[!htbp]
\begin{center}
\caption{Branching fractions of charmless radiative and FCNC decays with leptons of $B^0$ mesons in units of $\times 10^{-6}$. Upper limits are
at 90\% CL. 
Where values are shown in \red{red} (\blue{blue}), this indicates that
they are new \red{published} (\blue{preliminary}) results since PDG2017.}
\label{tab:radll_Bz}
\resizebox{\textwidth}{!}{

}
\end{center}
\scriptsize{
Results for LHCb are relative BFs converted to absolute BFs.\\     %
CLEO upper limits that have been greatly superseded are not shown. \\     %
$^\dag$~$1.25~\gevcc < M_{K\pi} < 1.6$~\gevcc.\\     %
$^\ddag$ $M_{K\pi\pi} < 1.8$~\gevcc.\\     %
$^\S$ Average of \babar\ results from ~\cite{Sanchez:2015pxu} %
and ~\cite{Aubert:2005xk}.\\ %
$^\P$~$1.0 < M_{K\pi\pi} < 2.0$~\gevcc.\\     %
$^\diamond$~This result takes into account the S-wave fraction in the $K\pi$ system.\\     %
$^1$~Muon pairs do not originate from resonances and $0.5<m(\pi^+\pi^-)<1.3~\gevcc$.     %
}
\end{table}
\clearpage

\begin{table}[!htbp]
\begin{center}
\caption{Branching fractions of charmless radiative, FCNC decays with leptons
and LFV/LNV decays
of $\Bpm/\Bz$ admixture in units of $\times 10^{-6}$. Upper limits are
at 90\% CL. 
Where values are shown in \red{red} (\blue{blue}), this indicates that
they are new \red{published} (\blue{preliminary}) results since PDG2017.}
\label{tab:radll_B}
\resizebox{\textwidth}{!}{
\begin{tabular}{|lccc @{}c c @{}c c @{}c c @{}c c|} \hline
RPP\# &Mode & PDG2017 Avg. & \babar & & Belle & & CLEO & & CDF & & Our Avg. \\
\sglinespb

~67                                               & %
$K \eta \gamma$                                   & %
$\cerr{8.5}{1.8}{1.6}$                            & %
\nodata                                           & %
\phantom{.}                                       & %
{$\aerr{8.5}{1.3}{1.2}{0.9}$}                     & %
\ifref {\cite{Nishida:2004fk}} \fi \phantom{.}    & %
\nodata                                           & %
\phantom{.}                                       & %
\nodata                                           & %
\phantom{.}                                       & %
$\cerr{8.5}{1.6}{1.5}$                            \\

~68                                               & %
$K_1(1400) \gamma $                               & %
$<1.27$                                           & %
\nodata                                           & %
\phantom{.}                                       & %
\nodata                                           & %
\phantom{.}                                       & %
$<1.27$                                           & %
\ifref {\cite{Coan:1999kh}} \fi \phantom{.}       & %
\nodata                                           & %
\phantom{.}                                       & %
$<1.27$                                           \\

~69                                               & %
$K_2^*(1430) \gamma $                             & %
$\cerr{17}{6}{5}$                                 & %
\nodata                                           & %
\phantom{.}                                       & %
\nodata                                           & %
\phantom{.}                                       & %
$\err{17}{6}{1}$                                  & %
\ifref {\cite{Coan:1999kh}} \fi \phantom{.}       & %
\nodata                                           & %
\phantom{.}                                       & %
$17 \pm 6$                                        \\

~71                                               & %
$K_3^*(1780) \gamma $                             & %
$<37$                                             & %
\nodata                                           & %
\phantom{.}                                       & %
{$<37$}~$^\S$                                     & %
\ifref {\cite{Nishida:2004fk}} \fi \phantom{.}    & %
\nodata                                           & %
\phantom{.}                                       & %
\nodata                                           & %
\phantom{.}                                       & %
{$<37$}~$^\S$                                     \\

~78                                               & %
$s \gamma$~$^\dag$                                & %
$349 \pm 19$                                      & %
$\cerr{341}{28}{28}$~$^{\P}$                      & %
\ifref {\cite{Aubert:2007my,Lees:2012ym,Lees:2012wg}} \fi \phantom{.}& %
{$\cerr{328}{20}{20}$}~$^{\P}$                    & %
\ifref {\cite{Limosani:2009qg,Saito:2014das,Belle:2016ufb}} \fi \phantom{.}& %
$\err{329}{44}{29}$                               & %
\ifref {\cite{Chen:2001fja}} \fi \phantom{.}      & %
\nodata                                           & %
\phantom{.}                                       & %
$332 \pm 15$                                      \\

~78                                               & %
$s \gamma$~$^{\diamond}$                          & %
\nodata                                           & %
$308 \pm 22$~$^{\P}$                              & %
\ifref {\cite{Aubert:2007my,Lees:2012ym,Lees:2012wg}} \fi \phantom{.}& %
{$\cerr{305}{16}{16}$}~$^{\P}$                    & %
\ifref {\cite{Saito:2014das,Belle:2016ufb}} \fi \phantom{.}& %
\nodata                                           & %
\phantom{.}                                       & %
\nodata                                           & %
\phantom{.}                                       & %
$306 \pm 12$                                      \\

~79                                               & %
$d \gamma$                                        & %
$9.2 \pm 3.0$                                     & %
{$\err{9.2}{2.0}{2.3}$}                           & %
\ifref {\cite{delAmoSanchez:2010ae}} \fi \phantom{.}& %
\nodata                                           & %
\phantom{.}                                       & %
\nodata                                           & %
\phantom{.}                                       & %
\nodata                                           & %
\phantom{.}                                       & %
$9.2 \pm 3.0$                                     \\

~85                                               & %
$\rho \gamma $                                    & %
$1.39\pm 0.25$                                    & %
{$\aerr{1.73}{0.34}{0.32}{0.17}$}                 & %
\ifref {\cite{Aubert:2008al}} \fi \phantom{.}     & %
{$\aerr{1.21}{0.24}{0.22}{0.12}$}                 & %
\ifref {\cite{Taniguchi:2008ty}} \fi \phantom{.}  & %
\nodata                                           & %
\phantom{.}                                       & %
\nodata                                           & %
\phantom{.}                                       & %
$\cerr{1.39}{0.22}{0.21}$                         \\

~86                                               & %
$\rho/\omega \gamma $                             & %
$1.30\pm0.23$                                     & %
{$\aerr{1.63}{0.30}{0.28}{0.16}$}                 & %
\ifref {\cite{Aubert:2008al}} \fi \phantom{.}     & %
{$\berr{1.14}{0.20}{0.10}{0.12}$}                 & %
\ifref {\cite{Taniguchi:2008ty}} \fi \phantom{.}  & %
\nodata                                           & %
\phantom{.}                                       & %
\nodata                                           & %
\phantom{.}                                       & %
$\cerr{1.30}{0.18}{0.19}$                         \\

121                                               & %
$se^+ e^-$~$^\ddag$                               & %
$6.7\pm 1.7$                                      & %
{$\aerrsy{7.69}{0.82}{0.77}{0.71}{0.60}$}         & %
\ifref {\cite{Lees:2013nxa}} \fi \phantom{.}      & %
{$\berr{4.05}{1.30}{0.87}{0.83}$}                 & %
\ifref {\cite{Iwasaki:2005sy}} \fi \phantom{.}    & %
\nodata                                           & %
\phantom{.}                                       & %
\nodata                                           & %
\phantom{.}                                       & %
$6.67 \pm 0.82$                                   \\

120                                               & %
$s\mu^+ \mu^-$~$^\ddag$                           & %
$4.3\pm 1.0$                                      & %
{$\aerrsy{4.41}{1.31}{1.17}{0.63}{0.50}$}         & %
\ifref {\cite{Lees:2013nxa}} \fi \phantom{.}      & %
{$\berr{4.13}{1.05}{0.85}{0.81}$}                 & %
\ifref {\cite{Iwasaki:2005sy}} \fi \phantom{.}    & %
\nodata                                           & %
\phantom{.}                                       & %
\nodata                                           & %
\phantom{.}                                       & %
$\cerr{4.27}{0.98}{0.91}$                         \\

123                                               & %
$s \ell^+ \ell^-$~$^\ddag$                        & %
$5.8\pm 1.3$                                      & %
{$\aerrsy{6.73}{0.70}{0.64}{0.60}{0.56}$}         & %
\ifref {\cite{Lees:2013nxa}} \fi \phantom{.}      & %
{$\berr{4.11}{0.83}{0.85}{0.81}$}                 & %
\ifref {\cite{Iwasaki:2005sy}} \fi \phantom{.}    & %
\nodata                                           & %
\phantom{.}                                       & %
\nodata                                           & %
\phantom{.}                                       & %
$5.84 \pm 0.69$                                   \\

124                                               & %
$\pi \ell^+ \ell^-$                               & %
$<0.059$                                          & %
$<0.059$                                          & %
\ifref {\cite{Lees:2013lvs}} \fi \phantom{.}      & %
$<0.062$                                          & %
\ifref {\cite{Wei:2008nv}} \fi \phantom{.}        & %
\nodata                                           & %
\phantom{.}                                       & %
\nodata                                           & %
\phantom{.}                                       & %
$<0.059$                                          \\

125                                               & %
$\pi e^+ e^-$                                     & %
$<0.110$                                          & %
$<0.110$                                          & %
\ifref {\cite{Lees:2013lvs}} \fi \phantom{.}      & %
\nodata                                           & %
\phantom{.}                                       & %
\nodata                                           & %
\phantom{.}                                       & %
\nodata                                           & %
\phantom{.}                                       & %
$<0.110$                                          \\

126                                               & %
$\pi \mu^+ \mu^-$                                 & %
$<0.050$                                          & %
$<0.050$                                          & %
\ifref {\cite{Lees:2013lvs}} \fi \phantom{.}      & %
\nodata                                           & %
\phantom{.}                                       & %
\nodata                                           & %
\phantom{.}                                       & %
\nodata                                           & %
\phantom{.}                                       & %
$<0.050$                                          \\

127                                               & %
$K e^+ e^-$                                       & %
$0.44 \pm 0.06$                                   & %
{$\aerr{0.39}{0.09}{0.08}{0.02}$}                 & %
\ifref {\cite{Aubert:2008ps}} \fi \phantom{.}     & %
{$\aerr{0.48}{0.08}{0.07}{0.03}$}                 & %
\ifref {\cite{Wei:2009zv}} \fi \phantom{.}        & %
\nodata                                           & %
\phantom{.}                                       & %
\nodata                                           & %
\phantom{.}                                       & %
$0.44 \pm 0.06$                                   \\

128                                               & %
$K^* e^+ e^-$                                     & %
$1.19\pm0.20$                                     & %
{$\aerr{0.99}{0.23}{0.21}{0.06}$}                 & %
\ifref {\cite{Aubert:2008ps}} \fi \phantom{.}     & %
{$\aerr{1.39}{0.23}{0.20}{0.12}$}                 & %
\ifref {\cite{Wei:2009zv}} \fi \phantom{.}        & %
\nodata                                           & %
\phantom{.}                                       & %
\nodata                                           & %
\phantom{.}                                       & %
$\cerr{1.19}{0.17}{0.16}$                         \\

129                                               & %
$K \mu^+ \mu^-$                                   & %
$0.44 \pm 0.04$                                   & %
{$\aerr{0.41}{0.13}{0.12}{0.02}$}                 & %
\ifref {\cite{Aubert:2008ps}} \fi \phantom{.}     & %
{$\err{0.50}{0.06}{0.03}$}                        & %
\ifref {\cite{Wei:2009zv}} \fi \phantom{.}        & %
\nodata                                           & %
\phantom{.}                                       & %
$\err{0.42}{0.04}{0.02}$                          & %
\ifref {\cite{Aaltonen:2011qs}} \fi \phantom{.}   & %
$0.44 \pm 0.04$                                   \\

130                                               & %
$K^* \mu^+ \mu^-$                                 & %
$1.06 \pm 0.09$                                   & %
{$\aerr{1.35}{0.35}{0.33}{0.10}$}                 & %
\ifref {\cite{Aubert:2008ps}} \fi \phantom{.}     & %
{$\aerr{1.10}{0.16}{0.14}{0.08}$}                 & %
\ifref {\cite{Wei:2009zv}} \fi \phantom{.}        & %
\nodata                                           & %
\phantom{.}                                       & %
$\err{1.01}{0.10}{0.05}$                          & %
\ifref {\cite{Aaltonen:2011qs}} \fi \phantom{.}   & %
$1.06 \pm 0.09$                                   \\

131                                               & %
$K \ell^+ \ell^-$                                 & %
$0.48 \pm 0.04$                                   & %
$\err{0.47}{0.06}{0.02}$                          & %
\ifref {\cite{Lees:2012tva}} \fi \phantom{.}      & %
{$\aerr{0.48}{0.05}{0.04}{0.03}$}                 & %
\ifref {\cite{Wei:2009zv}} \fi \phantom{.}        & %
\nodata                                           & %
\phantom{.}                                       & %
\nodata                                           & %
\phantom{.}                                       & %
$0.48 \pm 0.04$                                   \\

132                                               & %
$K^* \ell^+ \ell^-$                               & %
$1.05\pm 0.10$                                    & %
$\aerr{1.02}{0.14}{0.13}{0.05}$                   & %
\ifref {\cite{Lees:2012tva}} \fi \phantom{.}      & %
{$\aerr{1.07}{0.11}{0.10}{0.09}$}                 & %
\ifref {\cite{Wei:2009zv}} \fi \phantom{.}        & %
\nodata                                           & %
\phantom{.}                                       & %
\nodata                                           & %
\phantom{.}                                       & %
$1.05 \pm 0.10$                                   \\

133                                               & %
$K \nu \overline \nu$                             & %
$<17$                                             & %
{$<17$}                                           & %
\ifref {\cite{Lees:2013kla}} \fi \phantom{.}      & %
\red{$<16$}                                       & %
\ifref {\cite{Grygier:2017tzo}} \fi \phantom{.}   & %
\nodata                                           & %
\phantom{.}                                       & %
\nodata                                           & %
\phantom{.}                                       & %
{$<16$}                                           \\

134                                               & %
$K^* \nu \overline \nu$                           & %
$<76$                                             & %
$<76$                                             & %
\ifref {\cite{Lees:2013kla}} \fi \phantom{.}      & %
\red{$<27$}                                       & %
\ifref {\cite{Grygier:2017tzo}} \fi \phantom{.}   & %
\nodata                                           & %
\phantom{.}                                       & %
\nodata                                           & %
\phantom{.}                                       & %
{$<27$}                                           \\

\nodata                                           & %
$\pi \nu \overline \nu$                           & %
\nodata                                           & %
\nodata                                           & %
\phantom{.}                                       & %
{$<9$}                                            & %
\ifref {\cite{Grygier:2017tzo}} \fi \phantom{.}   & %
\nodata                                           & %
\phantom{.}                                       & %
\nodata                                           & %
\phantom{.}                                       & %
{$<9$}                                            \\

\nodata                                           & %
$\rho \nu \overline \nu$                          & %
\nodata                                           & %
\nodata                                           & %
\phantom{.}                                       & %
{$<30$}                                           & %
\ifref {\cite{Grygier:2017tzo}} \fi \phantom{.}   & %
\nodata                                           & %
\phantom{.}                                       & %
\nodata                                           & %
\phantom{.}                                       & %
{$<30$}                                           \\

136                                               & %
$\pi e^\pm \mu^\mp$                               & %
$<0.092$                                          & %
{$<0.092$}                                        & %
\ifref {\cite{Aubert:2007mm}} \fi \phantom{.}     & %
\nodata                                           & %
\phantom{.}                                       & %
\nodata                                           & %
\phantom{.}                                       & %
\nodata                                           & %
\phantom{.}                                       & %
{$<0.092$}                                        \\

137                                               & %
$\rho e^\pm \mu^\mp$                              & %
$<3.2$                                            & %
\nodata                                           & %
\phantom{.}                                       & %
\nodata                                           & %
\phantom{.}                                       & %
$<3.2$                                            & %
\ifref {\cite{Edwards:2002kq}} \fi \phantom{.}    & %
\nodata                                           & %
\phantom{.}                                       & %
$<3.2$                                            \\

138                                               & %
$K e^\pm \mu^\mp$                                 & %
$<0.038$                                          & %
{$<0.038$}                                        & %
\ifref {\cite{Aubert:2006vb}} \fi \phantom{.}     & %
\nodata                                           & %
\phantom{.}                                       & %
\nodata                                           & %
\phantom{.}                                       & %
\nodata                                           & %
\phantom{.}                                       & %
{$<0.038$}                                        \\

139                                               & %
$K^* e^\pm \mu^\mp$                               & %
$<0.51$                                           & %
{$<0.51$}                                         & %
\ifref {\cite{Aubert:2006vb}} \fi \phantom{.}     & %
\nodata                                           & %
\phantom{.}                                       & %
\nodata                                           & %
\phantom{.}                                       & %
\nodata                                           & %
\phantom{.}                                       & %
{$<0.51$}                                         \\

\hline
\end{tabular}
}
\end{center}
\scriptsize{
Channels with no RPP\# were not included in PDG Live as of Dec. 31, 2017. \\ %
Results for CDF are relative BFs converted to absolute BFs.\\     %
CLEO upper limits that have been greatly superseded are not shown. \\     %
$^\dag$~Results extrapolated to $E_{\gamma}>1.6$~\gev, using the method of Ref.~\cite{Buchmuller:2005zv}.\\ %
$^\ddag$ ~Belle: $m(\ell^+\ell^-)>0.2~\gevcc$, \babar: $m^2(\ell^+\ell^-)>0.1~\gevgevcccc$.\\     %
$^\S$~The value quoted is   ${\cal B}(B \to K^*_3 \gamma) \times {\cal B}(K^*_3 \to K\eta)$. PDG gives the BF assuming ${\cal B}(K^*_3 \to K\eta)=\cerr{11}{5}{4}\%$.\\     %
$^\P$~Average of several results, obtained with different methods.\\     %
$^\diamond$~Only results originally measured in the interval $E_\gamma > 1.9$~\gev (also taken into account in the previous line).\\     %
}
\end{table}

\begin{table}[htbp!]
\begin{center}
\caption{Branching fractions of charmless leptonic and radiative-leptonic 
$\Bp$ and $\Bz$ decays in units of $\times 10^{-6}$. Upper limits are
at 90\% CL. 
Where values are shown in \red{red} (\blue{blue}), this indicates that
they are new \red{published} (\blue{preliminary}) results since PDG2017.}
\label{tab:radll_lep}
\resizebox{\textwidth}{!}{
\begin{tabular}{|lccc @{}c c @{}c c @{}c c @{}c c @{}c c @{}c c|} \hline
RPP\# &Mode & PDG2017 Avg. & \babar & & Belle & & CDF & & LHCb & & CMS & & ATLAS & & Our Avg.  \\ \sglinespb
~31                                               & %
$e^+ \nu$                                         & %
$<0.98$                                           & %
{$<1.9$}                                          & %
\ifref {\cite{Aubert:2009ar}} \fi \phantom{.}     & %
$<0.98$~$^\dag$                                   & %
\ifref {\cite{Satoyama:2006xn}} \fi \phantom{.}   & %
\nodata                                           & %
\phantom{.}                                       & %
\nodata                                           & %
\phantom{.}                                       & %
\nodata                                           & %
\phantom{.}                                       & %
\nodata                                           & %
\phantom{.}                                       & %
$<0.98$~$^\dag$                                   \\

~32                                               & %
$\mu^+ \nu$                                       & %
$<1.0$                                            & %
$<1.0$                                            & %
\ifref {\cite{Aubert:2009ar}} \fi \phantom{.}     & %
\blue{$<1.07$}                                    & %
\ifref {\cite{Sibidanov:2017vph}} \fi \phantom{.} & %
\nodata                                           & %
\phantom{.}                                       & %
\nodata                                           & %
\phantom{.}                                       & %
\nodata                                           & %
\phantom{.}                                       & %
\nodata                                           & %
\phantom{.}                                       & %
$<1.0$                                            \\

~33                                               & %
$\tau^+ \nu$                                      & %
$109\pm24$                                        & %
$179 \pm 48$                                      & %
\ifref {\cite{Lees:2012ju}} \fi \phantom{.}       & %
{$\err{91}{19}{11}$}                              & %
\ifref {\cite{Kronenbitter:2015kls}} \fi \phantom{.}& %
\nodata                                           & %
\phantom{.}                                       & %
\nodata                                           & %
\phantom{.}                                       & %
\nodata                                           & %
\phantom{.}                                       & %
\nodata                                           & %
\phantom{.}                                       & %
$106 \pm 19$                                      \\

~34                                               & %
$\ell^+ \nu_{\ell} \gamma$                        & %
$<3.5$                                            & %
$<15.6$                                           & %
\ifref {\cite{Aubert:2009ya}} \fi \phantom{.}     & %
\blue{$<3.0$}                                     & %
\ifref {\cite{Gelb:2018end}} \fi \phantom{.}      & %
\nodata                                           & %
\phantom{.}                                       & %
\nodata                                           & %
\phantom{.}                                       & %
\nodata                                           & %
\phantom{.}                                       & %
\nodata                                           & %
\phantom{.}                                       & %
{$<3.0$}                                          \\

~35                                               & %
$e^+ \nu_e \gamma$                                & %
$<6.1$                                            & %
{$<17$}                                           & %
\ifref {\cite{Aubert:2009ya}} \fi \phantom{.}     & %
{$<6.1$}                                          & %
\ifref {\cite{Heller:2015vvm}} \fi \phantom{.}    & %
\nodata                                           & %
\phantom{.}                                       & %
\nodata                                           & %
\phantom{.}                                       & %
\nodata                                           & %
\phantom{.}                                       & %
\nodata                                           & %
\phantom{.}                                       & %
{$<6.1$}                                          \\

~36                                               & %
$\mu^+ \nu_{\mu} \gamma$                          & %
$<3.4$                                            & %
{$<24$}                                           & %
\ifref {\cite{Aubert:2009ya}} \fi \phantom{.}     & %
{$<3.4$}                                          & %
\ifref {\cite{Heller:2015vvm}} \fi \phantom{.}    & %
\nodata                                           & %
\phantom{.}                                       & %
\nodata                                           & %
\phantom{.}                                       & %
\nodata                                           & %
\phantom{.}                                       & %
\nodata                                           & %
\phantom{.}                                       & %
{$<3.4$}                                          \\

495                                               & %
$\gamma \gamma$                                   & %
$<0.32$                                           & %
$<0.32$                                           & %
\ifref {\cite{delAmoSanchez:2010bx}} \fi \phantom{.}& %
{$<0.62$}                                         & %
\ifref {\cite{Abe:2005bs}} \fi \phantom{.}        & %
\nodata                                           & %
\phantom{.}                                       & %
\nodata                                           & %
\phantom{.}                                       & %
\nodata                                           & %
\phantom{.}                                       & %
\nodata                                           & %
\phantom{.}                                       & %
$<0.32$                                           \\

458                                               & %
$e^+ e^-$                                         & %
$<0.083$                                          & %
{$<0.113$}                                        & %
\ifref {\cite{Aubert:2007hb}} \fi \phantom{.}     & %
$<0.19$                                           & %
\ifref {\cite{Chang:2003yy}} \fi \phantom{.}      & %
$<0.083$                                          & %
\ifref {\cite{Aaltonen:2009vr}} \fi \phantom{.}   & %
\nodata                                           & %
\phantom{.}                                       & %
\nodata                                           & %
\phantom{.}                                       & %
\nodata                                           & %
\phantom{.}                                       & %
$<0.083$                                          \\

497                                               & %
$e^+ e^- \gamma$                                  & %
$<0.12$                                           & %
{$<0.12$}                                         & %
\ifref {\cite{Aubert:2007up}} \fi \phantom{.}     & %
\nodata                                           & %
\phantom{.}                                       & %
\nodata                                           & %
\phantom{.}                                       & %
\nodata                                           & %
\phantom{.}                                       & %
\nodata                                           & %
\phantom{.}                                       & %
\nodata                                           & %
\phantom{.}                                       & %
{$<0.12$}                                         \\

498                                               & %
$\mu^+ \mu^-$                                     & %
$0.00018\pm0.00031$                               & %
{$<0.052$}                                        & %
\ifref {\cite{Aubert:2007hb}} \fi \phantom{.}     & %
$<0.16$                                           & %
\ifref {\cite{Chang:2003yy}} \fi \phantom{.}      & %
$<0.0038$                                         & %
\ifref {\cite{Aaltonen:2013as}} \fi \phantom{.}   & %
{$<0.00034$} $^\P$                                & %
\ifref {\cite{Aaij:2017vad}} \fi \phantom{.}      & %
$<0.00110$ $^\P$                                  & %
\ifref {\cite{Chatrchyan:2013bka}} \fi \phantom{.}& %
\blue{$<0.00021$} $^{\P}$                         & %
\ifref {\cite{Aaboud:2018mst}} \fi \phantom{.}    & %
$<0.00021$                                        \\

499                                               & %
$\mu^+ \mu^- \gamma$                              & %
$<0.16$                                           & %
{$<0.16$}                                         & %
\ifref {\cite{Aubert:2007up}} \fi \phantom{.}     & %
\nodata                                           & %
\phantom{.}                                       & %
\nodata                                           & %
\phantom{.}                                       & %
\nodata                                           & %
\phantom{.}                                       & %
\nodata                                           & %
\phantom{.}                                       & %
\nodata                                           & %
\phantom{.}                                       & %
{$<0.16$}                                         \\

500                                               & %
$\mu^+ \mu^- \mu^+ \mu^-$                         & %
$<0.0053$                                         & %
\nodata                                           & %
\phantom{.}                                       & %
\nodata                                           & %
\phantom{.}                                       & %
\nodata                                           & %
\phantom{.}                                       & %
{$<0.0053$} $^{\P}$                               & %
\ifref {\cite{Aaij:2016kfs}} \fi \phantom{.}      & %
\nodata                                           & %
\phantom{.}                                       & %
\nodata                                           & %
\phantom{.}                                       & %
{$<0.0053$} $^{\P}$                               \\

501                                               & %
$SP,  S\to \mu^+ \mu^-, P\to \mu^+ \mu^-$         & %
$<0.0051$                                         & %
\nodata                                           & %
\phantom{.}                                       & %
\nodata                                           & %
\phantom{.}                                       & %
\nodata                                           & %
\phantom{.}                                       & %
{$<0.0051$} $^\P$                                 & %
\ifref {\cite{Aaij:2016kfs}} \fi \phantom{.}      & %
\nodata                                           & %
\phantom{.}                                       & %
\nodata                                           & %
\phantom{.}                                       & %
{$<0.0051$} $^\P$                                 \\

502                                               & %
$\tau^+ \tau^-$                                   & %
$<4100$                                           & %
{$<4100$}                                         & %
\ifref {\cite{Aubert:2005qw}} \fi \phantom{.}     & %
\nodata                                           & %
\phantom{.}                                       & %
\nodata                                           & %
\phantom{.}                                       & %
{$<1600$}                                         & %
\ifref {\cite{Aaij:2017xqt}} \fi \phantom{.}      & %
\nodata                                           & %
\phantom{.}                                       & %
\nodata                                           & %
\phantom{.}                                       & %
{$<1600$}                                         \\

524                                               & %
$e^\pm \mu^\mp$                                   & %
$<0.0028$                                         & %
{$<0.092$}                                        & %
\ifref {\cite{Aubert:2007hb}} \fi \phantom{.}     & %
$<0.17$                                           & %
\ifref {\cite{Chang:2003yy}} \fi \phantom{.}      & %
$<0.064$                                          & %
\ifref {\cite{Aaltonen:2009vr}} \fi \phantom{.}   & %
\red{$<0.001$}                                    & %
\ifref {\cite{Aaij:2017cza}} \fi \phantom{.}      & %
\nodata                                           & %
\phantom{.}                                       & %
\nodata                                           & %
\phantom{.}                                       & %
{$<0.001$}                                        \\

530                                               & %
$e^{\pm} \tau^{\mp}$                              & %
{$<28$}                                           & %
{$<28$}                                           & %
\ifref {\cite{Aubert:2008cu}} \fi \phantom{.}     & %
\nodata                                           & %
\phantom{.}                                       & %
\nodata                                           & %
\phantom{.}                                       & %
\nodata                                           & %
\phantom{.}                                       & %
\nodata                                           & %
\phantom{.}                                       & %
\nodata                                           & %
\phantom{.}                                       & %
{$<28$}                                           \\

532                                               & %
$\mu^\pm \tau^\mp$                                & %
$<22$                                             & %
{$<22$}                                           & %
\ifref {\cite{Aubert:2008cu}} \fi \phantom{.}     & %
\nodata                                           & %
\phantom{.}                                       & %
\nodata                                           & %
\phantom{.}                                       & %
\nodata                                           & %
\phantom{.}                                       & %
\nodata                                           & %
\phantom{.}                                       & %
\nodata                                           & %
\phantom{.}                                       & %
{$<22$}                                           \\

521                                               & %
$\nu \bar\nu$                                     & %
$<24$                                             & %
$<24$                                             & %
\ifref {\cite{Lees:2012wv}} \fi \phantom{.}       & %
$<130$                                            & %
\ifref {\cite{Hsu:2012uh}} \fi \phantom{.}        & %
\nodata                                           & %
\phantom{.}                                       & %
\nodata                                           & %
\phantom{.}                                       & %
\nodata                                           & %
\phantom{.}                                       & %
\nodata                                           & %
\phantom{.}                                       & %
$<24$                                             \\

522                                               & %
$\nu \bar\nu\gamma$                               & %
{$<17$}                                           & %
{$<17$}                                           & %
\ifref {\cite{Lees:2012wv}} \fi \phantom{.}       & %
\nodata                                           & %
\phantom{.}                                       & %
\nodata                                           & %
\phantom{.}                                       & %
\nodata                                           & %
\phantom{.}                                       & %
\nodata                                           & %
\phantom{.}                                       & %
\nodata                                           & %
\phantom{.}                                       & %
{$<17$}                                           \\

\sglinespt
\end{tabular}
}
\end{center}
\scriptsize{
Results for CDF, LHCb, CMS and ATLAS are relative BFs converted to absolute BFs.\\     %
$^\dag$~More recent results exist, with hadronic tagging~\cite{Yook:2014kga},
that do not improve the limits ({$<3.5$} and {$<2.7$}) for $e^+\nu$ and $\mu^+\nu$, respectively).\\ %
$^\P$~UL at 95\% CL.      %
}
\end{table}

\begin{table}[!htbp]
\begin{center}
\caption{Relative branching fractions of charmless radiative and FCNC decays with leptons of 
$\Bp$ and $\Bz$ mesons. 
Where values are shown in \red{red} (\blue{blue}), this indicates that
they are new \red{published} (\blue{preliminary}) results since PDG2017.}
\label{tab:radll_rel}
\resizebox{\textwidth}{!}{

\begin{tabular}{|lccc @{}c c @{}c c @{}c c|} \hline
RPP\# &Mode & PDG2017 AVG. & Belle & & \babar & & LHCb & & Our Avg.  \\ \sglinespb
548/298                                           & %
$10^4\times\mathcal{B}(K^+\pi^+\pi^-\mu^+\mu^-)/\mathcal{B}(\psi(2S)K^+)$& %
{$\aerr{6.95}{0.46}{0.43}{0.34}$}                 & %
\nodata                                           & %
\phantom{.}                                       & %
\nodata                                           & %
\phantom{.}                                       & %
{$\aerr{6.95}{0.46}{0.43}{0.34}$}                 & %
\ifref {\cite{Aaij:2014kwa}} \fi \phantom{.}      & %
$\cerr{6.95}{0.57}{0.55}$                         \\

549/274                                           & %
$10^4\times\mathcal{B}(K^+\phi\mu^+\mu^-)/\mathcal{B}(\psi(2S)K^+)$& %
{$\aerrsy{1.58}{0.36}{0.32}{0.19}{0.07}$}         & %
\nodata                                           & %
\phantom{.}                                       & %
\nodata                                           & %
\phantom{.}                                       & %
{$\aerrsy{1.58}{0.36}{0.32}{0.19}{0.07}$}         & %
\ifref {\cite{Aaij:2014kwa}} \fi \phantom{.}      & %
$\cerr{1.58}{0.41}{0.33}$                         \\

536/540                                           & %
$\mathcal{B}(\pi^+\mu^+\mu^-)/\mathcal{B}(K^+\mu^+\mu^-)$~$^\dag$& %
$\err{0.053}{0.014}{0.01}$                        & %
\nodata                                           & %
\phantom{.}                                       & %
\nodata                                           & %
\phantom{.}                                       & %
{$\err{0.038}{0.009}{0.001}$}                     & %
\ifref {\cite{Aaij:2015nea}} \fi \phantom{.}      & %
$0.038 \pm 0.009$                                 \\

\nodata                                           & %
$\mathcal{B}(K^+\mu^+\mu^-)/\mathcal{B}(K^+e^+e^-)$~$^\ddag$& %
\nodata                                           & %
\nodata                                           & %
\phantom{.}                                       & %
\nodata                                           & %
\phantom{.}                                       & %
$\aerr{0.745}{0.090}{0.074}{0.036}$               & %
\ifref {\cite{Aaij:2014ora}} \fi \phantom{.}     & %
$\cerr{0.745}{0.097}{0.082}$                      \\

\nodata                                           & %
$\mathcal{B}(K^+\mu^+\mu^-)/\mathcal{B}(K^+e^+e^-)$~$^\ddag$& %
\nodata                                           & %
\nodata                                           & %
\phantom{.}                                       & %
$\aerr{1.00}{0.31}{0.25}{0.07}$                   & %
\ifref {\cite{Lees:2012tva}} \fi \phantom{.}      & %
\nodata                                           & %
\phantom{.}                                       & %
$\cerr{1.00}{0.32}{0.26}$                         \\

\nodata                                           & %
$\mathcal{B}(K^+\mu^+\mu^-)/\mathcal{B}(K^+e^+e^-)$~$^\S$& %
\nodata                                           & %
$\err{1.03}{0.19}{0.06}$                          & %
\ifref {\cite{Wei:2009zv}} \fi \phantom{.}        & %
\nodata                                           & %
\phantom{.}                                       & %
\nodata                                           & %
\phantom{.}                                       & %
$1.03 \pm 0.20$                                   \\

\nodata                                           & %
$\mathcal{B}(K^*\mu^+\mu^-)/\mathcal{B}(K^*e^+e^-)$~$^\S$& %
\nodata                                           & %
$\err{0.83}{0.17}{0.08}$                          & %
\ifref {\cite{Wei:2009zv}} \fi \phantom{.}        & %
\nodata                                           & %
\phantom{.}                                       & %
\nodata                                           & %
\phantom{.}                                       & %
$0.83 \pm 0.19$                                   \\

\nodata                                           & %
$\mathcal{B}(K^*\mu^+\mu^-)/\mathcal{B}(K^*e^+e^-)$~$^\P$& %
\nodata                                           & %
\nodata                                           & %
\phantom{.}                                       & %
$\aerr{1.013}{0.34}{0.26}{0.010}$                 & %
\ifref {\cite{Lees:2012tva}} \fi \phantom{.}      & %
\nodata                                           & %
\phantom{.}                                       & %
$\cerr{1.013}{0.340}{0.260}$                      \\

\nodata                                           & %
$\mathcal{B}(K^{*0}\mu^+\mu^-)/\mathcal{B}(K^{*0}e^+e^-)$~$^\diamond$& %
\nodata                                           & %
\nodata                                           & %
\phantom{.}                                       & %
\nodata                                           & %
\phantom{.}                                       & %
{$\aerr{0.66}{0.11}{0.07}{0.03}$}                 & %
\ifref {\cite{Aaij:2017vbb}} \fi \phantom{.}      & %
$\cerr{0.66}{0.11}{0.08}$                         \\

\nodata                                           & %
$\mathcal{B}(K^{*0}\mu^+\mu^-)/\mathcal{B}(K^{*0}e^+e^-)$~$^1$& %
\nodata                                           & %
\nodata                                           & %
\phantom{.}                                       & %
\nodata                                           & %
\phantom{.}                                       & %
{$\aerr{0.69}{0.11}{0.07}{0.05}$}                 & %
\ifref {\cite{Aaij:2017vbb}} \fi \phantom{.}      & %
$\cerr{0.69}{0.12}{0.09}$                         \\

\nodata                                           & %
$\mathcal{B}(B^0\to K^{*0}\gamma)/\mathcal{B}(B^0_s \to \phi\gamma)$& %
\nodata                                           & %
$\gerrsyt{1.10}{0.16}{0.09}{0.18}$                & %
\ifref {\cite{Horiguchi:2017ntw}} \fi \phantom{.} & %
\nodata                                           & %
\phantom{.}                                       & %
$\err{1.23}{0.06}{0.11}$                          & %
\ifref {\cite{Aaij:2012ita}} \fi \phantom{.}      & %
$1.21 \pm 0.11$                                   \\

\hline
\end{tabular}
}
\end{center}
\scriptsize{
Channels with no RPP\# were not included in PDG Live as of Dec. 31, 2017. \\ %
$^\dag$~For $0.1<m^2(\ell^+\ell^-)<6.0~\gevgevcccc$.\\      %
$^\ddag$~For $1.0<m^2(\ell^+\ell^-)<6.0~\gevgevcccc$.\\      %
$^\S$~For the full $m^2(\ell^+\ell^-)$ range.\\      %
$^\P$~For $0.10<m^2(\ell^+\ell^-)<8.12~\gevgevcccc$ and $m^2(\ell^+\ell^-)>10.11~\gevgevcccc$.\\      %
$^\diamond$~For $0.045<m^2(\ell^+\ell^-)<1.1~\gevgevcccc$.\\      %
$^1$~For $1.1<m^2(\ell^+\ell^-)<6.0~\gevgevcccc$.\\      %
}
\end{table}

\begin{table}[!htbp]
\begin{center}
\caption{Branching fractions of $\Bp/\Bz\to\bar{q}$ gluon decays in units of $\times 10^{-6}$. Upper limits are
at 90\% CL. 
Where values are shown in \red{red} (\blue{blue}), this indicates that
they are new \red{published} (\blue{preliminary}) results since PDG2017.}
\label{tab:radll_gluon}
\resizebox{\textwidth}{!}{
\begin{tabular}{|lccc @{}c c @{}c c @{}c c|} \hline
RPP\# &Mode & PDG2017 Avg. & \babar & & Belle & & CLEO & & Our Avg. \\
\sglinespb
$~81$                                             & %
$\eta X$                                          & %
$\cerr{260}{50}{80}$                              & %
\nodata                                           & %
\phantom{.}                                       & %
{\berr{261}{30}{44}{74}}~$^\S$                    & %
\ifref {\cite{Nishimura:2009ae}} \fi \phantom{.}  & %
$<440$                                            & %
\ifref {\cite{Browder:1998yb}} \fi \phantom{.}    & %
$\cerr{261}{53}{79}$                              \\

$~82$                                             & %
$\etapr X$                                        & %
$420 \pm 90$                                      & %
\err{390}{80}{90}~$^\dag$                         & %
\ifref {\cite{Aubert:2004eq}} \fi \phantom{.}     & %
\nodata                                           & %
\phantom{.}                                       & %
\err{460}{110}{60}~$^\dag$                        & %
\ifref {\cite{Bonvicini:2003aw}} \fi \phantom{.}  & %
$423 \pm 86$                                      \\

$~83$                                             & %
$K^+ X$                                           & %
$<187$                                            & %
$<187$~$^\ddag$                                   & %
\ifref {\cite{delAmoSanchez:2010gx}} \fi \phantom{.}& %
\nodata                                           & %
\phantom{.}                                       & %
\nodata                                           & %
\phantom{.}                                       & %
$<187$~$^\ddag$                                   \\

$~84$                                             & %
$K^0 X$                                           & %
\cerr{190}{70}{70}                                & %
\aerr{195}{51}{45}{50}~$^\ddag$                   & %
\ifref {\cite{delAmoSanchez:2010gx}} \fi \phantom{.}& %
\nodata                                           & %
\phantom{.}                                       & %
\nodata                                           & %
\phantom{.}                                       & %
$\cerr{195}{71}{67}$                              \\

$~95$                                             & %
$\pi^+ X$                                         & %
$370\pm80$                                        & %
{\aerr{372}{50}{47}{59}}~$^\P$                    & %
\ifref {\cite{delAmoSanchez:2010gx}} \fi \phantom{.}& %
\nodata                                           & %
\phantom{.}                                       & %
\nodata                                           & %
\phantom{.}                                       & %
$\cerr{372}{77}{75}$                              \\

\hline
\end{tabular}
}
\end{center}
\scriptsize{
$^\dag$~$2.0 < p^*(\etapr) < 2.7~\gevc$. \\      %
$^\ddag$~$m_{X} < 1.69~\gevcc$. \\      %
$^\S$~$0.4 < m_{X} < 2.6~\gevcc$. \\      %
$^\P$~$m_{X} < 1.71~\gevcc$.      %
}
\end{table}

\begin{table}[!htbp]
\begin{center}
\caption{Isospin asymmetry in radiative and FCNC decays with leptons of $B$ mesons.
The notations are those adopted by the PDG.
Where values are shown in \red{red} (\blue{blue}), this indicates that
they are new \red{published} (\blue{preliminary}) results since PDG2017.}
\label{tab:radll_AI}
\resizebox{\textwidth}{!}{
\begin{tabular}{|cccc @{}c c @{}c c @{}c c|}
\sgline
Parameter & & PDG2017 Avg. & \babar & & Belle & & LHCb & & Our Avg. \\
\sglinespb
$\Delta_{0^-}(X_s\gamma)$                         & %
\phantom{.}                                       & %
$-0.01 \pm 0.06$                                  & %
$\cerr{-0.01}{0.06}{0.06}$~$^\ddag$               & %
\ifref {\cite{Aubert:2005cua,Aubert:2007my}} \fi \phantom{.}& %
\red{\gerrsyt{-0.0048}{0.0149}{0.0097}{0.0115}}   & %
\ifref {\cite{Watanuki:2018xxg}} \fi \phantom{.}  & %
\nodata                                           & %
\phantom{.}                                       & %
$-0.0055 \pm 0.0198$                              \\

$\Delta_{0^+}(K^* \gamma)$                        & %
\phantom{.}                                       & %
$0.052 \pm 0.026$                                 & %
\err{0.066}{0.021}{0.022}                         & %
\ifref {\cite{Aubert:2009ak}} \fi \phantom{.}     & %
\red{\gerrsyt{0.062}{0.015}{0.006}{0.012}}        & %
\ifref {\cite{Horiguchi:2017ntw}} \fi \phantom{.} & %
\nodata                                           & %
\phantom{.}                                       & %
$0.063 \pm 0.017$                                 \\

$\Delta_{\rho \gamma}$                            & %
\phantom{.}                                       & %
$-0.46 \pm 0.17$                                  & %
{$\aerr{-0.43}{0.25}{0.22}{0.10}$}                & %
\ifref {\cite{Aubert:2008al}} \fi \phantom{.}     & %
{$\aerrsy{-0.48}{0.21}{0.19}{0.08}{0.09}$}        & %
\ifref {\cite{Taniguchi:2008ty}} \fi \phantom{.}  & %
\nodata                                           & %
\phantom{.}                                       & %
$\cerr{-0.46}{0.17}{0.16}$                        \\

$\Delta_{0-}(K\ell\ell)~^\dag$                    & %
\phantom{.}                                       & %
$-0.13 \pm 0.06$                                  & %
{$\aerr{-0.58}{0.29}{0.37}{0.02}$}                & %
\ifref {\cite{Lees:2012tva}} \fi \phantom{.}      & %
{$\aerr{-0.31}{0.17}{0.14}{0.08}$}                & %
\ifref {\cite{Wei:2009zv}} \fi \phantom{.}        & %
{$\aerr{-0.10}{0.08}{0.09}{0.02}$}~$^\S$          & %
\ifref {\cite{Aaij:2014pli}} \fi \phantom{.}      & %
$-0.17 \pm 0.08$                                  \\

$\Delta_{0-}(K^*\ell\ell)~^\dag$                  & %
\phantom{.}                                       & %
$-0.45 \pm 0.17$                                  & %
$\aerr{-0.64}{0.15}{0.14}{0.03}$                  & %
\ifref {\cite{Lees:2012tva}} \fi \phantom{.}      & %
$\aerr{0.30}{0.12}{0.11}{0.08}$                   & %
\ifref {\cite{Wei:2009zv}} \fi \phantom{.}        & %
{$\aerr{0.00}{0.12}{0.10}{0.02}$}~$^\S$           & %
\ifref {\cite{Aaij:2014pli}} \fi \phantom{.}      & %
$-0.06 \pm 0.07$                                  \\

\sglinespt
\end{tabular}
}
\end{center}
\scriptsize{
In some of the $B$-factory results it is assumed that
$\mathcal{B}(\Upsilon(4S) \to B^+ B^-) = \mathcal{B}(\Upsilon(4S) \to B^0 \overline{B}{}^0)$,
and in others a measured value of the ratio of branching fractions is used.
See original papers for details.
The averages quoted above are computed naively and should be treated with caution.\\
$^\dag$~Results given for the bin $1<m^2(\ell^+\ell^-)<6~\gevgevcccc$, see references for the other bins.\\      %
$^\ddag$~Average of two independent measurements from \babar~\cite{Aubert:2005cua,Aubert:2007my}.\\      %
$^\S$~Only muons are used. \\      %
}
\end{table}

\begin{table}[!htbp]
\begin{center}
\caption{Branching fractions of charmless semileptonic
$B^+$ decays to LFV and LNV final states in units of $\times 10^{-6}$. Upper limits are
at 90\% CL. 
Where values are shown in \red{red} (\blue{blue}), this indicates that
they are new \red{published} (\blue{preliminary}) results since PDG2017.}
\label{tab:radll_Bp_lnv}
\resizebox{\textwidth}{!}{

}
\end{center}
\scriptsize{
Results for LHCb are relative BFs converted to absolute BFs.\\     %
CLEO upper limits that have been greatly superseded are not shown. \\     %
$^\dag$~UL at $95\%$ CL.     %
}
\end{table}

\begin{table}[!htbp]
\begin{center}
\caption{Branching fractions of charmless semileptonic
$B^0$ decays to LFV and LNV final states in units of $\times 10^{-6}$. Upper limits are
at 90\% CL. 
Where values are shown in \red{red} (\blue{blue}), this indicates that
they are new \red{published} (\blue{preliminary}) results since PDG2017.}
\label{tab:radll_Bz_lnv}
\resizebox{\textwidth}{!}{
\begin{tabular}{|lccc @{}c c @{}c c @{}c c|} \hline
RPP\# &Mode & PDG2017 Avg. & \babar & & BELLE & & LHCb & & Our Avg. \\
\sglinespb

525                                               & %
$\pi^0 e^\pm \mu^\mp$                             & %
{$<0.14$}                                         & %
{$<0.14$}                                         & %
\ifref {\cite{Aubert:2007mm}} \fi \phantom{.}     & %
\nodata                                           & %
\phantom{.}                                       & %
\nodata                                           & %
\phantom{.}                                       & %
{$<0.14$}                                         \\

526                                               & %
$K^0 e^\pm \mu^\mp$                               & %
{$<0.27$}                                         & %
{$<0.27$}                                         & %
\ifref {\cite{Aubert:2006vb}} \fi \phantom{.}     & %
\nodata                                           & %
\phantom{.}                                       & %
\nodata                                           & %
\phantom{.}                                       & %
{$<0.27$}                                         \\

527                                               & %
$K^{*0} e^+ \mu^-$                                & %
{$<0.53$}                                         & %
{$<0.53$}                                         & %
\ifref {\cite{Aubert:2006vb}} \fi \phantom{.}     & %
\red{$<0.16$}                                     & %
\ifref {\cite{Sandilya:2018pop}} \fi \phantom{.}  & %
\nodata                                           & %
\phantom{.}                                       & %
{$<0.16$}                                         \\

528                                               & %
$K^{*0} e^- \mu^+$                                & %
{$<0.34$}                                         & %
{$<0.34$}                                         & %
\ifref {\cite{Aubert:2006vb}} \fi \phantom{.}     & %
\red{$<0.12$}                                     & %
\ifref {\cite{Sandilya:2018pop}} \fi \phantom{.}  & %
\nodata                                           & %
\phantom{.}                                       & %
{$<0.12$}                                         \\

529                                               & %
$K^{*0} e^{\pm} \mu^{\mp}$                        & %
{$<0.58$}                                         & %
{$<0.58$}                                         & %
\ifref {\cite{Aubert:2006vb}} \fi \phantom{.}     & %
\red{$<0.18$}                                     & %
\ifref {\cite{Sandilya:2018pop}} \fi \phantom{.}  & %
\nodata                                           & %
\phantom{.}                                       & %
{$<0.18$}                                         \\

532                                               & %
$\Lambda^{+}_{c}\mu^{-}$                          & %
{$<1.4$}                                          & %
{$<1.4$}                                          & %
\ifref {\cite{BABAR:2011ac}} \fi \phantom{.}      & %
\nodata                                           & %
\phantom{.}                                       & %
\nodata                                           & %
\phantom{.}                                       & %
{$<1.4$}                                          \\

533                                               & %
$\Lambda^{+}_{c}e^{-}$                            & %
{$<4$}                                            & %
{$<4$}                                            & %
\ifref {\cite{BABAR:2011ac}} \fi \phantom{.}      & %
\nodata                                           & %
\phantom{.}                                       & %
\nodata                                           & %
\phantom{.}                                       & %
{$<4$}                                            \\

\hline
\end{tabular}
}
\end{center}
\scriptsize{
Channels with no RPP\# were not included in PDG Live as of Dec. 31, 2017. \\ %
}
\end{table}

\clearpage
\noindent List of other measurements that are not included in the tables:
\begin{itemize}
\item  $B^+ \to K^+ \pi^- \pi^+ \gamma$ : LHCb has measured the up-down asymmetries in bins of the $K\pi\pi\gamma$ mass \cite{Aaij:2014wgo}.
\item  In \cite{Aaij:2013hha}, LHCb has also measured the branching fraction of $B^+ \to K^+ e^- e^+$  in the $m^2(\ell \ell)$ bin $[1, 6]~\gevgevcccc$.
\item  In the $B^+ \to \pi^+ \mu^+ \mu^-$ paper \cite{Aaij:2015nea},
LHCb has also measured the differential branching fraction in bins of  $m^2(\ell \ell)$.
\item For $B \to K \ell^- \ell^+$, LHCb has measured $F_H$ and $A_{\rm FB}$ in 17 (5) bins of $m^2(\ell \ell)$  for the $K^+$ (\ks) final state \cite{Aaij:2014tfa}. %
Belle has measured $F_L$ and $A_{\rm FB}$ in 6 $m^2(\ell \ell)$ bins~[64].%
\item For the  $B \to K^{*} \ell^- \ell^+$ analyses, partial branching fractions and angular observables in bins of $m^2(\ell\ell)$ are also available:  
\begin{itemize}
\item   $B^0 \to K^{*0} e^- e^+$ : LHCb has measured $F_L$, $A_T^{(2)}$, $A_T^{\rm Im}$, $A_T^{\rm Re}$ in the $[0.002, 1.120]~\gevgevcccc$ bin of $m^2(\ell \ell)$ \cite{Aaij:2015dea},
and has also determined the branching fraction in the dilepton mass region $[10,1000]~\mevcc$ \cite{Aaij:2013hha}.
\item   $B \to K^{*} \ell^- \ell^+$ : Belle has measured $F_L$, $A_{\rm FB}$, isospin asymmetry in 6 $m^2(\ell \ell)$ bins \cite{Wei:2009zv}
[41]
and $P_4'$, $P_5'$, $P_6'$, $P_8'$ in 4 $m^2(\ell \ell)$ bins \cite{Abdesselam:2016llu}. In a more recent paper~\cite{Wehle:2016yoi}, they report measurements of $P_4'$ and $P_5'$, separately for $\ell=\mu$ or $e$, in 4 $m^2(\ell \ell)$ bins and in the region $[1, 6]~\gevgevcccc$ bin of $m^2(\ell \ell)$. The measurements use both $\Bz$ and $\Bp$ decays. They also measure the LFV observables $Q_i = P_i^{\mu}-P_i^e$, for $i=4,5$.
\babar\ has measured $F_L$, $A_{\rm FB}$, $P_2$ in 5 $m^2(\ell \ell)$ bins \cite{Lees:2015ymt}.
\item   $B^0 \to K^{*0} \mu^- \mu^+$ : LHCb has measured $F_L$, $A_{\rm FB}$, $S_3-S_9$, $A_3-A_9$, $P_1-P_3$, $P_4'-P_8'$ in 8 $m^2(\ell \ell)$ bins \cite{Aaij:2015oid}.
CMS has  measured $F_L$ and $A_{\rm FB}$ in 7 $m^2(\ell \ell)$ bins \cite{Khachatryan:2015isa}, and $P_1,P_5'$ in \cite{Sirunyan:2017dhj}. ATLAS has measured $F_L$, $S_{3,4,5,7,8}$ and $P_{1,4,5,6,8}'$ in  6 $m^2(\ell \ell)$ bins \cite{ATLAS-CONF-2017-023}.

\end{itemize}
\item For $B \to X_s \ell^- \ell^+$ ($X_s$ is a hadronic system with an $s$ quark), Belle has measured $A_{\rm FB}$ in bins of $m^2(\ell \ell)$ with a sum of 10 exclusive final states \cite{Sato:2014pjr}.

\item $B^0 \to K^+ \pi^- \mu^+ \mu^-$, with $1330 < m(K^+ \pi^-) < 1530~\gevcc$: LHCb has measured the partial branching fraction in bins of  $m^2(\mu^+ \mu^-)$ in the range $[0.1,8.0]~\gevgevcccc$, and has also determined angular moments \cite{Aaij:2016kqt}.
\item In~\cite{Aaij:2016cbx}, LHCb measures the phase difference between the short- and long-distance contributions to the $\Bp \to K^+\mu^+\mu^-$ decay. The measurement is based on the analysis of the dimuon mass distribution in the regions of the $J/\psi$ and $\psi(2S)$ resonances and far from their poles, to probe long and short distance effects, respectively.
\item In~\cite{Sirunyan:2018jll} CMS studies the angular distribution of $\Bp \to K^+\mu^+\mu^-$ and measures, in 7 $m^2(\mu^+ \mu^-)$ bins, $A_{\rm FB}$ and the contribution $F_{\rm H}$ from the pseudoscalar, scalar and tensor amplitudes to the decay.
\item In~\cite{Aaij:2016qsm} LHCb performs a search for a hypothetical new scalar particle $\chi$, assumed to have a narrow width, through the decay $\Bp \to \chi(\mu^+\mu^-)$ in the ranges of mass $250<m(\chi)<4700\ \mevcc$ and lifetime $0.1<\tau(\chi)<1000$~ps. Upper limits are given as a function of $m(\chi)$ and $\tau(\chi)$.

\end{itemize}

\begin{figure}[htbp!]
\centering
\includegraphics[width=0.5\textwidth]{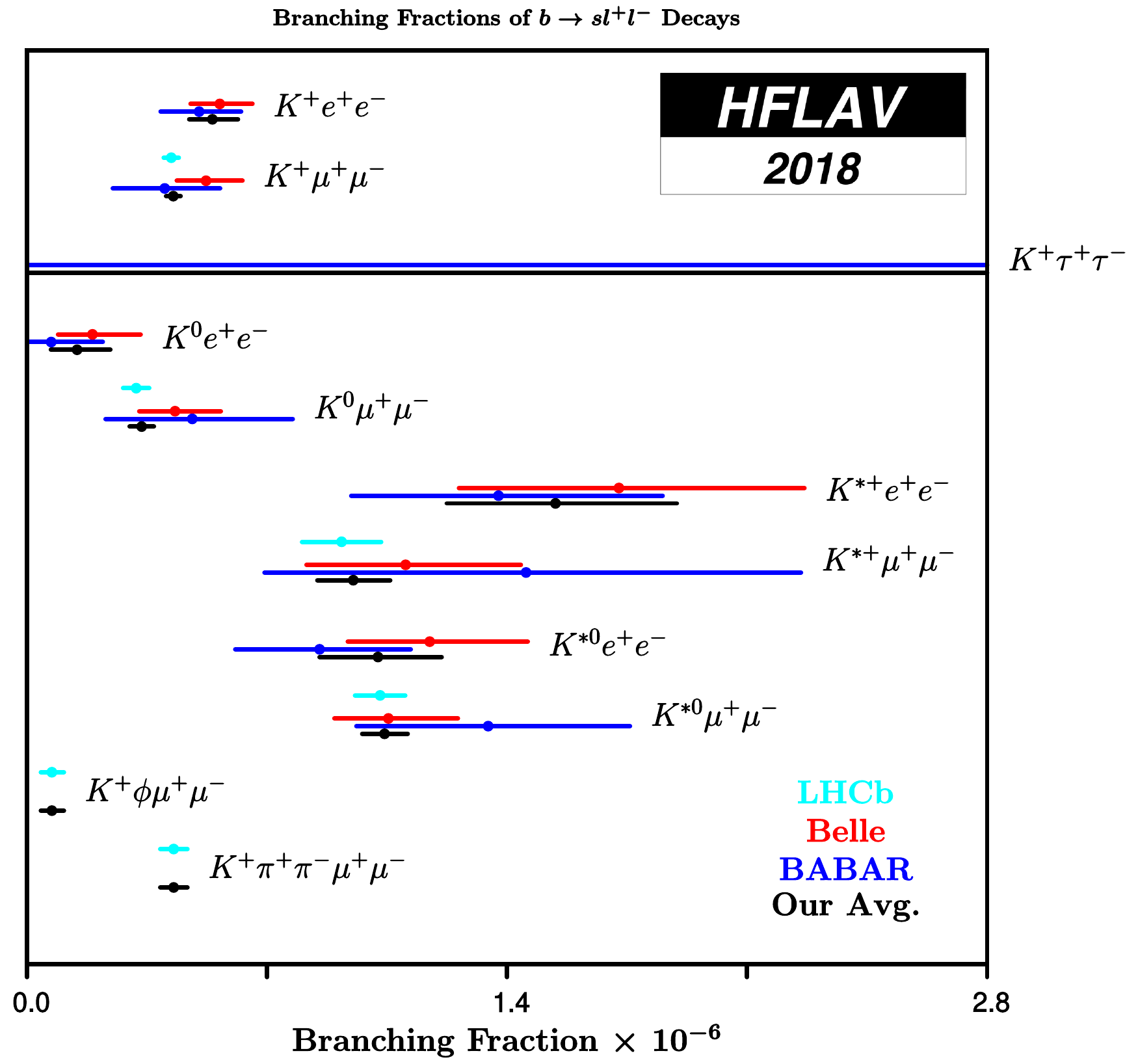}
\caption{Branching fractions of $b\to s\ell^{+}\ell^{-}$ decays.}
\label{fig:rare-btosll}
\end{figure}

\begin{figure}[htbp!]
\centering
\includegraphics[width=0.5\textwidth]{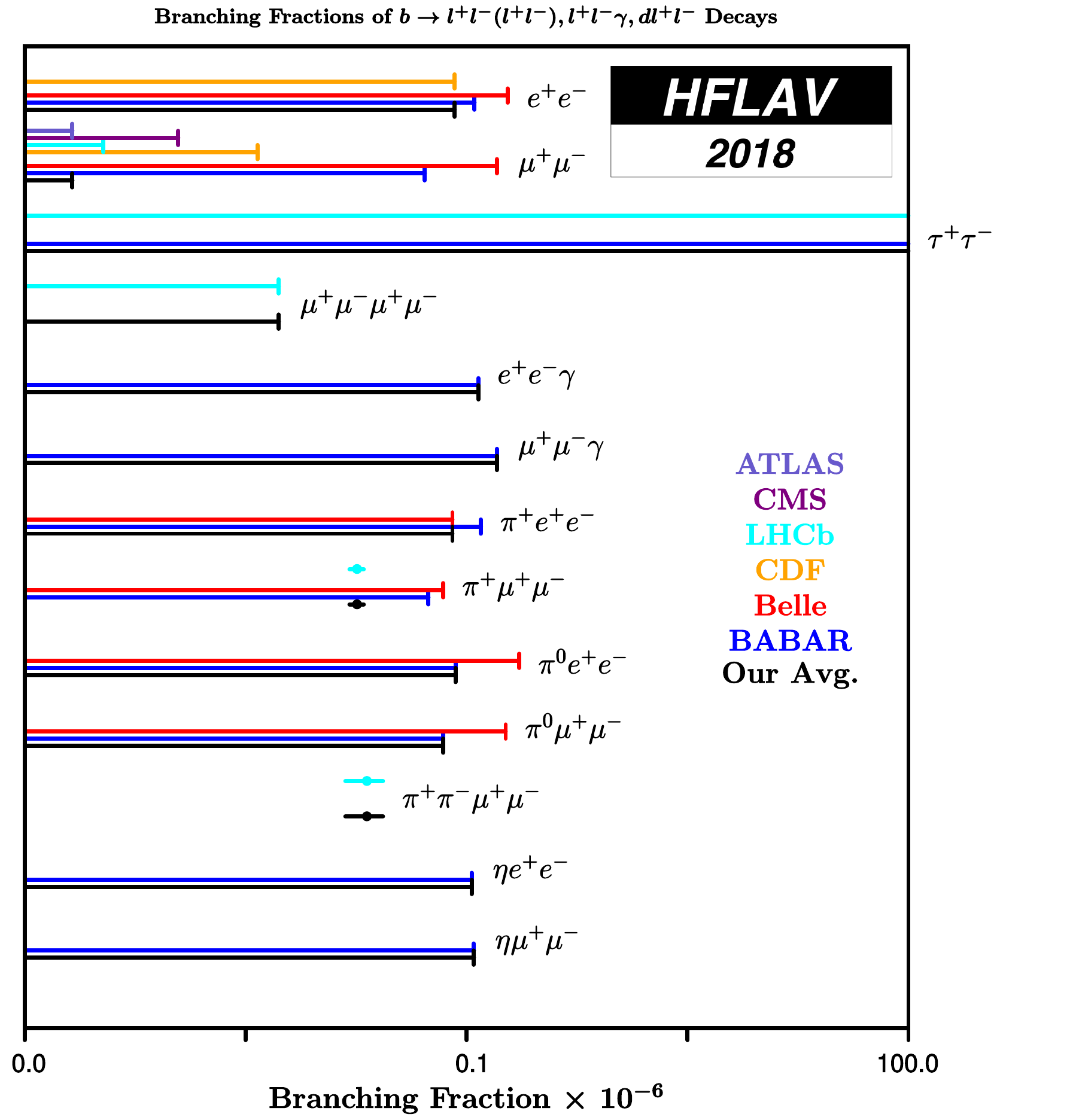}
\caption{Branching fractions of $b\to \ell^{+}\ell^{-}(\ell^{+}\ell^{-}), \ell^{+}\ell^{-}\gamma$ and $b\to d \ell^{+}\ell^{-}$ decays.}
\label{fig:rare-btodllg}
\end{figure}

\begin{figure}[htbp!]
\centering
\includegraphics[width=0.5\textwidth]{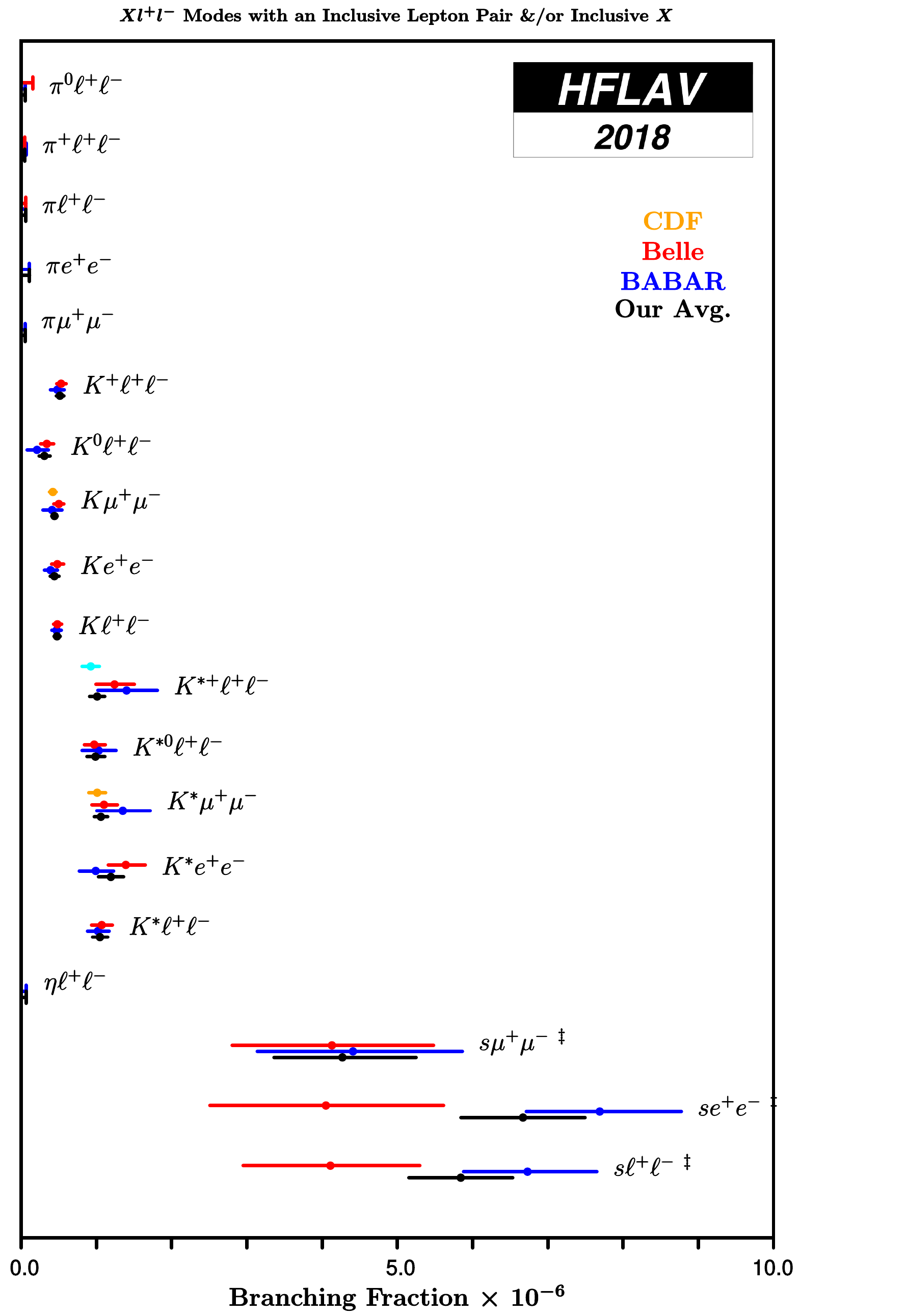}
\caption{$X \ell^+ \ell^-$ modes with an inclusive lepton pair and/or inclusive $X$.}
\label{fig:rare-kllsummary}
\end{figure}

\begin{figure}[htbp!]
\centering
\includegraphics[width=0.5\textwidth]{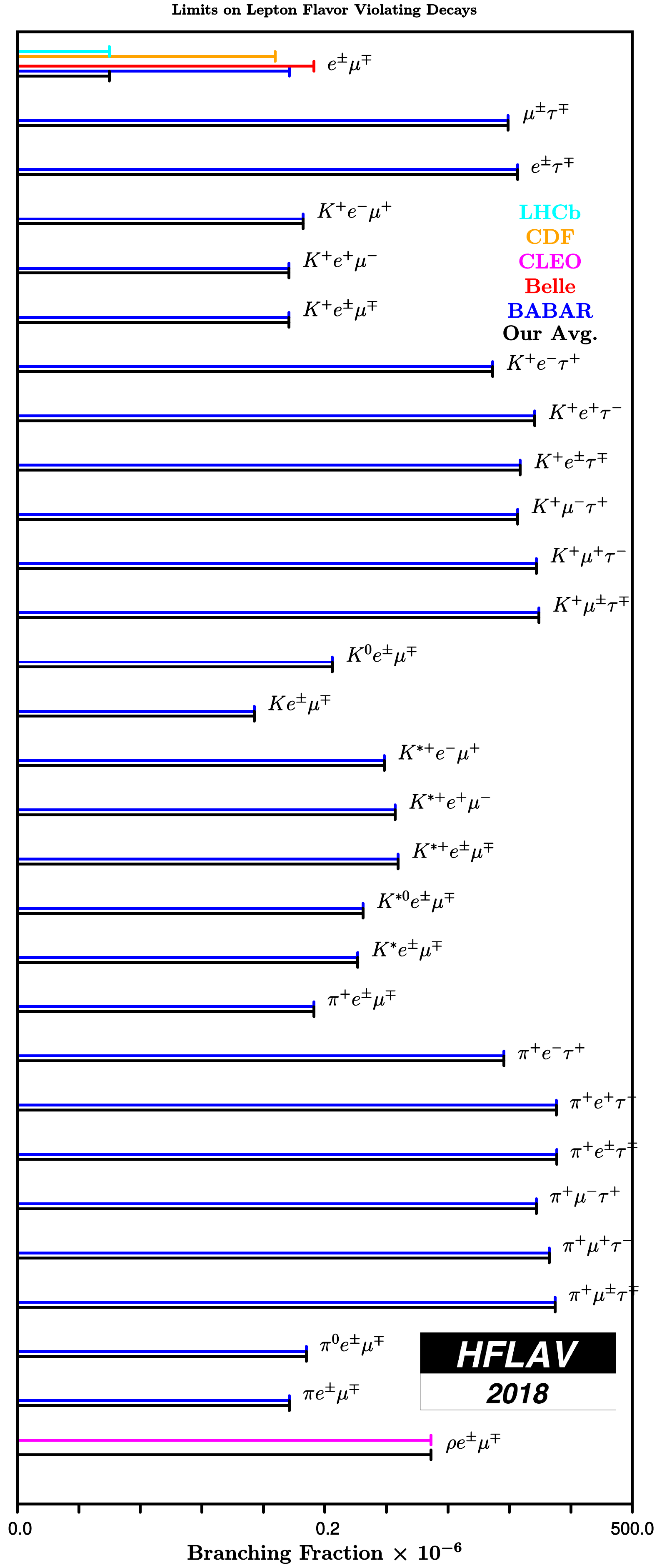}
\caption{Limits on lepton-flavour-violating decays.}
\label{fig:rare-leptonflavourviol}
\end{figure}

\begin{figure}[htbp!]
\centering
\includegraphics[width=0.5\textwidth]{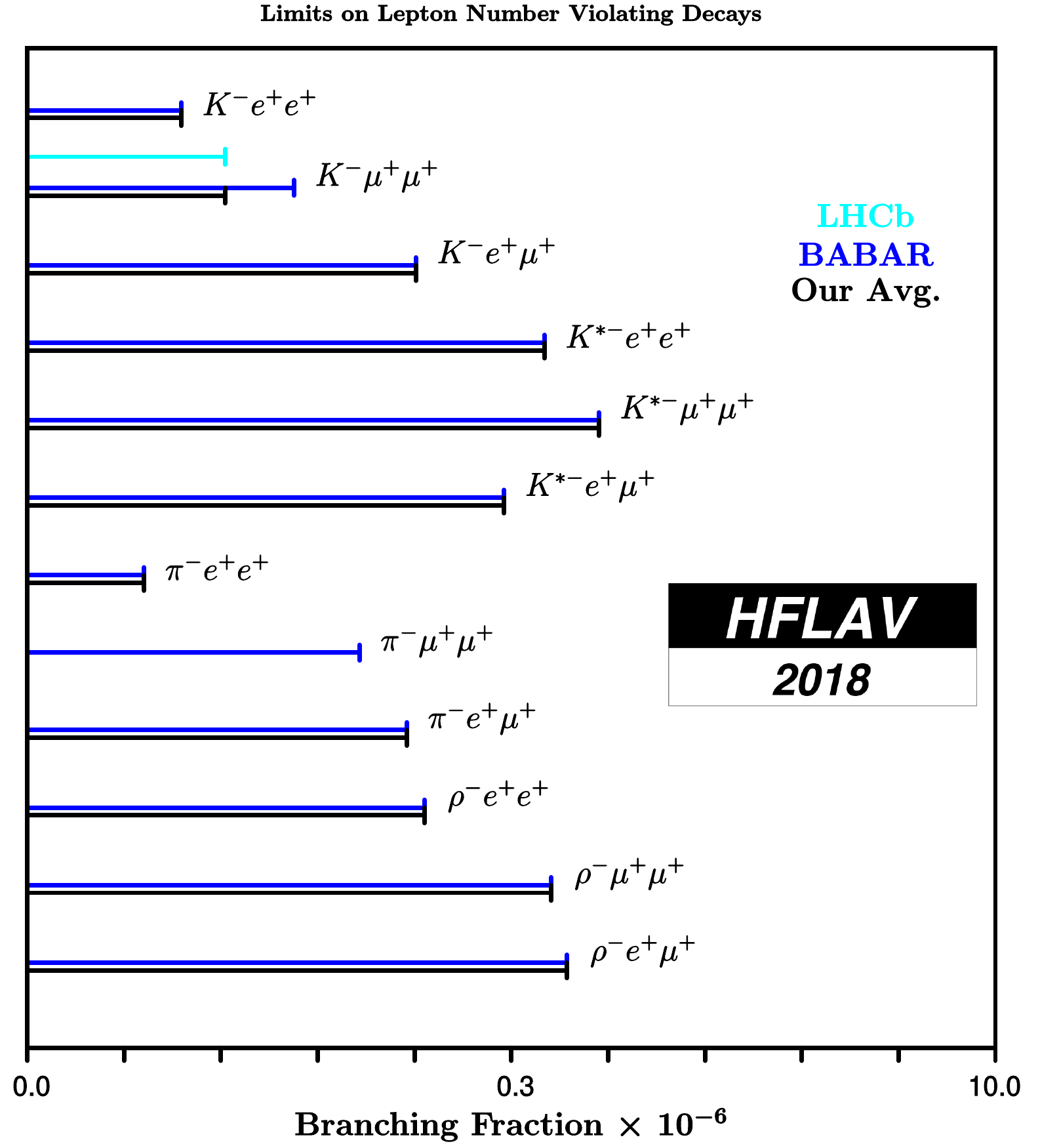}
\caption{Limits on lepton-number-violating decays.}
\label{fig:rare-leptonnumberviol}
\end{figure}

\clearpage

\begin{figure}[htbp!]
\centering
\includegraphics[width=0.5\textwidth]{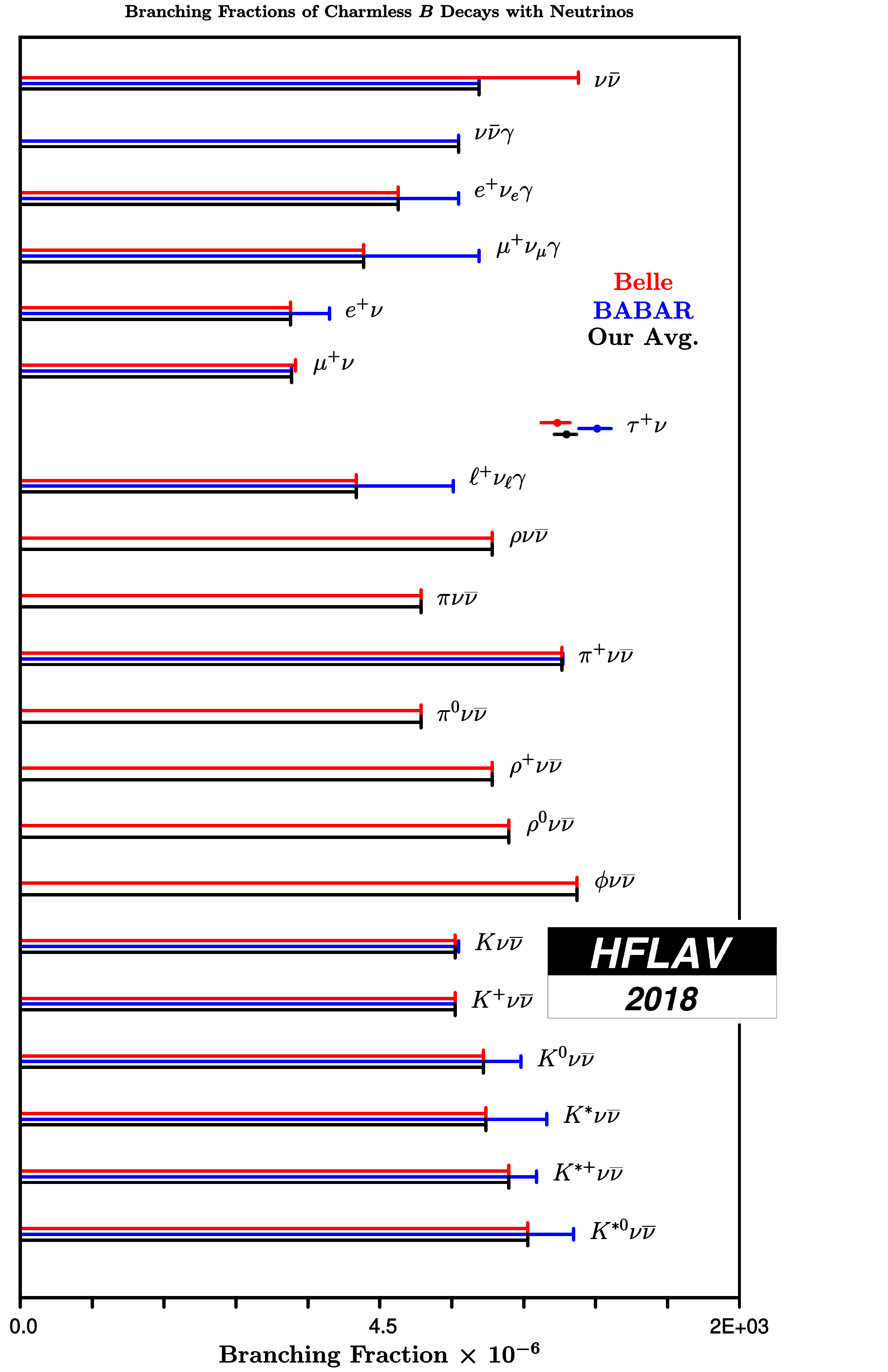}
\caption{Branching fractions of charmless $B$ decays with neutrinos.}
\label{fig:rare-neutrino}
\end{figure}

\mysubsection{Charge asymmetries in \b-hadron decays}
\label{sec:rare-acp}

This section contains, in Tables~\ref{tab:acp_Bp1} to~\ref{tab:acp_Lb},
compilations of \CP\ asymmetries in decays of various \b-hadrons: \Bp, \Bz
mesons, $\Bpm/\Bz$ admixtures, \Bs mesons and finally \Lb baryons.
Measurements of time-dependent \CP\ asymmetries are not listed here but are
discussed in Sec.~\ref{sec:cp_uta}.

Figure~\ref{fig:rare-acpselect} shows a graphic representation of a selection of results given in this section.
Footnote symbols indicate that the footnote in the corresponding table should be consulted.

\newpage

\begin{table}[!htbp]
\begin{center}
\caption{\CP\ asymmetries of charmless hadronic $\Bp$ decays (part 1).
Where values are shown in \red{red} (\blue{blue}), this indicates that
they are new \red{published} (\blue{preliminary}) results since PDG2017.}
\label{tab:acp_Bp1}
\resizebox{\textwidth}{!}
{

}
\end{center}
\scriptsize
$^*$~Errors from  PDG include a scale factor. \\     %
$^\dag$~PDG takes the value from the \babar\ amplitude analysis of $B^+ \to K^+ K^- K^+$, while our numbers are from amplitude analyses of $B^+ \to K^+ \pi^- \pi^+$.\\     %
$^\ddag$~PDG uses also a result from CLEO. \\    %
$^\S$~CP asymetry is also measured in different bins of $m_{K^+K^-}$.\\    %
$^\P$~CP asymetry is also measured in different bins of $m_{K_s K_s}$.
\end{table}

\begin{table}[!htbp]
\begin{center}
\caption{\CP\ asymmetries of charmless hadronic $\Bp$ decays (part 2).
Where values are shown in \red{red} (\blue{blue}), this indicates that
they are new \red{published} (\blue{preliminary}) results since PDG2017.}
\label{tab:acp_Bp2}
\resizebox{\textwidth}{!}
{

}
\end{center}
\scriptsize
$^*$~Errors from  PDG include a scale factor. \\     %
$^\dag$~PDG uses also a result from CLEO.\\     %
$^\ddag$~PDG swaps the Belle results corresponding to $A_{C\!P}(p\bar{p}\pi^+)$ and $A_{C\!P}(p\bar{p}K^+)$.\\
$^\S$~PDG uses also a previous result from \babar\ (\cite{Aubert:2008ps}).\\
$^\P$~LHCb also quotes results in bins of $m(\ell^+\ell^-)^2$.     %
\end{table}

\begin{table}[!htbp]
\begin{center}
\caption{\CP\ asymmetries of charmless hadronic $\Bz$ decays.
Where values are shown in \red{red} (\blue{blue}), this indicates that
they are new \red{published} (\blue{preliminary}) results since PDG2017.}
\label{tab:acp_Bz}
\resizebox{\textwidth}{!}
{

}
\end{center}

\scriptsize
Measurements of time-dependent $\CP$ asymmetries are listed in     %
the Unitarity Triangle home page. (http://www.slac.stanford.edu/xorg/hfag/triangle/index.html)\\ \\ %
$^\dag$~PDG uses also a result from CLEO.\\     %
$^\ddag$ Average of \babar\ results from     %
$B^0 \to K^+ \pi^- \pi^0$ and $B^0 \to K^0 \pi^+ \pi^-$.\\     %
$^\S$ PDG quotes the opposite asymmetry. \\     %
$^\P$~Extracted from measured $\Delta A_{\CP}=A_{\CP}     %
(\phi K^{*0})-A_{\CP}(J/\psi K^{*0})=     %
\err{0.015}{0.032}{0.005}$.\\     %
$^\diamond$~LHCb also quotes results in bins of $m(\ell^+\ell^-)^2$.  \\   %
$^1$~Last error comes from the Dalitz plot model.

\end{table}

\clearpage

\begin{table}[!htbp]
\begin{center}
\caption{\CP\ asymmetries of charmless hadronic decays of $B^\pm/B^0$ admixture.
Where values are shown in \red{red} (\blue{blue}), this indicates that
they are new \red{published} (\blue{preliminary}) results since PDG2017.}
\label{tab:acp_B}
\resizebox{\textwidth}{!}
{
\begin{tabular}{|@{}cccc @{}c c @{}c c|}
\hline
& Mode & PDG2017 Avg. & \babar & & Belle & & Our Avg. \\
\sglinespb

\nodata                                           & %
$K^* \gamma$                                      & %
$-0.003\pm 0.017$~$^\dag$                         & %
{$-0.003\pm 0.017\pm 0.007$}                      & %
\ifref {\cite{Aubert:2009ak}} \fi \phantom{.}     & %
\red{$\err{-0.004}{0.014}{0.003}$}~$^1$           & %
\ifref {\cite{Horiguchi:2017ntw}} \fi \phantom{.} & %
$-0.004 \pm 0.011$                                \\

\nodata                                           & %
$s \gamma$                                        & %
$0.015\pm0.020$                                   & %
$\err{0.017}{0.019}{0.010}$~$^\ddag$              & %
\ifref {\cite{Lees:2014uoa}} \fi \phantom{.}      & %
$\err{0.002}{0.050}{0.030}$                       & %
\ifref {\cite{Nishida:2003paa}} \fi \phantom{.}   & %
$0.015 \pm 0.020$                                 \\

\nodata                                           & %
$\Delta A_{CP}(s \gamma)$                         & %
\nodata                                           & %
\nodata                                           & %
\phantom{.}                                       & %
$\err{0.0369}{0.0265}{0.0076}$ $^2$               & %
\ifref {\cite{Watanuki:2018xxg}} \fi \phantom{.}  & %
$0.0370 \pm 0.0280$                               \\

\nodata                                           & %
$(s+d) \gamma$                                    & %
$0.010\pm0.031$                                   & %
$\err{0.057}{0.060}{0.018}$~$^\S$                 & %
\ifref {\cite{Lees:2012ym}} \fi \phantom{.}       & %
$\err{0.022}{0.039}{0.009}$~$^\diamond$           & %
\ifref {\cite{Pesantez:2015fza}} \fi \phantom{.}  & %
$0.032 \pm 0.034$                                 \\

\nodata                                           & %
$s \eta$                                          & %
$\cerr{-0.13}{0.04}{0.05}$                        & %
\nodata                                           & %
\phantom{.}                                       & %
{\berr{-0.13}{0.04}{0.02}{0.03}}                  & %
\ifref {\cite{Nishimura:2009ae}} \fi \phantom{.}  & %
$\cerr{-0.13}{0.04}{0.05}$                        \\

\nodata                                           & %
$\pi^+ X$                                         & %
$0.10 \pm 0.17$                                   & %
{\err{0.10}{0.16}{0.05}}                          & %
\ifref {\cite{delAmoSanchez:2010gx}} \fi \phantom{.}& %
\nodata                                           & %
\phantom{.}                                       & %
$0.10 \pm 0.17$                                   \\

\nodata                                           & %
$s \ell \ell$                                     & %
$0.04\pm0.11$                                     & %
$\err{0.04}{0.11}{0.01}$                          & %
\ifref {\cite{Lees:2013nxa}} \fi \phantom{.}      & %
\nodata                                           & %
\phantom{.}                                       & %
$0.04 \pm 0.11$                                   \\

\nodata                                           & %
$K^*e^+e^-$                                       & %
$-0.18 \pm 0.15$                                  & %
\nodata                                           & %
\phantom{.}                                       & %
$\err{-0.18}{0.15}{0.01}$                         & %
\ifref {\cite{Wei:2009zv}} \fi \phantom{.}        & %
$-0.18 \pm 0.15$                                  \\

\nodata                                           & %
$K^*\mu^+\mu^-$                                   & %
$-0.03 \pm 0.13$                                  & %
\nodata                                           & %
\phantom{.}                                       & %
$\err{-0.03}{0.13}{0.02}$                         & %
\ifref {\cite{Wei:2009zv}} \fi \phantom{.}        & %
$-0.03 \pm 0.13$                                  \\

\nodata                                           & %
$K \ell \ell$                                     & %
\nodata                                           & %
$\err{-0.03}{0.14}{0.01}$                         & %
\ifref {\cite{Lees:2012tva}} \fi \phantom{.}      & %
\nodata                                           & %
\phantom{.}                                       & %
$-0.03 \pm 0.14$                                  \\

\nodata                                           & %
$K^* \ell \ell$                                   & %
$-0.04 \pm 0.07$                                  & %
$\err{0.03}{0.13}{0.01}$~$^\P$                    & %
\ifref {\cite{Lees:2012tva}} \fi \phantom{.}      & %
$\err{-0.10}{0.10}{0.01}$                         & %
\ifref {\cite{Wei:2009zv}} \fi \phantom{.}        & %
$-0.05 \pm 0.08$                                  \\

\hline
\end{tabular}
}
\end{center}
\scriptsize
$^\dag$~PDG includes also a result from CLEO.\\     %
$^\ddag$~\babar\ also measures the difference in direct $C\!P$ asymmetry for charged and neutral $B$ mesons: $\Delta A_{C\!P}= +(5.0\pm3.9\pm1.5)\%$.\\     %
$^\S$~There is another \babar\ result using the recoil method (Phys. Rev. D 77, 051103),     %
and a CLEO result (Phys. Rev. Lett. 86, 5661)     %
that are used in the PDG average.\\     %
$^\P$~Previous \babar\ result is also included in the PDG Average.\\     %
$^\diamond$~Requires $E_\gamma >2.1$~GeV. \\     %
$^1$~Belle\ also measures the difference in direct $C\!P$ asymmetry for charged and neutral $B$ mesons: $\Delta A_{C\!P}= +(2.4\pm2.8\pm0.5)\%$.\\     %
$^2$~$\Delta A_{CP}(s \gamma)= A_{CP}(B^+\to X_s^+\gamma)-A_{CP}(B^0\to X_s^0\gamma) $
\end{table}

\begin{table}[!htbp]
\begin{center}
\caption{\CP\ asymmetries of charmless hadronic $\Bs$ decays.
Where values are shown in \red{red} (\blue{blue}), this indicates that
they are new \red{published} (\blue{preliminary}) results since PDG2017.}
\label{tab:acp_Bs}
\resizebox{\textwidth}{!}{
\begin{tabular}{|@{}cccc @{}c c @{}c c|}
\hline
& Mode & PDG2017 Avg. & CDF & & LHCb & & Our Avg. \\
\sglinespb

\nodata                                           & %
$\pi^+ K^-$                                       & %
$0.26 \pm 0.04$                                   & %
$\err{0.22}{0.07}{0.02}$                          & %
\ifref {\cite{Aaltonen:2014vra}} \fi \phantom{.}  & %
\red{$\err{0.213}{0.015}{0.007}$}                 & %
\ifref {\cite{Aaij:2018tfw}} \fi \phantom{.}      & %
$0.213 \pm 0.017$                                 \\

\hline
\end{tabular}
}
\end{center}
\end{table}

\begin{table}[!htbp]
\begin{center}
\caption{\CP\ asymmetries of charmless hadronic $\Lb$ decays.
Where values are shown in \red{red} (\blue{blue}), this indicates that
they are new \red{published} (\blue{preliminary}) results since PDG2017.}
\label{tab:acp_Lb}
\resizebox{\textwidth}{!}{
\begin{tabular}{|@{}cccc @{}c c @{}c c|}
\sgline
& Mode & PDG2017 Avg. & CDF & & LHCb & & Our Avg. \\
\hline

\nodata                                           & %
$p\pi^-$                                          & %
$0.06 \pm 0.08$                                   & %
$\err{0.06}{0.07}{0.03}$                          & %
\ifref {\cite{Aaltonen:2014vra}} \fi \phantom{.}  & %
\red{$\err{-0.035}{0.017}{0.020}$}~$^\dag$        & %
\ifref {\cite{Aaij:2018tlk}} \fi \phantom{.}      & %
$-0.025 \pm 0.024$                                \\

\nodata                                           & %
$p K^-$                                           & %
$-0.10 \pm 0.09$                                  & %
$\err{-0.10}{0.08}{0.04}$                         & %
\ifref {\cite{Aaltonen:2014vra}} \fi \phantom{.}  & %
\red{$\err{-0.020}{0.013}{0.019}$}~$^\dag$        & %
\ifref {\cite{Aaij:2018tlk}} \fi \phantom{.}      & %
$-0.025 \pm 0.022$                                \\

\nodata                                           & %
$\kzb p \pi^-$                                    & %
$0.22 \pm 0.13$                                   & %
\nodata                                           & %
\phantom{.}                                       & %
$\err{0.22}{0.13}{0.03}$                          & %
\ifref {\cite{Aaij:2014lpa}} \fi \phantom{.}      & %
$0.22 \pm 0.13$                                   \\

\nodata                                           & %
$\Lambda K^+\pi^-$                                & %
$-0.53 \pm 0.25$                                  & %
\nodata                                           & %
\phantom{.}                                       & %
$\err{-0.53}{0.23}{0.11}$                         & %
\ifref {\cite{Aaij:2016nrq}} \fi \phantom{.}      & %
$-0.53 \pm 0.26$                                  \\

\nodata                                           & %
$\Lambda K^+ K^-$                                 & %
$-0.28 \pm 0.12$                                  & %
\nodata                                           & %
\phantom{.}                                       & %
$\err{-0.28}{0.10}{0.07}$                         & %
\ifref {\cite{Aaij:2016nrq}} \fi \phantom{.}      & %
$-0.28 \pm 0.12$                                  \\

\nodata                                           & %
$p K^- \mu^+ \mu^-$                               & %
\nodata                                           & %
\nodata                                           & %
\phantom{.}                                       & %
$\err{-0.035}{0.05}{0.002}$                       & %
\ifref {\cite{Aaij:2017mib}} \fi \phantom{.}      & %
$-0.035 \pm 0.050$                                \\

\sglinespt
\end{tabular}
}
\end{center}
\scriptsize
$^\dag$~LHCb also reports $\Delta A_{CP}= A_{CP}(\Lb \to p K^-)-A_{CP}(\Lb \to p \pi^-) = \err{0.014}{0.022}{0.010}$. \\     %
\end{table}

\clearpage
\noindent List of other measurements that are not included in the tables:
\begin{itemize}

\item In \cite{Aaij:2016cla},
LHCb has measured the triple-product asymmetries for the decays $\Lb \to p\pi^-\pi^+\pi^-$ and $\Lb \to p\pi^- K^+ K^-$.
\item In \cite{Aaij:2017mib}, LHCb also measures $a_{CP}^{\hat T-odd}$ and $a_{P}^{\hat T-odd}$.
\item In \cite{Hsu:2017kir}, Belle also measure the partial branching fraction and CP asymmetry in different bins of $K^+K^-$ mass.
\item In \cite{Aaij:2018lsx},
LHCb has measured the triple-product asymmetries for the decays $\Lb \to p K^-\pi^+\pi^-$,  $\Lb \to p K^- K^+ K^-$ and $\Xi^{0}_{b} \to p K^- K^- \pi^+$.
\end{itemize}

\begin{figure}[htbp!]
\centering
\includegraphics[width=0.5\textwidth]{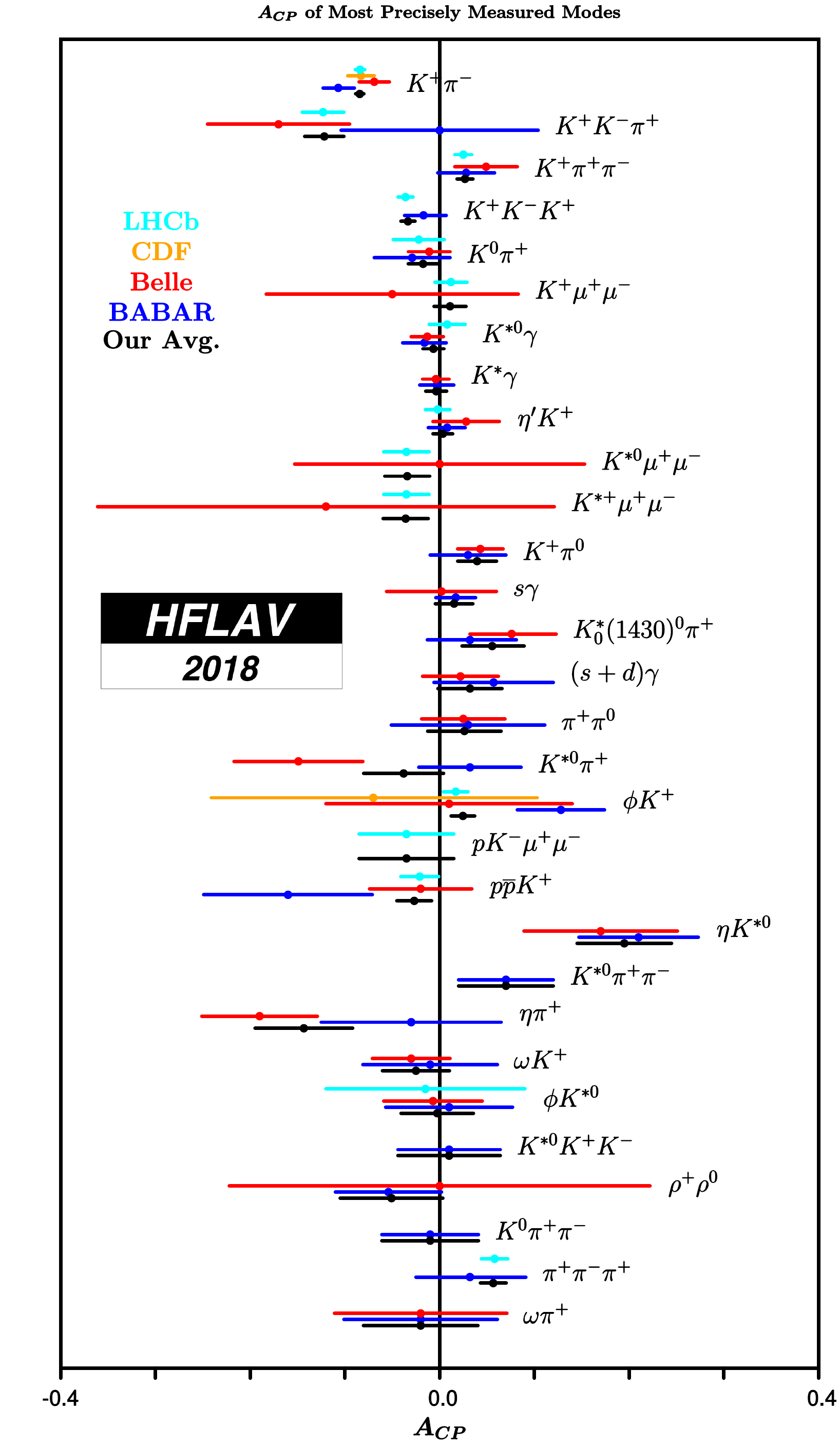}
\caption{$A_{\CP}$ of most precisely measured modes.}
\label{fig:rare-acpselect}
\end{figure}

\clearpage

\mysubsection{Polarization measurements in \b-hadron decays}
\label{sec:rare-polar}

In this section, compilations of polarization measurements in \b-hadron decays
are given. Table~\ref{tab:polar_Bp} (\ref{tab:polar_Bz}) details measurements
of the longitudinal fraction, $f_L$, in \Bp\ (\Bz) decays, and
Table~\ref{tab:polar_BpAng} (\ref{tab:polar_BzAng}) the results of the full
angular analyses of $\Bp$ (\Bz) $\to\phi\Kstar$ decays. 
Table~\ref{tab:polar_BzAng1430} gives results of the full angular analysis of
$\Bz\to\phi K_2^{\ast 0}(1430)$ decays. 
Tables~\ref{tab:polar_Bs} to~\ref{tab:polar_BsAng2} detail quantities of \Bs decays: $f_L$ measurements, and observables from full angular analyses of decays to $\phi\phi$ and $\phi\Kstarzb$.

Figures~\ref{fig:rare-polar} and~\ref{fig:rare-polarbs} show graphic representations of a selection of results shown in this section.
Footnote symbols indicate that the footnote in the corresponding table should be consulted.

\begin{table}[!htbp]
\begin{center}
\caption{Longitudinal polarization fraction $f_L$ for $B^+$ decays. 
Where values are shown in \red{red} (\blue{blue}), this indicates that
they are new \red{published} (\blue{preliminary}) results since PDG2017.}
\label{tab:polar_Bp}
\resizebox{\textwidth}{!}{

}
\end{center}
\end{table}

\begin{table}[!htbp]
\begin{center}
\caption{Longitudinal polarization fraction $f_L$ for $\Bz$ decays.
Where values are shown in \red{red} (\blue{blue}), this indicates that
they are new \red{published} (\blue{preliminary}) results since PDG2017.}
\label{tab:polar_Bz}
\resizebox{\textwidth}{!}{
%
}
\end{center}
\scriptsize
$\dag$~$0.002< q^2 < 1.120~GeV^{2}/c^{4}$
\end{table}

\begin{table}[!p]
\begin{center}
\caption{
Results of the full angular analyses of $B^+ \to \phi K^{*+}$ decays.
Where values are shown in \red{red} (\blue{blue}), this indicates that
they are new \red{published} (\blue{preliminary}) results since PDG2017.}
\label{tab:polar_BpAng}
\resizebox{\textwidth}{!}{
%
}
\end{center}
\scriptsize
Angles ($\phi$, $\delta$) are in radians. BF, $f_L$ and $A_{\CP}$ are tabulated separately.    %
\end{table}

\begin{table}[!p]
\begin{center}
\caption{
Results of the full angular analyses of $B^0 \to \phi K^{*0}$ decays.
Where values are shown in \red{red} (\blue{blue}), this indicates that
they are new \red{published} (\blue{preliminary}) results since PDG2017.}
\label{tab:polar_BzAng}
\resizebox{\textwidth}{!}{
%
}
\end{center}
\scriptsize
Angles ($\phi$, $\delta$) are in radians. BF, $f_L$ and $A_{\CP}$ are tabulated separately.\\     %
$^\dag$~Original LHCb notation adapted to match similar existing quantities.     %
\end{table}

\begin{table}[!p]
\begin{center}
\caption{
Results of the full angular analyses of $B^0 \to \phi K_2^{*0}(1430)$ decays.
Where values are shown in \red{red} (\blue{blue}), this indicates that
they are new \red{published} (\blue{preliminary}) results since PDG2017.}
\label{tab:polar_BzAng1430}
\resizebox{\textwidth}{!}{
%
}
\end{center}
\scriptsize
Angles ($\phi$, $\delta$) are in radians. BF, $f_L$ and $A_{\CP}$ are tabulated separately.     %
\end{table}

\begin{table}[!p]
\begin{center}
\caption{Longitudinal polarization fraction $f_L$ for $\Bs$ decays.
Where values are shown in \red{red} (\blue{blue}), this indicates that
they are new \red{published} (\blue{preliminary}) results since PDG2017.}
\label{tab:polar_Bs}
\resizebox{\textwidth}{!}{
%
}
\end{center}
\end{table}

\begin{table}[!p]
\begin{center}
\caption{
Results of the full angular analyses of $\Bs \to \phi\phi$ decays.
Where values are shown in \red{red} (\blue{blue}), this indicates that
they are new \red{published} (\blue{preliminary}) results since PDG2017.}
\label{tab:polar_BsAng1}
\resizebox{\textwidth}{!}{
%
}
\end{center}
\scriptsize
The parameter $\phi$ is in radians. BF, $f_L$ and $A_{\CP}$ are tabulated separately.     %
\end{table}

\begin{table}[!p]
\begin{center}
\caption{
Results of the full angular analyses of $\Bs \to \phi\overline{K}^{*0}$ decays.
Where values are shown in \red{red} (\blue{blue}), this indicates that
they are new \red{published} (\blue{preliminary}) results since PDG2017.}
\label{tab:polar_BsAng2}
%
\end{center}
\scriptsize
The parameter $\phi$ is in radians. BF, $f_L$ and $A_{\CP}$ are tabulated separately.\\     %
$^\dag$~Converted from the measurement of $\cos(\phi_\parallel)$. PDG takes the smallest resulting asymmetric error as parabolic.     %
\end{table}

\begin{table}[!p]
\begin{center}
\caption{
Results of the full angular analyses of $\Bs \to {K}^{*0}\overline{K}^{*0}$ decays.
Where values are shown in \red{red} (\blue{blue}), this indicates that
they are new \red{published} (\blue{preliminary}) results since PDG2017.}
\label{tab:polar_BsAng3}
%
\end{center}
\end{table}

\clearpage
\noindent List of other measurements that are not included in the tables:
\begin{itemize}
\item  In Ref.~\cite{Aaij:2017wgt}, LHCb presents a flavour-tagged, decay-time-dependent amplitude analysis of $\B^0_s \to (K^+\pi^-)(K^-\pi^+)$ decays in the $K^\pm\pi^\mp$ mass range from 750 to 1600~MeV/$c^2$. The paper includes measurements of
19 $\CP$-averaged amplitude parameters corresponding to scalar, vector and tensor final states. 
\end{itemize}

\begin{figure}[hp!]
\centering
\includegraphics[width=0.5\textwidth]{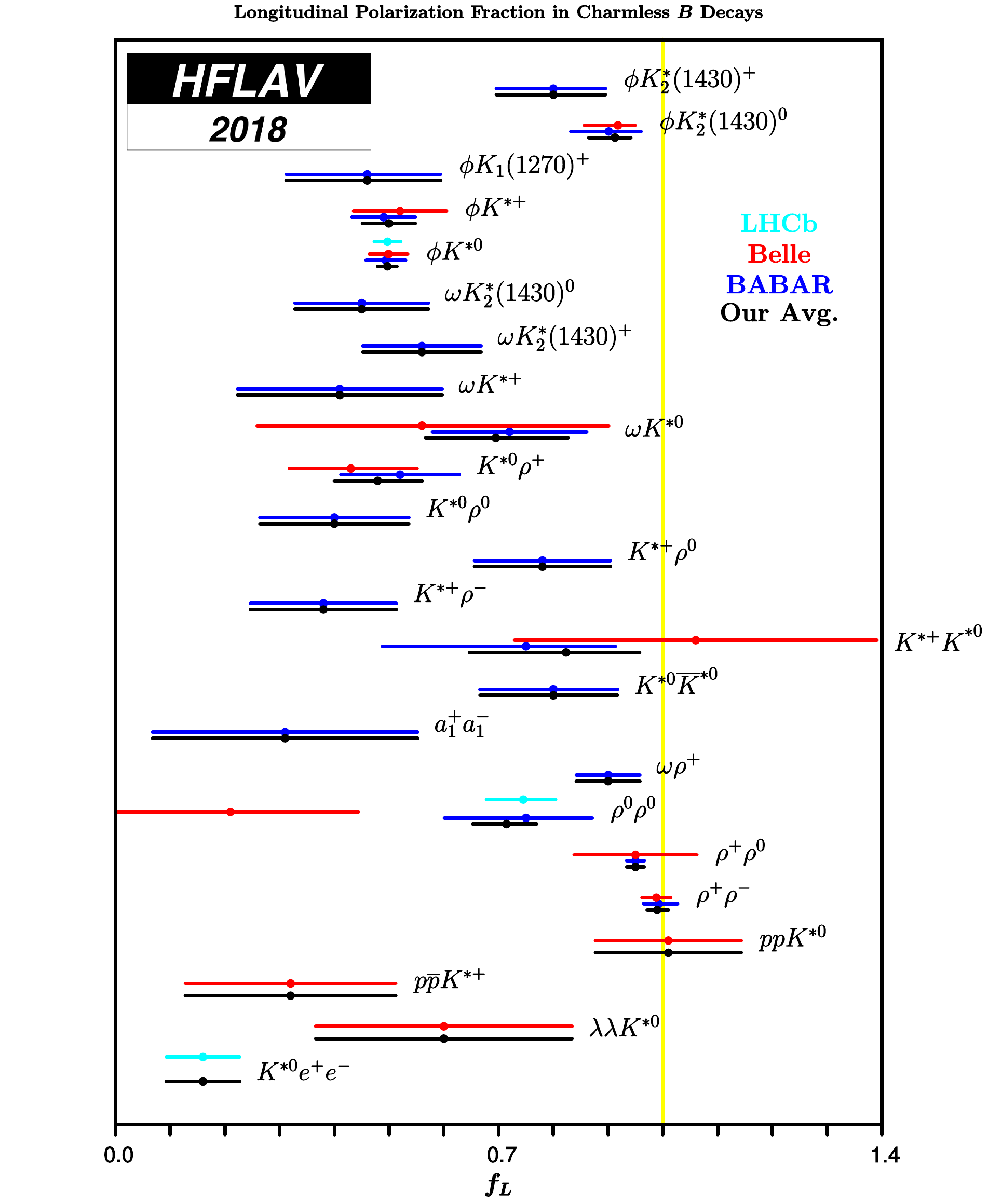}
\caption{Longitudinal polarization fraction in charmless $B$ decays.}
\label{fig:rare-polar}
\end{figure}

\begin{figure}[hbp!]
\centering
\includegraphics[width=0.5\textwidth]{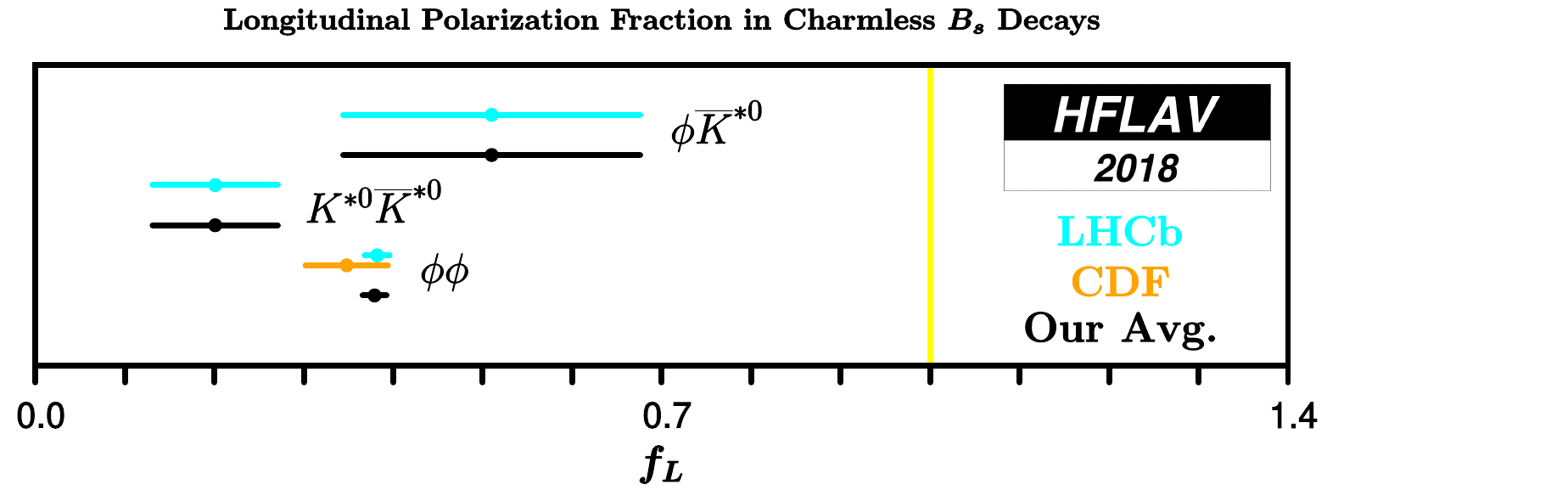}
\caption{Longitudinal polarization fraction in charmless $\Bs$ decays.}
\label{fig:rare-polarbs}
\end{figure}

\clearpage

\mysubsection{Decays of $\Bc$ mesons}
\label{sec:rare-bc}

Table~\ref{tab:Bc_BF} details branching fractions of $\Bc$ meson decays to
charmless hadronic final states.

\begin{table}[ht!]
\begin{center}
\caption{Relative branching fractions of
$\Bc$ decays.
Where values are shown in \red{red} (\blue{blue}), this indicates that
they are new \red{published} (\blue{preliminary}) results since PDG2017.}
\label{tab:Bc_BF}
\resizebox{\textwidth}{!}{
\begin{tabular}{|lccc @{}c c|} \hline
RPP\# & Mode & PDG2017 AVG. & LHCb & & Our Avg.  \\ \sglinespb
18                                                & %
$\it{f}_c\mathcal{B}(\Bc\rightarrow p \bar{p}\pi^+)/\it{f}_u$~$^\S$& %
$3.6 \times 10^{-8}$                              & %
$<2.8 \times 10^{-8}$                             & %
\ifref {\cite{Aaij:2016xxs}} \fi \phantom{.}      & %
$<2.8 \times 10^{-8}$                             \\

25                                                & %
$\it{f}_c\mathcal{B}(\Bc\rightarrow K^+ K^0)/\it{f}_u\mathcal{B}(B^+\rightarrow K_S^0\pi^+)$& %
$^\ddag$                                          & %
$<5.8\times 10^{-2}$                              & %
\ifref {\cite{Aaij:2013fja}} \fi \phantom{.}      & %
$<5.8\times 10^{-2}$                              \\

\nodata                                           & %
$\it{\sigma(\Bc)}\mathcal{B}(\Bc\rightarrow K^+ K^- \pi^+)/\it{\sigma(B^+)}$~$^\dag$& %
\nodata                                           & %
$<15 \times 10^{-8}$                              & %
\ifref {\cite{Aaij:2016xas}} \fi \phantom{.}      & %
$<15 \times 10^{-8}$                              \\

\sglinespt
\hline
\end{tabular}
}
\end{center}
\scriptsize
Channels with no RPP\# were not included in PDG Live as of Dec. 31, 2017. \\ %
$^\S$~PDG result at 95\% CL, LHCb at 90\% CL.\\
$^\dag$~Measured in the annihilation region m($K^-\pi^+)<1.834 $GeV$/c^2$.\\
$^\ddag$~PDG converts the LHCb result to ${f}_c\mathcal{B}(\Bc\rightarrow K^+ K^0)<4.6\times 10^{-7}$.
\end{table}

\clearpage
\section{Charm physics}
\label{sec:charm_physics}

\def\kbar{\overline{K}{}^{\,0}}
\def\dbar{\overline{D}{}^{\,0}}
\def\bbar{\overline{B}{}^{\,0}}
\def\cp{$\CP$}
\def\cpv{$CPV$}
\def\ra{\!\rightarrow\!}
\def\ddbar{$D^0$-$\dbar$}
\def\ycp{$y^{}_{\CP}$}

\def\dklnu{$D^0\ra K^+\ell^-\bar{\nu}$}
\def\dkpi{$D^0\ra K^+\pi^-$}
\def\dkk{$D^0\ra K^+K^-$}
\def\dpipi{$D^0\ra\pi^+\pi^-$}
\def\dkkpp{$D^0\ra K^+K^-/\pi^+\pi^-$}
\def\dkspp{$D^0\ra K^0_S\,\pi^+\pi^-$}
\def\dppp{$D^0\ra \pi^0\,\pi^+\pi^-$}
\def\dkskk{$D^0\ra K^0_S\,K^+ K^-$}
\def\dkppp{$D^0\ra K^+\pi^-\pi^+\pi^-$}

\def\dsphipi{$D^+_s\ra\phi\,\pi^+$}

\def\lcp{\Lambda_c^+}

\def\simge{\mathrel{%
   \rlap{\raise 0.511ex \hbox{$>$}}{\lower 0.511ex \hbox{$\sim$}}}}
\def\simle{\mathrel{
   \rlap{\raise 0.511ex \hbox{$<$}}{\lower 0.511ex \hbox{$\sim$}}}}

\newcommand{\Dnan}{\ensuremath{D_0^\ast(2400)^0}}
\newcommand{\Dtan}{\ensuremath{D_2^\ast(2460)^0}}
\newcommand{\Don}{\ensuremath{D_1(2420)^{0}}}
\newcommand{\Dopn}{\ensuremath{D_1(2430)^{0}}}
\newcommand{\Dnap}{\ensuremath{D_0^\ast(2400)^\pm}}
\newcommand{\Dtap}{\ensuremath{D_2^\ast(2460)^\pm}}
\newcommand{\Dop}{\ensuremath{D_1(2420)^{\pm}}}
\newcommand{\Dopp}{\ensuremath{D_1(2430)^{\pm}}}

\newcommand{\Dsa}{\ensuremath{D_s^{\ast\pm}}}
\newcommand{\Dsna}{\ensuremath{D_{s0}^\ast(2317)^{\pm}}}
\newcommand{\Dsop}{\ensuremath{D_{s1}(2460)^{\pm}}}
\newcommand{\Dso}{\ensuremath{D_{s1}(2536)^{\pm}}}
\newcommand{\Dst}{\ensuremath{D_{s2}(2573)^{\pm}}}
\newcommand{\Dsts}{\ensuremath{D_{sJ}(2700)^{\pm}}}
\newcommand{\Dste}{\ensuremath{D_{sJ}(2860)^{\pm}}}
\newcommand{\Dstsi}{\ensuremath{D_{sJ}(2632)^{\pm}}}

\newcommand{\citep}{\cite}

\newcommand{\kst}{K^*(892)^0}
\newcommand{\akst}{\overline{K}^*(892)^0}
\newcommand{\kstp}{K^*(1410)^0}
\newcommand{\akstp}{\overline{K}^*(1410)^0}
\newcommand{\kstd}{K^*_2(1430)^0}
\newcommand{\akstd}{\overline{K}^*_2(1430)^0}

\newcommand{\ksts}{K^*_0(1430)^0}
\newcommand{\aksts}{\overline{K}^*_0(1430)^0}

\newcommand{\ds}{D_{s}}
\newcommand{\dsp}{D_{s}^+}
\newcommand{\dsm}{D_{s}^-}
\newcommand{\dspm}{D_{s}^{\pm}}
\newcommand{\dsmunu}{\ds^+\to\mu^+\nu_{\mu}}
\newcommand{\dsellnu}{\ds^+\to\ell^+\nu_{\ell}}
\newcommand{\br}{{\cal B}}
\newcommand{\ellnu}{\ell^+\nu_{\ell}}
\newcommand{\enu}{e^+\nu_{e}}
\newcommand{\munu}{\mu^+\nu_{\mu}}
\newcommand{\taunu}{\tau^+\nu_{\tau}}
\newcommand{\taumunu}{\tau^+(\mu^+)\nu_{\tau}}
\newcommand{\tauenu}{\tau^+(e^+)\nu_{\tau}}
\newcommand{\taupinuCharm}{\tau^+(\pi^+)\nu_{\tau}}
\newcommand{\taurhonu}{\tau^+(\rho^+)\nu_{\tau}}

\subsection{\emph{$D^0$-$\dbar$} mixing and \emph{\cp}\ violation}
\label{sec:charm:mixcpv}

\subsubsection{Introduction}

The first evidence for $D^0$-$\dbar$ mixing was obtained in 2007 
by Belle~\cite{Staric:2007dt} and \babar~\cite{Aubert:2007wf}.
These results were confirmed by CDF~\cite{Aaltonen:2007uc}
and, much later, LHCb~\cite{Aaij:2013wda}.
There are now numerous measurements of $D^0$-$\dbar$ mixing 
with various levels of sensitivity. HFLAV performs a global fit
to all relevant measurements to determine world average values 
of mixing parameters, \cp-violation~(\cpv) parameters, 
and strong phase differences.

Our notation is as follows. We use the phase convention 
$\CP|D^0\rangle=-|\dbar\rangle$ and 
$\CP|\dbar\rangle=-|D^0\rangle$~\cite{Bergmann:2000id}
and denote the mass eigenstates as
\begin{eqnarray} 
D^{}_1 & = & p|D^0\rangle-q|\dbar\rangle \\ 
D^{}_2 & = & p|D^0\rangle+q|\dbar\rangle\,.
\end{eqnarray}
With this phase convention, in the absence of \cp\ violation ($p\!=\!q$)
$D^{}_1$ is \cp-even and $D^{}_2$ is \cp-odd.
The global fit determines central values and uncertainties
for ten underlying parameters. These consist of the following:
\begin{itemize}
\item mixing parameters $x$ and $y$, where
$x = 2(m^{}_1-m^{}_2)/(\Gamma^{}_1+\Gamma^{}_2)$,
$y= (\Gamma^{}_1-\Gamma^{}_2)/(\Gamma^{}_1+\Gamma^{}_2)$, and 
$m^{}_1,\,m^{}_2$ and $\Gamma^{}_1,\,\Gamma^{}_2$ are
the masses and decay widths of the mass eigenstates.
\item \cpv\ parameters $|q/p|$ and ${\rm Arg}(q/p)\equiv \phi$;
these give rise to {\it indirect\/} \cpv. Here we assume indirect \cpv\ is
``universal,'' i.e., independent of the final state in $D^0\ra f$ decays.
\item direct \cpv\ asymmetries 
\begin{eqnarray*}
A^{}_D & \equiv & \frac{\Gamma(D^0\ra K^+\pi^-)-\Gamma(\dbar\ra K^-\pi^+)}
{\Gamma(D^0\ra K^+\pi^-)+\Gamma(\dbar\ra K^-\pi^+)} \\ \\
A^{}_K & \equiv & \frac{\Gamma(D^0\ra K^+ K^-)-\Gamma(\dbar\ra K^- K^+)}
{\Gamma(D^0\ra K^+ K^-)+\Gamma(\dbar\ra K^- K^+)} \\ \\
A^{}_\pi & \equiv & \frac{\Gamma(D^0\ra\pi^+\pi^-)-\Gamma(\dbar\ra\pi^-\pi^+)}
{\Gamma(D^0\ra\pi^+\pi^-)+\Gamma(\dbar\ra\pi^-\pi^+)}\,,
\end{eqnarray*}
where the decay rates correspond to pure $D^0$ and $\dbar$ flavour eigenstates.
\item the ratio of doubly Cabibbo-suppressed to Cabibbo-favored decay rates
\begin{eqnarray*}
R^{}_D & \equiv & \frac{\Gamma(D^0\ra K^+\pi^-)+\Gamma(\dbar\ra K^-\pi^+)}
{\Gamma(D^0\ra K^-\pi^+)+\Gamma(\dbar\ra K^+\pi^-)}\,,
\end{eqnarray*}
where the decay rates correspond to pure $D^0$ and $\dbar$ flavour eigenstates.
\item the strong phase difference $\delta$ between the 
$\dbar\ra K^-\pi^+$ and $D^0\ra K^-\pi^+$ amplitudes; and 
\item the strong phase difference $\delta^{}_{K\pi\pi}$ between 
$\dbar\ra K^-\rho^+$ and $D^0\ra K^-\rho^+$ amplitudes.
\end{itemize}

The fit uses 49 measurements of observables from the 
following\footnote{Charge-conjugate modes are implicitly included.} decays:
\dklnu, \dkk, \dpipi, \dkpi, 
$D^0\ra K^+\pi^-\pi^0$, %
\dkspp, \dppp, \dkskk, and \dkppp.
The fit also uses measurements of mixing parameters and strong phases
obtained from double-tagged branching fractions measured at the 
$\psi(3770)$ resonance. The relationships between measured observables 
and fitted parameters are given in Table~\ref{tab:relationships}.
Correlations among observables are accounted for by using covariance 
matrices provided by the experimental collaborations. Uncertainties are 
assumed to be Gaussian, and systematic uncertainties among different 
experiments are assumed to be uncorrelated unless specific 
correlations have been identified.
We have compared this method with a second method that adds
together three-dimensional log-likelihood functions for $x$, 
$y$, and $\delta$ obtained from several independent measurements;
this combination accounts for non-Gaussian uncertainties.
When both methods are applied to the same set of 
measurements, equivalent results are obtained. 

\begin{table}
\renewcommand{\arraycolsep}{0.02in}
\renewcommand{\arraystretch}{1.3}
\begin{center}
\caption{\label{tab:relationships}
Left: decay modes used to determine the fitted parameters 
$x$, $y$, $\delta$, $\delta^{}_{K\pi\pi}$, $R^{}_D$, 
$A^{}_D$, $A^{}_K$, $A^{}_\pi$, $|q/p|$, and $\phi$.
Middle: measured observables for each decay mode. 
Right: relationships between the measured observables and the fitted
parameters. The symbol $\langle t\rangle$ denotes the mean reconstructed 
decay time for $D^0\ra K^+K^-$ or $D^0\ra\pi^+\pi^-$ decays.}
\vspace*{6pt}
\footnotesize
\resizebox{0.99\textwidth}{!}{
\begin{tabular}{l|c|l}
\hline
\textbf{Decay Mode} & \textbf{Observables} & \textbf{Relationship} \\
\hline
$D^0\ra K^+K^-/\pi^+\pi^-$  & 
\begin{tabular}{c}
 $y^{}_{\CP}$  \\
 $A^{}_{\Gamma}$
\end{tabular} & 
$\begin{array}{c}
2y^{}_{\CP} = 
\left(\left|q/p\right|+\left|p/q\right|\right)y\cos\phi - \\
\hskip0.50in \left(\left|q/p\right|-\left|p/q\right|\right)x\sin\phi \\
2A^{}_\Gamma = 
\left(\left|q/p\right|-\left|p/q\right|\right)y\cos\phi - \\
\hskip0.50in \left(\left|q/p\right|+\left|p/q\right|\right)x\sin\phi
\end{array}$   \\
\hline
$D^0\ra K^0_S\,\pi^+\pi^-$ & 
$\begin{array}{c}
x \\ 
y \\ 
|q/p| \\ 
\phi
\end{array}$ &   \\ 
\hline
$D^0\ra K^+\ell^-\bar{\nu}$ & $R^{}_M$  & $R^{}_M = (x^2 + y^2)/2$ \\
\hline \hskip-0.10in
\begin{tabular}{l}
$D^0\ra K^+\pi^-\pi^0$ \\
(Dalitz plot analysis)
\end{tabular} & 
$\begin{array}{c}
x'' \\ 
y''
\end{array}$ &
$\begin{array}{l}
x'' = x\cos\delta^{}_{K\pi\pi} + y\sin\delta^{}_{K\pi\pi} \\ 
y'' = y\cos\delta^{}_{K\pi\pi} - x\sin\delta^{}_{K\pi\pi}
\end{array}$ \\
\hline\hskip-0.10in
\begin{tabular}{l}
``Double-tagged'' \\
branching fractions \\
measured in \\
$\psi(3770)\ra DD$ decays
\end{tabular} & 
$\begin{array}{c}
R^{}_M \\
y \\
R^{}_D \\
\sqrt{R^{}_D}\cos\delta
\end{array}$ &   $R^{}_M = (x^2 + y^2)/2$ \\
\hline
$D^0\ra K^+\pi^-$ &
$\begin{array}{c}
x'^2,\ y' \\
x'^{2+},\ x'^{2-} \\
y'^+,\ y'^-
\end{array}$ & 
$\begin{array}{l}
x' = x\cos\delta + y\sin\delta \\ 
y' = y\cos\delta - x\sin\delta \\
A^{}_M\equiv (|q/p|^4-1)/(|q/p|^4+1) \\
x'^\pm = [(1\pm A^{}_M)/(1\mp A^{}_M)]^{1/4} \times \\
\hskip0.50in (x'\cos\phi\pm y'\sin\phi) \\
y'^\pm = [(1\pm A^{}_M)/(1\mp A^{}_M)]^{1/4} \times \\
\hskip0.50in (y'\cos\phi\mp x'\sin\phi) \\
\end{array}$ \\
\hline\hskip-0.10in
\begin{tabular}{l}
$D^0\ra K^+\pi^-/K^-\pi^+$ \\
(time-integrated)
\end{tabular} & 
\begin{tabular}{c}
$R^{}_D$ \\ 
$A^{}_D$ 
\end{tabular}   &  \\
\hline\hskip-0.10in
\begin{tabular}{l}
$D^0\ra K^+K^-/\pi^+\pi^-$ \\
(time-integrated)
\end{tabular} & 
\begin{tabular}{c}
$\frac{\displaystyle \Gamma(D^0\ra K^+K^-)-\Gamma(\dbar\ra K^+K^-)}
{\displaystyle \Gamma(D^0\ra K^+K^-)+\Gamma(\dbar\ra K^+K^-)}$    \\ \\
$\frac{\displaystyle \Gamma(D^0\ra\pi^+\pi^-)-\Gamma(\dbar\ra\pi^+\pi^-)}
{\displaystyle \Gamma(D^0\ra\pi^+\pi^-)+\Gamma(\dbar\ra\pi^+\pi^-)}$ 
\end{tabular} & 
\begin{tabular}{c}
$A^{}_K  + \frac{\displaystyle \langle t\rangle}
{\displaystyle \tau^{}_D}\,{\cal A}_{\CP}^{\rm indirect}$ 
\ \ (${\cal A}_{\CP}^{\rm indirect}\approx -A^{}_\Gamma$)
\\ \\ \\
$A^{}_\pi + \frac{\displaystyle \langle t\rangle}
{\displaystyle \tau^{}_D}\,{\cal A}_{\CP}^{\rm indirect}$ 
\ \ (${\cal A}_{\CP}^{\rm indirect}\approx -A^{}_\Gamma$)
\end{tabular} \\
\hline
\end{tabular}
}
\end{center}
\end{table}

Mixing in the $B^0$, and $B^0_s$ heavy flavour systems
is governed by a short-distance box diagram. In the $D^0$ 
system, this box diagram is both doubly-Cabibbo-suppressed 
and GIM-suppressed, and consequently the short-distance mixing rate 
is tiny. Thus, $D^0$-$\dbar$ mixing is expected to be dominated by 
long-distance processes. These are difficult to calculate, and 
theoretical estimates for $x$ and $y$ range over three orders 
of magnitude, up to the percent 
level~\cite{Bigi:2000wn,Petrov:2003un,Petrov:2004rf,Falk:2004wg}.

Almost all experimental analyses besides that of the 
$\psi(3770)\ra \overline{D}D$
measurements~\cite{Asner:2012xb} identify the flavour of the
$D^0$ or $\dbar$ when produced by reconstructing the decay
$D^{*+}\ra D^0\pi^+$ or $D^{*-}\ra\dbar\pi^-$. The charge
of the pion, which has low momentum in the lab frame relative 
to that of the $D^0$ and is often referred to as the ``soft'' 
pion, identifies the $D^0$ flavour. For $D^{*+}\ra D^0\pi^+$, 
$M^{}_{D^*}-M^{}_{D^0}-M^{}_{\pi^+}\equiv Q\approx 6~\mev$, 
which is close to the kinematic threshold; thus analyses 
typically require that the reconstructed $Q$ be small in order 
to suppress backgrounds. An LHCb measurement~\cite{Aaij:2014gsa} 
of the difference between time-integrated \cp\ asymmetries
$A_{\CP}(K^+K^-) - A_{\CP}(\pi^+\pi^-)$ identifies the flavour of
the $D^0$ by partially reconstructing $\overline{B}\ra D^0\mu^- X$
and $B\ra\dbar\mu^+ X$ decays; in this case the charge of the 
$\mu^\pm$ identifies the flavour of the $D^0$.

For time-dependent measurements, the $D^0$ decay time is 
calculated as 
$t = (\vec{d}\cdot\vec{p})\times M^{}_{D^0}/(cp^2)$, 
where $\vec{d}$ is the displacement vector from the
$D^{*+}$ vertex to the $D^0$ decay vertex; $\vec{p}$ is the
reconstructed $D^0$ momentum; and $p$ and $M^{}_{D^0}$ 
are in GeV. The $D^{*+}$ vertex position is 
taken as the intersection of the $D^0$ momentum vector 
with the beamspot profile for $e^+e^-$ experiments, and 
at the primary interaction vertex for $pp$ and $\bar{p}p$
experiments~\cite{Aaltonen:2007uc,Aaij:2013wda}.

\subsubsection{Input observables}

The global fit determines central values and errors for
ten parameters using a $\chi^2$ statistic.
The fitted parameters are $x$, $y$, $R^{}_D$, $A^{}_D$,
$|q/p|$, $\phi$, $\delta$, $\delta^{}_{K\pi\pi}$,
$A^{}_K$, and $A^{}_\pi$. In the $D\ra K^+\pi^-\pi^0$ 
Dalitz plot analysis~\cite{Aubert:2008zh}, 
the phases of intermediate resonances in the $\dbar\ra K^+\pi^-\pi^0$ 
decay amplitude are fitted relative to the phase for
${\cal A}(\dbar\ra K^+\rho^-)$, and the phases of intermediate
resonances for $D^0\ra K^+\pi^-\pi^0$ are fitted 
relative to the phase for ${\cal A}(D^0\ra K^+\rho^-)$. 
As the $\dbar$ and $D^0$ Dalitz plots are fitted independently, 
the phase difference 
$\delta^{}_{K\pi\pi} = 
{\rm Arg}[{\cal A}(\dbar\ra K^+\rho^-)/{\cal A}(D^0\ra K^+\rho^-)]$
between the reference amplitudes cannot be determined from these 
individual fits. However, this phase difference can be constrained 
in the global fit and thus is included as a fitted parameter.

All input measurements are listed in 
Tables~\ref{tab:observables1}-\ref{tab:observables3}.
There are three observables input to the fit that are world
average values:
\begin{eqnarray}
R^{}_M & = & \frac{x^2+y^2}{2} \\ 
 & & \nonumber \\
y^{}_{\CP} & = & 
\frac{1}{2}\left(\left|\frac{q}{p}\right| + \left|\frac{p}{q}\right|\right)
y\cos\phi - 
\frac{1}{2}\left(\left|\frac{q}{p}\right| - \left|\frac{p}{q}\right|\right)
x\sin\phi \\
 & & \nonumber \\
A^{}_\Gamma & = & 
\frac{1}{2}\left(\left|\frac{q}{p}\right| - \left|\frac{p}{q}\right|\right)
y\cos\phi - 
\frac{1}{2}\left(\left|\frac{q}{p}\right| + \left|\frac{p}{q}\right|\right)
x\sin\phi\,. 
\end{eqnarray} 
These world averages are calculated using the COMBOS 
program~\cite{Combos:1999}. The observable $R^{}_M$ 
is measured in both \dklnu\ and
$D^0\ra K^+\pi^-\pi^+\pi^-$~\cite{Aaij:2016rhq} decays, and 
it is for the first case (measured by several experiments) that 
the world average is used. The inputs used for 
this~\cite{Aitala:1996vz,Cawlfield:2005ze,Aubert:2007aa,Bitenc:2008bk}
are plotted in Fig.~\ref{fig:rm_semi}. 
The inputs used for world averages of $y^{}_{CP}$ and $A^{}_\Gamma$ are 
plotted in Figs.~\ref{fig:ycp} and \ref{fig:Agamma}, respectively. 

The \dkpi\ measurements used are from 
Belle~\cite{Zhang:2006dp,Ko:2014qvu}, 
\babar~\cite{Aubert:2007wf}, CDF~\cite{Aaltonen:2013pja}, 
and LHCb~\cite{Aaij:2017urz}; earlier measurements are either 
superseded or have much less precision and are not used.
The observables from \dkspp\ decays are measured in two ways:
assuming \cp\ conservation ($D^0$ and $\dbar$ decays combined),
and allowing for \cp\ violation ($D^0$ and $\dbar$ decays
fitted separately). The no-\cpv\ measurements are from 
Belle~\cite{Peng:2014oda}, \babar~\cite{delAmoSanchez:2010xz},
and LHCb~\cite{Aaij:2015xoa}; for the \cpv-allowed case, 
Belle~\cite{Peng:2014oda} and LHCb~\cite{Aaij:2019jot}
measurements are available. 
The $D^0\ra K^+\pi^-\pi^0$, $D^0\ra K^0_S K^+ K^-$, 
and $D^0\ra \pi^0\,\pi^+\pi^-$ results are from 
\babar~\cite{Aubert:2008zh,Lees:2016gom}; 
the $D^0\ra K^+\pi^-\pi^+\pi^-$ results are from LHCb~\cite{Aaij:2016rhq}; and
the $\psi(3770)\ra\overline{D}D$ results are from CLEOc~\cite{Asner:2012xb}.

\begin{table}
\renewcommand{\arraystretch}{1.4}
\renewcommand{\arraycolsep}{0.02in}
\renewcommand{\tabcolsep}{0.05in}
\caption{\label{tab:observables1}
Observables used in the global fit except those from
time-dependent \dkpi\ measurements, and those from direct 
\cpv\ measurements. The $D^0\ra K^+\pi^-\pi^0$ observables are
$x'' = x\cos\delta^{}_{K\pi\pi} + y\sin\delta^{}_{K\pi\pi}$ and 
$y'' = -x\sin\delta^{}_{K\pi\pi} + y\cos\delta^{}_{K\pi\pi}$.}
\vspace*{6pt}
\footnotesize
\resizebox{0.99\textwidth}{!}{
\begin{tabular}{l|ccc}
\hline
{\bf Mode} & \textbf{Observable} & {\bf Values} & {\bf Correlation coefficients} \\
\hline
\begin{tabular}{l}  
$D^0\ra K^+K^-/\pi^+\pi^-$, \\
\hskip0.30in $\phi\,K^0_S$
\end{tabular}
&
\begin{tabular}{c}
 $y^{}_{\CP}$  \\
 $A^{}_{\Gamma}$
\end{tabular} & 
$\begin{array}{c}
(0.715\pm 0.111)\% \\
(-0.032\pm 0.026)\% 
\end{array}$   & \\ 
\hline
\begin{tabular}{l}  
$D^0\ra K^0_S\,\pi^+\pi^-$~\cite{Peng:2014oda} \\
\ (Belle: no \cpv)
\end{tabular}
&
\begin{tabular}{c}
$x$ \\
$y$ 
\end{tabular} & 
\begin{tabular}{c}
 $(0.56\pm 0.19\,^{+0.067}_{-0.127})\%$ \\
 $(0.30\pm 0.15\,^{+0.050}_{-0.078})\%$ 
\end{tabular} & $+0.012$ \\ 
\begin{tabular}{l}  
$D^0\ra K^0_S\,\pi^+\pi^-$~\cite{Peng:2014oda} \\
\ (Belle: no direct \cpv)
\end{tabular}
&
\begin{tabular}{c}
$|q/p|$ \\
$\phi$  
\end{tabular} & 
\begin{tabular}{c}
 $0.90\,^{+0.16}_{-0.15}{}^{+0.078}_{-0.064}$ \\
 $(-6\pm 11\,^{+4.2}_{-5.0})$ degrees
\end{tabular} &
$\left\{ \begin{array}{cccc}
 1 &  0.054 & -0.074 & -0.031  \\
   &  1     &  0.034 & -0.019  \\
   &        &  1     &  0.044  \\
   &        &        &  1 
\end{array} \right\}$  \\
\begin{tabular}{l}  
$D^0\ra K^0_S\,\pi^+\pi^-$~\cite{Peng:2014oda} \\
\ (Belle: direct \\
\ \ \ \ \ \ \ \cpv\ allowed)
\end{tabular}
&
\begin{tabular}{c}
$x$ \\
$y$ \\
$|q/p|$ \\
$\phi$  
\end{tabular} & 
\begin{tabular}{c}
 $(0.58\pm 0.19^{+0.0734}_{-0.1177})\%$ \\
 $(0.27\pm 0.16^{+0.0546}_{-0.0854})\%$ \\
 $0.82\,^{+0.20}_{-0.18}{}^{+0.0807}_{-0.0645}$ \\
 $(-13\,^{+12}_{-13}\,^{+4.15}_{-4.77})$ degrees
\end{tabular} &  same as above \\
 & & \\
\begin{tabular}{l}  
$D^0\ra K^0_S\,\pi^+\pi^-$~\cite{Aaij:2015xoa} \\
\ (LHCb: 1~fb$^{-1}$ \\
\ \ \ \ \ \ \ \ no \cpv)
\end{tabular}
&
\begin{tabular}{c}
$x$ \\
$y$ 
\end{tabular} & 
\begin{tabular}{c}
 $(-0.86\,\pm 0.53\,\pm 0.17)\%$ \\
 $(0.03\,\pm 0.46\,\pm 0.13)\%$ 
\end{tabular} & $+0.37$ \\ 
\begin{tabular}{l}  
$D^0\ra K^0_S\,\pi^+\pi^-$~\cite{Aaij:2019jot} \\
\ (LHCb: 3~fb$^{-1}$ \\
\ \ \ \ \ \cpv\ allowed)
\end{tabular}
&
\begin{tabular}{c}
$x^{}_{CP}$ \\
$y^{}_{CP}$ \\
$\Delta x$  \\
$\Delta y$ 
\end{tabular} & 
\begin{tabular}{c}
 $(0.27\,\pm 0.16\,\pm 0.04)\%$ \\
 $(0.74\,\pm 0.36\,\pm 0.11)\%$ \\
 $(-0.053\,\pm 0.070\,\pm 0.022)\%$ \\
 $(0.06\,\pm 0.16\,\pm 0.03)\%$ 
\end{tabular} & 
\begin{tabular}{l}
\ \ \ $\left\{ \begin{array}{cccc}
 1 &  (-0.17+0.15) & (0.04+0.01) & (-0.02-0.02)  \\
   &  1     &  (-0.03-0.05) & (0.01-0.03)  \\
   &        &  1     &  (-0.13+0.14)  \\
   &        &        &  1 
\end{array} \right\}$  \\
Notation: above coefficients are (statistical+systematic). \\
For $(x, y, |q/p|, \phi)\rightarrow (x^{}_{CP}, y^{}_{CP}, \Delta x, \Delta y)$ 
mapping, see~\cite{DiCanto:2018tsd}. 
\end{tabular} \\
\begin{tabular}{l}  
$D^0\ra K^0_S\,\pi^+\pi^-$~\cite{delAmoSanchez:2010xz} \\
\hskip0.30in $K^0_S\,K^+ K^-$ \\
\ (\babar: no \cpv) 
\end{tabular}
&
\begin{tabular}{c}
$x$ \\
$y$ 
\end{tabular} & 
\begin{tabular}{c}
 $(0.16\pm 0.23\pm 0.12\pm 0.08)\%$ \\
 $(0.57\pm 0.20\pm 0.13\pm 0.07)\%$ 
\end{tabular} &  $+0.0615$ \\ 
 & & \\
\begin{tabular}{l}  
$D^0\ra \pi^0\,\pi^+\pi^-$~\cite{Lees:2016gom} \\
\ (\babar: no \cpv) 
\end{tabular}
&
\begin{tabular}{c}
$x$ \\
$y$ 
\end{tabular} & 
\begin{tabular}{c}
 $(1.5\pm 1.2\pm 0.6)\%$ \\
 $(0.2\pm 0.9\pm 0.5)\%$ 
\end{tabular} &  $-0.006$ \\ 
\hline
\begin{tabular}{l}  
$D^0\ra K^+\ell^-\bar{\nu}$
\end{tabular} 
  & $R^{}_M =(x^2+y^2)/2$ & $(0.0130\pm 0.0269)\%$  &  \\ 
\hline
\begin{tabular}{l}  
$D^0\ra K^+\pi^-\pi^0$~\cite{Aubert:2008zh}
\end{tabular} 
&
\begin{tabular}{c}
$x''$ \\ 
$y''$ 
\end{tabular} &
\begin{tabular}{c}
$(2.61\,^{+0.57}_{-0.68}\,\pm 0.39)\%$ \\ 
$(-0.06\,^{+0.55}_{-0.64}\,\pm 0.34)\%$ 
\end{tabular} & $-0.75$ \\
\hline
\begin{tabular}{l}  
$D^0\ra K^+\pi^-\pi^+\pi^-$~\cite{Aaij:2016rhq}
\end{tabular} 
  & $R^{}_M/2$  & $(4.8\pm 1.8)\times 10^{-5}$  &  \\ 
\hline
\begin{tabular}{c}  
$\psi(3770)\ra\overline{D}D$~\cite{Asner:2012xb} \\
(CLEOc)
\end{tabular}
&
\begin{tabular}{c}
$R^{}_D$ \\
$x^2$ \\
$y$ \\
$\cos\delta$ \\
$\sin\delta$ 
\end{tabular} & 
\begin{tabular}{c}
$(0.533 \pm 0.107 \pm 0.045)\%$ \\
$(0.06 \pm 0.23 \pm 0.11)\%$ \\
$(4.2 \pm 2.0 \pm 1.0)\%$ \\
$0.81\,^{+0.22}_{-0.18}\,^{+0.07}_{-0.05}$ \\
$-0.01\pm 0.41\pm 0.04$
\end{tabular} &
$\left\{ \begin{array}{ccccc}
1 & 0 &  0    & -0.42 &  0.01 \\
  & 1 & -0.73 &  0.39 &  0.02 \\
  &   &  1    & -0.53 & -0.03 \\
  &   &       &  1    &  0.04 \\
  &   &       &       &  1    
\end{array} \right\}$ \\
\hline
\end{tabular}
}
\end{table}

\begin{table}
\renewcommand{\arraystretch}{1.3}
\renewcommand{\arraycolsep}{0.02in}
\caption{\label{tab:observables2}
Time-dependent \dkpi\ observables used for the global fit.
The observables $R^+_D$ and $R^-_D$ are related to parameters 
$R^{}_D$ and $A^{}_D$ via $R^\pm_D = R^{}_D (1\pm A^{}_D)$.}
\vspace*{6pt}
\footnotesize
\begin{center}
\begin{tabular}{l|ccc}
\hline
{\bf Mode} & \textbf{Observable} & {\bf Values} & {\bf Correlation coefficients} \\
\hline
\begin{tabular}{l}  
$D^0\ra K^+\pi^-$~\cite{Aubert:2007wf} \\
(\babar~384~fb$^{-1}$)
\end{tabular}
&
\begin{tabular}{c}
$R^{}_D$ \\
$x'^{2+}$ \\
$y'^+$ 
\end{tabular} & 
\begin{tabular}{c}
 $(0.303\pm 0.0189)\%$ \\
 $(-0.024\pm 0.052)\%$ \\
 $(0.98\pm 0.78)\%$ 
\end{tabular} &
$\left\{ \begin{array}{ccc}
 1 &  0.77 &  -0.87 \\
   &  1    &  -0.94 \\
   &       &   1 
\end{array} \right\}$ \\ \\
\begin{tabular}{l}  
$\dbar\ra K^-\pi^+$~\cite{Aubert:2007wf} \\
(\babar~384~fb$^{-1}$)
\end{tabular}
&
\begin{tabular}{c}
$A^{}_D$ \\
$x'^{2-}$ \\
$y'^-$ 
\end{tabular} & 
\begin{tabular}{c}
 $(-2.1\pm 5.4)\%$ \\
 $(-0.020\pm 0.050)\%$ \\
 $(0.96\pm 0.75)\%$ 
\end{tabular} & same as above \\
\hline
\begin{tabular}{l}  
$D^0\ra K^+\pi^-$~\cite{Ko:2014qvu} \\
(Belle 976~fb$^{-1}$ No \cpv)
\end{tabular}
&
\begin{tabular}{c}
$R^{}_D$ \\
$x'^{2}$ \\
$y'$ 
\end{tabular} & 
\begin{tabular}{c}
 $(0.353\pm 0.013)\%$ \\
 $(0.009\pm 0.022)\%$ \\
 $(0.46\pm 0.34)\%$ 
\end{tabular} &
$\left\{ \begin{array}{ccc}
 1 &  0.737 &  -0.865 \\
   &  1     &  -0.948 \\
   &        &   1 
\end{array} \right\}$ \\ \\
\begin{tabular}{l}  
$D^0\ra K^+\pi^-$~\cite{Zhang:2006dp} \\
(Belle 400~fb$^{-1}$ \cpv-allowed)
\end{tabular}
&
\begin{tabular}{c}
$R^{}_D$ \\
$x'^{2+}$ \\
$y'^+$ 
\end{tabular} & 
\begin{tabular}{c}
 $(0.364\pm 0.018)\%$ \\
 $(0.032\pm 0.037)\%$ \\
 $(-0.12\pm 0.58)\%$ 
\end{tabular} &
$\left\{ \begin{array}{ccc}
 1 &  0.655 &  -0.834 \\
   &  1     &  -0.909 \\
   &        &   1 
\end{array} \right\}$ \\ \\
\begin{tabular}{l}  
$\dbar\ra K^-\pi^+$~\cite{Zhang:2006dp} \\
(Belle 400~fb$^{-1}$ \cpv-allowed)
\end{tabular}
&
\begin{tabular}{c}
$A^{}_D$ \\
$x'^{2-}$ \\
$y'^-$ 
\end{tabular} & 
\begin{tabular}{c}
 $(+2.3\pm 4.7)\%$ \\
 $(0.006\pm 0.034)\%$ \\
 $(0.20\pm 0.54)\%$ 
\end{tabular} & same as above \\
\hline
\begin{tabular}{l}  
$D^0\ra K^+\pi^-$~\cite{Aaltonen:2013pja} \\
(CDF 9.6~fb$^{-1}$ No \cpv)
\end{tabular}
&
\begin{tabular}{c}
$R^{}_D$ \\
$x'^{2}$ \\
$y'$ 
\end{tabular} & 
\begin{tabular}{c}
 $(0.351\pm 0.035)\%$ \\
 $(0.008\pm 0.018)\%$ \\
 $(0.43\pm 0.43)\%$ 
\end{tabular} & 
$\left\{ \begin{array}{ccc}
 1 &  0.90 &  -0.97 \\
   &  1    &  -0.98 \\
   &       &   1 
\end{array} \right\}$ \\ 
\hline
\begin{tabular}{l}  
$D^0\ra K^+\pi^-$~\cite{Aaij:2017urz} \\  
(LHCb 5.0~fb$^{-1}$ \cpv-allowed)
\end{tabular}
&
\begin{tabular}{c}
$R^{+}_D$ \\
$x'^{2+}$ \\
$y'^+$ 
\end{tabular} & 
\begin{tabular}{c}
 $(0.3454\pm 0.0045)\%$ \\
 $(0.0061\pm 0.0037)\%$ \\
 $(0.501\pm 0.074)\%$ 
\end{tabular} &
$\left\{ \begin{array}{ccc}
 1 &  0.843 &  -0.935 \\
   &  1     &  -0.963 \\
   &        &   1 
\end{array} \right\}$ \\ \\
\begin{tabular}{l}  
$\dbar\ra K^-\pi^+$~\cite{Aaij:2017urz} \\
(LHCb 5.0~fb$^{-1}$ \cpv-allowed)
\end{tabular}
&
\begin{tabular}{c}
$R^{-}_D$ \\
$x'^{2-}$ \\
$y'^-$ 
\end{tabular} & 
\begin{tabular}{c}
 $(0.3454\pm 0.0045)\%$ \\
 $(0.0016\pm 0.0039)\%$ \\
 $(0.554\pm 0.074)\%$ 
\end{tabular} & 
$\left\{ \begin{array}{ccc}
 1 &  0.846 &  -0.935 \\
   &  1     &  -0.964 \\
   &        &   1 
\end{array} \right\}$ \\
\hline
\end{tabular}
\end{center}
\end{table}

\begin{table}
\renewcommand{\arraystretch}{1.3}
\renewcommand{\arraycolsep}{0.02in}
\caption{\label{tab:observables3}
Measurements of time-integrated \cp\ asymmetries. The observable 
$A^{}_{\CP}(f)= [\Gamma(D^0\ra f)-\Gamma(\dbar\ra f)]/
[\Gamma(D^0\ra f)+\Gamma(\dbar\ra f)]$. The symbol
$\Delta\langle t\rangle$ denotes the difference
between the mean reconstructed decay times for 
$D^0\ra K^+K^-$ and $D^0\ra\pi^+\pi^-$ decays due to 
different trigger and reconstruction efficiencies.}
\vspace*{6pt}
\footnotesize
\begin{center}
\resizebox{\textwidth}{!}{
\begin{tabular}{l|ccc}
\hline
{\bf Mode} & \textbf{Observable} & {\bf Values} & 
                  {\boldmath $\Delta\langle t\rangle/\tau^{}_D$} \\
\hline
\begin{tabular}{c}
$D^0\ra h^+ h^-$~\cite{Aubert:2007if} \\
(\babar\ 386 fb$^{-1}$)
\end{tabular} & 
\begin{tabular}{c}
$A^{}_{\CP}(K^+K^-)$ \\
$A^{}_{\CP}(\pi^+\pi^-)$ 
\end{tabular} & 
\begin{tabular}{c}
$(+0.00 \pm 0.34 \pm 0.13)\%$ \\
$(-0.24 \pm 0.52 \pm 0.22)\%$ 
\end{tabular} &
0 \\
\hline
\begin{tabular}{c}
$D^0\ra h^+ h^-$~\cite{cdf_public_note_10784,Collaboration:2012qw} \\
(CDF 9.7~fb$^{-1}$)
\end{tabular} & 
\begin{tabular}{c}
$A^{}_{\CP}(K^+K^-)-A^{}_{\CP}(\pi^+\pi^-)$ \\
$A^{}_{\CP}(K^+K^-)$ \\
$A^{}_{\CP}(\pi^+\pi^-)$ 
\end{tabular} & 
\begin{tabular}{c}
$(-0.62 \pm 0.21 \pm 0.10)\%$ \\
$(-0.32 \pm 0.21)\%$ \\
$(+0.31 \pm 0.22)\%$ 
\end{tabular} &
$0.27 \pm 0.01$ \\
\hline
\begin{tabular}{c}
$D^0\ra h^+ h^-$~\cite{Aaij:2019kcg} \\
(LHCb 9.0~fb$^{-1}$, \\
$D^{*+}\ra D^0\pi^+$ + \\
\ $\overline{B}\ra D^0\mu^- X$ \\
\ \ tags combined)
\end{tabular} & 
$A^{}_{\CP}(K^+K^-)-A^{}_{\CP}(\pi^+\pi^-)$ &
$(-0.154 \pm 0.029)\%$ &
$0.115 \pm 0.002$ \\
\hline
\end{tabular}
}
\end{center}
\end{table}

\begin{figure}
\vskip-0.20in
\begin{center}
\includegraphics[width=5.0in]{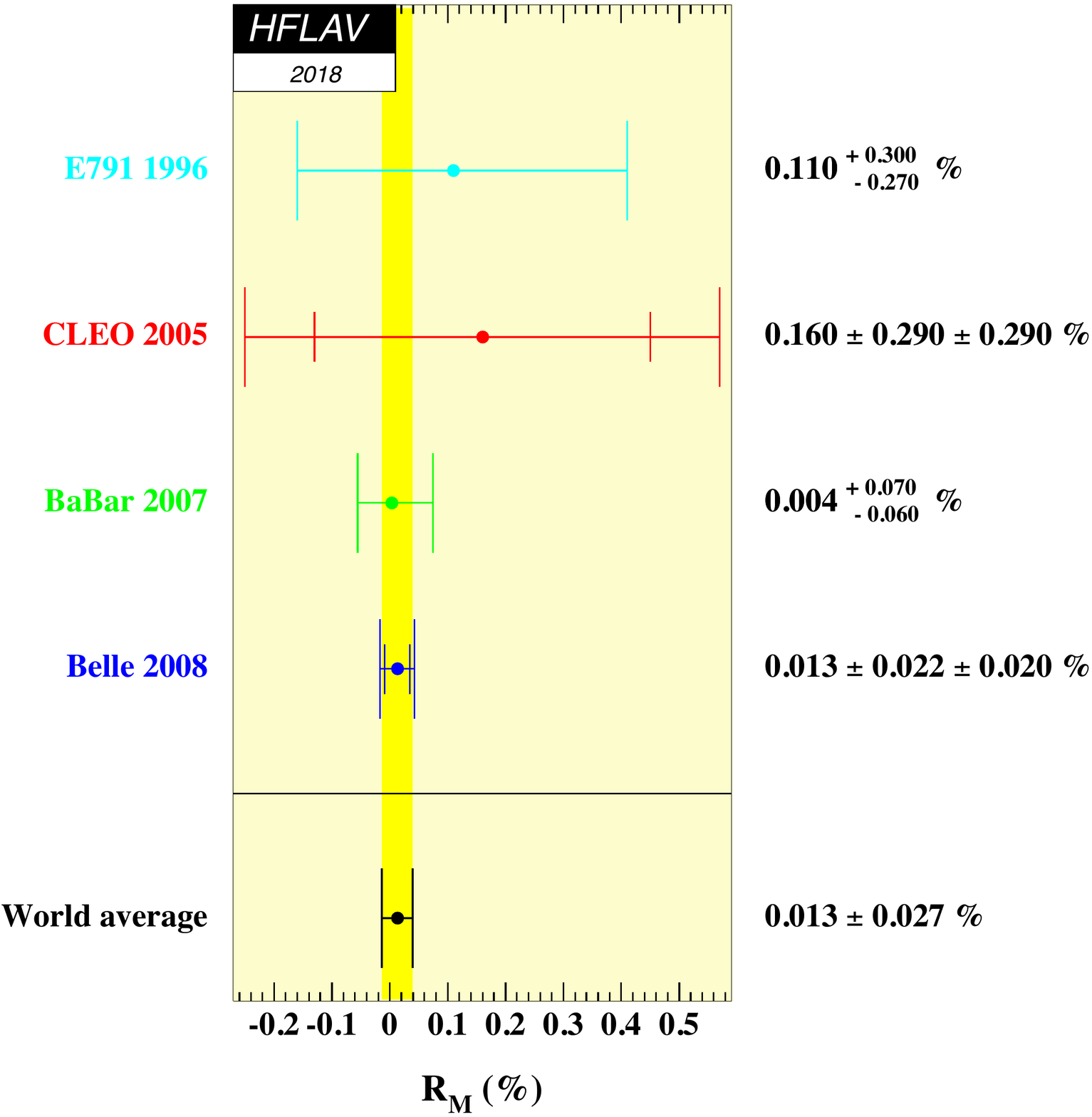}
\end{center}
\vskip-0.8in
\caption{\label{fig:rm_semi}
World average value of $R^{}_M=(x^2+y^2)/2$ 
as calculated from $D^0\ra K^+\ell^-\bar{\nu}$ 
measurements~\cite{Aitala:1996vz,Cawlfield:2005ze,Aubert:2007aa,Bitenc:2008bk}. }
\end{figure}

\begin{figure}
\vskip-0.20in
\begin{center}
\includegraphics[width=5.2in]{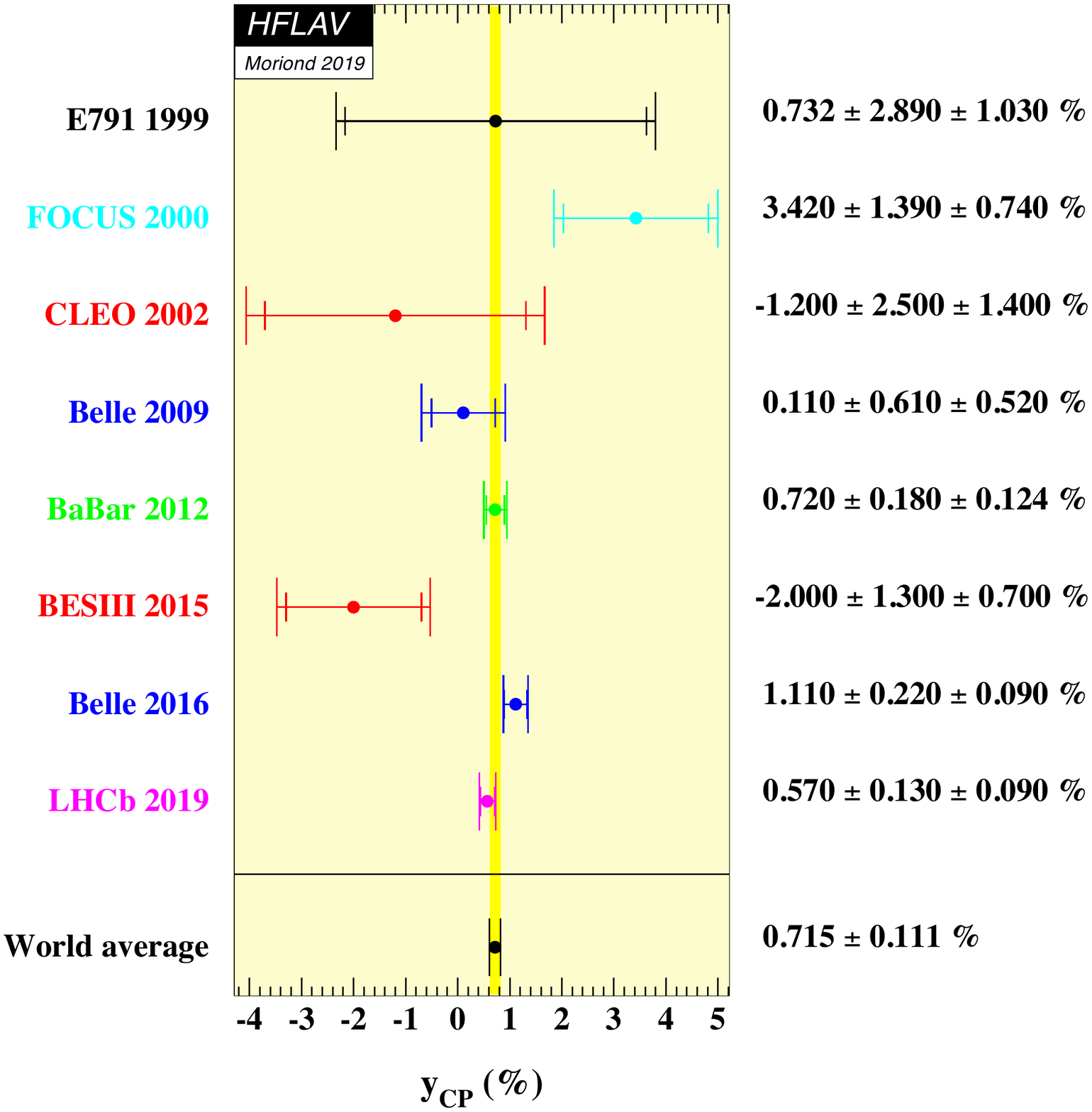}
\end{center}
\vskip-1.2in
\caption{\label{fig:ycp}
World average value of $y^{}_{\CP}$ 
as calculated from $D^0\ra K^+ K^-\!\!,\,\pi^+\pi^-$
measurements~\cite{Aitala:1999dt,Link:2000cu,Csorna:2001ww,Zupanc:2009sy,
Lees:2012qh,Ablikim:2015hih,Staric:2015sta,Aaij:2018qiw}.}
\end{figure}

\begin{figure}
\vskip-0.20in
\begin{center}
\includegraphics[width=5.4in]{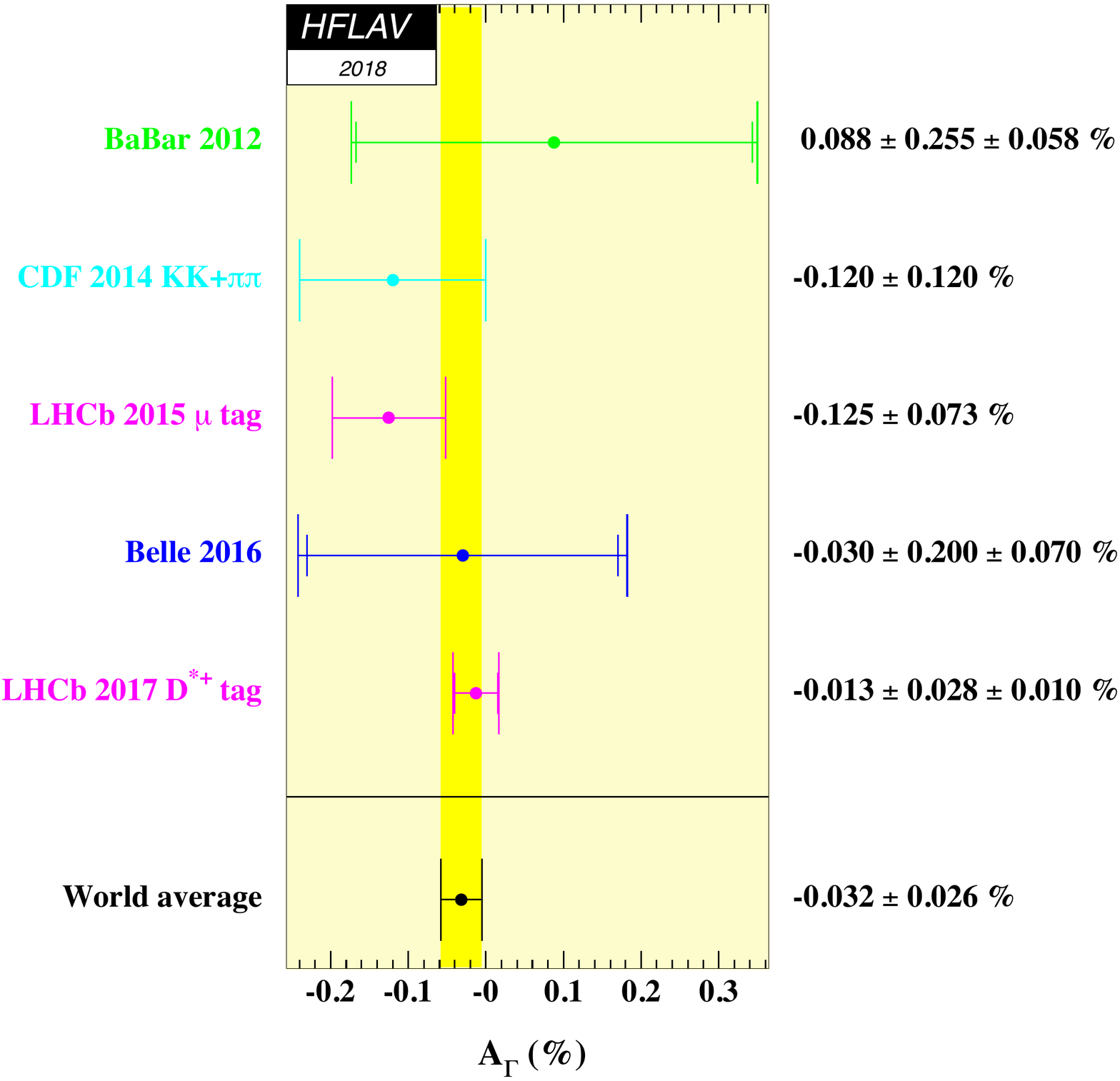}
\end{center}
\vskip-1.1in
\caption{\label{fig:Agamma}
World average value of $A^{}_\Gamma$ 
as calculated from $D^0\ra K^+ K^-\!\!,\,\pi^+\pi^-$
measurements~\cite{Lees:2012qh,Aaltonen:2014efa,Aaij:2015yda,
Staric:2015sta,Aaij:2017idz}.}
\end{figure}

As mentioned, Table~\ref{tab:relationships} lists the 
relationships between the observables and the global-fit parameters.
For each set of correlated observables, we use these relations
to construct a difference vector $\vec{V}$ between the measured 
values and those calculated from the fitted parameters.
For example, for $D^0\ra K^0_S\,\pi^+\pi^-$ decays,
$\vec{V}=(\Delta x,\Delta y,\Delta |q/p|,\Delta \phi)$,
where $\Delta x \equiv x^{}_{\rm measured} - x^{}_{\rm fitted}$
(and similarly for $\Delta y, \Delta |q/p|$, and $\Delta \phi$).
The contribution of a set of observables to the fit $\chi^2$ 
is calculated as $\vec{V}\cdot (M^{-1})\cdot\vec{V}^T$, 
where $M^{-1}$ is the inverse of the covariance matrix 
for the measured observables. Covariance matrices are 
constructed from the correlation coefficients among the 
observables. These correlation coefficients are furnished
by the experiments and listed in 
Tables~\ref{tab:observables1}-\ref{tab:observables3}.

\subsubsection{Fit results}

The global fitter uses MINUIT with the MIGRAD minimizer, 
and all uncertainties are obtained from MINOS~\cite{MINUIT:webpage}. 
Four separate fits are performed: 
\begin{enumerate}
\item 
assuming \cp\ conservation, \ie, fixing
$A^{}_D\!=\!0$, $A_K\!=\!0$, $A^{}_\pi\!=\!0$, $\phi\!=\!0$, 
and $|q/p|\!=\!1$;
\item  
assuming no direct \cpv\ in doubly Cabibbo-suppressed (DCS)
decays ($A^{}_D\!=\!0$) and fitting for the parameters 
$(x,y,|q/p|)$ or $(x,y,\phi)$; 
\item  
assuming no direct \cpv\ in DCS decays and fitting 
for alternative parameters~\cite{Grossman:2009mn,Kagan:2009gb}
$x^{}_{12}= 2|M^{}_{12}|/\Gamma$, 
$y^{}_{12}= |\Gamma^{}_{12}|/\Gamma$, and 
$\phi^{}_{12}= {\rm Arg}(M^{}_{12}/\Gamma^{}_{12})$,
where $M^{}_{12}$ and $\Gamma^{}_{12}$ are the off-diagonal
elements of the $D^0$-$\dbar$ mass and decay matrices, respectively.
The parameter $\phi^{}_{12}$ is a weak phase that is responsible for 
$\CP$ violation in mixing. The conventional parameters $(x, y, |q/p|, \phi)$ 
can be derived from $(x^{}_{12}, y^{}_{12}, \phi^{}_{12})$; see Ref.~\cite{Kagan:2009gb}.
\item  
allowing full \cpv, \ie, floating all parameters.
For this fit, we fit for $(x, y, |q/p|, \phi)$.
\end{enumerate}

For fit (2), in addition to fixing $A^{}_D\!=\!0$, we reduce
four independent parameters to three by imposing the 
relation~\cite{Ciuchini:2007cw,Kagan:2009gb}
$\tan\phi = (1-|q/p|^2)/(1+|q/p|^2)\times (x/y)$.\footnote{One can also use 
Eq.~(16) of Ref.~\cite{Grossman:2009mn} to reduce four parameters to three.} 
This constraint is imposed in two ways: 
in the first way we float parameters
$x$, $y$, and $\phi$ and from these derive $|q/p|$; 
and in the second way we float $x$, $y$, and $|q/p|$ and from these 
derive $\phi$. The central values returned by the two fits are identical,
but the first fit yields MINOS errors for $\phi$ while the second
fit yields MINOS errors for $|q/p|$. 
For the no-direct-\cpv\ fit (3),
we fit for parameters $x^{}_{12}$, $y^{}_{12}$, and $\phi^{}_{12}$
and from these derive $x$, $y$, $|q/p|$, and $\phi$; the latter
parameters are compared to measured observables to calculate the fit
$\chi^2$. All results are listed in Table~\ref{tab:results}. The
$\chi^2$ for the \cpv-allowed fit~(4) is 60.7 for 
$49-10=39$ degrees of freedom. Table~\ref{tab:results_chi2} 
lists the individual contributions to this~$\chi^2$.

\begin{table}
\renewcommand{\arraystretch}{1.4}
\begin{center}
\caption{\label{tab:results}
Results of the global fit for different assumptions regarding \cpv.}
\vspace*{6pt}
\footnotesize
\begin{tabular}{c|cccc}
\hline
\textbf{Parameter} & \textbf{\boldmath No \cpv} & \textbf{\boldmath No direct \cpv} 
& \textbf{\boldmath \cpv-allowed} & \textbf{\boldmath \cpv-allowed} \\
 & & \textbf{\boldmath in DCS decays} & & \textbf{\boldmath 95\% C.L.\ Interval} \\
\hline
$\begin{array}{c}
x\ (\%) \\ 
y\ (\%) \\ 
\delta^{}_{K\pi}\ (^\circ) \\ 
R^{}_D\ (\%) \\ 
A^{}_D\ (\%) \\ 
|q/p| \\ 
\phi\ (^\circ) \\
\delta^{}_{K\pi\pi}\ (^\circ)  \\
A^{}_{\pi} (\%) \\
A^{}_K (\%) \\
x^{}_{12}\ (\%) \\ 
y^{}_{12}\ (\%) \\ 
\phi^{}_{12} (^\circ)
\end{array}$ & 
$\begin{array}{c}
0.50\,^{+0.13}_{-0.14} \\
0.62\,\pm 0.07 \\
8.9\,^{+8.2}_{-8.9} \\
0.344\,\pm 0.002 \\
- \\
- \\
- \\
18.5\,^{+22.7}_{-23.4} \\
- \\
- \\
- \\
- \\
- 
\end{array}$ &
$\begin{array}{c}
0.43\,^{+0.10}_{-0.11}\\
0.63\,\,\pm 0.06 \\
9.3\,^{+8.3}_{-9.2}\\
0.344\,\pm 0.002 \\
- \\
0.998\,\pm 0.008 \\
0.08\,\pm 0.31 \\ 
22.1\,^{+22.6}_{-23.4} \\ 
0.05\,\pm 0.16 \\
-0.11\,\pm 0.16 \\
0.43\,^{+0.10}_{-0.11}\\
0.63\,\pm 0.06 \\
-0.25\,^{+0.96}_{-0.99}\\
\end{array}$ &
$\begin{array}{c}
0.39\,^{+0.11}_{-0.12}\\
0.651\,\,^{+0.063}_{-0.069}\\
12.1\,^{+8.6}_{-10.2} \\
0.344\,\pm 0.002 \\
-0.55\,^{+0.49}_{-0.51} \\
0.969\,^{+0.050}_{-0.045} \\ 
-3.9\,^{+4.5}_{-4.6} \\ 
25.8\,^{+23.0}_{-23.8} \\
0.06\,\pm 0.16 \\
-0.09\,\pm 0.16 \\
 \\
 \\
 \\
\end{array}$ &
$\begin{array}{c}
\left[ 0.16,\, 0.61\right] \\
\left[ 0.51,\, 0.77\right] \\
\left[ -10.4,\, 28.2\right] \\
\left[ 0.339,\, 0.348\right] \\
\left[ -1.5,\, 0.4\right] \\
\left[ 0.89,\, 1.07\right] \\\
\left[ -13.2,\, 5.1\right] \\
\left[ -21.3,\, 70.3\right] \\
\left[ -0.25,\, 0.38\right] \\
\left[ -0.40,\, 0.22\right] \\
\left[ 0.22,\, 0.63\right] \\
\left[ 0.50,\, 0.75\right] \\
\left[ -2.5,\, 1.8\right] \\
\end{array}$ \\
\hline
$\chi^2$ & 89.2 (Fit 1) & 61.1 (Fits 2, 3) & 60.7 (Fit 4) &  \\
\hline
\end{tabular}
\end{center}
\end{table}

\begin{table}
\renewcommand{\arraystretch}{1.3}
\begin{center}
\caption{\label{tab:results_chi2}
Individual contributions to the $\chi^2$ for the \cpv-allowed fit.}
\vspace*{6pt}
\footnotesize
\begin{tabular}{l|c|rr}
\hline
\textbf{Observable} & \textbf{\boldmath degrees of} & 
  \textbf{\boldmath $\chi^2$} & \textbf{\boldmath $\sum\chi^2$} \\
 & \textbf{\boldmath freedom} &  &  \\
\hline
$y^{}_{CP}$ World Average (Fig.~\ref{fig:ycp})       & 1 & 0.35 & 0.35 \\
$A^{}_\Gamma$ World Average (Fig.~\ref{fig:Agamma})  & 1 & 2.07 & 2.41 \\
\hline
$x^{}_{K^0\pi^+\pi^-}$ Belle~\cite{Peng:2014oda}       & 1 & 0.71 & 3.12 \\
$y^{}_{K^0\pi^+\pi^-}$ Belle~\cite{Peng:2014oda}       & 1 & 4.42 & 7.54 \\
$|q/p|^{}_{K^0\pi^+\pi^-}$ Belle~\cite{Peng:2014oda}   & 1 & 0.48 & 8.02 \\
$\phi^{}_{K^0\pi^+\pi^-}$  Belle~\cite{Peng:2014oda}   & 1 & 0.53 & 8.55 \\
\hline
$x^{}_{CP}\,(K^0\pi^+\pi^-)$ LHCb~\cite{Aaij:2019jot} & 1 & 0.55 & 9.10 \\
$y^{}_{CP}\,(K^0\pi^+\pi^-)$ LHCb~\cite{Aaij:2019jot} & 1 & 0.06 & 9.16 \\
$\Delta x\,(K^0\pi^+\pi^-)$ LHCb~\cite{Aaij:2019jot} & 1 & 0.00 & 9.16 \\
$\Delta y\,(K^0\pi^+\pi^-)$ LHCb~\cite{Aaij:2019jot} & 1 & 0.09 & 9.26 \\
\hline
$x^{}_{K^0 h^+ h^-}$ \babar~\cite{delAmoSanchez:2010xz} & 1 & 0.73 & 9.98 \\
$y^{}_{K^0 h^+ h^-}$ \babar~\cite{delAmoSanchez:2010xz} & 1 & 0.08 & 10.06 \\
\hline
$x^{}_{\pi^0\pi^+\pi^-}$ \babar~\cite{Lees:2016gom} & 1 & 0.68 & 10.74 \\
$y^{}_{\pi^0\pi^+\pi^-}$ \babar~\cite{Lees:2016gom} & 1 & 0.19 & 10.93 \\
\hline
$(x^2+y^2)^{}_{K^+\ell^-\nu}$ World Average 
                         (Fig.~\ref{fig:rm_semi}) & 1 & 0.14 & 11.07 \\
\hline
$x^{}_{K^+\pi^-\pi^0}$ \babar~\cite{Aubert:2008zh}     & 1 & 7.10 & 18.17 \\
$y^{}_{K^+\pi^-\pi^0}$ \babar~\cite{Aubert:2008zh}     & 1 & 3.91 & 22.08 \\
\hline
CLEOc~\cite{Asner:2012xb}                          &      &       \\
($x/y/R^{}_D/\cos\delta/\sin\delta$) 
                                & 5 & 10.53 & 32.60 \\
\hline
$R^+_D/x'{}^{2+}/y'{}^+$ \babar~\cite{Aubert:2007wf}  &  3 & 8.69 & 41.30    \\
$R^-_D/x'{}^{2-}/y'{}^-$ \babar~\cite{Aubert:2007wf}  &  3 & 4.02 & 45.32    \\
$R^+_D/x'{}^{2+}/y'{}^+$ Belle~\cite{Ko:2014qvu}      &  3 & 1.88 & 47.20    \\
$R^-_D/x'{}^{2-}/y'{}^-$ Belle~\cite{Ko:2014qvu}      &  3 & 2.36 & 49.56    \\
$R^{}_D/x'{}^{2}/y'$ CDF~\cite{Aaltonen:2013pja}      &  3 & 1.20 & 50.76    \\
$R^+_D/x'{}^{2+}/y'{}^+$ LHCb~\cite{Aaij:2017urz}     &  3 & 1.29 & 52.05    \\
$R^-_D/x'{}^{2-}/y'{}^-$ LHCb~\cite{Aaij:2017urz}     &  3 & 0.67 & 52.72    \\
\hline
$A^{}_{KK}/A^{}_{\pi\pi}$  \babar~\cite{Aubert:2007if} & 2 & 0.35 & 53.08  \\
$A^{}_{KK}/A^{}_{\pi\pi}$  CDF~\cite{cdf_public_note_10784} & 2 & 4.07 & 57.14  \\
$A^{}_{KK}-A^{}_{\pi\pi}$  LHCb~\cite{Aaij:2019kcg} ($D^*$, $B^0\ra D^0\mu X$ tags)   
                                & 1 & 0.05 & 57.19  \\
\hline
$(x^2+y^2)^{}_{K^+\pi^-\pi^+\pi^-}$ LHCb~\cite{Aaij:2016rhq} & 1 & 3.47 & 60.67 \\
\hline
\end{tabular}
\end{center}
\end{table}

Confidence contours in the two dimensions $(x,y)$ or 
$(|q/p|,\phi)$ are obtained by finding the minimum $\chi^2$
for each fixed point in the two-dimensional plane.
The resulting $1\sigma$-$5\sigma$ contours 
are shown 
in Fig.~\ref{fig:contours_ncpv} for the \cp-conserving fit~(1); 
in Fig.~\ref{fig:contours_ndcpv} for the no-direct-\cpv\ fit~(3); 
and in Fig.~\ref{fig:contours_cpv} for the \cpv-allowed fit~(4).
The contours are determined from the increase of the
$\chi^2$ above the minimum value.
One observes that the $(x,y)$ contours for the no-\cpv\ fit 
are very similar to those for the \cpv-allowed fit. In the \cpv-allowed
fit, the $\chi^2$ at the no-mixing point $(x,y)\!=\!(0,0)$ is 2028
units above the minimum value, which, for two degrees of freedom,
corresponds to a confidence level (C.L.) greater than $10\sigma$.
Thus, the no-mixing hypothesis is excluded at this high level. 
In the $(|q/p|,\phi)$ plot (Fig.~\ref{fig:contours_cpv} bottom), 
the no-\cpv\ point $(1,0)$ is within the $1\sigma$ contour, and
thus the data is consistent with \cp\ conservation.

One-dimensional likelihood curves for individual parameters 
are obtained by finding, for a fixed value of a parameter,
the minimum $\chi^2$. The resulting
functions $\Delta\chi^2=\chi^2-\chi^2_{\rm min}$, where $\chi^2_{\rm min}$
is the minimum value, are shown in Fig.~\ref{fig:1dlikelihood}.
The points where $\Delta\chi^2=3.84$ determine 95\% C.L. intervals 
for the parameters. These intervals are listed in Table~\ref{tab:results}.

\begin{figure}
\vskip-0.20in
\begin{center}
\includegraphics[width=4.0in]{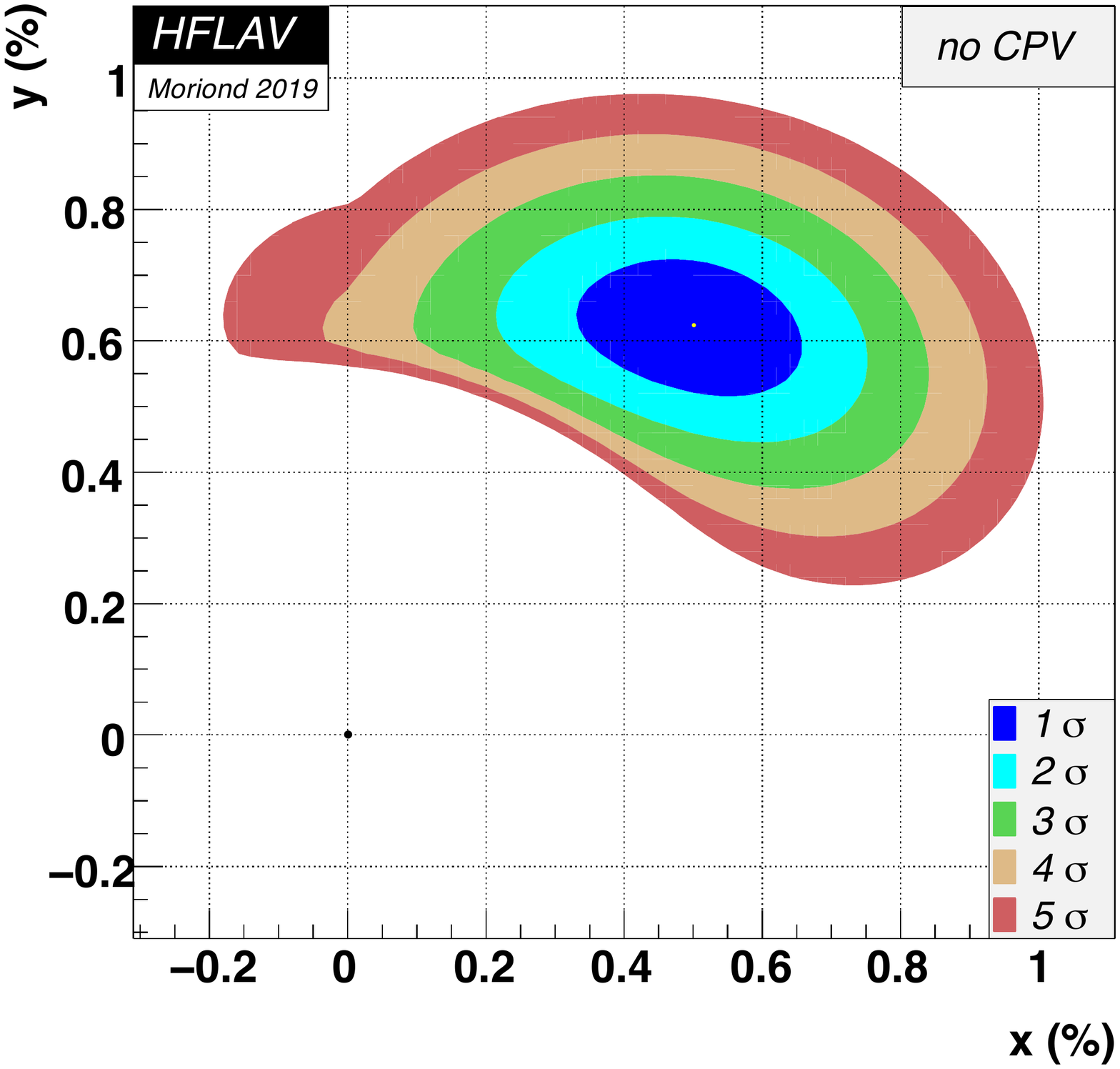}
\end{center}
\vskip-1.1in
\caption{\label{fig:contours_ncpv}
Two-dimensional contours for mixing parameters $(x,y)$, for no \cpv\ (fit 1).}
\end{figure}

\begin{figure}
\begin{center}
\vbox{
\includegraphics[width=84mm]{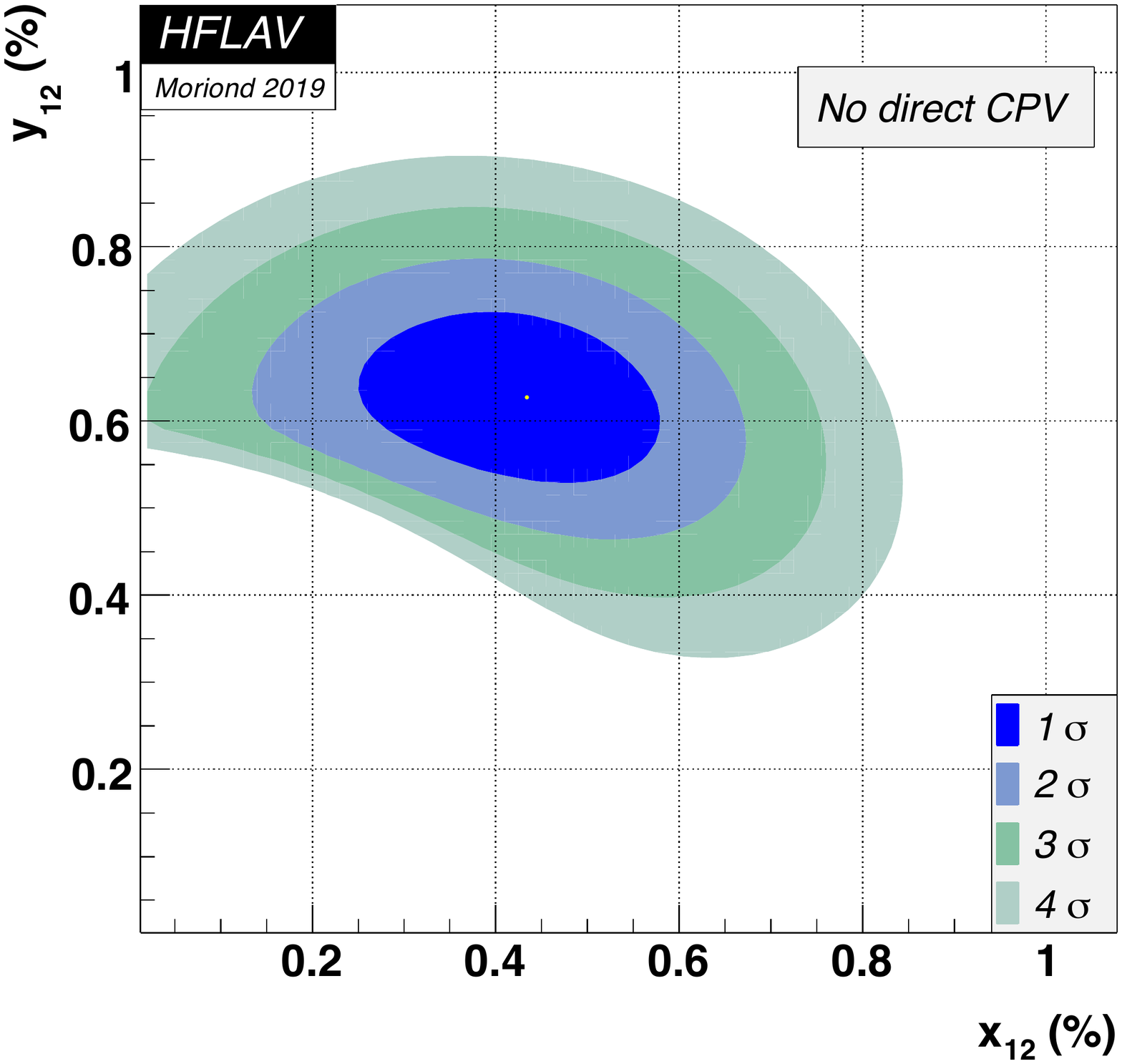}
\includegraphics[width=84mm]{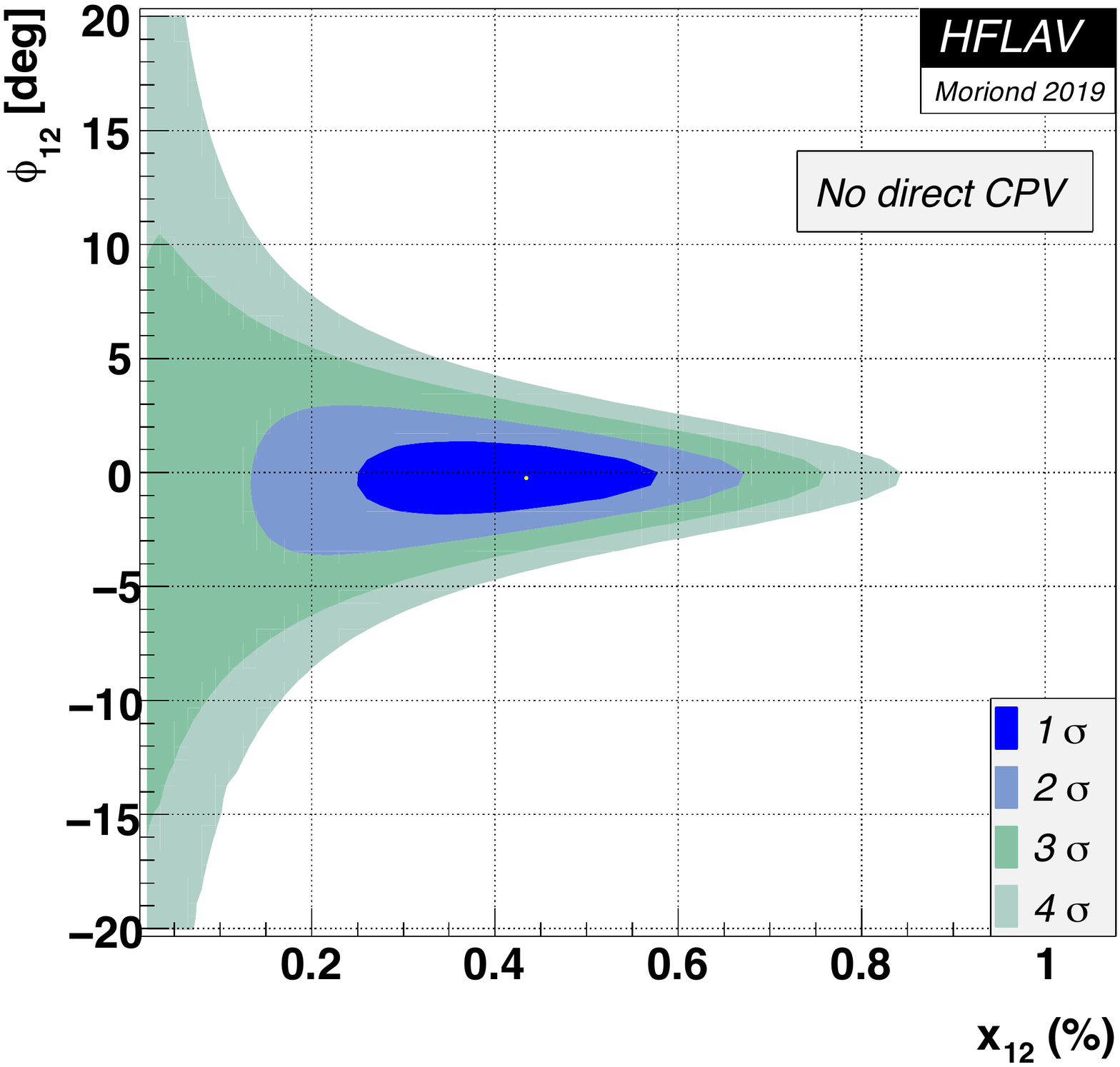}
\vskip-1.2in
\includegraphics[width=84mm]{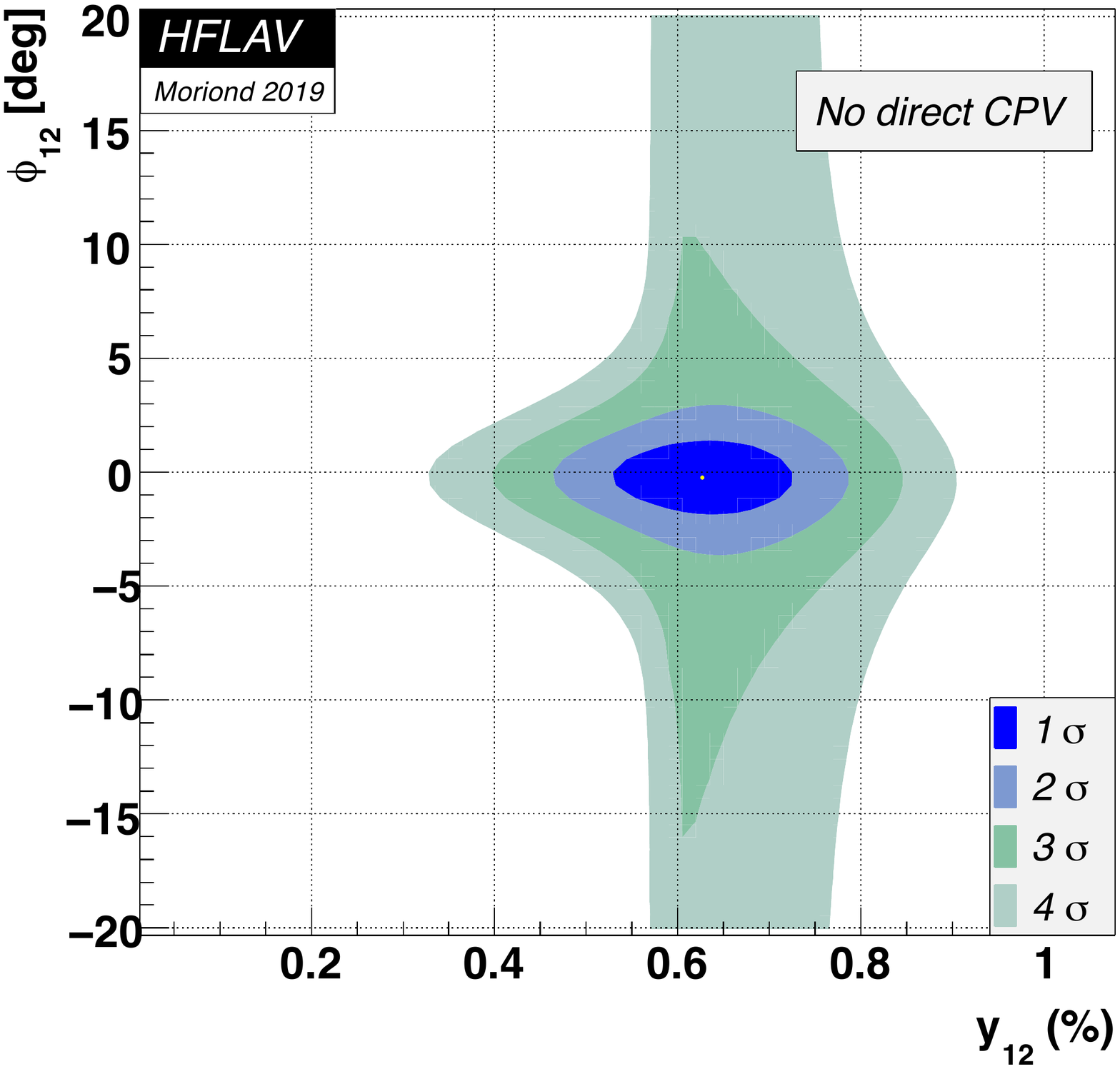}
}
\end{center}
\vskip-0.90in
\caption{\label{fig:contours_ndcpv}
Two-dimensional contours for theoretical parameters 
$(x^{}_{12},y^{}_{12})$ (top left), 
$(x^{}_{12},\phi^{}_{12})$ (top right), and 
$(y^{}_{12},\phi^{}_{12})$ (bottom), 
for no direct \cpv\ in DCS decays (fit 3).}
\end{figure}

\begin{figure}
\vskip-0.40in
\begin{center}
\vbox{
\includegraphics[width=4.0in]{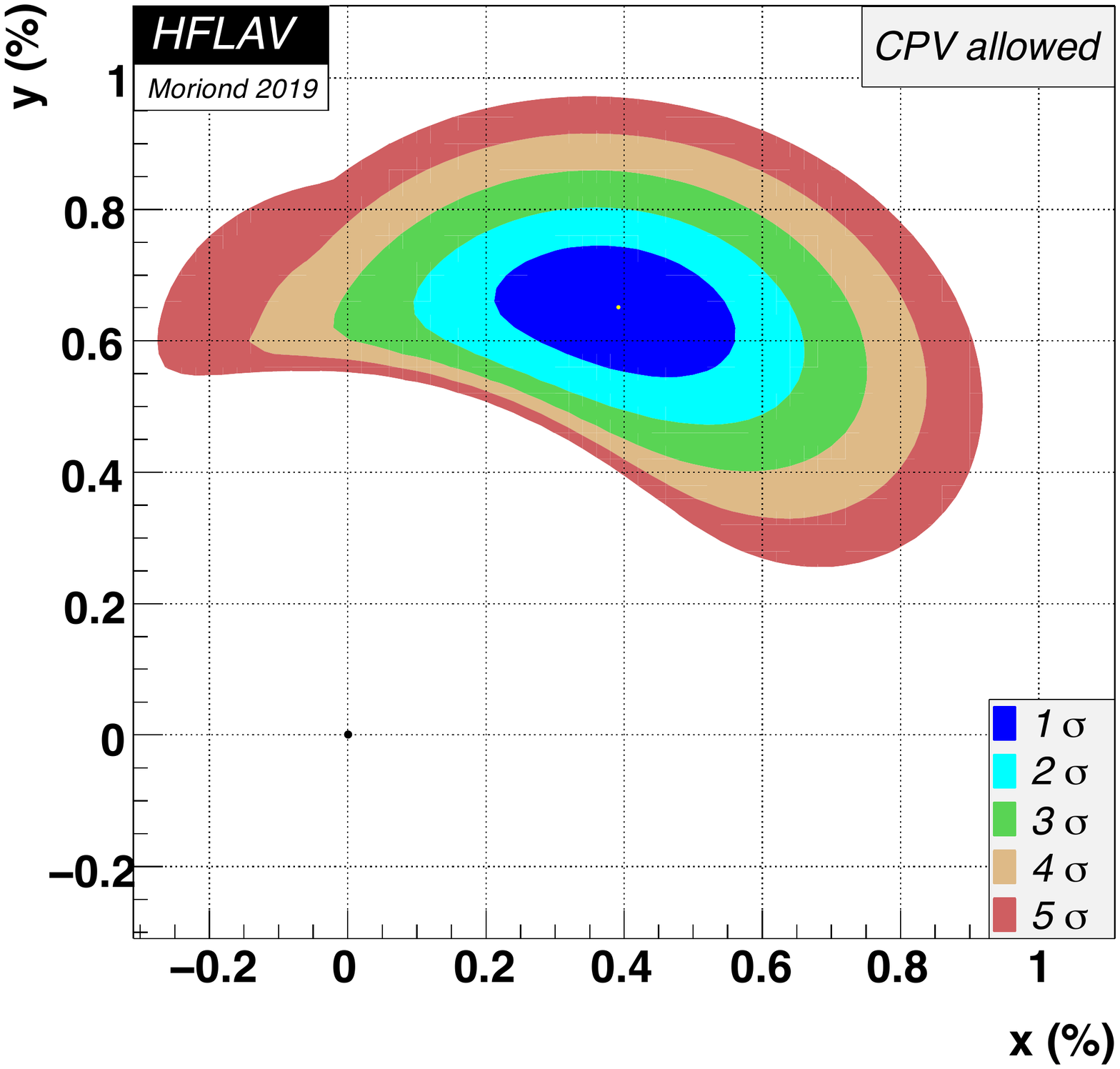}
\vskip-1.6in
\includegraphics[width=4.0in]{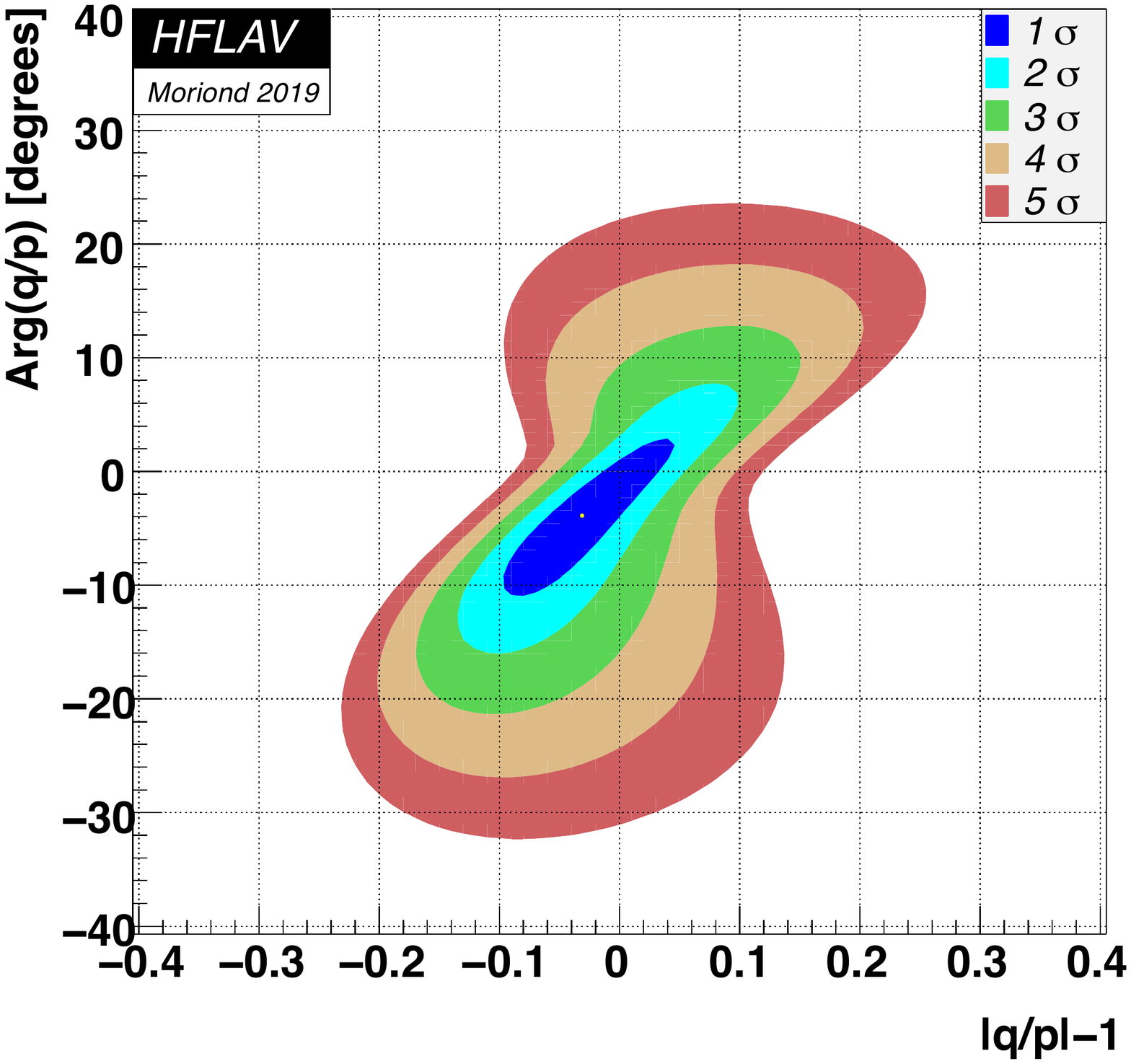}
}
\end{center}
\vskip-1.1in
\caption{\label{fig:contours_cpv}
Two-dimensional contours for parameters $(x,y)$ (top) 
and $(|q/p|-1,\phi)$ (bottom), allowing for \cpv\ (fit 4).}
\end{figure}

\begin{figure}
\centering
\vskip-0.40in
\includegraphics[width=68mm]{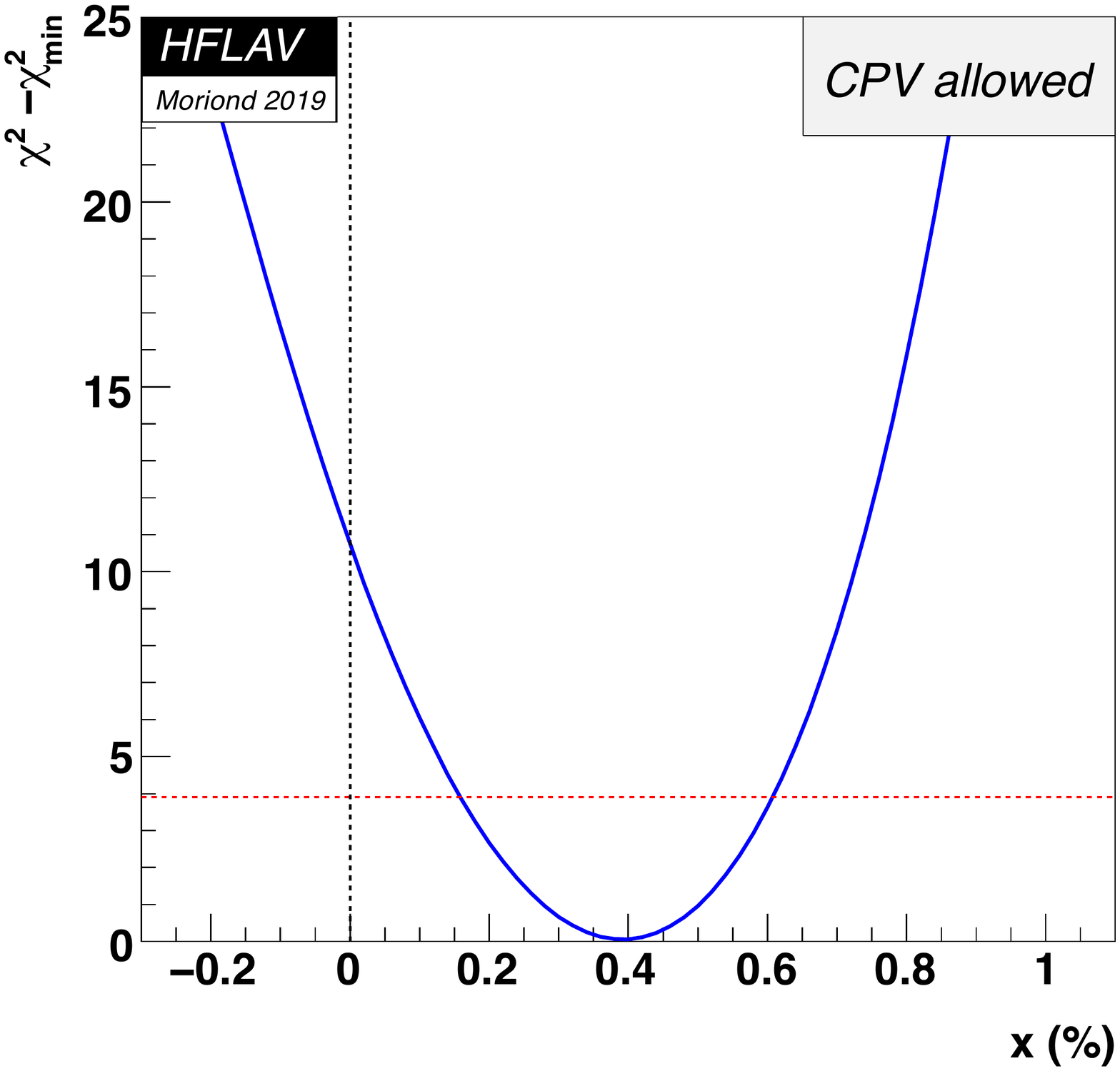}
\includegraphics[width=68mm]{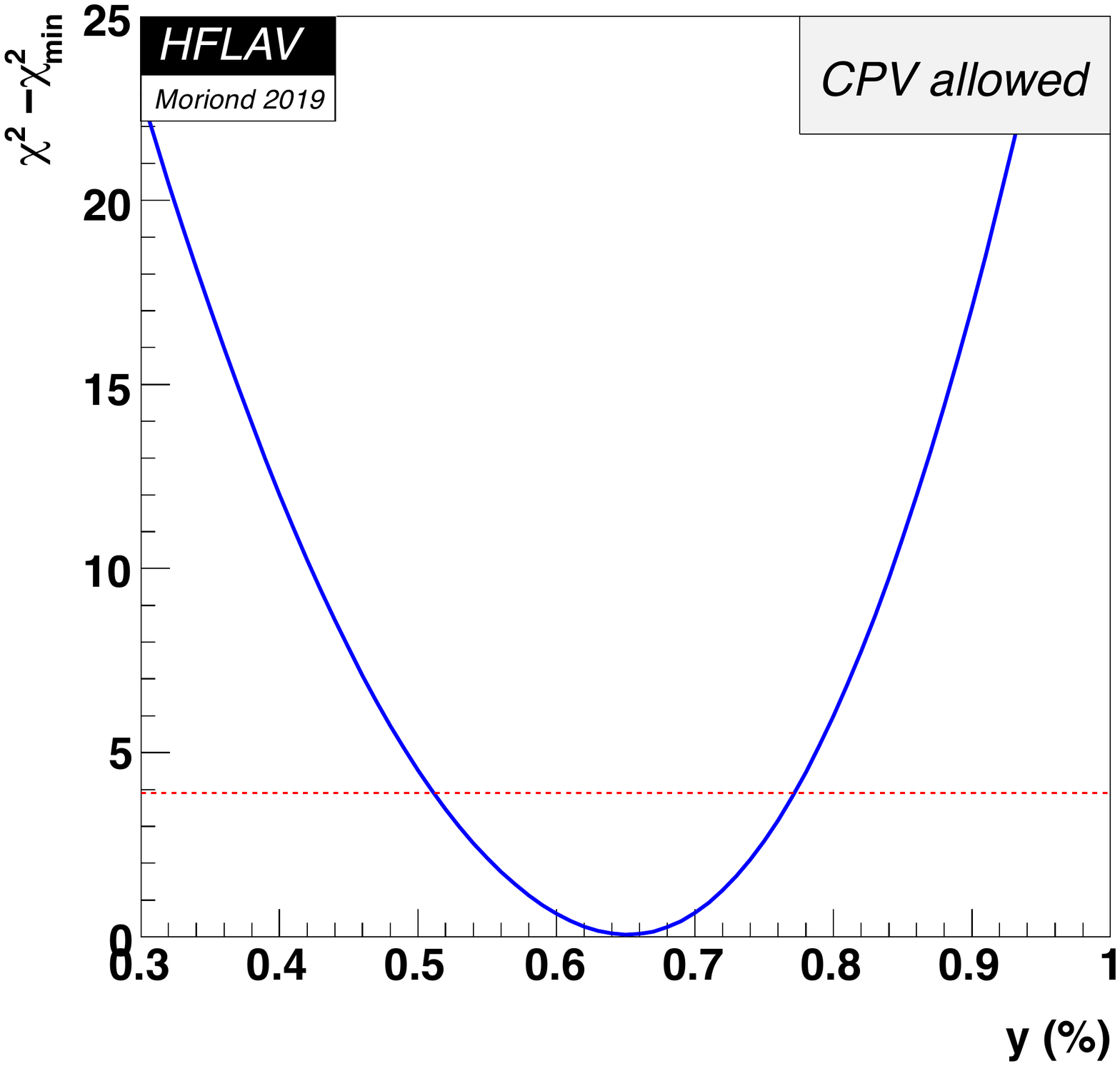}
\vskip-1.1in
\includegraphics[width=68mm]{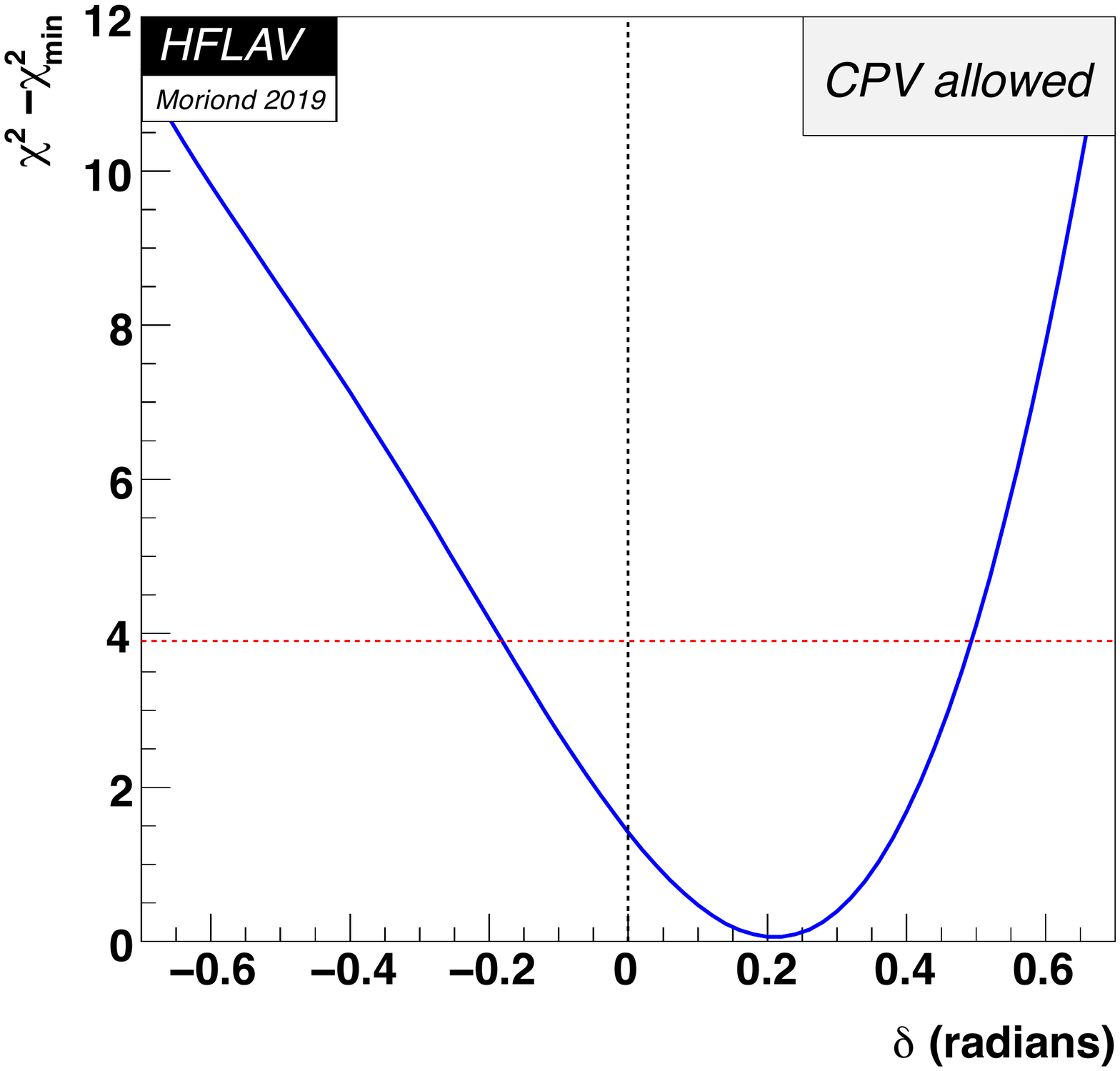}
\includegraphics[width=68mm]{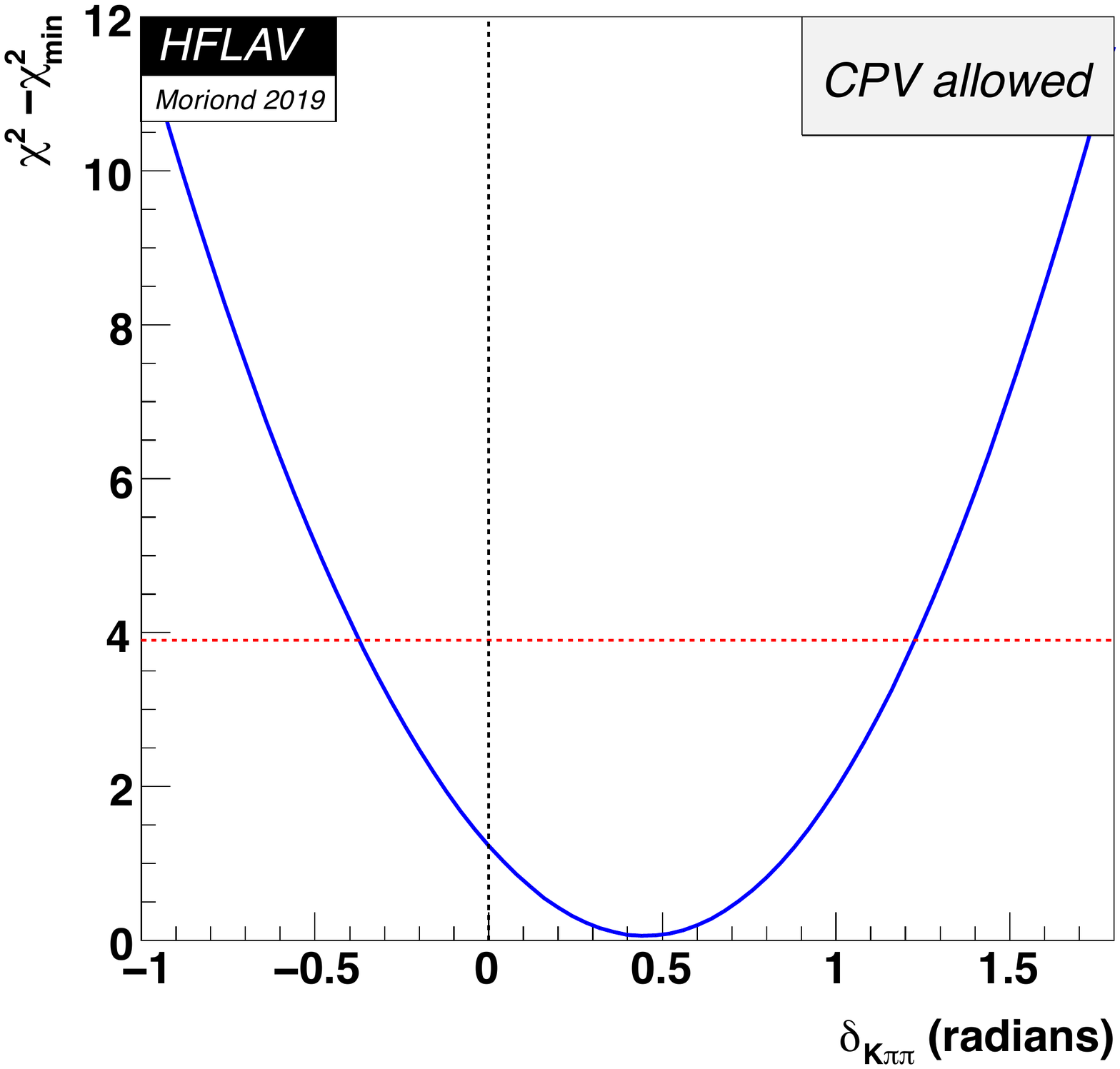}
\vskip-1.1in
\includegraphics[width=68mm]{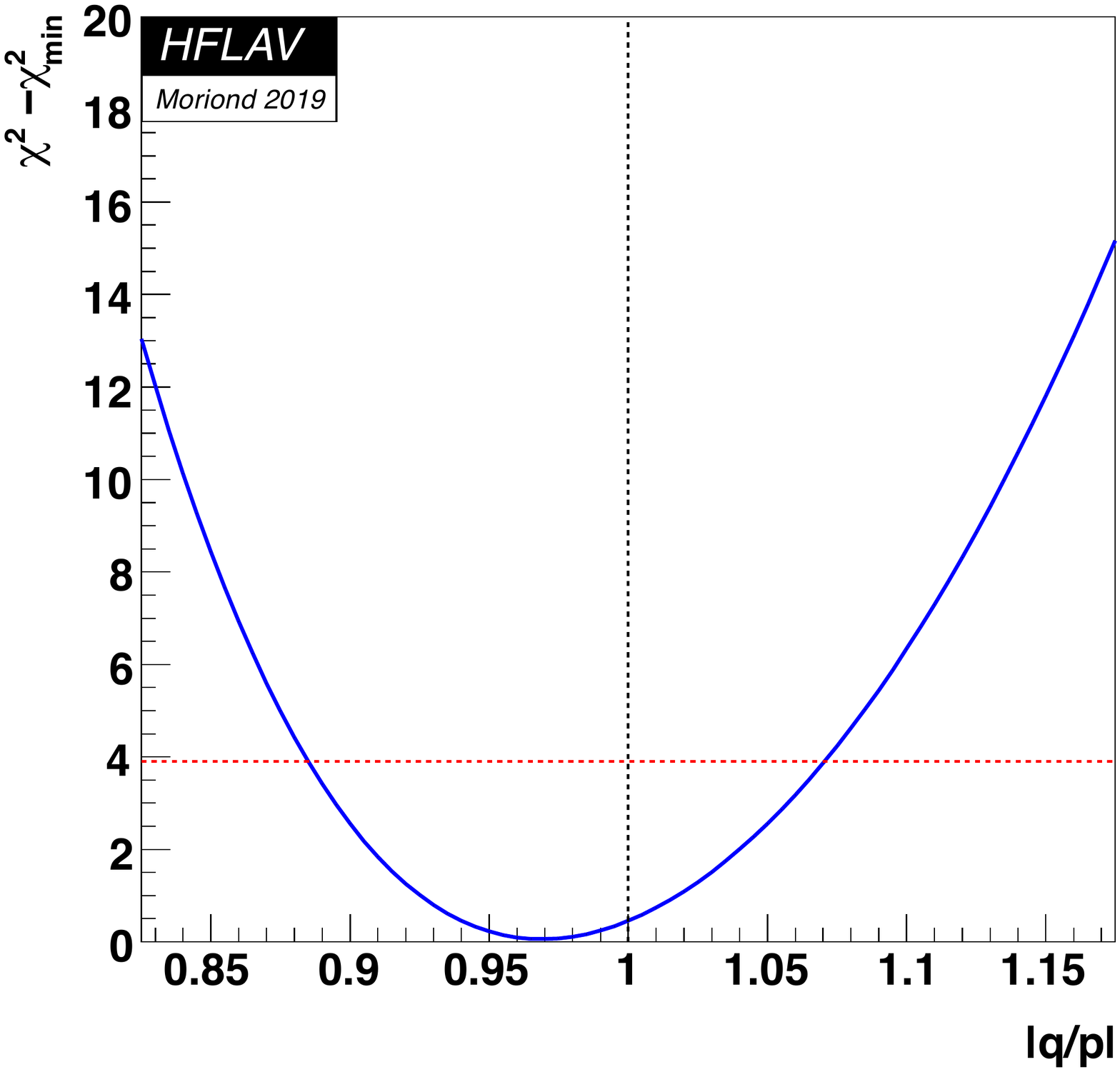}
\includegraphics[width=68mm]{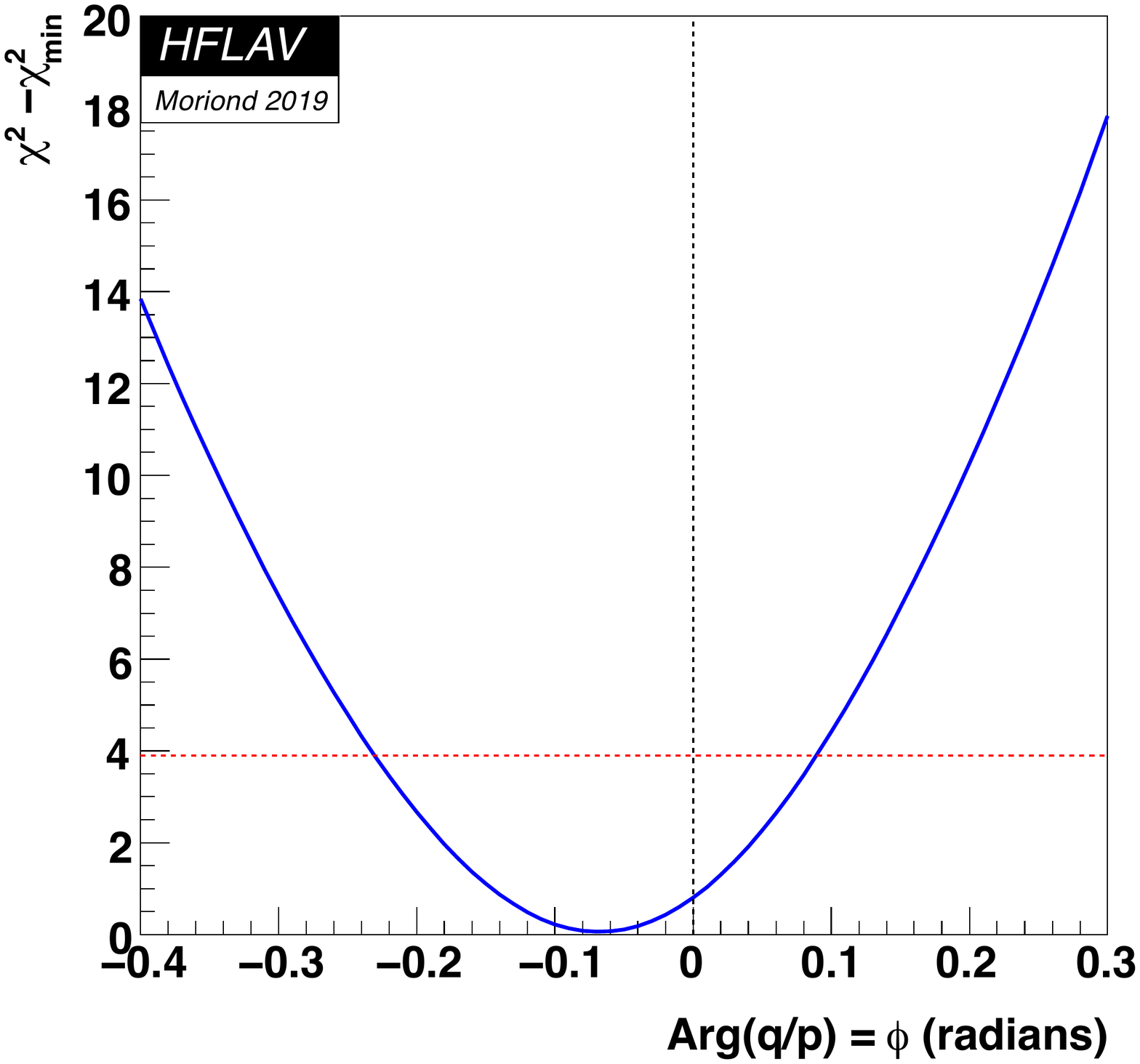}
\vskip-0.60in
\caption{\label{fig:1dlikelihood}
The function $\Delta\chi^2=\chi^2-\chi^2_{\rm min}$ 
for fitted parameters
$x,\,y,\,\delta,\,\delta^{}_{K\pi\pi},\,|q/p|$, and $\phi$.
The points where $\Delta\chi^2=3.84$ (denoted by dashed 
horizontal lines) determine 95\% C.L. intervals. }
\end{figure}

\subsubsection{Conclusions}

From the results listed in Table~\ref{tab:results}
and shown in Figs.~\ref{fig:contours_cpv} and \ref{fig:1dlikelihood},
we conclude the following:
\begin{itemize}
\item 
The experimental data consistently indicate $D^0$-$\dbar$ mixing. 
The no-mixing point $x=y=0$
is excluded at $>11.5\sigma$. The parameter $x$ differs
from zero by $3.1\sigma$, and $y$ differs from zero
by $>11.4\sigma$. This mixing is presumably dominated 
by long-distance processes, which are difficult to calculate.
\item 
Since \ycp\ is positive, the \cp-even state is shorter-lived,
as in the $K^0$-$\kbar$ system. However, since $x$ also appears
to be positive, the \cp-even state is heavier, unlike in the 
$K^0$-$\kbar$ system.
\item 
There is no evidence for \cpv\ arising from $D^0$-$\dbar$
mixing ($|q/p|\neq 1$) or from a phase difference between
the mixing amplitude and a direct decay amplitude ($\phi\neq 0$). 
However, {\it direct\/} \cp\ violation has recently been observed
by LHCb in time-integrated $D^0\ra K^+K^-\!, \pi^+\pi^-$ decays -- 
see Ref.~\cite{Aaij:2019kcg}. The measured \cp\ asymmetry is small, 
0.15\%. Several theory calculations indicate that this value is 
consistent with Standard Model expectations, while new physics 
contributions cannot be excluded.
\end{itemize}

\clearpage
\mysubsection{\CP\ asymmetries}\label{sec:cp_asym}

One way \CP\ violation manifests itself is in a difference between 
the decay rate for a particle and that of its \CP-conjugate\cite{Bigi:2000yz}. 
Such phenomena can be classified into two broad categories, 
termed {\it direct\/} \CP\ violation and 
{\it indirect\/} \CP\ violation~\cite{Nir:1999mg}. 
Direct \CP\ violation refers to charm changing $\Delta C\!=\!1$
processes and can occur in both charged and neutral charm hadron decays. 
It results from interference between two different decay amplitudes, \eg, a 
penguin amplitude and a tree amplitude, that have different weak and strong phases. 
The weak phase difference ($\Delta\phi$) will have opposite sign for $D\ra f$ 
and $\bar{D}\ra\bar{f}$ decays, while the strong phase difference ($\Delta\delta$)
will have the same sign. As a result, squaring the total amplitudes 
to obtain the decay rates gives an interference term proportional to
$\cos(\Delta\phi + \Delta\delta)$ for $D\ra f$ decays, and proportional to
$\cos(-\Delta\phi + \Delta\delta)$ for $\bar{D}\ra\bar{f}$ decays. Thus the 
decay rates will differ. This difference is unaffected when time-integrating
the decay rates, and the overall branching fractions will differ.

In the Standard Model (SM), a difference in strong phases can arise
due to final-state interactions (FSI)\cite{Buccella:1994nf}, different isospin 
amplitudes, intermediate resonance contributions, or different partial waves.
A difference in weak phases arises from different CKM vertex couplings, as 
is often the case for tree and penguin diagrams. Within the SM, direct 
\CP\ violation is expected only in singly Cabibbo-suppressed (SCS) charm 
decays, as only these decays receive a non-negligible contribution from the 
penguin amplitude. This type of \CP\ violation depends on the decay mode, 
and the \CP\ asymmetries can reach the percent level. Indirect \CP\ violation 
refers to $\Delta C\!=\!2$ processes and arises in $D^0$ decays due to 
$D^0$-$\dbar$ mixing. It can occur as an asymmetry in the mixing itself, or it 
can result from interference between a decay amplitude following mixing and a 
non-mixed amplitude. Within the SM, charm indirect \CP\ violation is expected 
to be universal, \ie, independent of final state. Current experimental limits 
on indirect \CP violation are discussed in Sec.~\ref{sec:charm:mixcpv}. 

The time-integrated \CP\ asymmetry $A_{\CP}$ is defined as the difference 
between $D$ and $\bar{D}$ partial widths divided by their sum:
\begin{eqnarray}  
A_{\CP} & = & \frac{\Gamma(D)-\Gamma(\bar{D})}
{\Gamma(D)+\Gamma(\bar{D})}\,.
\end{eqnarray}
In the case of $D^+$ and $D^+_s$ decays, $A^{}_{\CP}$ measures 
direct \CP\ violation; in the case of $D^0$ decays, $A^{}_{\CP}$ 
measures direct and indirect \CP\ violation combined 
(see also Sec.~\ref{sec:charm:cpvdir}). Given experimental 
constraints on $A_\Gamma$, a contribution from indirect \CP\ violation 
would be negligible compared to current $A_{\CP}$ sensitivities. 
Values of $A^{}_{\CP}$ for $D^+$, $D^0$ and $D_s^+$ decays are 
listed in Tables~\ref{tab:cp_charged}, \ref{tab:cp_charged2}, 
\ref{tab:cp_neutral}, \ref{tab:cp_neutral2} and 
\ref{tab:cp_ds} respectively. Modes with a single $K^{}_S$ meson 
in the final state can exhibit a \CP\ asymmetry due to \CP\ violation
in $K^0$-$\kbar$ mixing~\cite{Grossman:2012aa}; \ie, the rate for 
$\kbar\ra K^{}_S$ differs slightly from that for $K^0\ra K^{}_S$. 
This small effect is visible thus far only in $D^+ \to K^{}_S\pi^+$ 
decays (see Table~\ref{tab:cp_charged}). Modes with a $K^0$ or $\kbar$ 
in the final state have this effect already corrected for. 
The asymmetry for the DCS decay $D^0 \to K^+\pi^-$ is not included 
in these tables, as it is a by-product of charm-mixing measurements 
and thus is discussed in Sec.~\ref{sec:charm:mixcpv} (where it is 
referred to as $A_D$).

In each experiment, care must be taken to correct for production 
and detection asymmetries, as they can reach the percent level. 
To take into account differences in production rates between 
$D$ and $\bar{D}$, which would affect the number of respective 
decays observed, some experiments (such as E791 and FOCUS) normalize 
$A_{\CP}$ to that measured in a Cabibbo-favored mode. 
This method assumes there is negligible \CP\ violation in the 
normalization mode. Explicitly, the \CP\ asymmetry is calculated as
\begin{eqnarray}
A_{\CP} &=&\frac{\eta(D)-\eta(\bar{D})}{\eta(D)+\eta(\bar{D})}\,,
\end{eqnarray}
where (considering, for example, $D^0 \to K^-K^+$)
\begin{eqnarray}
 \eta(D) &=& \frac{N(D^0 \rightarrow K^-K^+)}{N(D^0 \rightarrow K^-\pi^+)}\,, \\
 \eta(\bar{D}) & = & \frac{N(\dbar\rightarrow K^-K^+)}
{N(\dbar\rightarrow K^+\pi^-)}\,.
\end{eqnarray}
In this method there is the additional advantage that most corrections due to 
reconstruction inefficiencies cancel out, reducing systematic uncertainties. 

Other experiments (such as Belle and LHCb) determine $A_{\CP}$ via the relation
\begin{eqnarray}
A_{\rm meas} & = & A_{\CP} + A_{\rm prod} + A_{\rm det}\,,
\end{eqnarray}
where $A_{\rm meas}$ is the measured (raw) asymmetry, $A_{\rm prod}$ is 
the asymmetry in the charm hadron production, and $A_{\rm det}$ 
is due to a difference in detection efficiencies between positively 
and negatively charged hadrons.
The production asymmetry at the LHC arises from a charge asymmetry of 
the colliding particles: in $pp$ collisions more charm baryons are
produced than anti-baryons, and, as a result, charm mesons are 
less abundantly produced than anti-charm mesons. 
Such a production asymmetry is expected to be dependent on kinematics 
of the produced charm hadrons. 
The production asymmetry in $e^+e^-$ 
collisions appears as a forward-backward (FB) asymmetry caused by an 
interference of the photon and off-shell $Z^0$ contributions. The detection 
asymmetries typically arise from differences in hadron interactions 
with detector material. In particular, the interaction cross sections 
for $K^+$ and $K^-$ significantly differ, with the differences being 
dependent on the kaon momentum. 

The $B$-factory strategy to separate the production and \CP\ 
asymmetries relies on the former being odd, while the latter is even, 
with respect to the center-of-mass production polar angle ($\theta^*$). 
The $A_{\rm meas}$ is measured in $|\cos \theta^*|$ bins and subsequently 
averaged; this removes the $A_{\rm prod}$ contribution. 
At LHCb, the production asymmetry is removed by measuring $A_{\CP}$ for
$D^*$-tagged $D^0 \to K^- \pi^+$ decays; this also corrects for the 
soft $\pi$ detection asymmetry.  
Subsequently, $D^+ \to K^- \pi^+ \pi^+$ decays are used to correct for 
the detection asymmetry introduced by the $K^- \pi^+$ system itself, and  
$D^+ \to K_S \pi^+$ decays are then used to remove the asymmetries in $D^+$ 
production and $\pi^+$ detection. 
Finally, the asymmetry related to the neutral kaon, 
\ie, from regeneration and different interactions of $K^0$ and $\kbar$ 
with the detector, as well as from \CP\ violation occurring in 
the $K^0 \text{-} \kbar$ mixing, is calculated. Put together, this gives
\begin{equation*}
A_{\CP} (K^+K^-) = A_{\rm meas}(K^+K^-) - A_{\rm meas} (K^- \pi^+) 
+ A_{\rm meas}(K^- \pi^+ \pi^+) - A_{\rm meas}(K_S \pi^+) 
+ A(\kbar \text{-} K^0).
\end{equation*}
For some decays, typically the ones with lower statistics, one corrects 
for nuisance asymmetries by measuring $A_{\CP}$ relative to 
some well-measured reference channel, for instance
\begin{equation*}
A_{\CP} (D_s^+\to \eta^{'}\pi^+) = A_{\rm meas}(D_s^+\to \eta^{'}\pi^+) 
- A_{\rm meas} (D_s^+\to \phi \pi^+)+ A_{\CP}(D_s^+\to \phi \pi^+).
\end{equation*}
The uncertainty of the reference $A_{\CP}$ is treated as an external input error.

Much easier than individual $A_{\CP}$ measurements, and often
easier for theoretical interpretation, are measurements of $A_{\CP}$ 
{\it differences}, denoted $\Delta A_{\CP}$. The most important one
is that for $D^0\to K^+K^-$ and $D^0\to \pi^+\pi^-$ decays, which is discussed 
in Sec.~\ref{sec:charm:cpvdir}. The difference $\Delta A_{\CP}$ in the
baryon asymmetries for $\Lambda_c^+\to p K^+K^-$ and $\Lambda_c^+\to p \pi^+\pi^-$ 
decays was recently measured by LHCb~\cite{Aaij:2017xva}. We note that, in the
limit of U-spin symmetry, direct \CP\ violation in $D^0\to K^+K^-$ and 
$D^0\to \pi^+\pi^-$ decays is expected to have equal magnitude and opposite
sign\cite{Brod:2012ud}; 
thus the measurement of $\Delta A_{\CP}$ ``doubles'' the effect. 
However, no such argument holds for baryonic $\Lambda_c^+\to p K^+K^-$ and
$\Lambda_c^+\to p \pi^+\pi^-$ decays.

\CP\ asymmetries arise from the interplay between weak and strong phases, 
and the latter change over the phase space of multi-body decays, which
usually proceed via intermediate states. Therefore local \CP\ asymmetries, 
\ie, measured in the phase space of the multi-body decays, or asymmetries 
for individual strong amplitudes, can offer better sensitivity 
than a global asymmetry measurement, in which an effect can be 
diluted. Probing the multi-body phase space is often done 
in a model-dependent way by employing a Dalitz analysis or more general 
amplitude analysis separately for $D$ and $\bar{D}$ decays; a \CP\ asymmetry 
is then measured for each contributing amplitude. The \cp-violating 
observables are asymmetries in magnitudes and phases of \cp-conjugate 
amplitudes, as well as asymmetries in the amplitude fit fractions. 

For multi-body decays, some experiments use model-independent 
techniques to search for local \CP\ asymmetries. One technique 
(see Refs.~\cite{Aubert:2008yd, Bediaga:2009tr}) uses a binned 
$\chi^2$ approach to compare the relative density in a bin 
of phase space for $B\to f$ with that of the \CP-conjugate decay.
Another technique (the ``Energy Test technique''~\cite{doi:10.1080/00949650410001661440})
uses a test statistic variable ($T$) to determine the average distance between
events in phase space. If the distribution of events in two \CP-conjugate samples
are identical (the \CP-symmetric case), $T$ will fluctuate around a value close to zero.
This technique yields a $p$-value for the no-\CP\ violation hypothesis and
localizes any \CP-asymmetric phase space regions.

In Tables~\ref{tab:cp_charged}, \ref{tab:cp_charged2}, \ref{tab:cp_neutral}, 
\ref{tab:cp_neutral2}, and \ref{tab:cp_ds}, asymmetries for three- 
and four-body decays are reported for their observed final state, \ie, 
resonant substructure is implicitly included but not considered 
separately. Most asymmetries measured for three- and four-body 
channels are still only global asymmetries. 
The reported model-independent tests, which attempt to probe 
the decay phase space, yield $p$-values typically at the level of 
a few percent or higher and thus consistent with no \CP\ violation. 
The lowest $p$-value of 0.6\%, corresponding to a significance for 
\CP\ violation of $2.7\sigma$, is obtained for the $P$-odd (parity-odd)
test of $D^0 \to \pi^+\pi^-\pi^+\pi^-$ decays~\cite{Aaij:2013aa}. This
implies that the effect, if not a statistical fluctuation, originates
in a $P$-odd amplitude such as $D^0 \to [\rho^0 \rho^0]_{L=1}$.
For $D^0 \to K^+K^-\pi^+\pi^-$ decays~\cite{Aaij:2018nis},
a model-dependent amplitude analysis was performed, and
\CP\ asymmetries were measured for 25 intermediate amplitudes.
The uncertainties on these asymmetries ranged from 1\% to 15\%
and were dominated by statistical errors.
No significant \CP\ violation was observed, and the most
significant asymmetry of $2.8\sigma$ was observed for the
phase of the $P$-odd amplitude $D^0 \to [\phi(1020) \rho(1450)^0]_{L=1}$. 
\cp\ violation arising through $P$ violation is 
discussed further in Sec.~\ref{sec:todd_asym}.

For the first time, $A_{\CP}$ has been measured for decays classified as rare:
radiative modes $D^0 \to V \gamma$, with $V=\ \bar{K}^{*0}, \ \phi(1020), 
\ \rho^{0}$, as well as di-muon decays $D^0 \to \pi^+ \pi^- \mu^+ \mu^-$
and $D^0 \to K^+ K^- \mu^+ \mu^-$. 
For the di-muon modes, in addition to their global asymmetries listed 
in Table~\ref{tab:cp_neutral2}, $A^{}_{\CP}$ was measured in bins of
di-muon invariant mass. Asymmetries for mass regions away from 
$\mu^+\mu^-$ production via 
$\eta$, $\rho\text{-}\omega$ or $\phi$ decays still have
very limited sensitivities, ranging from 12\% to 26\%. These 
non-resonance regions are particularly important for New Physics 
searches (see Sec.~\ref{sec:charm:rare}).
Overall, \CP\ asymmetries have been measured for more than 50 charm 
decay modes, and in several modes the sensitivity is well below 
$5 \times 10^{-3}$. There is currently no evidence 
for \CP\ violation in the charm meson sector.
The \CP\ asymmetry observed for the mode $D^+ \to K_S\pi^+$ 
is consistent with that expected due to $K^0\text{-}\kbar$ 
mixing~\cite{Grossman:2012aa}, and thus it is not attributed 
to charm.

In the charm baryon sector, there is also no evidence of \CP\ violation.
Until recently, there had been only two measurements 
for $\Lambda_c^+$; these were performed by CLEO~\cite{Hinson:2004pj} 
and FOCUS~\cite{Link:2005ft} and had limited sensitivity.
The former used the semileptonic 
$\Lambda_c^+ \to \Lambda e^+ \nu_{e}$ decays, while the latter 
the CF $\Lambda_c^+ \to \Lambda \pi^+$ decays; both have searched 
for \cp\ violation through an angular analysis exploiting 
the $\Lambda$ helicity angle. 
\CP\ asymmetry is accessed through comparison of $P$ asymmetry 
in decays of $\Lambda_c^+$ and $\Lambda_c^-$, measured with 
the weak-asymmetry parameters, respectively $\alpha_{\Lambda_c}$ 
and $\alpha_{\bar{\Lambda}_c}$.
As $\alpha_{\Lambda_c} = - \alpha_{\bar{\Lambda}_c}$ under 
$P$-parity conservation, the \CP-violating asymmetry is defined as
\begin{eqnarray}  
A_{\CP}^{\alpha} & = & 
\frac{ \alpha_{\Lambda_c} + \alpha_{\bar{\Lambda}_c}}
{\alpha_{\Lambda_c} - \alpha_{\bar{\Lambda}_c}} \, .
\end{eqnarray}
The CLEO measurement~\cite{Hinson:2004pj} gives
\begin{equation*}
A_{\CP}^{\alpha} (\Lambda_c^+ \to \Lambda e^+ \nu_{e}) 
= 0.00 \pm 0.03 \pm 0.01 \pm 0.02,
\end{equation*}
where the third error is related to the uncertainty of the $\Lambda$ 
weak-asymmetry parameter. 
The asymmetry measured by FOCUS~\cite{Link:2005ft} is
\begin{equation*}
A_{\CP}^{\alpha} (\Lambda_c^+ \to \Lambda \pi^+) 
= -0.07 \pm 0.19 \pm 0.12. 
\end{equation*}
The first high-statistics \cpv\ measurement of charm baryons comes 
from LHCb in the form of $\Delta A_{\CP}$ for 
the $\Lambda_c^+\to p K^+K^-$ and $\Lambda_c^+\to p \pi^+\pi^-$ SCS 
decays~\cite{Aaij:2017xva}, where the result is
\begin{equation*}
\Delta A_{\CP}(\Lambda_c^+ \to p h^+h^-) 
\equiv A_{\CP}(p K^+K^-) - A_{\CP}(p \pi^+\pi^-) 
= 0.003 \pm 0.009 \pm 0.006.
\end{equation*}
The measurement, performed in a phase-space integrated manner, 
has limited sensitivity and does not facilitate an interpretation.
However, the production asymmetry between $\Lambda_c^+$ and $\Lambda_c^-$
baryons cancels in this difference.
Given the potentially rich dynamics of these decays in their 
five-dimensional phase space\footnote{For unpolarized $\Lambda_c$, 
its decay phase space reduces to a two-dimensional Dalitz distribution.}, 
$\Delta A_{\CP}$ measured in phase-space regions or a model-dependent
measurement of intermediate amplitude asymmetries would be very desirable. 

For charm decays one can build various SU(3)-based sum rules which, 
in addition to testing SU(3) symmetry itself, are also useful for 
performing model-independent tests of the SM. Particularly interesting 
are sums exploiting the SU(3) subgroups, U-spin or isospin (I), 
as they involve less decays and offer more 
precise tests. While U-spin symmetry in charm decays is broken by 
a non-negligible amount due to the s-quark mass, isospin symmetry holds 
at the $(m_u-m_d)$ level and thus is very precise. 
Important for our considerations are isospin sum rules that relate 
individual \cp\ asymmetries of the isospin-related processes. 
Verifying such rules allows for tests to be performed with reduced 
uncertainty due to strong interaction effects. Such a sum rule
has been proposed for $D \to \pi \pi$ decays in 
Ref.~\cite{Grossman:2012eb}.

The isospin decomposition of $D \to \pi \pi$  amplitudes gives
\begin{eqnarray*}
A_{\pi^+\pi^-} & = & \sqrt{2} {\cal A}_3  + \sqrt{2} {\cal A}_1, \\
A_{\pi^0\pi^0} & = & 2 {\cal A}_3  - {\cal A}_1, \\
A_{\pi^+\pi^0} & = & 3 {\cal A}_3,  
\end{eqnarray*}
where ${\cal A}_1$ and ${\cal A}_3$ are amplitudes corresponding to 
the $\Delta I = 1/2$ and $\Delta I = 3/2$ transitions, respectively
(\ie, transitions to $\pi\pi$ final states with $I=0$ and $I=2$).
From this, one can get an amplitude isospin sum rule
\begin{equation}
\frac{1}{\sqrt{2}}A_{\pi^+\pi^-} + A_{\pi^0\pi^0} - A_{\pi^+\pi^0} \ =\ 0.
\end{equation}
Probing such a sum requires knowledge of strong phases, which are 
accessible only at charm-threshold experiments. However, without 
this knowledge the sum of differences of decay rates for $D$ and 
$\bar{D}$ decays can be measured:
\begin{equation}
|A_{\pi^+\pi^-}|^2 - |\bar{A}_{\pi^+\pi^-}|^2 + |A_{\pi^0\pi^0}|^2 - |\bar{A}_{\pi^0\pi^0}|^2  
- \frac{2}{3}(|A_{\pi^+\pi^0}|^2 - |\bar{A}_{\pi^-\pi^0}|^2)  \ =\ 
3( |{\cal A}_1|^2 -|{\cal \bar{A} }_1|^2 )\,.
\label{eq:rate_sum_rule}
\end{equation}
This equation suggests several SM tests. As the penguin amplitude is, 
to excellent approximation within the SM, purely $\Delta I = 1/2$, 
any \CP asymmetry observed in $D^+ \to \pi^+\pi^0$ 
would be a sign of New Physics in the $\Delta I = 3/2$ amplitude. 
If the sum in Eq.~(\ref{eq:rate_sum_rule}), 
depending only on ${\cal A}_1$, is found to be non-zero,
this would mean that \CP\ violation arises from the $\Delta I = 1/2$ 
transitions. Moreover, a scenario in which the sum in 
Eq.~(\ref{eq:rate_sum_rule}) is zero and individual asymmetries 
are non-zero would suggest New Physics contributing 
to the $\Delta I = 3/2$ amplitude. 

To facilitate an experimental test, the left-hand side of 
Eq.~(\ref{eq:rate_sum_rule}) is rewritten as a ratio~\cite{Babu:2017bjn}:
\begin{equation}
R\ \equiv\ \frac{|A_{\pi^+\pi^-}|^2 - |\bar{A}_{\pi^+\pi^-}|^2 
+ |A_{\pi^0\pi^0}|^2 - |\bar{A}_{\pi^0\pi^0}|^2  
- \frac{2}{3}(|A_{\pi^+\pi^0}|^2 - |\bar{A}_{\pi^-\pi^0}|^2)}
{|A_{\pi^+\pi^-}|^2 + |\bar{A}_{\pi^+\pi^-}|^2 
+ |A_{\pi^0\pi^0}|^2 + |\bar{A}_{\pi^0\pi^0}|^2  
+ \frac{2}{3}(|A_{\pi^+\pi^0}|^2 + |\bar{A}_{\pi^-\pi^0}|^2)}\,.
\label{eq:rate_sum_ratio}
\end{equation}
Using the relations
$|A|^2 \propto {\cal B}/\tau^{}_D$ and
$|A|^2 - |\bar{A}|^2 = A_{CP}(|A|^2 + |\bar{A}|^2)$,
we rewrite Eq.~(\ref{eq:rate_sum_ratio}) as
\begin{equation}
R\ =\  
\frac{A_{CP}(D^0 \to \pi^+ \pi^-)}
{1+\frac{\tau_{D^0}}{{\cal B}_{+-}}
\large(\frac{{\cal B}_{00}}{\tau_{D^0}} 
+\frac{2}{3}\frac{{\cal B}_{+0}}{\tau_{D^+}}\large) } 
  + 
\frac{A_{CP}(D^0 \to \pi^0 \pi^0)}
{1+\frac{\tau_{D^0}}{{\cal B}_{00}}
\large(\frac{{\cal B}_{+-}}{\tau_{D^0}} 
+\frac{2}{3}\frac{{\cal B}_{+0}}{\tau_{D^+}}\large) } 
  + 
\frac{A_{CP}(D^+ \to \pi^+ \pi^0)}
{1+\frac{3}{2}\frac{\tau_{D^+}}{{\cal B}_{+0}}
\large(\frac{{\cal B}_{00}}{\tau_{D^0}} 
+\frac{{\cal B}_{+-}}{\tau_{D^0}}\large) }\,,
\label{eq:rate4test}
\end{equation}
where ${\cal B}_{+-}$, ${\cal B}_{00}$, and ${\cal B}_{+0}$
denote the branching fractions for $D^0 \to \pi^+ \pi^-$,
$D^0 \to \pi^0 \pi^0$, and $D^+ \to \pi^+ \pi^0$, respectively.
The sum $R$ is calculated using our averages for \cp\ asymmetries 
(Tables~\ref{tab:cp_charged} and \ref{tab:cp_neutral}), and
PDG averages~\cite{PDG_2018} for branching fractions and 
lifetimes. The result is
\begin{equation}
R\ =\ (+0.01 \pm 2.65) \times 10^{-3},
\end{equation}
which is consistent with zero. In addition, all the individual 
asymmetries contributing to $R$ are consistent with zero. The 
uncertainty on $R$ is dominated by the uncertainties on 
individual asymmetries.

The sum rule for $D \to \bar{K} K$ decays involves full SU(3) 
considerations and thus is imprecise. 
Ref.~\cite{Grossman:2012eb} proposes a set of isospin sum rules 
for $D\to \rho \pi$ or $D\to \bar{K}^{(*)} K^{(*)}\pi$, but to test
these sum rules requires a number of not-yet-performed experimental
measurements.

\begin{table}[!htb]
\renewcommand{\arraystretch}{1.4}
\caption{\CP\ asymmetries 
$A^{}_{\CP}= [\Gamma(D^+)-\Gamma(D^-)]/[\Gamma(D^+)+\Gamma(D^-)]$
for two-body $D^\pm$ decays. In the individual asymmetries listed, the first uncertainties are statistical, and the second systematic, whereas the third uncertainty in $A^{}_{\CP}(D^+ \to \pi^+ \eta^\prime)$ from LHCb is due to $A^{}_{\CP}(D^+ \to \pi^+ K_S)$ used for calibration. 
\label{tab:cp_charged}}
\footnotesize
\begin{center}
\begin{tabular}{|l|c|c|c|} 
\hline
{\bf Mode} & {\bf Year} & {\bf Collaboration} & {\boldmath $A^{}_{\CP}$} \\
\hline
{\boldmath $D^+ \to \mu^+ \nu$} &
  2008 & CLEO~\cite{Eisenstein:2008aa} &  $ +0.08  \pm 0.08 $ \\
\hline
{\boldmath $D^+ \to \pi^+ \pi^0$} &
 2018 & Belle~\cite{Babu:2017bjn} & $+0.0231 \pm 0.0124 \pm 0.0023 $ \\
 & 2010 & CLEO~\cite{Mendez:2009aa} &  $ +0.029  \pm 0.029 \pm 0.003 $ \\
&      & HFLAV average             &  $ +0.024   \pm 0.012 $ \\
\hline
{\boldmath $D^+ \to \pi^+ \eta$} &
  2011 & Belle~\cite{Won:2011ng}    &  $ +0.0174  \pm 0.0113 \pm 0.0019 $ \\
& 2010 & CLEO~\cite{Mendez:2009aa} &  $ -0.020   \pm 0.023  \pm 0.003 $ \\
&      & HFLAV average             &  $ +0.010   \pm 0.010 $ \\
\hline
{\boldmath $D^+ \to \pi^+ \eta^\prime$} &
    2017 & LHCb~\cite{Aaij:2017eux} & $-0.0061 \pm 0.0072 \pm 0.0053 \pm 0.0012$ \\
&  2011 & Belle~\cite{Won:2011ng}    &  $ -0.0012  \pm 0.0112 \pm 0.0017 $ \\
& 2010 & CLEO~\cite{Mendez:2009aa} &  $ -0.040   \pm 0.034  \pm 0.003  $ \\
&      & HFLAV average             &  $ -0.006   \pm 0.007 $ \\  
\hline
{\boldmath $D^+ \to K^+ \pi^0$} &
  2010 & CLEO~\cite{Mendez:2009aa} &  $ -0.035  \pm 0.107 \pm 0.009 $ \\
\hline
{\boldmath $D^+ \to K_S\pi^+$}   &
   2014 & CLEO~\cite{Bonvicini:2013vxi} &  $ -0.011   \pm 0.006   \pm 0.002   $ \\
&  2012 & Belle~\cite{Ko:2012pe}        &  $ -0.00363 \pm 0.00094 \pm 0.00067 $ \\
&  2011 & \babar~\cite{Amo:2011ab}      &  $ -0.0044  \pm 0.0013  \pm 0.0010  $ \\
&  2002 & FOCUS~\cite{Link:2001zj}      &  $ -0.016   \pm 0.015   \pm 0.009   $ \\
&       & HFLAV average                &  $ -0.0041  \pm 0.0009 $ \\
\hline
{\boldmath $D^+ \to K_SK^+$} &
  2013 & \babar~\cite{Lees:2013aa}      &  $ +0.0013 \pm 0.0036 \pm 0.0025 $ \\
& 2013 & Belle~\cite{Ko:2013aa}         &  $ -0.0025 \pm 0.0028 \pm 0.0014 $ \\
& 2010 & CLEO~\cite{Mendez:2009aa}      &  $ -0.002  \pm 0.015  \pm 0.009  $ \\  
& 2002 & FOCUS~\cite{Link:2001zj}       &  $ +0.071  \pm 0.061  \pm 0.012  $ \\
&      & HFLAV average                 &  $ -0.0011 \pm 0.0025 $ \\
\hline
{\boldmath $D^+ \to (\kbar/K^0)K^+$} &
  2014 & LHCb~\cite{Aaij:2014ac}        &  $ +0.0003 \pm 0.0017 \pm 0.0014 $ \\
& 2013 & \babar~\cite{Lees:2013aa}      &  $ +0.0046 \pm 0.0036 \pm 0.0025 $ \\
& 2013 & Belle~\cite{Ko:2013aa}         &  $ -0.0008 \pm 0.0028 \pm 0.0014 $ \\
&      & HFLAV average                 &  $ +0.0011 \pm 0.0017 $ \\
\hline
\end{tabular}
\end{center} 
\end{table}

\begin{table}[!htb]
\renewcommand{\arraystretch}{1.4}
\caption{\CP\ asymmetries 
$A^{}_{\CP}= [\Gamma(D^+)-\Gamma(D^-)]/[\Gamma(D^+)+\Gamma(D^-)]$
for three- and four-body $D^\pm$ decays. In the individual asymmetries listed, the first uncertainties are statistical, and the second (if quoted) are systematic.
\label{tab:cp_charged2}}
\footnotesize
\begin{center}
\begin{tabular}{|l|c|c|c|} 
\hline
{\bf Mode} & {\bf Year} & {\bf Collaboration} & {\boldmath $A^{}_{\CP}$} \\
\hline
{\boldmath $D^+ \to \pi^+\pi^-\pi^+$} &
  2014 & LHCb~\cite{Aaij:2014aa}        &  Model independent technique, no 
evidence for \cpv \\
& 1997 & E791~\cite{Aitala:1996sh}      &  $ -0.017  \pm 0.042  $ (stat.) \\
\hline
{\boldmath $D^+ \to K^-\pi^+\pi^+$} &
   2014 & D0~\cite{Abazov:2014wga}       &  $ -0.0016 \pm 0.0015 \pm 0.0009 $ \\
&  2014 & CLEO~\cite{Bonvicini:2013vxi}  &  $ -0.003  \pm 0.002  \pm 0.004  $ \\
&       & HFLAV average                 &  $ -0.0018 \pm 0.0016            $ \\        
\hline
{\boldmath $D^+ \to K_S\pi^+\pi^0$} &
  2014 & CLEO~\cite{Bonvicini:2013vxi} &  $ -0.001  \pm 0.007 \pm 0.002  $ \\
\hline
{\boldmath $D^+ \to K^+K^-\pi^+$} &
   2014 & CLEO~\cite{Bonvicini:2013vxi}  &  $ -0.001  \pm 0.009  \pm 0.004  $ \\
&  2013 & \babar~\cite{Lees:2013ab}      &  $ +0.0037 \pm 0.0030 \pm 0.0015 $ \\
&  2008 & CLEO~\cite{Rubin:2008zi}     &  Dalitz plot analysis, no evidence 
for \cpv\\
&  2000 & FOCUS~\cite{Link:2000aw}       &  $ +0.006  \pm 0.011  \pm 0.005  $ \\
&  1997 & E791~\cite{Aitala:1996sh}      &  $ -0.014  \pm 0.029  $ (stat.)    \\
&       & HFLAV average                 &  $ +0.0032 \pm 0.0031 $            \\
\hline
{\boldmath $D^+ \to K^-\pi^+\pi^+\pi^0$} &
  2014 & CLEO~\cite{Bonvicini:2013vxi}   &  $ -0.003  \pm 0.006  \pm 0.004  $ \\
\hline
{\boldmath $D^+ \to K_S\pi^+\pi^+\pi^-$} &
  2014 & CLEO~\cite{Bonvicini:2013vxi}   &  $ +0.000  \pm 0.012  \pm 0.003  $ \\
\hline
{\boldmath $D^+ \to K_S K^+\pi^+\pi^-$} &
  2005 & FOCUS~\cite{Link:2005th}  &  $ -0.042  \pm 0.064  \pm 0.022  $ \\
\hline 
\end{tabular}
\end{center} 
\end{table}

\begin{table}[!htb]
\renewcommand{\arraystretch}{1.3}
\caption{\CP\ asymmetries 
$A^{}_{\CP}=[\Gamma(D^0)-\Gamma(\dbar)]/[\Gamma(D^0)+\Gamma(\dbar)]$
for two-body $D^0,\dbar$ decays. In the individual asymmetries listed, 
the first uncertainties are statistical, and the second are systematic, 
unless explicitly stated that they have been combined. The third 
uncertainty in Belle $A^{}_{\CP}(D^0 \to K_S K_S)$ is due to $A^{}_{\CP}(D^0 \to K_S \pi^0)$ used for normalization.
\label{tab:cp_neutral}}
\footnotesize
\begin{center}
\begin{tabular}{|l|c|c|c|} 
\hline
{\bf Mode} & {\bf Year} & {\bf Collaboration} & {\boldmath $A^{}_{\CP}$} \\
\hline
{\boldmath $D^0 \to \pi^+\pi^-$} &
   2017 & LHCb~\cite{Aaij:2016dfb} & $+0.0007 \pm 0.0014 \pm 0.0011 $ \\
& 2012 & CDF~\cite{Aaltonen:2012ab}  & $ +0.0022 \pm 0.0024 \pm 0.0011  $ \\
& 2008 & \babar~\cite{Aubert:2007if} & $ -0.0024 \pm 0.0052 \pm 0.0022  $ \\
& 2012 & Belle~\cite{Staric:2008rx}  & $ +0.0043 \pm 0.0052 \pm 0.0012  $ \\
& 2002 & CLEO~\cite{Csorna:2001ww}   & $ +0.019  \pm 0.032  \pm 0.008   $ \\
& 2000 & FOCUS~\cite{Link:2000aw}    & $ +0.048  \pm 0.039  \pm 0.025   $ \\
& 1998 & E791~\cite{Aitala:1997ff}   & $ -0.049  \pm 0.078  \pm 0.030   $ \\
&      & HFLAV average               & $ +0.0012 \pm 0.0014 $ \\
\hline
{\boldmath $D^0 \to \pi^0\pi^0$} &
  2014 & Belle~\cite{Nisar:2014fkc}     & $ -0.0003 \pm 0.0064 \pm 0.0010  $ \\
& 2001 & CLEO~\cite{Bonvicini:2000qm}  & $ +0.001  \pm 0.048 $ (stat. and syst. 
combined) \\
&      & HFLAV average                & $ -0.0003 \pm 0.0064 $ \\ 
\hline
{\boldmath $D^0 \to K_S\pi^0$} &
  2014 & Belle~\cite{Nisar:2014fkc}     & $ -0.0021 \pm 0.0016 \pm 0.0007 $ \\
& 2001 & CLEO~\cite{Bonvicini:2000qm}  & $ +0.001  \pm 0.013 $ (stat. and syst. 
combined) \\
&      & HFLAV average                & $ -0.0020 \pm 0.0017 $ \\
\hline
{\boldmath $D^0 \to K_S\eta$} &
  2011 & Belle~\cite{Ko:2011ab}        & $ +0.0054 \pm 0.0051 \pm 0.0016 $ \\
\hline
{\boldmath $D^0 \to K_S\eta^\prime$} &
  2011 & Belle~\cite{Ko:2011ab}        & $ +0.0098 \pm 0.0067 \pm 0.0014 $ \\  
\hline
{\boldmath $D^0 \to K_S K_S$} &
2018 & LHCb~\cite{Aaij:2018jud} & $+0.023 \pm 0.028 \pm 0.009$ \\
& 2017 & Belle~\cite{Dash:2017heu} & $-0.0002 \pm 0.0153 \pm 0.0002 \pm 0.0017$ \\
& 2001 & CLEO~\cite{Bonvicini:2000qm}   & $ -0.23  \pm 0.19  $ (stat. and syst. 
combined) \\
&      & HFLAV average                 & $ +0.004 \pm 0.014           $ \\ 
\hline
{\boldmath $D^0 \to K^-\pi^+$} &
 2014 & CLEO~\cite{Bonvicini:2013vxi} & $ +0.003  \pm 0.003  \pm 0.006 $  \\
\hline
{\boldmath $D^0 \to K^+K^-$} &
2017 & LHCb~\cite{Aaij:2016dfb} & $+0.0004 \pm 0.0012 \pm 0.0010 $ \\
& 2012 & CDF~\cite{Aaltonen:2012ab}  & $ -0.0024 \pm 0.0022 \pm 0.0009 $ \\
& 2008 & \babar~\cite{Aubert:2007if} & $ +0.0000 \pm 0.0034 \pm 0.0013 $ \\
& 2012 & Belle~\cite{Staric:2008rx}  & $ -0.0043 \pm 0.0030 \pm 0.0011 $ \\ 
& 2002 & CLEO~\cite{Csorna:2001ww}   & $ +0.000  \pm 0.022  \pm 0.008  $ \\
& 2000 & FOCUS~\cite{Link:2000aw}    & $ -0.001  \pm 0.022  \pm 0.015  $ \\
& 1998 & E791~\cite{Aitala:1997ff}   & $ -0.010  \pm 0.049  \pm 0.012  $ \\
&      & HFLAV average              & $ -0.0009 \pm 0.0011            $ \\
\hline
\end{tabular}
\end{center} 
\end{table}

\begin{table}[!htb]
\renewcommand{\arraystretch}{1.3}
\caption{\CP\ asymmetries 
$A^{}_{\CP}=[\Gamma(D^0)-\Gamma(\dbar)]/[\Gamma(D^0)+\Gamma(\dbar)]$
for three- and four-body $D^0,\dbar$ decays. In the individual 
asymmetries listed, the first uncertainties are statistical, and the 
second are systematic, unless only the former is given, or explicitly 
stated that these two have been combined. 
The Belle study of $D^0 \to K^+K^-\pi^+\pi^-$~\cite{Kim:2018mtf} 
employs a T-odd method for P-even variables, which corresponds to 
measuring a global $A_{\CP}$.
\label{tab:cp_neutral2}}
\footnotesize
\begin{center}
\begin{tabular}{|l|c|c|c|} 
\hline
{\bf Mode} & {\bf Year} & {\bf Collaboration} & {\boldmath $A^{}_{\CP}$} \\
\hline
{\boldmath $D^0 \to \pi^+\pi^-\pi^0$} &
   2015 & LHCb~\cite{Aaij:2014afa}          & Model-independent method, no evidence for 
\cpv \\
&  2008 & \babar~\cite{Aubert:2008yd}       & $ +0.0031 \pm  0.0041 \pm  0.0017$ \\
&  2008 & Belle~\cite{Arinstein:2008zh}     & $ +0.0043 \pm  0.0130 $ (stat. and 
syst. combined) \\
&  2005 & CLEO~\cite{CroninHennessy:2005sy} & $ +0.01^{+0.09}_{-0.07} \pm  0.05 $ \\
&       & HFLAV average                    & $ +0.0032 \pm 0.0042 $ \\
\hline
{\boldmath $D^0 \to K^-\pi^+\pi^0$} &
  2014 & CLEO~\cite{Bonvicini:2013vxi} & $ +0.001  \pm 0.003 \pm 0.004   $ \\
\hline   
{\boldmath $D^0 \to K^+\pi^-\pi^0$} &
  2005 & Belle~\cite{Tian:2005ik}        & $ -0.006  \pm 0.053  $ (stat.) \\
& 2001 & CLEO~\cite{Brandenburg:2001ze}  & $ +0.09^{+0.25}_{-0.22}  $ (stat.) \\
&      & HFLAV average                  & $ -0.0014 \pm 0.0517 $ \\
\hline
{\boldmath $D^0 \to K_S\pi^+\pi^-$} &
  2012 & CDF~\cite{Aaltonen:2012ac}  & $ -0.0005 \pm 0.0057 \pm 0.0054 $ \\
& 2004 & CLEO~\cite{Asner:2003uz}    & $ -0.009  \pm 0.021^{+0.016}_{-0.057} $ \\
&      & HFLAV average              & $ -0.0008 \pm 0.0077 $ \\
\hline
{\boldmath $D^0 \to K_S K^-\pi^+$} &
   2016 & LHCb~\cite{Aaij:2015lsa} &  Amplitude analysis, no evidence for 
\cpv \\ 
\hline
{\boldmath $D^0 \to K_S K^+\pi^-$} &
   2016 & LHCb~\cite{Aaij:2015lsa} &  Amplitude analysis, no evidence for 
\cpv \\ 
\hline
{\boldmath $D^0 \to K^+ K^-\pi^0$} &
   2008 & \babar~\cite{Aubert:2008yd} & $ -0.0100 \pm  0.0167 \pm  0.0025$ \\ 
\hline
{\boldmath $D^0 \to \pi^-\pi^-\pi^+\pi^+$} &
  2013 & LHCb~\cite{Aaij:2013aa}  & Model-independent method, no evidence for 
\cpv \\
\hline
{\boldmath $D^0 \to K^-\pi^+\pi^+\pi^-$} &
  2014 & CLEO~\cite{Bonvicini:2013vxi} & $ +0.002  \pm 0.003 \pm 0.004   $ \\
\hline 
{\boldmath $D^0 \to K^+\pi^-\pi^+\pi^-$} &
  2005 & Belle~\cite{Tian:2005ik} & $ -0.018  \pm 0.044  $ (stat.) \\
\hline
{\boldmath $D^0 \to K^+K^-\pi^+\pi^-$} &
 2018 & Belle~\cite{Kim:2018mtf} & $ +0.0034 \pm 0.0036 \pm 0.0006 $  \\
& 2018 & LHCb~\cite{Aaij:2018nis} & Amplitude analysis, no evidence for 
\cpv \\
& 2013 & LHCb~\cite{Aaij:2013aa}   & Model-independent method, no evidence for 
\cpv \\
& 2012 & CLEO~\cite{Artuso:2012df} & Amplitude analysis, no evidence for 
\cpv \\  
& 2005 & FOCUS~\cite{Link:2005th}  & $ -0.082  \pm 0.056  \pm 0.047  $ \\
&      & HFLAV average              & $ +0.0032 \pm 0.0036 $ \\
\hline  \hline                 
{\boldmath $D^0 \to \bar{K}^{*0} [\to K^-\pi^+] \gamma$} &
  2016 & Belle~\cite{Abdesselam:2016yvr} & $  -0.003 \pm 0.020 \pm 0.000  $ \\  
\hline
{\boldmath $D^0 \to \phi [\to K^+K^-] \gamma$} &
2016 & Belle~\cite{Abdesselam:2016yvr} & $ -0.094 \pm 0.066 \pm 0.001$ \\
\hline
{\boldmath $D^0 \to \rho^{0} [\to \pi^+ \pi^-] \gamma$} &
2016 & Belle~\cite{Abdesselam:2016yvr} & $ +0.056 \pm 0.152 \pm 0.006$ \\
\hline
{\boldmath $D^0 \to K^+ K^- \mu^+ \mu^-$} &
2018 & LHCb~\cite{Aaij:2018fpa} & $ +0.00 \pm 0.11 \pm 0.02$ \\
\hline
{\boldmath $D^0 \to \pi^+ \pi^- \mu^+ \mu^-$} &
2018 & LHCb~\cite{Aaij:2018fpa} & $ +0.049 \pm 0.038 \pm 0.007$ \\
\hline
\end{tabular}
\end{center} 
\end{table}

\begin{table}[!htb]
\renewcommand{\arraystretch}{1.4}
\caption{\CP\ asymmetries 
$A^{}_{\CP}=[\Gamma(D_s^+)-\Gamma(D_s^-)]/[\Gamma(D_s^+)+\Gamma(D_s^-)]$ for $D_s^\pm$ decays. In the individual asymmetries listed, the first
uncertainties are statistical, and the second systematic, whereas the
third uncertainty in $A^{}_{\CP}(D_s^+ \to \pi^+ \eta^\prime)$ from 
LHCb is due to $A^{}_{\CP}(D^+ \to \pi^+ \phi)$ used for calibration. 
\label{tab:cp_ds}}
\footnotesize
\begin{center}
\begin{tabular}{|l|c|c|c|} 
\hline
{\bf Mode} & {\bf Year} & {\bf Collaboration} & {\boldmath $A^{}_{\CP}$} \\
\hline
{\boldmath $D_s^+ \to \mu^+ \nu$} &
  2009 & CLEO~\cite{Alexander:2009ux} & $ +0.048 \pm 0.061 $ \\
\hline
{\boldmath $D_s^+ \to \pi^+ \eta$} &
  2013 & CLEO~\cite{Onyisi:2013bjt}     & $ +0.011 \pm 0.030 \pm 0.008 $ \\
\hline
{\boldmath $D_s^+ \to \pi^+ \eta^\prime$} &
    2017 & LHCb~\cite{Aaij:2017eux} & $-0.0082 \pm 0.0036 \pm 0.0022 \pm 0.002$ \\
&  2013 & CLEO~\cite{Onyisi:2013bjt}     & $ -0.022 \pm 0.022 \pm 0.006 $ \\
&      & HFLAV average             & $ -0.0088 \pm 0.0049 $            \\
\hline
{\boldmath $D_s^+ \to K_S\pi^+$}  &
  2013 & \babar~\cite{Lees:2013aa}  & $ +0.006  \pm 0.020  \pm 0.003  $ \\  
& 2010 & Belle~\cite{Ko:2010ng}     & $ +0.0545 \pm 0.0250 \pm 0.0033 $ \\
& 2010 & CLEO~\cite{Mendez:2009aa}  & $ +0.163  \pm 0.073  \pm 0.003  $ \\
&      & HFLAV average             & $ +0.0311 \pm 0.0154 $            \\
\hline
{\boldmath $D_s^+ \to (\kbar/K^0)\pi^+$}  &
  2014 & LHCb~\cite{Aaij:2014ac}    & $ +0.0038 \pm 0.0046 \pm 0.0017 $ \\
& 2013 & \babar~\cite{Lees:2013aa}  & $ +0.003  \pm 0.020  \pm 0.003  $ \\  
&      & HFLAV average             & $ +0.0038 \pm 0.0048 $            \\
\hline
{\boldmath $D_s^+ \to K_S K^+$}   &
  2013 & CLEO~\cite{Onyisi:2013bjt}  & $ +0.026  \pm 0.015  \pm 0.006  $ \\
& 2013 & \babar~\cite{Lees:2013aa}  & $ -0.0005 \pm 0.0023 \pm 0.0024 $ \\  
& 2010 & Belle~\cite{Ko:2010ng}     & $ +0.0012 \pm 0.0036 \pm 0.0022 $ \\
&      & HFLAV average             & $ +0.0008 \pm 0.0026 $            \\
\hline
{\boldmath $D_s^+ \to K^+ \pi^0$}   &
  2010 & CLEO~\cite{Mendez:2009aa} &  $ +0.266 \pm 0.228 \pm 0.009 $ \\
\hline
{\boldmath $D_s^+ \to K^+ \eta$}    &
  2010 & CLEO~\cite{Mendez:2009aa} &  $ +0.093 \pm 0.152 \pm 0.009 $ \\
\hline
{\boldmath $D_s^+ \to K^+ \eta^\prime$}  &
  2010 & CLEO~\cite{Mendez:2009aa}      &  $ +0.060 \pm 0.189 \pm 0.009 $ \\
\hline
{\boldmath $D_s^+ \to \pi^+ \pi^+ \pi^-$} &
  2013 & CLEO~\cite{Onyisi:2013bjt}        & $ -0.007 \pm 0.030 \pm 0.006 $ \\
\hline
{\boldmath $D_s^+ \to \pi^+ \pi^0 \eta$}  &
  2013 & CLEO~\cite{Onyisi:2013bjt}        & $ -0.005 \pm 0.039 \pm 0.020 $ \\
\hline
{\boldmath $D_s^+ \to \pi^+ \pi^0 \eta^\prime$} &
  2013 & CLEO~\cite{Onyisi:2013bjt}        & $ -0.004 \pm 0.074 \pm 0.019 $ \\
\hline
{\boldmath $D_s^+ \to K_S K^+ \pi^0$}   &
  2013 & CLEO~\cite{Onyisi:2013bjt}        & $ -0.016 \pm 0.060 \pm 0.011 $ \\
\hline
{\boldmath $D_s^+ \to K_S K_S \pi^+$} &
  2013 & CLEO~\cite{Onyisi:2013bjt}        & $ +0.031 \pm 0.052 \pm 0.006 $ \\
\hline
{\boldmath $D_s^+ \to K^+ \pi^+ \pi^-$} &
  2013 & CLEO~\cite{Onyisi:2013bjt}        & $ +0.045 \pm 0.048 \pm 0.006 $ \\
\hline
{\boldmath $D_s^+ \to K^+ K^- \pi^+$} &
  2013 & CLEO~\cite{Onyisi:2013bjt}        & $ -0.005 \pm 0.008 \pm 0.004 $ \\
\hline
{\boldmath $D_s^+ \to K_S K^- \pi^+\pi^+$} &
  2013 & CLEO~\cite{Onyisi:2013bjt}        & $ +0.041 \pm 0.027 \pm 0.009 $ \\
\hline
{\boldmath $D_s^+ \to K_S K^+ \pi^+\pi^-$} &
  2013 & CLEO~\cite{Onyisi:2013bjt}        & $ -0.057 \pm 0.053 \pm 0.009 $ \\
\hline
{\boldmath $D_s^+ \to K^+ K^- \pi^+\pi^0$} &
  2013 & CLEO~\cite{Onyisi:2013bjt}        & $ +0.000 \pm 0.027 \pm 0.012 $ \\ 
\hline 
\end{tabular}
\end{center} 
\end{table}

\clearpage
\mysubsection{$T$-odd asymmetries}\label{sec:todd_asym}
                                               
Measuring $T$-odd asymmetries provides a complementary way 
to search for \CP\ violation in the charm sector; 
it exploits ${\CP}T$ invariance. 
$T$-odd asymmetries are measured using triple-product correlations
of the form $\vec{a}\cdot(\vec{b}\times\vec{c})$, where $a$, $b$, 
and $c$ are spins or momenta; this combination is odd under time 
reversal~($T$). If a triple product is formed using {\it both\/} 
spin and momenta, \ie, 
\begin{equation*}
\vec{s_1} \cdot(\vec{p_2} \times \vec{p}_3),
\end{equation*}
it can be even under $P$-conjugation. 
However, if only momenta are used, \ie, 
\begin{equation*}
\vec{p_1} \cdot(\vec{p_2} \times \vec{p}_3),
\end{equation*}
it is odd under $P$-conjugation. 
Thus, in this case the $T$-odd method becomes $P$-odd and 
allows one to probe \CP\ violation occurring via $P$-violation.
This type of \cpv, arising in $P$-odd amplitudes, can be 
studied in decays of mesons into final states with at least 
four spinless particles. Two- and three-body hadronic decays 
of charm mesons to spinless particles involve only $P$-even
amplitudes\footnote{$P$-even 
amplitudes are accessed with $P$-even variables, like invariant 
masses or helicity angles.}, for which \CP\ violation can arise 
only through $C$-violation. 

\vspace{0.2cm}
Taking as an example the decay mode $D^0 \to K^+K^-\pi^+\pi^-$, 
involving spinless particles only, one forms a triple-product 
correlation using momenta of the final-state particles in 
the $D^0$ center-of-mass frame.\footnote{For momentum-only 
triple products, at least four-daughter final states are 
required to give a nonzero correlation, as only three out 
of four momenta are independent. For three-body decays, the 
daughters are in a plane and the triple product is zero.}
Defining the $T$-odd (and $P$-odd) correlation for $D^0$
\begin{equation} 
C_T \equiv \vec{p}^{}_{K^+}\cdot(\vec{p}_{\pi^+}\times \vec{p}_{\pi^-}),
\end{equation}  
and the corresponding quantity for $\dbar$
\begin{equation}
\overline{C}_T \equiv 
      \vec{p}^{}_{K^-}\cdot(\vec{p}_{\pi^-}\times \vec{p}_{\pi^+}),
\end{equation}      
one can construct the asymmetry for the $D^0$ decays as
\begin{equation}
 A_{T} = 
    \frac{\Gamma(C_T>0)-\Gamma(C_T<0)}{\Gamma(C_T>0)+\Gamma(C_T<0)},
\end{equation}
while for their \CP-conjugate decays as 
\begin{equation}
\overline{A}_{T} = 
   \frac{\Gamma(-\overline{C}_T>0)-\Gamma(-\overline{C}_T<0)}
                        {\Gamma(-\overline{C}_T>0)+\Gamma(-\overline{C}_T<0)}.
\end{equation} 
In these expressions, $\Gamma$ represents a partial width, 
and the following applies: 
\begin{equation}
P(C_T)=-C_T, \quad C(C_T)=\overline{C_T}, \quad C\!P(A_T) = \overline{A}_T.
\label{eq:parities}
\end{equation}
The asymmetries $A_T$ and $\overline{A}_T$ depend on angular 
distributions of the daughter particles and may be nonzero due to 
final-state interactions or $P$-violation in weak decays. 
Given Eq.~(\ref{eq:parities}), one can construct the \CP-violating, 
\ie\ \CP-odd (and $P$-odd, $T$-odd) asymmetry 
\begin{equation}
{\cal A}^{}_{T} \equiv \frac{A_{T}-\overline{A}_{T}}{2};
\label{eqn:atodd}
\end{equation}
where a nonzero value indicates \CP\ violation
(see Refs.~\cite{Golowich:1988ig,Bigi:2001sg,Bensalem:2002ys,
Bensalem:2000hq,Bensalem:2002pz,Gronau:2011cf}).
This asymmetry is referred to in the literature by 
several names: $A^{}_{T\,{\rm viol}}$, $a^P_{\CP}$, and $a^{T-{\rm odd}}_{\CP}$.

\vspace{0.2cm}
Values of ${\cal A}^{}_{T}$ for $D^+$, $D^+_s$, and
$D^0$ decay modes are listed in Table~\ref{tab:t_odd}. 
The first measurements were made by FOCUS, and subsequent 
\babar\ measurements reached a sensitivity of $\sim 1\%$. 
Currently the best sensitivity is from LHCb. 
However, despite relatively high precision ($<1\%$), 
there is no evidence for \CP\ violation. 
\begin{table}[htb]
\renewcommand{\arraystretch}{1.4}
\caption{Measurements of the $T$-odd \CP\ asymmetry 
${\cal A}^{}_{T} = (A_{T}-\overline{A}_{T})/2$.
\label{tab:t_odd}}
\footnotesize
\begin{center}
\begin{tabular}{|l|c|c|c|} 
\hline
{\bf Mode} & {\bf Year} & {\bf Collaboration} & {\boldmath ${\cal A}^{}_{T}$} \\
\hline
{\boldmath $D^0 \to K^+K^-\pi^+\pi^-$} &
   2018 & Belle~\cite{Kim:2018mtf}     &  $ +0.0052 \pm 0.0037 \pm 0.0007 $ \\
&  2014 & LHCb~\cite{Aaij:2014qwa}     &  $ +0.0018 \pm 0.0029 \pm 0.0004 $ \\
&  2010 & \babar~\cite{Sanchez:2010xj} &  $ +0.0010 \pm 0.0051 \pm 0.0044 $ \\
&  2005 & FOCUS~\cite{Link:2005th}     &  $ +0.010  \pm 0.057  \pm 0.037  $ \\
&       & HFLAV average               &  $ +0.0035 \pm 0.0021            $ \\  
\hline
{\boldmath $D^0 \to K_S \pi^+\pi^-\pi^0$} &
   2017 & Belle~\cite{Prasanth:2017beu}     &  $ -0.00028 \pm 0.00138 ^{+0.00023}_{-0.00076} $ \\
\hline
{\boldmath $D^+ \to K_SK^+\pi^+\pi^-$} &
  2011 & \babar~\cite{Lees:2011ab} &  $ -0.0120 \pm 0.0100 \pm 0.0046 $ \\
& 2005 & FOCUS~\cite{Link:2005th}  &  $ +0.023  \pm 0.062  \pm 0.022  $ \\
&      & HFLAV average            &  $ -0.0110 \pm 0.0109            $ \\
\hline
{\boldmath $D^+_s \to K_SK^+\pi^+\pi^-$} &
  2011 & \babar~\cite{Lees:2011ab} &  $ -0.0136 \pm 0.0077 \pm 0.0034 $ \\
& 2005 & FOCUS~\cite{Link:2005th}  &  $ -0.036  \pm 0.067  \pm 0.023  $ \\
&      & HFLAV average            &  $ -0.0139 \pm 0.0084            $ \\
\hline                    
\end{tabular}
\end{center} 
\end{table}

All $P$-even contributions contributing to ${\cal A}^{}_{T}$ cancel out 
in the difference; thus it is only sensitive to $P$-odd amplitudes or 
interference between $P$-odd and $P$-even ones. 
The cancellation typically applies also to detection asymmetries 
and, at LHCb, the production asymmetry, and
this is a significant advantage of the $T$-odd method.
Another way to probe $P$-odd amplitudes is through amplitude analysis 
using $P$-odd variables. One example is $\sin{\Phi}$, 
where $\Phi$ is the angle between the $K^+K^-$ decay plane and
the $\pi^+\pi^-$ decay plane in the decay
$D^0 \to K^+K^-\pi^+\pi^-$~\cite{Aaij:2018nis}.
It can be shown that $\sin{\Phi}$ is proportional to the triple product. 
The model-independent technique used for 
$D^0 \to \pi^+\pi^-\pi^+\pi^-$decays~\cite{Aaij:2013aa}
has been carried out separately for $P$-odd and $P$-even contributions, 
separated out using a triple product. The largest 
$P$-odd amplitudes in four-body decays of charm mesons 
are $D \to [VV]_{L=1}$, \ie, a final state with 
two vector mesons in a $P$-wave state. However, these
amplitudes are quite suppressed ($<10\%$)~\cite{Aaij:2018nis,Aaij:2017kbo}.

Decays of charm baryons also offer access to $P$-odd amplitudes, \eg, 
$\Lambda_c^+$ decays with a weakly-decaying baryon in the final state
such as $\Lambda_c^+ \to \Lambda\,\pi^+$.
Moreover, for polarized charm baryons, \eg, $\Lambda_c$ 
produced weakly in $\Lambda_b$ decays, one can build a triple 
product using the $\Lambda_c$ spin.
Recently, the topic of symmetries has been revisited (see 
Refs.~\cite{Bevan:2015xra,Durieux:2015zwa}), with the suggestion 
to exploit additional asymmetries constructed from triple products
in multi-body decays.

\clearpage
\subsection{Interplay of direct and indirect \cp\ violation}
\label{sec:charm:cpvdir}

In decays of $D^0$ mesons, \cp\ asymmetry measurements have contributions from 
both direct and indirect \cp\ violation as discussed in Sec.~\ref{sec:charm:mixcpv}.
The contribution from indirect \cp\ violation depends on the decay-time distribution 
of the data sample~\cite{Kagan:2009gb}. This section describes a combination of 
measurements that allows the determination of the individual contributions of the 
two types of \cp\ violation.
At the same time, the level of agreement for a no-\cp-violation hypothesis is 
tested. The observables are: 
\begin{equation}
A_{\Gamma} \equiv \frac{\tau(\dbar \ra h^+ h^-) - \tau(D^0 \ra h^+ h^- )}
{\tau(\dbar \ra h^+ h^-) + \tau(D^0 \ra h^+ h^- )},
\end{equation}
where $h^+ h^-$ can be $K^+ K^-$ or $\pi^+\pi^-$, and 
\begin{equation}
\Delta A_{\CP}   \equiv A_{\CP}(K^+K^-) - A_{\CP}(\pi^+\pi^-),
\end{equation}
where $A_{\CP}$ are time-integrated \cp\ asymmetries. The underlying 
theoretical parameters are: 
\begin{eqnarray}
a_{\CP}^{\rm dir} & \equiv & 
\frac{|{\cal A}_{D^0\rightarrow f} |^2 - |{\cal A}_{\dbar\rightarrow f} |^2} 
{|{\cal A}_{D^0\rightarrow f} |^2 + |{\cal A}_{\dbar\rightarrow f} |^2} ,\nonumber\\ 
a_{\CP}^{\rm ind}  & \equiv & \frac{1}{2} 
\left[ \left(\left|\frac{q}{p}\right| + \left|\frac{p}{q}\right|\right) x \sin \phi - 
\left(\left|\frac{q}{p}\right| - \left|\frac{p}{q}\right|\right) y \cos \phi \right] ,
\end{eqnarray}
where ${\cal A}_{D\rightarrow f}$ is the amplitude for $D\ra f$~\cite{Grossman:2006jg}. 
We use the relations~\cite{Gersabeck:2011xj}
\begin{eqnarray}
A_{\Gamma} & = & - a_{\CP}^{\rm ind} - a_{\CP}^{\rm dir} y_{\CP},\label{eqn:charm_MG_AGamma}\\ 
\Delta A_{\CP} & = &  \Delta a_{\CP}^{\rm dir} \left(1 + y_{\CP} 
\frac{\overline{\langle t\rangle}}{\tau} \right)   +   
   a_{\CP}^{\rm ind} \frac{\Delta\langle t\rangle}{\tau}   +   
  \overline{a_{\CP}^{\rm dir}} y_{\CP} \frac{\Delta\langle t\rangle}{\tau},\nonumber\\ 
& \approx & \Delta a_{\CP}^{\rm dir} \left(1 + y_{\CP} 
\frac{\overline{\langle t\rangle}}{\tau} \right)   +   a_{\CP}^{\rm ind} 
\frac{\Delta\langle t\rangle}{\tau}.\label{eqn:charm_MG_DACP}
\end{eqnarray}
between the observables and the underlying parameters.
Equation~(\ref{eqn:charm_MG_AGamma}) constrains mostly indirect \cp\ violation, and the 
direct \cp\ violation contribution can differ for different final states. 
In Eq.~(\ref{eqn:charm_MG_DACP}), $\langle t\rangle/\tau$ denotes the mean decay 
time in units of the $D^0$ lifetime; $\Delta X$ denotes the difference 
in quantity $X$ between $K^+K^-$ and $\pi^+\pi^-$ final states; and $\overline{X}$ 
denotes the average for quantity $X$. 
We neglect the last term in this relation as all three factors are 
$\mathcal{O}(10^{-2})$ or smaller, and thus this term is negligible 
with respect to the other two terms. 
Note that $\Delta\langle t\rangle/\tau \ll\langle t\rangle/\tau$, and 
it is expected that $|a_{\CP}^{\rm dir}| < |\Delta a_{\CP}^{\rm dir}|$ 
because $a_{\CP}^{\rm dir}(K^+K^-)$ and $a_{\CP}^{\rm dir}(\pi^+\pi^-)$ 
are expected to have opposite signs in the Standard Model~\cite{Grossman:2006jg}. 

A $\chi^2$ fit is performed in the plane $\Delta a_{\CP}^{\rm dir}$ 
vs. $a_{\CP}^{\rm ind}$. 
For the \babar result, the difference of the quoted values for 
$A_{\CP}(K^+K^-)$ and $A_{\CP}(\pi^+\pi^-)$ is calculated, 
adding all uncertainties in quadrature. 
This may overestimate the systematic uncertainty for the difference 
as it neglects correlated uncertainties; however, the result is conservative 
and the effect is small as all measurements are statistically limited. 
For all measurements, statistical and systematic uncertainties are added 
in quadrature when calculating the $\chi^2$. 
We use the HFLAV average value $y_{\CP} = (0.715 \pm 0.111)\%$ 
(see Sec.~\ref{sec:charm:mixcpv}) and the measurements listed in 
Table~\ref{tab:charm:dir_indir_comb}. In this fit, $A_\Gamma(KK)$ and $A_\Gamma(\pi\pi)$ are assumed to be identical.
This assumption, which is expected in the SM, is supported by all measurements to date.
A significant relative shift due to final-state dependent $A_\Gamma$ values between $\Delta A_{\CP}$ measurements with different mean decay times is excluded by these measurements.

\begin{table}
\centering 
\caption{Inputs to the fit for direct and indirect \cp\ violation. 
The first uncertainty listed is statistical and the second is systematic.}
\label{tab:charm:dir_indir_comb}
\vspace{3pt}
\begin{tabular}{ll|ccccc}
\hline \hline
Year & 	Experiment	& Results
& $\Delta \langle t\rangle/\tau$ & $\langle t\rangle/\tau$ & Reference\\
\hline
2012	& \babar	& $A_\Gamma = (+0.09 \pm 0.26 \pm 0.06 )\%$ &	-&	-&	 
\cite{Lees:2012qh}\\
2016	& LHCb	prompt & $A_\Gamma(KK) = (-0.030 \pm 0.032 \pm 0.010 )\%$ &	-&	-&	 
\cite{Aaij:2017idz}\\
    	&     	& $A_\Gamma(\pi\pi) = (+0.046 \pm 0.058 \pm 0.012 )\%$ &  -&	-&	 
                   \\
2014	& CDF & $A_\Gamma = (-0.12 \pm 0.12 )\%$ &	-&	-&	 
\cite{Aaltonen:2014efa}\\
2015	& LHCb SL & $A_\Gamma = (-0.125 \pm 0.073 )\%$ &	-&	-&	 
\cite{Aaij:2015yda}\\
2015	& Belle	& $A_\Gamma = (-0.03 \pm 0.20 \pm 0.07 )\%$ &	-&	-&	 
\cite{Staric:2015sta}\\
2008	& \babar	& $A_{\CP}(KK) = (+0.00 \pm 0.34 \pm 0.13 )\%$&&&\\ 
& & $A_{\CP}(\pi\pi) = (-0.24 \pm 0.52 \pm 0.22 )\%$ &	$0.00$ &	
$1.00$ &	 \cite{Aubert:2007if}\\
2012	& CDF	 & $\Delta A_{\CP} = (-0.62 \pm 0.21 \pm 0.10 )\%$ &	
$0.25$ &	$2.58$ &	 \cite{Collaboration:2012qw}\\
2014	& LHCb	SL & $\Delta A_{\CP} = (+0.14 \pm 0.16 \pm 0.08 )\%$ &	
$0.01$ &	$1.07$ &	 \cite{Aaij:2014gsa}\\
2016	& LHCb	prompt & $\Delta A_{\CP} = (-0.10 \pm 0.08 \pm 0.03 )\%$ &	
$0.12$ &	$2.10$ &	 \cite{Aaij:2016cfh}\\
2019	& LHCb	SL2 & $\Delta A_{\CP} = (-0.09 \pm 0.08 \pm 0.05 )\%$ &	
$0.00$ &	$1.21$ &	 \cite{Aaij:2019kcg}\\
2019	& LHCb	prompt2 & $\Delta A_{\CP} = (-0.18 \pm 0.03 \pm 0.09 )\%$ &	
$0.13$ &	$1.74$ &	 \cite{Aaij:2019kcg}\\
\hline
\end{tabular}
\end{table}

The combination plot (see Fig.~\ref{fig:charm:dir_indir_comb}) shows the measurements listed in 
Table~\ref{tab:charm:dir_indir_comb} for
$\Delta A_{\CP}$ and $A_\Gamma$.
\begin{figure}
\begin{center}
\includegraphics[width=0.90\textwidth]{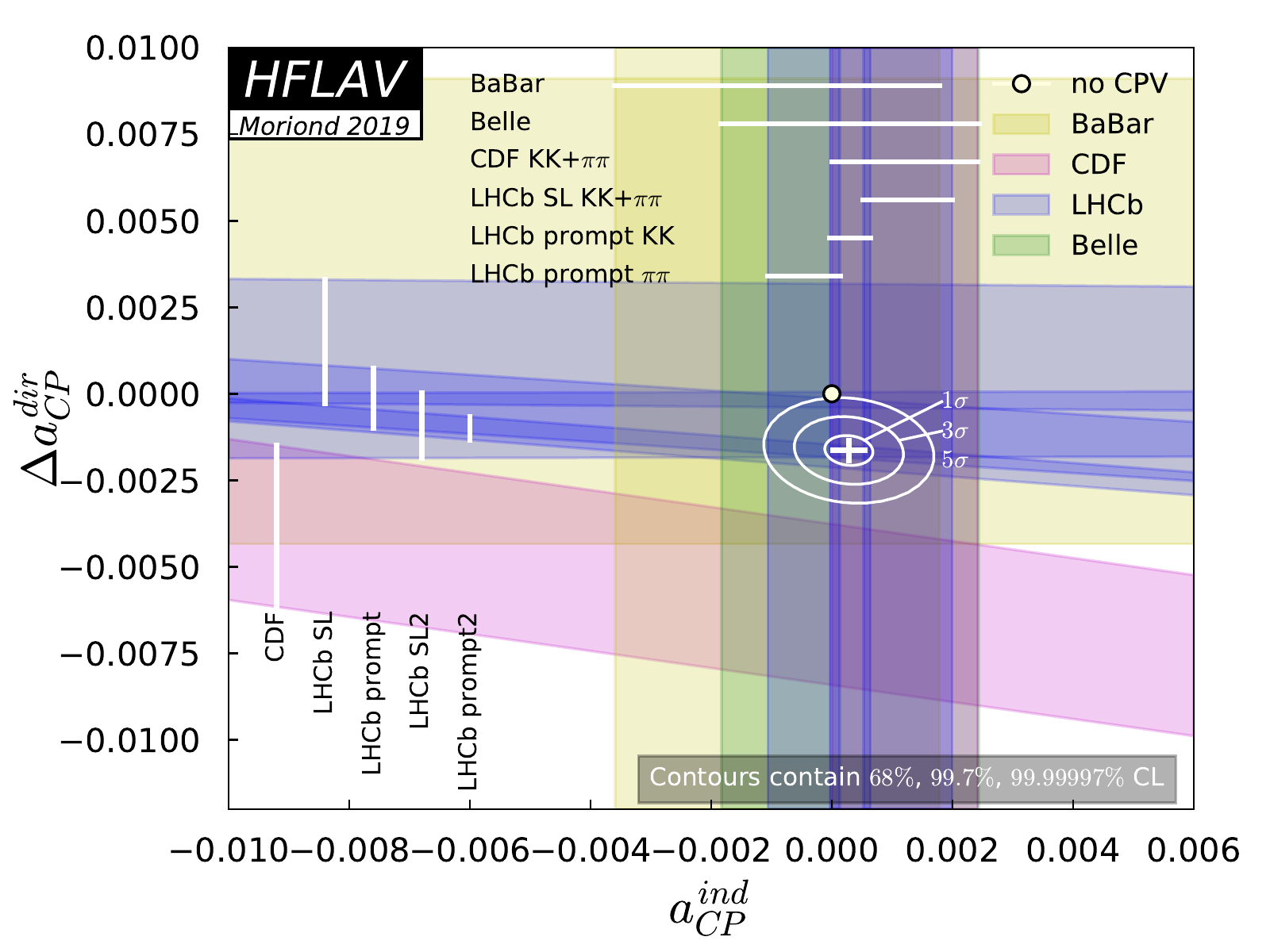}
\caption{Plot of all data and the fit result. Individual 
measurements are plotted as bands showing their $\pm1\sigma$ range. 
The no-\cpv\ point (0,0) is shown as a filled circle, and the best 
fit value is indicated by a cross showing the one-dimensional uncertainties. 
Two-dimensional $68\%$ C.L., $99.7\%$ C.L., and $99.99997\%$ C.L.\ regions are 
plotted as ellipses. }
\label{fig:charm:dir_indir_comb}
\end{center}
\end{figure}
From the fit, the change in $\chi^2$ from the minimum value for the no-\cpv\ 
point (0,0) is $33.5$, which corresponds to a C.L.\ of $5.4\times 10^{-8}$ for 
two degrees of freedom or $5.4$ standard deviations. The central
values and $\pm1\sigma$ uncertainties for the individual parameters are
\begin{eqnarray}
a_{\CP}^{\rm ind} & = & (+0.028 \pm 0.026 )\% \nonumber\\
\Delta a_{\CP}^{\rm dir} & = & (-0.164 \pm 0.028 )\%.
\end{eqnarray}
This constitutes the first time that the average rejects the hypothesis of CP symmetry with a significance exceeding $5\sigma$.
The average clearly points at CP violation in the decays to two charged hadrons.
\clearpage

\subsection{Semileptonic decays}
\label{sec:charm:semileptonic}

\subsubsection{Introduction}

Semileptonic decays of $D$ mesons involve the interaction of a leptonic
current with a hadronic current. The latter is nonperturbative
and cannot be calculated from first principles; thus it is usually
parameterized in terms of form factors. The transition matrix element 
is written
\begin{eqnarray}
  {\cal M} & = & -i\,\frac{G_F}{\sqrt{2}}\,V^{}_{cq}\,L^\mu H_\mu\,,
  \label{Melem}
\end{eqnarray}
where $G_F$ is the Fermi constant and $V^{}_{cq}$ is a CKM matrix element.
The leptonic current $L^\mu$ is evaluated directly from the lepton spinors 
and has a simple structure; this allows one to extract information about 
the form factors (in $H^{}_\mu$) from data on semileptonic decays~\cite{Becher:2005bg}.  
Conversely, because there are no strong final-state interactions between the
leptonic and hadronic systems, semileptonic decays for which the form 
factors can be calculated allow one to 
determine~$|V^{}_{cq}|$~\cite{Kobayashi:1973fv}.

\subsubsection{$D\ra P \overline \ell \nu_{\ell}$ decays}

When the final state hadron is a pseudoscalar, the hadronic 
current is given by
\begin{eqnarray}
\hspace{-1cm}
H_\mu & = & \left< P(p) | \bar{q}\gamma_\mu c | D(p') \right> \ =\  
f_+(q^2)\left[ (p' + p)_\mu -\frac{m_D^2-m_P^2}{q^2}q_\mu\right] + 
 f_0(q^2)\frac{m_D^2-m_P^2}{q^2}q_\mu\,,
\label{eq:hadronic}
\end{eqnarray}
where $m_D$ and $p'$ are the mass and four momentum of the 
parent $D$ meson, $m_P$ and $p$ are those of the daughter meson, 
$f_+(q^2)$ and $f_0(q^2)$ are form factors, and $q = p' - p$.  
Kinematics require that $f_+(0) = f_0(0)$.
The contraction $q_\mu L^\mu$ results in terms proportional 
to $m^{}_\ell$\cite{Gilman:1989uy}, and thus for $\ell=e $
the terms proportional to $q_\mu$ in Eq.~(\ref{eq:hadronic}) are negligible and only the $f_+(q^2)$ vector form factor 
is relevant. The corresponding differential partial width is
\begin{eqnarray}
\frac{d\Gamma(D \to P e \nu_e)}{dq^2\, d\cos\theta_e} & = & 
   \frac{G_F^2|V_{cq}|^2}{32\pi^3} p^{*\,3}|f_{+}(q^2)|^2\sin\theta^2_e\,,
\label{eq:dGamma}
\end{eqnarray}
where ${p^*}$ is the magnitude of the momentum of the final state hadron
in the $D$ rest frame, and $\theta_e$ is the angle of the electron in the 
$e\nu$ rest frame with respect to the direction of the pseudoscalar meson 
in the $D$ rest frame.

\subsubsection{Form factor parameterizations} 

The form factor is traditionally parameterized with an explicit pole 
and a sum of effective poles:
\begin{eqnarray}
f_+(q^2) & = & \frac{f_+(0)}{(1-\alpha)}\Bigg [
\left(\frac{1}{1- q^2/m^2_{\rm pole}}\right)\ +\ 
\sum_{k=1}^{N}\frac{\rho_k}{1- q^2/(\gamma_k\,m^2_{\rm pole})}\Bigg ]\,,
\label{eqn:expansion}
\end{eqnarray}
where $\rho_k$ and $\gamma_k$ are expansion parameters and $\alpha$ is 
a parameter that normalizes the form factor at $q^2=0$, $f_+(0)$. 
The parameter $m_{{\rm pole}}$ is the mass of the lowest-lying $c\bar{q}$ 
resonance with the vector quantum numbers; this is expected to 
provide the largest contribution to the form factor for the $c\ra q$ 
transition. The sum over $N$ gives the contribution of higher mass states.  
For example, for $D\to\pi$ transitions the dominant resonance is
expected to be the $D^*(2010)$, and thus $m^{}_{\rm pole}=m^{}_{D^*(2010)}$. 
For $D\to K$ transitions, the dominant resonance is expected to be the
$D^{*}_s(2112)$, and thus $m^{}_{\rm pole}=m^{}_{D^{*}_{s}(2112)}$.

\subsubsection{Simple pole}

Equation~(\ref{eqn:expansion}) can be simplified by neglecting the 
sum over effective poles, leaving only the explicit vector meson pole. 
This approximation is referred to as ``nearest pole dominance'' or 
``vector-meson dominance.''  The resulting parameterization is
\begin{eqnarray}
  f_+(q^2) & = & \frac{f_+(0)}{(1-q^2/m^2_{\rm pole})}\,. 
\label{SimplePole}
\end{eqnarray}
However, values of $m_{{\rm pole}}$ that give a good fit to the data 
do not agree with the expected vector meson masses~\cite{Hill:2006ub}. 
To address this problem, the ``modified pole'' or Becirevic-Kaidalov~(BK) 
parameterization~\cite{Becirevic:1999kt} was introduced.
In this parameterizatio $m_{\rm pole} /\sqrt{\alpha_{\rm BK}}$
is interpreted as the mass of an effective pole higher than 
$m_{\rm pole}$, i.e., it is expected that $\alpha_{\rm BK}<1$.
The parameterization takes the form
\begin{eqnarray}
f_+(q^2) & = & \frac{f_+(0)}{(1-q^2/m^2_{\rm pole})}
\frac{1}{\left(1-\alpha^{}_{\rm BK}\frac{q^2}{m^2_{\rm pole}}\right)}\,.
\end{eqnarray}
This parameterization is used by several experiments to 
determine form factor parameters.
Measured values of $m^{}_{\rm pole}$ and $\alpha^{}_{\rm BK}$ are
listed in Tables~\ref{kPseudoPole} and~\ref{piPseudoPole} for
$D\to K\ell\nu_{\ell}$ and $D\to\pi\ell\nu_{\ell}$ decays, respectively.

\subsubsection{$z$ expansion}

Alternatively, a power series expansion around some value $q^2=t_0$ can be used to 
parameterize $f^{}_+(q^2)$~\cite{Boyd:1994tt,Boyd:1997qw,Arnesen:2005ez,Becher:2005bg}. 
This parameterization is model-independent and satisfies general QCD constraints. 
The expansion is given in terms of a complex parameter $z$, which is 
the analytic continuation of $q^2$ into the complex plane:
\begin{eqnarray}
z(q^2,t_0) & = & \frac{\sqrt{t_+ - q^2} - \sqrt{t_+ - t_0}}{\sqrt{t_+ - q^2}
	  + \sqrt{t_+ - t_0}}\,, 
\end{eqnarray}
where $t_{0}= t_{+} (1-\sqrt{1-t_{-}/t_{+}})$ and $t_\pm \equiv (m_D \pm m_P)^2$. 
In this parameterization, $q^2=t_0$ corresponds to $z=0$, and the physical 
region extends in either direction up to $\pm|z|_{\rm max} = \pm 0.051$ for 
$D\to K \ell \nu_\ell$ decays, and up to $\pm 0.17$ for $D\to \pi \ell \nu_\ell$ 
decays. 

The form factor is expressed as
\begin{eqnarray}
f_+(q^2) & = & \frac{1}{P(q^2)\,\phi(q^2,t_0)}\sum_{k=0}^\infty
a_k(t_0)[z(q^2,t_0)]^k\,,
\label{z_expansion}
\end{eqnarray}
where the $P(q^2)$ factor accommodates sub-threshold resonances via
\begin{eqnarray}
P(q^2) & \equiv & 
\begin{cases} 
z(q^2,M^2_{D^*_s}) & (D\to K) \\
1 & (D\to \pi)\,.
\end{cases}
\end{eqnarray}
The ``outer'' function $\phi(t,t_0)$ can be any analytic function, but a preferred 
choice (see, \eg, Refs.~\cite{Boyd:1994tt,Boyd:1997qw,Bourrely:1980gp}), obtained
from the Operator Product Expansion (OPE), is
\begin{eqnarray}
\phi(q^2,t_0) & =  & \alpha 
\left(\sqrt{t_+ - q^2}+\sqrt{t_+ - t_0}\right) \times  \nonumber \\
 & & \hskip0.20in \frac{t_+ - q^2}{(t_+ - t_0)^{1/4}}\  
\frac{(\sqrt{t_+ - q^2}\ +\ \sqrt{t_+ - t_-})^{3/2}}
     {(\sqrt{t_+ - q^2}+\sqrt{t_+})^5}\,,
\label{eqn:outer}
\end{eqnarray}
with $\alpha = \sqrt{\pi m_c^2/3}$.
The OPE analysis provides a constraint upon the 
expansion coefficients, $\sum_{k=0}^{N}a_k^2 \leq 1$.
These coefficients receive $1/M_D$ corrections, and thus
the constraint is only approximate. However, the
expansion is expected to converge rapidly since 
$|z|<0.051\ (0.17)$ for $D\ra K$ ($D\ra\pi$) over 
the entire physical $q^2$ range, and Eq.~(\ref{z_expansion}) 
remains a useful parameterization. The main disadvantage as compared to 
phenomenological approaches is that there is no physical interpretation 
of the fitted coefficients~$a_K$.

\subsubsection{Three-pole formalism}
An update of the vector pole dominance model has been developed for 
the $D \to \pi \ell \nu_\ell$ channel~\cite{Becirevic:2014kaa}. It uses 
information of the residues of the semileptonic form factor at its first 
two poles, the $D^\ast(2010)$ and $D^{\ast '}(2600)$ resonances.  The form 
factor is expressed as an infinite sum of residues from $J^P =1^-$ states 
with masses $m^{}_{D^\ast_n}$: 
\begin{eqnarray}
f_+(q^2) = \sum_{n=0}^\infty 
\frac{\displaystyle{\underset{q^2= m_{D^\ast_n}^2}{\rm Res}} f_+(q^2)}{m_{D^\ast_n}^2-q^2} \,,
\label{ThreePole}
\end{eqnarray}
with the residues given by 
\begin{eqnarray}
\displaystyle{\underset{q^2=m_{D_n^\ast}^2}{\rm Res}} 
f_+(q^2)= \frac{1}{2} m^{}_{D_n^\ast}\,f^{}_{D_n^\ast}\,g^{}_{D_n^\ast D\pi}\,. 
\label{Residua}
\end{eqnarray}
Values of the $f_{D^\ast}$ and $f_{D^{\ast '}}$ decay constants have been 
calculated relative to $f_{D}$ via lattice QCD, with 2$\%$ 
and 28$\%$ precision, respectively~\cite{Becirevic:2014kaa}. 
The couplings to the $D\pi$ state, $g^{}_{D^\ast D\pi}$ and $g^{}_{D^{\ast '} D\pi}$, 
are extracted from measurements of the $D^\ast(2010)$ and  $D^{\ast '}(2600)$ 
widths by the \babar and LHCb 
experiments~\cite{Lees:2013uxa,delAmoSanchez:2010vq,Aaij:2013sza}. 
This results in the contribution from the first pole being determined
with $3\%$ accuracy. The contribution from the $D^{\ast '}(2600)$ pole 
is determined with poorer accuracy, $\sim 30\%$, mainly due to lattice 
uncertainties. A {\it superconvergence} condition~\cite{Burdman:1996kr}
\begin{eqnarray}
\sum_{n=0}^\infty 
{\displaystyle{\underset{q^2=m_{D^\ast_n}^2}{\rm Res}} f_+(q^2) }= 0
\label{superconvergence}
\end{eqnarray}
 is applied, protecting the form factor behavior at large $q^2$. Within this model, 
the first two poles are not sufficient to describe the data, and a third 
effective pole needs to be included. 

One of the advantages of this phenomenological model is that it can 
be extrapolated outside the charm physical region, providing a method 
to extract the magnitude of the CKM matrix element $V_{ub}$ using the ratio of the form 
factors of the $D\to \pi\ell \nu$ and $B\to \pi\ell \nu$ decay channels. 
It will be used once lattice calculations provide the form factor ratio 
$f^{+}_{B\pi}(q^2)/f^{+}_{D\pi}(q^2)$ at the same pion energy. 

This form factor description can be extended to the $D\to K \ell \nu$ 
decay channel, considering the contribution of several $c\bar s$ 
resonances with $J^P = 1^-$. The first two pole masses contributing 
to the form factor correspond to the $D^{*}_s(2112)$ and $D^{*}_{s1}(2700)$ 
resonant states \cite{PDG_2018}. A constraint on the first residue can be 
obtained using information of the $f_K$ decay constant~\cite{PDG_2018} and 
the $g$ coupling extracted from the $D^{\ast +}$ width~\cite{Lees:2013uxa}. 
The contribution from the second pole can be evaluated using the decay 
constants from~\cite{Becirevic:2012te}, the measured total width, and 
the ratio of $D^{\ast} K$ and $D K$ decay branching fractions~\cite{PDG_2018}.

\subsubsection{Experimental techniques and results}

Various techniques have been used by several experiments to measure 
$D$ semileptonic decays with a pseudoscalar particle in the 
final state. The most recent results are provided by the \babar~\cite{Lees:2014ihu} 
and BES III~\cite{Ablikim:2015ixa, Ablikim:2015qgt} collaborations.
Belle~\cite{Widhalm:2006wz}, \babar~\cite{Aubert:2007wg}, and 
\mbox{CLEO-c}~\cite{Besson:2009uv,Dobbs:2007aa} have all  
previously reported results. Belle fully 
reconstructs $e^+e^- \to D \bar D X$ events from the continuum 
under the $\Upsilon(4S)$ resonance, achieving very good $q^2$ 
resolution (15${\rm~MeV}^2$) and a low background level but with 
a low efficiency. Using 282~$\fb^{-1}$ of data, about 
1300 $D\to K\ell^+\nu$ (Cabibbo-favored) and 115 $D\to\pi\ell^+\nu$ 
(Cabibbo-suppressed) decays are reconstructed, considering the electron 
and muon channels together. The \babar experiment uses a partial 
reconstruction technique in which the semileptonic decays 
are tagged via $ D^{\ast +}\to D^0\pi^+$ decays. 
The $D$ direction and neutrino energy are obtained 
using information from the rest of the event. 
With 75~$\fb^{-1}$ of data, 74000 signal events in the 
$D^0 \to {K}^- e^+ \nu$ mode are obtained. This technique 
provides a large signal yield but also a high background level 
and a poor $q^2$ resolution (ranging from 66 to 219 MeV$^2$). In this 
case, the measurement of the branching fraction is obtained by normalizing 
to the $D^0 \to K^- \pi^+$ decay channel; thus the measurement would
benefit from future improvements in the determination of the branching fraction for this
reference channel. The Cabibbo-suppressed mode has been recently 
measured using the same technique and 350~fb$^{-1}$ data. For
this measurement, 5000 $D^0 \to {\pi}^- e^+ \nu$ 
signal events were reconstructed~\cite{Lees:2014ihu}.  

The CLEO-c experiment uses two different methods to measure charm semileptonic 
decays. The {\it tagged\/} analyses~\cite{Besson:2009uv} rely on the full 
reconstruction of 
$\Psi(3770)\to D {\overline D}$ events. One of the $D$ mesons is reconstructed 
in a hadronic decay mode, and the other in the semileptonic channel. The only missing 
particle is the neutrino, and thus the $q^2$ resolution is very good and the 
background level very low.   
With the entire CLEO-c data sample of 818 $\pb^{-1}$, 14123 and 1374 signal 
events are reconstructed for the $D^0 \to K^{-} e^+\nu$ and $D^0\to \pi^{-} e^+\nu$ 
channels, respectively, and 8467 and 838 are reconstructed for the 
$D^+\to {\overline K}^{0} e^+\nu$ and $D^+\to \pi^{0} e^+\nu$ decays, 
respectively. An alternative method that does not tag the $D$ decay in a 
hadronic mode (referred to as {\it untagged\/} analyses) has also been 
used by CLEO-c~\cite{Dobbs:2007aa}. In this method, the entire missing 
energy and momentum in an event are associated with the neutrino four 
momentum, with the penalty of larger backgrounds as compared to the 
tagged method. 

Using the tagged method, the BES III experiment measures the 
$D^0 \to {K}^- e^+ \nu$ and $D^0 \to {\pi}^- e^+ \nu$ decay channels. 
With 2.9~fb$^{-1}$ of data, they fully reconstruct 70700 and 6300 signal 
events, respectively, for the two channels~\cite{Ablikim:2015ixa}. In 
a separate analysis, BES~III measures the semileptonic decay 
$D^+ \to K^{0}_{L} e^+ \nu$ \cite{Ablikim:2015qgt}, with about 
20100 semileptonic candidates. 
Since 2016, BES III has reported additional measurements of 
$D\to \bar K\ell^+\nu_\ell$ and $\pi\ell^+\nu_\ell$. The signal yields 
are 26008, 5013, 47100, 20714, 3402, 2265, and 1335 events for 
$D^+\to \bar K^0(\pi^+\pi^-)e^+\nu_e$, $D^+\to \bar K^0(\pi^0\pi^0)e^+\nu_e$, 
$D^0\to K^-\mu^+\nu_\mu$, $D^+\to \bar K^0(\pi\pi)\mu^+\nu_\mu$, 
$D^+\to \pi^0 e^+\nu_e$, $D^0\to \pi^-\mu^+\nu_\mu$, and 
$D^+\to \pi^0\mu^+\nu_\mu$~\cite{bes3:2017fay, bes3:2019zsf, 
bes3:2016hzl, bes3:2016wy, bes3:2018wy}, respectively. The 
corresponding branching fractions are determined with good 
precision. In Refs.~\cite{bes3:2017fay, bes3:2019zsf}, the products 
of the $c\to s(d)$ CKM matrix element and the semileptonic form factor 
are measured to be 
$|V_{cs}|f_+^{D\to K}(0)=0.7053\pm0.0040\pm0.0112$, 
$|V_{cs}|f_+^{D\to K}(0)=0.7133\pm0.0038\pm0.0030$, and 
$|V_{cd}|f_+^{D\to \pi}(0)=0.1400\pm0.0026\pm0.0007$, respectively, 
based on a two-parameter series expansion.

Results of the hadronic form factor parameters, $m_{\rm pole}$ and $\alpha_{\rm BK}$,
obtained from the measurements discussed above, are given in 
Tables~\ref{kPseudoPole} and~\ref{piPseudoPole}.
The $z$-expansion formalism has been used by \babar~\cite{Aubert:2007wg,Lees:2014ihu}, 
BES III\cite{BESIII-new} and CLEO-c~\cite{Besson:2009uv},~\cite{Dobbs:2007aa}.
Their fits use the first three terms of the expansion, %
and the results for the ratios $r_1\equiv a_1/a_0$ and $r_2\equiv a_2/a_0$ are 
listed in Tables~\ref{KPseudoZ} and~\ref{piPseudoZ}. 

\begin{table}[tp]
\centering
\caption{Results for $m_{\rm pole}$ and $\alpha_{\rm BK}$ from various 
  experiments for $D^0\to K^-\ell^+\nu$ and $D^+\to \bar{K}{}^0\ell^+\nu$ decays.
  The last two rows list results for other $c\to s e^+\nu_e$ decays, for comparison.
\label{kPseudoPole}}
\resizebox{\textwidth}{!}{
\begin{tabular}{ccccc}
\hline
\vspace*{-10pt} & \\
 $D\to K\ell\nu_\ell$ Expt. &  Mode & Ref.  & $m_{\rm pole}$ ($\gevcc$) 
& $\alpha^{}_{\rm BK}$       \\
\vspace*{-10pt} & \\
\hline
 CLEO III   &  ($D^0$; $\ell=e,\mu$) & \cite{Huang:2004fra}         & $1.89\pm0.05^{+0.04}_{-0.03}$          & $0.36\pm0.10^{+0.03}_{-0.07}$ \\
 FOCUS      &  ($D^0$; $\ell=\mu$)& \cite{Link:2004dh}            & $1.93\pm0.05\pm0.03$                   & $0.28\pm0.08\pm0.07$     \\
 Belle      &  ($D^0$; $\ell=e,\mu$)& \cite{Widhalm:2006wz}         & $1.82\pm0.04\pm0.03$                   & $0.52\pm0.08\pm0.06$     \\
 \babar     &  ($D^0$; $\ell=e$) & \cite{Aubert:2007wg}          & $1.889\pm0.012\pm0.015$                & $0.366\pm0.023\pm0.029$  \\

 CLEO-c (tagged)  & ($D^0,D^+$; $\ell=e$) &\cite{Besson:2009uv}      & $1.93\pm0.02\pm0.01$                   & $0.30\pm0.03\pm0.01$     \\
 CLEO-c (untagged) & ($D^0$; $\ell=e$) &\cite{Dobbs:2007aa}       & $1.97 \pm0.03 \pm 0.01 $ & $0.21 \pm 0.05 \pm 0.03 $  \\
 CLEO-c (untagged) & ($D^+$; $\ell=e$) &\cite{Dobbs:2007aa}       & $1.96 \pm0.04 \pm 0.02 $ & $0.22 \pm 0.08 \pm 0.03$  \\
  BES III      & ($D^0$; $\ell=e$)          &\cite{Ablikim:2015ixa}                & $1.921 \pm 0.010 \pm 0.007$ & $ 0.309 \pm 0.020 \pm 0.013$   \\ %
  BES III      & ($D^+$; $\ell=e$)          &\cite{Ablikim:2015qgt}                & $1.953 \pm 0.044 \pm 0.036$ & $ 0.239 \pm 0.077 \pm 0.065$   \\ %
  BES III&$D^+\to \bar{K}{}^0_{\pi^+\pi^-}e^+\nu_e$&\cite{bes3:2017fay}&$1.935\pm0.017\pm0.006$&$0.294\pm0.031\pm0.010$\\
 \hline
BES III&$D^+_s\to \eta e^+\nu_e$              &\cite{bes3:2019yyh}&$3.759\pm0.084\pm0.045$&$0.304\pm0.044\pm0.022$\\
BES III&$D^+_s\to \eta^\prime e^+\nu_e$       &\cite{bes3:2019yyh}&$1.88\pm0.60\pm0.08$   &$1.62\pm0.90\pm0.13$   \\
\vspace*{-10pt} & \\
\hline
\end{tabular}
}
\end{table}

\begin{table}[htbp]
\centering
\caption{Results for $m_{\rm pole}$ and $\alpha_{\rm BK}$ from various experiments for 
  $D^0\to \pi^-\ell^+\nu$ and $D^+\to \pi^0\ell^+\nu$ decays.
  The last two rows list results for other $c\to d e^+\nu_e$ decays, for comparison.
\label{piPseudoPole}}
\resizebox{\textwidth}{!}{
\begin{tabular}{ccccc}
\hline
\vspace*{-10pt} & \\
 $D\to \pi\ell\nu_\ell$ Expt. &  Mode & Ref.               & $m_{\rm pole}$ ($\gevcc$) & $\alpha_{\rm BK}$ \\
\vspace*{-10pt} & \\
\hline
 \omit        & \omit         & \omit                & \omit                                  & \omit                  \\
 CLEO III     &   ($D^0$; $\ell=e,\mu$) & \cite{Huang:2004fra}      & $1.86^{+0.10+0.07}_{-0.06-0.03}$       & $0.37^{+0.20}_{-0.31}\pm0.15$         \\
 FOCUS        &    ($D^0$; $\ell=\mu$)  & \cite{Link:2004dh}      & $1.91^{+0.30}_{-0.15}\pm0.07$          & --                                    \\
 Belle        &  ($D^0$; $\ell=e,\mu$)  & \cite{Widhalm:2006wz}      & $1.97\pm0.08\pm0.04$                   & $0.10\pm0.21\pm0.10$                  \\
 CLEO-c (tagged)  & ($D^0,D^+$; $\ell=e$)  &\cite{Besson:2009uv}   & $1.91\pm0.02\pm0.01$                   & $0.21\pm0.07\pm0.02$     \\
 CLEO-c (untagged) &  ($D^0$; $\ell=e$)  &\cite{Dobbs:2007aa}   & $1.87 \pm0.03 \pm 0.01 $ & $0.37 \pm 0.08 \pm 0.03 $  \\
 CLEO-c (untagged) & ($D^+$; $\ell=e$) &\cite{Dobbs:2007aa}     & $1.97 \pm0.07 \pm 0.02 $ & $0.14 \pm 0.16 \pm 0.04$  \\
 BES III   &  ($D^0$; $\ell=e$) &\cite{Ablikim:2015ixa}       & $1.911 \pm 0.012 \pm 0.004$ & $ 0.279 \pm 0.035 \pm 0.011$   \\ %
  \babar  &($D^0$; $\ell=e$) &\cite{Lees:2014ihu}  & $1.906 \pm 0.029 \pm 0.023$ & $ 0.268 \pm 0.074 \pm 0.059$   \\ 
BES III&$D^+\to \pi^0e^+\nu_e$                &\cite{bes3:2017fay}&$1.898\pm0.020\pm0.003$&$0.285\pm0.057\pm0.010$\\ \hline
CLEO-c&$D^+\to \eta e^+\nu_e$                &\cite{cleo:2011jy} &$1.87\pm0.24\pm0.00$   &$0.21\pm0.44\pm0.05$   \\
BES III&$D^+\to \eta e^+\nu_e$                &\cite{bes3:2018zhy}&$1.73\pm0.17\pm0.03$   &$0.50\pm0.54\pm0.08$   \\
\vspace*{-10pt} & \\
\hline
\end{tabular}
}
\end{table}

\begin{table}[tbp]
\caption{Results for $r_1$ and $r_2$ from various experiments for $D\to K\ell\nu_{\ell}$ decays.
  The correlation coefficient between these parameters is larger than 0.9.
  For comparison, the last four rows list results for $c\to s e^+\nu_e$ decays in which
  only the first two terms of the $z$ expansion were used.} 
\label{KPseudoZ}
\begin{center}
\begin{tabular}{ccccc}
\hline
\vspace*{-10pt} & \\
Expt. $D\to K\ell\nu_{\ell}$    & Mode &  Ref.                         & $r_1$               & $r_2$   \\
\hline
 \omit    & \omit         & \omit                & \omit               & \omit         \\
 \babar              & ($D^0$; $\ell=e$)  & \cite{Aubert:2007wg}   & $-2.5\pm0.2\pm0.2$  & $0.6\pm6.0\pm5.0$    \\
 CLEO-c (tagged)     & ($D^0$; $\ell=e$)  & \cite{Besson:2009uv}   & $-2.65\pm0.34\pm0.08$  & $13\pm9\pm1$   \\
 CLEO-c (tagged)     & ($D^+$; $\ell=e$)  & \cite{Besson:2009uv}   & $-1.66\pm0.44\pm0.10$  & $-14\pm11\pm1$  \\
 CLEO-c (untagged)   & ($D^0$; $\ell=e$)  &\cite{Dobbs:2007aa}     & $-2.4\pm0.4\pm0.1$  & $21\pm11\pm2$      \\
 CLEO-c (untagged)   & ($D^+$; $\ell=e$)  & \cite{Dobbs:2007aa}    & $-2.8\pm6\pm2$      & $32\pm18\pm4$       \\
 BES III             & ($D^0$; $\ell=e$)  & \cite{Ablikim:2015ixa}      & $-2.334\pm0.159\pm0.080$ & $3.42\pm 3.91\pm 2.41$  \\
 BES III             & ($D^+$; $\ell=e$) & \cite{Ablikim:2015qgt}     & $ -2.23\pm 0.42 \pm 0.53 $ & $ 11.3\pm 8.5 \pm 8.7$  \\
\hline
BES III&$D^0 \to K^-\mu^+\nu_\mu$             &\cite{bes3:2019zsf}&$-1.90\pm0.21\pm0.07$  &--                     \\
BES III&$D^+\to \bar K^0_{\pi^+\pi^-}e^+\nu_e$&\cite{bes3:2017fay}&$-1.76\pm0.25\pm0.06$  &--                     \\
BES III&$D^+_s\to \eta e^+\nu_e$              &\cite{bes3:2019yyh}&$-7.3\pm1.7\pm0.4$     &--                     \\
BES III&$D^+_s\to \eta^\prime e^+\nu_e$       &\cite{bes3:2019yyh}&$-13.1\pm7.6\pm1.0$    &--                     \\
\hline
\end{tabular}
\end{center}
\end{table}

\begin{table}[htbp]
  \caption{Results for $r_1$ and $r_2$ from various experiments for $D\to \pi \ell\nu_{\ell}$
    decays. The correlation coefficient between these parameters is larger than 0.9.
    For comparison, the last three rows list results for $c\to d e^+\nu_e$ decays in
    which only the first two terms of the $z$ expansion were used.} 
\label{piPseudoZ}
\begin{center}
\begin{tabular}{cccccc}
\hline
\vspace*{-10pt} & \\
Expt. $D\to \pi\ell\nu_{\ell}$     & Mode &  Ref.                         & $r_1$               & $r_2$    \\
\hline
 \omit    & \omit         & \omit                & \omit               & \omit              \\
 CLEO-c (tagged)     & $(D^0$; $\ell=e$) & \cite{Besson:2009uv}      &  $-2.80\pm0.49\pm0.04$ & $6\pm3\pm0$\\            
 CLEO-c (tagged)     & $(D^+$; $\ell=e$) & \cite{Besson:2009uv}      &  $-1.37\pm0.88\pm0.24$ & $-4\pm5\pm1$\\            
 CLEO-c  (untagged)  & $(D^0$; $\ell=e$) & \cite{Dobbs:2007aa}  & $-2.1\pm0.7\pm0.3$      & $-1.2\pm4.8\pm1.7$ \\
 CLEO-c   (untagged) & $(D^+$; $\ell=e$) & \cite{Dobbs:2007aa}  & $-0.2\pm1.5\pm0.4$    & $-9.8\pm9.1\pm2.1$ \\
 BES III             & $(D^0$; $\ell=e$)  & \cite{Ablikim:2015ixa}                    & $-1.85 \pm 0.22 \pm 0.07$ & $-1.4 \pm 1.5 \pm 0.5$ \\
 \babar              & $(D^0$; $\ell=e$)  & \cite{Lees:2014ihu}                   & $ -1.31 \pm 0.70 \pm 0.43 $ & $-4.2 \pm 4.0 \pm 1.9$ \\
\hline
BES III&$D^+\to \pi^0e^+\nu_e$                &\cite{bes3:2017fay}&$-2.23\pm0.42\pm0.06$  &--                     \\
CLEO-c&$D^+\to \eta e^+\nu_e$                &\cite{cleo:2011jy} &$1.83\pm2.23\pm0.28$   &--                     \\
BES III&$D^+\to \eta e^+\nu_e$                &\cite{bes3:2018zhy}&$1.88\pm0.60\pm0.08$   &--                     \\
\hline
\end{tabular}
\end{center}
\end{table}

\subsubsection{Combined results for the $D\to K\ell\nu_\ell$ and $D\to \pi\ell\nu_\ell$ channels}
 
Results and world averages for the products $f_+^K(0)|V_{cs}|$ and 
$f_+^\pi(0)|V_{cd}|$ as measured by CLEO-c, Belle, BaBar, and BES III 
are summarized in Tables~\ref{tab:aver_FF_D_K} and \ref{tab:aver_FF_D_pi}, 
respectively, and plotted in Figs.~\ref{fig:aver_FF_D_K} and \ref{fig:aver_FF_D_pi}. 
When calculating these world averages, the systematic uncertainties of the
BES~III analyses are conservatively taken to be fully correlated. 

\begin{figure}[hbt!]
\centering
\includegraphics[width=0.9\textwidth]{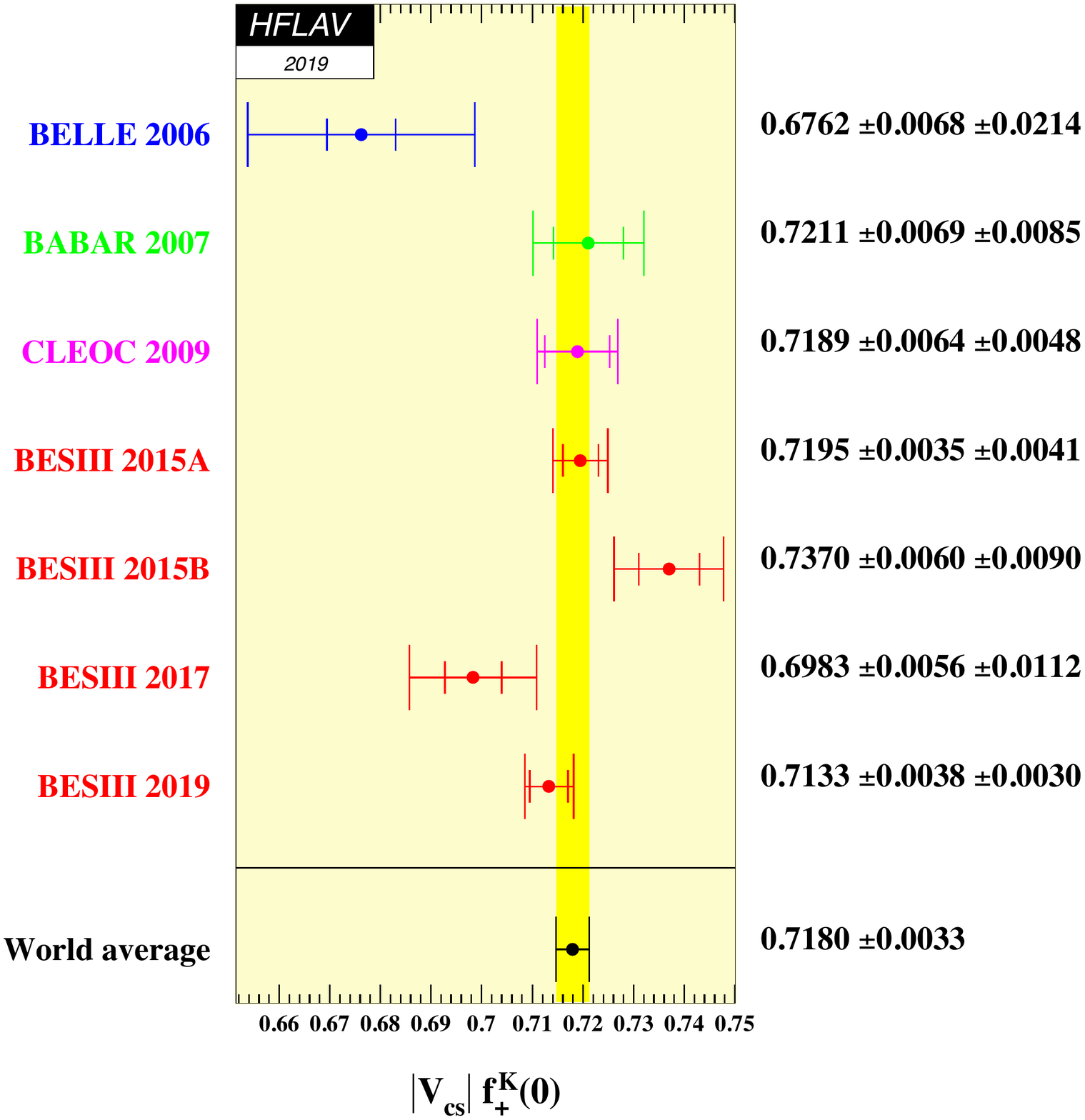}
\vskip-0.60in
\caption{
Comparison of the results of $f_+^K(0)|V_{cs}|$ measured by 
the Belle~\cite{Widhalm:2006wz}, BaBar~\cite{Aubert:2007wg}, 
CLEO-c~\cite{Besson:2009uv}, and 
BES~III~\cite{Ablikim:2015ixa,Ablikim:2015qgt,bes3:2017fay,bes3:2019zsf} experiments.
\label{fig:aver_FF_D_K}
}
\end{figure}

\begin{figure}[hbt!]
\centering
\includegraphics[width=0.9\textwidth]{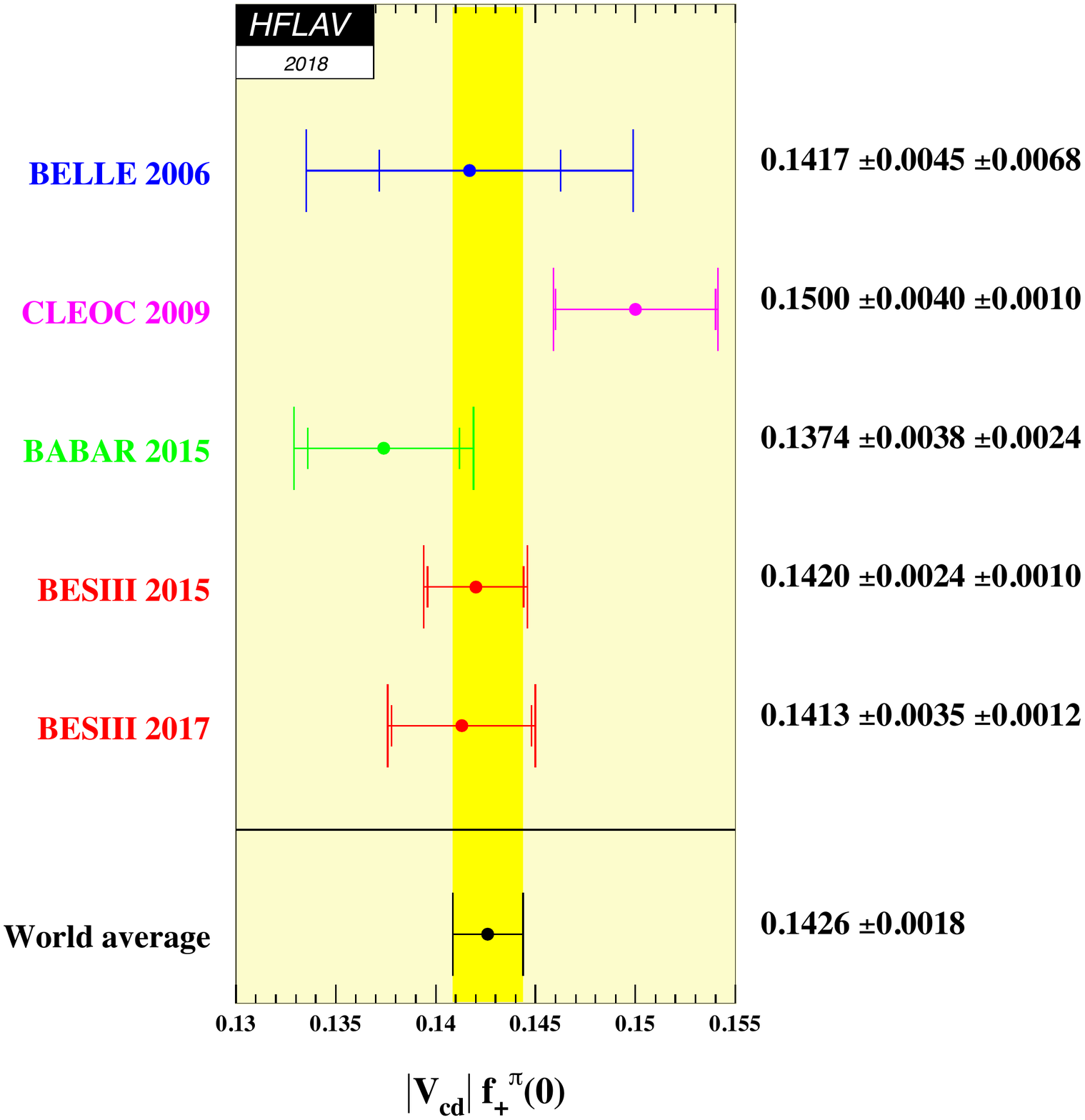}
\vskip-0.60in
\caption{
Comparison of the results of $f_+^\pi(0)|V_{cd}|$ measured by 
the Belle~\cite{Widhalm:2006wz}, BaBar~\cite{Lees:2014ihu}, 
CLEO-c~\cite{Besson:2009uv}, and 
BES~III~\cite{Ablikim:2015ixa,bes3:2017fay} experiments.
\label{fig:aver_FF_D_pi}
}
\end{figure}

\begin{table}[htbp]
\centering
\caption{Results for $f_+^K(0)|V_{cs}|$ from various experiments.}
\label{tab:aver_FF_D_K}
\begin{tabular}{l|cc|l}
\hline
\vspace*{-10pt} & \\
$D\to K \ell\nu_{\ell}$ Measurement  & Mode  &  $|V_{cs}| f_{+}^{K}(0)$ & Comment   \\
\hline
  BES III 2019~\cite{bes3:2019zsf} & ($D^0;\ell=\mu$)   & 0.7133(38)(30)  &
      $z$ expansion, 2 terms \\
  BES III 2017~\cite{bes3:2017fay} & ($D^+;\ell=e$)     & 0.6983(56)(112) & 
      $z$ expansion, 3 terms \\
  BES III 2015B~\cite{Ablikim:2015qgt} & ($D^+;\ell=e$) & 0.7370(60)(90)  & 
      $z$ expansion, 3 terms \\
  BES III 2015A~\cite{Ablikim:2015ixa} & ($D^0;\ell=e$) & 0.7195(35)(41)  &
      $z$ expansion, 3 terms \\
  CLEOc 2009~\cite{Besson:2009uv} & ($D^0$, $D^+;\ell=e$) & 0.7189(64)(48) &
      $z$ expansion, 3 terms \\
\hline
 \babar~2007~\cite{Aubert:2007wg}  &  ($D^0;\ell=e$)    &  0.7211(69)(85) &
\begin{tabular}{l}
Fitted pole mass + \\
\ \ modified pole ansatze; \\
$|V^{}_{cs}| = 0.9729\pm 0.0003$; \\
corrected for $\br(D^0\to K^-\pi^+)$
\end{tabular} \\
\hline
  Belle 2006~\cite{Widhalm:2006wz} &  ($D^0;\ell=e,\,\mu$)     & 0.6762(68)(214) & 
\begin{tabular}{l}
$|V^{}_{cs}| = 0.97296\pm 0.00024$ \\
(PDG 2006 w/unitarity)
\end{tabular} \\
 \hline 
{\bf World average} &    & {\bf 0.7180(33)} &  BES~III syst.~fully correlated \\
\hline
\end{tabular}
\end{table}

\begin{table}[htbp]
\centering
\caption{Results for $f_+^\pi(0)|V_{cd}|$ from various experiments.}
\label{tab:aver_FF_D_pi}
\begin{tabular}{l|cc|l}
\hline
$D\to \pi \ell\nu_{\ell}$ Measurement & Mode & $|V_{cd}| f_{+}^{\pi}(0)$ & Comment \\
\hline
BES III 2017~\cite{bes3:2017fay} & ($D^+;\ell=e$)            & 0.1413(35)(12)  &
      $z$ expansion, 3 terms \\
BES III 2015A~\cite{Ablikim:2015ixa} & ($D^0;\ell=e$)        & 0.1420(24)(10)  &
      $z$ expansion, 3 terms \\
CLEOc 2009~\cite{Besson:2009uv} & ($D^0$, $D^+;\ell=e$)      & 0.1500(40)(10)  &
      $z$ expansion, 3 terms \\
\babar~2015~\cite{Lees:2014ihu}   &  ($D^0;\ell=e$)          &  0.1374(38)(24) &
      $z$ expansion, 3 terms \\     
\hline
Belle 2006~\cite{Widhalm:2006wz} &  ($D^0;\ell=e,\,\mu$) &  0.1417(45)(68) &
\begin{tabular}{l}
$|V^{}_{cd}| = 0.2271\pm 0.0010$ \\
(PDG 2006 w/unitarity)
\end{tabular} \\
 \hline 
{\bf World average}   &    & {\bf 0.1426(18)} & BES~III syst.~fully correlated \\
\hline
\end{tabular}
\end{table}

\subsubsection{Form factors of other $D_{(s)}\to P\ell\nu_\ell$ decays}

In the past two decades, rapid progress in lattice QCD calculations of 
$f_+^{D\to K\,(\pi)}(0)$ has been achieved, motivated by much improved 
experimental measurements of $D\to \bar K\ell\nu_\ell$ and $D\to \pi\ell\nu_\ell$. 
However, in contrast, progress in theoretical calculations of form 
factors in other $D_{(s)}\to P\ell^+\nu_\ell$ decays has been slow, 
and experimental measurements sparse. Before BES~III, only CLEO reported 
a measurement, that of $f_+^{D\to\eta}(0)$~\cite{cleo:2011jy}. For this 
analysis both tagged and untagged methods were used.
Recently, BES III reported measurements of $f_+^{D\to\eta}(0)$, 
$f_+^{D_s\to\eta}(0)$, $f_+^{D_s\to\eta^\prime}(0)$ and $f_+^{D_s\to K}(0)$ using
a tagged method~\cite{bes3:2018zhy,bes3:2018sll,bes3:2019yyh}. These 
measurements greatly expand experimental knowledge of hadronic form factors 
in $D\to P\ell^+\nu_\ell$ decays. To date, there is still no measurement 
of $f_+^{D\to\eta^\prime}(0)$ due to the small amount of data available.

On the theory side, lattice QCD calculations of $f_+^{D_s\to\eta^{(\prime)}}(0)$ 
for $D^+_s\to \eta^{(\prime)} e^+\nu_e$ were presented in Ref.~\cite{dse:2015gsb},
but with no systematic uncertainties included. Other calculations of
$f_+^{D^+_{(s)}\to\eta^{(\prime)}}(0)$ and $f_+^{D_s\to K}(0)$ have been reported 
based on QCD light-cone sum rules (LCSR)~\cite{dse:2015gdu,dse:2013nof,dse:2011kaz}, 
three-point QCD sum rules (3PSR)~\cite{dse:2001pco}, 
a light-front quark model (LFQM)~\cite{dse:2012rcv,dse:2009ztw}, 
a constituent quark model (CQM)~\cite{dse:2000dme}, and 
a covariant confined quark model (CCQM)~\cite{dse:2018nrs}.

Table \ref{table:FF_otherD} summarizes both experimental measurements 
and theoretical calculations of these form factors.
The $f_+^{D_s\to K}(0)$ value measured by BES~III is 
consistent with current theoretical calculations.
The $f_+^{D_s\to\eta}(0)$ and $f_+^{D_s\to\eta^\prime}(0)$ values measured by
BES~III are consistent with the LCSR calculations of Refs.~\cite{dse:2015gdu, dse:2013nof}; 
however, the calculation of Ref.~\cite{dse:2013nof} is inconsistent with the measured
value of $f_+^{D\to\eta}(0)$.
More robust theoretical calculations of these form factors 
for both $D^+$ and $D^+_s$ semileptonic decays are desired.

\begin{table*}[tbp]
\centering
\caption{\label{table:FF_otherD}
  Comparison between theory and experiment for hadronic form factors of other $D_{(s)}\to P$ transitions.
  The BES III result for $f_+^{D\to\eta}(0)$ is obtained by dividing the measured product $f_+^{D\to\eta}(0)|V_{cd}|$
  by the world average value for $|V_{cd}|$. The uncertainties listed in the first and second parentheses are
  statistical and systematic uncertainties, respectively. }
\tiny
\begin{tabular}{lccccc} \hline
&$f_+^{D_s\to\eta}(0)$&$f_+^{D_s\to\eta^\prime}(0)$&$f_+^{D\to\eta}(0)$&$f_+^{D\to\eta^\prime}(0)$&$f_+^{D_s\to K}(0)$\\ \hline
CLEO                       & --          & --        &$0.38(03)(01)$\cite{cleo:2011jy}& --          & --          \\
BES III                     &$0.458(05)(04)$\cite{bes3:2019yyh}&$0.49(05)(01)$\cite{bes3:2019yyh}&$0.35(03)(01)$\cite{bes3:2018zhy}&--&$0.72(08)(01)$\cite{bes3:2018sll}\\ \hline
LQCD$_{m_\pi=470\,\rm MeV}$~\cite{dse:2015gsb}&$0.564\pm0.011$&$0.437\pm0.018$&--     & --          & --          \\
LQCD$_{m_\pi=370\,\rm MeV}$~\cite{dse:2015gsb}&$0.542\pm0.013$&$0.404\pm0.025$&--     & --          & --          \\
LCSR~\cite{dse:2015gdu}    &$0.495^{+0.030}_{-0.029}$&$0.558^{+0.047}_{-0.045}$&$0.429^{+0.165}_{-0.141}$&$0.292^{+0.113}_{-0.104}$& \\
LCSR~\cite{dse:2013nof}    &$0.432\pm0.033$&$0.520\pm0.080$&$0.552\pm0.051$&$0.458\pm0.105$& -- \\
LCSR~\cite{dse:2011kaz}    &$0.45\pm0.14$&$0.55\pm0.18$& --          & --          & --          \\
3PSR~\cite{dse:2001pco}    &$0.50\pm0.04$& --        & --          & --          & --          \\
LFQM~\cite{dse:2012rcv}    &0.76         & --        & 0.71        & --          & 0.66        \\
LFQM(I)~\cite{dse:2009ztw} &0.50         &0.62       & --          & --          & --          \\
LFQM(II)~\cite{dse:2009ztw}&0.48         &0.60       & --          & --          & --          \\
CQM \cite{dse:2000dme}     &0.78         &0.78       & --          & --          & 0.72        \\
CCQM\cite{dse:2018nrs}     &$0.78\pm0.12$&$0.73\pm11$&$0.67\pm0.11$&$0.76\pm0.11$&$0.60\pm0.09$\\ \hline
\end{tabular}
\end{table*}

\subsubsection{Determinations of $|V_{cs}|$ and $|V_{cd}|$}

Assuming unitarity of the CKM matrix, the values of the CKM matrix elements 
entering in charm semileptonic decays are evaluated from the $V_{ud}$, 
$V_{td}$ and $V_{cb}$ elements as~\cite{PDG_2018}
\begin{equation}
\label{eq:charm:ckm}
\begin{aligned}
|V_{cs}| & = 0.97359^{+0.00010}_{-0.00011} \, ,\\
|V_{cd}| & = 0.22438 \pm 0.00044 \, .
\end{aligned}
\end {equation}
Using the world average values of $f_+^K(0)|V_{cs}|$ and $f_+^{\pi}(0)|V_{cd}|$
from Tables~\ref{tab:aver_FF_D_K} and \ref{tab:aver_FF_D_pi} leads to the 
form factor values 
\begin{equation}
\begin{aligned}
 f_+^K(0) & = 0.7361 \pm 0.0034  \, , \\ 
 f_+^{\pi}(0) &= 0.6351 \pm 0.0081 \,, 
\end{aligned}\nonumber 
\end {equation}
which are in agreement with present averages of lattice QCD calculations.
Table \ref{table:FF_DKpi} summarizes $f^{D\to\pi}_+(0)$ and $f^{D\to K}_+(0)$ 
results based on $N_f=2+1+1$ flavour lattice QCD of the ETM collaboration~\cite{etm:2017lub},
and earlier results based on $N_f=2+1$ flavour lattice QCD of the HPQCD 
collaboration~\cite{hpqcd:2010hna,hpqcd:2011hna}. Recently, the Fermilab Lattice and MILC Collaborations released their preliminary results of $f^{D\to K}_+(0)$ and $f^{D\to \pi}_+(0)$ based on $N_f=2+1+1$ flavour lattice QCD calculations~\cite{fm:2019rui}. The weighted averages 
are $f^{D\to\pi}_+(0)=0.634\pm0.015$ and $f^{D\to K}_+(0)=0.760\pm0.011$, 
respectively. The experimental accuracy is at present better than that 
from lattice calculations.  

\begin{table*}[htp]
\centering
\caption{\label{table:FF_DKpi}
\small Summary of the latest LQCD calculations of $f^{D\to\pi}_+(0)$ and 
$f^{D\to K}_+(0)$ from the Fermilab/MILC, ETM, and HPQCD collaborations.}
\begin{tabular}{lcc}
\hline
Collaboration    &$f^{D\to\pi}_+(0)$&$f^{D\to K}_+(0)$\\ \hline
Fermilab Lattice and MILC~\cite{fm:2019rui}&$0.625\pm0.017\pm0.013$&$0.768\pm0.012\pm0.011$\\
ETM(2+1+1)~\cite{etm:2017lub}&$0.612\pm0.035$&$0.765\pm0.031$\\
HPQCD(2+1)~\cite{hpqcd:2010hna,hpqcd:2011hna}&$0.666\pm0.029$&$0.747\pm0.019$\\ \hline
Average&$0.634\pm0.015$&$0.760\pm0.011$\\ \hline
\end{tabular}
\end{table*}

Alternatively, if one assumes the lattice QCD form factor values,
the averages in Tables~\ref{tab:aver_FF_D_K} and \ref{tab:aver_FF_D_pi} give
\begin{equation}
\begin{aligned}
|V_{cs}| &= 0.943 \pm 0.004({\rm exp.})\pm 0.014({\rm LQCD})  \, , \\ 
|V_{cd}| &= 0.2249 \pm 0.0028({\rm exp.})\pm 0.0055({\rm LQCD})\,.
\end{aligned}\nonumber 
\end {equation} 
Here, the uncertainties are dominated by the lattice QCD calculations.
These values are consistent within $2.1\sigma$ and $0.1\sigma$, respectively,
with those obtained from the PDG global fit assuming CKM unitarity~\cite{PDG_2018}.

\subsubsection{Test of $e$-$\mu$ lepton flavour universality}

In the Standard Model (SM), the couplings between the three families 
of leptons and gauge bosons are expected to be equal; this is known 
as lepton flavour universality (LFU). The semileptonic decays of 
pseudoscalar mesons are well understood in the SM and thus offer
a robust way to test LFU and search for new physics. Various tests 
of LFU with $B$ semileptonic decays have been reported by BaBar, Belle, 
and LHCb. The average of the ratio of the branching fractions 
${\mathcal B}_{B\to \bar D^{(*)}\tau^+\nu_\tau}/
{\mathcal B}_{B\to \bar D^{(*)}\ell^+\nu_\ell}$~($\ell=\mu,\,e$) 
deviates from the SM prediction by $3.1\sigma$ 
(see section~\ref{slbdecays_b2dtaunu}).  Precision 
measurements of $D$ semileptonic decays also tests LFU, and in 
a manner complimentary to that of $B$ decays~\cite{th:2017svj}. 
Within the SM, the ratios
${\mathcal B}_{D\to \bar K\mu^+\nu_\mu}/{\mathcal B}_{D\to \bar K e^+\nu_e}$
and 
${\mathcal B}_{D\to \pi\mu^+\nu_\mu}/{\mathcal B}_{D\to \pi e^+\nu_e}$
are predicted to be $0.975\pm0.001$ and 
$0.985\pm0.002$, respectively~\cite{epjc:2018rig}.
Above $q^2=0.1$\,GeV$^2/c^4$, where $q$ is the four momentum of the
$\ell^+\nu_\ell$ system, the branching fraction ratios are expected to 
be close to unity with negligible uncertainty. This is due to the 
high correlation of the corresponding hadronic form factors~\cite{epjc:2018rig}.

In 2016, BES~III presented improved measurements of the branching 
fractions of $D^+\to\bar K^0\mu^+\nu_\mu$~\cite{bes3:2016hzl}, 
$D^0\to \pi^-\mu^+\nu_\mu$~\cite{bes3:2018wy}, and
$D^0\to K^-\mu^+\nu_\mu$~\cite{bes3:2019zsf}, and the 
first measurement of $D^+\to \pi^0\mu^+\nu_\mu$~\cite{bes3:2018wy}. 
All these analyses used the tagged method and 2.9 fb$^{-1}$ of data 
taken at 3.773 GeV. Combining these results with previous BES~III 
measurements of $\br(D^0\to \pi^-e^+\nu_e)$, $\br(D^+\to \pi^0e^+\nu_e)$, 
and $\br(D^0\to K^-e^+\nu_e)$ using the same data sample, the 
ratios of branching fractions are
\begin{eqnarray}
\frac{\br(D^0\to \pi^-\mu^+\nu_\mu)}{\br(D^0\to \pi^-e^+\nu_e)}
 & = & 0.922\pm0.030\pm0.022\,, \\
 & & \nonumber \\
\frac{\br(D^+\to \pi^0\mu^+\nu_\mu}{\br(D^+\to \pi^0e^+\nu_e)}
 & = & 0.964\pm0.037\pm0.026\,, \\
 & & \nonumber \\
\frac{\br(D^0\to K^-\mu^+\nu_\mu)}{\br(D^0\to K^-e^+\nu_e)}
 & = & 0.974\pm0.007\pm0.012\,.
\end{eqnarray}
In addition, using the world average for 
$\br(D^+\to \bar K^0 e^+\nu_e)$~\cite{PDG_2018} gives
\begin{eqnarray}
\frac{\br(D^+\to \bar K^0\mu^+\nu_\mu)}{\br(D^+\to \bar K^0 e^+\nu_e)}
 & = & 1.00\pm0.03\,.
\end{eqnarray}
These results indicate that any $e$-$\mu$ LFU violation in $D$ semileptonic decays 
has to be at the level of a few percent or less. BES~III also tested $e$-$\mu$ LFU
in separate $q^2$ intervals using 
$D^{0(+)}\to \pi^{-(0)}\ell^+\nu_\ell$~\cite{bes3:2018wy} and 
$D^0\to K^-\ell^+\nu_\ell$~\cite{bes3:2019zsf} decays.
No indication of LFU above the $2\sigma$ level was found.

In 2018, using 0.482 fb$^{-1}$ of data taken at a center-of-mass energy 
of 4.009~GeV, BES~III reported measurements of the branching fractions 
for semileptonic decays $D^+_s\to \phi\,\mu^+\nu_\mu$, 
$D^+_s\to \eta \mu^+\nu_\mu$, and 
$D^+_s\to \eta^\prime \mu^+\nu_\mu$~\cite{bes3:2018grp}. 
Combining these results with previous measurements of 
$D^+_s\to \phi\,e^+\nu_e$~\cite{bes3:2018grp}, 
$D^+_s\to \eta e^+\nu_e$, and $D^+_s\to \eta^\prime e^+\nu_e$~\cite{bes3:2016grp}
gives the ratios
\begin{eqnarray}
\frac{\br(D^+_s\to \phi\,\mu^+\nu_\mu)}{\br(D^+_s\to \phi\,e^+\nu_e)}
 & = & 0.86\pm0.29\,, \\
 & & \nonumber \\
\frac{\br(D^+_s\to \eta\mu^+\nu_\mu)}{\br(D^+_s\to \eta e^+\nu_e)}
 & = & 1.05\pm0.24\,, \\
 & & \nonumber \\
\frac{\br(D^+_s\to \eta'\mu^+\nu_\mu)}{\br(D^+_s\to \eta' e^+\nu_e)}
 & = & 1.14\pm0.68\,.
\end{eqnarray}
These values are all consistent with unity. The uncertainties include 
both statistical and systematic uncertainties, the former of which dominates.

\subsubsection{$D\ra V \ell \nu_\ell$ decays}

When the final state hadron is a vector meson, the decay can proceed through
both vector and axial vector currents, and four form factors are needed.
The hadronic current is $H^{}_\mu = V^{}_\mu + A^{}_\mu$, 
where~\cite{Gilman:1989uy} 
\begin{eqnarray}
V_\mu & = & \left< V(p,\varepsilon) | \bar{q}\gamma_\mu c | D(p') \right> \ =\  
\frac{2V(q^2)}{m_D+m_V} 
\varepsilon_{\mu\nu\rho\sigma}\varepsilon^{*\nu}p^{\prime\rho}p^\sigma \\
 & & \nonumber\\
A_\mu & = & \left< V(p,\varepsilon) | -\bar{q}\gamma_\mu\gamma_5 c | D(p') \right> 
 \ =\  -i\,(m_D+m_V)A_1(q^2)\varepsilon^*_\mu \nonumber \\
 & & \hskip2.10in 
  +\ i \frac{A_2(q^2)}{m_D+m_V}(\varepsilon^*\cdot q)(p' + p)_\mu \\
 & & \hskip2.10in 
+\ i\,\frac{2m_V}{q^2}\left(A_3(q^2)-A_0(q^2)\right)[\varepsilon^*\cdot (p' +
p)] q_\mu\,. \nonumber 
\end{eqnarray}
In this expression, $m_V$ is the daughter meson mass and
\begin{equation}
  A_3(q^2) = \frac{m_D + m_V}{2m_V}A_1(q^2)\ -\ \frac{m_D - m_V}{2m_V}A_2(q^2)\,.
\end{equation}
Kinematics require that $A_3(0) = A_0(0)$. Terms proportional to $q_\mu$ are 
only important for the case of $\tau$ leptons. Thus, only the three form factors 
$A_1(q^2)$, $A_2(q^2)$ and $V(q^2)$ are relevant in the decays involving muons and 
electrons. 

The differential decay rate is
\begin{eqnarray}
\frac{d\Gamma(D \to V \overline \ell \nu_\ell)}{dq^2\, d\cos\theta_\ell} & = & 
  \frac{G_F^2\,|V_{cq}|^2}{128\pi^3m_D^2}\,p^*\,q^2 \times \nonumber \\
 & &  
\left[\frac{(1-\cos\theta_\ell)^2}{2}|H_-|^2\ +\  
\frac{(1+\cos\theta_\ell)^2}{2}|H_+|^2\ +\ \sin^2\theta_\ell|H_0|^2\right]\,,
\end{eqnarray}
where $H^{}_\pm$ and $H^{}_0$ are helicity amplitudes, corresponding to 
helicities of the vector ($V$) meson or virtual $W$. The helicity amplitudes
can be expressed in terms of the form factors as
\begin{eqnarray}
H_\pm & = & \frac{1}{m_D + m_V}\left[(m_D+m_V)^2A_1(q^2)\ \mp\ 
      2m^{}_D\,p^* V(q^2)\right] \\
 & & \nonumber \\
H_0 & = & \frac{1}{|q|}\frac{m_D^2}{2m_V(m_D + m_V)}\ \times\ \nonumber \\
 & & \hskip0.01in \left[
    \left(1- \frac{m_V^2 - q^2}{m_D^2}\right)(m_D + m_V)^2 A_1(q^2) 
    \ -\ 4{p^*}^2 A_2(q^2) \right]\,.
\label{HelDef}
\end{eqnarray}
Here $p^*$ is the magnitude of the three-momentum of the $V$ system as measured 
in the $D$ rest frame, and $\theta_\ell$ is the angle of the lepton momentum 
with respect to the direction opposite that of the $D$ in the $W$ rest frame 
(see Fig.~\ref{DecayAngles} for the electron case, $\theta_e$).
The left-handed nature of the quark current manifests itself as
$|H_-|>|H_+|$. The differential decay rate for $D\ra V\ell\nu$ 
followed by the vector meson decaying into two pseudoscalars is
\begin{eqnarray}
\frac{d\Gamma(D\ra V \overline \ell\nu, V\ra P_1P_2)}{dq^2 d\cos\theta_V d\cos\theta_\ell d\chi} 
 &  = & \frac{3G_F^2}{2048\pi^4}
       |V_{cq}|^2 \frac{p^*(q^2)q^2}{m_D^2} {\cal B}(V\to P_1P_2)\ \times \nonumber \\ 
 & & \hskip0.10in \Big\{ (1 + \cos\theta_\ell)^2 \sin^2\theta_V |H_+(q^2)|^2 \nonumber \\
 & & \hskip0.20in +\ (1 - \cos\theta_\ell)^2 \sin^2\theta_V |H_-(q^2)|^2 \nonumber \\
 & & \hskip0.30in +\ 4\sin^2\theta_\ell\cos^2\theta_V|H_0(q^2)|^2 \nonumber \\
 & & \hskip0.40in -\ 4\sin\theta_\ell (1 + \cos\theta_\ell) 
             \sin\theta_V \cos\theta_V \cos\chi H_+(q^2) H_0(q^2) \nonumber \\
 & & \hskip0.50in +\ 4\sin\theta_\ell (1 - \cos\theta_\ell) 
          \sin\theta_V \cos\theta_V \cos\chi H_-(q^2) H_0(q^2) \nonumber \\
 & & \hskip0.60in -\ 2\sin^2\theta_\ell \sin^2\theta_V 
                \cos 2\chi H_+(q^2) H_-(q^2) \Big\}\,,
\label{eq:dGammaVector}
\end{eqnarray}
where the helicity angles $\theta^{}_\ell$, $\theta^{}_V$, and 
acoplanarity angle $\chi$ are defined as shown in Fig.~\ref{DecayAngles}. 
Typically, the ratios of the form factors at $q^2=0$ are defined as
\begin{eqnarray}
r^{}_V  & \equiv & \frac{V(0)}{A_1(0)}\,, \\
 & & \nonumber \\
r^{}_2 & \equiv & \frac{A_2(0)}{A_1(0)}\,. \label{rVr2_eq}
\end{eqnarray}

\begin{figure}[htbp]
  \begin{center}
\includegraphics[width=2.5in, viewport=0 0 320 200]{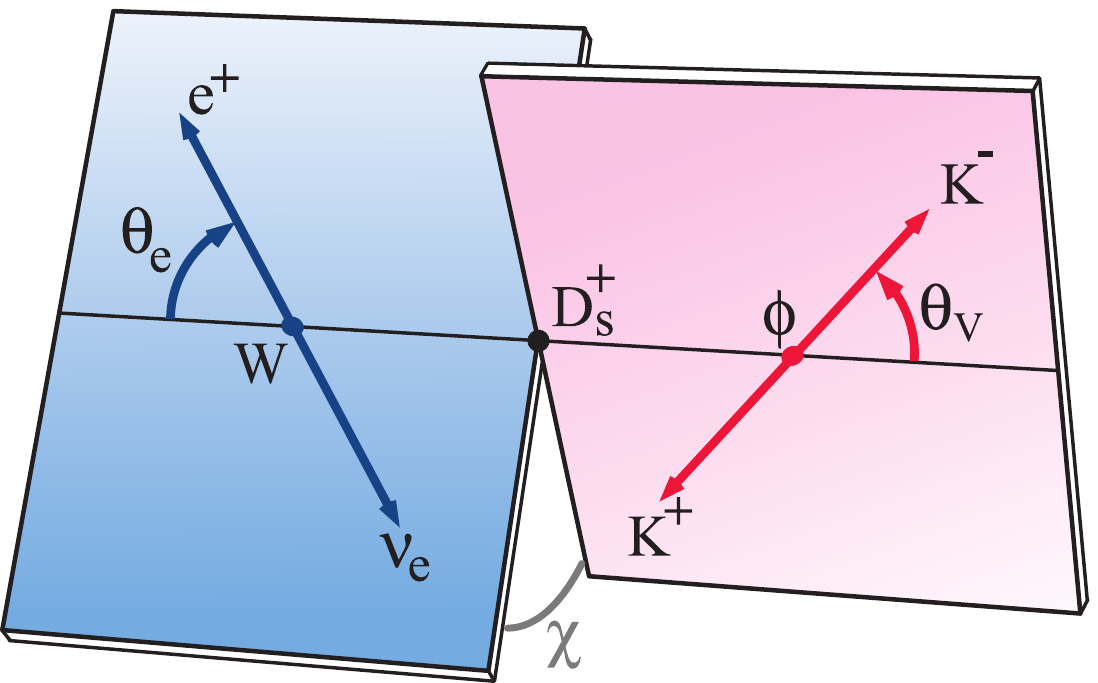}
  \end{center}
  \caption{
    Decay angles $\theta_V$, $\theta_\ell$ 
    and $\chi$. Note that the angle $\chi$ between the decay
    planes is defined in the $D$-meson reference frame, whereas
    the angles $\theta^{}_V$ and $\theta^{}_\ell$ are defined
    in the $V$ meson and $W$ reference frames, respectively.}
  \label{DecayAngles}
\end{figure}

\subsubsection{Vector form factor measurements}

In 2002 FOCUS reported an asymmetry in the observed $\cos(\theta_V)$ 
distribution in $D^+\to K^-\pi^+\mu^+\nu$ decays~\cite{Link:2002ev}. 
This was interpreted as
evidence for an $S$-wave $K^-\pi^+$ component in the decay amplitude. 
Since $H_0$ typically dominates over $H_{\pm}$, the distribution given 
by Eq.~(\ref{eq:dGammaVector}) is, after integration over $\chi$,
roughly proportional to $\cos^2\theta_V$. 
Inclusion of a constant $S$-wave amplitude of the form $A\,e^{i\delta}$ 
leads to an interference term proportional to 
$|A H_0 \sin\theta_\ell \cos\theta_V|$ which then causes an asymmetry 
in $\cos(\theta_V)$.
When FOCUS fit their data including this $S$-wave amplitude, 
they obtained $A = 0.330 \pm 0.022 \pm 0.015~\gev^{-1}$ and 
$\delta = 0.68 \pm 0.07 \pm 0.05$~\cite{Link:2002wg}. 
Both \babar~\cite{Aubert:2008rs} and CLEO-c~\cite{Ecklund:2009fia} 
have also found evidence for an $f^{}_0 \to K^+ K^-$ component in semileptonic $D^{}_s$ decays.

The CLEO-c collaboration extracted the form factors $H_+(q^2)$, $H_-(q^2)$, 
and $H_0(q^2)$ from 11000 $D^+ \rightarrow K^- \pi^+ \ell^+ \nu_\ell$ events 
in a model-independent fashion directly as functions of $q^2$~\cite{Briere:2010zc}. 
They also determined the $S$-wave form factor $h_0(q^2)$ via the interference term, despite the
fact that the $K\pi$ mass distribution appears dominated by the vector
$K^*(892)$ state. 
It is observed that $H_0(q^2)$ dominates over a wide range of $q^2$, especially at 
low $q^2$. The transverse form factor $H_t(q^2)$, which can be related 
to $A_3(q^2)$, is small compared to LQCD calculations 
and suggests that the form factor ratio $r_3 \equiv A_3(0) / A_1(0)$ is large and negative.

The \babar collaboration selected a large sample of 
$244\times 10^3$ $D^+ \rightarrow K^- \pi^+ e^+ \nu_e$ candidates 
with a ratio $S/B\sim 2.3$ from an integrated luminosity of 
$347~\fb^{-1}$~\cite{delAmoSanchez:2010fd}. With four particles emitted 
in the final state, the differential decay rate depends on five variables.
In addition to the four variables defined in previous sections there is 
also $m^2$, the mass squared of the $K\pi$ system.
To analyze the $D^+ \rightarrow K^- \pi^+ e^+ \nu_e$ decay channel, 
it was assumed that all form factors have a $q^2$ variation given by 
the simple pole model, and an effective pole mass of 
$m_A=(2.63 \pm 0.10 \pm 0.13)~\gevcc$ is fitted. This value is compatible
with expectations when comparing to the mass of $J^P=1^+$ charm mesons. 
For the mass dependence of the form factors, a Breit-Wigner with a
mass-dependent width and a Blatt-Weisskopf damping factor is used. For the 
$S$-wave amplitude, a polynomial below the $\overline{K}^*_0(1430)$, and 
a Breit-Wigner distribution above, are used. These are consistent with
measurements of $D^+ \rightarrow K^- \pi^+\pi^+$ decays.
For the polynomial part, a linear term is sufficient to fit the data.
It is verified that the variation of the $S$-wave phase is compatible 
with expectations from elastic $K\pi$ 
scattering~\cite{Estabrooks:1977xe,Aston:1987ir} (after correcting for
$\delta^{3/2}$) according to the Watson theorem~\cite{Watson:1954uc}.
As compared with elastic $K^-\pi^+$ scattering, there is an additional
negative sign between the $S$ and $P$ waves.
Contributions from other spin-1 and spin-2 resonances decaying into $K^-\pi^+$ 
are also considered.

Since 2016, several new measurements of form factors in
$D_{(s)}\to V e^+\nu_e$ decays have been reported by BES~III.
These measurements greatly increase the information available 
on $D\to V \ell^+\nu_e$ decays.
The BES~III data was recorded at center-of-mass energies of 3.773~GeV (2.9 fb$^{-1}$)
and 4.178~GeV (3.2 fb$^{-1}$). The $D\to V e^+\nu_e$ samples are reconstructed using 
a tagged method, and 
18262, 3112, 978, 491, and 155 signal events, respectively, are obtained for the 
$D^+\to \bar K^{*0}e^+\nu_e$, $D^0\to K^{*-}e^+\nu_e$, $D^{0,+}\to \rho e^+\nu_e$, 
$D^+\to \omega e^+\nu_e$, and $D^+_s\to K^{*0}e^+\nu_e$ decay 
modes~\cite{bes3:2016aff,bes3:2018dll,bes3:2016hy,bes3:2018zhl,bes3:2018sll}.
The form factor ratios $r_V$ and $r_2$ are subsequently extracted.

Table \ref{Table1} lists measurements of $r_V$ and $r_2$ from several
experiments. Most of the measurements assume that the $q^2$ dependence 
of the form factors is given by the simple pole ansatz. Some of these 
measurements do not consider a separate $S$-wave contribution; in this 
case such a contribution is implicitly included in the measured values. 

\begin{table}[htbp]
\caption{Results for $r_V$ and $r_2$ from various experiments. 
\label{Table1}}
\begin{center}
\begin{tabular}{cccc}
\hline
\vspace*{-10pt} & \\
Experiment & Ref. & $r_V$ & $r_2$ \\
\vspace*{-10pt} & \\
\hline
\vspace*{-10pt} & \\
$D^+\to \overline{K}^{*0}\ell^+\nu_\ell$ & \omit & \omit & \omit         \\
E691         & \cite{Anjos:1990pn}     & 2.0$\pm$  0.6$\pm$  0.3  & 0.0$\pm$  0.5$\pm$  0.2    \\
E653         & \cite{Kodama:1992tn}     & 2.00$\pm$ 0.33$\pm$ 0.16 & 0.82$\pm$ 0.22$\pm$ 0.11   \\
E687         & \cite{Frabetti:1993jq}     & 1.74$\pm$ 0.27$\pm$ 0.28 & 0.78$\pm$ 0.18$\pm$ 0.11   \\
E791 (e)     & \cite{Aitala:1997cm}    & 1.90$\pm$ 0.11$\pm$ 0.09 & 0.71$\pm$ 0.08$\pm$ 0.09   \\
E791 ($\mu$) & \cite{Aitala:1998ey}    & 1.84$\pm$0.11$\pm$0.09   & 0.75$\pm$0.08$\pm$0.09     \\
Beatrice     & \cite{Adamovich:1998ia} & 1.45$\pm$ 0.23$\pm$ 0.07 & 1.00$\pm$ 0.15$\pm$ 0.03   \\
FOCUS        & \cite{Link:2002wg}   & 1.504$\pm$0.057$\pm$0.039& 0.875$\pm$0.049$\pm$0.064  \\
BES III ($e$) &\cite{bes3:2016aff}&$1.406\pm0.058\pm0.022$&$0.784\pm0.041\pm0.024$\\
\hline
$D^0\to \overline{K}^0\pi^-\ell^+\nu_\ell$ & \omit & \omit & \omit         \\
FOCUS ($\mu$) & \cite{Link:2004uk}    & 1.706$\pm$0.677$\pm$0.342& 0.912$\pm$0.370$\pm$0.104 \\
\babar ($\mu$)& \cite{delAmoSanchez:2010fd} & $1.493 \pm 0.014 \pm 0.021$ & $0.775 \pm 0.011 \pm 0.011$ \\
BES III ($e$)  &\cite{bes3:2018dll} &$1.46\pm0.07\pm0.02$   &$0.67\pm0.06\pm0.01$\\
\hline
$D^+\to \omega e^+\nu_e$ & \omit & \omit & \omit         \\
BES III        &\cite{bes3:2016hy}  &$1.24\pm0.09\pm0.06$   &$1.06\pm0.15\pm0.05$\\
\hline
$D^0, D^+\to \rho\,e \nu_e$ & \omit  & \omit    & \omit                 \\
CLEO         & \cite{Mahlke:2007uf}    & 1.40$\pm$0.25$\pm$0.03   & 0.57$\pm$0.18$\pm$0.06    \\
BES III       &\cite{bes3:2018zhl}&$1.695\pm0.083\pm0.051$&$0.845\pm0.056\pm0.039$\\
\hline
$D_s^+ \to \phi\,e^+ \nu_e$ &\omit  &\omit     & \omit                  \\
\babar        & \cite{Aubert:2008rs}    & 1.849$\pm$0.060$\pm$0.095& 0.763$\pm$0.071$\pm$0.065\\ \hline
$D_s^+ \to K^{*0}\,e^+ \nu_e$ &\omit  &\omit     & \omit                  \\
BES III      &\cite{bes3:2018sll} &$1.67\pm0.34\pm0.16$   &$0.77\pm0.28\pm0.07$\\
\hline
\end{tabular}
\end{center}
\end{table}

\subsubsection{$D\to S \ell^+\nu_\ell$ decays}

In 2018, BES III reported measurements of semileptonic $D$ decays into a 
scalar meson. The experiment measured $D\to a_0(980) e^+\nu_e$, with 
$a_0(980)\to\eta\pi$. Signal yields of $25.7^{+6.4}_{-5.7}$ events for
$D^0\to a_0(980)^-e^+\nu_e$, and $10.2^{+5.0}_{-4.1}$ events for 
$D^+\to a_0(980)^0e^+\nu_e$, were obtained, resulting in statistical
significances of greater than 6.5$\sigma$ and 3.0$\sigma$, 
respectively~\cite{bes3:2018dzl}. As the branching fraction for
$a_0(980)\to\eta\pi$ is not well-measured, BES~III reports the
product branching fractions
\begin{eqnarray}
\br[D^0\to a_0(980)^-e^+\nu_e]\times \br[a_0(980)^-\to\eta\pi^-] 
   & = & (1.33^{+0.33}_{-0.29}\pm0.09)\times 10^{-4}\,, \\
\br[D^+\to a_0(980)^0e^+\nu_e]\times \br[a_0(980)^0\to\eta\pi^0]
   & = & (1.66^{+0.81}_{-0.66}\pm0.11)\times 10^{-4}\,.
\end{eqnarray}
The ratio of these values can be compared to a prediction based on
QCD light-cone sum rules~\cite{Cheng:2017fkw}, after relating the
$a_0(980)\to\eta\pi$ branching fractions via isospin.
The result is a difference of more than $2\sigma$.
Taking the lifetimes of the $D^0$ and $D^+$ into account, and assuming 
$\br[a_0(980)^-\to\eta\pi^-] = \br[a_0(980)^0\to\eta\pi^0]$, the 
ratio of the partial widths is
\begin{eqnarray}
\frac{\Gamma[D^0\to a_0(980)^-e^+\nu_e]}{\Gamma[D^+\to a_0(980)^0e^+\nu_e]}
 & = & 2.03\pm 0.95\pm 0.06 \,.
\end{eqnarray}
This value is consistent with isospin symmetry.

\subsubsection{$D\to A \ell^+\nu_\ell$ decays}

While semileptonic $D$ decays into $S$-wave states have been
studied in both theory and experiment, there is a long-standing 
puzzle whether transitions into $P$-wave states have
been established. Previously, CLEO-c reported evidence 
for $D^0\to K_1(1270)^-e^+\nu_e$ with a statistical significance of 
$4\sigma$~\cite{cleo:2007Artuso}. The branching fraction was measured
to be $\br[D^0\to K_1(1270)^-e^+\nu_e] = 
(7.6^{+4.1}_{-3.0}\pm0.6\pm0.7)\times 10^{-4}$. Recently, BES~III reported 
the first observation of $D^+\to \bar K_1(1270)^0e^+\nu_e$, with a statistical 
significance greater than $10\sigma$~\cite{bes3:2019liuk}. The branching fraction 
was measured to be 
$\br[D^+\to \bar K_1(1270)^0e^+\nu_e] = (23.0\pm2.6\pm1.8\pm2.5)\times 10^{-4}$,
which is notably higher than the CLEO result. The third error listed arises from 
the branching fraction for $K_1(1270)\to K\pi\pi$. Taking the lifetimes of 
the $D^0$ and $D^+$ into account, the ratio of the partial widths is
\begin{eqnarray}
\frac{\Gamma[D^+\to \bar K_1(1270)^0e^+\nu_e]}{\Gamma[D^0\to K_1(1270)^-e^+\nu_e]}
  & = & 1.2\,^{+0.7}_{-0.5}\,.
\end{eqnarray}
This value, like that for $D\to a_0(980)\ell^+\nu_\ell$ decays, is consistent 
with isospin symmetry.

\clearpage
\subsection{Leptonic decays}

Purely leptonic decays of $\Dp$ and $\dsp$ mesons are among the 
simplest and best understood probes of $c\to d$ and $c\to s$ 
quark flavour-changing transitions. The amplitude of purely leptonic 
decays consists of the annihilation of the initial quark-antiquark 
pair ($c\overline{d}$ or $c\overline{s}$) into a virtual $W^+$ that 
subsequently materializes as an antilepton-neutrino pair ($\ellnu$). 
The Standard Model branching fraction is given by
\begin{equation}
  \br(D_{q}^+\to \ell^+\nu_{\ell})=
  \frac{G_F^2}{8\pi}\tau^{}_{D_q} f_{D_{q}}^2 |V_{cq}|^2 m_{D_{q}}m_{\ell}^2
  \left(1-\frac{m_{\ell}^2}{m_{D_{q}}^2} \right)^2\,,
  \label{eq:brCharmLeptonicSM}
\end{equation}
where $m_{D_{q}}$ is the $D_{q}$ meson mass, 
$\tau^{}_{D_q}$ is its lifetime, 
$m_{\ell}$ is the charged lepton mass, 
$|V_{cq}|$ is the magnitude of the relevant CKM matrix element, and 
$G_F$ is the Fermi coupling constant. The parameter $f_{D_{q}}$ is the 
$D_q$ meson decay constant and parameterizes the overlap of the wave 
functions of the constituent quark and anti-quark. The decay constants 
have been calculated using several theory methods, the most accurate 
and robust being that of lattice QCD (LQCD). 
Using the $N_f=2\pm1\pm1$ flavour LQCD calculations of $f_{D^+}$ and 
$f_{D^+_s}$ from the ETM~\cite{etm:2015lqcd} and FNAL/MILC~\cite{Bazavov:2018lqcd} 
Collaborations, the Flavour Lattice Averaging Group (FLAG) calculates world 
average values~\cite{flag:2019}
\begin{eqnarray}
f^{\rm FLAG}_{D^+} & = & 212.0\pm0.7~{\rm MeV}\,, \label{eqn:fDplus_FLAG}  \\
 & & \nonumber \\
f^{\rm FLAG}_{D^+_s} & = & 249.9\pm0.5~{\rm MeV}\,, \label{eqn:fDs_FLAG}  
\end{eqnarray}
and the ratio
\begin{eqnarray}
\left(\frac{f_{D^+_s}}{f_{D^+}}\right)^{\rm FLAG} & = & 1.1783\pm0.0016\,. \label{eqn:ratio_FLAG}  
\end{eqnarray}
These values are used within this section to determine the magnitudes 
$|V_{cd}|$ and $|V_{cs}|$ from the measured branching fractions of 
$D^+\to \ell^+\nu_{\ell}$ and $D_s^+\to \ell^+\nu_{\ell}$.

The leptonic decays of pseudoscalar mesons 
are helicity-suppressed, and thus their decay rates are 
proportional to the square of the charged lepton mass. 
Thus, decays to $\tau^+\nu^{}_\tau$ are favored over decays 
to $\mu^+\nu^{}_\mu$, and decays to $e^+\nu^{}_e$, with an 
expected $\br\lesssim 10^{-7}$, are not yet experimentally 
observable. The ratio of $\tau^+\nu^{}_\tau$ to $\mu^+\nu^{}_\tau$ 
decays is given by
\begin{eqnarray}
R^{D_q}_{\tau/\mu}\ \equiv\  
\frac{\br(D^+_q\to\tau^+\nu_{\tau})}{\br(D^+_q\to\mu^+\nu_{\mu})} &  = &
\left(\frac{m_{\tau}^2}{m_{\mu}^2}\right)
\frac{(m^2_{D_q}-m^2_{\tau})^2}{(m^2_{D_q}-m^2_{\mu})^2}\,,
\end{eqnarray}
and equals $9.74\pm0.03$ for $D_s^+$ decays and 
$2.67\pm0.01$ for $D^+$ decays, based on the well-measured 
values of $m^{}_\mu$, $m^{}_\tau$, and $m^{}_{D_{(s)}}$~\cite{PDG_2018}. 
A significant deviation from this expectation would be 
interpreted as LFU violation in charged currents, which
signifies new physics~\cite{Filipuzzi:2012mg}.

In this section we present world average values for the product 
$f_{D_{q}} |V_{cq}|$, where $q=d,s$. For these averages, 
correlations between measurements and dependencies on 
input parameters are taken into account. 
Since our last report from 2016, there is one new experimental
measurement: that of $\br(D_s^+\to\munu)$ by BES~III~\cite{Ablikim:2018jun}.
In addition, lattice QCD calculations of $f^{}_D$ and $f^{}_{D_s}$ have improved.

\subsubsection{$D^+\to \ell^+\nu_{\ell}$ decays and $|V_{cd}|$}

We use measurements of the branching fraction $\br(D^+\to\munu)$ 
from \mbox{CLEO-c}~\cite{Eisenstein:2008aa} and 
BES~III~\cite{Ablikim:2013uvu} to calculate the 
world average (WA) value. We obtain
\begin{equation}
 \br^{\rm WA}(D^+\to\munu) = (3.77\pm0.17)\times10^{-4},
 \label{eq:Br:WA:DtoMuNu}
\end{equation}
from which we determine the product of the decay constant and the 
CKM matrix element to be
\begin{equation}
 f_{D}|V_{cd}| = \left(46.1\pm1.1\right)~\mbox{MeV}\,.
 \label{eq:expFDVCD}
\end{equation}
The uncertainty listed includes the uncertainty on $\br^{\rm WA}(D^+\to\munu)$, 
and also uncertainties on the external parameters $m^{}_\mu$, $m^{}_D$, and 
$\tau^{}_D$~\cite{PDG_2018} needed to extract $f_{D}|V_{cd}|$ 
from the branching fraction via Eq.~(\ref{eq:brCharmLeptonicSM}). 
Using the LQCD value for $f_D$ from FLAG [Eq.~(\ref{eqn:fDplus_FLAG})], 
we calculate the magnitude of the CKM matrix element $V_{cd}$ to be
\begin{equation}
 |V_{cd}| = 0.2173\pm0.0051\,(\rm exp.)\pm0.0007\,(\rm LQCD),
 \label{eq:Vcd:WA:Leptonic}
\end{equation}
where the uncertainties are from experiment and from LQCD, 
respectively. All input values and the resulting world average are 
summarized in Table~\ref{tab:DExpLeptonic} 
and plotted in Fig.~\ref{fig:ExpDLeptonic}.
The upper limit on the ratio of branching fractions $R_{\tau/\mu}^D$
is 3.2 at 90\%~C.L.; this is slightly above the SM expected value.

\begin{table}[t!]
\caption{Experimental results and world averages for 
${\cal{B}}(D^+\to \ell^+\nu_{\ell})$ and $f_{D}|V_{cd}|$.
The first uncertainty is statistical and the second is experimental 
systematic. The third uncertainty in the case of $f_{D^+}|V_{cd}|$ is 
due to external inputs (dominated by the uncertainty on $\tau_D$). 
Here, we take the unconstrained result from CLEO-c.
\label{tab:DExpLeptonic}}
\vskip0.15in
\begin{center}
\begin{tabular}{lccll}
\toprule
Mode 	& ${\cal{B}}$ ($10^{-4}$)	& $f_{D}|V_{cd}|$ (MeV)		& Reference & \\ 
\midrule
\multirow{2}{*}{$\munu$} & $3.95\pm0.35\pm 0.09$ 	& $47.1\pm2.1\pm0.5\pm0.2$	& CLEO-c & \cite{Eisenstein:2008aa}\\ 
			& $3.71\pm0.19\pm 0.06$ 	& $45.7\pm1.2\pm0.4\pm0.2$	& BES III & \cite{Ablikim:2013uvu}\\
\midrule
  & {\boldmath $3.77\pm0.17\pm 0.05$}  & {\boldmath $46.1\pm1.0\pm0.3\pm0.2$} 
 & {\bf Average} & \\
 & & & & \\
\midrule
$\enu$	 		& {$<0.088$ at 90\% C.L.}	&& CLEO-c & \cite{Eisenstein:2008aa}\\
\midrule
$\taunu$ 		& {$<12$ at 90\% C.L.}		&& CLEO-c & \cite{Eisenstein:2008aa}
\\ \bottomrule
\end{tabular}
\end{center}
\end{table}

\begin{figure}[hbt]
\centering
\includegraphics[width=0.6\textwidth]{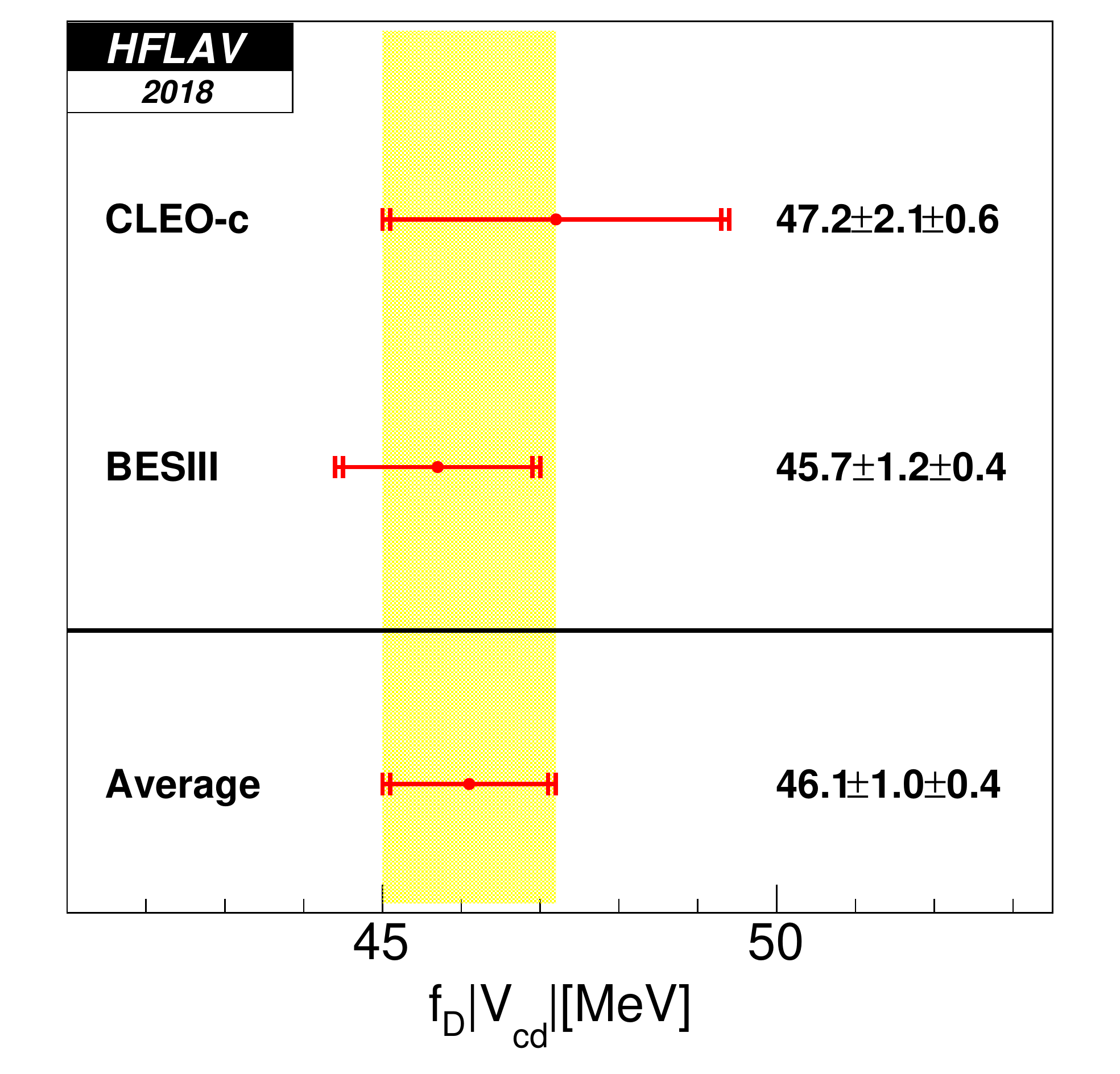}
\caption{
WA value for $f_{D}|V_{cd}|$. For each point, the first error listed is
statistical and the second error is systematic.
\label{fig:ExpDLeptonic}
}
\end{figure}

\subsubsection{$D_s^+\to \ell^+\nu_{\ell}$ decays and $|V_{cs}|$}

We use measurements of the branching fraction $\br(D_s^+\to\munu)$ 
from CLEO-c~\cite{Alexander:2009ux}, \babar~\cite{delAmoSanchez:2010jg},
Belle~\cite{Zupanc:2013byn}, and BES~III~\cite{Ablikim:2016duz,Ablikim:2018jun}
to obtain a WA value of
\begin{equation}
 \br^{\rm WA}(\dsmunu) = (5.51\pm0.16)\times10^{-3}.
 \label{eq:Br:WA:DstoMuNu}
\end{equation}
The WA value for $\br(D_s^+\to\taunu)$ is also calculated from 
CLEO-c, \babar, Belle, and BES~III measurements. 
CLEO-c made separate measurements using
$\tau^+\to e^+\nu_e\overline{\nu}{}_{\tau}$~\cite{Naik:2009tk},
$\tau^+\to\pi^+\overline{\nu}{}_{\tau}$~\cite{Alexander:2009ux}, and
$\tau^+\to\rho^+\overline{\nu}{}_{\tau}$ decays~\cite{Onyisi:2009th};
\babar made separate measurements using
$\tau^+\to e^+\nu_e\overline{\nu}{}_{\tau}$ and 
$\tau^+\to \mu^+\nu_{\mu}\overline{\nu}{}_{\tau}$ decays~\cite{delAmoSanchez:2010jg};
Belle made separate measurements using
$\tau^+\to e^+\nu_e\overline{\nu}{}_{\tau}$, 
$\tau^+\to \mu^+\nu_{\mu}\overline{\nu}{}_{\tau}$, 
and $\tau^+\to\pi^+\overline{\nu}{}_{\tau}$ decays~\cite{Zupanc:2013byn}; 
and BES~III made measurements using only
$\tau^+\to\pi^+\overline{\nu}{}_{\tau}$~\cite{Ablikim:2016duz} decays.
Combining all these results and accounting for correlations, we obtain 
a WA value of
\begin{equation}
 \br^{\rm WA}(\dsp\to\taunu) = (5.52\pm0.24)\times10^{-2}.
 \label{eq:Br:WA:DstoTauNu}
\end{equation}
The ratio of branching fractions is found to be
\begin{equation}
R_{\tau/\mu}^{\ds} = 10.02\pm0.52\,,
\label{eq:R:WA:Leptonic}
\end{equation}
which is consistent with the ratio expected in the SM.

Taking the average of 
$\br^{\rm WA}(D^+_s\to\mu^+\nu)$ and $\br^{\rm WA}(D^+_s\to\tau^+\nu)$
[Eqs.~(\ref{eq:Br:WA:DstoMuNu}) and (\ref{eq:Br:WA:DstoTauNu})],
and using the most recent values for $m^{}_\tau$, $m^{}_{D_s}$, 
and $\tau^{}_D$~\cite{PDG_2018}, we calculate the product of 
the $D_s$ decay constant and $|V_{cs}|$. The result is
\begin{equation}
 \fds|V_{cs}|=\left(247.8\pm3.1\right)~\mbox{MeV},
 \label{eq:expFDSVCS}
\end{equation}
where the uncertainty is due to the uncertainties on 
$\br^{\rm WA}(D_s^+\to\munu)$, $\br^{\rm WA}(D_s^+\to\taunu)$,
and the external inputs. 
All input values and the resulting world average are 
summarized in Table~\ref{tab:DsLeptonic} and plotted in 
Fig.~\ref{fig:ExpDsLeptonic}. To calculate this average,
we take into account correlations within each 
experiment\footnote{In the case of \babar, we use 
the covariance matrix from 
the Errata of~Ref.\cite{delAmoSanchez:2010jg}.} for 
uncertainties related to normalization, tracking, 
particle identification, signal and background 
parameterizations, and peaking background contributions.

Using the LQCD value for $\fds$ from FLAG [Eq.~(\ref{eqn:fDs_FLAG})], 
we calculate the magnitude of the CKM matrix element $V_{cs}$ to be
\begin{equation}
 |V_{cs}| = 0.991\pm0.013\,(\rm exp.)\pm0.002\,(\rm LQCD),
 \label{eq:Vcs:WA:Leptonic}
\end{equation}
where the uncertainties are from experiment and from 
lattice calculations, respectively.

\begin{table}[t!]
\caption{Experimental results and world averages for ${\cal{B}}(\dsellnu)$ 
and $f_{D_s}|V_{cs}|$. The first uncertainty is statistical and the second 
is experimental systematic. The third uncertainty in the case of 
$f_{D_s}|V_{cs}|$ is due to external inputs (dominated by the uncertainty 
on $\tau_{D_s}$). We have adjusted the $\br(\dsp\to\taunu)$ values quoted 
by CLEO-c and \babar\ to account for the most recent values of
$\br(\tau^+\to\pi^+\bar{\nu}^{}_\tau)$, $\br(\tau^+\to\mu^+\nu^{}_\mu\bar{\nu}^{}_\tau)$, 
and $\br(\tau^+\to e^+\nu^{}_e\bar{\nu}^{}_\tau)$~\cite{PDG_2018}.
CLEO-c and \babar include the uncertainty in the number of $\ds$ tags 
(denominator in the calculation of the
branching fraction) in the statistical uncertainty of $\br$; however,
we subtract this uncertainty from the statistical one and include it 
in the systematic uncertainty. 
\label{tab:DsLeptonic}}
\vskip0.15in
\begin{center}
\begin{tabular}{lccll}
\toprule
Mode 		& ${\cal{B}}$ ($10^{-2}$) 	& $f_{D_s}|V_{cs}|$ (MeV) 		& Reference & 
\\ \midrule
\multirow{4}{*}{$\munu$}	
&$0.565\pm0.044\pm 0.020$ 	& $249.8 \pm 9.7 \pm 4.4 \pm 1.0$	& CLEO-c &\cite{Alexander:2009ux}\\		
&$0.602\pm0.037\pm 0.032$ 	& $257.8 \pm 7.9 \pm 6.9 \pm 1.0$	& \babar  &\cite{delAmoSanchez:2010jg}\\
&$0.531\pm0.028\pm 0.020$ 	& $242.2 \pm 6.4 \pm 4.6 \pm 1.0$ 	& Belle  &\cite{Zupanc:2013byn}\\
&$0.517\pm0.075\pm 0.021$   & $238.9 \pm 17.3 \pm 4.9 \pm 0.9$      & BES III &\cite{Ablikim:2016duz}\\
&$0.549\pm0.016\pm 0.015$   & $246.2\pm 3.6 \pm 3.4 \pm 1.0$      & BES III &\cite{Ablikim:2018jun}\\
\midrule
    & {\boldmath $0.551\pm0.012\pm0.010$} & {\boldmath $246.7 \pm 2.8 \pm 2.3 \pm 1.0$} & {\bf Average} & \\
 & & & & \\
\midrule
$\tauenu$ 			& $5.32\pm0.47\pm0.22$ 		& $245.4 \pm 10.9 \pm 5.1 \pm 1.0$ 	& CLEO-c &\cite{Onyisi:2009th}\\
$\taupinuCharm$ & $6.47\pm0.80\pm0.22$ 		& $270.1 \pm 16.8 \pm 4.6 \pm 1.1$  & CLEO-c &\cite{Alexander:2009ux}\\
$\taurhonu$ 			& $5.50\pm0.54\pm0.24$ 		& $249.8 \pm 12.3 \pm 5.5 \pm 1.0$  & CLEO-c &\cite{Naik:2009tk}\\
\midrule
$\taunu$			& $5.59\pm0.32\pm0.14$		& $251.7 \pm 7.2 \pm 3.2 \pm 1.0$   & CLEO-c & \\
\midrule
$\tauenu$ 			& $5.09\pm0.52\pm0.68$ 		& $240.1 \pm 12.3 \pm 16.1 \pm 1.0$	& \multirow{2}{*}{\babar} & \multirow{2}{*}{\cite{delAmoSanchez:2010jg}}\\
$\taumunu$ 			& $4.90\pm0.46\pm0.54$ 		& $235.7 \pm 11.1 \pm 13.0 \pm 1.0$	&  & \\
\midrule
$\taunu$			& $4.96\pm0.37\pm0.57$		& $237.1 \pm 8.8 \pm 13.6 \pm 1.0$   & \babar &\\
\midrule
$\tauenu$  			& $5.38\pm0.33^{+0.35}_{-0.31}$ & $246.8 \pm 7.6^{+8.1}_{-7.1} \pm 1.0$  & \multirow{3}{*}{Belle} & \multirow{3}{*}{\cite{Zupanc:2013byn}} \\
$\taumunu$ 		 	& $5.86\pm0.37^{+0.34}_{-0.59}$ & $257.8 \pm 8.1^{+7.5}_{-13.0} \pm 1.0$  & &\\ 
$\taupinuCharm$  			& $6.05\pm0.43^{+0.46}_{-0.40}$ & $261.7 \pm 9.3^{+10.0}_{-8.7} \pm 1.0$  & &\\
\midrule
$\taunu$			& $5.70\pm0.21\pm0.31$		& $254.1 \pm 4.7 \pm 6.9 \pm 1.0$   & Belle & \\
\midrule
$\taupinuCharm$ 	        & $3.28\pm1.83\pm0.37$ 		& $193 \pm 54 \pm 11 \pm 1$  & BES III &\cite{Ablikim:2016duz}\\
\midrule
  & {\boldmath $5.52\pm0.16\pm0.18$} & {\boldmath $250.1 \pm 3.6 \pm 4.0 \pm 1.0$} & {\bf Average} & \\
 & & & & \\
\midrule
{\boldmath $\munu\,+\,\taunu$}  &  & {\boldmath $247.8\pm 2.2\pm 2.0\pm1.0$} & {\bf Average} & \\
 & & & & \\
\midrule
$\enu$			& $<0.0083$ at 90\% C.L.	& 	& Belle & \cite{Zupanc:2013byn} \\
\bottomrule
\end{tabular}
\end{center}
\end{table}
\begin{figure}[hbt!]
\centering
\includegraphics[width=0.8\textwidth]{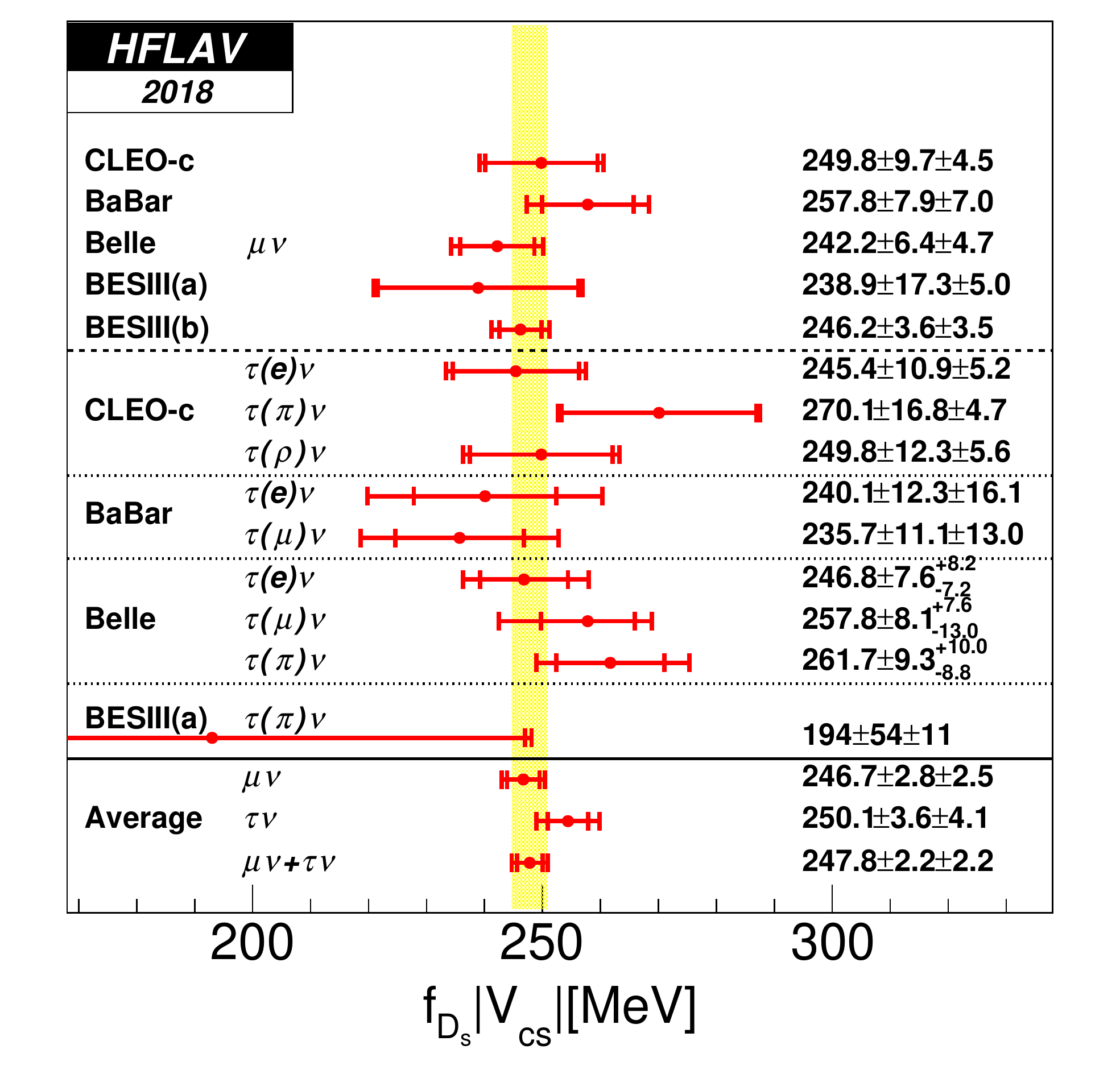}
\caption{
WA value for $f_{D_s}|V_{cs}|$. For each point, the first error listed is 
statistical and the second error listed is systematic. 
BES III(a) represents results based on 0.48~fb$^{-1}$ of 
data recorded at $\sqrt{s}=4.009$~GeV~\cite{Ablikim:2016duz},
and BES III(b) represents results based on 3.19~fb$^{-1}$ of 
data recorded at $\sqrt{s}=4.178$~GeV~\cite{Ablikim:2018jun}.
\label{fig:ExpDsLeptonic}
}
\end{figure}

\subsubsection{Comparison with other determinations of $|V_{cd}|$ and $|V_{cs}|$}

Table~\ref{tab:CKMVcdVcs} summarizes, and Fig.~\ref{fig:VcdVcsComparions} 
displays, all determinations of the magnitudes $|V_{cd}|$ and $|V_{cs}|$. The
table and figure show that, currently, the most precise direct determinations 
are from leptonic $D^+$ and $D^+_s$ decays. The values obtained are in agreement 
within uncertainties with those obtained from a global fit assuming 
CKM unitarity~\cite{Charles:2004jd}.

\begin{table}[htb]
\centering
\caption{Averages of the magnitudes of CKM matrix elements $|V_{cd}|$ and 
$|V_{cs}|$, as determined from leptonic and semileptonic $D^+_{(s)}$ decays.
In calculating these averages, we conservatively assume that uncertainties 
due to LQCD are fully correlated. For comparison, values determined from 
neutrino scattering, from $W$ decays, and from a global fit to the 
CKM matrix assuming unitarity~\cite{Charles:2004jd} are also listed. 
\label{tab:CKMVcdVcs}}
\vskip0.15in
\begin{tabular}{lcc}
\toprule
Method & Reference & Value \\ 
\midrule
&&{$|V_{cd}|$}\\
\cline{3-3}
$D\to\ell\nu_{\ell}$ 	 & This section	                 & $0.2173\pm0.0051(\rm exp.)\pm0.0007(\rm LQCD)$\\
$D\to\pi\ell\nu_{\ell}$  & Section~\ref{sec:charm:semileptonic}	& $0.2249\pm0.0028(\rm exp.)\pm0.0055(\rm LQCD)$\\
\midrule
$D\to\ell\nu_{\ell}$ 	& \multirow{2}{*}{Average}	& \multirow{2}{*}{$0.2204\pm0.0040$}\\
$D\to\pi\ell\nu_{\ell}$ & \multirow{-2}{*}{Average}	& \multirow{-2}{*}{$0.2204\pm0.0040$}\\
\midrule
$\nu N$			& PDG~\cite{PDG_2018}	& $0.230\pm0.011$\\
Global CKM Fit		& CKMFitter~\cite{Charles:2004jd}	& $0.22529^{+0.00041}_{-0.00032}$\\
\midrule
\midrule
&&{$|V_{cs}|$}\\
\cline{3-3}
$D_s\to\ell\nu_{\ell}$ 	 & This section            	& $0.991\pm0.013(\rm exp.)\pm0.002(\rm LQCD)$\\
$D\to K\ell\nu_{\ell}$   & Section~\ref{sec:charm:semileptonic}	& $0.943\pm0.004(\rm exp.)\pm0.014(\rm LQCD)$\\
\midrule
$D_s\to\ell\nu_{\ell}$ 	& \multirow{2}{*}{Average}	& \multirow{2}{*}{$0.969\pm0.010$}\\
$D\to K\ell\nu_{\ell}$ & \multirow{-2}{*}{Average}	& \multirow{-2}{*}{$0.969\pm0.010$}\\
\midrule
$W\to c\overline{s}$	& PDG~\cite{PDG_2018}	& $0.94^{+0.32}_{-0.26}\pm0.13$\\
Global CKM Fit		& CKMFitter~\cite{Charles:2004jd} & $0.973394^{+0.000074}_{-0.000096}$\\
\bottomrule
\end{tabular}
\end{table}

\begin{figure}[hbt]
\centering
\includegraphics[width=0.45\textwidth]{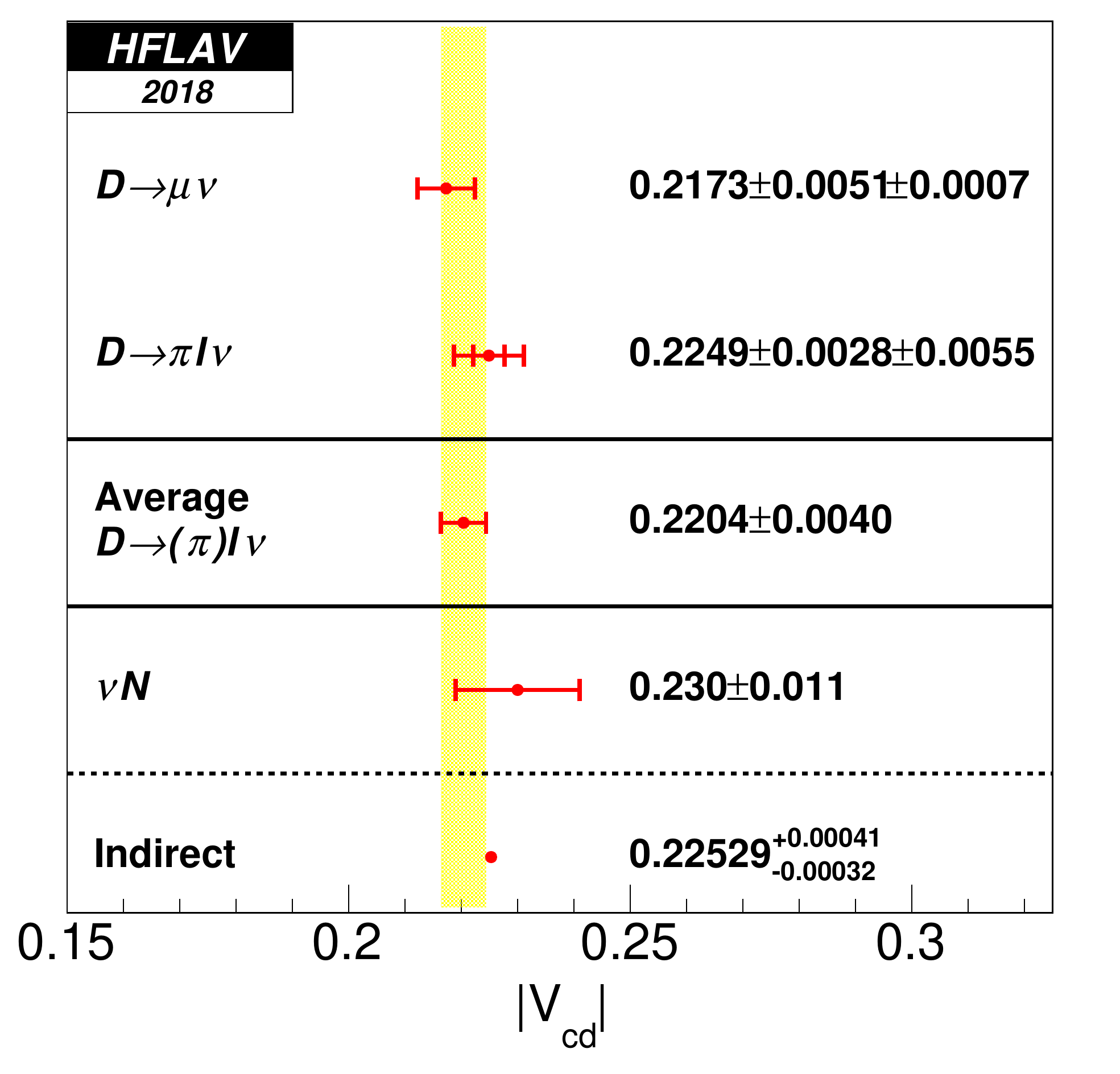}~
\includegraphics[width=0.45\textwidth]{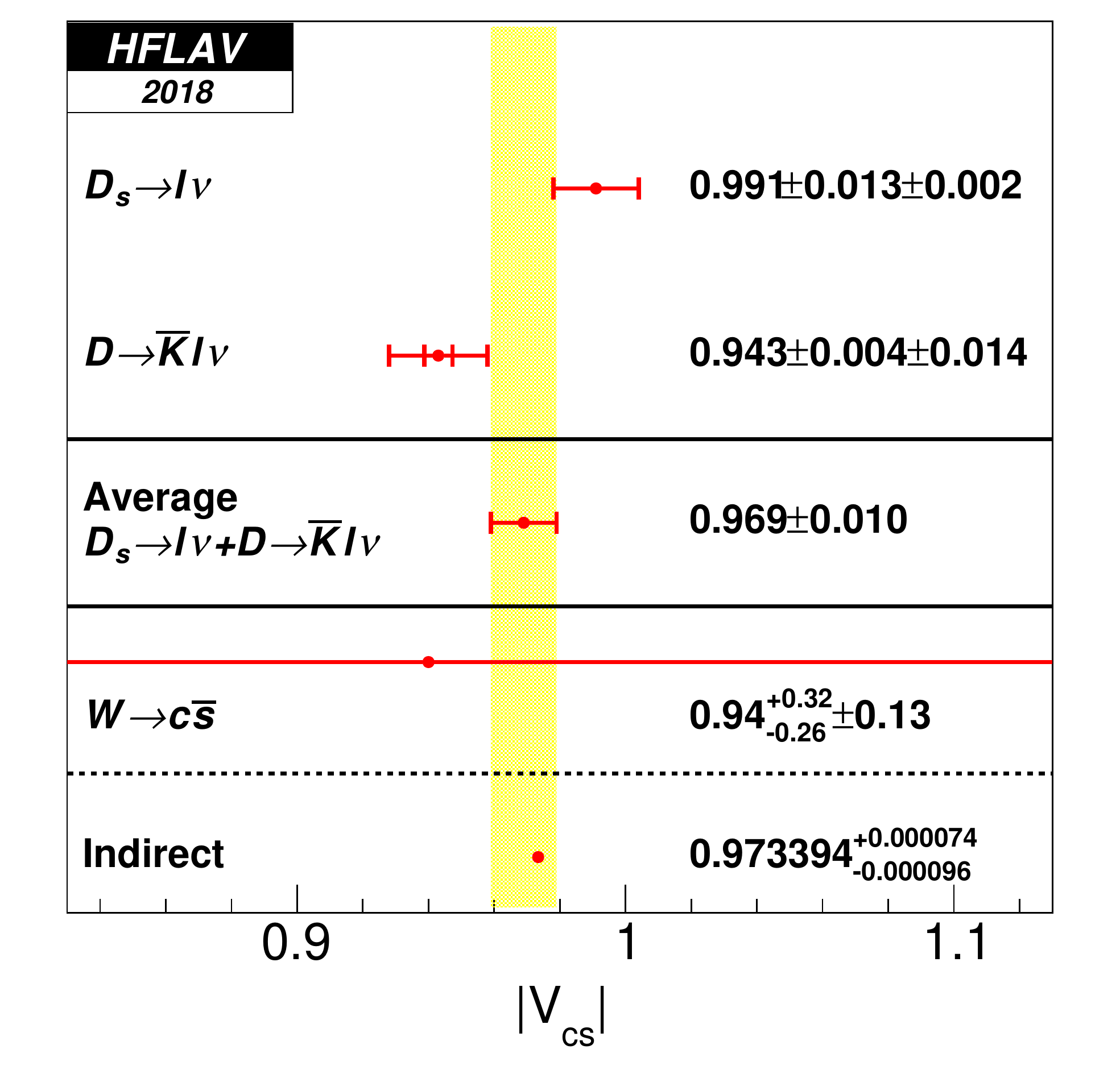}
\vskip-0.10in
\caption{
Comparison of magnitudes of CKM matrix elements $|V_{cd}|$ (left) and 
$|V_{cs}|$ (right), as determined from leptonic and semileptonic $D^+_{(s)}$ 
decays. Also listed are results from neutrino scattering, from $W$ decays,
and from a global fit of the CKM matrix assuming unitarity~\cite{Charles:2004jd}.
\label{fig:VcdVcsComparions}
}
\end{figure}

\subsubsection{Extraction of $D_{(s)}$ meson decay constants}

As listed in Table~\ref{tab:CKMVcdVcs} (and plotted in 
Fig.~\ref{fig:VcdVcsComparions}), the values of $|V^{}_{cs}|$ 
and $|V^{}_{cd}|$ can be determined from a global fit of the
CKM matrix assuming unitarity~\cite{Charles:2004jd}. 
These values can be used to extract the $D^+$ and $D^+_s$ decay constants 
from the world average values of $f_{D}|V_{cd}|$ and $f_{D_s}|V_{cs}|$ 
given in Eqs.~(\ref{eq:expFDVCD}) and~(\ref{eq:expFDSVCS}). 
The results are
\begin{eqnarray}
f_D^{\rm exp} & = & (205.4\pm4.8)~{\rm MeV,}\\ 
f_{D_s}^{\rm exp} & = & (254.5\pm3.2)~{\rm MeV,}
\end{eqnarray}
and the ratio of the decay constants is
\begin{equation}
\frac{f_{D_s}^{\rm exp}}{f_D^{\rm exp}} = 1.239\pm0.033\,.
\label{eq:fDsfDRatio:WA}
\end{equation}
These values are in agreement within their uncertainties 
with the LQCD values given by FLAG [Eqs.~(\ref{eqn:fDplus_FLAG})-(\ref{eqn:ratio_FLAG})]. 
The only discrepancy is in the ratio of decay constants; 
in this case the measurement is higher by $2.1\sigma$ than 
the LQCD prediction.

\clearpage
\subsection{Hadronic $D^0$ decays and final state radiation}

Measurements of the branching fractions for the decays $D^0\to K^\mp\pi^\pm$,
$D^0\to \pi^+\pi^-$, and $D^0\to K^+ K^-$ have reached sufficient precision to
allow averages with ${\cal O}(1\%)$ relative uncertainties. 
At this precision, Final 
State Radiation (FSR) must be treated correctly and consistently across 
the input measurements for the accuracy of the averages to match the 
precision.  The sensitivity of measurements to FSR arises because of 
a tail in the distribution of radiated energy that extends to the 
kinematic limit.  The tail beyond $\sum{E_\gamma} \approx 30$ MeV causes 
typical selection variables like the hadronic invariant mass to 
shift outside the selection range dictated by experimental 
resolution, as shown in Fig.~\ref{fig:FSR_mass_shift}.  While the 
differential rate for the tail is small, the integrated rate 
amounts to several percent of the total $h^+ h^-(n\gamma)$ 
rate because of the tail's extent.  The tail therefore 
translates directly into a several percent loss in 
experimental efficiency.

All measurements that include an FSR correction 
have a correction based on the use of 
PHOTOS~\cite{Barberio:1990ms,Barberio:1993qi,Golonka:2005pn,Golonka:2003xt,Golonka:2006tw} 
within the experiment's Monte Carlo simulation.  
PHOTOS itself, however, has evolved, over the period spanning the set of
measurements~\cite{Golonka:2003xt}.  
In particular, the incorporation of interference between
radiation from 
the two separate mesons has proceeded in stages: it was first
available for particle--antiparticle pairs in version 2.00 (1993), extended 
to any two-body, all-charged, final states in version 2.02 (1999), and 
further extended to multi-body final states in version 2.15 (2005).
The effects of interference are clearly visible, as shown in
Figure~\ref{fig:FSR_mass_shift}, and cause a 
roughly 30\% increase in the integrated rate into 
the high energy photon tail.  To evaluate the FSR 
correction incorporated into a given measurement, 
we must therefore note whether any correction was 
made, the version of PHOTOS used in correction, 
and whether the interference terms in PHOTOS were 
turned on. Also worth noting, an exponentiated multiple-photon mode was introduced in PHOTOS version 2.09, which allows PHOTOS to also simulate photons with low energies; this mode can be switched on or off. %
\begin{figure}[bh]
\begin{center}
\includegraphics[width=0.48\textwidth,angle=0.]{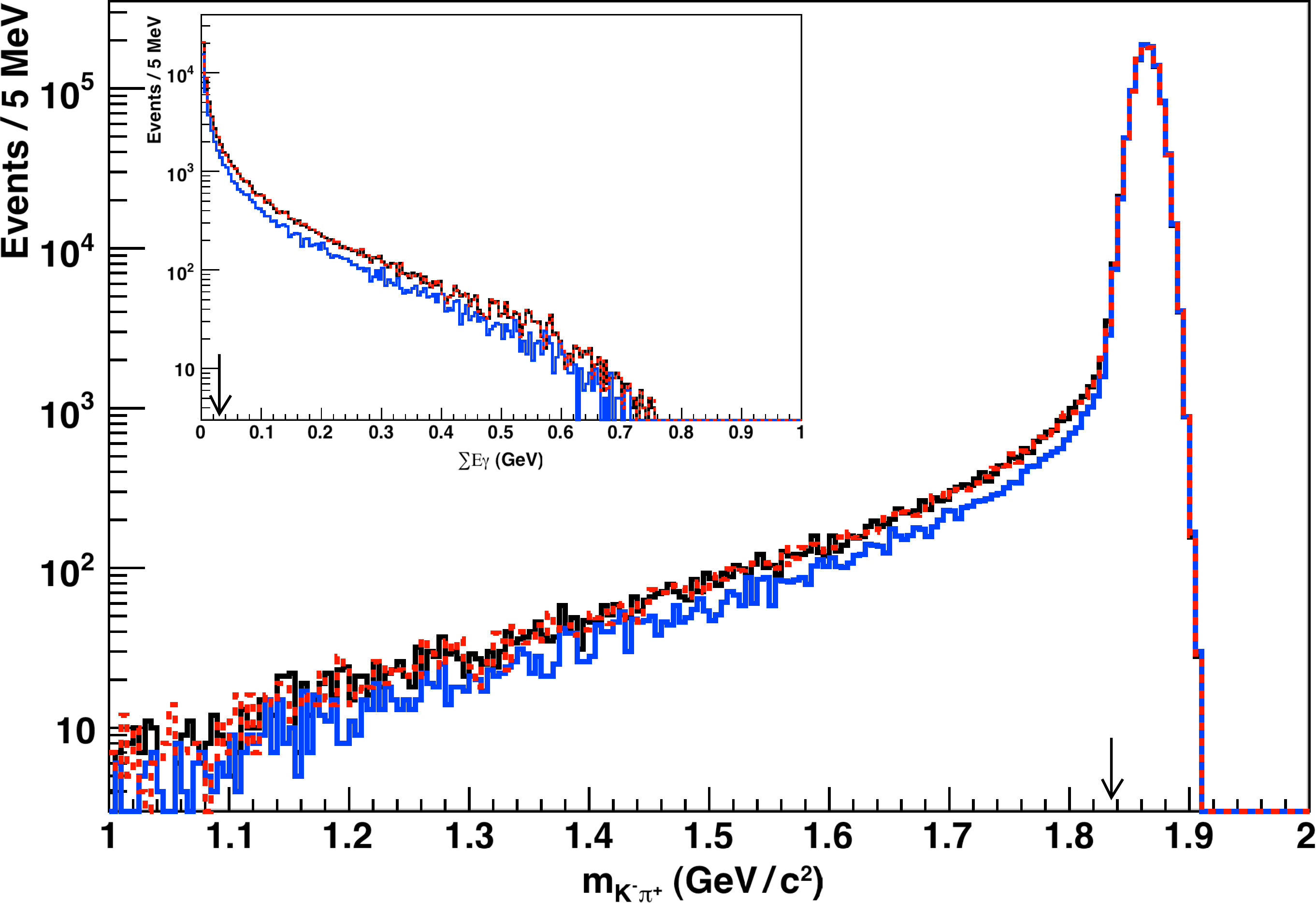}
\caption{The $K\pi$ invariant mass distribution for 
$D^0\to K^-\pi^+ (n\gamma)$ decays. The three curves correspond 
to three different configurations of PHOTOS for modeling FSR: 
version 2.02 without interference (blue/grey), version 2.02 with 
interference (red dashed) and version 2.15 with interference (black).  
The true invariant mass has been smeared with a typical experimental 
resolution of 10 MeV${}/c^2$.  Inset: The corresponding spectrum of 
total energy radiated per event.  The arrow indicates the $\sum E_\gamma$ 
value that begins to shift kinematic quantities outside of the range 
typically accepted in a measurement.}
\label{fig:FSR_mass_shift}
\end{center}
\end{figure}

\subsubsection{Updates to the branching fractions} %

Before averaging the measured branching fractions, the published 
results are updated, as necessary, to the FSR prediction of 
PHOTOS~2.15 with interference included and exponentiated multiple-photon mode turned on.  The update %
will always shift a branching fraction to a higher value: with no 
FSR correction or an FSR correction suboptimally modeled, %
the experimental efficiency determination will be biased high, 
and therefore the branching fraction will be biased low. 

Most of the branching fraction analyses used the kinematic quantity 
sensitive to FSR in the candidate selection criteria. For the 
analyses at the $\psi(3770)$, this variable was $\Delta E$, the 
difference between the candidate $D^0$ energy and the beam energy 
(\eg, 
$E_K + E_\pi - E_{\rm beam}$ 
for $D^0\to K^-\pi^+$).  
In the remainder of the analyses, the relevant quantity was the 
reconstructed hadronic two-body mass $m_{h^+h^-}$. To make an FSR 
correction, 
we need to evaluate the fraction of decays that FSR moves 
outside of the range accepted for the analysis. 
The corrections were evaluated using an event generator ({\sc EvtGen} 
\cite{Ryd:2005zz, Lange:2001uf}) that incorporates PHOTOS to simulate the 
portions of the decay process most relevant to the correction. 

We compared corrections determined both with and without smearing 
to account for experimental resolution; for the analyses using $m_{h^+h^-}$ as the kinematic quantity sensitive to FSR, the differences were 
negligible, typically of ${\cal O}(1\%)$ of the correction itself. 
The immunity of the correction to resolution effects comes about because 
most of the long FSR-induced tail in %
the $m_{h^+h^-}$ 
distribution resides well away from the selection boundaries.  The 
smearing from resolution, on the other hand, mainly affects the 
distribution of events right at the boundary. 
For the analyses using $\Delta E$ however, events with low energy photons are found to substantially move events across the selection boundary; thus PHOTOS versions with exponentiated multiple-photon mode turned on and off, respectively, can give substantially different FSR corrections. In the case that this mode is on, smearing of the events with low energy photons increases the amount of the FSR correction by about 10\%. This is well within the uncertainty on the FSR correction, as discussed later in this section, and thus ignored.

For measurements incorporating an FSR correction that did not 
include interference and/or use exponentiated multiple-photon mode, 
we update by assessing the FSR-induced 
efficiency loss for both the PHOTOS version and configuration 
used in the analysis and our nominal version 2.15 (with interference included and exponentiated multiple-photon mode turned on).  
For measurements that published their sensitivity to FSR, our 
generator-level predictions for the original efficiency loss 
agreed to within a few percent of the correction. 
This agreement lends additional credence to the procedure.

Once the event loss from FSR in the most sensitive kinematic 
quantity is accounted for, the event loss in other quantities 
is typically very small. For example, analyses using $D^{*+}$ tags show 
very little sensitivity to FSR in the reconstructed $D^{*+}\!-D^0$ 
mass difference, \ie, in $m^{}_{h^+h^-\pi^+}\!-m^{}_{h^+h^-}$. 
In this case, the effect of FSR tends to cancel in the difference 
of reconstructed masses. 
In the $\psi(3770)$ analyses, the 
beam-constrained mass distributions 
(\eg $\sqrt{E_{\rm beam}^2 - |\vec{p}_K + \vec{p}_\pi|^2}$)  
have some sensitivity, but provide negligible independent sensitivity after the  %
$\Delta E$ selection. %
\begin{figure}[t]
\begin{center}
\includegraphics[width=1.00\textwidth]{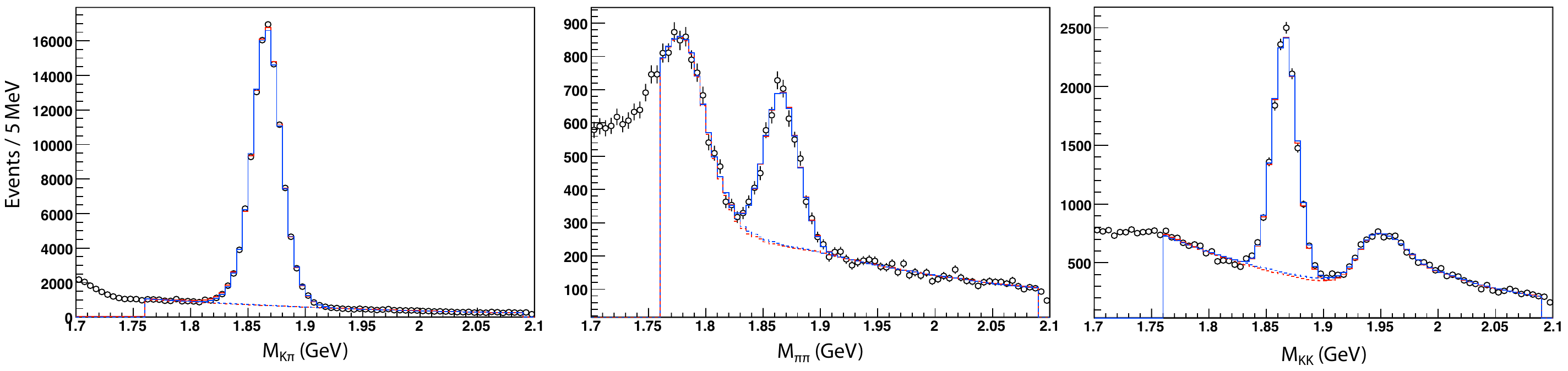}
\caption{FOCUS data (dots), original fits (blue) and 
toy MC parameterization (red) for $D^0\to K^-\pi^+$ (left), 
$D^0\to \pi^+\pi^-$ (center), and $D^0\to \pi^+\pi^-$ (right).}\label{fig:FocusFits}
\end{center}
\end{figure}

The FOCUS~\cite{Link:2002hi} analysis of the branching fraction
ratios ${\cal B}(D^0 \to \pi^+\pi^-)/{\cal B}(D^0\to K^- \pi^+)$ and 
${\cal B}(D^0 \to K^+ K^-)/{\cal B}(D^0 \to K^- \pi^+)$ obtained 
yields using fits to the two-body mass distributions.  FSR will 
both distort the low end of the signal mass peak, and will 
contribute a signal component to the low side tail used to 
estimate the background.  The fitting procedure is not sensitive 
to signal events out in the FSR tail, which would be counted as 
part of the background.

A more complex toy Monte Carlo procedure was required to analyze 
the effect of FSR on the fitted yields, which were published with 
no FSR corrections applied.  
Determining the update %
involved an iterative 
procedure in which samples of similar size to the FOCUS sample were 
generated and then fit using the FOCUS signal and background 
parameterizations. The MC parameterizations were tuned based 
on differences between the fits to the toy MC data and the FOCUS 
fits, and the procedure was repeated. These steps were iterated until 
the fit parameters matched the original FOCUS parameters.  
\begin{table}[!htb]
  \centering 
  \caption{The experimental measurements relating to ${\cal B}(D^0\to K^-\pi^+)$, ${\cal B}(D^0\to \pi^+\pi^-)$, and ${\cal B}(D^0\to K^+ K^-)$ after updating %
  them to the common version and configuration of PHOTOS.  The uncertainties are statistical and total systematic, with the FSR-related systematic estimated in this procedure shown in parentheses.  Also listed are the percent shifts in the results from those with the original correction (if any), in the case an update is applied here, as well as the original PHOTOS and interference configuration for each publication. %
  }
  \label{tab:FSR_corrections}
\begin{tabular}{lccc}
\hline \hline
Experiment\,(acronym) & Result\,(rescaled) & Update shift\,[\%] & PHOTOS \\ \hline %
\multicolumn{4}{l}{$D^{0} \to K^{-} \pi^{+}$} \\
      BES III 18 (BE18) \cite{Ablikim:2018ydy} & $3.931 \pm 0.006 \pm 0.067(44)\%$ & 1.25   & 2.03/Yes \\
      CLEO-c 14  (CC14) \cite{Bonvicini:2013vxi} & $3.934 \pm 0.021 \pm 0.061(31)\%$ & --   & 2.15/Yes \\
      \babar 07  (BA07) \cite{Aubert:2007wn}   & $4.035 \pm 0.037 \pm 0.074(24)\%$ & 0.69 & 2.02/No \\
      CLEO II 98 (CL98) \cite{Artuso:1997mc}   & $3.917 \pm 0.154 \pm 0.167(27)\%$ & 2.80 & none \\
      ALEPH 97   (AL97) \cite{Barate:1997mm}   & $3.931 \pm 0.091 \pm 0.124(27)\%$ & 0.79 & 2.0/No \\
      ARGUS 94   (AR94) \cite{Albrecht:1994nb} & $3.490 \pm 0.123 \pm 0.287(20)\%$ & 2.33 & none \\
      CLEO II 93 (CL93) \cite{Akerib:1993pm}   & $3.965 \pm 0.080 \pm 0.171(13)\%$ & 0.38 & 2.0/No \\
      ALEPH 91   (AL91) \cite{Decamp:1991jw}   & $3.733 \pm 0.351 \pm 0.455(28)\%$ & 3.12 & none \\
\hline
\multicolumn{4}{l}{$D^{0} \to \pi^{+}\pi^{-} $} \\
      BES III 18 \cite{Ablikim:2018ydy} & $ 0.1529  \pm 0.0018   \pm 0.0032(23)\%$  & 1.39   & 2.03/Yes \\
\hline
\multicolumn{4}{l}{$D^{0} \to \pi^{+}\pi^{-} / D^{0} \to K^{-} \pi^{+}$} \\
      CLEO-c 10  (CC10) \cite{Mendez:2009aa}   & $0.0370  \pm 0.0006  \pm 0.0009(02)$  & --   & 2.15/Yes \\
      CDF 05     (CD05) \cite{Acosta:2004ts}   & $0.03594 \pm 0.00054 \pm 0.00043(15)$ & --   & 2.15/Yes \\
      FOCUS 02   (FO02) \cite{Link:2002hi}     & $0.0364  \pm 0.0012  \pm 0.0006(02)$  & 3.10 & none \\
\hline
\multicolumn{4}{l}{$D^{0} \to K^{+}K^{-}$ } \\
      BES III 18 \cite{Ablikim:2018ydy} & $ 0.4271 \pm 0.0021 \pm 0.0069(27)\%$ & 0.89    & 2.03/Yes \\  
\hline
\multicolumn{4}{l}{$D^{0} \to K^{+}K^{-} / D^{0} \to K^{-} \pi^{+}$} \\
      CLEO-c 10   \cite{Mendez:2009aa}         & $0.1041 \pm 0.0011 \pm 0.0012(03)$ & --    & 2.15/Yes \\ 
      CDF 05      \cite{Acosta:2004ts}         & $0.0992 \pm 0.0011 \pm 0.0012(01)$ & --    & 2.15/Yes \\
      FOCUS 02    \cite{Link:2002hi}           & $0.0982 \pm 0.0014 \pm 0.0014(01)$ & -1.12 & none \\ \hline
\end{tabular}
\end{table}

The toy MC samples for the first iteration were based on the generator-level 
distributions of $m_{K^-\pi^+}$, $m_{\pi^+\pi^-}$, and $m_{K^+K^-}$, including 
the effects of FSR, smeared according to the original FOCUS resolution 
function,  and on backgrounds 
generated 
using the parameterization from the final
FOCUS fits.  For each iteration, 400 to 1600 individual 
data-sized samples were 
generated 
and fit. The central values %
of the parameters from these fits determined the 
corrections to the generator parameters for the following iteration.  The 
ratio between the number of signal events generated and the final signal 
yield provides the required FSR correction in the final iteration.  Only a 
few iterations were required in each mode.  Figure~\ref{fig:FocusFits} 
shows the FOCUS data, the published FOCUS fits, and the final toy MC 
parameterizations.  The toy MC provides an excellent description of the 
data.

The corrections obtained to the individual FOCUS yields were 
$1.0298 \pm 0.0001$ for $K^-\pi^+$, $1.062 \pm 0.001$ for $\pi^+\pi^-$, 
and $1.0183 \pm 0.0003$ for $K^+K^-$.  These corrections tend to 
cancel in the branching ratios, leading to corrections (update shifts)  
of 
$1.031 \pm 0.001$ 
($3.10\%$)
for 
${\cal B}(D^0\to \pi^+\pi^-)/{\cal B}(D^0\to K^-\pi^+)$, and 
$0.9888 \pm 0.0003$ 
($-1.12\%$)
for 
${\cal B}(D^0\to K^+ K^-)/{\cal B}(D^0\to K^-\pi^+)$.

Table~\ref{tab:FSR_corrections} summarizes the updated %
branching fractions. 
The published FSR-related modeling uncertainties have been replaced with a
new, common estimate; this estimate is based on the assumption that the dominant
uncertainty in the FSR corrections comes from the fact that the mesons are treated 
as structureless particles. No contributions from structure-dependent terms in 
the decay process (\eg, radiation from individual quarks) are included in PHOTOS. 
Internal studies performed by various experiments have indicated that in $K\pi$
decays, the PHOTOS corrections agree with data at the 20-30\% level. We therefore
attribute a 25\% uncertainty to the (updated) FSR correction from potential 
structure-dependent contributions. For the other two modes, the only difference 
in structure is the final state valence quark content. While radiative corrections 
typically enter with a $1/M$ dependence, the additional contribution from the
structure terms enters on a time scale shorter than the hadronization time scale.
Thus, this contribution corresponds to $M\!\sim\!\Lambda_{\rm QCD}$ rather than that of
the quark masses and would be the same for all three modes. We make this
assumption when treating the correlations among measurements. We also assume
that the PHOTOS amplitudes and any missing structure amplitudes interfere
constructively.
The uncertainties largely cancel 
in the branching fraction ratios. For the final average branching 
fractions, the FSR uncertainty on $K\pi$ %
is as large as the uncertainty due to other systematic effects. 
Note that because 
of the relative sizes of FSR in the different modes, the $\pi\pi/K\pi$ 
branching ratio uncertainty from FSR is 
positively correlated with that 
for the $K\pi$ branching fraction, while the $KK/K\pi$ branching ratio FSR
uncertainty is negatively correlated.

The ${\cal B}(D^0\to K^-\pi^+)$ measurement of reference~\cite{Coan:1997ye}, the  
${\cal B}(D^0\to \pi^+\pi^-)/{\cal B}(D^0\to K^-\pi^+)$ measurements of 
references~\cite{Aitala:1997ff} 
and~\cite{Csorna:2001ww}, and the 
${\cal B}(D^0\to K^+ K^-)/{\cal B}(D^0\to K^-\pi^+)$ measurement
of reference~\cite{Csorna:2001ww} are excluded from the branching 
fraction averages presented here.
These measurements appear not to have incorporated any FSR corrections, 
and insufficient information
is available to determine the 2-3\% update shifts %
that would be required.
\begin{sidewaystable}[p]
  \centering 
  \caption{The correlation matrix corresponding to the full covariance matrix. 
  Subscripts $h \in \{\pi, K\}$ denote which of the $D^0 \to h^+ h^-$ decay results from a single experiment
  is represented in that row or column.}\label{tab:correlations}
  \scriptsize
\begin{tabular}{lr@{.}lr@{.}lr@{.}lr@{.}lr@{.}lr@{.}lr@{.}lr@{.}lr@{.}lr@{.}lr@{.}lr@{.}lr@{.}lr@{.}lr@{.}lr@{.}l}
\hline\hline
           & \multicolumn{2}{c}{BE18}
           & \multicolumn{2}{c}{CC14}
                   & \multicolumn{2}{c}{BA07}
                           & \multicolumn{2}{c}{CL98}
                                   & \multicolumn{2}{c}{AL97}
                                           & \multicolumn{2}{c}{AR94} 
                                                   & \multicolumn{2}{c}{CL93} 
                                                           & \multicolumn{2}{c}{AL91} 
                                                                   & \multicolumn{2}{c}{BE18$_\pi$} 
                                                                   & \multicolumn{2}{c}{CC10$_\pi$} 
                                                                           & \multicolumn{2}{c}{CD05$_\pi$} 
                                                                                   & \multicolumn{2}{c}{FO02$_\pi$} 
                                                                                           & \multicolumn{2}{c}{BE18$_K$}
                                                                                           & \multicolumn{2}{c}{CC10$_K$}
                                                                                                    & \multicolumn{2}{c}{CD05$_K$} 
                                                                                                            & \multicolumn{2}{c}{FO02$_K$} \\ \hline
\\
BE18       & 1&0000  & 0&3143  & 0&1897  & 0&0777  & 0&1148  & 0&0419  & 0&0450  & 0&0319  & 0&6534  & 0&0930  & 0&1401  & 0&0948  & 0&5839  &-0&1153  &-0&0437  &-0&0259 \\ \\
CC14       & 0&3143  & 1&0000  & 0&1394  & 0&0571  & 0&0844  & 0&0308  & 0&0331  & 0&0234  & 0&3023  & 0&0683  & 0&1029  & 0&0697  & 0&1788  &-0&0847  &-0&0321  &-0&0191 \\ \\
BA07       & 0&1897  & 0&1394  & 1&0000  & 0&0345  & 0&0509  & 0&0186  & 0&0200  & 0&0141  & 0&1825  & 0&0413  & 0&0621  & 0&0421  & 0&1079  &-0&0511  &-0&0194  &-0&0115 \\ \\
CL98       & 0&0777  & 0&0571  & 0&0345  & 1&0000  & 0&0209  & 0&0076  & 0&0082  & 0&0058  & 0&0748  & 0&0169  & 0&0255  & 0&0172  & 0&0442  &-0&0209  &-0&0079  &-0&0047 \\ \\
AL97       & 0&1148  & 0&0844  & 0&0509  & 0&0209  & 1&0000  & 0&0112  & 0&0121  & 0&1156  & 0&1104  & 0&0250  & 0&0376  & 0&0254  & 0&0653  &-0&0309  &-0&0117  &-0&0070 \\ \\
AR94       & 0&0419  & 0&0308  & 0&0186  & 0&0076  & 0&0112  & 1&0000  & 0&0044  & 0&0031  & 0&0403  & 0&0091  & 0&0137  & 0&0093  & 0&0238  &-0&0113  &-0&0043  &-0&0025 \\ \\
CL93       & 0&0450  & 0&0331  & 0&0200  & 0&0082  & 0&0121  & 0&0044  & 1&0000  & 0&0034  & 0&0433  & 0&0098  & 0&0147  & 0&0100  & 0&0256  &-0&0121  &-0&0046  &-0&0027 \\ \\
AL91       & 0&0319  & 0&0234  & 0&0141  & 0&0058  & 0&1156  & 0&0031  & 0&0034  & 1&0000  & 0&0306  & 0&0069  & 0&0104  & 0&0071  & 0&0181  &-0&0086  &-0&0033  &-0&0019 \\ \\
BE18$_\pi$ & 0&6534  & 0&3023  & 0&1825  & 0&0748  & 0&1104  & 0&0403  & 0&0433  & 0&0306  & 1&0000  & 0&0895  & 0&1347  & 0&0912  & 0&4334  &-0&1109  &-0&0421  &-0&0249 \\ \\
CC10$_\pi$ & 0&0930  & 0&0683  & 0&0413  & 0&0169  & 0&0250  & 0&0091  & 0&0098  & 0&0069  & 0&0895  & 1&0000  & 0&0305  & 0&0206  & 0&0529  &-0&0251  &-0&0095  &-0&0056 \\ \\
CD05$_\pi$ & 0&1401  & 0&1029  & 0&0621  & 0&0255  & 0&0376  & 0&0137  & 0&0147  & 0&0104  & 0&1347  & 0&0305  & 1&0000  & 0&0310  & 0&0797  &-0&0378  &-0&0143  &-0&0085 \\ \\
FO02$_\pi$ & 0&0948  & 0&0697  & 0&0421  & 0&0172  & 0&0254  & 0&0093  & 0&0100  & 0&0071  & 0&0912  & 0&0206  & 0&0310  & 1&0000  & 0&0539  &-0&0255  &-0&0097  &-0&0057 \\ \\
BE18$_K$   & 0&5839  & 0&1788  & 0&1079  & 0&0442  & 0&0653  & 0&0238  & 0&0256  & 0&0181  & 0&4334  & 0&0529  & 0&0797  & 0&0539  & 1&0000  &-0&0656  &-0&0249  &-0&0148 \\ \\
CC10$_K$   &-0&1153  &-0&0847  &-0&0511  &-0&0209  &-0&0309  &-0&0113  &-0&0121  &-0&0086  &-0&1109  &-0&0251  &-0&0378  &-0&0255  &-0&0656  & 1&0000  & 0&0118  & 0&0070 \\ \\
CD05$_K$   &-0&0437  &-0&0321  &-0&0194  &-0&0079  &-0&0117  &-0&0043  &-0&0046  &-0&0033  &-0&0421  &-0&0095  &-0&0143  &-0&0097  &-0&0249  & 0&0118  & 1&0000  & 0&0027 \\ \\
FO02$_K$   &-0&0259  &-0&0191  &-0&0115  &-0&0047  &-0&0070  &-0&0025  &-0&0027  &-0&0019  &-0&0249  &-0&0056  &-0&0085  &-0&0057  &-0&0148  & 0&0070  & 0&0027  & 1&0000 \\ \\
\hline
\end{tabular}
\end{sidewaystable}

\subsubsection{Average branching fractions for
\texorpdfstring{$D^0\to K^-\pi^+$, $D^0\to \pi^+\pi^-$ and $D^0\to K^+ K^-$}{D0 to K-pi+, D0 to pi+pi-, D0 to K+K-}} %
\begin{figure}[b]
\begin{center}
\includegraphics[width=0.6\textwidth,angle=0.]{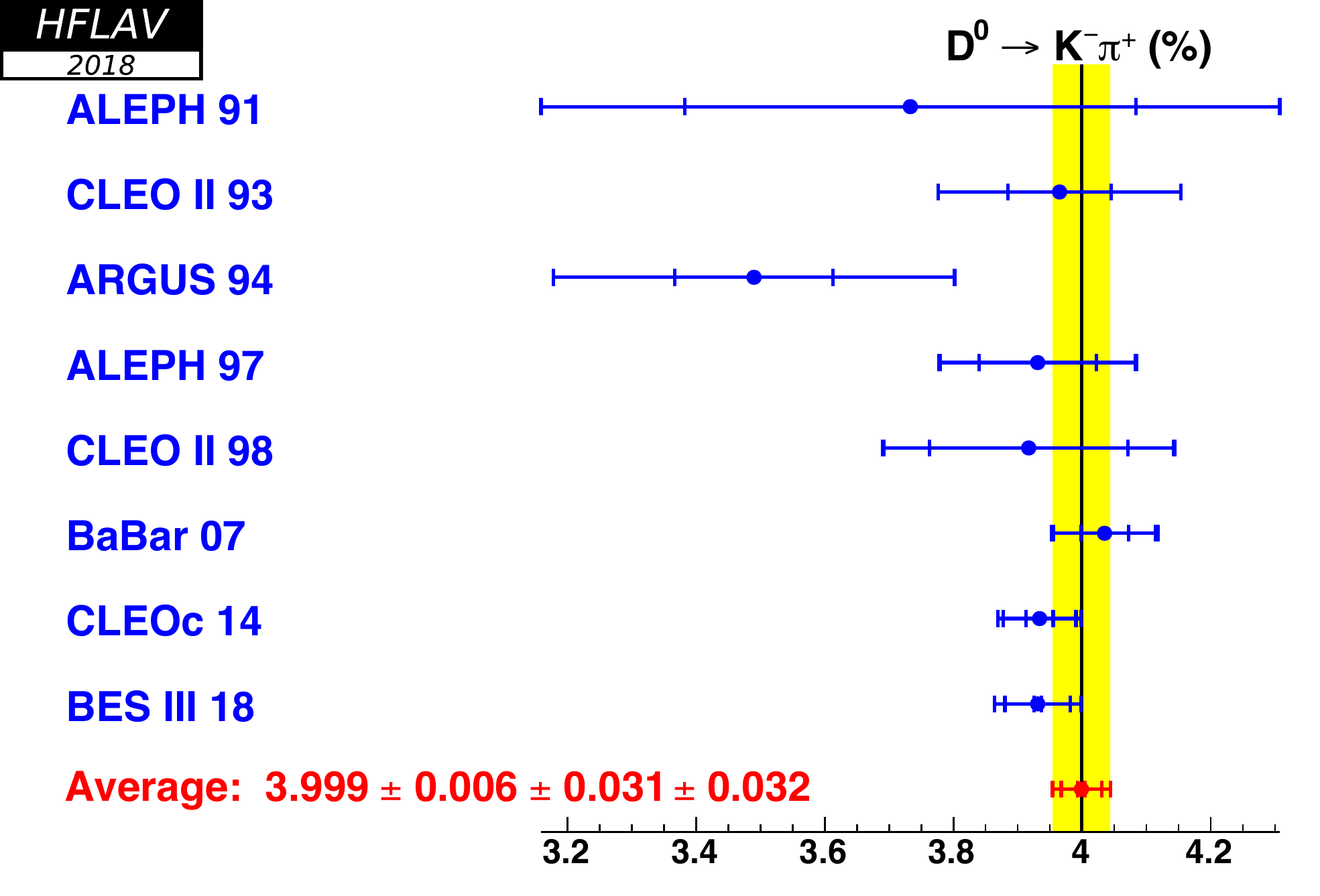}
\caption{Comparison of measurements of 
${\cal B}(D^0\to K^-\pi^+)$ (blue) with the average 
branching fraction obtained here (red, and yellow band).  For these measurements only, the partial $\chi^2$ is 4.9 in the final fit.}
\label{D0bfs}
\end{center}
\end{figure}

The average branching fractions for 
$D^0\to K^-\pi^+$, $D^0\to \pi^+\pi^-$ and $D^0\to K^+ K^-$ decays
are obtained from a single $\chi^2$ minimization procedure, 
in which the three branching fractions are floating parameters. 
The central values are obtained from a fit in which the full covariance matrix, accounting for all statistical, systematic (excluding FSR), and FSR 
measurement uncertainties, is used.  
Table~\ref{tab:correlations} presents the correlation matrix for 
this nominal fit. %
We then obtain the three reported uncertainties on those central values as follows:
The statistical uncertainties are obtained from a fit using only the statistical 
covariance matrix.  
The systematic uncertainties are obtained by subtracting (in quadrature) the 
statistical uncertainties 
from the uncertainties determined via a fit using a covariance matrix that 
accounts for both statistical and systematic measurement uncertainties.  
The FSR uncertainties are obtained by subtracting (in quadrature)
the uncertainties determined via a fit using a covariance matrix that accounts 
for both statistical and systematic measurement uncertainties
from the uncertainties determined via the fit using the full covariance matrix.

In forming the full covariance matrix, the FSR
uncertainties are treated as fully correlated (or anti-correlated) as 
described above.  %
For the covariance matrices involving systematic measurement uncertainties, ALEPH's systematic %
uncertainties in the $\theta_{D^*}$ parameter are treated
as fully correlated between the ALEPH 97 and ALEPH 91 measurements.  Similarly,
the tracking efficiency uncertainties in the CLEO II 98 and the
CLEO II 93 measurements are treated as fully correlated. For the three BES III 18 results, both tracking and particle identification efficiencies for any particles shared between decay modes are treated as fully correlated. Finally, the BES III 18 results also have a fully correlated statistical dependence on the number %
of ${D^0\bar{D}^0}$ pairs produced.

The averaging procedure results in a 
final $\chi^2$ of $36.0$ for $13$ ($16-3$) degrees 
of freedom.  The branching
fractions obtained are
\begin{eqnarray}
\label{DHad_results}
  {\cal B}(D^0\to K^-\pi^+)   & = & ( 3.999~\, \pm 0.006~\, \pm 0.031~\, \pm 0.032~\, )\,\%, \\
  {\cal B}(D^0\to \pi^+\pi^-) & = & ( 0.1490 \pm 0.0012 \pm 0.0015 \pm 0.0019 )\,\%, \\
  {\cal B}(D^0\to K^+ K^-)    & = & ( 0.4113 \pm 0.0017 \pm 0.0041 \pm 0.0025 )\,\%\,. 
\end{eqnarray}The uncertainties, estimated as described above, are statistical, 
systematic (excluding FSR), and
FSR modeling.  The correlation coefficients from the fit using the 
total uncertainties are 
\begin{center}
\begin{tabular}{llll}
               & $K^-\pi^+$ & $\pi^+\pi^-$ & $K^+ K^-$ \\
$K^-\pi^+$     &  1.00 & 0.77 & 0.76  \\
$\pi^+\pi^-$   &  0.77 & 1.00 & 0.58  \\
$K^+ K^-$      &  0.76 & 0.58 & 1.00  \\
\end{tabular}
\end{center}

\begin{table}[!htb]
  \centering 
  \caption{Evolution of the $D^0\to K^-\pi^+$ branching fraction from a fit with
  no FSR updates %
  or correlations (similar to the average in the 
  PDG 2018 update~\cite{PDG_2018}) to the nominal fit presented
here.} \label{tab:fit_evolution}
\resizebox{\textwidth}{!}{
\begin{tabular}{cccll}
\hline\hline
Modes &  Description & ${\cal B}(D^0\to K^-\pi^+)$ (\%)  & $\chi^2$/(deg.~of freedom) \\
fit   &              &                                   &     \\ \hline
$K^-\pi^+$ & PDG 2018\,\cite{PDG_2018} equivalent    %
     & $3.931 \pm 0.017 \pm 0.041$ & $~4.5/(8-1)=0.64$ \\
$K^-\pi^+$ & drop Ref.~\cite{Coan:1997ye}  & $3.937 \pm 0.017 \pm 0.041$ & $~4.4/(7-1)=0.73$ \\
$K^-\pi^+$ & add Ref.~\cite{Ablikim:2018ydy}  & $3.913 \pm 0.006 \pm 0.033$ & $~5.1/(8-1)=0.73$ \\
$K^-\pi^+$ & add FSR updates           & $3.948 \pm 0.006 \pm 0.032 \pm 0.019$ & $~3.5/(8-1)=0.50$  \\ %
$K^-\pi^+$ & add FSR correlations          & $3.949 \pm 0.006 \pm 0.032 \pm 0.033$ & $~3.7/(8-1)=0.53$  \\
all        & add CLEO-c, CDF, and FOCUS $h^+ h^-$   & $3.956 \pm 0.006 \pm 0.032 \pm 0.033$ &   $11.1/(14-3)=1.01$  \\
all        & add BES III $h^+ h^-$  & $3.999 \pm 0.006 \pm 0.031 \pm 0.032$ &   $36.0/(16-3)=2.77$  \\
\hline
\end{tabular}  
}
\end{table}
As %
Fig.~\ref{D0bfs} shows, the average 
value for ${\cal B}(D^0\to K^-\pi^+)$ and
the input branching fractions agree very well. 
For the ${\cal B}(D^0\to K^-\pi^+)$ measurements only, the partial $\chi^2$ is 4.9 in the final fit.
With the estimated 
uncertainty in the FSR modeling used here,
the FSR uncertainty dominates the statistical uncertainty 
in the average, suggesting that experimental
work in the near future should focus on verification of FSR with 
$\sum E_\gamma \simge 100$ MeV.  
Note that the systematic uncertainty excluding FSR
has now approached the level of 
the FSR uncertainty; in the most 
precise measurements of these branching fractions, the 
competing
uncertainty is the uncertainty on the tracking efficiency. 

The ${\cal B}(D^0\to
K^+K^-)$ and ${\cal B}(D^0\to \pi^+\pi^-)$ measurements inferred
from the branching ratio measurements %
do not agree as well
(Fig.~\ref{fig:kkpipi}). There is some tension among the results 
when all measurements related to ${\cal B}(D^0 \to K^+K^-)$
and ${\cal B}(D^0\to \pi^+\pi^-)$ are included in the average together.
For the measurements related to ${\cal B}(D^0\to K^+ K^-)$ [${\cal B}(D^0\to \pi^+\pi^-)$] only, the partial $\chi^2$ is 15.7 [6.0] in the final fit.

The ${\cal B}(D^0\to K^-\pi^+)$ average obtained here 
is 
approximately 
four statistical standard deviations 
higher than 
the PDG 2018 update average~\cite{PDG_2018}.
Table~\ref{tab:fit_evolution} shows the evolution from a
fit similar to the PDG fit (no FSR updates %
or correlations, 
reference~\cite{Coan:1997ye} 
included, reference~\cite{Ablikim:2018ydy} not
included)
 to the average presented here.
There are three main contributions to the difference. 
The
branching fraction in reference~\cite{Coan:1997ye} is
low, and its exclusion shifts the result upwards. 
A large shift 
($-0.024\%$) is due to the precision 
of reference~\cite{Ablikim:2018ydy} as it is added; reference~\cite{Ablikim:2018ydy} is a 
considerably lower result than the PDG average before the FSR update.
A subsequently larger shift
($+0.035\%$) is due to the FSR updates, %
which as
expected shift the result upwards, and coincidentally back to compatible with the PDG average. 
The largest shift ($+0.050\%$) occurs as all of the measurements related to ${\cal B}(D^0 \to K^+K^-)$
and ${\cal B}(D^0 \to \pi^+\pi^-)$ are included in the average together with the ${\cal B}(D^0\to K^-\pi^+)$ measurements. 
\begin{figure}
\begin{center}
\includegraphics[width=0.47\textwidth,angle=0.]{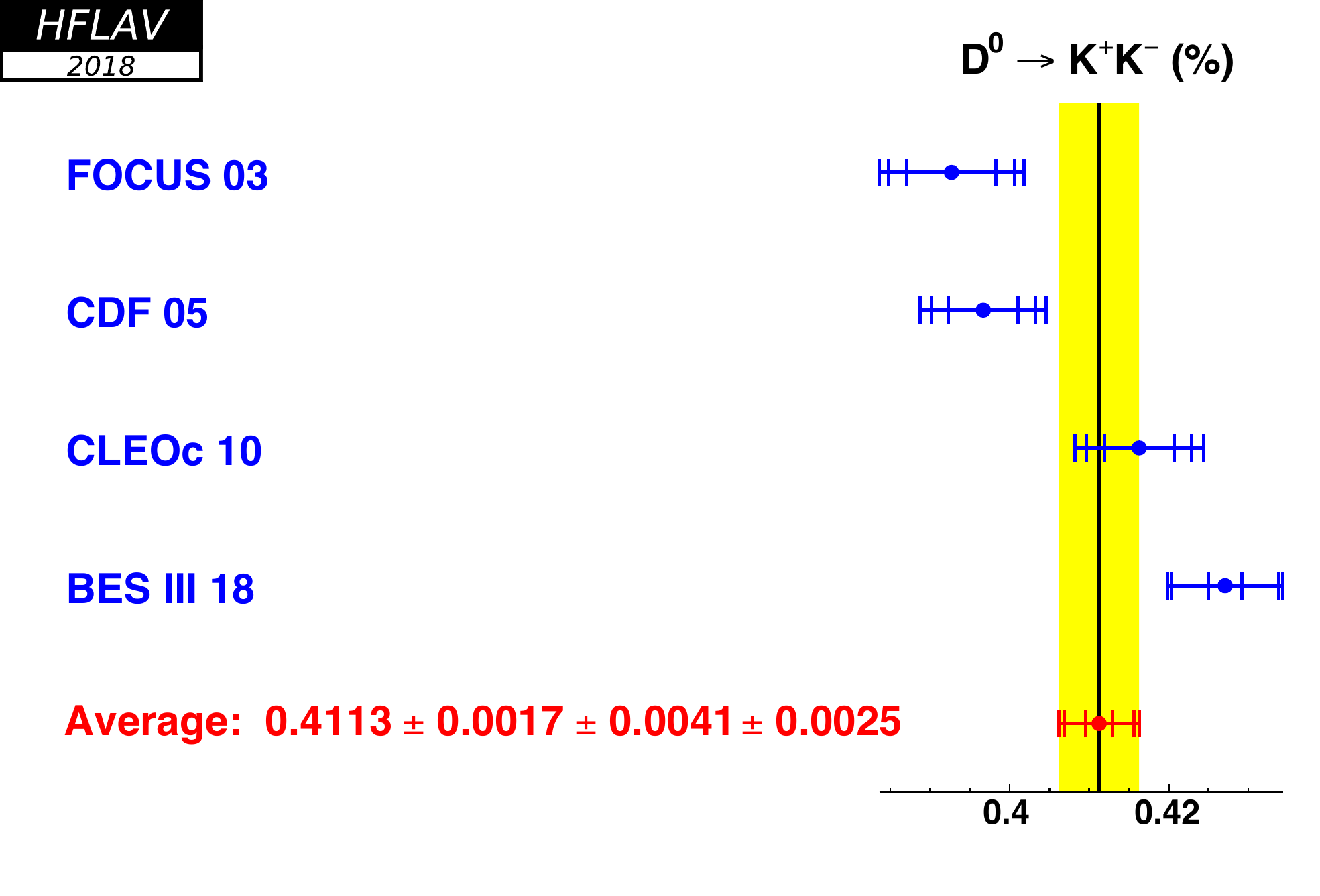}\hfill
\includegraphics[width=0.47\textwidth,angle=0.]{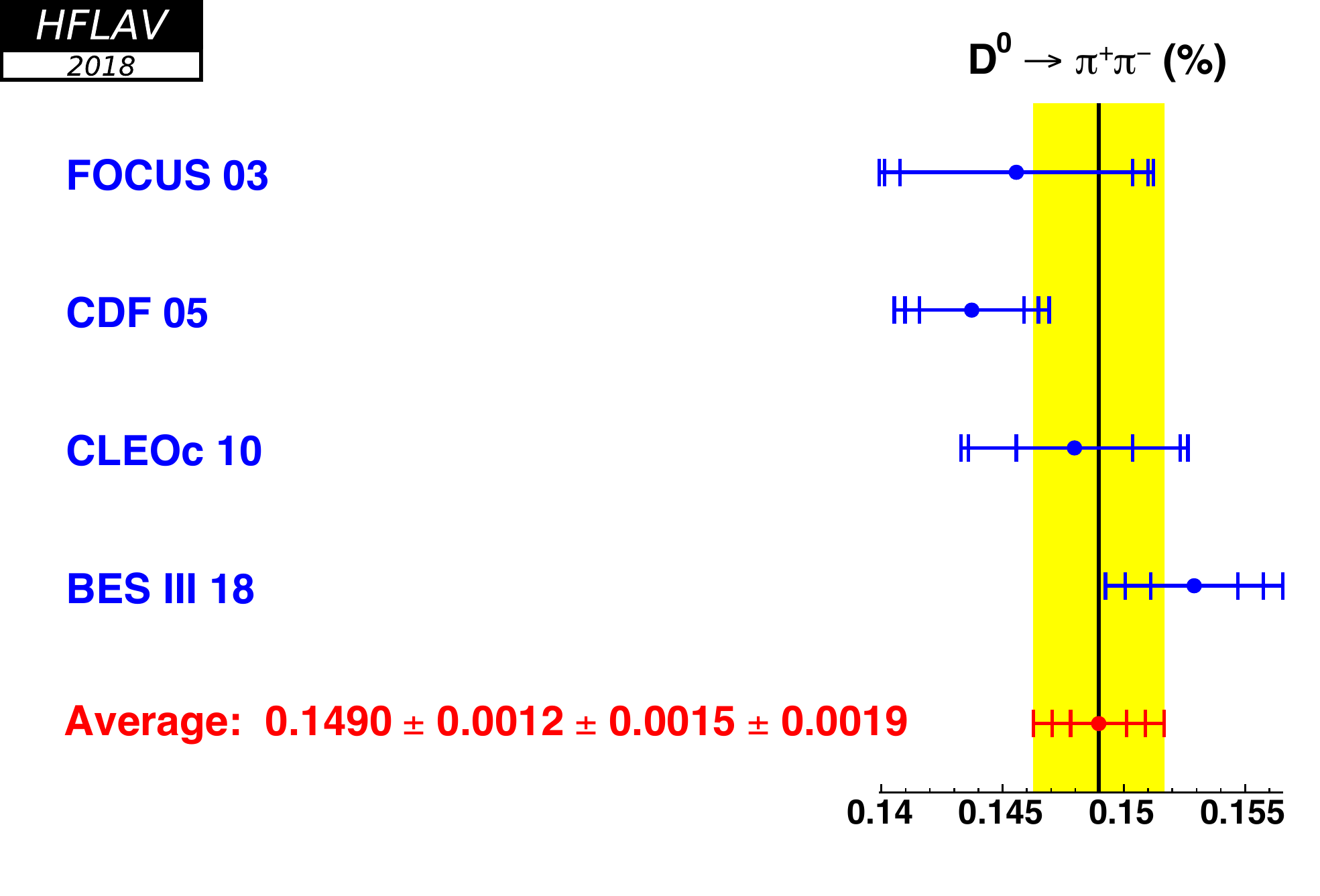}
\caption{The ${\cal B}(D^0\to K^+K^-)$ (left) and ${\cal B}(D^0\to \pi^+\pi^-)$ (right) 
values obtained either from absolute measurements or by scaling the measured branching ratios with the ${\cal B}(D^0\to K^-\pi^+)$ branching fraction
average obtained here.  For the measurements (blue points), the error bars correspond to the statistical, systematic
and either the $K\pi$ normalization uncertainties or, in case of an absolute measurement, the FSR modeling uncertainty.  The average obtained here (red point, yellow band) lists the statistical,
systematics excluding FSR, and the FSR systematic. 
 For the measurements related to ${\cal B}(D^0\to K^+K^-)$ [${\cal B}(D^0 \to \pi^+\pi^-)$] only, the partial $\chi^2$ is 15.7 [6.0] in the final fit.
\label{fig:kkpipi}}
\end{center}
\end{figure}

\subsubsection{Average branching fraction for 
\texorpdfstring{$D^0\to K^+\pi^-$}{D0 to K+ pi-}} 

There is no reason to presume that the effects of FSR should be
different in $D^0\to K^+\pi^-$ and $D^0\to K^-\pi^+$ decays, as both decay to
one charged kaon and one charged pion; indeed, for the same version of PHOTOS
the FSR simulations of these decays are identical. Measurements of the relative
branching fraction ratio between the doubly Cabibbo-suppressed decay
$D^0\to K^+\pi^-$ and the Cabibbo-favored decay $D^0\to K^-\pi^+$
($R_D$, determined in Section~\ref{sec:charm:mixcpv})
have now approached ${\cal O}(1\%)$ relative uncertainties. 
This makes it worthwhile to combine our $R_D$ average with the
${\cal B}(D^0\to K^-\pi^+)$ average obtained in Eq.~(\ref{DHad_results}),
to provide a measurement of the branching fraction:
\begin{eqnarray}
 {\cal B}(D^0\to K^+\pi^-)   & = & ( 1.376 \pm 0.017 ) \times 10^{-4}.%
\end{eqnarray} 
Note that, by definition of $R_D$, these branching fractions do not include
any contribution from Cabibbo-favored $\Dzb \to K^+\pi^-$ decays. 
Our result is more precise than the PDG 2018 value of 
$(1.366 \pm 0.028 ) \times 10^{-4}$~\cite{PDG_2018} due to our using
a more precise value for the ratio $R^{}_D$ (obtained from a global fit
to a range of mixing data, see Section~\ref{sec:charm:mixcpv}).

\subsubsection{Consideration of PHOTOS++}

The versions of PHOTOS that existing measurements were performed with are now well
over a decade out of date. The newest version, PHOTOS++ 3.61~\cite{Davidson:2010ew},
is now fully based on C++ instead of the original FORTRAN. 
None of the measurements used in our branching fraction averages use PHOTOS++, so we have
not yet undertaken an effort to update all results to this newest version. However, at this
time it is worth continuing our procedure to evaluate whether there is any continued low
bias in the the branching fractions, due to sub-optimal modeling of FSR.

We find that the FSR spectra for PHOTOS 2.15, with interference included and exponentiated
multiple-photon mode turned on, and PHOTOS++ (in its default mode) are compatible. The
distributions of $m_{K\pi}$ for simulated $D$ mesons from $B \to D^* X$ decays produced
at $\Upsilon(4S)$ threshold appear to be identical. As an example, the \babar 07 selection
criteria was applied to decays simulated with PHOTOS++ and our nominal version of PHOTOS~2.15;
both produce identical FSR corrections to within 0.01\%. 

The distributions of $\Delta E$ for simulated $D$ mesons produced at $\psi(3770)$ threshold
also appear to be identical. As an example, for the BES III 18
$D^0 \to K^-\pi^+$, $D^0 \to \pi^+\pi^-$, and $D^0 \to K^+K^-$ branching fraction results,
the additional update shifts required to correct from our nominal version of PHOTOS 2.15
to PHOTOS++ are less than or equal to 0.02\%. However, if smearing is applied with the
BES III 18 $\Delta E$ resolution, while the update for $D^0 \to K^-\pi^+$ remains negligible,
the update shifts for $D^0 \to \pi^+\pi^-$ and $D^0 \to K^+K^-$ are modest at -0.25\% and 0.19\%,
respectively; this level of shifts are well within the systematic uncertainty of our averages.

\clearpage
\subsection{Hadronic $D_s$ decays}

For $\dsp$ mesons, most branching fractions are measured relative to
the normalizing channels $\dsp\to K^-K^+\pi^+$ and $\dsp\to \overline{K}{}^0 K^+$.
Thus, it is important to know the absolute branching fractions for 
these modes as precisely as possible. To achieve that, we calculate world 
average values using all relevant measurements and accounting for 
correlations among measurements. In addition, we calculate a world 
average branching fraction for $\dsp \to \eta\pi^+$, for which absolute
branching fraction measurements exist. Other $\dsp$ decay modes 
are either measured relative to one of the normalization modes 
above, or only a single measurement exists (e.g., Ref.~\cite{Ablikim:2019pit}).
We note that the well-known two-body decay modes $\dsp\to \phi\pi^+$ and 
$\dsp\to \overline{K}{}^{*0} K^+$ are subsets of $\dsp\to K^-K^+\pi^+$.

All measurements used are listed in Table~\ref{tab:Dshadronic} and plotted 
along with the resulting world averages in Figs.~\ref{fig:DsKKp}, \ref{fig:DsKK}, 
and \ref{fig:Dsetap}. The measurements of $\br(\dsp\to K^-K^+\pi^+)$ are 
integrated over phase space and thus have uncertainties arising from the 
contributions of intermediate resonances. These are accounted for 
in the systematic uncertainties. For $\dsp\to \overline{K}{}^0 K^+$, 
we use measurements of $\br(\dsp\to K^0_S\,K^+)$ from CLEOc and BESIII, 
and a measurement of $\br(\dsp\to K^0_L\,K^+)$ from BESIII, assuming
$\br(\dsp\to \overline{K}{}^0K^+) = 2\times\br(\dsp\to K^0_S\,K^+)= 
2\times\br(\dsp\to K^0_L\,K^+)$.
The two BESIII measurements are statistically independent but have 
correlated systematic uncertainties; we take these correlations into 
account when calculating the world average. We perform our averaging 
using COMBOS~\cite{Combos:1999}, and the results are
\begin{eqnarray}
\br^{\rm WA}(\dsp\to K^-K^+\pi^+) & = & (5.44\pm0.14)\%\,,\\
\br^{\rm WA}(\dsp\to \overline{K}{}^0 K^+) & = & (2.94\pm0.05)\%\,,\\
\br^{\rm WA}(\dsp\to \eta\pi^+) & = & (1.71\pm0.08)\%\,.
\end{eqnarray}
The uncertainties listed are total uncertainties, i.e., statistical 
plus systematic combined.

\begin{table}[!htb]
\caption{Experimental measurements and world averages for the branching 
fractions $\br(\dsp\to K^-K^+\pi^+)$, $\br(\dsp\to \overline{K}{}^0K^+)$, 
and $\br(\dsp\to \eta\pi^+)$. The first uncertainty listed is 
statistical, and the second is systematic. 
\label{tab:Dshadronic}}
\begin{center}
\begin{tabular}{lclc}
\toprule
Mode & Branching fraction (\%)  & Reference  &	\\ 
\hline
\multirow{3}{*}{$K^-K^+\pi^+$}  
   & $5.78\pm0.20\pm 0.30$     & \babar & \cite{delAmoSanchez:2010jg}  \\ 
   & $5.06\pm0.15\pm 0.21$     & Belle  & \cite{Zupanc:2013byn}        \\
   & $5.55\pm0.14\pm 0.13$     & CLEOc  & \cite{Onyisi:2013bjt}        \\ 
\hline
   & {\boldmath $5.44\pm0.09\pm 0.11$}  & {\bf Average}  &            \\
 & & & \\
\hline
\multirow{2}{*}{$\overline{K}{}^0K^+$}
   & $2.95\pm0.11\pm 0.09$     & Belle  & \cite{Zupanc:2013byn}        \\
   & $3.04\pm0.10\pm 0.06$     & CLEOc  $\dsp\to K^0_S\,K^+$ & \cite{Onyisi:2013bjt}  \\ 
   & $2.850\pm0.076\pm 0.038$  & BESIII $\dsp\to K^0_S\,K^+$ & \cite{Ablikim:2019whl} \\ 
   & $2.970\pm0.078\pm 0.041$  & BESIII $\dsp\to K^0_L\,K^+$ & \cite{Ablikim:2019whl} \\ 
\hline
   & {\boldmath $2.94\pm0.04\pm 0.03$}	& {\bf Average} &             \\
 & & & \\
\hline
\multirow{2}{*}{$\eta\pi^+$}
   & $1.67\pm0.08\pm 0.06$     & CLEOc  & \cite{Onyisi:2013bjt}        \\ 
   & $1.82\pm0.14\pm 0.07$     & Belle  & \cite{Zupanc:2013byn}        \\
\hline
   & {\boldmath $1.71\pm0.07\pm 0.05$}	& {\bf Average} &             \\ 
\bottomrule
\end{tabular}
\end{center}
\end{table}

\begin{figure}
\centering
\vskip-0.70in
\includegraphics[width=0.60\textwidth]{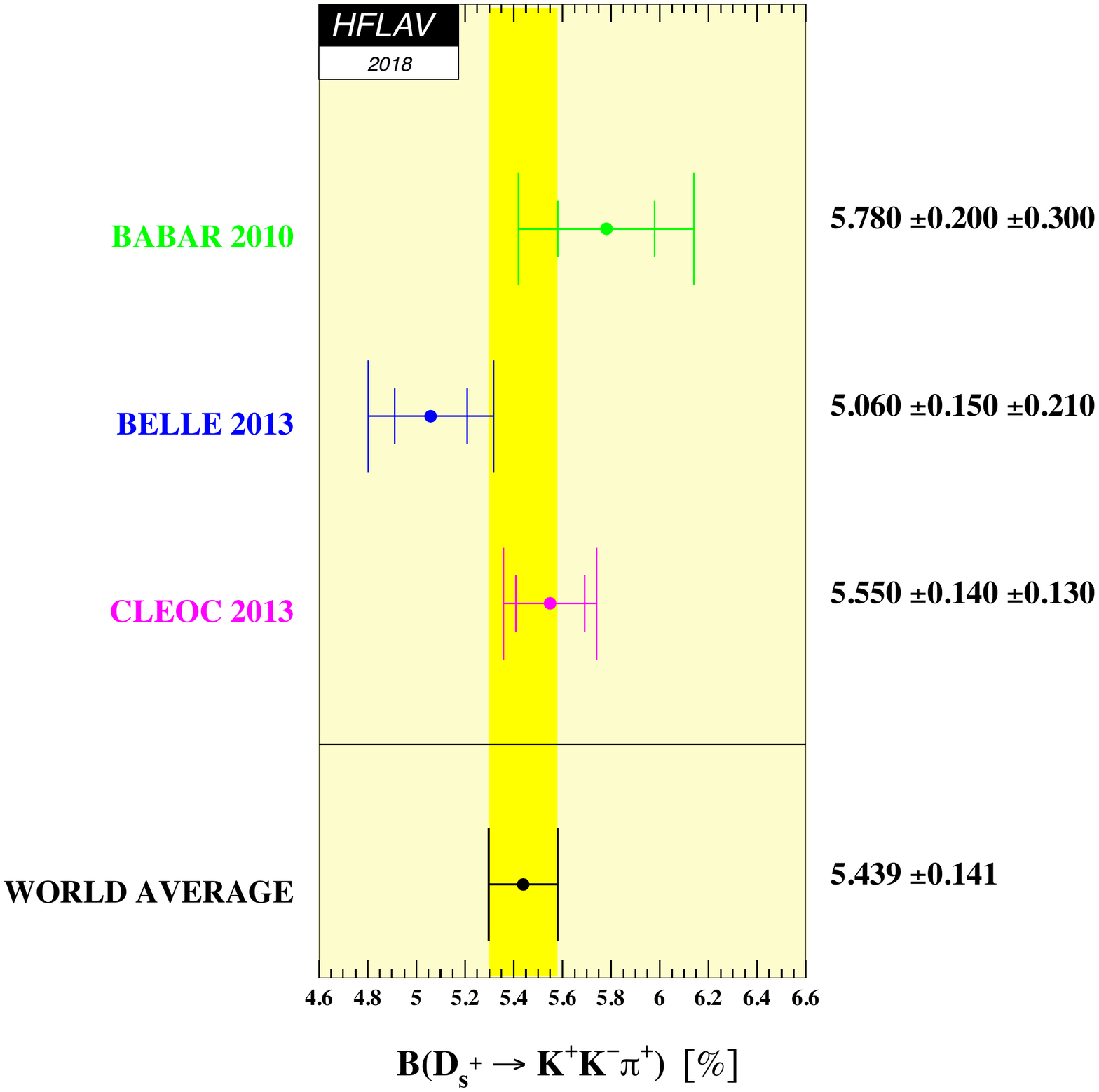}
\vskip-0.60in
\caption{Input values and world average for $\br(\dsp\to K^-K^+\pi^+)$.
The first uncertainty listed is statistical, and the second is systematic.
\label{fig:DsKKp}}
\end{figure}

\begin{figure}
\centering
\vskip-0.50in
\includegraphics[width=0.60\textwidth]{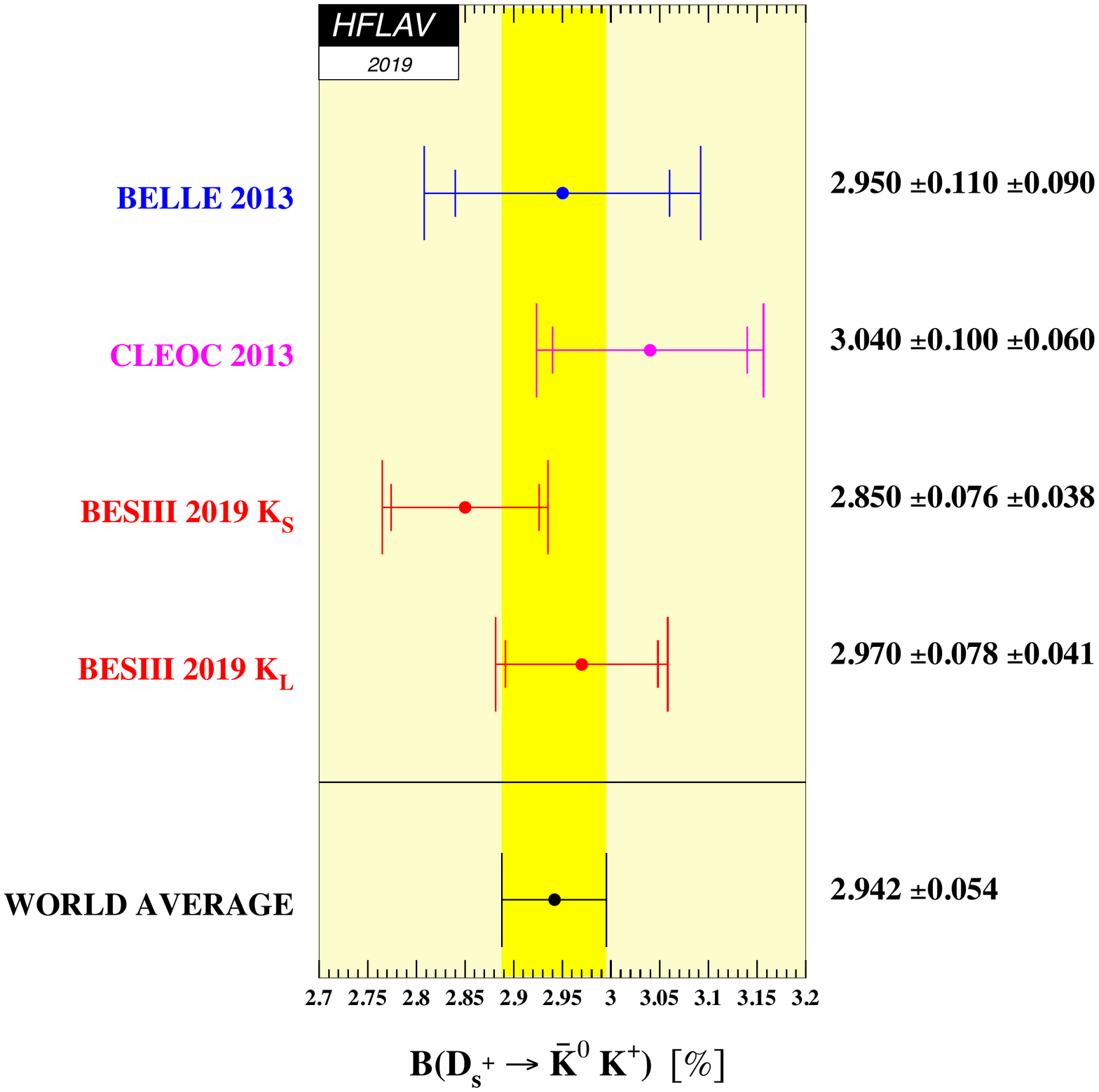}
\vskip-0.60in
\caption{Input values and world average for $\br(\dsp\to \overline{K}{}^0 K^+)$.
The first uncertainty listed is statistical, and the second is systematic.
\label{fig:DsKK}}
\end{figure}

\begin{figure}
\centering
\vskip-0.70in
\includegraphics[width=0.60\textwidth]{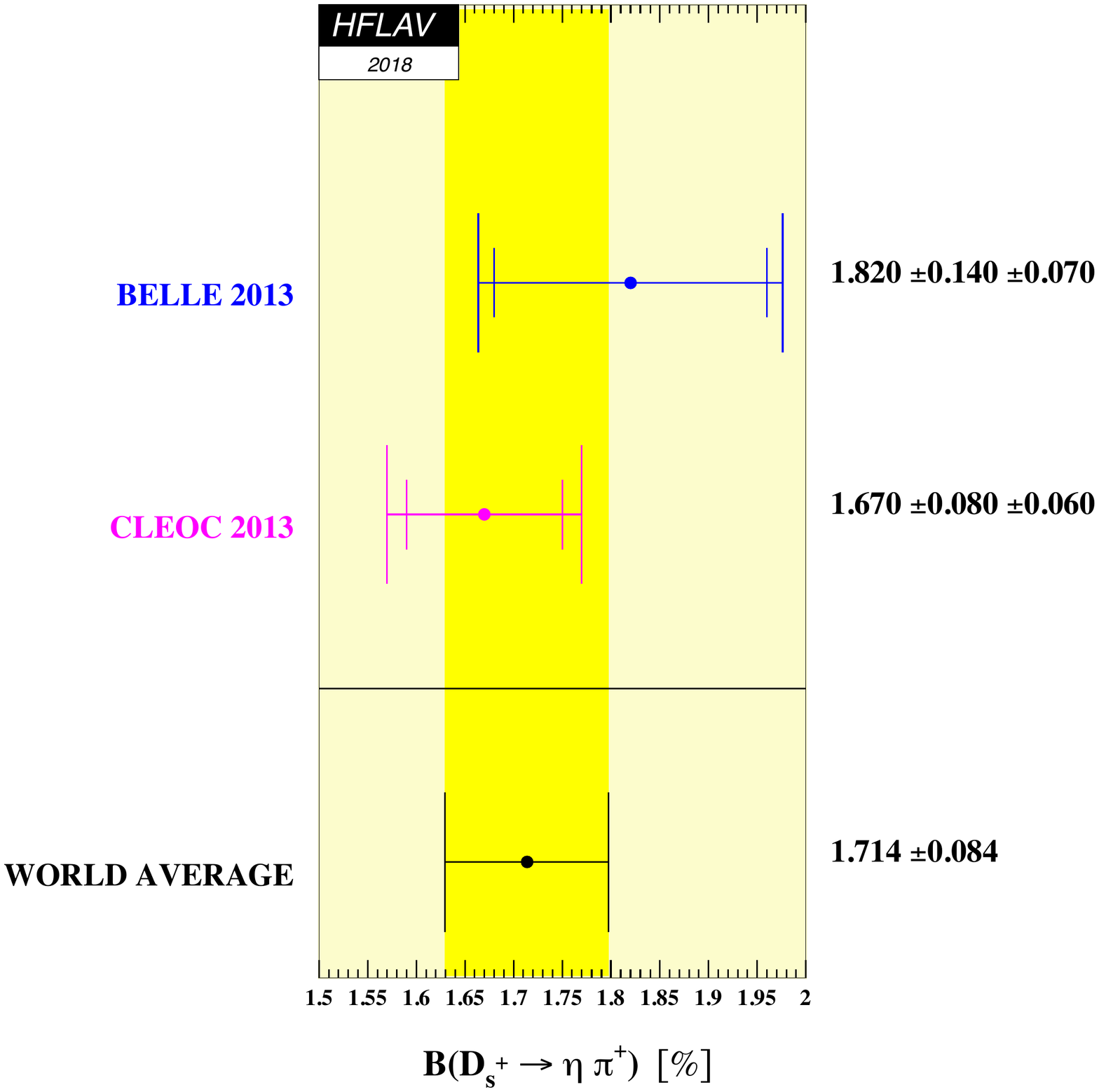}
\vskip-0.60in
\caption{Input values and world average for $\br(\dsp\to \eta\pi^+)$.
The first uncertainty listed is statistical, and the second is systematic.
\label{fig:Dsetap}}
\end{figure}

\clearpage
\mysubsection{Excited $D_{(s)}$ mesons}

Excited ``open'' charm mesons have received increased attention
since the first observation of states that were inconsistent with
QCD predictions~\cite{Aubert:2003fg,Besson:2003cp,Abe:2003jk,Aubert:2003pe}.
Their properties can be measured in both prompt analyses as well as in
amplitude analyses of multi-body $B$ decays.
Tables~\ref{table:charm:spect:D1}, \ref{table:charm:spect:D2}, and 
\ref{table:charm:spect:Ds} summarize the measurements of masses and
widths of excited $D$ and $D_{s}$ states.
If a preferred assignment of spin and parity was measured, 
it is listed in the column $J^{P}$, where the label ``natural'' denotes
$P = (-1)^J$ ($J^{P}=0^{+},1^{-},2^{+}\ldots$) and ``unnatural'' denotes
$P = (-1)^{J+1}$ ($J^{P}=0^{-},1^{+},2^{-}\ldots$). 
In some studies, it was possible to identify only whether the state has
natural or unnatural spin-parity, but not the values of the quantum numbers. 

\begin{table}[htb!]
\caption{\label{table:charm:spect:D1} Measurements of masses 
and widthes for excited $D$ mesons. The column $J^{P}$ lists 
the most significant assignment of spin and parity. If possible,
an average mass or width is calculated.}
\begin{adjustbox}{width=\textwidth,center}
{\setlength\tabcolsep{0pt}
 \begin{tabular}{cp{5pt}cp{5pt}cp{5pt}r@{}lp{10pt}r@{}lp{5pt}cp{5pt}c}
 \toprule
 Resonance &&$J^{P}$ && Decay mode && \multicolumn{2}{c}{Mass [MeV$/c^{2}$]} & & \multicolumn{2}{c}{Width [MeV]} & & \multicolumn{1}{c}{Measured by} && \multicolumn{1}{c}{Reference}
 \\ \midrule 
 \multirow{4}{*}{$D_{0}^{*}(2400)^{0}$} & & \multirow{4}{*}{$0^{+}$} & &$D^{+}\pi^{-}$ & &$ 2297$ &${}\pm8\pm20 $ & &$ 273$ &${}\pm12\pm48 $ & & \babar{} & & \cite{Aubert:2009wg} \\
 & & & &$D^{+}\pi^{-}$ & &$ 2308$ &${}\pm17\pm32 $ & &$ 276$ &${}\pm21\pm63 $ & & Belle & & \cite{Abe:2003zm} \\
 & & & &$D^{+}\pi^{-}$ & &$ 2407$ &${} \pm 21 \pm 35 $ & &$ 240$ &${}\pm55 \pm 59 $ & & FOCUS & & \cite{Link:2003bd} \\ \cmidrule{6-14}
 & & & & & \cellcolor{Gray} & \cellcolor{Gray}$2318$ & \cellcolor{Gray}$.2 \pm 16.9 $ & \cellcolor{Gray} &$ \cellcolor{Gray} 267$ & \cellcolor{Gray}$.4 \pm 35.6 $ & \cellcolor{Gray} & \cellcolor{Gray} Our average & \cellcolor{Gray} & \\ \midrule
 \multirow{4}{*}[-3pt]{$D_{0}^{*}(2400)^{\pm}$} & & \multirow{4}{*}[-3pt]{$0^{+}$} & &$D^{0}\pi^{+}$ & &$ 2349$ &${}\pm6\pm1\pm4 $ & &$ 217$ &${}\pm13\pm5\pm12 $ & & LHCb & & \cite{Aaij:2015sqa} \\
 & & & &$D^{0}\pi^{+}$ &   &$ 2360$ &${}\pm15\pm12\pm28 $ & &$ 255$ &${}\pm26\pm20\pm47 $ & & LHCb & & \cite{Aaij:2015kqa} \\
 & & & &$D^{0}\pi^{+}$ &   &$  2403$ &  ${} \pm 14 \pm35 $ &   &$  283$ &${} \pm24 \pm34 $ &   &   FOCUS($m$ \&$\Gamma$) + Belle($J^{P}$) &   & \cite{Link:2003bd} + \cite{Kuzmin:2006mw} \\ \cmidrule{6-14}
 & & & & & \cellcolor{Gray} & \cellcolor{Gray}$2350$ & \cellcolor{Gray}$.6 \pm 5.9 $ & \cellcolor{Gray} &$ \cellcolor{Gray} 233$ & \cellcolor{Gray}$.7 \pm 15.5 $ & \cellcolor{Gray} & \cellcolor{Gray} Our average & \cellcolor{Gray} & \\ \midrule
 \multirow{11}{*}{$D_{1}^{}(2420)^{0}$} & & \multirow{11}{*}{$1^{+}$} & &$D^{*+}\pi^{-}$ & &$ 2419$ &$.6\pm0.1\pm0.7 $ & &$ 35$ &$.2\pm0.4\pm0.9 $ & & LHCb & & \cite{Aaij:2013sza} \\
 & & & &$D^{*+}\pi^{-}$ & &$ 2423$ &$.1 \pm 1.5 ^{+0.4}_{-1.0} $ & &$ 38$ &$.8\pm5^{+1.9}_{-5.4} $ & & Zeus & & \cite{Abramowicz:2012ys} \\
 & & & &$D^{*+}\pi^{-}$ & &$ 2420$ &$.1 \pm0.1 \pm0.8 $ & &$ 31$ &$.4\pm0.5\pm1.3 $ & & \babar{} & & \cite{delAmoSanchez:2010vq} \\
 & & & &$D^{*+}\pi^{-}$ & &$ $ & & &$ 20$ &$.0\pm1.7\pm1.3 $ & & CDF & & \cite{Abulencia:2005ry} \\
 & & & &$D^{0}\pi^{+}\pi^{-}$ & &$ 2426$ &${}\pm3 \pm1 $ & &$ 24$ &${}\pm7\pm8 $ & & Belle & & \cite{Abe:2004sm} \\
 & & & &$D^{*+}\pi^{-}$ & &$ 2421$ &$.4 \pm1.5 \pm 0.9 $ & &$ 23$ &$.7\pm2.7\pm4.0 $ & & Belle & & \cite{Abe:2003zm} \\
 & & & &$D^{*+}\pi^{-}$ & &$ 2421$ &${}^{+1}_{-2}\pm2 $ & &$ 20$ &${}^{+6}_{-5}{}^{+3}_{-3} $ & & CLEO & & \cite{Avery:1994yc} \\
 & & & &$D^{*+}\pi^{-}$ & &$ 2422$ &${} \pm2 \pm2 $ & &$ 15$ &${}\pm8\pm4 $ & & E687 & & \cite{Frabetti:1993vv} \\
 & & & &$D^{*+}\pi^{-}$ & &$ 2428$ &${}\pm3\pm2 $ & &$ 23$ &${}^{+8}_{-6}{}^{+10}_{-4} $ & & CLEO & & \cite{Avery:1989ui} \\
 & & & &$D^{*+}\pi^{-}$ & &$ 2414$ &${}\pm2\pm5 $ & &$ 13$ &${}\pm6^{+10}_{-5} $ & & ARGUS & & \cite{Albrecht:1989pa} \\
 & & & &$D^{*+}\pi^{-}$ & &$ 2428$ &${} \pm 8 \pm5 $ & &$ 58$ &${}\pm14 \pm10 $ & & TPS & & \cite{Anjos:1988uf} \\ \cmidrule{6-14}
 & & & & & \cellcolor{Gray} &$ \cellcolor{Gray} 2420$ & \cellcolor{Gray}$.5 \pm 0.5 $ & \cellcolor{Gray} &$ \cellcolor{Gray} 31$ & \cellcolor{Gray}$.7 \pm 0.7 $ & \cellcolor{Gray} & \cellcolor{Gray} Our average & \cellcolor{Gray} & \\ \midrule
 \multirow{5}{*}{$D_{1}^{}(2420)^{\pm}$} & & \multirow{5}{*}{$1^{+}$} & &$D^{*0}\pi^{+}$ & &$ 2421$ &$.9\pm4.7^{+3.4}_{-1.2} $ & & & & & Zeus & & \cite{Abramowicz:2012ys} \\
 & & & &$D^{+}\pi^{-}\pi^{+}$ & &$ 2421$ &${}\pm2\pm1 $ & &$ 21$ &${}\pm5\pm8 $ & & Belle & & \cite{Abe:2004sm} \\
 & & & &$D^{*0}\pi^{+}$ & &$ 2425$ &${}\pm2\pm2 $ & &$ 26$ &${}^{+8}_{-7}\pm4 $ & & CLEO & & \cite{Bergfeld:1994af} \\
 & & & &$D^{*0}\pi^{+}$ & &$ 2443$ &${} \pm7\pm5 $ & &$ 41$ &${}\pm19\pm8 $ & & TPS & & \cite{Anjos:1988uf} \\ \cmidrule{6-14}
 & & & & & \cellcolor{Gray} &$ \cellcolor{Gray} 2423$ & \cellcolor{Gray}$.2 \pm 1.6 $\cellcolor{Gray} & \cellcolor{Gray} &$ \cellcolor{Gray} 25$ & \cellcolor{Gray}$.2 \pm 6.0 $ & \cellcolor{Gray} & \cellcolor{Gray} Our average & \cellcolor{Gray} & \\ \midrule
 \multirow{1}{*}{$D_{1}(2430)^{0}$} & & \multirow{1}{*}{$1^{+}$} & &$D^{*+}\pi^{-}$ & \cellcolor{LightGray} &$ \cellcolor{LightGray}2427$ & \cellcolor{LightGray}${}\pm26\pm25 $ & \cellcolor{LightGray} &$ \cellcolor{LightGray} 384$ & \cellcolor{LightGray}${}^{+107}_{-75}\pm74 $ & \cellcolor{LightGray} & \cellcolor{LightGray} Belle & \cellcolor{LightGray} & \cite{Abe:2003zm} \\ \midrule
 \multirow{15}{*}{$D_{2}^{*}(2460)^{0}$} & & \multirow{15}{*}{$2^{+}$} & &$D^{*+}\pi^{-}$ & &$ 2464$ &$.0\pm1.4\pm0.5\pm0.2 $ & &$ 43$ &$.8\pm2.9\pm1.7\pm0.6 $ & & LHCb & & \cite{Aaij:2015vea} \\
 & & & &$D^{*+}\pi^{-}$ & &$ 2460$ &$.4\pm0.4\pm1.2 $ & &$ 43$ &$.2\pm1.2\pm3.0 $ & & LHCb & & \cite{Aaij:2013sza} \\
 & & & &$D^{+}\pi^{-}$ & &$ 2460$ &$.4\pm0.1\pm0.1 $ & &$ 45$ &$.6\pm0.4\pm1.1 $ & & LHCb & & \cite{Aaij:2013sza} \\
 & & & &$D^{*+}\pi^{-}$, $D^{+}\pi^{-}$ & &$ 2462$ &$.5\pm2.4^{+1.3}_{-1.1} $ & &$ 46$ &$.6\pm8.1^{+5.9}_{-3.8} $ & & Zeus & & \cite{Abramowicz:2012ys} \\
 & & & &$D^{+}\pi^{-}$ & &$ 2462$ &$.2\pm0.1\pm0.8 $ & &$ 50$ &$.5\pm0.6\pm0.7 $ & & \babar{} & & \cite{delAmoSanchez:2010vq} \\
 & & & &$D^{+}\pi^{-}$ & &$ 2460$ &$.4\pm1.2\pm2.2 $ & &$ 41$ &$.8\pm2.5\pm2.9 $ & & \babar{} & & \cite{Aubert:2009wg} \\
 & & & &$D^{+}\pi^{-}$ & &$ $ & & &$ 49$ &$.2\pm2.3\pm1.3 $ & & CDF & & \cite{Abulencia:2005ry} \\
 & & & &$D^{+}\pi^{-}$ & &$ 2461$ &$.6\pm2.1\pm3.3 $ & &$ 45$ &$.6\pm4.4\pm6.7 $ & & Belle & & \cite{Abe:2003zm} \\
 & & & &$D^{+}\pi^{-}$ & &$ 2464$ &$.5\pm1.1\pm1.9 $ & &$ 38$ &$.7\pm5.3\pm2.9 $ & & FOCUS & & \cite{Link:2003bd} \\
 & & & &$D^{+}\pi^{-}$ & &$ 2465$ &${}\pm3\pm3 $ & &$ 28$ &${}^{+8}_{-7}\pm6 $ & & CLEO & & \cite{Avery:1994yc} \\
 & & & &$D^{+}\pi^{-}$ & &$ 2453$ &${}\pm3\pm2 $ & &$ 25$ &${}\pm10\pm5 $ & & E687 & & \cite{Frabetti:1993vv} \\
 & & & &$D^{*+}\pi^{-}$ & &$ 2461$ &${}\pm3\pm1 $ & &$ 20$ &${}^{+9}_{-12}{}^{+9}_{-10} $ & & CLEO & & \cite{Avery:1989ui} \\
 & & & &$D^{+}\pi^{-}$ & &$ 2455$ &${}\pm3\pm5 $ & &$ 15$ &${}^{+13}_{-10}{}^{+5}_{-10} $ & & ARGUS & & \cite{Albrecht:1988dj} \\
 & & & &$D^{+}\pi^{-}$ & &$ 2459$ &${}\pm3\pm2$ & &$ 20$ &${}\pm10\pm5$ & & TPS & & \cite{Anjos:1988uf} \\ 
 & & & &$D^{+}\pi^{-}$  & &$ 2463$&$.7\pm0.4\pm0.4\pm0.6$ & &$47$&$.0\pm0.8\pm0.9\pm0.3$ & & LHCb & & \cite{Aaij:2016fma} \\ 
 \cmidrule{6-14} 
 & & & & & \cellcolor{Gray} &$ \cellcolor{Gray} 2460$ & \cellcolor{Gray}$.58 \pm 0.14 $ & \cellcolor{Gray} &$ \cellcolor{Gray} 47$ & \cellcolor{Gray}$.42 \pm 0.57 $ & \cellcolor{Gray} & \cellcolor{Gray} Our average & \cellcolor{Gray} & \\ \bottomrule
 \end{tabular}
}
\end{adjustbox}

\end{table} 

\begin{table}[htb!]
\caption{\label{table:charm:spect:D2} Measurements of masses 
and widthes for excited $D$ mesons. The column $J^{P}$ lists 
the most significant assignment of spin and parity. If possible,
an average mass or width is calculated.}
\begin{adjustbox}{width=\textwidth,center}
{\setlength\tabcolsep{0pt}
 \begin{tabular}{cp{5pt}cp{5pt}cp{5pt}r@{}lp{10pt}r@{}lp{5pt}cp{5pt}c}
 \toprule
 Resonance &&$J^{P}$ && Decay mode && \multicolumn{2}{c}{Mass [MeV$/c^{2}$]} & & \multicolumn{2}{c}{Width [MeV]} & & \multicolumn{1}{c}{Measured by} && \multicolumn{1}{c}{Reference}
 \\ \midrule 
 \multirow{11}{*}{$D_{2}^{*}(2460)^{\pm}$} & & \multirow{11}{*}{$2^{+}$} & &$D^{0}\pi^{+}$ & &$ 2468$ &$.6\pm0.6\pm0.0\pm0.3 $ & &$ 47$ &$.3\pm1.5\pm0.3\pm0.6 $ & & LHCb & & \cite{Aaij:2015sqa} \\
 & & & &$D^{0}\pi^{+}$ & &$ 2465$ &$.6\pm1.8\pm0.5\pm1.2 $ & &$ 46$ &$.0\pm3.4\pm1.4\pm2.9 $ & & LHCb & & \cite{Aaij:2015kqa} \\
 & & & &$D^{0}\pi^{+}$ & &$ 2463$ &$.1\pm0.2\pm0.6 $ & &$ 48$ &$.6\pm1.3\pm1.9 $ & & LHCb & & \cite{Aaij:2013sza} \\
 & & & &$D^{*0}\pi^{+}$, $D^{0}\pi^{+}$ & &$ 2460$ &$.6\pm4.4^{+3.6}_{-0.8} $ & & & & & Zeus & & \cite{Abramowicz:2012ys} \\
 & & & &$D^{0}\pi^{+}$ & &$ 2465$ &$.4\pm0.2\pm1.1 $ & & & & & \babar{} & & \cite{delAmoSanchez:2010vq} \\
 & & & &$D^{0}\pi^{+}$ & &$ 2465$ &$.7\pm1.8^{+1.4}_{-4.8} $ & &$ 49$ &$.7\pm3.8\pm6.4 $ & & Belle & & \cite{Kuzmin:2006mw} \\
 & & & &$D^{0}\pi^{+}$ & &$ 2467$ &$.6\pm1.5\pm0.8 $ & &$ 34$ &$.1\pm6.5\pm4.2 $ & & FOCUS & & \cite{Link:2003bd} \\
 & & & &$D^{0}\pi^{+}$ & &$ 2463$ &${}\pm3\pm3 $ & &$ 27$ &${}^{+11}_{-8}\pm5 $ & & CLEO & & \cite{Bergfeld:1994af} \\
 & & & &$D^{0}\pi^{+}$ & &$ 2453$ &${}\pm3\pm2 $ & &$ 23$ &${}\pm9\pm5 $ & & E687 & & \cite{Frabetti:1993vv} \\
 & & & &$D^{0}\pi^{+}$ & &$ 2469$ &${}\pm4\pm6 $ & & & & & ARGUS & & \cite{Albrecht:1989gb} \\ \cmidrule{6-14}
 & & & & & \cellcolor{Gray} &$ \cellcolor{Gray} 2465$ & \cellcolor{Gray}$.55 \pm 0.40 $ & \cellcolor{Gray} &$ \cellcolor{Gray}46$ & \cellcolor{Gray}$.7 \pm 1.2 $ & \cellcolor{Gray} & \cellcolor{Gray} Our average & \cellcolor{Gray} & \\ \midrule
 \multirow{1}{*}{$D(2550)^{0}$} & & \multirow{1}{*}{$0^{-}$} & &$D^{*+}\pi^{-}$ & \cellcolor{LightGray} &$ \cellcolor{LightGray}2539$ & \cellcolor{LightGray}$.4\pm4.5\pm6.8 $ & \cellcolor{LightGray} &$ \cellcolor{LightGray}130$ & \cellcolor{LightGray}${}\pm12\pm13 $ & \cellcolor{LightGray} & \cellcolor{LightGray} \babar{} & \cellcolor{LightGray} & \cite{delAmoSanchez:2010vq} \\ \midrule
 \multirow{1}{*}{$D(2580)^{0}$} & & \multirow{1}{*}{Unnatural} & &$D^{*+}\pi^{-}$ & \cellcolor{LightGray} &$ \cellcolor{LightGray}2579$ & \cellcolor{LightGray}$.5\pm3.4\pm5.5 $ & \cellcolor{LightGray} &$ \cellcolor{LightGray} 117$ & \cellcolor{LightGray}$.5\pm17.8\pm46.0 $ & \cellcolor{LightGray} & \cellcolor{LightGray} LHCb & \cellcolor{LightGray} & \cite{Aaij:2013sza} \\ \midrule
 \multirow{1}{*}{$D(2600)^{0}$} & & \multirow{1}{*}{Natural} & &$D^{+}\pi^{-}$ & \cellcolor{LightGray} &$ \cellcolor{LightGray}2608$ & \cellcolor{LightGray}$.7\pm2.4\pm2.5 $ & \cellcolor{LightGray} &$ \cellcolor{LightGray} 93$ & \cellcolor{LightGray}${}\pm6\pm13 $ & \cellcolor{LightGray} & \cellcolor{LightGray} \babar{} & \cellcolor{LightGray} & \cite{delAmoSanchez:2010vq} \\ \midrule
 \multirow{1}{*}{$D(2600)^{\pm}$} & & \multirow{1}{*}{Natural} & &$D^{0}\pi^{+}$ & \cellcolor{LightGray} &$ \cellcolor{LightGray}2621$ & \cellcolor{LightGray}$.3\pm3.7\pm4.2 $ & \cellcolor{LightGray} & \cellcolor{LightGray} & \cellcolor{LightGray} & \cellcolor{LightGray} & \cellcolor{LightGray} \babar{} & \cellcolor{LightGray} & \cite{delAmoSanchez:2010vq} \\ \midrule
 \multirow{1}{*}{$D^{*}(2640)^{\pm}$} & & \multirow{1}{*}{$1^{-}$} & &$D^{*+}\pi^{+}\pi^{-}$ & \cellcolor{LightGray} &$ \cellcolor{LightGray}2637$ & \cellcolor{LightGray}${}\pm2\pm6 $ & \cellcolor{LightGray} & \cellcolor{LightGray} & \cellcolor{LightGray} & \cellcolor{LightGray} & \cellcolor{LightGray} Delphi & \cellcolor{LightGray} & \cite{Abreu:1998vk} \\ \midrule
 \multirow{1}{*}{$D^{*}(2650)^{0}$} & & \multirow{1}{*}{Natural} & &$D^{*+}\pi^{-}$ & \cellcolor{LightGray} &$ \cellcolor{LightGray}2649$ & \cellcolor{LightGray}$.2\pm3.5\pm3.5 $ & \cellcolor{LightGray} & \cellcolor{LightGray}$ \cellcolor{LightGray}140$ & \cellcolor{LightGray}$.2\pm17.1\pm18.6 $ & \cellcolor{LightGray} & \cellcolor{LightGray} LHCb & \cellcolor{LightGray} & \cite{Aaij:2013sza} \\ \midrule
 \multirow{1}{*}{$D^{*}_{1}(2680)^{0}$} & & \multirow{1}{*}{$1^{-}$} & &$D^{+}\pi^{-}$ & \cellcolor{LightGray} &$ \cellcolor{LightGray}2681$ & \cellcolor{LightGray}$.1\pm5.6\pm4.9\pm13.1 $ & \cellcolor{LightGray} & \cellcolor{LightGray}$ \cellcolor{LightGray}186$ & \cellcolor{LightGray}$.7\pm8.5\pm8.6\pm8.2 $ & \cellcolor{LightGray} & \cellcolor{LightGray} LHCb & \cellcolor{LightGray} & \cite{Aaij:2016fma} \\ \midrule
 \multirow{1}{*}{$D(2740)^{0}$} & & \multirow{1}{*}{Unnatural} & &$D^{*+}\pi^{-}$ & \cellcolor{LightGray} &$ \cellcolor{LightGray} 2737$ & \cellcolor{LightGray}$.0\pm3.5\pm11.2 $ & \cellcolor{LightGray} & \cellcolor{LightGray}$ \cellcolor{LightGray} 73$ & \cellcolor{LightGray}$.2\pm13.4\pm25.0 $ & \cellcolor{LightGray} & \cellcolor{LightGray} LHCb & \cellcolor{LightGray} & \cite{Aaij:2013sza} \\ \midrule
 \multirow{1}{*}{$D(2750)^{0}$} & & \multirow{1}{*}{} & &$D^{*+}\pi^{-}$ & \cellcolor{LightGray} &$ \cellcolor{LightGray} 2752$ & \cellcolor{LightGray}$.4\pm1.7\pm2.7 $ & \cellcolor{LightGray} & \cellcolor{LightGray}$ \cellcolor{LightGray} 71$ & \cellcolor{LightGray}${}\pm6\pm11 $ & \cellcolor{LightGray} & \cellcolor{LightGray} \babar{} & \cellcolor{LightGray} & \cite{delAmoSanchez:2010vq} \\ \midrule
 \multirow{5}{*}{$D^{*}_{1}(2760)^{0}$} & & \multirow{5}{*}{$1^+$} & &$D^{+}\pi^{-}$ & &$ 2781$ &${}\pm18\pm11\pm6 $ & &$ 177$ &${}\pm32\pm20\pm7 $ & & LHCb & & \cite{Aaij:2015vea} \\
 & & & &$D^{*+}\pi^{-}$ & &$ 2761$ &$.1\pm5.1\pm6.5 $ & &$ 74$ &$.4\pm3.4\pm37.0 $ & & LHCb & & \cite{Aaij:2013sza} \\
 & & & &$D^{+}\pi^{-}$ & &$ 2760$ &$.1\pm1.1\pm3.7 $ & &$ 74$ &$.4\pm3.4\pm19.1 $ & & LHCb & & \cite{Aaij:2013sza} \\
 & & & &$D^{+}\pi^{-}$ & &$ 2763$ &$.3\pm2.3\pm2.3 $ & &$ 60$ &$.9\pm5.1\pm3.6 $ & & \babar{} & & \cite{delAmoSanchez:2010vq} \\ \cmidrule{6-14}
 & & & & & \cellcolor{Gray} &$ \cellcolor{Gray} 2762$ & \cellcolor{Gray}$.1 \pm 2.4 $ & \cellcolor{Gray} &$ \cellcolor{Gray} 65$ & \cellcolor{Gray}$.1 \pm 5.8 $ & \cellcolor{Gray} & \cellcolor{Gray} Our average & \cellcolor{Gray} & \\ \midrule
 \multirow{1}{*}{$D^{*}_{3}(2760)^{0}$} & &  \multirow{1}{*}{$3^-$} & &$D^{+}\pi^{-}$ &\cellcolor{LightGray} &$\cellcolor{LightGray} 2775$&\cellcolor{LightGray}$.5\pm4.5\pm4.5\pm4.7$ & \cellcolor{LightGray}&$ \cellcolor{LightGray} 95$&\cellcolor{LightGray}$.3\pm9.6\pm7.9\pm33.1$ & \cellcolor{LightGray}&\cellcolor{LightGray} LHCb &\cellcolor{LightGray} & \cite{Aaij:2016fma}\\ \midrule
 \multirow{4}{*}{$D^{*}_{3}(2760)^{\pm}$} & & \multirow{4}{*}{$3^-$} & &$D^{0}\pi^{+}$ & &$ 2798$ &${}\pm7\pm1\pm7$ & &$ 105$ &${}\pm18\pm6\pm23 $ & & LHCb & & \cite{Aaij:2015sqa} \\
 & & & &$D^{0}\pi^{+}$ & &$ 2771$ &$.7\pm1.7\pm3.8 $ & &$ 66$ &$.7\pm6.6\pm10.5 $ & & LHCb & & \cite{Aaij:2013sza} \\
 & & & &$D^{0}\pi^{+}$ & &$ 2769$ &$.7\pm3.8\pm1.5 $ & &$ $ & & & \babar{} & & \cite{delAmoSanchez:2010vq} \\ \cmidrule{6-14}
 & & & & & \cellcolor{Gray} &$ \cellcolor{Gray} 2773$ & \cellcolor{Gray}$.9 \pm 3.3 $ & \cellcolor{Gray} &$ \cellcolor{Gray} 72$ & \cellcolor{Gray}$.3\pm 11.5 $ & \cellcolor{Gray} & \cellcolor{Gray} Our average & \cellcolor{Gray} & \\ \midrule
  \multirow{1}{*}{$D^{*}_{2}(3000)^{0}$} & &  \multirow{1}{*}{$2^+$} & &$D^{+}\pi^{-}$ &\cellcolor{LightGray} &$\cellcolor{LightGray} 3214$&\cellcolor{LightGray}${}\pm29\pm33\pm36$ & \cellcolor{LightGray}&$ \cellcolor{LightGray} 186$&\cellcolor{LightGray}${}\pm38\pm34\pm63$ & \cellcolor{LightGray}&\cellcolor{LightGray} LHCb &\cellcolor{LightGray} & \cite{Aaij:2016fma}\\
 \bottomrule
 \end{tabular}
}
\end{adjustbox}

\end{table} 

\begin{table}[htb!]
\caption{\label{table:charm:spect:Ds} Measurements of masses 
and widthes for excited $D^{}_s$ mesons. The column $J^{P}$ lists 
the most significant assignment of spin and parity. If possible,
an average mass or width is calculated.}
\begin{adjustbox}{width=\textwidth,center}
{\setlength\tabcolsep{0pt}
 \begin{tabular}{cp{5pt}cp{5pt}cp{5pt}r@{}lp{5pt}r@{}lp{5pt}cp{5pt}c}
 \toprule
 Resonance && $J^{P}$ && Decay mode && \multicolumn{2}{c}{Mass [MeV$/c^{2}$]} && \multicolumn{2}{c}{Width [MeV]} && \multicolumn{1}{c}{Measured by} && \multicolumn{1}{c}{Reference} 
 \\ \midrule
 \multirow{3}{*}{$D_{s0}^{*}(2317)^{\pm}$} && \multirow{3}{*}{$0^{+}$} && $D_{s}^{+}\pi^{0}$ &&$ 2319$&$.6\pm0.2\pm1.4 $&&$ $&$ $&& \babar{} && \cite{Aubert:2006bk} \\
 && && $D_{s}^{+}\pi^{0}$ &&$ 2317$&$.3\pm0.4\pm0.8 $&&$ $&$ $&& \babar{} && \cite{Aubert:2003pe} \\ 
 && && $D_{s}^{+}\pi^{0}$ &&$ 2318$&$.3\pm1.2\pm1.2 $&&$ $&$ $&& BESIII && \cite{Ablikim:2017rrr} \\   
 \cmidrule{6-14}
 && && &\cellcolor{Gray}&$ \cellcolor{Gray} 2318$&\cellcolor{Gray}$.01 \pm 0.69 $&\cellcolor{Gray}& \cellcolor{Gray}&\cellcolor{Gray} &\cellcolor{Gray}& \cellcolor{Gray} Our average &\cellcolor{Gray}& \\ \midrule

 \multirow{3}{*}{$D_{s1}(2460)^{\pm}$} && \multirow{3}{*}{$1^{+}$} && $D_{s}^{\ast}{}^{+}\pi^{0}, D_{s}^{+}\pi^{0}\gamma, D_{s}^{+}\gamma, D_{s}^{+}\pi^{+}\pi^{-}$ &&$ 2460$&$.1\pm0.2\pm0.8 $&&$ $&$ $&& \babar{} && \cite{Aubert:2006bk} \\
 && && $D_{s}^{+}\pi^{0}\gamma$ &&$ 2458$&${}\pm1.0\pm1.0 $&&$ $&$ $&& \babar{} && \cite{Aubert:2003pe} \\ \cmidrule{6-14}
 && && &\cellcolor{Gray}&$ \cellcolor{Gray} 2459$&\cellcolor{Gray}$.6 \pm 0.7 $&\cellcolor{Gray}& \cellcolor{Gray}&\cellcolor{Gray} &\cellcolor{Gray}& \cellcolor{Gray} Our average &\cellcolor{Gray}& \\ \midrule
 \multirow{11}{*}{$D_{s1}(2536)^{\pm}$} && \multirow{11}{*}{$1^{+}$} && $D^{*}{}^{+}K_{S}^{0}$ &&$ 2535$&$.7\pm0.6\pm0.5 $&&$ $&$ $&& D\O\ &&\cite{Abazov:2007wg} \\
 && && $D^{*}{}^{+}K_{S}^{0}, D^{*}{}^{0}K^{+}$ &&$ 2534$&$.78\pm0.31\pm0.40 $&&$ $&$ $&& \babar{} && \cite{Aubert:2007rva} \\
 && && $D_{s}^{+}\pi^{+}\pi^{-}$ &&$ 2534$&$.6\pm0.3\pm0.7 $&&$ $&$ $&& \babar{} && \cite{Aubert:2006bk} \\
 && && $D^{*}{}^{+}K_{S}^{0}, D^{*}{}^{0}K^{+}$ &&$ 2535$&$.0\pm0.6\pm1.0 $&&$ $&$ $&& E687 && \cite{Frabetti:1993vv} \\
 && && $D^{*}{}^{0}K^{+}$ &&$ 2535$&$.3\pm0.2\pm0.5 $&&$ $&$ $&& CLEO && \cite{Alexander:1993nq} \\
 && && $D^{*}{}^{+}K_{S}^{0}$ &&$ 2534$&$.8\pm0.6\pm0.6 $&&$ $&$ $&& CLEO && \cite{Alexander:1993nq} \\
 && && $D^{*}{}^{0}K^{+}$ &&$ 2535$&$.2\pm0.5\pm1.5 $&&$ $&$ $&& ARGUS && \cite{Albrecht:1992zh} \\
 && && $D^{*}{}^{+}K_{S}^{0}$ &&$ 2535$&$.6\pm0.7\pm0.4 $&&$ $&$ $&& CLEO && \cite{Avery:1989ui} \\
 && && $D^{*}{}^{+}K_{S}^{0}$ &&$ 2535$&$.9\pm0.6\pm2.0 $&&$ $&$ $&& ARGUS && \cite{Albrecht:1989yi} \\
 && && $D^{*}{}^{+}K_{S}^{0}$ &&$ $&$ $&&$ 0$&$.92\pm0.03\pm0.04 $&& \babar{} && \cite{Lees:2011um} \\ 
 && && %
    &&$ 2537$&$.7\pm0.5\pm3.1 $&&$ 1$&$.7\pm1.2\pm0.6$ && BESIII && \cite{Ablikim:2018qjv} \\ 
 \cmidrule{6-14}
 && && &\cellcolor{Gray}&$ \cellcolor{Gray}2535$&\cellcolor{Gray}$.12 \pm 0.26 $&\cellcolor{Gray}&$ \cellcolor{Gray}0$&\cellcolor{Gray}$.92\pm0.05 $&\cellcolor{Gray}& \cellcolor{Gray} Our average &\cellcolor{Gray}& \\ \midrule
 
 \multirow{6}{*}[-5pt]{$D_{s2}^{*}(2573)^{\pm}$} && \multirow{6}{*}[-5pt]{$2^{+}$} && $D^{0}K^{+},D^{*+}K_{S}^{0}$ &&$ 2568$&$.39\pm0.29\pm0.26 $&&$ 16$&$.9\pm0.5\pm0.6 $&& LHCb && \cite{Aaij:2014baa} \\
 && && $D^{+}K_{S}^{0}, D^{0}K^{+}$ &&$ 2569$&$.4\pm1.6\pm0.5 $&&$ 12$&$.1\pm4.5\pm1.6 $&& LHCb && \cite{Aaij:2011ju} \\
 && && $D^{+}K_{S}^{0}, D^{0}K^{+}$ &&$ 2572$&$.2\pm0.3\pm1.0 $&&$ 27$&$.1\pm0.6\pm5.6 $&& \babar{} && \cite{Aubert:2006mh} \\
 && && $D^{0}K^{+}$ &&$ 2574$&$.25\pm3.3\pm1.6 $&&$ 10$&$.4\pm8.3\pm3.0 $&& ARGUS && \cite{Albrecht:1995qx} \\
 && && $D^{0}K^{+}$ &&$ 2573$&$.2 _{-1.6}^{+1.7}\pm0.9 $&&$ 16$&${}_{-4}^{+5}\pm3 $&& CLEO && \cite{Kubota:1994gn} \\ 
 && && %
    &&$ 2570$&$.7\pm2.0\pm1.7$ &&$ 17$&$.2\pm3.6\pm1.1 $ && BESIII && \cite{Ablikim:2018qjv} \\
 \cmidrule{6-14}
 && && &\cellcolor{Gray}&$ \cellcolor{Gray}2569$&\cellcolor{Gray}$.10 \pm 0.35 $&\cellcolor{Gray}&$ \cellcolor{Gray} 16$&\cellcolor{Gray}$.91 \pm 0.74 $&\cellcolor{Gray}& \cellcolor{Gray}Our average &\cellcolor{Gray}& \\ \midrule
 
 \multirow{6}{*}[-2pt]{$D_{s1}^{*}(2700)^{\pm}$} && \multirow{6}{*}[-2pt]{$1^{-}$} && $D^{*}{}^{+}K_{S}^{0}, D^{*}{}^{0}K^{+}$ &&$ 2732$&$.3\pm4.3\pm5.8 $&&$ 136$&${}\pm19\pm24 $&& LHCb && \cite{Aaij:2016utb} \\
  && && $D^{0}K^{+}$ &&$ 2699$&${}^{+14}_{-7} $&&$ 127$&$_{-19}^{+24} $&& \babar{}  && \cite{Lees:2014abp} \\
 && && $D^{*}{}^{+}K_{S}^{0}, D^{*}{}^{0}K^{+}$ &&$ 2709$&$.2\pm1.9\pm4.5 $&&$ 115$&$.8\pm7.3\pm12.1 $&& LHCb && \cite{Aaij:2012pc} \\
 && && $DK, D^{*}K$ &&$ 2710$&${}\pm 2 _{-7}^{+12} $&&$ 149$&${}\pm7 _{-52}^{+39} $&& \babar{} && \cite{Aubert:2009ah} \\
 && && $D^{0}K^{+}$ &&$ 2708$&${}\pm9 _{-10}^{+11} $&&$ 108$&${}\pm2_{-31}^{+36} $&& Belle && \cite{Brodzicka:2007aa} \\ \cmidrule{6-14}
 && && &\cellcolor{Gray}&$ \cellcolor{Gray} 2712$&\cellcolor{Gray}$.0 \pm 1.5 $&\cellcolor{Gray}&$ \cellcolor{Gray} 121$&\cellcolor{Gray}$.5 \pm 10.2 $&\cellcolor{Gray}& \cellcolor{Gray} Our average &\cellcolor{Gray}& \\ \midrule
 \multirow{1}{*}{$D_{s1}^{*}(2860)^{\pm}$} && \multirow{1}{*}{1} && $ D^{0}K^{+}$ &\cellcolor{LightGray} &$ \cellcolor{LightGray} 2859$&\cellcolor{LightGray}${}\pm12\pm24 $&\cellcolor{LightGray}&$ \cellcolor{LightGray} 159$&\cellcolor{LightGray}${}\pm23\pm77 $&\cellcolor{LightGray}& \cellcolor{LightGray} LHCb &\cellcolor{LightGray}& \cite{Aaij:2014xza} \\ \midrule
 \multirow{3}{*}{$D_{s3}^{*}(2860)^{\pm}$} && \multirow{3}{*}{$3^-$}  &&  $D^{*}{}^{+}K_{S}^{0}, D^{*}{}^{0}K^{+}$ &&$  2867$&$.1\pm4.3\pm1.9 $&&$  50$&${}\pm11\pm13 $&&  LHCb && \cite{Aaij:2016utb} \\ 
 && && $ D^{0}K^{+}$ &&$  2860$&$.5\pm2.6\pm6.5 $&&$  53$&${}\pm7\pm7 $&&  LHCb && \cite{Aaij:2014xza} \\  \cmidrule{6-14}
 && && &\cellcolor{Gray}&$ \cellcolor{Gray} 2865$&\cellcolor{Gray}$.0 \pm 3.9 $&\cellcolor{Gray}&$ \cellcolor{Gray} 52$&\cellcolor{Gray}$.2 \pm 8.6 $&\cellcolor{Gray}& \cellcolor{Gray} Our average &\cellcolor{Gray}& \\ \midrule
 \multirow{1}{*}{$D_{sJ}(3040)^{\pm}$} && \multirow{1}{*}{Unnatural} && $D^{*}K$ &\cellcolor{LightGray}&$ \cellcolor{LightGray} 3044$&\cellcolor{LightGray}${}\pm8 _{-5}^{+30} $&\cellcolor{LightGray}&$ \cellcolor{LightGray}239$&\cellcolor{LightGray}${}\pm35 _{-42}^{+46} $&\cellcolor{LightGray}& \cellcolor{LightGray} \babar{} ($m$ \& $\Gamma$) + LHCb($J^{P}$)&\cellcolor{LightGray}& \cite{Aubert:2009ah}+\cite{Aaij:2016utb} \\ \bottomrule
\end{tabular}
}
\end{adjustbox}

\end{table}

For states in which multiple measurements are available, an average mass and width
are calculated; these are listed in the gray shaded rows. For simplicity, when calculating
averages we neglect possible correlations among individual measurements. All averaged masses
and widths are summarized in Figure~\ref{fig:charm:spect:1}. The resonances listed in the
tables and figures are as they appear in the respective publications. In some cases, it
is unclear whether separately listed states are in fact distinct or are the same resonance.
An example is the recently observed $D^{*}_{1}(2680)^{0}$ state~\cite{Aaij:2016fma}, which
has parameters close to those of the $D^{*}(2650)^{0}$. Further measurements are needed
to resolve these ambiguities.

\begin{figure}[htb!]
\begin{centering}
\includegraphics[width=0.49\textwidth]{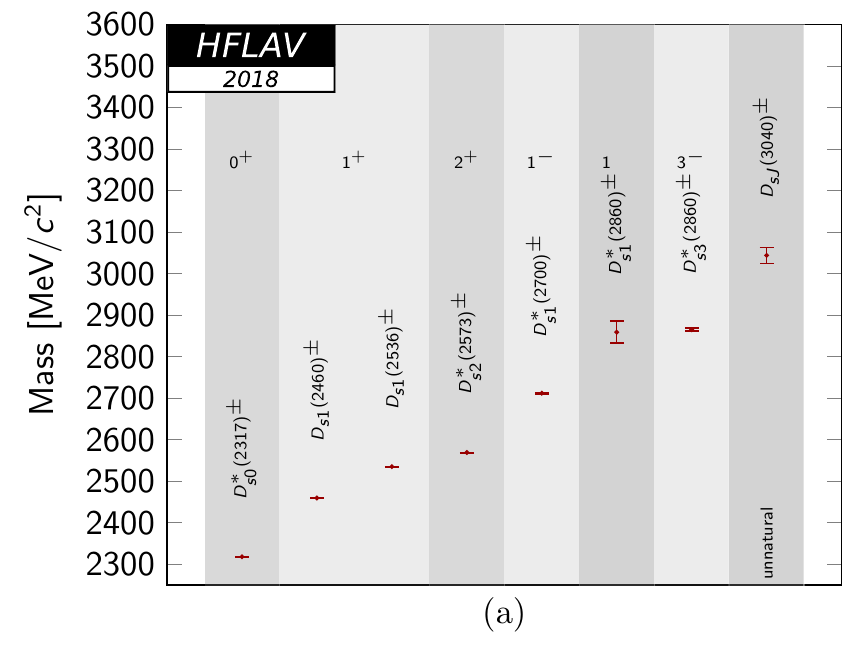}
\includegraphics[width=0.49\textwidth]{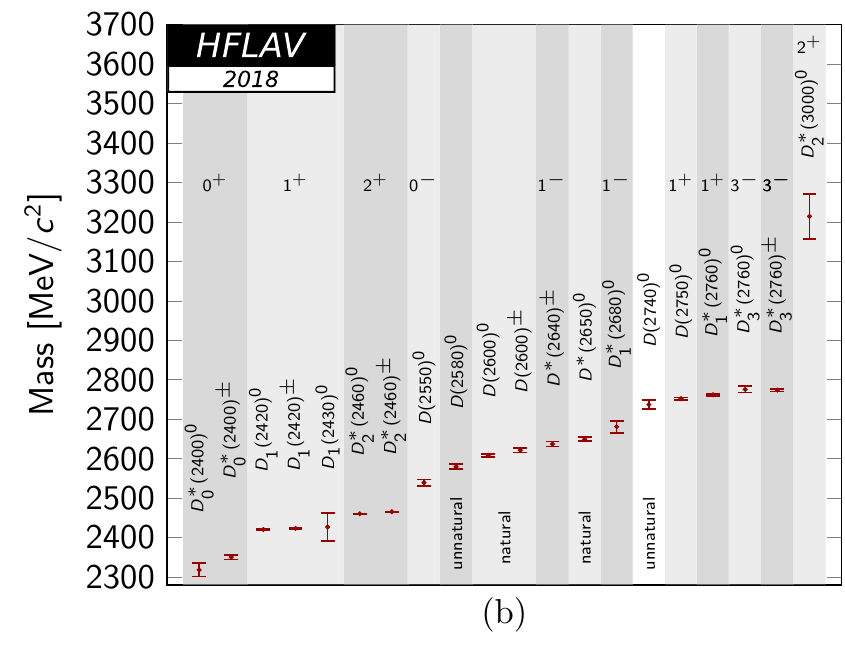}\\
\includegraphics[width=0.49\textwidth]{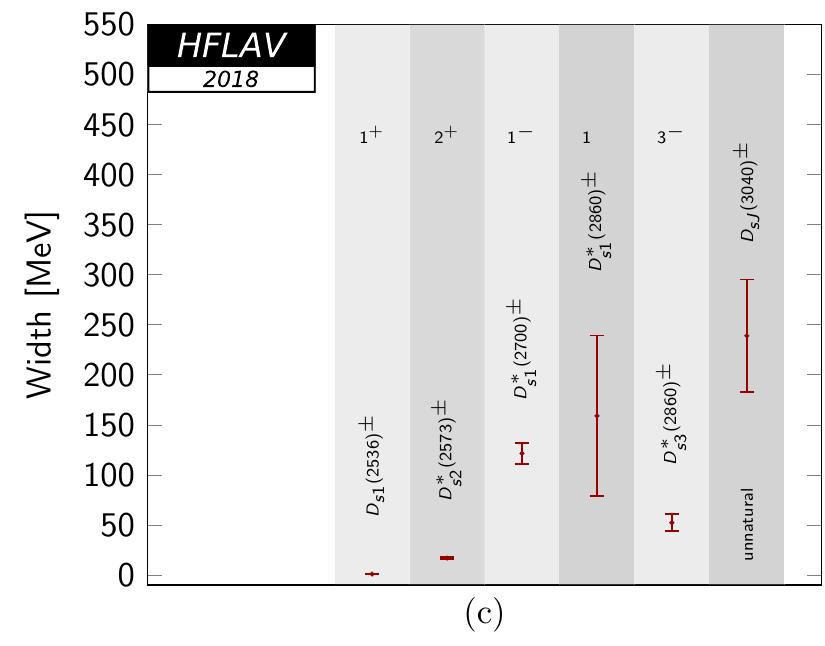}
\includegraphics[width=0.49\textwidth]{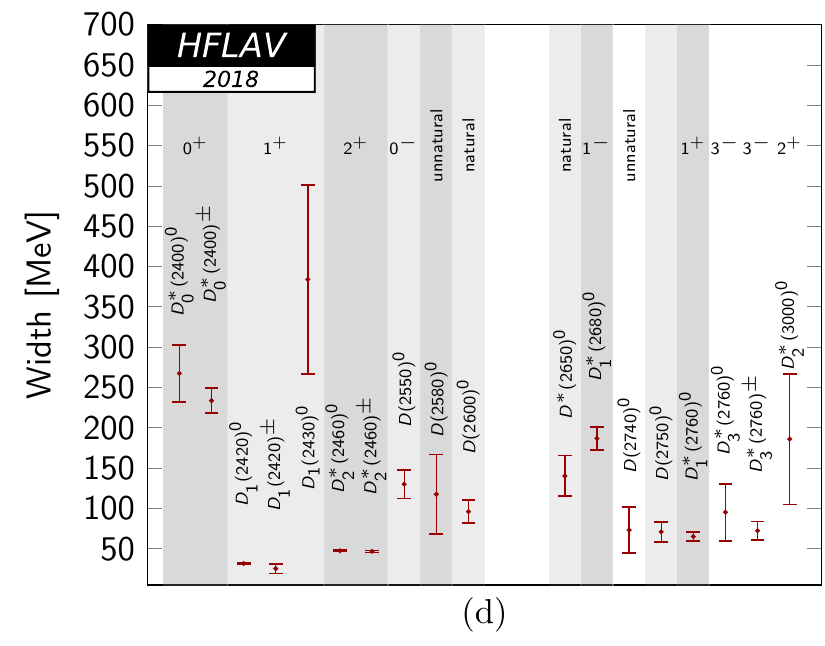}
\caption{\label{fig:charm:spect:1}
(a) Average masses for excited $D_{s}$ mesons;
(b) average masses for excited $D$ mesons;
(c) average widths for excited $D_{s}$ mesons;
(d) average widths for excited $D$ mesons.
The vertical shaded regions distinguish between different spin-parity states.}
\end{centering}
\end{figure}

The masses and widths of narrow ($\Gamma<50$~MeV) orbitally excited 
$D$ mesons (1P states), both neutral and charged, are 
well-established. Measurements of broad states ($\Gamma\sim$ 200--400~MeV) 
are less abundant, as identifying the signal is more challenging. There is 
a $\sim\!2\sigma$ difference between the $D_0^\ast(2400)^0$ masses measured by 
Belle~\cite{Abe:2003zm} and \babar~\cite{Aubert:2009wg} (which are in
good agreement) and that measured by the FOCUS~\cite{Link:2003bd} experiment.
No data exist yet for the $D_1(2430)^{\pm}$ state. Dalitz plot analyses of 
$B\to \overline{D}{}^{(\ast)}\pi\pi$ decays strongly favor the assignments $0^+$ 
and $1^+$ for the spin-parity quantum numbers of the 
$D_0^\ast(2400)^0/D_0^\ast(2400)^\pm$ and $D_1(2430)^{0}$ states, 
respectively. The measured masses and widths, as well as the $J^P$ 
values, are in agreement with theoretical predictions based on potential 
models~\cite{Godfrey:1985xj, Godfrey:1986wj, Isgur:1991wq, Schweitzer:2002nm}. 

The spectroscopic assignment of heavier states remains less clear. Further
theoretical studies suggest the identity of some 2S and 1D states~\cite{Chen:2015lpa,Godfrey:2015dva}
and tentatively discuss possible 1F, 3S and 2P states. Possible new states to
be found in the future are suggested in Ref.~\cite{Godfrey:2015dva}.

Tables~\ref{table:charm:spect:3} and \ref{table:charm:spect:4} summarize 
branching fractions of $B$ meson decays to excited $D$ and $D_{s}$ 
states, respectively. The measurements listed are the products of the
$B$ meson branching fraction and the daughter $D$ meson branching fraction.
It is notable that the branching fractions for 
$B$ mesons decaying to a narrow $D^{\ast}$ state and a pion are similar 
for charged and neutral $B$ initial states, while the branching fractions 
to a broad $D^{\ast}$ state and $\pi^+$ are much larger for $B^+$ than 
for $B^0$. This may be due to the fact that color-suppressed amplitudes 
contribute only to the $B^+$ decay and not to the $B^0$ decay (for a 
theoretical discussion, see Refs.~\cite{Jugeau:2005yr,Colangelo:2004vu}). 
Values for the branching fractions of the $D$ mesons are difficult
to extract due to 
the unknown (and difficult to calculate) $B \to D^{\ast} X$ branching fractions.

The discoveries of the $D_{s0}^\ast(2317)^{\pm}$ and $D_{s1}(2460)^{\pm}$ 
have triggered increased interest in properties of, and searches for, 
excited $D_s$ mesons. While 
the masses and widths of $D_{s1}(2536)^{\pm}$ and $D_{s2}^{\ast}(2573)^{\pm}$ 
states are in relatively good agreement with potential model predictions, 
the masses of $D_{s0}^\ast(2317)^{\pm}$ and $D_{s1}(2460)^{\pm}$ states are 
significantly lower than expected (see Ref.~\cite{Cahn:2003cw} for a 
discussion of $c\bar{s}$ models). Moreover, the mass splitting between 
these two states greatly exceeds that between the $D_{s1}(2536)^{\pm}$ 
and $D_{s2}(2573)^{\pm}$. These unexpected properties have led to 
interpretations of the $D_{s0}^\ast(2317)^{\pm}$ and $D_{s1}(2460)^{\pm}$ 
as exotic four-quark states~\cite{Barnes:2003dj,Lipkin:2003zk}.

While there are few measurements of the $J^P$ values of 
$D_{s0}^\ast(2317)^{\pm}$ and $D_{s1}(2460)^{\pm}$, the available data 
favor $0^+$ and $1^+$, respectively. A molecule-like ($DK$) interpretation 
of the $D_{s0}^\ast(2317)^{\pm}$ and 
$D_{s1}(2460)^{\pm}$~\cite{Barnes:2003dj,Lipkin:2003zk} that can account 
for their low masses and isospin-breaking decay modes is tested by searching 
for charged and neutral isospin partners of these states; thus far such 
searches have yielded negative results. Therefore the models 
that predict equal production rates for different charged states are
excluded. The molecular picture can also be tested by measuring the 
rates for the radiative processes 
$D_{s0}^\ast(2317)^{\pm}/D_{s1}(2460)^{\pm}\to D_s^{(\ast)}\gamma$ and 
comparing to theoretical predictions. The predicted rates, however, 
are below the sensitivity of current experiments. 

Another model successful in explaining the total widths and the 
$D_{s0}^\ast(2317)^{\pm}$ -- $D_{s1}(2460)^{\pm}$ mass splitting is based 
on the assumption that these states are chiral partners of the ground 
states $D_{s}^{+}$ and~$D_{s}^{*}$~\cite{Bardeen:2003kt}. While some 
measured branching fraction ratios agree with predicted values, further 
experimental tests with better sensitivity are needed to confirm or 
refute this scenario. A summary of the mass difference measurements 
is given in Table~\ref{table:charm:spect:5}.

Measurements by \babar{}~\cite{Aubert:2009ah} and LHCb~\cite{Aaij:2012pc} 
first indicated the existence of a strange-charm $D^{*}_{sJ}(2860)^{\pm}$ 
meson. An LHCb study of $B_{s}^{0}\to \overline{D}{}^{0}K^{-}\pi^{+}$ decays, 
in which they searched for excited $D_{s}$ mesons~\cite{Aaij:2014xza}, showed 
with $10\sigma$ significance that this state comprises two different 
particles,  one of spin 1 and one of spin 3. This represents the first 
measurement of a heavy flavoured spin-3 particle, and the first observation 
of $B$ meson decays to spin 3 particles. A subsequent study of $D_{sJ}$ mesons 
by the LHCb collaboration~\cite{Aaij:2016utb} supports the natural parity 
assignment for this state ($J^P=3^-$). This study also shows weak evidence 
for a further structure at a mass around 3040 MeV$/c^2$ with unnatural 
parity, which was first hinted at by a \babar\ analysis~\cite{Aubert:2009ah}.
The second observation of a spin-3 charm meson was a subsequent LHCb analysis
of $B^0 \to \overline{D}{}^0\pi^+\pi^-$ decays, which measured the spin-parity
assignment of the state $D^*_3(2760)^{\pm}$ to be $J^P=3^-$. This resonance was
in fact observed previously by \babar{}~\cite{delAmoSanchez:2010vq} and
LHCb~\cite{Aaij:2013sza}. The measurement suggests a spectroscopic assignment
of ${}^3D_3$. Recently, also the corresponding neutral state was observed by
LHCb, the $D^{*}_{3}(2760)^{0}$~\cite{Aaij:2016fma}.

Other observed excited $D_s$ states include $D_{s1}^{\ast}(2700)^{\pm}$  and 
$D_{s2}^{\ast}(2573)^{\pm}$. The properties of both (mass, width, $J^P$) have 
been measured and determined in several analyses. A theoretical 
discussion~\cite{Matsuki:2006rz} investigates the possibility that 
the $D_{s1}(2700)^{\pm}$ could represent radial excitations of the 
$D_s^{\ast\pm}$. Similarly, the  $D_{s1}^{\ast}(2860)^{\pm}$ and $D_{sJ}(3040)^{\pm}$ 
could be excitations of $D_{s0}^\ast(2317)^{\pm}$ and $D_{s1}(2460)^{\pm}$ or 
$D_{s1}(2536)^{\pm}$, respectively.

Table~\ref{table:charm:spect:6} summarizes measurements of the helicity parameter
 $A_{D}$ (also referred to as polarization amplitude). In $D^{\ast\ast}$ meson 
decays to $D^{\ast\ast} \to D^{\ast}\pi$, $D^{\ast} \to D \pi$, the helicity
distribution varies like $1 + A_{D}\cos^{2}\theta_{H}$, where $\theta_{H}$
is the angle in the $D^{\ast}$ rest frame between the two pions emitted
by decay $D^{\ast\ast} \to D^{\ast}\pi$ and the $D^{\ast} \to D \pi$. The
parameter is sensitive to possible S-wave contributions in the decay. 
In the case of a $D$ meson decay decaying purely via D-wave, 
the helicity parameter is predicted to give $A_{D}=3$. Studies of the 
$D_{1}(2420)^{0}$ meson by the ZEUS and \babar{} collaborations suggest 
that there is an S-wave admixture in the decay, which is  contrary to 
the expectation based on Heavy Quark Effective Theory~\cite{Isgur:1989vq,Neubert:1993mb}.

\begin{table}[htb!]
\begin{center}
\caption{\label{table:charm:spect:3} 
Product of the $B$ meson branching fraction and the
daughter (excited) $D$ meson branching fraction.}
\begin{adjustbox}{width=\textwidth,center}
{\setlength\tabcolsep{0pt}
 \begin{tabular}{cp{5pt}cp{5pt}r@{}lp{5pt}cp{5pt}c}
 \toprule
 Resonance&&Decay &&\multicolumn{2}{c}{$\mathcal{B}~[10^{-4}]$} & & \multicolumn{1}{c}{Measured by} && \multicolumn{1}{c}{Reference}
 \\ \midrule
 \multirow{5}{*}{$D_{0}^{*}(2400)^{0}$} & & \multirow{3}{*}{$B^{-}\to D_{0}^{*}(2400)^{0}(\to D^{+}\pi^{-})\pi^{-}$} & &$6$ &$.1\pm0.6\pm1.8$ & &Belle & & \cite{Abe:2003zm} \\ 
 & & & &$6$ &$.8\pm0.3\pm2.0$ &&\babar{} && \cite{Aubert:2009wg} \\ \cmidrule{4-9}
 & & & & \cellcolor{Gray}$6$ & \cellcolor{Gray}$.4\pm 1.4$ & \cellcolor{Gray} & \cellcolor{Gray} Our average & \cellcolor{Gray} & \\ \cmidrule{3-9}
 & & \multirow{1}{*}{$B^{-}\to D_{0}^{*}(2400)^{0}(\to D^{+}\pi^{-})K^{-}$} &  \cellcolor{LightGray}& \cellcolor{LightGray} $0$ & \cellcolor{LightGray}$.061\pm0.019\pm0.005\pm0.014\pm0.004$ & \cellcolor{LightGray} & \cellcolor{LightGray} LHCb & \cellcolor{LightGray} & \cite{Aaij:2015vea} \\ \midrule
 \multirow{5}{*}{$D_{0}^{*}(2400)^{\pm}$} & & \multirow{3}{*}{$\overline{B}{}^{0}\to D_{0}^{*}(2400)^{+}(\to D^{0}\pi^{+})\pi^{-}$} & &$0$ &$.77\pm0.05\pm0.03\pm0.03\pm0.04$ &  &LHCb & & \cite{Aaij:2015sqa} \\ 
 & & &  &$0$ &$.60\pm0.13\pm0.27$ &  &Belle & & \cite{Kuzmin:2006mw} \\\cmidrule{4-9}
 & & & \cellcolor{Gray} & \cellcolor{Gray} $0\cellcolor{Gray}$ & \cellcolor{Gray}$.76 \pm 0.07$ & \cellcolor{Gray} & \cellcolor{Gray} Our average & \cellcolor{Gray} & \\ \cmidrule{3-9}
 & & \multirow{1}{*}{$\overline{B}{}^{0}\to D_{0}^{*}(2400)^{+}(\to D^{0}\pi^{+})K^{-}$} & \cellcolor{LightGray} & \cellcolor{LightGray} $0$ & \cellcolor{LightGray}$.177\pm0.026\pm0.019\pm0.067\pm0.20$ & \cellcolor{LightGray} & \cellcolor{LightGray} LHCb & \cellcolor{LightGray} & \cite{Aaij:2015kqa} \\ \midrule 
 \multirow{3}{*}[-5pt]{$D_{1}^{}(2420)^{0}$} & & \multirow{1}{*}{$B^{-}\to D_{1}^{}(2420)^{0}(\to D^{*+}\pi^{-})\pi^{-}$} & \cellcolor{LightGray} & \cellcolor{LightGray} $6$ & \cellcolor{LightGray}$.8\pm0.7\pm1.3$ & \cellcolor{LightGray} & \cellcolor{LightGray} Belle & \cellcolor{LightGray} & \cite{Abe:2003zm} \\ \cmidrule{4-9}
 & & \multirow{1}{*}{$B^{-}\to D_{1}^{}(2420)^{0}(\to D^{0}\pi^{+}\pi^{-})\pi^{-}$} & \cellcolor{LightGray} & \cellcolor{LightGray} $1$ & \cellcolor{LightGray}$.85\pm0.29\pm0.27\pm0.41$ & \cellcolor{LightGray} & \cellcolor{LightGray} Belle & \cellcolor{LightGray} & \cite{Abe:2004sm} \\ \cmidrule{4-9}
  & & \multirow{1}{*}{$\overline{B}{}^{0}\to D_{1}^{}(2420)^{0}(\to D^{*+}\pi^{-})\omega$} & \cellcolor{LightGray} & \cellcolor{LightGray} $0$ & \cellcolor{LightGray}$.7\pm0.2^{+0.1}_{-0.0}\pm0.1$ & \cellcolor{LightGray} & \cellcolor{LightGray} Belle & \cellcolor{LightGray} & \cite{Matvienko:2015gqa} \\ \midrule
 \multirow{1}{*}{$D_{1}^{}(2420)^{\pm}$} & & \multirow{1}{*}{$\overline{B}{}^{0}\to D_{1}(2420)^{+}(\to D^{+}\pi^{-}\pi^{+})\pi^{-}$} & \cellcolor{LightGray} & \cellcolor{LightGray} $0$ & \cellcolor{LightGray}$.89\pm0.15\pm0.22$ & \cellcolor{LightGray} & \cellcolor{LightGray} Belle & \cellcolor{LightGray} & \cite{Abe:2004sm} \\ \midrule
 \multirow{2}{*}[-2pt]{$D_{1}(2430)^{0}$} & &$B^{-}\to D_{1}(2430)^{0}(\to D^{*+}\pi^{-})\pi^{-}$ & \cellcolor{LightGray} & \cellcolor{LightGray}$5$ & \cellcolor{LightGray}$.0\pm0.4\pm1.08$ & \cellcolor{LightGray} & \cellcolor{LightGray} Belle & \cellcolor{LightGray} & \cite{Abe:2003zm} \\ \cmidrule{4-9}
 & &$\overline{B}{}^{0}\to D_{1}(2430)^{0}(\to D^{*+}\pi^{-})\omega$ & \cellcolor{LightGray} & \cellcolor{LightGray}$2$ & \cellcolor{LightGray}$.5\pm0.4^{+0.7}_{-0.2}{}^{+0.4}_{-0.1}$ & \cellcolor{LightGray} & \cellcolor{LightGray} Belle & \cellcolor{LightGray} & \cite{Matvienko:2015gqa} \\ \midrule
 \multirow{8}{*}[-5pt]{$D_{2}^{*}(2460)^{0}$} & & \multirow{4}{*}{$B^{-}\to D_{2}^{*}(2460)^{0}(\to D^{+}\pi^{-})\pi^{-}$} & &$3$ &$.4\pm0.3\pm0.7$ & & Belle & & \cite{Abe:2003zm} \\ 
 & & & &$3$ &$.5\pm0.2\pm0.5$ & & \babar{} & & \cite{Aubert:2009wg} \\
 & & & &$3$ &$.62\pm0.06\pm0.14\pm0.09\pm0.25$ & & LHCb & & \cite{Aaij:2016fma}\\
 \cmidrule{4-9}
 & & & \cellcolor{Gray} & \cellcolor{Gray} $3\cellcolor{Gray}$ & \cellcolor{Gray}$.58 \pm 0.23$ & \cellcolor{Gray} & \cellcolor{Gray} Our average & \cellcolor{Gray} & \\ \cmidrule{3-9}
 & & \multirow{1}{*}{$B^{-}\to D_{2}^{*}(2460)^{0}(\to D^{*+}\pi^{-})\pi^{-}$} & \cellcolor{LightGray} & \cellcolor{LightGray} $1$ & \cellcolor{LightGray}$.8\pm0.3\pm0.4$ & \cellcolor{LightGray} & \cellcolor{LightGray} Belle & \cellcolor{LightGray} & \cite{Abe:2003zm} \\ \cmidrule{3-9}
  & & \multirow{1}{*}{$B^{-}\to D_{2}^{*}(2460)^{0}(\to D^{*+}\pi^{-})\omega$} & \cellcolor{LightGray} & \cellcolor{LightGray} $0$ & \cellcolor{LightGray}$.4\pm0.1^{+0.0}_{-0.1}\pm0.1$ & \cellcolor{LightGray} & \cellcolor{LightGray} Belle & \cellcolor{LightGray} & \cite{Matvienko:2015gqa} \\  \cmidrule{3-9}
   & & \multirow{1}{*}{$B^{-}\to D_{2}^{*}(2460)^{0}(\to D^{+}\pi^{-})K^{-}$} &  \cellcolor{LightGray}& \cellcolor{LightGray} $0$ & \cellcolor{LightGray}$.232\pm0.011\pm0.006\pm0.010\pm0.016$ & \cellcolor{LightGray} & \cellcolor{LightGray} LHCb & \cellcolor{LightGray} & \cite{Aaij:2015vea} \\ \midrule
 \multirow{4}{*}[-5pt]{$D_{2}^{*}(2460)^{\pm}$} & & \multirow{3}{*}{$\overline{B}{}^{0}\to D_{2}^{*}(2460)^{+}(\to D^{0}\pi^{+})\pi^{-}$} & &$2$ &$.44\pm0.07\pm0.10\pm0.04\pm0.12$ &  &LHCb & & \cite{Aaij:2015sqa} \\ 
 & & & &$2$ &$.15\pm0.17\pm0.31$ & &  Belle & & \cite{Kuzmin:2006mw} \\\cmidrule{4-9}
 & & & \cellcolor{Gray} & \cellcolor{Gray} $2\cellcolor{Gray}$ & \cellcolor{Gray}$.38 \pm 0.16$ & \cellcolor{Gray} & \cellcolor{Gray} Our average & \cellcolor{Gray} & \\ \cmidrule{3-9}
 & & \multirow{1}{*}{$\overline{B}{}^{0}\to D_{2}^{*}(2460)^{+}(\to D^{0}\pi^{+})K^{-}$} & \cellcolor{LightGray} & \cellcolor{LightGray} $0$ & \cellcolor{LightGray}$.212\pm0.010\pm0.011\pm0.011\pm0.25$ & \cellcolor{LightGray} & \cellcolor{LightGray} LHCb & \cellcolor{LightGray} & \cite{Aaij:2015kqa} \\ \midrule 
  \multirow{1}{*}{$D_{1}^{*}(2680)^{0}$}    & & \multirow{1}{*}{$B^{-}\to D_{1}^{*}(2680)^{0}(\to D^{+}\pi^{-})\pi^{-}$} &  \cellcolor{LightGray}& \cellcolor{LightGray} $0$ & \cellcolor{LightGray}$.84\pm0.06\pm0.07\pm0.18\pm0.06$ & \cellcolor{LightGray} & \cellcolor{LightGray} LHCb & \cellcolor{LightGray} & \cite{Aaij:2016fma} \\ \midrule
 \multirow{1}{*}{$D_{1}^{*}(2760)^{0}$}    & & \multirow{1}{*}{$B^{-}\to D_{1}^{*}(2760)^{0}(\to D^{+}\pi^{-})K^{-}$} &  \cellcolor{LightGray}& \cellcolor{LightGray} $0$ & \cellcolor{LightGray}$.036\pm0.009\pm0.003\pm0.007\pm0.002$ & \cellcolor{LightGray} & \cellcolor{LightGray} LHCb & \cellcolor{LightGray} & \cite{Aaij:2015vea} \\ \midrule
  \multirow{1}{*}{$D_{3}^{*}(2760)^{0}$}    & & \multirow{1}{*}{$\overline{B}{}^{-}\to D_{3}^{*}(2760)^{0}(\to D^{+}\pi^{-})\pi^{-}$} &  \cellcolor{LightGray}& \cellcolor{LightGray} $0$ & \cellcolor{LightGray}$.10\pm0.01\pm0.01\pm0.02\pm0.01$ & \cellcolor{LightGray} & \cellcolor{LightGray} LHCb & \cellcolor{LightGray} & \cite{Aaij:2016fma} \\  \midrule
  \multirow{1}{*}{$D_{3}^{*}(2760)^{\pm}$}    & & \multirow{1}{*}{$\overline{B}{}^{0}\to D_{3}^{*}(2760)^{+}(\to D^{0}\pi^{+})\pi^{-}$} &  \cellcolor{LightGray}& \cellcolor{LightGray} $0$ & \cellcolor{LightGray}$.103\pm0.016\pm0.007\pm0.008\pm0.005$ & \cellcolor{LightGray} & \cellcolor{LightGray} LHCb & \cellcolor{LightGray} & \cite{Aaij:2015sqa} \\  \midrule 
  \multirow{1}{*}{$D_{2}^{*}(3000)^{0}$}    & & \multirow{1}{*}{$\overline{B}{}^{0}\to D_{2}^{*}(3000)^{0}(\to D^{+}\pi^{-})\pi^{-}$} &  \cellcolor{LightGray}& \cellcolor{LightGray} $0$ & \cellcolor{LightGray}$.02\pm0.01\pm0.01\pm0.01\pm0.00$ & \cellcolor{LightGray} & \cellcolor{LightGray} LHCb & \cellcolor{LightGray} & \cite{Aaij:2016fma} \\ 
  \bottomrule 
 \end{tabular}
}
\end{adjustbox}

\end{center}
\end{table} 

\begin{table}[htb!]
\begin{center}
\caption{\label{table:charm:spect:4} 
Product of the $B$ meson branching fraction and the
daughter (excited) $D^{}_s$ meson branching fraction.}
\begin{adjustbox}{width=\textwidth,center}
{\setlength\tabcolsep{0pt}
 \begin{tabular}{cp{5pt}cp{5pt}r@{}lp{5pt}cp{5pt}c}
 \toprule
 Resonance & & Decay & & \multicolumn{2}{c}{$\mathcal{B}~[10^{-4}]$} & & \multicolumn{1}{c}{Measured by} & & \multicolumn{1}{c}{Reference} \\ \midrule 
 \multirow{6}{*}[-5pt]{$D_{s0}^{*}(2317)^{\pm}$} && \multirow{4}{*}[-2pt]{$B^{0}\to D_{s0}^{*}(2317)^{+}(\to D^{+}_{s}\pi^{0})D^{-}$} &&$ 8$&$.6^{+3.3}_{-2.6}\pm2.6 $&& Belle && \cite{Krokovny:2003zq} \\
 && &&$ 18$&$.0\pm 4.0{}^{+6.7}_{-5.0} $&& \babar{} && \cite{Aubert:2004pw} \\
  && &&$ 10$&$.1^{+1.3}_{-1.2}\pm 1.0\pm0.4 $&& Belle && \cite{Choi:2015lpc} \\ \cmidrule{4-9}
 && &\cellcolor{Gray}&$ \cellcolor{Gray}10$&\cellcolor{Gray}$.2 \pm 1.5 $&\cellcolor{Gray}& \cellcolor{Gray} Our average &\cellcolor{Gray}& \\ \cmidrule{3-9}
 && \multirow{1}{*}{$B^{+}\to D_{s0}^{*}(2317)^{+}(\to D^{+}_{s}\pi^{0})\overline{D}{}^{0}$} &\cellcolor{LightGray} &$ \cellcolor{LightGray} 8$&\cellcolor{LightGray}$.0^{+1.3}_{-1.2}\pm1.0\pm0.4 $&\cellcolor{LightGray}& \cellcolor{LightGray} Belle &\cellcolor{LightGray}& \cite{Choi:2015lpc} \\ \cmidrule{3-9}
  && \multirow{1}{*}{$B^{0}\to D_{s0}^{*}(2317)^{+}(\to D^{+}_{s}\pi^{0})K^{-}$} &\cellcolor{LightGray} &$ \cellcolor{LightGray} 0$&\cellcolor{LightGray}$.53^{+0.15}_{-0.13}\pm0.16 $&\cellcolor{LightGray}& \cellcolor{LightGray} Belle &\cellcolor{LightGray}& \cite{Abe:2004wz} \\ \midrule

 \multirow{9}{*}[-5pt]{$D_{s1}(2460)^{\pm}$} && \multirow{3}{*}{$B^{0}\to D_{s1}(2460)^{+}(\to D^{*+}_{s}\pi^{0})D^{-}$} &&$ 22$&$.7^{+7.3}_{-6.2}\pm6.8 $&& Belle && \cite{Krokovny:2003zq} \\
 && &&$ 28$&$.0\pm 8.0{}^{+11.2}_{-7.8} $&& \babar{} && \cite{Aubert:2004pw} \\ \cmidrule{4-9}
 && &\cellcolor{Gray}&$ \cellcolor{Gray}24$&\cellcolor{Gray}$.7 \pm 7.6 $&\cellcolor{Gray}& \cellcolor{Gray} Our average &\cellcolor{Gray}& \\ \cmidrule{3-9}
 && \multirow{3}{*}{$B^{0}\to D_{s1}(2460)^{+}(\to D^{*+}_{s}\gamma)D^{-}$} &&$ 8$&$.2^{+2.2}_{-1.9}\pm2.5 $&& Belle && \cite{Krokovny:2003zq} \\
 && &&$ 8$&$.0\pm 2.0{}^{+3.2}_{-2.3} $&& \babar{} && \cite{Aubert:2004pw} \\ \cmidrule{4-9}
 && &\cellcolor{Gray}&$ \cellcolor{Gray}8$&\cellcolor{Gray}$.1 \pm 2.3 $&\cellcolor{Gray}& \cellcolor{Gray} Our average &\cellcolor{Gray}& \\ \cmidrule{3-9}
 && $D_{s1}(2460)^{+}\to D^{*+}_{s}\pi^{0}$ &\cellcolor{LightGray} &$ \cellcolor{LightGray} (56$&\cellcolor{LightGray}${}\pm13\pm9)\% $&\cellcolor{LightGray}& \cellcolor{LightGray} \babar{} &\cellcolor{LightGray}& \cite{Aubert:2006nm} \\ \cmidrule{3-9}
 && $D_{s1}(2460)^{+}\to D^{*+}_{s}\gamma$ &\cellcolor{LightGray} &$ \cellcolor{LightGray} (16$&\cellcolor{LightGray}${}\pm4\pm3)\% $&\cellcolor{LightGray}& \cellcolor{LightGray} \babar{} &\cellcolor{LightGray}& \cite{Aubert:2006nm} \\ \midrule
 && \multirow{1}{*}{$B^{0}\to D_{s1}(2536)^{+}(\to D^{*0}K^{+})D^{-}$} &\cellcolor{LightGray} &$ \cellcolor{LightGray} 1$&\cellcolor{LightGray}$.71\pm0.48\pm0.32 $&\cellcolor{LightGray}& \cellcolor{LightGray} \babar{} &\cellcolor{LightGray}& \cite{Aubert:2007rva} \\ \cmidrule{3-9}
 && \multirow{1}{*}{$B^{0}\to D_{s1}(2536)^{+}(\to D^{*+}K^{0})D^{-}$} &\cellcolor{LightGray} &$ \cellcolor{LightGray} 2$&\cellcolor{LightGray}$.61\pm1.03\pm0.31 $&\cellcolor{LightGray}& \cellcolor{LightGray} \babar{} &\cellcolor{LightGray}& \cite{Aubert:2007rva} \\ \cmidrule{3-9}
 && \multirow{1}{*}{$B^{0}\to D_{s1}(2536)^{+}(\to D^{*0}K^{+})D^{*-}$} &\cellcolor{LightGray} &$ \cellcolor{LightGray} 3$&\cellcolor{LightGray}$.32\pm0.88\pm0.66 $&\cellcolor{LightGray}& \cellcolor{LightGray} \babar{} &\cellcolor{LightGray}& \cite{Aubert:2007rva} \\ \cmidrule{3-9}
 \multirow{2}{*}{$D_{s1}(2536)^{\pm}$} && \multirow{1}{*}{$B^{0}\to D_{s1}(2536)^{+}(\to D^{*+}K^{0})D^{*-}$} &\cellcolor{LightGray} &$ \cellcolor{LightGray} 5$&\cellcolor{LightGray}$.00\pm1.51\pm0.67 $&\cellcolor{LightGray}& \cellcolor{LightGray} \babar{} &\cellcolor{LightGray}& \cite{Aubert:2007rva} \\ \cmidrule{3-9}
 && \multirow{1}{*}{$B^{+}\to D_{s1}(2536)^{+}(\to D^{*0}K^{+})\overline{D}^{0}$} &\cellcolor{LightGray} &$ \cellcolor{LightGray} 2$&\cellcolor{LightGray}$.16\pm0.52\pm0.45 $&\cellcolor{LightGray}& \cellcolor{LightGray} \babar{} &\cellcolor{LightGray}& \cite{Aubert:2007rva} \\ \cmidrule{3-9}
 && \multirow{1}{*}{$B^{+}\to D_{s1}(2536)^{+}(\to D^{*+}K^{0})\overline{D}^{0}$} &\cellcolor{LightGray} &$ \cellcolor{LightGray} 2$&\cellcolor{LightGray}$.30\pm0.98\pm0.43 $&\cellcolor{LightGray}& \cellcolor{LightGray} \babar{} &\cellcolor{LightGray}& \cite{Aubert:2007rva} \\ \cmidrule{3-9}
 && \multirow{1}{*}{$B^{+}\to D_{s1}(2536)^{+}(\to D^{*0}K^{+})\overline{D}^{*0}$} &\cellcolor{LightGray} &$ \cellcolor{LightGray} 5$&\cellcolor{LightGray}$.46\pm1.17\pm1.04 $&\cellcolor{LightGray}& \cellcolor{LightGray} \babar{} &\cellcolor{LightGray}& \cite{Aubert:2007rva} \\ \cmidrule{3-9}
 && \multirow{1}{*}{$B^{+}\to D_{s1}(2536)^{+}(\to D^{*+}K^{0})\overline{D}^{*0}$} &\cellcolor{LightGray} &$ \cellcolor{LightGray} 3$&\cellcolor{LightGray}$.92\pm2.46\pm0.83 $&\cellcolor{LightGray}& \cellcolor{LightGray} \babar{} &\cellcolor{LightGray}& \cite{Aubert:2007rva} \\ \midrule
  \multirow{2}{*}[-2pt]{$D_{s2}^{\ast}(2573)^{\pm}$} && \multirow{1}{*}{$B^{0}\to D_{s2}^{\ast}(2573)(\to D^{0}K^{+})D^{-}$} &\cellcolor{LightGray} &$ \cellcolor{LightGray} 0$&\cellcolor{LightGray}$.34\pm0.17\pm0.05$&\cellcolor{LightGray}& \cellcolor{LightGray} \babar{} &\cellcolor{LightGray}& \cite{Lees:2014abp} \\ \cmidrule{3-9} 
  && \multirow{1}{*}{$B^{+}\to D_{s2}^{\ast}(2573)(\to D^{0}K^{+})\overline{D}{}^{0}$} &\cellcolor{LightGray} &$ \cellcolor{LightGray} 0$&\cellcolor{LightGray}$.08\pm14\pm0.05 $&\cellcolor{LightGray}& \cellcolor{LightGray} \babar{} &\cellcolor{LightGray}& \cite{Lees:2014abp} \\ \midrule
 \multirow{4}{*}[-3pt]{$D_{s1}{}^{\ast}(2700)^{\pm}$} && \multirow{4}{*}[3pt]{$B^{+}\to D_{s1}{}^{\ast}(2700)^{+}(\to D^{0}K^{+})\overline{D}{}^{0}$} &&$ 11$&$.3\pm2.2{}^{+1.4}_{-2.8} $&& Belle && \cite{Brodzicka:2007aa} \\
 && &&$ 5$&$.02\pm0.71\pm0.93 $&& \babar{} && \cite{Lees:2014abp} \\ \cmidrule{4-9}
 && &\cellcolor{Gray}&$ \cellcolor{Gray}5$&\cellcolor{Gray}$.83 \pm 1.09 $&\cellcolor{Gray}& \cellcolor{Gray} Our average &\cellcolor{Gray}& \\  \cmidrule{3-9}
 && \multirow{1}{*}{$B^{0}\to D_{s1}{}^{\ast}(2700)^{+}(\to D^{0}K^{+})D{}^{-}$} &\cellcolor{LightGray} &$ \cellcolor{LightGray} 7$&\cellcolor{LightGray}$.14\pm0.96\pm0.69 $&\cellcolor{LightGray}& \cellcolor{LightGray} \babar{} &\cellcolor{LightGray}& \cite{Lees:2014abp} \\  \bottomrule
\end{tabular}
}
\end{adjustbox}

\end{center}
\end{table}

\begin{table}[htb!]
\begin{center}
\caption{\label{table:charm:spect:5} 
Measurements of mass differences for excited $D$ mesons.}
{\setlength\tabcolsep{0pt}
 \begin{tabular}{cp{5pt}cp{5pt}r@{}lp{5pt}cp{5pt}c}
 \toprule
 Resonance & & Relative to && \multicolumn{2}{c}{$\Delta m$ [MeV$/c^{2}$]} & & \multicolumn{1}{c}{Measured by} && \multicolumn{1}{c}{Reference} 
 \\ \midrule
 \multirow{3}{*}{$D_{1}^{*}(2420)^{0}$} & & \multirow{3}{*}{$D^{*+}$} & & $ 410$ & $.2\pm2.1\pm0.9 $ & & Zeus & & \cite{Chekanov:2008ac} \\
 & & & & $ 411$ & $.7\pm0.7\pm0.4 $ & & CDF & & \cite{Abulencia:2005ry} \\ \cmidrule{4-9}
 & & & \cellcolor{Gray} & \cellcolor{Gray} $411$ & \cellcolor{Gray}$.5 \pm 0.8 $ & \cellcolor{Gray} & \cellcolor{Gray} Our average & \cellcolor{Gray} & \\ \midrule

 \multirow{1}{*}{$D_{1}^{}(2420)^{\pm}$} & & \multirow{1}{*}{$D_{1}^{*}(2420)^{0}$} & \cellcolor{LightGray} & $ \cellcolor{LightGray}4$ & \cellcolor{LightGray}${}^{+2}_{-3}\pm3 $ & \cellcolor{LightGray} & \cellcolor{LightGray} CLEO & \cellcolor{LightGray} & \cite{Bergfeld:1994af} \\ \midrule
 
 \multirow{2}{*}{$D_{2}^{*}(2460)^{0}$} & & $D^{+}$ & \cellcolor{LightGray} & $ \cellcolor{LightGray}593$ & \cellcolor{LightGray}$.9\pm0.6\pm0.5 $ & \cellcolor{LightGray} & \cellcolor{LightGray}CDF & \cellcolor{LightGray} & \cite{Abulencia:2005ry} \\ \cmidrule{3-9}
 & & $D^{*+}$ & \cellcolor{LightGray} & $ \cellcolor{LightGray} 458$ & \cellcolor{LightGray}$.8\pm3.7^{+1.2}_{-1.3} $ & \cellcolor{LightGray} & \cellcolor{LightGray} Zeus & \cellcolor{LightGray} & \cite{Chekanov:2008ac} \\ \midrule
 \multirow{4}{*}[-2pt]{$D_{2}^{*}(2460)^{\pm}$} & & \multirow{4}{*}[-2pt]{$D_{2}^{*}(2460)^{0}$} & & $ 3$ & $.1\pm1.9\pm0.9 $ & & FOCUS & & \cite{Link:2003bd} \\
 & & & & $-2$ & ${}\pm4\pm4 $ & & CLEO & & \cite{Bergfeld:1994af} \\
 & & & & $ 14$ & ${}\pm5\pm8 $ & & ARGUS & & \cite{Albrecht:1989gb} \\ \cmidrule{4-9}
 & & & \cellcolor{Gray} & $ \cellcolor{Gray} 3$ & \cellcolor{Gray}$.0 \pm 1.9 $ & \cellcolor{Gray} & \cellcolor{Gray} Our average & \cellcolor{Gray} & \\ \midrule

 \multirow{4}{*}[-2pt]{$D_{s0}^{*}(2317)^{\pm}$} & & \multirow{4}{*}[-2pt]{$D_{s}^{\pm}$} & & $ 348$ & $.7\pm0.5\pm0.7 $ & & Belle & & \cite{Abe:2003jk} \\
 & & & & $ 350$ & $.0\pm1.2\pm1.0 $ & & CLEO & & \cite{Besson:2003cp} \\
 & & & & $ 351$ & $.3\pm2.1\pm1.9 $ & & Belle & & \cite{Krokovny:2003zq} \\ \cmidrule{4-9}
 & & & \cellcolor{Gray} & $ \cellcolor{Gray} 349$ & \cellcolor{Gray}$.2 \pm 0.7 $ & \cellcolor{Gray} & \cellcolor{Gray} Our average & \cellcolor{Gray} & \\ \midrule
 \multirow{7}{*}[-5pt]{$D_{s1}(2460)^{\pm}$} & & \multirow{4}{*}[-2pt]{$D_{s}^{*\pm}$} & & $ 344$ & $.1\pm1.3\pm1.1 $ & & Belle & & \cite{Abe:2003jk} \\
 & & & & $ 351$ & $.2\pm1.7\pm1.0 $ & & CLEO & & \cite{Besson:2003cp} \\
 & & & & $ 346$ & $.8\pm1.6\pm1.9 $ & & Belle & & \cite{Krokovny:2003zq} \\ \cmidrule{4-9}
 & & & \cellcolor{Gray} & $ \cellcolor{Gray} 347$ & \cellcolor{Gray}$.1 \pm 1.1 $ & \cellcolor{Gray} & \cellcolor{Gray} Our average & \cellcolor{Gray} & \\ \cmidrule{3-9}
 & & \multirow{3}{*}[-2pt]{ $D_{s}^{\pm}$ } & & $ 491$ & $.0\pm1.3\pm1.9 $ & & Belle & & \cite{Abe:2003jk} \\
 & & & & $ 491$ & $.4\pm0.9\pm1.5 $ & & Belle & & \cite{Abe:2003jk} \\ \cmidrule{4-9}
 & & & \cellcolor{Gray} & $ \cellcolor{Gray} 491$ & \cellcolor{Gray}$.3 \pm 1.4 $ & \cellcolor{Gray} & \cellcolor{Gray} Our average & \cellcolor{Gray} & \\ \midrule
 
 \multirow{5}{*}[-4pt]{$D_{s1}(2536)^{\pm}$} & & \multirow{4}{*}[-2pt]{ $D^{*}(2010)^{\pm}$} & & $ 524$ & $.83\pm0.01\pm0.04 $ & & \babar{} & & \cite{Lees:2011um} \\
 & & & & $ 525$ & $.30_{-0.41}^{+0.44}\pm0.10 $ & & Zeus & & \cite{Chekanov:2008ac} \\
 & & & & $ 525$ & $.3\pm0.6\pm0.1 $ & & ALEPH & & \cite{Heister:2001nj} \\ \cmidrule{4-9}
 & & & \cellcolor{Gray} & $ \cellcolor{Gray} 524$ & \cellcolor{Gray}$.84 \pm 0.04 $ & \cellcolor{Gray} & \cellcolor{Gray} Our average & \cellcolor{Gray} & \\ \cmidrule{3-9}
 & & \multirow{1}{*}{ $D^{*}(2007)^{0}$} & \cellcolor{LightGray} & $ \cellcolor{LightGray} 528$ & \cellcolor{LightGray}$.7\pm1.9\pm0.5 $ & \cellcolor{LightGray} & \cellcolor{LightGray}ALEPH & \cellcolor{LightGray} & \cite{Heister:2001nj} \\ \midrule
 
 \multirow{1}{*}{$D_{s2}^{*}(2573)^{\pm}$} & & $D^{0}$ & \cellcolor{LightGray} & $ \cellcolor{LightGray} 704$ & \cellcolor{LightGray}${}\pm3\pm1 $ & \cellcolor{LightGray} & \cellcolor{LightGray}ALEPH & \cellcolor{LightGray} & \cite{Heister:2001nj} \\ \bottomrule
 \end{tabular}}

\end{center}
\end{table} 

\begin{table}[htb!]
\begin{center}
\caption{\label{table:charm:spect:6}
Measurements of polarization amplitudes for excited $D$ mesons.}
{\setlength\tabcolsep{0pt}
 \begin{tabular}{cp{5pt}r@{}lp{5pt}cp{5pt}c}
 \toprule
 Resonance &&\multicolumn{2}{c}{$A_{D}$} && \multicolumn{1}{c}{Measured by} && \multicolumn{1}{c}{Reference}
 \\ \midrule
 \multirow{5}{*}{$D_{1}^{}(2420)^{0}$} & &$7$ &$.8_{-2.7}^{+6.7}{}_{-1.8}^{+4.6}$ & & ZEUS & & \cite{Abramowicz:2012ys} \\
 & &$ 5$ &$.72\pm0.25$ & & \babar{} & & \cite{delAmoSanchez:2010vq} \\ 
 & &$ 5$ &$.9_{-1.7}^{+3.0}{}_{-1.0}^{+2.4}$ & & ZEUS & & \cite{Chekanov:2008ac} \\ 
 & &$ 3$ &$.8\pm0.6\pm0.8$ & & \babar{} & & \cite{Aubert:2008zc} \\ \cmidrule{3-7}
 & & \cellcolor{Gray}$5$ & \cellcolor{Gray}{$.61 \pm 0.24$} & \cellcolor{Gray} & \cellcolor{Gray} Our average & \cellcolor{Gray} & \\ \midrule
 \multirow{1}{*}{$D_{1}^{}(2420)^{\pm}$} & & \cellcolor{LightGray} $3$ & \cellcolor{LightGray}$.8\pm0.6\pm0.8$ & \cellcolor{LightGray} & \cellcolor{LightGray} \babar{} & \cellcolor{LightGray} & \cite{Aubert:2008zc} \\ \midrule 
 \multirow{1}{*}{$D_{2}^{*}(2460)^{0}$} & & \cellcolor{LightGray} $-1$ & \cellcolor{LightGray}$.16\pm0.35$ & \cellcolor{LightGray} & \cellcolor{LightGray} ZEUS & \cellcolor{LightGray} & \cite{Abramowicz:2012ys} \\ \midrule
 \multirow{1}{*}{$D_{}^{}(2750)^{0}$} & & \cellcolor{LightGray} $-0$ & \cellcolor{LightGray}$.33\pm0.28$ & \cellcolor{LightGray} & \cellcolor{LightGray} \babar{} & \cellcolor{LightGray} & \cite{delAmoSanchez:2010vq} \\ 
 \bottomrule
 \end{tabular}}

\end{center}
\end{table}

\clearpage
\subsection{Excited charm baryons}

In this section we summarize the present status of excited charmed
baryons, decaying strongly or electromagnetically. We list their
masses (or the mass difference between the excited baryon and the
corresponding ground state), natural widths, decay modes, and 
assigned quantum numbers. 
The present ground-state measurements are: $M(\Lambda_c^+)=2286.46\pm0.14$~MeV/$c^2$
measured by \babar~\cite{Aubert:2005gt},
$M(\Xi_c^0)=(2470.85^{+0.28}_{-0.04}$)~MeV/$c^2$ and
$M(\Xi_c^+)=(2467.93^{+0.28}_{-0.40}$)~MeV/$c^2$, both 
dominated by CDF~\cite{Aaltonen:2014wfa}, and
$M(\Omega_c^0)=(2695.2\pm1.7$)~MeV/$c^2$, dominated 
by Belle~\cite{Solovieva:2008fw}. Should these values 
change, so will some of the values for the masses of the excited states.

Table~\ref{sumtable1} summarizes the excited $\Lambda_c^+$ baryons.  
The first two states listed, namely the $\Lambda_c(2595)^+$ and $\Lambda_c(2625)^+$,
are well-established. 
The measured masses and decay patterns suggest that 
they are orbitally 
excited $\Lambda_c^+$ baryons with total angular momentum of the
light quarks $L=1$. Thus their quantum numbers are assigned to be 
$J^P=(\frac{1}{2})^-$ and $J^P=(\frac{3}{2})^-$, respectively. 
Their mass measurements are  
dominated by CDF~\cite{Aaltonen:2011sf}: 
$M(\Lambda_c(2595)^+)=(2592.25\pm 0.24\pm 0.14$)~MeV/$c^2$ and
$M(\Lambda_c(2625)^+)=(2628.11\pm 0.13\pm 0.14$)~MeV/$c^2$. 
Earlier measurements did not fully take into account the restricted
phase-space of the $\Lambda_c(2595)^+$ decays.

The next two states, $\Lambda_c(2765)^+$ and $\Lambda_c(2880)^+$, 
were discovered by CLEO~\cite{Artuso:2000xy} in the $\Lambda_c^+\pi^+\pi^-$ 
final state. CLEO found that a significant fraction of the $\Lambda_c(2880)^+$ decays 
proceeds via an intermediate $\Sigma_c(2445)^{++/0}\pi^{-/+}$.  
Later, \babar~\cite{Aubert:2006sp} 
observed that this state has also a $D^0 p$ decay mode. This was the 
first example of an excited charmed baryon decaying into a charm meson 
plus a baryon; previously all excited charmed baryon were found in their 
hadronic transitions into lower lying charmed 
baryons. In the same analysis, \babar observed for the
first time an additional state, $\Lambda_c(2940)^+$, 
decaying into $D^0 p$. Studying the $D^+ p$ final state,
\babar found no signal; this implies that the $\Lambda_c(2880)^+$ 
and $\Lambda_c(2940)^+$ are $\Lambda_c^+$ excited states
rather than $\Sigma_c$ excitations. 
Belle reported the result of an angular analysis that favors
$5/2$ for the $\Lambda_c(2880)^+$ spin hypothesis. 
Moreover, the measured ratio of branching fractions 
${\cal B}(\Lambda_c(2880)^+\rightarrow \Sigma_c(2520)\pi^{\pm})/{\cal B}(\Lambda_c(2880)^+\rightarrow \Sigma_c(2455)\pi^{\pm})
=(0.225\pm 0.062\pm 0.025)$, combined 
with theoretical predictions based on HQS~\cite{Isgur:1991wq,Cheng:2006dk}, 
favor even parity. However this prediction is only valid if the P-wave 
portion of $\Sigma_c(2520)\pi$ is suppressed. 
LHCb~\cite{Aaij:2017vbw} have analyzed the $D^0p$ system in the resonant substructure of $\Lambda_b$
decays. They confirm the $5/2$ identification of the $\Lambda_c(2880)^+$. In addition they find 
evidence for a further, wider, state they name the $\Lambda_c(2860)^+$, with $J^P=3/2^+$ (with the
parity measured with respect to that of the $\Lambda_c(2880)^+$.) The explanation for these states
in the heavy quark-light diquark model is that they are a pair of orbital D-wave excitations.
Furthermore, LHCb~\cite{Aaij:2017vbw} find evidence for the spin-parity of the $\Lambda_c(2940)^+$
to be $3/2^-$, and improve the world average measurements of both the mass and width of this particle.

A current open question concerns the nature of the $\Lambda_c(2765)^+$ state, or even whether it is
an excited $\Sigma_c^+$ or $\Lambda_c^+$. However, there is no doubt that 
the state exists, as it is clearly visible in Belle~\cite{Joo:2014fka} and LHCb~\cite{Aaij:2017svr} data.

\begin{table}[htb]
\caption{Summary of excited $\Lambda_c^+$ baryons.} 
\vskip0.15in
\begin{center}
\renewcommand{\arraystretch}{1.2}
\begin{tabular}{c|c|c|c|c}
\hline
Charmed baryon   & Mode  & Mass & Natural width  & $J^P$  \\
excited state &  &  (MeV/$c^2$) & (MeV)  \\
\hline
$\Lambda_c(2595)^+$ & $\Lambda_c^+\pi^+\pi^-$, $\Sigma_c(2455)\pi$ &  $2592.25\pm 0.28$ &
$2.59\pm 0.30 \pm 0.47$  & $1/2^-$  \\
\hline
$\Lambda_c(2625)^+$ & $\Lambda_c^+\pi^+\pi^-$   & $2628.11\pm 0.19$ & $<0.97$ & $3/2^-$  \\
\hline
$\Lambda_c(2765)^+$ & $\Lambda_c^+\pi^+\pi^-$, $\Sigma_c(2455)\pi$ & $2766.6\pm 2.4$ & $50$ & ?  \\
\hline
$\Lambda_c(2860)^+$ & $D^0p$ & $2856.1\,^{+2.0}_{-1.7}\pm 0.5\,^{+1.1}_{-5.6}$ &
$67.6\,^{+10.1}_{-8.1}\pm 1.4\,^{+5.9}_{-20.0}$ & $3/2^+$ \\
\hline
$\Lambda_c(2880)^+$ & $\Lambda_c^+\pi^+\pi^-$, $\Sigma_c(2455)\pi$,  &$2881.63\pm 0.24$ &
$5.6\,^{+0.8}_{-0.6} \pm 0.8$ & $5/2^+$ \\
 &  $\Sigma_c(2520)\pi$, $D^0p$     & & &  \\
\hline
$\Lambda_c(2940)^+$ & $D^0p$, $\Sigma_c(2455)\pi$ & $2939.6\,^{+1.3}_{-1.5}$ & $20\,^{+6}_{-5}$  & ?  \\
\hline 
\end{tabular}
\end{center}
\label{sumtable1} 
\end{table}

Table~\ref{sumtable2} summarizes the excited $\Sigma_c^{++,+,0}$ baryons.
The ground iso-triplets of $\Sigma_c(2455)^{++,+,0}$ and
$\Sigma_c(2520)^{++,+,0}$ baryons are well-established. 
Belle~\cite{Lee:2014htd} 
precisely measured the mass differences  
and widths of the doubly charged and neutral members of this triplet.
The short list of excited $\Sigma_c$ baryons is completed by the triplet 
of $\Sigma_c(2800)$ states observed by Belle~\cite{Mizuk:2004yu}. Based 
on the measured masses and theoretical predictions~\cite{Copley:1979wj,Pirjol:1997nh}, 
these states are assumed to be members of the predicted $\Sigma_{c2}$ $3/2^-$
triplet. From a study of resonant substructure 
in $B^-\rightarrow \Lambda_c^+\bar{p}\pi^-$ decays, \babar found 
a significant signal in the $\Lambda_c^+\pi^-$ final state with a mean value 
higher than measured for the $\Sigma_c(2800)$ by Belle by about $3\sigma$
(Table~\ref{sumtable2}). The decay widths measured by
Belle and \babar are consistent, but it is an open question if the 
observed state is the same as the Belle state.

\begin{table}[!htb]
\caption{Summary of the excited $\Sigma_c^{++,+,0}$ baryon family.} 
\vskip0.15in
\begin{center}
\renewcommand{\arraystretch}{1.2}
\begin{tabular}{c|c|c|c|c}
\hline
Charmed baryon   & Mode  & $\Delta M$ & Natural width  & $J^P$  \\
excited state &  &  (MeV/$c^2$) & (MeV)  \\
\hline
$\Sigma_c(2455)^{++}$ &$\Lambda_c^+\pi^+$  & $167.510 \pm 0.17$ & $1.89\,^{+0.09}_{-0.18}$ & $1/2^+$   \\
$\Sigma_c(2455)^{+}$ &$\Lambda_c^+\pi^0$  & $166.4\pm 0.4$ & $<4.6$~@~90$\%$~C.L. & $1/2^+$ \\
$\Sigma_c(2455)^{0}$ &$\Lambda_c^+\pi^-$  & $167.29\pm 0.17$ & $1.83\,^{+0.11}_{-0.19}$ & $1/2^+$    \\
\hline
$\Sigma_c(2520)^{++}$ &$\Lambda_c^+\pi^+$  & $231.95\,^{+0.17}_{-0.12}$ & $14.78\,^{+0.30}_{-0.40}$ & $3/2^+$   \\
$\Sigma_c(2520)^{+}$ &$\Lambda_c^+\pi^0$  & $231.0\pm 2.3$ & $<17$~@~90$\%$~C.L. & $3/2^+$ \\
$\Sigma_c(2520)^{0}$ &$\Lambda_c^+\pi^-$  & $232.02\,^{+0.15}_{-0.14}$ & $15.3\,^{+0.4}_{-0.5}$ & $3/2^+$    \\
\hline
$\Sigma_c(2800)^{++}$ & $\Lambda_c^+\pi^{+}$ & $514\,^{+4}_{-6}$ & $75\,^{+18+12}_{-13-11}$ & $3/2^-$?     \\
$\Sigma_c(2800)^{+}$ & $\Lambda_c^+\pi^{0}$&$505\,^{+15}_{-5}$ &$62\,^{+37+52}_{-23-38}$ &   \\
$\Sigma_c(2800)^{0}$ & $\Lambda_c^+\pi^{-}$&$519\,^{+5}_{-7}$ & $72\,^{+22}_{-15}$ &   \\
 & $\Lambda_c^+\pi^{-}$ & $560\pm 8\pm 10$ & $86\,^{+33}_{-22}$  \\

\hline 
\end{tabular}
\end{center}
\label{sumtable2} 
\end{table}

Table~\ref{sumtable3} summarizes the excited $\Xi_c^{+,0}$. The list of excited $\Xi_c$
baryons has several states, of unknown quantum numbers, having masses 
above 2900~MeV/$c^2$ and decaying into three different types of decay modes:
$\Lambda_c/\Sigma_c n\pi$, $\Xi_c n\pi$ and the most recently observed $\Lambda D$.  
Some of these states ($\Xi_c(2970)^+$, $\Xi_c(3055)$ and $\Xi_c(3080)^{+,0}$) have been
observed by
both Belle~\cite{Chistov:2006zj,YKato:2014,YKato:2016} 
and \babar~\cite{Aubert:2007eb}, are produced in the charm continuum, 
and are considered well-established.
The $\Xi_c(2930)^0$ state decaying into $\Lambda_c^+ K^-$, first reported
by \babar~\cite{Aubert:2007bd} in $B$ decays, has also been observed by Belle~\cite{Li:2017uvv}.
The latter analysis includes a study of the effects of possible interference and other resonances
in the mass distribution, and these are reflected in the large negative systematic uncertainty. As
the \babar~\cite{Aubert:2007bd} paper only ``suggests the presence of a $\Xi_c^0$ resonance,'' we quote
the mass and width measured by Belle~\cite{Li:2017uvv} rather than a weighted sum of the two measurements.
It is unclear if the the fact that the $\Xi_c(2930)$ has been observed in $B$ decays (in contrast
to the charm continuum) can be used to help identify the state.
Belle~\cite{Li:2018fmq} has also reported evidence of its charged partner.

The $\Xi_c(3123)^+$ reported by \babar~\cite{Aubert:2007eb}
in the $\Sigma_c(2520)^{++}\pi^-$ final state has not been
confirmed by Belle~\cite{YKato:2014} with twice the statistics; 
thus its existence is in doubt and it is omitted from Tab.~\ref{sumtable3}.

Several of the width and mass measurements for the $\Xi_c(3055)$ and $\Xi_c(3080)$ 
iso-doublets are only in marginal agreement between experiments and 
decay modes. However, there seems little doubt that the differing 
measurements are of the same particle.

Belle~\cite{Yelton:2016fqw} has recently analyzed large samples of 
$\Xi_c^\prime$, $\Xi_c(2645)$, $\Xi_c(2790)$, $\Xi_c(2815)$ and 
$\Xi_c(2970)$ decays. From this analysis they obtain the most 
precise mass measurements of all five iso-doublets, and the first
significant width measurements of the $\Xi_c(2645)$, $\Xi_c(2790)$ and $\Xi_c(2815)$.
The level of agreement in the different measurements of the mass and width 
of the $\Xi_c(2970)$, formerly named by the PDG as the $\Xi_c(2980)$, is not 
satisfactory. This leaves open the possibility of there being other resonances 
nearby or that threshold effects have not been fully understood.
The present situation in the excited $\Xi_c$ sector is summarized
in Table~\ref{sumtable3}. 

\begin{table}[b]
\caption{Summary of excited $\Xi_c^{+,0}$ states. 
For the first four iso-doublets, the mass difference with respect to the 
ground state is given, as the uncertainties are dominated by the uncertainty
in the ground state mass. In the remaining cases, the uncertainty on the 
measurement of the excited state itself dominates.} 
\vskip0.15in
\resizebox{\textwidth}{!}{
\renewcommand{\arraystretch}{1.2}
\begin{tabular}{c|c|c|c|c}
\hline
Charmed baryon   & Mode  & Mass difference & Natural width  & $J^P$  \\
excited state &  & (MeV/$c^2$) & (MeV) & \\
\hline
$\Xi_c'^+$ & $\Xi_c^+\gamma$ & $110.5 \pm 0.4$  &  & $1/2^+$    \\
$\Xi_c'^0$ & $\Xi_c^0\gamma$ & $108.3\pm 0.4$   &  & $1/2^+$   \\
\hline
$\Xi_c(2645)^+$ & $\Xi_c^0\pi^+$ & $178.5 \pm 0.1$  & $2.1 \pm 0.2 $ & $3/2^+$   \\
$\Xi_c(2645)^0$ & $\Xi_c^+\pi^-$ & $174.7 \pm 0.1$  & $2.4 \pm 0.2 $ & $3/2^+$   \\
\hline
$\Xi_c(2790)^+$ &$\Xi_c'^0\pi^+$ & $320.7\pm 0.5$ & $9  \pm 1$ & $1/2^-$   \\
$\Xi_c(2790)^0$ &$\Xi_c'^+\pi^-$ & $323.8\pm 0.5$ & $10 \pm 1$ & $1/2^-$   \\
\hline
$\Xi_c(2815)^+$ &$\Xi_c(2645)^0\pi^+$ & $348.8\pm 0.1$ & $2.43\pm0.23$ &  $3/2^-$  \\
$\Xi_c(2815)^0$ &$\Xi_c(2645)^+\pi^-$ & $349.4\pm 0.1$  & $2.54\pm0.23$  & $3/2^-$   \\
\hline
\hline
Charmed baryon   & Mode  & Mass  & Natural width  & $J^P$  \\
excited state &  &  (MeV/$c^2$) & (MeV)  \\
\hline
$\Xi_c(2930)^+$ &  $\Lambda_c^+ K^0_S$ &  $2942.3\pm4.4\pm1.5$        &$14.8\pm8.8\pm2.5$& ? \\
$\Xi_c(2930)^0$ & $\Lambda_c^+ K^-$ & $2928.6\pm3^{+0.9}_{-12.0}$ & $19.5\pm8.4^{+5.9}_{-7.9}$ & ?     \\
\hline
$\Xi_c(2970)^+$ & $\Lambda_c^+K^-\pi^+$, $\Sigma_c^{++}K^-$, $\Xi_c(2645)^0\pi^+$
 &  $2967.2\pm 0.8$  & $21 \pm 3$ & ?     \\
$\Xi_c(2970)^0$ & $\Xi_c(2645)^+\pi^-$
&  $2970.4\pm 0.8$ &$28\pm 3$ & ?       \\
\hline
$\Xi_c(3055)^+$ & $\Sigma_c^{++}K^-$, $\Lambda D$ & 	$3055.7\pm 0.4$  &     	$8.0 \pm 1.9 $  & ?   \\
$\Xi_c(3055)^0$ & $\Lambda D$ &                         $3059.0\pm 0.8$   &     $6.2 \pm 2.4$ & ?    \\
\hline
$\Xi_c(3080)^+$ & $\Lambda_c^+K^-\pi^+$, $\Sigma_c^{++}K^-$, $\Sigma_c(2520)^{++}K^-$ , $\Lambda D$ & $3077.8\pm 0.3$ & $3.6\pm 0.7$ & ?   \\
$\Xi_c(3080)^0$ &$\Lambda_c^+ K^0_S\pi^-$, $\Sigma_c^0K^0_S$, $\Sigma_c(2520)^{0}K^0_S$ & $3079.9\pm 1.0$ & $5.6\pm 2.2$ & ?   \\
\hline

\hline 
\end{tabular}
}
\label{sumtable3} 
\end{table}
 The $\Omega_c^{*0}$ doubly-strange charmed baryon has been seen by both 
\babar~\cite{Aubert:2006je} and Belle~\cite{Solovieva:2008fw}.
The mass differences $\delta M=M(\Omega_c^{*0})-M(\Omega_c^0)$ 
measured by the experiments are in good agreement
and are also consistent with most theoretical 
predictions~\cite{Rosner:1995yu,Glozman:1995xy,Jenkins:1996de,
Burakovsky:1997vm}. 
Recently, LHCb~\cite{Aaij:2017nav} has found a family of five excited $\Omega_c^{0}$ baryons
decaying into $\Xi_c^+K^-$. The natural explanation is that they are the five states with $L=1$
between the heavy quark and the light ($ss$) di-quark; however, there is no consensus as to
which state is which, and this overall interpretation is controversial. Four of the five states
have been confirmed by Belle~\cite{Yelton:2017qxg} and, although the Belle dataset is much
smaller than that of LHCb, these mass measurements do contribute to the world averages.
There is evidence for a further, wider, state at higher mass in the LHCb data. Belle
data shows a small excess in the same region, but it is of low significance.  
 
\begin{table}[b]
\caption{Summary of excited $\Omega_c^{0}$ baryons. 
For the $\Omega_c(2770)^0$, the mass difference with respect to the 
ground state is given, as the uncertainty is dominated by the uncertainty
in the ground state mass. In the remaining cases the total mass is shown,
though the uncertainty in the $\Xi_c^+$ mass makes an important contribution
to the total uncertainty.} 
\vskip0.15in
\renewcommand{\arraystretch}{1.2}
\begin{center}
    
\begin{tabular}{c|c|c|c|c}
\hline
Charmed baryon   & Mode  & Mass difference & Natural width  & $J^P$  \\
excited state &  & (MeV/$c^2$)  & (MeV)  \\
\hline
$\Omega_c(2770)^0$ & $\Omega_c^0\gamma$ & $70.7^{+0.8}_{-0.9}$ &      &$3/2^+$  \\
\hline 
\hline
Charmed baryon   & Mode  & Mass  & Natural width  & $J^P$  \\
excited state &  & (MeV/$c^2)$  & (MeV) & \\
\hline
$\Omega_c(3000)^0$  & $\Xi_c^+K^-$  & $3000.4\pm0.4$   &   $4.5\pm0.7$ &? \\
\hline
$\Omega_c(3050)^0$  & $\Xi_c^+K^-$  & $3050.2\pm0.3$   &   $<1.2$& ? \\
\hline
$\Omega_c(3065)^0$  & $\Xi_c^+K^-$  & $3065.5\pm0.4$      &   $3.5\pm0.5$ & ? \\
\hline
$\Omega_c(3090)^0$  & $\Xi_c^+K^-$  & $3090.0\pm0.6$       &   $8.7\pm1.4$ & ? \\
\hline
$\Omega_c(3120)^0$  & $\Xi_c^+K^-$  &  $3119.1\pm1.0$     &   $<2.6$ & ? \\
\hline
\end{tabular}
\label{sumtable4} 
\end{center}
\end{table}

Figure~\ref{charm:leveldiagram} shows the levels of excited charm
baryons along with corresponding transitions between them, and
also transitions to the ground states.
\begin{figure}[!htb]
\includegraphics[width=1.0\textwidth]{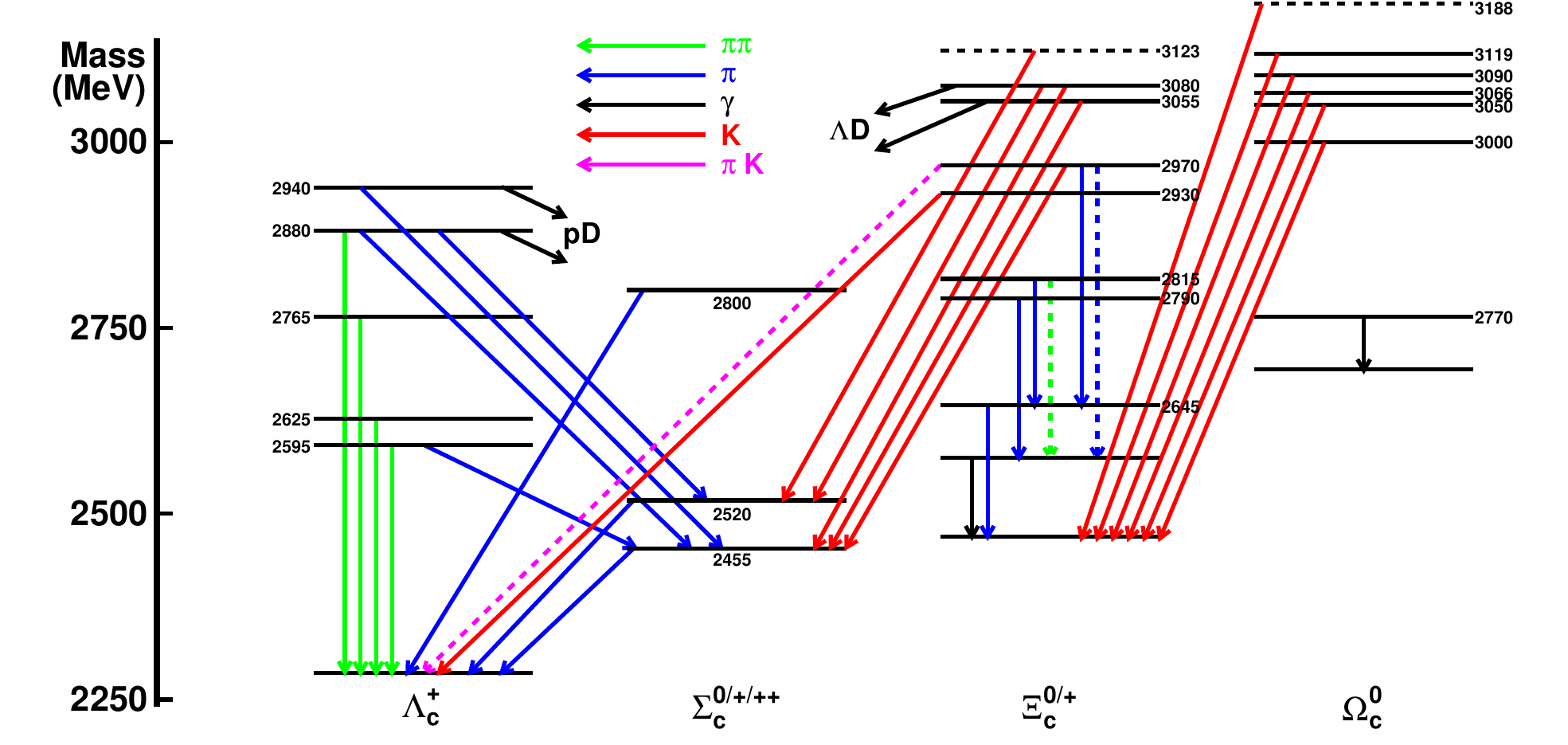}
\caption{Level diagram for multiplets and transitions for excited charm baryons.}
\label{charm:leveldiagram}
\end{figure} 
We note that Belle and \babar\ discovered
that transitions between families are possible, \ie, between 
the $\Xi_c$ and $\Lambda_c^+$ families of excited charmed 
baryons~\cite{Chistov:2006zj,Aubert:2007eb}, and that 
highly excited states are found to decay into a
non-charmed baryon and a $D$ meson\cite{Aubert:2006sp,YKato:2016}.

\clearpage
\subsection{Rare and forbidden decays}
\label{sec:charm:rare}

This section provides a summary of searches for rare and forbidden charm decays
in tabular form. The decay modes can be categorized as 
flavour-changing neutral currents and radiative, lepton-flavour-violating, 
lepton-number-violating, and both baryon- and lepton-number-violating decays.
Figures~\ref{fig:charm:rare_d0}-\ref{fig:charm:lambdac} plot the 
upper limits for $D^0$, $D^+$, $D_s^+$, and $\Lambda_c^+$ decays. 
Tables~\ref{tab:charm:rare_d0}-\ref{tab:charm:rare_lambdac} give the 
corresponding numerical results. Some theoretical predictions are given in 
Refs.~\cite{Burdman:2001tf,Fajfer:2002bu,Fajfer:2007dy,Golowich:2009ii,Paul:2010pq,Borisov:2011aa,Wang:2014dba,deBoer:2015boa}.

Some $D^0$ decay modes have been observed and are quoted as a branching fraction with uncertainties in the tables and shown as a symbol with a line representing the $68\%$ C.L. interval in the plots.

In several cases the rare-decay final states have been observed with 
the di-lepton pair being the decay product of a vector meson.
For these measurements the quoted limits are those expected for the 
non-resonant di-lepton spectrum.
For the extrapolation to the full spectrum a phase-space distribution 
of the non-resonant component has been assumed.
This applies to the CLEO measurement of the decays 
$D_{(s)}^+\to(K^+,\pi^+)e^+e^-$~\cite{Rubin:2010cq}, to the D0 measurements 
of the decays $D_{(s)}^+\to\pi^+\mu^+\mu^-$~\cite{Abazov:2007aj}, and to 
the \babar measurements of the decays $D_{(s)}^+\to(K^+,\pi^+)e^+e^-$ and 
$D_{(s)}^+\to(K^+,\pi^+)\mu^+\mu^-$, where the contribution from 
$\phi\to l^+l^-$ ($l=e,\mu$) has been excluded.
In the case of the LHCb measurements of the decays 
$D^0\to\pi^+\pi^-\mu^+\mu^-$~\cite{Aaij:2013uoa} as 
well as the decays $D_{(s)}^+\to\pi^+\mu^+\mu^-$~\cite{Aaij:2013sua} 
the contributions from $\phi\to l^+l^-$ as well as from 
$\rho,\omega\to l^+l^-$ ($l=e,\mu$) have been excluded. 

\begin{figure}
\begin{center}
\includegraphics[width=5.00in]{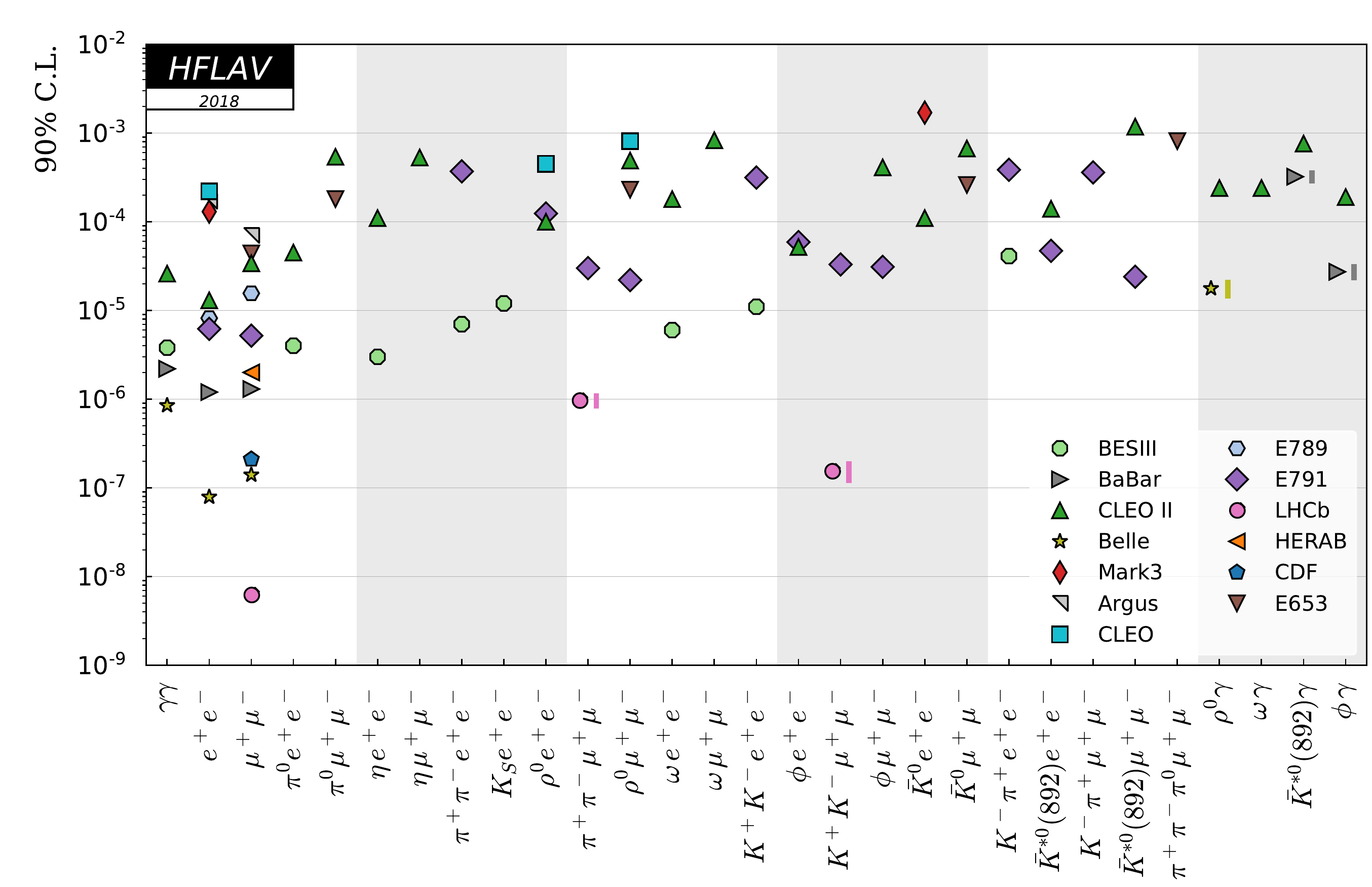}
\vskip0.10in
\includegraphics[width=5.00in]{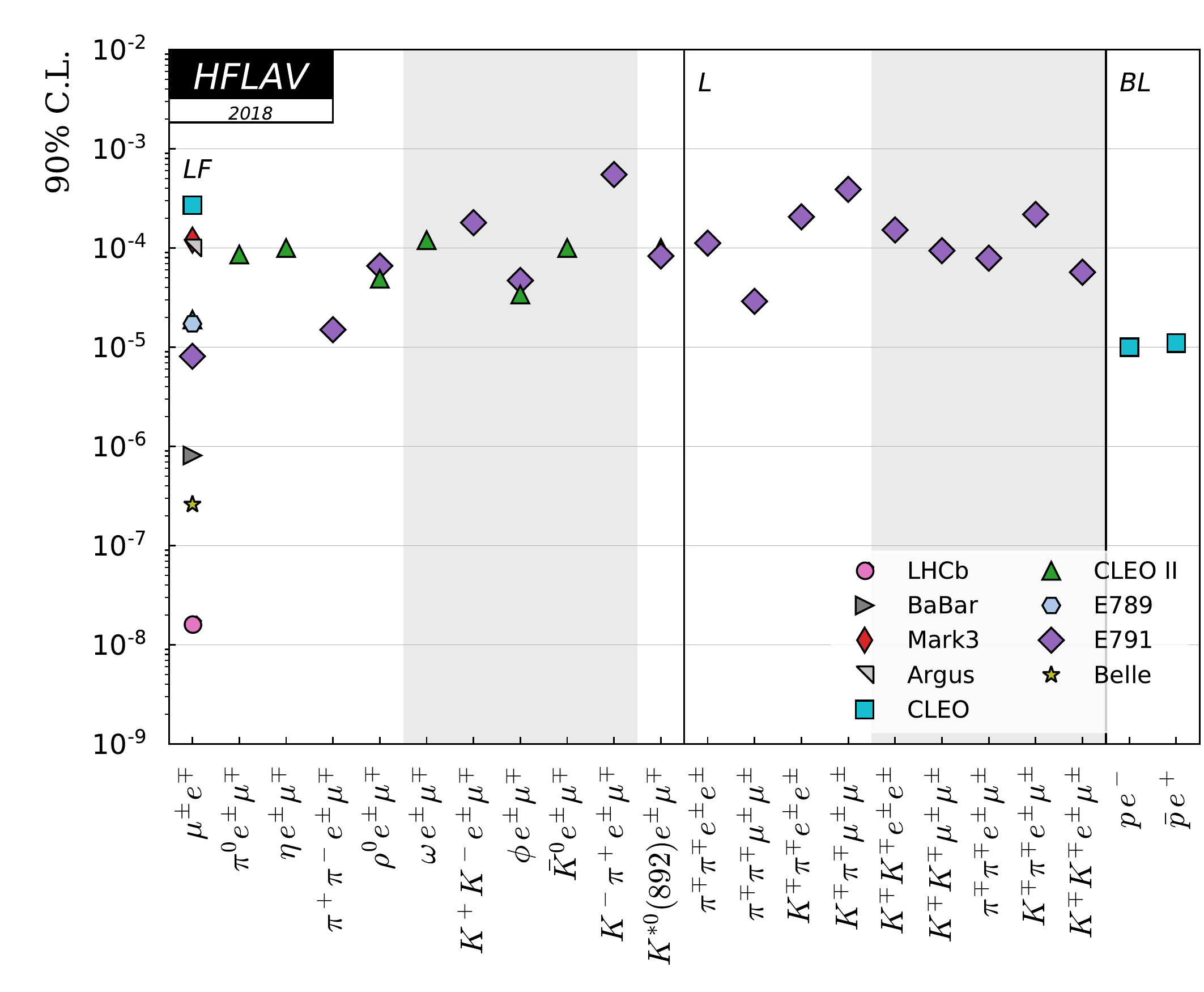}
\caption{Upper limits at $90\%$ C.L.\ for $D^0$ decays. The top plot
shows flavour-changing neutral current decays, and the bottom plot
shows lepton-flavour-changing (LF), lepton-number-changing (L), and 
both baryon- and lepton-number-changing (BL) decays.
}
\label{fig:charm:rare_d0}
\end{center}
\end{figure}

\begin{figure}
\begin{center}
\includegraphics[width=4.00in]{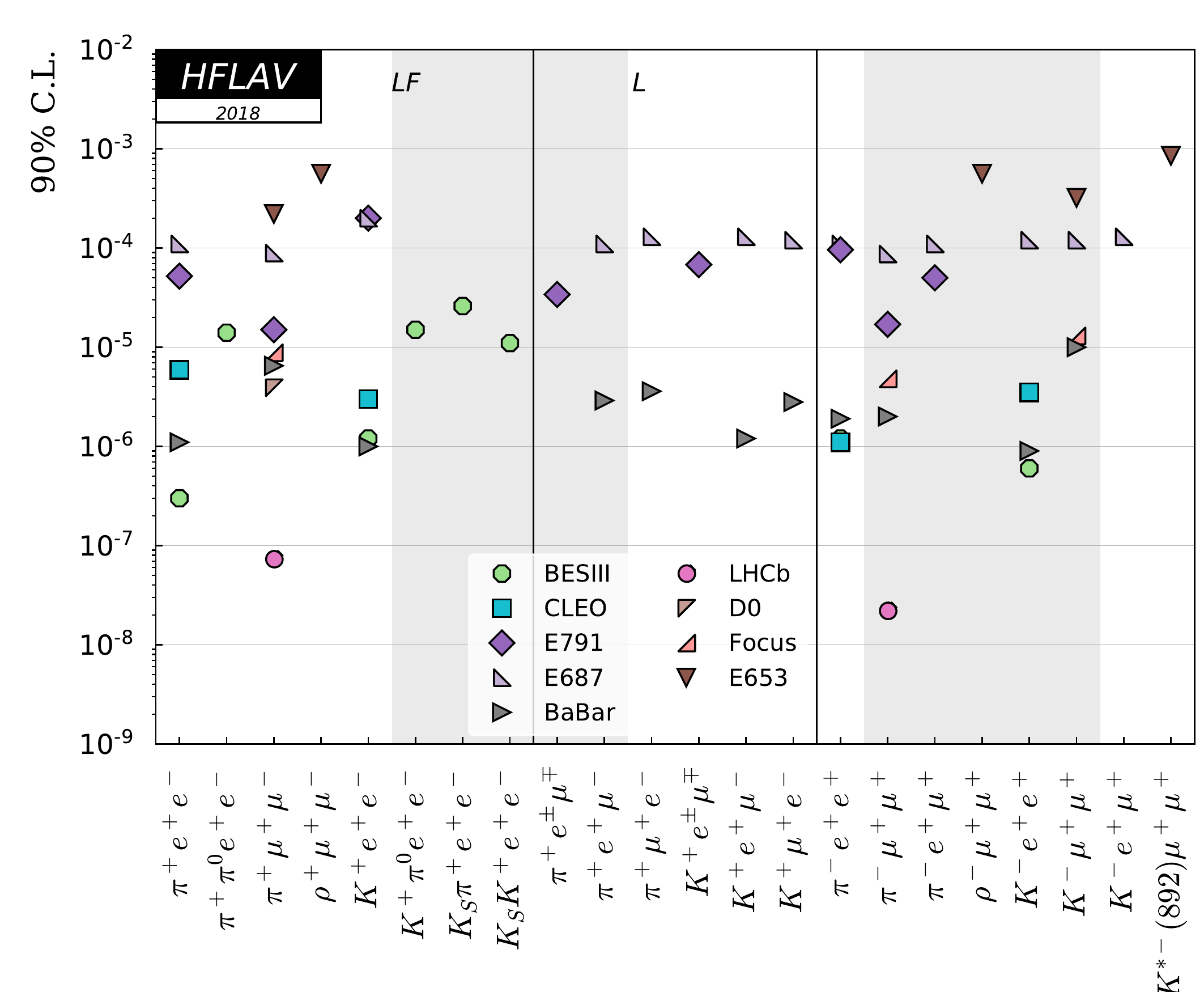}
\vskip0.10in
\includegraphics[width=4.00in]{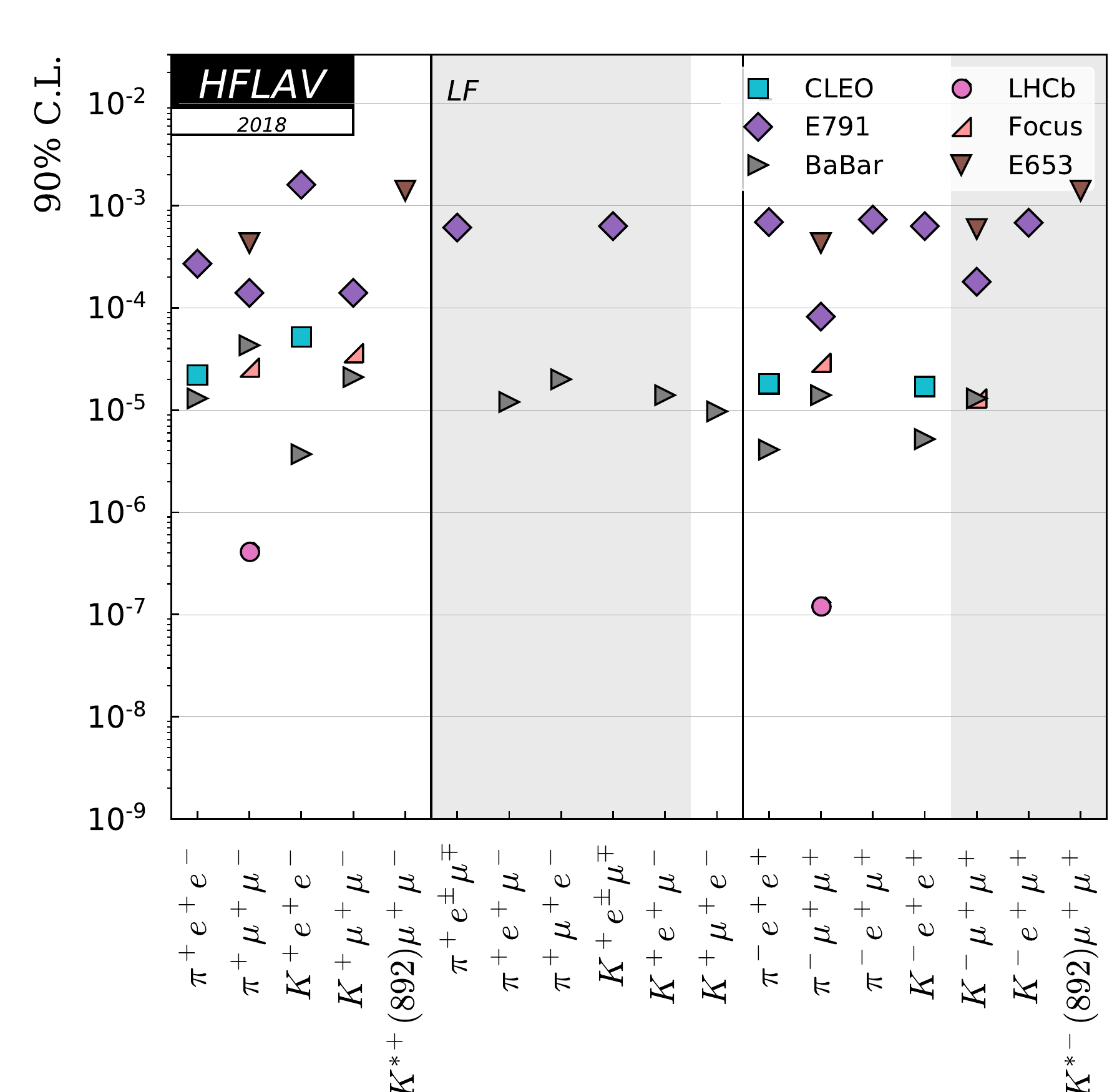}
\caption{Upper limits at $90\%$ C.L.\ for $D^+$ (top) and $D_s^+$ (bottom) 
decays. Each plot shows flavour-changing neutral current decays, 
lepton-flavour-changing decays (LF), and lepton-number-changing (L) decays. 
}
\label{fig:charm:rare_charged}
\end{center}
\end{figure}

\begin{figure}
\begin{center}
\includegraphics[width=3.0in]{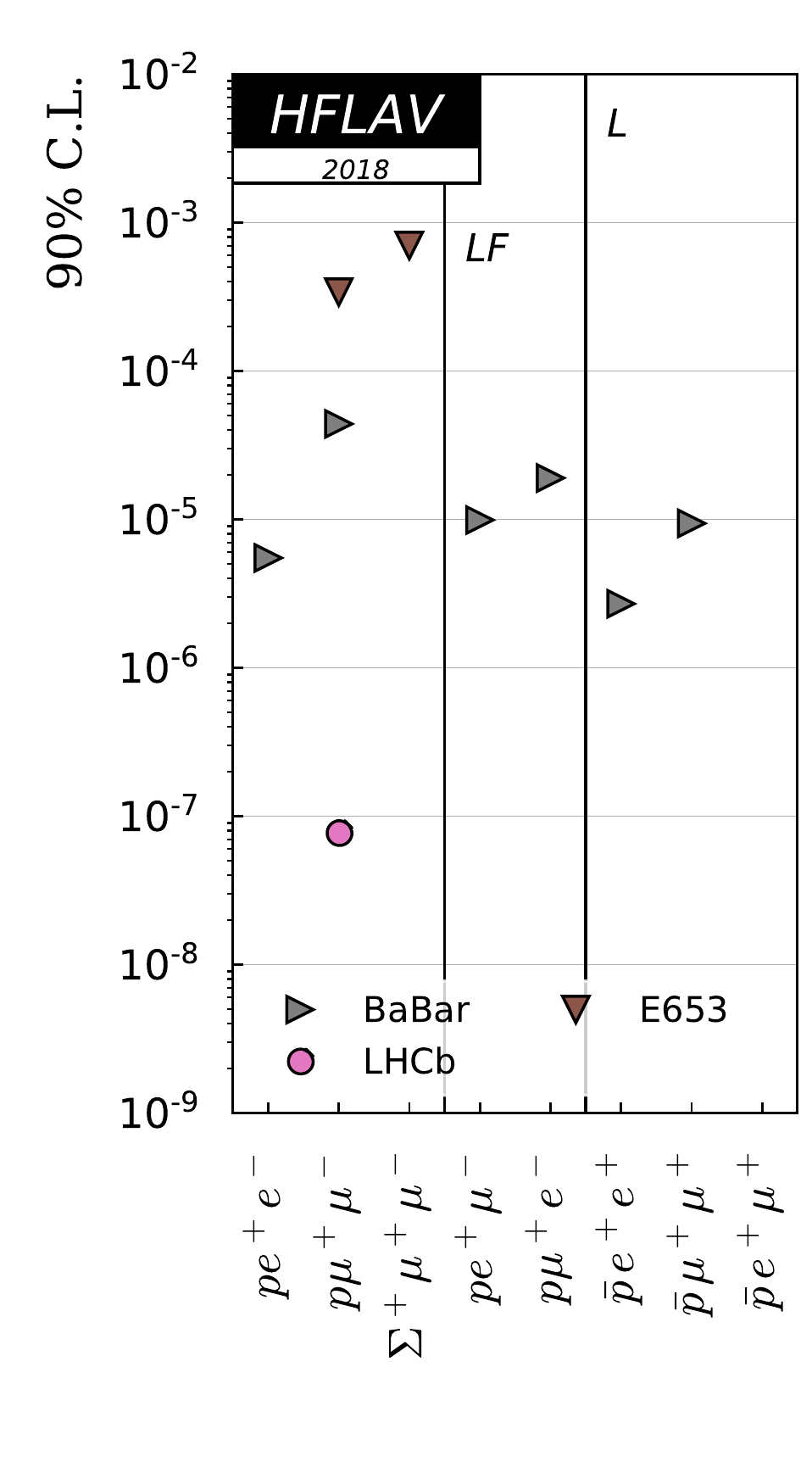}
\caption{Upper limits at $90\%$ C.L.\ for $\Lambda_c^+$ decays. Shown are 
flavour-changing neutral current decays, lepton-flavour-changing (LF) 
decays, and lepton-number-changing (L) decays. }
\label{fig:charm:lambdac}
\end{center}
\end{figure}

\begin{longtable}{l|ccc}
\caption{Upper limits for branching fractions at $90\%$ C.L.\ for $D^0$ decays. Where values are  quoted with uncertainties, these refer to observed branching fractions with the first uncertainty being statistical and all others systematic as detailed in the corresponding reference.
\label{tab:charm:rare_d0}
}\\

\hline\hline%
Decay & BF $\times10^6$ & Experiment & Reference\\
\endfirsthead
\multicolumn{4}{c}{\tablename\ \thetable{} -- continued from previous page} \\ \hline
Decay & BF $\times10^6$ & Experiment & Reference\\
\endhead
\hline
$\gamma{}\gamma{}$ & 26.0 & CLEO II & \cite{Coan:2002te}\\
& 3.8 & BESIII & \cite{Ablikim:2015djc}\\
& 2.2 & BaBar & \cite{Lees:2011qz}\\
& 0.85 & Belle & \cite{Nisar:2015gvd}\\
\hline
$e^+e^-$ & 220.0 & CLEO & \cite{Haas:1988bh}\\
& 170.0 & Argus & \cite{Albrecht:1988ge}\\
& 130.0 & Mark3 & \cite{Adler:1987cp}\\
& 13.0 & CLEO II & \cite{Freyberger:1996it}\\
& 8.19 & E789 & \cite{Pripstein:1999tq}\\
& 6.2 & E791 & \cite{Aitala:1999db}\\
& 1.2 & BaBar & \cite{Aubert:2004bs}\\
& 0.079 & Belle & \cite{Petric:2010yt}\\
\hline
$\mu{}^+\mu{}^-$ & 70.0 & Argus & \cite{Albrecht:1988ge}\\
& 44.0 & E653 & \cite{Kodama:1995ia}\\
& 34.0 & CLEO II & \cite{Freyberger:1996it}\\
& 15.6 & E789 & \cite{Pripstein:1999tq}\\
& 5.2 & E791 & \cite{Aitala:1999db}\\
& 2.0 & HERAb & \cite{Abt:2004hn}\\
& 1.3 & BaBar & \cite{Aubert:2004bs}\\
& 0.21 & CDF & \cite{Aaltonen:2010hz}\\
& 0.14 & Belle & \cite{Petric:2010yt}\\
& 0.0062 & LHCb & \cite{Aaij:2013cza}\\
\hline
$\pi{}^0e^+e^-$ & 45.0 & CLEO II & \cite{Freyberger:1996it}\\
& 4.0 & BESIII & \cite{Ablikim:2018gro}\\
\hline
$\pi{}^0\mu{}^+\mu{}^-$ & 540.0 & CLEO II & \cite{Freyberger:1996it}\\
& 180.0 & E653 & \cite{Kodama:1995ia}\\
\hline
$\eta\,{}e^+e^-$ & 110.0 & CLEO II & \cite{Freyberger:1996it}\\
& 3.0 & BESIII & \cite{Ablikim:2018gro}\\
\hline
$\eta\,{}\mu{}^+\mu{}^-$ & 530.0 & CLEO II & \cite{Freyberger:1996it}\\
\hline
$\pi{}^+\pi{}^-e^+e^-$ & 370.0 & E791 & \cite{Aitala:2000kk}\\
& 7.0 & BESIII & \cite{Ablikim:2018gro}\\
\hline
$K_Se^+e^-$ & 12.0 & BESIII & \cite{Ablikim:2018gro}\\
\hline
$\rho{}^0e^+e^-$ & 450.0 & CLEO & \cite{Haas:1988bh}\\
& 124.0 & E791 & \cite{Aitala:2000kk}\\
& 100.0 & CLEO II & \cite{Freyberger:1996it}\\
\hline
$\pi{}^+\pi{}^-\mu{}^+\mu{}^-$ & 30.0 & E791 & \cite{Aitala:2000kk}\\
& $0.964\pm0.048\pm0.051\pm0.097$ & LHCb & \cite{Aaij:2017iyr}\\
\hline
$\rho{}^0\mu{}^+\mu{}^-$ & 810.0 & CLEO & \cite{Haas:1988bh}\\
& 490.0 & CLEO II & \cite{Freyberger:1996it}\\
& 230.0 & E653 & \cite{Kodama:1995ia}\\
& 22.0 & E791 & \cite{Aitala:2000kk}\\
\hline
$\omega\,{}e^+e^-$ & 180.0 & CLEO II & \cite{Freyberger:1996it}\\
& 6.0 & BESIII & \cite{Ablikim:2018gro}\\
\hline
$\omega\,{}\mu{}^+\mu{}^-$ & 830.0 & CLEO II & \cite{Freyberger:1996it}\\
\hline
$K^+K^-e^+e^-$ & 315.0 & E791 & \cite{Aitala:2000kk}\\
& 11.0 & BESIII & \cite{Ablikim:2018gro}\\
\hline
$\phi\,{}e^+e^-$ & 59.0 & E791 & \cite{Aitala:2000kk}\\
& 52.0 & CLEO II & \cite{Freyberger:1996it}\\
\hline
$K^+K^-\mu{}^+\mu{}^-$ & 33.0 & E791 & \cite{Aitala:2000kk}\\
& $0.154\pm0.027\pm0.009\pm0.016$ & LHCb & \cite{Aaij:2017iyr}\\
\hline
$\phi\,{}\mu{}^+\mu{}^-$ & 410.0 & CLEO II & \cite{Freyberger:1996it}\\
& 31.0 & E791 & \cite{Aitala:2000kk}\\
\hline
$\overline{K}^0e^+e^-$ & 1700.0 & Mark3 & \cite{Adler:1988es}\\
& 110.0 & CLEO II & \cite{Freyberger:1996it}\\
\hline
$\overline{K}^0\mu{}^+\mu{}^-$ & 670.0 & CLEO II & \cite{Freyberger:1996it}\\
& 260.0 & E653 & \cite{Kodama:1995ia}\\
\hline
$K^-\pi{}^+e^+e^-$ & 385.0 & E791 & \cite{Aitala:2000kk}\\
& 41.0 & BESIII & \cite{Ablikim:2018gro}\\
\hline
$\overline{K}^*(892)^0e^+e^-$ & 140.0 & CLEO II & \cite{Freyberger:1996it}\\
& 47.0 & E791 & \cite{Aitala:2000kk}\\
\hline
$K^-\pi{}^+\mu{}^+\mu{}^-$ & 360.0 & E791 & \cite{Aitala:2000kk}\\
\hline
$\overline{K}^*(892)^0\mu{}^+\mu{}^-$ & 1180.0 & CLEO II & \cite{Freyberger:1996it}\\
& 24.0 & E791 & \cite{Aitala:2000kk}\\
\hline
$\pi{}^+\pi{}^-\pi{}^0\mu{}^+\mu{}^-$ & 810.0 & E653 & \cite{Kodama:1995ia}\\
\hline
$\rho{}^0\gamma{}$ & 240.0 & CLEO II & \cite{Asner:1998mv}\\
& $17.7\pm3.0\pm0.7$ & Belle & \cite{Abdesselam:2016yvr}\\
\hline
$\omega\,{}\gamma{}$ & 240.0 & CLEO II & \cite{Asner:1998mv}\\
\hline
$\overline{K}^*(892)^0\gamma{}$ & 760.0 & CLEO II & \cite{Asner:1998mv}\\
& $322.0\pm20.0\pm27.0$ & BaBar & \cite{Aubert:2008ai}\\
\hline
$\phi\,{}\gamma{}$ & 190.0 & CLEO II & \cite{Asner:1998mv}\\
& $27.3\pm3.0\pm2.6$ & BaBar & \cite{Aubert:2008ai}\\
\hline
$\mu{}^{\pm}e^{\mp}$ & 270.0 & CLEO & \cite{Haas:1988bh}\\
& 120.0 & Mark3 & \cite{Becker:1987mu}\\
& 100.0 & Argus & \cite{Albrecht:1988ge}\\
& 19.0 & CLEO II & \cite{Freyberger:1996it}\\
& 17.2 & E789 & \cite{Pripstein:1999tq}\\
& 8.1 & E791 & \cite{Aitala:1999db}\\
& 0.81 & BaBar & \cite{Aubert:2004bs}\\
& 0.26 & Belle & \cite{Petric:2010yt}\\
& 0.016 & LHCb & \cite{Aaij:2015qmj}\\
\hline
$\pi{}^0e^{\pm}\mu{}^{\mp}$ & 86.0 & CLEO II & \cite{Freyberger:1996it}\\
\hline
$\eta\,{}e^{\pm}\mu{}^{\mp}$ & 100.0 & CLEO II & \cite{Freyberger:1996it}\\
\hline
$\pi{}^+\pi{}^-e^{\pm}\mu{}^{\mp}$ & 15.0 & E791 & \cite{Aitala:2000kk}\\
\hline
$\rho{}^0e^{\pm}\mu{}^{\mp}$ & 66.0 & E791 & \cite{Aitala:2000kk}\\
& 49.0 & CLEO II & \cite{Freyberger:1996it}\\
\hline
$\omega\,{}e^{\pm}\mu{}^{\mp}$ & 120.0 & CLEO II & \cite{Freyberger:1996it}\\
\hline
$K^+K^-e^{\pm}\mu{}^{\mp}$ & 180.0 & E791 & \cite{Aitala:2000kk}\\
\hline
$\phi\,{}e^{\pm}\mu{}^{\mp}$ & 47.0 & E791 & \cite{Aitala:2000kk}\\
& 34.0 & CLEO II & \cite{Freyberger:1996it}\\
\hline
$\overline{K}^0e^{\pm}\mu{}^{\mp}$ & 100.0 & CLEO II & \cite{Freyberger:1996it}\\
\hline
$K^-\pi{}^+e^{\pm}\mu{}^{\mp}$ & 550.0 & E791 & \cite{Aitala:2000kk}\\
\hline
$K^*(892)^0e^{\pm}\mu{}^{\mp}$ & 100.0 & CLEO II & \cite{Freyberger:1996it}\\
& 83.0 & E791 & \cite{Aitala:2000kk}\\
\hline
$\pi{}^{\mp}\pi{}^{\mp}e^{\pm}e^{\pm}$ & 112.0 & E791 & \cite{Aitala:2000kk}\\
\hline
$\pi{}^{\mp}\pi{}^{\mp}\mu{}^{\pm}\mu{}^{\pm}$ & 29.0 & E791 & \cite{Aitala:2000kk}\\
\hline
$K^{\mp}\pi{}^{\mp}e^{\pm}e^{\pm}$ & 206.0 & E791 & \cite{Aitala:2000kk}\\
\hline
$K^{\mp}\pi{}^{\mp}\mu{}^{\pm}\mu{}^{\pm}$ & 390.0 & E791 & \cite{Aitala:2000kk}\\
\hline
$K^{\mp}K^{\mp}e^{\pm}e^{\pm}$ & 152.0 & E791 & \cite{Aitala:2000kk}\\
\hline
$K^{\mp}K^{\mp}\mu{}^{\pm}\mu{}^{\pm}$ & 94.0 & E791 & \cite{Aitala:2000kk}\\
\hline
$\pi{}^{\mp}\pi{}^{\mp}e^{\pm}\mu{}^{\pm}$ & 79.0 & E791 & \cite{Aitala:2000kk}\\
\hline
$K^{\mp}\pi{}^{\mp}e^{\pm}\mu{}^{\pm}$ & 218.0 & E791 & \cite{Aitala:2000kk}\\
\hline
$K^{\mp}K^{\mp}e^{\pm}\mu{}^{\pm}$ & 57.0 & E791 & \cite{Aitala:2000kk}\\
\hline
$p\,e^-$ & 10.0 & CLEO & \cite{Rubin:2009aa}\\
\hline
$\overline{p}\,{}e^+$ & 11.0 & CLEO & \cite{Rubin:2009aa}\\
\hline
\end{longtable}
\pagebreak

\begin{longtable}{l|ccc}
\caption{Upper limits at $90\%$ C.L.\ for $D^+$ decays.\label{tab:charm:rare_dplus}}\\
\hline\hline
Decay & Limit $\times10^6$ & Experiment & Reference\\
\endfirsthead
\multicolumn{4}{c}{\tablename\ \thetable{} -- continued from previous page} \\ \hline
Decay & Limit $\times10^6$ & Experiment & Reference\\
\endhead

\hline
$\pi{}^+e^+e^-$ & 110.0 & E687 & \cite{Frabetti:1997wp}\\
& 52.0 & E791 & \cite{Aitala:1999db}\\
& 5.9 & CLEO & \cite{Rubin:2010cq}\\
& 1.1 & BaBar & \cite{Lees:2011hb}\\
& 0.3 & BESIII & \cite{Zhao:2016jna}\\
\hline
$\pi{}^+\pi{}^0e^+e^-$ & 14.0 & BESIII & \cite{Ablikim:2018gro}\\
\hline
$\pi{}^+\mu{}^+\mu{}^-$ & 220.0 & E653 & \cite{Kodama:1995ia}\\
& 89.0 & E687 & \cite{Frabetti:1997wp}\\
& 15.0 & E791 & \cite{Aitala:1999db}\\
& 8.8 & Focus & \cite{Link:2003qp}\\
& 6.5 & BaBar & \cite{Lees:2011hb}\\
& 3.9 & D0 & \cite{Abazov:2007aj}\\
& 0.073 & LHCb & \cite{Aaij:2013sua}\\
\hline
$K^+\mu{}^+\mu{}^-$ & 97.0 & E687 & \cite{Frabetti:1997wp}\\
& 44.0 & E791 & \cite{Aitala:1999db}\\
& 9.2 & Focus & \cite{Link:2003qp}\\
& 4.3 & BaBar & \cite{Lees:2011hb}\\
\hline
$\rho{}^+\mu{}^+\mu{}^-$ & 560.0 & E653 & \cite{Kodama:1995ia}\\
\hline
$K^+e^+e^-$ & 200.0 & E687 & \cite{Frabetti:1997wp}\\
& 3.0 & CLEO & \cite{Rubin:2010cq}\\
& 1.2 & BESIII & \cite{Zhao:2016jna}\\
& 1.0 & BaBar & \cite{Lees:2011hb}\\
\hline
$K^+\pi{}^0e^+e^-$ & 15.0 & BESIII & \cite{Ablikim:2018gro}\\
\hline
$K_S\pi{}^+e^+e^-$ & 26.0 & BESIII & \cite{Ablikim:2018gro}\\
\hline
$K_SK^+e^+e^-$ & 11.0 & BESIII & \cite{Ablikim:2018gro}\\
\hline
$\pi{}^+e^{\pm}\mu{}^{\mp}$ & 34.0 & E791 & \cite{Aitala:1999db}\\
\hline
$\pi{}^+e^+\mu{}^-$ & 110.0 & E687 & \cite{Frabetti:1997wp}\\
& 2.9 & BaBar & \cite{Lees:2011hb}\\
\hline
$\pi{}^+\mu{}^+e^-$ & 130.0 & E687 & \cite{Frabetti:1997wp}\\
& 3.6 & BaBar & \cite{Lees:2011hb}\\
\hline
$K^+e^{\pm}\mu{}^{\mp}$ & 68.0 & E791 & \cite{Aitala:1999db}\\
\hline
$K^+e^+\mu{}^-$ & 130.0 & E687 & \cite{Frabetti:1997wp}\\
& 1.2 & BaBar & \cite{Lees:2011hb}\\
\hline
$K^+\mu{}^+e^-$ & 120.0 & E687 & \cite{Frabetti:1997wp}\\
& 2.8 & BaBar & \cite{Lees:2011hb}\\
\hline
$\pi{}^-e^+e^+$ & 110.0 & E687 & \cite{Frabetti:1997wp}\\
& 96.0 & E791 & \cite{Aitala:1999db}\\
& 1.9 & BaBar & \cite{Lees:2011hb}\\
& 1.2 & BESIII & \cite{Zhao:2016jna}\\
& 1.1 & CLEO & \cite{Rubin:2010cq}\\
\hline
$\pi{}^-\mu{}^+\mu{}^+$ & 87.0 & E687 & \cite{Frabetti:1997wp}\\
& 17.0 & E791 & \cite{Aitala:1999db}\\
& 4.8 & Focus & \cite{Link:2003qp}\\
& 2.0 & BaBar & \cite{Lees:2011hb}\\
& 0.022 & LHCb & \cite{Aaij:2013sua}\\
\hline
$\pi{}^-e^+\mu{}^+$ & 110.0 & E687 & \cite{Frabetti:1997wp}\\
& 50.0 & E791 & \cite{Aitala:1999db}\\
& 2.0 & BaBar & \cite{Lees:2011hb}\\
\hline
$\rho{}^-\mu{}^+\mu{}^+$ & 560.0 & E653 & \cite{Kodama:1995ia}\\
\hline
$K^-e^+e^+$ & 120.0 & E687 & \cite{Frabetti:1997wp}\\
& 3.5 & CLEO & \cite{Rubin:2010cq}\\
& 0.9 & BaBar & \cite{Lees:2011hb}\\
& 0.6 & BESIII & \cite{Zhao:2016jna}\\
\hline
$K^-\mu{}^+\mu{}^+$ & 320.0 & E653 & \cite{Kodama:1995ia}\\
& 120.0 & E687 & \cite{Frabetti:1997wp}\\
& 13.0 & Focus & \cite{Link:2003qp}\\
& 10.0 & BaBar & \cite{Lees:2011hb}\\
\hline
$K^-e^+\mu{}^+$ & 130.0 & E687 & \cite{Frabetti:1997wp}\\
\hline
$K^*(892)^-\mu{}^+\mu{}^+$ & 850.0 & E653 & \cite{Kodama:1995ia}\\
\hline
\end{longtable}

\begin{longtable}{l|ccc}
\caption{Upper limits at $90\%$ C.L.\ for $D_s^+$ decays.\label{tab:charm:rare_dsplus}}\\
\hline\hline
Decay & Limit $\times10^6$ & Experiment & Reference\\
\endfirsthead
\multicolumn{4}{c}{\tablename\ \thetable{} -- continued from previous page} \\ \hline
Decay & Limit $\times10^6$ & Experiment & Reference\\
\endhead

\hline
$\pi{}^+e^+e^-$ & 270.0 & E791 & \cite{Aitala:1999db}\\
& 22.0 & CLEO & \cite{Rubin:2010cq}\\
& 13.0 & BaBar & \cite{Lees:2011hb}\\
\hline
$\pi{}^+\mu{}^+\mu{}^-$ & 430.0 & E653 & \cite{Kodama:1995ia}\\
& 140.0 & E791 & \cite{Aitala:1999db}\\
& 43.0 & BaBar & \cite{Lees:2011hb}\\
& 26.0 & Focus & \cite{Link:2003qp}\\
& 0.41 & LHCb & \cite{Aaij:2013sua}\\
\hline
$K^+e^+e^-$ & 1600.0 & E791 & \cite{Aitala:1999db}\\
& 52.0 & CLEO & \cite{Rubin:2010cq}\\
& 3.7 & BaBar & \cite{Lees:2011hb}\\
\hline
$K^+\mu{}^+\mu{}^-$ & 140.0 & E791 & \cite{Aitala:1999db}\\
& 36.0 & Focus & \cite{Link:2003qp}\\
& 21.0 & BaBar & \cite{Lees:2011hb}\\
\hline
$K^*(892)^+\mu{}^+\mu{}^-$ & 1400.0 & E653 & \cite{Kodama:1995ia}\\
\hline
$\pi{}^+e^{\pm}\mu{}^{\mp}$ & 610.0 & E791 & \cite{Aitala:1999db}\\
\hline
$\pi{}^+e^+\mu{}^-$ & 12.0 & BaBar & \cite{Lees:2011hb}\\
\hline
$\pi{}^+\mu{}^+e^-$ & 20.0 & BaBar & \cite{Lees:2011hb}\\
\hline
$K^+e^{\pm}\mu{}^{\mp}$ & 630.0 & E791 & \cite{Aitala:1999db}\\
\hline
$K^+e^+\mu{}^-$ & 14.0 & BaBar & \cite{Lees:2011hb}\\
\hline
$K^+\mu{}^+e^-$ & 9.7 & BaBar & \cite{Lees:2011hb}\\
\hline
$\pi{}^-e^+e^+$ & 690.0 & E791 & \cite{Aitala:1999db}\\
& 18.0 & CLEO & \cite{Rubin:2010cq}\\
& 4.1 & BaBar & \cite{Lees:2011hb}\\
\hline
$\pi{}^-\mu{}^+\mu{}^+$ & 430.0 & E653 & \cite{Kodama:1995ia}\\
& 82.0 & E791 & \cite{Aitala:1999db}\\
& 29.0 & Focus & \cite{Link:2003qp}\\
& 14.0 & BaBar & \cite{Lees:2011hb}\\
& 0.12 & LHCb & \cite{Aaij:2013sua}\\
\hline
$\pi{}^-e^+\mu{}^+$ & 730.0 & E791 & \cite{Aitala:1999db}\\
& 8.4 & BaBar & \cite{Lees:2011hb}\\
\hline
$K^-e^+e^+$ & 630.0 & E791 & \cite{Aitala:1999db}\\
& 17.0 & CLEO & \cite{Rubin:2010cq}\\
& 5.2 & BaBar & \cite{Lees:2011hb}\\
\hline
$K^-\mu{}^+\mu{}^+$ & 590.0 & E653 & \cite{Kodama:1995ia}\\
& 180.0 & E791 & \cite{Aitala:1999db}\\
& 13.0 & BaBar & \cite{Lees:2011hb}\\
\hline
$K^-e^+\mu{}^+$ & 680.0 & E791 & \cite{Aitala:1999db}\\
& 6.1 & BaBar & \cite{Lees:2011hb}\\
\hline
$K^*(892)^-\mu{}^+\mu{}^+$ & 1400.0 & E653 & \cite{Kodama:1995ia}\\
\hline
\end{longtable}

\begin{table}
\centering
\caption{Upper limits at $90\%$ C.L.\ for $\Lambda_c^+$ decays.\label{tab:charm:rare_lambdac}}
\begin{tabular}{l|ccc}
\hline\hline
Decay & Limit $\times10^6$ & Experiment & Reference\\

\hline
$pe^+e^-$ & 5.5 & BaBar & \cite{Lees:2011hb}\\
\hline
$p\mu{}^+\mu{}^-$ & 340.0 & E653 & \cite{Kodama:1995ia}\\
& 44.0 & BaBar & \cite{Lees:2011hb}\\
& 0.077 & LHCb & \cite{Aaij:2017nsd}\\
\hline
$\Sigma{}^+\mu{}^+\mu{}^-$ & 700.0 & E653 & \cite{Kodama:1995ia}\\
\hline
$pe^+\mu{}^-$ & 9.9 & BaBar & \cite{Lees:2011hb}\\
\hline
$p\mu{}^+e^-$ & 19.0 & BaBar & \cite{Lees:2011hb}\\
\hline
$\overline{p}\,{}e^+e^+$ & 2.7 & BaBar & \cite{Lees:2011hb}\\
\hline
$\overline{p}\,{}\mu{}^+\mu{}^+$ & 9.4 & BaBar & \cite{Lees:2011hb}\\
\hline
$\overline{p}\,{}e^+\mu{}^+$ & 16.0 & BaBar & \cite{Lees:2011hb}\\
\hline
\end{tabular}
\end{table}

\clearpage
\section{Tau lepton properties}
\label{sec:tau}
\input{tau/tau_all_common}

\section{Acknowledgments}

We are grateful for the strong support of the 
ATLAS, \babar, \belle, BESIII, CLEO(c), CDF, CMS, 
\dzero\ and LHCb collaborations, without whom this 
compilation of results and world averages would not have  
been possible. The success of these experiments in turn would 
not have been possible without the excellent operations of the 
BEPC-II, CESR, KEKB, LHC, PEP-II, and Tevatron accelerators.
We also recognise the interplay between theoretical and experimental
communities that has provided a stimulus for many of the measurements 
in this document.
We thank the SLAC National Accelerator Laboratory for 
providing crucial computing resources and support to HFLAV in past years and CERN for taking this over from 2018.

Our averages and this compilation have benefitted greatly from 
contributions to the Heavy Flavour Averaging Group from numerous
individuals. In particular, we would like to thank Yinghui Guan for contributions to the charm physics chapter, and Paolo Gambino and Andreas Kronfeld for discussions regarding the averages in the semileptonic chapter. We also thank all past members of the HFLAV.
We especially thank the following for their careful review of the 
text in preparing this paper for publication: 
G.~Cowan, 
F.~Muheim,
U.~Nierst,
K-F~Chen,
Th.~Mannel,
A.~El-Khadra,
R.~Kowalewski,
H.-Y.~Cheng,
M.~Kreps,
G.~Mohanty,
G.~Hiller,
M.~Charles,
J.~Brod,
M.~Jamin,
K.~Moenig.

Members of HFLAV are supported by the following funding agencies:
Australian Research Council (Australia);
Ministry of Science and Technology, National Natural Science Foundation, Key Research Program of Frontier Sciences of the Chinese Academy of Sciences (China);  
CNRS/IN2P3 (France); 
Deutsche Forschungsgemeinschaft (Germany); 
German-Israeli Foundation for Scientific research and Development, Israel Science Foundation, Ministry of Science and Technology, US-Israel Binational Science Fund (Israel);
Istituto Nazionale di Fisica Nucleare (Italy);
Ministry of Education, Culture, Sports, Science and Technology, Japan Society for the Promotion of Science (Japan);
National Agency for Academic Exchange, Ministry of Science and Higher Education, National Science Centre (Poland);
Swiss National Science Foundation (Switzerland);
Science and Technology Facilities Council (UK);
Department of Energy (USA). 
\clearpage

\bibliographystyle{tex/HFLAV}
\raggedright
\setlength{\parskip}{0pt}
\setlength{\itemsep}{0pt plus 0.3ex}
\begin{small}
\bibliography{main,life_mix/life_mix,unitary/cp_uta,slbdecays/slb_ref,rare/RareDecaysBib,b2charm/b2charm,charm/charm_refs,tau/tau-refs,tau/tau-refs-pdg}

\ifx\mcitethebibliography\mciteundefinedmacro
\PackageError{unsrtM.bst}{mciteplus.sty has not been loaded}
{This bibstyle requires the use of the mciteplus package.}\fi
\begin{mcitethebibliography}{1000}

\bibitem{Amhis:2016xyh}
{Heavy Flavor Averaging Group}, Y.~Amhis {\em et al.},
  \href{http://dx.doi.org/10.1140/epjc/s10052-017-5058-4}{Eur. Phys. J. {\bf
  C77},  895 (2017)}, \href{http://arxiv.org/abs/1612.07233}{{\tt
  arXiv:1612.07233 [hep-ex]}}\relax
\mciteBstWouldAddEndPuncttrue
\mciteSetBstMidEndSepPunct{\mcitedefaultmidpunct}
{\mcitedefaultendpunct}{\mcitedefaultseppunct}\relax
\EndOfBibitem
\bibitem{Cabibbo:1963yz}
N.~Cabibbo, \href{http://dx.doi.org/10.1103/PhysRevLett.10.531}{Phys. Rev.
  Lett. {\bf 10},  531 (1963)}\relax
\mciteBstWouldAddEndPuncttrue
\mciteSetBstMidEndSepPunct{\mcitedefaultmidpunct}
{\mcitedefaultendpunct}{\mcitedefaultseppunct}\relax
\EndOfBibitem
\bibitem{Kobayashi:1973fv}
M.~Kobayashi and T.~Maskawa, \href{http://dx.doi.org/10.1143/PTP.49.652}{Prog.
  Theor. Phys. {\bf 49},  652 (1973)}\relax
\mciteBstWouldAddEndPuncttrue
\mciteSetBstMidEndSepPunct{\mcitedefaultmidpunct}
{\mcitedefaultendpunct}{\mcitedefaultseppunct}\relax
\EndOfBibitem
\bibitem{Abbaneo:2000ej_mod}
{ALEPH, CDF, DELPHI, L3, OPAL, and SLD collaborations}, D.~Abbaneo {\em et
  al.}, \href{http://arxiv.org/abs/hep-ex/0009052}{{\tt arXiv:hep-ex/0009052}}
  (2000), CERN-EP-2000-096\relax
\mciteBstWouldAddEndPuncttrue
\mciteSetBstMidEndSepPunct{\mcitedefaultmidpunct}
{\mcitedefaultendpunct}{\mcitedefaultseppunct}\relax
\EndOfBibitem
\bibitem{Abbaneo:2001bv_mod_cont}
\href{http://arxiv.org/abs/hep-ex/0112028}{{\tt arXiv:hep-ex/0112028}} (2001),
  CERN-EP-2001-050\relax
\mciteBstWouldAddEndPuncttrue
\mciteSetBstMidEndSepPunct{\mcitedefaultmidpunct}
{\mcitedefaultendpunct}{\mcitedefaultseppunct}\relax
\EndOfBibitem
\bibitem{PDG_2016}
{Particle Data Group}, C.~Patrignani {\em et al.},
  \href{http://dx.doi.org/10.1088/1674-1137/40/10/100001}{Chin. Phys. {\bf
  C40},  100001 (2016)}\relax
\mciteBstWouldAddEndPuncttrue
\mciteSetBstMidEndSepPunct{\mcitedefaultmidpunct}
{\mcitedefaultendpunct}{\mcitedefaultseppunct}\relax
\EndOfBibitem
\bibitem{Alexander:2000tb}
{CLEO} collaboration, J.~P. Alexander {\em et al.},
  \href{http://dx.doi.org/10.1103/PhysRevLett.86.2737}{Phys. Rev. Lett. {\bf
  86},  2737 (2001)}, \href{http://arxiv.org/abs/hep-ex/0006002}{{\tt
  arXiv:hep-ex/0006002 [hep-ex]}}\relax
\mciteBstWouldAddEndPuncttrue
\mciteSetBstMidEndSepPunct{\mcitedefaultmidpunct}
{\mcitedefaultendpunct}{\mcitedefaultseppunct}\relax
\EndOfBibitem
\bibitem{Athar:2002mr}
{CLEO} collaboration, S.~B. Athar {\em et al.},
  \href{http://dx.doi.org/10.1103/PhysRevD.66.052003}{Phys. Rev. {\bf D66},
  052003 (2002)}, \href{http://arxiv.org/abs/hep-ex/0202033}{{\tt
  arXiv:hep-ex/0202033 [hep-ex]}}\relax
\mciteBstWouldAddEndPuncttrue
\mciteSetBstMidEndSepPunct{\mcitedefaultmidpunct}
{\mcitedefaultendpunct}{\mcitedefaultseppunct}\relax
\EndOfBibitem
\bibitem{Hastings:2002ff}
{Belle} collaboration, N.~C. Hastings {\em et al.},
  \href{http://dx.doi.org/10.1103/PhysRevD.67.052004}{Phys. Rev. {\bf D67},
  052004 (2003)}, \href{http://arxiv.org/abs/hep-ex/0212033}{{\tt
  arXiv:hep-ex/0212033 [hep-ex]}}\relax
\mciteBstWouldAddEndPuncttrue
\mciteSetBstMidEndSepPunct{\mcitedefaultmidpunct}
{\mcitedefaultendpunct}{\mcitedefaultseppunct}\relax
\EndOfBibitem
\bibitem{Aubert:2004rz}
{\babar} collaboration, B.~Aubert {\em et al.},
  \href{http://dx.doi.org/10.1103/PhysRevLett.94.141801}{Phys. Rev. Lett. {\bf
  94},  141801 (2005)}, \href{http://arxiv.org/abs/hep-ex/0412062}{{\tt
  arXiv:hep-ex/0412062 [hep-ex]}}\relax
\mciteBstWouldAddEndPuncttrue
\mciteSetBstMidEndSepPunct{\mcitedefaultmidpunct}
{\mcitedefaultendpunct}{\mcitedefaultseppunct}\relax
\EndOfBibitem
\bibitem{Barish:1994mu}
{CLEO} collaboration, B.~Barish {\em et al.},
  \href{http://dx.doi.org/10.1103/PhysRevD.51.1014}{Phys. Rev. {\bf D51},  1014
  (1995)}, \href{http://arxiv.org/abs/hep-ex/9406005}{{\tt arXiv:hep-ex/9406005
  [hep-ex]}}\relax
\mciteBstWouldAddEndPuncttrue
\mciteSetBstMidEndSepPunct{\mcitedefaultmidpunct}
{\mcitedefaultendpunct}{\mcitedefaultseppunct}\relax
\EndOfBibitem
\bibitem{Aubert:2005bq}
{\babar} collaboration, B.~Aubert {\em et al.},
  \href{http://dx.doi.org/10.1103/PhysRevLett.95.042001}{Phys. Rev. Lett. {\bf
  95},  042001 (2005)}, \href{http://arxiv.org/abs/hep-ex/0504001}{{\tt
  arXiv:hep-ex/0504001 [hep-ex]}}\relax
\mciteBstWouldAddEndPuncttrue
\mciteSetBstMidEndSepPunct{\mcitedefaultmidpunct}
{\mcitedefaultendpunct}{\mcitedefaultseppunct}\relax
\EndOfBibitem
\bibitem{Aubert:2006bm}
{\babar} collaboration, B.~Aubert {\em et al.},
  \href{http://dx.doi.org/10.1103/PhysRevLett.96.232001}{Phys. Rev. Lett. {\bf
  96},  232001 (2006)}, \href{http://arxiv.org/abs/hep-ex/0604031}{{\tt
  arXiv:hep-ex/0604031 [hep-ex]}}\relax
\mciteBstWouldAddEndPuncttrue
\mciteSetBstMidEndSepPunct{\mcitedefaultmidpunct}
{\mcitedefaultendpunct}{\mcitedefaultseppunct}\relax
\EndOfBibitem
\bibitem{Sokolov:2006sd}
{Belle} collaboration, A.~Sokolov {\em et al.},
  \href{http://dx.doi.org/10.1103/PhysRevD.75.071103}{Phys. Rev. {\bf D75},
  071103 (2007)}, \href{http://arxiv.org/abs/hep-ex/0611026}{{\tt
  arXiv:hep-ex/0611026 [hep-ex]}}\relax
\mciteBstWouldAddEndPuncttrue
\mciteSetBstMidEndSepPunct{\mcitedefaultmidpunct}
{\mcitedefaultendpunct}{\mcitedefaultseppunct}\relax
\EndOfBibitem
\bibitem{Aubert:2008az}
{\babar} collaboration, B.~Aubert {\em et al.},
  \href{http://dx.doi.org/10.1103/PhysRevD.78.112002}{Phys. Rev. {\bf D78},
  112002 (2008)}, \href{http://arxiv.org/abs/0807.2014}{{\tt arXiv:0807.2014
  [hep-ex]}}\relax
\mciteBstWouldAddEndPuncttrue
\mciteSetBstMidEndSepPunct{\mcitedefaultmidpunct}
{\mcitedefaultendpunct}{\mcitedefaultseppunct}\relax
\EndOfBibitem
\bibitem{Guido:2018ywg}
{Belle} collaboration, E.~Guido {\em et al.},
  \href{http://dx.doi.org/10.1103/PhysRevLett.121.062001}{Phys. Rev. Lett. {\bf
  121},  062001 (2018)}, \href{http://arxiv.org/abs/1803.10303}{{\tt
  arXiv:1803.10303 [hep-ex]}}\relax
\mciteBstWouldAddEndPuncttrue
\mciteSetBstMidEndSepPunct{\mcitedefaultmidpunct}
{\mcitedefaultendpunct}{\mcitedefaultseppunct}\relax
\EndOfBibitem
\bibitem{Barish:1995cx}
{CLEO} collaboration, B.~Barish {\em et al.},
  \href{http://dx.doi.org/10.1103/PhysRevLett.76.1570}{Phys. Rev. Lett. {\bf
  76},  1570 (1996)}\relax
\mciteBstWouldAddEndPuncttrue
\mciteSetBstMidEndSepPunct{\mcitedefaultmidpunct}
{\mcitedefaultendpunct}{\mcitedefaultseppunct}\relax
\EndOfBibitem
\bibitem{Drutskoy:2010an}
{Belle} collaboration, A.~Drutskoy {\em et al.},
  \href{http://dx.doi.org/10.1103/PhysRevD.81.112003}{Phys. Rev. {\bf D81},
  112003 (2010)}, \href{http://arxiv.org/abs/1003.5885}{{\tt arXiv:1003.5885
  [hep-ex]}}\relax
\mciteBstWouldAddEndPuncttrue
\mciteSetBstMidEndSepPunct{\mcitedefaultmidpunct}
{\mcitedefaultendpunct}{\mcitedefaultseppunct}\relax
\EndOfBibitem
\bibitem{Esen:2012yz}
{Belle} collaboration, S.~Esen {\em et al.},
  \href{http://dx.doi.org/10.1103/PhysRevD.87.031101}{Phys. Rev. {\bf D87},
  031101 (2013)}, \href{http://arxiv.org/abs/1208.0323}{{\tt
  arXiv:1208.0323}}\relax
\mciteBstWouldAddEndPuncttrue
\mciteSetBstMidEndSepPunct{\mcitedefaultmidpunct}
{\mcitedefaultendpunct}{\mcitedefaultseppunct}\relax
\EndOfBibitem
\bibitem{Drutskoy:2006fg}
{Belle} collaboration, A.~Drutskoy {\em et al.},
  \href{http://dx.doi.org/10.1103/PhysRevLett.98.052001}{Phys. Rev. Lett. {\bf
  98},  052001 (2007)}, \href{http://arxiv.org/abs/hep-ex/0608015}{{\tt
  arXiv:hep-ex/0608015 [hep-ex]}}\relax
\mciteBstWouldAddEndPuncttrue
\mciteSetBstMidEndSepPunct{\mcitedefaultmidpunct}
{\mcitedefaultendpunct}{\mcitedefaultseppunct}\relax
\EndOfBibitem
\bibitem{Huang:2006em}
{CLEO} collaboration, G.~S. Huang {\em et al.},
  \href{http://dx.doi.org/10.1103/PhysRevD.75.012002}{Phys. Rev. {\bf D75},
  012002 (2007)}, \href{http://arxiv.org/abs/hep-ex/0610035}{{\tt
  arXiv:hep-ex/0610035 [hep-ex]}}\relax
\mciteBstWouldAddEndPuncttrue
\mciteSetBstMidEndSepPunct{\mcitedefaultmidpunct}
{\mcitedefaultendpunct}{\mcitedefaultseppunct}\relax
\EndOfBibitem
\bibitem{PDG_2018}
{Particle Data Group} collaboration, M.~Tanabashi {\em et al.},
  \href{http://dx.doi.org/10.1103/PhysRevD.98.030001}{Phys. Rev. {\bf D98},
  030001 (2018)}\relax
\mciteBstWouldAddEndPuncttrue
\mciteSetBstMidEndSepPunct{\mcitedefaultmidpunct}
{\mcitedefaultendpunct}{\mcitedefaultseppunct}\relax
\EndOfBibitem
\bibitem{Artuso:2005xw}
{CLEO} collaboration, M.~Artuso {\em et al.},
  \href{http://dx.doi.org/10.1103/PhysRevLett.95.261801}{Phys. Rev. Lett. {\bf
  95},  261801 (2005)}, \href{http://arxiv.org/abs/hep-ex/0508047}{{\tt
  arXiv:hep-ex/0508047}}\relax
\mciteBstWouldAddEndPuncttrue
\mciteSetBstMidEndSepPunct{\mcitedefaultmidpunct}
{\mcitedefaultendpunct}{\mcitedefaultseppunct}\relax
\EndOfBibitem
\bibitem{Abe:2007tk}
{Belle} collaboration, K.~F. Chen {\em et al.},
  \href{http://dx.doi.org/10.1103/PhysRevLett.100.112001}{Phys. Rev. Lett. {\bf
  100},  112001 (2008)}, \href{http://arxiv.org/abs/0710.2577}{{\tt
  arXiv:0710.2577 [hep-ex]}}\relax
\mciteBstWouldAddEndPuncttrue
\mciteSetBstMidEndSepPunct{\mcitedefaultmidpunct}
{\mcitedefaultendpunct}{\mcitedefaultseppunct}\relax
\EndOfBibitem
\bibitem{Garmash:2014dhx}
{Belle} collaboration, A.~Garmash {\em et al.},
  \href{http://dx.doi.org/10.1103/PhysRevD.91.072003}{Phys. Rev. {\bf D91},
  072003 (2015)}, \href{http://arxiv.org/abs/1403.0992}{{\tt arXiv:1403.0992
  [hep-ex]}}\relax
\mciteBstWouldAddEndPuncttrue
\mciteSetBstMidEndSepPunct{\mcitedefaultmidpunct}
{\mcitedefaultendpunct}{\mcitedefaultseppunct}\relax
\EndOfBibitem
\bibitem{Adachi:2011ji}
{Belle} collaboration, I.~Adachi {\em et al.},
  \href{http://dx.doi.org/10.1103/PhysRevLett.108.032001}{Phys. Rev. Lett. {\bf
  108},  032001 (2012)}, \href{http://arxiv.org/abs/1103.3419}{{\tt
  arXiv:1103.3419 [hep-ex]}}\relax
\mciteBstWouldAddEndPuncttrue
\mciteSetBstMidEndSepPunct{\mcitedefaultmidpunct}
{\mcitedefaultendpunct}{\mcitedefaultseppunct}\relax
\EndOfBibitem
\bibitem{Krokovny:2013mgx}
{Belle} collaboration, P.~Krokovny {\em et al.},
  \href{http://dx.doi.org/10.1103/PhysRevD.88.052016}{Phys. Rev. {\bf D88},
  052016 (2013)}, \href{http://arxiv.org/abs/1308.2646}{{\tt arXiv:1308.2646
  [hep-ex]}}\relax
\mciteBstWouldAddEndPuncttrue
\mciteSetBstMidEndSepPunct{\mcitedefaultmidpunct}
{\mcitedefaultendpunct}{\mcitedefaultseppunct}\relax
\EndOfBibitem
\bibitem{Tamponi:2018cuf}
{Belle} collaboration, U.~Tamponi {\em et al.},
  \href{http://dx.doi.org/10.1140/epjc/s10052-018-6086-4}{Eur. Phys. J. {\bf
  C78},  633 (2018)}, \href{http://arxiv.org/abs/1803.03225}{{\tt
  arXiv:1803.03225 [hep-ex]}}\relax
\mciteBstWouldAddEndPuncttrue
\mciteSetBstMidEndSepPunct{\mcitedefaultmidpunct}
{\mcitedefaultendpunct}{\mcitedefaultseppunct}\relax
\EndOfBibitem
\bibitem{thesis_Louvot}
 R.~Louvot, PhD thesis \#5213, EPFL, Lausanne, 2012,
  \url{{http://dx.doi.org/10.5075/epfl-thesis-5213}}\relax
\mciteBstWouldAddEndPuncttrue
\mciteSetBstMidEndSepPunct{\mcitedefaultmidpunct}
{\mcitedefaultendpunct}{\mcitedefaultseppunct}\relax
\EndOfBibitem
\bibitem{Lees:2011ji}
{\babar} collaboration, J.~P. Lees {\em et al.},
  \href{http://dx.doi.org/10.1103/PhysRevD.85.011101}{Phys. Rev. {\bf D85},
  011101 (2012)}, \href{http://arxiv.org/abs/1110.5600}{{\tt arXiv:1110.5600
  [hep-ex]}}\relax
\mciteBstWouldAddEndPuncttrue
\mciteSetBstMidEndSepPunct{\mcitedefaultmidpunct}
{\mcitedefaultendpunct}{\mcitedefaultseppunct}\relax
\EndOfBibitem
\bibitem{Li:2011pg}
{Belle} collaboration, J.~Li {\em et al.},
  \href{http://dx.doi.org/10.1103/PhysRevLett.106.121802}{Phys. Rev. Lett. {\bf
  106},  121802 (2011)}, \href{http://arxiv.org/abs/1102.2759}{{\tt
  arXiv:1102.2759 [hep-ex]}}\relax
\mciteBstWouldAddEndPuncttrue
\mciteSetBstMidEndSepPunct{\mcitedefaultmidpunct}
{\mcitedefaultendpunct}{\mcitedefaultseppunct}\relax
\EndOfBibitem
\bibitem{Louvot:2008sc}
{Belle} collaboration, R.~Louvot {\em et al.},
  \href{http://dx.doi.org/10.1103/PhysRevLett.102.021801}{Phys. Rev. Lett. {\bf
  102},  021801 (2009)}, \href{http://arxiv.org/abs/0809.2526}{{\tt
  arXiv:0809.2526 [hep-ex]}}\relax
\mciteBstWouldAddEndPuncttrue
\mciteSetBstMidEndSepPunct{\mcitedefaultmidpunct}
{\mcitedefaultendpunct}{\mcitedefaultseppunct}\relax
\EndOfBibitem
\bibitem{Abreu:1992rv}
{DELPHI} collaboration, P.~Abreu {\em et al.},
  \href{http://dx.doi.org/10.1016/0370-2693(92)91385-M}{Phys. Lett. {\bf B289},
   199 (1992)}\relax
\mciteBstWouldAddEndPuncttrue
\mciteSetBstMidEndSepPunct{\mcitedefaultmidpunct}
{\mcitedefaultendpunct}{\mcitedefaultseppunct}\relax
\EndOfBibitem
\bibitem{Acton:1992zq}
{OPAL} collaboration, P.~D. Acton {\em et al.},
  \href{http://dx.doi.org/10.1016/0370-2693(92)91578-W}{Phys. Lett. {\bf B295},
   357 (1992)}\relax
\mciteBstWouldAddEndPuncttrue
\mciteSetBstMidEndSepPunct{\mcitedefaultmidpunct}
{\mcitedefaultendpunct}{\mcitedefaultseppunct}\relax
\EndOfBibitem
\bibitem{Buskulic:1995bd}
{ALEPH} collaboration, D.~Buskulic {\em et al.},
  \href{http://dx.doi.org/10.1016/0370-2693(95)01173-N}{Phys. Lett. {\bf B361},
   221 (1995)}\relax
\mciteBstWouldAddEndPuncttrue
\mciteSetBstMidEndSepPunct{\mcitedefaultmidpunct}
{\mcitedefaultendpunct}{\mcitedefaultseppunct}\relax
\EndOfBibitem
\bibitem{Abreu:1995me}
{DELPHI} collaboration, P.~Abreu {\em et al.},
  \href{http://dx.doi.org/10.1007/BF01620713}{Z. Phys. {\bf C68},  375
  (1995)}\relax
\mciteBstWouldAddEndPuncttrue
\mciteSetBstMidEndSepPunct{\mcitedefaultmidpunct}
{\mcitedefaultendpunct}{\mcitedefaultseppunct}\relax
\EndOfBibitem
\bibitem{Barate:1997if}
{ALEPH} collaboration, R.~Barate {\em et al.},
  \href{http://dx.doi.org/10.1007/s100520050133}{Eur. Phys. J. {\bf C2},  197
  (1998)}\relax
\mciteBstWouldAddEndPuncttrue
\mciteSetBstMidEndSepPunct{\mcitedefaultmidpunct}
{\mcitedefaultendpunct}{\mcitedefaultseppunct}\relax
\EndOfBibitem
\bibitem{Buskulic:1996sm}
{ALEPH} collaboration, D.~Buskulic {\em et al.},
  \href{http://dx.doi.org/10.1016/0370-2693(96)00925-2}{Phys. Lett. {\bf B384},
   449 (1996)}\relax
\mciteBstWouldAddEndPuncttrue
\mciteSetBstMidEndSepPunct{\mcitedefaultmidpunct}
{\mcitedefaultendpunct}{\mcitedefaultseppunct}\relax
\EndOfBibitem
\bibitem{Abdallah:2005cw}
{DELPHI} collaboration, J.~Abdallah {\em et al.},
  \href{http://dx.doi.org/10.1140/epjc/s2005-02388-4}{Eur. Phys. J. {\bf C44},
  299 (2005)}, \href{http://arxiv.org/abs/hep-ex/0510023}{{\tt
  arXiv:hep-ex/0510023 [hep-ex]}}\relax
\mciteBstWouldAddEndPuncttrue
\mciteSetBstMidEndSepPunct{\mcitedefaultmidpunct}
{\mcitedefaultendpunct}{\mcitedefaultseppunct}\relax
\EndOfBibitem
\bibitem{Barate:1997ty}
{ALEPH} collaboration, R.~Barate {\em et al.},
  \href{http://dx.doi.org/10.1007/s100520050263}{Eur. Phys. J. {\bf C5},  205
  (1998)}\relax
\mciteBstWouldAddEndPuncttrue
\mciteSetBstMidEndSepPunct{\mcitedefaultmidpunct}
{\mcitedefaultendpunct}{\mcitedefaultseppunct}\relax
\EndOfBibitem
\bibitem{Abdallah:2003xp}
{DELPHI} collaboration, J.~Abdallah {\em et al.},
  \href{http://dx.doi.org/10.1016/j.physletb.2003.09.070}{Phys. Lett. {\bf
  B576},  29 (2003)}, \href{http://arxiv.org/abs/hep-ex/0311005}{{\tt
  arXiv:hep-ex/0311005 [hep-ex]}}\relax
\mciteBstWouldAddEndPuncttrue
\mciteSetBstMidEndSepPunct{\mcitedefaultmidpunct}
{\mcitedefaultendpunct}{\mcitedefaultseppunct}\relax
\EndOfBibitem
\bibitem{Affolder:1999iq}
{CDF} collaboration, T.~Affolder {\em et al.},
  \href{http://dx.doi.org/10.1103/PhysRevLett.84.1663}{Phys. Rev. Lett. {\bf
  84},  1663 (2000)}, \href{http://arxiv.org/abs/hep-ex/9909011}{{\tt
  arXiv:hep-ex/9909011 [hep-ex]}}\relax
\mciteBstWouldAddEndPuncttrue
\mciteSetBstMidEndSepPunct{\mcitedefaultmidpunct}
{\mcitedefaultendpunct}{\mcitedefaultseppunct}\relax
\EndOfBibitem
\bibitem{Aaltonen:2008zd}
{CDF} collaboration, T.~Aaltonen {\em et al.},
  \href{http://dx.doi.org/10.1103/PhysRevD.77.072003}{Phys. Rev. {\bf D77},
  072003 (2008)}, \href{http://arxiv.org/abs/0801.4375}{{\tt arXiv:0801.4375
  [hep-ex]}}\relax
\mciteBstWouldAddEndPuncttrue
\mciteSetBstMidEndSepPunct{\mcitedefaultmidpunct}
{\mcitedefaultendpunct}{\mcitedefaultseppunct}\relax
\EndOfBibitem
\bibitem{Aaltonen:2008eu}
{CDF} collaboration, T.~Aaltonen {\em et al.},
  \href{http://dx.doi.org/10.1103/PhysRevD.79.032001}{Phys. Rev. {\bf D79},
  032001 (2009)}, \href{http://arxiv.org/abs/0810.3213}{{\tt arXiv:0810.3213
  [hep-ex]}}\relax
\mciteBstWouldAddEndPuncttrue
\mciteSetBstMidEndSepPunct{\mcitedefaultmidpunct}
{\mcitedefaultendpunct}{\mcitedefaultseppunct}\relax
\EndOfBibitem
\bibitem{Abe:1999ta}
{CDF} collaboration, F.~Abe {\em et al.},
  \href{http://dx.doi.org/10.1103/PhysRevD.60.092005}{Phys. Rev. {\bf D60},
  092005 (1999)}\relax
\mciteBstWouldAddEndPuncttrue
\mciteSetBstMidEndSepPunct{\mcitedefaultmidpunct}
{\mcitedefaultendpunct}{\mcitedefaultseppunct}\relax
\EndOfBibitem
\bibitem{Abazov:2007am}
{\dzero} collaboration, V.~M. Abazov {\em et al.},
  \href{http://dx.doi.org/10.1103/PhysRevLett.99.052001}{Phys. Rev. Lett. {\bf
  99},  052001 (2007)}, \href{http://arxiv.org/abs/0706.1690}{{\tt
  arXiv:0706.1690 [hep-ex]}}\relax
\mciteBstWouldAddEndPuncttrue
\mciteSetBstMidEndSepPunct{\mcitedefaultmidpunct}
{\mcitedefaultendpunct}{\mcitedefaultseppunct}\relax
\EndOfBibitem
\bibitem{Abazov:2008qm}
{\dzero} collaboration, V.~M. Abazov {\em et al.},
  \href{http://dx.doi.org/10.1103/PhysRevLett.101.232002}{Phys. Rev. Lett. {\bf
  101},  232002 (2008)}, \href{http://arxiv.org/abs/0808.4142}{{\tt
  arXiv:0808.4142 [hep-ex]}}\relax
\mciteBstWouldAddEndPuncttrue
\mciteSetBstMidEndSepPunct{\mcitedefaultmidpunct}
{\mcitedefaultendpunct}{\mcitedefaultseppunct}\relax
\EndOfBibitem
\bibitem{Aaltonen:2009ny}
{CDF} collaboration, T.~Aaltonen {\em et al.},
  \href{http://dx.doi.org/10.1103/PhysRevD.80.072003}{Phys. Rev. {\bf D80},
  072003 (2009)}, \href{http://arxiv.org/abs/0905.3123}{{\tt arXiv:0905.3123
  [hep-ex]}}\relax
\mciteBstWouldAddEndPuncttrue
\mciteSetBstMidEndSepPunct{\mcitedefaultmidpunct}
{\mcitedefaultendpunct}{\mcitedefaultseppunct}\relax
\EndOfBibitem
\bibitem{Aaij:2011jp}
{LHCb} collaboration, R.~Aaij {\em et al.},
  \href{http://dx.doi.org/10.1103/PhysRevD.85.032008}{Phys. Rev. {\bf D85},
  032008 (2012)}, \href{http://arxiv.org/abs/1111.2357}{{\tt arXiv:1111.2357
  [hep-ex]}}, with numerical results and full covariance matrix available at
  \url{https://cdsweb.cern.ch/record/1390838}\relax
\mciteBstWouldAddEndPuncttrue
\mciteSetBstMidEndSepPunct{\mcitedefaultmidpunct}
{\mcitedefaultendpunct}{\mcitedefaultseppunct}\relax
\EndOfBibitem
\bibitem{Aaij:2013qqa}
{LHCb} collaboration, R.~Aaij {\em et al.},
  \href{http://dx.doi.org/10.1007/JHEP04(2013)001}{JHEP {\bf 04},  001 (2013)},
  \href{http://arxiv.org/abs/1301.5286}{{\tt arXiv:1301.5286 [hep-ex]}}, with
  numerical results and full covariance matrix available at
  \url{http://cdsweb.cern.ch/record/1507868}\relax
\mciteBstWouldAddEndPuncttrue
\mciteSetBstMidEndSepPunct{\mcitedefaultmidpunct}
{\mcitedefaultendpunct}{\mcitedefaultseppunct}\relax
\EndOfBibitem
\bibitem{Aaij:2014jyk}
{LHCb} collaboration, R.~Aaij {\em et al.},
  \href{http://dx.doi.org/10.1007/JHEP08(2014)143}{JHEP {\bf 08},  143 (2014)},
  \href{http://arxiv.org/abs/1405.6842}{{\tt arXiv:1405.6842 [hep-ex]}}\relax
\mciteBstWouldAddEndPuncttrue
\mciteSetBstMidEndSepPunct{\mcitedefaultmidpunct}
{\mcitedefaultendpunct}{\mcitedefaultseppunct}\relax
\EndOfBibitem
\bibitem{Aad:2015cda}
{ATLAS} collaboration, G.~Aad {\em et al.},
  \href{http://dx.doi.org/10.1103/PhysRevLett.115.262001}{Phys. Rev. Lett. {\bf
  115},  262001 (2015)}, \href{http://arxiv.org/abs/1507.08925}{{\tt
  arXiv:1507.08925 [hep-ex]}}\relax
\mciteBstWouldAddEndPuncttrue
\mciteSetBstMidEndSepPunct{\mcitedefaultmidpunct}
{\mcitedefaultendpunct}{\mcitedefaultseppunct}\relax
\EndOfBibitem
\bibitem{Liu:2013nea}
X.~Liu, W.~Wang, and Y.~Xie,
  \href{http://dx.doi.org/10.1103/PhysRevD.89.094010}{Phys. Rev. {\bf D89},
  094010 (2014)}, \href{http://arxiv.org/abs/1309.0313}{{\tt arXiv:1309.0313
  [hep-ph]}}\relax
\mciteBstWouldAddEndPuncttrue
\mciteSetBstMidEndSepPunct{\mcitedefaultmidpunct}
{\mcitedefaultendpunct}{\mcitedefaultseppunct}\relax
\EndOfBibitem
\bibitem{ALEPH:2005ab}
{ALEPH, CDF, DELPHI, L3, OPAL, and SLD collaborations, LEP electroweak working
  group, SLD electroweak and heavy flavour working groups}, S.~Schael {\em et
  al.}, \href{http://dx.doi.org/10.1016/j.physrep.2005.12.006}{Phys. Rept. {\bf
  427},  257 (2006)}, \href{http://arxiv.org/abs/hep-ex/0509008}{{\tt
  arXiv:hep-ex/0509008 [hep-ex]}}\relax
\mciteBstWouldAddEndPuncttrue
\mciteSetBstMidEndSepPunct{\mcitedefaultmidpunct}
{\mcitedefaultendpunct}{\mcitedefaultseppunct}\relax
\EndOfBibitem
\bibitem{Abazov:2006qw}
{\dzero} collaboration, V.~M. Abazov {\em et al.},
  \href{http://dx.doi.org/10.1103/PhysRevD.74.092001}{Phys. Rev. {\bf D74},
  092001 (2006)}, \href{http://arxiv.org/abs/hep-ex/0609014}{{\tt
  arXiv:hep-ex/0609014 [hep-ex]}}\relax
\mciteBstWouldAddEndPuncttrue
\mciteSetBstMidEndSepPunct{\mcitedefaultmidpunct}
{\mcitedefaultendpunct}{\mcitedefaultseppunct}\relax
\EndOfBibitem
\bibitem{Acosta:2003ie}
{CDF} collaboration, D.~Acosta {\em et al.},
  \href{http://dx.doi.org/10.1103/PhysRevD.69.012002}{Phys. Rev. {\bf D69},
  012002 (2004)}, \href{http://arxiv.org/abs/hep-ex/0309030}{{\tt
  arXiv:hep-ex/0309030 [hep-ex]}}\relax
\mciteBstWouldAddEndPuncttrue
\mciteSetBstMidEndSepPunct{\mcitedefaultmidpunct}
{\mcitedefaultendpunct}{\mcitedefaultseppunct}\relax
\EndOfBibitem
\bibitem{LHCb-CONF-2013-011}
{LHCb} collaboration, LHCb-CONF-2013-011, 2013,
  \url{{https://cdsweb.cern.ch/record/1559262}}\relax
\mciteBstWouldAddEndPuncttrue
\mciteSetBstMidEndSepPunct{\mcitedefaultmidpunct}
{\mcitedefaultendpunct}{\mcitedefaultseppunct}\relax
\EndOfBibitem
\bibitem{Shifman:1986mx}
M.~A. Shifman and M.~B. Voloshin, Sov. Phys. JETP {\bf 64},  698 (1986)\relax
\mciteBstWouldAddEndPuncttrue
\mciteSetBstMidEndSepPunct{\mcitedefaultmidpunct}
{\mcitedefaultendpunct}{\mcitedefaultseppunct}\relax
\EndOfBibitem
\bibitem{Chay:1990da}
J.~Chay, H.~Georgi, and B.~Grinstein,
  \href{http://dx.doi.org/10.1016/0370-2693(90)90916-T}{Phys. Lett. {\bf B247},
   399 (1990)}\relax
\mciteBstWouldAddEndPuncttrue
\mciteSetBstMidEndSepPunct{\mcitedefaultmidpunct}
{\mcitedefaultendpunct}{\mcitedefaultseppunct}\relax
\EndOfBibitem
\bibitem{Bigi:1992su}
I.~I. Bigi, N.~G. Uraltsev, and A.~I. Vainshtein,
  \href{http://dx.doi.org/10.1016/0370-2693(92)90908-M}{Phys. Lett. {\bf B293},
   430 (1992)}, \href{http://arxiv.org/abs/hep-ph/9207214}{{\tt
  arXiv:hep-ph/9207214 [hep-ph]}}, Erratum ibid.\
  \href{http://dx.doi.org/10.1016/0370-2693(92)91287-J}{{\bf B297}, 477}
  (1992)\relax
\mciteBstWouldAddEndPuncttrue
\mciteSetBstMidEndSepPunct{\mcitedefaultmidpunct}
{\mcitedefaultendpunct}{\mcitedefaultseppunct}\relax
\EndOfBibitem
\bibitem{Wilson:1969zs}
K.~G. Wilson, \href{http://dx.doi.org/10.1103/PhysRev.179.1499}{Phys. Rev. {\bf
  179},  1499 (1969)}\relax
\mciteBstWouldAddEndPuncttrue
\mciteSetBstMidEndSepPunct{\mcitedefaultmidpunct}
{\mcitedefaultendpunct}{\mcitedefaultseppunct}\relax
\EndOfBibitem
\bibitem{Shifman:2000jv}
M.~A. Shifman, \href{http://arxiv.org/abs/hep-ph/0009131}{{\tt
  arXiv:hep-ph/0009131 [hep-ph]}} (2000)\relax
\mciteBstWouldAddEndPuncttrue
\mciteSetBstMidEndSepPunct{\mcitedefaultmidpunct}
{\mcitedefaultendpunct}{\mcitedefaultseppunct}\relax
\EndOfBibitem
\bibitem{Bigi:2001ys}
I.~I.~Y. Bigi and N.~Uraltsev,
  \href{http://dx.doi.org/10.1142/S0217751X01005535}{Int. J. Mod. Phys. {\bf
  A16},  5201 (2001)}, \href{http://arxiv.org/abs/hep-ph/0106346}{{\tt
  arXiv:hep-ph/0106346 [hep-ph]}}\relax
\mciteBstWouldAddEndPuncttrue
\mciteSetBstMidEndSepPunct{\mcitedefaultmidpunct}
{\mcitedefaultendpunct}{\mcitedefaultseppunct}\relax
\EndOfBibitem
\bibitem{Jubb:2016mvq}
T.~Jubb, M.~Kirk, A.~Lenz, and G.~Tetlalmatzi-Xolocotzi,
  \href{http://dx.doi.org/10.1016/j.nuclphysb.2016.12.020}{Nucl. Phys. {\bf
  B915},  431 (2017)}, \href{http://arxiv.org/abs/1603.07770}{{\tt
  arXiv:1603.07770 [hep-ph]}}\relax
\mciteBstWouldAddEndPuncttrue
\mciteSetBstMidEndSepPunct{\mcitedefaultmidpunct}
{\mcitedefaultendpunct}{\mcitedefaultseppunct}\relax
\EndOfBibitem
\bibitem{Voloshin:1999pz}
M.~B. Voloshin, \href{http://dx.doi.org/10.1016/S0370-1573(99)00054-X}{Phys.
  Rept. {\bf 320},  275 (1999)},
  \href{http://arxiv.org/abs/hep-ph/9901445}{{\tt arXiv:hep-ph/9901445
  [hep-ph]}}\relax
\mciteBstWouldAddEndPuncttrue
\mciteSetBstMidEndSepPunct{\mcitedefaultmidpunct}
{\mcitedefaultendpunct}{\mcitedefaultseppunct}\relax
\EndOfBibitem
\bibitem{Guberina:1999bw}
B.~Guberina, B.~Melic, and H.~Stefancic,
  \href{http://dx.doi.org/10.1016/S0370-2693(99)01198-3}{Phys. Lett. {\bf
  B469},  253 (1999)}, \href{http://arxiv.org/abs/hep-ph/9907468}{{\tt
  arXiv:hep-ph/9907468 [hep-ph]}}\relax
\mciteBstWouldAddEndPuncttrue
\mciteSetBstMidEndSepPunct{\mcitedefaultmidpunct}
{\mcitedefaultendpunct}{\mcitedefaultseppunct}\relax
\EndOfBibitem
\bibitem{Neubert:1996we}
M.~Neubert and C.~T. Sachrajda,
  \href{http://dx.doi.org/10.1016/S0550-3213(96)00559-7}{Nucl. Phys. {\bf
  B483},  339 (1997)}, \href{http://arxiv.org/abs/hep-ph/9603202}{{\tt
  arXiv:hep-ph/9603202 [hep-ph]}}\relax
\mciteBstWouldAddEndPuncttrue
\mciteSetBstMidEndSepPunct{\mcitedefaultmidpunct}
{\mcitedefaultendpunct}{\mcitedefaultseppunct}\relax
\EndOfBibitem
\bibitem{Bigi:1997fj}
I.~I.~Y. Bigi, M.~A. Shifman, and N.~Uraltsev,
  \href{http://dx.doi.org/10.1146/annurev.nucl.47.1.591}{Ann. Rev. Nucl. Part.
  Sci. {\bf 47},  591 (1997)}, \href{http://arxiv.org/abs/hep-ph/9703290}{{\tt
  arXiv:hep-ph/9703290 [hep-ph]}}\relax
\mciteBstWouldAddEndPuncttrue
\mciteSetBstMidEndSepPunct{\mcitedefaultmidpunct}
{\mcitedefaultendpunct}{\mcitedefaultseppunct}\relax
\EndOfBibitem
\bibitem{Beneke:1996gn}
M.~Beneke, G.~Buchalla, and I.~Dunietz,
  \href{http://dx.doi.org/10.1103/PhysRevD.54.4419}{Phys. Rev. {\bf D54},  4419
  (1996)}, \href{http://arxiv.org/abs/hep-ph/9605259}{{\tt arXiv:hep-ph/9605259
  [hep-ph]}}, Erratum ibid.\
  \href{http://dx.doi.org/10.1103/PhysRevD.83.119902}{{\bf D83}, 119902},
  (2011)\relax
\mciteBstWouldAddEndPuncttrue
\mciteSetBstMidEndSepPunct{\mcitedefaultmidpunct}
{\mcitedefaultendpunct}{\mcitedefaultseppunct}\relax
\EndOfBibitem
\bibitem{Lenz:2011ti}
A.~Lenz and U.~Nierste, \href{http://arxiv.org/abs/1102.4274}{{\tt
  arXiv:1102.4274 [hep-ph]}} (2011)\relax
\mciteBstWouldAddEndPuncttrue
\mciteSetBstMidEndSepPunct{\mcitedefaultmidpunct}
{\mcitedefaultendpunct}{\mcitedefaultseppunct}\relax
\EndOfBibitem
\bibitem{Lenz:2006hd}
A.~Lenz and U.~Nierste,
  \href{http://dx.doi.org/10.1088/1126-6708/2007/06/072}{JHEP {\bf 06},  072
  (2007)}, \href{http://arxiv.org/abs/hep-ph/0612167}{{\tt arXiv:hep-ph/0612167
  [hep-ph]}}\relax
\mciteBstWouldAddEndPuncttrue
\mciteSetBstMidEndSepPunct{\mcitedefaultmidpunct}
{\mcitedefaultendpunct}{\mcitedefaultseppunct}\relax
\EndOfBibitem
\bibitem{Beneke:1998sy}
M.~Beneke, G.~Buchalla, C.~Greub, A.~Lenz, and U.~Nierste,
  \href{http://dx.doi.org/10.1016/S0370-2693(99)00684-X}{Phys. Lett. {\bf
  B459},  631 (1999)}, \href{http://arxiv.org/abs/hep-ph/9808385}{{\tt
  arXiv:hep-ph/9808385 [hep-ph]}}\relax
\mciteBstWouldAddEndPuncttrue
\mciteSetBstMidEndSepPunct{\mcitedefaultmidpunct}
{\mcitedefaultendpunct}{\mcitedefaultseppunct}\relax
\EndOfBibitem
\bibitem{Beneke:2002rj}
M.~Beneke, G.~Buchalla, C.~Greub, A.~Lenz, and U.~Nierste,
  \href{http://dx.doi.org/10.1016/S0550-3213(02)00561-8}{Nucl. Phys. {\bf
  B639},  389 (2002)}, \href{http://arxiv.org/abs/hep-ph/0202106}{{\tt
  arXiv:hep-ph/0202106 [hep-ph]}}\relax
\mciteBstWouldAddEndPuncttrue
\mciteSetBstMidEndSepPunct{\mcitedefaultmidpunct}
{\mcitedefaultendpunct}{\mcitedefaultseppunct}\relax
\EndOfBibitem
\bibitem{Franco:2002fc}
E.~Franco, V.~Lubicz, F.~Mescia, and C.~Tarantino,
  \href{http://dx.doi.org/10.1016/S0550-3213(02)00262-6}{Nucl. Phys. {\bf
  B633},  212 (2002)}, \href{http://arxiv.org/abs/hep-ph/0203089}{{\tt
  arXiv:hep-ph/0203089 [hep-ph]}}\relax
\mciteBstWouldAddEndPuncttrue
\mciteSetBstMidEndSepPunct{\mcitedefaultmidpunct}
{\mcitedefaultendpunct}{\mcitedefaultseppunct}\relax
\EndOfBibitem
\bibitem{Ciuchini:2003ww}
M.~Ciuchini, E.~Franco, V.~Lubicz, F.~Mescia, and C.~Tarantino,
  \href{http://dx.doi.org/10.1088/1126-6708/2003/08/031}{JHEP {\bf 08},  031
  (2003)}, \href{http://arxiv.org/abs/hep-ph/0308029}{{\tt arXiv:hep-ph/0308029
  [hep-ph]}}\relax
\mciteBstWouldAddEndPuncttrue
\mciteSetBstMidEndSepPunct{\mcitedefaultmidpunct}
{\mcitedefaultendpunct}{\mcitedefaultseppunct}\relax
\EndOfBibitem
\bibitem{Beneke:2003az}
M.~Beneke, G.~Buchalla, A.~Lenz, and U.~Nierste,
  \href{http://dx.doi.org/10.1016/j.physletb.2003.09.089}{Phys. Lett. {\bf
  B576},  173 (2003)}, \href{http://arxiv.org/abs/hep-ph/0307344}{{\tt
  arXiv:hep-ph/0307344 [hep-ph]}}\relax
\mciteBstWouldAddEndPuncttrue
\mciteSetBstMidEndSepPunct{\mcitedefaultmidpunct}
{\mcitedefaultendpunct}{\mcitedefaultseppunct}\relax
\EndOfBibitem
\bibitem{Ciuchini:2001vx}
M.~Ciuchini, E.~Franco, V.~Lubicz, and F.~Mescia,
  \href{http://dx.doi.org/10.1016/S0550-3213(02)00006-8}{Nucl. Phys. {\bf
  B625},  211 (2002)}, \href{http://arxiv.org/abs/hep-ph/0110375}{{\tt
  arXiv:hep-ph/0110375 [hep-ph]}}\relax
\mciteBstWouldAddEndPuncttrue
\mciteSetBstMidEndSepPunct{\mcitedefaultmidpunct}
{\mcitedefaultendpunct}{\mcitedefaultseppunct}\relax
\EndOfBibitem
\bibitem{Tarantino:2003qw}
C.~Tarantino, \href{http://dx.doi.org/10.1140/epjcd/s2003-03-1006-y}{Eur. Phys.
  J. {\bf C33},  S895 (2004)}, \href{http://arxiv.org/abs/hep-ph/0310241}{{\tt
  arXiv:hep-ph/0310241 [hep-ph]}}\relax
\mciteBstWouldAddEndPuncttrue
\mciteSetBstMidEndSepPunct{\mcitedefaultmidpunct}
{\mcitedefaultendpunct}{\mcitedefaultseppunct}\relax
\EndOfBibitem
\bibitem{Gabbiani:2003pq}
F.~Gabbiani, A.~I. Onishchenko, and A.~A. Petrov,
  \href{http://dx.doi.org/10.1103/PhysRevD.68.114006}{Phys. Rev. {\bf D68},
  114006 (2003)}, \href{http://arxiv.org/abs/hep-ph/0303235}{{\tt
  arXiv:hep-ph/0303235 [hep-ph]}}\relax
\mciteBstWouldAddEndPuncttrue
\mciteSetBstMidEndSepPunct{\mcitedefaultmidpunct}
{\mcitedefaultendpunct}{\mcitedefaultseppunct}\relax
\EndOfBibitem
\bibitem{Gabbiani:2004tp}
F.~Gabbiani, A.~I. Onishchenko, and A.~A. Petrov,
  \href{http://dx.doi.org/10.1103/PhysRevD.70.094031}{Phys. Rev. {\bf D70},
  094031 (2004)}, \href{http://arxiv.org/abs/hep-ph/0407004}{{\tt
  arXiv:hep-ph/0407004 [hep-ph]}}\relax
\mciteBstWouldAddEndPuncttrue
\mciteSetBstMidEndSepPunct{\mcitedefaultmidpunct}
{\mcitedefaultendpunct}{\mcitedefaultseppunct}\relax
\EndOfBibitem
\bibitem{Lenz:2015dra}
A.~Lenz, \href{http://dx.doi.org/10.1142/S0217751X15430058}{Int. J. Mod. Phys.
  {\bf A30},  1543005 (2015)}, \href{http://arxiv.org/abs/1405.3601}{{\tt
  arXiv:1405.3601 [hep-ph]}}\relax
\mciteBstWouldAddEndPuncttrue
\mciteSetBstMidEndSepPunct{\mcitedefaultmidpunct}
{\mcitedefaultendpunct}{\mcitedefaultseppunct}\relax
\EndOfBibitem
\bibitem{Kirk:2017juj}
M.~Kirk, A.~Lenz, and T.~Rauh,
  \href{http://dx.doi.org/10.1007/JHEP12(2017)068}{JHEP {\bf 12},  068 (2017)},
  \href{http://arxiv.org/abs/1711.02100}{{\tt arXiv:1711.02100 [hep-ph]}}\relax
\mciteBstWouldAddEndPuncttrue
\mciteSetBstMidEndSepPunct{\mcitedefaultmidpunct}
{\mcitedefaultendpunct}{\mcitedefaultseppunct}\relax
\EndOfBibitem
\bibitem{Buskulic:1993gj}
{ALEPH} collaboration, D.~Buskulic {\em et al.},
  \href{http://dx.doi.org/10.1016/0370-2693(93)91265-O}{Phys. Lett. {\bf B314},
   459 (1993)}\relax
\mciteBstWouldAddEndPuncttrue
\mciteSetBstMidEndSepPunct{\mcitedefaultmidpunct}
{\mcitedefaultendpunct}{\mcitedefaultseppunct}\relax
\EndOfBibitem
\bibitem{Abreu:1994dr}
{DELPHI} collaboration, P.~Abreu {\em et al.},
  \href{http://dx.doi.org/10.1007/BF01577539}{Z. Phys. {\bf C63},  3
  (1994)}\relax
\mciteBstWouldAddEndPuncttrue
\mciteSetBstMidEndSepPunct{\mcitedefaultmidpunct}
{\mcitedefaultendpunct}{\mcitedefaultseppunct}\relax
\EndOfBibitem
\bibitem{Abreu:1996hv}
{DELPHI} collaboration, P.~Abreu {\em et al.},
  \href{http://dx.doi.org/10.1016/0370-2693(96)00452-2}{Phys. Lett. {\bf B377},
   195 (1996)}\relax
\mciteBstWouldAddEndPuncttrue
\mciteSetBstMidEndSepPunct{\mcitedefaultmidpunct}
{\mcitedefaultendpunct}{\mcitedefaultseppunct}\relax
\EndOfBibitem
\bibitem{Abdallah:2003sb}
{DELPHI} collaboration, J.~Abdallah {\em et al.},
  \href{http://dx.doi.org/10.1140/epjc/s2004-01599-5}{Eur. Phys. J. {\bf C33},
  307 (2004)}, \href{http://arxiv.org/abs/hep-ex/0401025}{{\tt
  arXiv:hep-ex/0401025 [hep-ex]}}\relax
\mciteBstWouldAddEndPuncttrue
\mciteSetBstMidEndSepPunct{\mcitedefaultmidpunct}
{\mcitedefaultendpunct}{\mcitedefaultseppunct}\relax
\EndOfBibitem
\bibitem{Acciarri:1997tt}
{L3} collaboration, M.~Acciarri {\em et al.},
  \href{http://dx.doi.org/10.1016/S0370-2693(97)01379-8}{Phys. Lett. {\bf
  B416},  220 (1998)}\relax
\mciteBstWouldAddEndPuncttrue
\mciteSetBstMidEndSepPunct{\mcitedefaultmidpunct}
{\mcitedefaultendpunct}{\mcitedefaultseppunct}\relax
\EndOfBibitem
\bibitem{Ackerstaff:1996as}
{OPAL} collaboration, K.~Ackerstaff {\em et al.},
  \href{http://dx.doi.org/10.1007/s002880050329}{Z. Phys. {\bf C73},  397
  (1997)}\relax
\mciteBstWouldAddEndPuncttrue
\mciteSetBstMidEndSepPunct{\mcitedefaultmidpunct}
{\mcitedefaultendpunct}{\mcitedefaultseppunct}\relax
\EndOfBibitem
\bibitem{Abe:1995rm}
{SLD} collaboration, K.~Abe {\em et al.},
  \href{http://dx.doi.org/10.1103/PhysRevLett.75.3624}{Phys. Rev. Lett. {\bf
  75},  3624 (1995)}, \href{http://arxiv.org/abs/hep-ex/9511005}{{\tt
  arXiv:hep-ex/9511005 [hep-ex]}}\relax
\mciteBstWouldAddEndPuncttrue
\mciteSetBstMidEndSepPunct{\mcitedefaultmidpunct}
{\mcitedefaultendpunct}{\mcitedefaultseppunct}\relax
\EndOfBibitem
\bibitem{Buskulic:1995rw}
{ALEPH} collaboration, D.~Buskulic {\em et al.},
  \href{http://dx.doi.org/10.1016/0370-2693(95)01586-8}{Phys. Lett. {\bf B369},
   151 (1996)}\relax
\mciteBstWouldAddEndPuncttrue
\mciteSetBstMidEndSepPunct{\mcitedefaultmidpunct}
{\mcitedefaultendpunct}{\mcitedefaultseppunct}\relax
\EndOfBibitem
\bibitem{Acton:1993xk}
{OPAL} collaboration, P.~D. Acton {\em et al.},
  \href{http://dx.doi.org/10.1007/BF01474617}{Z. Phys. {\bf C60},  217
  (1993)}\relax
\mciteBstWouldAddEndPuncttrue
\mciteSetBstMidEndSepPunct{\mcitedefaultmidpunct}
{\mcitedefaultendpunct}{\mcitedefaultseppunct}\relax
\EndOfBibitem
\bibitem{Abe:1997bd}
{CDF} collaboration, F.~Abe {\em et al.},
  \href{http://dx.doi.org/10.1103/PhysRevD.57.5382}{Phys. Rev. {\bf D57},  5382
  (1998)}\relax
\mciteBstWouldAddEndPuncttrue
\mciteSetBstMidEndSepPunct{\mcitedefaultmidpunct}
{\mcitedefaultendpunct}{\mcitedefaultseppunct}\relax
\EndOfBibitem
\bibitem{Barate:2000bs}
{ALEPH} collaboration, R.~Barate {\em et al.},
  \href{http://dx.doi.org/10.1016/S0370-2693(00)01093-5}{Phys. Lett. {\bf
  B492},  275 (2000)}, \href{http://arxiv.org/abs/hep-ex/0008016}{{\tt
  arXiv:hep-ex/0008016 [hep-ex]}}\relax
\mciteBstWouldAddEndPuncttrue
\mciteSetBstMidEndSepPunct{\mcitedefaultmidpunct}
{\mcitedefaultendpunct}{\mcitedefaultseppunct}\relax
\EndOfBibitem
\bibitem{Buskulic:1996hy}
{ALEPH} collaboration, D.~Buskulic {\em et al.},
  \href{http://dx.doi.org/10.1007/s002880050145}{Z. Phys. {\bf C71},  31
  (1996)}\relax
\mciteBstWouldAddEndPuncttrue
\mciteSetBstMidEndSepPunct{\mcitedefaultmidpunct}
{\mcitedefaultendpunct}{\mcitedefaultseppunct}\relax
\EndOfBibitem
\bibitem{Abreu:1995mc}
{DELPHI} collaboration, P.~Abreu {\em et al.},
  \href{http://dx.doi.org/10.1007/BF01579800}{Z. Phys. {\bf C68},  13
  (1995)}\relax
\mciteBstWouldAddEndPuncttrue
\mciteSetBstMidEndSepPunct{\mcitedefaultmidpunct}
{\mcitedefaultendpunct}{\mcitedefaultseppunct}\relax
\EndOfBibitem
\bibitem{Adam:1995mb}
{DELPHI} collaboration, W.~Adam {\em et al.},
  \href{http://dx.doi.org/10.1007/BF01620712}{Z. Phys. {\bf C68},  363
  (1995)}\relax
\mciteBstWouldAddEndPuncttrue
\mciteSetBstMidEndSepPunct{\mcitedefaultmidpunct}
{\mcitedefaultendpunct}{\mcitedefaultseppunct}\relax
\EndOfBibitem
\bibitem{Abreu:1996gb}
{DELPHI} collaboration, P.~Abreu {\em et al.},
  \href{http://dx.doi.org/10.1007/s002880050367}{Z. Phys. {\bf C74},  19
  (1997)}\relax
\mciteBstWouldAddEndPuncttrue
\mciteSetBstMidEndSepPunct{\mcitedefaultmidpunct}
{\mcitedefaultendpunct}{\mcitedefaultseppunct}\relax
\EndOfBibitem
\bibitem{Acciarri:1998uv}
{L3} collaboration, M.~Acciarri {\em et al.},
  \href{http://dx.doi.org/10.1016/S0370-2693(98)01114-9}{Phys. Lett. {\bf
  B438},  417 (1998)}\relax
\mciteBstWouldAddEndPuncttrue
\mciteSetBstMidEndSepPunct{\mcitedefaultmidpunct}
{\mcitedefaultendpunct}{\mcitedefaultseppunct}\relax
\EndOfBibitem
\bibitem{Akers:1995pa}
{OPAL} collaboration, R.~Akers {\em et al.},
  \href{http://dx.doi.org/10.1007/BF01624581}{Z. Phys. {\bf C67},  379
  (1995)}\relax
\mciteBstWouldAddEndPuncttrue
\mciteSetBstMidEndSepPunct{\mcitedefaultmidpunct}
{\mcitedefaultendpunct}{\mcitedefaultseppunct}\relax
\EndOfBibitem
\bibitem{Abbiendi:1998av}
{OPAL} collaboration, G.~Abbiendi {\em et al.},
  \href{http://dx.doi.org/10.1007/s100520000322}{Eur. Phys. J. {\bf C12},  609
  (2000)}, \href{http://arxiv.org/abs/hep-ex/9901017}{{\tt arXiv:hep-ex/9901017
  [hep-ex]}}\relax
\mciteBstWouldAddEndPuncttrue
\mciteSetBstMidEndSepPunct{\mcitedefaultmidpunct}
{\mcitedefaultendpunct}{\mcitedefaultseppunct}\relax
\EndOfBibitem
\bibitem{Abbiendi:2000ec}
{OPAL} collaboration, G.~Abbiendi {\em et al.},
  \href{http://dx.doi.org/10.1016/S0370-2693(00)01145-X}{Phys. Lett. {\bf
  B493},  266 (2000)}, \href{http://arxiv.org/abs/hep-ex/0010013}{{\tt
  arXiv:hep-ex/0010013 [hep-ex]}}\relax
\mciteBstWouldAddEndPuncttrue
\mciteSetBstMidEndSepPunct{\mcitedefaultmidpunct}
{\mcitedefaultendpunct}{\mcitedefaultseppunct}\relax
\EndOfBibitem
\bibitem{Abe:1997ys}
{SLD} collaboration, K.~Abe {\em et al.},
  \href{http://dx.doi.org/10.1103/PhysRevLett.79.590}{Phys. Rev. Lett. {\bf
  79},  590 (1997)}\relax
\mciteBstWouldAddEndPuncttrue
\mciteSetBstMidEndSepPunct{\mcitedefaultmidpunct}
{\mcitedefaultendpunct}{\mcitedefaultseppunct}\relax
\EndOfBibitem
\bibitem{Abe:1998wt}
{CDF} collaboration, F.~Abe {\em et al.},
  \href{http://dx.doi.org/10.1103/PhysRevD.58.092002}{Phys. Rev. {\bf D58},
  092002 (1998)}, \href{http://arxiv.org/abs/hep-ex/9806018}{{\tt
  arXiv:hep-ex/9806018 [hep-ex]}}\relax
\mciteBstWouldAddEndPuncttrue
\mciteSetBstMidEndSepPunct{\mcitedefaultmidpunct}
{\mcitedefaultendpunct}{\mcitedefaultseppunct}\relax
\EndOfBibitem
\bibitem{Acosta:2002nd}
{CDF} collaboration, D.~Acosta {\em et al.},
  \href{http://dx.doi.org/10.1103/PhysRevD.65.092009}{Phys. Rev. {\bf D65},
  092009 (2002)}\relax
\mciteBstWouldAddEndPuncttrue
\mciteSetBstMidEndSepPunct{\mcitedefaultmidpunct}
{\mcitedefaultendpunct}{\mcitedefaultseppunct}\relax
\EndOfBibitem
\bibitem{Aaltonen:2010pj}
{CDF} collaboration, T.~Aaltonen {\em et al.},
  \href{http://dx.doi.org/10.1103/PhysRevLett.106.121804}{Phys. Rev. Lett. {\bf
  106},  121804 (2011)}, \href{http://arxiv.org/abs/1012.3138}{{\tt
  arXiv:1012.3138 [hep-ex]}}\relax
\mciteBstWouldAddEndPuncttrue
\mciteSetBstMidEndSepPunct{\mcitedefaultmidpunct}
{\mcitedefaultendpunct}{\mcitedefaultseppunct}\relax
\EndOfBibitem
\bibitem{Abazov:2008jz}
{\dzero} collaboration, V.~M. Abazov {\em et al.},
  \href{http://dx.doi.org/10.1103/PhysRevLett.102.032001}{Phys. Rev. Lett. {\bf
  102},  032001 (2009)}, \href{http://arxiv.org/abs/0810.0037}{{\tt
  arXiv:0810.0037 [hep-ex]}}\relax
\mciteBstWouldAddEndPuncttrue
\mciteSetBstMidEndSepPunct{\mcitedefaultmidpunct}
{\mcitedefaultendpunct}{\mcitedefaultseppunct}\relax
\EndOfBibitem
\bibitem{Abazov:2012iy}
{\dzero} collaboration, V.~M. Abazov {\em et al.},
  \href{http://dx.doi.org/10.1103/PhysRevD.85.112003}{Phys. Rev. {\bf D85},
  112003 (2012)}, \href{http://arxiv.org/abs/1204.2340}{{\tt arXiv:1204.2340
  [hep-ex]}}\relax
\mciteBstWouldAddEndPuncttrue
\mciteSetBstMidEndSepPunct{\mcitedefaultmidpunct}
{\mcitedefaultendpunct}{\mcitedefaultseppunct}\relax
\EndOfBibitem
\bibitem{Abazov:2014rua}
{\dzero} collaboration, V.~M. Abazov {\em et al.},
  \href{http://dx.doi.org/10.1103/PhysRevLett.114.062001}{Phys. Rev. Lett. {\bf
  114},  062001 (2015)}, \href{http://arxiv.org/abs/1410.1568}{{\tt
  arXiv:1410.1568 [hep-ex]}}\relax
\mciteBstWouldAddEndPuncttrue
\mciteSetBstMidEndSepPunct{\mcitedefaultmidpunct}
{\mcitedefaultendpunct}{\mcitedefaultseppunct}\relax
\EndOfBibitem
\bibitem{Aubert:2001uw}
{\babar} collaboration, B.~Aubert {\em et al.},
  \href{http://dx.doi.org/10.1103/PhysRevLett.87.201803}{Phys. Rev. Lett. {\bf
  87},  201803 (2001)}, \href{http://arxiv.org/abs/hep-ex/0107019}{{\tt
  arXiv:hep-ex/0107019 [hep-ex]}}\relax
\mciteBstWouldAddEndPuncttrue
\mciteSetBstMidEndSepPunct{\mcitedefaultmidpunct}
{\mcitedefaultendpunct}{\mcitedefaultseppunct}\relax
\EndOfBibitem
\bibitem{Aubert:2002gi}
{\babar} collaboration, B.~Aubert {\em et al.},
  \href{http://dx.doi.org/10.1103/PhysRevLett.89.011802}{Phys. Rev. Lett. {\bf
  89},  011802 (2002)}, \href{http://arxiv.org/abs/hep-ex/0202005}{{\tt
  arXiv:hep-ex/0202005 [hep-ex]}}, Erratum ibid.\
  \href{http://dx.doi.org/10.1103/PhysRevLett.89.169903}{{\bf 89}, 169903},
  (2002)\relax
\mciteBstWouldAddEndPuncttrue
\mciteSetBstMidEndSepPunct{\mcitedefaultmidpunct}
{\mcitedefaultendpunct}{\mcitedefaultseppunct}\relax
\EndOfBibitem
\bibitem{Aubert:2002sh}
{\babar} collaboration, B.~Aubert {\em et al.},
  \href{http://dx.doi.org/10.1103/PhysRevD.67.072002}{Phys. Rev. {\bf D67},
  072002 (2003)}, \href{http://arxiv.org/abs/hep-ex/0212017}{{\tt
  arXiv:hep-ex/0212017 [hep-ex]}}\relax
\mciteBstWouldAddEndPuncttrue
\mciteSetBstMidEndSepPunct{\mcitedefaultmidpunct}
{\mcitedefaultendpunct}{\mcitedefaultseppunct}\relax
\EndOfBibitem
\bibitem{Aubert:2002ms}
{\babar} collaboration, B.~Aubert {\em et al.},
  \href{http://dx.doi.org/10.1103/PhysRevD.67.091101}{Phys. Rev. {\bf D67},
  091101 (2003)}, \href{http://arxiv.org/abs/hep-ex/0212012}{{\tt
  arXiv:hep-ex/0212012 [hep-ex]}}\relax
\mciteBstWouldAddEndPuncttrue
\mciteSetBstMidEndSepPunct{\mcitedefaultmidpunct}
{\mcitedefaultendpunct}{\mcitedefaultseppunct}\relax
\EndOfBibitem
\bibitem{Aubert:2005kf}
{\babar} collaboration, B.~Aubert {\em et al.},
  \href{http://dx.doi.org/10.1103/PhysRevD.73.012004}{Phys. Rev. {\bf D73},
  012004 (2006)}, \href{http://arxiv.org/abs/hep-ex/0507054}{{\tt
  arXiv:hep-ex/0507054 [hep-ex]}}\relax
\mciteBstWouldAddEndPuncttrue
\mciteSetBstMidEndSepPunct{\mcitedefaultmidpunct}
{\mcitedefaultendpunct}{\mcitedefaultseppunct}\relax
\EndOfBibitem
\bibitem{Abe:2004mz}
{Belle} collaboration, K.~Abe {\em et al.},
  \href{http://dx.doi.org/10.1103/PhysRevD.71.072003}{Phys. Rev. {\bf D71},
  072003 (2005)}, \href{http://arxiv.org/abs/hep-ex/0408111}{{\tt
  arXiv:hep-ex/0408111 [hep-ex]}}, Erratum ibid.\
  \href{http://dx.doi.org/10.1103/PhysRevD.71.079903}{{\bf D71}, 079903},
  (2005)\relax
\mciteBstWouldAddEndPuncttrue
\mciteSetBstMidEndSepPunct{\mcitedefaultmidpunct}
{\mcitedefaultendpunct}{\mcitedefaultseppunct}\relax
\EndOfBibitem
\bibitem{Aad:2012bpa}
{ATLAS} collaboration, G.~Aad {\em et al.},
  \href{http://dx.doi.org/10.1103/PhysRevD.87.032002}{Phys. Rev. {\bf D87},
  032002 (2013)}, \href{http://arxiv.org/abs/1207.2284}{{\tt arXiv:1207.2284
  [hep-ex]}}\relax
\mciteBstWouldAddEndPuncttrue
\mciteSetBstMidEndSepPunct{\mcitedefaultmidpunct}
{\mcitedefaultendpunct}{\mcitedefaultseppunct}\relax
\EndOfBibitem
\bibitem{Sirunyan:2017nbv}
{CMS} collaboration, A.~M. Sirunyan {\em et al.},
  \href{http://dx.doi.org/10.1140/epjc/s10052-018-5929-3,
  10.1140/epjc/s10052-018-6014-7}{Eur. Phys. J. {\bf C78},  457 (2018)},
  \href{http://arxiv.org/abs/1710.08949}{{\tt arXiv:1710.08949 [hep-ex]}},
  Erratum ibid.\ \href{http://dx.doi.org/10.1140/epjc/s10052-018-6014-7}{{\bf
  C78}, 561} (2018)\relax
\mciteBstWouldAddEndPuncttrue
\mciteSetBstMidEndSepPunct{\mcitedefaultmidpunct}
{\mcitedefaultendpunct}{\mcitedefaultseppunct}\relax
\EndOfBibitem
\bibitem{Aaij:2014owa}
{LHCb} collaboration, R.~Aaij {\em et al.},
  \href{http://dx.doi.org/10.1007/JHEP04(2014)114}{JHEP {\bf 04},  114 (2014)},
  \href{http://arxiv.org/abs/1402.2554}{{\tt arXiv:1402.2554 [hep-ex]}}\relax
\mciteBstWouldAddEndPuncttrue
\mciteSetBstMidEndSepPunct{\mcitedefaultmidpunct}
{\mcitedefaultendpunct}{\mcitedefaultseppunct}\relax
\EndOfBibitem
\bibitem{Aaij:2014fia}
{LHCb} collaboration, R.~Aaij {\em et al.},
  \href{http://dx.doi.org/10.1016/j.physletb.2014.07.051}{Phys. Lett. {\bf
  B736},  446 (2014)}, \href{http://arxiv.org/abs/1406.7204}{{\tt
  arXiv:1406.7204 [hep-ex]}}\relax
\mciteBstWouldAddEndPuncttrue
\mciteSetBstMidEndSepPunct{\mcitedefaultmidpunct}
{\mcitedefaultendpunct}{\mcitedefaultseppunct}\relax
\EndOfBibitem
\bibitem{Aaltonen:2010ta}
{CDF} collaboration, T.~Aaltonen {\em et al.},
  \href{http://dx.doi.org/10.1103/PhysRevD.83.032008}{Phys. Rev. {\bf D83},
  032008 (2011)}, \href{http://arxiv.org/abs/1004.4855}{{\tt arXiv:1004.4855
  [hep-ex]}}\relax
\mciteBstWouldAddEndPuncttrue
\mciteSetBstMidEndSepPunct{\mcitedefaultmidpunct}
{\mcitedefaultendpunct}{\mcitedefaultseppunct}\relax
\EndOfBibitem
\bibitem{Abazov:2004sa}
{\dzero} collaboration, V.~M. Abazov {\em et al.},
  \href{http://dx.doi.org/10.1103/PhysRevLett.94.182001}{Phys. Rev. Lett. {\bf
  94},  182001 (2005)}, \href{http://arxiv.org/abs/hep-ex/0410052}{{\tt
  arXiv:hep-ex/0410052 [hep-ex]}}\relax
\mciteBstWouldAddEndPuncttrue
\mciteSetBstMidEndSepPunct{\mcitedefaultmidpunct}
{\mcitedefaultendpunct}{\mcitedefaultseppunct}\relax
\EndOfBibitem
\bibitem{Aaij:2012eq}
{LHCb} collaboration, R.~Aaij {\em et al.},
  \href{http://dx.doi.org/10.1103/PhysRevLett.108.241801}{Phys. Rev. Lett. {\bf
  108},  241801 (2012)}, \href{http://arxiv.org/abs/1202.4717}{{\tt
  arXiv:1202.4717 [hep-ex]}}\relax
\mciteBstWouldAddEndPuncttrue
\mciteSetBstMidEndSepPunct{\mcitedefaultmidpunct}
{\mcitedefaultendpunct}{\mcitedefaultseppunct}\relax
\EndOfBibitem
\bibitem{Artuso:2015swg}
M.~Artuso, G.~Borissov, and A.~Lenz,
  \href{http://dx.doi.org/10.1103/RevModPhys.88.045002}{Rev. Mod. Phys. {\bf
  88},  045002 (2016)}, \href{http://arxiv.org/abs/1511.09466}{{\tt
  arXiv:1511.09466 [hep-ph]}}\relax
\mciteBstWouldAddEndPuncttrue
\mciteSetBstMidEndSepPunct{\mcitedefaultmidpunct}
{\mcitedefaultendpunct}{\mcitedefaultseppunct}\relax
\EndOfBibitem
\bibitem{Laplace:2002ik}
S.~Laplace, Z.~Ligeti, Y.~Nir, and G.~Perez,
  \href{http://dx.doi.org/10.1103/PhysRevD.65.094040}{Phys. Rev. {\bf D65},
  094040 (2002)}, \href{http://arxiv.org/abs/hep-ph/0202010}{{\tt
  arXiv:hep-ph/0202010 [hep-ph]}}\relax
\mciteBstWouldAddEndPuncttrue
\mciteSetBstMidEndSepPunct{\mcitedefaultmidpunct}
{\mcitedefaultendpunct}{\mcitedefaultseppunct}\relax
\EndOfBibitem
\bibitem{Hartkorn:1999ga}
K.~Hartkorn and H.~G. Moser,
  \href{http://dx.doi.org/10.1007/s100520050472}{Eur. Phys. J. {\bf C8},  381
  (1999)}\relax
\mciteBstWouldAddEndPuncttrue
\mciteSetBstMidEndSepPunct{\mcitedefaultmidpunct}
{\mcitedefaultendpunct}{\mcitedefaultseppunct}\relax
\EndOfBibitem
\bibitem{Dunietz:2000cr}
I.~Dunietz, R.~Fleischer, and U.~Nierste,
  \href{http://dx.doi.org/10.1103/PhysRevD.63.114015}{Phys. Rev. {\bf D63},
  114015 (2001)}, \href{http://arxiv.org/abs/hep-ph/0012219}{{\tt
  arXiv:hep-ph/0012219 [hep-ph]}}\relax
\mciteBstWouldAddEndPuncttrue
\mciteSetBstMidEndSepPunct{\mcitedefaultmidpunct}
{\mcitedefaultendpunct}{\mcitedefaultseppunct}\relax
\EndOfBibitem
\bibitem{Fleischer:2011cw}
R.~Fleischer and R.~Knegjens,
  \href{http://dx.doi.org/10.1140/epjc/s10052-011-1789-9}{Eur. Phys. J. {\bf
  C71},  1789 (2011)}, \href{http://arxiv.org/abs/1109.5115}{{\tt
  arXiv:1109.5115 [hep-ph]}}\relax
\mciteBstWouldAddEndPuncttrue
\mciteSetBstMidEndSepPunct{\mcitedefaultmidpunct}
{\mcitedefaultendpunct}{\mcitedefaultseppunct}\relax
\EndOfBibitem
\bibitem{Barate:1997ua}
{ALEPH} collaboration, R.~Barate {\em et al.},
  \href{http://dx.doi.org/10.1007/s100520050215}{Eur. Phys. J. {\bf C4},  367
  (1998)}\relax
\mciteBstWouldAddEndPuncttrue
\mciteSetBstMidEndSepPunct{\mcitedefaultmidpunct}
{\mcitedefaultendpunct}{\mcitedefaultseppunct}\relax
\EndOfBibitem
\bibitem{Abreu:2000ev}
{DELPHI} collaboration, P.~Abreu {\em et al.},
  \href{http://dx.doi.org/10.1007/s100520000531}{Eur. Phys. J. {\bf C18},  229
  (2000)}, \href{http://arxiv.org/abs/hep-ex/0105077}{{\tt arXiv:hep-ex/0105077
  [hep-ex]}}\relax
\mciteBstWouldAddEndPuncttrue
\mciteSetBstMidEndSepPunct{\mcitedefaultmidpunct}
{\mcitedefaultendpunct}{\mcitedefaultseppunct}\relax
\EndOfBibitem
\bibitem{Ackerstaff:1997ne}
{OPAL} collaboration, K.~Ackerstaff {\em et al.},
  \href{http://dx.doi.org/10.1007/s100520050150}{Eur. Phys. J. {\bf C2},  407
  (1998)}, \href{http://arxiv.org/abs/hep-ex/9708023}{{\tt arXiv:hep-ex/9708023
  [hep-ex]}}\relax
\mciteBstWouldAddEndPuncttrue
\mciteSetBstMidEndSepPunct{\mcitedefaultmidpunct}
{\mcitedefaultendpunct}{\mcitedefaultseppunct}\relax
\EndOfBibitem
\bibitem{Buskulic:1996ei}
{ALEPH} collaboration, D.~Buskulic {\em et al.},
  \href{http://dx.doi.org/10.1016/0370-2693(96)00451-0}{Phys. Lett. {\bf B377},
   205 (1996)}\relax
\mciteBstWouldAddEndPuncttrue
\mciteSetBstMidEndSepPunct{\mcitedefaultmidpunct}
{\mcitedefaultendpunct}{\mcitedefaultseppunct}\relax
\EndOfBibitem
\bibitem{Abe:1998cj}
{CDF} collaboration, F.~Abe {\em et al.},
  \href{http://dx.doi.org/10.1103/PhysRevD.59.032004}{Phys. Rev. {\bf D59},
  032004 (1999)}, \href{http://arxiv.org/abs/hep-ex/9808003}{{\tt
  arXiv:hep-ex/9808003 [hep-ex]}}\relax
\mciteBstWouldAddEndPuncttrue
\mciteSetBstMidEndSepPunct{\mcitedefaultmidpunct}
{\mcitedefaultendpunct}{\mcitedefaultseppunct}\relax
\EndOfBibitem
\bibitem{Abreu:2000sh}
{DELPHI} collaboration, P.~Abreu {\em et al.},
  \href{http://dx.doi.org/10.1007/s100520000415}{Eur. Phys. J. {\bf C16},  555
  (2000)}, \href{http://arxiv.org/abs/hep-ex/0107077}{{\tt arXiv:hep-ex/0107077
  [hep-ex]}}\relax
\mciteBstWouldAddEndPuncttrue
\mciteSetBstMidEndSepPunct{\mcitedefaultmidpunct}
{\mcitedefaultendpunct}{\mcitedefaultseppunct}\relax
\EndOfBibitem
\bibitem{Ackerstaff:1997qi}
{OPAL} collaboration, K.~Ackerstaff {\em et al.},
  \href{http://dx.doi.org/10.1016/S0370-2693(98)00289-5}{Phys. Lett. {\bf
  B426},  161 (1998)}, \href{http://arxiv.org/abs/hep-ex/9802002}{{\tt
  arXiv:hep-ex/9802002 [hep-ex]}}\relax
\mciteBstWouldAddEndPuncttrue
\mciteSetBstMidEndSepPunct{\mcitedefaultmidpunct}
{\mcitedefaultendpunct}{\mcitedefaultseppunct}\relax
\EndOfBibitem
\bibitem{Aaltonen:2011qsa}
{CDF} collaboration, T.~Aaltonen {\em et al.},
  \href{http://dx.doi.org/10.1103/PhysRevLett.107.272001}{Phys. Rev. Lett. {\bf
  107},  272001 (2011)}, \href{http://arxiv.org/abs/1103.1864}{{\tt
  arXiv:1103.1864 [hep-ex]}}\relax
\mciteBstWouldAddEndPuncttrue
\mciteSetBstMidEndSepPunct{\mcitedefaultmidpunct}
{\mcitedefaultendpunct}{\mcitedefaultseppunct}\relax
\EndOfBibitem
\bibitem{Aaij:2013bvd}
{LHCb} collaboration, R.~Aaij {\em et al.},
  \href{http://dx.doi.org/10.1103/PhysRevLett.112.111802}{Phys. Rev. Lett. {\bf
  112},  111802 (2014)}, \href{http://arxiv.org/abs/1312.1217}{{\tt
  arXiv:1312.1217 [hep-ex]}}\relax
\mciteBstWouldAddEndPuncttrue
\mciteSetBstMidEndSepPunct{\mcitedefaultmidpunct}
{\mcitedefaultendpunct}{\mcitedefaultseppunct}\relax
\EndOfBibitem
\bibitem{Aaij:2014sua}
{LHCb} collaboration, R.~Aaij {\em et al.},
  \href{http://dx.doi.org/10.1103/PhysRevLett.113.172001}{Phys. Rev. Lett. {\bf
  113},  172001 (2014)}, \href{http://arxiv.org/abs/1407.5873}{{\tt
  arXiv:1407.5873 [hep-ex]}}\relax
\mciteBstWouldAddEndPuncttrue
\mciteSetBstMidEndSepPunct{\mcitedefaultmidpunct}
{\mcitedefaultendpunct}{\mcitedefaultseppunct}\relax
\EndOfBibitem
\bibitem{Aaij:2017vqj}
{LHCb} collaboration, R.~Aaij {\em et al.},
  \href{http://dx.doi.org/10.1103/PhysRevLett.119.101801}{Phys. Rev. Lett. {\bf
  119},  101801 (2017)}, \href{http://arxiv.org/abs/1705.03475}{{\tt
  arXiv:1705.03475 [hep-ex]}}\relax
\mciteBstWouldAddEndPuncttrue
\mciteSetBstMidEndSepPunct{\mcitedefaultmidpunct}
{\mcitedefaultendpunct}{\mcitedefaultseppunct}\relax
\EndOfBibitem
\bibitem{Abazov:2004ce}
{\dzero} collaboration, V.~M. Abazov {\em et al.},
  \href{http://dx.doi.org/10.1103/PhysRevLett.94.042001}{Phys. Rev. Lett. {\bf
  94},  042001 (2005)}, \href{http://arxiv.org/abs/hep-ex/0409043}{{\tt
  arXiv:hep-ex/0409043 [hep-ex]}}\relax
\mciteBstWouldAddEndPuncttrue
\mciteSetBstMidEndSepPunct{\mcitedefaultmidpunct}
{\mcitedefaultendpunct}{\mcitedefaultseppunct}\relax
\EndOfBibitem
\bibitem{Aaij:2017vad}
{LHCb} collaboration, R.~Aaij {\em et al.},
  \href{http://dx.doi.org/10.1103/PhysRevLett.118.191801}{Phys. Rev. Lett. {\bf
  118},  191801 (2017)}, \href{http://arxiv.org/abs/1703.05747}{{\tt
  arXiv:1703.05747 [hep-ex]}}\relax
\mciteBstWouldAddEndPuncttrue
\mciteSetBstMidEndSepPunct{\mcitedefaultmidpunct}
{\mcitedefaultendpunct}{\mcitedefaultseppunct}\relax
\EndOfBibitem
\bibitem{Barate:2000kd}
{ALEPH} collaboration, R.~Barate {\em et al.},
  \href{http://dx.doi.org/10.1016/S0370-2693(00)00750-4}{Phys. Lett. {\bf
  B486},  286 (2000)}\relax
\mciteBstWouldAddEndPuncttrue
\mciteSetBstMidEndSepPunct{\mcitedefaultmidpunct}
{\mcitedefaultendpunct}{\mcitedefaultseppunct}\relax
\EndOfBibitem
\bibitem{Aaij:2012kn}
{LHCb} collaboration, R.~Aaij {\em et al.},
  \href{http://dx.doi.org/10.1016/j.physletb.2011.12.058}{Phys. Lett. {\bf
  B707},  349 (2012)}, \href{http://arxiv.org/abs/1111.0521}{{\tt
  arXiv:1111.0521 [hep-ex]}}\relax
\mciteBstWouldAddEndPuncttrue
\mciteSetBstMidEndSepPunct{\mcitedefaultmidpunct}
{\mcitedefaultendpunct}{\mcitedefaultseppunct}\relax
\EndOfBibitem
\bibitem{Aaij:2016dzn}
{LHCb} collaboration, R.~Aaij {\em et al.},
  \href{http://dx.doi.org/10.1016/j.physletb.2016.10.006}{Phys. Lett. {\bf
  B762},  484 (2016)}, \href{http://arxiv.org/abs/1607.06314}{{\tt
  arXiv:1607.06314 [hep-ex]}}\relax
\mciteBstWouldAddEndPuncttrue
\mciteSetBstMidEndSepPunct{\mcitedefaultmidpunct}
{\mcitedefaultendpunct}{\mcitedefaultseppunct}\relax
\EndOfBibitem
\bibitem{Aaij:2013eia}
{LHCb} collaboration, R.~Aaij {\em et al.},
  \href{http://dx.doi.org/10.1016/j.nuclphysb.2013.04.021}{Nucl. Phys. {\bf
  B873},  275 (2013)}, \href{http://arxiv.org/abs/1304.4500}{{\tt
  arXiv:1304.4500 [hep-ex]}}\relax
\mciteBstWouldAddEndPuncttrue
\mciteSetBstMidEndSepPunct{\mcitedefaultmidpunct}
{\mcitedefaultendpunct}{\mcitedefaultseppunct}\relax
\EndOfBibitem
\bibitem{Aaltonen:2011nk}
{CDF} collaboration, T.~Aaltonen {\em et al.},
  \href{http://dx.doi.org/10.1103/PhysRevD.84.052012}{Phys. Rev. {\bf D84},
  052012 (2011)}, \href{http://arxiv.org/abs/1106.3682}{{\tt arXiv:1106.3682
  [hep-ex]}}\relax
\mciteBstWouldAddEndPuncttrue
\mciteSetBstMidEndSepPunct{\mcitedefaultmidpunct}
{\mcitedefaultendpunct}{\mcitedefaultseppunct}\relax
\EndOfBibitem
\bibitem{Abazov:2016oqi}
{\dzero} collaboration, V.~M. Abazov {\em et al.},
  \href{http://dx.doi.org/10.1103/PhysRevD.94.012001}{Phys. Rev. {\bf D94},
  012001 (2016)}, \href{http://arxiv.org/abs/1603.01302}{{\tt arXiv:1603.01302
  [hep-ex]}}\relax
\mciteBstWouldAddEndPuncttrue
\mciteSetBstMidEndSepPunct{\mcitedefaultmidpunct}
{\mcitedefaultendpunct}{\mcitedefaultseppunct}\relax
\EndOfBibitem
\bibitem{Aaij:2013oba}
{LHCb} collaboration, R.~Aaij {\em et al.},
  \href{http://dx.doi.org/10.1103/PhysRevD.87.112010}{Phys. Rev. {\bf D87},
  112010 (2013)}, \href{http://arxiv.org/abs/1304.2600}{{\tt arXiv:1304.2600
  [hep-ex]}}\relax
\mciteBstWouldAddEndPuncttrue
\mciteSetBstMidEndSepPunct{\mcitedefaultmidpunct}
{\mcitedefaultendpunct}{\mcitedefaultseppunct}\relax
\EndOfBibitem
\bibitem{Aaij:2014bba}
{LHCb} collaboration, R.~Aaij {\em et al.},
  \href{http://dx.doi.org/10.1016/j.physletb.2014.10.005}{Phys. Lett. {\bf
  B739},  218 (2014)}, \href{http://arxiv.org/abs/1408.0275}{{\tt
  arXiv:1408.0275 [hep-ex]}}\relax
\mciteBstWouldAddEndPuncttrue
\mciteSetBstMidEndSepPunct{\mcitedefaultmidpunct}
{\mcitedefaultendpunct}{\mcitedefaultseppunct}\relax
\EndOfBibitem
\bibitem{Abe:1998wi}
{CDF} collaboration, F.~Abe {\em et al.},
  \href{http://dx.doi.org/10.1103/PhysRevLett.81.2432}{Phys. Rev. Lett. {\bf
  81},  2432 (1998)}, \href{http://arxiv.org/abs/hep-ex/9805034}{{\tt
  arXiv:hep-ex/9805034 [hep-ex]}}\relax
\mciteBstWouldAddEndPuncttrue
\mciteSetBstMidEndSepPunct{\mcitedefaultmidpunct}
{\mcitedefaultendpunct}{\mcitedefaultseppunct}\relax
\EndOfBibitem
\bibitem{Abulencia:2006zu}
{CDF} collaboration, A.~Abulencia {\em et al.},
  \href{http://dx.doi.org/10.1103/PhysRevLett.97.012002}{Phys. Rev. Lett. {\bf
  97},  012002 (2006)}, \href{http://arxiv.org/abs/hep-ex/0603027}{{\tt
  arXiv:hep-ex/0603027 [hep-ex]}}\relax
\mciteBstWouldAddEndPuncttrue
\mciteSetBstMidEndSepPunct{\mcitedefaultmidpunct}
{\mcitedefaultendpunct}{\mcitedefaultseppunct}\relax
\EndOfBibitem
\bibitem{Abazov:2008rba}
{\dzero} collaboration, V.~M. Abazov {\em et al.},
  \href{http://dx.doi.org/10.1103/PhysRevLett.102.092001}{Phys. Rev. Lett. {\bf
  102},  092001 (2009)}, \href{http://arxiv.org/abs/0805.2614}{{\tt
  arXiv:0805.2614 [hep-ex]}}\relax
\mciteBstWouldAddEndPuncttrue
\mciteSetBstMidEndSepPunct{\mcitedefaultmidpunct}
{\mcitedefaultendpunct}{\mcitedefaultseppunct}\relax
\EndOfBibitem
\bibitem{Aaltonen:2012yb}
{CDF} collaboration, T.~Aaltonen {\em et al.},
  \href{http://dx.doi.org/10.1103/PhysRevD.87.011101}{Phys. Rev. {\bf D87},
  011101 (2013)}, \href{http://arxiv.org/abs/1210.2366}{{\tt arXiv:1210.2366
  [hep-ex]}}\relax
\mciteBstWouldAddEndPuncttrue
\mciteSetBstMidEndSepPunct{\mcitedefaultmidpunct}
{\mcitedefaultendpunct}{\mcitedefaultseppunct}\relax
\EndOfBibitem
\bibitem{Aaij:2014bva}
{LHCb} collaboration, R.~Aaij {\em et al.},
  \href{http://dx.doi.org/10.1140/epjc/s10052-014-2839-x}{Eur. Phys. J. {\bf
  C74},  2839 (2014)}, \href{http://arxiv.org/abs/1401.6932}{{\tt
  arXiv:1401.6932 [hep-ex]}}\relax
\mciteBstWouldAddEndPuncttrue
\mciteSetBstMidEndSepPunct{\mcitedefaultmidpunct}
{\mcitedefaultendpunct}{\mcitedefaultseppunct}\relax
\EndOfBibitem
\bibitem{Aaij:2014gka}
{LHCb} collaboration, R.~Aaij {\em et al.},
  \href{http://dx.doi.org/10.1016/j.physletb.2015.01.010}{Phys. Lett. {\bf
  B742},  29 (2015)}, \href{http://arxiv.org/abs/1411.6899}{{\tt
  arXiv:1411.6899 [hep-ex]}}\relax
\mciteBstWouldAddEndPuncttrue
\mciteSetBstMidEndSepPunct{\mcitedefaultmidpunct}
{\mcitedefaultendpunct}{\mcitedefaultseppunct}\relax
\EndOfBibitem
\bibitem{Abreu:1999hu}
{DELPHI} collaboration, P.~Abreu {\em et al.},
  \href{http://dx.doi.org/10.1007/s100520050582}{Eur. Phys. J. {\bf C10},  185
  (1999)}\relax
\mciteBstWouldAddEndPuncttrue
\mciteSetBstMidEndSepPunct{\mcitedefaultmidpunct}
{\mcitedefaultendpunct}{\mcitedefaultseppunct}\relax
\EndOfBibitem
\bibitem{Abreu:1996nt}
{DELPHI} collaboration, P.~Abreu {\em et al.},
  \href{http://dx.doi.org/10.1007/s002880050164}{Z. Phys. {\bf C71},  199
  (1996)}\relax
\mciteBstWouldAddEndPuncttrue
\mciteSetBstMidEndSepPunct{\mcitedefaultmidpunct}
{\mcitedefaultendpunct}{\mcitedefaultseppunct}\relax
\EndOfBibitem
\bibitem{Akers:1995ui}
{OPAL} collaboration, R.~Akers {\em et al.},
  \href{http://dx.doi.org/10.1007/s002880050020}{Z. Phys. {\bf C69},  195
  (1996)}\relax
\mciteBstWouldAddEndPuncttrue
\mciteSetBstMidEndSepPunct{\mcitedefaultmidpunct}
{\mcitedefaultendpunct}{\mcitedefaultseppunct}\relax
\EndOfBibitem
\bibitem{Abe:1996df}
{CDF} collaboration, F.~Abe {\em et al.},
  \href{http://dx.doi.org/10.1103/PhysRevLett.77.1439}{Phys. Rev. Lett. {\bf
  77},  1439 (1996)}\relax
\mciteBstWouldAddEndPuncttrue
\mciteSetBstMidEndSepPunct{\mcitedefaultmidpunct}
{\mcitedefaultendpunct}{\mcitedefaultseppunct}\relax
\EndOfBibitem
\bibitem{Abazov:2007al}
{\dzero} collaboration, V.~M. Abazov {\em et al.},
  \href{http://dx.doi.org/10.1103/PhysRevLett.99.182001}{Phys. Rev. Lett. {\bf
  99},  182001 (2007)}, \href{http://arxiv.org/abs/0706.2358}{{\tt
  arXiv:0706.2358 [hep-ex]}}\relax
\mciteBstWouldAddEndPuncttrue
\mciteSetBstMidEndSepPunct{\mcitedefaultmidpunct}
{\mcitedefaultendpunct}{\mcitedefaultseppunct}\relax
\EndOfBibitem
\bibitem{Aaltonen:2009zn}
{CDF} collaboration, T.~Aaltonen {\em et al.},
  \href{http://dx.doi.org/10.1103/PhysRevLett.104.102002}{Phys. Rev. Lett. {\bf
  104},  102002 (2010)}, \href{http://arxiv.org/abs/0912.3566}{{\tt
  arXiv:0912.3566 [hep-ex]}}\relax
\mciteBstWouldAddEndPuncttrue
\mciteSetBstMidEndSepPunct{\mcitedefaultmidpunct}
{\mcitedefaultendpunct}{\mcitedefaultseppunct}\relax
\EndOfBibitem
\bibitem{Aaltonen:2014wfa}
{CDF} collaboration, T.~A. Aaltonen {\em et al.},
  \href{http://dx.doi.org/10.1103/PhysRevD.89.072014}{Phys. Rev. {\bf D89},
  072014 (2014)}, \href{http://arxiv.org/abs/1403.8126}{{\tt arXiv:1403.8126
  [hep-ex]}}\relax
\mciteBstWouldAddEndPuncttrue
\mciteSetBstMidEndSepPunct{\mcitedefaultmidpunct}
{\mcitedefaultendpunct}{\mcitedefaultseppunct}\relax
\EndOfBibitem
\bibitem{Chatrchyan:2013sxa}
{CMS} collaboration, S.~Chatrchyan {\em et al.},
  \href{http://dx.doi.org/10.1007/JHEP07(2013)163}{JHEP {\bf 07},  163 (2013)},
  \href{http://arxiv.org/abs/1304.7495}{{\tt arXiv:1304.7495 [hep-ex]}}\relax
\mciteBstWouldAddEndPuncttrue
\mciteSetBstMidEndSepPunct{\mcitedefaultmidpunct}
{\mcitedefaultendpunct}{\mcitedefaultseppunct}\relax
\EndOfBibitem
\bibitem{Aaij:2014zyy}
{LHCb} collaboration, R.~Aaij {\em et al.},
  \href{http://dx.doi.org/10.1016/j.physletb.2014.05.021}{Phys. Lett. {\bf
  B734},  122 (2014)}, \href{http://arxiv.org/abs/1402.6242}{{\tt
  arXiv:1402.6242 [hep-ex]}}\relax
\mciteBstWouldAddEndPuncttrue
\mciteSetBstMidEndSepPunct{\mcitedefaultmidpunct}
{\mcitedefaultendpunct}{\mcitedefaultseppunct}\relax
\EndOfBibitem
\bibitem{Abreu:1995kt}
{DELPHI} collaboration, P.~Abreu {\em et al.},
  \href{http://dx.doi.org/10.1007/BF01565255}{Z. Phys. {\bf C68},  541
  (1995)}\relax
\mciteBstWouldAddEndPuncttrue
\mciteSetBstMidEndSepPunct{\mcitedefaultmidpunct}
{\mcitedefaultendpunct}{\mcitedefaultseppunct}\relax
\EndOfBibitem
\bibitem{Aaij:2014sia}
{LHCb} collaboration, R.~Aaij {\em et al.},
  \href{http://dx.doi.org/10.1016/j.physletb.2014.06.064}{Phys. Lett. {\bf
  B736},  154 (2014)}, \href{http://arxiv.org/abs/1405.1543}{{\tt
  arXiv:1405.1543 [hep-ex]}}\relax
\mciteBstWouldAddEndPuncttrue
\mciteSetBstMidEndSepPunct{\mcitedefaultmidpunct}
{\mcitedefaultendpunct}{\mcitedefaultseppunct}\relax
\EndOfBibitem
\bibitem{Aaij:2014lxa}
{LHCb} collaboration, R.~Aaij {\em et al.},
  \href{http://dx.doi.org/10.1103/PhysRevLett.113.242002}{Phys. Rev. Lett. {\bf
  113},  242002 (2014)}, \href{http://arxiv.org/abs/1409.8568}{{\tt
  arXiv:1409.8568 [hep-ex]}}\relax
\mciteBstWouldAddEndPuncttrue
\mciteSetBstMidEndSepPunct{\mcitedefaultmidpunct}
{\mcitedefaultendpunct}{\mcitedefaultseppunct}\relax
\EndOfBibitem
\bibitem{Aaij:2014esa}
{LHCb} collaboration, R.~Aaij {\em et al.},
  \href{http://dx.doi.org/10.1103/PhysRevLett.113.032001}{Phys. Rev. Lett. {\bf
  113},  032001 (2014)}, \href{http://arxiv.org/abs/1405.7223}{{\tt
  arXiv:1405.7223 [hep-ex]}}\relax
\mciteBstWouldAddEndPuncttrue
\mciteSetBstMidEndSepPunct{\mcitedefaultmidpunct}
{\mcitedefaultendpunct}{\mcitedefaultseppunct}\relax
\EndOfBibitem
\bibitem{Aaij:2016dls}
{LHCb} collaboration, R.~Aaij {\em et al.},
  \href{http://dx.doi.org/10.1103/PhysRevD.93.092007}{Phys. Rev. {\bf D93},
  092007 (2016)}, \href{http://arxiv.org/abs/1604.01412}{{\tt arXiv:1604.01412
  [hep-ex]}}\relax
\mciteBstWouldAddEndPuncttrue
\mciteSetBstMidEndSepPunct{\mcitedefaultmidpunct}
{\mcitedefaultendpunct}{\mcitedefaultseppunct}\relax
\EndOfBibitem
\bibitem{Keum:1998fd}
Y.-Y. Keum and U.~Nierste,
  \href{http://dx.doi.org/10.1103/PhysRevD.57.4282}{Phys. Rev. {\bf D57},  4282
  (1998)}, \href{http://arxiv.org/abs/hep-ph/9710512}{{\tt arXiv:hep-ph/9710512
  [hep-ph]}}\relax
\mciteBstWouldAddEndPuncttrue
\mciteSetBstMidEndSepPunct{\mcitedefaultmidpunct}
{\mcitedefaultendpunct}{\mcitedefaultseppunct}\relax
\EndOfBibitem
\bibitem{Uraltsev:1996ta}
N.~G. Uraltsev, \href{http://dx.doi.org/10.1016/0370-2693(96)00305-X}{Phys.
  Lett. {\bf B376},  303 (1996)},
  \href{http://arxiv.org/abs/hep-ph/9602324}{{\tt arXiv:hep-ph/9602324
  [hep-ph]}}\relax
\mciteBstWouldAddEndPuncttrue
\mciteSetBstMidEndSepPunct{\mcitedefaultmidpunct}
{\mcitedefaultendpunct}{\mcitedefaultseppunct}\relax
\EndOfBibitem
\bibitem{Pirjol:1998ur}
D.~Pirjol and N.~Uraltsev,
  \href{http://dx.doi.org/10.1103/PhysRevD.59.034012}{Phys. Rev. {\bf D59},
  034012 (1999)}, \href{http://arxiv.org/abs/hep-ph/9805488}{{\tt
  arXiv:hep-ph/9805488 [hep-ph]}}\relax
\mciteBstWouldAddEndPuncttrue
\mciteSetBstMidEndSepPunct{\mcitedefaultmidpunct}
{\mcitedefaultendpunct}{\mcitedefaultseppunct}\relax
\EndOfBibitem
\bibitem{Colangelo:1996ta}
P.~Colangelo and F.~De~Fazio,
  \href{http://dx.doi.org/10.1016/0370-2693(96)01049-0}{Phys. Lett. {\bf B387},
   371 (1996)}, \href{http://arxiv.org/abs/hep-ph/9604425}{{\tt
  arXiv:hep-ph/9604425 [hep-ph]}}\relax
\mciteBstWouldAddEndPuncttrue
\mciteSetBstMidEndSepPunct{\mcitedefaultmidpunct}
{\mcitedefaultendpunct}{\mcitedefaultseppunct}\relax
\EndOfBibitem
\bibitem{DiPierro:1999tb}
{UKQCD} collaboration, M.~Di~Pierro, C.~T. Sachrajda, and C.~Michael,
  \href{http://dx.doi.org/10.1016/S0370-2693(99)01166-1}{Phys. Lett. {\bf
  B468},  143 (1999)}, \href{http://arxiv.org/abs/hep-lat/9906031}{{\tt
  arXiv:hep-lat/9906031 [hep-lat]}}, Erratum ibid.\
  \href{http://dx.doi.org/10.1016/S0370-2693(01)01458-7}{{\bf D525}, 360},
  (2002)\relax
\mciteBstWouldAddEndPuncttrue
\mciteSetBstMidEndSepPunct{\mcitedefaultmidpunct}
{\mcitedefaultendpunct}{\mcitedefaultseppunct}\relax
\EndOfBibitem
\bibitem{Buskulic:1996qt}
{ALEPH} collaboration, D.~Buskulic {\em et al.},
  \href{http://dx.doi.org/10.1007/s002880050483}{Z. Phys. {\bf C75},  397
  (1997)}\relax
\mciteBstWouldAddEndPuncttrue
\mciteSetBstMidEndSepPunct{\mcitedefaultmidpunct}
{\mcitedefaultendpunct}{\mcitedefaultseppunct}\relax
\EndOfBibitem
\bibitem{Abreu:1997xq}
{DELPHI} collaboration, P.~Abreu {\em et al.},
  \href{http://dx.doi.org/10.1007/s002880050582}{Z. Phys. {\bf C76},  579
  (1997)}\relax
\mciteBstWouldAddEndPuncttrue
\mciteSetBstMidEndSepPunct{\mcitedefaultmidpunct}
{\mcitedefaultendpunct}{\mcitedefaultseppunct}\relax
\EndOfBibitem
\bibitem{Abdallah:2002mr}
{DELPHI} collaboration, J.~Abdallah {\em et al.},
  \href{http://dx.doi.org/10.1140/epjc/s2003-01183-7}{Eur. Phys. J. {\bf C28},
  155 (2003)}, \href{http://arxiv.org/abs/hep-ex/0303032}{{\tt
  arXiv:hep-ex/0303032 [hep-ex]}}\relax
\mciteBstWouldAddEndPuncttrue
\mciteSetBstMidEndSepPunct{\mcitedefaultmidpunct}
{\mcitedefaultendpunct}{\mcitedefaultseppunct}\relax
\EndOfBibitem
\bibitem{Acciarri:1998pq}
{L3} collaboration, M.~Acciarri {\em et al.},
  \href{http://dx.doi.org/10.1007/s100520050262}{Eur. Phys. J. {\bf C5},  195
  (1998)}\relax
\mciteBstWouldAddEndPuncttrue
\mciteSetBstMidEndSepPunct{\mcitedefaultmidpunct}
{\mcitedefaultendpunct}{\mcitedefaultseppunct}\relax
\EndOfBibitem
\bibitem{Ackerstaff:1997iw}
{OPAL} collaboration, K.~Ackerstaff {\em et al.},
  \href{http://dx.doi.org/10.1007/s002880050565}{Z. Phys. {\bf C76},  417
  (1997)}, \href{http://arxiv.org/abs/hep-ex/9707010}{{\tt arXiv:hep-ex/9707010
  [hep-ex]}}\relax
\mciteBstWouldAddEndPuncttrue
\mciteSetBstMidEndSepPunct{\mcitedefaultmidpunct}
{\mcitedefaultendpunct}{\mcitedefaultseppunct}\relax
\EndOfBibitem
\bibitem{Ackerstaff:1997vd}
{OPAL} collaboration, K.~Ackerstaff {\em et al.},
  \href{http://dx.doi.org/10.1007/s002880050564}{Z. Phys. {\bf C76},  401
  (1997)}, \href{http://arxiv.org/abs/hep-ex/9707009}{{\tt arXiv:hep-ex/9707009
  [hep-ex]}}\relax
\mciteBstWouldAddEndPuncttrue
\mciteSetBstMidEndSepPunct{\mcitedefaultmidpunct}
{\mcitedefaultendpunct}{\mcitedefaultseppunct}\relax
\EndOfBibitem
\bibitem{Alexander:1996id}
{OPAL} collaboration, G.~Alexander {\em et al.},
  \href{http://dx.doi.org/10.1007/s002880050258}{Z. Phys. {\bf C72},  377
  (1996)}\relax
\mciteBstWouldAddEndPuncttrue
\mciteSetBstMidEndSepPunct{\mcitedefaultmidpunct}
{\mcitedefaultendpunct}{\mcitedefaultseppunct}\relax
\EndOfBibitem
\bibitem{Abe:1997qf}
{CDF} collaboration, F.~Abe {\em et al.},
  \href{http://dx.doi.org/10.1103/PhysRevLett.80.2057}{Phys. Rev. Lett. {\bf
  80},  2057 (1998)}, \href{http://arxiv.org/abs/hep-ex/9712004}{{\tt
  arXiv:hep-ex/9712004 [hep-ex]}}\relax
\mciteBstWouldAddEndPuncttrue
\mciteSetBstMidEndSepPunct{\mcitedefaultmidpunct}
{\mcitedefaultendpunct}{\mcitedefaultseppunct}\relax
\EndOfBibitem
\bibitem{Abe:1998sq}
{CDF} collaboration, F.~Abe {\em et al.},
  \href{http://dx.doi.org/10.1103/PhysRevD.59.032001}{Phys. Rev. {\bf D59},
  032001 (1999)}, \href{http://arxiv.org/abs/hep-ex/9806026}{{\tt
  arXiv:hep-ex/9806026 [hep-ex]}}\relax
\mciteBstWouldAddEndPuncttrue
\mciteSetBstMidEndSepPunct{\mcitedefaultmidpunct}
{\mcitedefaultendpunct}{\mcitedefaultseppunct}\relax
\EndOfBibitem
\bibitem{Abe:1999pv}
{CDF} collaboration, F.~Abe {\em et al.},
  \href{http://dx.doi.org/10.1103/PhysRevD.60.051101}{Phys. Rev. {\bf D60},
  051101 (1999)}\relax
\mciteBstWouldAddEndPuncttrue
\mciteSetBstMidEndSepPunct{\mcitedefaultmidpunct}
{\mcitedefaultendpunct}{\mcitedefaultseppunct}\relax
\EndOfBibitem
\bibitem{Abe:1999ds}
{CDF} collaboration, F.~Abe {\em et al.},
  \href{http://dx.doi.org/10.1103/PhysRevD.60.072003}{Phys. Rev. {\bf D60},
  072003 (1999)}, \href{http://arxiv.org/abs/hep-ex/9903011}{{\tt
  arXiv:hep-ex/9903011 [hep-ex]}}\relax
\mciteBstWouldAddEndPuncttrue
\mciteSetBstMidEndSepPunct{\mcitedefaultmidpunct}
{\mcitedefaultendpunct}{\mcitedefaultseppunct}\relax
\EndOfBibitem
\bibitem{Affolder:1999cn}
{CDF} collaboration, T.~Affolder {\em et al.},
  \href{http://dx.doi.org/10.1103/PhysRevD.60.112004}{Phys. Rev. {\bf D60},
  112004 (1999)}, \href{http://arxiv.org/abs/hep-ex/9907053}{{\tt
  arXiv:hep-ex/9907053 [hep-ex]}}\relax
\mciteBstWouldAddEndPuncttrue
\mciteSetBstMidEndSepPunct{\mcitedefaultmidpunct}
{\mcitedefaultendpunct}{\mcitedefaultseppunct}\relax
\EndOfBibitem
\bibitem{Abazov:2006qp}
{\dzero} collaboration, V.~M. Abazov {\em et al.},
  \href{http://dx.doi.org/10.1103/PhysRevD.74.112002}{Phys. Rev. {\bf D74},
  112002 (2006)}, \href{http://arxiv.org/abs/hep-ex/0609034}{{\tt
  arXiv:hep-ex/0609034 [hep-ex]}}\relax
\mciteBstWouldAddEndPuncttrue
\mciteSetBstMidEndSepPunct{\mcitedefaultmidpunct}
{\mcitedefaultendpunct}{\mcitedefaultseppunct}\relax
\EndOfBibitem
\bibitem{Aubert:2001te}
{\babar} collaboration, B.~Aubert {\em et al.},
  \href{http://dx.doi.org/10.1103/PhysRevLett.88.221802}{Phys. Rev. Lett. {\bf
  88},  221802 (2002)}, \href{http://arxiv.org/abs/hep-ex/0112044}{{\tt
  arXiv:hep-ex/0112044 [hep-ex]}}\relax
\mciteBstWouldAddEndPuncttrue
\mciteSetBstMidEndSepPunct{\mcitedefaultmidpunct}
{\mcitedefaultendpunct}{\mcitedefaultseppunct}\relax
\EndOfBibitem
\bibitem{Aubert:2002rg}
{\babar} collaboration, B.~Aubert {\em et al.},
  \href{http://dx.doi.org/10.1103/PhysRevD.66.032003}{Phys. Rev. {\bf D66},
  032003 (2002)}, \href{http://arxiv.org/abs/hep-ex/0201020}{{\tt
  arXiv:hep-ex/0201020 [hep-ex]}}\relax
\mciteBstWouldAddEndPuncttrue
\mciteSetBstMidEndSepPunct{\mcitedefaultmidpunct}
{\mcitedefaultendpunct}{\mcitedefaultseppunct}\relax
\EndOfBibitem
\bibitem{Aubert:2001tf}
{\babar} collaboration, B.~Aubert {\em et al.},
  \href{http://dx.doi.org/10.1103/PhysRevLett.88.221803}{Phys. Rev. Lett. {\bf
  88},  221803 (2002)}, \href{http://arxiv.org/abs/hep-ex/0112045}{{\tt
  arXiv:hep-ex/0112045 [hep-ex]}}\relax
\mciteBstWouldAddEndPuncttrue
\mciteSetBstMidEndSepPunct{\mcitedefaultmidpunct}
{\mcitedefaultendpunct}{\mcitedefaultseppunct}\relax
\EndOfBibitem
\bibitem{Zheng:2002jv}
{Belle} collaboration, Y.~Zheng {\em et al.},
  \href{http://dx.doi.org/10.1103/PhysRevD.67.092004}{Phys. Rev. {\bf D67},
  092004 (2003)}, \href{http://arxiv.org/abs/hep-ex/0211065}{{\tt
  arXiv:hep-ex/0211065 [hep-ex]}}\relax
\mciteBstWouldAddEndPuncttrue
\mciteSetBstMidEndSepPunct{\mcitedefaultmidpunct}
{\mcitedefaultendpunct}{\mcitedefaultseppunct}\relax
\EndOfBibitem
\bibitem{Aaij:2011qx}
{LHCb} collaboration, R.~Aaij {\em et al.},
  \href{http://dx.doi.org/10.1016/j.physletb.2012.02.031}{Phys. Lett. {\bf
  B709},  177 (2012)}, \href{http://arxiv.org/abs/1112.4311}{{\tt
  arXiv:1112.4311 [hep-ex]}}\relax
\mciteBstWouldAddEndPuncttrue
\mciteSetBstMidEndSepPunct{\mcitedefaultmidpunct}
{\mcitedefaultendpunct}{\mcitedefaultseppunct}\relax
\EndOfBibitem
\bibitem{Aaij:2012nt}
{LHCb} collaboration, R.~Aaij {\em et al.},
  \href{http://dx.doi.org/10.1016/j.physletb.2013.01.019}{Phys. Lett. {\bf
  B719},  318 (2013)}, \href{http://arxiv.org/abs/1210.6750}{{\tt
  arXiv:1210.6750 [hep-ex]}}\relax
\mciteBstWouldAddEndPuncttrue
\mciteSetBstMidEndSepPunct{\mcitedefaultmidpunct}
{\mcitedefaultendpunct}{\mcitedefaultseppunct}\relax
\EndOfBibitem
\bibitem{Aaij:2013gja}
{LHCb} collaboration, R.~Aaij {\em et al.},
  \href{http://dx.doi.org/10.1140/epjc/s10052-013-2655-8}{Eur. Phys. J. {\bf
  C73},  2655 (2013)}, \href{http://arxiv.org/abs/1308.1302}{{\tt
  arXiv:1308.1302 [hep-ex]}}\relax
\mciteBstWouldAddEndPuncttrue
\mciteSetBstMidEndSepPunct{\mcitedefaultmidpunct}
{\mcitedefaultendpunct}{\mcitedefaultseppunct}\relax
\EndOfBibitem
\bibitem{Aaij:2016fdk}
{LHCb} collaboration, R.~Aaij {\em et al.},
  \href{http://dx.doi.org/10.1140/epjc/s10052-016-4250-2}{Eur. Phys. J. {\bf
  C76},  412 (2016)}, \href{http://arxiv.org/abs/1604.03475}{{\tt
  arXiv:1604.03475 [hep-ex]}}\relax
\mciteBstWouldAddEndPuncttrue
\mciteSetBstMidEndSepPunct{\mcitedefaultmidpunct}
{\mcitedefaultendpunct}{\mcitedefaultseppunct}\relax
\EndOfBibitem
\bibitem{Albrecht:1992yd}
{ARGUS} collaboration, H.~Albrecht {\em et al.},
  \href{http://dx.doi.org/10.1007/BF01565092}{Z. Phys. {\bf C55},  357
  (1992)}\relax
\mciteBstWouldAddEndPuncttrue
\mciteSetBstMidEndSepPunct{\mcitedefaultmidpunct}
{\mcitedefaultendpunct}{\mcitedefaultseppunct}\relax
\EndOfBibitem
\bibitem{Albrecht:1993gr}
{ARGUS} collaboration, H.~Albrecht {\em et al.},
  \href{http://dx.doi.org/10.1016/0370-2693(94)90415-4}{Phys. Lett. {\bf B324},
   249 (1994)}\relax
\mciteBstWouldAddEndPuncttrue
\mciteSetBstMidEndSepPunct{\mcitedefaultmidpunct}
{\mcitedefaultendpunct}{\mcitedefaultseppunct}\relax
\EndOfBibitem
\bibitem{Bartelt:1993cf}
{CLEO} collaboration, J.~E. Bartelt {\em et al.},
  \href{http://dx.doi.org/10.1103/PhysRevLett.71.1680}{Phys. Rev. Lett. {\bf
  71},  1680 (1993)}\relax
\mciteBstWouldAddEndPuncttrue
\mciteSetBstMidEndSepPunct{\mcitedefaultmidpunct}
{\mcitedefaultendpunct}{\mcitedefaultseppunct}\relax
\EndOfBibitem
\bibitem{Behrens:2000qu}
{CLEO} collaboration, B.~H. Behrens {\em et al.},
  \href{http://dx.doi.org/10.1016/S0370-2693(00)00990-4}{Phys. Lett. {\bf
  B490},  36 (2000)}, \href{http://arxiv.org/abs/hep-ex/0005013}{{\tt
  arXiv:hep-ex/0005013 [hep-ex]}}\relax
\mciteBstWouldAddEndPuncttrue
\mciteSetBstMidEndSepPunct{\mcitedefaultmidpunct}
{\mcitedefaultendpunct}{\mcitedefaultseppunct}\relax
\EndOfBibitem
\bibitem{Aubert:2003hd}
{\babar} collaboration, B.~Aubert {\em et al.},
  \href{http://dx.doi.org/10.1103/PhysRevLett.92.181801}{Phys. Rev. Lett. {\bf
  92},  181801 (2004)}, \href{http://arxiv.org/abs/hep-ex/0311037}{{\tt
  arXiv:hep-ex/0311037 [hep-ex]}}\relax
\mciteBstWouldAddEndPuncttrue
\mciteSetBstMidEndSepPunct{\mcitedefaultmidpunct}
{\mcitedefaultendpunct}{\mcitedefaultseppunct}\relax
\EndOfBibitem
\bibitem{Aubert:2004xga}
{\babar} collaboration, B.~Aubert {\em et al.},
  \href{http://dx.doi.org/10.1103/PhysRevD.70.012007}{Phys. Rev. {\bf D70},
  012007 (2004)}, \href{http://arxiv.org/abs/hep-ex/0403002}{{\tt
  arXiv:hep-ex/0403002 [hep-ex]}}\relax
\mciteBstWouldAddEndPuncttrue
\mciteSetBstMidEndSepPunct{\mcitedefaultmidpunct}
{\mcitedefaultendpunct}{\mcitedefaultseppunct}\relax
\EndOfBibitem
\bibitem{Higuchi:2012kx}
{Belle} collaboration, T.~Higuchi {\em et al.},
  \href{http://dx.doi.org/10.1103/PhysRevD.85.071105}{Phys. Rev. {\bf D85},
  071105 (2012)}, \href{http://arxiv.org/abs/1203.0930}{{\tt arXiv:1203.0930
  [hep-ex]}}\relax
\mciteBstWouldAddEndPuncttrue
\mciteSetBstMidEndSepPunct{\mcitedefaultmidpunct}
{\mcitedefaultendpunct}{\mcitedefaultseppunct}\relax
\EndOfBibitem
\bibitem{Gershon:2010wx}
T.~Gershon, \href{http://dx.doi.org/10.1088/0954-3899/38/1/015007}{J. Phys.
  {\bf G38},  015007 (2011)}, \href{http://arxiv.org/abs/1007.5135}{{\tt
  arXiv:1007.5135 [hep-ph]}}, Erratum ibid.\
  \href{http://dx.doi.org/10.1088/0954-3899/42/11/119501}{{\bf G42}, 119501},
  (2015)\relax
\mciteBstWouldAddEndPuncttrue
\mciteSetBstMidEndSepPunct{\mcitedefaultmidpunct}
{\mcitedefaultendpunct}{\mcitedefaultseppunct}\relax
\EndOfBibitem
\bibitem{Aaboud:2016bro}
{ATLAS} collaboration, M.~Aaboud {\em et al.},
  \href{http://dx.doi.org/10.1007/JHEP06(2016)081}{JHEP {\bf 06},  081 (2016)},
  \href{http://arxiv.org/abs/1605.07485}{{\tt arXiv:1605.07485 [hep-ex]}}\relax
\mciteBstWouldAddEndPuncttrue
\mciteSetBstMidEndSepPunct{\mcitedefaultmidpunct}
{\mcitedefaultendpunct}{\mcitedefaultseppunct}\relax
\EndOfBibitem
\bibitem{Charles:2011va_mod}
{CKMfitter group}, J.~Charles {\em et al.},
  \href{http://dx.doi.org/10.1103/PhysRevD.84.033005}{Phys. Rev. {\bf D84},
  033005 (2011)}, \href{http://arxiv.org/abs/1106.4041}{{\tt arXiv:1106.4041
  [hep-ph]}}, with updated results and plots available at
  \url{http://ckmfitter.in2p3.fr}\relax
\mciteBstWouldAddEndPuncttrue
\mciteSetBstMidEndSepPunct{\mcitedefaultmidpunct}
{\mcitedefaultendpunct}{\mcitedefaultseppunct}\relax
\EndOfBibitem
\bibitem{Bona:2006ah_mod}
{UTfit} collaboration, M.~Bona {\em et al.},
  \href{http://dx.doi.org/10.1088/1126-6708/2006/10/081}{JHEP {\bf 10},  081
  (2006)}, \href{http://arxiv.org/abs/hep-ph/0606167}{{\tt arXiv:hep-ph/0606167
  [hep-ph]}}, with similar updated results and plots available at
  \url{http://www.utfit.org}\relax
\mciteBstWouldAddEndPuncttrue
\mciteSetBstMidEndSepPunct{\mcitedefaultmidpunct}
{\mcitedefaultendpunct}{\mcitedefaultseppunct}\relax
\EndOfBibitem
\bibitem{Abazov:2013uma}
{\dzero} collaboration, V.~M. Abazov {\em et al.},
  \href{http://dx.doi.org/10.1103/PhysRevD.89.012002}{Phys. Rev. {\bf D89},
  012002 (2014)}, \href{http://arxiv.org/abs/1310.0447}{{\tt arXiv:1310.0447
  [hep-ex]}}\relax
\mciteBstWouldAddEndPuncttrue
\mciteSetBstMidEndSepPunct{\mcitedefaultmidpunct}
{\mcitedefaultendpunct}{\mcitedefaultseppunct}\relax
\EndOfBibitem
\bibitem{Nierste_CKM2014}
 {U.~Nierste, talk presented at the 8th International Workshop on the CKM
  unitarity Triangle (CKM 2014)}, 2014, {\small
  \url{{http://indico.cern.ch/event/253826/contributions/567426/}}}\relax
\mciteBstWouldAddEndPuncttrue
\mciteSetBstMidEndSepPunct{\mcitedefaultmidpunct}
{\mcitedefaultendpunct}{\mcitedefaultseppunct}\relax
\EndOfBibitem
\bibitem{Aaltonen:2012ie}
{CDF} collaboration, T.~Aaltonen {\em et al.},
  \href{http://dx.doi.org/10.1103/PhysRevLett.109.171802}{Phys. Rev. Lett. {\bf
  109},  171802 (2012)}, \href{http://arxiv.org/abs/1208.2967}{{\tt
  arXiv:1208.2967 [hep-ex]}}\relax
\mciteBstWouldAddEndPuncttrue
\mciteSetBstMidEndSepPunct{\mcitedefaultmidpunct}
{\mcitedefaultendpunct}{\mcitedefaultseppunct}\relax
\EndOfBibitem
\bibitem{Abazov:2011ry}
{\dzero} collaboration, V.~M. Abazov {\em et al.},
  \href{http://dx.doi.org/10.1103/PhysRevD.85.032006}{Phys. Rev. {\bf D85},
  032006 (2012)}, \href{http://arxiv.org/abs/1109.3166}{{\tt arXiv:1109.3166
  [hep-ex]}}\relax
\mciteBstWouldAddEndPuncttrue
\mciteSetBstMidEndSepPunct{\mcitedefaultmidpunct}
{\mcitedefaultendpunct}{\mcitedefaultseppunct}\relax
\EndOfBibitem
\bibitem{Aad:2014cqa}
{ATLAS} collaboration, G.~Aad {\em et al.},
  \href{http://dx.doi.org/10.1103/PhysRevD.90.052007}{Phys. Rev. {\bf D90},
  052007 (2014)}, \href{http://arxiv.org/abs/1407.1796}{{\tt arXiv:1407.1796
  [hep-ex]}}\relax
\mciteBstWouldAddEndPuncttrue
\mciteSetBstMidEndSepPunct{\mcitedefaultmidpunct}
{\mcitedefaultendpunct}{\mcitedefaultseppunct}\relax
\EndOfBibitem
\bibitem{Aad:2016tdj}
{ATLAS} collaboration, G.~Aad {\em et al.},
  \href{http://dx.doi.org/10.1007/JHEP08(2016)147}{JHEP {\bf 08},  147 (2016)},
  \href{http://arxiv.org/abs/1601.03297}{{\tt arXiv:1601.03297 [hep-ex]}}\relax
\mciteBstWouldAddEndPuncttrue
\mciteSetBstMidEndSepPunct{\mcitedefaultmidpunct}
{\mcitedefaultendpunct}{\mcitedefaultseppunct}\relax
\EndOfBibitem
\bibitem{Khachatryan:2015nza}
{CMS} collaboration, V.~Khachatryan {\em et al.},
  \href{http://dx.doi.org/10.1016/j.physletb.2016.03.046}{Phys. Lett. {\bf
  B757},  97 (2016)}, \href{http://arxiv.org/abs/1507.07527}{{\tt
  arXiv:1507.07527 [hep-ex]}}\relax
\mciteBstWouldAddEndPuncttrue
\mciteSetBstMidEndSepPunct{\mcitedefaultmidpunct}
{\mcitedefaultendpunct}{\mcitedefaultseppunct}\relax
\EndOfBibitem
\bibitem{Aaij:2014zsa}
{LHCb} collaboration, R.~Aaij {\em et al.},
  \href{http://dx.doi.org/10.1103/PhysRevLett.114.041801}{Phys. Rev. Lett. {\bf
  114},  041801 (2015)}, \href{http://arxiv.org/abs/1411.3104}{{\tt
  arXiv:1411.3104 [hep-ex]}}\relax
\mciteBstWouldAddEndPuncttrue
\mciteSetBstMidEndSepPunct{\mcitedefaultmidpunct}
{\mcitedefaultendpunct}{\mcitedefaultseppunct}\relax
\EndOfBibitem
\bibitem{Aaij:2017zgz}
{LHCb} collaboration, R.~Aaij {\em et al.},
  \href{http://dx.doi.org/10.1007/JHEP08(2017)037}{JHEP {\bf 08},  037 (2017)},
  \href{http://arxiv.org/abs/1704.08217}{{\tt arXiv:1704.08217 [hep-ex]}}\relax
\mciteBstWouldAddEndPuncttrue
\mciteSetBstMidEndSepPunct{\mcitedefaultmidpunct}
{\mcitedefaultendpunct}{\mcitedefaultseppunct}\relax
\EndOfBibitem
\bibitem{Aaij:2016ohx}
{LHCb} collaboration, R.~Aaij {\em et al.},
  \href{http://dx.doi.org/10.1016/j.physletb.2016.09.028}{Phys. Lett. {\bf
  B762},  253 (2016)}, \href{http://arxiv.org/abs/1608.04855}{{\tt
  arXiv:1608.04855 [hep-ex]}}\relax
\mciteBstWouldAddEndPuncttrue
\mciteSetBstMidEndSepPunct{\mcitedefaultmidpunct}
{\mcitedefaultendpunct}{\mcitedefaultseppunct}\relax
\EndOfBibitem
\bibitem{Lenz:2012mb}
A.~Lenz, \href{http://arxiv.org/abs/1205.1444}{{\tt arXiv:1205.1444 [hep-ph]}}
  (2012)\relax
\mciteBstWouldAddEndPuncttrue
\mciteSetBstMidEndSepPunct{\mcitedefaultmidpunct}
{\mcitedefaultendpunct}{\mcitedefaultseppunct}\relax
\EndOfBibitem
\bibitem{Esen:2010jq}
{Belle} collaboration, S.~Esen {\em et al.},
  \href{http://dx.doi.org/10.1103/PhysRevLett.105.201802}{Phys. Rev. Lett. {\bf
  105},  201802 (2010)}, \href{http://arxiv.org/abs/1005.5177}{{\tt
  arXiv:1005.5177 [hep-ex]}}\relax
\mciteBstWouldAddEndPuncttrue
\mciteSetBstMidEndSepPunct{\mcitedefaultmidpunct}
{\mcitedefaultendpunct}{\mcitedefaultseppunct}\relax
\EndOfBibitem
\bibitem{Abazov:2008ig}
{\dzero} collaboration, V.~Abazov {\em et al.},
  \href{http://dx.doi.org/10.1103/PhysRevLett.102.091801}{Phys. Rev. Lett. {\bf
  102},  091801 (2009)}, \href{http://arxiv.org/abs/0811.2173}{{\tt
  arXiv:0811.2173 [hep-ex]}}\relax
\mciteBstWouldAddEndPuncttrue
\mciteSetBstMidEndSepPunct{\mcitedefaultmidpunct}
{\mcitedefaultendpunct}{\mcitedefaultseppunct}\relax
\EndOfBibitem
\bibitem{Abulencia:2007zz}
{CDF} collaboration, T.~Aaltonen {\em et al.},
  \href{http://dx.doi.org/10.1103/PhysRevLett.100.021803}{Phys. Rev. Lett. {\bf
  100},  021803 (2008)}\relax
\mciteBstWouldAddEndPuncttrue
\mciteSetBstMidEndSepPunct{\mcitedefaultmidpunct}
{\mcitedefaultendpunct}{\mcitedefaultseppunct}\relax
\EndOfBibitem
\bibitem{Heister:2002gk}
{ALEPH} collaboration, A.~Heister {\em et al.},
  \href{http://dx.doi.org/10.1140/epjc/s2003-01230-5}{Eur. Phys. J. {\bf C29},
  143 (2003)}\relax
\mciteBstWouldAddEndPuncttrue
\mciteSetBstMidEndSepPunct{\mcitedefaultmidpunct}
{\mcitedefaultendpunct}{\mcitedefaultseppunct}\relax
\EndOfBibitem
\bibitem{Abdallah:2003qga}
{DELPHI} collaboration, J.~Abdallah {\em et al.},
  \href{http://dx.doi.org/10.1140/epjc/s2004-01827-0}{Eur. Phys. J. {\bf C35},
  35 (2004)}, \href{http://arxiv.org/abs/hep-ex/0404013}{{\tt
  arXiv:hep-ex/0404013 [hep-ex]}}\relax
\mciteBstWouldAddEndPuncttrue
\mciteSetBstMidEndSepPunct{\mcitedefaultmidpunct}
{\mcitedefaultendpunct}{\mcitedefaultseppunct}\relax
\EndOfBibitem
\bibitem{Abbiendi:1999gm}
{OPAL} collaboration, G.~Abbiendi {\em et al.},
  \href{http://dx.doi.org/10.1007/s100520050658}{Eur. Phys. J. {\bf C11},  587
  (1999)}, \href{http://arxiv.org/abs/hep-ex/9907061}{{\tt arXiv:hep-ex/9907061
  [hep-ex]}}\relax
\mciteBstWouldAddEndPuncttrue
\mciteSetBstMidEndSepPunct{\mcitedefaultmidpunct}
{\mcitedefaultendpunct}{\mcitedefaultseppunct}\relax
\EndOfBibitem
\bibitem{Abbiendi:2000bh}
{OPAL} collaboration, G.~Abbiendi {\em et al.},
  \href{http://dx.doi.org/10.1007/s100520100591}{Eur. Phys. J. {\bf C19},  241
  (2001)}, \href{http://arxiv.org/abs/hep-ex/0011052}{{\tt arXiv:hep-ex/0011052
  [hep-ex]}}\relax
\mciteBstWouldAddEndPuncttrue
\mciteSetBstMidEndSepPunct{\mcitedefaultmidpunct}
{\mcitedefaultendpunct}{\mcitedefaultseppunct}\relax
\EndOfBibitem
\bibitem{Abe:2002ua}
{SLD} collaboration, K.~Abe {\em et al.},
  \href{http://dx.doi.org/10.1103/PhysRevD.67.012006}{Phys. Rev. {\bf D67},
  012006 (2003)}, \href{http://arxiv.org/abs/hep-ex/0209002}{{\tt
  arXiv:hep-ex/0209002 [hep-ex]}}\relax
\mciteBstWouldAddEndPuncttrue
\mciteSetBstMidEndSepPunct{\mcitedefaultmidpunct}
{\mcitedefaultendpunct}{\mcitedefaultseppunct}\relax
\EndOfBibitem
\bibitem{Abe:2002wfa}
{SLD} collaboration, K.~Abe {\em et al.},
  \href{http://dx.doi.org/10.1103/PhysRevD.66.032009}{Phys. Rev. {\bf D66},
  032009 (2002)}, \href{http://arxiv.org/abs/hep-ex/0207048}{{\tt
  arXiv:hep-ex/0207048 [hep-ex]}}\relax
\mciteBstWouldAddEndPuncttrue
\mciteSetBstMidEndSepPunct{\mcitedefaultmidpunct}
{\mcitedefaultendpunct}{\mcitedefaultseppunct}\relax
\EndOfBibitem
\bibitem{Abe:1998qj}
{CDF} collaboration, F.~Abe {\em et al.},
  \href{http://dx.doi.org/10.1103/PhysRevLett.82.3576}{Phys. Rev. Lett. {\bf
  82},  3576 (1999)}\relax
\mciteBstWouldAddEndPuncttrue
\mciteSetBstMidEndSepPunct{\mcitedefaultmidpunct}
{\mcitedefaultendpunct}{\mcitedefaultseppunct}\relax
\EndOfBibitem
\bibitem{Abazov:2006dm}
{\dzero} collaboration, V.~M. Abazov {\em et al.},
  \href{http://dx.doi.org/10.1103/PhysRevLett.97.021802}{Phys. Rev. Lett. {\bf
  97},  021802 (2006)}, \href{http://arxiv.org/abs/hep-ex/0603029}{{\tt
  arXiv:hep-ex/0603029 [hep-ex]}}\relax
\mciteBstWouldAddEndPuncttrue
\mciteSetBstMidEndSepPunct{\mcitedefaultmidpunct}
{\mcitedefaultendpunct}{\mcitedefaultseppunct}\relax
\EndOfBibitem
\bibitem{Abulencia:2006ze}
{CDF} collaboration, A.~Abulencia {\em et al.},
  \href{http://dx.doi.org/10.1103/PhysRevLett.97.242003}{Phys. Rev. Lett. {\bf
  97},  242003 (2006)}, \href{http://arxiv.org/abs/hep-ex/0609040}{{\tt
  arXiv:hep-ex/0609040 [hep-ex]}}\relax
\mciteBstWouldAddEndPuncttrue
\mciteSetBstMidEndSepPunct{\mcitedefaultmidpunct}
{\mcitedefaultendpunct}{\mcitedefaultseppunct}\relax
\EndOfBibitem
\bibitem{Aaij:2013mpa}
{LHCb} collaboration, R.~Aaij {\em et al.},
  \href{http://dx.doi.org/10.1088/1367-2630/15/5/053021}{New J. Phys. {\bf 15},
   053021 (2013)}, \href{http://arxiv.org/abs/1304.4741}{{\tt arXiv:1304.4741
  [hep-ex]}}\relax
\mciteBstWouldAddEndPuncttrue
\mciteSetBstMidEndSepPunct{\mcitedefaultmidpunct}
{\mcitedefaultendpunct}{\mcitedefaultseppunct}\relax
\EndOfBibitem
\bibitem{Aoki:2019cca}
{FLAG} collaboration, S.~Aoki {\em et al.},
  \href{http://arxiv.org/abs/1902.08191}{{\tt arXiv:1902.08191
  [hep-lat]}}\relax
\mciteBstWouldAddEndPuncttrue
\mciteSetBstMidEndSepPunct{\mcitedefaultmidpunct}
{\mcitedefaultendpunct}{\mcitedefaultseppunct}\relax
\EndOfBibitem
\bibitem{Bazavov:2016nty}
{Fermilab Lattice and MILC} collaborations, A.~Bazavov {\em et al.},
  \href{http://dx.doi.org/10.1103/PhysRevD.93.113016}{Phys. Rev. {\bf D93},
  113016 (2016)}, \href{http://arxiv.org/abs/1602.03560}{{\tt arXiv:1602.03560
  [hep-lat]}}\relax
\mciteBstWouldAddEndPuncttrue
\mciteSetBstMidEndSepPunct{\mcitedefaultmidpunct}
{\mcitedefaultendpunct}{\mcitedefaultseppunct}\relax
\EndOfBibitem
\bibitem{Jaffe:2001hz}
{CLEO} collaboration, D.~E. Jaffe {\em et al.},
  \href{http://dx.doi.org/10.1103/PhysRevLett.86.5000}{Phys. Rev. Lett. {\bf
  86},  5000 (2001)}, \href{http://arxiv.org/abs/hep-ex/0101006}{{\tt
  arXiv:hep-ex/0101006 [hep-ex]}}\relax
\mciteBstWouldAddEndPuncttrue
\mciteSetBstMidEndSepPunct{\mcitedefaultmidpunct}
{\mcitedefaultendpunct}{\mcitedefaultseppunct}\relax
\EndOfBibitem
\bibitem{Lees:2014kep}
{\babar} collaboration, J.~P. Lees {\em et al.},
  \href{http://dx.doi.org/10.1103/PhysRevLett.114.081801}{Phys. Rev. Lett. {\bf
  114},  081801 (2015)}, \href{http://arxiv.org/abs/1411.1842}{{\tt
  arXiv:1411.1842 [hep-ex]}}\relax
\mciteBstWouldAddEndPuncttrue
\mciteSetBstMidEndSepPunct{\mcitedefaultmidpunct}
{\mcitedefaultendpunct}{\mcitedefaultseppunct}\relax
\EndOfBibitem
\bibitem{Abe:1996zt}
{CDF} collaboration, F.~Abe {\em et al.},
  \href{http://dx.doi.org/10.1103/PhysRevD.55.2546}{Phys. Rev. {\bf D55},  2546
  (1997)}\relax
\mciteBstWouldAddEndPuncttrue
\mciteSetBstMidEndSepPunct{\mcitedefaultmidpunct}
{\mcitedefaultendpunct}{\mcitedefaultseppunct}\relax
\EndOfBibitem
\bibitem{Barate:2000uk}
{ALEPH} collaboration, R.~Barate {\em et al.},
  \href{http://dx.doi.org/10.1007/s100520100644}{Eur. Phys. J. {\bf C20},  431
  (2001)}\relax
\mciteBstWouldAddEndPuncttrue
\mciteSetBstMidEndSepPunct{\mcitedefaultmidpunct}
{\mcitedefaultendpunct}{\mcitedefaultseppunct}\relax
\EndOfBibitem
\bibitem{Lees:2013sua}
{\babar} collaboration, J.~P. Lees {\em et al.},
  \href{http://dx.doi.org/10.1103/PhysRevLett.111.101802}{Phys. Rev. Lett. {\bf
  111},  101802 (2013)}, \href{http://arxiv.org/abs/1305.1575}{{\tt
  arXiv:1305.1575 [hep-ex]}}, Erratum ibid.\
  \href{http://dx.doi.org/10.1103/PhysRevLett.111.159901}{{\bf 111}, 159901},
  (2013)\relax
\mciteBstWouldAddEndPuncttrue
\mciteSetBstMidEndSepPunct{\mcitedefaultmidpunct}
{\mcitedefaultendpunct}{\mcitedefaultseppunct}\relax
\EndOfBibitem
\bibitem{Aubert:2006nf}
{\babar} collaboration, B.~Aubert {\em et al.},
  \href{http://dx.doi.org/10.1103/PhysRevLett.96.251802}{Phys. Rev. Lett. {\bf
  96},  251802 (2006)}, \href{http://arxiv.org/abs/hep-ex/0603053}{{\tt
  arXiv:hep-ex/0603053 [hep-ex]}}\relax
\mciteBstWouldAddEndPuncttrue
\mciteSetBstMidEndSepPunct{\mcitedefaultmidpunct}
{\mcitedefaultendpunct}{\mcitedefaultseppunct}\relax
\EndOfBibitem
\bibitem{Nakano:2005jb}
{Belle} collaboration, E.~Nakano {\em et al.},
  \href{http://dx.doi.org/10.1103/PhysRevD.73.112002}{Phys. Rev. {\bf D73},
  112002 (2006)}, \href{http://arxiv.org/abs/hep-ex/0505017}{{\tt
  arXiv:hep-ex/0505017 [hep-ex]}}\relax
\mciteBstWouldAddEndPuncttrue
\mciteSetBstMidEndSepPunct{\mcitedefaultmidpunct}
{\mcitedefaultendpunct}{\mcitedefaultseppunct}\relax
\EndOfBibitem
\bibitem{Beneke:1996hv}
M.~Beneke, G.~Buchalla, and I.~Dunietz,
  \href{http://dx.doi.org/10.1016/S0370-2693(96)01648-6}{Phys. Lett. {\bf
  B393},  132 (1997)}, \href{http://arxiv.org/abs/hep-ph/9609357}{{\tt
  arXiv:hep-ph/9609357 [hep-ph]}}\relax
\mciteBstWouldAddEndPuncttrue
\mciteSetBstMidEndSepPunct{\mcitedefaultmidpunct}
{\mcitedefaultendpunct}{\mcitedefaultseppunct}\relax
\EndOfBibitem
\bibitem{Dunietz:1998av}
I.~Dunietz, \href{http://dx.doi.org/10.1007/s100529801005}{Eur. Phys. J. {\bf
  C7},  197 (1999)}, \href{http://arxiv.org/abs/hep-ph/9806521}{{\tt
  arXiv:hep-ph/9806521 [hep-ph]}}\relax
\mciteBstWouldAddEndPuncttrue
\mciteSetBstMidEndSepPunct{\mcitedefaultmidpunct}
{\mcitedefaultendpunct}{\mcitedefaultseppunct}\relax
\EndOfBibitem
\bibitem{Abazov:2012hha}
{\dzero} collaboration, V.~M. Abazov {\em et al.},
  \href{http://dx.doi.org/10.1103/PhysRevD.86.072009}{Phys. Rev. {\bf D86},
  072009 (2012)}, \href{http://arxiv.org/abs/1208.5813}{{\tt arXiv:1208.5813
  [hep-ex]}}\relax
\mciteBstWouldAddEndPuncttrue
\mciteSetBstMidEndSepPunct{\mcitedefaultmidpunct}
{\mcitedefaultendpunct}{\mcitedefaultseppunct}\relax
\EndOfBibitem
\bibitem{Aaij:2014nxa}
{LHCb} collaboration, R.~Aaij {\em et al.},
  \href{http://dx.doi.org/10.1103/PhysRevLett.114.041601}{Phys. Rev. Lett. {\bf
  114},  041601 (2015)}, \href{http://arxiv.org/abs/1409.8586}{{\tt
  arXiv:1409.8586 [hep-ex]}}\relax
\mciteBstWouldAddEndPuncttrue
\mciteSetBstMidEndSepPunct{\mcitedefaultmidpunct}
{\mcitedefaultendpunct}{\mcitedefaultseppunct}\relax
\EndOfBibitem
\bibitem{Amhis:2012bh}
{Heavy Flavor Averaging Group}, Y.~Amhis {\em et al.},
  \href{http://arxiv.org/abs/1207.1158}{{\tt arXiv:1207.1158 [hep-ex]}}
  (2012)\relax
\mciteBstWouldAddEndPuncttrue
\mciteSetBstMidEndSepPunct{\mcitedefaultmidpunct}
{\mcitedefaultendpunct}{\mcitedefaultseppunct}\relax
\EndOfBibitem
\bibitem{Abazov:2012zz}
{\dzero} collaboration, V.~M. Abazov {\em et al.},
  \href{http://dx.doi.org/10.1103/PhysRevLett.110.011801}{Phys. Rev. Lett. {\bf
  110},  011801 (2013)}, \href{http://arxiv.org/abs/1207.1769}{{\tt
  arXiv:1207.1769 [hep-ex]}}\relax
\mciteBstWouldAddEndPuncttrue
\mciteSetBstMidEndSepPunct{\mcitedefaultmidpunct}
{\mcitedefaultendpunct}{\mcitedefaultseppunct}\relax
\EndOfBibitem
\bibitem{Aaij:2016yze}
{LHCb} collaboration, R.~Aaij {\em et al.},
  \href{http://dx.doi.org/10.1103/PhysRevLett.117.061803}{Phys. Rev. Lett. {\bf
  117},  061803 (2016)}, \href{http://arxiv.org/abs/1605.09768}{{\tt
  arXiv:1605.09768 [hep-ex]}}\relax
\mciteBstWouldAddEndPuncttrue
\mciteSetBstMidEndSepPunct{\mcitedefaultmidpunct}
{\mcitedefaultendpunct}{\mcitedefaultseppunct}\relax
\EndOfBibitem
\bibitem{Lenz_private_communication}
 A.~Lenz, private communication, 2017\relax
\mciteBstWouldAddEndPuncttrue
\mciteSetBstMidEndSepPunct{\mcitedefaultmidpunct}
{\mcitedefaultendpunct}{\mcitedefaultseppunct}\relax
\EndOfBibitem
\bibitem{DescotesGenon:2012kr}
S.~Descotes-Genon and J.~F. Kamenik,
  \href{http://dx.doi.org/10.1103/PhysRevD.87.074036}{Phys. Rev. {\bf D87},
  074036 (2013)}, \href{http://arxiv.org/abs/1207.4483}{{\tt arXiv:1207.4483
  [hep-ph]}}, Erratum ibid.\
  \href{http://dx.doi.org/10.1103/PhysRevD.92.079903}{{\bf D92}, 079903},
  (2015)\relax
\mciteBstWouldAddEndPuncttrue
\mciteSetBstMidEndSepPunct{\mcitedefaultmidpunct}
{\mcitedefaultendpunct}{\mcitedefaultseppunct}\relax
\EndOfBibitem
\bibitem{Aaboud:2016bmk}
{ATLAS} collaboration, M.~Aaboud {\em et al.},
  \href{http://dx.doi.org/10.1007/JHEP02(2017)071}{JHEP {\bf 02},  071 (2017)},
  \href{http://arxiv.org/abs/1610.07869}{{\tt arXiv:1610.07869 [hep-ex]}}\relax
\mciteBstWouldAddEndPuncttrue
\mciteSetBstMidEndSepPunct{\mcitedefaultmidpunct}
{\mcitedefaultendpunct}{\mcitedefaultseppunct}\relax
\EndOfBibitem
\bibitem{Aaij:2014dka}
{LHCb} collaboration, R.~Aaij {\em et al.},
  \href{http://dx.doi.org/10.1016/j.physletb.2014.06.079}{Phys. Lett. {\bf
  B736},  186 (2014)}, \href{http://arxiv.org/abs/1405.4140}{{\tt
  arXiv:1405.4140 [hep-ex]}}\relax
\mciteBstWouldAddEndPuncttrue
\mciteSetBstMidEndSepPunct{\mcitedefaultmidpunct}
{\mcitedefaultendpunct}{\mcitedefaultseppunct}\relax
\EndOfBibitem
\bibitem{LHCb:2012ae}
{LHCb} collaboration, R.~Aaij {\em et al.},
  \href{http://dx.doi.org/10.1103/PhysRevD.86.052006}{Phys. Rev. {\bf D86},
  052006 (2012)}, \href{http://arxiv.org/abs/1204.5643}{{\tt arXiv:1204.5643
  [hep-ex]}}\relax
\mciteBstWouldAddEndPuncttrue
\mciteSetBstMidEndSepPunct{\mcitedefaultmidpunct}
{\mcitedefaultendpunct}{\mcitedefaultseppunct}\relax
\EndOfBibitem
\bibitem{Aaij:2014ywt}
{LHCb} collaboration, R.~Aaij {\em et al.},
  \href{http://dx.doi.org/10.1103/PhysRevLett.113.211801}{Phys. Rev. Lett. {\bf
  113},  211801 (2014)}, \href{http://arxiv.org/abs/1409.4619}{{\tt
  arXiv:1409.4619 [hep-ex]}}\relax
\mciteBstWouldAddEndPuncttrue
\mciteSetBstMidEndSepPunct{\mcitedefaultmidpunct}
{\mcitedefaultendpunct}{\mcitedefaultseppunct}\relax
\EndOfBibitem
\bibitem{Frings:2015eva}
P.~Frings, U.~Nierste, and M.~Wiebusch,
  \href{http://dx.doi.org/10.1103/PhysRevLett.115.061802}{Phys. Rev. Lett. {\bf
  115},  061802 (2015)}, \href{http://arxiv.org/abs/1503.00859}{{\tt
  arXiv:1503.00859 [hep-ph]}}\relax
\mciteBstWouldAddEndPuncttrue
\mciteSetBstMidEndSepPunct{\mcitedefaultmidpunct}
{\mcitedefaultendpunct}{\mcitedefaultseppunct}\relax
\EndOfBibitem
\bibitem{Aaij:2014xba}
{LHCb} collaboration, R.~Aaij {\em et al.},
  \href{http://dx.doi.org/10.1016/j.physletb.2014.12.015}{Phys. Lett. {\bf
  B741},  1 (2015)}, \href{http://arxiv.org/abs/1408.4368}{{\tt arXiv:1408.4368
  [hep-ex]}}\relax
\mciteBstWouldAddEndPuncttrue
\mciteSetBstMidEndSepPunct{\mcitedefaultmidpunct}
{\mcitedefaultendpunct}{\mcitedefaultseppunct}\relax
\EndOfBibitem
\bibitem{Chau:1984fp}
L.-L. Chau and W.-Y. Keung,
  \href{http://dx.doi.org/10.1103/PhysRevLett.53.1802}{Phys. Rev. Lett. {\bf
  53},  1802 (1984)}\relax
\mciteBstWouldAddEndPuncttrue
\mciteSetBstMidEndSepPunct{\mcitedefaultmidpunct}
{\mcitedefaultendpunct}{\mcitedefaultseppunct}\relax
\EndOfBibitem
\bibitem{Wolfenstein:1983yz}
L.~Wolfenstein, \href{http://dx.doi.org/10.1103/PhysRevLett.51.1945}{Phys. Rev.
  Lett. {\bf 51},  1945 (1983)}\relax
\mciteBstWouldAddEndPuncttrue
\mciteSetBstMidEndSepPunct{\mcitedefaultmidpunct}
{\mcitedefaultendpunct}{\mcitedefaultseppunct}\relax
\EndOfBibitem
\bibitem{Buras:1994ec}
A.~J. Buras, M.~E. Lautenbacher, and G.~Ostermaier,
  \href{http://dx.doi.org/10.1103/PhysRevD.50.3433}{Phys. Rev. {\bf D50},  3433
  (1994)}, \href{http://arxiv.org/abs/hep-ph/9403384}{{\tt
  arXiv:hep-ph/9403384}}\relax
\mciteBstWouldAddEndPuncttrue
\mciteSetBstMidEndSepPunct{\mcitedefaultmidpunct}
{\mcitedefaultendpunct}{\mcitedefaultseppunct}\relax
\EndOfBibitem
\bibitem{Jarlskog:1985ht}
C.~Jarlskog, \href{http://dx.doi.org/10.1103/PhysRevLett.55.1039}{Phys. Rev.
  Lett. {\bf 55},  1039 (1985)}\relax
\mciteBstWouldAddEndPuncttrue
\mciteSetBstMidEndSepPunct{\mcitedefaultmidpunct}
{\mcitedefaultendpunct}{\mcitedefaultseppunct}\relax
\EndOfBibitem
\bibitem{Jarlskog:2005uq}
C.~Jarlskog, \href{http://dx.doi.org/10.1016/j.physletb.2005.04.033}{Phys.
  Lett. {\bf B615},  207 (2005)},
  \href{http://arxiv.org/abs/hep-ph/0503199}{{\tt arXiv:hep-ph/0503199}}\relax
\mciteBstWouldAddEndPuncttrue
\mciteSetBstMidEndSepPunct{\mcitedefaultmidpunct}
{\mcitedefaultendpunct}{\mcitedefaultseppunct}\relax
\EndOfBibitem
\bibitem{Harrison:2009bz}
P.~F. Harrison, S.~Dallison, and W.~G. Scott,
  \href{http://dx.doi.org/10.1016/j.physletb.2009.09.004}{Phys. Lett. {\bf
  B680},  328 (2009)}, \href{http://arxiv.org/abs/0904.3077}{{\tt
  arXiv:0904.3077 [hep-ph]}}\relax
\mciteBstWouldAddEndPuncttrue
\mciteSetBstMidEndSepPunct{\mcitedefaultmidpunct}
{\mcitedefaultendpunct}{\mcitedefaultseppunct}\relax
\EndOfBibitem
\bibitem{Frampton:2010ii}
P.~H. Frampton and X.-G. He,
  \href{http://dx.doi.org/10.1016/j.physletb.2010.03.077}{Phys. Lett. {\bf
  B688},  67 (2010)}, \href{http://arxiv.org/abs/1003.0310}{{\tt
  arXiv:1003.0310 [hep-ph]}}\relax
\mciteBstWouldAddEndPuncttrue
\mciteSetBstMidEndSepPunct{\mcitedefaultmidpunct}
{\mcitedefaultendpunct}{\mcitedefaultseppunct}\relax
\EndOfBibitem
\bibitem{Frampton:2010uq}
P.~H. Frampton and X.-G. He,
  \href{http://dx.doi.org/10.1103/PhysRevD.82.017301}{Phys. Rev. {\bf D82},
  017301 (2010)}, \href{http://arxiv.org/abs/1004.3679}{{\tt arXiv:1004.3679
  [hep-ph]}}\relax
\mciteBstWouldAddEndPuncttrue
\mciteSetBstMidEndSepPunct{\mcitedefaultmidpunct}
{\mcitedefaultendpunct}{\mcitedefaultseppunct}\relax
\EndOfBibitem
\bibitem{Charles:2004jd}
{CKMfitter group}, J.~Charles {\em et al.},
  \href{http://dx.doi.org/10.1140/epjc/s2005-02169-1}{Eur. Phys. J. {\bf C41},
  1 (2005)}, \href{http://arxiv.org/abs/hep-ph/0406184}{{\tt
  arXiv:hep-ph/0406184}}, see also online updates,
  \url{http://ckmfitter.in2p3.fr/}\relax
\mciteBstWouldAddEndPuncttrue
\mciteSetBstMidEndSepPunct{\mcitedefaultmidpunct}
{\mcitedefaultendpunct}{\mcitedefaultseppunct}\relax
\EndOfBibitem
\bibitem{Aubert:2001sp}
{\babar} collaboration, B.~Aubert {\em et al.},
  \href{http://dx.doi.org/10.1103/PhysRevLett.86.2515}{Phys. Rev. Lett. {\bf
  86},  2515 (2001)}, \href{http://arxiv.org/abs/hep-ex/0102030}{{\tt
  arXiv:hep-ex/0102030}}\relax
\mciteBstWouldAddEndPuncttrue
\mciteSetBstMidEndSepPunct{\mcitedefaultmidpunct}
{\mcitedefaultendpunct}{\mcitedefaultseppunct}\relax
\EndOfBibitem
\bibitem{Aaij:2012ke}
{LHCb} collaboration, R.~Aaij {\em et al.},
  \href{http://dx.doi.org/10.1016/j.physletb.2013.02.054}{Phys. Lett. {\bf
  B721},  24 (2013)}, \href{http://arxiv.org/abs/1211.6093}{{\tt
  arXiv:1211.6093 [hep-ex]}}\relax
\mciteBstWouldAddEndPuncttrue
\mciteSetBstMidEndSepPunct{\mcitedefaultmidpunct}
{\mcitedefaultendpunct}{\mcitedefaultseppunct}\relax
\EndOfBibitem
\bibitem{Abe:2001xe}
{Belle} collaboration, K.~Abe {\em et al.},
  \href{http://dx.doi.org/10.1103/PhysRevLett.87.091802}{Phys. Rev. Lett. {\bf
  87},  091802 (2001)}, \href{http://arxiv.org/abs/hep-ex/0107061}{{\tt
  arXiv:hep-ex/0107061}}\relax
\mciteBstWouldAddEndPuncttrue
\mciteSetBstMidEndSepPunct{\mcitedefaultmidpunct}
{\mcitedefaultendpunct}{\mcitedefaultseppunct}\relax
\EndOfBibitem
\bibitem{Carter:1980tk}
A.~B. Carter and A.~I. Sanda,
  \href{http://dx.doi.org/10.1103/PhysRevD.23.1567}{Phys. Rev. {\bf D23},  1567
  (1981)}\relax
\mciteBstWouldAddEndPuncttrue
\mciteSetBstMidEndSepPunct{\mcitedefaultmidpunct}
{\mcitedefaultendpunct}{\mcitedefaultseppunct}\relax
\EndOfBibitem
\bibitem{Bigi:1981qs}
I.~I.~Y. Bigi and A.~I. Sanda,
  \href{http://dx.doi.org/10.1016/0550-3213(81)90519-8}{Nucl. Phys. {\bf B193},
   85 (1981)}\relax
\mciteBstWouldAddEndPuncttrue
\mciteSetBstMidEndSepPunct{\mcitedefaultmidpunct}
{\mcitedefaultendpunct}{\mcitedefaultseppunct}\relax
\EndOfBibitem
\bibitem{Aaij:2014vda}
{LHCb} collaboration, R.~Aaij {\em et al.},
  \href{http://dx.doi.org/10.1016/j.physletb.2015.01.008}{Phys. Lett. {\bf
  B742},  38 (2015)}, \href{http://arxiv.org/abs/1411.1634}{{\tt
  arXiv:1411.1634 [hep-ex]}}\relax
\mciteBstWouldAddEndPuncttrue
\mciteSetBstMidEndSepPunct{\mcitedefaultmidpunct}
{\mcitedefaultendpunct}{\mcitedefaultseppunct}\relax
\EndOfBibitem
\bibitem{Aubert:2008ah}
{\babar} collaboration, B.~Aubert {\em et al.},
  \href{http://dx.doi.org/10.1103/PhysRevD.79.032002}{Phys. Rev. {\bf D79},
  032002 (2009)}, \href{http://arxiv.org/abs/0808.1866}{{\tt arXiv:0808.1866
  [hep-ex]}}\relax
\mciteBstWouldAddEndPuncttrue
\mciteSetBstMidEndSepPunct{\mcitedefaultmidpunct}
{\mcitedefaultendpunct}{\mcitedefaultseppunct}\relax
\EndOfBibitem
\bibitem{Krokovny:2006sv}
{Belle} collaboration, P.~Krokovny {\em et al.},
  \href{http://dx.doi.org/10.1103/PhysRevLett.97.081801}{Phys. Rev. Lett. {\bf
  97},  081801 (2006)}, \href{http://arxiv.org/abs/hep-ex/0605023}{{\tt
  arXiv:hep-ex/0605023}}\relax
\mciteBstWouldAddEndPuncttrue
\mciteSetBstMidEndSepPunct{\mcitedefaultmidpunct}
{\mcitedefaultendpunct}{\mcitedefaultseppunct}\relax
\EndOfBibitem
\bibitem{Aubert:2007rp}
{\babar} collaboration, B.~Aubert {\em et al.},
  \href{http://dx.doi.org/10.1103/PhysRevLett.99.231802}{Phys. Rev. Lett. {\bf
  99},  231802 (2007)}, \href{http://arxiv.org/abs/0708.1544}{{\tt
  arXiv:0708.1544 [hep-ex]}}\relax
\mciteBstWouldAddEndPuncttrue
\mciteSetBstMidEndSepPunct{\mcitedefaultmidpunct}
{\mcitedefaultendpunct}{\mcitedefaultseppunct}\relax
\EndOfBibitem
\bibitem{Adachi:2018itz}
{\babar\ and Belle} collaborations, I.~Adachi {\em et al.},
  \href{http://dx.doi.org/10.1103/PhysRevLett.121.261801}{Phys. Rev. Lett. {\bf
  121},  261801 (2018)}, \href{http://arxiv.org/abs/1804.06152}{{\tt
  arXiv:1804.06152 [hep-ex]}}\relax
\mciteBstWouldAddEndPuncttrue
\mciteSetBstMidEndSepPunct{\mcitedefaultmidpunct}
{\mcitedefaultendpunct}{\mcitedefaultseppunct}\relax
\EndOfBibitem
\bibitem{Adachi:2018jqe}
{\babar\ and Belle} collaborations, I.~Adachi {\em et al.},
  \href{http://dx.doi.org/10.1103/PhysRevD.98.112012}{Phys. Rev. {\bf D98},
  112012 (2018)}, \href{http://arxiv.org/abs/1804.06153}{{\tt arXiv:1804.06153
  [hep-ex]}}\relax
\mciteBstWouldAddEndPuncttrue
\mciteSetBstMidEndSepPunct{\mcitedefaultmidpunct}
{\mcitedefaultendpunct}{\mcitedefaultseppunct}\relax
\EndOfBibitem
\bibitem{Vorobyev:2016npn}
{Belle} collaboration, V.~Vorobyev {\em et al.},
  \href{http://dx.doi.org/10.1103/PhysRevD.94.052004}{Phys. Rev. {\bf D94},
  052004 (2016)}, \href{http://arxiv.org/abs/1607.05813}{{\tt arXiv:1607.05813
  [hep-ex]}}\relax
\mciteBstWouldAddEndPuncttrue
\mciteSetBstMidEndSepPunct{\mcitedefaultmidpunct}
{\mcitedefaultendpunct}{\mcitedefaultseppunct}\relax
\EndOfBibitem
\bibitem{Libby:2010nu}
{CLEO} collaboration, J.~Libby {\em et al.},
  \href{http://dx.doi.org/10.1103/PhysRevD.82.112006}{Phys. Rev. {\bf D82},
  112006 (2010)}, \href{http://arxiv.org/abs/1010.2817}{{\tt arXiv:1010.2817
  [hep-ex]}}\relax
\mciteBstWouldAddEndPuncttrue
\mciteSetBstMidEndSepPunct{\mcitedefaultmidpunct}
{\mcitedefaultendpunct}{\mcitedefaultseppunct}\relax
\EndOfBibitem
\bibitem{Browder:1999ng}
T.~E. Browder, A.~Datta, P.~J. O'Donnell, and S.~Pakvasa,
  \href{http://dx.doi.org/10.1103/PhysRevD.61.054009}{Phys. Rev. {\bf D61},
  054009 (2000)}, \href{http://arxiv.org/abs/hep-ph/9905425}{{\tt
  arXiv:hep-ph/9905425}}\relax
\mciteBstWouldAddEndPuncttrue
\mciteSetBstMidEndSepPunct{\mcitedefaultmidpunct}
{\mcitedefaultendpunct}{\mcitedefaultseppunct}\relax
\EndOfBibitem
\bibitem{Aubert:2006fh}
{\babar} collaboration, B.~Aubert {\em et al.},
  \href{http://dx.doi.org/10.1103/PhysRevD.74.091101}{Phys. Rev. {\bf D74},
  091101 (2006)}, \href{http://arxiv.org/abs/hep-ex/0608016}{{\tt
  arXiv:hep-ex/0608016 [hep-ex]}}\relax
\mciteBstWouldAddEndPuncttrue
\mciteSetBstMidEndSepPunct{\mcitedefaultmidpunct}
{\mcitedefaultendpunct}{\mcitedefaultseppunct}\relax
\EndOfBibitem
\bibitem{Dalseno:2007hx}
{Belle} collaboration, J.~Dalseno {\em et al.},
  \href{http://dx.doi.org/10.1103/PhysRevD.76.072004}{Phys. Rev. {\bf D76},
  072004 (2007)}, \href{http://arxiv.org/abs/0706.2045}{{\tt arXiv:0706.2045
  [hep-ex]}}\relax
\mciteBstWouldAddEndPuncttrue
\mciteSetBstMidEndSepPunct{\mcitedefaultmidpunct}
{\mcitedefaultendpunct}{\mcitedefaultseppunct}\relax
\EndOfBibitem
\bibitem{Aaij:2014siy}
{LHCb} collaboration, R.~Aaij {\em et al.},
  \href{http://dx.doi.org/10.1103/PhysRevD.90.012003}{Phys. Rev. {\bf D90},
  012003 (2014)}, \href{http://arxiv.org/abs/1404.5673}{{\tt arXiv:1404.5673
  [hep-ex]}}\relax
\mciteBstWouldAddEndPuncttrue
\mciteSetBstMidEndSepPunct{\mcitedefaultmidpunct}
{\mcitedefaultendpunct}{\mcitedefaultseppunct}\relax
\EndOfBibitem
\bibitem{Pelaez:2015qba}
J.~R. Pelaez, \href{http://dx.doi.org/10.1016/j.physrep.2016.09.001}{Phys.
  Rept. {\bf 658},  1 (2016)}, \href{http://arxiv.org/abs/1510.00653}{{\tt
  arXiv:1510.00653 [hep-ph]}}\relax
\mciteBstWouldAddEndPuncttrue
\mciteSetBstMidEndSepPunct{\mcitedefaultmidpunct}
{\mcitedefaultendpunct}{\mcitedefaultseppunct}\relax
\EndOfBibitem
\bibitem{Aubert:2007sd}
{\babar} collaboration, B.~Aubert {\em et al.},
  \href{http://dx.doi.org/10.1103/PhysRevLett.99.161802}{Phys. Rev. Lett. {\bf
  99},  161802 (2007)}, \href{http://arxiv.org/abs/0706.3885}{{\tt
  arXiv:0706.3885 [hep-ex]}}\relax
\mciteBstWouldAddEndPuncttrue
\mciteSetBstMidEndSepPunct{\mcitedefaultmidpunct}
{\mcitedefaultendpunct}{\mcitedefaultseppunct}\relax
\EndOfBibitem
\bibitem{Nakahama:2010nj}
{Belle} collaboration, Y.~Nakahama {\em et al.},
  \href{http://dx.doi.org/10.1103/PhysRevD.82.073011}{Phys. Rev. {\bf D82},
  073011 (2010)}, \href{http://arxiv.org/abs/1007.3848}{{\tt arXiv:1007.3848
  [hep-ex]}}\relax
\mciteBstWouldAddEndPuncttrue
\mciteSetBstMidEndSepPunct{\mcitedefaultmidpunct}
{\mcitedefaultendpunct}{\mcitedefaultseppunct}\relax
\EndOfBibitem
\bibitem{Lees:2012kxa}
{\babar} collaboration, J.~P. Lees {\em et al.},
  \href{http://dx.doi.org/10.1103/PhysRevD.85.112010}{Phys. Rev. {\bf D85},
  112010 (2012)}, \href{http://arxiv.org/abs/1201.5897}{{\tt arXiv:1201.5897
  [hep-ex]}}\relax
\mciteBstWouldAddEndPuncttrue
\mciteSetBstMidEndSepPunct{\mcitedefaultmidpunct}
{\mcitedefaultendpunct}{\mcitedefaultseppunct}\relax
\EndOfBibitem
\bibitem{Garmash:2004wa}
{Belle} collaboration, A.~Garmash {\em et al.},
  \href{http://dx.doi.org/10.1103/PhysRevD.71.092003}{Phys. Rev. {\bf D71},
  092003 (2005)}, \href{http://arxiv.org/abs/hep-ex/0412066}{{\tt
  arXiv:hep-ex/0412066 [hep-ex]}}\relax
\mciteBstWouldAddEndPuncttrue
\mciteSetBstMidEndSepPunct{\mcitedefaultmidpunct}
{\mcitedefaultendpunct}{\mcitedefaultseppunct}\relax
\EndOfBibitem
\bibitem{Aubert:2006nu}
{\babar} collaboration, B.~Aubert {\em et al.},
  \href{http://dx.doi.org/10.1103/PhysRevD.74.032003}{Phys. Rev. {\bf D74},
  032003 (2006)}, \href{http://arxiv.org/abs/hep-ex/0605003}{{\tt
  arXiv:hep-ex/0605003 [hep-ex]}}\relax
\mciteBstWouldAddEndPuncttrue
\mciteSetBstMidEndSepPunct{\mcitedefaultmidpunct}
{\mcitedefaultendpunct}{\mcitedefaultseppunct}\relax
\EndOfBibitem
\bibitem{Aubert:2009me}
{\babar} collaboration, B.~Aubert {\em et al.},
  \href{http://dx.doi.org/10.1103/PhysRevD.80.112001}{Phys. Rev. {\bf D80},
  112001 (2009)}, \href{http://arxiv.org/abs/0905.3615}{{\tt arXiv:0905.3615
  [hep-ex]}}\relax
\mciteBstWouldAddEndPuncttrue
\mciteSetBstMidEndSepPunct{\mcitedefaultmidpunct}
{\mcitedefaultendpunct}{\mcitedefaultseppunct}\relax
\EndOfBibitem
\bibitem{Dalseno:2008wwa}
{Belle} collaboration, J.~Dalseno {\em et al.},
  \href{http://dx.doi.org/10.1103/PhysRevD.79.072004}{Phys. Rev. {\bf D79},
  072004 (2009)}, \href{http://arxiv.org/abs/0811.3665}{{\tt arXiv:0811.3665
  [hep-ex]}}\relax
\mciteBstWouldAddEndPuncttrue
\mciteSetBstMidEndSepPunct{\mcitedefaultmidpunct}
{\mcitedefaultendpunct}{\mcitedefaultseppunct}\relax
\EndOfBibitem
\bibitem{Garmash:2005rv}
{Belle} collaboration, A.~Garmash {\em et al.},
  \href{http://dx.doi.org/10.1103/PhysRevLett.96.251803}{Phys. Rev. Lett. {\bf
  96},  251803 (2006)}, \href{http://arxiv.org/abs/hep-ex/0512066}{{\tt
  arXiv:hep-ex/0512066 [hep-ex]}}\relax
\mciteBstWouldAddEndPuncttrue
\mciteSetBstMidEndSepPunct{\mcitedefaultmidpunct}
{\mcitedefaultendpunct}{\mcitedefaultseppunct}\relax
\EndOfBibitem
\bibitem{Aubert:2005ce}
{\babar} collaboration, B.~Aubert {\em et al.},
  \href{http://dx.doi.org/10.1103/PhysRevD.72.072003}{Phys. Rev. {\bf D72},
  072003 (2005)}, \href{http://arxiv.org/abs/hep-ex/0507004}{{\tt
  arXiv:hep-ex/0507004 [hep-ex]}}, Erratum ibid.\
  \href{http://dx.doi.org/10.1103/PhysRevD.74.099903}{{\bf D74}, 099903},
  (2006)\relax
\mciteBstWouldAddEndPuncttrue
\mciteSetBstMidEndSepPunct{\mcitedefaultmidpunct}
{\mcitedefaultendpunct}{\mcitedefaultseppunct}\relax
\EndOfBibitem
\bibitem{Aubert:2008bj}
{\babar} collaboration, B.~Aubert {\em et al.},
  \href{http://dx.doi.org/10.1103/PhysRevD.78.012004}{Phys. Rev. {\bf D78},
  012004 (2008)}, \href{http://arxiv.org/abs/0803.4451}{{\tt arXiv:0803.4451
  [hep-ex]}}\relax
\mciteBstWouldAddEndPuncttrue
\mciteSetBstMidEndSepPunct{\mcitedefaultmidpunct}
{\mcitedefaultendpunct}{\mcitedefaultseppunct}\relax
\EndOfBibitem
\bibitem{Snyder:1993mx}
A.~E. Snyder and H.~R. Quinn,
  \href{http://dx.doi.org/10.1103/PhysRevD.48.2139}{Phys. Rev. {\bf D48},  2139
  (1993)}\relax
\mciteBstWouldAddEndPuncttrue
\mciteSetBstMidEndSepPunct{\mcitedefaultmidpunct}
{\mcitedefaultendpunct}{\mcitedefaultseppunct}\relax
\EndOfBibitem
\bibitem{Quinn:2000by}
H.~R. Quinn and J.~P. Silva,
  \href{http://dx.doi.org/10.1103/PhysRevD.62.054002}{Phys. Rev. {\bf D62},
  054002 (2000)}, \href{http://arxiv.org/abs/hep-ph/0001290}{{\tt
  arXiv:hep-ph/0001290}}\relax
\mciteBstWouldAddEndPuncttrue
\mciteSetBstMidEndSepPunct{\mcitedefaultmidpunct}
{\mcitedefaultendpunct}{\mcitedefaultseppunct}\relax
\EndOfBibitem
\bibitem{Aubert:2007jn}
{\babar} collaboration, B.~Aubert {\em et al.},
  \href{http://dx.doi.org/10.1103/PhysRevD.76.012004}{Phys. Rev. {\bf D76},
  012004 (2007)}, \href{http://arxiv.org/abs/hep-ex/0703008}{{\tt
  arXiv:hep-ex/0703008}}\relax
\mciteBstWouldAddEndPuncttrue
\mciteSetBstMidEndSepPunct{\mcitedefaultmidpunct}
{\mcitedefaultendpunct}{\mcitedefaultseppunct}\relax
\EndOfBibitem
\bibitem{Lees:2013nwa}
{\babar} collaboration, J.~P. Lees {\em et al.},
  \href{http://dx.doi.org/10.1103/PhysRevD.88.012003}{Phys. Rev. {\bf D88},
  012003 (2013)}, \href{http://arxiv.org/abs/1304.3503}{{\tt arXiv:1304.3503
  [hep-ex]}}\relax
\mciteBstWouldAddEndPuncttrue
\mciteSetBstMidEndSepPunct{\mcitedefaultmidpunct}
{\mcitedefaultendpunct}{\mcitedefaultseppunct}\relax
\EndOfBibitem
\bibitem{Kusaka:2007dv}
{Belle} collaboration, A.~Kusaka {\em et al.},
  \href{http://dx.doi.org/10.1103/PhysRevLett.98.221602}{Phys. Rev. Lett. {\bf
  98},  221602 (2007)}, \href{http://arxiv.org/abs/hep-ex/0701015}{{\tt
  arXiv:hep-ex/0701015}}\relax
\mciteBstWouldAddEndPuncttrue
\mciteSetBstMidEndSepPunct{\mcitedefaultmidpunct}
{\mcitedefaultendpunct}{\mcitedefaultseppunct}\relax
\EndOfBibitem
\bibitem{Kusaka:2007mj}
{Belle} collaboration, A.~Kusaka {\em et al.},
  \href{http://dx.doi.org/10.1103/PhysRevD.77.072001}{Phys. Rev. {\bf D77},
  072001 (2008)}, \href{http://arxiv.org/abs/0710.4974}{{\tt arXiv:0710.4974
  [hep-ex]}}\relax
\mciteBstWouldAddEndPuncttrue
\mciteSetBstMidEndSepPunct{\mcitedefaultmidpunct}
{\mcitedefaultendpunct}{\mcitedefaultseppunct}\relax
\EndOfBibitem
\bibitem{Aubert:2007pa}
{\babar} collaboration, B.~Aubert {\em et al.},
  \href{http://dx.doi.org/10.1103/PhysRevLett.99.071801}{Phys. Rev. Lett. {\bf
  99},  071801 (2007)}, \href{http://arxiv.org/abs/0705.1190}{{\tt
  arXiv:0705.1190 [hep-ex]}}\relax
\mciteBstWouldAddEndPuncttrue
\mciteSetBstMidEndSepPunct{\mcitedefaultmidpunct}
{\mcitedefaultendpunct}{\mcitedefaultseppunct}\relax
\EndOfBibitem
\bibitem{Aushev:2004uc}
{Belle} collaboration, T.~Aushev {\em et al.},
  \href{http://dx.doi.org/10.1103/PhysRevLett.93.201802}{Phys. Rev. Lett. {\bf
  93},  201802 (2004)}, \href{http://arxiv.org/abs/hep-ex/0408051}{{\tt
  arXiv:hep-ex/0408051}}\relax
\mciteBstWouldAddEndPuncttrue
\mciteSetBstMidEndSepPunct{\mcitedefaultmidpunct}
{\mcitedefaultendpunct}{\mcitedefaultseppunct}\relax
\EndOfBibitem
\bibitem{Rohrken:2012ta}
{Belle} collaboration, M.~Rohrken {\em et al.},
  \href{http://dx.doi.org/10.1103/PhysRevD.85.091106}{Phys. Rev. {\bf D85},
  091106 (2012)}, \href{http://arxiv.org/abs/1203.6647}{{\tt arXiv:1203.6647
  [hep-ex]}}\relax
\mciteBstWouldAddEndPuncttrue
\mciteSetBstMidEndSepPunct{\mcitedefaultmidpunct}
{\mcitedefaultendpunct}{\mcitedefaultseppunct}\relax
\EndOfBibitem
\bibitem{Aubert:2003wr}
{\babar} collaboration, B.~Aubert {\em et al.},
  \href{http://dx.doi.org/10.1103/PhysRevLett.91.201802}{Phys. Rev. Lett. {\bf
  91},  201802 (2003)}, \href{http://arxiv.org/abs/hep-ex/0306030}{{\tt
  arXiv:hep-ex/0306030 [hep-ex]}}\relax
\mciteBstWouldAddEndPuncttrue
\mciteSetBstMidEndSepPunct{\mcitedefaultmidpunct}
{\mcitedefaultendpunct}{\mcitedefaultseppunct}\relax
\EndOfBibitem
\bibitem{Wang:2004va}
{Belle} collaboration, C.~C. Wang {\em et al.},
  \href{http://dx.doi.org/10.1103/PhysRevLett.94.121801}{Phys. Rev. Lett. {\bf
  94},  121801 (2005)}, \href{http://arxiv.org/abs/hep-ex/0408003}{{\tt
  arXiv:hep-ex/0408003}}\relax
\mciteBstWouldAddEndPuncttrue
\mciteSetBstMidEndSepPunct{\mcitedefaultmidpunct}
{\mcitedefaultendpunct}{\mcitedefaultseppunct}\relax
\EndOfBibitem
\bibitem{Aubert:2006tw}
{\babar} collaboration, B.~Aubert {\em et al.},
  \href{http://dx.doi.org/10.1103/PhysRevD.73.111101}{Phys. Rev. {\bf D73},
  111101 (2006)}, \href{http://arxiv.org/abs/hep-ex/0602049}{{\tt
  arXiv:hep-ex/0602049}}\relax
\mciteBstWouldAddEndPuncttrue
\mciteSetBstMidEndSepPunct{\mcitedefaultmidpunct}
{\mcitedefaultendpunct}{\mcitedefaultseppunct}\relax
\EndOfBibitem
\bibitem{Aubert:2005yf}
{\babar} collaboration, B.~Aubert {\em et al.},
  \href{http://dx.doi.org/10.1103/PhysRevD.71.112003}{Phys. Rev. {\bf D71},
  112003 (2005)}, \href{http://arxiv.org/abs/hep-ex/0504035}{{\tt
  arXiv:hep-ex/0504035}}\relax
\mciteBstWouldAddEndPuncttrue
\mciteSetBstMidEndSepPunct{\mcitedefaultmidpunct}
{\mcitedefaultendpunct}{\mcitedefaultseppunct}\relax
\EndOfBibitem
\bibitem{Long:2003wq}
O.~Long, M.~Baak, R.~N. Cahn, and D.~Kirkby,
  \href{http://dx.doi.org/10.1103/PhysRevD.68.034010}{Phys. Rev. {\bf D68},
  034010 (2003)}, \href{http://arxiv.org/abs/hep-ex/0303030}{{\tt
  arXiv:hep-ex/0303030}}\relax
\mciteBstWouldAddEndPuncttrue
\mciteSetBstMidEndSepPunct{\mcitedefaultmidpunct}
{\mcitedefaultendpunct}{\mcitedefaultseppunct}\relax
\EndOfBibitem
\bibitem{Bahinipati:2011yq}
{Belle} collaboration, S.~Bahinipati {\em et al.},
  \href{http://dx.doi.org/10.1103/PhysRevD.84.021101}{Phys. Rev. {\bf D84},
  021101 (2011)}, \href{http://arxiv.org/abs/1102.0888}{{\tt arXiv:1102.0888
  [hep-ex]}}\relax
\mciteBstWouldAddEndPuncttrue
\mciteSetBstMidEndSepPunct{\mcitedefaultmidpunct}
{\mcitedefaultendpunct}{\mcitedefaultseppunct}\relax
\EndOfBibitem
\bibitem{Ronga:2006hv}
{Belle} collaboration, F.~J. Ronga {\em et al.},
  \href{http://dx.doi.org/10.1103/PhysRevD.73.092003}{Phys. Rev. {\bf D73},
  092003 (2006)}, \href{http://arxiv.org/abs/hep-ex/0604013}{{\tt
  arXiv:hep-ex/0604013}}\relax
\mciteBstWouldAddEndPuncttrue
\mciteSetBstMidEndSepPunct{\mcitedefaultmidpunct}
{\mcitedefaultendpunct}{\mcitedefaultseppunct}\relax
\EndOfBibitem
\bibitem{Fleischer:2003yb}
R.~Fleischer, \href{http://dx.doi.org/10.1016/j.nuclphysb.2003.08.010}{Nucl.
  Phys. {\bf B671},  459 (2003)},
  \href{http://arxiv.org/abs/hep-ph/0304027}{{\tt arXiv:hep-ph/0304027}}\relax
\mciteBstWouldAddEndPuncttrue
\mciteSetBstMidEndSepPunct{\mcitedefaultmidpunct}
{\mcitedefaultendpunct}{\mcitedefaultseppunct}\relax
\EndOfBibitem
\bibitem{Aaij:2018kpq}
{LHCb} collaboration, R.~Aaij {\em et al.},
  \href{http://dx.doi.org/10.1007/JHEP06(2018)084}{JHEP {\bf 06},  084 (2018)},
  \href{http://arxiv.org/abs/1805.03448}{{\tt arXiv:1805.03448 [hep-ex]}}\relax
\mciteBstWouldAddEndPuncttrue
\mciteSetBstMidEndSepPunct{\mcitedefaultmidpunct}
{\mcitedefaultendpunct}{\mcitedefaultseppunct}\relax
\EndOfBibitem
\bibitem{Aaij:2014fba}
{LHCb} collaboration, R.~Aaij {\em et al.},
  \href{http://dx.doi.org/10.1007/JHEP11(2014)060}{JHEP {\bf 11},  060 (2014)},
  \href{http://arxiv.org/abs/1407.6127}{{\tt arXiv:1407.6127 [hep-ex]}}\relax
\mciteBstWouldAddEndPuncttrue
\mciteSetBstMidEndSepPunct{\mcitedefaultmidpunct}
{\mcitedefaultendpunct}{\mcitedefaultseppunct}\relax
\EndOfBibitem
\bibitem{Aaij:2017lff}
{LHCb} collaboration, R.~Aaij {\em et al.},
  \href{http://dx.doi.org/10.1007/JHEP03(2018)059}{JHEP {\bf 03},  059 (2018)},
  \href{http://arxiv.org/abs/1712.07428}{{\tt arXiv:1712.07428 [hep-ex]}}\relax
\mciteBstWouldAddEndPuncttrue
\mciteSetBstMidEndSepPunct{\mcitedefaultmidpunct}
{\mcitedefaultendpunct}{\mcitedefaultseppunct}\relax
\EndOfBibitem
\bibitem{Atwood:1997zr}
D.~Atwood, M.~Gronau, and A.~Soni,
  \href{http://dx.doi.org/10.1103/PhysRevLett.79.185}{Phys. Rev. Lett. {\bf
  79},  185 (1997)}, \href{http://arxiv.org/abs/hep-ph/9704272}{{\tt
  arXiv:hep-ph/9704272}}\relax
\mciteBstWouldAddEndPuncttrue
\mciteSetBstMidEndSepPunct{\mcitedefaultmidpunct}
{\mcitedefaultendpunct}{\mcitedefaultseppunct}\relax
\EndOfBibitem
\bibitem{Atwood:2004jj}
D.~Atwood, T.~Gershon, M.~Hazumi, and A.~Soni,
  \href{http://dx.doi.org/10.1103/PhysRevD.71.076003}{Phys. Rev. {\bf D71},
  076003 (2005)}, \href{http://arxiv.org/abs/hep-ph/0410036}{{\tt
  arXiv:hep-ph/0410036}}\relax
\mciteBstWouldAddEndPuncttrue
\mciteSetBstMidEndSepPunct{\mcitedefaultmidpunct}
{\mcitedefaultendpunct}{\mcitedefaultseppunct}\relax
\EndOfBibitem
\bibitem{Grinstein:2004uu}
B.~Grinstein, Y.~Grossman, Z.~Ligeti, and D.~Pirjol,
  \href{http://dx.doi.org/10.1103/PhysRevD.71.011504}{Phys. Rev. {\bf D71},
  011504 (2005)}, \href{http://arxiv.org/abs/hep-ph/0412019}{{\tt
  arXiv:hep-ph/0412019}}\relax
\mciteBstWouldAddEndPuncttrue
\mciteSetBstMidEndSepPunct{\mcitedefaultmidpunct}
{\mcitedefaultendpunct}{\mcitedefaultseppunct}\relax
\EndOfBibitem
\bibitem{Grinstein:2005nu}
B.~Grinstein and D.~Pirjol,
  \href{http://dx.doi.org/10.1103/PhysRevD.73.014013}{Phys. Rev. {\bf D73},
  014013 (2006)}, \href{http://arxiv.org/abs/hep-ph/0510104}{{\tt
  arXiv:hep-ph/0510104}}\relax
\mciteBstWouldAddEndPuncttrue
\mciteSetBstMidEndSepPunct{\mcitedefaultmidpunct}
{\mcitedefaultendpunct}{\mcitedefaultseppunct}\relax
\EndOfBibitem
\bibitem{Matsumori:2005ax}
M.~Matsumori and A.~I. Sanda,
  \href{http://dx.doi.org/10.1103/PhysRevD.73.114022}{Phys. Rev. {\bf D73},
  114022 (2006)}, \href{http://arxiv.org/abs/hep-ph/0512175}{{\tt
  arXiv:hep-ph/0512175}}\relax
\mciteBstWouldAddEndPuncttrue
\mciteSetBstMidEndSepPunct{\mcitedefaultmidpunct}
{\mcitedefaultendpunct}{\mcitedefaultseppunct}\relax
\EndOfBibitem
\bibitem{Ball:2006cva}
P.~Ball and R.~Zwicky,
  \href{http://dx.doi.org/10.1016/j.physletb.2006.10.013}{Phys. Lett. {\bf
  B642},  478 (2006)}, \href{http://arxiv.org/abs/hep-ph/0609037}{{\tt
  arXiv:hep-ph/0609037}}\relax
\mciteBstWouldAddEndPuncttrue
\mciteSetBstMidEndSepPunct{\mcitedefaultmidpunct}
{\mcitedefaultendpunct}{\mcitedefaultseppunct}\relax
\EndOfBibitem
\bibitem{Muheim:2008vu}
F.~Muheim, Y.~Xie, and R.~Zwicky,
  \href{http://dx.doi.org/10.1016/j.physletb.2008.05.032}{Phys. Lett. {\bf
  B664},  174 (2008)}, \href{http://arxiv.org/abs/0802.0876}{{\tt
  arXiv:0802.0876 [hep-ph]}}\relax
\mciteBstWouldAddEndPuncttrue
\mciteSetBstMidEndSepPunct{\mcitedefaultmidpunct}
{\mcitedefaultendpunct}{\mcitedefaultseppunct}\relax
\EndOfBibitem
\bibitem{Bigi:1988ym}
I.~I.~Y. Bigi and A.~I. Sanda,
  \href{http://dx.doi.org/10.1016/0370-2693(88)90836-2}{Phys. Lett. {\bf B211},
   213 (1988)}\relax
\mciteBstWouldAddEndPuncttrue
\mciteSetBstMidEndSepPunct{\mcitedefaultmidpunct}
{\mcitedefaultendpunct}{\mcitedefaultseppunct}\relax
\EndOfBibitem
\bibitem{Gronau:1990ra}
M.~Gronau and D.~London.,
  \href{http://dx.doi.org/10.1016/0370-2693(91)91756-L}{Phys. Lett. {\bf B253},
   483 (1991)}\relax
\mciteBstWouldAddEndPuncttrue
\mciteSetBstMidEndSepPunct{\mcitedefaultmidpunct}
{\mcitedefaultendpunct}{\mcitedefaultseppunct}\relax
\EndOfBibitem
\bibitem{Gronau:1991dp}
M.~Gronau and D.~Wyler,
  \href{http://dx.doi.org/10.1016/0370-2693(91)90034-N}{Phys. Lett. {\bf B265},
   172 (1991)}\relax
\mciteBstWouldAddEndPuncttrue
\mciteSetBstMidEndSepPunct{\mcitedefaultmidpunct}
{\mcitedefaultendpunct}{\mcitedefaultseppunct}\relax
\EndOfBibitem
\bibitem{Atwood:1996ci}
D.~Atwood, I.~Dunietz, and A.~Soni,
  \href{http://dx.doi.org/10.1103/PhysRevLett.78.3257}{Phys. Rev. Lett. {\bf
  78},  3257 (1997)}, \href{http://arxiv.org/abs/hep-ph/9612433}{{\tt
  arXiv:hep-ph/9612433}}\relax
\mciteBstWouldAddEndPuncttrue
\mciteSetBstMidEndSepPunct{\mcitedefaultmidpunct}
{\mcitedefaultendpunct}{\mcitedefaultseppunct}\relax
\EndOfBibitem
\bibitem{Atwood:2000ck}
D.~Atwood, I.~Dunietz, and A.~Soni,
  \href{http://dx.doi.org/10.1103/PhysRevD.63.036005}{Phys. Rev. {\bf D63},
  036005 (2001)}, \href{http://arxiv.org/abs/hep-ph/0008090}{{\tt
  arXiv:hep-ph/0008090}}\relax
\mciteBstWouldAddEndPuncttrue
\mciteSetBstMidEndSepPunct{\mcitedefaultmidpunct}
{\mcitedefaultendpunct}{\mcitedefaultseppunct}\relax
\EndOfBibitem
\bibitem{Giri:2003ty}
A.~Giri, Y.~Grossman, A.~Soffer, and J.~Zupan,
  \href{http://dx.doi.org/10.1103/PhysRevD.68.054018}{Phys. Rev. {\bf D68},
  054018 (2003)}, \href{http://arxiv.org/abs/hep-ph/0303187}{{\tt
  arXiv:hep-ph/0303187}}\relax
\mciteBstWouldAddEndPuncttrue
\mciteSetBstMidEndSepPunct{\mcitedefaultmidpunct}
{\mcitedefaultendpunct}{\mcitedefaultseppunct}\relax
\EndOfBibitem
\bibitem{Poluektov:2004mf}
{Belle} collaboration, A.~Poluektov {\em et al.},
  \href{http://dx.doi.org/10.1103/PhysRevD.70.072003}{Phys. Rev. {\bf D70},
  072003 (2004)}, \href{http://arxiv.org/abs/hep-ex/0406067}{{\tt
  arXiv:hep-ex/0406067}}\relax
\mciteBstWouldAddEndPuncttrue
\mciteSetBstMidEndSepPunct{\mcitedefaultmidpunct}
{\mcitedefaultendpunct}{\mcitedefaultseppunct}\relax
\EndOfBibitem
\bibitem{Brod:2013sga}
J.~Brod and J.~Zupan, \href{http://dx.doi.org/10.1007/JHEP01(2014)051}{JHEP
  {\bf 01},  051 (2014)}, \href{http://arxiv.org/abs/1308.5663}{{\tt
  arXiv:1308.5663 [hep-ph]}}\relax
\mciteBstWouldAddEndPuncttrue
\mciteSetBstMidEndSepPunct{\mcitedefaultmidpunct}
{\mcitedefaultendpunct}{\mcitedefaultseppunct}\relax
\EndOfBibitem
\bibitem{Gronau:2002mu}
M.~Gronau, \href{http://dx.doi.org/10.1016/S0370-2693(03)00192-8}{Phys. Lett.
  {\bf B557},  198 (2003)}, \href{http://arxiv.org/abs/hep-ph/0211282}{{\tt
  arXiv:hep-ph/0211282}}\relax
\mciteBstWouldAddEndPuncttrue
\mciteSetBstMidEndSepPunct{\mcitedefaultmidpunct}
{\mcitedefaultendpunct}{\mcitedefaultseppunct}\relax
\EndOfBibitem
\bibitem{Aleksan:2002mh}
R.~Aleksan, T.~C. Petersen, and A.~Soffer,
  \href{http://dx.doi.org/10.1103/PhysRevD.67.096002}{Phys. Rev. {\bf D67},
  096002 (2003)}, \href{http://arxiv.org/abs/hep-ph/0209194}{{\tt
  arXiv:hep-ph/0209194 [hep-ph]}}\relax
\mciteBstWouldAddEndPuncttrue
\mciteSetBstMidEndSepPunct{\mcitedefaultmidpunct}
{\mcitedefaultendpunct}{\mcitedefaultseppunct}\relax
\EndOfBibitem
\bibitem{Gershon:2008pe}
T.~Gershon, \href{http://dx.doi.org/10.1103/PhysRevD.79.051301}{Phys. Rev. {\bf
  D79},  051301 (2009)}, \href{http://arxiv.org/abs/0810.2706}{{\tt
  arXiv:0810.2706 [hep-ph]}}\relax
\mciteBstWouldAddEndPuncttrue
\mciteSetBstMidEndSepPunct{\mcitedefaultmidpunct}
{\mcitedefaultendpunct}{\mcitedefaultseppunct}\relax
\EndOfBibitem
\bibitem{Gershon:2009qc}
T.~Gershon and M.~Williams,
  \href{http://dx.doi.org/10.1103/PhysRevD.80.092002}{Phys. Rev. {\bf D80},
  092002 (2009)}, \href{http://arxiv.org/abs/0909.1495}{{\tt arXiv:0909.1495
  [hep-ph]}}\relax
\mciteBstWouldAddEndPuncttrue
\mciteSetBstMidEndSepPunct{\mcitedefaultmidpunct}
{\mcitedefaultendpunct}{\mcitedefaultseppunct}\relax
\EndOfBibitem
\bibitem{Bondar:2004bi}
A.~Bondar and T.~Gershon,
  \href{http://dx.doi.org/10.1103/PhysRevD.70.091503}{Phys. Rev. {\bf D70},
  091503 (2004)}, \href{http://arxiv.org/abs/hep-ph/0409281}{{\tt
  arXiv:hep-ph/0409281}}\relax
\mciteBstWouldAddEndPuncttrue
\mciteSetBstMidEndSepPunct{\mcitedefaultmidpunct}
{\mcitedefaultendpunct}{\mcitedefaultseppunct}\relax
\EndOfBibitem
\bibitem{Atwood:2003mj}
D.~Atwood and A.~Soni,
  \href{http://dx.doi.org/10.1103/PhysRevD.68.033003}{Phys. Rev. {\bf D68},
  033003 (2003)}, \href{http://arxiv.org/abs/hep-ph/0304085}{{\tt
  arXiv:hep-ph/0304085}}\relax
\mciteBstWouldAddEndPuncttrue
\mciteSetBstMidEndSepPunct{\mcitedefaultmidpunct}
{\mcitedefaultendpunct}{\mcitedefaultseppunct}\relax
\EndOfBibitem
\bibitem{Grossman:2002aq}
Y.~Grossman, Z.~Ligeti, and A.~Soffer,
  \href{http://dx.doi.org/10.1103/PhysRevD.67.071301}{Phys. Rev. {\bf D67},
  071301 (2003)}, \href{http://arxiv.org/abs/hep-ph/0210433}{{\tt
  arXiv:hep-ph/0210433 [hep-ph]}}\relax
\mciteBstWouldAddEndPuncttrue
\mciteSetBstMidEndSepPunct{\mcitedefaultmidpunct}
{\mcitedefaultendpunct}{\mcitedefaultseppunct}\relax
\EndOfBibitem
\bibitem{Nayak:2014tea}
M.~Nayak {\em et al.},
  \href{http://dx.doi.org/10.1016/j.physletb.2014.11.022}{Phys. Lett. {\bf
  B740},  1 (2014)}, \href{http://arxiv.org/abs/1410.3964}{{\tt arXiv:1410.3964
  [hep-ex]}}\relax
\mciteBstWouldAddEndPuncttrue
\mciteSetBstMidEndSepPunct{\mcitedefaultmidpunct}
{\mcitedefaultendpunct}{\mcitedefaultseppunct}\relax
\EndOfBibitem
\bibitem{Bondar:2005ki}
A.~Bondar and A.~Poluektov,
  \href{http://dx.doi.org/10.1140/epjc/s2006-02590-x}{Eur. Phys. J. {\bf C47},
  347 (2006)}, \href{http://arxiv.org/abs/hep-ph/0510246}{{\tt
  arXiv:hep-ph/0510246}}\relax
\mciteBstWouldAddEndPuncttrue
\mciteSetBstMidEndSepPunct{\mcitedefaultmidpunct}
{\mcitedefaultendpunct}{\mcitedefaultseppunct}\relax
\EndOfBibitem
\bibitem{Bondar:2008hh}
A.~Bondar and A.~Poluektov,
  \href{http://dx.doi.org/10.1140/epjc/s10052-008-0600-z}{Eur. Phys. J. {\bf
  C55},  51 (2008)}, \href{http://arxiv.org/abs/0801.0840}{{\tt arXiv:0801.0840
  [hep-ex]}}\relax
\mciteBstWouldAddEndPuncttrue
\mciteSetBstMidEndSepPunct{\mcitedefaultmidpunct}
{\mcitedefaultendpunct}{\mcitedefaultseppunct}\relax
\EndOfBibitem
\bibitem{Gershon:2015xra}
T.~Gershon, J.~Libby, and G.~Wilkinson,
  \href{http://dx.doi.org/10.1016/j.physletb.2015.08.063}{Phys. Lett. {\bf
  B750},  338 (2015)}, \href{http://arxiv.org/abs/1506.08594}{{\tt
  arXiv:1506.08594 [hep-ph]}}\relax
\mciteBstWouldAddEndPuncttrue
\mciteSetBstMidEndSepPunct{\mcitedefaultmidpunct}
{\mcitedefaultendpunct}{\mcitedefaultseppunct}\relax
\EndOfBibitem
\bibitem{Aubert:2007ii}
{\babar} collaboration, B.~Aubert {\em et al.},
  \href{http://dx.doi.org/10.1103/PhysRevLett.99.251801}{Phys. Rev. Lett. {\bf
  99},  251801 (2007)}, \href{http://arxiv.org/abs/hep-ex/0703037}{{\tt
  arXiv:hep-ex/0703037}}\relax
\mciteBstWouldAddEndPuncttrue
\mciteSetBstMidEndSepPunct{\mcitedefaultmidpunct}
{\mcitedefaultendpunct}{\mcitedefaultseppunct}\relax
\EndOfBibitem
\bibitem{Aubert:2007hz}
{\babar} collaboration, B.~Aubert {\em et al.},
  \href{http://dx.doi.org/10.1103/PhysRevD.76.031102}{Phys. Rev. {\bf D76},
  031102 (2007)}, \href{http://arxiv.org/abs/0704.0522}{{\tt arXiv:0704.0522
  [hep-ex]}}\relax
\mciteBstWouldAddEndPuncttrue
\mciteSetBstMidEndSepPunct{\mcitedefaultmidpunct}
{\mcitedefaultendpunct}{\mcitedefaultseppunct}\relax
\EndOfBibitem
\bibitem{Itoh:2005ks}
{Belle} collaboration, R.~Itoh {\em et al.},
  \href{http://dx.doi.org/10.1103/PhysRevLett.95.091601}{Phys. Rev. Lett. {\bf
  95},  091601 (2005)}, \href{http://arxiv.org/abs/hep-ex/0504030}{{\tt
  arXiv:hep-ex/0504030 [hep-ex]}}\relax
\mciteBstWouldAddEndPuncttrue
\mciteSetBstMidEndSepPunct{\mcitedefaultmidpunct}
{\mcitedefaultendpunct}{\mcitedefaultseppunct}\relax
\EndOfBibitem
\bibitem{Acosta:2004gt}
{CDF} collaboration, D.~Acosta {\em et al.},
  \href{http://dx.doi.org/10.1103/PhysRevLett.94.101803}{Phys. Rev. Lett. {\bf
  94},  101803 (2005)}, \href{http://arxiv.org/abs/hep-ex/0412057}{{\tt
  arXiv:hep-ex/0412057}}\relax
\mciteBstWouldAddEndPuncttrue
\mciteSetBstMidEndSepPunct{\mcitedefaultmidpunct}
{\mcitedefaultendpunct}{\mcitedefaultseppunct}\relax
\EndOfBibitem
\bibitem{Aaij:2013cma}
{LHCb} collaboration, R.~Aaij {\em et al.},
  \href{http://dx.doi.org/10.1103/PhysRevD.88.052002}{Phys. Rev. {\bf D88},
  052002 (2013)}, \href{http://arxiv.org/abs/1307.2782}{{\tt arXiv:1307.2782
  [hep-ex]}}\relax
\mciteBstWouldAddEndPuncttrue
\mciteSetBstMidEndSepPunct{\mcitedefaultmidpunct}
{\mcitedefaultendpunct}{\mcitedefaultseppunct}\relax
\EndOfBibitem
\bibitem{Jung:2012mp}
M.~Jung, \href{http://dx.doi.org/10.1103/PhysRevD.86.053008}{Phys. Rev. {\bf
  D86},  053008 (2012)}, \href{http://arxiv.org/abs/1206.2050}{{\tt
  arXiv:1206.2050 [hep-ph]}}\relax
\mciteBstWouldAddEndPuncttrue
\mciteSetBstMidEndSepPunct{\mcitedefaultmidpunct}
{\mcitedefaultendpunct}{\mcitedefaultseppunct}\relax
\EndOfBibitem
\bibitem{DeBruyn:2014oga}
K.~De~Bruyn and R.~Fleischer,
  \href{http://dx.doi.org/10.1007/JHEP03(2015)145}{JHEP {\bf 03},  145 (2015)},
  \href{http://arxiv.org/abs/1412.6834}{{\tt arXiv:1412.6834 [hep-ph]}}\relax
\mciteBstWouldAddEndPuncttrue
\mciteSetBstMidEndSepPunct{\mcitedefaultmidpunct}
{\mcitedefaultendpunct}{\mcitedefaultseppunct}\relax
\EndOfBibitem
\bibitem{:2009yr}
{\babar} collaboration, B.~Aubert {\em et al.},
  \href{http://dx.doi.org/10.1103/PhysRevD.79.072009}{Phys. Rev. {\bf D79},
  072009 (2009)}, \href{http://arxiv.org/abs/0902.1708}{{\tt arXiv:0902.1708
  [hep-ex]}}\relax
\mciteBstWouldAddEndPuncttrue
\mciteSetBstMidEndSepPunct{\mcitedefaultmidpunct}
{\mcitedefaultendpunct}{\mcitedefaultseppunct}\relax
\EndOfBibitem
\bibitem{Adachi:2012et}
{Belle} collaboration, I.~Adachi {\em et al.},
  \href{http://dx.doi.org/10.1103/PhysRevLett.108.171802}{Phys. Rev. Lett. {\bf
  108},  171802 (2012)}, \href{http://arxiv.org/abs/1201.4643}{{\tt
  arXiv:1201.4643 [hep-ex]}}\relax
\mciteBstWouldAddEndPuncttrue
\mciteSetBstMidEndSepPunct{\mcitedefaultmidpunct}
{\mcitedefaultendpunct}{\mcitedefaultseppunct}\relax
\EndOfBibitem
\bibitem{Aaij:2015vza}
{LHCb} collaboration, R.~Aaij {\em et al.},
  \href{http://dx.doi.org/10.1103/PhysRevLett.115.031601}{Phys. Rev. Lett. {\bf
  115},  031601 (2015)}, \href{http://arxiv.org/abs/1503.07089}{{\tt
  arXiv:1503.07089 [hep-ex]}}\relax
\mciteBstWouldAddEndPuncttrue
\mciteSetBstMidEndSepPunct{\mcitedefaultmidpunct}
{\mcitedefaultendpunct}{\mcitedefaultseppunct}\relax
\EndOfBibitem
\bibitem{Aaij:2017yld}
{LHCb} collaboration, R.~Aaij {\em et al.},
  \href{http://dx.doi.org/10.1007/JHEP11(2017)170}{JHEP {\bf 11},  170 (2017)},
  \href{http://arxiv.org/abs/1709.03944}{{\tt arXiv:1709.03944 [hep-ex]}}\relax
\mciteBstWouldAddEndPuncttrue
\mciteSetBstMidEndSepPunct{\mcitedefaultmidpunct}
{\mcitedefaultendpunct}{\mcitedefaultseppunct}\relax
\EndOfBibitem
\bibitem{Aubert:2003xn}
{\babar} collaboration, B.~Aubert {\em et al.},
  \href{http://dx.doi.org/10.1103/PhysRevD.69.052001}{Phys. Rev. {\bf D69},
  052001 (2004)}, \href{http://arxiv.org/abs/hep-ex/0309039}{{\tt
  arXiv:hep-ex/0309039}}\relax
\mciteBstWouldAddEndPuncttrue
\mciteSetBstMidEndSepPunct{\mcitedefaultmidpunct}
{\mcitedefaultendpunct}{\mcitedefaultseppunct}\relax
\EndOfBibitem
\bibitem{Barate:2000tf}
{ALEPH} collaboration, R.~Barate {\em et al.},
  \href{http://dx.doi.org/10.1016/S0370-2693(00)01091-1}{Phys. Lett. {\bf
  B492},  259 (2000)}, \href{http://arxiv.org/abs/hep-ex/0009058}{{\tt
  arXiv:hep-ex/0009058}}\relax
\mciteBstWouldAddEndPuncttrue
\mciteSetBstMidEndSepPunct{\mcitedefaultmidpunct}
{\mcitedefaultendpunct}{\mcitedefaultseppunct}\relax
\EndOfBibitem
\bibitem{Ackerstaff:1998xz}
{OPAL} collaboration, K.~Ackerstaff {\em et al.},
  \href{http://dx.doi.org/10.1007/s100520050284}{Eur. Phys. J. {\bf C5},  379
  (1998)}, \href{http://arxiv.org/abs/hep-ex/9801022}{{\tt
  arXiv:hep-ex/9801022}}\relax
\mciteBstWouldAddEndPuncttrue
\mciteSetBstMidEndSepPunct{\mcitedefaultmidpunct}
{\mcitedefaultendpunct}{\mcitedefaultseppunct}\relax
\EndOfBibitem
\bibitem{Affolder:1999gg}
{CDF} collaboration, A.~A. Affolder {\em et al.},
  \href{http://dx.doi.org/10.1103/PhysRevD.61.072005}{Phys. Rev. {\bf D61},
  072005 (2000)}, \href{http://arxiv.org/abs/hep-ex/9909003}{{\tt
  arXiv:hep-ex/9909003}}\relax
\mciteBstWouldAddEndPuncttrue
\mciteSetBstMidEndSepPunct{\mcitedefaultmidpunct}
{\mcitedefaultendpunct}{\mcitedefaultseppunct}\relax
\EndOfBibitem
\bibitem{Sato:2012hu}
{Belle} collaboration, Y.~Sato {\em et al.},
  \href{http://dx.doi.org/10.1103/PhysRevLett.108.171801}{Phys. Rev. Lett. {\bf
  108},  171801 (2012)}, \href{http://arxiv.org/abs/1201.3502}{{\tt
  arXiv:1201.3502 [hep-ex]}}\relax
\mciteBstWouldAddEndPuncttrue
\mciteSetBstMidEndSepPunct{\mcitedefaultmidpunct}
{\mcitedefaultendpunct}{\mcitedefaultseppunct}\relax
\EndOfBibitem
\bibitem{Bona:2005vz}
{UTfit} collaboration, M.~Bona {\em et al.},
  \href{http://dx.doi.org/10.1088/1126-6708/2005/07/028}{JHEP {\bf 07},  028
  (2005)}, \href{http://arxiv.org/abs/hep-ph/0501199}{{\tt
  arXiv:hep-ph/0501199}}, see also online updates,
  \url{http://www.utfit.org/}\relax
\mciteBstWouldAddEndPuncttrue
\mciteSetBstMidEndSepPunct{\mcitedefaultmidpunct}
{\mcitedefaultendpunct}{\mcitedefaultseppunct}\relax
\EndOfBibitem
\bibitem{Lunghi:2008aa}
E.~Lunghi and A.~Soni,
  \href{http://dx.doi.org/10.1016/j.physletb.2008.07.015}{Phys. Lett. {\bf
  B666},  162 (2008)}, \href{http://arxiv.org/abs/0803.4340}{{\tt
  arXiv:0803.4340 [hep-ph]}}\relax
\mciteBstWouldAddEndPuncttrue
\mciteSetBstMidEndSepPunct{\mcitedefaultmidpunct}
{\mcitedefaultendpunct}{\mcitedefaultseppunct}\relax
\EndOfBibitem
\bibitem{Eigen:2013cv}
G.~Eigen, G.~Dubois-Felsmann, D.~Hitlin, and F.~Porter,
  \href{http://dx.doi.org/10.1103/PhysRevD.89.033004}{Phys. Rev. {\bf D89},
  033004 (2014)}, \href{http://arxiv.org/abs/1301.5867}{{\tt arXiv:1301.5867
  [hep-ex]}}\relax
\mciteBstWouldAddEndPuncttrue
\mciteSetBstMidEndSepPunct{\mcitedefaultmidpunct}
{\mcitedefaultendpunct}{\mcitedefaultseppunct}\relax
\EndOfBibitem
\bibitem{Dunietz:1990cj}
I.~Dunietz, H.~R. Quinn, A.~Snyder, W.~Toki, and H.~J. Lipkin,
  \href{http://dx.doi.org/10.1103/PhysRevD.43.2193}{Phys. Rev. {\bf D43},  2193
  (1991)}\relax
\mciteBstWouldAddEndPuncttrue
\mciteSetBstMidEndSepPunct{\mcitedefaultmidpunct}
{\mcitedefaultendpunct}{\mcitedefaultseppunct}\relax
\EndOfBibitem
\bibitem{Aston:1987ir}
D.~Aston {\em et al.},
  \href{http://dx.doi.org/10.1016/0550-3213(88)90028-4}{Nucl. Phys. {\bf B296},
   493 (1988)}\relax
\mciteBstWouldAddEndPuncttrue
\mciteSetBstMidEndSepPunct{\mcitedefaultmidpunct}
{\mcitedefaultendpunct}{\mcitedefaultseppunct}\relax
\EndOfBibitem
\bibitem{Suzuki:2001za}
M.~Suzuki, \href{http://dx.doi.org/10.1103/PhysRevD.64.117503}{Phys. Rev. {\bf
  D64},  117503 (2001)}, \href{http://arxiv.org/abs/hep-ph/0106354}{{\tt
  arXiv:hep-ph/0106354}}\relax
\mciteBstWouldAddEndPuncttrue
\mciteSetBstMidEndSepPunct{\mcitedefaultmidpunct}
{\mcitedefaultendpunct}{\mcitedefaultseppunct}\relax
\EndOfBibitem
\bibitem{Aubert:2004cp}
{\babar} collaboration, B.~Aubert {\em et al.},
  \href{http://dx.doi.org/10.1103/PhysRevD.71.032005}{Phys. Rev. {\bf D71},
  032005 (2005)}, \href{http://arxiv.org/abs/hep-ex/0411016}{{\tt
  arXiv:hep-ex/0411016}}\relax
\mciteBstWouldAddEndPuncttrue
\mciteSetBstMidEndSepPunct{\mcitedefaultmidpunct}
{\mcitedefaultendpunct}{\mcitedefaultseppunct}\relax
\EndOfBibitem
\bibitem{Grossman:1996ke}
Y.~Grossman and M.~P. Worah,
  \href{http://dx.doi.org/10.1016/S0370-2693(97)00068-3}{Phys. Lett. {\bf
  B395},  241 (1997)}, \href{http://arxiv.org/abs/hep-ph/9612269}{{\tt
  arXiv:hep-ph/9612269}}\relax
\mciteBstWouldAddEndPuncttrue
\mciteSetBstMidEndSepPunct{\mcitedefaultmidpunct}
{\mcitedefaultendpunct}{\mcitedefaultseppunct}\relax
\EndOfBibitem
\bibitem{Fleischer:2003ai}
R.~Fleischer, \href{http://dx.doi.org/10.1016/S0370-2693(03)00582-3}{Phys.
  Lett. {\bf B562},  234 (2003)},
  \href{http://arxiv.org/abs/hep-ph/0301255}{{\tt arXiv:hep-ph/0301255}}\relax
\mciteBstWouldAddEndPuncttrue
\mciteSetBstMidEndSepPunct{\mcitedefaultmidpunct}
{\mcitedefaultendpunct}{\mcitedefaultseppunct}\relax
\EndOfBibitem
\bibitem{Fleischer:2003aj}
R.~Fleischer, \href{http://dx.doi.org/10.1016/S0550-3213(03)00225-6}{Nucl.
  Phys. {\bf B659},  321 (2003)},
  \href{http://arxiv.org/abs/hep-ph/0301256}{{\tt arXiv:hep-ph/0301256}}\relax
\mciteBstWouldAddEndPuncttrue
\mciteSetBstMidEndSepPunct{\mcitedefaultmidpunct}
{\mcitedefaultendpunct}{\mcitedefaultseppunct}\relax
\EndOfBibitem
\bibitem{Abdesselam:2015gha}
{\babar\ and Belle} collaborations, A.~Abdesselam {\em et al.},
  \href{http://dx.doi.org/10.1103/PhysRevLett.115.121604}{Phys. Rev. Lett. {\bf
  115},  121604 (2015)}, \href{http://arxiv.org/abs/1505.04147}{{\tt
  arXiv:1505.04147 [hep-ex]}}\relax
\mciteBstWouldAddEndPuncttrue
\mciteSetBstMidEndSepPunct{\mcitedefaultmidpunct}
{\mcitedefaultendpunct}{\mcitedefaultseppunct}\relax
\EndOfBibitem
\bibitem{Bondar:2005gk}
A.~Bondar, T.~Gershon, and P.~Krokovny,
  \href{http://dx.doi.org/10.1016/j.physletb.2005.07.053}{Phys. Lett. {\bf
  B624},  1 (2005)}, \href{http://arxiv.org/abs/hep-ph/0503174}{{\tt
  arXiv:hep-ph/0503174}}\relax
\mciteBstWouldAddEndPuncttrue
\mciteSetBstMidEndSepPunct{\mcitedefaultmidpunct}
{\mcitedefaultendpunct}{\mcitedefaultseppunct}\relax
\EndOfBibitem
\bibitem{Botella:2005ks}
F.~Botella and J.~Silva,
  \href{http://dx.doi.org/10.1103/PhysRevD.71.094008}{Phys. Rev. {\bf D71},
  094008 (2005)}, \href{http://arxiv.org/abs/hep-ph/0503136}{{\tt
  arXiv:hep-ph/0503136 [hep-ph]}}\relax
\mciteBstWouldAddEndPuncttrue
\mciteSetBstMidEndSepPunct{\mcitedefaultmidpunct}
{\mcitedefaultendpunct}{\mcitedefaultseppunct}\relax
\EndOfBibitem
\bibitem{Aubert:2008bs}
{\babar} collaboration, B.~Aubert {\em et al.},
  \href{http://dx.doi.org/10.1103/PhysRevLett.101.021801}{Phys. Rev. Lett. {\bf
  101},  021801 (2008)}, \href{http://arxiv.org/abs/0804.0896}{{\tt
  arXiv:0804.0896 [hep-ex]}}\relax
\mciteBstWouldAddEndPuncttrue
\mciteSetBstMidEndSepPunct{\mcitedefaultmidpunct}
{\mcitedefaultendpunct}{\mcitedefaultseppunct}\relax
\EndOfBibitem
\bibitem{Pal:2018olx}
{Belle} collaboration, B.~Pal {\em et al.},
  \href{http://dx.doi.org/10.1103/PhysRevD.98.112008}{Phys. Rev. {\bf D98},
  112008 (2018)}, \href{http://arxiv.org/abs/1810.01356}{{\tt arXiv:1810.01356
  [hep-ex]}}\relax
\mciteBstWouldAddEndPuncttrue
\mciteSetBstMidEndSepPunct{\mcitedefaultmidpunct}
{\mcitedefaultendpunct}{\mcitedefaultseppunct}\relax
\EndOfBibitem
\bibitem{Aaij:2016yip}
{LHCb} collaboration, R.~Aaij {\em et al.},
  \href{http://dx.doi.org/10.1103/PhysRevLett.117.261801}{Phys. Rev. Lett. {\bf
  117},  261801 (2016)}, \href{http://arxiv.org/abs/1608.06620}{{\tt
  arXiv:1608.06620 [hep-ex]}}\relax
\mciteBstWouldAddEndPuncttrue
\mciteSetBstMidEndSepPunct{\mcitedefaultmidpunct}
{\mcitedefaultendpunct}{\mcitedefaultseppunct}\relax
\EndOfBibitem
\bibitem{Lees:2012px}
{\babar} collaboration, J.~P. Lees {\em et al.},
  \href{http://dx.doi.org/10.1103/PhysRevD.86.112006}{Phys. Rev. {\bf D86},
  112006 (2012)}, \href{http://arxiv.org/abs/1208.1282}{{\tt arXiv:1208.1282
  [hep-ex]}}\relax
\mciteBstWouldAddEndPuncttrue
\mciteSetBstMidEndSepPunct{\mcitedefaultmidpunct}
{\mcitedefaultendpunct}{\mcitedefaultseppunct}\relax
\EndOfBibitem
\bibitem{Kronenbitter:2012ha}
{Belle} collaboration, B.~Kronenbitter {\em et al.},
  \href{http://dx.doi.org/10.1103/PhysRevD.86.071103}{Phys. Rev. {\bf D86},
  071103 (2012)}, \href{http://arxiv.org/abs/1207.5611}{{\tt arXiv:1207.5611
  [hep-ex]}}\relax
\mciteBstWouldAddEndPuncttrue
\mciteSetBstMidEndSepPunct{\mcitedefaultmidpunct}
{\mcitedefaultendpunct}{\mcitedefaultseppunct}\relax
\EndOfBibitem
\bibitem{Fratina:2007zk}
{Belle} collaboration, S.~Fratina {\em et al.},
  \href{http://dx.doi.org/10.1103/PhysRevLett.98.221802}{Phys. Rev. Lett. {\bf
  98},  221802 (2007)}, \href{http://arxiv.org/abs/hep-ex/0702031}{{\tt
  arXiv:hep-ex/0702031}}\relax
\mciteBstWouldAddEndPuncttrue
\mciteSetBstMidEndSepPunct{\mcitedefaultmidpunct}
{\mcitedefaultendpunct}{\mcitedefaultseppunct}\relax
\EndOfBibitem
\bibitem{Fleischer:1999nz}
R.~Fleischer, \href{http://dx.doi.org/10.1007/s100529900099}{Eur. Phys. J. {\bf
  C10},  299 (1999)}, \href{http://arxiv.org/abs/hep-ph/9903455}{{\tt
  arXiv:hep-ph/9903455 [hep-ph]}}\relax
\mciteBstWouldAddEndPuncttrue
\mciteSetBstMidEndSepPunct{\mcitedefaultmidpunct}
{\mcitedefaultendpunct}{\mcitedefaultseppunct}\relax
\EndOfBibitem
\bibitem{DeBruyn:2010hh}
K.~De~Bruyn, R.~Fleischer, and P.~Koppenburg,
  \href{http://dx.doi.org/10.1140/epjc/s10052-010-1495-z}{Eur. Phys. J. {\bf
  C70},  1025 (2010)}, \href{http://arxiv.org/abs/1010.0089}{{\tt
  arXiv:1010.0089 [hep-ph]}}\relax
\mciteBstWouldAddEndPuncttrue
\mciteSetBstMidEndSepPunct{\mcitedefaultmidpunct}
{\mcitedefaultendpunct}{\mcitedefaultseppunct}\relax
\EndOfBibitem
\bibitem{Aaij:2015tza}
{LHCb} collaboration, R.~Aaij {\em et al.},
  \href{http://dx.doi.org/10.1007/JHEP06(2015)131}{JHEP {\bf 06},  131 (2015)},
  \href{http://arxiv.org/abs/1503.07055}{{\tt arXiv:1503.07055 [hep-ex]}}\relax
\mciteBstWouldAddEndPuncttrue
\mciteSetBstMidEndSepPunct{\mcitedefaultmidpunct}
{\mcitedefaultendpunct}{\mcitedefaultseppunct}\relax
\EndOfBibitem
\bibitem{Fleischer:1996bv}
R.~Fleischer, \href{http://dx.doi.org/10.1142/S0217751X97001432}{Int. J. Mod.
  Phys. {\bf A12},  2459 (1997)},
  \href{http://arxiv.org/abs/hep-ph/9612446}{{\tt arXiv:hep-ph/9612446}}\relax
\mciteBstWouldAddEndPuncttrue
\mciteSetBstMidEndSepPunct{\mcitedefaultmidpunct}
{\mcitedefaultendpunct}{\mcitedefaultseppunct}\relax
\EndOfBibitem
\bibitem{London:1997zk}
D.~London and A.~Soni,
  \href{http://dx.doi.org/10.1016/S0370-2693(97)00695-3}{Phys. Lett. {\bf
  B407},  61 (1997)}, \href{http://arxiv.org/abs/hep-ph/9704277}{{\tt
  arXiv:hep-ph/9704277}}\relax
\mciteBstWouldAddEndPuncttrue
\mciteSetBstMidEndSepPunct{\mcitedefaultmidpunct}
{\mcitedefaultendpunct}{\mcitedefaultseppunct}\relax
\EndOfBibitem
\bibitem{Ciuchini:1997zp}
M.~Ciuchini, E.~Franco, G.~Martinelli, A.~Masiero, and L.~Silvestrini,
  \href{http://dx.doi.org/10.1103/PhysRevLett.79.978}{Phys. Rev. Lett. {\bf
  79},  978 (1997)}, \href{http://arxiv.org/abs/hep-ph/9704274}{{\tt
  arXiv:hep-ph/9704274}}\relax
\mciteBstWouldAddEndPuncttrue
\mciteSetBstMidEndSepPunct{\mcitedefaultmidpunct}
{\mcitedefaultendpunct}{\mcitedefaultseppunct}\relax
\EndOfBibitem
\bibitem{Okubo:1963fa}
S.~Okubo, \href{http://dx.doi.org/10.1016/S0375-9601(63)92548-9}{Phys. Lett.
  {\bf 5},  165--168 (1963)}\relax
\mciteBstWouldAddEndPuncttrue
\mciteSetBstMidEndSepPunct{\mcitedefaultmidpunct}
{\mcitedefaultendpunct}{\mcitedefaultseppunct}\relax
\EndOfBibitem
\bibitem{Zweig:1964jf}
 G.~Zweig, CERN-TH-412, 1964\relax
\mciteBstWouldAddEndPuncttrue
\mciteSetBstMidEndSepPunct{\mcitedefaultmidpunct}
{\mcitedefaultendpunct}{\mcitedefaultseppunct}\relax
\EndOfBibitem
\bibitem{Iizuka:1966fk}
J.~Iizuka, \href{http://dx.doi.org/10.1143/PTPS.37.21}{Prog. Theor. Phys.
  Suppl. {\bf 37},  21--34 (1966)}\relax
\mciteBstWouldAddEndPuncttrue
\mciteSetBstMidEndSepPunct{\mcitedefaultmidpunct}
{\mcitedefaultendpunct}{\mcitedefaultseppunct}\relax
\EndOfBibitem
\bibitem{Gershon:2004tk}
T.~Gershon and M.~Hazumi,
  \href{http://dx.doi.org/10.1016/j.physletb.2004.06.095}{Phys. Lett. {\bf
  B596},  163 (2004)}, \href{http://arxiv.org/abs/hep-ph/0402097}{{\tt
  arXiv:hep-ph/0402097}}\relax
\mciteBstWouldAddEndPuncttrue
\mciteSetBstMidEndSepPunct{\mcitedefaultmidpunct}
{\mcitedefaultendpunct}{\mcitedefaultseppunct}\relax
\EndOfBibitem
\bibitem{Grossman:2003qp}
Y.~Grossman, Z.~Ligeti, Y.~Nir, and H.~Quinn,
  \href{http://dx.doi.org/10.1103/PhysRevD.68.015004}{Phys. Rev. {\bf D68},
  015004 (2003)}, \href{http://arxiv.org/abs/hep-ph/0303171}{{\tt
  arXiv:hep-ph/0303171}}\relax
\mciteBstWouldAddEndPuncttrue
\mciteSetBstMidEndSepPunct{\mcitedefaultmidpunct}
{\mcitedefaultendpunct}{\mcitedefaultseppunct}\relax
\EndOfBibitem
\bibitem{Gronau:2003ep}
M.~Gronau and J.~L. Rosner,
  \href{http://dx.doi.org/10.1016/S0370-2693(03)00702-0}{Phys. Lett. {\bf
  B564},  90 (2003)}, \href{http://arxiv.org/abs/hep-ph/0304178}{{\tt
  arXiv:hep-ph/0304178}}\relax
\mciteBstWouldAddEndPuncttrue
\mciteSetBstMidEndSepPunct{\mcitedefaultmidpunct}
{\mcitedefaultendpunct}{\mcitedefaultseppunct}\relax
\EndOfBibitem
\bibitem{Gronau:2003kx}
M.~Gronau, Y.~Grossman, and J.~L. Rosner,
  \href{http://dx.doi.org/10.1016/j.physletb.2003.11.015}{Phys. Lett. {\bf
  B579},  331 (2004)}, \href{http://arxiv.org/abs/hep-ph/0310020}{{\tt
  arXiv:hep-ph/0310020}}\relax
\mciteBstWouldAddEndPuncttrue
\mciteSetBstMidEndSepPunct{\mcitedefaultmidpunct}
{\mcitedefaultendpunct}{\mcitedefaultseppunct}\relax
\EndOfBibitem
\bibitem{Gronau:2004hp}
M.~Gronau, J.~L. Rosner, and J.~Zupan,
  \href{http://dx.doi.org/10.1016/j.physletb.2004.06.086}{Phys. Lett. {\bf
  B596},  107 (2004)}, \href{http://arxiv.org/abs/hep-ph/0403287}{{\tt
  arXiv:hep-ph/0403287}}\relax
\mciteBstWouldAddEndPuncttrue
\mciteSetBstMidEndSepPunct{\mcitedefaultmidpunct}
{\mcitedefaultendpunct}{\mcitedefaultseppunct}\relax
\EndOfBibitem
\bibitem{Cheng:2005bg}
H.-Y. Cheng, C.-K. Chua, and A.~Soni,
  \href{http://dx.doi.org/10.1103/PhysRevD.72.014006}{Phys. Rev. {\bf D72},
  014006 (2005)}, \href{http://arxiv.org/abs/hep-ph/0502235}{{\tt
  arXiv:hep-ph/0502235}}\relax
\mciteBstWouldAddEndPuncttrue
\mciteSetBstMidEndSepPunct{\mcitedefaultmidpunct}
{\mcitedefaultendpunct}{\mcitedefaultseppunct}\relax
\EndOfBibitem
\bibitem{Gronau:2005gz}
M.~Gronau and J.~L. Rosner,
  \href{http://dx.doi.org/10.1103/PhysRevD.71.074019}{Phys. Rev. {\bf D71},
  074019 (2005)}, \href{http://arxiv.org/abs/hep-ph/0503131}{{\tt
  arXiv:hep-ph/0503131}}\relax
\mciteBstWouldAddEndPuncttrue
\mciteSetBstMidEndSepPunct{\mcitedefaultmidpunct}
{\mcitedefaultendpunct}{\mcitedefaultseppunct}\relax
\EndOfBibitem
\bibitem{Buchalla:2005us}
G.~Buchalla, G.~Hiller, Y.~Nir, and G.~Raz,
  \href{http://dx.doi.org/10.1088/1126-6708/2005/09/074}{JHEP {\bf 09},  074
  (2005)}, \href{http://arxiv.org/abs/hep-ph/0503151}{{\tt
  arXiv:hep-ph/0503151}}\relax
\mciteBstWouldAddEndPuncttrue
\mciteSetBstMidEndSepPunct{\mcitedefaultmidpunct}
{\mcitedefaultendpunct}{\mcitedefaultseppunct}\relax
\EndOfBibitem
\bibitem{Beneke:2005pu}
M.~Beneke, \href{http://dx.doi.org/10.1016/j.physletb.2005.06.045}{Phys. Lett.
  {\bf B620},  143 (2005)}, \href{http://arxiv.org/abs/hep-ph/0505075}{{\tt
  arXiv:hep-ph/0505075}}\relax
\mciteBstWouldAddEndPuncttrue
\mciteSetBstMidEndSepPunct{\mcitedefaultmidpunct}
{\mcitedefaultendpunct}{\mcitedefaultseppunct}\relax
\EndOfBibitem
\bibitem{Engelhard:2005hu}
G.~Engelhard, Y.~Nir, and G.~Raz,
  \href{http://dx.doi.org/10.1103/PhysRevD.72.075013}{Phys. Rev. {\bf D72},
  075013 (2005)}, \href{http://arxiv.org/abs/hep-ph/0505194}{{\tt
  arXiv:hep-ph/0505194}}\relax
\mciteBstWouldAddEndPuncttrue
\mciteSetBstMidEndSepPunct{\mcitedefaultmidpunct}
{\mcitedefaultendpunct}{\mcitedefaultseppunct}\relax
\EndOfBibitem
\bibitem{Cheng:2005ug}
H.-Y. Cheng, C.-K. Chua, and A.~Soni,
  \href{http://dx.doi.org/10.1103/PhysRevD.72.094003}{Phys. Rev. {\bf D72},
  094003 (2005)}, \href{http://arxiv.org/abs/hep-ph/0506268}{{\tt
  arXiv:hep-ph/0506268}}\relax
\mciteBstWouldAddEndPuncttrue
\mciteSetBstMidEndSepPunct{\mcitedefaultmidpunct}
{\mcitedefaultendpunct}{\mcitedefaultseppunct}\relax
\EndOfBibitem
\bibitem{Engelhard:2005ky}
G.~Engelhard and G.~Raz,
  \href{http://dx.doi.org/10.1103/PhysRevD.72.114017}{Phys. Rev. {\bf D72},
  114017 (2005)}, \href{http://arxiv.org/abs/hep-ph/0508046}{{\tt
  arXiv:hep-ph/0508046}}\relax
\mciteBstWouldAddEndPuncttrue
\mciteSetBstMidEndSepPunct{\mcitedefaultmidpunct}
{\mcitedefaultendpunct}{\mcitedefaultseppunct}\relax
\EndOfBibitem
\bibitem{Gronau:2006qh}
M.~Gronau, J.~L. Rosner, and J.~Zupan,
  \href{http://dx.doi.org/10.1103/PhysRevD.74.093003}{Phys. Rev. {\bf D74},
  093003 (2006)}, \href{http://arxiv.org/abs/hep-ph/0608085}{{\tt
  arXiv:hep-ph/0608085}}\relax
\mciteBstWouldAddEndPuncttrue
\mciteSetBstMidEndSepPunct{\mcitedefaultmidpunct}
{\mcitedefaultendpunct}{\mcitedefaultseppunct}\relax
\EndOfBibitem
\bibitem{Silvestrini:2007yf}
L.~Silvestrini,
  \href{http://dx.doi.org/10.1146/annurev.nucl.57.090506.123007}{Ann. Rev.
  Nucl. Part. Sci. {\bf 57},  405 (2007)},
  \href{http://arxiv.org/abs/0705.1624}{{\tt arXiv:0705.1624 [hep-ph]}}\relax
\mciteBstWouldAddEndPuncttrue
\mciteSetBstMidEndSepPunct{\mcitedefaultmidpunct}
{\mcitedefaultendpunct}{\mcitedefaultseppunct}\relax
\EndOfBibitem
\bibitem{Dutta:2008xw}
R.~Dutta and S.~Gardner,
  \href{http://dx.doi.org/10.1103/PhysRevD.78.034021}{Phys. Rev. {\bf D78},
  034021 (2008)}, \href{http://arxiv.org/abs/0805.1963}{{\tt arXiv:0805.1963
  [hep-ph]}}\relax
\mciteBstWouldAddEndPuncttrue
\mciteSetBstMidEndSepPunct{\mcitedefaultmidpunct}
{\mcitedefaultendpunct}{\mcitedefaultseppunct}\relax
\EndOfBibitem
\bibitem{Fujikawa:2008pk}
{Belle} collaboration, M.~Fujikawa {\em et al.},
  \href{http://dx.doi.org/10.1103/PhysRevD.81.011101}{Phys. Rev. {\bf D81},
  011101 (2010)}, \href{http://arxiv.org/abs/0809.4366}{{\tt arXiv:0809.4366
  [hep-ex]}}\relax
\mciteBstWouldAddEndPuncttrue
\mciteSetBstMidEndSepPunct{\mcitedefaultmidpunct}
{\mcitedefaultendpunct}{\mcitedefaultseppunct}\relax
\EndOfBibitem
\bibitem{Abe:2006gy}
{Belle} collaboration, K.~Abe {\em et al.},
  \href{http://dx.doi.org/10.1103/PhysRevD.76.091103}{Phys. Rev. {\bf D76},
  091103 (2007)}, \href{http://arxiv.org/abs/hep-ex/0609006}{{\tt
  arXiv:hep-ex/0609006}}\relax
\mciteBstWouldAddEndPuncttrue
\mciteSetBstMidEndSepPunct{\mcitedefaultmidpunct}
{\mcitedefaultendpunct}{\mcitedefaultseppunct}\relax
\EndOfBibitem
\bibitem{Aubert:2005ja}
{\babar} collaboration, B.~Aubert {\em et al.},
  \href{http://dx.doi.org/10.1103/PhysRevD.71.091102}{Phys. Rev. {\bf D71},
  091102 (2005)}, \href{http://arxiv.org/abs/hep-ex/0502019}{{\tt
  arXiv:hep-ex/0502019}}\relax
\mciteBstWouldAddEndPuncttrue
\mciteSetBstMidEndSepPunct{\mcitedefaultmidpunct}
{\mcitedefaultendpunct}{\mcitedefaultseppunct}\relax
\EndOfBibitem
\bibitem{:2008se}
{\babar} collaboration, B.~Aubert {\em et al.},
  \href{http://dx.doi.org/10.1103/PhysRevD.79.052003}{Phys. Rev. {\bf D79},
  052003 (2009)}, \href{http://arxiv.org/abs/0809.1174}{{\tt arXiv:0809.1174
  [hep-ex]}}\relax
\mciteBstWouldAddEndPuncttrue
\mciteSetBstMidEndSepPunct{\mcitedefaultmidpunct}
{\mcitedefaultendpunct}{\mcitedefaultseppunct}\relax
\EndOfBibitem
\bibitem{Santelj:2014sja}
{Belle} collaboration, L.~\v{S}antelj {\em et al.},
  \href{http://dx.doi.org/10.1007/JHEP10(2014)165}{JHEP {\bf 10},  165 (2014)},
  \href{http://arxiv.org/abs/1408.5991}{{\tt arXiv:1408.5991 [hep-ex]}}\relax
\mciteBstWouldAddEndPuncttrue
\mciteSetBstMidEndSepPunct{\mcitedefaultmidpunct}
{\mcitedefaultendpunct}{\mcitedefaultseppunct}\relax
\EndOfBibitem
\bibitem{Lees:2011nf}
{\babar} collaboration, J.~P. Lees {\em et al.},
  \href{http://dx.doi.org/10.1103/PhysRevD.85.054023}{Phys. Rev. {\bf D85},
  054023 (2012)}, \href{http://arxiv.org/abs/1111.3636}{{\tt arXiv:1111.3636
  [hep-ex]}}\relax
\mciteBstWouldAddEndPuncttrue
\mciteSetBstMidEndSepPunct{\mcitedefaultmidpunct}
{\mcitedefaultendpunct}{\mcitedefaultseppunct}\relax
\EndOfBibitem
\bibitem{Chen:2006nk}
{Belle} collaboration, K.~F. Chen {\em et al.},
  \href{http://dx.doi.org/10.1103/PhysRevLett.98.031802}{Phys. Rev. Lett. {\bf
  98},  031802 (2007)}, \href{http://arxiv.org/abs/hep-ex/0608039}{{\tt
  arXiv:hep-ex/0608039}}\relax
\mciteBstWouldAddEndPuncttrue
\mciteSetBstMidEndSepPunct{\mcitedefaultmidpunct}
{\mcitedefaultendpunct}{\mcitedefaultseppunct}\relax
\EndOfBibitem
\bibitem{Chobanova:2013ddr}
{Belle} collaboration, V.~Chobanova {\em et al.},
  \href{http://dx.doi.org/10.1103/PhysRevD.90.012002}{Phys. Rev. {\bf D90},
  012002 (2014)}, \href{http://arxiv.org/abs/1311.6666}{{\tt arXiv:1311.6666
  [hep-ex]}}\relax
\mciteBstWouldAddEndPuncttrue
\mciteSetBstMidEndSepPunct{\mcitedefaultmidpunct}
{\mcitedefaultendpunct}{\mcitedefaultseppunct}\relax
\EndOfBibitem
\bibitem{Aubert:2007ub}
{\babar} collaboration, B.~Aubert {\em et al.},
  \href{http://dx.doi.org/10.1103/PhysRevD.76.071101}{Phys. Rev. {\bf D76},
  071101 (2007)}, \href{http://arxiv.org/abs/hep-ex/0702010}{{\tt
  arXiv:hep-ex/0702010}}\relax
\mciteBstWouldAddEndPuncttrue
\mciteSetBstMidEndSepPunct{\mcitedefaultmidpunct}
{\mcitedefaultendpunct}{\mcitedefaultseppunct}\relax
\EndOfBibitem
\bibitem{Yusa:2018hmz}
{Belle} collaboration, Y.~Yusa {\em et al.},
  \href{http://dx.doi.org/10.1103/PhysRevD.99.011102}{Phys. Rev. {\bf D99},
  011102 (2019)}, \href{http://arxiv.org/abs/1810.03336}{{\tt arXiv:1810.03336
  [hep-ex]}}\relax
\mciteBstWouldAddEndPuncttrue
\mciteSetBstMidEndSepPunct{\mcitedefaultmidpunct}
{\mcitedefaultendpunct}{\mcitedefaultseppunct}\relax
\EndOfBibitem
\bibitem{Aubert:2008zza}
{\babar} collaboration, B.~Aubert {\em et al.},
  \href{http://dx.doi.org/10.1103/PhysRevD.78.092008}{Phys. Rev. {\bf D78},
  092008 (2008)}, \href{http://arxiv.org/abs/0808.3586}{{\tt arXiv:0808.3586
  [hep-ex]}}\relax
\mciteBstWouldAddEndPuncttrue
\mciteSetBstMidEndSepPunct{\mcitedefaultmidpunct}
{\mcitedefaultendpunct}{\mcitedefaultseppunct}\relax
\EndOfBibitem
\bibitem{Dunietz:1993rm}
 I.~Dunietz, FERMILAB-CONF-93-090-T, 1993\relax
\mciteBstWouldAddEndPuncttrue
\mciteSetBstMidEndSepPunct{\mcitedefaultmidpunct}
{\mcitedefaultendpunct}{\mcitedefaultseppunct}\relax
\EndOfBibitem
\bibitem{Fleischer:1999pa}
R.~Fleischer, \href{http://dx.doi.org/10.1016/S0370-2693(99)00640-1}{Phys.
  Lett. {\bf B459},  306 (1999)},
  \href{http://arxiv.org/abs/hep-ph/9903456}{{\tt arXiv:hep-ph/9903456
  [hep-ph]}}\relax
\mciteBstWouldAddEndPuncttrue
\mciteSetBstMidEndSepPunct{\mcitedefaultmidpunct}
{\mcitedefaultendpunct}{\mcitedefaultseppunct}\relax
\EndOfBibitem
\bibitem{Aaij:2018tfw}
{LHCb} collaboration, R.~Aaij {\em et al.},
  \href{http://dx.doi.org/10.1103/PhysRevD.98.032004}{Phys. Rev. {\bf D98},
  032004 (2018)}, \href{http://arxiv.org/abs/1805.06759}{{\tt arXiv:1805.06759
  [hep-ex]}}\relax
\mciteBstWouldAddEndPuncttrue
\mciteSetBstMidEndSepPunct{\mcitedefaultmidpunct}
{\mcitedefaultendpunct}{\mcitedefaultseppunct}\relax
\EndOfBibitem
\bibitem{Aaij:2013tna}
{LHCb} collaboration, R.~Aaij {\em et al.},
  \href{http://dx.doi.org/10.1007/JHEP10(2013)183}{JHEP {\bf 10},  183 (2013)},
  \href{http://arxiv.org/abs/1308.1428}{{\tt arXiv:1308.1428 [hep-ex]}}\relax
\mciteBstWouldAddEndPuncttrue
\mciteSetBstMidEndSepPunct{\mcitedefaultmidpunct}
{\mcitedefaultendpunct}{\mcitedefaultseppunct}\relax
\EndOfBibitem
\bibitem{Raidal:2002ph}
M.~Raidal, \href{http://dx.doi.org/10.1103/PhysRevLett.89.231803}{Phys. Rev.
  Lett. {\bf 89},  231803 (2002)},
  \href{http://arxiv.org/abs/hep-ph/0208091}{{\tt arXiv:hep-ph/0208091
  [hep-ph]}}\relax
\mciteBstWouldAddEndPuncttrue
\mciteSetBstMidEndSepPunct{\mcitedefaultmidpunct}
{\mcitedefaultendpunct}{\mcitedefaultseppunct}\relax
\EndOfBibitem
\bibitem{Aaij:2019uld}
{LHCb} collaboration, R.~Aaij {\em et al.},
  \href{http://arxiv.org/abs/1907.10003}{{\tt arXiv:1907.10003 [hep-ex]}}\relax
\mciteBstWouldAddEndPuncttrue
\mciteSetBstMidEndSepPunct{\mcitedefaultmidpunct}
{\mcitedefaultendpunct}{\mcitedefaultseppunct}\relax
\EndOfBibitem
\bibitem{Aubert:2006gm}
{\babar} collaboration, B.~Aubert {\em et al.},
  \href{http://dx.doi.org/10.1103/PhysRevLett.97.171805}{Phys. Rev. Lett. {\bf
  97},  171805 (2006)}, \href{http://arxiv.org/abs/hep-ex/0608036}{{\tt
  arXiv:hep-ex/0608036 [hep-ex]}}\relax
\mciteBstWouldAddEndPuncttrue
\mciteSetBstMidEndSepPunct{\mcitedefaultmidpunct}
{\mcitedefaultendpunct}{\mcitedefaultseppunct}\relax
\EndOfBibitem
\bibitem{Nakahama:2007dg}
{Belle} collaboration, Y.~Nakahama {\em et al.},
  \href{http://dx.doi.org/10.1103/PhysRevLett.100.121601}{Phys. Rev. Lett. {\bf
  100},  121601 (2008)}, \href{http://arxiv.org/abs/0712.4234}{{\tt
  arXiv:0712.4234 [hep-ex]}}\relax
\mciteBstWouldAddEndPuncttrue
\mciteSetBstMidEndSepPunct{\mcitedefaultmidpunct}
{\mcitedefaultendpunct}{\mcitedefaultseppunct}\relax
\EndOfBibitem
\bibitem{Akar:2018zhv}
S.~Akar, E.~Ben-Haim, J.~Hebinger, E.~Kou, and F.-S. Yu,
  \href{http://dx.doi.org/10.1007/JHEP09(2019)034}{JHEP {\bf 09},  034 (2019)},
  \href{http://arxiv.org/abs/1802.09433}{{\tt arXiv:1802.09433 [hep-ph]}}\relax
\mciteBstWouldAddEndPuncttrue
\mciteSetBstMidEndSepPunct{\mcitedefaultmidpunct}
{\mcitedefaultendpunct}{\mcitedefaultseppunct}\relax
\EndOfBibitem
\bibitem{Akar:2013ima}
 S.~Akar, PhD thesis, LPNHE, Universit\'e Pierre et Marie Curie - Paris VI,
  2013, \url{https://tel.archives-ouvertes.fr/tel-00998252}\relax
\mciteBstWouldAddEndPuncttrue
\mciteSetBstMidEndSepPunct{\mcitedefaultmidpunct}
{\mcitedefaultendpunct}{\mcitedefaultseppunct}\relax
\EndOfBibitem
\bibitem{Sanchez:2015pxu}
{\babar} collaboration, P.~del Amo~Sanchez {\em et al.},
  \href{http://dx.doi.org/10.1103/PhysRevD.93.052013}{Phys. Rev. {\bf D93},
  052013 (2016)}, \href{http://arxiv.org/abs/1512.03579}{{\tt arXiv:1512.03579
  [hep-ex]}}\relax
\mciteBstWouldAddEndPuncttrue
\mciteSetBstMidEndSepPunct{\mcitedefaultmidpunct}
{\mcitedefaultendpunct}{\mcitedefaultseppunct}\relax
\EndOfBibitem
\bibitem{Li:2008qma}
{Belle} collaboration, J.~Li {\em et al.},
  \href{http://dx.doi.org/10.1103/PhysRevLett.101.251601}{Phys. Rev. Lett. {\bf
  101},  251601 (2008)}, \href{http://arxiv.org/abs/0806.1980}{{\tt
  arXiv:0806.1980 [hep-ex]}}\relax
\mciteBstWouldAddEndPuncttrue
\mciteSetBstMidEndSepPunct{\mcitedefaultmidpunct}
{\mcitedefaultendpunct}{\mcitedefaultseppunct}\relax
\EndOfBibitem
\bibitem{Aubert:2008gy}
{\babar} collaboration, B.~Aubert {\em et al.},
  \href{http://dx.doi.org/10.1103/PhysRevD.78.071102}{Phys. Rev. {\bf D78},
  071102 (2008)}, \href{http://arxiv.org/abs/0807.3103}{{\tt arXiv:0807.3103
  [hep-ex]}}\relax
\mciteBstWouldAddEndPuncttrue
\mciteSetBstMidEndSepPunct{\mcitedefaultmidpunct}
{\mcitedefaultendpunct}{\mcitedefaultseppunct}\relax
\EndOfBibitem
\bibitem{Ushiroda:2006fi}
{Belle} collaboration, Y.~Ushiroda {\em et al.},
  \href{http://dx.doi.org/10.1103/PhysRevD.74.111104}{Phys. Rev. {\bf D74},
  111104 (2006)}, \href{http://arxiv.org/abs/hep-ex/0608017}{{\tt
  arXiv:hep-ex/0608017}}\relax
\mciteBstWouldAddEndPuncttrue
\mciteSetBstMidEndSepPunct{\mcitedefaultmidpunct}
{\mcitedefaultendpunct}{\mcitedefaultseppunct}\relax
\EndOfBibitem
\bibitem{Aubert:2008js}
{\babar} collaboration, B.~Aubert {\em et al.},
  \href{http://dx.doi.org/10.1103/PhysRevD.79.011102}{Phys. Rev. {\bf D79},
  011102 (2009)}, \href{http://arxiv.org/abs/0805.1317}{{\tt arXiv:0805.1317
  [hep-ex]}}\relax
\mciteBstWouldAddEndPuncttrue
\mciteSetBstMidEndSepPunct{\mcitedefaultmidpunct}
{\mcitedefaultendpunct}{\mcitedefaultseppunct}\relax
\EndOfBibitem
\bibitem{Nakano:2018lqo}
{Belle} collaboration, H.~Nakano {\em et al.},
  \href{http://dx.doi.org/10.1103/PhysRevD.97.092003}{Phys. Rev. {\bf D97},
  092003 (2018)}, \href{http://arxiv.org/abs/1803.07774}{{\tt arXiv:1803.07774
  [hep-ex]}}\relax
\mciteBstWouldAddEndPuncttrue
\mciteSetBstMidEndSepPunct{\mcitedefaultmidpunct}
{\mcitedefaultendpunct}{\mcitedefaultseppunct}\relax
\EndOfBibitem
\bibitem{Sahoo:2011zd}
{Belle} collaboration, H.~Sahoo {\em et al.},
  \href{http://dx.doi.org/10.1103/PhysRevD.84.071101}{Phys. Rev. {\bf D84},
  071101 (2011)}, \href{http://arxiv.org/abs/1104.5590}{{\tt arXiv:1104.5590
  [hep-ex]}}\relax
\mciteBstWouldAddEndPuncttrue
\mciteSetBstMidEndSepPunct{\mcitedefaultmidpunct}
{\mcitedefaultendpunct}{\mcitedefaultseppunct}\relax
\EndOfBibitem
\bibitem{Aaij:2016ofv}
{LHCb} collaboration, R.~Aaij {\em et al.},
  \href{http://dx.doi.org/10.1103/PhysRevLett.118.021801}{Phys. Rev. Lett. {\bf
  118},  021801 (2017)}, \href{http://arxiv.org/abs/1609.02032}{{\tt
  arXiv:1609.02032 [hep-ex]}}\relax
\mciteBstWouldAddEndPuncttrue
\mciteSetBstMidEndSepPunct{\mcitedefaultmidpunct}
{\mcitedefaultendpunct}{\mcitedefaultseppunct}\relax
\EndOfBibitem
\bibitem{Ushiroda:2007jf}
{Belle} collaboration, Y.~Ushiroda {\em et al.},
  \href{http://dx.doi.org/10.1103/PhysRevLett.100.021602}{Phys. Rev. Lett. {\bf
  100},  021602 (2008)}, \href{http://arxiv.org/abs/0709.2769}{{\tt
  arXiv:0709.2769 [hep-ex]}}\relax
\mciteBstWouldAddEndPuncttrue
\mciteSetBstMidEndSepPunct{\mcitedefaultmidpunct}
{\mcitedefaultendpunct}{\mcitedefaultseppunct}\relax
\EndOfBibitem
\bibitem{Aubert:2007nua}
{\babar} collaboration, B.~Aubert {\em et al.},
  \href{http://dx.doi.org/10.1103/PhysRevD.76.052007}{Phys. Rev. {\bf D76},
  052007 (2007)}, \href{http://arxiv.org/abs/0705.2157}{{\tt arXiv:0705.2157
  [hep-ex]}}\relax
\mciteBstWouldAddEndPuncttrue
\mciteSetBstMidEndSepPunct{\mcitedefaultmidpunct}
{\mcitedefaultendpunct}{\mcitedefaultseppunct}\relax
\EndOfBibitem
\bibitem{Vanhoefer:2015ijw}
{Belle} collaboration, P.~Vanhoefer {\em et al.},
  \href{http://dx.doi.org/10.1103/PhysRevD.94.099903,
  10.1103/PhysRevD.93.032010}{Phys. Rev. {\bf D93},  032010 (2016)},
  \href{http://arxiv.org/abs/1510.01245}{{\tt arXiv:1510.01245 [hep-ex]}},
  [addendum: Phys. Rev.D94,no.9,099903(2016)]\relax
\mciteBstWouldAddEndPuncttrue
\mciteSetBstMidEndSepPunct{\mcitedefaultmidpunct}
{\mcitedefaultendpunct}{\mcitedefaultseppunct}\relax
\EndOfBibitem
\bibitem{Aubert:2008au}
{\babar} collaboration, B.~Aubert {\em et al.},
  \href{http://dx.doi.org/10.1103/PhysRevD.78.071104}{Phys. Rev. {\bf D78},
  071104 (2008)}, \href{http://arxiv.org/abs/0807.4977}{{\tt arXiv:0807.4977
  [hep-ex]}}\relax
\mciteBstWouldAddEndPuncttrue
\mciteSetBstMidEndSepPunct{\mcitedefaultmidpunct}
{\mcitedefaultendpunct}{\mcitedefaultseppunct}\relax
\EndOfBibitem
\bibitem{Adachi:2012cz}
{Belle} collaboration, I.~Adachi {\em et al.},
  \href{http://dx.doi.org/10.1103/PhysRevD.89.072008}{Phys. Rev. {\bf D89},
  072008 (2014)}, \href{http://arxiv.org/abs/1212.4015}{{\tt arXiv:1212.4015
  [hep-ex]}}, addendum ibid.\
  \href{http://dx.doi.org/10.1103/PhysRevD.89.119903}{{\bf D89}, 119903},
  (2014)\relax
\mciteBstWouldAddEndPuncttrue
\mciteSetBstMidEndSepPunct{\mcitedefaultmidpunct}
{\mcitedefaultendpunct}{\mcitedefaultseppunct}\relax
\EndOfBibitem
\bibitem{Aaij:2015ria}
{LHCb} collaboration, R.~Aaij {\em et al.},
  \href{http://dx.doi.org/10.1016/j.physletb.2015.06.027}{Phys. Lett. {\bf
  B747},  468 (2015)}, \href{http://arxiv.org/abs/1503.07770}{{\tt
  arXiv:1503.07770 [hep-ex]}}\relax
\mciteBstWouldAddEndPuncttrue
\mciteSetBstMidEndSepPunct{\mcitedefaultmidpunct}
{\mcitedefaultendpunct}{\mcitedefaultseppunct}\relax
\EndOfBibitem
\bibitem{Aubert:2006gb}
{\babar} collaboration, B.~Aubert {\em et al.},
  \href{http://dx.doi.org/10.1103/PhysRevLett.98.181803}{Phys. Rev. Lett. {\bf
  98},  181803 (2007)}, \href{http://arxiv.org/abs/hep-ex/0612050}{{\tt
  arXiv:hep-ex/0612050}}\relax
\mciteBstWouldAddEndPuncttrue
\mciteSetBstMidEndSepPunct{\mcitedefaultmidpunct}
{\mcitedefaultendpunct}{\mcitedefaultseppunct}\relax
\EndOfBibitem
\bibitem{Dalseno:2012hp}
{Belle} collaboration, J.~Dalseno {\em et al.},
  \href{http://dx.doi.org/10.1103/PhysRevD.86.092012}{Phys. Rev. {\bf D86},
  092012 (2012)}, \href{http://arxiv.org/abs/1205.5957}{{\tt arXiv:1205.5957
  [hep-ex]}}\relax
\mciteBstWouldAddEndPuncttrue
\mciteSetBstMidEndSepPunct{\mcitedefaultmidpunct}
{\mcitedefaultendpunct}{\mcitedefaultseppunct}\relax
\EndOfBibitem
\bibitem{Aubert:2009ab}
{\babar} collaboration, B.~Aubert {\em et al.},
  \href{http://dx.doi.org/10.1103/PhysRevD.81.052009}{Phys. Rev. {\bf D81},
  052009 (2010)}, \href{http://arxiv.org/abs/0909.2171}{{\tt arXiv:0909.2171
  [hep-ex]}}\relax
\mciteBstWouldAddEndPuncttrue
\mciteSetBstMidEndSepPunct{\mcitedefaultmidpunct}
{\mcitedefaultendpunct}{\mcitedefaultseppunct}\relax
\EndOfBibitem
\bibitem{Lees:2012mma}
{\babar} collaboration, J.~P. Lees {\em et al.},
  \href{http://dx.doi.org/10.1103/PhysRevD.87.052009}{Phys. Rev. {\bf D87},
  052009 (2013)}, \href{http://arxiv.org/abs/1206.3525}{{\tt arXiv:1206.3525
  [hep-ex]}}\relax
\mciteBstWouldAddEndPuncttrue
\mciteSetBstMidEndSepPunct{\mcitedefaultmidpunct}
{\mcitedefaultendpunct}{\mcitedefaultseppunct}\relax
\EndOfBibitem
\bibitem{Adachi:2013mae}
{Belle} collaboration, I.~Adachi {\em et al.},
  \href{http://dx.doi.org/10.1103/PhysRevD.88.092003}{Phys. Rev. {\bf D88},
  092003 (2013)}, \href{http://arxiv.org/abs/1302.0551}{{\tt arXiv:1302.0551
  [hep-ex]}}\relax
\mciteBstWouldAddEndPuncttrue
\mciteSetBstMidEndSepPunct{\mcitedefaultmidpunct}
{\mcitedefaultendpunct}{\mcitedefaultseppunct}\relax
\EndOfBibitem
\bibitem{Gronau:1990ka}
M.~Gronau and D.~London,
  \href{http://dx.doi.org/10.1103/PhysRevLett.65.3381}{Phys. Rev. Lett. {\bf
  65},  3381 (1990)}\relax
\mciteBstWouldAddEndPuncttrue
\mciteSetBstMidEndSepPunct{\mcitedefaultmidpunct}
{\mcitedefaultendpunct}{\mcitedefaultseppunct}\relax
\EndOfBibitem
\bibitem{Asner:2010qj}
{Heavy Flavor Averaging Group}, D.~Asner {\em et al.},
  \href{http://arxiv.org/abs/1010.1589}{{\tt arXiv:1010.1589 [hep-ex]}}
  (2010)\relax
\mciteBstWouldAddEndPuncttrue
\mciteSetBstMidEndSepPunct{\mcitedefaultmidpunct}
{\mcitedefaultendpunct}{\mcitedefaultseppunct}\relax
\EndOfBibitem
\bibitem{Aubert:2009it}
{\babar} collaboration, B.~Aubert {\em et al.},
  \href{http://dx.doi.org/10.1103/PhysRevLett.102.141802}{Phys. Rev. Lett. {\bf
  102},  141802 (2009)}, \href{http://arxiv.org/abs/0901.3522}{{\tt
  arXiv:0901.3522 [hep-ex]}}\relax
\mciteBstWouldAddEndPuncttrue
\mciteSetBstMidEndSepPunct{\mcitedefaultmidpunct}
{\mcitedefaultendpunct}{\mcitedefaultseppunct}\relax
\EndOfBibitem
\bibitem{Lipkin:1991st}
H.~J. Lipkin, Y.~Nir, H.~R. Quinn, and A.~Snyder,
  \href{http://dx.doi.org/10.1103/PhysRevD.44.1454}{Phys. Rev. {\bf D44},  1454
  (1991)}\relax
\mciteBstWouldAddEndPuncttrue
\mciteSetBstMidEndSepPunct{\mcitedefaultmidpunct}
{\mcitedefaultendpunct}{\mcitedefaultseppunct}\relax
\EndOfBibitem
\bibitem{Gronau:2005kw}
M.~Gronau and J.~Zupan,
  \href{http://dx.doi.org/10.1103/PhysRevD.73.057502}{Phys. Rev. {\bf D73},
  057502 (2006)}, \href{http://arxiv.org/abs/hep-ph/0512148}{{\tt
  arXiv:hep-ph/0512148}}\relax
\mciteBstWouldAddEndPuncttrue
\mciteSetBstMidEndSepPunct{\mcitedefaultmidpunct}
{\mcitedefaultendpunct}{\mcitedefaultseppunct}\relax
\EndOfBibitem
\bibitem{Bevan:2014iga}
{\babar\ and Belle} collaborations, A.~Bevan {\em et al.},
  \href{http://dx.doi.org/10.1140/epjc/s10052-014-3026-9}{Eur. Phys. J. {\bf
  C74},  3026 (2014)}, \href{http://arxiv.org/abs/1406.6311}{{\tt
  arXiv:1406.6311 [hep-ex]}}\relax
\mciteBstWouldAddEndPuncttrue
\mciteSetBstMidEndSepPunct{\mcitedefaultmidpunct}
{\mcitedefaultendpunct}{\mcitedefaultseppunct}\relax
\EndOfBibitem
\bibitem{Gronau:2005pq}
M.~Gronau and J.~Zupan,
  \href{http://dx.doi.org/10.1103/PhysRevD.71.074017}{Phys. Rev. {\bf D71},
  074017 (2005)}, \href{http://arxiv.org/abs/hep-ph/0502139}{{\tt
  arXiv:hep-ph/0502139 [hep-ph]}}\relax
\mciteBstWouldAddEndPuncttrue
\mciteSetBstMidEndSepPunct{\mcitedefaultmidpunct}
{\mcitedefaultendpunct}{\mcitedefaultseppunct}\relax
\EndOfBibitem
\bibitem{Gardner:2005pq}
S.~Gardner, \href{http://dx.doi.org/10.1103/PhysRevD.72.034015}{Phys. Rev. {\bf
  D72},  034015 (2005)}, \href{http://arxiv.org/abs/hep-ph/0505071}{{\tt
  arXiv:hep-ph/0505071 [hep-ph]}}\relax
\mciteBstWouldAddEndPuncttrue
\mciteSetBstMidEndSepPunct{\mcitedefaultmidpunct}
{\mcitedefaultendpunct}{\mcitedefaultseppunct}\relax
\EndOfBibitem
\bibitem{Charles:2017evz}
J.~Charles, O.~Deschamps, S.~Descotes-Genon, and V.~Niess,
  \href{http://dx.doi.org/10.1140/epjc/s10052-017-5126-9}{Eur. Phys. J. {\bf
  C77},  574 (2017)}, \href{http://arxiv.org/abs/1705.02981}{{\tt
  arXiv:1705.02981 [hep-ph]}}\relax
\mciteBstWouldAddEndPuncttrue
\mciteSetBstMidEndSepPunct{\mcitedefaultmidpunct}
{\mcitedefaultendpunct}{\mcitedefaultseppunct}\relax
\EndOfBibitem
\bibitem{gammacombo}
 {{\tt GammaCombo} framework for combinations of measurements and computation
  of confidence intervals, CERN}, \url{{http://gammacombo.github.io/}}\relax
\mciteBstWouldAddEndPuncttrue
\mciteSetBstMidEndSepPunct{\mcitedefaultmidpunct}
{\mcitedefaultendpunct}{\mcitedefaultseppunct}\relax
\EndOfBibitem
\bibitem{Julius:2017jso}
{Belle} collaboration, T.~Julius {\em et al.},
  \href{http://arxiv.org/abs/1705.02083}{{\tt arXiv:1705.02083 [hep-ex]}}
  (2017)\relax
\mciteBstWouldAddEndPuncttrue
\mciteSetBstMidEndSepPunct{\mcitedefaultmidpunct}
{\mcitedefaultendpunct}{\mcitedefaultseppunct}\relax
\EndOfBibitem
\bibitem{Dunietz:1997in}
I.~Dunietz, \href{http://dx.doi.org/10.1016/S0370-2693(98)00304-9}{Phys. Lett.
  {\bf B427},  179 (1998)}, \href{http://arxiv.org/abs/hep-ph/9712401}{{\tt
  arXiv:hep-ph/9712401 [hep-ph]}}\relax
\mciteBstWouldAddEndPuncttrue
\mciteSetBstMidEndSepPunct{\mcitedefaultmidpunct}
{\mcitedefaultendpunct}{\mcitedefaultseppunct}\relax
\EndOfBibitem
\bibitem{Baak:2007gp}
 M.A.~Baak, PhD thesis, Vrije U., Amsterdam, 2007, {\small
  \url{http://www-public.slac.stanford.edu/sciDoc/docMeta.aspx?slacPubNumber=slac-r-858}}\relax
\mciteBstWouldAddEndPuncttrue
\mciteSetBstMidEndSepPunct{\mcitedefaultmidpunct}
{\mcitedefaultendpunct}{\mcitedefaultseppunct}\relax
\EndOfBibitem
\bibitem{DeBruyn:2012jp}
K.~De~Bruyn, R.~Fleischer, R.~Knegjens, M.~Merk, M.~Schiller, and N.~Tuning,
  \href{http://dx.doi.org/10.1016/j.nuclphysb.2012.11.012}{Nucl. Phys. {\bf
  B868},  351 (2013)}, \href{http://arxiv.org/abs/1208.6463}{{\tt
  arXiv:1208.6463 [hep-ph]}}\relax
\mciteBstWouldAddEndPuncttrue
\mciteSetBstMidEndSepPunct{\mcitedefaultmidpunct}
{\mcitedefaultendpunct}{\mcitedefaultseppunct}\relax
\EndOfBibitem
\bibitem{Kenzie:2016yee}
M.~Kenzie, M.~Martinelli, and N.~Tuning,
  \href{http://dx.doi.org/10.1103/PhysRevD.94.054021}{Phys. Rev. {\bf D94},
  054021 (2016)}, \href{http://arxiv.org/abs/1606.09129}{{\tt arXiv:1606.09129
  [hep-ph]}}\relax
\mciteBstWouldAddEndPuncttrue
\mciteSetBstMidEndSepPunct{\mcitedefaultmidpunct}
{\mcitedefaultendpunct}{\mcitedefaultseppunct}\relax
\EndOfBibitem
\bibitem{Aubert:2007qe}
{\babar} collaboration, B.~Aubert {\em et al.},
  \href{http://dx.doi.org/10.1103/PhysRevD.77.071102}{Phys. Rev. {\bf D77},
  071102 (2008)}, \href{http://arxiv.org/abs/0712.3469}{{\tt arXiv:0712.3469
  [hep-ex]}}\relax
\mciteBstWouldAddEndPuncttrue
\mciteSetBstMidEndSepPunct{\mcitedefaultmidpunct}
{\mcitedefaultendpunct}{\mcitedefaultseppunct}\relax
\EndOfBibitem
\bibitem{Dunietz:1987bv}
I.~Dunietz and R.~G. Sachs,
  \href{http://dx.doi.org/10.1103/PhysRevD.37.3186}{Phys. Rev. {\bf D37},  3186
  (1988)}, Erratum ibid.\
  \href{http://dx.doi.org/10.1103/PhysRevD.39.3515}{{\bf D39}, 3515},
  (1989)\relax
\mciteBstWouldAddEndPuncttrue
\mciteSetBstMidEndSepPunct{\mcitedefaultmidpunct}
{\mcitedefaultendpunct}{\mcitedefaultseppunct}\relax
\EndOfBibitem
\bibitem{Aleksan:1991nh}
R.~Aleksan, I.~Dunietz, and B.~Kayser,
  \href{http://dx.doi.org/10.1007/BF01559494}{Z. Phys. {\bf C54},  653
  (1992)}\relax
\mciteBstWouldAddEndPuncttrue
\mciteSetBstMidEndSepPunct{\mcitedefaultmidpunct}
{\mcitedefaultendpunct}{\mcitedefaultseppunct}\relax
\EndOfBibitem
\bibitem{Rama:2013voa}
M.~Rama, \href{http://dx.doi.org/10.1103/PhysRevD.89.014021}{Phys. Rev. {\bf
  D89},  014021 (2014)}, \href{http://arxiv.org/abs/1307.4384}{{\tt
  arXiv:1307.4384 [hep-ex]}}\relax
\mciteBstWouldAddEndPuncttrue
\mciteSetBstMidEndSepPunct{\mcitedefaultmidpunct}
{\mcitedefaultendpunct}{\mcitedefaultseppunct}\relax
\EndOfBibitem
\bibitem{delAmoSanchez:2010ji}
{\babar} collaboration, P.~del Amo~Sanchez {\em et al.},
  \href{http://dx.doi.org/10.1103/PhysRevD.82.072004}{Phys. Rev. {\bf D82},
  072004 (2010)}, \href{http://arxiv.org/abs/1007.0504}{{\tt arXiv:1007.0504
  [hep-ex]}}\relax
\mciteBstWouldAddEndPuncttrue
\mciteSetBstMidEndSepPunct{\mcitedefaultmidpunct}
{\mcitedefaultendpunct}{\mcitedefaultseppunct}\relax
\EndOfBibitem
\bibitem{Abe:2006hc}
{Belle} collaboration, K.~Abe {\em et al.},
  \href{http://dx.doi.org/10.1103/PhysRevD.73.051106}{Phys. Rev. {\bf D73},
  051106 (2006)}, \href{http://arxiv.org/abs/hep-ex/0601032}{{\tt
  arXiv:hep-ex/0601032}}\relax
\mciteBstWouldAddEndPuncttrue
\mciteSetBstMidEndSepPunct{\mcitedefaultmidpunct}
{\mcitedefaultendpunct}{\mcitedefaultseppunct}\relax
\EndOfBibitem
\bibitem{Aaltonen:2009hz}
{CDF} collaboration, T.~Aaltonen {\em et al.},
  \href{http://dx.doi.org/10.1103/PhysRevD.81.031105}{Phys. Rev. {\bf D81},
  031105 (2010)}, \href{http://arxiv.org/abs/0911.0425}{{\tt arXiv:0911.0425
  [hep-ex]}}\relax
\mciteBstWouldAddEndPuncttrue
\mciteSetBstMidEndSepPunct{\mcitedefaultmidpunct}
{\mcitedefaultendpunct}{\mcitedefaultseppunct}\relax
\EndOfBibitem
\bibitem{Aaij:2017ryw}
{LHCb} collaboration, R.~Aaij {\em et al.},
  \href{http://dx.doi.org/10.1016/j.physletb.2017.11.070}{Phys. Lett. {\bf
  B777},  16--30 (2018)}, \href{http://arxiv.org/abs/1708.06370}{{\tt
  arXiv:1708.06370 [hep-ex]}}\relax
\mciteBstWouldAddEndPuncttrue
\mciteSetBstMidEndSepPunct{\mcitedefaultmidpunct}
{\mcitedefaultendpunct}{\mcitedefaultseppunct}\relax
\EndOfBibitem
\bibitem{:2008jd}
{\babar} collaboration, B.~Aubert {\em et al.},
  \href{http://dx.doi.org/10.1103/PhysRevD.78.092002}{Phys. Rev. {\bf D78},
  092002 (2008)}, \href{http://arxiv.org/abs/0807.2408}{{\tt arXiv:0807.2408
  [hep-ex]}}\relax
\mciteBstWouldAddEndPuncttrue
\mciteSetBstMidEndSepPunct{\mcitedefaultmidpunct}
{\mcitedefaultendpunct}{\mcitedefaultseppunct}\relax
\EndOfBibitem
\bibitem{Aubert:2009yw}
{\babar} collaboration, B.~Aubert {\em et al.},
  \href{http://dx.doi.org/10.1103/PhysRevD.80.092001}{Phys. Rev. {\bf D80},
  092001 (2009)}, \href{http://arxiv.org/abs/0909.3981}{{\tt arXiv:0909.3981
  [hep-ex]}}\relax
\mciteBstWouldAddEndPuncttrue
\mciteSetBstMidEndSepPunct{\mcitedefaultmidpunct}
{\mcitedefaultendpunct}{\mcitedefaultseppunct}\relax
\EndOfBibitem
\bibitem{Aaij:2017glf}
{LHCb} collaboration, R.~Aaij {\em et al.},
  \href{http://dx.doi.org/10.1007/JHEP11(2017)156}{JHEP {\bf 11},  156 (2017)},
  \href{http://arxiv.org/abs/1709.05855}{{\tt arXiv:1709.05855 [hep-ex]}},
  Erratum ibid.\ \href{http://dx.doi.org/10.1007/JHEP05(2018)067}{{\bf 05},
  067}, (2018)\relax
\mciteBstWouldAddEndPuncttrue
\mciteSetBstMidEndSepPunct{\mcitedefaultmidpunct}
{\mcitedefaultendpunct}{\mcitedefaultseppunct}\relax
\EndOfBibitem
\bibitem{Aaij:2015ina}
{LHCb} collaboration, R.~Aaij {\em et al.},
  \href{http://dx.doi.org/10.1103/PhysRevD.92.112005}{Phys. Rev. {\bf D92},
  112005 (2015)}, \href{http://arxiv.org/abs/1505.07044}{{\tt arXiv:1505.07044
  [hep-ex]}}\relax
\mciteBstWouldAddEndPuncttrue
\mciteSetBstMidEndSepPunct{\mcitedefaultmidpunct}
{\mcitedefaultendpunct}{\mcitedefaultseppunct}\relax
\EndOfBibitem
\bibitem{Aaij:2014eha}
{LHCb} collaboration, R.~Aaij {\em et al.},
  \href{http://dx.doi.org/10.1103/PhysRevD.90.112002}{Phys. Rev. {\bf D90},
  112002 (2014)}, \href{http://arxiv.org/abs/1407.8136}{{\tt arXiv:1407.8136
  [hep-ex]}}\relax
\mciteBstWouldAddEndPuncttrue
\mciteSetBstMidEndSepPunct{\mcitedefaultmidpunct}
{\mcitedefaultendpunct}{\mcitedefaultseppunct}\relax
\EndOfBibitem
\bibitem{Aaij:2016bqv}
{LHCb} collaboration, R.~Aaij {\em et al.},
  \href{http://dx.doi.org/10.1103/PhysRevD.93.112018}{Phys. Rev. {\bf D93},
  112018 (2016)}, \href{http://arxiv.org/abs/1602.03455}{{\tt arXiv:1602.03455
  [hep-ex]}}\relax
\mciteBstWouldAddEndPuncttrue
\mciteSetBstMidEndSepPunct{\mcitedefaultmidpunct}
{\mcitedefaultendpunct}{\mcitedefaultseppunct}\relax
\EndOfBibitem
\bibitem{Malde:2015mha}
S.~Malde {\em et al.},
  \href{http://dx.doi.org/10.1016/j.physletb.2015.05.043}{Phys. Lett. {\bf
  B747},  9 (2015)}, \href{http://arxiv.org/abs/1504.05878}{{\tt
  arXiv:1504.05878 [hep-ex]}}\relax
\mciteBstWouldAddEndPuncttrue
\mciteSetBstMidEndSepPunct{\mcitedefaultmidpunct}
{\mcitedefaultendpunct}{\mcitedefaultseppunct}\relax
\EndOfBibitem
\bibitem{Aaij:2015jna}
{LHCb} collaboration, R.~Aaij {\em et al.},
  \href{http://dx.doi.org/10.1103/PhysRevD.91.112014}{Phys. Rev. {\bf D91},
  112014 (2015)}, \href{http://arxiv.org/abs/1504.05442}{{\tt arXiv:1504.05442
  [hep-ex]}}\relax
\mciteBstWouldAddEndPuncttrue
\mciteSetBstMidEndSepPunct{\mcitedefaultmidpunct}
{\mcitedefaultendpunct}{\mcitedefaultseppunct}\relax
\EndOfBibitem
\bibitem{Aaij:2016oso}
{LHCb} collaboration, R.~Aaij {\em et al.},
  \href{http://dx.doi.org/10.1016/j.physletb.2016.06.022}{Phys. Lett. {\bf
  B760},  117 (2016)}, \href{http://arxiv.org/abs/1603.08993}{{\tt
  arXiv:1603.08993 [hep-ex]}}\relax
\mciteBstWouldAddEndPuncttrue
\mciteSetBstMidEndSepPunct{\mcitedefaultmidpunct}
{\mcitedefaultendpunct}{\mcitedefaultseppunct}\relax
\EndOfBibitem
\bibitem{delAmoSanchez:2010dz}
{\babar} collaboration, P.~del Amo~Sanchez {\em et al.},
  \href{http://dx.doi.org/10.1103/PhysRevD.82.072006}{Phys. Rev. {\bf D82},
  072006 (2010)}, \href{http://arxiv.org/abs/1006.4241}{{\tt arXiv:1006.4241
  [hep-ex]}}\relax
\mciteBstWouldAddEndPuncttrue
\mciteSetBstMidEndSepPunct{\mcitedefaultmidpunct}
{\mcitedefaultendpunct}{\mcitedefaultseppunct}\relax
\EndOfBibitem
\bibitem{Belle:2011ac}
{Belle} collaboration, Y.~Horii {\em et al.},
  \href{http://dx.doi.org/10.1103/PhysRevLett.106.231803}{Phys. Rev. Lett. {\bf
  106},  231803 (2011)}, \href{http://arxiv.org/abs/1103.5951}{{\tt
  arXiv:1103.5951 [hep-ex]}}\relax
\mciteBstWouldAddEndPuncttrue
\mciteSetBstMidEndSepPunct{\mcitedefaultmidpunct}
{\mcitedefaultendpunct}{\mcitedefaultseppunct}\relax
\EndOfBibitem
\bibitem{Aaltonen:2011uu}
{CDF} collaboration, T.~Aaltonen {\em et al.},
  \href{http://dx.doi.org/10.1103/PhysRevD.84.091504}{Phys. Rev. {\bf D84},
  091504 (2011)}, \href{http://arxiv.org/abs/1108.5765}{{\tt arXiv:1108.5765
  [hep-ex]}}\relax
\mciteBstWouldAddEndPuncttrue
\mciteSetBstMidEndSepPunct{\mcitedefaultmidpunct}
{\mcitedefaultendpunct}{\mcitedefaultseppunct}\relax
\EndOfBibitem
\bibitem{Lees:2011up}
{\babar} collaboration, J.~P. Lees {\em et al.},
  \href{http://dx.doi.org/10.1103/PhysRevD.84.012002}{Phys. Rev. {\bf D84},
  012002 (2011)}, \href{http://arxiv.org/abs/1104.4472}{{\tt arXiv:1104.4472
  [hep-ex]}}\relax
\mciteBstWouldAddEndPuncttrue
\mciteSetBstMidEndSepPunct{\mcitedefaultmidpunct}
{\mcitedefaultendpunct}{\mcitedefaultseppunct}\relax
\EndOfBibitem
\bibitem{Nayak:2013tgg}
{Belle} collaboration, M.~Nayak {\em et al.},
  \href{http://dx.doi.org/10.1103/PhysRevD.88.091104}{Phys. Rev. {\bf D88},
  091104 (2013)}, \href{http://arxiv.org/abs/1310.1741}{{\tt arXiv:1310.1741
  [hep-ex]}}\relax
\mciteBstWouldAddEndPuncttrue
\mciteSetBstMidEndSepPunct{\mcitedefaultmidpunct}
{\mcitedefaultendpunct}{\mcitedefaultseppunct}\relax
\EndOfBibitem
\bibitem{:2009au}
{\babar} collaboration, B.~Aubert {\em et al.},
  \href{http://dx.doi.org/10.1103/PhysRevD.80.031102}{Phys. Rev. {\bf D80},
  031102 (2009)}, \href{http://arxiv.org/abs/0904.2112}{{\tt arXiv:0904.2112
  [hep-ex]}}\relax
\mciteBstWouldAddEndPuncttrue
\mciteSetBstMidEndSepPunct{\mcitedefaultmidpunct}
{\mcitedefaultendpunct}{\mcitedefaultseppunct}\relax
\EndOfBibitem
\bibitem{Negishi:2012uxa}
{Belle} collaboration, K.~Negishi {\em et al.},
  \href{http://dx.doi.org/10.1103/PhysRevD.86.011101}{Phys. Rev. {\bf D86},
  011101 (2012)}, \href{http://arxiv.org/abs/1205.0422}{{\tt arXiv:1205.0422
  [hep-ex]}}\relax
\mciteBstWouldAddEndPuncttrue
\mciteSetBstMidEndSepPunct{\mcitedefaultmidpunct}
{\mcitedefaultendpunct}{\mcitedefaultseppunct}\relax
\EndOfBibitem
\bibitem{Asner:2008ft}
{CLEO} collaboration, D.~M. Asner {\em et al.},
  \href{http://dx.doi.org/10.1103/PhysRevD.78.012001}{Phys. Rev. {\bf D78},
  012001 (2008)}, \href{http://arxiv.org/abs/0802.2268}{{\tt arXiv:0802.2268
  [hep-ex]}}\relax
\mciteBstWouldAddEndPuncttrue
\mciteSetBstMidEndSepPunct{\mcitedefaultmidpunct}
{\mcitedefaultendpunct}{\mcitedefaultseppunct}\relax
\EndOfBibitem
\bibitem{Lowery:2009id}
{CLEO} collaboration, N.~Lowrey {\em et al.},
  \href{http://dx.doi.org/10.1103/PhysRevD.80.031105}{Phys. Rev. {\bf D80},
  031105 (2009)}, \href{http://arxiv.org/abs/0903.4853}{{\tt arXiv:0903.4853
  [hep-ex]}}\relax
\mciteBstWouldAddEndPuncttrue
\mciteSetBstMidEndSepPunct{\mcitedefaultmidpunct}
{\mcitedefaultendpunct}{\mcitedefaultseppunct}\relax
\EndOfBibitem
\bibitem{Poluektov:2010wz}
{Belle} collaboration, A.~Poluektov {\em et al.},
  \href{http://dx.doi.org/10.1103/PhysRevD.81.112002}{Phys. Rev. {\bf D81},
  112002 (2010)}, \href{http://arxiv.org/abs/1003.3360}{{\tt arXiv:1003.3360
  [hep-ex]}}\relax
\mciteBstWouldAddEndPuncttrue
\mciteSetBstMidEndSepPunct{\mcitedefaultmidpunct}
{\mcitedefaultendpunct}{\mcitedefaultseppunct}\relax
\EndOfBibitem
\bibitem{delAmoSanchez:2010rq}
{\babar} collaboration, P.~del Amo~Sanchez {\em et al.},
  \href{http://dx.doi.org/10.1103/PhysRevLett.105.121801}{Phys. Rev. Lett. {\bf
  105},  121801 (2010)}, \href{http://arxiv.org/abs/1005.1096}{{\tt
  arXiv:1005.1096 [hep-ex]}}\relax
\mciteBstWouldAddEndPuncttrue
\mciteSetBstMidEndSepPunct{\mcitedefaultmidpunct}
{\mcitedefaultendpunct}{\mcitedefaultseppunct}\relax
\EndOfBibitem
\bibitem{Aaij:2014iba}
{LHCb} collaboration, R.~Aaij {\em et al.},
  \href{http://dx.doi.org/10.1016/j.nuclphysb.2014.09.015}{Nucl. Phys. {\bf
  B888},  169 (2014)}, \href{http://arxiv.org/abs/1407.6211}{{\tt
  arXiv:1407.6211 [hep-ex]}}\relax
\mciteBstWouldAddEndPuncttrue
\mciteSetBstMidEndSepPunct{\mcitedefaultmidpunct}
{\mcitedefaultendpunct}{\mcitedefaultseppunct}\relax
\EndOfBibitem
\bibitem{Poluektov:2006ia}
{Belle} collaboration, A.~Poluektov {\em et al.},
  \href{http://dx.doi.org/10.1103/PhysRevD.73.112009}{Phys. Rev. {\bf D73},
  112009 (2006)}, \href{http://arxiv.org/abs/hep-ex/0604054}{{\tt
  arXiv:hep-ex/0604054}}\relax
\mciteBstWouldAddEndPuncttrue
\mciteSetBstMidEndSepPunct{\mcitedefaultmidpunct}
{\mcitedefaultendpunct}{\mcitedefaultseppunct}\relax
\EndOfBibitem
\bibitem{Aaij:2016zlt}
{LHCb} collaboration, R.~Aaij {\em et al.},
  \href{http://dx.doi.org/10.1007/JHEP08(2016)137}{JHEP {\bf 08},  137 (2016)},
  \href{http://arxiv.org/abs/1605.01082}{{\tt arXiv:1605.01082 [hep-ex]}}\relax
\mciteBstWouldAddEndPuncttrue
\mciteSetBstMidEndSepPunct{\mcitedefaultmidpunct}
{\mcitedefaultendpunct}{\mcitedefaultseppunct}\relax
\EndOfBibitem
\bibitem{Aubert:2008yn}
{\babar} collaboration, B.~Aubert {\em et al.},
  \href{http://dx.doi.org/10.1103/PhysRevD.79.072003}{Phys. Rev. {\bf D79},
  072003 (2009)}, \href{http://arxiv.org/abs/0805.2001}{{\tt arXiv:0805.2001
  [hep-ex]}}\relax
\mciteBstWouldAddEndPuncttrue
\mciteSetBstMidEndSepPunct{\mcitedefaultmidpunct}
{\mcitedefaultendpunct}{\mcitedefaultseppunct}\relax
\EndOfBibitem
\bibitem{Briere:2009aa}
{CLEO} collaboration, R.~A. Briere {\em et al.},
  \href{http://dx.doi.org/10.1103/PhysRevD.80.032002}{Phys. Rev. {\bf D80},
  032002 (2009)}, \href{http://arxiv.org/abs/0903.1681}{{\tt arXiv:0903.1681
  [hep-ex]}}\relax
\mciteBstWouldAddEndPuncttrue
\mciteSetBstMidEndSepPunct{\mcitedefaultmidpunct}
{\mcitedefaultendpunct}{\mcitedefaultseppunct}\relax
\EndOfBibitem
\bibitem{Aihara:2012aw}
{Belle} collaboration, H.~Aihara {\em et al.},
  \href{http://dx.doi.org/10.1103/PhysRevD.85.112014}{Phys. Rev. {\bf D85},
  112014 (2012)}, \href{http://arxiv.org/abs/1204.6561}{{\tt arXiv:1204.6561
  [hep-ex]}}\relax
\mciteBstWouldAddEndPuncttrue
\mciteSetBstMidEndSepPunct{\mcitedefaultmidpunct}
{\mcitedefaultendpunct}{\mcitedefaultseppunct}\relax
\EndOfBibitem
\bibitem{Aaij:2014uva}
{LHCb} collaboration, R.~Aaij {\em et al.},
  \href{http://dx.doi.org/10.1007/JHEP10(2014)097}{JHEP {\bf 10},  97 (2014)},
  \href{http://arxiv.org/abs/1408.2748}{{\tt arXiv:1408.2748 [hep-ex]}}\relax
\mciteBstWouldAddEndPuncttrue
\mciteSetBstMidEndSepPunct{\mcitedefaultmidpunct}
{\mcitedefaultendpunct}{\mcitedefaultseppunct}\relax
\EndOfBibitem
\bibitem{Aaij:2018uns}
{LHCb} collaboration, R.~Aaij {\em et al.},
  \href{http://dx.doi.org/10.1007/JHEP08(2018)176}{JHEP {\bf 08},  176 (2018)},
  \href{http://arxiv.org/abs/1806.01202}{{\tt arXiv:1806.01202 [hep-ex]}},
  Erratum ibid.\ \href{http://dx.doi.org/10.1007/JHEP10(2018)107}{{\bf
  10},107}, (2018)\relax
\mciteBstWouldAddEndPuncttrue
\mciteSetBstMidEndSepPunct{\mcitedefaultmidpunct}
{\mcitedefaultendpunct}{\mcitedefaultseppunct}\relax
\EndOfBibitem
\bibitem{Negishi:2015vqa}
{Belle} collaboration, K.~Negishi {\em et al.},
  \href{http://dx.doi.org/10.1093/ptep/ptw030}{PTEP {\bf 2016},  043C01
  (2016)}, \href{http://arxiv.org/abs/1509.01098}{{\tt arXiv:1509.01098
  [hep-ex]}}\relax
\mciteBstWouldAddEndPuncttrue
\mciteSetBstMidEndSepPunct{\mcitedefaultmidpunct}
{\mcitedefaultendpunct}{\mcitedefaultseppunct}\relax
\EndOfBibitem
\bibitem{Aaij:2016nao}
{LHCb} collaboration, R.~Aaij {\em et al.},
  \href{http://dx.doi.org/10.1007/JHEP06(2016)131}{JHEP {\bf 06},  131 (2016)},
  \href{http://arxiv.org/abs/1604.01525}{{\tt arXiv:1604.01525 [hep-ex]}}\relax
\mciteBstWouldAddEndPuncttrue
\mciteSetBstMidEndSepPunct{\mcitedefaultmidpunct}
{\mcitedefaultendpunct}{\mcitedefaultseppunct}\relax
\EndOfBibitem
\bibitem{Aaij:2014dia}
{LHCb} collaboration, R.~Aaij {\em et al.},
  \href{http://dx.doi.org/10.1016/j.physletb.2014.03.051}{Phys. Lett. {\bf
  B733},  36 (2014)}, \href{http://arxiv.org/abs/1402.2982}{{\tt
  arXiv:1402.2982 [hep-ex]}}\relax
\mciteBstWouldAddEndPuncttrue
\mciteSetBstMidEndSepPunct{\mcitedefaultmidpunct}
{\mcitedefaultendpunct}{\mcitedefaultseppunct}\relax
\EndOfBibitem
\bibitem{Insler:2012pm}
{CLEO} collaboration, J.~Insler {\em et al.},
  \href{http://dx.doi.org/10.1103/PhysRevD.85.092016}{Phys. Rev. {\bf D85},
  092016 (2012)}, \href{http://arxiv.org/abs/1203.3804}{{\tt arXiv:1203.3804
  [hep-ex]}}\relax
\mciteBstWouldAddEndPuncttrue
\mciteSetBstMidEndSepPunct{\mcitedefaultmidpunct}
{\mcitedefaultendpunct}{\mcitedefaultseppunct}\relax
\EndOfBibitem
\bibitem{Lees:2013nha}
{\babar} collaboration, J.~P. Lees {\em et al.},
  \href{http://dx.doi.org/10.1103/PhysRevD.87.052015}{Phys. Rev. {\bf D87},
  052015 (2013)}, \href{http://arxiv.org/abs/1301.1029}{{\tt arXiv:1301.1029
  [hep-ex]}}\relax
\mciteBstWouldAddEndPuncttrue
\mciteSetBstMidEndSepPunct{\mcitedefaultmidpunct}
{\mcitedefaultendpunct}{\mcitedefaultseppunct}\relax
\EndOfBibitem
\bibitem{Aaij:2016kjh}
{LHCb} collaboration, R.~Aaij {\em et al.},
  \href{http://dx.doi.org/10.1007/JHEP12(2016)087}{JHEP {\bf 12},  087 (2016)},
  \href{http://arxiv.org/abs/1611.03076}{{\tt arXiv:1611.03076 [hep-ex]}}\relax
\mciteBstWouldAddEndPuncttrue
\mciteSetBstMidEndSepPunct{\mcitedefaultmidpunct}
{\mcitedefaultendpunct}{\mcitedefaultseppunct}\relax
\EndOfBibitem
\bibitem{LHCb-CONF-2018-002}
{LHCb} collaboration,
  \href{http://cds.cern.ch/record/2319289}{LHCb-CONF-2018-002}, {2018}\relax
\mciteBstWouldAddEndPuncttrue
\mciteSetBstMidEndSepPunct{\mcitedefaultmidpunct}
{\mcitedefaultendpunct}{\mcitedefaultseppunct}\relax
\EndOfBibitem
\bibitem{Aaij:2013zfa}
{LHCb} collaboration, R.~Aaij {\em et al.},
  \href{http://dx.doi.org/10.1016/j.physletb.2013.08.020}{Phys. Lett. {\bf
  B726},  151 (2013)}, \href{http://arxiv.org/abs/1305.2050}{{\tt
  arXiv:1305.2050 [hep-ex]}}\relax
\mciteBstWouldAddEndPuncttrue
\mciteSetBstMidEndSepPunct{\mcitedefaultmidpunct}
{\mcitedefaultendpunct}{\mcitedefaultseppunct}\relax
\EndOfBibitem
\bibitem{Evans:2016tlp}
T.~Evans, S.~Harnew, J.~Libby, S.~Malde, J.~Rademacker, and G.~Wilkinson,
  \href{http://dx.doi.org/10.1016/j.physletb.2016.04.037}{Phys. Lett. {\bf
  B757},  520 (2016)}, \href{http://arxiv.org/abs/1602.07430}{{\tt
  arXiv:1602.07430 [hep-ex]}}\relax
\mciteBstWouldAddEndPuncttrue
\mciteSetBstMidEndSepPunct{\mcitedefaultmidpunct}
{\mcitedefaultendpunct}{\mcitedefaultseppunct}\relax
\EndOfBibitem
\bibitem{Aaij:2015lsa}
{LHCb} collaboration, R.~Aaij {\em et al.},
  \href{http://dx.doi.org/10.1103/PhysRevD.93.052018}{Phys. Rev. {\bf D93},
  052018 (2016)}, \href{http://arxiv.org/abs/1509.06628}{{\tt arXiv:1509.06628
  [hep-ex]}}\relax
\mciteBstWouldAddEndPuncttrue
\mciteSetBstMidEndSepPunct{\mcitedefaultmidpunct}
{\mcitedefaultendpunct}{\mcitedefaultseppunct}\relax
\EndOfBibitem
\bibitem{Silva:1999bd}
J.~P. Silva and A.~Soffer,
  \href{http://dx.doi.org/10.1103/PhysRevD.61.112001}{Phys. Rev. {\bf D61},
  112001 (2000)}, \href{http://arxiv.org/abs/hep-ph/9912242}{{\tt
  arXiv:hep-ph/9912242 [hep-ph]}}\relax
\mciteBstWouldAddEndPuncttrue
\mciteSetBstMidEndSepPunct{\mcitedefaultmidpunct}
{\mcitedefaultendpunct}{\mcitedefaultseppunct}\relax
\EndOfBibitem
\bibitem{Grossman:2005rp}
Y.~Grossman, A.~Soffer, and J.~Zupan,
  \href{http://dx.doi.org/10.1103/PhysRevD.72.031501}{Phys. Rev. {\bf D72},
  031501 (2005)}, \href{http://arxiv.org/abs/hep-ph/0505270}{{\tt
  arXiv:hep-ph/0505270 [hep-ph]}}\relax
\mciteBstWouldAddEndPuncttrue
\mciteSetBstMidEndSepPunct{\mcitedefaultmidpunct}
{\mcitedefaultendpunct}{\mcitedefaultseppunct}\relax
\EndOfBibitem
\bibitem{HFLAV_sl:webpage}
 HFLAV averages of semileptonic $B$ decays,
  \url{https://hflav.web.cern.ch/content/semileptonic-b-decays}\relax
\mciteBstWouldAddEndPuncttrue
\mciteSetBstMidEndSepPunct{\mcitedefaultmidpunct}
{\mcitedefaultendpunct}{\mcitedefaultseppunct}\relax
\EndOfBibitem
\bibitem{Neubert:1993mb}
M.~Neubert, \href{http://dx.doi.org/10.1016/0370-1573(94)90091-4}{Phys. Rept.
  {\bf 245},  259 (1994)}, \href{http://arxiv.org/abs/hep-ph/9306320}{{\tt
  arXiv:hep-ph/9306320 [hep-ph]}}\relax
\mciteBstWouldAddEndPuncttrue
\mciteSetBstMidEndSepPunct{\mcitedefaultmidpunct}
{\mcitedefaultendpunct}{\mcitedefaultseppunct}\relax
\EndOfBibitem
\bibitem{Sirlin:1981ie}
A.~Sirlin, \href{http://dx.doi.org/10.1016/0550-3213(82)90303-0}{Nucl. Phys.
  {\bf B196},  83 (1982)}\relax
\mciteBstWouldAddEndPuncttrue
\mciteSetBstMidEndSepPunct{\mcitedefaultmidpunct}
{\mcitedefaultendpunct}{\mcitedefaultseppunct}\relax
\EndOfBibitem
\bibitem{HFLAV_sl:inputparams}
 $B$ semileptonic decays common input parameters,
  \url{https://hflav-eos.web.cern.ch/hflav-eos/semi/spring19/common/common.param.writeup18}\relax
\mciteBstWouldAddEndPuncttrue
\mciteSetBstMidEndSepPunct{\mcitedefaultmidpunct}
{\mcitedefaultendpunct}{\mcitedefaultseppunct}\relax
\EndOfBibitem
\bibitem{Buskulic:1996yq}
{ALEPH} collaboration, D.~Buskulic {\em et al.},
  \href{http://dx.doi.org/10.1016/S0370-2693(97)00071-3}{Phys. Lett. {\bf
  B395},  373 (1997)}\relax
\mciteBstWouldAddEndPuncttrue
\mciteSetBstMidEndSepPunct{\mcitedefaultmidpunct}
{\mcitedefaultendpunct}{\mcitedefaultseppunct}\relax
\EndOfBibitem
\bibitem{Abbiendi:2000hk}
{OPAL} collaboration, G.~Abbiendi {\em et al.},
  \href{http://dx.doi.org/10.1016/S0370-2693(00)00457-3}{Phys. Lett. {\bf
  B482},  15 (2000)}, \href{http://arxiv.org/abs/hep-ex/0003013}{{\tt
  arXiv:hep-ex/0003013}}\relax
\mciteBstWouldAddEndPuncttrue
\mciteSetBstMidEndSepPunct{\mcitedefaultmidpunct}
{\mcitedefaultendpunct}{\mcitedefaultseppunct}\relax
\EndOfBibitem
\bibitem{Abreu:2001ic}
{DELPHI} collaboration, P.~Abreu {\em et al.},
  \href{http://dx.doi.org/10.1016/S0370-2693(01)00569-X}{Phys. Lett. {\bf
  B510},  55 (2001)}, \href{http://arxiv.org/abs/hep-ex/0104026}{{\tt
  arXiv:hep-ex/0104026}}\relax
\mciteBstWouldAddEndPuncttrue
\mciteSetBstMidEndSepPunct{\mcitedefaultmidpunct}
{\mcitedefaultendpunct}{\mcitedefaultseppunct}\relax
\EndOfBibitem
\bibitem{Abdallah:2004rz}
{DELPHI} collaboration, J.~Abdallah {\em et al.},
  \href{http://dx.doi.org/10.1140/epjc/s2004-01598-6}{Eur. Phys. J. {\bf C33},
  213 (2004)}, \href{http://arxiv.org/abs/hep-ex/0401023}{{\tt
  arXiv:hep-ex/0401023}}\relax
\mciteBstWouldAddEndPuncttrue
\mciteSetBstMidEndSepPunct{\mcitedefaultmidpunct}
{\mcitedefaultendpunct}{\mcitedefaultseppunct}\relax
\EndOfBibitem
\bibitem{Adam:2002uw}
{CLEO} collaboration, N.~E. Adam {\em et al.},
  \href{http://dx.doi.org/10.1103/PhysRevD.67.032001}{Phys. Rev. {\bf D67},
  032001 (2003)}, \href{http://arxiv.org/abs/hep-ex/0210040}{{\tt
  arXiv:hep-ex/0210040}}\relax
\mciteBstWouldAddEndPuncttrue
\mciteSetBstMidEndSepPunct{\mcitedefaultmidpunct}
{\mcitedefaultendpunct}{\mcitedefaultseppunct}\relax
\EndOfBibitem
\bibitem{Waheed:2018djm}
{Belle} collaboration, E.~Waheed {\em et al.},
  \href{http://dx.doi.org/10.1103/PhysRevD.100.052007}{Phys. Rev. D {\bf 100},
  052007 (2019)}, \href{http://arxiv.org/abs/1809.03290}{{\tt arXiv:1809.03290
  [hep-ex]}}\relax
\mciteBstWouldAddEndPuncttrue
\mciteSetBstMidEndSepPunct{\mcitedefaultmidpunct}
{\mcitedefaultendpunct}{\mcitedefaultseppunct}\relax
\EndOfBibitem
\bibitem{Abdesselam:2017kjf}
{Belle} collaboration, A.~Abdesselam {\em et al.},
  \href{http://arxiv.org/abs/1702.01521}{{\tt arXiv:1702.01521 [hep-ex]}}\relax
\mciteBstWouldAddEndPuncttrue
\mciteSetBstMidEndSepPunct{\mcitedefaultmidpunct}
{\mcitedefaultendpunct}{\mcitedefaultseppunct}\relax
\EndOfBibitem
\bibitem{Aubert:2006mb}
{\babar} collaboration, B.~Aubert {\em et al.},
  \href{http://dx.doi.org/10.1103/PhysRevD.77.032002}{Phys. Rev. {\bf D77},
  032002 (2008)}, \href{http://arxiv.org/abs/0705.4008}{{\tt arXiv:0705.4008
  [hep-ex]}}\relax
\mciteBstWouldAddEndPuncttrue
\mciteSetBstMidEndSepPunct{\mcitedefaultmidpunct}
{\mcitedefaultendpunct}{\mcitedefaultseppunct}\relax
\EndOfBibitem
\bibitem{Aubert:vcbExcl}
{\babar} collaboration, B.~Aubert {\em et al.},
  \href{http://dx.doi.org/10.1103/PhysRevLett.100.151802}{Phys. Rev. Lett. {\bf
  100},  151802 (2008)}, \href{http://arxiv.org/abs/0712.3503}{{\tt
  arXiv:0712.3503 [hep-ex]}}\relax
\mciteBstWouldAddEndPuncttrue
\mciteSetBstMidEndSepPunct{\mcitedefaultmidpunct}
{\mcitedefaultendpunct}{\mcitedefaultseppunct}\relax
\EndOfBibitem
\bibitem{Aubert:2009_3}
{\babar} collaboration, B.~Aubert {\em et al.},
  \href{http://dx.doi.org/10.1103/PhysRevLett.100.231803}{Phys. Rev. Lett. {\bf
  100},  231803 (2008)}, \href{http://arxiv.org/abs/0712.3493}{{\tt
  arXiv:0712.3493 [hep-ex]}}\relax
\mciteBstWouldAddEndPuncttrue
\mciteSetBstMidEndSepPunct{\mcitedefaultmidpunct}
{\mcitedefaultendpunct}{\mcitedefaultseppunct}\relax
\EndOfBibitem
\bibitem{CLN}
I.~Caprini, L.~Lellouch, and M.~Neubert,
  \href{http://dx.doi.org/10.1016/S0550-3213(98)00350-2}{Nucl. Phys. {\bf
  B530},  153 (1998)}, \href{http://arxiv.org/abs/hep-ph/9712417}{{\tt
  arXiv:hep-ph/9712417}}\relax
\mciteBstWouldAddEndPuncttrue
\mciteSetBstMidEndSepPunct{\mcitedefaultmidpunct}
{\mcitedefaultendpunct}{\mcitedefaultseppunct}\relax
\EndOfBibitem
\bibitem{Aubert:2009_1}
{\babar} collaboration, B.~Aubert {\em et al.},
  \href{http://dx.doi.org/10.1103/PhysRevD.79.012002}{Phys. Rev. {\bf D79},
  012002 (2009)}, \href{http://arxiv.org/abs/0809.0828}{{\tt arXiv:0809.0828
  [hep-ex]}}\relax
\mciteBstWouldAddEndPuncttrue
\mciteSetBstMidEndSepPunct{\mcitedefaultmidpunct}
{\mcitedefaultendpunct}{\mcitedefaultseppunct}\relax
\EndOfBibitem
\bibitem{Bailey:2014tva}
{Fermilab Lattice and MILC} collaborations, J.~A. Bailey {\em et al.},
  \href{http://dx.doi.org/10.1103/PhysRevD.89.114504}{Phys. Rev. {\bf D89},
  114504 (2014)}, \href{http://arxiv.org/abs/1403.0635}{{\tt arXiv:1403.0635
  [hep-lat]}}\relax
\mciteBstWouldAddEndPuncttrue
\mciteSetBstMidEndSepPunct{\mcitedefaultmidpunct}
{\mcitedefaultendpunct}{\mcitedefaultseppunct}\relax
\EndOfBibitem
\bibitem{Harrison:2017fmw}
{HPQCD} collaboration, J.~Harrison, C.~Davies, and M.~Wingate,
  \href{http://dx.doi.org/10.1103/PhysRevD.97.054502}{Phys.\ Rev.\ D {\bf 97},
  054502 (2018)}, \href{http://arxiv.org/abs/1711.11013}{{\tt arXiv:1711.11013
  [hep-lat]}}\relax
\mciteBstWouldAddEndPuncttrue
\mciteSetBstMidEndSepPunct{\mcitedefaultmidpunct}
{\mcitedefaultendpunct}{\mcitedefaultseppunct}\relax
\EndOfBibitem
\bibitem{Lattice:2015rga}
{MILC} collaboration, J.~A. Bailey {\em et al.},
  \href{http://dx.doi.org/10.1103/PhysRevD.92.034506}{Phys. Rev. {\bf D92},
  034506 (2015)}, \href{http://arxiv.org/abs/1503.07237}{{\tt arXiv:1503.07237
  [hep-lat]}}\relax
\mciteBstWouldAddEndPuncttrue
\mciteSetBstMidEndSepPunct{\mcitedefaultmidpunct}
{\mcitedefaultendpunct}{\mcitedefaultseppunct}\relax
\EndOfBibitem
\bibitem{Boyd:1997kz}
C.~Boyd, B.~Grinstein, and R.~F. Lebed,
  \href{http://dx.doi.org/10.1103/PhysRevD.56.6895}{Phys. Rev. D {\bf 56},
  6895--6911 (1997)}, \href{http://arxiv.org/abs/hep-ph/9705252}{{\tt
  arXiv:hep-ph/9705252}}\relax
\mciteBstWouldAddEndPuncttrue
\mciteSetBstMidEndSepPunct{\mcitedefaultmidpunct}
{\mcitedefaultendpunct}{\mcitedefaultseppunct}\relax
\EndOfBibitem
\bibitem{Grinstein:2017nlq}
B.~Grinstein and A.~Kobach,
  \href{http://dx.doi.org/10.1016/j.physletb.2017.05.078}{Phys. Lett. B {\bf
  771},  359--364 (2017)}, \href{http://arxiv.org/abs/1703.08170}{{\tt
  arXiv:1703.08170 [hep-ph]}}\relax
\mciteBstWouldAddEndPuncttrue
\mciteSetBstMidEndSepPunct{\mcitedefaultmidpunct}
{\mcitedefaultendpunct}{\mcitedefaultseppunct}\relax
\EndOfBibitem
\bibitem{Bigi:2017njr}
D.~Bigi, P.~Gambino, and S.~Schacht,
  \href{http://dx.doi.org/10.1016/j.physletb.2017.04.022}{Phys. Lett. B {\bf
  769},  441--445 (2017)}, \href{http://arxiv.org/abs/1703.06124}{{\tt
  arXiv:1703.06124 [hep-ph]}}\relax
\mciteBstWouldAddEndPuncttrue
\mciteSetBstMidEndSepPunct{\mcitedefaultmidpunct}
{\mcitedefaultendpunct}{\mcitedefaultseppunct}\relax
\EndOfBibitem
\bibitem{Dey:2019bgc}
{BaBar} collaboration, J.~P. Lees {\em et al.},
  \href{http://arxiv.org/abs/1903.10002}{{\tt arXiv:1903.10002 [hep-ex]}}\relax
\mciteBstWouldAddEndPuncttrue
\mciteSetBstMidEndSepPunct{\mcitedefaultmidpunct}
{\mcitedefaultendpunct}{\mcitedefaultseppunct}\relax
\EndOfBibitem
\bibitem{Bartelt:1998dq}
{CLEO} collaboration, J.~E. Bartelt {\em et al.},
  \href{http://dx.doi.org/10.1103/PhysRevLett.82.3746}{Phys. Rev. Lett. {\bf
  82},  3746 (1999)}, \href{http://arxiv.org/abs/hep-ex/9811042}{{\tt
  arXiv:hep-ex/9811042}}\relax
\mciteBstWouldAddEndPuncttrue
\mciteSetBstMidEndSepPunct{\mcitedefaultmidpunct}
{\mcitedefaultendpunct}{\mcitedefaultseppunct}\relax
\EndOfBibitem
\bibitem{Aubert:2009_2}
{\babar} collaboration, B.~Aubert {\em et al.},
  \href{http://dx.doi.org/10.1103/PhysRevLett.104.011802}{Phys. Rev. Lett. {\bf
  104},  011802 (2010)}, \href{http://arxiv.org/abs/0904.4063}{{\tt
  arXiv:0904.4063 [hep-ex]}}\relax
\mciteBstWouldAddEndPuncttrue
\mciteSetBstMidEndSepPunct{\mcitedefaultmidpunct}
{\mcitedefaultendpunct}{\mcitedefaultseppunct}\relax
\EndOfBibitem
\bibitem{Glattauer:2015teq}
{Belle} collaboration, R.~Glattauer {\em et al.},
  \href{http://dx.doi.org/10.1103/PhysRevD.93.032006}{Phys. Rev. {\bf D93},
  032006 (2016)}, \href{http://arxiv.org/abs/1510.03657}{{\tt arXiv:1510.03657
  [hep-ex]}}\relax
\mciteBstWouldAddEndPuncttrue
\mciteSetBstMidEndSepPunct{\mcitedefaultmidpunct}
{\mcitedefaultendpunct}{\mcitedefaultseppunct}\relax
\EndOfBibitem
\bibitem{Na:2015kha}
{HPQCD} collaboration, H.~Na, C.~M. Bouchard, G.~P. Lepage, C.~Monahan, and
  J.~Shigemitsu, \href{http://dx.doi.org/10.1103/PhysRevD.93.119906}{Phys. Rev.
  {\bf D92},  054510 (2015)}, \href{http://arxiv.org/abs/1505.03925}{{\tt
  arXiv:1505.03925 [hep-lat]}}, erratum ibid.\
  \href{http://dx.doi.org/10.1103/PhysRevD.92.054510}{{\bf D93}, 119906},
  (2016)\relax
\mciteBstWouldAddEndPuncttrue
\mciteSetBstMidEndSepPunct{\mcitedefaultmidpunct}
{\mcitedefaultendpunct}{\mcitedefaultseppunct}\relax
\EndOfBibitem
\bibitem{Aubert:2009ac}
{BaBar} collaboration, B.~Aubert {\em et al.},
  \href{http://dx.doi.org/10.1103/PhysRevLett.104.011802}{Phys. Rev. Lett. {\bf
  104},  011802 (2010)}, \href{http://arxiv.org/abs/0904.4063}{{\tt
  arXiv:0904.4063 [hep-ex]}}\relax
\mciteBstWouldAddEndPuncttrue
\mciteSetBstMidEndSepPunct{\mcitedefaultmidpunct}
{\mcitedefaultendpunct}{\mcitedefaultseppunct}\relax
\EndOfBibitem
\bibitem{Vossen:2018zeg}
{Belle} collaboration, A.~Vossen {\em et al.},
  \href{http://dx.doi.org/10.1103/PhysRevD.98.012005}{Phys. Rev. {\bf D98},
  012005 (2018)}, \href{http://arxiv.org/abs/1803.06444}{{\tt arXiv:1803.06444
  [hep-ex]}}\relax
\mciteBstWouldAddEndPuncttrue
\mciteSetBstMidEndSepPunct{\mcitedefaultmidpunct}
{\mcitedefaultendpunct}{\mcitedefaultseppunct}\relax
\EndOfBibitem
\bibitem{Isgur:1991wq}
N.~Isgur and M.~B. Wise,
  \href{http://dx.doi.org/10.1103/PhysRevLett.66.1130}{Phys. Rev. Lett. {\bf
  66},  1130 (1991)}\relax
\mciteBstWouldAddEndPuncttrue
\mciteSetBstMidEndSepPunct{\mcitedefaultmidpunct}
{\mcitedefaultendpunct}{\mcitedefaultseppunct}\relax
\EndOfBibitem
\bibitem{Aubert:2009_4}
{\babar} collaboration, B.~Aubert {\em et al.},
  \href{http://dx.doi.org/10.1103/PhysRevLett.101.261802}{Phys. Rev. Lett. {\bf
  101},  261802 (2008)}, \href{http://arxiv.org/abs/0808.0528}{{\tt
  arXiv:0808.0528 [hep-ex]}}\relax
\mciteBstWouldAddEndPuncttrue
\mciteSetBstMidEndSepPunct{\mcitedefaultmidpunct}
{\mcitedefaultendpunct}{\mcitedefaultseppunct}\relax
\EndOfBibitem
\bibitem{Live:Dss}
{Belle} collaboration, D.~Liventsev {\em et al.},
  \href{http://dx.doi.org/10.1103/PhysRevD.77.091503}{Phys. Rev. {\bf D77},
  091503 (2008)}, \href{http://arxiv.org/abs/0711.3252}{{\tt arXiv:0711.3252
  [hep-ex]}}\relax
\mciteBstWouldAddEndPuncttrue
\mciteSetBstMidEndSepPunct{\mcitedefaultmidpunct}
{\mcitedefaultendpunct}{\mcitedefaultseppunct}\relax
\EndOfBibitem
\bibitem{Abdallah:2005cx}
{DELPHI} collaboration, J.~Abdallah {\em et al.},
  \href{http://dx.doi.org/10.1140/epjc/s2005-02406-7}{Eur. Phys. J. {\bf C45},
  35 (2006)}, \href{http://arxiv.org/abs/hep-ex/0510024}{{\tt
  arXiv:hep-ex/0510024}}\relax
\mciteBstWouldAddEndPuncttrue
\mciteSetBstMidEndSepPunct{\mcitedefaultmidpunct}
{\mcitedefaultendpunct}{\mcitedefaultseppunct}\relax
\EndOfBibitem
\bibitem{Aleph:Dss}
{ALEPH} collaboration, D.~Buskulic {\em et al.},
  \href{http://dx.doi.org/10.1007/s002880050351}{Z. Phys. {\bf C73},  601
  (1997)}\relax
\mciteBstWouldAddEndPuncttrue
\mciteSetBstMidEndSepPunct{\mcitedefaultmidpunct}
{\mcitedefaultendpunct}{\mcitedefaultseppunct}\relax
\EndOfBibitem
\bibitem{opal:Dss}
{OPAL} collaboration, G.~Abbiendi {\em et al.},
  \href{http://dx.doi.org/10.1140/epjc/s2003-01322-2}{Eur. Phys. J. {\bf C30},
  467 (2003)}, \href{http://arxiv.org/abs/hep-ex/0301018}{{\tt
  arXiv:hep-ex/0301018}}\relax
\mciteBstWouldAddEndPuncttrue
\mciteSetBstMidEndSepPunct{\mcitedefaultmidpunct}
{\mcitedefaultendpunct}{\mcitedefaultseppunct}\relax
\EndOfBibitem
\bibitem{cleo:Dss}
{CLEO} collaboration, A.~Anastassov {\em et al.},
  \href{http://dx.doi.org/10.1103/PhysRevLett.80.4127}{Phys. Rev. Lett. {\bf
  80},  4127 (1998)}, \href{http://arxiv.org/abs/hep-ex/9708035}{{\tt
  arXiv:hep-ex/9708035}}\relax
\mciteBstWouldAddEndPuncttrue
\mciteSetBstMidEndSepPunct{\mcitedefaultmidpunct}
{\mcitedefaultendpunct}{\mcitedefaultseppunct}\relax
\EndOfBibitem
\bibitem{D0:Dss}
{\dzero} collaboration, V.~M. Abazov {\em et al.},
  \href{http://dx.doi.org/10.1103/PhysRevLett.95.171803}{Phys. Rev. Lett. {\bf
  95},  171803 (2005)}, \href{http://arxiv.org/abs/hep-ex/0507046}{{\tt
  arXiv:hep-ex/0507046}}\relax
\mciteBstWouldAddEndPuncttrue
\mciteSetBstMidEndSepPunct{\mcitedefaultmidpunct}
{\mcitedefaultendpunct}{\mcitedefaultseppunct}\relax
\EndOfBibitem
\bibitem{Aubert:2008zc}
{\babar} collaboration, B.~Aubert {\em et al.},
  \href{http://dx.doi.org/10.1103/PhysRevLett.103.051803}{Phys. Rev. Lett. {\bf
  103},  051803 (2009)}, \href{http://arxiv.org/abs/0808.0333}{{\tt
  arXiv:0808.0333 [hep-ex]}}\relax
\mciteBstWouldAddEndPuncttrue
\mciteSetBstMidEndSepPunct{\mcitedefaultmidpunct}
{\mcitedefaultendpunct}{\mcitedefaultseppunct}\relax
\EndOfBibitem
\bibitem{Benson:2003kp}
D.~Benson, I.~I. Bigi, T.~Mannel, and N.~Uraltsev,
  \href{http://dx.doi.org/10.1016/S0550-3213(03)00452-8}{Nucl. Phys. {\bf
  B665},  367 (2003)}, \href{http://arxiv.org/abs/hep-ph/0302262}{{\tt
  arXiv:hep-ph/0302262}}\relax
\mciteBstWouldAddEndPuncttrue
\mciteSetBstMidEndSepPunct{\mcitedefaultmidpunct}
{\mcitedefaultendpunct}{\mcitedefaultseppunct}\relax
\EndOfBibitem
\bibitem{Gambino:2004qm}
P.~Gambino and N.~Uraltsev,
  \href{http://dx.doi.org/10.1140/epjc/s2004-01671-2}{Eur. Phys. J. {\bf C34},
  181 (2004)}, \href{http://arxiv.org/abs/hep-ph/0401063}{{\tt
  arXiv:hep-ph/0401063}}\relax
\mciteBstWouldAddEndPuncttrue
\mciteSetBstMidEndSepPunct{\mcitedefaultmidpunct}
{\mcitedefaultendpunct}{\mcitedefaultseppunct}\relax
\EndOfBibitem
\bibitem{Gambino:2011cq}
P.~Gambino, \href{http://dx.doi.org/10.1007/JHEP09(2011)055}{JHEP {\bf 09},
  055 (2011)}, \href{http://arxiv.org/abs/1107.3100}{{\tt arXiv:1107.3100
  [hep-ph]}}\relax
\mciteBstWouldAddEndPuncttrue
\mciteSetBstMidEndSepPunct{\mcitedefaultmidpunct}
{\mcitedefaultendpunct}{\mcitedefaultseppunct}\relax
\EndOfBibitem
\bibitem{Alberti:2014yda}
A.~Alberti, P.~Gambino, K.~J. Healey, and S.~Nandi,
  \href{http://dx.doi.org/10.1103/PhysRevLett.114.061802}{Phys. Rev. Lett. {\bf
  114},  061802 (2015)}, \href{http://arxiv.org/abs/1411.6560}{{\tt
  arXiv:1411.6560 [hep-ph]}}\relax
\mciteBstWouldAddEndPuncttrue
\mciteSetBstMidEndSepPunct{\mcitedefaultmidpunct}
{\mcitedefaultendpunct}{\mcitedefaultseppunct}\relax
\EndOfBibitem
\bibitem{Bauer:2004ve}
C.~W. Bauer, Z.~Ligeti, M.~Luke, A.~V. Manohar, and M.~Trott,
  \href{http://dx.doi.org/10.1103/PhysRevD.70.094017}{Phys. Rev. {\bf D70},
  094017 (2004)}, \href{http://arxiv.org/abs/hep-ph/0408002}{{\tt
  arXiv:hep-ph/0408002}}\relax
\mciteBstWouldAddEndPuncttrue
\mciteSetBstMidEndSepPunct{\mcitedefaultmidpunct}
{\mcitedefaultendpunct}{\mcitedefaultseppunct}\relax
\EndOfBibitem
\bibitem{Aubert:2009qda}
{\babar} collaboration, B.~Aubert {\em et al.},
  \href{http://dx.doi.org/10.1103/PhysRevD.81.032003}{Phys. Rev. {\bf D81},
  032003 (2010)}, \href{http://arxiv.org/abs/0908.0415}{{\tt arXiv:0908.0415
  [hep-ex]}}\relax
\mciteBstWouldAddEndPuncttrue
\mciteSetBstMidEndSepPunct{\mcitedefaultmidpunct}
{\mcitedefaultendpunct}{\mcitedefaultseppunct}\relax
\EndOfBibitem
\bibitem{Aubert:2004td}
{\babar} collaboration, B.~Aubert {\em et al.},
  \href{http://dx.doi.org/10.1103/PhysRevD.69.111104}{Phys. Rev. {\bf D69},
  111104 (2004)}, \href{http://arxiv.org/abs/hep-ex/0403030}{{\tt
  arXiv:hep-ex/0403030}}\relax
\mciteBstWouldAddEndPuncttrue
\mciteSetBstMidEndSepPunct{\mcitedefaultmidpunct}
{\mcitedefaultendpunct}{\mcitedefaultseppunct}\relax
\EndOfBibitem
\bibitem{Schwanda:2006nf}
{Belle} collaboration, C.~Schwanda {\em et al.},
  \href{http://dx.doi.org/10.1103/PhysRevD.75.032005}{Phys. Rev. {\bf D75},
  032005 (2007)}, \href{http://arxiv.org/abs/hep-ex/0611044}{{\tt
  arXiv:hep-ex/0611044}}\relax
\mciteBstWouldAddEndPuncttrue
\mciteSetBstMidEndSepPunct{\mcitedefaultmidpunct}
{\mcitedefaultendpunct}{\mcitedefaultseppunct}\relax
\EndOfBibitem
\bibitem{Urquijo:2006wd}
{Belle} collaboration, P.~Urquijo {\em et al.},
  \href{http://dx.doi.org/10.1103/PhysRevD.75.032001}{Phys. Rev. {\bf D75},
  032001 (2007)}, \href{http://arxiv.org/abs/hep-ex/0610012}{{\tt
  arXiv:hep-ex/0610012}}\relax
\mciteBstWouldAddEndPuncttrue
\mciteSetBstMidEndSepPunct{\mcitedefaultmidpunct}
{\mcitedefaultendpunct}{\mcitedefaultseppunct}\relax
\EndOfBibitem
\bibitem{Acosta:2005qh}
{CDF} collaboration, D.~E. Acosta {\em et al.},
  \href{http://dx.doi.org/10.1103/PhysRevD.71.051103}{Phys. Rev. {\bf D71},
  051103 (2005)}, \href{http://arxiv.org/abs/hep-ex/0502003}{{\tt
  arXiv:hep-ex/0502003}}\relax
\mciteBstWouldAddEndPuncttrue
\mciteSetBstMidEndSepPunct{\mcitedefaultmidpunct}
{\mcitedefaultendpunct}{\mcitedefaultseppunct}\relax
\EndOfBibitem
\bibitem{Csorna:2004kp}
{CLEO} collaboration, S.~E. Csorna {\em et al.},
  \href{http://dx.doi.org/10.1103/PhysRevD.70.032002}{Phys. Rev. {\bf D70},
  032002 (2004)}, \href{http://arxiv.org/abs/hep-ex/0403052}{{\tt
  arXiv:hep-ex/0403052}}\relax
\mciteBstWouldAddEndPuncttrue
\mciteSetBstMidEndSepPunct{\mcitedefaultmidpunct}
{\mcitedefaultendpunct}{\mcitedefaultseppunct}\relax
\EndOfBibitem
\bibitem{Chetyrkin:2009fv}
K.~Chetyrkin {\em et al.},
  \href{http://dx.doi.org/10.1103/PhysRevD.80.074010}{Phys. Rev. {\bf D80},
  074010 (2009)}, \href{http://arxiv.org/abs/0907.2110}{{\tt arXiv:0907.2110
  [hep-ph]}}\relax
\mciteBstWouldAddEndPuncttrue
\mciteSetBstMidEndSepPunct{\mcitedefaultmidpunct}
{\mcitedefaultendpunct}{\mcitedefaultseppunct}\relax
\EndOfBibitem
\bibitem{Aubert:2005cua}
{\babar} collaboration, B.~Aubert {\em et al.},
  \href{http://dx.doi.org/10.1103/PhysRevD.72.052004}{Phys. Rev. {\bf D72},
  052004 (2005)}, \href{http://arxiv.org/abs/hep-ex/0508004}{{\tt
  arXiv:hep-ex/0508004 [hep-ex]}}\relax
\mciteBstWouldAddEndPuncttrue
\mciteSetBstMidEndSepPunct{\mcitedefaultmidpunct}
{\mcitedefaultendpunct}{\mcitedefaultseppunct}\relax
\EndOfBibitem
\bibitem{Aubert:2006gg}
{\babar} collaboration, B.~Aubert {\em et al.},
  \href{http://dx.doi.org/10.1103/PhysRevLett.97.171803}{Phys. Rev. Lett. {\bf
  97},  171803 (2006)}, \href{http://arxiv.org/abs/hep-ex/0607071}{{\tt
  arXiv:hep-ex/0607071 [hep-ex]}}\relax
\mciteBstWouldAddEndPuncttrue
\mciteSetBstMidEndSepPunct{\mcitedefaultmidpunct}
{\mcitedefaultendpunct}{\mcitedefaultseppunct}\relax
\EndOfBibitem
\bibitem{Limosani:2009qg}
{Belle} collaboration, A.~Limosani {\em et al.},
  \href{http://dx.doi.org/10.1103/PhysRevLett.103.241801}{Phys. Rev. Lett. {\bf
  103},  241801 (2009)}, \href{http://arxiv.org/abs/0907.1384}{{\tt
  arXiv:0907.1384 [hep-ex]}}\relax
\mciteBstWouldAddEndPuncttrue
\mciteSetBstMidEndSepPunct{\mcitedefaultmidpunct}
{\mcitedefaultendpunct}{\mcitedefaultseppunct}\relax
\EndOfBibitem
\bibitem{Chen:2001fja}
{CLEO} collaboration, S.~Chen {\em et al.},
  \href{http://dx.doi.org/10.1103/PhysRevLett.87.251807}{Phys. Rev. Lett. {\bf
  87},  251807 (2001)}, \href{http://arxiv.org/abs/hep-ex/0108032}{{\tt
  arXiv:hep-ex/0108032 [hep-ex]}}\relax
\mciteBstWouldAddEndPuncttrue
\mciteSetBstMidEndSepPunct{\mcitedefaultmidpunct}
{\mcitedefaultendpunct}{\mcitedefaultseppunct}\relax
\EndOfBibitem
\bibitem{Gambino:2013rza}
P.~Gambino and C.~Schwanda,
  \href{http://dx.doi.org/10.1103/PhysRevD.89.014022}{Phys. Rev. {\bf D89},
  014022 (2014)}, \href{http://arxiv.org/abs/1307.4551}{{\tt arXiv:1307.4551
  [hep-ph]}}\relax
\mciteBstWouldAddEndPuncttrue
\mciteSetBstMidEndSepPunct{\mcitedefaultmidpunct}
{\mcitedefaultendpunct}{\mcitedefaultseppunct}\relax
\EndOfBibitem
\bibitem{Schwanda:2008kw}
{Belle} collaboration, C.~Schwanda {\em et al.},
  \href{http://dx.doi.org/10.1103/PhysRevD.78.032016}{Phys. Rev. {\bf D78},
  032016 (2008)}, \href{http://arxiv.org/abs/0803.2158}{{\tt arXiv:0803.2158
  [hep-ex]}}\relax
\mciteBstWouldAddEndPuncttrue
\mciteSetBstMidEndSepPunct{\mcitedefaultmidpunct}
{\mcitedefaultendpunct}{\mcitedefaultseppunct}\relax
\EndOfBibitem
\bibitem{Aaij:2015bfa}
{LHCb} collaboration, R.~Aaij {\em et al.},
  \href{http://dx.doi.org/10.1038/nphys3415}{Nature Phys. {\bf 11},  743
  (2015)}, \href{http://arxiv.org/abs/1504.01568}{{\tt arXiv:1504.01568
  [hep-ex]}}\relax
\mciteBstWouldAddEndPuncttrue
\mciteSetBstMidEndSepPunct{\mcitedefaultmidpunct}
{\mcitedefaultendpunct}{\mcitedefaultseppunct}\relax
\EndOfBibitem
\bibitem{Ha:2010rf}
{Belle} collaboration, H.~Ha {\em et al.},
  \href{http://dx.doi.org/10.1103/PhysRevD.83.071101}{Phys. Rev. {\bf D83},
  071101 (2011)}, \href{http://arxiv.org/abs/1012.0090}{{\tt arXiv:1012.0090
  [hep-ex]}}\relax
\mciteBstWouldAddEndPuncttrue
\mciteSetBstMidEndSepPunct{\mcitedefaultmidpunct}
{\mcitedefaultendpunct}{\mcitedefaultseppunct}\relax
\EndOfBibitem
\bibitem{Sibidanov:2013rkk}
{Belle} collaboration, A.~Sibidanov {\em et al.},
  \href{http://dx.doi.org/10.1103/PhysRevD.88.032005}{Phys. Rev. {\bf D88},
  032005 (2013)}, \href{http://arxiv.org/abs/1306.2781}{{\tt arXiv:1306.2781
  [hep-ex]}}\relax
\mciteBstWouldAddEndPuncttrue
\mciteSetBstMidEndSepPunct{\mcitedefaultmidpunct}
{\mcitedefaultendpunct}{\mcitedefaultseppunct}\relax
\EndOfBibitem
\bibitem{delAmoSanchez:2010af}
{\babar} collaboration, P.~del Amo~Sanchez {\em et al.},
  \href{http://dx.doi.org/10.1103/PhysRevD.83.032007}{Phys. Rev. {\bf D83},
  032007 (2011)}, \href{http://arxiv.org/abs/1005.3288}{{\tt arXiv:1005.3288
  [hep-ex]}}\relax
\mciteBstWouldAddEndPuncttrue
\mciteSetBstMidEndSepPunct{\mcitedefaultmidpunct}
{\mcitedefaultendpunct}{\mcitedefaultseppunct}\relax
\EndOfBibitem
\bibitem{Lees:2012vv}
{\babar} collaboration, J.~P. Lees {\em et al.},
  \href{http://dx.doi.org/10.1103/PhysRevD.86.092004}{Phys. Rev. {\bf D86},
  092004 (2012)}, \href{http://arxiv.org/abs/1208.1253}{{\tt arXiv:1208.1253
  [hep-ex]}}\relax
\mciteBstWouldAddEndPuncttrue
\mciteSetBstMidEndSepPunct{\mcitedefaultmidpunct}
{\mcitedefaultendpunct}{\mcitedefaultseppunct}\relax
\EndOfBibitem
\bibitem{Bourrely:2008za}
C.~Bourrely, I.~Caprini, and L.~Lellouch,
  \href{http://dx.doi.org/10.1103/PhysRevD.79.013008}{Phys. Rev. {\bf D79},
  013008 (2009)}, \href{http://arxiv.org/abs/0807.2722}{{\tt arXiv:0807.2722
  [hep-ph]}}, erratum ibid.\
  \href{http://dx.doi.org/10.1103/PhysRevD.82.099902}{{\bf D82}, 099902},
  (2010)\relax
\mciteBstWouldAddEndPuncttrue
\mciteSetBstMidEndSepPunct{\mcitedefaultmidpunct}
{\mcitedefaultendpunct}{\mcitedefaultseppunct}\relax
\EndOfBibitem
\bibitem{Aoki:2016frl}
{FLAG}, S.~Aoki {\em et al.},
  \href{http://dx.doi.org/10.1140/epjc/s10052-016-4509-7}{Eur. Phys. J. {\bf
  C77},  112 (2017)}, \href{http://arxiv.org/abs/1607.00299}{{\tt
  arXiv:1607.00299 [hep-lat]}}, see also \href{http://flag.unibe.ch/}{{\tt
  http://flag.unibe.ch/}}\relax
\mciteBstWouldAddEndPuncttrue
\mciteSetBstMidEndSepPunct{\mcitedefaultmidpunct}
{\mcitedefaultendpunct}{\mcitedefaultseppunct}\relax
\EndOfBibitem
\bibitem{Lattice:2015tia}
{Fermilab Lattice and MILC} collaborations, J.~A. Bailey {\em et al.},
  \href{http://dx.doi.org/10.1103/PhysRevD.92.014024}{Phys. Rev. {\bf D92},
  014024 (2015)}, \href{http://arxiv.org/abs/1503.07839}{{\tt arXiv:1503.07839
  [hep-lat]}}\relax
\mciteBstWouldAddEndPuncttrue
\mciteSetBstMidEndSepPunct{\mcitedefaultmidpunct}
{\mcitedefaultendpunct}{\mcitedefaultseppunct}\relax
\EndOfBibitem
\bibitem{Flynn:2015mha}
{RBC and UKQCD} collaborations, J.~M. Flynn, T.~Izubuchi, T.~Kawanai,
  C.~Lehner, A.~Soni, R.~S. Van~de Water, and O.~Witzel,
  \href{http://dx.doi.org/10.1103/PhysRevD.91.074510}{Phys. Rev. {\bf D91},
  074510 (2015)}, \href{http://arxiv.org/abs/1501.05373}{{\tt arXiv:1501.05373
  [hep-lat]}}\relax
\mciteBstWouldAddEndPuncttrue
\mciteSetBstMidEndSepPunct{\mcitedefaultmidpunct}
{\mcitedefaultendpunct}{\mcitedefaultseppunct}\relax
\EndOfBibitem
\bibitem{Bharucha:2012wy}
A.~Bharucha, \href{http://dx.doi.org/10.1007/JHEP05(2012)092}{JHEP {\bf 05},
  092 (2012)}, \href{http://arxiv.org/abs/1203.1359}{{\tt arXiv:1203.1359
  [hep-ph]}}\relax
\mciteBstWouldAddEndPuncttrue
\mciteSetBstMidEndSepPunct{\mcitedefaultmidpunct}
{\mcitedefaultendpunct}{\mcitedefaultseppunct}\relax
\EndOfBibitem
\bibitem{Detmold:2015aaa}
W.~Detmold, C.~Lehner, and S.~Meinel,
  \href{http://dx.doi.org/10.1103/PhysRevD.92.034503}{Phys. Rev. {\bf D92},
  034503 (2015)}, \href{http://arxiv.org/abs/1503.01421}{{\tt arXiv:1503.01421
  [hep-lat]}}\relax
\mciteBstWouldAddEndPuncttrue
\mciteSetBstMidEndSepPunct{\mcitedefaultmidpunct}
{\mcitedefaultendpunct}{\mcitedefaultseppunct}\relax
\EndOfBibitem
\bibitem{Faustov:2016pal}
R.~N. Faustov and V.~O. Galkin,
  \href{http://dx.doi.org/10.1103/PhysRevD.94.073008}{Phys. Rev. {\bf D94},
  073008 (2016)}, \href{http://arxiv.org/abs/1609.00199}{{\tt arXiv:1609.00199
  [hep-ph]}}\relax
\mciteBstWouldAddEndPuncttrue
\mciteSetBstMidEndSepPunct{\mcitedefaultmidpunct}
{\mcitedefaultendpunct}{\mcitedefaultseppunct}\relax
\EndOfBibitem
\bibitem{Ball:2004ye}
P.~Ball and R.~Zwicky,
  \href{http://dx.doi.org/10.1103/PhysRevD.71.014015}{Phys. Rev. {\bf D71},
  014015 (2005)}, \href{http://arxiv.org/abs/hep-ph/0406232}{{\tt
  arXiv:hep-ph/0406232}}\relax
\mciteBstWouldAddEndPuncttrue
\mciteSetBstMidEndSepPunct{\mcitedefaultmidpunct}
{\mcitedefaultendpunct}{\mcitedefaultseppunct}\relax
\EndOfBibitem
\bibitem{Straub:2015ica}
A.~Bharucha, D.~M. Straub, and R.~Zwicky,
  \href{http://dx.doi.org/10.1007/JHEP08(2016)098}{JHEP {\bf 08},  098 (2016)},
  \href{http://arxiv.org/abs/1503.05534}{{\tt arXiv:1503.05534 [hep-ph]}}\relax
\mciteBstWouldAddEndPuncttrue
\mciteSetBstMidEndSepPunct{\mcitedefaultmidpunct}
{\mcitedefaultendpunct}{\mcitedefaultseppunct}\relax
\EndOfBibitem
\bibitem{Behrens:1999vv}
{CLEO} collaboration, B.~H. Behrens {\em et al.},
  \href{http://dx.doi.org/10.1103/PhysRevD.61.052001}{Phys. Rev. {\bf D61},
  052001 (2000)}, \href{http://arxiv.org/abs/hep-ex/9905056}{{\tt
  arXiv:hep-ex/9905056}}\relax
\mciteBstWouldAddEndPuncttrue
\mciteSetBstMidEndSepPunct{\mcitedefaultmidpunct}
{\mcitedefaultendpunct}{\mcitedefaultseppunct}\relax
\EndOfBibitem
\bibitem{Adam:2007pv}
{CLEO} collaboration, N.~E. Adam {\em et al.},
  \href{http://dx.doi.org/10.1103/PhysRevLett.99.041802}{Phys. Rev. Lett. {\bf
  99},  041802 (2007)}, \href{http://arxiv.org/abs/hep-ex/0703041}{{\tt
  arXiv:hep-ex/0703041}}\relax
\mciteBstWouldAddEndPuncttrue
\mciteSetBstMidEndSepPunct{\mcitedefaultmidpunct}
{\mcitedefaultendpunct}{\mcitedefaultseppunct}\relax
\EndOfBibitem
\bibitem{Hokuue:2006nr}
{Belle} collaboration, T.~Hokuue {\em et al.},
  \href{http://dx.doi.org/10.1016/j.physletb.2007.02.067}{Phys. Lett. {\bf
  B648},  139 (2007)}, \href{http://arxiv.org/abs/hep-ex/0604024}{{\tt
  arXiv:hep-ex/0604024}}\relax
\mciteBstWouldAddEndPuncttrue
\mciteSetBstMidEndSepPunct{\mcitedefaultmidpunct}
{\mcitedefaultendpunct}{\mcitedefaultseppunct}\relax
\EndOfBibitem
\bibitem{Schwanda:2004fa}
{Belle} collaboration, C.~Schwanda {\em et al.},
  \href{http://dx.doi.org/10.1103/PhysRevLett.93.131803}{Phys. Rev. Lett. {\bf
  93},  131803 (2004)}, \href{http://arxiv.org/abs/hep-ex/0402023}{{\tt
  arXiv:hep-ex/0402023 [hep-ex]}}\relax
\mciteBstWouldAddEndPuncttrue
\mciteSetBstMidEndSepPunct{\mcitedefaultmidpunct}
{\mcitedefaultendpunct}{\mcitedefaultseppunct}\relax
\EndOfBibitem
\bibitem{Lees:2012mq}
{\babar} collaboration, J.~P. Lees {\em et al.},
  \href{http://dx.doi.org/10.1103/PhysRevD.87.032004}{Phys. Rev. {\bf D87},
  032004 (2013)}, \href{http://arxiv.org/abs/1205.6245}{{\tt arXiv:1205.6245
  [hep-ex]}}\relax
\mciteBstWouldAddEndPuncttrue
\mciteSetBstMidEndSepPunct{\mcitedefaultmidpunct}
{\mcitedefaultendpunct}{\mcitedefaultseppunct}\relax
\EndOfBibitem
\bibitem{Lees:2013gja}
{\babar} collaboration, J.~P. Lees {\em et al.},
  \href{http://dx.doi.org/10.1103/PhysRevD.88.072006}{Phys. Rev. {\bf D88},
  072006 (2013)}, \href{http://arxiv.org/abs/1308.2589}{{\tt arXiv:1308.2589
  [hep-ex]}}\relax
\mciteBstWouldAddEndPuncttrue
\mciteSetBstMidEndSepPunct{\mcitedefaultmidpunct}
{\mcitedefaultendpunct}{\mcitedefaultseppunct}\relax
\EndOfBibitem
\bibitem{Gray:2007pw}
{CLEO} collaboration, R.~Gray {\em et al.},
  \href{http://dx.doi.org/10.1103/PhysRevD.76.012007}{Phys. Rev. {\bf D76},
  012007 (2007)}, \href{http://arxiv.org/abs/hep-ex/0703042}{{\tt
  arXiv:hep-ex/0703042}}\relax
\mciteBstWouldAddEndPuncttrue
\mciteSetBstMidEndSepPunct{\mcitedefaultmidpunct}
{\mcitedefaultendpunct}{\mcitedefaultseppunct}\relax
\EndOfBibitem
\bibitem{Aubert:2008ct}
{\babar} collaboration, B.~Aubert {\em et al.},
  \href{http://dx.doi.org/10.1103/PhysRevD.79.052011}{Phys. Rev. {\bf D79},
  052011 (2009)}, \href{http://arxiv.org/abs/0808.3524}{{\tt arXiv:0808.3524
  [hep-ex]}}\relax
\mciteBstWouldAddEndPuncttrue
\mciteSetBstMidEndSepPunct{\mcitedefaultmidpunct}
{\mcitedefaultendpunct}{\mcitedefaultseppunct}\relax
\EndOfBibitem
\bibitem{Aubert:2008bf}
{\babar} collaboration, B.~Aubert {\em et al.},
  \href{http://dx.doi.org/10.1103/PhysRevLett.101.081801}{Phys. Rev. Lett. {\bf
  101},  081801 (2008)}, \href{http://arxiv.org/abs/0805.2408}{{\tt
  arXiv:0805.2408 [hep-ex]}}\relax
\mciteBstWouldAddEndPuncttrue
\mciteSetBstMidEndSepPunct{\mcitedefaultmidpunct}
{\mcitedefaultendpunct}{\mcitedefaultseppunct}\relax
\EndOfBibitem
\bibitem{Beleno:2017cao}
{Belle} collaboration, C.~Beleño {\em et al.},
  \href{http://dx.doi.org/10.1103/PhysRevD.96.091102}{Phys. Rev. {\bf D96},
  091102 (2017)}, \href{http://arxiv.org/abs/1703.10216}{{\tt arXiv:1703.10216
  [hep-ex]}}\relax
\mciteBstWouldAddEndPuncttrue
\mciteSetBstMidEndSepPunct{\mcitedefaultmidpunct}
{\mcitedefaultendpunct}{\mcitedefaultseppunct}\relax
\EndOfBibitem
\bibitem{ref:belle-multivariate}
{Belle} collaboration, P.~Urquijo {\em et al.},
  \href{http://dx.doi.org/10.1103/PhysRevLett.104.021801}{Phys. Rev. Lett. {\bf
  104},  021801 (2010)}, \href{http://arxiv.org/abs/0907.0379}{{\tt
  arXiv:0907.0379 [hep-ex]}}\relax
\mciteBstWouldAddEndPuncttrue
\mciteSetBstMidEndSepPunct{\mcitedefaultmidpunct}
{\mcitedefaultendpunct}{\mcitedefaultseppunct}\relax
\EndOfBibitem
\bibitem{Lees:2011fv}
{\babar} collaboration, J.~P. Lees {\em et al.},
  \href{http://dx.doi.org/10.1103/PhysRevD.86.032004}{Phys. Rev. {\bf D86},
  032004 (2012)}, \href{http://arxiv.org/abs/1112.0702}{{\tt arXiv:1112.0702
  [hep-ex]}}\relax
\mciteBstWouldAddEndPuncttrue
\mciteSetBstMidEndSepPunct{\mcitedefaultmidpunct}
{\mcitedefaultendpunct}{\mcitedefaultseppunct}\relax
\EndOfBibitem
\bibitem{ref:BLL}
C.~W. Bauer, Z.~Ligeti, and M.~E. Luke,
  \href{http://dx.doi.org/10.1103/PhysRevD.64.113004}{Phys. Rev. {\bf D64},
  113004 (2001)}, \href{http://arxiv.org/abs/hep-ph/0107074}{{\tt
  arXiv:hep-ph/0107074}}\relax
\mciteBstWouldAddEndPuncttrue
\mciteSetBstMidEndSepPunct{\mcitedefaultmidpunct}
{\mcitedefaultendpunct}{\mcitedefaultseppunct}\relax
\EndOfBibitem
\bibitem{Neubert:1993um}
M.~Neubert, \href{http://dx.doi.org/10.1103/PhysRevD.49.4623}{Phys. Rev. {\bf
  D49},  4623 (1994)}, \href{http://arxiv.org/abs/hep-ph/9312311}{{\tt
  arXiv:hep-ph/9312311}}\relax
\mciteBstWouldAddEndPuncttrue
\mciteSetBstMidEndSepPunct{\mcitedefaultmidpunct}
{\mcitedefaultendpunct}{\mcitedefaultseppunct}\relax
\EndOfBibitem
\bibitem{Leibovich:1999xf}
A.~K. Leibovich, I.~Low, and I.~Z. Rothstein,
  \href{http://dx.doi.org/10.1103/PhysRevD.61.053006}{Phys. Rev. {\bf D61},
  053006 (2000)}, \href{http://arxiv.org/abs/hep-ph/9909404}{{\tt
  arXiv:hep-ph/9909404}}\relax
\mciteBstWouldAddEndPuncttrue
\mciteSetBstMidEndSepPunct{\mcitedefaultmidpunct}
{\mcitedefaultendpunct}{\mcitedefaultseppunct}\relax
\EndOfBibitem
\bibitem{Lange:2005qn}
B.~O. Lange, M.~Neubert, and G.~Paz,
  \href{http://dx.doi.org/10.1088/1126-6708/2005/10/084}{JHEP {\bf 10},  084
  (2005)}, \href{http://arxiv.org/abs/hep-ph/0508178}{{\tt
  arXiv:hep-ph/0508178}}\relax
\mciteBstWouldAddEndPuncttrue
\mciteSetBstMidEndSepPunct{\mcitedefaultmidpunct}
{\mcitedefaultendpunct}{\mcitedefaultseppunct}\relax
\EndOfBibitem
\bibitem{TheBABAR:2016lja}
{\babar} collaboration, J.~P. Lees {\em et al.},
  \href{http://dx.doi.org/10.1103/PhysRevD.95.072001}{Phys. Rev. {\bf D95},
  072001 (2017)}, \href{http://arxiv.org/abs/1611.05624}{{\tt arXiv:1611.05624
  [hep-ex]}}\relax
\mciteBstWouldAddEndPuncttrue
\mciteSetBstMidEndSepPunct{\mcitedefaultmidpunct}
{\mcitedefaultendpunct}{\mcitedefaultseppunct}\relax
\EndOfBibitem
\bibitem{ref:babar-endpoint}
{\babar} collaboration, B.~Aubert {\em et al.},
  \href{http://dx.doi.org/10.1103/PhysRevD.73.012006}{Phys. Rev. {\bf D73},
  012006 (2006)}, \href{http://arxiv.org/abs/hep-ex/0509040}{{\tt
  arXiv:hep-ex/0509040}}\relax
\mciteBstWouldAddEndPuncttrue
\mciteSetBstMidEndSepPunct{\mcitedefaultmidpunct}
{\mcitedefaultendpunct}{\mcitedefaultseppunct}\relax
\EndOfBibitem
\bibitem{ref:shmax}
R.~V. Kowalewski and S.~Menke,
  \href{http://dx.doi.org/10.1016/S0370-2693(02)02181-0}{Phys. Lett. {\bf
  B541},  29 (2002)}, \href{http://arxiv.org/abs/hep-ex/0205038}{{\tt
  arXiv:hep-ex/0205038}}\relax
\mciteBstWouldAddEndPuncttrue
\mciteSetBstMidEndSepPunct{\mcitedefaultmidpunct}
{\mcitedefaultendpunct}{\mcitedefaultseppunct}\relax
\EndOfBibitem
\bibitem{ref:babar-elq2}
{\babar} collaboration, B.~Aubert {\em et al.},
  \href{http://dx.doi.org/10.1103/PhysRevLett.95.111801}{Phys. Rev. Lett. {\bf
  95},  111801 (2005)}, \href{http://arxiv.org/abs/hep-ex/0506036}{{\tt
  arXiv:hep-ex/0506036}}\relax
\mciteBstWouldAddEndPuncttrue
\mciteSetBstMidEndSepPunct{\mcitedefaultmidpunct}
{\mcitedefaultendpunct}{\mcitedefaultseppunct}\relax
\EndOfBibitem
\bibitem{ref:cleo-endpoint}
{CLEO} collaboration, A.~Bornheim {\em et al.},
  \href{http://dx.doi.org/10.1103/PhysRevLett.88.231803}{Phys. Rev. Lett. {\bf
  88},  231803 (2002)}, \href{http://arxiv.org/abs/hep-ex/0202019}{{\tt
  arXiv:hep-ex/0202019}}\relax
\mciteBstWouldAddEndPuncttrue
\mciteSetBstMidEndSepPunct{\mcitedefaultmidpunct}
{\mcitedefaultendpunct}{\mcitedefaultseppunct}\relax
\EndOfBibitem
\bibitem{ref:belle-endpoint}
{Belle} collaboration, A.~Limosani {\em et al.},
  \href{http://dx.doi.org/10.1016/j.physletb.2005.06.011}{Phys. Lett. {\bf
  B621},  28 (2005)}, \href{http://arxiv.org/abs/hep-ex/0504046}{{\tt
  arXiv:hep-ex/0504046}}\relax
\mciteBstWouldAddEndPuncttrue
\mciteSetBstMidEndSepPunct{\mcitedefaultmidpunct}
{\mcitedefaultendpunct}{\mcitedefaultseppunct}\relax
\EndOfBibitem
\bibitem{ref:belle-mxq2Anneal}
{Belle} collaboration, H.~Kakuno {\em et al.},
  \href{http://dx.doi.org/10.1103/PhysRevLett.92.101801}{Phys. Rev. Lett. {\bf
  92},  101801 (2004)}, \href{http://arxiv.org/abs/hep-ex/0311048}{{\tt
  arXiv:hep-ex/0311048}}\relax
\mciteBstWouldAddEndPuncttrue
\mciteSetBstMidEndSepPunct{\mcitedefaultmidpunct}
{\mcitedefaultendpunct}{\mcitedefaultseppunct}\relax
\EndOfBibitem
\bibitem{ref:belle-mx}
{Belle} collaboration, I.~Bizjak {\em et al.},
  \href{http://dx.doi.org/10.1103/PhysRevLett.95.241801}{Phys. Rev. Lett. {\bf
  95},  241801 (2005)}, \href{http://arxiv.org/abs/hep-ex/0505088}{{\tt
  arXiv:hep-ex/0505088}}\relax
\mciteBstWouldAddEndPuncttrue
\mciteSetBstMidEndSepPunct{\mcitedefaultmidpunct}
{\mcitedefaultendpunct}{\mcitedefaultseppunct}\relax
\EndOfBibitem
\bibitem{ref:BLNP}
B.~O. Lange, M.~Neubert, and G.~Paz,
  \href{http://dx.doi.org/10.1103/PhysRevD.72.073006}{Phys. Rev. {\bf D72},
  073006 (2005)}, \href{http://arxiv.org/abs/hep-ph/0504071}{{\tt
  arXiv:hep-ph/0504071}}\relax
\mciteBstWouldAddEndPuncttrue
\mciteSetBstMidEndSepPunct{\mcitedefaultmidpunct}
{\mcitedefaultendpunct}{\mcitedefaultseppunct}\relax
\EndOfBibitem
\bibitem{ref:Neubert-new-1}
S.~W. Bosch, B.~O. Lange, M.~Neubert, and G.~Paz,
  \href{http://dx.doi.org/10.1016/j.nuclphysb.2004.07.041}{Nucl. Phys. {\bf
  B699},  335 (2004)}, \href{http://arxiv.org/abs/hep-ph/0402094}{{\tt
  arXiv:hep-ph/0402094}}\relax
\mciteBstWouldAddEndPuncttrue
\mciteSetBstMidEndSepPunct{\mcitedefaultmidpunct}
{\mcitedefaultendpunct}{\mcitedefaultseppunct}\relax
\EndOfBibitem
\bibitem{ref:Neubert-new-2}
S.~W. Bosch, M.~Neubert, and G.~Paz,
  \href{http://dx.doi.org/10.1088/1126-6708/2004/11/073}{JHEP {\bf 11},  073
  (2004)}, \href{http://arxiv.org/abs/hep-ph/0409115}{{\tt
  arXiv:hep-ph/0409115}}\relax
\mciteBstWouldAddEndPuncttrue
\mciteSetBstMidEndSepPunct{\mcitedefaultmidpunct}
{\mcitedefaultendpunct}{\mcitedefaultseppunct}\relax
\EndOfBibitem
\bibitem{ref:Neubert-new-3}
M.~Neubert, \href{http://dx.doi.org/10.1140/epjc/s2005-02360-4}{Eur. Phys. J.
  {\bf C44},  205 (2005)}, \href{http://arxiv.org/abs/hep-ph/0411027}{{\tt
  arXiv:hep-ph/0411027}}\relax
\mciteBstWouldAddEndPuncttrue
\mciteSetBstMidEndSepPunct{\mcitedefaultmidpunct}
{\mcitedefaultendpunct}{\mcitedefaultseppunct}\relax
\EndOfBibitem
\bibitem{Neubert:2004sp}
M.~Neubert, \href{http://dx.doi.org/10.1016/j.physletb.2005.02.055}{Phys. Lett.
  {\bf B612},  13 (2005)}, \href{http://arxiv.org/abs/hep-ph/0412241}{{\tt
  arXiv:hep-ph/0412241 [hep-ph]}}\relax
\mciteBstWouldAddEndPuncttrue
\mciteSetBstMidEndSepPunct{\mcitedefaultmidpunct}
{\mcitedefaultendpunct}{\mcitedefaultseppunct}\relax
\EndOfBibitem
\bibitem{Neubert:2005nt}
M.~Neubert, \href{http://dx.doi.org/10.1103/PhysRevD.72.074025}{Phys. Rev. {\bf
  D72},  074025 (2005)}, \href{http://arxiv.org/abs/hep-ph/0506245}{{\tt
  arXiv:hep-ph/0506245 [hep-ph]}}\relax
\mciteBstWouldAddEndPuncttrue
\mciteSetBstMidEndSepPunct{\mcitedefaultmidpunct}
{\mcitedefaultendpunct}{\mcitedefaultseppunct}\relax
\EndOfBibitem
\bibitem{ref:DGE}
J.~R. Andersen and E.~Gardi,
  \href{http://dx.doi.org/10.1088/1126-6708/2006/01/097}{JHEP {\bf 01},  097
  (2006)}, \href{http://arxiv.org/abs/hep-ph/0509360}{{\tt
  arXiv:hep-ph/0509360}}\relax
\mciteBstWouldAddEndPuncttrue
\mciteSetBstMidEndSepPunct{\mcitedefaultmidpunct}
{\mcitedefaultendpunct}{\mcitedefaultseppunct}\relax
\EndOfBibitem
\bibitem{Aglietti:2006yb}
U.~Aglietti, G.~Ferrera, and G.~Ricciardi,
  \href{http://dx.doi.org/10.1016/j.nuclphysb.2007.01.014}{Nucl. Phys. {\bf
  B768},  85 (2007)}, \href{http://arxiv.org/abs/hep-ph/0608047}{{\tt
  arXiv:hep-ph/0608047}}\relax
\mciteBstWouldAddEndPuncttrue
\mciteSetBstMidEndSepPunct{\mcitedefaultmidpunct}
{\mcitedefaultendpunct}{\mcitedefaultseppunct}\relax
\EndOfBibitem
\bibitem{Blokland:2005uk}
I.~R. Blokland, A.~Czarnecki, M.~Misiak, M.~Slusarczyk, and F.~Tkachov,
  \href{http://dx.doi.org/10.1103/PhysRevD.72.033014}{Phys. Rev. {\bf D72},
  033014 (2005)}, \href{http://arxiv.org/abs/hep-ph/0506055}{{\tt
  arXiv:hep-ph/0506055 [hep-ph]}}\relax
\mciteBstWouldAddEndPuncttrue
\mciteSetBstMidEndSepPunct{\mcitedefaultmidpunct}
{\mcitedefaultendpunct}{\mcitedefaultseppunct}\relax
\EndOfBibitem
\bibitem{Gambino:2008fj}
P.~Gambino and P.~Giordano,
  \href{http://dx.doi.org/10.1016/j.physletb.2008.09.046}{Phys. Lett. {\bf
  B669},  69--73 (2008)}, \href{http://arxiv.org/abs/0805.0271}{{\tt
  arXiv:0805.0271 [hep-ph]}}\relax
\mciteBstWouldAddEndPuncttrue
\mciteSetBstMidEndSepPunct{\mcitedefaultmidpunct}
{\mcitedefaultendpunct}{\mcitedefaultseppunct}\relax
\EndOfBibitem
\bibitem{Gambino:2007rp}
P.~Gambino, P.~Giordano, G.~Ossola, and N.~Uraltsev,
  \href{http://dx.doi.org/10.1088/1126-6708/2007/10/058}{JHEP {\bf 10},  058
  (2007)}, \href{http://arxiv.org/abs/0707.2493}{{\tt arXiv:0707.2493
  [hep-ph]}}\relax
\mciteBstWouldAddEndPuncttrue
\mciteSetBstMidEndSepPunct{\mcitedefaultmidpunct}
{\mcitedefaultendpunct}{\mcitedefaultseppunct}\relax
\EndOfBibitem
\bibitem{Aglietti:2007ik}
U.~Aglietti, F.~Di~Lodovico, G.~Ferrera, and G.~Ricciardi,
  \href{http://dx.doi.org/10.1140/epjc/s10052-008-0817-x}{Eur. Phys. J. {\bf
  C59},  831 (2009)}, \href{http://arxiv.org/abs/0711.0860}{{\tt
  arXiv:0711.0860 [hep-ph]}}\relax
\mciteBstWouldAddEndPuncttrue
\mciteSetBstMidEndSepPunct{\mcitedefaultmidpunct}
{\mcitedefaultendpunct}{\mcitedefaultseppunct}\relax
\EndOfBibitem
\bibitem{Aglietti:2005mb}
U.~Aglietti, G.~Ricciardi, and G.~Ferrera,
  \href{http://dx.doi.org/10.1103/PhysRevD.74.034004}{Phys. Rev. {\bf D74},
  034004 (2006)}, \href{http://arxiv.org/abs/hep-ph/0507285}{{\tt
  arXiv:hep-ph/0507285}}\relax
\mciteBstWouldAddEndPuncttrue
\mciteSetBstMidEndSepPunct{\mcitedefaultmidpunct}
{\mcitedefaultendpunct}{\mcitedefaultseppunct}\relax
\EndOfBibitem
\bibitem{Aglietti:2005bm}
U.~Aglietti, G.~Ricciardi, and G.~Ferrera,
  \href{http://dx.doi.org/10.1103/PhysRevD.74.034005}{Phys. Rev. {\bf D74},
  034005 (2006)}, \href{http://arxiv.org/abs/hep-ph/0509095}{{\tt
  arXiv:hep-ph/0509095}}\relax
\mciteBstWouldAddEndPuncttrue
\mciteSetBstMidEndSepPunct{\mcitedefaultmidpunct}
{\mcitedefaultendpunct}{\mcitedefaultseppunct}\relax
\EndOfBibitem
\bibitem{Aglietti:2005eq}
U.~Aglietti, G.~Ricciardi, and G.~Ferrera,
  \href{http://dx.doi.org/10.1103/PhysRevD.74.034006}{Phys. Rev. {\bf D74},
  034006 (2006)}, \href{http://arxiv.org/abs/hep-ph/0509271}{{\tt
  arXiv:hep-ph/0509271}}\relax
\mciteBstWouldAddEndPuncttrue
\mciteSetBstMidEndSepPunct{\mcitedefaultmidpunct}
{\mcitedefaultendpunct}{\mcitedefaultseppunct}\relax
\EndOfBibitem
\bibitem{Duraisamy:2014sna}
M.~Duraisamy, P.~Sharma, and A.~Datta,
  \href{http://dx.doi.org/10.1103/PhysRevD.90.074013}{Phys. Rev. {\bf D90},
  074013 (2014)}, \href{http://arxiv.org/abs/1405.3719}{{\tt arXiv:1405.3719
  [hep-ph]}}\relax
\mciteBstWouldAddEndPuncttrue
\mciteSetBstMidEndSepPunct{\mcitedefaultmidpunct}
{\mcitedefaultendpunct}{\mcitedefaultseppunct}\relax
\EndOfBibitem
\bibitem{Bigi:2016mdz}
D.~Bigi and P.~Gambino,
  \href{http://dx.doi.org/10.1103/PhysRevD.94.094008}{Phys. Rev. {\bf D94},
  094008 (2016)}, \href{http://arxiv.org/abs/1606.08030}{{\tt arXiv:1606.08030
  [hep-ph]}}\relax
\mciteBstWouldAddEndPuncttrue
\mciteSetBstMidEndSepPunct{\mcitedefaultmidpunct}
{\mcitedefaultendpunct}{\mcitedefaultseppunct}\relax
\EndOfBibitem
\bibitem{Bernlochner:2017jka}
F.~U. Bernlochner, Z.~Ligeti, M.~Papucci, and D.~J. Robinson,
  \href{http://dx.doi.org/10.1103/PhysRevD.95.115008,
  10.1103/PhysRevD.97.059902}{Phys. Rev. {\bf D95},  115008 (2017)},
  \href{http://arxiv.org/abs/1703.05330}{{\tt arXiv:1703.05330 [hep-ph]}},
  erratum ibid {\bf D97},059902 (2018)]\relax
\mciteBstWouldAddEndPuncttrue
\mciteSetBstMidEndSepPunct{\mcitedefaultmidpunct}
{\mcitedefaultendpunct}{\mcitedefaultseppunct}\relax
\EndOfBibitem
\bibitem{Jaiswal:2017rve}
S.~Jaiswal, S.~Nandi, and S.~K. Patra,
  \href{http://dx.doi.org/10.1007/JHEP12(2017)060}{JHEP {\bf 12},  060 (2017)},
  \href{http://arxiv.org/abs/1707.09977}{{\tt arXiv:1707.09977 [hep-ph]}}\relax
\mciteBstWouldAddEndPuncttrue
\mciteSetBstMidEndSepPunct{\mcitedefaultmidpunct}
{\mcitedefaultendpunct}{\mcitedefaultseppunct}\relax
\EndOfBibitem
\bibitem{Bigi:2017jbd}
D.~Bigi, P.~Gambino, and S.~Schacht,
  \href{http://dx.doi.org/10.1007/JHEP11(2017)061}{JHEP {\bf 11},  061 (2017)},
  \href{http://arxiv.org/abs/1707.09509}{{\tt arXiv:1707.09509 [hep-ph]}}\relax
\mciteBstWouldAddEndPuncttrue
\mciteSetBstMidEndSepPunct{\mcitedefaultmidpunct}
{\mcitedefaultendpunct}{\mcitedefaultseppunct}\relax
\EndOfBibitem
\bibitem{Fajfer:2012vx}
S.~Fajfer, J.~F. Kamenik, and I.~Nisandzic,
  \href{http://dx.doi.org/10.1103/PhysRevD.85.094025}{Phys. Rev. {\bf D85},
  094025 (2012)}, \href{http://arxiv.org/abs/1203.2654}{{\tt arXiv:1203.2654
  [hep-ph]}}\relax
\mciteBstWouldAddEndPuncttrue
\mciteSetBstMidEndSepPunct{\mcitedefaultmidpunct}
{\mcitedefaultendpunct}{\mcitedefaultseppunct}\relax
\EndOfBibitem
\bibitem{Gambino:2019sif}
P.~Gambino, M.~Jung, and S.~Schacht,
  \href{http://dx.doi.org/10.1016/j.physletb.2019.06.039}{Phys. Lett. {\bf
  B795},  386--390 (2019)}, \href{http://arxiv.org/abs/1905.08209}{{\tt
  arXiv:1905.08209 [hep-ph]}}\relax
\mciteBstWouldAddEndPuncttrue
\mciteSetBstMidEndSepPunct{\mcitedefaultmidpunct}
{\mcitedefaultendpunct}{\mcitedefaultseppunct}\relax
\EndOfBibitem
\bibitem{Bordone:2019vic}
M.~Bordone, M.~Jung, and D.~van Dyk,
  \href{http://dx.doi.org/10.1140/epjc/s10052-020-7616-4}{Eur.\ Phys.\ J.\ C
  {\bf 80},  74 (2020)}, \href{http://arxiv.org/abs/1908.09398}{{\tt
  arXiv:1908.09398 [hep-ph]}}\relax
\mciteBstWouldAddEndPuncttrue
\mciteSetBstMidEndSepPunct{\mcitedefaultmidpunct}
{\mcitedefaultendpunct}{\mcitedefaultseppunct}\relax
\EndOfBibitem
\bibitem{Matyja:2007kt}
{Belle} collaboration, A.~Matyja {\em et al.},
  \href{http://dx.doi.org/10.1103/PhysRevLett.99.191807}{Phys. Rev. Lett. {\bf
  99},  191807 (2007)}, \href{http://arxiv.org/abs/0706.4429}{{\tt
  arXiv:0706.4429 [hep-ex]}}\relax
\mciteBstWouldAddEndPuncttrue
\mciteSetBstMidEndSepPunct{\mcitedefaultmidpunct}
{\mcitedefaultendpunct}{\mcitedefaultseppunct}\relax
\EndOfBibitem
\bibitem{Aubert:2007dsa}
{\babar} collaboration, B.~Aubert {\em et al.},
  \href{http://dx.doi.org/10.1103/PhysRevLett.100.021801}{Phys. Rev. Lett. {\bf
  100},  021801 (2008)}, \href{http://arxiv.org/abs/0709.1698}{{\tt
  arXiv:0709.1698 [hep-ex]}}\relax
\mciteBstWouldAddEndPuncttrue
\mciteSetBstMidEndSepPunct{\mcitedefaultmidpunct}
{\mcitedefaultendpunct}{\mcitedefaultseppunct}\relax
\EndOfBibitem
\bibitem{Bozek:2010xy}
{Belle} collaboration, A.~Bozek {\em et al.},
  \href{http://dx.doi.org/10.1103/PhysRevD.82.072005}{Phys. Rev. {\bf D82},
  072005 (2010)}, \href{http://arxiv.org/abs/1005.2302}{{\tt arXiv:1005.2302
  [hep-ex]}}\relax
\mciteBstWouldAddEndPuncttrue
\mciteSetBstMidEndSepPunct{\mcitedefaultmidpunct}
{\mcitedefaultendpunct}{\mcitedefaultseppunct}\relax
\EndOfBibitem
\bibitem{Lees:2012xj}
{\babar} collaboration, J.~P. Lees {\em et al.},
  \href{http://dx.doi.org/10.1103/PhysRevLett.109.101802}{Phys. Rev. Lett. {\bf
  109},  101802 (2012)}, \href{http://arxiv.org/abs/1205.5442}{{\tt
  arXiv:1205.5442 [hep-ex]}}\relax
\mciteBstWouldAddEndPuncttrue
\mciteSetBstMidEndSepPunct{\mcitedefaultmidpunct}
{\mcitedefaultendpunct}{\mcitedefaultseppunct}\relax
\EndOfBibitem
\bibitem{Huschle:2015rga}
{Belle} collaboration, M.~Huschle {\em et al.},
  \href{http://dx.doi.org/10.1103/PhysRevD.92.072014}{Phys. Rev. {\bf D92},
  072014 (2015)}, \href{http://arxiv.org/abs/1507.03233}{{\tt arXiv:1507.03233
  [hep-ex]}}\relax
\mciteBstWouldAddEndPuncttrue
\mciteSetBstMidEndSepPunct{\mcitedefaultmidpunct}
{\mcitedefaultendpunct}{\mcitedefaultseppunct}\relax
\EndOfBibitem
\bibitem{Belle:2019rba}
{Belle} collaboration, G.~Caria {\em et al.},
  \href{http://dx.doi.org/10.1103/PhysRevLett.124.161803}{Phys. Rev. Lett. {\bf
  124}, no.~16, 161803 (2020)}, \href{http://arxiv.org/abs/1910.05864}{{\tt
  arXiv:1910.05864 [hep-ex]}}\relax
\mciteBstWouldAddEndPuncttrue
\mciteSetBstMidEndSepPunct{\mcitedefaultmidpunct}
{\mcitedefaultendpunct}{\mcitedefaultseppunct}\relax
\EndOfBibitem
\bibitem{Hirose:2016wfn}
{Belle} collaboration, S.~Hirose {\em et al.},
  \href{http://dx.doi.org/10.1103/PhysRevLett.118.211801}{Phys. Rev. Lett. {\bf
  118},  211801 (2017)}, \href{http://arxiv.org/abs/1612.00529}{{\tt
  arXiv:1612.00529 [hep-ex]}}\relax
\mciteBstWouldAddEndPuncttrue
\mciteSetBstMidEndSepPunct{\mcitedefaultmidpunct}
{\mcitedefaultendpunct}{\mcitedefaultseppunct}\relax
\EndOfBibitem
\bibitem{Aaij:2015yra}
{LHCb} collaboration, R.~Aaij {\em et al.},
  \href{http://dx.doi.org/10.1103/PhysRevLett.115.111803}{Phys. Rev. Lett. {\bf
  115},  111803 (2015)}, \href{http://arxiv.org/abs/1506.08614}{{\tt
  arXiv:1506.08614 [hep-ex]}}, addendum ibid.\
  \href{http://dx.doi.org/10.1103/PhysRevLett.115.159901}{{\bf 115}, 159901},
  (2015)\relax
\mciteBstWouldAddEndPuncttrue
\mciteSetBstMidEndSepPunct{\mcitedefaultmidpunct}
{\mcitedefaultendpunct}{\mcitedefaultseppunct}\relax
\EndOfBibitem
\bibitem{Aaij:2017uff}
{LHCb} collaboration, R.~Aaij {\em et al.},
  \href{http://dx.doi.org/10.1103/PhysRevLett.120.171802}{Phys. Rev. Lett. {\bf
  120},  171802 (2018)}, \href{http://arxiv.org/abs/1708.08856}{{\tt
  arXiv:1708.08856 [hep-ex]}}\relax
\mciteBstWouldAddEndPuncttrue
\mciteSetBstMidEndSepPunct{\mcitedefaultmidpunct}
{\mcitedefaultendpunct}{\mcitedefaultseppunct}\relax
\EndOfBibitem
\bibitem{Lees:2013uzd}
{\babar} collaboration, J.~P. Lees {\em et al.},
  \href{http://dx.doi.org/10.1103/PhysRevD.88.072012}{Phys. Rev. {\bf D88},
  072012 (2013)}, \href{http://arxiv.org/abs/1303.0571}{{\tt arXiv:1303.0571
  [hep-ex]}}\relax
\mciteBstWouldAddEndPuncttrue
\mciteSetBstMidEndSepPunct{\mcitedefaultmidpunct}
{\mcitedefaultendpunct}{\mcitedefaultseppunct}\relax
\EndOfBibitem
\bibitem{Aaij:2017deq}
{LHCb} collaboration, R.~Aaij {\em et al.},
  \href{http://dx.doi.org/10.1103/PhysRevD.97.072013}{Phys. Rev. {\bf D97},
  072013 (2018)}, \href{http://arxiv.org/abs/1711.02505}{{\tt arXiv:1711.02505
  [hep-ex]}}\relax
\mciteBstWouldAddEndPuncttrue
\mciteSetBstMidEndSepPunct{\mcitedefaultmidpunct}
{\mcitedefaultendpunct}{\mcitedefaultseppunct}\relax
\EndOfBibitem
\bibitem{Lenz:2014nka}
A.~J. Lenz, \href{http://dx.doi.org/10.1088/0954-3899/41/10/103001}{J. Phys.
  {\bf G41},  103001 (2014)}, \href{http://arxiv.org/abs/1404.6197}{{\tt
  arXiv:1404.6197 [hep-ph]}}\relax
\mciteBstWouldAddEndPuncttrue
\mciteSetBstMidEndSepPunct{\mcitedefaultmidpunct}
{\mcitedefaultendpunct}{\mcitedefaultseppunct}\relax
\EndOfBibitem
\bibitem{Choi:2003ue}
{Belle} collaboration, S.~Choi {\em et al.},
  \href{http://dx.doi.org/10.1103/PhysRevLett.91.262001}{Phys. Rev. Lett. {\bf
  91},  262001 (2003)}, \href{http://arxiv.org/abs/hep-ex/0309032}{{\tt
  arXiv:hep-ex/0309032 [hep-ex]}}\relax
\mciteBstWouldAddEndPuncttrue
\mciteSetBstMidEndSepPunct{\mcitedefaultmidpunct}
{\mcitedefaultendpunct}{\mcitedefaultseppunct}\relax
\EndOfBibitem
\bibitem{Aaij:2015eva}
{LHCb} collaboration, R.~Aaij {\em et al.},
  \href{http://dx.doi.org/10.1103/PhysRevD.92.011102}{Phys. Rev. {\bf D92},
  011102 (2015)}, \href{http://arxiv.org/abs/1504.06339}{{\tt arXiv:1504.06339
  [hep-ex]}}\relax
\mciteBstWouldAddEndPuncttrue
\mciteSetBstMidEndSepPunct{\mcitedefaultmidpunct}
{\mcitedefaultendpunct}{\mcitedefaultseppunct}\relax
\EndOfBibitem
\bibitem{Choi:2007wga}
{Belle} collaboration, S.~Choi {\em et al.},
  \href{http://dx.doi.org/10.1103/PhysRevLett.100.142001}{Phys. Rev. Lett. {\bf
  100},  142001 (2008)}, \href{http://arxiv.org/abs/0708.1790}{{\tt
  arXiv:0708.1790 [hep-ex]}}\relax
\mciteBstWouldAddEndPuncttrue
\mciteSetBstMidEndSepPunct{\mcitedefaultmidpunct}
{\mcitedefaultendpunct}{\mcitedefaultseppunct}\relax
\EndOfBibitem
\bibitem{Aaij:2014jqa}
{LHCb} collaboration, R.~Aaij {\em et al.},
  \href{http://dx.doi.org/10.1103/PhysRevLett.112.222002}{Phys. Rev. Lett. {\bf
  112},  222002 (2014)}, \href{http://arxiv.org/abs/1404.1903}{{\tt
  arXiv:1404.1903 [hep-ex]}}\relax
\mciteBstWouldAddEndPuncttrue
\mciteSetBstMidEndSepPunct{\mcitedefaultmidpunct}
{\mcitedefaultendpunct}{\mcitedefaultseppunct}\relax
\EndOfBibitem
\bibitem{Aaij:2015tga}
{LHCb} collaboration, R.~Aaij {\em et al.},
  \href{http://dx.doi.org/10.1103/PhysRevLett.115.072001}{Phys. Rev. Lett. {\bf
  115},  072001 (2015)}, \href{http://arxiv.org/abs/1507.03414}{{\tt
  arXiv:1507.03414 [hep-ex]}}\relax
\mciteBstWouldAddEndPuncttrue
\mciteSetBstMidEndSepPunct{\mcitedefaultmidpunct}
{\mcitedefaultendpunct}{\mcitedefaultseppunct}\relax
\EndOfBibitem
\bibitem{Barlow:2004wg}
R.~Barlow, \href{http://arxiv.org/abs/physics/0406120}{{\tt
  arXiv:physics/0406120 [physics]}} (2004)\relax
\mciteBstWouldAddEndPuncttrue
\mciteSetBstMidEndSepPunct{\mcitedefaultmidpunct}
{\mcitedefaultendpunct}{\mcitedefaultseppunct}\relax
\EndOfBibitem
\bibitem{DeBruyn:2012wj}
K.~De~Bruyn {\em et al.},
  \href{http://dx.doi.org/10.1103/PhysRevD.86.014027}{Phys. Rev. {\bf D86},
  014027 (2012)}, \href{http://arxiv.org/abs/1204.1735}{{\tt arXiv:1204.1735
  [hep-ph]}}\relax
\mciteBstWouldAddEndPuncttrue
\mciteSetBstMidEndSepPunct{\mcitedefaultmidpunct}
{\mcitedefaultendpunct}{\mcitedefaultseppunct}\relax
\EndOfBibitem
\bibitem{Jung:2015yma}
M.~Jung, \href{http://dx.doi.org/10.1016/j.physletb.2015.12.024}{Phys. Lett.
  {\bf B753},  187 (2016)}, \href{http://arxiv.org/abs/1510.03423}{{\tt
  arXiv:1510.03423 [hep-ph]}}\relax
\mciteBstWouldAddEndPuncttrue
\mciteSetBstMidEndSepPunct{\mcitedefaultmidpunct}
{\mcitedefaultendpunct}{\mcitedefaultseppunct}\relax
\EndOfBibitem
\bibitem{Aubert:2006cd}
{\babar} collaboration, B.~Aubert {\em et al.},
  \href{http://dx.doi.org/10.1103/PhysRevD.75.031101}{Phys. Rev. {\bf D75},
  031101 (2007)}, \href{http://arxiv.org/abs/hep-ex/0610027}{{\tt
  arXiv:hep-ex/0610027 [hep-ex]}}\relax
\mciteBstWouldAddEndPuncttrue
\mciteSetBstMidEndSepPunct{\mcitedefaultmidpunct}
{\mcitedefaultendpunct}{\mcitedefaultseppunct}\relax
\EndOfBibitem
\bibitem{Aubert:2006jc}
{\babar} collaboration, B.~Aubert {\em et al.},
  \href{http://dx.doi.org/10.1103/PhysRevD.74.111102}{Phys. Rev. {\bf D74},
  111102 (2006)}, \href{http://arxiv.org/abs/hep-ex/0609033}{{\tt
  arXiv:hep-ex/0609033 [hep-ex]}}\relax
\mciteBstWouldAddEndPuncttrue
\mciteSetBstMidEndSepPunct{\mcitedefaultmidpunct}
{\mcitedefaultendpunct}{\mcitedefaultseppunct}\relax
\EndOfBibitem
\bibitem{Kuzmin:2006mw}
{Belle} collaboration, A.~Kuzmin {\em et al.},
  \href{http://dx.doi.org/10.1103/PhysRevD.76.012006}{Phys. Rev. {\bf D76},
  012006 (2007)}, \href{http://arxiv.org/abs/hep-ex/0611054}{{\tt
  arXiv:hep-ex/0611054 [hep-ex]}}\relax
\mciteBstWouldAddEndPuncttrue
\mciteSetBstMidEndSepPunct{\mcitedefaultmidpunct}
{\mcitedefaultendpunct}{\mcitedefaultseppunct}\relax
\EndOfBibitem
\bibitem{Majumder:2004su}
{Belle} collaboration, G.~Majumder {\em et al.},
  \href{http://dx.doi.org/10.1103/PhysRevD.70.111103}{Phys. Rev. {\bf D70},
  111103 (2004)}, \href{http://arxiv.org/abs/hep-ex/0409008}{{\tt
  arXiv:hep-ex/0409008 [hep-ex]}}\relax
\mciteBstWouldAddEndPuncttrue
\mciteSetBstMidEndSepPunct{\mcitedefaultmidpunct}
{\mcitedefaultendpunct}{\mcitedefaultseppunct}\relax
\EndOfBibitem
\bibitem{TheBABAR:2016vzj}
{\babar} collaboration, J.~P. Lees {\em et al.},
  \href{http://dx.doi.org/10.1103/PhysRevD.94.091101}{Phys. Rev. {\bf D94},
  091101 (2016)}, \href{http://arxiv.org/abs/1609.06802}{{\tt arXiv:1609.06802
  [hep-ex]}}\relax
\mciteBstWouldAddEndPuncttrue
\mciteSetBstMidEndSepPunct{\mcitedefaultmidpunct}
{\mcitedefaultendpunct}{\mcitedefaultseppunct}\relax
\EndOfBibitem
\bibitem{Matvienko:2015gqa}
{Belle} collaboration, D.~Matvienko {\em et al.},
  \href{http://dx.doi.org/10.1103/PhysRevD.92.012013}{Phys. Rev. {\bf D92},
  012013 (2015)}, \href{http://arxiv.org/abs/1505.03362}{{\tt arXiv:1505.03362
  [hep-ex]}}\relax
\mciteBstWouldAddEndPuncttrue
\mciteSetBstMidEndSepPunct{\mcitedefaultmidpunct}
{\mcitedefaultendpunct}{\mcitedefaultseppunct}\relax
\EndOfBibitem
\bibitem{Aubert:2006zb}
{\babar} collaboration, B.~Aubert {\em et al.},
  \href{http://dx.doi.org/10.1103/PhysRevD.74.012001}{Phys. Rev. {\bf D74},
  012001 (2006)}, \href{http://arxiv.org/abs/hep-ex/0604009}{{\tt
  arXiv:hep-ex/0604009 [hep-ex]}}\relax
\mciteBstWouldAddEndPuncttrue
\mciteSetBstMidEndSepPunct{\mcitedefaultmidpunct}
{\mcitedefaultendpunct}{\mcitedefaultseppunct}\relax
\EndOfBibitem
\bibitem{Blyth:2006at}
{Belle} collaboration, S.~Blyth {\em et al.},
  \href{http://dx.doi.org/10.1103/PhysRevD.74.092002}{Phys. Rev. {\bf D74},
  092002 (2006)}, \href{http://arxiv.org/abs/hep-ex/0607029}{{\tt
  arXiv:hep-ex/0607029 [hep-ex]}}\relax
\mciteBstWouldAddEndPuncttrue
\mciteSetBstMidEndSepPunct{\mcitedefaultmidpunct}
{\mcitedefaultendpunct}{\mcitedefaultseppunct}\relax
\EndOfBibitem
\bibitem{Lees:2011gw}
{\babar} collaboration, J.~P. Lees {\em et al.},
  \href{http://dx.doi.org/10.1103/PhysRevD.84.112007}{Phys. Rev. {\bf D84},
  112007 (2011)}, \href{http://arxiv.org/abs/1107.5751}{{\tt arXiv:1107.5751
  [hep-ex]}}, Erratum ibid.\
  \href{http://dx.doi.org/10.1103/PhysRevD.87.039901}{{\bf D87}, 039901},
  (2013)\relax
\mciteBstWouldAddEndPuncttrue
\mciteSetBstMidEndSepPunct{\mcitedefaultmidpunct}
{\mcitedefaultendpunct}{\mcitedefaultseppunct}\relax
\EndOfBibitem
\bibitem{Aaij:2015sqa}
{LHCb} collaboration, R.~Aaij {\em et al.},
  \href{http://dx.doi.org/10.1103/PhysRevD.92.032002}{Phys. Rev. {\bf D92},
  032002 (2015)}, \href{http://arxiv.org/abs/1505.01710}{{\tt arXiv:1505.01710
  [hep-ex]}}\relax
\mciteBstWouldAddEndPuncttrue
\mciteSetBstMidEndSepPunct{\mcitedefaultmidpunct}
{\mcitedefaultendpunct}{\mcitedefaultseppunct}\relax
\EndOfBibitem
\bibitem{Satpathy:2002js}
{Belle} collaboration, A.~Satpathy {\em et al.},
  \href{http://dx.doi.org/10.1016/S0370-2693(02)03198-2}{Phys. Lett. {\bf
  B553},  159 (2003)}, \href{http://arxiv.org/abs/hep-ex/0211022}{{\tt
  arXiv:hep-ex/0211022 [hep-ex]}}\relax
\mciteBstWouldAddEndPuncttrue
\mciteSetBstMidEndSepPunct{\mcitedefaultmidpunct}
{\mcitedefaultendpunct}{\mcitedefaultseppunct}\relax
\EndOfBibitem
\bibitem{Schumann:2005ej}
{Belle} collaboration, J.~Schumann {\em et al.},
  \href{http://dx.doi.org/10.1103/PhysRevD.72.011103}{Phys. Rev. {\bf D72},
  011103 (2005)}, \href{http://arxiv.org/abs/hep-ex/0501013}{{\tt
  arXiv:hep-ex/0501013 [hep-ex]}}\relax
\mciteBstWouldAddEndPuncttrue
\mciteSetBstMidEndSepPunct{\mcitedefaultmidpunct}
{\mcitedefaultendpunct}{\mcitedefaultseppunct}\relax
\EndOfBibitem
\bibitem{Abe:2001waa}
{Belle} collaboration, K.~Abe {\em et al.},
  \href{http://dx.doi.org/10.1103/PhysRevLett.87.111801}{Phys. Rev. Lett. {\bf
  87},  111801 (2001)}, \href{http://arxiv.org/abs/hep-ex/0104051}{{\tt
  arXiv:hep-ex/0104051 [hep-ex]}}\relax
\mciteBstWouldAddEndPuncttrue
\mciteSetBstMidEndSepPunct{\mcitedefaultmidpunct}
{\mcitedefaultendpunct}{\mcitedefaultseppunct}\relax
\EndOfBibitem
\bibitem{Aubert:2004at}
{\babar} collaboration, B.~Aubert {\em et al.},
  \href{http://dx.doi.org/10.1103/PhysRevLett.95.171802}{Phys. Rev. Lett. {\bf
  95},  171802 (2005)}, \href{http://arxiv.org/abs/hep-ex/0412040}{{\tt
  arXiv:hep-ex/0412040 [hep-ex]}}\relax
\mciteBstWouldAddEndPuncttrue
\mciteSetBstMidEndSepPunct{\mcitedefaultmidpunct}
{\mcitedefaultendpunct}{\mcitedefaultseppunct}\relax
\EndOfBibitem
\bibitem{Drutskoy:2002ib}
{Belle} collaboration, A.~Drutskoy {\em et al.},
  \href{http://dx.doi.org/10.1016/S0370-2693(02)02373-0}{Phys. Lett. {\bf
  B542},  171 (2002)}, \href{http://arxiv.org/abs/hep-ex/0207041}{{\tt
  arXiv:hep-ex/0207041 [hep-ex]}}\relax
\mciteBstWouldAddEndPuncttrue
\mciteSetBstMidEndSepPunct{\mcitedefaultmidpunct}
{\mcitedefaultendpunct}{\mcitedefaultseppunct}\relax
\EndOfBibitem
\bibitem{Krokovny:2002ua}
{Belle} collaboration, P.~Krokovny {\em et al.},
  \href{http://dx.doi.org/10.1103/PhysRevLett.90.141802}{Phys. Rev. Lett. {\bf
  90},  141802 (2003)}, \href{http://arxiv.org/abs/hep-ex/0212066}{{\tt
  arXiv:hep-ex/0212066 [hep-ex]}}\relax
\mciteBstWouldAddEndPuncttrue
\mciteSetBstMidEndSepPunct{\mcitedefaultmidpunct}
{\mcitedefaultendpunct}{\mcitedefaultseppunct}\relax
\EndOfBibitem
\bibitem{Aubert:2006qn}
{\babar} collaboration, B.~Aubert {\em et al.},
  \href{http://dx.doi.org/10.1103/PhysRevD.74.031101}{Phys. Rev. {\bf D74},
  031101 (2006)}, \href{http://arxiv.org/abs/hep-ex/0604016}{{\tt
  arXiv:hep-ex/0604016 [hep-ex]}}\relax
\mciteBstWouldAddEndPuncttrue
\mciteSetBstMidEndSepPunct{\mcitedefaultmidpunct}
{\mcitedefaultendpunct}{\mcitedefaultseppunct}\relax
\EndOfBibitem
\bibitem{Aubert:2005yt}
{\babar} collaboration, B.~Aubert {\em et al.},
  \href{http://dx.doi.org/10.1103/PhysRevLett.96.011803}{Phys. Rev. Lett. {\bf
  96},  011803 (2006)}, \href{http://arxiv.org/abs/hep-ex/0509036}{{\tt
  arXiv:hep-ex/0509036 [hep-ex]}}\relax
\mciteBstWouldAddEndPuncttrue
\mciteSetBstMidEndSepPunct{\mcitedefaultmidpunct}
{\mcitedefaultendpunct}{\mcitedefaultseppunct}\relax
\EndOfBibitem
\bibitem{Aaij:2018rol}
{LHCb} collaboration, R.~Aaij {\em et al.},
  \href{http://arxiv.org/abs/1807.01891}{{\tt arXiv:1807.01891 [hep-ex]}}\relax
\mciteBstWouldAddEndPuncttrue
\mciteSetBstMidEndSepPunct{\mcitedefaultmidpunct}
{\mcitedefaultendpunct}{\mcitedefaultseppunct}\relax
\EndOfBibitem
\bibitem{Aaij:2018jqv}
{LHCb} collaboration, R.~Aaij {\em et al.},
  \href{http://arxiv.org/abs/1807.01892}{{\tt arXiv:1807.01892 [hep-ex]}}\relax
\mciteBstWouldAddEndPuncttrue
\mciteSetBstMidEndSepPunct{\mcitedefaultmidpunct}
{\mcitedefaultendpunct}{\mcitedefaultseppunct}\relax
\EndOfBibitem
\bibitem{Das:2010be}
{Belle} collaboration, A.~Das {\em et al.},
  \href{http://dx.doi.org/10.1103/PhysRevD.82.051103}{Phys. Rev. {\bf D82},
  051103 (2010)}, \href{http://arxiv.org/abs/1007.4619}{{\tt arXiv:1007.4619
  [hep-ex]}}\relax
\mciteBstWouldAddEndPuncttrue
\mciteSetBstMidEndSepPunct{\mcitedefaultmidpunct}
{\mcitedefaultendpunct}{\mcitedefaultseppunct}\relax
\EndOfBibitem
\bibitem{Aubert:2008zi}
{\babar} collaboration, B.~Aubert {\em et al.},
  \href{http://dx.doi.org/10.1103/PhysRevD.78.032005}{Phys. Rev. {\bf D78},
  032005 (2008)}, \href{http://arxiv.org/abs/0803.4296}{{\tt arXiv:0803.4296
  [hep-ex]}}\relax
\mciteBstWouldAddEndPuncttrue
\mciteSetBstMidEndSepPunct{\mcitedefaultmidpunct}
{\mcitedefaultendpunct}{\mcitedefaultseppunct}\relax
\EndOfBibitem
\bibitem{Joshi:2009yv}
{Belle} collaboration, N.~Joshi {\em et al.},
  \href{http://dx.doi.org/10.1103/PhysRevD.81.031101}{Phys. Rev. {\bf D81},
  031101 (2010)}, \href{http://arxiv.org/abs/0912.2594}{{\tt arXiv:0912.2594
  [hep-ex]}}\relax
\mciteBstWouldAddEndPuncttrue
\mciteSetBstMidEndSepPunct{\mcitedefaultmidpunct}
{\mcitedefaultendpunct}{\mcitedefaultseppunct}\relax
\EndOfBibitem
\bibitem{Aubert:2005qt}
{\babar} collaboration, B.~Aubert {\em et al.},
  \href{http://dx.doi.org/10.1103/PhysRevD.73.071103}{Phys. Rev. {\bf D73},
  071103 (2006)}, \href{http://arxiv.org/abs/hep-ex/0512031}{{\tt
  arXiv:hep-ex/0512031 [hep-ex]}}\relax
\mciteBstWouldAddEndPuncttrue
\mciteSetBstMidEndSepPunct{\mcitedefaultmidpunct}
{\mcitedefaultendpunct}{\mcitedefaultseppunct}\relax
\EndOfBibitem
\bibitem{Aubert:2007xma}
{\babar} collaboration, B.~Aubert {\em et al.},
  \href{http://dx.doi.org/10.1103/PhysRevLett.100.171803}{Phys. Rev. Lett. {\bf
  100},  171803 (2008)}, \href{http://arxiv.org/abs/0707.1043}{{\tt
  arXiv:0707.1043 [hep-ex]}}\relax
\mciteBstWouldAddEndPuncttrue
\mciteSetBstMidEndSepPunct{\mcitedefaultmidpunct}
{\mcitedefaultendpunct}{\mcitedefaultseppunct}\relax
\EndOfBibitem
\bibitem{Aaij:2011rj}
{LHCb} collaboration, R.~Aaij {\em et al.},
  \href{http://dx.doi.org/10.1103/PhysRevD.84.092001}{Phys. Rev. {\bf D84},
  092001 (2011)}, \href{http://arxiv.org/abs/1109.6831}{{\tt arXiv:1109.6831
  [hep-ex]}}, Erratum ibid.\
  \href{http://dx.doi.org/10.1103/PhysRevD.85.039904}{{\bf D85}, 039904},
  (2011)\relax
\mciteBstWouldAddEndPuncttrue
\mciteSetBstMidEndSepPunct{\mcitedefaultmidpunct}
{\mcitedefaultendpunct}{\mcitedefaultseppunct}\relax
\EndOfBibitem
\bibitem{Aaij:2012mra}
{LHCb} collaboration, R.~Aaij {\em et al.},
  \href{http://dx.doi.org/10.1103/PhysRevD.86.112005}{Phys. Rev. {\bf D86},
  112005 (2012)}, \href{http://arxiv.org/abs/1211.1541}{{\tt arXiv:1211.1541
  [hep-ex]}}\relax
\mciteBstWouldAddEndPuncttrue
\mciteSetBstMidEndSepPunct{\mcitedefaultmidpunct}
{\mcitedefaultendpunct}{\mcitedefaultseppunct}\relax
\EndOfBibitem
\bibitem{Aaij:2013xca}
{LHCb} collaboration, R.~Aaij {\em et al.},
  \href{http://dx.doi.org/10.1103/PhysRevD.87.092001}{Phys. Rev. {\bf D87},
  092001 (2013)}, \href{http://arxiv.org/abs/1303.6861}{{\tt arXiv:1303.6861
  [hep-ex]}}\relax
\mciteBstWouldAddEndPuncttrue
\mciteSetBstMidEndSepPunct{\mcitedefaultmidpunct}
{\mcitedefaultendpunct}{\mcitedefaultseppunct}\relax
\EndOfBibitem
\bibitem{Aaij:2013pua}
{LHCb} collaboration, R.~Aaij {\em et al.},
  \href{http://dx.doi.org/10.1103/PhysRevD.87.112009}{Phys. Rev. {\bf D87},
  112009 (2013)}, \href{http://arxiv.org/abs/1304.6317}{{\tt arXiv:1304.6317
  [hep-ex]}}\relax
\mciteBstWouldAddEndPuncttrue
\mciteSetBstMidEndSepPunct{\mcitedefaultmidpunct}
{\mcitedefaultendpunct}{\mcitedefaultseppunct}\relax
\EndOfBibitem
\bibitem{Aaij:2012bw}
{LHCb} collaboration, R.~Aaij {\em et al.},
  \href{http://dx.doi.org/10.1103/PhysRevLett.108.161801}{Phys. Rev. Lett. {\bf
  108},  161801 (2012)}, \href{http://arxiv.org/abs/1201.4402}{{\tt
  arXiv:1201.4402 [hep-ex]}}\relax
\mciteBstWouldAddEndPuncttrue
\mciteSetBstMidEndSepPunct{\mcitedefaultmidpunct}
{\mcitedefaultendpunct}{\mcitedefaultseppunct}\relax
\EndOfBibitem
\bibitem{Aaij:2014jpa}
{LHCb} collaboration, R.~Aaij {\em et al.},
  \href{http://dx.doi.org/10.1007/JHEP05(2015)019}{JHEP {\bf 05},  019 (2015)},
  \href{http://arxiv.org/abs/1412.7654}{{\tt arXiv:1412.7654 [hep-ex]}}\relax
\mciteBstWouldAddEndPuncttrue
\mciteSetBstMidEndSepPunct{\mcitedefaultmidpunct}
{\mcitedefaultendpunct}{\mcitedefaultseppunct}\relax
\EndOfBibitem
\bibitem{Abe:2004sm}
{Belle} collaboration, K.~Abe {\em et al.},
  \href{http://dx.doi.org/10.1103/PhysRevLett.94.221805}{Phys. Rev. Lett. {\bf
  94},  221805 (2005)}, \href{http://arxiv.org/abs/hep-ex/0410091}{{\tt
  arXiv:hep-ex/0410091 [hep-ex]}}\relax
\mciteBstWouldAddEndPuncttrue
\mciteSetBstMidEndSepPunct{\mcitedefaultmidpunct}
{\mcitedefaultendpunct}{\mcitedefaultseppunct}\relax
\EndOfBibitem
\bibitem{Abe:2004wz}
{Belle} collaboration, A.~Drutskoy {\em et al.},
  \href{http://dx.doi.org/10.1103/PhysRevLett.94.061802}{Phys. Rev. Lett. {\bf
  94},  061802 (2005)}, \href{http://arxiv.org/abs/hep-ex/0409026}{{\tt
  arXiv:hep-ex/0409026 [hep-ex]}}\relax
\mciteBstWouldAddEndPuncttrue
\mciteSetBstMidEndSepPunct{\mcitedefaultmidpunct}
{\mcitedefaultendpunct}{\mcitedefaultseppunct}\relax
\EndOfBibitem
\bibitem{delAmoSanchez:2011gi}
{\babar} collaboration, P.~del Amo~Sanchez {\em et al.},
  \href{http://dx.doi.org/10.1103/PhysRevD.85.092017}{Phys. Rev. {\bf D85},
  092017 (2012)}, \href{http://arxiv.org/abs/1111.4387}{{\tt arXiv:1111.4387
  [hep-ex]}}\relax
\mciteBstWouldAddEndPuncttrue
\mciteSetBstMidEndSepPunct{\mcitedefaultmidpunct}
{\mcitedefaultendpunct}{\mcitedefaultseppunct}\relax
\EndOfBibitem
\bibitem{Abe:2002tw}
{Belle} collaboration, K.~Abe {\em et al.},
  \href{http://dx.doi.org/10.1103/PhysRevLett.89.151802}{Phys. Rev. Lett. {\bf
  89},  151802 (2002)}, \href{http://arxiv.org/abs/hep-ex/0205083}{{\tt
  arXiv:hep-ex/0205083 [hep-ex]}}\relax
\mciteBstWouldAddEndPuncttrue
\mciteSetBstMidEndSepPunct{\mcitedefaultmidpunct}
{\mcitedefaultendpunct}{\mcitedefaultseppunct}\relax
\EndOfBibitem
\bibitem{Medvedeva:2007af}
{Belle} collaboration, T.~Medvedeva {\em et al.},
  \href{http://dx.doi.org/10.1103/PhysRevD.76.051102}{Phys. Rev. {\bf D76},
  051102 (2007)}, \href{http://arxiv.org/abs/0704.2652}{{\tt arXiv:0704.2652
  [hep-ex]}}\relax
\mciteBstWouldAddEndPuncttrue
\mciteSetBstMidEndSepPunct{\mcitedefaultmidpunct}
{\mcitedefaultendpunct}{\mcitedefaultseppunct}\relax
\EndOfBibitem
\bibitem{Chang:2008yw}
{Belle} collaboration, Y.~W. Chang {\em et al.},
  \href{http://dx.doi.org/10.1103/PhysRevD.79.052006}{Phys. Rev. {\bf D79},
  052006 (2009)}, \href{http://arxiv.org/abs/0811.3826}{{\tt arXiv:0811.3826
  [hep-ex]}}\relax
\mciteBstWouldAddEndPuncttrue
\mciteSetBstMidEndSepPunct{\mcitedefaultmidpunct}
{\mcitedefaultendpunct}{\mcitedefaultseppunct}\relax
\EndOfBibitem
\bibitem{Lees:2014mka}
{\babar} collaboration, J.~P. Lees {\em et al.},
  \href{http://dx.doi.org/10.1103/PhysRevD.89.112002}{Phys. Rev. {\bf D89},
  112002 (2014)}, \href{http://arxiv.org/abs/1401.5990}{{\tt arXiv:1401.5990
  [hep-ex]}}\relax
\mciteBstWouldAddEndPuncttrue
\mciteSetBstMidEndSepPunct{\mcitedefaultmidpunct}
{\mcitedefaultendpunct}{\mcitedefaultseppunct}\relax
\EndOfBibitem
\bibitem{Chang:2015fja}
{Belle} collaboration, Y.~Y. Chang {\em et al.},
  \href{http://dx.doi.org/10.1103/PhysRevLett.115.221803}{Phys. Rev. Lett. {\bf
  115},  221803 (2015)}, \href{http://arxiv.org/abs/1509.03034}{{\tt
  arXiv:1509.03034 [hep-ex]}}\relax
\mciteBstWouldAddEndPuncttrue
\mciteSetBstMidEndSepPunct{\mcitedefaultmidpunct}
{\mcitedefaultendpunct}{\mcitedefaultseppunct}\relax
\EndOfBibitem
\bibitem{Aubert:2006ia}
{\babar} collaboration, B.~Aubert {\em et al.},
  \href{http://dx.doi.org/10.1103/PhysRevD.73.112004}{Phys. Rev. {\bf D73},
  112004 (2006)}, \href{http://arxiv.org/abs/hep-ex/0604037}{{\tt
  arXiv:hep-ex/0604037 [hep-ex]}}\relax
\mciteBstWouldAddEndPuncttrue
\mciteSetBstMidEndSepPunct{\mcitedefaultmidpunct}
{\mcitedefaultendpunct}{\mcitedefaultseppunct}\relax
\EndOfBibitem
\bibitem{Adachi:2008cj}
{Belle} collaboration, I.~Adachi {\em et al.},
  \href{http://dx.doi.org/10.1103/PhysRevD.77.091101}{Phys. Rev. {\bf D77},
  091101 (2008)}, \href{http://arxiv.org/abs/0802.2988}{{\tt arXiv:0802.2988
  [hep-ex]}}\relax
\mciteBstWouldAddEndPuncttrue
\mciteSetBstMidEndSepPunct{\mcitedefaultmidpunct}
{\mcitedefaultendpunct}{\mcitedefaultseppunct}\relax
\EndOfBibitem
\bibitem{delAmoSanchez:2010pg}
{\babar} collaboration, P.~del Amo~Sanchez {\em et al.},
  \href{http://dx.doi.org/10.1103/PhysRevD.83.032004}{Phys. Rev. {\bf D83},
  032004 (2011)}, \href{http://arxiv.org/abs/1011.3929}{{\tt arXiv:1011.3929
  [hep-ex]}}\relax
\mciteBstWouldAddEndPuncttrue
\mciteSetBstMidEndSepPunct{\mcitedefaultmidpunct}
{\mcitedefaultendpunct}{\mcitedefaultseppunct}\relax
\EndOfBibitem
\bibitem{Gokhroo:2006bt}
{Belle} collaboration, G.~Gokhroo {\em et al.},
  \href{http://dx.doi.org/10.1103/PhysRevLett.97.162002}{Phys. Rev. Lett. {\bf
  97},  162002 (2006)}, \href{http://arxiv.org/abs/hep-ex/0606055}{{\tt
  arXiv:hep-ex/0606055 [hep-ex]}}\relax
\mciteBstWouldAddEndPuncttrue
\mciteSetBstMidEndSepPunct{\mcitedefaultmidpunct}
{\mcitedefaultendpunct}{\mcitedefaultseppunct}\relax
\EndOfBibitem
\bibitem{Zupanc:2007pu}
{Belle} collaboration, A.~Zupanc {\em et al.},
  \href{http://dx.doi.org/10.1103/PhysRevD.75.091102}{Phys. Rev. {\bf D75},
  091102 (2007)}, \href{http://arxiv.org/abs/hep-ex/0703040}{{\tt
  arXiv:hep-ex/0703040 [hep-ex]}}\relax
\mciteBstWouldAddEndPuncttrue
\mciteSetBstMidEndSepPunct{\mcitedefaultmidpunct}
{\mcitedefaultendpunct}{\mcitedefaultseppunct}\relax
\EndOfBibitem
\bibitem{Aubert:2006nm}
{\babar} collaboration, B.~Aubert {\em et al.},
  \href{http://dx.doi.org/10.1103/PhysRevD.74.031103}{Phys. Rev. {\bf D74},
  031103 (2006)}, \href{http://arxiv.org/abs/hep-ex/0605036}{{\tt
  arXiv:hep-ex/0605036 [hep-ex]}}\relax
\mciteBstWouldAddEndPuncttrue
\mciteSetBstMidEndSepPunct{\mcitedefaultmidpunct}
{\mcitedefaultendpunct}{\mcitedefaultseppunct}\relax
\EndOfBibitem
\bibitem{Aubert:2003jj}
{\babar} collaboration, B.~Aubert {\em et al.},
  \href{http://dx.doi.org/10.1103/PhysRevD.67.092003}{Phys. Rev. {\bf D67},
  092003 (2003)}, \href{http://arxiv.org/abs/hep-ex/0302015}{{\tt
  arXiv:hep-ex/0302015 [hep-ex]}}\relax
\mciteBstWouldAddEndPuncttrue
\mciteSetBstMidEndSepPunct{\mcitedefaultmidpunct}
{\mcitedefaultendpunct}{\mcitedefaultseppunct}\relax
\EndOfBibitem
\bibitem{Aubert:2005xu}
{\babar} collaboration, B.~Aubert {\em et al.},
  \href{http://dx.doi.org/10.1103/PhysRevD.71.091104}{Phys. Rev. {\bf D71},
  091104 (2005)}, \href{http://arxiv.org/abs/hep-ex/0502041}{{\tt
  arXiv:hep-ex/0502041 [hep-ex]}}\relax
\mciteBstWouldAddEndPuncttrue
\mciteSetBstMidEndSepPunct{\mcitedefaultmidpunct}
{\mcitedefaultendpunct}{\mcitedefaultseppunct}\relax
\EndOfBibitem
\bibitem{Aubert:2005jv}
{\babar} collaboration, B.~Aubert {\em et al.},
  \href{http://dx.doi.org/10.1103/PhysRevD.72.111101}{Phys. Rev. {\bf D72},
  111101 (2005)}, \href{http://arxiv.org/abs/hep-ex/0510051}{{\tt
  arXiv:hep-ex/0510051 [hep-ex]}}\relax
\mciteBstWouldAddEndPuncttrue
\mciteSetBstMidEndSepPunct{\mcitedefaultmidpunct}
{\mcitedefaultendpunct}{\mcitedefaultseppunct}\relax
\EndOfBibitem
\bibitem{Aaij:2013fha}
{LHCb} collaboration, R.~Aaij {\em et al.},
  \href{http://dx.doi.org/10.1103/PhysRevD.87.092007}{Phys. Rev. {\bf D87},
  092007 (2013)}, \href{http://arxiv.org/abs/1302.5854}{{\tt arXiv:1302.5854
  [hep-ex]}}\relax
\mciteBstWouldAddEndPuncttrue
\mciteSetBstMidEndSepPunct{\mcitedefaultmidpunct}
{\mcitedefaultendpunct}{\mcitedefaultseppunct}\relax
\EndOfBibitem
\bibitem{Krokovny:2003zq}
{Belle} collaboration, P.~Krokovny {\em et al.},
  \href{http://dx.doi.org/10.1103/PhysRevLett.91.262002}{Phys. Rev. Lett. {\bf
  91},  262002 (2003)}, \href{http://arxiv.org/abs/hep-ex/0308019}{{\tt
  arXiv:hep-ex/0308019 [hep-ex]}}\relax
\mciteBstWouldAddEndPuncttrue
\mciteSetBstMidEndSepPunct{\mcitedefaultmidpunct}
{\mcitedefaultendpunct}{\mcitedefaultseppunct}\relax
\EndOfBibitem
\bibitem{Aubert:2004pw}
{\babar} collaboration, B.~Aubert {\em et al.},
  \href{http://dx.doi.org/10.1103/PhysRevLett.93.181801}{Phys. Rev. Lett. {\bf
  93},  181801 (2004)}, \href{http://arxiv.org/abs/hep-ex/0408041}{{\tt
  arXiv:hep-ex/0408041 [hep-ex]}}\relax
\mciteBstWouldAddEndPuncttrue
\mciteSetBstMidEndSepPunct{\mcitedefaultmidpunct}
{\mcitedefaultendpunct}{\mcitedefaultseppunct}\relax
\EndOfBibitem
\bibitem{Choi:2015lpc}
{Belle} collaboration, S.~K. Choi {\em et al.},
  \href{http://dx.doi.org/10.1103/PhysRevD.91.092011}{Phys. Rev. {\bf D91},
  092011 (2015)}, \href{http://arxiv.org/abs/1504.02637}{{\tt arXiv:1504.02637
  [hep-ex]}}, Addendum ibid.\
  \href{http://dx.doi.org/10.1103/PhysRevD.92.039905}{{\bf D92}, 039905},
  (2015)\relax
\mciteBstWouldAddEndPuncttrue
\mciteSetBstMidEndSepPunct{\mcitedefaultmidpunct}
{\mcitedefaultendpunct}{\mcitedefaultseppunct}\relax
\EndOfBibitem
\bibitem{Aubert:2007rva}
{\babar} collaboration, B.~Aubert {\em et al.},
  \href{http://dx.doi.org/10.1103/PhysRevD.77.011102}{Phys. Rev. {\bf D77},
  011102 (2008)}, \href{http://arxiv.org/abs/0708.1565}{{\tt arXiv:0708.1565
  [hep-ex]}}\relax
\mciteBstWouldAddEndPuncttrue
\mciteSetBstMidEndSepPunct{\mcitedefaultmidpunct}
{\mcitedefaultendpunct}{\mcitedefaultseppunct}\relax
\EndOfBibitem
\bibitem{Belle:2011ad}
{Belle} collaboration, T.~Aushev {\em et al.},
  \href{http://dx.doi.org/10.1103/PhysRevD.83.059902}{Phys. Rev. {\bf D83},
  051102 (2011)}, \href{http://arxiv.org/abs/1102.0935}{{\tt arXiv:1102.0935
  [hep-ex]}}, Erratum ibid.\
  \href{http://dx.doi.org/10.1103/PhysRevD.83.051102}{{\bf D83}, 051102},
  (2011)\relax
\mciteBstWouldAddEndPuncttrue
\mciteSetBstMidEndSepPunct{\mcitedefaultmidpunct}
{\mcitedefaultendpunct}{\mcitedefaultseppunct}\relax
\EndOfBibitem
\bibitem{Abe:1995aw}
{CDF} collaboration, F.~Abe {\em et al.},
  \href{http://dx.doi.org/10.1103/PhysRevLett.76.2015}{Phys. Rev. Lett. {\bf
  76},  2015 (1996)}\relax
\mciteBstWouldAddEndPuncttrue
\mciteSetBstMidEndSepPunct{\mcitedefaultmidpunct}
{\mcitedefaultendpunct}{\mcitedefaultseppunct}\relax
\EndOfBibitem
\bibitem{Abe:2002rc}
{Belle} collaboration, K.~Abe {\em et al.},
  \href{http://dx.doi.org/10.1103/PhysRevD.67.032003}{Phys. Rev. {\bf D67},
  032003 (2003)}, \href{http://arxiv.org/abs/hep-ex/0211047}{{\tt
  arXiv:hep-ex/0211047 [hep-ex]}}\relax
\mciteBstWouldAddEndPuncttrue
\mciteSetBstMidEndSepPunct{\mcitedefaultmidpunct}
{\mcitedefaultendpunct}{\mcitedefaultseppunct}\relax
\EndOfBibitem
\bibitem{Chilikin:2014bkk}
{Belle} collaboration, K.~Chilikin {\em et al.},
  \href{http://dx.doi.org/10.1103/PhysRevD.90.112009}{Phys. Rev. {\bf D90},
  112009 (2014)}, \href{http://arxiv.org/abs/1408.6457}{{\tt arXiv:1408.6457
  [hep-ex]}}\relax
\mciteBstWouldAddEndPuncttrue
\mciteSetBstMidEndSepPunct{\mcitedefaultmidpunct}
{\mcitedefaultendpunct}{\mcitedefaultseppunct}\relax
\EndOfBibitem
\bibitem{Abe:1998yu}
{CDF} collaboration, F.~Abe {\em et al.},
  \href{http://dx.doi.org/10.1103/PhysRevD.58.072001}{Phys. Rev. {\bf D58},
  072001 (1998)}, \href{http://arxiv.org/abs/hep-ex/9803013}{{\tt
  arXiv:hep-ex/9803013 [hep-ex]}}\relax
\mciteBstWouldAddEndPuncttrue
\mciteSetBstMidEndSepPunct{\mcitedefaultmidpunct}
{\mcitedefaultendpunct}{\mcitedefaultseppunct}\relax
\EndOfBibitem
\bibitem{Aaij:2014naa}
{LHCb} collaboration, R.~Aaij {\em et al.},
  \href{http://dx.doi.org/10.1007/JHEP07(2014)140}{JHEP {\bf 07},  140 (2014)},
  \href{http://arxiv.org/abs/1405.3219}{{\tt arXiv:1405.3219 [hep-ex]}}\relax
\mciteBstWouldAddEndPuncttrue
\mciteSetBstMidEndSepPunct{\mcitedefaultmidpunct}
{\mcitedefaultendpunct}{\mcitedefaultseppunct}\relax
\EndOfBibitem
\bibitem{Affolder:2001qi}
{CDF} collaboration, T.~Affolder {\em et al.},
  \href{http://dx.doi.org/10.1103/PhysRevLett.88.071801}{Phys. Rev. Lett. {\bf
  88},  071801 (2002)}, \href{http://arxiv.org/abs/hep-ex/0108022}{{\tt
  arXiv:hep-ex/0108022 [hep-ex]}}\relax
\mciteBstWouldAddEndPuncttrue
\mciteSetBstMidEndSepPunct{\mcitedefaultmidpunct}
{\mcitedefaultendpunct}{\mcitedefaultseppunct}\relax
\EndOfBibitem
\bibitem{delAmoSanchez:2010jr}
{\babar} collaboration, P.~del Amo~Sanchez {\em et al.},
  \href{http://dx.doi.org/10.1103/PhysRevD.82.011101}{Phys. Rev. {\bf D82},
  011101 (2010)}, \href{http://arxiv.org/abs/1005.5190}{{\tt arXiv:1005.5190
  [hep-ex]}}\relax
\mciteBstWouldAddEndPuncttrue
\mciteSetBstMidEndSepPunct{\mcitedefaultmidpunct}
{\mcitedefaultendpunct}{\mcitedefaultseppunct}\relax
\EndOfBibitem
\bibitem{Aubert:2003ii}
{\babar} collaboration, B.~Aubert {\em et al.},
  \href{http://dx.doi.org/10.1103/PhysRevLett.91.071801}{Phys. Rev. Lett. {\bf
  91},  071801 (2003)}, \href{http://arxiv.org/abs/hep-ex/0304014}{{\tt
  arXiv:hep-ex/0304014 [hep-ex]}}\relax
\mciteBstWouldAddEndPuncttrue
\mciteSetBstMidEndSepPunct{\mcitedefaultmidpunct}
{\mcitedefaultendpunct}{\mcitedefaultseppunct}\relax
\EndOfBibitem
\bibitem{Abe:2001wa}
{Belle} collaboration, K.~Abe {\em et al.},
  \href{http://dx.doi.org/10.1103/PhysRevLett.87.161601}{Phys. Rev. Lett. {\bf
  87},  161601 (2001)}, \href{http://arxiv.org/abs/hep-ex/0105014}{{\tt
  arXiv:hep-ex/0105014 [hep-ex]}}\relax
\mciteBstWouldAddEndPuncttrue
\mciteSetBstMidEndSepPunct{\mcitedefaultmidpunct}
{\mcitedefaultendpunct}{\mcitedefaultseppunct}\relax
\EndOfBibitem
\bibitem{Iwashita:2013wnn}
{Belle} collaboration, T.~Iwashita {\em et al.},
  \href{http://dx.doi.org/10.1093/ptep/ptu043}{PTEP {\bf 2014},  043C01
  (2014)}, \href{http://arxiv.org/abs/1310.2704}{{\tt arXiv:1310.2704
  [hep-ex]}}\relax
\mciteBstWouldAddEndPuncttrue
\mciteSetBstMidEndSepPunct{\mcitedefaultmidpunct}
{\mcitedefaultendpunct}{\mcitedefaultseppunct}\relax
\EndOfBibitem
\bibitem{Aubert:2004fc}
{\babar} collaboration, B.~Aubert {\em et al.},
  \href{http://dx.doi.org/10.1103/PhysRevLett.93.041801}{Phys. Rev. Lett. {\bf
  93},  041801 (2004)}, \href{http://arxiv.org/abs/hep-ex/0402025}{{\tt
  arXiv:hep-ex/0402025 [hep-ex]}}\relax
\mciteBstWouldAddEndPuncttrue
\mciteSetBstMidEndSepPunct{\mcitedefaultmidpunct}
{\mcitedefaultendpunct}{\mcitedefaultseppunct}\relax
\EndOfBibitem
\bibitem{Mizuk:2009da}
{Belle} collaboration, R.~Mizuk {\em et al.},
  \href{http://dx.doi.org/10.1103/PhysRevD.80.031104}{Phys. Rev. {\bf D80},
  031104 (2009)}, \href{http://arxiv.org/abs/0905.2869}{{\tt arXiv:0905.2869
  [hep-ex]}}\relax
\mciteBstWouldAddEndPuncttrue
\mciteSetBstMidEndSepPunct{\mcitedefaultmidpunct}
{\mcitedefaultendpunct}{\mcitedefaultseppunct}\relax
\EndOfBibitem
\bibitem{Bhardwaj:2013rmw}
{Belle} collaboration, V.~Bhardwaj {\em et al.},
  \href{http://dx.doi.org/10.1103/PhysRevLett.111.032001}{Phys. Rev. Lett. {\bf
  111},  032001 (2013)}, \href{http://arxiv.org/abs/1304.3975}{{\tt
  arXiv:1304.3975 [hep-ex]}}\relax
\mciteBstWouldAddEndPuncttrue
\mciteSetBstMidEndSepPunct{\mcitedefaultmidpunct}
{\mcitedefaultendpunct}{\mcitedefaultseppunct}\relax
\EndOfBibitem
\bibitem{Aubert:2005vwa}
{\babar} collaboration, B.~Aubert {\em et al.},
  \href{http://dx.doi.org/10.1103/PhysRevLett.94.171801}{Phys. Rev. Lett. {\bf
  94},  171801 (2005)}, \href{http://arxiv.org/abs/hep-ex/0501061}{{\tt
  arXiv:hep-ex/0501061 [hep-ex]}}\relax
\mciteBstWouldAddEndPuncttrue
\mciteSetBstMidEndSepPunct{\mcitedefaultmidpunct}
{\mcitedefaultendpunct}{\mcitedefaultseppunct}\relax
\EndOfBibitem
\bibitem{Aubert:2008ak}
{\babar} collaboration, B.~Aubert {\em et al.},
  \href{http://dx.doi.org/10.1103/PhysRevD.78.091101}{Phys. Rev. {\bf D78},
  091101 (2008)}, \href{http://arxiv.org/abs/0808.1487}{{\tt arXiv:0808.1487
  [hep-ex]}}\relax
\mciteBstWouldAddEndPuncttrue
\mciteSetBstMidEndSepPunct{\mcitedefaultmidpunct}
{\mcitedefaultendpunct}{\mcitedefaultseppunct}\relax
\EndOfBibitem
\bibitem{Bhardwaj:2011dj}
{Belle} collaboration, V.~Bhardwaj {\em et al.},
  \href{http://dx.doi.org/10.1103/PhysRevLett.107.091803}{Phys. Rev. Lett. {\bf
  107},  091803 (2011)}, \href{http://arxiv.org/abs/1105.0177}{{\tt
  arXiv:1105.0177 [hep-ex]}}\relax
\mciteBstWouldAddEndPuncttrue
\mciteSetBstMidEndSepPunct{\mcitedefaultmidpunct}
{\mcitedefaultendpunct}{\mcitedefaultseppunct}\relax
\EndOfBibitem
\bibitem{Aubert:2008ae}
{\babar} collaboration, B.~Aubert {\em et al.},
  \href{http://dx.doi.org/10.1103/PhysRevLett.102.132001}{Phys. Rev. Lett. {\bf
  102},  132001 (2009)}, \href{http://arxiv.org/abs/0809.0042}{{\tt
  arXiv:0809.0042 [hep-ex]}}\relax
\mciteBstWouldAddEndPuncttrue
\mciteSetBstMidEndSepPunct{\mcitedefaultmidpunct}
{\mcitedefaultendpunct}{\mcitedefaultseppunct}\relax
\EndOfBibitem
\bibitem{Bhardwaj:2015rju}
{Belle} collaboration, V.~Bhardwaj {\em et al.},
  \href{http://dx.doi.org/10.1103/PhysRevD.93.052016}{Phys. Rev. {\bf D93},
  052016 (2016)}, \href{http://arxiv.org/abs/1512.02672}{{\tt arXiv:1512.02672
  [hep-ex]}}\relax
\mciteBstWouldAddEndPuncttrue
\mciteSetBstMidEndSepPunct{\mcitedefaultmidpunct}
{\mcitedefaultendpunct}{\mcitedefaultseppunct}\relax
\EndOfBibitem
\bibitem{Lees:2011ik}
{\babar} collaboration, J.~P. Lees {\em et al.},
  \href{http://dx.doi.org/10.1103/PhysRevD.85.052003}{Phys. Rev. {\bf D85},
  052003 (2012)}, \href{http://arxiv.org/abs/1111.5919}{{\tt arXiv:1111.5919
  [hep-ex]}}\relax
\mciteBstWouldAddEndPuncttrue
\mciteSetBstMidEndSepPunct{\mcitedefaultmidpunct}
{\mcitedefaultendpunct}{\mcitedefaultseppunct}\relax
\EndOfBibitem
\bibitem{Soni:2005fw}
{Belle} collaboration, N.~Soni {\em et al.},
  \href{http://dx.doi.org/10.1016/j.physletb.2006.01.013}{Phys. Lett. {\bf
  B634},  155 (2006)}, \href{http://arxiv.org/abs/hep-ex/0508032}{{\tt
  arXiv:hep-ex/0508032 [hep-ex]}}\relax
\mciteBstWouldAddEndPuncttrue
\mciteSetBstMidEndSepPunct{\mcitedefaultmidpunct}
{\mcitedefaultendpunct}{\mcitedefaultseppunct}\relax
\EndOfBibitem
\bibitem{Fang:2002gi}
{Belle} collaboration, F.~Fang {\em et al.},
  \href{http://dx.doi.org/10.1103/PhysRevLett.90.071801}{Phys. Rev. Lett. {\bf
  90},  071801 (2003)}, \href{http://arxiv.org/abs/hep-ex/0208047}{{\tt
  arXiv:hep-ex/0208047 [hep-ex]}}\relax
\mciteBstWouldAddEndPuncttrue
\mciteSetBstMidEndSepPunct{\mcitedefaultmidpunct}
{\mcitedefaultendpunct}{\mcitedefaultseppunct}\relax
\EndOfBibitem
\bibitem{Aubert:2007qea}
{\babar} collaboration, B.~Aubert {\em et al.},
  \href{http://dx.doi.org/10.1103/PhysRevD.76.092004}{Phys. Rev. {\bf D76},
  092004 (2007)}, \href{http://arxiv.org/abs/0707.1648}{{\tt arXiv:0707.1648
  [hep-ex]}}\relax
\mciteBstWouldAddEndPuncttrue
\mciteSetBstMidEndSepPunct{\mcitedefaultmidpunct}
{\mcitedefaultendpunct}{\mcitedefaultseppunct}\relax
\EndOfBibitem
\bibitem{Aubert:2004gc}
{\babar} collaboration, B.~Aubert {\em et al.},
  \href{http://dx.doi.org/10.1103/PhysRevD.70.011101}{Phys. Rev. {\bf D70},
  011101 (2004)}, \href{http://arxiv.org/abs/hep-ex/0403007}{{\tt
  arXiv:hep-ex/0403007 [hep-ex]}}\relax
\mciteBstWouldAddEndPuncttrue
\mciteSetBstMidEndSepPunct{\mcitedefaultmidpunct}
{\mcitedefaultendpunct}{\mcitedefaultseppunct}\relax
\EndOfBibitem
\bibitem{Aubert:2008kp}
{\babar} collaboration, B.~Aubert {\em et al.},
  \href{http://dx.doi.org/10.1103/PhysRevD.78.012006}{Phys. Rev. {\bf D78},
  012006 (2008)}, \href{http://arxiv.org/abs/0804.1208}{{\tt arXiv:0804.1208
  [hep-ex]}}\relax
\mciteBstWouldAddEndPuncttrue
\mciteSetBstMidEndSepPunct{\mcitedefaultmidpunct}
{\mcitedefaultendpunct}{\mcitedefaultseppunct}\relax
\EndOfBibitem
\bibitem{Aaij:2018bla}
{LHCb} collaboration, R.~Aaij {\em et al.}, Submitted to: Eur. Phys. J.
  (2018), \href{http://arxiv.org/abs/1809.07416}{{\tt arXiv:1809.07416
  [hep-ex]}}\relax
\mciteBstWouldAddEndPuncttrue
\mciteSetBstMidEndSepPunct{\mcitedefaultmidpunct}
{\mcitedefaultendpunct}{\mcitedefaultseppunct}\relax
\EndOfBibitem
\bibitem{Aubert:2007xw}
{\babar} collaboration, B.~Aubert {\em et al.},
  \href{http://dx.doi.org/10.1103/PhysRevD.76.031101}{Phys. Rev. {\bf D76},
  031101 (2007)}, \href{http://arxiv.org/abs/0704.1266}{{\tt arXiv:0704.1266
  [hep-ex]}}\relax
\mciteBstWouldAddEndPuncttrue
\mciteSetBstMidEndSepPunct{\mcitedefaultmidpunct}
{\mcitedefaultendpunct}{\mcitedefaultseppunct}\relax
\EndOfBibitem
\bibitem{Chang:2012gnb}
{Belle} collaboration, M.~Chang {\em et al.},
  \href{http://dx.doi.org/10.1103/PhysRevD.85.091102}{Phys. Rev. {\bf D85},
  091102 (2012)}, \href{http://arxiv.org/abs/1203.3399}{{\tt arXiv:1203.3399
  [hep-ex]}}\relax
\mciteBstWouldAddEndPuncttrue
\mciteSetBstMidEndSepPunct{\mcitedefaultmidpunct}
{\mcitedefaultendpunct}{\mcitedefaultseppunct}\relax
\EndOfBibitem
\bibitem{Aaij:2013rja}
{LHCb} collaboration, R.~Aaij {\em et al.},
  \href{http://dx.doi.org/10.1103/PhysRevLett.112.091802}{Phys. Rev. Lett. {\bf
  112},  091802 (2014)}, \href{http://arxiv.org/abs/1310.2145}{{\tt
  arXiv:1310.2145 [hep-ex]}}\relax
\mciteBstWouldAddEndPuncttrue
\mciteSetBstMidEndSepPunct{\mcitedefaultmidpunct}
{\mcitedefaultendpunct}{\mcitedefaultseppunct}\relax
\EndOfBibitem
\bibitem{Kumar:2008ir}
{Belle} collaboration, R.~Kumar {\em et al.},
  \href{http://dx.doi.org/10.1103/PhysRevD.78.091104}{Phys. Rev. {\bf D78},
  091104 (2008)}, \href{http://arxiv.org/abs/0809.1778}{{\tt arXiv:0809.1778
  [hep-ex]}}\relax
\mciteBstWouldAddEndPuncttrue
\mciteSetBstMidEndSepPunct{\mcitedefaultmidpunct}
{\mcitedefaultendpunct}{\mcitedefaultseppunct}\relax
\EndOfBibitem
\bibitem{Aaij:2013mtm}
{LHCb} collaboration, R.~Aaij {\em et al.},
  \href{http://dx.doi.org/10.1103/PhysRevD.88.072005}{Phys. Rev. {\bf D88},
  072005 (2013)}, \href{http://arxiv.org/abs/1308.5916}{{\tt arXiv:1308.5916
  [hep-ex]}}\relax
\mciteBstWouldAddEndPuncttrue
\mciteSetBstMidEndSepPunct{\mcitedefaultmidpunct}
{\mcitedefaultendpunct}{\mcitedefaultseppunct}\relax
\EndOfBibitem
\bibitem{Aaij:2013zpt}
{LHCb} collaboration, R.~Aaij {\em et al.},
  \href{http://dx.doi.org/10.1103/PhysRevD.87.052001}{Phys. Rev. {\bf D87},
  052001 (2013)}, \href{http://arxiv.org/abs/1301.5347}{{\tt arXiv:1301.5347
  [hep-ex]}}\relax
\mciteBstWouldAddEndPuncttrue
\mciteSetBstMidEndSepPunct{\mcitedefaultmidpunct}
{\mcitedefaultendpunct}{\mcitedefaultseppunct}\relax
\EndOfBibitem
\bibitem{Liu:2008bta}
{Belle} collaboration, Y.~Liu {\em et al.},
  \href{http://dx.doi.org/10.1103/PhysRevD.78.011106}{Phys. Rev. {\bf D78},
  011106 (2008)}, \href{http://arxiv.org/abs/0805.3225}{{\tt arXiv:0805.3225
  [hep-ex]}}\relax
\mciteBstWouldAddEndPuncttrue
\mciteSetBstMidEndSepPunct{\mcitedefaultmidpunct}
{\mcitedefaultendpunct}{\mcitedefaultseppunct}\relax
\EndOfBibitem
\bibitem{Aaij:2015uoa}
{LHCb} collaboration, R.~Aaij {\em et al.},
  \href{http://dx.doi.org/10.1103/PhysRevD.92.112002}{Phys. Rev. {\bf D92},
  112002 (2015)}, \href{http://arxiv.org/abs/1510.04866}{{\tt arXiv:1510.04866
  [hep-ex]}}\relax
\mciteBstWouldAddEndPuncttrue
\mciteSetBstMidEndSepPunct{\mcitedefaultmidpunct}
{\mcitedefaultendpunct}{\mcitedefaultseppunct}\relax
\EndOfBibitem
\bibitem{Aubert:2004xd}
{\babar} collaboration, B.~Aubert {\em et al.},
  \href{http://dx.doi.org/10.1103/PhysRevD.70.091104}{Phys. Rev. {\bf D70},
  091104 (2004)}, \href{http://arxiv.org/abs/hep-ex/0408018}{{\tt
  arXiv:hep-ex/0408018 [hep-ex]}}\relax
\mciteBstWouldAddEndPuncttrue
\mciteSetBstMidEndSepPunct{\mcitedefaultmidpunct}
{\mcitedefaultendpunct}{\mcitedefaultseppunct}\relax
\EndOfBibitem
\bibitem{Aaij:2013yba}
{LHCb} collaboration, R.~Aaij {\em et al.},
  \href{http://dx.doi.org/10.1007/JHEP09(2013)006}{JHEP {\bf 09},  006 (2013)},
  \href{http://arxiv.org/abs/1306.4489}{{\tt arXiv:1306.4489 [hep-ex]}}\relax
\mciteBstWouldAddEndPuncttrue
\mciteSetBstMidEndSepPunct{\mcitedefaultmidpunct}
{\mcitedefaultendpunct}{\mcitedefaultseppunct}\relax
\EndOfBibitem
\bibitem{Xie:2005tf}
{Belle} collaboration, Q.~Xie {\em et al.},
  \href{http://dx.doi.org/10.1103/PhysRevD.72.051105}{Phys. Rev. {\bf D72},
  051105 (2005)}, \href{http://arxiv.org/abs/hep-ex/0508011}{{\tt
  arXiv:hep-ex/0508011 [hep-ex]}}\relax
\mciteBstWouldAddEndPuncttrue
\mciteSetBstMidEndSepPunct{\mcitedefaultmidpunct}
{\mcitedefaultendpunct}{\mcitedefaultseppunct}\relax
\EndOfBibitem
\bibitem{Aubert:2003ww}
{\babar} collaboration, B.~Aubert {\em et al.},
  \href{http://dx.doi.org/10.1103/PhysRevLett.90.231801}{Phys. Rev. Lett. {\bf
  90},  231801 (2003)}, \href{http://arxiv.org/abs/hep-ex/0303036}{{\tt
  arXiv:hep-ex/0303036 [hep-ex]}}\relax
\mciteBstWouldAddEndPuncttrue
\mciteSetBstMidEndSepPunct{\mcitedefaultmidpunct}
{\mcitedefaultendpunct}{\mcitedefaultseppunct}\relax
\EndOfBibitem
\bibitem{Zhang:2005bs}
{Belle} collaboration, L.~Zhang {\em et al.},
  \href{http://dx.doi.org/10.1103/PhysRevD.71.091107}{Phys. Rev. {\bf D71},
  091107 (2005)}, \href{http://arxiv.org/abs/hep-ex/0503037}{{\tt
  arXiv:hep-ex/0503037 [hep-ex]}}\relax
\mciteBstWouldAddEndPuncttrue
\mciteSetBstMidEndSepPunct{\mcitedefaultmidpunct}
{\mcitedefaultendpunct}{\mcitedefaultseppunct}\relax
\EndOfBibitem
\bibitem{Aubert:2005tr}
{\babar} collaboration, B.~Aubert {\em et al.},
  \href{http://dx.doi.org/10.1103/PhysRevD.71.091103}{Phys. Rev. {\bf D71},
  091103 (2005)}, \href{http://arxiv.org/abs/hep-ex/0503021}{{\tt
  arXiv:hep-ex/0503021 [hep-ex]}}\relax
\mciteBstWouldAddEndPuncttrue
\mciteSetBstMidEndSepPunct{\mcitedefaultmidpunct}
{\mcitedefaultendpunct}{\mcitedefaultseppunct}\relax
\EndOfBibitem
\bibitem{Abe:1996kc}
{CDF} collaboration, F.~Abe {\em et al.},
  \href{http://dx.doi.org/10.1103/PhysRevD.54.6596}{Phys. Rev. {\bf D54},  6596
  (1996)}, \href{http://arxiv.org/abs/hep-ex/9607003}{{\tt arXiv:hep-ex/9607003
  [hep-ex]}}\relax
\mciteBstWouldAddEndPuncttrue
\mciteSetBstMidEndSepPunct{\mcitedefaultmidpunct}
{\mcitedefaultendpunct}{\mcitedefaultseppunct}\relax
\EndOfBibitem
\bibitem{LHCb:2012cw}
{LHCb} collaboration, R.~Aaij {\em et al.},
  \href{http://dx.doi.org/10.1016/j.nuclphysb.2012.10.021}{Nucl. Phys. {\bf
  B867},  547 (2013)}, \href{http://arxiv.org/abs/1210.2631}{{\tt
  arXiv:1210.2631 [hep-ex]}}\relax
\mciteBstWouldAddEndPuncttrue
\mciteSetBstMidEndSepPunct{\mcitedefaultmidpunct}
{\mcitedefaultendpunct}{\mcitedefaultseppunct}\relax
\EndOfBibitem
\bibitem{Aaij:2013cpa}
{LHCb} collaboration, R.~Aaij {\em et al.},
  \href{http://dx.doi.org/10.1016/j.nuclphysb.2013.03.004}{Nucl. Phys. {\bf
  B871},  403 (2013)}, \href{http://arxiv.org/abs/1302.6354}{{\tt
  arXiv:1302.6354 [hep-ex]}}\relax
\mciteBstWouldAddEndPuncttrue
\mciteSetBstMidEndSepPunct{\mcitedefaultmidpunct}
{\mcitedefaultendpunct}{\mcitedefaultseppunct}\relax
\EndOfBibitem
\bibitem{Aaij:2012dda}
{LHCb} collaboration, R.~Aaij {\em et al.},
  \href{http://dx.doi.org/10.1140/epjc/s10052-012-2118-7}{Eur. Phys. J. {\bf
  C72},  2118 (2012)}, \href{http://arxiv.org/abs/1205.0918}{{\tt
  arXiv:1205.0918 [hep-ex]}}\relax
\mciteBstWouldAddEndPuncttrue
\mciteSetBstMidEndSepPunct{\mcitedefaultmidpunct}
{\mcitedefaultendpunct}{\mcitedefaultseppunct}\relax
\EndOfBibitem
\bibitem{Aaij:2014jna}
{LHCb} collaboration, R.~Aaij {\em et al.},
  \href{http://dx.doi.org/10.1007/JHEP01(2015)024}{JHEP {\bf 01},  024 (2015)},
  \href{http://arxiv.org/abs/1411.0943}{{\tt arXiv:1411.0943 [hep-ex]}}\relax
\mciteBstWouldAddEndPuncttrue
\mciteSetBstMidEndSepPunct{\mcitedefaultmidpunct}
{\mcitedefaultendpunct}{\mcitedefaultseppunct}\relax
\EndOfBibitem
\bibitem{Aubert:2004ei}
{\babar} collaboration, B.~Aubert {\em et al.},
  \href{http://dx.doi.org/10.1103/PhysRevLett.93.081801}{Phys. Rev. Lett. {\bf
  93},  081801 (2004)}, \href{http://arxiv.org/abs/hep-ex/0404005}{{\tt
  arXiv:hep-ex/0404005 [hep-ex]}}\relax
\mciteBstWouldAddEndPuncttrue
\mciteSetBstMidEndSepPunct{\mcitedefaultmidpunct}
{\mcitedefaultendpunct}{\mcitedefaultseppunct}\relax
\EndOfBibitem
\bibitem{Aubert:2010zv}
{\babar} collaboration, B.~Aubert {\em et al.},
  \href{http://dx.doi.org/10.1103/PhysRevD.82.031102}{Phys. Rev. {\bf D82},
  031102 (2010)}, \href{http://arxiv.org/abs/1007.1370}{{\tt arXiv:1007.1370
  [hep-ex]}}\relax
\mciteBstWouldAddEndPuncttrue
\mciteSetBstMidEndSepPunct{\mcitedefaultmidpunct}
{\mcitedefaultendpunct}{\mcitedefaultseppunct}\relax
\EndOfBibitem
\bibitem{Gabyshev:2002zq}
{Belle} collaboration, N.~Gabyshev {\em et al.},
  \href{http://dx.doi.org/10.1103/PhysRevD.66.091102}{Phys. Rev. {\bf D66},
  091102 (2002)}, \href{http://arxiv.org/abs/hep-ex/0208041}{{\tt
  arXiv:hep-ex/0208041 [hep-ex]}}\relax
\mciteBstWouldAddEndPuncttrue
\mciteSetBstMidEndSepPunct{\mcitedefaultmidpunct}
{\mcitedefaultendpunct}{\mcitedefaultseppunct}\relax
\EndOfBibitem
\bibitem{Lees:2013bya}
{\babar} collaboration, J.~P. Lees {\em et al.},
  \href{http://dx.doi.org/10.1103/PhysRevD.87.092004}{Phys. Rev. {\bf D87},
  092004 (2013)}, \href{http://arxiv.org/abs/1302.0191}{{\tt arXiv:1302.0191
  [hep-ex]}}\relax
\mciteBstWouldAddEndPuncttrue
\mciteSetBstMidEndSepPunct{\mcitedefaultmidpunct}
{\mcitedefaultendpunct}{\mcitedefaultseppunct}\relax
\EndOfBibitem
\bibitem{Park:2006uj}
{Belle} collaboration, K.~Park {\em et al.},
  \href{http://dx.doi.org/10.1103/PhysRevD.75.011101}{Phys. Rev. {\bf D75},
  011101 (2007)}, \href{http://arxiv.org/abs/hep-ex/0608025}{{\tt
  arXiv:hep-ex/0608025 [hep-ex]}}\relax
\mciteBstWouldAddEndPuncttrue
\mciteSetBstMidEndSepPunct{\mcitedefaultmidpunct}
{\mcitedefaultendpunct}{\mcitedefaultseppunct}\relax
\EndOfBibitem
\bibitem{Abe:2005ib}
{Belle} collaboration, K.~Abe {\em et al.},
  \href{http://dx.doi.org/10.1103/PhysRevLett.97.202003}{Phys. Rev. Lett. {\bf
  97},  202003 (2006)}, \href{http://arxiv.org/abs/hep-ex/0508015}{{\tt
  arXiv:hep-ex/0508015 [hep-ex]}}\relax
\mciteBstWouldAddEndPuncttrue
\mciteSetBstMidEndSepPunct{\mcitedefaultmidpunct}
{\mcitedefaultendpunct}{\mcitedefaultseppunct}\relax
\EndOfBibitem
\bibitem{Aubert:2007eb}
{\babar} collaboration, B.~Aubert {\em et al.},
  \href{http://dx.doi.org/10.1103/PhysRevD.77.012002}{Phys. Rev. {\bf D77},
  012002 (2008)}, \href{http://arxiv.org/abs/0710.5763}{{\tt arXiv:0710.5763
  [hep-ex]}}\relax
\mciteBstWouldAddEndPuncttrue
\mciteSetBstMidEndSepPunct{\mcitedefaultmidpunct}
{\mcitedefaultendpunct}{\mcitedefaultseppunct}\relax
\EndOfBibitem
\bibitem{Lees:2014uta}
{\babar} collaboration, J.~P. Lees {\em et al.},
  \href{http://dx.doi.org/10.1103/PhysRevD.91.031102}{Phys. Rev. {\bf D91},
  031102 (2015)}, \href{http://arxiv.org/abs/1410.3644}{{\tt arXiv:1410.3644
  [hep-ex]}}\relax
\mciteBstWouldAddEndPuncttrue
\mciteSetBstMidEndSepPunct{\mcitedefaultmidpunct}
{\mcitedefaultendpunct}{\mcitedefaultseppunct}\relax
\EndOfBibitem
\bibitem{Gabyshev:2002dt}
{Belle} collaboration, N.~Gabyshev {\em et al.},
  \href{http://dx.doi.org/10.1103/PhysRevLett.90.121802}{Phys. Rev. Lett. {\bf
  90},  121802 (2003)}, \href{http://arxiv.org/abs/hep-ex/0212052}{{\tt
  arXiv:hep-ex/0212052 [hep-ex]}}\relax
\mciteBstWouldAddEndPuncttrue
\mciteSetBstMidEndSepPunct{\mcitedefaultmidpunct}
{\mcitedefaultendpunct}{\mcitedefaultseppunct}\relax
\EndOfBibitem
\bibitem{Aubert:2008ax}
{\babar} collaboration, B.~Aubert {\em et al.},
  \href{http://dx.doi.org/10.1103/PhysRevD.78.112003}{Phys. Rev. {\bf D78},
  112003 (2008)}, \href{http://arxiv.org/abs/0807.4974}{{\tt arXiv:0807.4974
  [hep-ex]}}\relax
\mciteBstWouldAddEndPuncttrue
\mciteSetBstMidEndSepPunct{\mcitedefaultmidpunct}
{\mcitedefaultendpunct}{\mcitedefaultseppunct}\relax
\EndOfBibitem
\bibitem{Aubert:2009aj}
{\babar} collaboration, B.~Aubert {\em et al.},
  \href{http://dx.doi.org/10.1103/PhysRevD.80.051105}{Phys. Rev. {\bf D80},
  051105 (2009)}, \href{http://arxiv.org/abs/0907.4566}{{\tt arXiv:0907.4566
  [hep-ex]}}\relax
\mciteBstWouldAddEndPuncttrue
\mciteSetBstMidEndSepPunct{\mcitedefaultmidpunct}
{\mcitedefaultendpunct}{\mcitedefaultseppunct}\relax
\EndOfBibitem
\bibitem{Chistov:2005zb}
{Belle} collaboration, R.~Chistov {\em et al.},
  \href{http://dx.doi.org/10.1103/PhysRevD.74.111105}{Phys. Rev. {\bf D74},
  111105 (2006)}, \href{http://arxiv.org/abs/hep-ex/0510074}{{\tt
  arXiv:hep-ex/0510074 [hep-ex]}}\relax
\mciteBstWouldAddEndPuncttrue
\mciteSetBstMidEndSepPunct{\mcitedefaultmidpunct}
{\mcitedefaultendpunct}{\mcitedefaultseppunct}\relax
\EndOfBibitem
\bibitem{Uchida:2007gx}
{Belle} collaboration, Y.~Uchida {\em et al.},
  \href{http://dx.doi.org/10.1103/PhysRevD.77.051101}{Phys. Rev. {\bf D77},
  051101 (2008)}, \href{http://arxiv.org/abs/0708.1105}{{\tt arXiv:0708.1105
  [hep-ex]}}\relax
\mciteBstWouldAddEndPuncttrue
\mciteSetBstMidEndSepPunct{\mcitedefaultmidpunct}
{\mcitedefaultendpunct}{\mcitedefaultseppunct}\relax
\EndOfBibitem
\bibitem{Lees:2011rf}
{\babar} collaboration, J.~P. Lees {\em et al.},
  \href{http://dx.doi.org/10.1103/PhysRevD.84.071102}{Phys. Rev. {\bf D84},
  071102 (2011)}, \href{http://arxiv.org/abs/1108.3211}{{\tt arXiv:1108.3211
  [hep-ex]}}, Erratum ibid.\
  \href{http://dx.doi.org/10.1103/PhysRevD.85.039903}{{\bf D85}, 039903},
  (2012)\relax
\mciteBstWouldAddEndPuncttrue
\mciteSetBstMidEndSepPunct{\mcitedefaultmidpunct}
{\mcitedefaultendpunct}{\mcitedefaultseppunct}\relax
\EndOfBibitem
\bibitem{TheBABAR:2013fda}
{\babar} collaboration, J.~P. Lees {\em et al.},
  \href{http://dx.doi.org/10.1103/PhysRevD.89.071102}{Phys. Rev. {\bf D89},
  071102 (2014)}, \href{http://arxiv.org/abs/1312.6800}{{\tt arXiv:1312.6800
  [hep-ex]}}\relax
\mciteBstWouldAddEndPuncttrue
\mciteSetBstMidEndSepPunct{\mcitedefaultmidpunct}
{\mcitedefaultendpunct}{\mcitedefaultseppunct}\relax
\EndOfBibitem
\bibitem{Aaij:2014pha}
{LHCb} collaboration, R.~Aaij {\em et al.},
  \href{http://dx.doi.org/10.1103/PhysRevLett.112.202001}{Phys. Rev. Lett. {\bf
  112},  202001 (2014)}, \href{http://arxiv.org/abs/1403.3606}{{\tt
  arXiv:1403.3606 [hep-ex]}}\relax
\mciteBstWouldAddEndPuncttrue
\mciteSetBstMidEndSepPunct{\mcitedefaultmidpunct}
{\mcitedefaultendpunct}{\mcitedefaultseppunct}\relax
\EndOfBibitem
\bibitem{Aubert:2008gu}
{\babar} collaboration, B.~Aubert {\em et al.},
  \href{http://dx.doi.org/10.1103/PhysRevD.77.111101}{Phys. Rev. {\bf D77},
  111101 (2008)}, \href{http://arxiv.org/abs/0803.2838}{{\tt arXiv:0803.2838
  [hep-ex]}}\relax
\mciteBstWouldAddEndPuncttrue
\mciteSetBstMidEndSepPunct{\mcitedefaultmidpunct}
{\mcitedefaultendpunct}{\mcitedefaultseppunct}\relax
\EndOfBibitem
\bibitem{Mizuk:2008me}
{Belle} collaboration, R.~Mizuk {\em et al.},
  \href{http://dx.doi.org/10.1103/PhysRevD.78.072004}{Phys. Rev. {\bf D78},
  072004 (2008)}, \href{http://arxiv.org/abs/0806.4098}{{\tt arXiv:0806.4098
  [hep-ex]}}\relax
\mciteBstWouldAddEndPuncttrue
\mciteSetBstMidEndSepPunct{\mcitedefaultmidpunct}
{\mcitedefaultendpunct}{\mcitedefaultseppunct}\relax
\EndOfBibitem
\bibitem{Aubert:2005vi}
{\babar} collaboration, B.~Aubert {\em et al.},
  \href{http://dx.doi.org/10.1103/PhysRevLett.96.052002}{Phys. Rev. Lett. {\bf
  96},  052002 (2006)}, \href{http://arxiv.org/abs/hep-ex/0510070}{{\tt
  arXiv:hep-ex/0510070 [hep-ex]}}\relax
\mciteBstWouldAddEndPuncttrue
\mciteSetBstMidEndSepPunct{\mcitedefaultmidpunct}
{\mcitedefaultendpunct}{\mcitedefaultseppunct}\relax
\EndOfBibitem
\bibitem{Aubert:2004zr}
{\babar} collaboration, B.~Aubert {\em et al.},
  \href{http://dx.doi.org/10.1103/PhysRevD.71.031501}{Phys. Rev. {\bf D71},
  031501 (2005)}, \href{http://arxiv.org/abs/hep-ex/0412051}{{\tt
  arXiv:hep-ex/0412051 [hep-ex]}}\relax
\mciteBstWouldAddEndPuncttrue
\mciteSetBstMidEndSepPunct{\mcitedefaultmidpunct}
{\mcitedefaultendpunct}{\mcitedefaultseppunct}\relax
\EndOfBibitem
\bibitem{Aubert:2008aa}
{\babar} collaboration, B.~Aubert {\em et al.},
  \href{http://dx.doi.org/10.1103/PhysRevD.79.112001}{Phys. Rev. {\bf D79},
  112001 (2009)}, \href{http://arxiv.org/abs/0811.0564}{{\tt arXiv:0811.0564
  [hep-ex]}}\relax
\mciteBstWouldAddEndPuncttrue
\mciteSetBstMidEndSepPunct{\mcitedefaultmidpunct}
{\mcitedefaultendpunct}{\mcitedefaultseppunct}\relax
\EndOfBibitem
\bibitem{Iwabuchi:2008av}
{Belle} collaboration, M.~Iwabuchi {\em et al.},
  \href{http://dx.doi.org/10.1103/PhysRevLett.101.041601}{Phys. Rev. Lett. {\bf
  101},  041601 (2008)}, \href{http://arxiv.org/abs/0804.0831}{{\tt
  arXiv:0804.0831 [hep-ex]}}\relax
\mciteBstWouldAddEndPuncttrue
\mciteSetBstMidEndSepPunct{\mcitedefaultmidpunct}
{\mcitedefaultendpunct}{\mcitedefaultseppunct}\relax
\EndOfBibitem
\bibitem{Kato:2017gfv}
{Belle} collaboration, Y.~Kato {\em et al.},
  \href{http://dx.doi.org/10.1103/PhysRevD.97.012005}{Phys. Rev. {\bf D97},
  012005 (2018)}, \href{http://arxiv.org/abs/1709.06108}{{\tt arXiv:1709.06108
  [hep-ex]}}\relax
\mciteBstWouldAddEndPuncttrue
\mciteSetBstMidEndSepPunct{\mcitedefaultmidpunct}
{\mcitedefaultendpunct}{\mcitedefaultseppunct}\relax
\EndOfBibitem
\bibitem{Abe:2003zm}
{Belle} collaboration, K.~Abe {\em et al.},
  \href{http://dx.doi.org/10.1103/PhysRevD.69.112002}{Phys. Rev. {\bf D69},
  112002 (2004)}, \href{http://arxiv.org/abs/hep-ex/0307021}{{\tt
  arXiv:hep-ex/0307021 [hep-ex]}}\relax
\mciteBstWouldAddEndPuncttrue
\mciteSetBstMidEndSepPunct{\mcitedefaultmidpunct}
{\mcitedefaultendpunct}{\mcitedefaultseppunct}\relax
\EndOfBibitem
\bibitem{Aubert:2009wg}
{\babar} collaboration, B.~Aubert {\em et al.},
  \href{http://dx.doi.org/10.1103/PhysRevD.79.112004}{Phys. Rev. {\bf D79},
  112004 (2009)}, \href{http://arxiv.org/abs/0901.1291}{{\tt arXiv:0901.1291
  [hep-ex]}}\relax
\mciteBstWouldAddEndPuncttrue
\mciteSetBstMidEndSepPunct{\mcitedefaultmidpunct}
{\mcitedefaultendpunct}{\mcitedefaultseppunct}\relax
\EndOfBibitem
\bibitem{Aubert:2003hm}
{\babar} collaboration, B.~Aubert {\em et al.},
  \href{http://arxiv.org/abs/hep-ex/0308026}{{\tt arXiv:hep-ex/0308026
  [hep-ex]}} (2003)\relax
\mciteBstWouldAddEndPuncttrue
\mciteSetBstMidEndSepPunct{\mcitedefaultmidpunct}
{\mcitedefaultendpunct}{\mcitedefaultseppunct}\relax
\EndOfBibitem
\bibitem{Swain:2003yu}
{Belle} collaboration, S.~Swain {\em et al.},
  \href{http://dx.doi.org/10.1103/PhysRevD.68.051101}{Phys. Rev. {\bf D68},
  051101 (2003)}, \href{http://arxiv.org/abs/hep-ex/0304032}{{\tt
  arXiv:hep-ex/0304032 [hep-ex]}}\relax
\mciteBstWouldAddEndPuncttrue
\mciteSetBstMidEndSepPunct{\mcitedefaultmidpunct}
{\mcitedefaultendpunct}{\mcitedefaultseppunct}\relax
\EndOfBibitem
\bibitem{Aubert:2006um}
{\babar} collaboration, B.~Aubert {\em et al.},
  \href{http://dx.doi.org/10.1103/PhysRevD.73.111104}{Phys. Rev. {\bf D73},
  111104 (2006)}, \href{http://arxiv.org/abs/hep-ex/0604017}{{\tt
  arXiv:hep-ex/0604017 [hep-ex]}}\relax
\mciteBstWouldAddEndPuncttrue
\mciteSetBstMidEndSepPunct{\mcitedefaultmidpunct}
{\mcitedefaultendpunct}{\mcitedefaultseppunct}\relax
\EndOfBibitem
\bibitem{Aubert:2003ae}
{\babar} collaboration, B.~Aubert {\em et al.},
  \href{http://dx.doi.org/10.1103/PhysRevLett.92.141801}{Phys. Rev. Lett. {\bf
  92},  141801 (2004)}, \href{http://arxiv.org/abs/hep-ex/0308057}{{\tt
  arXiv:hep-ex/0308057 [hep-ex]}}\relax
\mciteBstWouldAddEndPuncttrue
\mciteSetBstMidEndSepPunct{\mcitedefaultmidpunct}
{\mcitedefaultendpunct}{\mcitedefaultseppunct}\relax
\EndOfBibitem
\bibitem{Aaij:2015vea}
{LHCb} collaboration, R.~Aaij {\em et al.},
  \href{http://dx.doi.org/10.1103/PhysRevD.91.092002}{Phys. Rev. {\bf D91},
  092002 (2015)}, \href{http://arxiv.org/abs/1503.02995}{{\tt arXiv:1503.02995
  [hep-ex]}}, Erratum ibid.\
  \href{http://dx.doi.org/10.1103/PhysRevD.93.119901}{{\bf D93}, 119901},
  (2016)\relax
\mciteBstWouldAddEndPuncttrue
\mciteSetBstMidEndSepPunct{\mcitedefaultmidpunct}
{\mcitedefaultendpunct}{\mcitedefaultseppunct}\relax
\EndOfBibitem
\bibitem{delAmoSanchez:2010rf}
{\babar} collaboration, P.~del Amo~Sanchez {\em et al.},
  \href{http://dx.doi.org/10.1103/PhysRevD.82.092006}{Phys. Rev. {\bf D82},
  092006 (2010)}, \href{http://arxiv.org/abs/1005.0068}{{\tt arXiv:1005.0068
  [hep-ex]}}\relax
\mciteBstWouldAddEndPuncttrue
\mciteSetBstMidEndSepPunct{\mcitedefaultmidpunct}
{\mcitedefaultendpunct}{\mcitedefaultseppunct}\relax
\EndOfBibitem
\bibitem{Aaij:2015dwl}
{LHCb} collaboration, R.~Aaij {\em et al.},
  \href{http://dx.doi.org/10.1103/PhysRevD.93.051101}{Phys. Rev. {\bf D93},
  051101 (2016)}, \href{http://arxiv.org/abs/1512.02494}{{\tt arXiv:1512.02494
  [hep-ex]}}, Erratum ibid.\
  \href{http://dx.doi.org/10.1103/PhysRevD.93.119902}{{\bf D93}, 119902},
  (2016)\relax
\mciteBstWouldAddEndPuncttrue
\mciteSetBstMidEndSepPunct{\mcitedefaultmidpunct}
{\mcitedefaultendpunct}{\mcitedefaultseppunct}\relax
\EndOfBibitem
\bibitem{Aubert:2005ra}
{\babar} collaboration, B.~Aubert {\em et al.},
  \href{http://dx.doi.org/10.1103/PhysRevD.72.011102}{Phys. Rev. {\bf D72},
  011102 (2005)}, \href{http://arxiv.org/abs/hep-ex/0505099}{{\tt
  arXiv:hep-ex/0505099 [hep-ex]}}\relax
\mciteBstWouldAddEndPuncttrue
\mciteSetBstMidEndSepPunct{\mcitedefaultmidpunct}
{\mcitedefaultendpunct}{\mcitedefaultseppunct}\relax
\EndOfBibitem
\bibitem{Abulencia:2005ia}
{CDF} collaboration, A.~Abulencia {\em et al.},
  \href{http://dx.doi.org/10.1103/PhysRevLett.96.191801}{Phys. Rev. Lett. {\bf
  96},  191801 (2006)}, \href{http://arxiv.org/abs/hep-ex/0508014}{{\tt
  arXiv:hep-ex/0508014 [hep-ex]}}\relax
\mciteBstWouldAddEndPuncttrue
\mciteSetBstMidEndSepPunct{\mcitedefaultmidpunct}
{\mcitedefaultendpunct}{\mcitedefaultseppunct}\relax
\EndOfBibitem
\bibitem{Horii:2008as}
{Belle} collaboration, Y.~Horii {\em et al.},
  \href{http://dx.doi.org/10.1103/PhysRevD.78.071901}{Phys. Rev. {\bf D78},
  071901 (2008)}, \href{http://arxiv.org/abs/0804.2063}{{\tt arXiv:0804.2063
  [hep-ex]}}\relax
\mciteBstWouldAddEndPuncttrue
\mciteSetBstMidEndSepPunct{\mcitedefaultmidpunct}
{\mcitedefaultendpunct}{\mcitedefaultseppunct}\relax
\EndOfBibitem
\bibitem{Aubert:2003uy}
{\babar} collaboration, B.~Aubert {\em et al.},
  \href{http://dx.doi.org/10.1103/PhysRevLett.92.202002}{Phys. Rev. Lett. {\bf
  92},  202002 (2004)}, \href{http://arxiv.org/abs/hep-ex/0311032}{{\tt
  arXiv:hep-ex/0311032 [hep-ex]}}\relax
\mciteBstWouldAddEndPuncttrue
\mciteSetBstMidEndSepPunct{\mcitedefaultmidpunct}
{\mcitedefaultendpunct}{\mcitedefaultseppunct}\relax
\EndOfBibitem
\bibitem{Aubert:2004hu}
{\babar} collaboration, B.~Aubert {\em et al.},
  \href{http://dx.doi.org/10.1103/PhysRevD.71.031102}{Phys. Rev. {\bf D71},
  031102 (2005)}, \href{http://arxiv.org/abs/hep-ex/0411091}{{\tt
  arXiv:hep-ex/0411091 [hep-ex]}}\relax
\mciteBstWouldAddEndPuncttrue
\mciteSetBstMidEndSepPunct{\mcitedefaultmidpunct}
{\mcitedefaultendpunct}{\mcitedefaultseppunct}\relax
\EndOfBibitem
\bibitem{Aaij:2017dtg}
{LHCb} collaboration, R.~Aaij {\em et al.},
  \href{http://dx.doi.org/10.1103/PhysRevD.96.011101}{Phys. Rev. {\bf D96},
  011101 (2017)}, \href{http://arxiv.org/abs/1704.07581}{{\tt arXiv:1704.07581
  [hep-ex]}}\relax
\mciteBstWouldAddEndPuncttrue
\mciteSetBstMidEndSepPunct{\mcitedefaultmidpunct}
{\mcitedefaultendpunct}{\mcitedefaultseppunct}\relax
\EndOfBibitem
\bibitem{Wiechczynski:2009rg}
{Belle} collaboration, J.~Wiechczynski {\em et al.},
  \href{http://dx.doi.org/10.1103/PhysRevD.80.052005}{Phys. Rev. {\bf D80},
  052005 (2009)}, \href{http://arxiv.org/abs/0903.4956}{{\tt arXiv:0903.4956
  [hep-ex]}}\relax
\mciteBstWouldAddEndPuncttrue
\mciteSetBstMidEndSepPunct{\mcitedefaultmidpunct}
{\mcitedefaultendpunct}{\mcitedefaultseppunct}\relax
\EndOfBibitem
\bibitem{Aubert:2006xy}
{\babar} collaboration, B.~Aubert {\em et al.},
  \href{http://dx.doi.org/10.1103/PhysRevLett.98.171801}{Phys. Rev. Lett. {\bf
  98},  171801 (2007)}, \href{http://arxiv.org/abs/hep-ex/0611030}{{\tt
  arXiv:hep-ex/0611030 [hep-ex]}}\relax
\mciteBstWouldAddEndPuncttrue
\mciteSetBstMidEndSepPunct{\mcitedefaultmidpunct}
{\mcitedefaultendpunct}{\mcitedefaultseppunct}\relax
\EndOfBibitem
\bibitem{Aaij:2017vxc}
{LHCb} collaboration, R.~Aaij {\em et al.},
  \href{http://dx.doi.org/10.1007/JHEP01(2018)131}{JHEP {\bf 01},  131 (2018)},
  \href{http://arxiv.org/abs/1711.05637}{{\tt arXiv:1711.05637 [hep-ex]}}\relax
\mciteBstWouldAddEndPuncttrue
\mciteSetBstMidEndSepPunct{\mcitedefaultmidpunct}
{\mcitedefaultendpunct}{\mcitedefaultseppunct}\relax
\EndOfBibitem
\bibitem{Aubert:2005qq}
{\babar} collaboration, B.~Aubert {\em et al.},
  \href{http://dx.doi.org/10.1103/PhysRevD.73.011103}{Phys. Rev. {\bf D73},
  011103 (2006)}, \href{http://arxiv.org/abs/hep-ex/0512028}{{\tt
  arXiv:hep-ex/0512028 [hep-ex]}}\relax
\mciteBstWouldAddEndPuncttrue
\mciteSetBstMidEndSepPunct{\mcitedefaultmidpunct}
{\mcitedefaultendpunct}{\mcitedefaultseppunct}\relax
\EndOfBibitem
\bibitem{Chen:2011hy}
{Belle} collaboration, P.~Chen {\em et al.},
  \href{http://dx.doi.org/10.1103/PhysRevD.84.071501}{Phys. Rev. {\bf D84},
  071501 (2011)}, \href{http://arxiv.org/abs/1108.4271}{{\tt arXiv:1108.4271
  [hep-ex]}}\relax
\mciteBstWouldAddEndPuncttrue
\mciteSetBstMidEndSepPunct{\mcitedefaultmidpunct}
{\mcitedefaultendpunct}{\mcitedefaultseppunct}\relax
\EndOfBibitem
\bibitem{Seon:2011ni}
{BELLE} collaboration, O.~Seon {\em et al.},
  \href{http://dx.doi.org/10.1103/PhysRevD.84.071106}{Phys. Rev. {\bf D84},
  071106 (2011)}, \href{http://arxiv.org/abs/1107.0642}{{\tt arXiv:1107.0642
  [hep-ex]}}\relax
\mciteBstWouldAddEndPuncttrue
\mciteSetBstMidEndSepPunct{\mcitedefaultmidpunct}
{\mcitedefaultendpunct}{\mcitedefaultseppunct}\relax
\EndOfBibitem
\bibitem{Majumder:2005kp}
{Belle} collaboration, G.~Majumder {\em et al.},
  \href{http://dx.doi.org/10.1103/PhysRevLett.95.041803}{Phys. Rev. Lett. {\bf
  95},  041803 (2005)}, \href{http://arxiv.org/abs/hep-ex/0502038}{{\tt
  arXiv:hep-ex/0502038 [hep-ex]}}\relax
\mciteBstWouldAddEndPuncttrue
\mciteSetBstMidEndSepPunct{\mcitedefaultmidpunct}
{\mcitedefaultendpunct}{\mcitedefaultseppunct}\relax
\EndOfBibitem
\bibitem{Brodzicka:2007aa}
{Belle} collaboration, J.~Brodzicka {\em et al.},
  \href{http://dx.doi.org/10.1103/PhysRevLett.100.092001}{Phys. Rev. Lett. {\bf
  100},  092001 (2008)}, \href{http://arxiv.org/abs/0707.3491}{{\tt
  arXiv:0707.3491 [hep-ex]}}\relax
\mciteBstWouldAddEndPuncttrue
\mciteSetBstMidEndSepPunct{\mcitedefaultmidpunct}
{\mcitedefaultendpunct}{\mcitedefaultseppunct}\relax
\EndOfBibitem
\bibitem{Abe:2003zv}
{Belle} collaboration, K.~Abe {\em et al.},
  \href{http://dx.doi.org/10.1103/PhysRevLett.93.051803}{Phys. Rev. Lett. {\bf
  93},  051803 (2004)}, \href{http://arxiv.org/abs/hep-ex/0307061}{{\tt
  arXiv:hep-ex/0307061 [hep-ex]}}\relax
\mciteBstWouldAddEndPuncttrue
\mciteSetBstMidEndSepPunct{\mcitedefaultmidpunct}
{\mcitedefaultendpunct}{\mcitedefaultseppunct}\relax
\EndOfBibitem
\bibitem{Abe:2002haa}
{Belle} collaboration, K.~Abe {\em et al.},
  \href{http://dx.doi.org/10.1016/S0370-2693(02)01969-X}{Phys. Lett. {\bf
  B538},  11 (2002)}, \href{http://arxiv.org/abs/hep-ex/0205021}{{\tt
  arXiv:hep-ex/0205021 [hep-ex]}}\relax
\mciteBstWouldAddEndPuncttrue
\mciteSetBstMidEndSepPunct{\mcitedefaultmidpunct}
{\mcitedefaultendpunct}{\mcitedefaultseppunct}\relax
\EndOfBibitem
\bibitem{Acosta:2002pw}
{CDF} collaboration, D.~Acosta {\em et al.},
  \href{http://dx.doi.org/10.1103/PhysRevD.66.052005}{Phys. Rev. {\bf D66},
  052005 (2002)}\relax
\mciteBstWouldAddEndPuncttrue
\mciteSetBstMidEndSepPunct{\mcitedefaultmidpunct}
{\mcitedefaultendpunct}{\mcitedefaultseppunct}\relax
\EndOfBibitem
\bibitem{Guler:2010if}
{Belle} collaboration, H.~Guler {\em et al.},
  \href{http://dx.doi.org/10.1103/PhysRevD.83.032005}{Phys. Rev. {\bf D83},
  032005 (2011)}, \href{http://arxiv.org/abs/1009.5256}{{\tt arXiv:1009.5256
  [hep-ex]}}\relax
\mciteBstWouldAddEndPuncttrue
\mciteSetBstMidEndSepPunct{\mcitedefaultmidpunct}
{\mcitedefaultendpunct}{\mcitedefaultseppunct}\relax
\EndOfBibitem
\bibitem{Aubert:2004ns}
{\babar} collaboration, B.~Aubert {\em et al.},
  \href{http://dx.doi.org/10.1103/PhysRevD.71.071103}{Phys. Rev. {\bf D71},
  071103 (2005)}, \href{http://arxiv.org/abs/hep-ex/0406022}{{\tt
  arXiv:hep-ex/0406022 [hep-ex]}}\relax
\mciteBstWouldAddEndPuncttrue
\mciteSetBstMidEndSepPunct{\mcitedefaultmidpunct}
{\mcitedefaultendpunct}{\mcitedefaultseppunct}\relax
\EndOfBibitem
\bibitem{Abe:2001mw}
{Belle} collaboration, K.~Abe {\em et al.},
  \href{http://dx.doi.org/10.1103/PhysRevLett.88.031802}{Phys. Rev. Lett. {\bf
  88},  031802 (2002)}, \href{http://arxiv.org/abs/hep-ex/0111069}{{\tt
  arXiv:hep-ex/0111069 [hep-ex]}}\relax
\mciteBstWouldAddEndPuncttrue
\mciteSetBstMidEndSepPunct{\mcitedefaultmidpunct}
{\mcitedefaultendpunct}{\mcitedefaultseppunct}\relax
\EndOfBibitem
\bibitem{Khachatryan:2016npv}
{CMS} collaboration, V.~Khachatryan {\em et al.},
  \href{http://dx.doi.org/10.1016/j.physletb.2016.11.001}{Phys. Lett. {\bf
  B764},  66 (2017)}, \href{http://arxiv.org/abs/1607.02638}{{\tt
  arXiv:1607.02638 [hep-ex]}}\relax
\mciteBstWouldAddEndPuncttrue
\mciteSetBstMidEndSepPunct{\mcitedefaultmidpunct}
{\mcitedefaultendpunct}{\mcitedefaultseppunct}\relax
\EndOfBibitem
\bibitem{Vinokurova:2011dy}
{Belle} collaboration, A.~Vinokurova {\em et al.},
  \href{http://dx.doi.org/10.1016/j.physletb.2011.11.014}{Phys. Lett. {\bf
  B706},  139 (2011)}, \href{http://arxiv.org/abs/1105.0978}{{\tt
  arXiv:1105.0978 [hep-ex]}}\relax
\mciteBstWouldAddEndPuncttrue
\mciteSetBstMidEndSepPunct{\mcitedefaultmidpunct}
{\mcitedefaultendpunct}{\mcitedefaultseppunct}\relax
\EndOfBibitem
\bibitem{Wu:2006vx}
{Belle} collaboration, C.-H. Wu {\em et al.},
  \href{http://dx.doi.org/10.1103/PhysRevLett.97.162003}{Phys. Rev. Lett. {\bf
  97},  162003 (2006)}, \href{http://arxiv.org/abs/hep-ex/0606022}{{\tt
  arXiv:hep-ex/0606022 [hep-ex]}}\relax
\mciteBstWouldAddEndPuncttrue
\mciteSetBstMidEndSepPunct{\mcitedefaultmidpunct}
{\mcitedefaultendpunct}{\mcitedefaultseppunct}\relax
\EndOfBibitem
\bibitem{Aubert:2005gw}
{\babar} collaboration, B.~Aubert {\em et al.},
  \href{http://dx.doi.org/10.1103/PhysRevD.72.051101}{Phys. Rev. {\bf D72},
  051101 (2005)}, \href{http://arxiv.org/abs/hep-ex/0507012}{{\tt
  arXiv:hep-ex/0507012 [hep-ex]}}\relax
\mciteBstWouldAddEndPuncttrue
\mciteSetBstMidEndSepPunct{\mcitedefaultmidpunct}
{\mcitedefaultendpunct}{\mcitedefaultseppunct}\relax
\EndOfBibitem
\bibitem{Fang:2006bz}
{Belle} collaboration, F.~Fang {\em et al.},
  \href{http://dx.doi.org/10.1103/PhysRevD.74.012007}{Phys. Rev. {\bf D74},
  012007 (2006)}, \href{http://arxiv.org/abs/hep-ex/0605007}{{\tt
  arXiv:hep-ex/0605007 [hep-ex]}}\relax
\mciteBstWouldAddEndPuncttrue
\mciteSetBstMidEndSepPunct{\mcitedefaultmidpunct}
{\mcitedefaultendpunct}{\mcitedefaultseppunct}\relax
\EndOfBibitem
\bibitem{Aaij:2012jw}
{LHCb} collaboration, R.~Aaij {\em et al.},
  \href{http://dx.doi.org/10.1103/PhysRevD.85.091105}{Phys. Rev. {\bf D85},
  091105 (2012)}, \href{http://arxiv.org/abs/1203.3592}{{\tt arXiv:1203.3592
  [hep-ex]}}\relax
\mciteBstWouldAddEndPuncttrue
\mciteSetBstMidEndSepPunct{\mcitedefaultmidpunct}
{\mcitedefaultendpunct}{\mcitedefaultseppunct}\relax
\EndOfBibitem
\bibitem{Aubert:2004pra}
{\babar} collaboration, B.~Aubert {\em et al.},
  \href{http://dx.doi.org/10.1103/PhysRevLett.92.241802}{Phys. Rev. Lett. {\bf
  92},  241802 (2004)}, \href{http://arxiv.org/abs/hep-ex/0401035}{{\tt
  arXiv:hep-ex/0401035 [hep-ex]}}\relax
\mciteBstWouldAddEndPuncttrue
\mciteSetBstMidEndSepPunct{\mcitedefaultmidpunct}
{\mcitedefaultendpunct}{\mcitedefaultseppunct}\relax
\EndOfBibitem
\bibitem{Aubert:2005sk}
{\babar} collaboration, B.~Aubert {\em et al.},
  \href{http://dx.doi.org/10.1103/PhysRevD.72.052002}{Phys. Rev. {\bf D72},
  052002 (2005)}, \href{http://arxiv.org/abs/hep-ex/0507025}{{\tt
  arXiv:hep-ex/0507025 [hep-ex]}}\relax
\mciteBstWouldAddEndPuncttrue
\mciteSetBstMidEndSepPunct{\mcitedefaultmidpunct}
{\mcitedefaultendpunct}{\mcitedefaultseppunct}\relax
\EndOfBibitem
\bibitem{Kumar:2006sg}
{Belle} collaboration, R.~Kumar {\em et al.},
  \href{http://dx.doi.org/10.1103/PhysRevD.74.051103}{Phys. Rev. {\bf D74},
  051103 (2006)}, \href{http://arxiv.org/abs/hep-ex/0607008}{{\tt
  arXiv:hep-ex/0607008 [hep-ex]}}\relax
\mciteBstWouldAddEndPuncttrue
\mciteSetBstMidEndSepPunct{\mcitedefaultmidpunct}
{\mcitedefaultendpunct}{\mcitedefaultseppunct}\relax
\EndOfBibitem
\bibitem{Abazov:2008jk}
{\dzero} collaboration, V.~Abazov {\em et al.},
  \href{http://dx.doi.org/10.1103/PhysRevD.79.111102}{Phys. Rev. {\bf D79},
  111102 (2009)}, \href{http://arxiv.org/abs/0805.2576}{{\tt arXiv:0805.2576
  [hep-ex]}}\relax
\mciteBstWouldAddEndPuncttrue
\mciteSetBstMidEndSepPunct{\mcitedefaultmidpunct}
{\mcitedefaultendpunct}{\mcitedefaultseppunct}\relax
\EndOfBibitem
\bibitem{Aaij:2013rha}
{LHCb} collaboration, R.~Aaij {\em et al.},
  \href{http://dx.doi.org/10.1140/epjc/s10052-013-2462-2}{Eur. Phys. J. {\bf
  C73},  2462 (2013)}, \href{http://arxiv.org/abs/1303.7133}{{\tt
  arXiv:1303.7133 [hep-ex]}}\relax
\mciteBstWouldAddEndPuncttrue
\mciteSetBstMidEndSepPunct{\mcitedefaultmidpunct}
{\mcitedefaultendpunct}{\mcitedefaultseppunct}\relax
\EndOfBibitem
\bibitem{Abe:1996yya}
{CDF} collaboration, F.~Abe {\em et al.},
  \href{http://dx.doi.org/10.1103/PhysRevLett.77.5176}{Phys. Rev. Lett. {\bf
  77},  5176 (1996)}\relax
\mciteBstWouldAddEndPuncttrue
\mciteSetBstMidEndSepPunct{\mcitedefaultmidpunct}
{\mcitedefaultendpunct}{\mcitedefaultseppunct}\relax
\EndOfBibitem
\bibitem{Abulencia:2007zzb}
{CDF} collaboration, A.~Abulencia,
  \href{http://dx.doi.org/10.1103/PhysRevD.79.112003}{Phys. Rev. {\bf D79},
  112003 (2009)}, \href{http://arxiv.org/abs/0905.2146}{{\tt arXiv:0905.2146
  [hep-ex]}}\relax
\mciteBstWouldAddEndPuncttrue
\mciteSetBstMidEndSepPunct{\mcitedefaultmidpunct}
{\mcitedefaultendpunct}{\mcitedefaultseppunct}\relax
\EndOfBibitem
\bibitem{Abazov:2013sqa}
{\dzero} collaboration, V.~M. Abazov {\em et al.},
  \href{http://dx.doi.org/10.1103/PhysRevLett.110.241801}{Phys. Rev. Lett. {\bf
  110},  241801 (2013)}, \href{http://arxiv.org/abs/1304.1655}{{\tt
  arXiv:1304.1655 [hep-ex]}}\relax
\mciteBstWouldAddEndPuncttrue
\mciteSetBstMidEndSepPunct{\mcitedefaultmidpunct}
{\mcitedefaultendpunct}{\mcitedefaultseppunct}\relax
\EndOfBibitem
\bibitem{Li:2017uvv}
{Belle} collaboration, Y.~B. Li {\em et al.},
  \href{http://dx.doi.org/10.1140/epjc/s10052-018-5720-5}{Eur. Phys. J. {\bf
  C78},  252 (2018)}, \href{http://arxiv.org/abs/1712.03612}{{\tt
  arXiv:1712.03612 [hep-ex]}}\relax
\mciteBstWouldAddEndPuncttrue
\mciteSetBstMidEndSepPunct{\mcitedefaultmidpunct}
{\mcitedefaultendpunct}{\mcitedefaultseppunct}\relax
\EndOfBibitem
\bibitem{Lees:2012kc}
{\babar} collaboration, J.~P. Lees {\em et al.},
  \href{http://dx.doi.org/10.1103/PhysRevD.86.091102}{Phys. Rev. {\bf D86},
  091102 (2012)}, \href{http://arxiv.org/abs/1208.3086}{{\tt arXiv:1208.3086
  [hep-ex]}}\relax
\mciteBstWouldAddEndPuncttrue
\mciteSetBstMidEndSepPunct{\mcitedefaultmidpunct}
{\mcitedefaultendpunct}{\mcitedefaultseppunct}\relax
\EndOfBibitem
\bibitem{Choi:2011fc}
{Belle} collaboration, S.~K. Choi {\em et al.},
  \href{http://dx.doi.org/10.1103/PhysRevD.84.052004}{Phys. Rev. {\bf D84},
  052004 (2011)}, \href{http://arxiv.org/abs/1107.0163}{{\tt arXiv:1107.0163
  [hep-ex]}}\relax
\mciteBstWouldAddEndPuncttrue
\mciteSetBstMidEndSepPunct{\mcitedefaultmidpunct}
{\mcitedefaultendpunct}{\mcitedefaultseppunct}\relax
\EndOfBibitem
\bibitem{Aubert:2006aj}
{\babar} collaboration, B.~Aubert {\em et al.},
  \href{http://dx.doi.org/10.1103/PhysRevD.74.071101}{Phys. Rev. {\bf D74},
  071101 (2006)}, \href{http://arxiv.org/abs/hep-ex/0607050}{{\tt
  arXiv:hep-ex/0607050 [hep-ex]}}\relax
\mciteBstWouldAddEndPuncttrue
\mciteSetBstMidEndSepPunct{\mcitedefaultmidpunct}
{\mcitedefaultendpunct}{\mcitedefaultseppunct}\relax
\EndOfBibitem
\bibitem{Aubert:2005zh}
{\babar} collaboration, B.~Aubert {\em et al.},
  \href{http://dx.doi.org/10.1103/PhysRevD.73.011101}{Phys. Rev. {\bf D73},
  011101 (2006)}, \href{http://arxiv.org/abs/hep-ex/0507090}{{\tt
  arXiv:hep-ex/0507090 [hep-ex]}}\relax
\mciteBstWouldAddEndPuncttrue
\mciteSetBstMidEndSepPunct{\mcitedefaultmidpunct}
{\mcitedefaultendpunct}{\mcitedefaultseppunct}\relax
\EndOfBibitem
\bibitem{Aaij:2016iza}
{LHCb} collaboration, R.~Aaij {\em et al.},
  \href{http://dx.doi.org/10.1103/PhysRevLett.118.022003}{Phys. Rev. Lett. {\bf
  118},  022003 (2017)}, \href{http://arxiv.org/abs/1606.07895}{{\tt
  arXiv:1606.07895 [hep-ex]}}\relax
\mciteBstWouldAddEndPuncttrue
\mciteSetBstMidEndSepPunct{\mcitedefaultmidpunct}
{\mcitedefaultendpunct}{\mcitedefaultseppunct}\relax
\EndOfBibitem
\bibitem{Abazov:2013xda}
{\dzero} collaboration, V.~M. Abazov {\em et al.},
  \href{http://dx.doi.org/10.1103/PhysRevD.89.012004}{Phys. Rev. {\bf D89},
  012004 (2014)}, \href{http://arxiv.org/abs/1309.6580}{{\tt arXiv:1309.6580
  [hep-ex]}}\relax
\mciteBstWouldAddEndPuncttrue
\mciteSetBstMidEndSepPunct{\mcitedefaultmidpunct}
{\mcitedefaultendpunct}{\mcitedefaultseppunct}\relax
\EndOfBibitem
\bibitem{Adachi:2008sua}
{Belle} collaboration, T.~Aushev {\em et al.},
  \href{http://dx.doi.org/10.1103/PhysRevD.81.031103}{Phys. Rev. {\bf D81},
  031103 (2010)}, \href{http://arxiv.org/abs/0810.0358}{{\tt arXiv:0810.0358
  [hep-ex]}}\relax
\mciteBstWouldAddEndPuncttrue
\mciteSetBstMidEndSepPunct{\mcitedefaultmidpunct}
{\mcitedefaultendpunct}{\mcitedefaultseppunct}\relax
\EndOfBibitem
\bibitem{Abe:2004zs}
{Belle} collaboration, K.~Abe {\em et al.},
  \href{http://dx.doi.org/10.1103/PhysRevLett.94.182002}{Phys. Rev. Lett. {\bf
  94},  182002 (2005)}, \href{http://arxiv.org/abs/hep-ex/0408126}{{\tt
  arXiv:hep-ex/0408126 [hep-ex]}}\relax
\mciteBstWouldAddEndPuncttrue
\mciteSetBstMidEndSepPunct{\mcitedefaultmidpunct}
{\mcitedefaultendpunct}{\mcitedefaultseppunct}\relax
\EndOfBibitem
\bibitem{Aaij:2012zz}
{LHCb} collaboration, R.~Aaij {\em et al.},
  \href{http://dx.doi.org/10.1007/JHEP06(2012)115}{JHEP {\bf 06},  115 (2012)},
  \href{http://arxiv.org/abs/1204.1237}{{\tt arXiv:1204.1237 [hep-ex]}}\relax
\mciteBstWouldAddEndPuncttrue
\mciteSetBstMidEndSepPunct{\mcitedefaultmidpunct}
{\mcitedefaultendpunct}{\mcitedefaultseppunct}\relax
\EndOfBibitem
\bibitem{Louvot:2010rd}
{Belle} collaboration, R.~Louvot {\em et al.},
  \href{http://dx.doi.org/10.1103/PhysRevLett.104.231801}{Phys. Rev. Lett. {\bf
  104},  231801 (2010)}, \href{http://arxiv.org/abs/1003.5312}{{\tt
  arXiv:1003.5312 [hep-ex]}}\relax
\mciteBstWouldAddEndPuncttrue
\mciteSetBstMidEndSepPunct{\mcitedefaultmidpunct}
{\mcitedefaultendpunct}{\mcitedefaultseppunct}\relax
\EndOfBibitem
\bibitem{Aaij:2015dsa}
{LHCb} collaboration, R.~Aaij {\em et al.},
  \href{http://dx.doi.org/10.1007/JHEP06(2015)130}{JHEP {\bf 06},  130 (2015)},
  \href{http://arxiv.org/abs/1503.09086}{{\tt arXiv:1503.09086 [hep-ex]}}\relax
\mciteBstWouldAddEndPuncttrue
\mciteSetBstMidEndSepPunct{\mcitedefaultmidpunct}
{\mcitedefaultendpunct}{\mcitedefaultseppunct}\relax
\EndOfBibitem
\bibitem{Aaij:2016amk}
{LHCb} collaboration, R.~Aaij {\em et al.},
  \href{http://dx.doi.org/10.1103/PhysRevLett.116.161802}{Phys. Rev. Lett. {\bf
  116},  161802 (2016)}, \href{http://arxiv.org/abs/1603.02408}{{\tt
  arXiv:1603.02408 [hep-ex]}}\relax
\mciteBstWouldAddEndPuncttrue
\mciteSetBstMidEndSepPunct{\mcitedefaultmidpunct}
{\mcitedefaultendpunct}{\mcitedefaultseppunct}\relax
\EndOfBibitem
\bibitem{Aaij:2011tz}
{LHCb} collaboration, R.~Aaij {\em et al.},
  \href{http://dx.doi.org/10.1016/j.physletb.2011.10.073}{Phys. Lett. {\bf
  B706},  32 (2011)}, \href{http://arxiv.org/abs/1110.3676}{{\tt
  arXiv:1110.3676 [hep-ex]}}\relax
\mciteBstWouldAddEndPuncttrue
\mciteSetBstMidEndSepPunct{\mcitedefaultmidpunct}
{\mcitedefaultendpunct}{\mcitedefaultseppunct}\relax
\EndOfBibitem
\bibitem{Aaij:2013fpa}
{LHCb} collaboration, R.~Aaij {\em et al.},
  \href{http://dx.doi.org/10.1103/PhysRevD.87.071101}{Phys. Rev. {\bf D87},
  071101 (2013)}, \href{http://arxiv.org/abs/1302.6446}{{\tt arXiv:1302.6446
  [hep-ex]}}\relax
\mciteBstWouldAddEndPuncttrue
\mciteSetBstMidEndSepPunct{\mcitedefaultmidpunct}
{\mcitedefaultendpunct}{\mcitedefaultseppunct}\relax
\EndOfBibitem
\bibitem{Aaij:2015rqa}
{LHCb} collaboration, R.~Aaij {\em et al.},
  \href{http://dx.doi.org/10.1007/JHEP08(2015)005}{JHEP {\bf 08},  005 (2015)},
  \href{http://arxiv.org/abs/1505.01654}{{\tt arXiv:1505.01654 [hep-ex]}}\relax
\mciteBstWouldAddEndPuncttrue
\mciteSetBstMidEndSepPunct{\mcitedefaultmidpunct}
{\mcitedefaultendpunct}{\mcitedefaultseppunct}\relax
\EndOfBibitem
\bibitem{Abulencia:2006aa}
{CDF} collaboration, A.~Abulencia {\em et al.},
  \href{http://dx.doi.org/10.1103/PhysRevLett.98.061802}{Phys. Rev. Lett. {\bf
  98},  061802 (2007)}, \href{http://arxiv.org/abs/hep-ex/0610045}{{\tt
  arXiv:hep-ex/0610045 [hep-ex]}}\relax
\mciteBstWouldAddEndPuncttrue
\mciteSetBstMidEndSepPunct{\mcitedefaultmidpunct}
{\mcitedefaultendpunct}{\mcitedefaultseppunct}\relax
\EndOfBibitem
\bibitem{Aaij:2013dda}
{LHCb} collaboration, R.~Aaij {\em et al.},
  \href{http://dx.doi.org/10.1016/j.physletb.2013.10.057}{Phys. Lett. {\bf
  B727},  403 (2013)}, \href{http://arxiv.org/abs/1308.4583}{{\tt
  arXiv:1308.4583 [hep-ex]}}\relax
\mciteBstWouldAddEndPuncttrue
\mciteSetBstMidEndSepPunct{\mcitedefaultmidpunct}
{\mcitedefaultendpunct}{\mcitedefaultseppunct}\relax
\EndOfBibitem
\bibitem{Aaltonen:2008ab}
{CDF} collaboration, T.~Aaltonen {\em et al.},
  \href{http://dx.doi.org/10.1103/PhysRevLett.103.191802}{Phys. Rev. Lett. {\bf
  103},  191802 (2009)}, \href{http://arxiv.org/abs/0809.0080}{{\tt
  arXiv:0809.0080 [hep-ex]}}\relax
\mciteBstWouldAddEndPuncttrue
\mciteSetBstMidEndSepPunct{\mcitedefaultmidpunct}
{\mcitedefaultendpunct}{\mcitedefaultseppunct}\relax
\EndOfBibitem
\bibitem{Aaltonen:2012mg}
{CDF} collaboration, T.~Aaltonen {\em et al.},
  \href{http://dx.doi.org/10.1103/PhysRevLett.108.201801}{Phys. Rev. Lett. {\bf
  108},  201801 (2012)}, \href{http://arxiv.org/abs/1204.0536}{{\tt
  arXiv:1204.0536 [hep-ex]}}\relax
\mciteBstWouldAddEndPuncttrue
\mciteSetBstMidEndSepPunct{\mcitedefaultmidpunct}
{\mcitedefaultendpunct}{\mcitedefaultseppunct}\relax
\EndOfBibitem
\bibitem{Aaij:2016rja}
{LHCb} collaboration, R.~Aaij {\em et al.},
  \href{http://dx.doi.org/10.1103/PhysRevD.93.092008}{Phys. Rev. {\bf D93},
  092008 (2016)}, \href{http://arxiv.org/abs/1602.07543}{{\tt arXiv:1602.07543
  [hep-ex]}}\relax
\mciteBstWouldAddEndPuncttrue
\mciteSetBstMidEndSepPunct{\mcitedefaultmidpunct}
{\mcitedefaultendpunct}{\mcitedefaultseppunct}\relax
\EndOfBibitem
\bibitem{Belle:2012aa}
{Belle} collaboration, J.~Li {\em et al.},
  \href{http://dx.doi.org/10.1103/PhysRevLett.108.181808}{Phys. Rev. Lett. {\bf
  108},  181808 (2012)}, \href{http://arxiv.org/abs/1202.0103}{{\tt
  arXiv:1202.0103 [hep-ex]}}\relax
\mciteBstWouldAddEndPuncttrue
\mciteSetBstMidEndSepPunct{\mcitedefaultmidpunct}
{\mcitedefaultendpunct}{\mcitedefaultseppunct}\relax
\EndOfBibitem
\bibitem{Aaij:2013orb}
{LHCb} collaboration, R.~Aaij {\em et al.},
  \href{http://dx.doi.org/10.1103/PhysRevD.87.072004}{Phys. Rev. {\bf D87},
  072004 (2013)}, \href{http://arxiv.org/abs/1302.1213}{{\tt arXiv:1302.1213
  [hep-ex]}}\relax
\mciteBstWouldAddEndPuncttrue
\mciteSetBstMidEndSepPunct{\mcitedefaultmidpunct}
{\mcitedefaultendpunct}{\mcitedefaultseppunct}\relax
\EndOfBibitem
\bibitem{Thorne:2013llu}
{Belle} collaboration, F.~Thorne {\em et al.},
  \href{http://dx.doi.org/10.1103/PhysRevD.88.114006}{Phys. Rev. {\bf D88},
  114006 (2013)}, \href{http://arxiv.org/abs/1309.0704}{{\tt arXiv:1309.0704
  [hep-ex]}}\relax
\mciteBstWouldAddEndPuncttrue
\mciteSetBstMidEndSepPunct{\mcitedefaultmidpunct}
{\mcitedefaultendpunct}{\mcitedefaultseppunct}\relax
\EndOfBibitem
\bibitem{Aaij:2012di}
{LHCb} collaboration, R.~Aaij {\em et al.},
  \href{http://dx.doi.org/10.1016/j.physletb.2012.05.062}{Phys. Lett. {\bf
  B713},  172 (2012)}, \href{http://arxiv.org/abs/1205.0934}{{\tt
  arXiv:1205.0934 [hep-ex]}}\relax
\mciteBstWouldAddEndPuncttrue
\mciteSetBstMidEndSepPunct{\mcitedefaultmidpunct}
{\mcitedefaultendpunct}{\mcitedefaultseppunct}\relax
\EndOfBibitem
\bibitem{Aaltonen:2011sy}
{CDF} collaboration, T.~Aaltonen {\em et al.},
  \href{http://dx.doi.org/10.1103/PhysRevD.83.052012}{Phys. Rev. {\bf D83},
  052012 (2011)}, \href{http://arxiv.org/abs/1102.1961}{{\tt arXiv:1102.1961
  [hep-ex]}}\relax
\mciteBstWouldAddEndPuncttrue
\mciteSetBstMidEndSepPunct{\mcitedefaultmidpunct}
{\mcitedefaultendpunct}{\mcitedefaultseppunct}\relax
\EndOfBibitem
\bibitem{Aaij:2015mea}
{LHCb} collaboration, R.~Aaij {\em et al.},
  \href{http://dx.doi.org/10.1007/JHEP11(2015)082}{JHEP {\bf 11},  082 (2015)},
  \href{http://arxiv.org/abs/1509.00400}{{\tt arXiv:1509.00400 [hep-ex]}}\relax
\mciteBstWouldAddEndPuncttrue
\mciteSetBstMidEndSepPunct{\mcitedefaultmidpunct}
{\mcitedefaultendpunct}{\mcitedefaultseppunct}\relax
\EndOfBibitem
\bibitem{Aaij:2011ac}
{LHCb} collaboration, R.~Aaij {\em et al.},
  \href{http://dx.doi.org/10.1103/PhysRevLett.108.151801}{Phys. Rev. Lett. {\bf
  108},  151801 (2012)}, \href{http://arxiv.org/abs/1112.4695}{{\tt
  arXiv:1112.4695 [hep-ex]}}\relax
\mciteBstWouldAddEndPuncttrue
\mciteSetBstMidEndSepPunct{\mcitedefaultmidpunct}
{\mcitedefaultendpunct}{\mcitedefaultseppunct}\relax
\EndOfBibitem
\bibitem{Abazov:2012dz}
{\dzero} collaboration, V.~M. Abazov {\em et al.},
  \href{http://dx.doi.org/10.1103/PhysRevD.86.092011}{Phys. Rev. {\bf D86},
  092011 (2012)}, \href{http://arxiv.org/abs/1204.5723}{{\tt arXiv:1204.5723
  [hep-ex]}}\relax
\mciteBstWouldAddEndPuncttrue
\mciteSetBstMidEndSepPunct{\mcitedefaultmidpunct}
{\mcitedefaultendpunct}{\mcitedefaultseppunct}\relax
\EndOfBibitem
\bibitem{Aaij:2011fx}
{LHCb} collaboration, R.~Aaij {\em et al.},
  \href{http://dx.doi.org/10.1016/j.physletb.2011.03.006}{Phys. Lett. {\bf
  B698},  115 (2011)}, \href{http://arxiv.org/abs/1102.0206}{{\tt
  arXiv:1102.0206 [hep-ex]}}\relax
\mciteBstWouldAddEndPuncttrue
\mciteSetBstMidEndSepPunct{\mcitedefaultmidpunct}
{\mcitedefaultendpunct}{\mcitedefaultseppunct}\relax
\EndOfBibitem
\bibitem{Abulencia:2006jp}
{CDF} collaboration, A.~Abulencia {\em et al.},
  \href{http://dx.doi.org/10.1103/PhysRevLett.96.231801}{Phys. Rev. Lett. {\bf
  96},  231801 (2006)}, \href{http://arxiv.org/abs/hep-ex/0602005}{{\tt
  arXiv:hep-ex/0602005 [hep-ex]}}\relax
\mciteBstWouldAddEndPuncttrue
\mciteSetBstMidEndSepPunct{\mcitedefaultmidpunct}
{\mcitedefaultendpunct}{\mcitedefaultseppunct}\relax
\EndOfBibitem
\bibitem{Abazov:2011hv}
{\dzero} collaboration, V.~M. Abazov {\em et al.},
  \href{http://dx.doi.org/10.1103/PhysRevD.85.011103}{Phys. Rev. {\bf D85},
  011103 (2012)}, \href{http://arxiv.org/abs/1110.4272}{{\tt arXiv:1110.4272
  [hep-ex]}}\relax
\mciteBstWouldAddEndPuncttrue
\mciteSetBstMidEndSepPunct{\mcitedefaultmidpunct}
{\mcitedefaultendpunct}{\mcitedefaultseppunct}\relax
\EndOfBibitem
\bibitem{Khachatryan:2015lua}
{CMS} collaboration, V.~Khachatryan {\em et al.},
  \href{http://dx.doi.org/10.1016/j.physletb.2016.02.047}{Phys. Lett. {\bf
  B756},  84 (2016)}, \href{http://arxiv.org/abs/1501.06089}{{\tt
  arXiv:1501.06089 [hep-ex]}}\relax
\mciteBstWouldAddEndPuncttrue
\mciteSetBstMidEndSepPunct{\mcitedefaultmidpunct}
{\mcitedefaultendpunct}{\mcitedefaultseppunct}\relax
\EndOfBibitem
\bibitem{Aaij:2018jlf}
{LHCb} collaboration, R.~Aaij {\em et al.},
  \href{http://dx.doi.org/10.1007/JHEP08(2018)131}{JHEP {\bf 08},  131 (2018)},
  \href{http://arxiv.org/abs/1806.08084}{{\tt arXiv:1806.08084 [hep-ex]}}\relax
\mciteBstWouldAddEndPuncttrue
\mciteSetBstMidEndSepPunct{\mcitedefaultmidpunct}
{\mcitedefaultendpunct}{\mcitedefaultseppunct}\relax
\EndOfBibitem
\bibitem{Aaij:2016qim}
{LHCb} collaboration, R.~Aaij {\em et al.},
  \href{http://dx.doi.org/10.1007/JHEP03(2016)040}{JHEP {\bf 03},  040 (2016)},
  \href{http://arxiv.org/abs/1601.05284}{{\tt arXiv:1601.05284 [hep-ex]}}\relax
\mciteBstWouldAddEndPuncttrue
\mciteSetBstMidEndSepPunct{\mcitedefaultmidpunct}
{\mcitedefaultendpunct}{\mcitedefaultseppunct}\relax
\EndOfBibitem
\bibitem{Aaij:2015wza}
{LHCb} collaboration, R.~Aaij {\em et al.},
  \href{http://dx.doi.org/10.1016/j.physletb.2015.06.038}{Phys. Lett. {\bf
  B747},  484 (2015)}, \href{http://arxiv.org/abs/1503.07112}{{\tt
  arXiv:1503.07112 [hep-ex]}}\relax
\mciteBstWouldAddEndPuncttrue
\mciteSetBstMidEndSepPunct{\mcitedefaultmidpunct}
{\mcitedefaultendpunct}{\mcitedefaultseppunct}\relax
\EndOfBibitem
\bibitem{Aaij:2014emv}
{LHCb} collaboration, R.~Aaij {\em et al.},
  \href{http://dx.doi.org/10.1103/PhysRevD.89.092006}{Phys. Rev. {\bf D89},
  092006 (2014)}, \href{http://arxiv.org/abs/1402.6248}{{\tt arXiv:1402.6248
  [hep-ex]}}\relax
\mciteBstWouldAddEndPuncttrue
\mciteSetBstMidEndSepPunct{\mcitedefaultmidpunct}
{\mcitedefaultendpunct}{\mcitedefaultseppunct}\relax
\EndOfBibitem
\bibitem{Solovieva:2013rhq}
{Belle} collaboration, E.~Solovieva {\em et al.},
  \href{http://dx.doi.org/10.1016/j.physletb.2013.08.057}{Phys. Lett. {\bf
  B726},  206 (2013)}, \href{http://arxiv.org/abs/1304.6931}{{\tt
  arXiv:1304.6931 [hep-ex]}}\relax
\mciteBstWouldAddEndPuncttrue
\mciteSetBstMidEndSepPunct{\mcitedefaultmidpunct}
{\mcitedefaultendpunct}{\mcitedefaultseppunct}\relax
\EndOfBibitem
\bibitem{Aaij:2017kea}
{LHCb} collaboration, R.~Aaij {\em et al.},
  \href{http://dx.doi.org/10.1103/PhysRevLett.118.111803}{Phys. Rev. Lett. {\bf
  118},  111803 (2017)}, \href{http://arxiv.org/abs/1701.01856}{{\tt
  arXiv:1701.01856 [hep-ex]}}\relax
\mciteBstWouldAddEndPuncttrue
\mciteSetBstMidEndSepPunct{\mcitedefaultmidpunct}
{\mcitedefaultendpunct}{\mcitedefaultseppunct}\relax
\EndOfBibitem
\bibitem{Aaij:2017gon}
{LHCb} collaboration, R.~Aaij {\em et al.},
  \href{http://arxiv.org/abs/1712.04702}{{\tt arXiv:1712.04702 [hep-ex]}}\relax
\mciteBstWouldAddEndPuncttrue
\mciteSetBstMidEndSepPunct{\mcitedefaultmidpunct}
{\mcitedefaultendpunct}{\mcitedefaultseppunct}\relax
\EndOfBibitem
\bibitem{Aaij:2013gia}
{LHCb} collaboration, R.~Aaij {\em et al.},
  \href{http://dx.doi.org/10.1103/PhysRevD.87.112012}{Phys. Rev. {\bf D87},
  112012 (2013)}, \href{http://arxiv.org/abs/1304.4530}{{\tt arXiv:1304.4530
  [hep-ex]}}\relax
\mciteBstWouldAddEndPuncttrue
\mciteSetBstMidEndSepPunct{\mcitedefaultmidpunct}
{\mcitedefaultendpunct}{\mcitedefaultseppunct}\relax
\EndOfBibitem
\bibitem{Aad:2015eza}
{ATLAS} collaboration, G.~Aad {\em et al.},
  \href{http://dx.doi.org/10.1140/epjc/s10052-015-3743-8}{Eur. Phys. J. {\bf
  C76},  4 (2016)}, \href{http://arxiv.org/abs/1507.07099}{{\tt
  arXiv:1507.07099 [hep-ex]}}\relax
\mciteBstWouldAddEndPuncttrue
\mciteSetBstMidEndSepPunct{\mcitedefaultmidpunct}
{\mcitedefaultendpunct}{\mcitedefaultseppunct}\relax
\EndOfBibitem
\bibitem{LHCb:2012ag}
{LHCb} collaboration, R.~Aaij {\em et al.},
  \href{http://dx.doi.org/10.1103/PhysRevLett.108.251802}{Phys. Rev. Lett. {\bf
  108},  251802 (2012)}, \href{http://arxiv.org/abs/1204.0079}{{\tt
  arXiv:1204.0079 [hep-ex]}}\relax
\mciteBstWouldAddEndPuncttrue
\mciteSetBstMidEndSepPunct{\mcitedefaultmidpunct}
{\mcitedefaultendpunct}{\mcitedefaultseppunct}\relax
\EndOfBibitem
\bibitem{Khachatryan:2014nfa}
{CMS} collaboration, V.~Khachatryan {\em et al.},
  \href{http://dx.doi.org/10.1007/JHEP01(2015)063}{JHEP {\bf 01},  063 (2015)},
  \href{http://arxiv.org/abs/1410.5729}{{\tt arXiv:1410.5729 [hep-ex]}}\relax
\mciteBstWouldAddEndPuncttrue
\mciteSetBstMidEndSepPunct{\mcitedefaultmidpunct}
{\mcitedefaultendpunct}{\mcitedefaultseppunct}\relax
\EndOfBibitem
\bibitem{Aaij:2016qlz}
{LHCb} collaboration, R.~Aaij {\em et al.},
  \href{http://dx.doi.org/10.1103/PhysRevD.95.032005}{Phys. Rev. {\bf D95},
  032005 (2017)}, \href{http://arxiv.org/abs/1612.07421}{{\tt arXiv:1612.07421
  [hep-ex]}}\relax
\mciteBstWouldAddEndPuncttrue
\mciteSetBstMidEndSepPunct{\mcitedefaultmidpunct}
{\mcitedefaultendpunct}{\mcitedefaultseppunct}\relax
\EndOfBibitem
\bibitem{Aaij:2013vcx}
{LHCb} collaboration, R.~Aaij {\em et al.},
  \href{http://dx.doi.org/10.1007/JHEP09(2013)075}{JHEP {\bf 09},  075 (2013)},
  \href{http://arxiv.org/abs/1306.6723}{{\tt arXiv:1306.6723 [hep-ex]}}\relax
\mciteBstWouldAddEndPuncttrue
\mciteSetBstMidEndSepPunct{\mcitedefaultmidpunct}
{\mcitedefaultendpunct}{\mcitedefaultseppunct}\relax
\EndOfBibitem
\bibitem{Aaij:2013gxa}
{LHCb} collaboration, R.~Aaij {\em et al.},
  \href{http://dx.doi.org/10.1007/JHEP11(2013)094}{JHEP {\bf 11},  094 (2013)},
  \href{http://arxiv.org/abs/1309.0587}{{\tt arXiv:1309.0587 [hep-ex]}}\relax
\mciteBstWouldAddEndPuncttrue
\mciteSetBstMidEndSepPunct{\mcitedefaultmidpunct}
{\mcitedefaultendpunct}{\mcitedefaultseppunct}\relax
\EndOfBibitem
\bibitem{Aaij:2015xga}
{LHCb} collaboration, R.~Aaij {\em et al.},
  \href{http://dx.doi.org/10.1103/PhysRevD.92.072007}{Phys. Rev. {\bf D92},
  072007 (2015)}, \href{http://arxiv.org/abs/1507.03516}{{\tt arXiv:1507.03516
  [hep-ex]}}\relax
\mciteBstWouldAddEndPuncttrue
\mciteSetBstMidEndSepPunct{\mcitedefaultmidpunct}
{\mcitedefaultendpunct}{\mcitedefaultseppunct}\relax
\EndOfBibitem
\bibitem{Aaij:2014ija}
{LHCb} collaboration, R.~Aaij {\em et al.},
  \href{http://dx.doi.org/10.1103/PhysRevLett.114.132001}{Phys. Rev. Lett. {\bf
  114},  132001 (2015)}, \href{http://arxiv.org/abs/1411.2943}{{\tt
  arXiv:1411.2943 [hep-ex]}}\relax
\mciteBstWouldAddEndPuncttrue
\mciteSetBstMidEndSepPunct{\mcitedefaultmidpunct}
{\mcitedefaultendpunct}{\mcitedefaultseppunct}\relax
\EndOfBibitem
\bibitem{Aaij:2012dd}
{LHCb} collaboration, R.~Aaij {\em et al.},
  \href{http://dx.doi.org/10.1103/PhysRevLett.109.232001}{Phys. Rev. Lett. {\bf
  109},  232001 (2012)}, \href{http://arxiv.org/abs/1209.5634}{{\tt
  arXiv:1209.5634 [hep-ex]}}\relax
\mciteBstWouldAddEndPuncttrue
\mciteSetBstMidEndSepPunct{\mcitedefaultmidpunct}
{\mcitedefaultendpunct}{\mcitedefaultseppunct}\relax
\EndOfBibitem
\bibitem{Aaij:2016xas}
{LHCb} collaboration, R.~Aaij {\em et al.},
  \href{http://dx.doi.org/10.1103/PhysRevD.94.091102}{Phys. Rev. {\bf D94},
  091102 (2016)}, \href{http://arxiv.org/abs/1607.06134}{{\tt arXiv:1607.06134
  [hep-ex]}}\relax
\mciteBstWouldAddEndPuncttrue
\mciteSetBstMidEndSepPunct{\mcitedefaultmidpunct}
{\mcitedefaultendpunct}{\mcitedefaultseppunct}\relax
\EndOfBibitem
\bibitem{Aaij:2013cda}
{LHCb} collaboration, R.~Aaij {\em et al.},
  \href{http://dx.doi.org/10.1103/PhysRevLett.111.181801}{Phys. Rev. Lett. {\bf
  111},  181801 (2013)}, \href{http://arxiv.org/abs/1308.4544}{{\tt
  arXiv:1308.4544 [hep-ex]}}\relax
\mciteBstWouldAddEndPuncttrue
\mciteSetBstMidEndSepPunct{\mcitedefaultmidpunct}
{\mcitedefaultendpunct}{\mcitedefaultseppunct}\relax
\EndOfBibitem
\bibitem{Aaij:2013pka}
{LHCb} collaboration, R.~Aaij {\em et al.},
  \href{http://dx.doi.org/10.1103/PhysRevD.89.032001}{Phys. Rev. {\bf D89},
  032001 (2014)}, \href{http://arxiv.org/abs/1311.4823}{{\tt arXiv:1311.4823
  [hep-ex]}}\relax
\mciteBstWouldAddEndPuncttrue
\mciteSetBstMidEndSepPunct{\mcitedefaultmidpunct}
{\mcitedefaultendpunct}{\mcitedefaultseppunct}\relax
\EndOfBibitem
\bibitem{Aaij:2015fea}
{LHCb} collaboration, R.~Aaij {\em et al.},
  \href{http://dx.doi.org/10.1088/1674-1137/40/1/011001}{Chin. Phys. {\bf C40},
   011001 (2016)}, \href{http://arxiv.org/abs/1509.00292}{{\tt arXiv:1509.00292
  [hep-ex]}}\relax
\mciteBstWouldAddEndPuncttrue
\mciteSetBstMidEndSepPunct{\mcitedefaultmidpunct}
{\mcitedefaultendpunct}{\mcitedefaultseppunct}\relax
\EndOfBibitem
\bibitem{Abe:1996tr}
{CDF} collaboration, F.~Abe {\em et al.},
  \href{http://dx.doi.org/10.1103/PhysRevD.55.1142}{Phys. Rev. {\bf D55},  1142
  (1997)}\relax
\mciteBstWouldAddEndPuncttrue
\mciteSetBstMidEndSepPunct{\mcitedefaultmidpunct}
{\mcitedefaultendpunct}{\mcitedefaultseppunct}\relax
\EndOfBibitem
\bibitem{Abazov:2011wt}
{\dzero} collaboration, V.~M. Abazov {\em et al.},
  \href{http://dx.doi.org/10.1103/PhysRevD.84.031102}{Phys. Rev. {\bf D84},
  031102 (2011)}, \href{http://arxiv.org/abs/1105.0690}{{\tt arXiv:1105.0690
  [hep-ex]}}\relax
\mciteBstWouldAddEndPuncttrue
\mciteSetBstMidEndSepPunct{\mcitedefaultmidpunct}
{\mcitedefaultendpunct}{\mcitedefaultseppunct}\relax
\EndOfBibitem
\bibitem{Aad:2015msa}
{ATLAS} collaboration, G.~Aad {\em et al.},
  \href{http://dx.doi.org/10.1016/j.physletb.2015.10.009}{Phys. Lett. {\bf
  B751},  63 (2015)}, \href{http://arxiv.org/abs/1507.08202}{{\tt
  arXiv:1507.08202 [hep-ex]}}\relax
\mciteBstWouldAddEndPuncttrue
\mciteSetBstMidEndSepPunct{\mcitedefaultmidpunct}
{\mcitedefaultendpunct}{\mcitedefaultseppunct}\relax
\EndOfBibitem
\bibitem{Aaij:2014zoa}
{LHCb} collaboration, R.~Aaij {\em et al.},
  \href{http://dx.doi.org/10.1007/JHEP07(2014)103}{JHEP {\bf 07},  103 (2014)},
  \href{http://arxiv.org/abs/1406.0755}{{\tt arXiv:1406.0755 [hep-ex]}}\relax
\mciteBstWouldAddEndPuncttrue
\mciteSetBstMidEndSepPunct{\mcitedefaultmidpunct}
{\mcitedefaultendpunct}{\mcitedefaultseppunct}\relax
\EndOfBibitem
\bibitem{Aaij:2016wxd}
{LHCb} collaboration, R.~Aaij {\em et al.},
  \href{http://dx.doi.org/10.1007/JHEP05(2016)132}{JHEP {\bf 05},  132 (2016)},
  \href{http://arxiv.org/abs/1603.06961}{{\tt arXiv:1603.06961 [hep-ex]}}\relax
\mciteBstWouldAddEndPuncttrue
\mciteSetBstMidEndSepPunct{\mcitedefaultmidpunct}
{\mcitedefaultendpunct}{\mcitedefaultseppunct}\relax
\EndOfBibitem
\bibitem{Aaij:2017awb}
{LHCb} collaboration, R.~Aaij {\em et al.},
  \href{http://dx.doi.org/10.1103/PhysRevLett.119.062001}{Phys. Rev. Lett. {\bf
  119},  062001 (2017)}, \href{http://arxiv.org/abs/1704.07900}{{\tt
  arXiv:1704.07900 [hep-ex]}}\relax
\mciteBstWouldAddEndPuncttrue
\mciteSetBstMidEndSepPunct{\mcitedefaultmidpunct}
{\mcitedefaultendpunct}{\mcitedefaultseppunct}\relax
\EndOfBibitem
\bibitem{Aaij:2013oxa}
{LHCb} collaboration, R.~Aaij {\em et al.},
  \href{http://dx.doi.org/10.1016/j.physletb.2013.05.041}{Phys. Lett. {\bf
  B724},  27 (2013)}, \href{http://arxiv.org/abs/1302.5578}{{\tt
  arXiv:1302.5578 [hep-ex]}}\relax
\mciteBstWouldAddEndPuncttrue
\mciteSetBstMidEndSepPunct{\mcitedefaultmidpunct}
{\mcitedefaultendpunct}{\mcitedefaultseppunct}\relax
\EndOfBibitem
\bibitem{Sirunyan:2018bfd}
{CMS} collaboration, A.~M. Sirunyan {\em et al.},
  \href{http://dx.doi.org/10.1103/PhysRevD.97.072010}{Phys. Rev. {\bf D97},
  072010 (2018)}, \href{http://arxiv.org/abs/1802.04867}{{\tt arXiv:1802.04867
  [hep-ex]}}\relax
\mciteBstWouldAddEndPuncttrue
\mciteSetBstMidEndSepPunct{\mcitedefaultmidpunct}
{\mcitedefaultendpunct}{\mcitedefaultseppunct}\relax
\EndOfBibitem
\bibitem{Aad:2014iba}
{ATLAS} collaboration, G.~Aad {\em et al.},
  \href{http://dx.doi.org/10.1103/PhysRevD.89.092009}{Phys. Rev. {\bf D89},
  092009 (2014)}, \href{http://arxiv.org/abs/1404.1071}{{\tt arXiv:1404.1071
  [hep-ex]}}\relax
\mciteBstWouldAddEndPuncttrue
\mciteSetBstMidEndSepPunct{\mcitedefaultmidpunct}
{\mcitedefaultendpunct}{\mcitedefaultseppunct}\relax
\EndOfBibitem
\bibitem{CDF:2011aa}
{CDF} collaboration, T.~Aaltonen {\em et al.},
  \href{http://dx.doi.org/10.1103/PhysRevD.85.032003}{Phys. Rev. {\bf D85},
  032003 (2012)}, \href{http://arxiv.org/abs/1112.3334}{{\tt arXiv:1112.3334
  [hep-ex]}}\relax
\mciteBstWouldAddEndPuncttrue
\mciteSetBstMidEndSepPunct{\mcitedefaultmidpunct}
{\mcitedefaultendpunct}{\mcitedefaultseppunct}\relax
\EndOfBibitem
\bibitem{Abulencia:2006df}
{CDF} collaboration, A.~Abulencia {\em et al.},
  \href{http://dx.doi.org/10.1103/PhysRevLett.98.122002}{Phys. Rev. Lett. {\bf
  98},  122002 (2007)}, \href{http://arxiv.org/abs/hep-ex/0601003}{{\tt
  arXiv:hep-ex/0601003 [hep-ex]}}\relax
\mciteBstWouldAddEndPuncttrue
\mciteSetBstMidEndSepPunct{\mcitedefaultmidpunct}
{\mcitedefaultendpunct}{\mcitedefaultseppunct}\relax
\EndOfBibitem
\bibitem{Aaij:2017vbw}
{LHCb} collaboration, R.~Aaij {\em et al.},
  \href{http://dx.doi.org/10.1007/JHEP05(2017)030}{JHEP {\bf 05},  030 (2017)},
  \href{http://arxiv.org/abs/1701.07873}{{\tt arXiv:1701.07873 [hep-ex]}}\relax
\mciteBstWouldAddEndPuncttrue
\mciteSetBstMidEndSepPunct{\mcitedefaultmidpunct}
{\mcitedefaultendpunct}{\mcitedefaultseppunct}\relax
\EndOfBibitem
\bibitem{Aaij:2018bre}
{LHCb} collaboration, R.~Aaij {\em et al.}, Submitted to: Phys. Lett.   (2018),
  \href{http://arxiv.org/abs/1804.09617}{{\tt arXiv:1804.09617 [hep-ex]}}\relax
\mciteBstWouldAddEndPuncttrue
\mciteSetBstMidEndSepPunct{\mcitedefaultmidpunct}
{\mcitedefaultendpunct}{\mcitedefaultseppunct}\relax
\EndOfBibitem
\bibitem{Aaij:2015yoy}
{LHCb} collaboration, R.~Aaij {\em et al.},
  \href{http://dx.doi.org/10.1103/PhysRevLett.115.241801}{Phys. Rev. Lett. {\bf
  115},  241801 (2015)}, \href{http://arxiv.org/abs/1510.03829}{{\tt
  arXiv:1510.03829 [hep-ex]}}\relax
\mciteBstWouldAddEndPuncttrue
\mciteSetBstMidEndSepPunct{\mcitedefaultmidpunct}
{\mcitedefaultendpunct}{\mcitedefaultseppunct}\relax
\EndOfBibitem
\bibitem{Aaij:2016ymb}
{LHCb} collaboration, R.~Aaij {\em et al.},
  \href{http://dx.doi.org/10.1103/PhysRevLett.118.119901,
  10.1103/PhysRevLett.117.082003, 10.1103/PhysRevLett.117.109902}{Phys. Rev.
  Lett. {\bf 117},  082003 (2016)}, \href{http://arxiv.org/abs/1606.06999}{{\tt
  arXiv:1606.06999 [hep-ex]}}, addendum: Phys. Rev. Lett. {\bf 118}, 119901
  (2017)\relax
\mciteBstWouldAddEndPuncttrue
\mciteSetBstMidEndSepPunct{\mcitedefaultmidpunct}
{\mcitedefaultendpunct}{\mcitedefaultseppunct}\relax
\EndOfBibitem
\bibitem{Duh:2012ie}
{Belle} collaboration, Y.~T. Duh {\em et al.},
  \href{http://dx.doi.org/10.1103/PhysRevD.87.031103}{Phys. Rev. {\bf D87},
  031103 (2013)}, \href{http://arxiv.org/abs/1210.1348}{{\tt arXiv:1210.1348
  [hep-ex]}}\relax
\mciteBstWouldAddEndPuncttrue
\mciteSetBstMidEndSepPunct{\mcitedefaultmidpunct}
{\mcitedefaultendpunct}{\mcitedefaultseppunct}\relax
\EndOfBibitem
\bibitem{Bornheim:2003bv}
{CLEO} collaboration, A.~Bornheim {\em et al.},
  \href{http://dx.doi.org/10.1103/PhysRevD.68.052002}{Phys. Rev. {\bf D68},
  052002 (2003)}, \href{http://arxiv.org/abs/hep-ex/0302026}{{\tt
  arXiv:hep-ex/0302026 [hep-ex]}}, Erratum ibid.\
  \href{http://dx.doi.org/10.1103/PhysRevD.75.119907}{{\bf D75}, 119907},
  (2007)\relax
\mciteBstWouldAddEndPuncttrue
\mciteSetBstMidEndSepPunct{\mcitedefaultmidpunct}
{\mcitedefaultendpunct}{\mcitedefaultseppunct}\relax
\EndOfBibitem
\bibitem{Aubert:2007hh}
{\babar} collaboration, B.~Aubert {\em et al.},
  \href{http://dx.doi.org/10.1103/PhysRevD.76.091102}{Phys. Rev. {\bf D76},
  091102 (2007)}, \href{http://arxiv.org/abs/0707.2798}{{\tt arXiv:0707.2798
  [hep-ex]}}\relax
\mciteBstWouldAddEndPuncttrue
\mciteSetBstMidEndSepPunct{\mcitedefaultmidpunct}
{\mcitedefaultendpunct}{\mcitedefaultseppunct}\relax
\EndOfBibitem
\bibitem{Aubert:2009yx}
{\babar} collaboration, B.~Aubert {\em et al.},
  \href{http://dx.doi.org/10.1103/PhysRevD.80.112002}{Phys. Rev. {\bf D80},
  112002 (2009)}, \href{http://arxiv.org/abs/0907.1743}{{\tt arXiv:0907.1743
  [hep-ex]}}\relax
\mciteBstWouldAddEndPuncttrue
\mciteSetBstMidEndSepPunct{\mcitedefaultmidpunct}
{\mcitedefaultendpunct}{\mcitedefaultseppunct}\relax
\EndOfBibitem
\bibitem{Schumann:2006bg}
{Belle} collaboration, J.~Schumann {\em et al.},
  \href{http://dx.doi.org/10.1103/PhysRevLett.97.061802}{Phys. Rev. Lett. {\bf
  97},  061802 (2006)}, \href{http://arxiv.org/abs/hep-ex/0603001}{{\tt
  arXiv:hep-ex/0603001 [hep-ex]}}\relax
\mciteBstWouldAddEndPuncttrue
\mciteSetBstMidEndSepPunct{\mcitedefaultmidpunct}
{\mcitedefaultendpunct}{\mcitedefaultseppunct}\relax
\EndOfBibitem
\bibitem{delAmoSanchez:2010qa}
{\babar} collaboration, P.~del Amo~Sanchez {\em et al.},
  \href{http://dx.doi.org/10.1103/PhysRevD.82.011502}{Phys. Rev. {\bf D82},
  011502 (2010)}, \href{http://arxiv.org/abs/1004.0240}{{\tt arXiv:1004.0240
  [hep-ex]}}\relax
\mciteBstWouldAddEndPuncttrue
\mciteSetBstMidEndSepPunct{\mcitedefaultmidpunct}
{\mcitedefaultendpunct}{\mcitedefaultseppunct}\relax
\EndOfBibitem
\bibitem{Schumann:2007ae}
{Belle} collaboration, J.~Schumann {\em et al.},
  \href{http://dx.doi.org/10.1103/PhysRevD.75.092002}{Phys. Rev. {\bf D75},
  092002 (2007)}, \href{http://arxiv.org/abs/hep-ex/0701046}{{\tt
  arXiv:hep-ex/0701046 [hep-ex]}}\relax
\mciteBstWouldAddEndPuncttrue
\mciteSetBstMidEndSepPunct{\mcitedefaultmidpunct}
{\mcitedefaultendpunct}{\mcitedefaultseppunct}\relax
\EndOfBibitem
\bibitem{Hoi:2011gv}
{Belle} collaboration, C.~T. Hoi {\em et al.},
  \href{http://dx.doi.org/10.1103/PhysRevLett.108.031801}{Phys. Rev. Lett. {\bf
  108},  031801 (2012)}, \href{http://arxiv.org/abs/1110.2000}{{\tt
  arXiv:1110.2000 [hep-ex]}}\relax
\mciteBstWouldAddEndPuncttrue
\mciteSetBstMidEndSepPunct{\mcitedefaultmidpunct}
{\mcitedefaultendpunct}{\mcitedefaultseppunct}\relax
\EndOfBibitem
\bibitem{Richichi:1999kj}
{CLEO} collaboration, S.~J. Richichi {\em et al.},
  \href{http://dx.doi.org/10.1103/PhysRevLett.85.520}{Phys. Rev. Lett. {\bf
  85},  520 (2000)}, \href{http://arxiv.org/abs/hep-ex/9912059}{{\tt
  arXiv:hep-ex/9912059 [hep-ex]}}\relax
\mciteBstWouldAddEndPuncttrue
\mciteSetBstMidEndSepPunct{\mcitedefaultmidpunct}
{\mcitedefaultendpunct}{\mcitedefaultseppunct}\relax
\EndOfBibitem
\bibitem{Aubert:2006fj}
{\babar} collaboration, B.~Aubert {\em et al.},
  \href{http://dx.doi.org/10.1103/PhysRevLett.97.201802}{Phys. Rev. Lett. {\bf
  97},  201802 (2006)}, \href{http://arxiv.org/abs/hep-ex/0608005}{{\tt
  arXiv:hep-ex/0608005 [hep-ex]}}\relax
\mciteBstWouldAddEndPuncttrue
\mciteSetBstMidEndSepPunct{\mcitedefaultmidpunct}
{\mcitedefaultendpunct}{\mcitedefaultseppunct}\relax
\EndOfBibitem
\bibitem{Wang:2007rzb}
{Belle} collaboration, C.~H. Wang {\em et al.},
  \href{http://dx.doi.org/10.1103/PhysRevD.75.092005}{Phys. Rev. {\bf D75},
  092005 (2007)}, \href{http://arxiv.org/abs/hep-ex/0701057}{{\tt
  arXiv:hep-ex/0701057 [hep-ex]}}\relax
\mciteBstWouldAddEndPuncttrue
\mciteSetBstMidEndSepPunct{\mcitedefaultmidpunct}
{\mcitedefaultendpunct}{\mcitedefaultseppunct}\relax
\EndOfBibitem
\bibitem{Aubert:2008bk}
{\babar} collaboration, B.~Aubert {\em et al.},
  \href{http://dx.doi.org/10.1103/PhysRevLett.101.091801}{Phys. Rev. Lett. {\bf
  101},  091801 (2008)}, \href{http://arxiv.org/abs/0804.0411}{{\tt
  arXiv:0804.0411 [hep-ex]}}\relax
\mciteBstWouldAddEndPuncttrue
\mciteSetBstMidEndSepPunct{\mcitedefaultmidpunct}
{\mcitedefaultendpunct}{\mcitedefaultseppunct}\relax
\EndOfBibitem
\bibitem{Aubert:2007si}
{\babar} collaboration, B.~Aubert {\em et al.},
  \href{http://dx.doi.org/10.1103/PhysRevD.76.031103}{Phys. Rev. {\bf D76},
  031103 (2007)}, \href{http://arxiv.org/abs/0706.3893}{{\tt arXiv:0706.3893
  [hep-ex]}}\relax
\mciteBstWouldAddEndPuncttrue
\mciteSetBstMidEndSepPunct{\mcitedefaultmidpunct}
{\mcitedefaultendpunct}{\mcitedefaultseppunct}\relax
\EndOfBibitem
\bibitem{Jessop:2000bv}
{CLEO} collaboration, C.~P. Jessop {\em et al.},
  \href{http://dx.doi.org/10.1103/PhysRevLett.85.2881}{Phys. Rev. Lett. {\bf
  85},  2881 (2000)}, \href{http://arxiv.org/abs/hep-ex/0006008}{{\tt
  arXiv:hep-ex/0006008 [hep-ex]}}\relax
\mciteBstWouldAddEndPuncttrue
\mciteSetBstMidEndSepPunct{\mcitedefaultmidpunct}
{\mcitedefaultendpunct}{\mcitedefaultseppunct}\relax
\EndOfBibitem
\bibitem{Aubert:2009sx}
{\babar} collaboration, B.~Aubert {\em et al.},
  \href{http://dx.doi.org/10.1103/PhysRevD.79.052005}{Phys. Rev. {\bf D79},
  052005 (2009)}, \href{http://arxiv.org/abs/0901.3703}{{\tt arXiv:0901.3703
  [hep-ex]}}\relax
\mciteBstWouldAddEndPuncttrue
\mciteSetBstMidEndSepPunct{\mcitedefaultmidpunct}
{\mcitedefaultendpunct}{\mcitedefaultseppunct}\relax
\EndOfBibitem
\bibitem{Aubert:2004hs}
{\babar} collaboration, B.~Aubert {\em et al.},
  \href{http://dx.doi.org/10.1103/PhysRevD.70.111102}{Phys. Rev. {\bf D70},
  111102 (2004)}, \href{http://arxiv.org/abs/hep-ex/0407013}{{\tt
  arXiv:hep-ex/0407013 [hep-ex]}}\relax
\mciteBstWouldAddEndPuncttrue
\mciteSetBstMidEndSepPunct{\mcitedefaultmidpunct}
{\mcitedefaultendpunct}{\mcitedefaultseppunct}\relax
\EndOfBibitem
\bibitem{BABAR:2011aaa}
{\babar} collaboration, J.~P. Lees {\em et al.},
  \href{http://dx.doi.org/10.1103/PhysRevD.84.092007}{Phys. Rev. {\bf D84},
  092007 (2011)}, \href{http://arxiv.org/abs/1109.0143}{{\tt arXiv:1109.0143
  [hep-ex]}}\relax
\mciteBstWouldAddEndPuncttrue
\mciteSetBstMidEndSepPunct{\mcitedefaultmidpunct}
{\mcitedefaultendpunct}{\mcitedefaultseppunct}\relax
\EndOfBibitem
\bibitem{Aubert:2008rr}
{\babar} collaboration, B.~Aubert {\em et al.},
  \href{http://dx.doi.org/10.1103/PhysRevD.78.091102}{Phys. Rev. {\bf D78},
  091102 (2008)}, \href{http://arxiv.org/abs/0808.0900}{{\tt arXiv:0808.0900
  [hep-ex]}}\relax
\mciteBstWouldAddEndPuncttrue
\mciteSetBstMidEndSepPunct{\mcitedefaultmidpunct}
{\mcitedefaultendpunct}{\mcitedefaultseppunct}\relax
\EndOfBibitem
\bibitem{Garmash:2003er}
{Belle} collaboration, A.~Garmash {\em et al.},
  \href{http://dx.doi.org/10.1103/PhysRevD.69.012001}{Phys. Rev. {\bf D69},
  012001 (2004)}, \href{http://arxiv.org/abs/hep-ex/0307082}{{\tt
  arXiv:hep-ex/0307082 [hep-ex]}}\relax
\mciteBstWouldAddEndPuncttrue
\mciteSetBstMidEndSepPunct{\mcitedefaultmidpunct}
{\mcitedefaultendpunct}{\mcitedefaultseppunct}\relax
\EndOfBibitem
\bibitem{LHCb:2016rul}
{LHCb} collaboration, R.~Aaij {\em et al.},
  \href{http://dx.doi.org/10.1016/j.physletb.2016.11.053}{Phys. Lett. {\bf
  B765},  307 (2017)}, \href{http://arxiv.org/abs/1608.01478}{{\tt
  arXiv:1608.01478 [hep-ex]}}\relax
\mciteBstWouldAddEndPuncttrue
\mciteSetBstMidEndSepPunct{\mcitedefaultmidpunct}
{\mcitedefaultendpunct}{\mcitedefaultseppunct}\relax
\EndOfBibitem
\bibitem{Bergfeld:1996dd}
{CLEO} collaboration, T.~Bergfeld {\em et al.},
  \href{http://dx.doi.org/10.1103/PhysRevLett.77.4503}{Phys. Rev. Lett. {\bf
  77},  4503 (1996)}\relax
\mciteBstWouldAddEndPuncttrue
\mciteSetBstMidEndSepPunct{\mcitedefaultmidpunct}
{\mcitedefaultendpunct}{\mcitedefaultseppunct}\relax
\EndOfBibitem
\bibitem{Eckhart:2002qr}
{CLEO} collaboration, E.~Eckhart {\em et al.},
  \href{http://dx.doi.org/10.1103/PhysRevLett.89.251801}{Phys. Rev. Lett. {\bf
  89},  251801 (2002)}, \href{http://arxiv.org/abs/hep-ex/0206024}{{\tt
  arXiv:hep-ex/0206024 [hep-ex]}}\relax
\mciteBstWouldAddEndPuncttrue
\mciteSetBstMidEndSepPunct{\mcitedefaultmidpunct}
{\mcitedefaultendpunct}{\mcitedefaultseppunct}\relax
\EndOfBibitem
\bibitem{Aubert:2007mb}
{\babar} collaboration, B.~Aubert {\em et al.},
  \href{http://dx.doi.org/10.1103/PhysRevD.76.011103}{Phys. Rev. {\bf D76},
  011103 (2007)}, \href{http://arxiv.org/abs/hep-ex/0702043}{{\tt
  arXiv:hep-ex/0702043 [hep-ex]}}\relax
\mciteBstWouldAddEndPuncttrue
\mciteSetBstMidEndSepPunct{\mcitedefaultmidpunct}
{\mcitedefaultendpunct}{\mcitedefaultseppunct}\relax
\EndOfBibitem
\bibitem{Aubert:2006aw}
{\babar} collaboration, B.~Aubert {\em et al.},
  \href{http://dx.doi.org/10.1103/PhysRevD.74.051104}{Phys. Rev. {\bf D74},
  051104 (2006)}, \href{http://arxiv.org/abs/hep-ex/0607113}{{\tt
  arXiv:hep-ex/0607113 [hep-ex]}}\relax
\mciteBstWouldAddEndPuncttrue
\mciteSetBstMidEndSepPunct{\mcitedefaultmidpunct}
{\mcitedefaultendpunct}{\mcitedefaultseppunct}\relax
\EndOfBibitem
\bibitem{delAmoSanchez:2010mz}
{\babar} collaboration, P.~del Amo~Sanchez {\em et al.},
  \href{http://dx.doi.org/10.1103/PhysRevD.83.051101}{Phys. Rev. {\bf D83},
  051101 (2011)}, \href{http://arxiv.org/abs/1012.4044}{{\tt arXiv:1012.4044
  [hep-ex]}}\relax
\mciteBstWouldAddEndPuncttrue
\mciteSetBstMidEndSepPunct{\mcitedefaultmidpunct}
{\mcitedefaultendpunct}{\mcitedefaultseppunct}\relax
\EndOfBibitem
\bibitem{Aubert:2007ds}
{\babar} collaboration, B.~Aubert {\em et al.},
  \href{http://dx.doi.org/10.1103/PhysRevLett.100.051803}{Phys. Rev. Lett. {\bf
  100},  051803 (2008)}, \href{http://arxiv.org/abs/0709.4165}{{\tt
  arXiv:0709.4165 [hep-ex]}}\relax
\mciteBstWouldAddEndPuncttrue
\mciteSetBstMidEndSepPunct{\mcitedefaultmidpunct}
{\mcitedefaultendpunct}{\mcitedefaultseppunct}\relax
\EndOfBibitem
\bibitem{Aubert:2008bg}
{\babar} collaboration, B.~Aubert {\em et al.},
  \href{http://dx.doi.org/10.1103/PhysRevD.78.011104}{Phys. Rev. {\bf D78},
  011104 (2008)}, \href{http://arxiv.org/abs/0805.1217}{{\tt arXiv:0805.1217
  [hep-ex]}}\relax
\mciteBstWouldAddEndPuncttrue
\mciteSetBstMidEndSepPunct{\mcitedefaultmidpunct}
{\mcitedefaultendpunct}{\mcitedefaultseppunct}\relax
\EndOfBibitem
\bibitem{Aubert:2006fs}
{\babar} collaboration, B.~Aubert {\em et al.},
  \href{http://dx.doi.org/10.1103/PhysRevLett.97.201801}{Phys. Rev. Lett. {\bf
  97},  201801 (2006)}, \href{http://arxiv.org/abs/hep-ex/0607057}{{\tt
  arXiv:hep-ex/0607057 [hep-ex]}}\relax
\mciteBstWouldAddEndPuncttrue
\mciteSetBstMidEndSepPunct{\mcitedefaultmidpunct}
{\mcitedefaultendpunct}{\mcitedefaultseppunct}\relax
\EndOfBibitem
\bibitem{Abe:2004mq}
{Belle} collaboration, J.~Zhang {\em et al.},
  \href{http://dx.doi.org/10.1103/PhysRevLett.95.141801}{Phys. Rev. Lett. {\bf
  95},  141801 (2005)}, \href{http://arxiv.org/abs/hep-ex/0408102}{{\tt
  arXiv:hep-ex/0408102 [hep-ex]}}\relax
\mciteBstWouldAddEndPuncttrue
\mciteSetBstMidEndSepPunct{\mcitedefaultmidpunct}
{\mcitedefaultendpunct}{\mcitedefaultseppunct}\relax
\EndOfBibitem
\bibitem{Albrecht:1989ny}
{ARGUS} collaboration, H.~Albrecht {\em et al.},
  \href{http://dx.doi.org/10.1016/0370-2693(91)90436-T}{Phys. Lett. {\bf B254},
   288 (1991)}\relax
\mciteBstWouldAddEndPuncttrue
\mciteSetBstMidEndSepPunct{\mcitedefaultmidpunct}
{\mcitedefaultendpunct}{\mcitedefaultseppunct}\relax
\EndOfBibitem
\bibitem{Aubert:2007xd}
{\babar} collaboration, B.~Aubert {\em et al.},
  \href{http://dx.doi.org/10.1103/PhysRevLett.99.241803}{Phys. Rev. Lett. {\bf
  99},  241803 (2007)}, \href{http://arxiv.org/abs/0707.4561}{{\tt
  arXiv:0707.4561 [hep-ex]}}\relax
\mciteBstWouldAddEndPuncttrue
\mciteSetBstMidEndSepPunct{\mcitedefaultmidpunct}
{\mcitedefaultendpunct}{\mcitedefaultseppunct}\relax
\EndOfBibitem
\bibitem{Aubert:2009qb}
{\babar} collaboration, B.~Aubert {\em et al.},
  \href{http://dx.doi.org/10.1103/PhysRevD.80.051101}{Phys. Rev. {\bf D80},
  051101 (2009)}, \href{http://arxiv.org/abs/0907.3485}{{\tt arXiv:0907.3485
  [hep-ex]}}\relax
\mciteBstWouldAddEndPuncttrue
\mciteSetBstMidEndSepPunct{\mcitedefaultmidpunct}
{\mcitedefaultendpunct}{\mcitedefaultseppunct}\relax
\EndOfBibitem
\bibitem{Aaij:2013fja}
{LHCb} collaboration, R.~Aaij {\em et al.},
  \href{http://dx.doi.org/10.1016/j.physletb.2013.09.046}{Phys. Lett. {\bf
  B726},  646 (2013)}, \href{http://arxiv.org/abs/1308.1277}{{\tt
  arXiv:1308.1277 [hep-ex]}}\relax
\mciteBstWouldAddEndPuncttrue
\mciteSetBstMidEndSepPunct{\mcitedefaultmidpunct}
{\mcitedefaultendpunct}{\mcitedefaultseppunct}\relax
\EndOfBibitem
\bibitem{Abdesselam:2018rbc}
{Belle} collaboration, A.~Abdesselam {\em et al.},
  \href{http://arxiv.org/abs/1808.00174}{{\tt arXiv:1808.00174 [hep-ex]}}\relax
\mciteBstWouldAddEndPuncttrue
\mciteSetBstMidEndSepPunct{\mcitedefaultmidpunct}
{\mcitedefaultendpunct}{\mcitedefaultseppunct}\relax
\EndOfBibitem
\bibitem{Aubert:2008aw}
{\babar} collaboration, B.~Aubert {\em et al.},
  \href{http://dx.doi.org/10.1103/PhysRevD.79.051101}{Phys. Rev. {\bf D79},
  051101 (2009)}, \href{http://arxiv.org/abs/0811.1979}{{\tt arXiv:0811.1979
  [hep-ex]}}\relax
\mciteBstWouldAddEndPuncttrue
\mciteSetBstMidEndSepPunct{\mcitedefaultmidpunct}
{\mcitedefaultendpunct}{\mcitedefaultseppunct}\relax
\EndOfBibitem
\bibitem{Aubert:2007xb}
{\babar} collaboration, B.~Aubert {\em et al.},
  \href{http://dx.doi.org/10.1103/PhysRevLett.99.221801}{Phys. Rev. Lett. {\bf
  99},  221801 (2007)}, \href{http://arxiv.org/abs/0708.0376}{{\tt
  arXiv:0708.0376 [hep-ex]}}\relax
\mciteBstWouldAddEndPuncttrue
\mciteSetBstMidEndSepPunct{\mcitedefaultmidpunct}
{\mcitedefaultendpunct}{\mcitedefaultseppunct}\relax
\EndOfBibitem
\bibitem{Hsu:2017kir}
{Belle} collaboration, C.~L. Hsu {\em et al.},
  \href{http://dx.doi.org/10.1103/PhysRevD.96.031101}{Phys. Rev. {\bf D96},
  031101 (2017)}, \href{http://arxiv.org/abs/1705.02640}{{\tt arXiv:1705.02640
  [hep-ex]}}\relax
\mciteBstWouldAddEndPuncttrue
\mciteSetBstMidEndSepPunct{\mcitedefaultmidpunct}
{\mcitedefaultendpunct}{\mcitedefaultseppunct}\relax
\EndOfBibitem
\bibitem{Aubert:2007ua}
{\babar} collaboration, B.~Aubert {\em et al.},
  \href{http://dx.doi.org/10.1103/PhysRevD.76.071103}{Phys. Rev. {\bf D76},
  071103 (2007)}, \href{http://arxiv.org/abs/0706.1059}{{\tt arXiv:0706.1059
  [hep-ex]}}\relax
\mciteBstWouldAddEndPuncttrue
\mciteSetBstMidEndSepPunct{\mcitedefaultmidpunct}
{\mcitedefaultendpunct}{\mcitedefaultseppunct}\relax
\EndOfBibitem
\bibitem{Huang:2003dr}
{Belle} collaboration, H.~C. Huang {\em et al.},
  \href{http://dx.doi.org/10.1103/PhysRevLett.91.241802}{Phys. Rev. Lett. {\bf
  91},  241802 (2003)}, \href{http://arxiv.org/abs/hep-ex/0305068}{{\tt
  arXiv:hep-ex/0305068 [hep-ex]}}\relax
\mciteBstWouldAddEndPuncttrue
\mciteSetBstMidEndSepPunct{\mcitedefaultmidpunct}
{\mcitedefaultendpunct}{\mcitedefaultseppunct}\relax
\EndOfBibitem
\bibitem{Aubert:2009ax}
{\babar} collaboration, B.~Aubert {\em et al.},
  \href{http://dx.doi.org/10.1103/PhysRevD.79.051102}{Phys. Rev. {\bf D79},
  051102 (2009)}, \href{http://arxiv.org/abs/0901.1223}{{\tt arXiv:0901.1223
  [hep-ex]}}\relax
\mciteBstWouldAddEndPuncttrue
\mciteSetBstMidEndSepPunct{\mcitedefaultmidpunct}
{\mcitedefaultendpunct}{\mcitedefaultseppunct}\relax
\EndOfBibitem
\bibitem{Goh:2015kaa}
{Belle} collaboration, Y.~M. Goh {\em et al.},
  \href{http://dx.doi.org/10.1103/PhysRevD.91.071101}{Phys. Rev. {\bf D91},
  071101 (2015)}, \href{http://arxiv.org/abs/1502.00381}{{\tt arXiv:1502.00381
  [hep-ex]}}\relax
\mciteBstWouldAddEndPuncttrue
\mciteSetBstMidEndSepPunct{\mcitedefaultmidpunct}
{\mcitedefaultendpunct}{\mcitedefaultseppunct}\relax
\EndOfBibitem
\bibitem{Briere:2001ue}
{CLEO} collaboration, R.~A. Briere {\em et al.},
  \href{http://dx.doi.org/10.1103/PhysRevLett.86.3718}{Phys. Rev. Lett. {\bf
  86},  3718 (2001)}, \href{http://arxiv.org/abs/hep-ex/0101032}{{\tt
  arXiv:hep-ex/0101032 [hep-ex]}}\relax
\mciteBstWouldAddEndPuncttrue
\mciteSetBstMidEndSepPunct{\mcitedefaultmidpunct}
{\mcitedefaultendpunct}{\mcitedefaultseppunct}\relax
\EndOfBibitem
\bibitem{Acosta:2005eu}
{CDF} collaboration, D.~Acosta {\em et al.},
  \href{http://dx.doi.org/10.1103/PhysRevLett.95.031801}{Phys. Rev. Lett. {\bf
  95},  031801 (2005)}, \href{http://arxiv.org/abs/hep-ex/0502044}{{\tt
  arXiv:hep-ex/0502044 [hep-ex]}}\relax
\mciteBstWouldAddEndPuncttrue
\mciteSetBstMidEndSepPunct{\mcitedefaultmidpunct}
{\mcitedefaultendpunct}{\mcitedefaultseppunct}\relax
\EndOfBibitem
\bibitem{Abulencia:2005aj}
{CDF} collaboration, A.~Abulencia {\em et al.},
  \href{http://dx.doi.org/10.1103/PhysRevD.73.032003}{Phys. Rev. {\bf D73},
  032003 (2006)}, \href{http://arxiv.org/abs/hep-ex/0510048}{{\tt
  arXiv:hep-ex/0510048 [hep-ex]}}\relax
\mciteBstWouldAddEndPuncttrue
\mciteSetBstMidEndSepPunct{\mcitedefaultmidpunct}
{\mcitedefaultendpunct}{\mcitedefaultseppunct}\relax
\EndOfBibitem
\bibitem{Aubert:2007ac}
{\babar} collaboration, B.~Aubert {\em et al.},
  \href{http://dx.doi.org/10.1103/PhysRevLett.99.201802}{Phys. Rev. Lett. {\bf
  99},  201802 (2007)}, \href{http://arxiv.org/abs/0705.1798}{{\tt
  arXiv:0705.1798 [hep-ex]}}\relax
\mciteBstWouldAddEndPuncttrue
\mciteSetBstMidEndSepPunct{\mcitedefaultmidpunct}
{\mcitedefaultendpunct}{\mcitedefaultseppunct}\relax
\EndOfBibitem
\bibitem{Chen:2003jfa}
{Belle} collaboration, K.~F. Chen {\em et al.},
  \href{http://dx.doi.org/10.1103/PhysRevLett.91.201801}{Phys. Rev. Lett. {\bf
  91},  201801 (2003)}, \href{http://arxiv.org/abs/hep-ex/0307014}{{\tt
  arXiv:hep-ex/0307014 [hep-ex]}}\relax
\mciteBstWouldAddEndPuncttrue
\mciteSetBstMidEndSepPunct{\mcitedefaultmidpunct}
{\mcitedefaultendpunct}{\mcitedefaultseppunct}\relax
\EndOfBibitem
\bibitem{Aubert:2008bc}
{\babar} collaboration, B.~Aubert {\em et al.},
  \href{http://dx.doi.org/10.1103/PhysRevLett.101.161801}{Phys. Rev. Lett. {\bf
  101},  161801 (2008)}, \href{http://arxiv.org/abs/0806.4419}{{\tt
  arXiv:0806.4419 [hep-ex]}}\relax
\mciteBstWouldAddEndPuncttrue
\mciteSetBstMidEndSepPunct{\mcitedefaultmidpunct}
{\mcitedefaultendpunct}{\mcitedefaultseppunct}\relax
\EndOfBibitem
\bibitem{Sanchez:2010qm}
{\babar} collaboration, P.~del Amo~Sanchez {\em et al.},
  \href{http://dx.doi.org/10.1103/PhysRevD.82.091101}{Phys. Rev. {\bf D82},
  091101 (2010)}, \href{http://arxiv.org/abs/1007.2732}{{\tt arXiv:1007.2732
  [hep-ex]}}\relax
\mciteBstWouldAddEndPuncttrue
\mciteSetBstMidEndSepPunct{\mcitedefaultmidpunct}
{\mcitedefaultendpunct}{\mcitedefaultseppunct}\relax
\EndOfBibitem
\bibitem{Lees:2011zh}
{\babar} collaboration, J.~P. Lees {\em et al.},
  \href{http://dx.doi.org/10.1103/PhysRevD.84.012001}{Phys. Rev. {\bf D84},
  012001 (2011)}, \href{http://arxiv.org/abs/1105.5159}{{\tt arXiv:1105.5159
  [hep-ex]}}\relax
\mciteBstWouldAddEndPuncttrue
\mciteSetBstMidEndSepPunct{\mcitedefaultmidpunct}
{\mcitedefaultendpunct}{\mcitedefaultseppunct}\relax
\EndOfBibitem
\bibitem{Aubert:2006caa}
{\babar} collaboration, B.~Aubert {\em et al.},
  \href{http://dx.doi.org/10.1103/PhysRevD.74.031105}{Phys. Rev. {\bf D74},
  031105 (2006)}, \href{http://arxiv.org/abs/hep-ex/0605008}{{\tt
  arXiv:hep-ex/0605008 [hep-ex]}}\relax
\mciteBstWouldAddEndPuncttrue
\mciteSetBstMidEndSepPunct{\mcitedefaultmidpunct}
{\mcitedefaultendpunct}{\mcitedefaultseppunct}\relax
\EndOfBibitem
\bibitem{Liu:2009kca}
{Belle} collaboration, C.~Liu {\em et al.},
  \href{http://dx.doi.org/10.1103/PhysRevD.79.071102}{Phys. Rev. {\bf D79},
  071102 (2009)}, \href{http://arxiv.org/abs/0902.4757}{{\tt arXiv:0902.4757
  [hep-ex]}}\relax
\mciteBstWouldAddEndPuncttrue
\mciteSetBstMidEndSepPunct{\mcitedefaultmidpunct}
{\mcitedefaultendpunct}{\mcitedefaultseppunct}\relax
\EndOfBibitem
\bibitem{Aubert:2009av}
{\babar} collaboration, B.~Aubert {\em et al.},
  \href{http://dx.doi.org/10.1103/PhysRevD.79.072006}{Phys. Rev. {\bf D79},
  072006 (2009)}, \href{http://arxiv.org/abs/0902.2051}{{\tt arXiv:0902.2051
  [hep-ex]}}\relax
\mciteBstWouldAddEndPuncttrue
\mciteSetBstMidEndSepPunct{\mcitedefaultmidpunct}
{\mcitedefaultendpunct}{\mcitedefaultseppunct}\relax
\EndOfBibitem
\bibitem{Gordon:2002yt}
{Belle} collaboration, A.~Gordon {\em et al.},
  \href{http://dx.doi.org/10.1016/S0370-2693(02)02374-2}{Phys. Lett. {\bf
  B542},  183 (2002)}, \href{http://arxiv.org/abs/hep-ex/0207007}{{\tt
  arXiv:hep-ex/0207007 [hep-ex]}}\relax
\mciteBstWouldAddEndPuncttrue
\mciteSetBstMidEndSepPunct{\mcitedefaultmidpunct}
{\mcitedefaultendpunct}{\mcitedefaultseppunct}\relax
\EndOfBibitem
\bibitem{Albrecht:1990am}
{ARGUS} collaboration, H.~Albrecht {\em et al.},
  \href{http://dx.doi.org/10.1016/0370-2693(90)91293-K}{Phys. Lett. {\bf B241},
   278 (1990)}\relax
\mciteBstWouldAddEndPuncttrue
\mciteSetBstMidEndSepPunct{\mcitedefaultmidpunct}
{\mcitedefaultendpunct}{\mcitedefaultseppunct}\relax
\EndOfBibitem
\bibitem{Aubert:2007py}
{\babar} collaboration, B.~Aubert {\em et al.},
  \href{http://dx.doi.org/10.1103/PhysRevD.75.091103}{Phys. Rev. {\bf D75},
  091103 (2007)}, \href{http://arxiv.org/abs/hep-ex/0701035}{{\tt
  arXiv:hep-ex/0701035 [hep-ex]}}\relax
\mciteBstWouldAddEndPuncttrue
\mciteSetBstMidEndSepPunct{\mcitedefaultmidpunct}
{\mcitedefaultendpunct}{\mcitedefaultseppunct}\relax
\EndOfBibitem
\bibitem{Zhang:2004wza}
{Belle} collaboration, J.~Zhang {\em et al.},
  \href{http://dx.doi.org/10.1103/PhysRevLett.94.031801}{Phys. Rev. Lett. {\bf
  94},  031801 (2005)}, \href{http://arxiv.org/abs/hep-ex/0406006}{{\tt
  arXiv:hep-ex/0406006 [hep-ex]}}\relax
\mciteBstWouldAddEndPuncttrue
\mciteSetBstMidEndSepPunct{\mcitedefaultmidpunct}
{\mcitedefaultendpunct}{\mcitedefaultseppunct}\relax
\EndOfBibitem
\bibitem{Zhang:2003up}
{Belle} collaboration, J.~Zhang {\em et al.},
  \href{http://dx.doi.org/10.1103/PhysRevLett.91.221801}{Phys. Rev. Lett. {\bf
  91},  221801 (2003)}, \href{http://arxiv.org/abs/hep-ex/0306007}{{\tt
  arXiv:hep-ex/0306007 [hep-ex]}}\relax
\mciteBstWouldAddEndPuncttrue
\mciteSetBstMidEndSepPunct{\mcitedefaultmidpunct}
{\mcitedefaultendpunct}{\mcitedefaultseppunct}\relax
\EndOfBibitem
\bibitem{Aubert:2007kpb}
{\babar} collaboration, B.~Aubert {\em et al.},
  \href{http://dx.doi.org/10.1103/PhysRevLett.99.261801}{Phys. Rev. Lett. {\bf
  99},  261801 (2007)}, \href{http://arxiv.org/abs/0708.0050}{{\tt
  arXiv:0708.0050 [hep-ex]}}\relax
\mciteBstWouldAddEndPuncttrue
\mciteSetBstMidEndSepPunct{\mcitedefaultmidpunct}
{\mcitedefaultendpunct}{\mcitedefaultseppunct}\relax
\EndOfBibitem
\bibitem{Jen:2006in}
{Belle} collaboration, C.~M. Jen {\em et al.},
  \href{http://dx.doi.org/10.1103/PhysRevD.74.111101}{Phys. Rev. {\bf D74},
  111101 (2006)}, \href{http://arxiv.org/abs/hep-ex/0609022}{{\tt
  arXiv:hep-ex/0609022 [hep-ex]}}\relax
\mciteBstWouldAddEndPuncttrue
\mciteSetBstMidEndSepPunct{\mcitedefaultmidpunct}
{\mcitedefaultendpunct}{\mcitedefaultseppunct}\relax
\EndOfBibitem
\bibitem{Aubert:2008fu}
{\babar} collaboration, B.~Aubert {\em et al.},
  \href{http://dx.doi.org/10.1103/PhysRevD.78.011107}{Phys. Rev. {\bf D78},
  011107 (2008)}, \href{http://arxiv.org/abs/0804.2422}{{\tt arXiv:0804.2422
  [hep-ex]}}\relax
\mciteBstWouldAddEndPuncttrue
\mciteSetBstMidEndSepPunct{\mcitedefaultmidpunct}
{\mcitedefaultendpunct}{\mcitedefaultseppunct}\relax
\EndOfBibitem
\bibitem{Aubert:2006nn}
{\babar} collaboration, B.~Aubert {\em et al.},
  \href{http://dx.doi.org/10.1103/PhysRevD.74.011102}{Phys. Rev. {\bf D74},
  011102 (2006)}, \href{http://arxiv.org/abs/hep-ex/0605037}{{\tt
  arXiv:hep-ex/0605037 [hep-ex]}}\relax
\mciteBstWouldAddEndPuncttrue
\mciteSetBstMidEndSepPunct{\mcitedefaultmidpunct}
{\mcitedefaultendpunct}{\mcitedefaultseppunct}\relax
\EndOfBibitem
\bibitem{Kim:2012gt}
{Belle} collaboration, J.~H. Kim {\em et al.},
  \href{http://dx.doi.org/10.1103/PhysRevD.86.031101}{Phys. Rev. {\bf D86},
  031101 (2012)}, \href{http://arxiv.org/abs/1206.4760}{{\tt arXiv:1206.4760
  [hep-ex]}}\relax
\mciteBstWouldAddEndPuncttrue
\mciteSetBstMidEndSepPunct{\mcitedefaultmidpunct}
{\mcitedefaultendpunct}{\mcitedefaultseppunct}\relax
\EndOfBibitem
\bibitem{Aaij:2013lja}
{LHCb} collaboration, R.~Aaij {\em et al.},
  \href{http://dx.doi.org/10.1016/j.physletb.2013.11.036}{Phys. Lett. {\bf
  B728},  85 (2014)}, \href{http://arxiv.org/abs/1309.3742}{{\tt
  arXiv:1309.3742 [hep-ex]}}\relax
\mciteBstWouldAddEndPuncttrue
\mciteSetBstMidEndSepPunct{\mcitedefaultmidpunct}
{\mcitedefaultendpunct}{\mcitedefaultseppunct}\relax
\EndOfBibitem
\bibitem{Aubert:2008fq}
{\babar} collaboration, B.~Aubert {\em et al.},
  \href{http://dx.doi.org/10.1103/PhysRevLett.101.201801}{Phys. Rev. Lett. {\bf
  101},  201801 (2008)}, \href{http://arxiv.org/abs/0807.3935}{{\tt
  arXiv:0807.3935 [hep-ex]}}\relax
\mciteBstWouldAddEndPuncttrue
\mciteSetBstMidEndSepPunct{\mcitedefaultmidpunct}
{\mcitedefaultendpunct}{\mcitedefaultseppunct}\relax
\EndOfBibitem
\bibitem{Aubert:2007vf}
{\babar} collaboration, B.~Aubert {\em et al.},
  \href{http://dx.doi.org/10.1103/PhysRevD.77.011101}{Phys. Rev. {\bf D77},
  011101 (2008)}, \href{http://arxiv.org/abs/0708.0963}{{\tt arXiv:0708.0963
  [hep-ex]}}\relax
\mciteBstWouldAddEndPuncttrue
\mciteSetBstMidEndSepPunct{\mcitedefaultmidpunct}
{\mcitedefaultendpunct}{\mcitedefaultseppunct}\relax
\EndOfBibitem
\bibitem{Bortoletto:1989mu}
{CLEO} collaboration, D.~Bortoletto {\em et al.},
  \href{http://dx.doi.org/10.1103/PhysRevLett.62.2436}{Phys. Rev. Lett. {\bf
  62},  2436 (1989)}\relax
\mciteBstWouldAddEndPuncttrue
\mciteSetBstMidEndSepPunct{\mcitedefaultmidpunct}
{\mcitedefaultendpunct}{\mcitedefaultseppunct}\relax
\EndOfBibitem
\bibitem{Aubert:2006fha}
{\babar} collaboration, B.~Aubert {\em et al.},
  \href{http://dx.doi.org/10.1103/PhysRevD.75.012008}{Phys. Rev. {\bf D75},
  012008 (2007)}, \href{http://arxiv.org/abs/hep-ex/0608003}{{\tt
  arXiv:hep-ex/0608003 [hep-ex]}}\relax
\mciteBstWouldAddEndPuncttrue
\mciteSetBstMidEndSepPunct{\mcitedefaultmidpunct}
{\mcitedefaultendpunct}{\mcitedefaultseppunct}\relax
\EndOfBibitem
\bibitem{Sato:2014gcp}
{Belle} collaboration, S.~Sato {\em et al.},
  \href{http://dx.doi.org/10.1103/PhysRevD.90.072009}{Phys. Rev. {\bf D90},
  072009 (2014)}, \href{http://arxiv.org/abs/1408.6343}{{\tt arXiv:1408.6343
  [hep-ex]}}\relax
\mciteBstWouldAddEndPuncttrue
\mciteSetBstMidEndSepPunct{\mcitedefaultmidpunct}
{\mcitedefaultendpunct}{\mcitedefaultseppunct}\relax
\EndOfBibitem
\bibitem{Aubert:2007ij}
{\babar} collaboration, B.~Aubert {\em et al.},
  \href{http://dx.doi.org/10.1103/PhysRevD.75.111102}{Phys. Rev. {\bf D75},
  111102 (2007)}, \href{http://arxiv.org/abs/hep-ex/0703038}{{\tt
  arXiv:hep-ex/0703038 [hep-ex]}}\relax
\mciteBstWouldAddEndPuncttrue
\mciteSetBstMidEndSepPunct{\mcitedefaultmidpunct}
{\mcitedefaultendpunct}{\mcitedefaultseppunct}\relax
\EndOfBibitem
\bibitem{Ammar:2001gi}
{CLEO} collaboration, R.~Ammar {\em et al.},
  \href{http://dx.doi.org/10.1103/PhysRevLett.87.271801}{Phys. Rev. Lett. {\bf
  87},  271801 (2001)}, \href{http://arxiv.org/abs/hep-ex/0106038}{{\tt
  arXiv:hep-ex/0106038 [hep-ex]}}\relax
\mciteBstWouldAddEndPuncttrue
\mciteSetBstMidEndSepPunct{\mcitedefaultmidpunct}
{\mcitedefaultendpunct}{\mcitedefaultseppunct}\relax
\EndOfBibitem
\bibitem{Goldenzweig:2008sz}
{Belle} collaboration, P.~Goldenzweig {\em et al.},
  \href{http://dx.doi.org/10.1103/PhysRevLett.101.231801}{Phys. Rev. Lett. {\bf
  101},  231801 (2008)}, \href{http://arxiv.org/abs/0807.4271}{{\tt
  arXiv:0807.4271 [hep-ex]}}\relax
\mciteBstWouldAddEndPuncttrue
\mciteSetBstMidEndSepPunct{\mcitedefaultmidpunct}
{\mcitedefaultendpunct}{\mcitedefaultseppunct}\relax
\EndOfBibitem
\bibitem{BABAR:2011ae}
{\babar} collaboration, J.~P. Lees {\em et al.},
  \href{http://dx.doi.org/10.1103/PhysRevD.83.112010}{Phys. Rev. {\bf D83},
  112010 (2011)}, \href{http://arxiv.org/abs/1105.0125}{{\tt arXiv:1105.0125
  [hep-ex]}}\relax
\mciteBstWouldAddEndPuncttrue
\mciteSetBstMidEndSepPunct{\mcitedefaultmidpunct}
{\mcitedefaultendpunct}{\mcitedefaultseppunct}\relax
\EndOfBibitem
\bibitem{Chang:2004um}
{Belle} collaboration, P.~Chang {\em et al.},
  \href{http://dx.doi.org/10.1016/j.physletb.2004.07.063}{Phys. Lett. {\bf
  B599},  148 (2004)}, \href{http://arxiv.org/abs/hep-ex/0406075}{{\tt
  arXiv:hep-ex/0406075 [hep-ex]}}\relax
\mciteBstWouldAddEndPuncttrue
\mciteSetBstMidEndSepPunct{\mcitedefaultmidpunct}
{\mcitedefaultendpunct}{\mcitedefaultseppunct}\relax
\EndOfBibitem
\bibitem{Aubert:2007bs}
{\babar} collaboration, B.~Aubert {\em et al.},
  \href{http://dx.doi.org/10.1103/PhysRevD.78.052005}{Phys. Rev. {\bf D78},
  052005 (2008)}, \href{http://arxiv.org/abs/0711.4417}{{\tt arXiv:0711.4417
  [hep-ex]}}\relax
\mciteBstWouldAddEndPuncttrue
\mciteSetBstMidEndSepPunct{\mcitedefaultmidpunct}
{\mcitedefaultendpunct}{\mcitedefaultseppunct}\relax
\EndOfBibitem
\bibitem{Garmash:2006fh}
{Belle} collaboration, A.~Garmash {\em et al.},
  \href{http://dx.doi.org/10.1103/PhysRevD.75.012006}{Phys. Rev. {\bf D75},
  012006 (2007)}, \href{http://arxiv.org/abs/hep-ex/0610081}{{\tt
  arXiv:hep-ex/0610081 [hep-ex]}}\relax
\mciteBstWouldAddEndPuncttrue
\mciteSetBstMidEndSepPunct{\mcitedefaultmidpunct}
{\mcitedefaultendpunct}{\mcitedefaultseppunct}\relax
\EndOfBibitem
\bibitem{Aaij:2017zpx}
{LHCb} collaboration, R.~Aaij {\em et al.},
  \href{http://dx.doi.org/10.1007/JHEP11(2017)027}{JHEP {\bf 11},  027 (2017)},
  \href{http://arxiv.org/abs/1707.01665}{{\tt arXiv:1707.01665 [hep-ex]}}\relax
\mciteBstWouldAddEndPuncttrue
\mciteSetBstMidEndSepPunct{\mcitedefaultmidpunct}
{\mcitedefaultendpunct}{\mcitedefaultseppunct}\relax
\EndOfBibitem
\bibitem{Adam:1996ts}
{DELPHI} collaboration, W.~Adam {\em et al.},
  \href{http://dx.doi.org/10.1007/s002880050238}{Z. Phys. {\bf C72},  207--220
  (1996)}\relax
\mciteBstWouldAddEndPuncttrue
\mciteSetBstMidEndSepPunct{\mcitedefaultmidpunct}
{\mcitedefaultendpunct}{\mcitedefaultseppunct}\relax
\EndOfBibitem
\bibitem{Kyeong:2009qx}
{Belle} collaboration, S.~H. Kyeong {\em et al.},
  \href{http://dx.doi.org/10.1103/PhysRevD.80.051103}{Phys. Rev. {\bf D80},
  051103 (2009)}, \href{http://arxiv.org/abs/0905.0763}{{\tt arXiv:0905.0763
  [hep-ex]}}\relax
\mciteBstWouldAddEndPuncttrue
\mciteSetBstMidEndSepPunct{\mcitedefaultmidpunct}
{\mcitedefaultendpunct}{\mcitedefaultseppunct}\relax
\EndOfBibitem
\bibitem{Aubert:2007fm}
{\babar} collaboration, B.~Aubert {\em et al.},
  \href{http://dx.doi.org/10.1103/PhysRevD.76.071104}{Phys. Rev. {\bf D76},
  071104 (2007)}, \href{http://arxiv.org/abs/0708.2543}{{\tt arXiv:0708.2543
  [hep-ex]}}\relax
\mciteBstWouldAddEndPuncttrue
\mciteSetBstMidEndSepPunct{\mcitedefaultmidpunct}
{\mcitedefaultendpunct}{\mcitedefaultseppunct}\relax
\EndOfBibitem
\bibitem{Lees:2011dq}
{\babar} collaboration, J.~P. Lees {\em et al.},
  \href{http://dx.doi.org/10.1103/PhysRevD.85.072005}{Phys. Rev. {\bf D85},
  072005 (2012)}, \href{http://arxiv.org/abs/1112.3896}{{\tt arXiv:1112.3896
  [hep-ex]}}\relax
\mciteBstWouldAddEndPuncttrue
\mciteSetBstMidEndSepPunct{\mcitedefaultmidpunct}
{\mcitedefaultendpunct}{\mcitedefaultseppunct}\relax
\EndOfBibitem
\bibitem{Aaltonen:2011jv}
{CDF} collaboration, T.~Aaltonen {\em et al.},
  \href{http://dx.doi.org/10.1103/PhysRevLett.108.211803}{Phys. Rev. Lett. {\bf
  108},  211803 (2012)}, \href{http://arxiv.org/abs/1111.0485}{{\tt
  arXiv:1111.0485 [hep-ex]}}\relax
\mciteBstWouldAddEndPuncttrue
\mciteSetBstMidEndSepPunct{\mcitedefaultmidpunct}
{\mcitedefaultendpunct}{\mcitedefaultseppunct}\relax
\EndOfBibitem
\bibitem{Aaij:2016elb}
{LHCb} collaboration, R.~Aaij {\em et al.},
  \href{http://dx.doi.org/10.1103/PhysRevLett.118.081801}{Phys. Rev. Lett. {\bf
  118},  081801 (2017)}, \href{http://arxiv.org/abs/1610.08288}{{\tt
  arXiv:1610.08288 [hep-ex]}}\relax
\mciteBstWouldAddEndPuncttrue
\mciteSetBstMidEndSepPunct{\mcitedefaultmidpunct}
{\mcitedefaultendpunct}{\mcitedefaultseppunct}\relax
\EndOfBibitem
\bibitem{delAmoSanchez:2010ur}
{\babar} collaboration, P.~del Amo~Sanchez {\em et al.},
  \href{http://dx.doi.org/10.1103/PhysRevD.82.031101}{Phys. Rev. {\bf D82},
  031101 (2010)}, \href{http://arxiv.org/abs/1003.0640}{{\tt arXiv:1003.0640
  [hep-ex]}}\relax
\mciteBstWouldAddEndPuncttrue
\mciteSetBstMidEndSepPunct{\mcitedefaultmidpunct}
{\mcitedefaultendpunct}{\mcitedefaultseppunct}\relax
\EndOfBibitem
\bibitem{Abdesselam:2018fin}
{Belle} collaboration, A.~Abdesselam {\em et al.},
  \href{http://arxiv.org/abs/1807.06782}{{\tt arXiv:1807.06782 [hep-ex]}}
  (2018)\relax
\mciteBstWouldAddEndPuncttrue
\mciteSetBstMidEndSepPunct{\mcitedefaultmidpunct}
{\mcitedefaultendpunct}{\mcitedefaultseppunct}\relax
\EndOfBibitem
\bibitem{Aaij:2014aaa}
{LHCb} collaboration, R.~Aaij {\em et al.},
  \href{http://dx.doi.org/10.1088/1367-2630/16/12/123001}{New J. Phys. {\bf
  16},  123001 (2014)}, \href{http://arxiv.org/abs/1407.7704}{{\tt
  arXiv:1407.7704 [hep-ex]}}\relax
\mciteBstWouldAddEndPuncttrue
\mciteSetBstMidEndSepPunct{\mcitedefaultmidpunct}
{\mcitedefaultendpunct}{\mcitedefaultseppunct}\relax
\EndOfBibitem
\bibitem{Aubert:2006wu}
{\babar} collaboration, B.~Aubert {\em et al.},
  \href{http://dx.doi.org/10.1103/PhysRevD.74.072008}{Phys. Rev. {\bf D74},
  072008 (2006)}, \href{http://arxiv.org/abs/hep-ex/0606050}{{\tt
  arXiv:hep-ex/0606050 [hep-ex]}}\relax
\mciteBstWouldAddEndPuncttrue
\mciteSetBstMidEndSepPunct{\mcitedefaultmidpunct}
{\mcitedefaultendpunct}{\mcitedefaultseppunct}\relax
\EndOfBibitem
\bibitem{Aaij:2015asa}
{LHCb} collaboration, R.~Aaij {\em et al.},
  \href{http://dx.doi.org/10.1007/JHEP01(2016)012}{JHEP {\bf 01},  012 (2016)},
  \href{http://arxiv.org/abs/1506.08634}{{\tt arXiv:1506.08634 [hep-ex]}}\relax
\mciteBstWouldAddEndPuncttrue
\mciteSetBstMidEndSepPunct{\mcitedefaultmidpunct}
{\mcitedefaultendpunct}{\mcitedefaultseppunct}\relax
\EndOfBibitem
\bibitem{Gaur:2013uou}
{Belle} collaboration, V.~Gaur {\em et al.},
  \href{http://dx.doi.org/10.1103/PhysRevD.87.091101}{Phys. Rev. {\bf D87},
  091101 (2013)}, \href{http://arxiv.org/abs/1304.5312}{{\tt arXiv:1304.5312
  [hep-ex]}}\relax
\mciteBstWouldAddEndPuncttrue
\mciteSetBstMidEndSepPunct{\mcitedefaultmidpunct}
{\mcitedefaultendpunct}{\mcitedefaultseppunct}\relax
\EndOfBibitem
\bibitem{Aubert:2009al}
{\babar} collaboration, B.~Aubert {\em et al.},
  \href{http://dx.doi.org/10.1103/PhysRevD.80.011101}{Phys. Rev. {\bf D80},
  011101 (2009)}, \href{http://arxiv.org/abs/0905.0868}{{\tt arXiv:0905.0868
  [hep-ex]}}\relax
\mciteBstWouldAddEndPuncttrue
\mciteSetBstMidEndSepPunct{\mcitedefaultmidpunct}
{\mcitedefaultendpunct}{\mcitedefaultseppunct}\relax
\EndOfBibitem
\bibitem{Aubert:2006zy}
{\babar} collaboration, B.~Aubert {\em et al.},
  \href{http://dx.doi.org/10.1103/PhysRevD.74.032005}{Phys. Rev. {\bf D74},
  032005 (2006)}, \href{http://arxiv.org/abs/hep-ex/0606031}{{\tt
  arXiv:hep-ex/0606031 [hep-ex]}}\relax
\mciteBstWouldAddEndPuncttrue
\mciteSetBstMidEndSepPunct{\mcitedefaultmidpunct}
{\mcitedefaultendpunct}{\mcitedefaultseppunct}\relax
\EndOfBibitem
\bibitem{Prim:2013nmy}
{Belle} collaboration, M.~Prim {\em et al.},
  \href{http://dx.doi.org/10.1103/PhysRevD.88.072004}{Phys. Rev. {\bf D88},
  072004 (2013)}, \href{http://arxiv.org/abs/1308.1830}{{\tt arXiv:1308.1830
  [hep-ex]}}\relax
\mciteBstWouldAddEndPuncttrue
\mciteSetBstMidEndSepPunct{\mcitedefaultmidpunct}
{\mcitedefaultendpunct}{\mcitedefaultseppunct}\relax
\EndOfBibitem
\bibitem{Chiang:2010ga}
{Belle} collaboration, C.~C. Chiang {\em et al.},
  \href{http://dx.doi.org/10.1103/PhysRevD.81.071101}{Phys. Rev. {\bf D81},
  071101 (2010)}, \href{http://arxiv.org/abs/1001.4595}{{\tt arXiv:1001.4595
  [hep-ex]}}\relax
\mciteBstWouldAddEndPuncttrue
\mciteSetBstMidEndSepPunct{\mcitedefaultmidpunct}
{\mcitedefaultendpunct}{\mcitedefaultseppunct}\relax
\EndOfBibitem
\bibitem{Aubert:2007xc}
{\babar} collaboration, B.~Aubert {\em et al.},
  \href{http://dx.doi.org/10.1103/PhysRevLett.100.081801}{Phys. Rev. Lett. {\bf
  100},  081801 (2008)}, \href{http://arxiv.org/abs/0708.2248}{{\tt
  arXiv:0708.2248 [hep-ex]}}\relax
\mciteBstWouldAddEndPuncttrue
\mciteSetBstMidEndSepPunct{\mcitedefaultmidpunct}
{\mcitedefaultendpunct}{\mcitedefaultseppunct}\relax
\EndOfBibitem
\bibitem{Aubert:2008bb}
{\babar} collaboration, B.~Aubert {\em et al.},
  \href{http://dx.doi.org/10.1103/PhysRevD.78.051103}{Phys. Rev. {\bf D78},
  051103 (2008)}, \href{http://arxiv.org/abs/0806.4467}{{\tt arXiv:0806.4467
  [hep-ex]}}\relax
\mciteBstWouldAddEndPuncttrue
\mciteSetBstMidEndSepPunct{\mcitedefaultmidpunct}
{\mcitedefaultendpunct}{\mcitedefaultseppunct}\relax
\EndOfBibitem
\bibitem{Aubert:2007nw}
{\babar} collaboration, B.~Aubert {\em et al.},
  \href{http://dx.doi.org/10.1103/PhysRevD.76.051103}{Phys. Rev. {\bf D76},
  051103 (2007)}, \href{http://arxiv.org/abs/0705.0398}{{\tt arXiv:0705.0398
  [hep-ex]}}\relax
\mciteBstWouldAddEndPuncttrue
\mciteSetBstMidEndSepPunct{\mcitedefaultmidpunct}
{\mcitedefaultendpunct}{\mcitedefaultseppunct}\relax
\EndOfBibitem
\bibitem{Aaltonen:2011qt}
{CDF} collaboration, T.~Aaltonen {\em et al.},
  \href{http://dx.doi.org/10.1103/PhysRevLett.106.181802}{Phys. Rev. Lett. {\bf
  106},  181802 (2011)}, \href{http://arxiv.org/abs/1103.5762}{{\tt
  arXiv:1103.5762 [hep-ex]}}\relax
\mciteBstWouldAddEndPuncttrue
\mciteSetBstMidEndSepPunct{\mcitedefaultmidpunct}
{\mcitedefaultendpunct}{\mcitedefaultseppunct}\relax
\EndOfBibitem
\bibitem{Aaij:2012as}
{LHCb} collaboration, R.~Aaij {\em et al.},
  \href{http://dx.doi.org/10.1007/JHEP10(2012)037}{JHEP {\bf 10},  037 (2012)},
  \href{http://arxiv.org/abs/1206.2794}{{\tt arXiv:1206.2794 [hep-ex]}}\relax
\mciteBstWouldAddEndPuncttrue
\mciteSetBstMidEndSepPunct{\mcitedefaultmidpunct}
{\mcitedefaultendpunct}{\mcitedefaultseppunct}\relax
\EndOfBibitem
\bibitem{Pal:2015ewa}
{Belle} collaboration, B.~Pal {\em et al.},
  \href{http://dx.doi.org/10.1103/PhysRevD.92.011101}{Phys. Rev. {\bf D92},
  011101 (2015)}, \href{http://arxiv.org/abs/1504.00957}{{\tt arXiv:1504.00957
  [hep-ex]}}\relax
\mciteBstWouldAddEndPuncttrue
\mciteSetBstMidEndSepPunct{\mcitedefaultmidpunct}
{\mcitedefaultendpunct}{\mcitedefaultseppunct}\relax
\EndOfBibitem
\bibitem{Abdesselam:2016tpr}
{Belle} collaboration, A.~Abdesselam {\em et al.},
  \href{http://arxiv.org/abs/1609.03267}{{\tt arXiv:1609.03267 [hep-ex]}}
  (2016)\relax
\mciteBstWouldAddEndPuncttrue
\mciteSetBstMidEndSepPunct{\mcitedefaultmidpunct}
{\mcitedefaultendpunct}{\mcitedefaultseppunct}\relax
\EndOfBibitem
\bibitem{Lees:2013yea}
{\babar} collaboration, J.~P. Lees {\em et al.},
  \href{http://dx.doi.org/10.1103/PhysRevD.89.051101}{Phys. Rev. {\bf D89},
  051101 (2014)}, \href{http://arxiv.org/abs/1312.0056}{{\tt arXiv:1312.0056
  [hep-ex]}}\relax
\mciteBstWouldAddEndPuncttrue
\mciteSetBstMidEndSepPunct{\mcitedefaultmidpunct}
{\mcitedefaultendpunct}{\mcitedefaultseppunct}\relax
\EndOfBibitem
\bibitem{Aaij:2016qnm}
{LHCb} collaboration, R.~Aaij {\em et al.},
  \href{http://dx.doi.org/10.1103/PhysRevD.95.012006}{Phys. Rev. {\bf D95},
  012006 (2017)}, \href{http://arxiv.org/abs/1610.05187}{{\tt arXiv:1610.05187
  [hep-ex]}}\relax
\mciteBstWouldAddEndPuncttrue
\mciteSetBstMidEndSepPunct{\mcitedefaultmidpunct}
{\mcitedefaultendpunct}{\mcitedefaultseppunct}\relax
\EndOfBibitem
\bibitem{Aubert:2003fm}
{\babar} collaboration, B.~Aubert {\em et al.},
  \href{http://dx.doi.org/10.1103/PhysRevLett.93.051802}{Phys. Rev. Lett. {\bf
  93},  051802 (2004)}, \href{http://arxiv.org/abs/hep-ex/0311049}{{\tt
  arXiv:hep-ex/0311049 [hep-ex]}}\relax
\mciteBstWouldAddEndPuncttrue
\mciteSetBstMidEndSepPunct{\mcitedefaultmidpunct}
{\mcitedefaultendpunct}{\mcitedefaultseppunct}\relax
\EndOfBibitem
\bibitem{Aubert:2006dd}
{\babar} collaboration, B.~Aubert {\em et al.},
  \href{http://dx.doi.org/10.1103/PhysRevLett.97.051802}{Phys. Rev. Lett. {\bf
  97},  051802 (2006)}, \href{http://arxiv.org/abs/hep-ex/0603050}{{\tt
  arXiv:hep-ex/0603050 [hep-ex]}}\relax
\mciteBstWouldAddEndPuncttrue
\mciteSetBstMidEndSepPunct{\mcitedefaultmidpunct}
{\mcitedefaultendpunct}{\mcitedefaultseppunct}\relax
\EndOfBibitem
\bibitem{Aubert:2006sw}
{\babar} collaboration, B.~Aubert {\em et al.},
  \href{http://dx.doi.org/10.1103/PhysRevD.74.031104}{Phys. Rev. {\bf D74},
  031104 (2006)}, \href{http://arxiv.org/abs/hep-ex/0605024}{{\tt
  arXiv:hep-ex/0605024 [hep-ex]}}\relax
\mciteBstWouldAddEndPuncttrue
\mciteSetBstMidEndSepPunct{\mcitedefaultmidpunct}
{\mcitedefaultendpunct}{\mcitedefaultseppunct}\relax
\EndOfBibitem
\bibitem{Aubert:2009zr}
{\babar} collaboration, B.~Aubert {\em et al.},
  \href{http://dx.doi.org/10.1103/PhysRevD.80.092007}{Phys. Rev. {\bf D80},
  092007 (2009)}, \href{http://arxiv.org/abs/0907.1776}{{\tt arXiv:0907.1776
  [hep-ex]}}\relax
\mciteBstWouldAddEndPuncttrue
\mciteSetBstMidEndSepPunct{\mcitedefaultmidpunct}
{\mcitedefaultendpunct}{\mcitedefaultseppunct}\relax
\EndOfBibitem
\bibitem{Wei:2007fg}
{Belle} collaboration, J.~T. Wei {\em et al.},
  \href{http://dx.doi.org/10.1016/j.physletb.2007.11.063}{Phys. Lett. {\bf
  B659},  80 (2008)}, \href{http://arxiv.org/abs/0706.4167}{{\tt
  arXiv:0706.4167 [hep-ex]}}\relax
\mciteBstWouldAddEndPuncttrue
\mciteSetBstMidEndSepPunct{\mcitedefaultmidpunct}
{\mcitedefaultendpunct}{\mcitedefaultseppunct}\relax
\EndOfBibitem
\bibitem{Aaij:2014tua}
{LHCb} collaboration, R.~Aaij {\em et al.},
  \href{http://dx.doi.org/10.1103/PhysRevLett.113.141801}{Phys. Rev. Lett. {\bf
  113},  141801 (2014)}, \href{http://arxiv.org/abs/1407.5907}{{\tt
  arXiv:1407.5907 [hep-ex]}}\relax
\mciteBstWouldAddEndPuncttrue
\mciteSetBstMidEndSepPunct{\mcitedefaultmidpunct}
{\mcitedefaultendpunct}{\mcitedefaultseppunct}\relax
\EndOfBibitem
\bibitem{Wang:2005fc}
{Belle} collaboration, M.~Z. Wang {\em et al.},
  \href{http://dx.doi.org/10.1016/j.physletb.2005.05.008}{Phys. Lett. {\bf
  B617},  141 (2005)}, \href{http://arxiv.org/abs/hep-ex/0503047}{{\tt
  arXiv:hep-ex/0503047 [hep-ex]}}\relax
\mciteBstWouldAddEndPuncttrue
\mciteSetBstMidEndSepPunct{\mcitedefaultmidpunct}
{\mcitedefaultendpunct}{\mcitedefaultseppunct}\relax
\EndOfBibitem
\bibitem{Chen:2008jy}
{Belle} collaboration, J.~H. Chen {\em et al.},
  \href{http://dx.doi.org/10.1103/PhysRevLett.100.251801}{Phys. Rev. Lett. {\bf
  100},  251801 (2008)}, \href{http://arxiv.org/abs/0802.0336}{{\tt
  arXiv:0802.0336 [hep-ex]}}\relax
\mciteBstWouldAddEndPuncttrue
\mciteSetBstMidEndSepPunct{\mcitedefaultmidpunct}
{\mcitedefaultendpunct}{\mcitedefaultseppunct}\relax
\EndOfBibitem
\bibitem{Tsai:2007pp}
{Belle} collaboration, Y.~T. Tsai {\em et al.},
  \href{http://dx.doi.org/10.1103/PhysRevD.75.111101}{Phys. Rev. {\bf D75},
  111101 (2007)}, \href{http://arxiv.org/abs/hep-ex/0703048}{{\tt
  arXiv:hep-ex/0703048 [hep-ex]}}\relax
\mciteBstWouldAddEndPuncttrue
\mciteSetBstMidEndSepPunct{\mcitedefaultmidpunct}
{\mcitedefaultendpunct}{\mcitedefaultseppunct}\relax
\EndOfBibitem
\bibitem{Aaij:2016xfa}
{LHCb} collaboration, R.~Aaij {\em et al.},
  \href{http://dx.doi.org/10.1007/JHEP04(2017)162}{JHEP {\bf 04},  162 (2017)},
  \href{http://arxiv.org/abs/1611.07805}{{\tt arXiv:1611.07805 [hep-ex]}}\relax
\mciteBstWouldAddEndPuncttrue
\mciteSetBstMidEndSepPunct{\mcitedefaultmidpunct}
{\mcitedefaultendpunct}{\mcitedefaultseppunct}\relax
\EndOfBibitem
\bibitem{Wang:2007as}
{Belle} collaboration, M.~Z. Wang {\em et al.},
  \href{http://dx.doi.org/10.1103/PhysRevD.76.052004}{Phys. Rev. {\bf D76},
  052004 (2007)}, \href{http://arxiv.org/abs/0704.2672}{{\tt arXiv:0704.2672
  [hep-ex]}}\relax
\mciteBstWouldAddEndPuncttrue
\mciteSetBstMidEndSepPunct{\mcitedefaultmidpunct}
{\mcitedefaultendpunct}{\mcitedefaultseppunct}\relax
\EndOfBibitem
\bibitem{Chen:2009xg}
{Belle} collaboration, P.~Chen {\em et al.},
  \href{http://dx.doi.org/10.1103/PhysRevD.80.111103}{Phys. Rev. {\bf D80},
  111103 (2009)}, \href{http://arxiv.org/abs/0910.5817}{{\tt arXiv:0910.5817
  [hep-ex]}}\relax
\mciteBstWouldAddEndPuncttrue
\mciteSetBstMidEndSepPunct{\mcitedefaultmidpunct}
{\mcitedefaultendpunct}{\mcitedefaultseppunct}\relax
\EndOfBibitem
\bibitem{Lu:2018qbw}
{Belle} collaboration, P.~C. Lu {\em et al.},
  \href{http://dx.doi.org/10.1103/PhysRevD.99.032003}{Phys. Rev. {\bf D99},
  032003 (2019)}, \href{http://arxiv.org/abs/1807.10503}{{\tt arXiv:1807.10503
  [hep-ex]}}\relax
\mciteBstWouldAddEndPuncttrue
\mciteSetBstMidEndSepPunct{\mcitedefaultmidpunct}
{\mcitedefaultendpunct}{\mcitedefaultseppunct}\relax
\EndOfBibitem
\bibitem{Aubert:2004fy}
{\babar} collaboration, B.~Aubert {\em et al.},
  \href{http://dx.doi.org/10.1103/PhysRevD.69.091503}{Phys. Rev. {\bf D69},
  091503 (2004)}, \href{http://arxiv.org/abs/hep-ex/0403003}{{\tt
  arXiv:hep-ex/0403003 [hep-ex]}}\relax
\mciteBstWouldAddEndPuncttrue
\mciteSetBstMidEndSepPunct{\mcitedefaultmidpunct}
{\mcitedefaultendpunct}{\mcitedefaultseppunct}\relax
\EndOfBibitem
\bibitem{Aaij:2017gum}
{LHCb} collaboration, R.~Aaij {\em et al.},
  \href{http://dx.doi.org/10.1103/PhysRevLett.119.232001}{Phys. Rev. Lett. {\bf
  119},  232001 (2017)}, \href{http://arxiv.org/abs/1709.01156}{{\tt
  arXiv:1709.01156 [hep-ex]}}\relax
\mciteBstWouldAddEndPuncttrue
\mciteSetBstMidEndSepPunct{\mcitedefaultmidpunct}
{\mcitedefaultendpunct}{\mcitedefaultseppunct}\relax
\EndOfBibitem
\bibitem{Aaij:2017pgn}
{LHCb} collaboration, R.~Aaij {\em et al.},
  \href{http://dx.doi.org/10.1103/PhysRevD.96.051103}{Phys. Rev. {\bf D96},
  051103 (2017)}, \href{http://arxiv.org/abs/1704.08497}{{\tt arXiv:1704.08497
  [hep-ex]}}\relax
\mciteBstWouldAddEndPuncttrue
\mciteSetBstMidEndSepPunct{\mcitedefaultmidpunct}
{\mcitedefaultendpunct}{\mcitedefaultseppunct}\relax
\EndOfBibitem
\bibitem{Aubert:2009am}
{\babar} collaboration, B.~Aubert {\em et al.},
  \href{http://dx.doi.org/10.1103/PhysRevD.79.112009}{Phys. Rev. {\bf D79},
  112009 (2009)}, \href{http://arxiv.org/abs/0904.4724}{{\tt arXiv:0904.4724
  [hep-ex]}}\relax
\mciteBstWouldAddEndPuncttrue
\mciteSetBstMidEndSepPunct{\mcitedefaultmidpunct}
{\mcitedefaultendpunct}{\mcitedefaultseppunct}\relax
\EndOfBibitem
\bibitem{Wang:2003yi}
{Belle} collaboration, M.~Z. Wang {\em et al.},
  \href{http://dx.doi.org/10.1103/PhysRevLett.90.201802}{Phys. Rev. Lett. {\bf
  90},  201802 (2003)}, \href{http://arxiv.org/abs/hep-ex/0302024}{{\tt
  arXiv:hep-ex/0302024 [hep-ex]}}\relax
\mciteBstWouldAddEndPuncttrue
\mciteSetBstMidEndSepPunct{\mcitedefaultmidpunct}
{\mcitedefaultendpunct}{\mcitedefaultseppunct}\relax
\EndOfBibitem
\bibitem{BABAR:2018erd}
{BaBar} collaboration, J.~P. Lees {\em et al.},
  \href{http://dx.doi.org/10.1103/PhysRevD.98.071102}{Phys. Rev. {\bf D98},
  071102 (2018)}, \href{http://arxiv.org/abs/1803.10378}{{\tt arXiv:1803.10378
  [hep-ex]}}\relax
\mciteBstWouldAddEndPuncttrue
\mciteSetBstMidEndSepPunct{\mcitedefaultmidpunct}
{\mcitedefaultendpunct}{\mcitedefaultseppunct}\relax
\EndOfBibitem
\bibitem{Aaij:2014lpa}
{LHCb} collaboration, R.~Aaij {\em et al.},
  \href{http://dx.doi.org/10.1007/JHEP04(2014)087}{JHEP {\bf 04},  087 (2014)},
  \href{http://arxiv.org/abs/1402.0770}{{\tt arXiv:1402.0770 [hep-ex]}}\relax
\mciteBstWouldAddEndPuncttrue
\mciteSetBstMidEndSepPunct{\mcitedefaultmidpunct}
{\mcitedefaultendpunct}{\mcitedefaultseppunct}\relax
\EndOfBibitem
\bibitem{Aaltonen:2008hg}
{CDF} collaboration, T.~Aaltonen {\em et al.},
  \href{http://dx.doi.org/10.1103/PhysRevLett.103.031801}{Phys. Rev. Lett. {\bf
  103},  031801 (2009)}, \href{http://arxiv.org/abs/0812.4271}{{\tt
  arXiv:0812.4271 [hep-ex]}}\relax
\mciteBstWouldAddEndPuncttrue
\mciteSetBstMidEndSepPunct{\mcitedefaultmidpunct}
{\mcitedefaultendpunct}{\mcitedefaultseppunct}\relax
\EndOfBibitem
\bibitem{Aaltonen:2011qs}
{CDF} collaboration, T.~Aaltonen {\em et al.},
  \href{http://dx.doi.org/10.1103/PhysRevLett.107.201802}{Phys. Rev. Lett. {\bf
  107},  201802 (2011)}, \href{http://arxiv.org/abs/1107.3753}{{\tt
  arXiv:1107.3753 [hep-ex]}}\relax
\mciteBstWouldAddEndPuncttrue
\mciteSetBstMidEndSepPunct{\mcitedefaultmidpunct}
{\mcitedefaultendpunct}{\mcitedefaultseppunct}\relax
\EndOfBibitem
\bibitem{Aaij:2013mna}
{LHCb} collaboration, R.~Aaij {\em et al.},
  \href{http://dx.doi.org/10.1016/j.physletb.2013.06.060}{Phys. Lett. {\bf
  B725},  25 (2013)}, \href{http://arxiv.org/abs/1306.2577}{{\tt
  arXiv:1306.2577 [hep-ex]}}\relax
\mciteBstWouldAddEndPuncttrue
\mciteSetBstMidEndSepPunct{\mcitedefaultmidpunct}
{\mcitedefaultendpunct}{\mcitedefaultseppunct}\relax
\EndOfBibitem
\bibitem{Acosta:2002fh}
{CDF} collaboration, D.~Acosta {\em et al.},
  \href{http://dx.doi.org/10.1103/PhysRevD.66.112002}{Phys. Rev. {\bf D66},
  112002 (2002)}, \href{http://arxiv.org/abs/hep-ex/0208035}{{\tt
  arXiv:hep-ex/0208035 [hep-ex]}}\relax
\mciteBstWouldAddEndPuncttrue
\mciteSetBstMidEndSepPunct{\mcitedefaultmidpunct}
{\mcitedefaultendpunct}{\mcitedefaultseppunct}\relax
\EndOfBibitem
\bibitem{Aaij:2015eqa}
{LHCb} collaboration, R.~Aaij {\em et al.},
  \href{http://dx.doi.org/10.1007/JHEP09(2015)006}{JHEP {\bf 09},  006 (2015)},
  \href{http://arxiv.org/abs/1505.03295}{{\tt arXiv:1505.03295 [hep-ex]}}\relax
\mciteBstWouldAddEndPuncttrue
\mciteSetBstMidEndSepPunct{\mcitedefaultmidpunct}
{\mcitedefaultendpunct}{\mcitedefaultseppunct}\relax
\EndOfBibitem
\bibitem{Aaij:2016nrq}
{LHCb} collaboration, R.~Aaij {\em et al.},
  \href{http://dx.doi.org/10.1007/JHEP05(2016)081}{JHEP {\bf 05},  081 (2016)},
  \href{http://arxiv.org/abs/1603.00413}{{\tt arXiv:1603.00413 [hep-ex]}}\relax
\mciteBstWouldAddEndPuncttrue
\mciteSetBstMidEndSepPunct{\mcitedefaultmidpunct}
{\mcitedefaultendpunct}{\mcitedefaultseppunct}\relax
\EndOfBibitem
\bibitem{Aaij:2016zhm}
{LHCb} collaboration, R.~Aaij {\em et al.},
  \href{http://dx.doi.org/10.1016/j.physletb.2016.05.077}{Phys. Lett. {\bf
  B759},  282 (2016)}, \href{http://arxiv.org/abs/1603.02870}{{\tt
  arXiv:1603.02870 [hep-ex]}}\relax
\mciteBstWouldAddEndPuncttrue
\mciteSetBstMidEndSepPunct{\mcitedefaultmidpunct}
{\mcitedefaultendpunct}{\mcitedefaultseppunct}\relax
\EndOfBibitem
\bibitem{Aaij:2017ewm}
{LHCb} collaboration, R.~Aaij {\em et al.},
  \href{http://dx.doi.org/10.1007/JHEP04(2017)029}{JHEP {\bf 04},  029 (2017)},
  \href{http://arxiv.org/abs/1701.08705}{{\tt arXiv:1701.08705 [hep-ex]}}\relax
\mciteBstWouldAddEndPuncttrue
\mciteSetBstMidEndSepPunct{\mcitedefaultmidpunct}
{\mcitedefaultendpunct}{\mcitedefaultseppunct}\relax
\EndOfBibitem
\bibitem{Aaij:2017pgy}
{LHCb} collaboration, R.~Aaij {\em et al.},
  \href{http://dx.doi.org/10.1007/JHEP02(2018)098}{JHEP {\bf 02},  098 (2018)},
  \href{http://arxiv.org/abs/1711.05490}{{\tt arXiv:1711.05490 [hep-ex]}}\relax
\mciteBstWouldAddEndPuncttrue
\mciteSetBstMidEndSepPunct{\mcitedefaultmidpunct}
{\mcitedefaultendpunct}{\mcitedefaultseppunct}\relax
\EndOfBibitem
\bibitem{Aaij:2015xza}
{LHCb} collaboration, R.~Aaij {\em et al.},
  \href{http://dx.doi.org/10.1007/JHEP06(2015)115}{JHEP {\bf 06},  115 (2015)},
  \href{http://arxiv.org/abs/1503.07138}{{\tt arXiv:1503.07138 [hep-ex]}}\relax
\mciteBstWouldAddEndPuncttrue
\mciteSetBstMidEndSepPunct{\mcitedefaultmidpunct}
{\mcitedefaultendpunct}{\mcitedefaultseppunct}\relax
\EndOfBibitem
\bibitem{Aaij:2016zab}
{LHCb} collaboration, R.~Aaij {\em et al.},
  \href{http://dx.doi.org/10.1103/PhysRevLett.118.071801}{Phys. Rev. Lett. {\bf
  118},  071801 (2017)}, \href{http://arxiv.org/abs/1612.02244}{{\tt
  arXiv:1612.02244 [hep-ex]}}\relax
\mciteBstWouldAddEndPuncttrue
\mciteSetBstMidEndSepPunct{\mcitedefaultmidpunct}
{\mcitedefaultendpunct}{\mcitedefaultseppunct}\relax
\EndOfBibitem
\bibitem{Aaij:2018gwm}
{LHCb} collaboration, R.~Aaij {\em et al.},
  \href{http://dx.doi.org/10.1007/JHEP09(2018)146}{JHEP {\bf 09},  146 (2018)},
  \href{http://arxiv.org/abs/1808.00264}{{\tt arXiv:1808.00264 [hep-ex]}}\relax
\mciteBstWouldAddEndPuncttrue
\mciteSetBstMidEndSepPunct{\mcitedefaultmidpunct}
{\mcitedefaultendpunct}{\mcitedefaultseppunct}\relax
\EndOfBibitem
\bibitem{Aaij:2014yka}
{LHCb} collaboration, R.~Aaij {\em et al.},
  \href{http://dx.doi.org/10.1103/PhysRevLett.114.062004}{Phys. Rev. Lett. {\bf
  114},  062004 (2015)}, \href{http://arxiv.org/abs/1411.4849}{{\tt
  arXiv:1411.4849 [hep-ex]}}\relax
\mciteBstWouldAddEndPuncttrue
\mciteSetBstMidEndSepPunct{\mcitedefaultmidpunct}
{\mcitedefaultendpunct}{\mcitedefaultseppunct}\relax
\EndOfBibitem
\bibitem{Aaij:2016jnn}
{LHCb} collaboration, R.~Aaij {\em et al.},
  \href{http://dx.doi.org/10.1007/JHEP05(2016)161}{JHEP {\bf 05},  161 (2016)},
  \href{http://arxiv.org/abs/1604.03896}{{\tt arXiv:1604.03896 [hep-ex]}}\relax
\mciteBstWouldAddEndPuncttrue
\mciteSetBstMidEndSepPunct{\mcitedefaultmidpunct}
{\mcitedefaultendpunct}{\mcitedefaultseppunct}\relax
\EndOfBibitem
\bibitem{Aaij:2017inn}
{LHCb} collaboration, R.~Aaij {\em et al.},
  \href{http://arxiv.org/abs/1708.05808}{{\tt arXiv:1708.05808 [hep-ex]}}
  (2017)\relax
\mciteBstWouldAddEndPuncttrue
\mciteSetBstMidEndSepPunct{\mcitedefaultmidpunct}
{\mcitedefaultendpunct}{\mcitedefaultseppunct}\relax
\EndOfBibitem
\bibitem{Peng:2010ze}
{Belle} collaboration, C.~C. Peng {\em et al.},
  \href{http://dx.doi.org/10.1103/PhysRevD.82.072007}{Phys. Rev. {\bf D82},
  072007 (2010)}, \href{http://arxiv.org/abs/1006.5115}{{\tt arXiv:1006.5115
  [hep-ex]}}\relax
\mciteBstWouldAddEndPuncttrue
\mciteSetBstMidEndSepPunct{\mcitedefaultmidpunct}
{\mcitedefaultendpunct}{\mcitedefaultseppunct}\relax
\EndOfBibitem
\bibitem{Aaij:2015qga}
{LHCb} collaboration, R.~Aaij {\em et al.},
  \href{http://dx.doi.org/10.1103/PhysRevLett.115.051801}{Phys. Rev. Lett. {\bf
  115},  051801 (2015)}, \href{http://arxiv.org/abs/1503.07483}{{\tt
  arXiv:1503.07483 [hep-ex]}}\relax
\mciteBstWouldAddEndPuncttrue
\mciteSetBstMidEndSepPunct{\mcitedefaultmidpunct}
{\mcitedefaultendpunct}{\mcitedefaultseppunct}\relax
\EndOfBibitem
\bibitem{Aaltonen:2011rs}
{CDF} collaboration, T.~Aaltonen {\em et al.},
  \href{http://dx.doi.org/10.1103/PhysRevLett.107.261802}{Phys. Rev. Lett. {\bf
  107},  261802 (2011)}, \href{http://arxiv.org/abs/1107.4999}{{\tt
  arXiv:1107.4999 [hep-ex]}}\relax
\mciteBstWouldAddEndPuncttrue
\mciteSetBstMidEndSepPunct{\mcitedefaultmidpunct}
{\mcitedefaultendpunct}{\mcitedefaultseppunct}\relax
\EndOfBibitem
\bibitem{Aaij:2015cxj}
{LHCb} collaboration, R.~Aaij {\em et al.},
  \href{http://dx.doi.org/10.1007/JHEP10(2015)053}{JHEP {\bf 10},  053 (2015)},
  \href{http://arxiv.org/abs/1508.00788}{{\tt arXiv:1508.00788 [hep-ex]}}\relax
\mciteBstWouldAddEndPuncttrue
\mciteSetBstMidEndSepPunct{\mcitedefaultmidpunct}
{\mcitedefaultendpunct}{\mcitedefaultseppunct}\relax
\EndOfBibitem
\bibitem{Pal:2015ghq}
{Belle} collaboration, B.~Pal {\em et al.},
  \href{http://dx.doi.org/10.1103/PhysRevLett.116.161801}{Phys. Rev. Lett. {\bf
  116},  161801 (2016)}, \href{http://arxiv.org/abs/1512.02145}{{\tt
  arXiv:1512.02145 [hep-ex]}}\relax
\mciteBstWouldAddEndPuncttrue
\mciteSetBstMidEndSepPunct{\mcitedefaultmidpunct}
{\mcitedefaultendpunct}{\mcitedefaultseppunct}\relax
\EndOfBibitem
\bibitem{Aaij:2015kba}
{LHCb} collaboration, R.~Aaij {\em et al.},
  \href{http://dx.doi.org/10.1007/JHEP07(2015)166}{JHEP {\bf 07},  166 (2015)},
  \href{http://arxiv.org/abs/1503.05362}{{\tt arXiv:1503.05362 [hep-ex]}}\relax
\mciteBstWouldAddEndPuncttrue
\mciteSetBstMidEndSepPunct{\mcitedefaultmidpunct}
{\mcitedefaultendpunct}{\mcitedefaultseppunct}\relax
\EndOfBibitem
\bibitem{Aaij:2013gga}
{LHCb} collaboration, R.~Aaij {\em et al.},
  \href{http://dx.doi.org/10.1007/JHEP11(2013)092}{JHEP {\bf 11},  092 (2013)},
  \href{http://arxiv.org/abs/1306.2239}{{\tt arXiv:1306.2239 [hep-ex]}}\relax
\mciteBstWouldAddEndPuncttrue
\mciteSetBstMidEndSepPunct{\mcitedefaultmidpunct}
{\mcitedefaultendpunct}{\mcitedefaultseppunct}\relax
\EndOfBibitem
\bibitem{Dutta:2014sxo}
{Belle} collaboration, D.~Dutta {\em et al.},
  \href{http://dx.doi.org/10.1103/PhysRevD.91.011101}{Phys. Rev. {\bf D91},
  011101 (2015)}, \href{http://arxiv.org/abs/1411.7771}{{\tt arXiv:1411.7771
  [hep-ex]}}\relax
\mciteBstWouldAddEndPuncttrue
\mciteSetBstMidEndSepPunct{\mcitedefaultmidpunct}
{\mcitedefaultendpunct}{\mcitedefaultseppunct}\relax
\EndOfBibitem
\bibitem{Aaij:2012ita}
{LHCb} collaboration, R.~Aaij {\em et al.},
  \href{http://dx.doi.org/10.1016/j.nuclphysb.2012.09.013}{Nucl. Phys. {\bf
  B867},  1 (2013)}, \href{http://arxiv.org/abs/1209.0313}{{\tt arXiv:1209.0313
  [hep-ex]}}\relax
\mciteBstWouldAddEndPuncttrue
\mciteSetBstMidEndSepPunct{\mcitedefaultmidpunct}
{\mcitedefaultendpunct}{\mcitedefaultseppunct}\relax
\EndOfBibitem
\bibitem{Aaltonen:2013as}
{CDF} collaboration, T.~Aaltonen {\em et al.},
  \href{http://dx.doi.org/10.1103/PhysRevD.87.072003}{Phys. Rev. {\bf D87},
  072003 (2013)}, \href{http://arxiv.org/abs/1301.7048}{{\tt arXiv:1301.7048
  [hep-ex]}}\relax
\mciteBstWouldAddEndPuncttrue
\mciteSetBstMidEndSepPunct{\mcitedefaultmidpunct}
{\mcitedefaultendpunct}{\mcitedefaultseppunct}\relax
\EndOfBibitem
\bibitem{Abazov:2013wjb}
{D0} collaboration, V.~M. Abazov {\em et al.},
  \href{http://dx.doi.org/10.1103/PhysRevD.87.072006}{Phys. Rev. {\bf D87},
  072006 (2013)}, \href{http://arxiv.org/abs/1301.4507}{{\tt arXiv:1301.4507
  [hep-ex]}}\relax
\mciteBstWouldAddEndPuncttrue
\mciteSetBstMidEndSepPunct{\mcitedefaultmidpunct}
{\mcitedefaultendpunct}{\mcitedefaultseppunct}\relax
\EndOfBibitem
\bibitem{CMS:2014xfa}
{CMS and LHCb} collaborations, V.~Khachatryan {\em et al.},
  \href{http://dx.doi.org/10.1038/nature14474}{Nature {\bf 522},  68 (2015)},
  \href{http://arxiv.org/abs/1411.4413}{{\tt arXiv:1411.4413 [hep-ex]}}\relax
\mciteBstWouldAddEndPuncttrue
\mciteSetBstMidEndSepPunct{\mcitedefaultmidpunct}
{\mcitedefaultendpunct}{\mcitedefaultseppunct}\relax
\EndOfBibitem
\bibitem{Aaboud:2018mst}
{ATLAS} collaboration, M.~Aaboud {\em et al.}, Submitted to: JHEP   (2018),
  \href{http://arxiv.org/abs/1812.03017}{{\tt arXiv:1812.03017 [hep-ex]}}\relax
\mciteBstWouldAddEndPuncttrue
\mciteSetBstMidEndSepPunct{\mcitedefaultmidpunct}
{\mcitedefaultendpunct}{\mcitedefaultseppunct}\relax
\EndOfBibitem
\bibitem{Aaltonen:2009vr}
{CDF} collaboration, T.~Aaltonen {\em et al.},
  \href{http://dx.doi.org/10.1103/PhysRevLett.102.201801}{Phys. Rev. Lett. {\bf
  102},  201801 (2009)}, \href{http://arxiv.org/abs/0901.3803}{{\tt
  arXiv:0901.3803 [hep-ex]}}\relax
\mciteBstWouldAddEndPuncttrue
\mciteSetBstMidEndSepPunct{\mcitedefaultmidpunct}
{\mcitedefaultendpunct}{\mcitedefaultseppunct}\relax
\EndOfBibitem
\bibitem{Aaij:2017xqt}
{LHCb} collaboration, R.~Aaij {\em et al.},
  \href{http://dx.doi.org/10.1103/PhysRevLett.118.251802}{Phys. Rev. Lett. {\bf
  118},  251802 (2017)}, \href{http://arxiv.org/abs/1703.02508}{{\tt
  arXiv:1703.02508 [hep-ex]}}\relax
\mciteBstWouldAddEndPuncttrue
\mciteSetBstMidEndSepPunct{\mcitedefaultmidpunct}
{\mcitedefaultendpunct}{\mcitedefaultseppunct}\relax
\EndOfBibitem
\bibitem{Aaij:2016kfs}
{LHCb} collaboration, R.~Aaij {\em et al.},
  \href{http://dx.doi.org/10.1007/JHEP03(2017)001}{JHEP {\bf 03},  001 (2017)},
  \href{http://arxiv.org/abs/1611.07704}{{\tt arXiv:1611.07704 [hep-ex]}}\relax
\mciteBstWouldAddEndPuncttrue
\mciteSetBstMidEndSepPunct{\mcitedefaultmidpunct}
{\mcitedefaultendpunct}{\mcitedefaultseppunct}\relax
\EndOfBibitem
\bibitem{Abazov:2006qm}
{D0} collaboration, V.~M. Abazov {\em et al.},
  \href{http://dx.doi.org/10.1103/PhysRevD.74.031107}{Phys. Rev. {\bf D74},
  031107 (2006)}, \href{http://arxiv.org/abs/hep-ex/0604015}{{\tt
  arXiv:hep-ex/0604015 [hep-ex]}}\relax
\mciteBstWouldAddEndPuncttrue
\mciteSetBstMidEndSepPunct{\mcitedefaultmidpunct}
{\mcitedefaultendpunct}{\mcitedefaultseppunct}\relax
\EndOfBibitem
\bibitem{Aaij:2015esa}
{LHCb} collaboration, R.~Aaij {\em et al.},
  \href{http://dx.doi.org/10.1007/JHEP09(2015)179}{JHEP {\bf 09},  179 (2015)},
  \href{http://arxiv.org/abs/1506.08777}{{\tt arXiv:1506.08777 [hep-ex]}}\relax
\mciteBstWouldAddEndPuncttrue
\mciteSetBstMidEndSepPunct{\mcitedefaultmidpunct}
{\mcitedefaultendpunct}{\mcitedefaultseppunct}\relax
\EndOfBibitem
\bibitem{Aaij:2014lba}
{LHCb} collaboration, R.~Aaij {\em et al.},
  \href{http://dx.doi.org/10.1016/j.physletb.2015.02.010}{Phys. Lett. {\bf
  B743},  46 (2015)}, \href{http://arxiv.org/abs/1412.6433}{{\tt
  arXiv:1412.6433 [hep-ex]}}\relax
\mciteBstWouldAddEndPuncttrue
\mciteSetBstMidEndSepPunct{\mcitedefaultmidpunct}
{\mcitedefaultendpunct}{\mcitedefaultseppunct}\relax
\EndOfBibitem
\bibitem{Aaij:2017cza}
{LHCb} collaboration, R.~Aaij {\em et al.},
  \href{http://dx.doi.org/10.1007/JHEP03(2018)078}{JHEP {\bf 03},  078 (2018)},
  \href{http://arxiv.org/abs/1710.04111}{{\tt arXiv:1710.04111 [hep-ex]}}\relax
\mciteBstWouldAddEndPuncttrue
\mciteSetBstMidEndSepPunct{\mcitedefaultmidpunct}
{\mcitedefaultendpunct}{\mcitedefaultseppunct}\relax
\EndOfBibitem
\bibitem{Aaij:2017vnw}
{LHCb} collaboration, R.~Aaij {\em et al.},
  \href{http://dx.doi.org/10.1103/PhysRevLett.119.041802}{Phys. Rev. Lett. {\bf
  119},  041802 (2017)}, \href{http://arxiv.org/abs/1704.07908}{{\tt
  arXiv:1704.07908 [hep-ex]}}\relax
\mciteBstWouldAddEndPuncttrue
\mciteSetBstMidEndSepPunct{\mcitedefaultmidpunct}
{\mcitedefaultendpunct}{\mcitedefaultseppunct}\relax
\EndOfBibitem
\bibitem{Aaij:2016yuv}
{LHCb} collaboration, R.~Aaij {\em et al.},
  \href{http://dx.doi.org/10.1007/JHEP05(2017)158}{JHEP {\bf 05},  158 (2017)},
  \href{http://arxiv.org/abs/1612.08110}{{\tt arXiv:1612.08110 [hep-ex]}}\relax
\mciteBstWouldAddEndPuncttrue
\mciteSetBstMidEndSepPunct{\mcitedefaultmidpunct}
{\mcitedefaultendpunct}{\mcitedefaultseppunct}\relax
\EndOfBibitem
\bibitem{Aaij:2018jhg}
{LHCb} collaboration, R.~Aaij {\em et al.},
  \href{http://dx.doi.org/10.1007/JHEP07(2018)020}{JHEP {\bf 07},  020 (2018)},
  \href{http://arxiv.org/abs/1804.07167}{{\tt arXiv:1804.07167 [hep-ex]}}\relax
\mciteBstWouldAddEndPuncttrue
\mciteSetBstMidEndSepPunct{\mcitedefaultmidpunct}
{\mcitedefaultendpunct}{\mcitedefaultseppunct}\relax
\EndOfBibitem
\bibitem{Aubert:2009ak}
{\babar} collaboration, B.~Aubert {\em et al.},
  \href{http://dx.doi.org/10.1103/PhysRevLett.103.211802}{Phys. Rev. Lett. {\bf
  103},  211802 (2009)}, \href{http://arxiv.org/abs/0906.2177}{{\tt
  arXiv:0906.2177 [hep-ex]}}\relax
\mciteBstWouldAddEndPuncttrue
\mciteSetBstMidEndSepPunct{\mcitedefaultmidpunct}
{\mcitedefaultendpunct}{\mcitedefaultseppunct}\relax
\EndOfBibitem
\bibitem{Horiguchi:2017ntw}
{Belle} collaboration, T.~Horiguchi {\em et al.},
  \href{http://dx.doi.org/10.1103/PhysRevLett.119.191802}{Phys. Rev. Lett. {\bf
  119},  191802 (2017)}, \href{http://arxiv.org/abs/1707.00394}{{\tt
  arXiv:1707.00394 [hep-ex]}}\relax
\mciteBstWouldAddEndPuncttrue
\mciteSetBstMidEndSepPunct{\mcitedefaultmidpunct}
{\mcitedefaultendpunct}{\mcitedefaultseppunct}\relax
\EndOfBibitem
\bibitem{Yang:2004as}
{Belle} collaboration, H.~Yang {\em et al.},
  \href{http://dx.doi.org/10.1103/PhysRevLett.94.111802}{Phys. Rev. Lett. {\bf
  94},  111802 (2005)}, \href{http://arxiv.org/abs/hep-ex/0412039}{{\tt
  arXiv:hep-ex/0412039 [hep-ex]}}\relax
\mciteBstWouldAddEndPuncttrue
\mciteSetBstMidEndSepPunct{\mcitedefaultmidpunct}
{\mcitedefaultendpunct}{\mcitedefaultseppunct}\relax
\EndOfBibitem
\bibitem{Nishida:2004fk}
{Belle} collaboration, S.~Nishida {\em et al.},
  \href{http://dx.doi.org/10.1016/j.physletb.2005.01.097}{Phys. Lett. {\bf
  B610},  23 (2005)}, \href{http://arxiv.org/abs/hep-ex/0411065}{{\tt
  arXiv:hep-ex/0411065 [hep-ex]}}\relax
\mciteBstWouldAddEndPuncttrue
\mciteSetBstMidEndSepPunct{\mcitedefaultmidpunct}
{\mcitedefaultendpunct}{\mcitedefaultseppunct}\relax
\EndOfBibitem
\bibitem{Aubert:2006vs}
{\babar} collaboration, B.~Aubert {\em et al.},
  \href{http://dx.doi.org/10.1103/PhysRevD.74.031102}{Phys. Rev. {\bf D74},
  031102 (2006)}, \href{http://arxiv.org/abs/hep-ex/0603054}{{\tt
  arXiv:hep-ex/0603054 [hep-ex]}}\relax
\mciteBstWouldAddEndPuncttrue
\mciteSetBstMidEndSepPunct{\mcitedefaultmidpunct}
{\mcitedefaultendpunct}{\mcitedefaultseppunct}\relax
\EndOfBibitem
\bibitem{Wedd:2008ru}
{Belle} collaboration, R.~Wedd {\em et al.},
  \href{http://dx.doi.org/10.1103/PhysRevD.81.111104}{Phys. Rev. {\bf D81},
  111104 (2010)}, \href{http://arxiv.org/abs/0810.0804}{{\tt arXiv:0810.0804
  [hep-ex]}}\relax
\mciteBstWouldAddEndPuncttrue
\mciteSetBstMidEndSepPunct{\mcitedefaultmidpunct}
{\mcitedefaultendpunct}{\mcitedefaultseppunct}\relax
\EndOfBibitem
\bibitem{Aubert:2006he}
{\babar} collaboration, B.~Aubert {\em et al.},
  \href{http://dx.doi.org/10.1103/PhysRevD.75.051102}{Phys. Rev. {\bf D75},
  051102 (2007)}, \href{http://arxiv.org/abs/hep-ex/0611037}{{\tt
  arXiv:hep-ex/0611037 [hep-ex]}}\relax
\mciteBstWouldAddEndPuncttrue
\mciteSetBstMidEndSepPunct{\mcitedefaultmidpunct}
{\mcitedefaultendpunct}{\mcitedefaultseppunct}\relax
\EndOfBibitem
\bibitem{Aubert:2005xk}
{\babar} collaboration, B.~Aubert {\em et al.},
  \href{http://dx.doi.org/10.1103/PhysRevLett.100.189903}{Phys. Rev. Lett. {\bf
  98},  211804 (2007)}, \href{http://arxiv.org/abs/hep-ex/0507031}{{\tt
  arXiv:hep-ex/0507031 [hep-ex]}}, Erratum ibid.\
  \href{http://dx.doi.org/10.1103/PhysRevLett.100.199905}{{\bf 100}, 199905},
  (2008)\relax
\mciteBstWouldAddEndPuncttrue
\mciteSetBstMidEndSepPunct{\mcitedefaultmidpunct}
{\mcitedefaultendpunct}{\mcitedefaultseppunct}\relax
\EndOfBibitem
\bibitem{Nishida:2002me}
{Belle} collaboration, S.~Nishida {\em et al.},
  \href{http://dx.doi.org/10.1103/PhysRevLett.89.231801}{Phys. Rev. Lett. {\bf
  89},  231801 (2002)}, \href{http://arxiv.org/abs/hep-ex/0205025}{{\tt
  arXiv:hep-ex/0205025 [hep-ex]}}\relax
\mciteBstWouldAddEndPuncttrue
\mciteSetBstMidEndSepPunct{\mcitedefaultmidpunct}
{\mcitedefaultendpunct}{\mcitedefaultseppunct}\relax
\EndOfBibitem
\bibitem{Aubert:2003zs}
{\babar} collaboration, B.~Aubert {\em et al.},
  \href{http://dx.doi.org/10.1103/PhysRevD.70.091105}{Phys. Rev. {\bf D70},
  091105 (2004)}, \href{http://arxiv.org/abs/hep-ex/0409035}{{\tt
  arXiv:hep-ex/0409035 [hep-ex]}}\relax
\mciteBstWouldAddEndPuncttrue
\mciteSetBstMidEndSepPunct{\mcitedefaultmidpunct}
{\mcitedefaultendpunct}{\mcitedefaultseppunct}\relax
\EndOfBibitem
\bibitem{Albrecht:1989ga}
{ARGUS} collaboration, H.~Albrecht {\em et al.},
  \href{http://dx.doi.org/10.1016/0370-2693(89)91177-5}{Phys. Lett. {\bf B229},
   304--308 (1989)}\relax
\mciteBstWouldAddEndPuncttrue
\mciteSetBstMidEndSepPunct{\mcitedefaultmidpunct}
{\mcitedefaultendpunct}{\mcitedefaultseppunct}\relax
\EndOfBibitem
\bibitem{Aubert:2008al}
{\babar} collaboration, B.~Aubert {\em et al.},
  \href{http://dx.doi.org/10.1103/PhysRevD.78.112001}{Phys. Rev. {\bf D78},
  112001 (2008)}, \href{http://arxiv.org/abs/0808.1379}{{\tt arXiv:0808.1379
  [hep-ex]}}\relax
\mciteBstWouldAddEndPuncttrue
\mciteSetBstMidEndSepPunct{\mcitedefaultmidpunct}
{\mcitedefaultendpunct}{\mcitedefaultseppunct}\relax
\EndOfBibitem
\bibitem{Taniguchi:2008ty}
{Belle} collaboration, N.~Taniguchi {\em et al.},
  \href{http://dx.doi.org/10.1103/PhysRevLett.101.111801}{Phys. Rev. Lett. {\bf
  101},  111801 (2008)}, \href{http://arxiv.org/abs/0804.4770}{{\tt
  arXiv:0804.4770 [hep-ex]}}, Erratum ibid.\
  \href{http://dx.doi.org/10.1103/PhysRevLett.101.129904}{{\bf 101}, 129904},
  (2008)\relax
\mciteBstWouldAddEndPuncttrue
\mciteSetBstMidEndSepPunct{\mcitedefaultmidpunct}
{\mcitedefaultendpunct}{\mcitedefaultseppunct}\relax
\EndOfBibitem
\bibitem{Lee:2005fba}
{Belle} collaboration, Y.~J. Lee {\em et al.},
  \href{http://dx.doi.org/10.1103/PhysRevLett.95.061802}{Phys. Rev. Lett. {\bf
  95},  061802 (2005)}, \href{http://arxiv.org/abs/hep-ex/0503046}{{\tt
  arXiv:hep-ex/0503046 [hep-ex]}}\relax
\mciteBstWouldAddEndPuncttrue
\mciteSetBstMidEndSepPunct{\mcitedefaultmidpunct}
{\mcitedefaultendpunct}{\mcitedefaultseppunct}\relax
\EndOfBibitem
\bibitem{Lees:2013lvs}
{\babar} collaboration, J.~P. Lees {\em et al.},
  \href{http://dx.doi.org/10.1103/PhysRevD.88.032012}{Phys. Rev. {\bf D88},
  032012 (2013)}, \href{http://arxiv.org/abs/1303.6010}{{\tt arXiv:1303.6010
  [hep-ex]}}\relax
\mciteBstWouldAddEndPuncttrue
\mciteSetBstMidEndSepPunct{\mcitedefaultmidpunct}
{\mcitedefaultendpunct}{\mcitedefaultseppunct}\relax
\EndOfBibitem
\bibitem{Wei:2008nv}
{Belle} collaboration, J.~T. Wei {\em et al.},
  \href{http://dx.doi.org/10.1103/PhysRevD.78.011101}{Phys. Rev. {\bf D78},
  011101 (2008)}, \href{http://arxiv.org/abs/0804.3656}{{\tt arXiv:0804.3656
  [hep-ex]}}\relax
\mciteBstWouldAddEndPuncttrue
\mciteSetBstMidEndSepPunct{\mcitedefaultmidpunct}
{\mcitedefaultendpunct}{\mcitedefaultseppunct}\relax
\EndOfBibitem
\bibitem{Aaij:2015nea}
{LHCb} collaboration, R.~Aaij {\em et al.},
  \href{http://dx.doi.org/10.1007/JHEP10(2015)034}{JHEP {\bf 10},  034 (2015)},
  \href{http://arxiv.org/abs/1509.00414}{{\tt arXiv:1509.00414 [hep-ex]}}\relax
\mciteBstWouldAddEndPuncttrue
\mciteSetBstMidEndSepPunct{\mcitedefaultmidpunct}
{\mcitedefaultendpunct}{\mcitedefaultseppunct}\relax
\EndOfBibitem
\bibitem{Aubert:2004ws}
{\babar} collaboration, B.~Aubert {\em et al.},
  \href{http://dx.doi.org/10.1103/PhysRevLett.94.101801}{Phys. Rev. Lett. {\bf
  94},  101801 (2005)}, \href{http://arxiv.org/abs/hep-ex/0411061}{{\tt
  arXiv:hep-ex/0411061 [hep-ex]}}\relax
\mciteBstWouldAddEndPuncttrue
\mciteSetBstMidEndSepPunct{\mcitedefaultmidpunct}
{\mcitedefaultendpunct}{\mcitedefaultseppunct}\relax
\EndOfBibitem
\bibitem{Lutz:2013ftz}
{Belle} collaboration, O.~Lutz {\em et al.},
  \href{http://dx.doi.org/10.1103/PhysRevD.87.111103}{Phys. Rev. {\bf D87},
  111103 (2013)}, \href{http://arxiv.org/abs/1303.3719}{{\tt arXiv:1303.3719
  [hep-ex]}}\relax
\mciteBstWouldAddEndPuncttrue
\mciteSetBstMidEndSepPunct{\mcitedefaultmidpunct}
{\mcitedefaultendpunct}{\mcitedefaultseppunct}\relax
\EndOfBibitem
\bibitem{Aubert:2008ps}
{\babar} collaboration, B.~Aubert {\em et al.},
  \href{http://dx.doi.org/10.1103/PhysRevLett.102.091803}{Phys. Rev. Lett. {\bf
  102},  091803 (2009)}, \href{http://arxiv.org/abs/0807.4119}{{\tt
  arXiv:0807.4119 [hep-ex]}}\relax
\mciteBstWouldAddEndPuncttrue
\mciteSetBstMidEndSepPunct{\mcitedefaultmidpunct}
{\mcitedefaultendpunct}{\mcitedefaultseppunct}\relax
\EndOfBibitem
\bibitem{Wei:2009zv}
{Belle} collaboration, J.~T. Wei {\em et al.},
  \href{http://dx.doi.org/10.1103/PhysRevLett.103.171801}{Phys. Rev. Lett. {\bf
  103},  171801 (2009)}, \href{http://arxiv.org/abs/0904.0770}{{\tt
  arXiv:0904.0770 [hep-ex]}}\relax
\mciteBstWouldAddEndPuncttrue
\mciteSetBstMidEndSepPunct{\mcitedefaultmidpunct}
{\mcitedefaultendpunct}{\mcitedefaultseppunct}\relax
\EndOfBibitem
\bibitem{Aaij:2012vr}
{LHCb} collaboration, R.~Aaij {\em et al.},
  \href{http://dx.doi.org/10.1007/JHEP02(2013)105}{JHEP {\bf 02},  105 (2013)},
  \href{http://arxiv.org/abs/1209.4284}{{\tt arXiv:1209.4284 [hep-ex]}}\relax
\mciteBstWouldAddEndPuncttrue
\mciteSetBstMidEndSepPunct{\mcitedefaultmidpunct}
{\mcitedefaultendpunct}{\mcitedefaultseppunct}\relax
\EndOfBibitem
\bibitem{TheBaBar:2016xwe}
{\babar} collaboration, J.~P. Lees {\em et al.},
  \href{http://dx.doi.org/10.1103/PhysRevLett.118.031802}{Phys. Rev. Lett. {\bf
  118},  031802 (2017)}, \href{http://arxiv.org/abs/1605.09637}{{\tt
  arXiv:1605.09637 [hep-ex]}}\relax
\mciteBstWouldAddEndPuncttrue
\mciteSetBstMidEndSepPunct{\mcitedefaultmidpunct}
{\mcitedefaultendpunct}{\mcitedefaultseppunct}\relax
\EndOfBibitem
\bibitem{Lees:2013kla}
{\babar} collaboration, J.~P. Lees {\em et al.},
  \href{http://dx.doi.org/10.1103/PhysRevD.87.112005}{Phys. Rev. {\bf D87},
  112005 (2013)}, \href{http://arxiv.org/abs/1303.7465}{{\tt arXiv:1303.7465
  [hep-ex]}}\relax
\mciteBstWouldAddEndPuncttrue
\mciteSetBstMidEndSepPunct{\mcitedefaultmidpunct}
{\mcitedefaultendpunct}{\mcitedefaultseppunct}\relax
\EndOfBibitem
\bibitem{Grygier:2017tzo}
{Belle} collaboration, J.~Grygier {\em et al.},
  \href{http://dx.doi.org/10.1103/PhysRevD.96.091101}{Phys. Rev. {\bf D96},
  091101 (2017)}, \href{http://arxiv.org/abs/1702.03224}{{\tt arXiv:1702.03224
  [hep-ex]}}\relax
\mciteBstWouldAddEndPuncttrue
\mciteSetBstMidEndSepPunct{\mcitedefaultmidpunct}
{\mcitedefaultendpunct}{\mcitedefaultseppunct}\relax
\EndOfBibitem
\bibitem{Aaij:2014pli}
{LHCb} collaboration, R.~Aaij {\em et al.},
  \href{http://dx.doi.org/10.1007/JHEP06(2014)133}{JHEP {\bf 06},  133 (2014)},
  \href{http://arxiv.org/abs/1403.8044}{{\tt arXiv:1403.8044 [hep-ex]}}\relax
\mciteBstWouldAddEndPuncttrue
\mciteSetBstMidEndSepPunct{\mcitedefaultmidpunct}
{\mcitedefaultendpunct}{\mcitedefaultseppunct}\relax
\EndOfBibitem
\bibitem{Aaij:2014kwa}
{LHCb} collaboration, R.~Aaij {\em et al.},
  \href{http://dx.doi.org/10.1007/JHEP10(2014)064}{JHEP {\bf 10},  064 (2014)},
  \href{http://arxiv.org/abs/1408.1137}{{\tt arXiv:1408.1137 [hep-ex]}}\relax
\mciteBstWouldAddEndPuncttrue
\mciteSetBstMidEndSepPunct{\mcitedefaultmidpunct}
{\mcitedefaultendpunct}{\mcitedefaultseppunct}\relax
\EndOfBibitem
\bibitem{BaBar-LambdaPNuNu-CKM2018}
{\babar} collaboration,  (CKM, 2018),
  \url{https://indico.cern.ch/event/686555/contributions/2986950/attachments/1680649/2699988/talk.pdf}\relax
\mciteBstWouldAddEndPuncttrue
\mciteSetBstMidEndSepPunct{\mcitedefaultmidpunct}
{\mcitedefaultendpunct}{\mcitedefaultseppunct}\relax
\EndOfBibitem
\bibitem{Hyun:2010an}
{Belle} collaboration, H.~J. Hyun {\em et al.},
  \href{http://dx.doi.org/10.1103/PhysRevLett.105.091801}{Phys. Rev. Lett. {\bf
  105},  091801 (2010)}, \href{http://arxiv.org/abs/1005.1450}{{\tt
  arXiv:1005.1450 [hep-ex]}}\relax
\mciteBstWouldAddEndPuncttrue
\mciteSetBstMidEndSepPunct{\mcitedefaultmidpunct}
{\mcitedefaultendpunct}{\mcitedefaultseppunct}\relax
\EndOfBibitem
\bibitem{Aubert:2005qc}
{\babar} collaboration, B.~Aubert {\em et al.},
  \href{http://dx.doi.org/10.1103/PhysRevD.72.091103}{Phys. Rev. {\bf D72},
  091103 (2005)}, \href{http://arxiv.org/abs/hep-ex/0501038}{{\tt
  arXiv:hep-ex/0501038 [hep-ex]}}\relax
\mciteBstWouldAddEndPuncttrue
\mciteSetBstMidEndSepPunct{\mcitedefaultmidpunct}
{\mcitedefaultendpunct}{\mcitedefaultseppunct}\relax
\EndOfBibitem
\bibitem{King:2016cxv}
{Belle} collaboration, Z.~King {\em et al.},
  \href{http://dx.doi.org/10.1103/PhysRevD.93.111101}{Phys. Rev. {\bf D93},
  111101 (2016)}, \href{http://arxiv.org/abs/1603.06546}{{\tt arXiv:1603.06546
  [hep-ex]}}\relax
\mciteBstWouldAddEndPuncttrue
\mciteSetBstMidEndSepPunct{\mcitedefaultmidpunct}
{\mcitedefaultendpunct}{\mcitedefaultseppunct}\relax
\EndOfBibitem
\bibitem{Lai:2013xht}
{Belle} collaboration, Y.~T. Lai {\em et al.},
  \href{http://dx.doi.org/10.1103/PhysRevD.89.051103}{Phys. Rev. {\bf D89},
  051103 (2014)}, \href{http://arxiv.org/abs/1312.4228}{{\tt arXiv:1312.4228
  [hep-ex]}}\relax
\mciteBstWouldAddEndPuncttrue
\mciteSetBstMidEndSepPunct{\mcitedefaultmidpunct}
{\mcitedefaultendpunct}{\mcitedefaultseppunct}\relax
\EndOfBibitem
\bibitem{Aaij:2016flj}
{LHCb} collaboration, R.~Aaij {\em et al.},
  \href{http://dx.doi.org/10.1007/JHEP11(2016)047}{JHEP {\bf 11},  047 (2016)},
  \href{http://arxiv.org/abs/1606.04731}{{\tt arXiv:1606.04731 [hep-ex]}}\relax
\mciteBstWouldAddEndPuncttrue
\mciteSetBstMidEndSepPunct{\mcitedefaultmidpunct}
{\mcitedefaultendpunct}{\mcitedefaultseppunct}\relax
\EndOfBibitem
\bibitem{Aaij:2015tna}
{LHCb} collaboration, R.~Aaij {\em et al.},
  \href{http://dx.doi.org/10.1103/PhysRevLett.115.161802}{Phys. Rev. Lett. {\bf
  115},  161802 (2015)}, \href{http://arxiv.org/abs/1508.04094}{{\tt
  arXiv:1508.04094 [hep-ex]}}\relax
\mciteBstWouldAddEndPuncttrue
\mciteSetBstMidEndSepPunct{\mcitedefaultmidpunct}
{\mcitedefaultendpunct}{\mcitedefaultseppunct}\relax
\EndOfBibitem
\bibitem{Coan:1999kh}
{CLEO} collaboration, T.~E. Coan {\em et al.},
  \href{http://dx.doi.org/10.1103/PhysRevLett.84.5283}{Phys. Rev. Lett. {\bf
  84},  5283 (2000)}, \href{http://arxiv.org/abs/hep-ex/9912057}{{\tt
  arXiv:hep-ex/9912057 [hep-ex]}}\relax
\mciteBstWouldAddEndPuncttrue
\mciteSetBstMidEndSepPunct{\mcitedefaultmidpunct}
{\mcitedefaultendpunct}{\mcitedefaultseppunct}\relax
\EndOfBibitem
\bibitem{Aubert:2007my}
{\babar} collaboration, B.~Aubert {\em et al.},
  \href{http://dx.doi.org/10.1103/PhysRevD.77.051103}{Phys. Rev. {\bf D77},
  051103 (2008)}, \href{http://arxiv.org/abs/0711.4889}{{\tt arXiv:0711.4889
  [hep-ex]}}\relax
\mciteBstWouldAddEndPuncttrue
\mciteSetBstMidEndSepPunct{\mcitedefaultmidpunct}
{\mcitedefaultendpunct}{\mcitedefaultseppunct}\relax
\EndOfBibitem
\bibitem{Lees:2012ym}
{\babar} collaboration, J.~P. Lees {\em et al.},
  \href{http://dx.doi.org/10.1103/PhysRevLett.109.191801}{Phys. Rev. Lett. {\bf
  109},  191801 (2012)}, \href{http://arxiv.org/abs/1207.2690}{{\tt
  arXiv:1207.2690 [hep-ex]}}\relax
\mciteBstWouldAddEndPuncttrue
\mciteSetBstMidEndSepPunct{\mcitedefaultmidpunct}
{\mcitedefaultendpunct}{\mcitedefaultseppunct}\relax
\EndOfBibitem
\bibitem{Lees:2012wg}
{\babar} collaboration, J.~P. Lees {\em et al.},
  \href{http://dx.doi.org/10.1103/PhysRevD.86.052012}{Phys. Rev. {\bf D86},
  052012 (2012)}, \href{http://arxiv.org/abs/1207.2520}{{\tt arXiv:1207.2520
  [hep-ex]}}\relax
\mciteBstWouldAddEndPuncttrue
\mciteSetBstMidEndSepPunct{\mcitedefaultmidpunct}
{\mcitedefaultendpunct}{\mcitedefaultseppunct}\relax
\EndOfBibitem
\bibitem{Saito:2014das}
{Belle} collaboration, T.~Saito {\em et al.},
  \href{http://dx.doi.org/10.1103/PhysRevD.91.052004}{Phys. Rev. {\bf D91},
  052004 (2015)}, \href{http://arxiv.org/abs/1411.7198}{{\tt arXiv:1411.7198
  [hep-ex]}}\relax
\mciteBstWouldAddEndPuncttrue
\mciteSetBstMidEndSepPunct{\mcitedefaultmidpunct}
{\mcitedefaultendpunct}{\mcitedefaultseppunct}\relax
\EndOfBibitem
\bibitem{Belle:2016ufb}
{Belle} collaboration, A.~Abdesselam {\em et al.},
  \href{http://arxiv.org/abs/1608.02344}{{\tt arXiv:1608.02344 [hep-ex]}}
  (2016)\relax
\mciteBstWouldAddEndPuncttrue
\mciteSetBstMidEndSepPunct{\mcitedefaultmidpunct}
{\mcitedefaultendpunct}{\mcitedefaultseppunct}\relax
\EndOfBibitem
\bibitem{delAmoSanchez:2010ae}
{\babar} collaboration, P.~del Amo~Sanchez {\em et al.},
  \href{http://dx.doi.org/10.1103/PhysRevD.82.051101}{Phys. Rev. {\bf D82},
  051101 (2010)}, \href{http://arxiv.org/abs/1005.4087}{{\tt arXiv:1005.4087
  [hep-ex]}}\relax
\mciteBstWouldAddEndPuncttrue
\mciteSetBstMidEndSepPunct{\mcitedefaultmidpunct}
{\mcitedefaultendpunct}{\mcitedefaultseppunct}\relax
\EndOfBibitem
\bibitem{Lees:2013nxa}
{\babar} collaboration, J.~P. Lees {\em et al.},
  \href{http://dx.doi.org/10.1103/PhysRevLett.112.211802}{Phys. Rev. Lett. {\bf
  112},  211802 (2014)}, \href{http://arxiv.org/abs/1312.5364}{{\tt
  arXiv:1312.5364 [hep-ex]}}\relax
\mciteBstWouldAddEndPuncttrue
\mciteSetBstMidEndSepPunct{\mcitedefaultmidpunct}
{\mcitedefaultendpunct}{\mcitedefaultseppunct}\relax
\EndOfBibitem
\bibitem{Iwasaki:2005sy}
{Belle} collaboration, M.~Iwasaki {\em et al.},
  \href{http://dx.doi.org/10.1103/PhysRevD.72.092005}{Phys. Rev. {\bf D72},
  092005 (2005)}, \href{http://arxiv.org/abs/hep-ex/0503044}{{\tt
  arXiv:hep-ex/0503044 [hep-ex]}}\relax
\mciteBstWouldAddEndPuncttrue
\mciteSetBstMidEndSepPunct{\mcitedefaultmidpunct}
{\mcitedefaultendpunct}{\mcitedefaultseppunct}\relax
\EndOfBibitem
\bibitem{Lees:2012tva}
{\babar} collaboration, J.~P. Lees {\em et al.},
  \href{http://dx.doi.org/10.1103/PhysRevD.86.032012}{Phys. Rev. {\bf D86},
  032012 (2012)}, \href{http://arxiv.org/abs/1204.3933}{{\tt arXiv:1204.3933
  [hep-ex]}}\relax
\mciteBstWouldAddEndPuncttrue
\mciteSetBstMidEndSepPunct{\mcitedefaultmidpunct}
{\mcitedefaultendpunct}{\mcitedefaultseppunct}\relax
\EndOfBibitem
\bibitem{Aubert:2007mm}
{\babar} collaboration, B.~Aubert {\em et al.},
  \href{http://dx.doi.org/10.1103/PhysRevLett.99.051801}{Phys. Rev. Lett. {\bf
  99},  051801 (2007)}, \href{http://arxiv.org/abs/hep-ex/0703018}{{\tt
  arXiv:hep-ex/0703018 [hep-ex]}}\relax
\mciteBstWouldAddEndPuncttrue
\mciteSetBstMidEndSepPunct{\mcitedefaultmidpunct}
{\mcitedefaultendpunct}{\mcitedefaultseppunct}\relax
\EndOfBibitem
\bibitem{Edwards:2002kq}
{CLEO} collaboration, K.~W. Edwards {\em et al.},
  \href{http://dx.doi.org/10.1103/PhysRevD.65.111102}{Phys. Rev. {\bf D65},
  111102 (2002)}, \href{http://arxiv.org/abs/hep-ex/0204017}{{\tt
  arXiv:hep-ex/0204017 [hep-ex]}}\relax
\mciteBstWouldAddEndPuncttrue
\mciteSetBstMidEndSepPunct{\mcitedefaultmidpunct}
{\mcitedefaultendpunct}{\mcitedefaultseppunct}\relax
\EndOfBibitem
\bibitem{Aubert:2006vb}
{\babar} collaboration, B.~Aubert {\em et al.},
  \href{http://dx.doi.org/10.1103/PhysRevD.73.092001}{Phys. Rev. {\bf D73},
  092001 (2006)}, \href{http://arxiv.org/abs/hep-ex/0604007}{{\tt
  arXiv:hep-ex/0604007 [hep-ex]}}\relax
\mciteBstWouldAddEndPuncttrue
\mciteSetBstMidEndSepPunct{\mcitedefaultmidpunct}
{\mcitedefaultendpunct}{\mcitedefaultseppunct}\relax
\EndOfBibitem
\bibitem{Buchmuller:2005zv}
O.~Buchmuller and H.~Flacher,
  \href{http://dx.doi.org/10.1103/PhysRevD.73.073008}{Phys. Rev. {\bf D73},
  073008 (2006)}, \href{http://arxiv.org/abs/hep-ph/0507253}{{\tt
  arXiv:hep-ph/0507253 [hep-ph]}}\relax
\mciteBstWouldAddEndPuncttrue
\mciteSetBstMidEndSepPunct{\mcitedefaultmidpunct}
{\mcitedefaultendpunct}{\mcitedefaultseppunct}\relax
\EndOfBibitem
\bibitem{Aubert:2009ar}
{\babar} collaboration, B.~Aubert {\em et al.},
  \href{http://dx.doi.org/10.1103/PhysRevD.79.091101}{Phys. Rev. {\bf D79},
  091101 (2009)}, \href{http://arxiv.org/abs/0903.1220}{{\tt arXiv:0903.1220
  [hep-ex]}}\relax
\mciteBstWouldAddEndPuncttrue
\mciteSetBstMidEndSepPunct{\mcitedefaultmidpunct}
{\mcitedefaultendpunct}{\mcitedefaultseppunct}\relax
\EndOfBibitem
\bibitem{Satoyama:2006xn}
{Belle} collaboration, N.~Satoyama {\em et al.},
  \href{http://dx.doi.org/10.1016/j.physletb.2007.01.068}{Phys. Lett. {\bf
  B647},  67 (2007)}, \href{http://arxiv.org/abs/hep-ex/0611045}{{\tt
  arXiv:hep-ex/0611045 [hep-ex]}}\relax
\mciteBstWouldAddEndPuncttrue
\mciteSetBstMidEndSepPunct{\mcitedefaultmidpunct}
{\mcitedefaultendpunct}{\mcitedefaultseppunct}\relax
\EndOfBibitem
\bibitem{Sibidanov:2017vph}
{Belle} collaboration, A.~Sibidanov {\em et al.},
  \href{http://arxiv.org/abs/1712.04123}{{\tt arXiv:1712.04123 [hep-ex]}}
  (2017)\relax
\mciteBstWouldAddEndPuncttrue
\mciteSetBstMidEndSepPunct{\mcitedefaultmidpunct}
{\mcitedefaultendpunct}{\mcitedefaultseppunct}\relax
\EndOfBibitem
\bibitem{Lees:2012ju}
{\babar} collaboration, J.~P. Lees {\em et al.},
  \href{http://dx.doi.org/10.1103/PhysRevD.88.031102}{Phys. Rev. {\bf D88},
  031102 (2013)}, \href{http://arxiv.org/abs/1207.0698}{{\tt arXiv:1207.0698
  [hep-ex]}}\relax
\mciteBstWouldAddEndPuncttrue
\mciteSetBstMidEndSepPunct{\mcitedefaultmidpunct}
{\mcitedefaultendpunct}{\mcitedefaultseppunct}\relax
\EndOfBibitem
\bibitem{Kronenbitter:2015kls}
{Belle} collaboration, B.~Kronenbitter {\em et al.},
  \href{http://dx.doi.org/10.1103/PhysRevD.92.051102}{Phys. Rev. {\bf D92},
  051102 (2015)}, \href{http://arxiv.org/abs/1503.05613}{{\tt arXiv:1503.05613
  [hep-ex]}}\relax
\mciteBstWouldAddEndPuncttrue
\mciteSetBstMidEndSepPunct{\mcitedefaultmidpunct}
{\mcitedefaultendpunct}{\mcitedefaultseppunct}\relax
\EndOfBibitem
\bibitem{Aubert:2009ya}
{\babar} collaboration, B.~Aubert {\em et al.},
  \href{http://dx.doi.org/10.1103/PhysRevD.80.111105}{Phys. Rev. {\bf D80},
  111105 (2009)}, \href{http://arxiv.org/abs/0907.1681}{{\tt arXiv:0907.1681
  [hep-ex]}}\relax
\mciteBstWouldAddEndPuncttrue
\mciteSetBstMidEndSepPunct{\mcitedefaultmidpunct}
{\mcitedefaultendpunct}{\mcitedefaultseppunct}\relax
\EndOfBibitem
\bibitem{Gelb:2018end}
{Belle} collaboration, M.~Gelb {\em et al.},
  \href{http://arxiv.org/abs/1810.12976}{{\tt arXiv:1810.12976 [hep-ex]}}\relax
\mciteBstWouldAddEndPuncttrue
\mciteSetBstMidEndSepPunct{\mcitedefaultmidpunct}
{\mcitedefaultendpunct}{\mcitedefaultseppunct}\relax
\EndOfBibitem
\bibitem{Heller:2015vvm}
{Belle} collaboration, A.~Heller {\em et al.},
  \href{http://dx.doi.org/10.1103/PhysRevD.91.112009}{Phys. Rev. {\bf D91},
  112009 (2015)}, \href{http://arxiv.org/abs/1504.05831}{{\tt arXiv:1504.05831
  [hep-ex]}}\relax
\mciteBstWouldAddEndPuncttrue
\mciteSetBstMidEndSepPunct{\mcitedefaultmidpunct}
{\mcitedefaultendpunct}{\mcitedefaultseppunct}\relax
\EndOfBibitem
\bibitem{delAmoSanchez:2010bx}
{\babar} collaboration, P.~del Amo~Sanchez {\em et al.},
  \href{http://dx.doi.org/10.1103/PhysRevD.83.032006}{Phys. Rev. {\bf D83},
  032006 (2011)}, \href{http://arxiv.org/abs/1010.2229}{{\tt arXiv:1010.2229
  [hep-ex]}}\relax
\mciteBstWouldAddEndPuncttrue
\mciteSetBstMidEndSepPunct{\mcitedefaultmidpunct}
{\mcitedefaultendpunct}{\mcitedefaultseppunct}\relax
\EndOfBibitem
\bibitem{Abe:2005bs}
{Belle} collaboration, S.~Villa {\em et al.},
  \href{http://dx.doi.org/10.1103/PhysRevD.73.051107}{Phys. Rev. {\bf D73},
  051107 (2006)}, \href{http://arxiv.org/abs/hep-ex/0507036}{{\tt
  arXiv:hep-ex/0507036 [hep-ex]}}\relax
\mciteBstWouldAddEndPuncttrue
\mciteSetBstMidEndSepPunct{\mcitedefaultmidpunct}
{\mcitedefaultendpunct}{\mcitedefaultseppunct}\relax
\EndOfBibitem
\bibitem{Aubert:2007hb}
{\babar} collaboration, B.~Aubert {\em et al.},
  \href{http://dx.doi.org/10.1103/PhysRevD.77.032007}{Phys. Rev. {\bf D77},
  032007 (2008)}, \href{http://arxiv.org/abs/0712.1516}{{\tt arXiv:0712.1516
  [hep-ex]}}\relax
\mciteBstWouldAddEndPuncttrue
\mciteSetBstMidEndSepPunct{\mcitedefaultmidpunct}
{\mcitedefaultendpunct}{\mcitedefaultseppunct}\relax
\EndOfBibitem
\bibitem{Chang:2003yy}
{Belle} collaboration, M.~C. Chang {\em et al.},
  \href{http://dx.doi.org/10.1103/PhysRevD.68.111101}{Phys. Rev. {\bf D68},
  111101 (2003)}, \href{http://arxiv.org/abs/hep-ex/0309069}{{\tt
  arXiv:hep-ex/0309069 [hep-ex]}}\relax
\mciteBstWouldAddEndPuncttrue
\mciteSetBstMidEndSepPunct{\mcitedefaultmidpunct}
{\mcitedefaultendpunct}{\mcitedefaultseppunct}\relax
\EndOfBibitem
\bibitem{Aubert:2007up}
{\babar} collaboration, B.~Aubert {\em et al.},
  \href{http://dx.doi.org/10.1103/PhysRevD.77.011104}{Phys. Rev. {\bf D77},
  011104 (2008)}, \href{http://arxiv.org/abs/0706.2870}{{\tt arXiv:0706.2870
  [hep-ex]}}\relax
\mciteBstWouldAddEndPuncttrue
\mciteSetBstMidEndSepPunct{\mcitedefaultmidpunct}
{\mcitedefaultendpunct}{\mcitedefaultseppunct}\relax
\EndOfBibitem
\bibitem{Chatrchyan:2013bka}
{CMS} collaboration, S.~Chatrchyan {\em et al.},
  \href{http://dx.doi.org/10.1103/PhysRevLett.111.101804}{Phys. Rev. Lett. {\bf
  111},  101804 (2013)}, \href{http://arxiv.org/abs/1307.5025}{{\tt
  arXiv:1307.5025 [hep-ex]}}\relax
\mciteBstWouldAddEndPuncttrue
\mciteSetBstMidEndSepPunct{\mcitedefaultmidpunct}
{\mcitedefaultendpunct}{\mcitedefaultseppunct}\relax
\EndOfBibitem
\bibitem{Aubert:2005qw}
{\babar} collaboration, B.~Aubert {\em et al.},
  \href{http://dx.doi.org/10.1103/PhysRevLett.96.241802}{Phys. Rev. Lett. {\bf
  96},  241802 (2006)}, \href{http://arxiv.org/abs/hep-ex/0511015}{{\tt
  arXiv:hep-ex/0511015 [hep-ex]}}\relax
\mciteBstWouldAddEndPuncttrue
\mciteSetBstMidEndSepPunct{\mcitedefaultmidpunct}
{\mcitedefaultendpunct}{\mcitedefaultseppunct}\relax
\EndOfBibitem
\bibitem{Aubert:2008cu}
{\babar} collaboration, B.~Aubert {\em et al.},
  \href{http://dx.doi.org/10.1103/PhysRevD.77.091104}{Phys. Rev. {\bf D77},
  091104 (2008)}, \href{http://arxiv.org/abs/0801.0697}{{\tt arXiv:0801.0697
  [hep-ex]}}\relax
\mciteBstWouldAddEndPuncttrue
\mciteSetBstMidEndSepPunct{\mcitedefaultmidpunct}
{\mcitedefaultendpunct}{\mcitedefaultseppunct}\relax
\EndOfBibitem
\bibitem{Lees:2012wv}
{\babar} collaboration, J.~P. Lees {\em et al.},
  \href{http://dx.doi.org/10.1103/PhysRevD.86.051105}{Phys. Rev. {\bf D86},
  051105 (2012)}, \href{http://arxiv.org/abs/1206.2543}{{\tt arXiv:1206.2543
  [hep-ex]}}\relax
\mciteBstWouldAddEndPuncttrue
\mciteSetBstMidEndSepPunct{\mcitedefaultmidpunct}
{\mcitedefaultendpunct}{\mcitedefaultseppunct}\relax
\EndOfBibitem
\bibitem{Hsu:2012uh}
{Belle} collaboration, C.~L. Hsu {\em et al.},
  \href{http://dx.doi.org/10.1103/PhysRevD.86.032002}{Phys. Rev. {\bf D86},
  032002 (2012)}, \href{http://arxiv.org/abs/1206.5948}{{\tt arXiv:1206.5948
  [hep-ex]}}\relax
\mciteBstWouldAddEndPuncttrue
\mciteSetBstMidEndSepPunct{\mcitedefaultmidpunct}
{\mcitedefaultendpunct}{\mcitedefaultseppunct}\relax
\EndOfBibitem
\bibitem{Yook:2014kga}
{Belle} collaboration, Y.~Yook {\em et al.},
  \href{http://dx.doi.org/10.1103/PhysRevD.91.052016}{Phys. Rev. {\bf D91},
  052016 (2015)}, \href{http://arxiv.org/abs/1406.6356}{{\tt arXiv:1406.6356
  [hep-ex]}}\relax
\mciteBstWouldAddEndPuncttrue
\mciteSetBstMidEndSepPunct{\mcitedefaultmidpunct}
{\mcitedefaultendpunct}{\mcitedefaultseppunct}\relax
\EndOfBibitem
\bibitem{Aaij:2014ora}
{LHCb} collaboration, R.~Aaij {\em et al.},
  \href{http://dx.doi.org/10.1103/PhysRevLett.113.151601}{Phys. Rev. Lett. {\bf
  113},  151601 (2014)}, \href{http://arxiv.org/abs/1406.6482}{{\tt
  arXiv:1406.6482 [hep-ex]}}\relax
\mciteBstWouldAddEndPuncttrue
\mciteSetBstMidEndSepPunct{\mcitedefaultmidpunct}
{\mcitedefaultendpunct}{\mcitedefaultseppunct}\relax
\EndOfBibitem
\bibitem{Aaij:2017vbb}
{LHCb} collaboration, R.~Aaij {\em et al.},
  \href{http://arxiv.org/abs/1705.05802}{{\tt arXiv:1705.05802 [hep-ex]}}
  (2017)\relax
\mciteBstWouldAddEndPuncttrue
\mciteSetBstMidEndSepPunct{\mcitedefaultmidpunct}
{\mcitedefaultendpunct}{\mcitedefaultseppunct}\relax
\EndOfBibitem
\bibitem{Nishimura:2009ae}
{Belle} collaboration, K.~Nishimura {\em et al.},
  \href{http://dx.doi.org/10.1103/PhysRevLett.105.191803}{Phys. Rev. Lett. {\bf
  105},  191803 (2010)}, \href{http://arxiv.org/abs/0910.4751}{{\tt
  arXiv:0910.4751 [hep-ex]}}\relax
\mciteBstWouldAddEndPuncttrue
\mciteSetBstMidEndSepPunct{\mcitedefaultmidpunct}
{\mcitedefaultendpunct}{\mcitedefaultseppunct}\relax
\EndOfBibitem
\bibitem{Browder:1998yb}
{CLEO} collaboration, T.~E. Browder {\em et al.},
  \href{http://dx.doi.org/10.1103/PhysRevLett.81.1786}{Phys. Rev. Lett. {\bf
  81},  1786 (1998)}, \href{http://arxiv.org/abs/hep-ex/9804018}{{\tt
  arXiv:hep-ex/9804018 [hep-ex]}}\relax
\mciteBstWouldAddEndPuncttrue
\mciteSetBstMidEndSepPunct{\mcitedefaultmidpunct}
{\mcitedefaultendpunct}{\mcitedefaultseppunct}\relax
\EndOfBibitem
\bibitem{Aubert:2004eq}
{\babar} collaboration, B.~Aubert {\em et al.},
  \href{http://dx.doi.org/10.1103/PhysRevLett.93.061801}{Phys. Rev. Lett. {\bf
  93},  061801 (2004)}, \href{http://arxiv.org/abs/hep-ex/0401006}{{\tt
  arXiv:hep-ex/0401006 [hep-ex]}}\relax
\mciteBstWouldAddEndPuncttrue
\mciteSetBstMidEndSepPunct{\mcitedefaultmidpunct}
{\mcitedefaultendpunct}{\mcitedefaultseppunct}\relax
\EndOfBibitem
\bibitem{Bonvicini:2003aw}
{CLEO} collaboration, G.~Bonvicini {\em et al.},
  \href{http://dx.doi.org/10.1103/PhysRevD.68.011101}{Phys. Rev. {\bf D68},
  011101 (2003)}, \href{http://arxiv.org/abs/hep-ex/0303009}{{\tt
  arXiv:hep-ex/0303009 [hep-ex]}}\relax
\mciteBstWouldAddEndPuncttrue
\mciteSetBstMidEndSepPunct{\mcitedefaultmidpunct}
{\mcitedefaultendpunct}{\mcitedefaultseppunct}\relax
\EndOfBibitem
\bibitem{delAmoSanchez:2010gx}
{\babar} collaboration, P.~del Amo~Sanchez {\em et al.},
  \href{http://dx.doi.org/10.1103/PhysRevD.83.031103}{Phys. Rev. {\bf D83},
  031103 (2011)}, \href{http://arxiv.org/abs/1012.5031}{{\tt arXiv:1012.5031
  [hep-ex]}}\relax
\mciteBstWouldAddEndPuncttrue
\mciteSetBstMidEndSepPunct{\mcitedefaultmidpunct}
{\mcitedefaultendpunct}{\mcitedefaultseppunct}\relax
\EndOfBibitem
\bibitem{Watanuki:2018xxg}
{Belle} collaboration, S.~Watanuki {\em et al.},
  \href{http://dx.doi.org/10.1103/PhysRevD.99.032012}{Phys. Rev. {\bf D99},
  032012 (2019)}, \href{http://arxiv.org/abs/1807.04236}{{\tt arXiv:1807.04236
  [hep-ex]}}\relax
\mciteBstWouldAddEndPuncttrue
\mciteSetBstMidEndSepPunct{\mcitedefaultmidpunct}
{\mcitedefaultendpunct}{\mcitedefaultseppunct}\relax
\EndOfBibitem
\bibitem{Lees:2012zz}
{\babar} collaboration, J.~P. Lees {\em et al.},
  \href{http://dx.doi.org/10.1103/PhysRevD.86.012004}{Phys. Rev. {\bf D86},
  012004 (2012)}, \href{http://arxiv.org/abs/1204.2852}{{\tt arXiv:1204.2852
  [hep-ex]}}\relax
\mciteBstWouldAddEndPuncttrue
\mciteSetBstMidEndSepPunct{\mcitedefaultmidpunct}
{\mcitedefaultendpunct}{\mcitedefaultseppunct}\relax
\EndOfBibitem
\bibitem{BABAR:2012aa}
{\babar} collaboration, J.~P. Lees {\em et al.},
  \href{http://dx.doi.org/10.1103/PhysRevD.85.071103}{Phys. Rev. {\bf D85},
  071103 (2012)}, \href{http://arxiv.org/abs/1202.3650}{{\tt arXiv:1202.3650
  [hep-ex]}}\relax
\mciteBstWouldAddEndPuncttrue
\mciteSetBstMidEndSepPunct{\mcitedefaultmidpunct}
{\mcitedefaultendpunct}{\mcitedefaultseppunct}\relax
\EndOfBibitem
\bibitem{Aaij:2014aba}
{LHCb} collaboration, R.~Aaij {\em et al.},
  \href{http://dx.doi.org/10.1103/PhysRevLett.112.131802}{Phys. Rev. Lett. {\bf
  112},  131802 (2014)}, \href{http://arxiv.org/abs/1401.5361}{{\tt
  arXiv:1401.5361 [hep-ex]}}\relax
\mciteBstWouldAddEndPuncttrue
\mciteSetBstMidEndSepPunct{\mcitedefaultmidpunct}
{\mcitedefaultendpunct}{\mcitedefaultseppunct}\relax
\EndOfBibitem
\bibitem{Lees:2013gdj}
{\babar} collaboration, J.~P. Lees {\em et al.},
  \href{http://dx.doi.org/10.1103/PhysRevD.89.011102}{Phys. Rev. {\bf D89},
  011102 (2014)}, \href{http://arxiv.org/abs/1310.8238}{{\tt arXiv:1310.8238
  [hep-ex]}}\relax
\mciteBstWouldAddEndPuncttrue
\mciteSetBstMidEndSepPunct{\mcitedefaultmidpunct}
{\mcitedefaultendpunct}{\mcitedefaultseppunct}\relax
\EndOfBibitem
\bibitem{Aaij:2011ex}
{LHCb} collaboration, R.~Aaij {\em et al.},
  \href{http://dx.doi.org/10.1103/PhysRevLett.108.101601}{Phys. Rev. Lett. {\bf
  108},  101601 (2012)}, \href{http://arxiv.org/abs/1110.0730}{{\tt
  arXiv:1110.0730 [hep-ex]}}\relax
\mciteBstWouldAddEndPuncttrue
\mciteSetBstMidEndSepPunct{\mcitedefaultmidpunct}
{\mcitedefaultendpunct}{\mcitedefaultseppunct}\relax
\EndOfBibitem
\bibitem{Aaij:2012zr}
{LHCb} collaboration, R.~Aaij {\em et al.},
  \href{http://dx.doi.org/10.1103/PhysRevD.85.112004}{Phys. Rev. {\bf D85},
  112004 (2012)}, \href{http://arxiv.org/abs/1201.5600}{{\tt arXiv:1201.5600
  [hep-ex]}}\relax
\mciteBstWouldAddEndPuncttrue
\mciteSetBstMidEndSepPunct{\mcitedefaultmidpunct}
{\mcitedefaultendpunct}{\mcitedefaultseppunct}\relax
\EndOfBibitem
\bibitem{BABAR:2011ac}
{BaBar} collaboration, P.~del Amo~Sanchez {\em et al.},
  \href{http://dx.doi.org/10.1103/PhysRevD.83.091101}{Phys. Rev. {\bf D83},
  091101 (2011)}, \href{http://arxiv.org/abs/1101.3830}{{\tt arXiv:1101.3830
  [hep-ex]}}\relax
\mciteBstWouldAddEndPuncttrue
\mciteSetBstMidEndSepPunct{\mcitedefaultmidpunct}
{\mcitedefaultendpunct}{\mcitedefaultseppunct}\relax
\EndOfBibitem
\bibitem{Sandilya:2018pop}
{Belle} collaboration, S.~Sandilya {\em et al.},
  \href{http://dx.doi.org/10.1103/PhysRevD.98.071101}{Phys. Rev. {\bf D98},
  071101 (2018)}, \href{http://arxiv.org/abs/1807.03267}{{\tt arXiv:1807.03267
  [hep-ex]}}\relax
\mciteBstWouldAddEndPuncttrue
\mciteSetBstMidEndSepPunct{\mcitedefaultmidpunct}
{\mcitedefaultendpunct}{\mcitedefaultseppunct}\relax
\EndOfBibitem
\bibitem{Aaij:2014wgo}
{LHCb} collaboration, R.~Aaij {\em et al.},
  \href{http://dx.doi.org/10.1103/PhysRevLett.112.161801}{Phys. Rev. Lett. {\bf
  112},  161801 (2014)}, \href{http://arxiv.org/abs/1402.6852}{{\tt
  arXiv:1402.6852 [hep-ex]}}\relax
\mciteBstWouldAddEndPuncttrue
\mciteSetBstMidEndSepPunct{\mcitedefaultmidpunct}
{\mcitedefaultendpunct}{\mcitedefaultseppunct}\relax
\EndOfBibitem
\bibitem{Aaij:2013hha}
{LHCb} collaboration, R.~Aaij {\em et al.},
  \href{http://dx.doi.org/10.1007/JHEP05(2013)159}{JHEP {\bf 05},  159 (2013)},
  \href{http://arxiv.org/abs/1304.3035}{{\tt arXiv:1304.3035 [hep-ex]}}\relax
\mciteBstWouldAddEndPuncttrue
\mciteSetBstMidEndSepPunct{\mcitedefaultmidpunct}
{\mcitedefaultendpunct}{\mcitedefaultseppunct}\relax
\EndOfBibitem
\bibitem{Aaij:2014tfa}
{LHCb} collaboration, R.~Aaij {\em et al.},
  \href{http://dx.doi.org/10.1007/JHEP05(2014)082}{JHEP {\bf 05},  082 (2014)},
  \href{http://arxiv.org/abs/1403.8045}{{\tt arXiv:1403.8045 [hep-ex]}}\relax
\mciteBstWouldAddEndPuncttrue
\mciteSetBstMidEndSepPunct{\mcitedefaultmidpunct}
{\mcitedefaultendpunct}{\mcitedefaultseppunct}\relax
\EndOfBibitem
\bibitem{Aaij:2015dea}
{LHCb} collaboration, R.~Aaij {\em et al.},
  \href{http://dx.doi.org/10.1007/JHEP04(2015)064}{JHEP {\bf 04},  064 (2015)},
  \href{http://arxiv.org/abs/1501.03038}{{\tt arXiv:1501.03038 [hep-ex]}}\relax
\mciteBstWouldAddEndPuncttrue
\mciteSetBstMidEndSepPunct{\mcitedefaultmidpunct}
{\mcitedefaultendpunct}{\mcitedefaultseppunct}\relax
\EndOfBibitem
\bibitem{Abdesselam:2016llu}
{Belle} collaboration, A.~Abdesselam {\em et al.},
  \href{http://arxiv.org/abs/1604.04042}{{\tt arXiv:1604.04042 [hep-ex]}}
  (2016)\relax
\mciteBstWouldAddEndPuncttrue
\mciteSetBstMidEndSepPunct{\mcitedefaultmidpunct}
{\mcitedefaultendpunct}{\mcitedefaultseppunct}\relax
\EndOfBibitem
\bibitem{Wehle:2016yoi}
{Belle} collaboration, S.~Wehle {\em et al.},
  \href{http://dx.doi.org/10.1103/PhysRevLett.118.111801}{Phys. Rev. Lett. {\bf
  118},  111801 (2017)}, \href{http://arxiv.org/abs/1612.05014}{{\tt
  arXiv:1612.05014 [hep-ex]}}\relax
\mciteBstWouldAddEndPuncttrue
\mciteSetBstMidEndSepPunct{\mcitedefaultmidpunct}
{\mcitedefaultendpunct}{\mcitedefaultseppunct}\relax
\EndOfBibitem
\bibitem{Lees:2015ymt}
{\babar} collaboration, J.~P. Lees {\em et al.},
  \href{http://dx.doi.org/10.1103/PhysRevD.93.052015}{Phys. Rev. {\bf D93},
  052015 (2016)}, \href{http://arxiv.org/abs/1508.07960}{{\tt arXiv:1508.07960
  [hep-ex]}}\relax
\mciteBstWouldAddEndPuncttrue
\mciteSetBstMidEndSepPunct{\mcitedefaultmidpunct}
{\mcitedefaultendpunct}{\mcitedefaultseppunct}\relax
\EndOfBibitem
\bibitem{Aaij:2015oid}
{LHCb} collaboration, R.~Aaij {\em et al.},
  \href{http://dx.doi.org/10.1007/JHEP02(2016)104}{JHEP {\bf 02},  104 (2016)},
  \href{http://arxiv.org/abs/1512.04442}{{\tt arXiv:1512.04442 [hep-ex]}}\relax
\mciteBstWouldAddEndPuncttrue
\mciteSetBstMidEndSepPunct{\mcitedefaultmidpunct}
{\mcitedefaultendpunct}{\mcitedefaultseppunct}\relax
\EndOfBibitem
\bibitem{Khachatryan:2015isa}
{CMS} collaboration, V.~Khachatryan {\em et al.},
  \href{http://dx.doi.org/10.1016/j.physletb.2015.12.020}{Phys. Lett. {\bf
  B753},  424 (2016)}, \href{http://arxiv.org/abs/1507.08126}{{\tt
  arXiv:1507.08126 [hep-ex]}}\relax
\mciteBstWouldAddEndPuncttrue
\mciteSetBstMidEndSepPunct{\mcitedefaultmidpunct}
{\mcitedefaultendpunct}{\mcitedefaultseppunct}\relax
\EndOfBibitem
\bibitem{Sirunyan:2017dhj}
{CMS} collaboration, A.~M. Sirunyan {\em et al.},
  \href{http://dx.doi.org/10.1016/j.physletb.2018.04.030}{Phys. Lett. {\bf
  B781},  517 (2018)}, \href{http://arxiv.org/abs/1710.02846}{{\tt
  arXiv:1710.02846 [hep-ex]}}\relax
\mciteBstWouldAddEndPuncttrue
\mciteSetBstMidEndSepPunct{\mcitedefaultmidpunct}
{\mcitedefaultendpunct}{\mcitedefaultseppunct}\relax
\EndOfBibitem
\bibitem{ATLAS-CONF-2017-023}
{ATLAS} collaboration,
  \href{http://cds.cern.ch/record/2258146}{ATLAS-CONF-2017-023}, {2017}\relax
\mciteBstWouldAddEndPuncttrue
\mciteSetBstMidEndSepPunct{\mcitedefaultmidpunct}
{\mcitedefaultendpunct}{\mcitedefaultseppunct}\relax
\EndOfBibitem
\bibitem{Sato:2014pjr}
{Belle} collaboration, Y.~Sato {\em et al.},
  \href{http://dx.doi.org/10.1103/PhysRevD.93.032008}{Phys. Rev. {\bf D93},
  032008 (2016)}, \href{http://arxiv.org/abs/1402.7134}{{\tt arXiv:1402.7134
  [hep-ex]}}, addendum ibid.\
  \href{http://dx.doi.org/10.1103/PhysRevD.93.059901}{{\bf D93}, 059901},
  (2016)\relax
\mciteBstWouldAddEndPuncttrue
\mciteSetBstMidEndSepPunct{\mcitedefaultmidpunct}
{\mcitedefaultendpunct}{\mcitedefaultseppunct}\relax
\EndOfBibitem
\bibitem{Aaij:2016kqt}
{LHCb} collaboration, R.~Aaij {\em et al.},
  \href{http://dx.doi.org/10.1007/JHEP12(2016)065}{JHEP {\bf 12},  065 (2016)},
  \href{http://arxiv.org/abs/1609.04736}{{\tt arXiv:1609.04736 [hep-ex]}}\relax
\mciteBstWouldAddEndPuncttrue
\mciteSetBstMidEndSepPunct{\mcitedefaultmidpunct}
{\mcitedefaultendpunct}{\mcitedefaultseppunct}\relax
\EndOfBibitem
\bibitem{Aaij:2016cbx}
{LHCb} collaboration, R.~Aaij {\em et al.},
  \href{http://dx.doi.org/10.1140/epjc/s10052-017-4703-2}{Eur. Phys. J. {\bf
  C77},  161 (2017)}, \href{http://arxiv.org/abs/1612.06764}{{\tt
  arXiv:1612.06764 [hep-ex]}}\relax
\mciteBstWouldAddEndPuncttrue
\mciteSetBstMidEndSepPunct{\mcitedefaultmidpunct}
{\mcitedefaultendpunct}{\mcitedefaultseppunct}\relax
\EndOfBibitem
\bibitem{Sirunyan:2018jll}
{CMS} collaboration, A.~M. Sirunyan {\em et al.},
  \href{http://dx.doi.org/10.1103/PhysRevD.98.112011}{Phys. Rev. {\bf D98},
  112011 (2018)}, \href{http://arxiv.org/abs/1806.00636}{{\tt arXiv:1806.00636
  [hep-ex]}}\relax
\mciteBstWouldAddEndPuncttrue
\mciteSetBstMidEndSepPunct{\mcitedefaultmidpunct}
{\mcitedefaultendpunct}{\mcitedefaultseppunct}\relax
\EndOfBibitem
\bibitem{Aaij:2016qsm}
{LHCb} collaboration, R.~Aaij {\em et al.},
  \href{http://dx.doi.org/10.1103/PhysRevD.95.071101}{Phys. Rev. {\bf D95},
  071101 (2017)}, \href{http://arxiv.org/abs/1612.07818}{{\tt arXiv:1612.07818
  [hep-ex]}}\relax
\mciteBstWouldAddEndPuncttrue
\mciteSetBstMidEndSepPunct{\mcitedefaultmidpunct}
{\mcitedefaultendpunct}{\mcitedefaultseppunct}\relax
\EndOfBibitem
\bibitem{Pesantez:2015fza}
{Belle} collaboration, L.~Pesántez {\em et al.},
  \href{http://dx.doi.org/10.1103/PhysRevLett.114.151601}{Phys. Rev. Lett. {\bf
  114},  151601 (2015)}, \href{http://arxiv.org/abs/1501.01702}{{\tt
  arXiv:1501.01702 [hep-ex]}}\relax
\mciteBstWouldAddEndPuncttrue
\mciteSetBstMidEndSepPunct{\mcitedefaultmidpunct}
{\mcitedefaultendpunct}{\mcitedefaultseppunct}\relax
\EndOfBibitem
\bibitem{Aaij:2014iva}
{LHCb} collaboration, R.~Aaij {\em et al.},
  \href{http://dx.doi.org/10.1103/PhysRevD.90.112004}{Phys. Rev. {\bf D90},
  112004 (2014)}, \href{http://arxiv.org/abs/1408.5373}{{\tt arXiv:1408.5373
  [hep-ex]}}\relax
\mciteBstWouldAddEndPuncttrue
\mciteSetBstMidEndSepPunct{\mcitedefaultmidpunct}
{\mcitedefaultendpunct}{\mcitedefaultseppunct}\relax
\EndOfBibitem
\bibitem{Chen:2005zv}
{Belle} collaboration, K.~F. Chen {\em et al.},
  \href{http://dx.doi.org/10.1103/PhysRevLett.94.221804}{Phys. Rev. Lett. {\bf
  94},  221804 (2005)}, \href{http://arxiv.org/abs/hep-ex/0503013}{{\tt
  arXiv:hep-ex/0503013 [hep-ex]}}\relax
\mciteBstWouldAddEndPuncttrue
\mciteSetBstMidEndSepPunct{\mcitedefaultmidpunct}
{\mcitedefaultendpunct}{\mcitedefaultseppunct}\relax
\EndOfBibitem
\bibitem{Aaij:2014bsa}
{LHCb} collaboration, R.~Aaij {\em et al.},
  \href{http://dx.doi.org/10.1007/JHEP09(2014)177}{JHEP {\bf 09},  177 (2014)},
  \href{http://arxiv.org/abs/1408.0978}{{\tt arXiv:1408.0978 [hep-ex]}}\relax
\mciteBstWouldAddEndPuncttrue
\mciteSetBstMidEndSepPunct{\mcitedefaultmidpunct}
{\mcitedefaultendpunct}{\mcitedefaultseppunct}\relax
\EndOfBibitem
\bibitem{Aaltonen:2014vra}
{CDF} collaboration, T.~A. Aaltonen {\em et al.},
  \href{http://dx.doi.org/10.1103/PhysRevLett.113.242001}{Phys. Rev. Lett. {\bf
  113},  242001 (2014)}, \href{http://arxiv.org/abs/1403.5586}{{\tt
  arXiv:1403.5586 [hep-ex]}}\relax
\mciteBstWouldAddEndPuncttrue
\mciteSetBstMidEndSepPunct{\mcitedefaultmidpunct}
{\mcitedefaultendpunct}{\mcitedefaultseppunct}\relax
\EndOfBibitem
\bibitem{Aaij:2017ngy}
{LHCb} collaboration, R.~Aaij {\em et al.},
  \href{http://dx.doi.org/10.1103/PhysRevLett.120.261801}{Phys. Rev. Lett. {\bf
  120},  261801 (2018)}, \href{http://arxiv.org/abs/1712.09320}{{\tt
  arXiv:1712.09320 [hep-ex]}}\relax
\mciteBstWouldAddEndPuncttrue
\mciteSetBstMidEndSepPunct{\mcitedefaultmidpunct}
{\mcitedefaultendpunct}{\mcitedefaultseppunct}\relax
\EndOfBibitem
\bibitem{Aaij:2014tpa}
{LHCb} collaboration, R.~Aaij {\em et al.},
  \href{http://dx.doi.org/10.1007/JHEP05(2014)069}{JHEP {\bf 05},  069 (2014)},
  \href{http://arxiv.org/abs/1403.2888}{{\tt arXiv:1403.2888 [hep-ex]}}\relax
\mciteBstWouldAddEndPuncttrue
\mciteSetBstMidEndSepPunct{\mcitedefaultmidpunct}
{\mcitedefaultendpunct}{\mcitedefaultseppunct}\relax
\EndOfBibitem
\bibitem{Lees:2014uoa}
{\babar} collaboration, J.~P. Lees {\em et al.},
  \href{http://dx.doi.org/10.1103/PhysRevD.90.092001}{Phys. Rev. {\bf D90},
  092001 (2014)}, \href{http://arxiv.org/abs/1406.0534}{{\tt arXiv:1406.0534
  [hep-ex]}}\relax
\mciteBstWouldAddEndPuncttrue
\mciteSetBstMidEndSepPunct{\mcitedefaultmidpunct}
{\mcitedefaultendpunct}{\mcitedefaultseppunct}\relax
\EndOfBibitem
\bibitem{Nishida:2003paa}
{Belle} collaboration, S.~Nishida {\em et al.},
  \href{http://dx.doi.org/10.1103/PhysRevLett.93.031803}{Phys. Rev. Lett. {\bf
  93},  031803 (2004)}, \href{http://arxiv.org/abs/hep-ex/0308038}{{\tt
  arXiv:hep-ex/0308038 [hep-ex]}}\relax
\mciteBstWouldAddEndPuncttrue
\mciteSetBstMidEndSepPunct{\mcitedefaultmidpunct}
{\mcitedefaultendpunct}{\mcitedefaultseppunct}\relax
\EndOfBibitem
\bibitem{Aaij:2018tlk}
{LHCb} collaboration, R.~Aaij {\em et al.},
  \href{http://dx.doi.org/10.1016/j.physletb.2018.10.039}{Phys. Lett. {\bf
  B787},  124--133 (2018)}, \href{http://arxiv.org/abs/1807.06544}{{\tt
  arXiv:1807.06544 [hep-ex]}}\relax
\mciteBstWouldAddEndPuncttrue
\mciteSetBstMidEndSepPunct{\mcitedefaultmidpunct}
{\mcitedefaultendpunct}{\mcitedefaultseppunct}\relax
\EndOfBibitem
\bibitem{Aaij:2017mib}
{LHCb} collaboration, R.~Aaij {\em et al.},
  \href{http://dx.doi.org/10.1007/JHEP06(2017)108}{JHEP {\bf 06},  108 (2017)},
  \href{http://arxiv.org/abs/1703.00256}{{\tt arXiv:1703.00256 [hep-ex]}}\relax
\mciteBstWouldAddEndPuncttrue
\mciteSetBstMidEndSepPunct{\mcitedefaultmidpunct}
{\mcitedefaultendpunct}{\mcitedefaultseppunct}\relax
\EndOfBibitem
\bibitem{Aaij:2016cla}
{LHCb} collaboration, R.~Aaij {\em et al.},
  \href{http://dx.doi.org/10.1038/nphys4021}{Nature Phys. {\bf 13},  391
  (2017)}, \href{http://arxiv.org/abs/1609.05216}{{\tt arXiv:1609.05216
  [hep-ex]}}\relax
\mciteBstWouldAddEndPuncttrue
\mciteSetBstMidEndSepPunct{\mcitedefaultmidpunct}
{\mcitedefaultendpunct}{\mcitedefaultseppunct}\relax
\EndOfBibitem
\bibitem{Aaij:2018lsx}
{LHCb} collaboration, R.~Aaij {\em et al.},
  \href{http://arxiv.org/abs/1805.03941}{{\tt arXiv:1805.03941 [hep-ex]}}\relax
\mciteBstWouldAddEndPuncttrue
\mciteSetBstMidEndSepPunct{\mcitedefaultmidpunct}
{\mcitedefaultendpunct}{\mcitedefaultseppunct}\relax
\EndOfBibitem
\bibitem{Aaij:2017wgt}
{LHCb} collaboration, R.~Aaij {\em et al.},
  \href{http://dx.doi.org/10.1007/JHEP03(2018)140}{JHEP {\bf 03},  140 (2018)},
  \href{http://arxiv.org/abs/1712.08683}{{\tt arXiv:1712.08683 [hep-ex]}}\relax
\mciteBstWouldAddEndPuncttrue
\mciteSetBstMidEndSepPunct{\mcitedefaultmidpunct}
{\mcitedefaultendpunct}{\mcitedefaultseppunct}\relax
\EndOfBibitem
\bibitem{LHCb-CONF-2018-001}
{LHCb} collaboration,
  \href{http://cdsweb.cern.ch/record/2314360}{LHCb-CONF-2018-001}, {2018}\relax
\mciteBstWouldAddEndPuncttrue
\mciteSetBstMidEndSepPunct{\mcitedefaultmidpunct}
{\mcitedefaultendpunct}{\mcitedefaultseppunct}\relax
\EndOfBibitem
\bibitem{Aaij:2016xxs}
{LHCb} collaboration, R.~Aaij {\em et al.},
  \href{http://dx.doi.org/10.1016/j.physletb.2016.05.074}{Phys. Lett. {\bf
  B759},  313 (2016)}, \href{http://arxiv.org/abs/1603.07037}{{\tt
  arXiv:1603.07037 [hep-ex]}}\relax
\mciteBstWouldAddEndPuncttrue
\mciteSetBstMidEndSepPunct{\mcitedefaultmidpunct}
{\mcitedefaultendpunct}{\mcitedefaultseppunct}\relax
\EndOfBibitem
\bibitem{Staric:2007dt}
{Belle} collaboration, M.~Staric {\em et al.},
  \href{http://dx.doi.org/10.1103/PhysRevLett.98.211803}{Phys. Rev. Lett. {\bf
  98},  211803 (2007)}, \href{http://arxiv.org/abs/hep-ex/0703036}{{\tt
  arXiv:hep-ex/0703036}}\relax
\mciteBstWouldAddEndPuncttrue
\mciteSetBstMidEndSepPunct{\mcitedefaultmidpunct}
{\mcitedefaultendpunct}{\mcitedefaultseppunct}\relax
\EndOfBibitem
\bibitem{Aubert:2007wf}
{\babar} collaboration, B.~Aubert {\em et al.},
  \href{http://dx.doi.org/10.1103/PhysRevLett.98.211802}{Phys. Rev. Lett. {\bf
  98},  211802 (2007)}, \href{http://arxiv.org/abs/hep-ex/0703020}{{\tt
  arXiv:hep-ex/0703020}}\relax
\mciteBstWouldAddEndPuncttrue
\mciteSetBstMidEndSepPunct{\mcitedefaultmidpunct}
{\mcitedefaultendpunct}{\mcitedefaultseppunct}\relax
\EndOfBibitem
\bibitem{Aaltonen:2007uc}
{CDF} collaboration, T.~Aaltonen {\em et al.},
  \href{http://dx.doi.org/10.1103/PhysRevLett.100.121802}{Phys. Rev. Lett. {\bf
  100},  121802 (2008)}, \href{http://arxiv.org/abs/0712.1567}{{\tt
  arXiv:0712.1567 [hep-ex]}}\relax
\mciteBstWouldAddEndPuncttrue
\mciteSetBstMidEndSepPunct{\mcitedefaultmidpunct}
{\mcitedefaultendpunct}{\mcitedefaultseppunct}\relax
\EndOfBibitem
\bibitem{Aaij:2013wda}
{LHCb} collaboration, R.~Aaij {\em et al.},
  \href{http://dx.doi.org/10.1103/PhysRevLett.111.251801}{Phys. Rev. Lett. {\bf
  111},  251801 (2013)}, \href{http://arxiv.org/abs/1309.6534}{{\tt
  arXiv:1309.6534 [hep-ex]}}\relax
\mciteBstWouldAddEndPuncttrue
\mciteSetBstMidEndSepPunct{\mcitedefaultmidpunct}
{\mcitedefaultendpunct}{\mcitedefaultseppunct}\relax
\EndOfBibitem
\bibitem{Bergmann:2000id}
S.~Bergmann, Y.~Grossman, Z.~Ligeti, Y.~Nir, and A.~A. Petrov,
  \href{http://dx.doi.org/10.1016/S0370-2693(00)00772-3}{Phys. Lett. {\bf
  B486},  418 (2000)}, \href{http://arxiv.org/abs/hep-ph/0005181}{{\tt
  arXiv:hep-ph/0005181 [hep-ph]}}\relax
\mciteBstWouldAddEndPuncttrue
\mciteSetBstMidEndSepPunct{\mcitedefaultmidpunct}
{\mcitedefaultendpunct}{\mcitedefaultseppunct}\relax
\EndOfBibitem
\bibitem{Bigi:2000wn}
I.~I.~Y. Bigi and N.~G. Uraltsev,
  \href{http://dx.doi.org/10.1016/S0550-3213(00)00604-0}{Nucl. Phys. {\bf
  B592},  92 (2001)}, \href{http://arxiv.org/abs/hep-ph/0005089}{{\tt
  arXiv:hep-ph/0005089}}\relax
\mciteBstWouldAddEndPuncttrue
\mciteSetBstMidEndSepPunct{\mcitedefaultmidpunct}
{\mcitedefaultendpunct}{\mcitedefaultseppunct}\relax
\EndOfBibitem
\bibitem{Petrov:2003un}
A.~A. Petrov, \href{http://arxiv.org/abs/hep-ph/0311371}{{\tt
  arXiv:hep-ph/0311371}} (2003)\relax
\mciteBstWouldAddEndPuncttrue
\mciteSetBstMidEndSepPunct{\mcitedefaultmidpunct}
{\mcitedefaultendpunct}{\mcitedefaultseppunct}\relax
\EndOfBibitem
\bibitem{Petrov:2004rf}
A.~A. Petrov, \href{http://dx.doi.org/10.1016/j.nuclphysbps.2005.01.057}{Nucl.
  Phys. Proc. Suppl. {\bf 142},  333 (2005)},
  \href{http://arxiv.org/abs/hep-ph/0409130}{{\tt arXiv:hep-ph/0409130}}\relax
\mciteBstWouldAddEndPuncttrue
\mciteSetBstMidEndSepPunct{\mcitedefaultmidpunct}
{\mcitedefaultendpunct}{\mcitedefaultseppunct}\relax
\EndOfBibitem
\bibitem{Falk:2004wg}
A.~F. Falk, Y.~Grossman, Z.~Ligeti, Y.~Nir, and A.~A. Petrov,
  \href{http://dx.doi.org/10.1103/PhysRevD.69.114021}{Phys. Rev. {\bf D69},
  114021 (2004)}, \href{http://arxiv.org/abs/hep-ph/0402204}{{\tt
  arXiv:hep-ph/0402204}}\relax
\mciteBstWouldAddEndPuncttrue
\mciteSetBstMidEndSepPunct{\mcitedefaultmidpunct}
{\mcitedefaultendpunct}{\mcitedefaultseppunct}\relax
\EndOfBibitem
\bibitem{Asner:2012xb}
{CLEO} collaboration, D.~Asner {\em et al.},
  \href{http://dx.doi.org/10.1103/PhysRevD.86.112001}{Phys. Rev. {\bf D86},
  112001 (2012)}, \href{http://arxiv.org/abs/1210.0939}{{\tt arXiv:1210.0939
  [hep-ex]}}\relax
\mciteBstWouldAddEndPuncttrue
\mciteSetBstMidEndSepPunct{\mcitedefaultmidpunct}
{\mcitedefaultendpunct}{\mcitedefaultseppunct}\relax
\EndOfBibitem
\bibitem{Aaij:2014gsa}
{LHCb} collaboration, R.~Aaij {\em et al.},
  \href{http://dx.doi.org/10.1007/JHEP07(2014)041}{JHEP {\bf 07},  041 (2014)},
  \href{http://arxiv.org/abs/1405.2797}{{\tt arXiv:1405.2797 [hep-ex]}}\relax
\mciteBstWouldAddEndPuncttrue
\mciteSetBstMidEndSepPunct{\mcitedefaultmidpunct}
{\mcitedefaultendpunct}{\mcitedefaultseppunct}\relax
\EndOfBibitem
\bibitem{Aubert:2008zh}
{\babar} collaboration, B.~Aubert {\em et al.},
  \href{http://dx.doi.org/10.1103/PhysRevLett.103.211801}{Phys. Rev. Lett. {\bf
  103},  211801 (2009)}, \href{http://arxiv.org/abs/0807.4544}{{\tt
  arXiv:0807.4544 [hep-ex]}}\relax
\mciteBstWouldAddEndPuncttrue
\mciteSetBstMidEndSepPunct{\mcitedefaultmidpunct}
{\mcitedefaultendpunct}{\mcitedefaultseppunct}\relax
\EndOfBibitem
\bibitem{Combos:1999}
 Documentation available at, 1999, {\small
  \url{http://www.slac.stanford.edu/xorg/hflav/docs/combos.ps}}\relax
\mciteBstWouldAddEndPuncttrue
\mciteSetBstMidEndSepPunct{\mcitedefaultmidpunct}
{\mcitedefaultendpunct}{\mcitedefaultseppunct}\relax
\EndOfBibitem
\bibitem{Aaij:2016rhq}
{LHCb} collaboration, R.~Aaij {\em et al.},
  \href{http://dx.doi.org/10.1103/PhysRevLett.116.241801}{Phys. Rev. Lett. {\bf
  116},  241801 (2016)}, \href{http://arxiv.org/abs/1602.07224}{{\tt
  arXiv:1602.07224 [hep-ex]}}\relax
\mciteBstWouldAddEndPuncttrue
\mciteSetBstMidEndSepPunct{\mcitedefaultmidpunct}
{\mcitedefaultendpunct}{\mcitedefaultseppunct}\relax
\EndOfBibitem
\bibitem{Aitala:1996vz}
{Fermilab E791} collaboration, E.~M. Aitala {\em et al.},
  \href{http://dx.doi.org/10.1103/PhysRevLett.77.2384}{Phys. Rev. Lett. {\bf
  77},  2384 (1996)}, \href{http://arxiv.org/abs/hep-ex/9606016}{{\tt
  arXiv:hep-ex/9606016}}\relax
\mciteBstWouldAddEndPuncttrue
\mciteSetBstMidEndSepPunct{\mcitedefaultmidpunct}
{\mcitedefaultendpunct}{\mcitedefaultseppunct}\relax
\EndOfBibitem
\bibitem{Cawlfield:2005ze}
{CLEO} collaboration, C.~Cawlfield {\em et al.},
  \href{http://dx.doi.org/10.1103/PhysRevD.71.077101}{Phys. Rev. {\bf D71},
  077101 (2005)}, \href{http://arxiv.org/abs/hep-ex/0502012}{{\tt
  arXiv:hep-ex/0502012}}\relax
\mciteBstWouldAddEndPuncttrue
\mciteSetBstMidEndSepPunct{\mcitedefaultmidpunct}
{\mcitedefaultendpunct}{\mcitedefaultseppunct}\relax
\EndOfBibitem
\bibitem{Aubert:2007aa}
{\babar} collaboration, B.~Aubert {\em et al.},
  \href{http://dx.doi.org/10.1103/PhysRevD.76.014018}{Phys. Rev. {\bf D76},
  014018 (2007)}, \href{http://arxiv.org/abs/0705.0704}{{\tt arXiv:0705.0704
  [hep-ex]}}\relax
\mciteBstWouldAddEndPuncttrue
\mciteSetBstMidEndSepPunct{\mcitedefaultmidpunct}
{\mcitedefaultendpunct}{\mcitedefaultseppunct}\relax
\EndOfBibitem
\bibitem{Bitenc:2008bk}
{Belle} collaboration, U.~Bitenc {\em et al.},
  \href{http://dx.doi.org/10.1103/PhysRevD.77.112003}{Phys. Rev. {\bf D77},
  112003 (2008)}, \href{http://arxiv.org/abs/0802.2952}{{\tt arXiv:0802.2952
  [hep-ex]}}\relax
\mciteBstWouldAddEndPuncttrue
\mciteSetBstMidEndSepPunct{\mcitedefaultmidpunct}
{\mcitedefaultendpunct}{\mcitedefaultseppunct}\relax
\EndOfBibitem
\bibitem{Zhang:2006dp}
{Belle} collaboration, L.~M. Zhang {\em et al.},
  \href{http://dx.doi.org/10.1103/PhysRevLett.96.151801}{Phys. Rev. Lett. {\bf
  96},  151801 (2006)}, \href{http://arxiv.org/abs/hep-ex/0601029}{{\tt
  arXiv:hep-ex/0601029}}\relax
\mciteBstWouldAddEndPuncttrue
\mciteSetBstMidEndSepPunct{\mcitedefaultmidpunct}
{\mcitedefaultendpunct}{\mcitedefaultseppunct}\relax
\EndOfBibitem
\bibitem{Ko:2014qvu}
{Belle} collaboration, B.~R. Ko {\em et al.},
  \href{http://dx.doi.org/10.1103/PhysRevLett.112.111801}{Phys. Rev. Lett. {\bf
  112},  111801 (2014)}, \href{http://arxiv.org/abs/1401.3402}{{\tt
  arXiv:1401.3402 [hep-ex]}}\relax
\mciteBstWouldAddEndPuncttrue
\mciteSetBstMidEndSepPunct{\mcitedefaultmidpunct}
{\mcitedefaultendpunct}{\mcitedefaultseppunct}\relax
\EndOfBibitem
\bibitem{Aaltonen:2013pja}
{CDF} collaboration, T.~A. Aaltonen {\em et al.},
  \href{http://dx.doi.org/10.1103/PhysRevLett.111.231802}{Phys. Rev. Lett. {\bf
  111},  231802 (2013)}, \href{http://arxiv.org/abs/1309.4078}{{\tt
  arXiv:1309.4078 [hep-ex]}}\relax
\mciteBstWouldAddEndPuncttrue
\mciteSetBstMidEndSepPunct{\mcitedefaultmidpunct}
{\mcitedefaultendpunct}{\mcitedefaultseppunct}\relax
\EndOfBibitem
\bibitem{Aaij:2017urz}
{LHCb} collaboration, R.~Aaij {\em et al.},
  \href{http://dx.doi.org/10.1103/PhysRevD.97.031101}{Phys. Rev. {\bf D97},
  031101 (2018)}, \href{http://arxiv.org/abs/1712.03220}{{\tt arXiv:1712.03220
  [hep-ex]}}\relax
\mciteBstWouldAddEndPuncttrue
\mciteSetBstMidEndSepPunct{\mcitedefaultmidpunct}
{\mcitedefaultendpunct}{\mcitedefaultseppunct}\relax
\EndOfBibitem
\bibitem{Peng:2014oda}
{Belle} collaboration, T.~Peng {\em et al.},
  \href{http://dx.doi.org/10.1103/PhysRevD.89.091103}{Phys. Rev. {\bf D89},
  091103 (2014)}, \href{http://arxiv.org/abs/1404.2412}{{\tt arXiv:1404.2412
  [hep-ex]}}\relax
\mciteBstWouldAddEndPuncttrue
\mciteSetBstMidEndSepPunct{\mcitedefaultmidpunct}
{\mcitedefaultendpunct}{\mcitedefaultseppunct}\relax
\EndOfBibitem
\bibitem{delAmoSanchez:2010xz}
{\babar} collaboration, P.~del Amo~Sanchez {\em et al.},
  \href{http://dx.doi.org/10.1103/PhysRevLett.105.081803}{Phys. Rev. Lett. {\bf
  105},  081803 (2010)}, \href{http://arxiv.org/abs/1004.5053}{{\tt
  arXiv:1004.5053 [hep-ex]}}\relax
\mciteBstWouldAddEndPuncttrue
\mciteSetBstMidEndSepPunct{\mcitedefaultmidpunct}
{\mcitedefaultendpunct}{\mcitedefaultseppunct}\relax
\EndOfBibitem
\bibitem{Aaij:2015xoa}
{LHCb} collaboration, R.~Aaij {\em et al.},
  \href{http://dx.doi.org/10.1007/JHEP04(2016)033}{JHEP {\bf 04},  033 (2016)},
  \href{http://arxiv.org/abs/1510.01664}{{\tt arXiv:1510.01664 [hep-ex]}}\relax
\mciteBstWouldAddEndPuncttrue
\mciteSetBstMidEndSepPunct{\mcitedefaultmidpunct}
{\mcitedefaultendpunct}{\mcitedefaultseppunct}\relax
\EndOfBibitem
\bibitem{Aaij:2019jot}
{LHCb} collaboration, R.~Aaij {\em et al.}, Submitted to: Phys. Rev. Lett.
  (2019), \href{http://arxiv.org/abs/1903.03074}{{\tt arXiv:1903.03074
  [hep-ex]}}\relax
\mciteBstWouldAddEndPuncttrue
\mciteSetBstMidEndSepPunct{\mcitedefaultmidpunct}
{\mcitedefaultendpunct}{\mcitedefaultseppunct}\relax
\EndOfBibitem
\bibitem{Lees:2016gom}
{\babar} collaboration, J.~P. Lees {\em et al.},
  \href{http://dx.doi.org/10.1103/PhysRevD.93.112014}{Phys. Rev. {\bf D93},
  112014 (2016)}, \href{http://arxiv.org/abs/1604.00857}{{\tt arXiv:1604.00857
  [hep-ex]}}\relax
\mciteBstWouldAddEndPuncttrue
\mciteSetBstMidEndSepPunct{\mcitedefaultmidpunct}
{\mcitedefaultendpunct}{\mcitedefaultseppunct}\relax
\EndOfBibitem
\bibitem{DiCanto:2018tsd}
A.~Di~Canto, J.~Garra~Ticó, T.~Gershon, N.~Jurik, M.~Martinelli, T.~Pilař,
  S.~Stahl, and D.~Tonelli,
  \href{http://dx.doi.org/10.1103/PhysRevD.99.012007}{Phys. Rev. {\bf D99},
  012007 (2019)}, \href{http://arxiv.org/abs/1811.01032}{{\tt arXiv:1811.01032
  [hep-ex]}}\relax
\mciteBstWouldAddEndPuncttrue
\mciteSetBstMidEndSepPunct{\mcitedefaultmidpunct}
{\mcitedefaultendpunct}{\mcitedefaultseppunct}\relax
\EndOfBibitem
\bibitem{Aubert:2007if}
{\babar} collaboration, B.~Aubert {\em et al.},
  \href{http://dx.doi.org/10.1103/PhysRevLett.100.061803}{Phys. Rev. Lett. {\bf
  100},  061803 (2008)}, \href{http://arxiv.org/abs/0709.2715}{{\tt
  arXiv:0709.2715 [hep-ex]}}\relax
\mciteBstWouldAddEndPuncttrue
\mciteSetBstMidEndSepPunct{\mcitedefaultmidpunct}
{\mcitedefaultendpunct}{\mcitedefaultseppunct}\relax
\EndOfBibitem
\bibitem{cdf_public_note_10784}
{CDF} collaboration, CDF note 10784, 2012, {\small
  \url{{http://www-cdf.fnal.gov/physics/new/bottom/120216.blessed-CPVcharm10fb/}}}\relax
\mciteBstWouldAddEndPuncttrue
\mciteSetBstMidEndSepPunct{\mcitedefaultmidpunct}
{\mcitedefaultendpunct}{\mcitedefaultseppunct}\relax
\EndOfBibitem
\bibitem{Collaboration:2012qw}
{CDF} collaboration, T.~Aaltonen {\em et al.},
  \href{http://dx.doi.org/10.1103/PhysRevLett.109.111801}{Phys. Rev. Lett. {\bf
  109},  111801 (2012)}, \href{http://arxiv.org/abs/1207.2158}{{\tt
  arXiv:1207.2158 [hep-ex]}}\relax
\mciteBstWouldAddEndPuncttrue
\mciteSetBstMidEndSepPunct{\mcitedefaultmidpunct}
{\mcitedefaultendpunct}{\mcitedefaultseppunct}\relax
\EndOfBibitem
\bibitem{Aaij:2019kcg}
{LHCb} collaboration, R.~Aaij {\em et al.},
  \href{http://arxiv.org/abs/1903.08726}{{\tt arXiv:1903.08726 [hep-ex]}}\relax
\mciteBstWouldAddEndPuncttrue
\mciteSetBstMidEndSepPunct{\mcitedefaultmidpunct}
{\mcitedefaultendpunct}{\mcitedefaultseppunct}\relax
\EndOfBibitem
\bibitem{Aitala:1999dt}
{Fermilab E791} collaboration, E.~M. Aitala {\em et al.},
  \href{http://dx.doi.org/10.1103/PhysRevLett.83.32}{Phys. Rev. Lett. {\bf 83},
   32 (1999)}, \href{http://arxiv.org/abs/hep-ex/9903012}{{\tt
  arXiv:hep-ex/9903012}}\relax
\mciteBstWouldAddEndPuncttrue
\mciteSetBstMidEndSepPunct{\mcitedefaultmidpunct}
{\mcitedefaultendpunct}{\mcitedefaultseppunct}\relax
\EndOfBibitem
\bibitem{Link:2000cu}
{FOCUS} collaboration, J.~M. Link {\em et al.},
  \href{http://dx.doi.org/10.1016/S0370-2693(00)00694-8}{Phys. Lett. {\bf
  B485},  62 (2000)}, \href{http://arxiv.org/abs/hep-ex/0004034}{{\tt
  arXiv:hep-ex/0004034}}\relax
\mciteBstWouldAddEndPuncttrue
\mciteSetBstMidEndSepPunct{\mcitedefaultmidpunct}
{\mcitedefaultendpunct}{\mcitedefaultseppunct}\relax
\EndOfBibitem
\bibitem{Csorna:2001ww}
{CLEO} collaboration, S.~E. Csorna {\em et al.},
  \href{http://dx.doi.org/10.1103/PhysRevD.65.092001}{Phys. Rev. {\bf D65},
  092001 (2002)}, \href{http://arxiv.org/abs/hep-ex/0111024}{{\tt
  arXiv:hep-ex/0111024}}\relax
\mciteBstWouldAddEndPuncttrue
\mciteSetBstMidEndSepPunct{\mcitedefaultmidpunct}
{\mcitedefaultendpunct}{\mcitedefaultseppunct}\relax
\EndOfBibitem
\bibitem{Zupanc:2009sy}
{Belle} collaboration, A.~Zupanc {\em et al.},
  \href{http://dx.doi.org/10.1103/PhysRevD.80.052006}{Phys. Rev. {\bf D80},
  052006 (2009)}, \href{http://arxiv.org/abs/0905.4185}{{\tt arXiv:0905.4185
  [hep-ex]}}\relax
\mciteBstWouldAddEndPuncttrue
\mciteSetBstMidEndSepPunct{\mcitedefaultmidpunct}
{\mcitedefaultendpunct}{\mcitedefaultseppunct}\relax
\EndOfBibitem
\bibitem{Lees:2012qh}
{\babar} collaboration, J.~P. Lees {\em et al.},
  \href{http://dx.doi.org/10.1103/PhysRevD.87.012004}{Phys. Rev. {\bf D87},
  012004 (2013)}, \href{http://arxiv.org/abs/1209.3896}{{\tt arXiv:1209.3896
  [hep-ex]}}\relax
\mciteBstWouldAddEndPuncttrue
\mciteSetBstMidEndSepPunct{\mcitedefaultmidpunct}
{\mcitedefaultendpunct}{\mcitedefaultseppunct}\relax
\EndOfBibitem
\bibitem{Ablikim:2015hih}
{BESIII} collaboration, M.~Ablikim {\em et al.},
  \href{http://dx.doi.org/10.1016/j.physletb.2015.04.008}{Phys. Lett. {\bf
  B744},  339 (2015)}, \href{http://arxiv.org/abs/1501.01378}{{\tt
  arXiv:1501.01378 [hep-ex]}}\relax
\mciteBstWouldAddEndPuncttrue
\mciteSetBstMidEndSepPunct{\mcitedefaultmidpunct}
{\mcitedefaultendpunct}{\mcitedefaultseppunct}\relax
\EndOfBibitem
\bibitem{Staric:2015sta}
{Belle} collaboration, M.~Staric {\em et al.},
  \href{http://dx.doi.org/10.1016/j.physletb.2015.12.025}{Phys. Lett. {\bf
  B753},  412--418 (2016)}, \href{http://arxiv.org/abs/1509.08266}{{\tt
  arXiv:1509.08266 [hep-ex]}}\relax
\mciteBstWouldAddEndPuncttrue
\mciteSetBstMidEndSepPunct{\mcitedefaultmidpunct}
{\mcitedefaultendpunct}{\mcitedefaultseppunct}\relax
\EndOfBibitem
\bibitem{Aaij:2018qiw}
{LHCb} collaboration, R.~Aaij {\em et al.},
  \href{http://dx.doi.org/10.1103/PhysRevLett.122.011802}{Phys. Rev. Lett. {\bf
  122},  011802 (2019)}, \href{http://arxiv.org/abs/1810.06874}{{\tt
  arXiv:1810.06874 [hep-ex]}}\relax
\mciteBstWouldAddEndPuncttrue
\mciteSetBstMidEndSepPunct{\mcitedefaultmidpunct}
{\mcitedefaultendpunct}{\mcitedefaultseppunct}\relax
\EndOfBibitem
\bibitem{Aaltonen:2014efa}
{CDF} collaboration, T.~A. Aaltonen {\em et al.},
  \href{http://dx.doi.org/10.1103/PhysRevD.90.111103}{Phys. Rev. {\bf D90},
  111103 (2014)}, \href{http://arxiv.org/abs/1410.5435}{{\tt arXiv:1410.5435
  [hep-ex]}}\relax
\mciteBstWouldAddEndPuncttrue
\mciteSetBstMidEndSepPunct{\mcitedefaultmidpunct}
{\mcitedefaultendpunct}{\mcitedefaultseppunct}\relax
\EndOfBibitem
\bibitem{Aaij:2015yda}
{LHCb} collaboration, R.~Aaij {\em et al.},
  \href{http://dx.doi.org/10.1007/JHEP04(2015)043}{JHEP {\bf 04},  043 (2015)},
  \href{http://arxiv.org/abs/1501.06777}{{\tt arXiv:1501.06777 [hep-ex]}}\relax
\mciteBstWouldAddEndPuncttrue
\mciteSetBstMidEndSepPunct{\mcitedefaultmidpunct}
{\mcitedefaultendpunct}{\mcitedefaultseppunct}\relax
\EndOfBibitem
\bibitem{Aaij:2017idz}
{LHCb} collaboration, R.~Aaij {\em et al.},
  \href{http://dx.doi.org/10.1103/PhysRevLett.118.261803}{Phys. Rev. Lett. {\bf
  118},  261803 (2017)}, \href{http://arxiv.org/abs/1702.06490}{{\tt
  arXiv:1702.06490 [hep-ex]}}\relax
\mciteBstWouldAddEndPuncttrue
\mciteSetBstMidEndSepPunct{\mcitedefaultmidpunct}
{\mcitedefaultendpunct}{\mcitedefaultseppunct}\relax
\EndOfBibitem
\bibitem{MINUIT:webpage}
 {CERN Program Library Long Writeup D506}, 1994,
  \url{http://wwwasdoc.web.cern.ch/wwwasdoc/minuit/minmain.html}\relax
\mciteBstWouldAddEndPuncttrue
\mciteSetBstMidEndSepPunct{\mcitedefaultmidpunct}
{\mcitedefaultendpunct}{\mcitedefaultseppunct}\relax
\EndOfBibitem
\bibitem{Grossman:2009mn}
Y.~Grossman, Y.~Nir, and G.~Perez,
  \href{http://dx.doi.org/10.1103/PhysRevLett.103.071602}{Phys. Rev. Lett. {\bf
  103},  071602 (2009)}, \href{http://arxiv.org/abs/0904.0305}{{\tt
  arXiv:0904.0305 [hep-ph]}}\relax
\mciteBstWouldAddEndPuncttrue
\mciteSetBstMidEndSepPunct{\mcitedefaultmidpunct}
{\mcitedefaultendpunct}{\mcitedefaultseppunct}\relax
\EndOfBibitem
\bibitem{Kagan:2009gb}
A.~L. Kagan and M.~D. Sokoloff,
  \href{http://dx.doi.org/10.1103/PhysRevD.80.076008}{Phys. Rev. {\bf D80},
  076008 (2009)}, \href{http://arxiv.org/abs/0907.3917}{{\tt arXiv:0907.3917
  [hep-ph]}}\relax
\mciteBstWouldAddEndPuncttrue
\mciteSetBstMidEndSepPunct{\mcitedefaultmidpunct}
{\mcitedefaultendpunct}{\mcitedefaultseppunct}\relax
\EndOfBibitem
\bibitem{Ciuchini:2007cw}
M.~Ciuchini {\em et al.},
  \href{http://dx.doi.org/10.1016/j.physletb.2007.08.055}{Phys. Lett. {\bf
  B655},  162 (2007)}, \href{http://arxiv.org/abs/hep-ph/0703204}{{\tt
  arXiv:hep-ph/0703204 [hep-ph]}}\relax
\mciteBstWouldAddEndPuncttrue
\mciteSetBstMidEndSepPunct{\mcitedefaultmidpunct}
{\mcitedefaultendpunct}{\mcitedefaultseppunct}\relax
\EndOfBibitem
\bibitem{Bigi:2000yz}
I.~I.~Y. Bigi and A.~I. Sanda, Camb. Monogr. Part. Phys. Nucl. Phys. Cosmol.
  {\bf 9},  1 (2000)\relax
\mciteBstWouldAddEndPuncttrue
\mciteSetBstMidEndSepPunct{\mcitedefaultmidpunct}
{\mcitedefaultendpunct}{\mcitedefaultseppunct}\relax
\EndOfBibitem
\bibitem{Nir:1999mg}
Y.~Nir, \href{http://arxiv.org/abs/hep-ph/9911321}{{\tt arXiv:hep-ph/9911321}}
  (1999)\relax
\mciteBstWouldAddEndPuncttrue
\mciteSetBstMidEndSepPunct{\mcitedefaultmidpunct}
{\mcitedefaultendpunct}{\mcitedefaultseppunct}\relax
\EndOfBibitem
\bibitem{Buccella:1994nf}
F.~Buccella, M.~Lusignoli, G.~Miele, A.~Pugliese, and P.~Santorelli,
  \href{http://dx.doi.org/10.1103/PhysRevD.51.3478}{Phys. Rev. {\bf D51},  3478
  (1995)}, \href{http://arxiv.org/abs/hep-ph/9411286}{{\tt
  arXiv:hep-ph/9411286}}\relax
\mciteBstWouldAddEndPuncttrue
\mciteSetBstMidEndSepPunct{\mcitedefaultmidpunct}
{\mcitedefaultendpunct}{\mcitedefaultseppunct}\relax
\EndOfBibitem
\bibitem{Grossman:2012aa}
Y.~Grossman and Y.~Nir, \href{http://dx.doi.org/10.1007/JHEP04(2012)002}{JHEP
  {\bf 04},  002 (2012)}, \href{http://arxiv.org/abs/1110.3790}{{\tt
  arXiv:1110.3790 [hep-ph]}}\relax
\mciteBstWouldAddEndPuncttrue
\mciteSetBstMidEndSepPunct{\mcitedefaultmidpunct}
{\mcitedefaultendpunct}{\mcitedefaultseppunct}\relax
\EndOfBibitem
\bibitem{Aaij:2017xva}
{LHCb} collaboration, R.~Aaij {\em et al.},
  \href{http://dx.doi.org/10.1007/JHEP03(2018)182}{JHEP {\bf 03},  182 (2018)},
  \href{http://arxiv.org/abs/1712.07051}{{\tt arXiv:1712.07051 [hep-ex]}}\relax
\mciteBstWouldAddEndPuncttrue
\mciteSetBstMidEndSepPunct{\mcitedefaultmidpunct}
{\mcitedefaultendpunct}{\mcitedefaultseppunct}\relax
\EndOfBibitem
\bibitem{Brod:2012ud}
J.~Brod, Y.~Grossman, A.~L. Kagan, and J.~Zupan,
  \href{http://dx.doi.org/10.1007/JHEP10(2012)161}{JHEP {\bf 10},  161 (2012)},
  \href{http://arxiv.org/abs/1203.6659}{{\tt arXiv:1203.6659 [hep-ph]}}\relax
\mciteBstWouldAddEndPuncttrue
\mciteSetBstMidEndSepPunct{\mcitedefaultmidpunct}
{\mcitedefaultendpunct}{\mcitedefaultseppunct}\relax
\EndOfBibitem
\bibitem{Aubert:2008yd}
{\babar} collaboration, B.~Aubert {\em et al.},
  \href{http://dx.doi.org/10.1103/PhysRevD.78.051102}{Phys. Rev. {\bf D78},
  051102 (2008)}, \href{http://arxiv.org/abs/0802.4035}{{\tt arXiv:0802.4035
  [hep-ex]}}\relax
\mciteBstWouldAddEndPuncttrue
\mciteSetBstMidEndSepPunct{\mcitedefaultmidpunct}
{\mcitedefaultendpunct}{\mcitedefaultseppunct}\relax
\EndOfBibitem
\bibitem{Bediaga:2009tr}
I.~Bediaga, I.~I. Bigi, A.~Gomes, G.~Guerrer, J.~Miranda, and A.~C.~d. Reis,
  \href{http://dx.doi.org/10.1103/PhysRevD.80.096006}{Phys. Rev. {\bf D80},
  096006 (2009)}, \href{http://arxiv.org/abs/0905.4233}{{\tt arXiv:0905.4233
  [hep-ph]}}\relax
\mciteBstWouldAddEndPuncttrue
\mciteSetBstMidEndSepPunct{\mcitedefaultmidpunct}
{\mcitedefaultendpunct}{\mcitedefaultseppunct}\relax
\EndOfBibitem
\bibitem{doi:10.1080/00949650410001661440}
B.~Aslan and G.~Zech,
  \href{http://dx.doi.org/10.1080/00949650410001661440}{Journal of Statistical
  Computation and Simulation {\bf 75},  109 (2005)}\relax
\mciteBstWouldAddEndPuncttrue
\mciteSetBstMidEndSepPunct{\mcitedefaultmidpunct}
{\mcitedefaultendpunct}{\mcitedefaultseppunct}\relax
\EndOfBibitem
\bibitem{Aaij:2013aa}
{LHCb} collaboration, R.~Aaij {\em et al.},
  \href{http://dx.doi.org/10.1016/j.physletb.2013.09.011}{Phys. Lett. {\bf
  B726},  623 (2013)}, \href{http://arxiv.org/abs/1308.3189}{{\tt
  arXiv:1308.3189 [hep-ex]}}\relax
\mciteBstWouldAddEndPuncttrue
\mciteSetBstMidEndSepPunct{\mcitedefaultmidpunct}
{\mcitedefaultendpunct}{\mcitedefaultseppunct}\relax
\EndOfBibitem
\bibitem{Aaij:2018nis}
{LHCb} collaboration, R.~Aaij {\em et al.},
  \href{http://dx.doi.org/10.1007/JHEP02(2019)126}{JHEP {\bf 02},  126 (2019)},
  \href{http://arxiv.org/abs/1811.08304}{{\tt arXiv:1811.08304 [hep-ex]}}\relax
\mciteBstWouldAddEndPuncttrue
\mciteSetBstMidEndSepPunct{\mcitedefaultmidpunct}
{\mcitedefaultendpunct}{\mcitedefaultseppunct}\relax
\EndOfBibitem
\bibitem{Hinson:2004pj}
{CLEO} collaboration, J.~W. Hinson {\em et al.},
  \href{http://dx.doi.org/10.1103/PhysRevLett.94.191801}{Phys. Rev. Lett. {\bf
  94},  191801 (2005)}, \href{http://arxiv.org/abs/hep-ex/0501002}{{\tt
  arXiv:hep-ex/0501002 [hep-ex]}}\relax
\mciteBstWouldAddEndPuncttrue
\mciteSetBstMidEndSepPunct{\mcitedefaultmidpunct}
{\mcitedefaultendpunct}{\mcitedefaultseppunct}\relax
\EndOfBibitem
\bibitem{Link:2005ft}
{FOCUS} collaboration, J.~M. Link {\em et al.},
  \href{http://dx.doi.org/10.1016/j.physletb.2006.01.017}{Phys. Lett. {\bf
  B634},  165 (2006)}, \href{http://arxiv.org/abs/hep-ex/0509042}{{\tt
  arXiv:hep-ex/0509042 [hep-ex]}}\relax
\mciteBstWouldAddEndPuncttrue
\mciteSetBstMidEndSepPunct{\mcitedefaultmidpunct}
{\mcitedefaultendpunct}{\mcitedefaultseppunct}\relax
\EndOfBibitem
\bibitem{Grossman:2012eb}
Y.~Grossman, A.~L. Kagan, and J.~Zupan,
  \href{http://dx.doi.org/10.1103/PhysRevD.85.114036}{Phys. Rev. {\bf D85},
  114036 (2012)}, \href{http://arxiv.org/abs/1204.3557}{{\tt arXiv:1204.3557
  [hep-ph]}}\relax
\mciteBstWouldAddEndPuncttrue
\mciteSetBstMidEndSepPunct{\mcitedefaultmidpunct}
{\mcitedefaultendpunct}{\mcitedefaultseppunct}\relax
\EndOfBibitem
\bibitem{Babu:2017bjn}
{Belle} collaboration, V.~Babu {\em et al.},
  \href{http://dx.doi.org/10.1103/PhysRevD.97.011101}{Phys. Rev. {\bf D97},
  011101 (2018)}, \href{http://arxiv.org/abs/1712.00619}{{\tt arXiv:1712.00619
  [hep-ex]}}\relax
\mciteBstWouldAddEndPuncttrue
\mciteSetBstMidEndSepPunct{\mcitedefaultmidpunct}
{\mcitedefaultendpunct}{\mcitedefaultseppunct}\relax
\EndOfBibitem
\bibitem{Eisenstein:2008aa}
{CLEO} collaboration, B.~I. Eisenstein {\em et al.},
  \href{http://dx.doi.org/10.1103/PhysRevD.78.052003}{Phys. Rev. {\bf D78},
  052003 (2008)}, \href{http://arxiv.org/abs/0806.2112}{{\tt arXiv:0806.2112
  [hep-ex]}}\relax
\mciteBstWouldAddEndPuncttrue
\mciteSetBstMidEndSepPunct{\mcitedefaultmidpunct}
{\mcitedefaultendpunct}{\mcitedefaultseppunct}\relax
\EndOfBibitem
\bibitem{Mendez:2009aa}
{CLEO} collaboration, H.~Mendez {\em et al.},
  \href{http://dx.doi.org/10.1103/PhysRevD.81.052013}{Phys. Rev. {\bf D81},
  052013 (2010)}, \href{http://arxiv.org/abs/0906.3198}{{\tt arXiv:0906.3198
  [hep-ex]}}\relax
\mciteBstWouldAddEndPuncttrue
\mciteSetBstMidEndSepPunct{\mcitedefaultmidpunct}
{\mcitedefaultendpunct}{\mcitedefaultseppunct}\relax
\EndOfBibitem
\bibitem{Won:2011ng}
{Belle} collaboration, E.~Won {\em et al.},
  \href{http://dx.doi.org/10.1103/PhysRevLett.107221801}{Phys. Rev. Lett. {\bf
  107},  221801 (2011)}, \href{http://arxiv.org/abs/1107.0553}{{\tt
  arXiv:1107.0553 [hep-ex]}}\relax
\mciteBstWouldAddEndPuncttrue
\mciteSetBstMidEndSepPunct{\mcitedefaultmidpunct}
{\mcitedefaultendpunct}{\mcitedefaultseppunct}\relax
\EndOfBibitem
\bibitem{Aaij:2017eux}
{LHCb} collaboration, R.~Aaij {\em et al.},
  \href{http://dx.doi.org/10.1016/j.physletb.2017.05.013}{Phys. Lett. {\bf
  B771},  21 (2017)}, \href{http://arxiv.org/abs/1701.01871}{{\tt
  arXiv:1701.01871 [hep-ex]}}\relax
\mciteBstWouldAddEndPuncttrue
\mciteSetBstMidEndSepPunct{\mcitedefaultmidpunct}
{\mcitedefaultendpunct}{\mcitedefaultseppunct}\relax
\EndOfBibitem
\bibitem{Bonvicini:2013vxi}
{CLEO} collaboration, G.~Bonvicini {\em et al.},
  \href{http://dx.doi.org/10.1103/PhysRevD.89.072002}{Phys. Rev. {\bf D89},
  072002 (2014)}, \href{http://arxiv.org/abs/1312.6775}{{\tt arXiv:1312.6775
  [hep-ex]}}\relax
\mciteBstWouldAddEndPuncttrue
\mciteSetBstMidEndSepPunct{\mcitedefaultmidpunct}
{\mcitedefaultendpunct}{\mcitedefaultseppunct}\relax
\EndOfBibitem
\bibitem{Ko:2012pe}
{Belle} collaboration, B.~R. Ko {\em et al.},
  \href{http://dx.doi.org/10.1103/PhysRevLett.109.021601}{Phys. Rev. Lett. {\bf
  109},  021601 (2012)}, \href{http://arxiv.org/abs/1203.6409}{{\tt
  arXiv:1203.6409 [hep-ex]}}\relax
\mciteBstWouldAddEndPuncttrue
\mciteSetBstMidEndSepPunct{\mcitedefaultmidpunct}
{\mcitedefaultendpunct}{\mcitedefaultseppunct}\relax
\EndOfBibitem
\bibitem{Amo:2011ab}
{\babar} collaboration, P.~del Amo~Sanchez {\em et al.},
  \href{http://dx.doi.org/10.1103/PhysRevD.83.071103}{Phys. Rev. {\bf D83},
  071103 (2011)}, \href{http://arxiv.org/abs/1011.5477}{{\tt arXiv:1011.5477
  [hep-ex]}}\relax
\mciteBstWouldAddEndPuncttrue
\mciteSetBstMidEndSepPunct{\mcitedefaultmidpunct}
{\mcitedefaultendpunct}{\mcitedefaultseppunct}\relax
\EndOfBibitem
\bibitem{Link:2001zj}
{FOCUS} collaboration, J.~M. Link {\em et al.},
  \href{http://dx.doi.org/10.1103/PhysRevLett.88.041602}{Phys. Rev. Lett. {\bf
  88},  041602; 159903(E) (2002)},
  \href{http://arxiv.org/abs/hep-ex/0109022}{{\tt arXiv:hep-ex/0109022}}\relax
\mciteBstWouldAddEndPuncttrue
\mciteSetBstMidEndSepPunct{\mcitedefaultmidpunct}
{\mcitedefaultendpunct}{\mcitedefaultseppunct}\relax
\EndOfBibitem
\bibitem{Lees:2013aa}
{\babar} collaboration, J.~P. Lees {\em et al.},
  \href{http://dx.doi.org/10.1103/PhysRevD.87.052012}{Phys. Rev. {\bf D87},
  052012 (2013)}, \href{http://arxiv.org/abs/1212.3003}{{\tt arXiv:1212.3003
  [hep-ex]}}\relax
\mciteBstWouldAddEndPuncttrue
\mciteSetBstMidEndSepPunct{\mcitedefaultmidpunct}
{\mcitedefaultendpunct}{\mcitedefaultseppunct}\relax
\EndOfBibitem
\bibitem{Ko:2013aa}
{Belle} collaboration, B.~R. Ko {\em et al.},
  \href{http://dx.doi.org/10.1007/JHEP02(2013)098}{JHEP {\bf 02},  098 (2013)},
  \href{http://arxiv.org/abs/1212.6112}{{\tt arXiv:1212.6112 [hep-ex]}}\relax
\mciteBstWouldAddEndPuncttrue
\mciteSetBstMidEndSepPunct{\mcitedefaultmidpunct}
{\mcitedefaultendpunct}{\mcitedefaultseppunct}\relax
\EndOfBibitem
\bibitem{Aaij:2014ac}
{LHCb} collaboration, R.~Aaij {\em et al.},
  \href{http://dx.doi.org/10.1007/JHEP10(2014)025}{JHEP {\bf 10},  25 (2014)},
  \href{http://arxiv.org/abs/1406.2624}{{\tt arXiv:1406.2624 [hep-ex]}}\relax
\mciteBstWouldAddEndPuncttrue
\mciteSetBstMidEndSepPunct{\mcitedefaultmidpunct}
{\mcitedefaultendpunct}{\mcitedefaultseppunct}\relax
\EndOfBibitem
\bibitem{Aaij:2014aa}
{LHCb} collaboration, R.~Aaij {\em et al.},
  \href{http://dx.doi.org/10.1016/j.physletb.2013.12.035}{Phys. Lett. {\bf
  B728},  585 (2014)}, \href{http://arxiv.org/abs/1310.7953}{{\tt
  arXiv:1310.7953 [hep-ex]}}\relax
\mciteBstWouldAddEndPuncttrue
\mciteSetBstMidEndSepPunct{\mcitedefaultmidpunct}
{\mcitedefaultendpunct}{\mcitedefaultseppunct}\relax
\EndOfBibitem
\bibitem{Aitala:1996sh}
{Fermilab E791} collaboration, E.~M. Aitala {\em et al.},
  \href{http://dx.doi.org/10.1016/S0370-2693(97)00565-0}{Phys. Lett. {\bf
  B403},  377 (1997)}, \href{http://arxiv.org/abs/hep-ex/9612005}{{\tt
  arXiv:hep-ex/9612005}}\relax
\mciteBstWouldAddEndPuncttrue
\mciteSetBstMidEndSepPunct{\mcitedefaultmidpunct}
{\mcitedefaultendpunct}{\mcitedefaultseppunct}\relax
\EndOfBibitem
\bibitem{Abazov:2014wga}
{\dzero} collaboration, V.~M. Abazov {\em et al.},
  \href{http://dx.doi.org/10.1103/PhysRevD.90.111102}{Phys. Rev. {\bf D90},
  111102 (2014)}, \href{http://arxiv.org/abs/1408.6848}{{\tt arXiv:1408.6848
  [hep-ex]}}\relax
\mciteBstWouldAddEndPuncttrue
\mciteSetBstMidEndSepPunct{\mcitedefaultmidpunct}
{\mcitedefaultendpunct}{\mcitedefaultseppunct}\relax
\EndOfBibitem
\bibitem{Lees:2013ab}
{\babar} collaboration, J.~P. Lees {\em et al.},
  \href{http://dx.doi.org/10.1103/PhysRevD.87.052010}{Phys. Rev. {\bf D87},
  052010 (2013)}, \href{http://arxiv.org/abs/1212.1856}{{\tt arXiv:1212.1856
  [hep-ex]}}\relax
\mciteBstWouldAddEndPuncttrue
\mciteSetBstMidEndSepPunct{\mcitedefaultmidpunct}
{\mcitedefaultendpunct}{\mcitedefaultseppunct}\relax
\EndOfBibitem
\bibitem{Rubin:2008zi}
{CLEO} collaboration, P.~Rubin {\em et al.},
  \href{http://dx.doi.org/10.1103/PhysRevD.78.072003}{Phys. Rev. {\bf D78},
  072003 (2008)}, \href{http://arxiv.org/abs/0807.4545}{{\tt arXiv:0807.4545
  [hep-ex]}}\relax
\mciteBstWouldAddEndPuncttrue
\mciteSetBstMidEndSepPunct{\mcitedefaultmidpunct}
{\mcitedefaultendpunct}{\mcitedefaultseppunct}\relax
\EndOfBibitem
\bibitem{Link:2000aw}
{FOCUS} collaboration, J.~M. Link {\em et al.},
  \href{http://dx.doi.org/10.1016/S0370-2693(00)01039-X}{Phys. Lett. {\bf
  B491},  232 (2000)}, \href{http://arxiv.org/abs/hep-ex/0005037}{{\tt
  arXiv:hep-ex/0005037}}\relax
\mciteBstWouldAddEndPuncttrue
\mciteSetBstMidEndSepPunct{\mcitedefaultmidpunct}
{\mcitedefaultendpunct}{\mcitedefaultseppunct}\relax
\EndOfBibitem
\bibitem{Link:2005th}
{FOCUS} collaboration, J.~M. Link {\em et al.},
  \href{http://dx.doi.org/10.1016/j.physletb.2005.07.024}{Phys. Lett. {\bf
  B622},  239 (2005)}, \href{http://arxiv.org/abs/hep-ex/0506012}{{\tt
  arXiv:hep-ex/0506012}}\relax
\mciteBstWouldAddEndPuncttrue
\mciteSetBstMidEndSepPunct{\mcitedefaultmidpunct}
{\mcitedefaultendpunct}{\mcitedefaultseppunct}\relax
\EndOfBibitem
\bibitem{Aaij:2016dfb}
{LHCb} collaboration, R.~Aaij {\em et al.},
  \href{http://dx.doi.org/10.1016/j.physletb.2017.01.061}{Phys. Lett. {\bf
  B767},  177--187 (2017)}, \href{http://arxiv.org/abs/1610.09476}{{\tt
  arXiv:1610.09476 [hep-ex]}}\relax
\mciteBstWouldAddEndPuncttrue
\mciteSetBstMidEndSepPunct{\mcitedefaultmidpunct}
{\mcitedefaultendpunct}{\mcitedefaultseppunct}\relax
\EndOfBibitem
\bibitem{Aaltonen:2012ab}
{CDF} collaboration, T.~Aaltonen {\em et al.},
  \href{http://dx.doi.org/10.1103/PhysRevD.85.012009}{Phys. Rev. {\bf D85},
  012009 (2012)}, \href{http://arxiv.org/abs/1111.5023}{{\tt arXiv:1111.5023
  [hep-ex]}}\relax
\mciteBstWouldAddEndPuncttrue
\mciteSetBstMidEndSepPunct{\mcitedefaultmidpunct}
{\mcitedefaultendpunct}{\mcitedefaultseppunct}\relax
\EndOfBibitem
\bibitem{Staric:2008rx}
{Belle} collaboration, M.~Staric {\em et al.},
  \href{http://dx.doi.org/10.1016/j.physletb.2008.10.052}{Phys. Lett. {\bf
  B670},  190 (2008)}, \href{http://arxiv.org/abs/0807.0148}{{\tt
  arXiv:0807.0148 [hep-ex]}}\relax
\mciteBstWouldAddEndPuncttrue
\mciteSetBstMidEndSepPunct{\mcitedefaultmidpunct}
{\mcitedefaultendpunct}{\mcitedefaultseppunct}\relax
\EndOfBibitem
\bibitem{Aitala:1997ff}
{Fermilab E791} collaboration, E.~M. Aitala {\em et al.},
  \href{http://dx.doi.org/10.1016/S0370-2693(97)01570-0}{Phys. Lett. {\bf
  B421},  405 (1998)}, \href{http://arxiv.org/abs/hep-ex/9711003}{{\tt
  arXiv:hep-ex/9711003}}\relax
\mciteBstWouldAddEndPuncttrue
\mciteSetBstMidEndSepPunct{\mcitedefaultmidpunct}
{\mcitedefaultendpunct}{\mcitedefaultseppunct}\relax
\EndOfBibitem
\bibitem{Nisar:2014fkc}
{Belle} collaboration, N.~K. Nisar {\em et al.},
  \href{http://dx.doi.org/10.1103/PhysRevLett.112.211601}{Phys. Rev. Lett. {\bf
  112},  211601 (2014)}, \href{http://arxiv.org/abs/1404.1266}{{\tt
  arXiv:1404.1266 [hep-ex]}}\relax
\mciteBstWouldAddEndPuncttrue
\mciteSetBstMidEndSepPunct{\mcitedefaultmidpunct}
{\mcitedefaultendpunct}{\mcitedefaultseppunct}\relax
\EndOfBibitem
\bibitem{Bonvicini:2000qm}
{CLEO} collaboration, G.~Bonvicini {\em et al.},
  \href{http://dx.doi.org/10.1103/PhysRevD.63.071101}{Phys. Rev. {\bf D63},
  071101 (2001)}, \href{http://arxiv.org/abs/hep-ex/0012054}{{\tt
  arXiv:hep-ex/0012054}}\relax
\mciteBstWouldAddEndPuncttrue
\mciteSetBstMidEndSepPunct{\mcitedefaultmidpunct}
{\mcitedefaultendpunct}{\mcitedefaultseppunct}\relax
\EndOfBibitem
\bibitem{Ko:2011ab}
{Belle} collaboration, B.~R. Ko {\em et al.},
  \href{http://dx.doi.org/10.1103/PhysRevLett.106.211801}{Phys. Rev. Lett. {\bf
  106},  211801 (2011)}, \href{http://arxiv.org/abs/1101.3365}{{\tt
  arXiv:1101.3365 [hep-ex]}}\relax
\mciteBstWouldAddEndPuncttrue
\mciteSetBstMidEndSepPunct{\mcitedefaultmidpunct}
{\mcitedefaultendpunct}{\mcitedefaultseppunct}\relax
\EndOfBibitem
\bibitem{Aaij:2018jud}
{LHCb} collaboration, R.~Aaij {\em et al.},
  \href{http://dx.doi.org/10.1007/JHEP11(2018)048}{JHEP {\bf 11},  048 (2018)},
  \href{http://arxiv.org/abs/1806.01642}{{\tt arXiv:1806.01642 [hep-ex]}}\relax
\mciteBstWouldAddEndPuncttrue
\mciteSetBstMidEndSepPunct{\mcitedefaultmidpunct}
{\mcitedefaultendpunct}{\mcitedefaultseppunct}\relax
\EndOfBibitem
\bibitem{Dash:2017heu}
N.~Dash {\em et al.},
  \href{http://dx.doi.org/10.1103/PhysRevLett.119.171801}{Phys. Rev. Lett. {\bf
  119},  171801 (2017)}, \href{http://arxiv.org/abs/1705.05966}{{\tt
  arXiv:1705.05966 [hep-ex]}}\relax
\mciteBstWouldAddEndPuncttrue
\mciteSetBstMidEndSepPunct{\mcitedefaultmidpunct}
{\mcitedefaultendpunct}{\mcitedefaultseppunct}\relax
\EndOfBibitem
\bibitem{Kim:2018mtf}
{Belle} collaboration, J.~B. Kim {\em et al.},
  \href{http://arxiv.org/abs/1810.06457}{{\tt arXiv:1810.06457 [hep-ex]}}\relax
\mciteBstWouldAddEndPuncttrue
\mciteSetBstMidEndSepPunct{\mcitedefaultmidpunct}
{\mcitedefaultendpunct}{\mcitedefaultseppunct}\relax
\EndOfBibitem
\bibitem{Aaij:2014afa}
{LHCb} collaboration, R.~Aaij {\em et al.},
  \href{http://dx.doi.org/10.1016/j.physletb.2014.11.043}{Phys. Lett. {\bf
  B740},  158 (2015)}, \href{http://arxiv.org/abs/1410.4170}{{\tt
  arXiv:1410.4170 [hep-ex]}}\relax
\mciteBstWouldAddEndPuncttrue
\mciteSetBstMidEndSepPunct{\mcitedefaultmidpunct}
{\mcitedefaultendpunct}{\mcitedefaultseppunct}\relax
\EndOfBibitem
\bibitem{Arinstein:2008zh}
{Belle} collaboration, K.~Arinstein,
  \href{http://dx.doi.org/10.1016/j.physletb.2008.02.054}{Phys. Lett. {\bf
  B662},  102 (2008)}, \href{http://arxiv.org/abs/0801.2439}{{\tt
  arXiv:0801.2439 [hep-ex]}}\relax
\mciteBstWouldAddEndPuncttrue
\mciteSetBstMidEndSepPunct{\mcitedefaultmidpunct}
{\mcitedefaultendpunct}{\mcitedefaultseppunct}\relax
\EndOfBibitem
\bibitem{CroninHennessy:2005sy}
{CLEO} collaboration, D.~Cronin-Hennessy {\em et al.},
  \href{http://dx.doi.org/10.1103/PhysRevD.72.031102}{Phys. Rev. {\bf D72},
  031102 (2005)}, \href{http://arxiv.org/abs/hep-ex/0503052}{{\tt
  arXiv:hep-ex/0503052}}\relax
\mciteBstWouldAddEndPuncttrue
\mciteSetBstMidEndSepPunct{\mcitedefaultmidpunct}
{\mcitedefaultendpunct}{\mcitedefaultseppunct}\relax
\EndOfBibitem
\bibitem{Tian:2005ik}
{Belle} collaboration, X.~C. Tian {\em et al.},
  \href{http://dx.doi.org/10.1103/PhysRevLett.95.231801}{Phys. Rev. Lett. {\bf
  95},  231801 (2005)}, \href{http://arxiv.org/abs/hep-ex/0507071}{{\tt
  arXiv:hep-ex/0507071}}\relax
\mciteBstWouldAddEndPuncttrue
\mciteSetBstMidEndSepPunct{\mcitedefaultmidpunct}
{\mcitedefaultendpunct}{\mcitedefaultseppunct}\relax
\EndOfBibitem
\bibitem{Brandenburg:2001ze}
{CLEO} collaboration, G.~Brandenburg {\em et al.},
  \href{http://dx.doi.org/10.1103/PhysRevLett.87.071802}{Phys. Rev. Lett. {\bf
  87},  071802 (2001)}, \href{http://arxiv.org/abs/hep-ex/0105002}{{\tt
  arXiv:hep-ex/0105002}}\relax
\mciteBstWouldAddEndPuncttrue
\mciteSetBstMidEndSepPunct{\mcitedefaultmidpunct}
{\mcitedefaultendpunct}{\mcitedefaultseppunct}\relax
\EndOfBibitem
\bibitem{Aaltonen:2012ac}
{CDF} collaboration, T.~Aaltonen {\em et al.},
  \href{http://dx.doi.org/10.1103/PhysRevD.86.032007}{Phys. Rev. {\bf D86},
  032007 (2012)}, \href{http://arxiv.org/abs/1207.0825}{{\tt arXiv:1207.0825
  [hep-ex]}}\relax
\mciteBstWouldAddEndPuncttrue
\mciteSetBstMidEndSepPunct{\mcitedefaultmidpunct}
{\mcitedefaultendpunct}{\mcitedefaultseppunct}\relax
\EndOfBibitem
\bibitem{Asner:2003uz}
{CLEO} collaboration, D.~M. Asner {\em et al.},
  \href{http://dx.doi.org/10.1103/PhysRevD.70.091101}{Phys. Rev. {\bf D70},
  091101 (2004)}, \href{http://arxiv.org/abs/hep-ex/0311033}{{\tt
  arXiv:hep-ex/0311033}}\relax
\mciteBstWouldAddEndPuncttrue
\mciteSetBstMidEndSepPunct{\mcitedefaultmidpunct}
{\mcitedefaultendpunct}{\mcitedefaultseppunct}\relax
\EndOfBibitem
\bibitem{Artuso:2012df}
{CLEO} collaboration, M.~Artuso {\em et al.},
  \href{http://dx.doi.org/10.1103/PhysRevD.85.122002}{Phys. Rev. {\bf D85},
  122002 (2012)}, \href{http://arxiv.org/abs/1201.5716}{{\tt arXiv:1201.5716
  [hep-ex]}}\relax
\mciteBstWouldAddEndPuncttrue
\mciteSetBstMidEndSepPunct{\mcitedefaultmidpunct}
{\mcitedefaultendpunct}{\mcitedefaultseppunct}\relax
\EndOfBibitem
\bibitem{Abdesselam:2016yvr}
{Belle} collaboration, A.~Abdesselam {\em et al.},
  \href{http://dx.doi.org/10.1103/PhysRevLett.118.051801}{Phys. Rev. Lett. {\bf
  118},  051801 (2017)}, \href{http://arxiv.org/abs/1603.03257}{{\tt
  arXiv:1603.03257 [hep-ex]}}\relax
\mciteBstWouldAddEndPuncttrue
\mciteSetBstMidEndSepPunct{\mcitedefaultmidpunct}
{\mcitedefaultendpunct}{\mcitedefaultseppunct}\relax
\EndOfBibitem
\bibitem{Aaij:2018fpa}
{LHCb} collaboration, R.~Aaij {\em et al.},
  \href{http://dx.doi.org/10.1103/PhysRevLett.121.091801}{Phys. Rev. Lett. {\bf
  121},  091801 (2018)}, \href{http://arxiv.org/abs/1806.10793}{{\tt
  arXiv:1806.10793 [hep-ex]}}\relax
\mciteBstWouldAddEndPuncttrue
\mciteSetBstMidEndSepPunct{\mcitedefaultmidpunct}
{\mcitedefaultendpunct}{\mcitedefaultseppunct}\relax
\EndOfBibitem
\bibitem{Alexander:2009ux}
{CLEO} collaboration, J.~Alexander {\em et al.},
  \href{http://dx.doi.org/10.1103/PhysRevD.79.052001}{Phys. Rev. {\bf D79},
  052001 (2009)}, \href{http://arxiv.org/abs/0901.1216}{{\tt arXiv:0901.1216
  [hep-ex]}}\relax
\mciteBstWouldAddEndPuncttrue
\mciteSetBstMidEndSepPunct{\mcitedefaultmidpunct}
{\mcitedefaultendpunct}{\mcitedefaultseppunct}\relax
\EndOfBibitem
\bibitem{Onyisi:2013bjt}
{CLEO} collaboration, P.~Onyisi {\em et al.},
  \href{http://dx.doi.org/10.1103/PhysRevD.88.032009}{Phys. Rev. {\bf D88},
  032009 (2013)}, \href{http://arxiv.org/abs/1306.5363}{{\tt arXiv:1306.5363
  [hep-ex]}}\relax
\mciteBstWouldAddEndPuncttrue
\mciteSetBstMidEndSepPunct{\mcitedefaultmidpunct}
{\mcitedefaultendpunct}{\mcitedefaultseppunct}\relax
\EndOfBibitem
\bibitem{Ko:2010ng}
{Belle} collaboration, B.~R. Ko {\em et al.},
  \href{http://dx.doi.org/10.1103/PhysRevLett.104.181602}{Phys. Rev. Lett. {\bf
  104},  181602 (2010)}, \href{http://arxiv.org/abs/1001.3202}{{\tt
  arXiv:1001.3202 [hep-ex]}}\relax
\mciteBstWouldAddEndPuncttrue
\mciteSetBstMidEndSepPunct{\mcitedefaultmidpunct}
{\mcitedefaultendpunct}{\mcitedefaultseppunct}\relax
\EndOfBibitem
\bibitem{Golowich:1988ig}
E.~Golowich and G.~Valencia,
  \href{http://dx.doi.org/10.1103/PhysRevD.40.112}{Phys. Rev. {\bf D40},  112
  (1989)}\relax
\mciteBstWouldAddEndPuncttrue
\mciteSetBstMidEndSepPunct{\mcitedefaultmidpunct}
{\mcitedefaultendpunct}{\mcitedefaultseppunct}\relax
\EndOfBibitem
\bibitem{Bigi:2001sg}
I.~I.~Y. Bigi, \href{http://arxiv.org/abs/hep-ph/0107102}{{\tt
  arXiv:hep-ph/0107102}} (2001)\relax
\mciteBstWouldAddEndPuncttrue
\mciteSetBstMidEndSepPunct{\mcitedefaultmidpunct}
{\mcitedefaultendpunct}{\mcitedefaultseppunct}\relax
\EndOfBibitem
\bibitem{Bensalem:2002ys}
W.~Bensalem, A.~Datta, and D.~London,
  \href{http://dx.doi.org/10.1103/PhysRevD.66.094004}{Phys. Rev. {\bf D66},
  094004 (2002)}, \href{http://arxiv.org/abs/hep-ph/0208054}{{\tt
  arXiv:hep-ph/0208054}}\relax
\mciteBstWouldAddEndPuncttrue
\mciteSetBstMidEndSepPunct{\mcitedefaultmidpunct}
{\mcitedefaultendpunct}{\mcitedefaultseppunct}\relax
\EndOfBibitem
\bibitem{Bensalem:2000hq}
W.~Bensalem and D.~London,
  \href{http://dx.doi.org/10.1103/PhysRevD.64.116003}{Phys. Rev. {\bf D64},
  116003 (2001)}, \href{http://arxiv.org/abs/hep-ph/0005018}{{\tt
  arXiv:hep-ph/0005018}}\relax
\mciteBstWouldAddEndPuncttrue
\mciteSetBstMidEndSepPunct{\mcitedefaultmidpunct}
{\mcitedefaultendpunct}{\mcitedefaultseppunct}\relax
\EndOfBibitem
\bibitem{Bensalem:2002pz}
W.~Bensalem, A.~Datta, and D.~London,
  \href{http://dx.doi.org/10.1016/S0370-2693(02)02028-2}{Phys. Lett. {\bf
  B538},  309 (2002)}, \href{http://arxiv.org/abs/hep-ph/0205009}{{\tt
  arXiv:hep-ph/0205009}}\relax
\mciteBstWouldAddEndPuncttrue
\mciteSetBstMidEndSepPunct{\mcitedefaultmidpunct}
{\mcitedefaultendpunct}{\mcitedefaultseppunct}\relax
\EndOfBibitem
\bibitem{Gronau:2011cf}
M.~Gronau and J.~L. Rosner,
  \href{http://dx.doi.org/10.1103/PhysRevD.84.096013}{Phys. Rev. {\bf D84},
  096013 (2011)}, \href{http://arxiv.org/abs/1107.1232}{{\tt arXiv:1107.1232
  [hep-ph]}}\relax
\mciteBstWouldAddEndPuncttrue
\mciteSetBstMidEndSepPunct{\mcitedefaultmidpunct}
{\mcitedefaultendpunct}{\mcitedefaultseppunct}\relax
\EndOfBibitem
\bibitem{Aaij:2014qwa}
{LHCb} collaboration, R.~Aaij {\em et al.},
  \href{http://dx.doi.org/10.1007/JHEP10(2014)005}{JHEP {\bf 10},  005 (2014)},
  \href{http://arxiv.org/abs/1408.1299}{{\tt arXiv:1408.1299 [hep-ex]}}\relax
\mciteBstWouldAddEndPuncttrue
\mciteSetBstMidEndSepPunct{\mcitedefaultmidpunct}
{\mcitedefaultendpunct}{\mcitedefaultseppunct}\relax
\EndOfBibitem
\bibitem{Sanchez:2010xj}
{\babar} collaboration, P.~del Amo~Sanchez {\em et al.},
  \href{http://dx.doi.org/10.1103/PhysRevD.81.111103}{Phys. Rev. {\bf D81},
  111103 (2010)}, \href{http://arxiv.org/abs/1003.3397}{{\tt arXiv:1003.3397
  [hep-ex]}}\relax
\mciteBstWouldAddEndPuncttrue
\mciteSetBstMidEndSepPunct{\mcitedefaultmidpunct}
{\mcitedefaultendpunct}{\mcitedefaultseppunct}\relax
\EndOfBibitem
\bibitem{Prasanth:2017beu}
{Belle} collaboration, K.~Prasanth {\em et al.},
  \href{http://dx.doi.org/10.1103/PhysRevD.95.091101}{Phys. Rev. {\bf D95},
  091101 (2017)}, \href{http://arxiv.org/abs/1703.05721}{{\tt arXiv:1703.05721
  [hep-ex]}}\relax
\mciteBstWouldAddEndPuncttrue
\mciteSetBstMidEndSepPunct{\mcitedefaultmidpunct}
{\mcitedefaultendpunct}{\mcitedefaultseppunct}\relax
\EndOfBibitem
\bibitem{Lees:2011ab}
{\babar} collaboration, J.~P. Lees {\em et al.},
  \href{http://dx.doi.org/10.1103/PhysRevD.84.031103}{Phys. Rev. {\bf D84},
  031103 (2011)}, \href{http://arxiv.org/abs/1105.4410}{{\tt arXiv:1105.4410
  [hep-ex]}}\relax
\mciteBstWouldAddEndPuncttrue
\mciteSetBstMidEndSepPunct{\mcitedefaultmidpunct}
{\mcitedefaultendpunct}{\mcitedefaultseppunct}\relax
\EndOfBibitem
\bibitem{Aaij:2017kbo}
{LHCb} collaboration, R.~Aaij {\em et al.},
  \href{http://dx.doi.org/10.1140/epjc/s10052-018-5758-4}{Eur. Phys. J. {\bf
  C78},  443 (2018)}, \href{http://arxiv.org/abs/1712.08609}{{\tt
  arXiv:1712.08609 [hep-ex]}}\relax
\mciteBstWouldAddEndPuncttrue
\mciteSetBstMidEndSepPunct{\mcitedefaultmidpunct}
{\mcitedefaultendpunct}{\mcitedefaultseppunct}\relax
\EndOfBibitem
\bibitem{Bevan:2015xra}
A.~J. Bevan, \href{http://arxiv.org/abs/1506.04246}{{\tt arXiv:1506.04246
  [hep-ex]}} (2015)\relax
\mciteBstWouldAddEndPuncttrue
\mciteSetBstMidEndSepPunct{\mcitedefaultmidpunct}
{\mcitedefaultendpunct}{\mcitedefaultseppunct}\relax
\EndOfBibitem
\bibitem{Durieux:2015zwa}
G.~Durieux and Y.~Grossman,
  \href{http://dx.doi.org/10.1103/PhysRevD.92.076013}{Phys. Rev. {\bf D92},
  076013 (2015)}, \href{http://arxiv.org/abs/1508.03054}{{\tt arXiv:1508.03054
  [hep-ph]}}\relax
\mciteBstWouldAddEndPuncttrue
\mciteSetBstMidEndSepPunct{\mcitedefaultmidpunct}
{\mcitedefaultendpunct}{\mcitedefaultseppunct}\relax
\EndOfBibitem
\bibitem{Grossman:2006jg}
Y.~Grossman, A.~L. Kagan, and Y.~Nir,
  \href{http://dx.doi.org/10.1103/PhysRevD.75.036008}{Phys. Rev. {\bf D75},
  036008 (2007)}, \href{http://arxiv.org/abs/hep-ph/0609178}{{\tt
  arXiv:hep-ph/0609178 [hep-ph]}}\relax
\mciteBstWouldAddEndPuncttrue
\mciteSetBstMidEndSepPunct{\mcitedefaultmidpunct}
{\mcitedefaultendpunct}{\mcitedefaultseppunct}\relax
\EndOfBibitem
\bibitem{Gersabeck:2011xj}
M.~Gersabeck, M.~Alexander, S.~Borghi, V.~Gligorov, and C.~Parkes,
  \href{http://dx.doi.org/10.1088/0954-3899/39/4/045005}{J. Phys. {\bf G39},
  045005 (2012)}, \href{http://arxiv.org/abs/1111.6515}{{\tt arXiv:1111.6515
  [hep-ex]}}\relax
\mciteBstWouldAddEndPuncttrue
\mciteSetBstMidEndSepPunct{\mcitedefaultmidpunct}
{\mcitedefaultendpunct}{\mcitedefaultseppunct}\relax
\EndOfBibitem
\bibitem{Aaij:2016cfh}
{LHCb} collaboration, R.~Aaij {\em et al.},
  \href{http://dx.doi.org/10.1103/PhysRevLett.116.191601}{Phys. Rev. Lett. {\bf
  116},  191601 (2016)}, \href{http://arxiv.org/abs/1602.03160}{{\tt
  arXiv:1602.03160 [hep-ex]}}\relax
\mciteBstWouldAddEndPuncttrue
\mciteSetBstMidEndSepPunct{\mcitedefaultmidpunct}
{\mcitedefaultendpunct}{\mcitedefaultseppunct}\relax
\EndOfBibitem
\bibitem{Becher:2005bg}
T.~Becher and R.~J. Hill,
  \href{http://dx.doi.org/10.1016/j.physletb.2005.11.063}{Phys. Lett. {\bf
  B633},  61 (2006)}, \href{http://arxiv.org/abs/hep-ph/0509090}{{\tt
  arXiv:hep-ph/0509090 [hep-ph]}}\relax
\mciteBstWouldAddEndPuncttrue
\mciteSetBstMidEndSepPunct{\mcitedefaultmidpunct}
{\mcitedefaultendpunct}{\mcitedefaultseppunct}\relax
\EndOfBibitem
\bibitem{Gilman:1989uy}
F.~J. Gilman and R.~L. Singleton,
  \href{http://dx.doi.org/10.1103/PhysRevD.41.142}{Phys. Rev. {\bf D41},  142
  (1990)}\relax
\mciteBstWouldAddEndPuncttrue
\mciteSetBstMidEndSepPunct{\mcitedefaultmidpunct}
{\mcitedefaultendpunct}{\mcitedefaultseppunct}\relax
\EndOfBibitem
\bibitem{Hill:2006ub}
R.~J. Hill, eConf {\bf C060409},  027 (2006),
  \href{http://arxiv.org/abs/hep-ph/0606023}{{\tt arXiv:hep-ph/0606023
  [hep-ph]}} (2006)\relax
\mciteBstWouldAddEndPuncttrue
\mciteSetBstMidEndSepPunct{\mcitedefaultmidpunct}
{\mcitedefaultendpunct}{\mcitedefaultseppunct}\relax
\EndOfBibitem
\bibitem{Becirevic:1999kt}
D.~Becirevic and A.~B. Kaidalov,
  \href{http://dx.doi.org/10.1016/S0370-2693(00)00290-2}{Phys. Lett. {\bf
  B478},  417 (2000)}, \href{http://arxiv.org/abs/hep-ph/9904490}{{\tt
  arXiv:hep-ph/9904490 [hep-ph]}}\relax
\mciteBstWouldAddEndPuncttrue
\mciteSetBstMidEndSepPunct{\mcitedefaultmidpunct}
{\mcitedefaultendpunct}{\mcitedefaultseppunct}\relax
\EndOfBibitem
\bibitem{Boyd:1994tt}
C.~G. Boyd, B.~Grinstein, and R.~F. Lebed,
  \href{http://dx.doi.org/10.1103/PhysRevLett.74.4603}{Phys. Rev. Lett. {\bf
  74},  4603 (1995)}, \href{http://arxiv.org/abs/hep-ph/9412324}{{\tt
  arXiv:hep-ph/9412324 [hep-ph]}}\relax
\mciteBstWouldAddEndPuncttrue
\mciteSetBstMidEndSepPunct{\mcitedefaultmidpunct}
{\mcitedefaultendpunct}{\mcitedefaultseppunct}\relax
\EndOfBibitem
\bibitem{Boyd:1997qw}
C.~G. Boyd and M.~J. Savage,
  \href{http://dx.doi.org/10.1103/PhysRevD.56.303}{Phys. Rev. {\bf D56},  303
  (1997)}, \href{http://arxiv.org/abs/hep-ph/9702300}{{\tt arXiv:hep-ph/9702300
  [hep-ph]}}\relax
\mciteBstWouldAddEndPuncttrue
\mciteSetBstMidEndSepPunct{\mcitedefaultmidpunct}
{\mcitedefaultendpunct}{\mcitedefaultseppunct}\relax
\EndOfBibitem
\bibitem{Arnesen:2005ez}
M.~C. Arnesen, B.~Grinstein, I.~Z. Rothstein, and I.~W. Stewart,
  \href{http://dx.doi.org/10.1103/PhysRevLett.95.071802}{Phys. Rev. Lett. {\bf
  95},  071802 (2005)}\relax
\mciteBstWouldAddEndPuncttrue
\mciteSetBstMidEndSepPunct{\mcitedefaultmidpunct}
{\mcitedefaultendpunct}{\mcitedefaultseppunct}\relax
\EndOfBibitem
\bibitem{Bourrely:1980gp}
C.~Bourrely, B.~Machet, and E.~de~Rafael,
  \href{http://dx.doi.org/10.1016/0550-3213(81)90086-9}{Nucl. Phys. {\bf B189},
   157 (1981)}\relax
\mciteBstWouldAddEndPuncttrue
\mciteSetBstMidEndSepPunct{\mcitedefaultmidpunct}
{\mcitedefaultendpunct}{\mcitedefaultseppunct}\relax
\EndOfBibitem
\bibitem{Becirevic:2014kaa}
D.~Becirevic, A.~L. Yaouanc, A.~Oyanguren, P.~Roudeau, and F.~Sanfilippo,
  \href{http://arxiv.org/abs/1407.1019}{{\tt arXiv:1407.1019 [hep-ph]}}
  (2014)\relax
\mciteBstWouldAddEndPuncttrue
\mciteSetBstMidEndSepPunct{\mcitedefaultmidpunct}
{\mcitedefaultendpunct}{\mcitedefaultseppunct}\relax
\EndOfBibitem
\bibitem{Lees:2013uxa}
{\babar} collaboration, J.~P. Lees {\em et al.},
  \href{http://dx.doi.org/10.1103/PhysRevD.88.052003}{Phys. Rev. {\bf D88},
  052003 (2013)}, \href{http://arxiv.org/abs/1304.5009}{{\tt arXiv:1304.5009
  [hep-ex]}}, Erratum ibid.\
  \href{http://dx.doi.org/10.1103/PhysRevD.88.079902}{{\bf D88}, 079902},
  (2013)\relax
\mciteBstWouldAddEndPuncttrue
\mciteSetBstMidEndSepPunct{\mcitedefaultmidpunct}
{\mcitedefaultendpunct}{\mcitedefaultseppunct}\relax
\EndOfBibitem
\bibitem{delAmoSanchez:2010vq}
{\babar} collaboration, P.~del Amo~Sanchez {\em et al.},
  \href{http://dx.doi.org/10.1103/PhysRevD.82.111101}{Phys. Rev. {\bf D82},
  111101 (2010)}, \href{http://arxiv.org/abs/1009.2076}{{\tt arXiv:1009.2076
  [hep-ex]}}\relax
\mciteBstWouldAddEndPuncttrue
\mciteSetBstMidEndSepPunct{\mcitedefaultmidpunct}
{\mcitedefaultendpunct}{\mcitedefaultseppunct}\relax
\EndOfBibitem
\bibitem{Aaij:2013sza}
{LHCb} collaboration, R.~Aaij {\em et al.},
  \href{http://dx.doi.org/10.1007/JHEP09(2013)145}{JHEP {\bf 09},  145 (2013)},
  \href{http://arxiv.org/abs/1307.4556}{{\tt arXiv:1307.4556}}\relax
\mciteBstWouldAddEndPuncttrue
\mciteSetBstMidEndSepPunct{\mcitedefaultmidpunct}
{\mcitedefaultendpunct}{\mcitedefaultseppunct}\relax
\EndOfBibitem
\bibitem{Burdman:1996kr}
G.~Burdman and J.~Kambor,
  \href{http://dx.doi.org/10.1103/PhysRevD.55.2817}{Phys. Rev. {\bf D55},  2817
  (1997)}, \href{http://arxiv.org/abs/hep-ph/9602353}{{\tt arXiv:hep-ph/9602353
  [hep-ph]}}\relax
\mciteBstWouldAddEndPuncttrue
\mciteSetBstMidEndSepPunct{\mcitedefaultmidpunct}
{\mcitedefaultendpunct}{\mcitedefaultseppunct}\relax
\EndOfBibitem
\bibitem{Becirevic:2012te}
D.~Becirevic, A.~Le~Yaouanc, L.~Oliver, J.-C. Raynal, P.~Roudeau, and
  J.~Serrano, \href{http://dx.doi.org/10.1103/PhysRevD.87.054007}{Phys. Rev.
  {\bf D87},  054007 (2013)}, \href{http://arxiv.org/abs/1206.5869}{{\tt
  arXiv:1206.5869 [hep-ph]}}\relax
\mciteBstWouldAddEndPuncttrue
\mciteSetBstMidEndSepPunct{\mcitedefaultmidpunct}
{\mcitedefaultendpunct}{\mcitedefaultseppunct}\relax
\EndOfBibitem
\bibitem{Lees:2014ihu}
{\babar} collaboration, J.~P. Lees {\em et al.},
  \href{http://dx.doi.org/10.1103/PhysRevD.91.052022}{Phys. Rev. {\bf D91},
  052022 (2015)}, \href{http://arxiv.org/abs/1412.5502}{{\tt arXiv:1412.5502
  [hep-ex]}}\relax
\mciteBstWouldAddEndPuncttrue
\mciteSetBstMidEndSepPunct{\mcitedefaultmidpunct}
{\mcitedefaultendpunct}{\mcitedefaultseppunct}\relax
\EndOfBibitem
\bibitem{Ablikim:2015ixa}
{BESIII} collaboration, M.~Ablikim {\em et al.},
  \href{http://dx.doi.org/10.1103/PhysRevD.92.072012}{Phys. Rev. {\bf D92},
  072012 (2015)}, \href{http://arxiv.org/abs/1508.07560}{{\tt arXiv:1508.07560
  [hep-ex]}}\relax
\mciteBstWouldAddEndPuncttrue
\mciteSetBstMidEndSepPunct{\mcitedefaultmidpunct}
{\mcitedefaultendpunct}{\mcitedefaultseppunct}\relax
\EndOfBibitem
\bibitem{Ablikim:2015qgt}
{BESIII} collaboration, M.~Ablikim {\em et al.},
  \href{http://dx.doi.org/10.1103/PhysRevD.92.112008}{Phys. Rev. {\bf D92},
  112008 (2015)}, \href{http://arxiv.org/abs/1510.00308}{{\tt arXiv:1510.00308
  [hep-ex]}}\relax
\mciteBstWouldAddEndPuncttrue
\mciteSetBstMidEndSepPunct{\mcitedefaultmidpunct}
{\mcitedefaultendpunct}{\mcitedefaultseppunct}\relax
\EndOfBibitem
\bibitem{Widhalm:2006wz}
{Belle} collaboration, L.~Widhalm {\em et al.},
  \href{http://dx.doi.org/10.1103/PhysRevLett.97.061804}{Phys. Rev. Lett. {\bf
  97},  061804 (2006)}, \href{http://arxiv.org/abs/hep-ex/0604049}{{\tt
  arXiv:hep-ex/0604049 [hep-ex]}}\relax
\mciteBstWouldAddEndPuncttrue
\mciteSetBstMidEndSepPunct{\mcitedefaultmidpunct}
{\mcitedefaultendpunct}{\mcitedefaultseppunct}\relax
\EndOfBibitem
\bibitem{Aubert:2007wg}
{\babar} collaboration, B.~Aubert {\em et al.},
  \href{http://dx.doi.org/10.1103/PhysRevD.76.052005}{Phys. Rev. {\bf D76},
  052005 (2007)}, \href{http://arxiv.org/abs/0704.0020}{{\tt arXiv:0704.0020
  [hep-ex]}}\relax
\mciteBstWouldAddEndPuncttrue
\mciteSetBstMidEndSepPunct{\mcitedefaultmidpunct}
{\mcitedefaultendpunct}{\mcitedefaultseppunct}\relax
\EndOfBibitem
\bibitem{Besson:2009uv}
{CLEO} collaboration, D.~Besson {\em et al.},
  \href{http://dx.doi.org/10.1103/PhysRevD.80.032005}{Phys. Rev. {\bf D80},
  032005 (2009)}, \href{http://arxiv.org/abs/0906.2983}{{\tt arXiv:0906.2983
  [hep-ex]}}\relax
\mciteBstWouldAddEndPuncttrue
\mciteSetBstMidEndSepPunct{\mcitedefaultmidpunct}
{\mcitedefaultendpunct}{\mcitedefaultseppunct}\relax
\EndOfBibitem
\bibitem{Dobbs:2007aa}
{CLEO} collaboration, S.~Dobbs {\em et al.},
  \href{http://dx.doi.org/10.1103/PhysRevD.77.112005}{Phys. Rev. {\bf D77},
  112005 (2008)}, \href{http://arxiv.org/abs/0712.1020}{{\tt arXiv:0712.1020
  [hep-ex]}}\relax
\mciteBstWouldAddEndPuncttrue
\mciteSetBstMidEndSepPunct{\mcitedefaultmidpunct}
{\mcitedefaultendpunct}{\mcitedefaultseppunct}\relax
\EndOfBibitem
\bibitem{bes3:2017fay}
{BESIII} collaboration, M.~Ablikim {\em et al.},
  \href{http://dx.doi.org/10.1103/PhysRevD.96.012002}{Phys. Rev. D {\bf 96},
  012002 (2017)}, \href{http://arxiv.org/abs/1703.09084}{{\tt arXiv:1703.09084
  [hep-ex]}}\relax
\mciteBstWouldAddEndPuncttrue
\mciteSetBstMidEndSepPunct{\mcitedefaultmidpunct}
{\mcitedefaultendpunct}{\mcitedefaultseppunct}\relax
\EndOfBibitem
\bibitem{bes3:2019zsf}
{BESIII} collaboration, M.~Ablikim {\em et al.},
  \href{http://dx.doi.org/10.1103/PhysRevLett.121.011804}{Phys. Rev. Lett. {\bf
  122},  011804 (2019)}, \href{http://arxiv.org/abs/1810.03127}{{\tt
  arXiv:1810.03127 [hep-ex]}}\relax
\mciteBstWouldAddEndPuncttrue
\mciteSetBstMidEndSepPunct{\mcitedefaultmidpunct}
{\mcitedefaultendpunct}{\mcitedefaultseppunct}\relax
\EndOfBibitem
\bibitem{bes3:2016hzl}
{BESIII} collaboration, M.~Ablikim {\em et al.},
  \href{http://dx.doi.org/10.1140/epjc/s10052-016-4198-2}{Eur. Phys. J. C {\bf
  76},  369 (2016)}, \href{http://arxiv.org/abs/1605.00068}{{\tt
  arXiv:1605.00068 [hep-ex]}}\relax
\mciteBstWouldAddEndPuncttrue
\mciteSetBstMidEndSepPunct{\mcitedefaultmidpunct}
{\mcitedefaultendpunct}{\mcitedefaultseppunct}\relax
\EndOfBibitem
\bibitem{bes3:2016wy}
{BESIII} collaboration, M.~Ablikim {\em et al.},
  \href{http://dx.doi.org/10.1088/1674-1137/40/11/113001}{Chin. Phys. C {\bf
  40},  113001 (2016)}, \href{http://arxiv.org/abs/1605.00208}{{\tt
  arXiv:1605.00208 [hep-ex]}}\relax
\mciteBstWouldAddEndPuncttrue
\mciteSetBstMidEndSepPunct{\mcitedefaultmidpunct}
{\mcitedefaultendpunct}{\mcitedefaultseppunct}\relax
\EndOfBibitem
\bibitem{bes3:2018wy}
{BESIII} collaboration, M.~Ablikim {\em et al.},
  \href{http://dx.doi.org/10.1103/PhysRevLett.121.171803}{Phys. Rev. Lett. {\bf
  121},  171803 (2018)}, \href{http://arxiv.org/abs/1802.05492}{{\tt
  arXiv:1802.05492 [hep-ex]}}\relax
\mciteBstWouldAddEndPuncttrue
\mciteSetBstMidEndSepPunct{\mcitedefaultmidpunct}
{\mcitedefaultendpunct}{\mcitedefaultseppunct}\relax
\EndOfBibitem
\bibitem{BESIII-new}
Y.~Zheng, {BESIII} collaboration. {presented at the 37th International
  Conference on High Energy Physics (ICHEP 2014)}, 2014\relax
\mciteBstWouldAddEndPuncttrue
\mciteSetBstMidEndSepPunct{\mcitedefaultmidpunct}
{\mcitedefaultendpunct}{\mcitedefaultseppunct}\relax
\EndOfBibitem
\bibitem{Huang:2004fra}
{CLEO} collaboration, G.~Huang {\em et al.},
  \href{http://dx.doi.org/10.1103/PhysRevLett.94.011802}{Phys. Rev. Lett. {\bf
  94},  011802 (2005)}, \href{http://arxiv.org/abs/hep-ex/0407035}{{\tt
  arXiv:hep-ex/0407035 [hep-ex]}}\relax
\mciteBstWouldAddEndPuncttrue
\mciteSetBstMidEndSepPunct{\mcitedefaultmidpunct}
{\mcitedefaultendpunct}{\mcitedefaultseppunct}\relax
\EndOfBibitem
\bibitem{Link:2004dh}
{FOCUS} collaboration, J.~Link {\em et al.},
  \href{http://dx.doi.org/10.1016/j.physletb.2004.12.036}{Phys. Lett. {\bf
  B607},  233 (2005)}, \href{http://arxiv.org/abs/hep-ex/0410037}{{\tt
  arXiv:hep-ex/0410037 [hep-ex]}}\relax
\mciteBstWouldAddEndPuncttrue
\mciteSetBstMidEndSepPunct{\mcitedefaultmidpunct}
{\mcitedefaultendpunct}{\mcitedefaultseppunct}\relax
\EndOfBibitem
\bibitem{bes3:2019yyh}
{BESIII} collaboration, M.~Ablikim {\em et al.},
  \href{http://dx.doi.org/10.1103/PhysRevLett.122.121801}{Phys. Rev. Lett. {\bf
  122},  121801 (2019)}, \href{http://arxiv.org/abs/1901.02133}{{\tt
  arXiv:1901.02133 [hep-ex]}}\relax
\mciteBstWouldAddEndPuncttrue
\mciteSetBstMidEndSepPunct{\mcitedefaultmidpunct}
{\mcitedefaultendpunct}{\mcitedefaultseppunct}\relax
\EndOfBibitem
\bibitem{cleo:2011jy}
{CLEO} collaboration, J.~Yelton {\em et al.},
  \href{http://dx.doi.org/10.1103/PhysRevD.84.032001}{Phys. Rev. D {\bf 84},
  032001 (2011)}, \href{http://arxiv.org/abs/1011.1195}{{\tt arXiv:1011.1195
  [hep-ex]}}\relax
\mciteBstWouldAddEndPuncttrue
\mciteSetBstMidEndSepPunct{\mcitedefaultmidpunct}
{\mcitedefaultendpunct}{\mcitedefaultseppunct}\relax
\EndOfBibitem
\bibitem{bes3:2018zhy}
{BESIII} collaboration, M.~Ablikim {\em et al.},
  \href{http://dx.doi.org/10.1103/PhysRevD.97.092009}{Phys. Rev. D {\bf 97},
  092009 (2018)}, \href{http://arxiv.org/abs/1512.08627}{{\tt arXiv:1512.08627
  [hep-ex]}}\relax
\mciteBstWouldAddEndPuncttrue
\mciteSetBstMidEndSepPunct{\mcitedefaultmidpunct}
{\mcitedefaultendpunct}{\mcitedefaultseppunct}\relax
\EndOfBibitem
\bibitem{bes3:2018sll}
{BESIII} collaboration, M.~Ablikim {\em et al.},
  \href{http://dx.doi.org/10.1103/PhysRevLett.122.061801}{Phys. Rev. Lett. {\bf
  122},  061801 (2019)}, \href{http://arxiv.org/abs/1811.02911}{{\tt
  arXiv:1811.02911 [hep-ex]}}\relax
\mciteBstWouldAddEndPuncttrue
\mciteSetBstMidEndSepPunct{\mcitedefaultmidpunct}
{\mcitedefaultendpunct}{\mcitedefaultseppunct}\relax
\EndOfBibitem
\bibitem{dse:2015gsb}
G.~S. Bali {\em et al.},
  \href{http://dx.doi.org/10.1103/PhysRevD.91.014503}{Phys. Rev. D {\bf 91},
  014503 (2015)}, \href{http://arxiv.org/abs/1406.5449}{{\tt arXiv:1406.5449
  [hep-ph]}}\relax
\mciteBstWouldAddEndPuncttrue
\mciteSetBstMidEndSepPunct{\mcitedefaultmidpunct}
{\mcitedefaultendpunct}{\mcitedefaultseppunct}\relax
\EndOfBibitem
\bibitem{dse:2015gdu}
G.~Duplancic {\em et al.},
  \href{http://dx.doi.org/10.1007/JHEP11(2015)138}{JHEP {\bf 1511},  138
  (2015)}, \href{http://arxiv.org/abs/1508.05287}{{\tt arXiv:1508.05287
  [hep-ph]}}\relax
\mciteBstWouldAddEndPuncttrue
\mciteSetBstMidEndSepPunct{\mcitedefaultmidpunct}
{\mcitedefaultendpunct}{\mcitedefaultseppunct}\relax
\EndOfBibitem
\bibitem{dse:2013nof}
N.~Offen {\em et al.},
  \href{http://dx.doi.org/10.1103/PhysRevD.88.034023}{Phys. Rev. D {\bf 88},
  034023 (2013)}, \href{http://arxiv.org/abs/1307.2797}{{\tt arXiv:1307.2797
  [hep-ph]}}\relax
\mciteBstWouldAddEndPuncttrue
\mciteSetBstMidEndSepPunct{\mcitedefaultmidpunct}
{\mcitedefaultendpunct}{\mcitedefaultseppunct}\relax
\EndOfBibitem
\bibitem{dse:2011kaz}
K.~Azizi {\em et al.},
  \href{http://dx.doi.org/10.1088/0954-3899/38/9/095001}{J. Phys. G {\bf 38},
  095001 (2011)}, \href{http://arxiv.org/abs/1011.6046}{{\tt arXiv:1011.6046
  [hep-ph]}}\relax
\mciteBstWouldAddEndPuncttrue
\mciteSetBstMidEndSepPunct{\mcitedefaultmidpunct}
{\mcitedefaultendpunct}{\mcitedefaultseppunct}\relax
\EndOfBibitem
\bibitem{dse:2001pco}
P.~Colangelo {\em et al.},
  \href{http://dx.doi.org/10.1016/S0370-2693(01)01112-1}{Phys. Lett. B {\bf
  520},  78 (2001)}, \href{http://arxiv.org/abs/0107137}{{\tt arXiv:0107137
  [hep-ph]}}\relax
\mciteBstWouldAddEndPuncttrue
\mciteSetBstMidEndSepPunct{\mcitedefaultmidpunct}
{\mcitedefaultendpunct}{\mcitedefaultseppunct}\relax
\EndOfBibitem
\bibitem{dse:2012rcv}
R.~C. Verma, \href{http://dx.doi.org/10.1088/0954-3899/39/2/025005}{J. Phys. G
  {\bf 39},  025005 (2012)}, \href{http://arxiv.org/abs/1103.2973}{{\tt
  arXiv:1103.2973 [hep-ph]}}\relax
\mciteBstWouldAddEndPuncttrue
\mciteSetBstMidEndSepPunct{\mcitedefaultmidpunct}
{\mcitedefaultendpunct}{\mcitedefaultseppunct}\relax
\EndOfBibitem
\bibitem{dse:2009ztw}
Z.~T. Wei {\em et al.},
  \href{http://dx.doi.org/10.1103/PhysRevD.80.015022}{Phys. Rev. D {\bf 80},
  015022 (2009)}, \href{http://arxiv.org/abs/0905.3069}{{\tt arXiv:0905.3069
  [hep-ph]}}\relax
\mciteBstWouldAddEndPuncttrue
\mciteSetBstMidEndSepPunct{\mcitedefaultmidpunct}
{\mcitedefaultendpunct}{\mcitedefaultseppunct}\relax
\EndOfBibitem
\bibitem{dse:2000dme}
D.~Melikhov {\em et al.},
  \href{http://dx.doi.org/10.1103/PhysRevD.62.014006}{Phys. Rev. D {\bf 62},
  014006 (2000)}, \href{http://arxiv.org/abs/0001113}{{\tt arXiv:0001113
  [hep-ph]}}\relax
\mciteBstWouldAddEndPuncttrue
\mciteSetBstMidEndSepPunct{\mcitedefaultmidpunct}
{\mcitedefaultendpunct}{\mcitedefaultseppunct}\relax
\EndOfBibitem
\bibitem{dse:2018nrs}
N.~R. Soni {\em et al.},
  \href{http://dx.doi.org/10.1103/PhysRevD.98.114031}{Phys. Rev. D {\bf 98},
  114031 (2018)}, \href{http://arxiv.org/abs/1810.11907}{{\tt arXiv:1810.11907
  [hep-ph]}}\relax
\mciteBstWouldAddEndPuncttrue
\mciteSetBstMidEndSepPunct{\mcitedefaultmidpunct}
{\mcitedefaultendpunct}{\mcitedefaultseppunct}\relax
\EndOfBibitem
\bibitem{etm:2017lub}
{ETM} collaboration, V.~Lubicz {\em et al.},
  \href{http://dx.doi.org/10.1103/PhysRevD.96.054514}{Phys. Rev. D {\bf 96},
  054514 (2017)}, \href{http://arxiv.org/abs/1706.03017}{{\tt arXiv:1706.03017
  [hep-lat]}}\relax
\mciteBstWouldAddEndPuncttrue
\mciteSetBstMidEndSepPunct{\mcitedefaultmidpunct}
{\mcitedefaultendpunct}{\mcitedefaultseppunct}\relax
\EndOfBibitem
\bibitem{hpqcd:2010hna}
{HPQCD} collaboration, H.~Na {\em et al.},
  \href{http://dx.doi.org/10.1103/PhysRevD.82.114506}{Phys. Rev. D {\bf 82},
  114506 (2010)}, \href{http://arxiv.org/abs/1008.4562}{{\tt arXiv:1008.4562
  [hep-lat]}}\relax
\mciteBstWouldAddEndPuncttrue
\mciteSetBstMidEndSepPunct{\mcitedefaultmidpunct}
{\mcitedefaultendpunct}{\mcitedefaultseppunct}\relax
\EndOfBibitem
\bibitem{hpqcd:2011hna}
{HPQCD} collaboration, H.~Na {\em et al.},
  \href{http://dx.doi.org/10.1103/PhysRevD.84.114505}{Phys. Rev. D {\bf 84},
  114505 (2011)}, \href{http://arxiv.org/abs/1109.1501}{{\tt arXiv:1109.1501
  [hep-lat]}}\relax
\mciteBstWouldAddEndPuncttrue
\mciteSetBstMidEndSepPunct{\mcitedefaultmidpunct}
{\mcitedefaultendpunct}{\mcitedefaultseppunct}\relax
\EndOfBibitem
\bibitem{fm:2019rui}
{Fermilab Lattice and MILC} collaboration, R.~Li {\em et al.},
  \href{http://dx.doi.org/http://www.pa.msu.edu/conf/Lattice2018/}{Lattice 2018
    (2019)}, \href{http://arxiv.org/abs/1901.08989}{{\tt arXiv:1901.08989
  [hep-lat]}}\relax
\mciteBstWouldAddEndPuncttrue
\mciteSetBstMidEndSepPunct{\mcitedefaultmidpunct}
{\mcitedefaultendpunct}{\mcitedefaultseppunct}\relax
\EndOfBibitem
\bibitem{th:2017svj}
S.~Fajfer {\em et al.},
  \href{http://dx.doi.org/10.1103/PhysRevD.91.094009}{Phys.Rev. D91 (2015)
  no.9, 094009 {\bf 91},  094009 (2015)},
  \href{http://arxiv.org/abs/1502.07488}{{\tt arXiv:1502.07488
  [hep-lat]}}\relax
\mciteBstWouldAddEndPuncttrue
\mciteSetBstMidEndSepPunct{\mcitedefaultmidpunct}
{\mcitedefaultendpunct}{\mcitedefaultseppunct}\relax
\EndOfBibitem
\bibitem{epjc:2018rig}
L.~Riggio {\em et al.},
  \href{http://dx.doi.org/10.1140/epjc/s10052-018-5943-5}{Eur. Phys. J. C {\bf
  78},  501 (2018)}, \href{http://arxiv.org/abs/1706.03657}{{\tt
  arXiv:1706.03657 [hep-lat]}}\relax
\mciteBstWouldAddEndPuncttrue
\mciteSetBstMidEndSepPunct{\mcitedefaultmidpunct}
{\mcitedefaultendpunct}{\mcitedefaultseppunct}\relax
\EndOfBibitem
\bibitem{bes3:2018grp}
{BESIII} collaboration, M.~Ablikim {\em et al.},
  \href{http://dx.doi.org/10.1103/PhysRevD.97.012006}{Phys. Rev. D {\bf 97},
  012006 (2018)}, \href{http://arxiv.org/abs/1709.03680}{{\tt arXiv:1709.03680
  [hep-ex]}}\relax
\mciteBstWouldAddEndPuncttrue
\mciteSetBstMidEndSepPunct{\mcitedefaultmidpunct}
{\mcitedefaultendpunct}{\mcitedefaultseppunct}\relax
\EndOfBibitem
\bibitem{bes3:2016grp}
{BESIII} collaboration, M.~Ablikim {\em et al.},
  \href{http://dx.doi.org/10.1103/PhysRevD.94.112003}{Phys. Rev. D {\bf 94},
  112003 (2016)}, \href{http://arxiv.org/abs/1608.06484}{{\tt arXiv:1608.06484
  [hep-ex]}}\relax
\mciteBstWouldAddEndPuncttrue
\mciteSetBstMidEndSepPunct{\mcitedefaultmidpunct}
{\mcitedefaultendpunct}{\mcitedefaultseppunct}\relax
\EndOfBibitem
\bibitem{Link:2002ev}
{FOCUS} collaboration, J.~Link {\em et al.},
  \href{http://dx.doi.org/10.1016/S0370-2693(02)01715-X}{Phys. Lett. {\bf
  B535},  43 (2002)}, \href{http://arxiv.org/abs/hep-ex/0203031}{{\tt
  arXiv:hep-ex/0203031 [hep-ex]}}\relax
\mciteBstWouldAddEndPuncttrue
\mciteSetBstMidEndSepPunct{\mcitedefaultmidpunct}
{\mcitedefaultendpunct}{\mcitedefaultseppunct}\relax
\EndOfBibitem
\bibitem{Link:2002wg}
{FOCUS} collaboration, J.~Link {\em et al.},
  \href{http://dx.doi.org/10.1016/S0370-2693(02)02386-9}{Phys. Lett. {\bf
  B544},  89 (2002)}, \href{http://arxiv.org/abs/hep-ex/0207049}{{\tt
  arXiv:hep-ex/0207049 [hep-ex]}}\relax
\mciteBstWouldAddEndPuncttrue
\mciteSetBstMidEndSepPunct{\mcitedefaultmidpunct}
{\mcitedefaultendpunct}{\mcitedefaultseppunct}\relax
\EndOfBibitem
\bibitem{Aubert:2008rs}
{\babar} collaboration, B.~Aubert {\em et al.},
  \href{http://dx.doi.org/10.1103/PhysRevD.78.051101}{Phys. Rev. {\bf D78},
  051101 (2008)}, \href{http://arxiv.org/abs/0807.1599}{{\tt arXiv:0807.1599
  [hep-ex]}}\relax
\mciteBstWouldAddEndPuncttrue
\mciteSetBstMidEndSepPunct{\mcitedefaultmidpunct}
{\mcitedefaultendpunct}{\mcitedefaultseppunct}\relax
\EndOfBibitem
\bibitem{Ecklund:2009fia}
{CLEO} collaboration, K.~M. Ecklund {\em et al.},
  \href{http://dx.doi.org/10.1103/PhysRevD.80.052009}{Phys. Rev. {\bf D80},
  052009 (2009)}, \href{http://arxiv.org/abs/0907.3201}{{\tt arXiv:0907.3201
  [hep-ex]}}\relax
\mciteBstWouldAddEndPuncttrue
\mciteSetBstMidEndSepPunct{\mcitedefaultmidpunct}
{\mcitedefaultendpunct}{\mcitedefaultseppunct}\relax
\EndOfBibitem
\bibitem{Briere:2010zc}
{CLEO} collaboration, R.~A. Briere {\em et al.},
  \href{http://dx.doi.org/10.1103/PhysRevD.81.112001}{Phys. Rev. {\bf D81},
  112001 (2010)}, \href{http://arxiv.org/abs/1004.1954}{{\tt arXiv:1004.1954
  [hep-ex]}}\relax
\mciteBstWouldAddEndPuncttrue
\mciteSetBstMidEndSepPunct{\mcitedefaultmidpunct}
{\mcitedefaultendpunct}{\mcitedefaultseppunct}\relax
\EndOfBibitem
\bibitem{delAmoSanchez:2010fd}
{\babar} collaboration, P.~del Amo~Sanchez {\em et al.},
  \href{http://dx.doi.org/10.1103/PhysRevD.83.072001}{Phys. Rev. {\bf D83},
  072001 (2011)}, \href{http://arxiv.org/abs/1012.1810}{{\tt arXiv:1012.1810
  [hep-ex]}}\relax
\mciteBstWouldAddEndPuncttrue
\mciteSetBstMidEndSepPunct{\mcitedefaultmidpunct}
{\mcitedefaultendpunct}{\mcitedefaultseppunct}\relax
\EndOfBibitem
\bibitem{Estabrooks:1977xe}
P.~Estabrooks {\em et al.},
  \href{http://dx.doi.org/10.1016/0550-3213(78)90238-9}{Nucl. Phys. {\bf B133},
   490 (1978)}\relax
\mciteBstWouldAddEndPuncttrue
\mciteSetBstMidEndSepPunct{\mcitedefaultmidpunct}
{\mcitedefaultendpunct}{\mcitedefaultseppunct}\relax
\EndOfBibitem
\bibitem{Watson:1954uc}
K.~M. Watson, \href{http://dx.doi.org/10.1103/PhysRev.95.228}{Phys. Rev. {\bf
  95},  228--236 (1954)}\relax
\mciteBstWouldAddEndPuncttrue
\mciteSetBstMidEndSepPunct{\mcitedefaultmidpunct}
{\mcitedefaultendpunct}{\mcitedefaultseppunct}\relax
\EndOfBibitem
\bibitem{bes3:2016aff}
{BESIII} collaboration, M.~Ablikim {\em et al.},
  \href{http://dx.doi.org/10.1103/PhysRevD.94.032001}{Phys. Rev. D {\bf 94},
  032001 (2016)}, \href{http://arxiv.org/abs/1512.08627}{{\tt arXiv:1512.08627
  [hep-ex]}}\relax
\mciteBstWouldAddEndPuncttrue
\mciteSetBstMidEndSepPunct{\mcitedefaultmidpunct}
{\mcitedefaultendpunct}{\mcitedefaultseppunct}\relax
\EndOfBibitem
\bibitem{bes3:2018dll}
{BESIII} collaboration, M.~Ablikim {\em et al.},
  \href{http://dx.doi.org/10.1103/PhysRevD.99.011103}{Phys. Rev. {\bf D99},
  011103 (2019)}, \href{http://arxiv.org/abs/1811.11349}{{\tt arXiv:1811.11349
  [hep-ex]}}\relax
\mciteBstWouldAddEndPuncttrue
\mciteSetBstMidEndSepPunct{\mcitedefaultmidpunct}
{\mcitedefaultendpunct}{\mcitedefaultseppunct}\relax
\EndOfBibitem
\bibitem{bes3:2016hy}
{BESIII} collaboration, M.~Ablikim {\em et al.},
  \href{http://dx.doi.org/10.1103/PhysRevD.92.071101}{Phys. Rev. D {\bf 92},
  071101 (2016)}, \href{http://arxiv.org/abs/1508.00151}{{\tt arXiv:1508.00151
  [hep-ex]}}\relax
\mciteBstWouldAddEndPuncttrue
\mciteSetBstMidEndSepPunct{\mcitedefaultmidpunct}
{\mcitedefaultendpunct}{\mcitedefaultseppunct}\relax
\EndOfBibitem
\bibitem{bes3:2018zhl}
{BESIII} collaboration, M.~Ablikim {\em et al.},
  \href{http://dx.doi.org/10.1103/PhysRevLett.122.062001}{Phys. Rev. Lett. {\bf
  122},  062001 (2019)}, \href{http://arxiv.org/abs/1809.06496}{{\tt
  arXiv:1809.06496 [hep-ex]}}\relax
\mciteBstWouldAddEndPuncttrue
\mciteSetBstMidEndSepPunct{\mcitedefaultmidpunct}
{\mcitedefaultendpunct}{\mcitedefaultseppunct}\relax
\EndOfBibitem
\bibitem{Anjos:1990pn}
{Fermilab E691} collaboration, J.~Anjos {\em et al.},
  \href{http://dx.doi.org/10.1103/PhysRevLett.65.2630}{Phys. Rev. Lett. {\bf
  65},  2630 (1990)}\relax
\mciteBstWouldAddEndPuncttrue
\mciteSetBstMidEndSepPunct{\mcitedefaultmidpunct}
{\mcitedefaultendpunct}{\mcitedefaultseppunct}\relax
\EndOfBibitem
\bibitem{Kodama:1992tn}
{Fermilab E653} collaboration, K.~Kodama {\em et al.},
  \href{http://dx.doi.org/10.1016/0370-2693(92)90530-H}{Phys. Lett. {\bf B274},
   246 (1992)}\relax
\mciteBstWouldAddEndPuncttrue
\mciteSetBstMidEndSepPunct{\mcitedefaultmidpunct}
{\mcitedefaultendpunct}{\mcitedefaultseppunct}\relax
\EndOfBibitem
\bibitem{Frabetti:1993jq}
{Fermilab E687} collaboration, P.~Frabetti {\em et al.},
  \href{http://dx.doi.org/10.1016/0370-2693(93)90216-5}{Phys. Lett. {\bf B307},
   262 (1993)}\relax
\mciteBstWouldAddEndPuncttrue
\mciteSetBstMidEndSepPunct{\mcitedefaultmidpunct}
{\mcitedefaultendpunct}{\mcitedefaultseppunct}\relax
\EndOfBibitem
\bibitem{Aitala:1997cm}
{Fermilab E791} collaboration, E.~Aitala {\em et al.},
  \href{http://dx.doi.org/10.1103/PhysRevLett.80.1393}{Phys. Rev. Lett. {\bf
  80},  1393 (1998)}, \href{http://arxiv.org/abs/hep-ph/9710216}{{\tt
  arXiv:hep-ph/9710216 [hep-ph]}}\relax
\mciteBstWouldAddEndPuncttrue
\mciteSetBstMidEndSepPunct{\mcitedefaultmidpunct}
{\mcitedefaultendpunct}{\mcitedefaultseppunct}\relax
\EndOfBibitem
\bibitem{Aitala:1998ey}
{Fermilab E791} collaboration, E.~Aitala {\em et al.},
  \href{http://dx.doi.org/10.1016/S0370-2693(98)01243-X}{Phys. Lett. {\bf
  B440},  435 (1998)}, \href{http://arxiv.org/abs/hep-ex/9809026}{{\tt
  arXiv:hep-ex/9809026 [hep-ex]}}\relax
\mciteBstWouldAddEndPuncttrue
\mciteSetBstMidEndSepPunct{\mcitedefaultmidpunct}
{\mcitedefaultendpunct}{\mcitedefaultseppunct}\relax
\EndOfBibitem
\bibitem{Adamovich:1998ia}
{BEATRICE} collaboration, M.~Adamovich {\em et al.},
  \href{http://dx.doi.org/10.1007/s100529801012}{Eur. Phys. J. {\bf C6},  35
  (1999)}\relax
\mciteBstWouldAddEndPuncttrue
\mciteSetBstMidEndSepPunct{\mcitedefaultmidpunct}
{\mcitedefaultendpunct}{\mcitedefaultseppunct}\relax
\EndOfBibitem
\bibitem{Link:2004uk}
{FOCUS} collaboration, J.~Link {\em et al.},
  \href{http://dx.doi.org/10.1016/j.physletb.2004.12.037}{Phys. Lett. {\bf
  B607},  67 (2005)}, \href{http://arxiv.org/abs/hep-ex/0410067}{{\tt
  arXiv:hep-ex/0410067 [hep-ex]}}\relax
\mciteBstWouldAddEndPuncttrue
\mciteSetBstMidEndSepPunct{\mcitedefaultmidpunct}
{\mcitedefaultendpunct}{\mcitedefaultseppunct}\relax
\EndOfBibitem
\bibitem{Mahlke:2007uf}
H.~Mahlke, eConf {\bf C0610161},  014 (2006),
  \href{http://arxiv.org/abs/hep-ex/0702014}{{\tt arXiv:hep-ex/0702014
  [hep-ex]}}\relax
\mciteBstWouldAddEndPuncttrue
\mciteSetBstMidEndSepPunct{\mcitedefaultmidpunct}
{\mcitedefaultendpunct}{\mcitedefaultseppunct}\relax
\EndOfBibitem
\bibitem{bes3:2018dzl}
{BESIII} collaboration, M.~Ablikim {\em et al.},
  \href{http://dx.doi.org/10.1103/PhysRevLett.121.081802}{Phys. Rev. Lett. {\bf
  121},  081802 (2018)}, \href{http://arxiv.org/abs/1803.02166}{{\tt
  arXiv:1803.02166 [hep-ex]}}\relax
\mciteBstWouldAddEndPuncttrue
\mciteSetBstMidEndSepPunct{\mcitedefaultmidpunct}
{\mcitedefaultendpunct}{\mcitedefaultseppunct}\relax
\EndOfBibitem
\bibitem{Cheng:2017fkw}
X.-D. Cheng, H.-B. Li, B.~Wei, Y.-G. Xu, and M.-Z. Yang,
  \href{http://dx.doi.org/10.1103/PhysRevD.96.033002}{Phys. Rev. {\bf D96},
  033002 (2017)}, \href{http://arxiv.org/abs/1706.01019}{{\tt arXiv:1706.01019
  [hep-ph]}}\relax
\mciteBstWouldAddEndPuncttrue
\mciteSetBstMidEndSepPunct{\mcitedefaultmidpunct}
{\mcitedefaultendpunct}{\mcitedefaultseppunct}\relax
\EndOfBibitem
\bibitem{cleo:2007Artuso}
{CLEO} collaboration, M.~Artuso {\em et al.},
  \href{http://dx.doi.org/10.1103/PhysRevLett.99.191801}{Phys. Rev. Lett. {\bf
  99},  191801 (2007)}, \href{http://arxiv.org/abs/0705.4276}{{\tt
  arXiv:0705.4276 [hep-ex]}}\relax
\mciteBstWouldAddEndPuncttrue
\mciteSetBstMidEndSepPunct{\mcitedefaultmidpunct}
{\mcitedefaultendpunct}{\mcitedefaultseppunct}\relax
\EndOfBibitem
\bibitem{bes3:2019liuk}
{BESIII} collaboration, M.~Ablikim {\em et al.},    (2019),
  \href{http://arxiv.org/abs/1907.11370}{{\tt arXiv:1907.11370 [hep-ex]}}\relax
\mciteBstWouldAddEndPuncttrue
\mciteSetBstMidEndSepPunct{\mcitedefaultmidpunct}
{\mcitedefaultendpunct}{\mcitedefaultseppunct}\relax
\EndOfBibitem
\bibitem{etm:2015lqcd}
{ETM} collaboration, N.~Carrasco {\em et al.},
  \href{http://dx.doi.org/10.1103/PhysRevD.91.054507}{Phys. Rev. {\bf D91},
  054507 (2015)}, \href{http://arxiv.org/abs/1411.7908}{{\tt arXiv:1411.7908
  [hep-lat]}}\relax
\mciteBstWouldAddEndPuncttrue
\mciteSetBstMidEndSepPunct{\mcitedefaultmidpunct}
{\mcitedefaultendpunct}{\mcitedefaultseppunct}\relax
\EndOfBibitem
\bibitem{Bazavov:2018lqcd}
{ETM} collaboration, A.~Bazavov {\em et al.},
  \href{http://dx.doi.org/10.1103/PhysRevD.98.074512}{Phys. Rev. {\bf D98},
  074512 (2018)}, \href{http://arxiv.org/abs/1712.09262}{{\tt arXiv:1712.09262
  [hep-lat]}}\relax
\mciteBstWouldAddEndPuncttrue
\mciteSetBstMidEndSepPunct{\mcitedefaultmidpunct}
{\mcitedefaultendpunct}{\mcitedefaultseppunct}\relax
\EndOfBibitem
\bibitem{flag:2019}
{FLAG} collaboration, S.~Aoki {\em et al.},    (2019),
  \href{http://arxiv.org/abs/1902.08191}{{\tt arXiv:1902.08191
  [hep-lat]}}\relax
\mciteBstWouldAddEndPuncttrue
\mciteSetBstMidEndSepPunct{\mcitedefaultmidpunct}
{\mcitedefaultendpunct}{\mcitedefaultseppunct}\relax
\EndOfBibitem
\bibitem{Filipuzzi:2012mg}
A.~Filipuzzi, J.~Portoles, and M.~Gonzalez-Alonso,
  \href{http://dx.doi.org/10.1103/PhysRevD.85.116010}{Phys. Rev. {\bf D85},
  116010 (2012)}, \href{http://arxiv.org/abs/1203.2092}{{\tt arXiv:1203.2092
  [hep-ph]}}\relax
\mciteBstWouldAddEndPuncttrue
\mciteSetBstMidEndSepPunct{\mcitedefaultmidpunct}
{\mcitedefaultendpunct}{\mcitedefaultseppunct}\relax
\EndOfBibitem
\bibitem{Ablikim:2018jun}
{BESIII} collaboration, M.~Ablikim {\em et al.},
  \href{http://dx.doi.org/10.1103/PhysRevLett.122.071802}{Phys. Rev. Lett. {\bf
  122},  071802 (2019)}, \href{http://arxiv.org/abs/1811.10890}{{\tt
  arXiv:1811.10890 [hep-ex]}}\relax
\mciteBstWouldAddEndPuncttrue
\mciteSetBstMidEndSepPunct{\mcitedefaultmidpunct}
{\mcitedefaultendpunct}{\mcitedefaultseppunct}\relax
\EndOfBibitem
\bibitem{Ablikim:2013uvu}
{BESIII} collaboration, M.~Ablikim {\em et al.},
  \href{http://dx.doi.org/10.1103/PhysRevD.89.051104}{Phys. Rev. {\bf D89},
  051104 (2014)}, \href{http://arxiv.org/abs/1312.0374}{{\tt arXiv:1312.0374
  [hep-ex]}}\relax
\mciteBstWouldAddEndPuncttrue
\mciteSetBstMidEndSepPunct{\mcitedefaultmidpunct}
{\mcitedefaultendpunct}{\mcitedefaultseppunct}\relax
\EndOfBibitem
\bibitem{delAmoSanchez:2010jg}
{\babar} collaboration, P.~del Amo~Sanchez {\em et al.},
  \href{http://dx.doi.org/10.1103/PhysRevD.82.091103}{Phys. Rev. {\bf D82},
  091103 (2010)}, \href{http://arxiv.org/abs/1008.4080}{{\tt arXiv:1008.4080
  [hep-ex]}}, Erratum ibid.\
  \href{http://dx.doi.org/10.1103/PhysRevD.91.019901}{{\bf D91}, 019901}
  (2015)\relax
\mciteBstWouldAddEndPuncttrue
\mciteSetBstMidEndSepPunct{\mcitedefaultmidpunct}
{\mcitedefaultendpunct}{\mcitedefaultseppunct}\relax
\EndOfBibitem
\bibitem{Zupanc:2013byn}
{Belle} collaboration, A.~Zupanc {\em et al.},
  \href{http://dx.doi.org/10.1007/JHEP09(2013)139}{JHEP {\bf 09},  139 (2013)},
  \href{http://arxiv.org/abs/1307.6240}{{\tt arXiv:1307.6240 [hep-ex]}}\relax
\mciteBstWouldAddEndPuncttrue
\mciteSetBstMidEndSepPunct{\mcitedefaultmidpunct}
{\mcitedefaultendpunct}{\mcitedefaultseppunct}\relax
\EndOfBibitem
\bibitem{Ablikim:2016duz}
{BESIII} collaboration, M.~Ablikim {\em et al.},
  \href{http://dx.doi.org/10.1103/PhysRevD.94.072004}{Phys. Rev. {\bf D94},
  072004 (2016)}, \href{http://arxiv.org/abs/1608.06732}{{\tt arXiv:1608.06732
  [hep-ex]}}\relax
\mciteBstWouldAddEndPuncttrue
\mciteSetBstMidEndSepPunct{\mcitedefaultmidpunct}
{\mcitedefaultendpunct}{\mcitedefaultseppunct}\relax
\EndOfBibitem
\bibitem{Naik:2009tk}
{CLEO} collaboration, P.~Naik {\em et al.},
  \href{http://dx.doi.org/10.1103/PhysRevD.80.112004}{Phys. Rev. {\bf D80},
  112004 (2009)}, \href{http://arxiv.org/abs/0910.3602}{{\tt arXiv:0910.3602
  [hep-ex]}}\relax
\mciteBstWouldAddEndPuncttrue
\mciteSetBstMidEndSepPunct{\mcitedefaultmidpunct}
{\mcitedefaultendpunct}{\mcitedefaultseppunct}\relax
\EndOfBibitem
\bibitem{Onyisi:2009th}
{CLEO} collaboration, P.~Onyisi {\em et al.},
  \href{http://dx.doi.org/10.1103/PhysRevD.79.052002}{Phys. Rev. {\bf D79},
  052002 (2009)}, \href{http://arxiv.org/abs/0901.1147}{{\tt arXiv:0901.1147
  [hep-ex]}}\relax
\mciteBstWouldAddEndPuncttrue
\mciteSetBstMidEndSepPunct{\mcitedefaultmidpunct}
{\mcitedefaultendpunct}{\mcitedefaultseppunct}\relax
\EndOfBibitem
\bibitem{Barberio:1990ms}
E.~Barberio, B.~van Eijk, and Z.~Was,
  \href{http://dx.doi.org/10.1016/0010-4655(91)90012-A}{Comput. Phys. Commun.
  {\bf 66},  115 (1991)}\relax
\mciteBstWouldAddEndPuncttrue
\mciteSetBstMidEndSepPunct{\mcitedefaultmidpunct}
{\mcitedefaultendpunct}{\mcitedefaultseppunct}\relax
\EndOfBibitem
\bibitem{Barberio:1993qi}
E.~Barberio and Z.~Was,
  \href{http://dx.doi.org/10.1016/0010-4655(94)90074-4}{Comput. Phys. Commun.
  {\bf 79},  291 (1994)}\relax
\mciteBstWouldAddEndPuncttrue
\mciteSetBstMidEndSepPunct{\mcitedefaultmidpunct}
{\mcitedefaultendpunct}{\mcitedefaultseppunct}\relax
\EndOfBibitem
\bibitem{Golonka:2005pn}
P.~Golonka and Z.~Was, \href{http://dx.doi.org/10.1140/epjc/s2005-02396-4}{Eur.
  Phys. J. {\bf C45},  97 (2006)},
  \href{http://arxiv.org/abs/hep-ph/0506026}{{\tt arXiv:hep-ph/0506026}}\relax
\mciteBstWouldAddEndPuncttrue
\mciteSetBstMidEndSepPunct{\mcitedefaultmidpunct}
{\mcitedefaultendpunct}{\mcitedefaultseppunct}\relax
\EndOfBibitem
\bibitem{Golonka:2003xt}
P.~Golonka, B.~Kersevan, T.~Pierzchala, E.~Richter-Was, Z.~Was, and M.~Worek,
  \href{http://dx.doi.org/10.1016/j.cpc.2005.12.018}{Comput. Phys. Commun. {\bf
  174},  818--835 (2006)}, \href{http://arxiv.org/abs/hep-ph/0312240}{{\tt
  arXiv:hep-ph/0312240 [hep-ph]}}\relax
\mciteBstWouldAddEndPuncttrue
\mciteSetBstMidEndSepPunct{\mcitedefaultmidpunct}
{\mcitedefaultendpunct}{\mcitedefaultseppunct}\relax
\EndOfBibitem
\bibitem{Golonka:2006tw}
P.~Golonka and Z.~Was,
  \href{http://dx.doi.org/10.1140/epjc/s10052-006-0205-3}{Eur. Phys. J. {\bf
  C50},  53 (2007)}, \href{http://arxiv.org/abs/hep-ph/0604232}{{\tt
  arXiv:hep-ph/0604232}}\relax
\mciteBstWouldAddEndPuncttrue
\mciteSetBstMidEndSepPunct{\mcitedefaultmidpunct}
{\mcitedefaultendpunct}{\mcitedefaultseppunct}\relax
\EndOfBibitem
\bibitem{Ryd:2005zz}
 EVTGEN-V00-11-07, 2005, \url{{http://inspirehep.net/record/707695}}\relax
\mciteBstWouldAddEndPuncttrue
\mciteSetBstMidEndSepPunct{\mcitedefaultmidpunct}
{\mcitedefaultendpunct}{\mcitedefaultseppunct}\relax
\EndOfBibitem
\bibitem{Lange:2001uf}
D.~J. Lange, \href{http://dx.doi.org/10.1016/S0168-9002(01)00089-4}{Nucl.
  Instrum. Meth. {\bf A462},  152--155 (2001)}\relax
\mciteBstWouldAddEndPuncttrue
\mciteSetBstMidEndSepPunct{\mcitedefaultmidpunct}
{\mcitedefaultendpunct}{\mcitedefaultseppunct}\relax
\EndOfBibitem
\bibitem{Link:2002hi}
{FOCUS} collaboration, J.~M. Link {\em et al.},
  \href{http://dx.doi.org/10.1016/S0370-2693(03)00053-4}{Phys. Lett. {\bf
  B555},  167 (2003)}, \href{http://arxiv.org/abs/hep-ex/0212058}{{\tt
  arXiv:hep-ex/0212058}}\relax
\mciteBstWouldAddEndPuncttrue
\mciteSetBstMidEndSepPunct{\mcitedefaultmidpunct}
{\mcitedefaultendpunct}{\mcitedefaultseppunct}\relax
\EndOfBibitem
\bibitem{Ablikim:2018ydy}
{BESIII} collaboration, M.~Ablikim {\em et al.},
  \href{http://dx.doi.org/10.1103/PhysRevD.97.072004}{Phys. Rev. {\bf D97},
  072004 (2018)}, \href{http://arxiv.org/abs/1802.03119}{{\tt arXiv:1802.03119
  [hep-ex]}}\relax
\mciteBstWouldAddEndPuncttrue
\mciteSetBstMidEndSepPunct{\mcitedefaultmidpunct}
{\mcitedefaultendpunct}{\mcitedefaultseppunct}\relax
\EndOfBibitem
\bibitem{Aubert:2007wn}
{\babar} collaboration, B.~Aubert {\em et al.},
  \href{http://dx.doi.org/10.1103/PhysRevLett.100.051802}{Phys. Rev. Lett. {\bf
  100},  051802 (2008)}, \href{http://arxiv.org/abs/0704.2080}{{\tt
  arXiv:0704.2080 [hep-ex]}}\relax
\mciteBstWouldAddEndPuncttrue
\mciteSetBstMidEndSepPunct{\mcitedefaultmidpunct}
{\mcitedefaultendpunct}{\mcitedefaultseppunct}\relax
\EndOfBibitem
\bibitem{Artuso:1997mc}
{CLEO} collaboration, M.~Artuso {\em et al.},
  \href{http://dx.doi.org/10.1103/PhysRevLett.80.3193}{Phys. Rev. Lett. {\bf
  80},  3193 (1998)}, \href{http://arxiv.org/abs/hep-ex/9712023}{{\tt
  arXiv:hep-ex/9712023}}\relax
\mciteBstWouldAddEndPuncttrue
\mciteSetBstMidEndSepPunct{\mcitedefaultmidpunct}
{\mcitedefaultendpunct}{\mcitedefaultseppunct}\relax
\EndOfBibitem
\bibitem{Barate:1997mm}
{ALEPH} collaboration, R.~Barate {\em et al.},
  \href{http://dx.doi.org/10.1016/S0370-2693(97)00585-6}{Phys. Lett. {\bf
  B403},  367 (1997)}\relax
\mciteBstWouldAddEndPuncttrue
\mciteSetBstMidEndSepPunct{\mcitedefaultmidpunct}
{\mcitedefaultendpunct}{\mcitedefaultseppunct}\relax
\EndOfBibitem
\bibitem{Albrecht:1994nb}
{ARGUS} collaboration, H.~Albrecht {\em et al.},
  \href{http://dx.doi.org/10.1016/0370-2693(94)91308-0}{Phys. Lett. {\bf B340},
   125 (1994)}\relax
\mciteBstWouldAddEndPuncttrue
\mciteSetBstMidEndSepPunct{\mcitedefaultmidpunct}
{\mcitedefaultendpunct}{\mcitedefaultseppunct}\relax
\EndOfBibitem
\bibitem{Akerib:1993pm}
{CLEO} collaboration, D.~S. Akerib {\em et al.},
  \href{http://dx.doi.org/10.1103/PhysRevLett.71.3070}{Phys. Rev. Lett. {\bf
  71},  3070 (1993)}\relax
\mciteBstWouldAddEndPuncttrue
\mciteSetBstMidEndSepPunct{\mcitedefaultmidpunct}
{\mcitedefaultendpunct}{\mcitedefaultseppunct}\relax
\EndOfBibitem
\bibitem{Decamp:1991jw}
{ALEPH} collaboration, D.~Decamp {\em et al.},
  \href{http://dx.doi.org/10.1016/0370-2693(91)90769-M}{Phys. Lett. {\bf B266},
   218 (1991)}\relax
\mciteBstWouldAddEndPuncttrue
\mciteSetBstMidEndSepPunct{\mcitedefaultmidpunct}
{\mcitedefaultendpunct}{\mcitedefaultseppunct}\relax
\EndOfBibitem
\bibitem{Acosta:2004ts}
{CDF} collaboration, D.~E. Acosta {\em et al.},
  \href{http://dx.doi.org/10.1103/PhysRevLett.94.122001}{Phys. Rev. Lett. {\bf
  94},  122001 (2005)}, \href{http://arxiv.org/abs/hep-ex/0504006}{{\tt
  arXiv:hep-ex/0504006}}\relax
\mciteBstWouldAddEndPuncttrue
\mciteSetBstMidEndSepPunct{\mcitedefaultmidpunct}
{\mcitedefaultendpunct}{\mcitedefaultseppunct}\relax
\EndOfBibitem
\bibitem{Coan:1997ye}
{CLEO} collaboration, T.~E. Coan {\em et al.},
  \href{http://dx.doi.org/10.1103/PhysRevLett.80.1150}{Phys. Rev. Lett. {\bf
  80},  1150 (1998)}, \href{http://arxiv.org/abs/hep-ex/9710028}{{\tt
  arXiv:hep-ex/9710028}}\relax
\mciteBstWouldAddEndPuncttrue
\mciteSetBstMidEndSepPunct{\mcitedefaultmidpunct}
{\mcitedefaultendpunct}{\mcitedefaultseppunct}\relax
\EndOfBibitem
\bibitem{Davidson:2010ew}
N.~Davidson, T.~Przedzinski, and Z.~Was,
  \href{http://dx.doi.org/10.1016/j.cpc.2015.09.013}{Comput. Phys. Commun. {\bf
  199},  86--101 (2016)}, \href{http://arxiv.org/abs/1011.0937}{{\tt
  arXiv:1011.0937 [hep-ph]}}\relax
\mciteBstWouldAddEndPuncttrue
\mciteSetBstMidEndSepPunct{\mcitedefaultmidpunct}
{\mcitedefaultendpunct}{\mcitedefaultseppunct}\relax
\EndOfBibitem
\bibitem{Ablikim:2019pit}
{BESIII} collaboration, M.~Ablikim {\em et al.},
  \href{http://arxiv.org/abs/1903.04118}{{\tt arXiv:1903.04118 [hep-ex]}}\relax
\mciteBstWouldAddEndPuncttrue
\mciteSetBstMidEndSepPunct{\mcitedefaultmidpunct}
{\mcitedefaultendpunct}{\mcitedefaultseppunct}\relax
\EndOfBibitem
\bibitem{Ablikim:2019whl}
{BESIII} collaboration, M.~Ablikim {\em et al.},
  \href{http://dx.doi.org/10.1103/PhysRevD.99.112005}{Phys. Rev. {\bf D99},
  112005 (2019)}, \href{http://arxiv.org/abs/1903.04164}{{\tt arXiv:1903.04164
  [hep-ex]}}\relax
\mciteBstWouldAddEndPuncttrue
\mciteSetBstMidEndSepPunct{\mcitedefaultmidpunct}
{\mcitedefaultendpunct}{\mcitedefaultseppunct}\relax
\EndOfBibitem
\bibitem{Aubert:2003fg}
{\babar} collaboration, B.~Aubert {\em et al.},
  \href{http://dx.doi.org/10.1103/PhysRevLett.90.242001}{Phys. Rev. Lett. {\bf
  90},  242001 (2003)}, \href{http://arxiv.org/abs/hep-ex/0304021}{{\tt
  arXiv:hep-ex/0304021 [hep-ex]}}\relax
\mciteBstWouldAddEndPuncttrue
\mciteSetBstMidEndSepPunct{\mcitedefaultmidpunct}
{\mcitedefaultendpunct}{\mcitedefaultseppunct}\relax
\EndOfBibitem
\bibitem{Besson:2003cp}
{CLEO} collaboration, D.~Besson {\em et al.},
  \href{http://dx.doi.org/10.1103/PhysRevD.68.032002}{Phys. Rev. {\bf D68},
  032002 (2003)}, \href{http://arxiv.org/abs/hep-ex/0305100}{{\tt
  arXiv:hep-ex/0305100 [hep-ex]}}, Erratum ibid.\
  \href{http://dx.doi.org/10.1103/PhysRevD.75.119908}{{\bf D75}, 119908}
  (2007)\relax
\mciteBstWouldAddEndPuncttrue
\mciteSetBstMidEndSepPunct{\mcitedefaultmidpunct}
{\mcitedefaultendpunct}{\mcitedefaultseppunct}\relax
\EndOfBibitem
\bibitem{Abe:2003jk}
{Belle} collaboration, K.~Abe {\em et al.},
  \href{http://dx.doi.org/10.1103/PhysRevLett.92.012002}{Phys. Rev. Lett. {\bf
  92},  012002 (2004)}, \href{http://arxiv.org/abs/hep-ex/0307052}{{\tt
  arXiv:hep-ex/0307052 [hep-ex]}}\relax
\mciteBstWouldAddEndPuncttrue
\mciteSetBstMidEndSepPunct{\mcitedefaultmidpunct}
{\mcitedefaultendpunct}{\mcitedefaultseppunct}\relax
\EndOfBibitem
\bibitem{Aubert:2003pe}
{\babar} collaboration, B.~Aubert {\em et al.},
  \href{http://dx.doi.org/10.1103/PhysRevD.69.031101}{Phys. Rev. {\bf D69},
  031101 (2004)}, \href{http://arxiv.org/abs/hep-ex/0310050}{{\tt
  arXiv:hep-ex/0310050 [hep-ex]}}\relax
\mciteBstWouldAddEndPuncttrue
\mciteSetBstMidEndSepPunct{\mcitedefaultmidpunct}
{\mcitedefaultendpunct}{\mcitedefaultseppunct}\relax
\EndOfBibitem
\bibitem{Link:2003bd}
{FOCUS} collaboration, J.~M. Link {\em et al.},
  \href{http://dx.doi.org/10.1016/j.physletb.2004.02.017}{Phys. Lett. {\bf
  B586},  11 (2004)}, \href{http://arxiv.org/abs/hep-ex/0312060}{{\tt
  arXiv:hep-ex/0312060 [hep-ex]}}\relax
\mciteBstWouldAddEndPuncttrue
\mciteSetBstMidEndSepPunct{\mcitedefaultmidpunct}
{\mcitedefaultendpunct}{\mcitedefaultseppunct}\relax
\EndOfBibitem
\bibitem{Aaij:2015kqa}
{LHCb} collaboration, R.~Aaij {\em et al.},
  \href{http://dx.doi.org/10.1103/PhysRevD.92.012012}{Phys. Rev. {\bf D92},
  012012 (2015)}, \href{http://arxiv.org/abs/1505.01505}{{\tt arXiv:1505.01505
  [hep-ex]}}\relax
\mciteBstWouldAddEndPuncttrue
\mciteSetBstMidEndSepPunct{\mcitedefaultmidpunct}
{\mcitedefaultendpunct}{\mcitedefaultseppunct}\relax
\EndOfBibitem
\bibitem{Abramowicz:2012ys}
{ZEUS} collaboration, H.~Abramowicz {\em et al.},
  \href{http://dx.doi.org/10.1016/j.nuclphysb.2012.09.007}{Nucl. Phys. {\bf
  B866},  229 (2013)}, \href{http://arxiv.org/abs/1208.4468}{{\tt
  arXiv:1208.4468 [hep-ex]}}\relax
\mciteBstWouldAddEndPuncttrue
\mciteSetBstMidEndSepPunct{\mcitedefaultmidpunct}
{\mcitedefaultendpunct}{\mcitedefaultseppunct}\relax
\EndOfBibitem
\bibitem{Abulencia:2005ry}
{CDF} collaboration, A.~Abulencia {\em et al.},
  \href{http://dx.doi.org/10.1103/PhysRevD.73.051104}{Phys. Rev. {\bf D73},
  051104 (2006)}, \href{http://arxiv.org/abs/hep-ex/0512069}{{\tt
  arXiv:hep-ex/0512069 [hep-ex]}}\relax
\mciteBstWouldAddEndPuncttrue
\mciteSetBstMidEndSepPunct{\mcitedefaultmidpunct}
{\mcitedefaultendpunct}{\mcitedefaultseppunct}\relax
\EndOfBibitem
\bibitem{Avery:1994yc}
{CLEO} collaboration, P.~Avery {\em et al.},
  \href{http://dx.doi.org/10.1016/0370-2693(94)90968-7}{Phys. Lett. {\bf B331},
   236 (1994)}, \href{http://arxiv.org/abs/hep-ph/9403359}{{\tt
  arXiv:hep-ph/9403359 [hep-ph]}}\relax
\mciteBstWouldAddEndPuncttrue
\mciteSetBstMidEndSepPunct{\mcitedefaultmidpunct}
{\mcitedefaultendpunct}{\mcitedefaultseppunct}\relax
\EndOfBibitem
\bibitem{Frabetti:1993vv}
{Fermilab E687} collaboration, P.~L. Frabetti {\em et al.},
  \href{http://dx.doi.org/10.1103/PhysRevLett.72.324}{Phys. Rev. Lett. {\bf
  72},  324 (1994)}\relax
\mciteBstWouldAddEndPuncttrue
\mciteSetBstMidEndSepPunct{\mcitedefaultmidpunct}
{\mcitedefaultendpunct}{\mcitedefaultseppunct}\relax
\EndOfBibitem
\bibitem{Avery:1989ui}
{CLEO} collaboration, P.~Avery {\em et al.},
  \href{http://dx.doi.org/10.1103/PhysRevD.41.774}{Phys. Rev. {\bf D41},  774
  (1990)}\relax
\mciteBstWouldAddEndPuncttrue
\mciteSetBstMidEndSepPunct{\mcitedefaultmidpunct}
{\mcitedefaultendpunct}{\mcitedefaultseppunct}\relax
\EndOfBibitem
\bibitem{Albrecht:1989pa}
{ARGUS} collaboration, H.~Albrecht {\em et al.},
  \href{http://dx.doi.org/10.1016/0370-2693(89)90764-8}{Phys. Lett. {\bf B232},
   398 (1989)}\relax
\mciteBstWouldAddEndPuncttrue
\mciteSetBstMidEndSepPunct{\mcitedefaultmidpunct}
{\mcitedefaultendpunct}{\mcitedefaultseppunct}\relax
\EndOfBibitem
\bibitem{Anjos:1988uf}
{Tagged Photon Spectrometer} collaboration, J.~C. Anjos {\em et al.},
  \href{http://dx.doi.org/10.1103/PhysRevLett.62.1717}{Phys. Rev. Lett. {\bf
  62},  1717 (1989)}\relax
\mciteBstWouldAddEndPuncttrue
\mciteSetBstMidEndSepPunct{\mcitedefaultmidpunct}
{\mcitedefaultendpunct}{\mcitedefaultseppunct}\relax
\EndOfBibitem
\bibitem{Bergfeld:1994af}
{CLEO} collaboration, T.~Bergfeld {\em et al.},
  \href{http://dx.doi.org/10.1016/0370-2693(94)01348-9}{Phys. Lett. {\bf B340},
   194 (1994)}\relax
\mciteBstWouldAddEndPuncttrue
\mciteSetBstMidEndSepPunct{\mcitedefaultmidpunct}
{\mcitedefaultendpunct}{\mcitedefaultseppunct}\relax
\EndOfBibitem
\bibitem{Albrecht:1988dj}
{ARGUS} collaboration, H.~Albrecht {\em et al.},
  \href{http://dx.doi.org/10.1016/0370-2693(89)91737-1}{Phys. Lett. {\bf B221},
   422 (1989)}\relax
\mciteBstWouldAddEndPuncttrue
\mciteSetBstMidEndSepPunct{\mcitedefaultmidpunct}
{\mcitedefaultendpunct}{\mcitedefaultseppunct}\relax
\EndOfBibitem
\bibitem{Aaij:2016fma}
{LHCb} collaboration, R.~Aaij {\em et al.},
  \href{http://dx.doi.org/10.1103/PhysRevD.94.072001}{Phys. Rev. {\bf D94},
  072001 (2016)}, \href{http://arxiv.org/abs/1608.01289}{{\tt arXiv:1608.01289
  [hep-ex]}}\relax
\mciteBstWouldAddEndPuncttrue
\mciteSetBstMidEndSepPunct{\mcitedefaultmidpunct}
{\mcitedefaultendpunct}{\mcitedefaultseppunct}\relax
\EndOfBibitem
\bibitem{Albrecht:1989gb}
{ARGUS} collaboration, H.~Albrecht {\em et al.},
  \href{http://dx.doi.org/10.1016/0370-2693(89)90141-X}{Phys. Lett. {\bf B231},
   208 (1989)}\relax
\mciteBstWouldAddEndPuncttrue
\mciteSetBstMidEndSepPunct{\mcitedefaultmidpunct}
{\mcitedefaultendpunct}{\mcitedefaultseppunct}\relax
\EndOfBibitem
\bibitem{Abreu:1998vk}
{DELPHI} collaboration, P.~Abreu {\em et al.},
  \href{http://dx.doi.org/10.1016/S0370-2693(98)00346-3}{Phys. Lett. {\bf
  B426},  231 (1998)}\relax
\mciteBstWouldAddEndPuncttrue
\mciteSetBstMidEndSepPunct{\mcitedefaultmidpunct}
{\mcitedefaultendpunct}{\mcitedefaultseppunct}\relax
\EndOfBibitem
\bibitem{Aubert:2006bk}
{\babar} collaboration, B.~Aubert {\em et al.},
  \href{http://dx.doi.org/10.1103/PhysRevD.74.032007}{Phys. Rev. {\bf D74},
  032007 (2006)}, \href{http://arxiv.org/abs/hep-ex/0604030}{{\tt
  arXiv:hep-ex/0604030 [hep-ex]}}\relax
\mciteBstWouldAddEndPuncttrue
\mciteSetBstMidEndSepPunct{\mcitedefaultmidpunct}
{\mcitedefaultendpunct}{\mcitedefaultseppunct}\relax
\EndOfBibitem
\bibitem{Ablikim:2017rrr}
{BESIII} collaboration, M.~Ablikim {\em et al.},
  \href{http://dx.doi.org/10.1103/PhysRevD.97.051103}{Phys. Rev. {\bf D97},
  051103 (2018)}, \href{http://arxiv.org/abs/1711.08293}{{\tt arXiv:1711.08293
  [hep-ex]}}\relax
\mciteBstWouldAddEndPuncttrue
\mciteSetBstMidEndSepPunct{\mcitedefaultmidpunct}
{\mcitedefaultendpunct}{\mcitedefaultseppunct}\relax
\EndOfBibitem
\bibitem{Abazov:2007wg}
{\dzero} collaboration, V.~M. Abazov {\em et al.},
  \href{http://dx.doi.org/10.1103/PhysRevLett.102.051801}{Phys. Rev. Lett. {\bf
  102},  051801 (2009)}, \href{http://arxiv.org/abs/0712.3789}{{\tt
  arXiv:0712.3789 [hep-ex]}}\relax
\mciteBstWouldAddEndPuncttrue
\mciteSetBstMidEndSepPunct{\mcitedefaultmidpunct}
{\mcitedefaultendpunct}{\mcitedefaultseppunct}\relax
\EndOfBibitem
\bibitem{Alexander:1993nq}
{CLEO} collaboration, J.~P. Alexander {\em et al.},
  \href{http://dx.doi.org/10.1016/0370-2693(93)91448-V}{Phys. Lett. {\bf B303},
   377 (1993)}\relax
\mciteBstWouldAddEndPuncttrue
\mciteSetBstMidEndSepPunct{\mcitedefaultmidpunct}
{\mcitedefaultendpunct}{\mcitedefaultseppunct}\relax
\EndOfBibitem
\bibitem{Albrecht:1992zh}
{ARGUS} collaboration, H.~Albrecht {\em et al.},
  \href{http://dx.doi.org/10.1016/0370-2693(92)91282-E}{Phys. Lett. {\bf B297},
   425 (1992)}\relax
\mciteBstWouldAddEndPuncttrue
\mciteSetBstMidEndSepPunct{\mcitedefaultmidpunct}
{\mcitedefaultendpunct}{\mcitedefaultseppunct}\relax
\EndOfBibitem
\bibitem{Albrecht:1989yi}
{ARGUS} collaboration, H.~Albrecht {\em et al.},
  \href{http://dx.doi.org/10.1016/0370-2693(89)91672-9}{Phys. Lett. {\bf B230},
   162 (1989)}\relax
\mciteBstWouldAddEndPuncttrue
\mciteSetBstMidEndSepPunct{\mcitedefaultmidpunct}
{\mcitedefaultendpunct}{\mcitedefaultseppunct}\relax
\EndOfBibitem
\bibitem{Lees:2011um}
{\babar} collaboration, J.~P. Lees {\em et al.},
  \href{http://dx.doi.org/10.1103/PhysRevD.83.072003}{Phys. Rev. {\bf D83},
  072003 (2011)}, \href{http://arxiv.org/abs/1103.2675}{{\tt arXiv:1103.2675
  [hep-ex]}}\relax
\mciteBstWouldAddEndPuncttrue
\mciteSetBstMidEndSepPunct{\mcitedefaultmidpunct}
{\mcitedefaultendpunct}{\mcitedefaultseppunct}\relax
\EndOfBibitem
\bibitem{Ablikim:2018qjv}
{BESIII} collaboration, M.~Ablikim {\em et al.},
  \href{http://arxiv.org/abs/1812.09800}{{\tt arXiv:1812.09800 [hep-ex]}}\relax
\mciteBstWouldAddEndPuncttrue
\mciteSetBstMidEndSepPunct{\mcitedefaultmidpunct}
{\mcitedefaultendpunct}{\mcitedefaultseppunct}\relax
\EndOfBibitem
\bibitem{Aaij:2014baa}
{LHCb} collaboration, R.~Aaij {\em et al.},
  \href{http://dx.doi.org/10.1103/PhysRevD.90.072003}{Phys. Rev. {\bf D90},
  072003 (2014)}, \href{http://arxiv.org/abs/1407.7712}{{\tt arXiv:1407.7712
  [hep-ex]}}\relax
\mciteBstWouldAddEndPuncttrue
\mciteSetBstMidEndSepPunct{\mcitedefaultmidpunct}
{\mcitedefaultendpunct}{\mcitedefaultseppunct}\relax
\EndOfBibitem
\bibitem{Aaij:2011ju}
{LHCb} collaboration, R.~Aaij {\em et al.},
  \href{http://dx.doi.org/10.1016/j.physletb.2011.02.039}{Phys. Lett. {\bf
  B698},  14 (2011)}, \href{http://arxiv.org/abs/1102.0348}{{\tt
  arXiv:1102.0348 [hep-ex]}}\relax
\mciteBstWouldAddEndPuncttrue
\mciteSetBstMidEndSepPunct{\mcitedefaultmidpunct}
{\mcitedefaultendpunct}{\mcitedefaultseppunct}\relax
\EndOfBibitem
\bibitem{Aubert:2006mh}
{\babar} collaboration, B.~Aubert {\em et al.},
  \href{http://dx.doi.org/10.1103/PhysRevLett.97.222001}{Phys. Rev. Lett. {\bf
  97},  222001 (2006)}, \href{http://arxiv.org/abs/hep-ex/0607082}{{\tt
  arXiv:hep-ex/0607082 [hep-ex]}}\relax
\mciteBstWouldAddEndPuncttrue
\mciteSetBstMidEndSepPunct{\mcitedefaultmidpunct}
{\mcitedefaultendpunct}{\mcitedefaultseppunct}\relax
\EndOfBibitem
\bibitem{Albrecht:1995qx}
{ARGUS} collaboration, H.~Albrecht {\em et al.},
  \href{http://dx.doi.org/10.1007/s002880050040}{Z. Phys. {\bf C69},  405
  (1996)}\relax
\mciteBstWouldAddEndPuncttrue
\mciteSetBstMidEndSepPunct{\mcitedefaultmidpunct}
{\mcitedefaultendpunct}{\mcitedefaultseppunct}\relax
\EndOfBibitem
\bibitem{Kubota:1994gn}
{CLEO} collaboration, Y.~Kubota {\em et al.},
  \href{http://dx.doi.org/10.1103/PhysRevLett.72.1972}{Phys. Rev. Lett. {\bf
  72},  1972 (1994)}, \href{http://arxiv.org/abs/hep-ph/9403325}{{\tt
  arXiv:hep-ph/9403325 [hep-ph]}}\relax
\mciteBstWouldAddEndPuncttrue
\mciteSetBstMidEndSepPunct{\mcitedefaultmidpunct}
{\mcitedefaultendpunct}{\mcitedefaultseppunct}\relax
\EndOfBibitem
\bibitem{Aaij:2016utb}
{LHCb} collaboration, R.~Aaij {\em et al.},
  \href{http://dx.doi.org/10.1007/JHEP02(2016)133}{JHEP {\bf 02},  133 (2016)},
  \href{http://arxiv.org/abs/1601.01495}{{\tt arXiv:1601.01495 [hep-ex]}}\relax
\mciteBstWouldAddEndPuncttrue
\mciteSetBstMidEndSepPunct{\mcitedefaultmidpunct}
{\mcitedefaultendpunct}{\mcitedefaultseppunct}\relax
\EndOfBibitem
\bibitem{Lees:2014abp}
{\babar} collaboration, J.~P. Lees {\em et al.},
  \href{http://dx.doi.org/10.1103/PhysRevD.91.052002}{Phys. Rev. {\bf D91},
  052002 (2015)}, \href{http://arxiv.org/abs/1412.6751}{{\tt arXiv:1412.6751
  [hep-ex]}}\relax
\mciteBstWouldAddEndPuncttrue
\mciteSetBstMidEndSepPunct{\mcitedefaultmidpunct}
{\mcitedefaultendpunct}{\mcitedefaultseppunct}\relax
\EndOfBibitem
\bibitem{Aaij:2012pc}
{LHCb} collaboration, R.~Aaij {\em et al.},
  \href{http://dx.doi.org/10.1007/JHEP10(2012)151}{JHEP {\bf 10},  151 (2012)},
  \href{http://arxiv.org/abs/1207.6016}{{\tt arXiv:1207.6016 [hep-ex]}}\relax
\mciteBstWouldAddEndPuncttrue
\mciteSetBstMidEndSepPunct{\mcitedefaultmidpunct}
{\mcitedefaultendpunct}{\mcitedefaultseppunct}\relax
\EndOfBibitem
\bibitem{Aubert:2009ah}
{\babar} collaboration, B.~Aubert {\em et al.},
  \href{http://dx.doi.org/10.1103/PhysRevD.80.092003}{Phys. Rev. {\bf D80},
  092003 (2009)}, \href{http://arxiv.org/abs/0908.0806}{{\tt arXiv:0908.0806
  [hep-ex]}}\relax
\mciteBstWouldAddEndPuncttrue
\mciteSetBstMidEndSepPunct{\mcitedefaultmidpunct}
{\mcitedefaultendpunct}{\mcitedefaultseppunct}\relax
\EndOfBibitem
\bibitem{Aaij:2014xza}
{LHCb} collaboration, R.~Aaij {\em et al.},
  \href{http://dx.doi.org/10.1103/PhysRevLett.113.162001}{Phys. Rev. Lett. {\bf
  113},  162001 (2014)}, \href{http://arxiv.org/abs/1407.7574}{{\tt
  arXiv:1407.7574 [hep-ex]}}\relax
\mciteBstWouldAddEndPuncttrue
\mciteSetBstMidEndSepPunct{\mcitedefaultmidpunct}
{\mcitedefaultendpunct}{\mcitedefaultseppunct}\relax
\EndOfBibitem
\bibitem{Godfrey:1985xj}
S.~Godfrey and N.~Isgur, \href{http://dx.doi.org/10.1103/PhysRevD.32.189}{Phys.
  Rev. {\bf D32},  189 (1985)}\relax
\mciteBstWouldAddEndPuncttrue
\mciteSetBstMidEndSepPunct{\mcitedefaultmidpunct}
{\mcitedefaultendpunct}{\mcitedefaultseppunct}\relax
\EndOfBibitem
\bibitem{Godfrey:1986wj}
S.~Godfrey and R.~Kokoski,
  \href{http://dx.doi.org/10.1103/PhysRevD.43.1679}{Phys. Rev. {\bf D43},  1679
  (1991)}\relax
\mciteBstWouldAddEndPuncttrue
\mciteSetBstMidEndSepPunct{\mcitedefaultmidpunct}
{\mcitedefaultendpunct}{\mcitedefaultseppunct}\relax
\EndOfBibitem
\bibitem{Schweitzer:2002nm}
P.~Schweitzer, S.~Boffi, and M.~Radici,
  \href{http://dx.doi.org/10.1103/PhysRevD.66.114004}{Phys. Rev. {\bf D66},
  114004 (2002)}, \href{http://arxiv.org/abs/hep-ph/0207230}{{\tt
  arXiv:hep-ph/0207230 [hep-ph]}}\relax
\mciteBstWouldAddEndPuncttrue
\mciteSetBstMidEndSepPunct{\mcitedefaultmidpunct}
{\mcitedefaultendpunct}{\mcitedefaultseppunct}\relax
\EndOfBibitem
\bibitem{Chen:2015lpa}
B.~Chen, X.~Liu, and A.~Zhang,
  \href{http://dx.doi.org/10.1103/PhysRevD.92.034005}{Phys. Rev. {\bf D92},
  034005 (2015)}, \href{http://arxiv.org/abs/1507.02339}{{\tt arXiv:1507.02339
  [hep-ph]}}\relax
\mciteBstWouldAddEndPuncttrue
\mciteSetBstMidEndSepPunct{\mcitedefaultmidpunct}
{\mcitedefaultendpunct}{\mcitedefaultseppunct}\relax
\EndOfBibitem
\bibitem{Godfrey:2015dva}
S.~Godfrey and K.~Moats,
  \href{http://dx.doi.org/10.1103/PhysRevD.93.034035}{Phys. Rev. {\bf D93},
  034035 (2016)}, \href{http://arxiv.org/abs/1510.08305}{{\tt arXiv:1510.08305
  [hep-ph]}}\relax
\mciteBstWouldAddEndPuncttrue
\mciteSetBstMidEndSepPunct{\mcitedefaultmidpunct}
{\mcitedefaultendpunct}{\mcitedefaultseppunct}\relax
\EndOfBibitem
\bibitem{Jugeau:2005yr}
F.~Jugeau, A.~Le~Yaouanc, L.~Oliver, and J.-C. Raynal,
  \href{http://dx.doi.org/10.1103/PhysRevD.72.094010}{Phys. Rev. {\bf D72},
  094010 (2005)}, \href{http://arxiv.org/abs/hep-ph/0504206}{{\tt
  arXiv:hep-ph/0504206 [hep-ph]}}\relax
\mciteBstWouldAddEndPuncttrue
\mciteSetBstMidEndSepPunct{\mcitedefaultmidpunct}
{\mcitedefaultendpunct}{\mcitedefaultseppunct}\relax
\EndOfBibitem
\bibitem{Colangelo:2004vu}
P.~Colangelo, F.~De~Fazio, and R.~Ferrandes,
  \href{http://dx.doi.org/10.1142/S0217732304015269}{Mod. Phys. Lett. {\bf
  A19},  2083 (2004)}, \href{http://arxiv.org/abs/hep-ph/0407137}{{\tt
  arXiv:hep-ph/0407137 [hep-ph]}}\relax
\mciteBstWouldAddEndPuncttrue
\mciteSetBstMidEndSepPunct{\mcitedefaultmidpunct}
{\mcitedefaultendpunct}{\mcitedefaultseppunct}\relax
\EndOfBibitem
\bibitem{Cahn:2003cw}
R.~N. Cahn and J.~D. Jackson,
  \href{http://dx.doi.org/10.1103/PhysRevD.68.037502}{Phys. Rev. {\bf D68},
  037502 (2003)}, \href{http://arxiv.org/abs/hep-ph/0305012}{{\tt
  arXiv:hep-ph/0305012 [hep-ph]}}\relax
\mciteBstWouldAddEndPuncttrue
\mciteSetBstMidEndSepPunct{\mcitedefaultmidpunct}
{\mcitedefaultendpunct}{\mcitedefaultseppunct}\relax
\EndOfBibitem
\bibitem{Barnes:2003dj}
T.~Barnes, F.~E. Close, and H.~J. Lipkin,
  \href{http://dx.doi.org/10.1103/PhysRevD.68.054006}{Phys. Rev. {\bf D68},
  054006 (2003)}, \href{http://arxiv.org/abs/hep-ph/0305025}{{\tt
  arXiv:hep-ph/0305025 [hep-ph]}}\relax
\mciteBstWouldAddEndPuncttrue
\mciteSetBstMidEndSepPunct{\mcitedefaultmidpunct}
{\mcitedefaultendpunct}{\mcitedefaultseppunct}\relax
\EndOfBibitem
\bibitem{Lipkin:2003zk}
H.~Lipkin, \href{http://dx.doi.org/10.1016/j.physletb.2003.10.117}{Phys. Lett.
  {\bf B580},  50 (2004)}, \href{http://arxiv.org/abs/hep-ph/0306204}{{\tt
  arXiv:hep-ph/0306204 [hep-ph]}}\relax
\mciteBstWouldAddEndPuncttrue
\mciteSetBstMidEndSepPunct{\mcitedefaultmidpunct}
{\mcitedefaultendpunct}{\mcitedefaultseppunct}\relax
\EndOfBibitem
\bibitem{Bardeen:2003kt}
W.~A. Bardeen, E.~J. Eichten, and C.~T. Hill,
  \href{http://dx.doi.org/10.1103/PhysRevD.68.054024}{Phys. Rev. {\bf D68},
  054024 (2003)}, \href{http://arxiv.org/abs/hep-ph/0305049}{{\tt
  arXiv:hep-ph/0305049 [hep-ph]}}\relax
\mciteBstWouldAddEndPuncttrue
\mciteSetBstMidEndSepPunct{\mcitedefaultmidpunct}
{\mcitedefaultendpunct}{\mcitedefaultseppunct}\relax
\EndOfBibitem
\bibitem{Matsuki:2006rz}
T.~Matsuki, T.~Morii, and K.~Sudoh,
  \href{http://dx.doi.org/10.1140/epja/i2006-10287-1}{Eur. Phys. J. {\bf A31},
  701 (2007)}, \href{http://arxiv.org/abs/hep-ph/0610186}{{\tt
  arXiv:hep-ph/0610186 [hep-ph]}}\relax
\mciteBstWouldAddEndPuncttrue
\mciteSetBstMidEndSepPunct{\mcitedefaultmidpunct}
{\mcitedefaultendpunct}{\mcitedefaultseppunct}\relax
\EndOfBibitem
\bibitem{Isgur:1989vq}
N.~Isgur and M.~B. Wise,
  \href{http://dx.doi.org/10.1016/0370-2693(89)90566-2}{Phys. Lett. {\bf B232},
   113 (1989)}\relax
\mciteBstWouldAddEndPuncttrue
\mciteSetBstMidEndSepPunct{\mcitedefaultmidpunct}
{\mcitedefaultendpunct}{\mcitedefaultseppunct}\relax
\EndOfBibitem
\bibitem{Chekanov:2008ac}
{ZEUS} collaboration, S.~Chekanov {\em et al.},
  \href{http://dx.doi.org/10.1140/epjc/s10052-009-0881-x}{Eur. Phys. J. {\bf
  C60},  25 (2009)}, \href{http://arxiv.org/abs/0807.1290}{{\tt arXiv:0807.1290
  [hep-ex]}}\relax
\mciteBstWouldAddEndPuncttrue
\mciteSetBstMidEndSepPunct{\mcitedefaultmidpunct}
{\mcitedefaultendpunct}{\mcitedefaultseppunct}\relax
\EndOfBibitem
\bibitem{Heister:2001nj}
{ALEPH} collaboration, A.~Heister {\em et al.},
  \href{http://dx.doi.org/10.1016/S0370-2693(01)01465-4}{Phys. Lett. {\bf
  B526},  34 (2002)}, \href{http://arxiv.org/abs/hep-ex/0112010}{{\tt
  arXiv:hep-ex/0112010 [hep-ex]}}\relax
\mciteBstWouldAddEndPuncttrue
\mciteSetBstMidEndSepPunct{\mcitedefaultmidpunct}
{\mcitedefaultendpunct}{\mcitedefaultseppunct}\relax
\EndOfBibitem
\bibitem{Aubert:2005gt}
{\babar} collaboration, B.~Aubert {\em et al.},
  \href{http://dx.doi.org/10.1103/PhysRevD.72.052006}{Phys. Rev. {\bf D72},
  052006 (2005)}, \href{http://arxiv.org/abs/hep-ex/0507009}{{\tt
  arXiv:hep-ex/0507009 [hep-ex]}}\relax
\mciteBstWouldAddEndPuncttrue
\mciteSetBstMidEndSepPunct{\mcitedefaultmidpunct}
{\mcitedefaultendpunct}{\mcitedefaultseppunct}\relax
\EndOfBibitem
\bibitem{Solovieva:2008fw}
{Belle} collaboration, E.~Solovieva {\em et al.},
  \href{http://dx.doi.org/10.1016/j.physletb.2008.12.062}{Phys. Lett. {\bf
  B672},  1 (2009)}, \href{http://arxiv.org/abs/0808.3677}{{\tt arXiv:0808.3677
  [hep-ex]}}\relax
\mciteBstWouldAddEndPuncttrue
\mciteSetBstMidEndSepPunct{\mcitedefaultmidpunct}
{\mcitedefaultendpunct}{\mcitedefaultseppunct}\relax
\EndOfBibitem
\bibitem{Aaltonen:2011sf}
{CDF} collaboration, T.~Aaltonen {\em et al.},
  \href{http://dx.doi.org/10.1103/PhysRevD.84.012003}{Phys. Rev. {\bf D84},
  012003 (2011)}, \href{http://arxiv.org/abs/1105.5995}{{\tt arXiv:1105.5995
  [hep-ex]}}\relax
\mciteBstWouldAddEndPuncttrue
\mciteSetBstMidEndSepPunct{\mcitedefaultmidpunct}
{\mcitedefaultendpunct}{\mcitedefaultseppunct}\relax
\EndOfBibitem
\bibitem{Artuso:2000xy}
{CLEO} collaboration, M.~Artuso {\em et al.},
  \href{http://dx.doi.org/10.1103/PhysRevLett.86.4479}{Phys. Rev. Lett. {\bf
  86},  4479 (2001)}, \href{http://arxiv.org/abs/hep-ex/0010080}{{\tt
  arXiv:hep-ex/0010080 [hep-ex]}}\relax
\mciteBstWouldAddEndPuncttrue
\mciteSetBstMidEndSepPunct{\mcitedefaultmidpunct}
{\mcitedefaultendpunct}{\mcitedefaultseppunct}\relax
\EndOfBibitem
\bibitem{Aubert:2006sp}
{\babar} collaboration, B.~Aubert {\em et al.},
  \href{http://dx.doi.org/10.1103/PhysRevLett.98.012001}{Phys. Rev. Lett. {\bf
  98},  012001 (2007)}, \href{http://arxiv.org/abs/hep-ex/0603052}{{\tt
  arXiv:hep-ex/0603052 [hep-ex]}}\relax
\mciteBstWouldAddEndPuncttrue
\mciteSetBstMidEndSepPunct{\mcitedefaultmidpunct}
{\mcitedefaultendpunct}{\mcitedefaultseppunct}\relax
\EndOfBibitem
\bibitem{Cheng:2006dk}
H.-Y. Cheng and C.-K. Chua,
  \href{http://dx.doi.org/10.1103/PhysRevD.75.014006}{Phys. Rev. {\bf D75},
  014006 (2007)}, \href{http://arxiv.org/abs/hep-ph/0610283}{{\tt
  arXiv:hep-ph/0610283 [hep-ph]}}\relax
\mciteBstWouldAddEndPuncttrue
\mciteSetBstMidEndSepPunct{\mcitedefaultmidpunct}
{\mcitedefaultendpunct}{\mcitedefaultseppunct}\relax
\EndOfBibitem
\bibitem{Joo:2014fka}
C.~W. Joo, Y.~Kato, K.~Tanida, and Y.~Kato,
  \href{http://dx.doi.org/10.22323/1.205.0201}{PoS {\bf Hadron2013},  201
  (2013)}\relax
\mciteBstWouldAddEndPuncttrue
\mciteSetBstMidEndSepPunct{\mcitedefaultmidpunct}
{\mcitedefaultendpunct}{\mcitedefaultseppunct}\relax
\EndOfBibitem
\bibitem{Aaij:2017svr}
{LHCb} collaboration, R.~Aaij {\em et al.},
  \href{http://dx.doi.org/10.1103/PhysRevD.96.112005}{Phys. Rev. {\bf D96},
  112005 (2017)}, \href{http://arxiv.org/abs/1709.01920}{{\tt arXiv:1709.01920
  [hep-ex]}}\relax
\mciteBstWouldAddEndPuncttrue
\mciteSetBstMidEndSepPunct{\mcitedefaultmidpunct}
{\mcitedefaultendpunct}{\mcitedefaultseppunct}\relax
\EndOfBibitem
\bibitem{Lee:2014htd}
{Belle} collaboration, S.~H. Lee {\em et al.},
  \href{http://dx.doi.org/10.1103/PhysRevD.89.091102}{Phys. Rev. {\bf D89},
  091102 (2014)}, \href{http://arxiv.org/abs/1404.5389}{{\tt arXiv:1404.5389
  [hep-ex]}}\relax
\mciteBstWouldAddEndPuncttrue
\mciteSetBstMidEndSepPunct{\mcitedefaultmidpunct}
{\mcitedefaultendpunct}{\mcitedefaultseppunct}\relax
\EndOfBibitem
\bibitem{Mizuk:2004yu}
{Belle} collaboration, R.~Mizuk {\em et al.},
  \href{http://dx.doi.org/10.1103/PhysRevLett.94.122002}{Phys. Rev. Lett. {\bf
  94},  122002 (2005)}, \href{http://arxiv.org/abs/hep-ex/0412069}{{\tt
  arXiv:hep-ex/0412069 [hep-ex]}}\relax
\mciteBstWouldAddEndPuncttrue
\mciteSetBstMidEndSepPunct{\mcitedefaultmidpunct}
{\mcitedefaultendpunct}{\mcitedefaultseppunct}\relax
\EndOfBibitem
\bibitem{Copley:1979wj}
L.~Copley, N.~Isgur, and G.~Karl,
  \href{http://dx.doi.org/10.1103/PhysRevD.20.768}{Phys. Rev. {\bf D20},  768
  (1979)}, Erratum ibid.\ \href{http://dx.doi.org/10.1103/PhysRevD.23.817}{{\bf
  D23}, 817}, (1981)\relax
\mciteBstWouldAddEndPuncttrue
\mciteSetBstMidEndSepPunct{\mcitedefaultmidpunct}
{\mcitedefaultendpunct}{\mcitedefaultseppunct}\relax
\EndOfBibitem
\bibitem{Pirjol:1997nh}
D.~Pirjol and T.-M. Yan,
  \href{http://dx.doi.org/10.1103/PhysRevD.56.5483}{Phys. Rev. {\bf D56},  5483
  (1997)}, \href{http://arxiv.org/abs/hep-ph/9701291}{{\tt arXiv:hep-ph/9701291
  [hep-ph]}}\relax
\mciteBstWouldAddEndPuncttrue
\mciteSetBstMidEndSepPunct{\mcitedefaultmidpunct}
{\mcitedefaultendpunct}{\mcitedefaultseppunct}\relax
\EndOfBibitem
\bibitem{Chistov:2006zj}
{Belle} collaboration, R.~Chistov {\em et al.},
  \href{http://dx.doi.org/10.1103/PhysRevLett.97.162001}{Phys. Rev. Lett. {\bf
  97},  162001 (2006)}, \href{http://arxiv.org/abs/hep-ex/0606051}{{\tt
  arXiv:hep-ex/0606051 [hep-ex]}}\relax
\mciteBstWouldAddEndPuncttrue
\mciteSetBstMidEndSepPunct{\mcitedefaultmidpunct}
{\mcitedefaultendpunct}{\mcitedefaultseppunct}\relax
\EndOfBibitem
\bibitem{YKato:2014}
{Belle} collaboration, Y.~Kato {\em et al.},
  \href{http://dx.doi.org/10.1103/PhysRevD.89.052003}{Phys. Rev. {\bf D89},
  052003 (2014)}, \href{http://arxiv.org/abs/1312.1026}{{\tt arXiv:1312.1026
  [hep-ex]}}\relax
\mciteBstWouldAddEndPuncttrue
\mciteSetBstMidEndSepPunct{\mcitedefaultmidpunct}
{\mcitedefaultendpunct}{\mcitedefaultseppunct}\relax
\EndOfBibitem
\bibitem{YKato:2016}
{Belle} collaboration, Y.~Kato {\em et al.},
  \href{http://dx.doi.org/10.1103/PhysRevD.94.032002}{Phys. Rev. {\bf D94},
  032002 (2016)}, \href{http://arxiv.org/abs/1605.09103}{{\tt arXiv:1605.09103
  [hep-ex]}}\relax
\mciteBstWouldAddEndPuncttrue
\mciteSetBstMidEndSepPunct{\mcitedefaultmidpunct}
{\mcitedefaultendpunct}{\mcitedefaultseppunct}\relax
\EndOfBibitem
\bibitem{Aubert:2007bd}
{\babar} collaboration, B.~Aubert {\em et al.},
  \href{http://dx.doi.org/10.1103/PhysRevD.77.031101}{Phys. Rev. {\bf D77},
  031101 (2008)}, \href{http://arxiv.org/abs/0710.5775}{{\tt arXiv:0710.5775
  [hep-ex]}}\relax
\mciteBstWouldAddEndPuncttrue
\mciteSetBstMidEndSepPunct{\mcitedefaultmidpunct}
{\mcitedefaultendpunct}{\mcitedefaultseppunct}\relax
\EndOfBibitem
\bibitem{Li:2018fmq}
{Belle} collaboration, Y.~B. Li {\em et al.},
  \href{http://dx.doi.org/10.1140/epjc/s10052-018-6425-5}{Eur. Phys. J. {\bf
  C78},  928 (2018)}, \href{http://arxiv.org/abs/1806.09182}{{\tt
  arXiv:1806.09182 [hep-ex]}}\relax
\mciteBstWouldAddEndPuncttrue
\mciteSetBstMidEndSepPunct{\mcitedefaultmidpunct}
{\mcitedefaultendpunct}{\mcitedefaultseppunct}\relax
\EndOfBibitem
\bibitem{Yelton:2016fqw}
{Belle} collaboration, J.~Yelton {\em et al.},
  \href{http://dx.doi.org/10.1103/PhysRevD.94.052011}{Phys. Rev. {\bf D94},
  052011 (2016)}, \href{http://arxiv.org/abs/1607.07123}{{\tt arXiv:1607.07123
  [hep-ex]}}\relax
\mciteBstWouldAddEndPuncttrue
\mciteSetBstMidEndSepPunct{\mcitedefaultmidpunct}
{\mcitedefaultendpunct}{\mcitedefaultseppunct}\relax
\EndOfBibitem
\bibitem{Aubert:2006je}
{\babar} collaboration, B.~Aubert {\em et al.},
  \href{http://dx.doi.org/10.1103/PhysRevLett.97.232001}{Phys. Rev. Lett. {\bf
  97},  232001 (2006)}, \href{http://arxiv.org/abs/hep-ex/0608055}{{\tt
  arXiv:hep-ex/0608055 [hep-ex]}}\relax
\mciteBstWouldAddEndPuncttrue
\mciteSetBstMidEndSepPunct{\mcitedefaultmidpunct}
{\mcitedefaultendpunct}{\mcitedefaultseppunct}\relax
\EndOfBibitem
\bibitem{Rosner:1995yu}
J.~L. Rosner, \href{http://dx.doi.org/10.1103/PhysRevD.52.6461}{Phys. Rev. {\bf
  D52},  6461 (1995)}, \href{http://arxiv.org/abs/hep-ph/9508252}{{\tt
  arXiv:hep-ph/9508252 [hep-ph]}}\relax
\mciteBstWouldAddEndPuncttrue
\mciteSetBstMidEndSepPunct{\mcitedefaultmidpunct}
{\mcitedefaultendpunct}{\mcitedefaultseppunct}\relax
\EndOfBibitem
\bibitem{Glozman:1995xy}
L.~Y. Glozman and D.~Riska,
  \href{http://dx.doi.org/10.1016/0375-9474(96)80005-C}{Nucl. Phys. {\bf A603},
   326 (1996)}, \href{http://arxiv.org/abs/hep-ph/9509269}{{\tt
  arXiv:hep-ph/9509269 [hep-ph]}}, Erratum ibid.\
  \href{http://dx.doi.org/10.1016/S0375-9474(97)00200-5}{{\bf A620}, 510},
  (1997)\relax
\mciteBstWouldAddEndPuncttrue
\mciteSetBstMidEndSepPunct{\mcitedefaultmidpunct}
{\mcitedefaultendpunct}{\mcitedefaultseppunct}\relax
\EndOfBibitem
\bibitem{Jenkins:1996de}
E.~E. Jenkins, \href{http://dx.doi.org/10.1103/PhysRevD.54.4515}{Phys. Rev.
  {\bf D54},  4515 (1996)}, \href{http://arxiv.org/abs/hep-ph/9603449}{{\tt
  arXiv:hep-ph/9603449 [hep-ph]}}\relax
\mciteBstWouldAddEndPuncttrue
\mciteSetBstMidEndSepPunct{\mcitedefaultmidpunct}
{\mcitedefaultendpunct}{\mcitedefaultseppunct}\relax
\EndOfBibitem
\bibitem{Burakovsky:1997vm}
L.~Burakovsky, J.~T. Goldman, and L.~Horwitz,
  \href{http://dx.doi.org/10.1103/PhysRevD.56.7124}{Phys. Rev. {\bf D56},  7124
  (1997)}, \href{http://arxiv.org/abs/hep-ph/9706464}{{\tt arXiv:hep-ph/9706464
  [hep-ph]}}\relax
\mciteBstWouldAddEndPuncttrue
\mciteSetBstMidEndSepPunct{\mcitedefaultmidpunct}
{\mcitedefaultendpunct}{\mcitedefaultseppunct}\relax
\EndOfBibitem
\bibitem{Aaij:2017nav}
{LHCb} collaboration, R.~Aaij {\em et al.},
  \href{http://dx.doi.org/10.1103/PhysRevLett.118.182001}{Phys. Rev. Lett. {\bf
  118},  182001 (2017)}, \href{http://arxiv.org/abs/1703.04639}{{\tt
  arXiv:1703.04639 [hep-ex]}}\relax
\mciteBstWouldAddEndPuncttrue
\mciteSetBstMidEndSepPunct{\mcitedefaultmidpunct}
{\mcitedefaultendpunct}{\mcitedefaultseppunct}\relax
\EndOfBibitem
\bibitem{Yelton:2017qxg}
{Belle} collaboration, J.~Yelton {\em et al.},
  \href{http://dx.doi.org/10.1103/PhysRevD.97.051102}{Phys. Rev. {\bf D97},
  051102 (2018)}, \href{http://arxiv.org/abs/1711.07927}{{\tt arXiv:1711.07927
  [hep-ex]}}\relax
\mciteBstWouldAddEndPuncttrue
\mciteSetBstMidEndSepPunct{\mcitedefaultmidpunct}
{\mcitedefaultendpunct}{\mcitedefaultseppunct}\relax
\EndOfBibitem
\bibitem{Burdman:2001tf}
G.~Burdman, E.~Golowich, J.~L. Hewett, and S.~Pakvasa,
  \href{http://dx.doi.org/10.1103/PhysRevD.66.014009}{Phys. Rev. {\bf D66},
  014009 (2002)}, \href{http://arxiv.org/abs/hep-ph/0112235}{{\tt
  arXiv:hep-ph/0112235 [hep-ph]}}\relax
\mciteBstWouldAddEndPuncttrue
\mciteSetBstMidEndSepPunct{\mcitedefaultmidpunct}
{\mcitedefaultendpunct}{\mcitedefaultseppunct}\relax
\EndOfBibitem
\bibitem{Fajfer:2002bu}
S.~Fajfer, A.~Prapotnik, S.~Prelovsek, P.~Singer, and J.~Zupan,
  \href{http://dx.doi.org/10.1016/S0920-5632(02)01961-8}{Nucl. Phys. Proc.
  Suppl. {\bf 115},  93 (2003)},
  \href{http://arxiv.org/abs/hep-ph/0208201}{{\tt arXiv:hep-ph/0208201
  [hep-ph]}}\relax
\mciteBstWouldAddEndPuncttrue
\mciteSetBstMidEndSepPunct{\mcitedefaultmidpunct}
{\mcitedefaultendpunct}{\mcitedefaultseppunct}\relax
\EndOfBibitem
\bibitem{Fajfer:2007dy}
S.~Fajfer, N.~Kosnik, and S.~Prelovsek,
  \href{http://dx.doi.org/10.1103/PhysRevD.76.074010}{Phys. Rev. {\bf D76},
  074010 (2007)}, \href{http://arxiv.org/abs/0706.1133}{{\tt arXiv:0706.1133
  [hep-ph]}}\relax
\mciteBstWouldAddEndPuncttrue
\mciteSetBstMidEndSepPunct{\mcitedefaultmidpunct}
{\mcitedefaultendpunct}{\mcitedefaultseppunct}\relax
\EndOfBibitem
\bibitem{Golowich:2009ii}
E.~Golowich, J.~Hewett, S.~Pakvasa, and A.~A. Petrov,
  \href{http://dx.doi.org/10.1103/PhysRevD.79.114030}{Phys. Rev. {\bf D79},
  114030 (2009)}, \href{http://arxiv.org/abs/0903.2830}{{\tt arXiv:0903.2830
  [hep-ph]}}\relax
\mciteBstWouldAddEndPuncttrue
\mciteSetBstMidEndSepPunct{\mcitedefaultmidpunct}
{\mcitedefaultendpunct}{\mcitedefaultseppunct}\relax
\EndOfBibitem
\bibitem{Paul:2010pq}
A.~Paul, I.~I. Bigi, and S.~Recksiegel,
  \href{http://dx.doi.org/10.1103/PhysRevD.82.094006}{Phys. Rev. {\bf D82},
  094006 (2010)}, \href{http://arxiv.org/abs/1008.3141}{{\tt arXiv:1008.3141
  [hep-ph]}}, Erratum ibid.\
  \href{http://dx.doi.org/10.1103/PhysRevD.83.019901}{{\bf D83}, 019901},
  (2011)\relax
\mciteBstWouldAddEndPuncttrue
\mciteSetBstMidEndSepPunct{\mcitedefaultmidpunct}
{\mcitedefaultendpunct}{\mcitedefaultseppunct}\relax
\EndOfBibitem
\bibitem{Borisov:2011aa}
A.~Borisov, \href{http://arxiv.org/abs/1112.3269}{{\tt arXiv:1112.3269
  [hep-ph]}} (2011)\relax
\mciteBstWouldAddEndPuncttrue
\mciteSetBstMidEndSepPunct{\mcitedefaultmidpunct}
{\mcitedefaultendpunct}{\mcitedefaultseppunct}\relax
\EndOfBibitem
\bibitem{Wang:2014dba}
R.-M. Wang, J.-H. Sheng, J.~Zhu, Y.-Y. Fan, and Y.~Gao,
  \href{http://dx.doi.org/10.1142/S0217751X14501693}{Int. J. Mod. Phys. {\bf
  A29},  1450169 (2014)}\relax
\mciteBstWouldAddEndPuncttrue
\mciteSetBstMidEndSepPunct{\mcitedefaultmidpunct}
{\mcitedefaultendpunct}{\mcitedefaultseppunct}\relax
\EndOfBibitem
\bibitem{deBoer:2015boa}
S.~de~Boer and G.~Hiller,
  \href{http://dx.doi.org/10.1103/PhysRevD.93.074001}{Phys. Rev. {\bf D93},
  074001 (2016)}, \href{http://arxiv.org/abs/1510.00311}{{\tt arXiv:1510.00311
  [hep-ph]}}\relax
\mciteBstWouldAddEndPuncttrue
\mciteSetBstMidEndSepPunct{\mcitedefaultmidpunct}
{\mcitedefaultendpunct}{\mcitedefaultseppunct}\relax
\EndOfBibitem
\bibitem{Rubin:2010cq}
{CLEO} collaboration, P.~Rubin {\em et al.},
  \href{http://dx.doi.org/10.1103/PhysRevD.82.092007}{Phys. Rev. {\bf D82},
  092007 (2010)}, \href{http://arxiv.org/abs/1009.1606}{{\tt arXiv:1009.1606
  [hep-ex]}}\relax
\mciteBstWouldAddEndPuncttrue
\mciteSetBstMidEndSepPunct{\mcitedefaultmidpunct}
{\mcitedefaultendpunct}{\mcitedefaultseppunct}\relax
\EndOfBibitem
\bibitem{Abazov:2007aj}
{\dzero} collaboration, V.~Abazov {\em et al.},
  \href{http://dx.doi.org/10.1103/PhysRevLett.100.101801}{Phys. Rev. Lett. {\bf
  100},  101801 (2008)}, \href{http://arxiv.org/abs/0708.2094}{{\tt
  arXiv:0708.2094 [hep-ex]}}\relax
\mciteBstWouldAddEndPuncttrue
\mciteSetBstMidEndSepPunct{\mcitedefaultmidpunct}
{\mcitedefaultendpunct}{\mcitedefaultseppunct}\relax
\EndOfBibitem
\bibitem{Aaij:2013uoa}
{LHCb} collaboration, R.~Aaij {\em et al.},
  \href{http://dx.doi.org/10.1016/j.physletb.2013.11.053}{Phys. Lett. {\bf
  B728},  234 (2014)}, \href{http://arxiv.org/abs/1310.2535}{{\tt
  arXiv:1310.2535 [hep-ex]}}\relax
\mciteBstWouldAddEndPuncttrue
\mciteSetBstMidEndSepPunct{\mcitedefaultmidpunct}
{\mcitedefaultendpunct}{\mcitedefaultseppunct}\relax
\EndOfBibitem
\bibitem{Aaij:2013sua}
{LHCb} collaboration, R.~Aaij {\em et al.},
  \href{http://dx.doi.org/10.1016/j.physletb.2013.06.010}{Phys. Lett. {\bf
  B724},  203 (2013)}, \href{http://arxiv.org/abs/1304.6365}{{\tt
  arXiv:1304.6365 [hep-ex]}}\relax
\mciteBstWouldAddEndPuncttrue
\mciteSetBstMidEndSepPunct{\mcitedefaultmidpunct}
{\mcitedefaultendpunct}{\mcitedefaultseppunct}\relax
\EndOfBibitem
\bibitem{Coan:2002te}
{CLEO} collaboration, T.~Coan {\em et al.},
  \href{http://dx.doi.org/10.1103/PhysRevLett.90.101801}{Phys. Rev. Lett. {\bf
  90},  101801 (2003)}, \href{http://arxiv.org/abs/hep-ex/0212045}{{\tt
  arXiv:hep-ex/0212045 [hep-ex]}}\relax
\mciteBstWouldAddEndPuncttrue
\mciteSetBstMidEndSepPunct{\mcitedefaultmidpunct}
{\mcitedefaultendpunct}{\mcitedefaultseppunct}\relax
\EndOfBibitem
\bibitem{Ablikim:2015djc}
{BESIII} collaboration, M.~Ablikim {\em et al.},
  \href{http://dx.doi.org/10.1103/PhysRevD.91.112015}{Phys. Rev. {\bf D91},
  112015 (2015)}, \href{http://arxiv.org/abs/1505.03087}{{\tt arXiv:1505.03087
  [hep-ex]}}\relax
\mciteBstWouldAddEndPuncttrue
\mciteSetBstMidEndSepPunct{\mcitedefaultmidpunct}
{\mcitedefaultendpunct}{\mcitedefaultseppunct}\relax
\EndOfBibitem
\bibitem{Lees:2011qz}
{\babar} collaboration, J.~P. Lees {\em et al.},
  \href{http://dx.doi.org/10.1103/PhysRevD.85.091107}{Phys. Rev. {\bf D85},
  091107 (2012)}, \href{http://arxiv.org/abs/1110.6480}{{\tt arXiv:1110.6480
  [hep-ex]}}\relax
\mciteBstWouldAddEndPuncttrue
\mciteSetBstMidEndSepPunct{\mcitedefaultmidpunct}
{\mcitedefaultendpunct}{\mcitedefaultseppunct}\relax
\EndOfBibitem
\bibitem{Nisar:2015gvd}
{Belle} collaboration, N.~K. Nisar {\em et al.},
  \href{http://dx.doi.org/10.1103/PhysRevD.93.051102}{Phys. Rev. {\bf D93},
  051102 (2016)}, \href{http://arxiv.org/abs/1512.02992}{{\tt arXiv:1512.02992
  [hep-ex]}}\relax
\mciteBstWouldAddEndPuncttrue
\mciteSetBstMidEndSepPunct{\mcitedefaultmidpunct}
{\mcitedefaultendpunct}{\mcitedefaultseppunct}\relax
\EndOfBibitem
\bibitem{Haas:1988bh}
{CLEO} collaboration, P.~Haas {\em et al.},
  \href{http://dx.doi.org/10.1103/PhysRevLett.60.1614}{Phys. Rev. Lett. {\bf
  60},  1614 (1988)}\relax
\mciteBstWouldAddEndPuncttrue
\mciteSetBstMidEndSepPunct{\mcitedefaultmidpunct}
{\mcitedefaultendpunct}{\mcitedefaultseppunct}\relax
\EndOfBibitem
\bibitem{Albrecht:1988ge}
{ARGUS} collaboration, H.~Albrecht {\em et al.},
  \href{http://dx.doi.org/10.1016/0370-2693(88)90967-7}{Phys. Lett. {\bf B209},
   380 (1988)}\relax
\mciteBstWouldAddEndPuncttrue
\mciteSetBstMidEndSepPunct{\mcitedefaultmidpunct}
{\mcitedefaultendpunct}{\mcitedefaultseppunct}\relax
\EndOfBibitem
\bibitem{Adler:1987cp}
{MARK-III} collaboration, J.~Adler {\em et al.},
  \href{http://dx.doi.org/10.1103/PhysRevD.37.2023}{Phys. Rev. {\bf D37},  2023
  (1988)}, Erratum ibid.\
  \href{http://dx.doi.org/10.1103/PhysRevD.40.3788}{{\bf D40}, 3788},
  (1989)\relax
\mciteBstWouldAddEndPuncttrue
\mciteSetBstMidEndSepPunct{\mcitedefaultmidpunct}
{\mcitedefaultendpunct}{\mcitedefaultseppunct}\relax
\EndOfBibitem
\bibitem{Freyberger:1996it}
{CLEO} collaboration, A.~Freyberger {\em et al.},
  \href{http://dx.doi.org/10.1103/PhysRevLett.76.3065}{Phys. Rev. Lett. {\bf
  76},  3065 (1996)}\relax
\mciteBstWouldAddEndPuncttrue
\mciteSetBstMidEndSepPunct{\mcitedefaultmidpunct}
{\mcitedefaultendpunct}{\mcitedefaultseppunct}\relax
\EndOfBibitem
\bibitem{Pripstein:1999tq}
{Fermilab E789} collaboration, D.~Pripstein {\em et al.},
  \href{http://dx.doi.org/10.1103/PhysRevD.61.032005}{Phys. Rev. {\bf D61},
  032005 (2000)}, \href{http://arxiv.org/abs/hep-ex/9906022}{{\tt
  arXiv:hep-ex/9906022 [hep-ex]}}\relax
\mciteBstWouldAddEndPuncttrue
\mciteSetBstMidEndSepPunct{\mcitedefaultmidpunct}
{\mcitedefaultendpunct}{\mcitedefaultseppunct}\relax
\EndOfBibitem
\bibitem{Aitala:1999db}
{Fermilab E791} collaboration, E.~Aitala {\em et al.},
  \href{http://dx.doi.org/10.1016/S0370-2693(99)00902-8}{Phys. Lett. {\bf
  B462},  401 (1999)}, \href{http://arxiv.org/abs/hep-ex/9906045}{{\tt
  arXiv:hep-ex/9906045 [hep-ex]}}\relax
\mciteBstWouldAddEndPuncttrue
\mciteSetBstMidEndSepPunct{\mcitedefaultmidpunct}
{\mcitedefaultendpunct}{\mcitedefaultseppunct}\relax
\EndOfBibitem
\bibitem{Aubert:2004bs}
{\babar} collaboration, B.~Aubert {\em et al.},
  \href{http://dx.doi.org/10.1103/PhysRevLett.93.191801}{Phys. Rev. Lett. {\bf
  93},  191801 (2004)}, \href{http://arxiv.org/abs/hep-ex/0408023}{{\tt
  arXiv:hep-ex/0408023 [hep-ex]}}\relax
\mciteBstWouldAddEndPuncttrue
\mciteSetBstMidEndSepPunct{\mcitedefaultmidpunct}
{\mcitedefaultendpunct}{\mcitedefaultseppunct}\relax
\EndOfBibitem
\bibitem{Petric:2010yt}
{Belle} collaboration, M.~Petric {\em et al.},
  \href{http://dx.doi.org/10.1103/PhysRevD.81.091102}{Phys. Rev. {\bf D81},
  091102 (2010)}, \href{http://arxiv.org/abs/1003.2345}{{\tt arXiv:1003.2345
  [hep-ex]}}\relax
\mciteBstWouldAddEndPuncttrue
\mciteSetBstMidEndSepPunct{\mcitedefaultmidpunct}
{\mcitedefaultendpunct}{\mcitedefaultseppunct}\relax
\EndOfBibitem
\bibitem{Kodama:1995ia}
{Fermilab E653} collaboration, K.~Kodama {\em et al.},
  \href{http://dx.doi.org/10.1016/0370-2693(94)01610-O}{Phys. Lett. {\bf B345},
   85 (1995)}\relax
\mciteBstWouldAddEndPuncttrue
\mciteSetBstMidEndSepPunct{\mcitedefaultmidpunct}
{\mcitedefaultendpunct}{\mcitedefaultseppunct}\relax
\EndOfBibitem
\bibitem{Abt:2004hn}
{HERA-B} collaboration, I.~Abt {\em et al.},
  \href{http://dx.doi.org/10.1016/j.physletb.2004.06.097}{Phys. Lett. {\bf
  B596},  173 (2004)}, \href{http://arxiv.org/abs/hep-ex/0405059}{{\tt
  arXiv:hep-ex/0405059 [hep-ex]}}\relax
\mciteBstWouldAddEndPuncttrue
\mciteSetBstMidEndSepPunct{\mcitedefaultmidpunct}
{\mcitedefaultendpunct}{\mcitedefaultseppunct}\relax
\EndOfBibitem
\bibitem{Aaltonen:2010hz}
{CDF} collaboration, T.~Aaltonen {\em et al.},
  \href{http://dx.doi.org/10.1103/PhysRevD.82.091105}{Phys. Rev. {\bf D82},
  091105 (2010)}, \href{http://arxiv.org/abs/1008.5077}{{\tt arXiv:1008.5077
  [hep-ex]}}\relax
\mciteBstWouldAddEndPuncttrue
\mciteSetBstMidEndSepPunct{\mcitedefaultmidpunct}
{\mcitedefaultendpunct}{\mcitedefaultseppunct}\relax
\EndOfBibitem
\bibitem{Aaij:2013cza}
{LHCb} collaboration, R.~Aaij {\em et al.},
  \href{http://dx.doi.org/10.1016/j.physletb.2013.06.037}{Phys. Lett. {\bf
  B725},  15 (2013)}, \href{http://arxiv.org/abs/1305.5059}{{\tt
  arXiv:1305.5059 [hep-ex]}}\relax
\mciteBstWouldAddEndPuncttrue
\mciteSetBstMidEndSepPunct{\mcitedefaultmidpunct}
{\mcitedefaultendpunct}{\mcitedefaultseppunct}\relax
\EndOfBibitem
\bibitem{Ablikim:2018gro}
{BESIII} collaboration, M.~Ablikim {\em et al.},
  \href{http://dx.doi.org/10.1103/PhysRevD.97.072015}{Phys. Rev. {\bf D97},
  072015 (2018)}, \href{http://arxiv.org/abs/1802.09752}{{\tt arXiv:1802.09752
  [hep-ex]}}\relax
\mciteBstWouldAddEndPuncttrue
\mciteSetBstMidEndSepPunct{\mcitedefaultmidpunct}
{\mcitedefaultendpunct}{\mcitedefaultseppunct}\relax
\EndOfBibitem
\bibitem{Aitala:2000kk}
{Fermilab E791} collaboration, E.~Aitala {\em et al.},
  \href{http://dx.doi.org/10.1103/PhysRevLett.86.3969}{Phys. Rev. Lett. {\bf
  86},  3969 (2001)}, \href{http://arxiv.org/abs/hep-ex/0011077}{{\tt
  arXiv:hep-ex/0011077 [hep-ex]}}\relax
\mciteBstWouldAddEndPuncttrue
\mciteSetBstMidEndSepPunct{\mcitedefaultmidpunct}
{\mcitedefaultendpunct}{\mcitedefaultseppunct}\relax
\EndOfBibitem
\bibitem{Aaij:2017iyr}
{LHCb} collaboration, R.~Aaij {\em et al.},
  \href{http://dx.doi.org/10.1103/PhysRevLett.119.181805}{Phys. Rev. Lett. {\bf
  119},  181805 (2017)}, \href{http://arxiv.org/abs/1707.08377}{{\tt
  arXiv:1707.08377 [hep-ex]}}\relax
\mciteBstWouldAddEndPuncttrue
\mciteSetBstMidEndSepPunct{\mcitedefaultmidpunct}
{\mcitedefaultendpunct}{\mcitedefaultseppunct}\relax
\EndOfBibitem
\bibitem{Adler:1988es}
{MARK-III} collaboration, J.~Adler {\em et al.},
  \href{http://dx.doi.org/10.1103/PhysRevD.40.906}{Phys. Rev. {\bf D40},  906
  (1989)}\relax
\mciteBstWouldAddEndPuncttrue
\mciteSetBstMidEndSepPunct{\mcitedefaultmidpunct}
{\mcitedefaultendpunct}{\mcitedefaultseppunct}\relax
\EndOfBibitem
\bibitem{Asner:1998mv}
{CLEO} collaboration, D.~M. Asner {\em et al.},
  \href{http://dx.doi.org/10.1103/PhysRevD.58.092001}{Phys. Rev. {\bf D58},
  092001 (1998)}, \href{http://arxiv.org/abs/hep-ex/9803022}{{\tt
  arXiv:hep-ex/9803022 [hep-ex]}}\relax
\mciteBstWouldAddEndPuncttrue
\mciteSetBstMidEndSepPunct{\mcitedefaultmidpunct}
{\mcitedefaultendpunct}{\mcitedefaultseppunct}\relax
\EndOfBibitem
\bibitem{Aubert:2008ai}
{BaBar} collaboration, B.~Aubert {\em et al.},
  \href{http://dx.doi.org/10.1103/PhysRevD.78.071101}{Phys. Rev. {\bf D78},
  071101 (2008)}, \href{http://arxiv.org/abs/0808.1838}{{\tt arXiv:0808.1838
  [hep-ex]}}\relax
\mciteBstWouldAddEndPuncttrue
\mciteSetBstMidEndSepPunct{\mcitedefaultmidpunct}
{\mcitedefaultendpunct}{\mcitedefaultseppunct}\relax
\EndOfBibitem
\bibitem{Becker:1987mu}
{MARK-III} collaboration, J.~Becker {\em et al.},
  \href{http://dx.doi.org/10.1016/0370-2693(87)90473-4}{Phys. Lett. {\bf B193},
   147 (1987)}\relax
\mciteBstWouldAddEndPuncttrue
\mciteSetBstMidEndSepPunct{\mcitedefaultmidpunct}
{\mcitedefaultendpunct}{\mcitedefaultseppunct}\relax
\EndOfBibitem
\bibitem{Aaij:2015qmj}
{LHCb} collaboration, R.~Aaij {\em et al.},
  \href{http://dx.doi.org/10.1016/j.physletb.2016.01.029}{Phys. Lett. {\bf
  B754},  167 (2016)}, \href{http://arxiv.org/abs/1512.00322}{{\tt
  arXiv:1512.00322 [hep-ex]}}\relax
\mciteBstWouldAddEndPuncttrue
\mciteSetBstMidEndSepPunct{\mcitedefaultmidpunct}
{\mcitedefaultendpunct}{\mcitedefaultseppunct}\relax
\EndOfBibitem
\bibitem{Rubin:2009aa}
{CLEO} collaboration, P.~Rubin {\em et al.},
  \href{http://dx.doi.org/10.1103/PhysRevD.79.097101}{Phys. Rev. {\bf D79},
  097101 (2009)}, \href{http://arxiv.org/abs/0904.1619}{{\tt arXiv:0904.1619
  [hep-ex]}}\relax
\mciteBstWouldAddEndPuncttrue
\mciteSetBstMidEndSepPunct{\mcitedefaultmidpunct}
{\mcitedefaultendpunct}{\mcitedefaultseppunct}\relax
\EndOfBibitem
\bibitem{Frabetti:1997wp}
{Fermilab E687} collaboration, P.~Frabetti {\em et al.},
  \href{http://dx.doi.org/10.1016/S0370-2693(97)00229-3}{Phys. Lett. {\bf
  B398},  239 (1997)}\relax
\mciteBstWouldAddEndPuncttrue
\mciteSetBstMidEndSepPunct{\mcitedefaultmidpunct}
{\mcitedefaultendpunct}{\mcitedefaultseppunct}\relax
\EndOfBibitem
\bibitem{Lees:2011hb}
{\babar} collaboration, J.~P. Lees {\em et al.},
  \href{http://dx.doi.org/10.1103/PhysRevD.84.072006}{Phys. Rev. {\bf D84},
  072006 (2011)}, \href{http://arxiv.org/abs/1107.4465}{{\tt arXiv:1107.4465
  [hep-ex]}}\relax
\mciteBstWouldAddEndPuncttrue
\mciteSetBstMidEndSepPunct{\mcitedefaultmidpunct}
{\mcitedefaultendpunct}{\mcitedefaultseppunct}\relax
\EndOfBibitem
\bibitem{Zhao:2016jna}
{BESIII} collaboration, M.-G. Zhao, \href{http://arxiv.org/abs/1605.08952}{{\tt
  arXiv:1605.08952 [hep-ex]}} (2016)\relax
\mciteBstWouldAddEndPuncttrue
\mciteSetBstMidEndSepPunct{\mcitedefaultmidpunct}
{\mcitedefaultendpunct}{\mcitedefaultseppunct}\relax
\EndOfBibitem
\bibitem{Link:2003qp}
{FOCUS} collaboration, J.~Link {\em et al.},
  \href{http://dx.doi.org/10.1016/j.physletb.2003.07.079}{Phys. Lett. {\bf
  B572},  21 (2003)}, \href{http://arxiv.org/abs/hep-ex/0306049}{{\tt
  arXiv:hep-ex/0306049 [hep-ex]}}\relax
\mciteBstWouldAddEndPuncttrue
\mciteSetBstMidEndSepPunct{\mcitedefaultmidpunct}
{\mcitedefaultendpunct}{\mcitedefaultseppunct}\relax
\EndOfBibitem
\bibitem{Aaij:2017nsd}
{LHCb} collaboration, R.~Aaij {\em et al.},
  \href{http://dx.doi.org/10.1103/PhysRevD.97.091101}{Phys. Rev. {\bf D97},
  091101 (2018)}, \href{http://arxiv.org/abs/1712.07938}{{\tt arXiv:1712.07938
  [hep-ex]}}\relax
\mciteBstWouldAddEndPuncttrue
\mciteSetBstMidEndSepPunct{\mcitedefaultmidpunct}
{\mcitedefaultendpunct}{\mcitedefaultseppunct}\relax
\EndOfBibitem
\bibitem{Davier:2005xq}
M.~Davier, A.~Hocker, and Z.~Zhang,
  \href{http://dx.doi.org/10.1103/RevModPhys.78.1043}{Rev. Mod. Phys. {\bf 78},
   1043 (2006)}, \href{http://arxiv.org/abs/hep-ph/0507078}{{\tt
  arXiv:hep-ph/0507078 [hep-ph]}}\relax
\mciteBstWouldAddEndPuncttrue
\mciteSetBstMidEndSepPunct{\mcitedefaultmidpunct}
{\mcitedefaultendpunct}{\mcitedefaultseppunct}\relax
\EndOfBibitem
\bibitem{BaBar:2018qry}
{{BaBar}} collaboration, J.~P. Lees {\em et al.},
  \href{http://dx.doi.org/10.1103/PhysRevD.98.032010}{Phys. Rev. {\bf D98},
  032010 (2018)}, \href{http://arxiv.org/abs/1806.10280}{{\tt arXiv:1806.10280
  [hep-ex]}}\relax
\mciteBstWouldAddEndPuncttrue
\mciteSetBstMidEndSepPunct{\mcitedefaultmidpunct}
{\mcitedefaultendpunct}{\mcitedefaultseppunct}\relax
\EndOfBibitem
\bibitem{Lusiani:2019mis}
A.~Lusiani, \href{http://dx.doi.org/10.1051/epjconf/201921208001}{EPJ Web Conf.
  {\bf 212},  08001 (2019)}, \href{http://arxiv.org/abs/1906.02626}{{\tt
  arXiv:1906.02626 [hep-ex]}}\relax
\mciteBstWouldAddEndPuncttrue
\mciteSetBstMidEndSepPunct{\mcitedefaultmidpunct}
{\mcitedefaultendpunct}{\mcitedefaultseppunct}\relax
\EndOfBibitem
\bibitem{Aubert:2007jh}
{\babar} collaboration, B.~Aubert {\em et al.},
  \href{http://dx.doi.org/10.1103/PhysRevD.76.051104}{Phys. Rev. {\bf D76},
  051104 (2007)}, \href{http://arxiv.org/abs/0707.2922}{{\tt arXiv:0707.2922
  [hep-ex]}}\relax
\mciteBstWouldAddEndPuncttrue
\mciteSetBstMidEndSepPunct{\mcitedefaultmidpunct}
{\mcitedefaultendpunct}{\mcitedefaultseppunct}\relax
\EndOfBibitem
\bibitem{Schael:2005am}
{ALEPH} collaboration, S.~Schael {\em et al.},
  \href{http://dx.doi.org/10.1016/j.physrep.2005.06.007}{Phys. Rept. {\bf 421},
   191 (2005)}, \href{http://arxiv.org/abs/hep-ex/0506072}{{\tt
  arXiv:hep-ex/0506072 [hep-ex]}}, HFLAV-tau uses measurements of $\tau \to h
  X$ and $\tau \to K X$ and obtains $\tau \to \pi X$ by difference; the
  measurement of ${\cal B}\left( \tau^- \to 3h^- 2h^+ \pi^0 \nu_{\tau} \ ({\rm
  ex.} K^0)\right)$ has been read as $(2.1 \pm 0.7 \pm 0.6) \times 10^{-4}$
  whereas PDG11 uses $(2.1 \pm 0.7 \pm 0.9) \times 10^{-4}$\relax
\mciteBstWouldAddEndPuncttrue
\mciteSetBstMidEndSepPunct{\mcitedefaultmidpunct}
{\mcitedefaultendpunct}{\mcitedefaultseppunct}\relax
\EndOfBibitem
\bibitem{Abreu:1999rb}
{DELPHI} collaboration, P.~Abreu {\em et al.},
  \href{http://dx.doi.org/10.1007/s100520050583}{Eur. Phys. J. {\bf C10},  201
  (1999)}\relax
\mciteBstWouldAddEndPuncttrue
\mciteSetBstMidEndSepPunct{\mcitedefaultmidpunct}
{\mcitedefaultendpunct}{\mcitedefaultseppunct}\relax
\EndOfBibitem
\bibitem{Acciarri:2001sg}
{L3} collaboration, M.~Acciarri {\em et al.},
  \href{http://dx.doi.org/10.1016/S0370-2693(01)00294-5}{Phys. Lett. {\bf
  B507},  47 (2001)}, \href{http://arxiv.org/abs/hep-ex/0102023}{{\tt
  arXiv:hep-ex/0102023 [hep-ex]}}\relax
\mciteBstWouldAddEndPuncttrue
\mciteSetBstMidEndSepPunct{\mcitedefaultmidpunct}
{\mcitedefaultendpunct}{\mcitedefaultseppunct}\relax
\EndOfBibitem
\bibitem{Abbiendi:2002jw}
{OPAL} collaboration, G.~Abbiendi {\em et al.},
  \href{http://dx.doi.org/10.1016/S0370-2693(02)03020-4}{Phys. Lett. {\bf
  B551},  35 (2003)}, \href{http://arxiv.org/abs/hep-ex/0211066}{{\tt
  arXiv:hep-ex/0211066 [hep-ex]}}\relax
\mciteBstWouldAddEndPuncttrue
\mciteSetBstMidEndSepPunct{\mcitedefaultmidpunct}
{\mcitedefaultendpunct}{\mcitedefaultseppunct}\relax
\EndOfBibitem
\bibitem{Albrecht:1991rh}
{ARGUS} collaboration, H.~Albrecht {\em et al.},
  \href{http://dx.doi.org/10.1007/BF01625895}{Z. Phys. {\bf C53},  367
  (1992)}\relax
\mciteBstWouldAddEndPuncttrue
\mciteSetBstMidEndSepPunct{\mcitedefaultmidpunct}
{\mcitedefaultendpunct}{\mcitedefaultseppunct}\relax
\EndOfBibitem
\bibitem{Aubert:2009qj}
{\babar} collaboration, B.~Aubert {\em et al.},
  \href{http://dx.doi.org/10.1103/PhysRevLett.105.051602}{Phys. Rev. Lett. {\bf
  105},  051602 (2010)}, \href{http://arxiv.org/abs/0912.0242}{{\tt
  arXiv:0912.0242 [hep-ex]}}\relax
\mciteBstWouldAddEndPuncttrue
\mciteSetBstMidEndSepPunct{\mcitedefaultmidpunct}
{\mcitedefaultendpunct}{\mcitedefaultseppunct}\relax
\EndOfBibitem
\bibitem{Anastassov:1996tc}
{CLEO} collaboration, A.~Anastassov {\em et al.},
  \href{http://dx.doi.org/10.1103/PhysRevD.55.2559}{Phys. Rev. {\bf D55},  2559
  (1997)}, Erratum ibid.\
  \href{http://dx.doi.org/10.1103/PhysRevD.58.119903}{{\bf D58}, 119903},
  (1998)\relax
\mciteBstWouldAddEndPuncttrue
\mciteSetBstMidEndSepPunct{\mcitedefaultmidpunct}
{\mcitedefaultendpunct}{\mcitedefaultseppunct}\relax
\EndOfBibitem
\bibitem{Abbiendi:1998cx}
{OPAL} collaboration, G.~Abbiendi {\em et al.},
  \href{http://dx.doi.org/10.1016/S0370-2693(98)01553-6}{Phys. Lett. {\bf
  B447},  134 (1999)}, \href{http://arxiv.org/abs/hep-ex/9812017}{{\tt
  arXiv:hep-ex/9812017 [hep-ex]}}\relax
\mciteBstWouldAddEndPuncttrue
\mciteSetBstMidEndSepPunct{\mcitedefaultmidpunct}
{\mcitedefaultendpunct}{\mcitedefaultseppunct}\relax
\EndOfBibitem
\bibitem{Abreu:1992gn}
{DELPHI} collaboration, P.~Abreu {\em et al.},
  \href{http://dx.doi.org/10.1007/BF01561293}{Z. Phys. {\bf C55},  555
  (1992)}\relax
\mciteBstWouldAddEndPuncttrue
\mciteSetBstMidEndSepPunct{\mcitedefaultmidpunct}
{\mcitedefaultendpunct}{\mcitedefaultseppunct}\relax
\EndOfBibitem
\bibitem{Acciarri:1994vr}
{L3} collaboration, M.~Acciarri {\em et al.},
  \href{http://dx.doi.org/10.1016/0370-2693(94)01587-3}{Phys. Lett. {\bf B345},
   93 (1995)}\relax
\mciteBstWouldAddEndPuncttrue
\mciteSetBstMidEndSepPunct{\mcitedefaultmidpunct}
{\mcitedefaultendpunct}{\mcitedefaultseppunct}\relax
\EndOfBibitem
\bibitem{Alexander:1991am}
{OPAL} collaboration, G.~Alexander {\em et al.},
  \href{http://dx.doi.org/10.1016/0370-2693(91)90768-L}{Phys. Lett. {\bf B266},
   201 (1991)}\relax
\mciteBstWouldAddEndPuncttrue
\mciteSetBstMidEndSepPunct{\mcitedefaultmidpunct}
{\mcitedefaultendpunct}{\mcitedefaultseppunct}\relax
\EndOfBibitem
\bibitem{Abdallah:2003cw}
{DELPHI} collaboration, J.~Abdallah {\em et al.},
  \href{http://dx.doi.org/10.1140/epjc/s2006-02494-9}{Eur. Phys. J. {\bf C46},
  1 (2006)}, \href{http://arxiv.org/abs/hep-ex/0603044}{{\tt
  arXiv:hep-ex/0603044 [hep-ex]}}\relax
\mciteBstWouldAddEndPuncttrue
\mciteSetBstMidEndSepPunct{\mcitedefaultmidpunct}
{\mcitedefaultendpunct}{\mcitedefaultseppunct}\relax
\EndOfBibitem
\bibitem{Ackerstaff:1997tx}
{OPAL} collaboration, K.~Ackerstaff {\em et al.},
  \href{http://dx.doi.org/10.1007/s100520050197}{Eur. Phys. J. {\bf C4},  193
  (1998)}, \href{http://arxiv.org/abs/hep-ex/9801029}{{\tt arXiv:hep-ex/9801029
  [hep-ex]}}\relax
\mciteBstWouldAddEndPuncttrue
\mciteSetBstMidEndSepPunct{\mcitedefaultmidpunct}
{\mcitedefaultendpunct}{\mcitedefaultseppunct}\relax
\EndOfBibitem
\bibitem{Barate:1999hi}
{ALEPH} collaboration, R.~Barate {\em et al.},
  \href{http://dx.doi.org/10.1007/s100529900146}{Eur. Phys. J. {\bf C10},  1
  (1999)}, \href{http://arxiv.org/abs/hep-ex/9903014}{{\tt arXiv:hep-ex/9903014
  [hep-ex]}}\relax
\mciteBstWouldAddEndPuncttrue
\mciteSetBstMidEndSepPunct{\mcitedefaultmidpunct}
{\mcitedefaultendpunct}{\mcitedefaultseppunct}\relax
\EndOfBibitem
\bibitem{Battle:1994by}
{CLEO} collaboration, M.~Battle {\em et al.},
  \href{http://dx.doi.org/10.1103/PhysRevLett.73.1079}{Phys. Rev. Lett. {\bf
  73},  1079 (1994)}, \href{http://arxiv.org/abs/hep-ph/9403329}{{\tt
  arXiv:hep-ph/9403329 [hep-ph]}}\relax
\mciteBstWouldAddEndPuncttrue
\mciteSetBstMidEndSepPunct{\mcitedefaultmidpunct}
{\mcitedefaultendpunct}{\mcitedefaultseppunct}\relax
\EndOfBibitem
\bibitem{Abreu:1994fi}
{DELPHI} collaboration, P.~Abreu {\em et al.},
  \href{http://dx.doi.org/10.1016/0370-2693(94)90711-0}{Phys. Lett. {\bf B334},
   435 (1994)}\relax
\mciteBstWouldAddEndPuncttrue
\mciteSetBstMidEndSepPunct{\mcitedefaultmidpunct}
{\mcitedefaultendpunct}{\mcitedefaultseppunct}\relax
\EndOfBibitem
\bibitem{Abbiendi:2000ee}
{OPAL} collaboration, G.~Abbiendi {\em et al.},
  \href{http://dx.doi.org/10.1007/s100520100632}{Eur. Phys. J. {\bf C19},  653
  (2001)}, \href{http://arxiv.org/abs/hep-ex/0009017}{{\tt arXiv:hep-ex/0009017
  [hep-ex]}}\relax
\mciteBstWouldAddEndPuncttrue
\mciteSetBstMidEndSepPunct{\mcitedefaultmidpunct}
{\mcitedefaultendpunct}{\mcitedefaultseppunct}\relax
\EndOfBibitem
\bibitem{Fujikawa:2008ma}
{Belle} collaboration, M.~Fujikawa {\em et al.},
  \href{http://dx.doi.org/10.1103/PhysRevD.78.072006}{Phys. Rev. {\bf D78},
  072006 (2008)}, \href{http://arxiv.org/abs/0805.3773}{{\tt arXiv:0805.3773
  [hep-ex]}}\relax
\mciteBstWouldAddEndPuncttrue
\mciteSetBstMidEndSepPunct{\mcitedefaultmidpunct}
{\mcitedefaultendpunct}{\mcitedefaultseppunct}\relax
\EndOfBibitem
\bibitem{Artuso:1994ii}
{CLEO} collaboration, M.~Artuso {\em et al.},
  \href{http://dx.doi.org/10.1103/PhysRevLett.72.3762}{Phys. Rev. Lett. {\bf
  72},  3762 (1994)}, \href{http://arxiv.org/abs/hep-ph/9404310}{{\tt
  arXiv:hep-ph/9404310 [hep-ph]}}\relax
\mciteBstWouldAddEndPuncttrue
\mciteSetBstMidEndSepPunct{\mcitedefaultmidpunct}
{\mcitedefaultendpunct}{\mcitedefaultseppunct}\relax
\EndOfBibitem
\bibitem{Abbiendi:2004xa}
{OPAL} collaboration, G.~Abbiendi {\em et al.},
  \href{http://dx.doi.org/10.1140/epjc/s2004-01877-2}{Eur. Phys. J. {\bf C35},
  437 (2004)}, \href{http://arxiv.org/abs/hep-ex/0406007}{{\tt
  arXiv:hep-ex/0406007 [hep-ex]}}\relax
\mciteBstWouldAddEndPuncttrue
\mciteSetBstMidEndSepPunct{\mcitedefaultmidpunct}
{\mcitedefaultendpunct}{\mcitedefaultseppunct}\relax
\EndOfBibitem
\bibitem{Procario:1992hd}
{CLEO} collaboration, M.~Procario {\em et al.},
  \href{http://dx.doi.org/10.1103/PhysRevLett.70.1207}{Phys. Rev. Lett. {\bf
  70},  1207 (1993)}\relax
\mciteBstWouldAddEndPuncttrue
\mciteSetBstMidEndSepPunct{\mcitedefaultmidpunct}
{\mcitedefaultendpunct}{\mcitedefaultseppunct}\relax
\EndOfBibitem
\bibitem{Barate:1997tt}
{ALEPH} collaboration, R.~Barate {\em et al.},
  \href{http://dx.doi.org/10.1007/s100529800879}{Eur. Phys. J. {\bf C4},  29
  (1998)}, \url{http://cdsweb.cern.ch/record/346304}\relax
\mciteBstWouldAddEndPuncttrue
\mciteSetBstMidEndSepPunct{\mcitedefaultmidpunct}
{\mcitedefaultendpunct}{\mcitedefaultseppunct}\relax
\EndOfBibitem
\bibitem{Akers:1994td}
{OPAL} collaboration, R.~Akers {\em et al.},
  \href{http://dx.doi.org/10.1016/0370-2693(94)90645-9}{Phys. Lett. {\bf B339},
   278 (1994)}\relax
\mciteBstWouldAddEndPuncttrue
\mciteSetBstMidEndSepPunct{\mcitedefaultmidpunct}
{\mcitedefaultendpunct}{\mcitedefaultseppunct}\relax
\EndOfBibitem
\bibitem{Coan:1996iu}
{CLEO} collaboration, T.~Coan {\em et al.},
  \href{http://dx.doi.org/10.1103/PhysRevD.53.6037}{Phys. Rev. {\bf D53},  6037
  (1996)}\relax
\mciteBstWouldAddEndPuncttrue
\mciteSetBstMidEndSepPunct{\mcitedefaultmidpunct}
{\mcitedefaultendpunct}{\mcitedefaultseppunct}\relax
\EndOfBibitem
\bibitem{Ryu:2014vpc}
{Belle} collaboration, S.~Ryu {\em et al.},
  \href{http://dx.doi.org/10.1103/PhysRevD.89.072009}{Phys. Rev. {\bf D89},
  072009 (2014)}, \href{http://arxiv.org/abs/1402.5213}{{\tt arXiv:1402.5213
  [hep-ex]}}\relax
\mciteBstWouldAddEndPuncttrue
\mciteSetBstMidEndSepPunct{\mcitedefaultmidpunct}
{\mcitedefaultendpunct}{\mcitedefaultseppunct}\relax
\EndOfBibitem
\bibitem{Acciarri:1995kx}
{L3} collaboration, M.~Acciarri {\em et al.},
  \href{http://dx.doi.org/10.1016/0370-2693(95)00509-J}{Phys. Lett. {\bf B352},
   487 (1995)}\relax
\mciteBstWouldAddEndPuncttrue
\mciteSetBstMidEndSepPunct{\mcitedefaultmidpunct}
{\mcitedefaultendpunct}{\mcitedefaultseppunct}\relax
\EndOfBibitem
\bibitem{Abbiendi:1999pm}
{OPAL} collaboration, G.~Abbiendi {\em et al.},
  \href{http://dx.doi.org/10.1007/s100520000317}{Eur. Phys. J. {\bf C13},  213
  (2000)}, \href{http://arxiv.org/abs/hep-ex/9911029}{{\tt arXiv:hep-ex/9911029
  [hep-ex]}}\relax
\mciteBstWouldAddEndPuncttrue
\mciteSetBstMidEndSepPunct{\mcitedefaultmidpunct}
{\mcitedefaultendpunct}{\mcitedefaultseppunct}\relax
\EndOfBibitem
\bibitem{Barate:1999hj}
{ALEPH} collaboration, R.~Barate {\em et al.},
  \href{http://dx.doi.org/10.1007/s100520050659}{Eur. Phys. J. {\bf C11},  599
  (1999)}, \href{http://arxiv.org/abs/hep-ex/9903015}{{\tt arXiv:hep-ex/9903015
  [hep-ex]}}\relax
\mciteBstWouldAddEndPuncttrue
\mciteSetBstMidEndSepPunct{\mcitedefaultmidpunct}
{\mcitedefaultendpunct}{\mcitedefaultseppunct}\relax
\EndOfBibitem
\bibitem{Lees:2012de}
{\babar} collaboration, J.~P. Lees {\em et al.},
  \href{http://dx.doi.org/10.1103/PhysRevD.86.092013}{Phys. Rev. {\bf D86},
  092013 (2012)}, \href{http://arxiv.org/abs/1208.0376}{{\tt arXiv:1208.0376
  [hep-ex]}}\relax
\mciteBstWouldAddEndPuncttrue
\mciteSetBstMidEndSepPunct{\mcitedefaultmidpunct}
{\mcitedefaultendpunct}{\mcitedefaultseppunct}\relax
\EndOfBibitem
\bibitem{Behrend:1989wc}
{CELLO} collaboration, H.~Behrend {\em et al.},
  \href{http://dx.doi.org/10.1016/0370-2693(89)90741-7}{Phys. Lett. {\bf B222},
   163 (1989)}\relax
\mciteBstWouldAddEndPuncttrue
\mciteSetBstMidEndSepPunct{\mcitedefaultmidpunct}
{\mcitedefaultendpunct}{\mcitedefaultseppunct}\relax
\EndOfBibitem
\bibitem{Adeva:1991qq}
{L3} collaboration, B.~Adeva {\em et al.},
  \href{http://dx.doi.org/10.1016/0370-2693(91)90081-Z}{Phys. Lett. {\bf B265},
   451 (1991)}\relax
\mciteBstWouldAddEndPuncttrue
\mciteSetBstMidEndSepPunct{\mcitedefaultmidpunct}
{\mcitedefaultendpunct}{\mcitedefaultseppunct}\relax
\EndOfBibitem
\bibitem{Aihara:1986mw}
{TPC/Two Gamma} collaboration, H.~Aihara {\em et al.},
  \href{http://dx.doi.org/10.1103/PhysRevD.35.1553}{Phys. Rev. {\bf D35},  1553
  (1987)}\relax
\mciteBstWouldAddEndPuncttrue
\mciteSetBstMidEndSepPunct{\mcitedefaultmidpunct}
{\mcitedefaultendpunct}{\mcitedefaultseppunct}\relax
\EndOfBibitem
\bibitem{Achard:2001pk}
{L3} collaboration, P.~Achard {\em et al.},
  \href{http://dx.doi.org/10.1016/S0370-2693(01)01099-1}{Phys. Lett. {\bf
  B519},  189 (2001)}, \href{http://arxiv.org/abs/hep-ex/0107055}{{\tt
  arXiv:hep-ex/0107055 [hep-ex]}}\relax
\mciteBstWouldAddEndPuncttrue
\mciteSetBstMidEndSepPunct{\mcitedefaultmidpunct}
{\mcitedefaultendpunct}{\mcitedefaultseppunct}\relax
\EndOfBibitem
\bibitem{Akers:1995ry}
{OPAL} collaboration, R.~Akers {\em et al.},
  \href{http://dx.doi.org/10.1007/BF01565256}{Z. Phys. {\bf C68},  555
  (1995)}\relax
\mciteBstWouldAddEndPuncttrue
\mciteSetBstMidEndSepPunct{\mcitedefaultmidpunct}
{\mcitedefaultendpunct}{\mcitedefaultseppunct}\relax
\EndOfBibitem
\bibitem{Balest:1995kq}
{CLEO} collaboration, R.~Balest {\em et al.},
  \href{http://dx.doi.org/10.1103/PhysRevLett.75.3809}{Phys. Rev. Lett. {\bf
  75},  3809 (1995)}\relax
\mciteBstWouldAddEndPuncttrue
\mciteSetBstMidEndSepPunct{\mcitedefaultmidpunct}
{\mcitedefaultendpunct}{\mcitedefaultseppunct}\relax
\EndOfBibitem
\bibitem{Aubert:2007mh}
{\babar} collaboration, B.~Aubert {\em et al.},
  \href{http://dx.doi.org/10.1103/PhysRevLett.100.011801}{Phys. Rev. Lett. {\bf
  100},  011801 (2008)}, \href{http://arxiv.org/abs/0707.2981}{{\tt
  arXiv:0707.2981 [hep-ex]}}\relax
\mciteBstWouldAddEndPuncttrue
\mciteSetBstMidEndSepPunct{\mcitedefaultmidpunct}
{\mcitedefaultendpunct}{\mcitedefaultseppunct}\relax
\EndOfBibitem
\bibitem{Lee:2010tc}
{Belle} collaboration, M.~Lee {\em et al.},
  \href{http://dx.doi.org/10.1103/PhysRevD.81.113007}{Phys. Rev. {\bf D81},
  113007 (2010)}, \href{http://arxiv.org/abs/1001.0083}{{\tt arXiv:1001.0083
  [hep-ex]}}\relax
\mciteBstWouldAddEndPuncttrue
\mciteSetBstMidEndSepPunct{\mcitedefaultmidpunct}
{\mcitedefaultendpunct}{\mcitedefaultseppunct}\relax
\EndOfBibitem
\bibitem{Briere:2003fr}
{CLEO} collaboration, R.~A. Briere {\em et al.},
  \href{http://dx.doi.org/10.1103/PhysRevLett.90.181802}{Phys. Rev. Lett. {\bf
  90},  181802 (2003)}, \href{http://arxiv.org/abs/hep-ex/0302028}{{\tt
  arXiv:hep-ex/0302028 [hep-ex]}}\relax
\mciteBstWouldAddEndPuncttrue
\mciteSetBstMidEndSepPunct{\mcitedefaultmidpunct}
{\mcitedefaultendpunct}{\mcitedefaultseppunct}\relax
\EndOfBibitem
\bibitem{Edwards:1999fj}
{CLEO} collaboration, K.~Edwards {\em et al.},
  \href{http://dx.doi.org/10.1103/PhysRevD.61.072003}{Phys. Rev. {\bf D61},
  072003 (2000)}, \href{http://arxiv.org/abs/hep-ex/9908024}{{\tt
  arXiv:hep-ex/9908024 [hep-ex]}}\relax
\mciteBstWouldAddEndPuncttrue
\mciteSetBstMidEndSepPunct{\mcitedefaultmidpunct}
{\mcitedefaultendpunct}{\mcitedefaultseppunct}\relax
\EndOfBibitem
\bibitem{Bortoletto:1993px}
{CLEO} collaboration, D.~Bortoletto {\em et al.},
  \href{http://dx.doi.org/10.1103/PhysRevLett.71.1791}{Phys. Rev. Lett. {\bf
  71},  1791 (1993)}\relax
\mciteBstWouldAddEndPuncttrue
\mciteSetBstMidEndSepPunct{\mcitedefaultmidpunct}
{\mcitedefaultendpunct}{\mcitedefaultseppunct}\relax
\EndOfBibitem
\bibitem{Anastassov:2000xu}
{CLEO} collaboration, A.~Anastassov {\em et al.},
  \href{http://dx.doi.org/10.1103/PhysRevLett.86.4467}{Phys. Rev. Lett. {\bf
  86},  4467 (2001)}, \href{http://arxiv.org/abs/hep-ex/0010025}{{\tt
  arXiv:hep-ex/0010025 [hep-ex]}}\relax
\mciteBstWouldAddEndPuncttrue
\mciteSetBstMidEndSepPunct{\mcitedefaultmidpunct}
{\mcitedefaultendpunct}{\mcitedefaultseppunct}\relax
\EndOfBibitem
\bibitem{Richichi:1998bc}
{CLEO} collaboration, S.~Richichi {\em et al.},
  \href{http://dx.doi.org/10.1103/PhysRevD.60.112002}{Phys. Rev. {\bf D60},
  112002 (1999)}, \href{http://arxiv.org/abs/hep-ex/9810026}{{\tt
  arXiv:hep-ex/9810026 [hep-ex]}}\relax
\mciteBstWouldAddEndPuncttrue
\mciteSetBstMidEndSepPunct{\mcitedefaultmidpunct}
{\mcitedefaultendpunct}{\mcitedefaultseppunct}\relax
\EndOfBibitem
\bibitem{Bauer:1993wn}
{TPC/Two Gamma} collaboration, D.~A. Bauer {\em et al.},
  \href{http://dx.doi.org/10.1103/PhysRevD.50.13}{Phys. Rev. {\bf D50},  13
  (1994)}\relax
\mciteBstWouldAddEndPuncttrue
\mciteSetBstMidEndSepPunct{\mcitedefaultmidpunct}
{\mcitedefaultendpunct}{\mcitedefaultseppunct}\relax
\EndOfBibitem
\bibitem{Barate:1997ma}
{ALEPH} collaboration, R.~Barate {\em et al.},
  \href{http://dx.doi.org/10.1007/s100520050062}{Eur. Phys. J. {\bf C1},  65
  (1998)}\relax
\mciteBstWouldAddEndPuncttrue
\mciteSetBstMidEndSepPunct{\mcitedefaultmidpunct}
{\mcitedefaultendpunct}{\mcitedefaultseppunct}\relax
\EndOfBibitem
\bibitem{Arms:2005qg}
{CLEO} collaboration, K.~E. Arms {\em et al.},
  \href{http://dx.doi.org/10.1103/PhysRevLett.94.241802}{Phys. Rev. Lett. {\bf
  94},  241802 (2005)}, \href{http://arxiv.org/abs/hep-ex/0501042}{{\tt
  arXiv:hep-ex/0501042 [hep-ex]}}\relax
\mciteBstWouldAddEndPuncttrue
\mciteSetBstMidEndSepPunct{\mcitedefaultmidpunct}
{\mcitedefaultendpunct}{\mcitedefaultseppunct}\relax
\EndOfBibitem
\bibitem{Abbiendi:1999cq}
{OPAL} collaboration, G.~Abbiendi {\em et al.},
  \href{http://dx.doi.org/10.1007/s100520000272}{Eur. Phys. J. {\bf C13},  197
  (2000)}, \href{http://arxiv.org/abs/hep-ex/9908013}{{\tt arXiv:hep-ex/9908013
  [hep-ex]}}\relax
\mciteBstWouldAddEndPuncttrue
\mciteSetBstMidEndSepPunct{\mcitedefaultmidpunct}
{\mcitedefaultendpunct}{\mcitedefaultseppunct}\relax
\EndOfBibitem
\bibitem{Gibaut:1994ik}
{CLEO} collaboration, D.~Gibaut {\em et al.},
  \href{http://dx.doi.org/10.1103/PhysRevLett.73.934}{Phys. Rev. Lett. {\bf
  73},  934 (1994)}\relax
\mciteBstWouldAddEndPuncttrue
\mciteSetBstMidEndSepPunct{\mcitedefaultmidpunct}
{\mcitedefaultendpunct}{\mcitedefaultseppunct}\relax
\EndOfBibitem
\bibitem{Bylsma:1986zy}
B.~Bylsma {\em et al.}, \href{http://dx.doi.org/10.1103/PhysRevD.35.2269}{Phys.
  Rev. {\bf D35},  2269 (1987)}\relax
\mciteBstWouldAddEndPuncttrue
\mciteSetBstMidEndSepPunct{\mcitedefaultmidpunct}
{\mcitedefaultendpunct}{\mcitedefaultseppunct}\relax
\EndOfBibitem
\bibitem{Albrecht:1987zf}
{ARGUS} collaboration, H.~Albrecht {\em et al.},
  \href{http://dx.doi.org/10.1016/0370-2693(88)90870-2}{Phys. Lett. {\bf B202},
   149 (1988)}\relax
\mciteBstWouldAddEndPuncttrue
\mciteSetBstMidEndSepPunct{\mcitedefaultmidpunct}
{\mcitedefaultendpunct}{\mcitedefaultseppunct}\relax
\EndOfBibitem
\bibitem{Ackerstaff:1998ia}
{OPAL} collaboration, K.~Ackerstaff {\em et al.},
  \href{http://dx.doi.org/10.1007/s100529901057}{Eur. Phys. J. {\bf C8},  183
  (1999)}, \href{http://arxiv.org/abs/hep-ex/9808011}{{\tt arXiv:hep-ex/9808011
  [hep-ex]}}\relax
\mciteBstWouldAddEndPuncttrue
\mciteSetBstMidEndSepPunct{\mcitedefaultmidpunct}
{\mcitedefaultendpunct}{\mcitedefaultseppunct}\relax
\EndOfBibitem
\bibitem{Buskulic:1996qs}
{ALEPH} collaboration, D.~Buskulic {\em et al.},
  \href{http://dx.doi.org/10.1007/s002880050387}{Z. Phys. {\bf C74},  263
  (1997)}\relax
\mciteBstWouldAddEndPuncttrue
\mciteSetBstMidEndSepPunct{\mcitedefaultmidpunct}
{\mcitedefaultendpunct}{\mcitedefaultseppunct}\relax
\EndOfBibitem
\bibitem{Inami:2008ar}
{Belle} collaboration, K.~Inami {\em et al.},
  \href{http://dx.doi.org/10.1016/j.physletb.2009.01.047}{Phys. Lett. {\bf
  B672},  209 (2009)}, \href{http://arxiv.org/abs/0811.0088}{{\tt
  arXiv:0811.0088 [hep-ex]}}\relax
\mciteBstWouldAddEndPuncttrue
\mciteSetBstMidEndSepPunct{\mcitedefaultmidpunct}
{\mcitedefaultendpunct}{\mcitedefaultseppunct}\relax
\EndOfBibitem
\bibitem{Artuso:1992qu}
{CLEO} collaboration, M.~Artuso {\em et al.},
  \href{http://dx.doi.org/10.1103/PhysRevLett.69.3278}{Phys. Rev. Lett. {\bf
  69},  3278 (1992)}\relax
\mciteBstWouldAddEndPuncttrue
\mciteSetBstMidEndSepPunct{\mcitedefaultmidpunct}
{\mcitedefaultendpunct}{\mcitedefaultseppunct}\relax
\EndOfBibitem
\bibitem{delAmoSanchez:2010pc}
{\babar} collaboration, P.~del Amo~Sanchez {\em et al.},
  \href{http://dx.doi.org/10.1103/PhysRevD.83.032002}{Phys. Rev. {\bf D83},
  032002 (2011)}, \href{http://arxiv.org/abs/1011.3917}{{\tt arXiv:1011.3917
  [hep-ex]}}\relax
\mciteBstWouldAddEndPuncttrue
\mciteSetBstMidEndSepPunct{\mcitedefaultmidpunct}
{\mcitedefaultendpunct}{\mcitedefaultseppunct}\relax
\EndOfBibitem
\bibitem{Bartelt:1996iv}
{CLEO} collaboration, J.~E. Bartelt {\em et al.},
  \href{http://dx.doi.org/10.1103/PhysRevLett.76.4119}{Phys. Rev. Lett. {\bf
  76},  4119 (1996)}\relax
\mciteBstWouldAddEndPuncttrue
\mciteSetBstMidEndSepPunct{\mcitedefaultmidpunct}
{\mcitedefaultendpunct}{\mcitedefaultseppunct}\relax
\EndOfBibitem
\bibitem{Bishai:1998gf}
{CLEO} collaboration, M.~Bishai {\em et al.},
  \href{http://dx.doi.org/10.1103/PhysRevLett.82.281}{Phys. Rev. Lett. {\bf
  82},  281 (1999)}, \href{http://arxiv.org/abs/hep-ex/9809012}{{\tt
  arXiv:hep-ex/9809012 [hep-ex]}}\relax
\mciteBstWouldAddEndPuncttrue
\mciteSetBstMidEndSepPunct{\mcitedefaultmidpunct}
{\mcitedefaultendpunct}{\mcitedefaultseppunct}\relax
\EndOfBibitem
\bibitem{Baringer:1987tr}
{CLEO} collaboration, P.~S. Baringer {\em et al.},
  \href{http://dx.doi.org/10.1103/PhysRevLett.59.1993}{Phys. Rev. Lett. {\bf
  59},  1993 (1987)}\relax
\mciteBstWouldAddEndPuncttrue
\mciteSetBstMidEndSepPunct{\mcitedefaultmidpunct}
{\mcitedefaultendpunct}{\mcitedefaultseppunct}\relax
\EndOfBibitem
\bibitem{Buskulic:1995ty}
{ALEPH} collaboration, D.~Buskulic {\em et al.},
  \href{http://dx.doi.org/10.1007/s002880050134}{Z. Phys. {\bf C70},  579
  (1996)}\relax
\mciteBstWouldAddEndPuncttrue
\mciteSetBstMidEndSepPunct{\mcitedefaultmidpunct}
{\mcitedefaultendpunct}{\mcitedefaultseppunct}\relax
\EndOfBibitem
\bibitem{Lees:2012ks}
{\babar} collaboration, J.~P. Lees {\em et al.},
  \href{http://dx.doi.org/10.1103/PhysRevD.86.092010}{Phys. Rev. {\bf D86},
  092010 (2012)}, \href{http://arxiv.org/abs/1209.2734}{{\tt arXiv:1209.2734
  [hep-ex]}}\relax
\mciteBstWouldAddEndPuncttrue
\mciteSetBstMidEndSepPunct{\mcitedefaultmidpunct}
{\mcitedefaultendpunct}{\mcitedefaultseppunct}\relax
\EndOfBibitem
\bibitem{Belous:2013dba}
{Belle} collaboration, K.~Belous {\em et al.},
  \href{http://dx.doi.org/10.1103/PhysRevLett.112.031801}{Phys. Rev. Lett. {\bf
  112},  031801 (2014)}, \href{http://arxiv.org/abs/1310.8503}{{\tt
  arXiv:1310.8503 [hep-ex]}}\relax
\mciteBstWouldAddEndPuncttrue
\mciteSetBstMidEndSepPunct{\mcitedefaultmidpunct}
{\mcitedefaultendpunct}{\mcitedefaultseppunct}\relax
\EndOfBibitem
\bibitem{Marciano:1988vm}
W.~Marciano and A.~Sirlin,
  \href{http://dx.doi.org/10.1103/PhysRevLett.61.1815}{Phys. Rev. Lett. {\bf
  61},  1815 (1988)}\relax
\mciteBstWouldAddEndPuncttrue
\mciteSetBstMidEndSepPunct{\mcitedefaultmidpunct}
{\mcitedefaultendpunct}{\mcitedefaultseppunct}\relax
\EndOfBibitem
\bibitem{Pich:2013lsa}
A.~Pich, \href{http://dx.doi.org/10.1016/j.ppnp.2013.11.002}{Prog. Part. Nucl.
  Phys. {\bf 75},  41 (2014)}, \href{http://arxiv.org/abs/1310.7922}{{\tt
  arXiv:1310.7922 [hep-ph]}}\relax
\mciteBstWouldAddEndPuncttrue
\mciteSetBstMidEndSepPunct{\mcitedefaultmidpunct}
{\mcitedefaultendpunct}{\mcitedefaultseppunct}\relax
\EndOfBibitem
\bibitem{Ferroglia:2013dga}
A.~Ferroglia, C.~Greub, A.~Sirlin, and Z.~Zhang,
  \href{http://dx.doi.org/10.1103/PhysRevD.88.033012}{Phys. Rev. {\bf D88},
  033012 (2013)}, \href{http://arxiv.org/abs/1307.6900}{{\tt arXiv:1307.6900
  [hep-ph]}}\relax
\mciteBstWouldAddEndPuncttrue
\mciteSetBstMidEndSepPunct{\mcitedefaultmidpunct}
{\mcitedefaultendpunct}{\mcitedefaultseppunct}\relax
\EndOfBibitem
\bibitem{Fael:2013pja}
M.~Fael, L.~Mercolli, and M.~Passera,
  \href{http://dx.doi.org/10.1103/PhysRevD.88.093011}{Phys. Rev. {\bf D88},
  093011 (2013)}, \href{http://arxiv.org/abs/1310.1081}{{\tt arXiv:1310.1081
  [hep-ph]}}\relax
\mciteBstWouldAddEndPuncttrue
\mciteSetBstMidEndSepPunct{\mcitedefaultmidpunct}
{\mcitedefaultendpunct}{\mcitedefaultseppunct}\relax
\EndOfBibitem
\bibitem{Marciano:1993sh}
W.~J. Marciano and A.~Sirlin,
  \href{http://dx.doi.org/10.1103/PhysRevLett.71.3629}{Phys. Rev. Lett. {\bf
  71},  3629 (1993)}\relax
\mciteBstWouldAddEndPuncttrue
\mciteSetBstMidEndSepPunct{\mcitedefaultmidpunct}
{\mcitedefaultendpunct}{\mcitedefaultseppunct}\relax
\EndOfBibitem
\bibitem{Decker:1994dd}
R.~Decker and M.~Finkemeier,
  \href{http://dx.doi.org/10.1016/0370-2693(94)90611-4}{Phys. Lett. {\bf B334},
   199 (1994)}\relax
\mciteBstWouldAddEndPuncttrue
\mciteSetBstMidEndSepPunct{\mcitedefaultmidpunct}
{\mcitedefaultendpunct}{\mcitedefaultseppunct}\relax
\EndOfBibitem
\bibitem{Decker:1994ea}
R.~Decker and M.~Finkemeier,
  \href{http://dx.doi.org/10.1016/0550-3213(95)00597-L}{Nucl. Phys. {\bf B438},
   17 (1995)}, \href{http://arxiv.org/abs/hep-ph/9403385}{{\tt
  arXiv:hep-ph/9403385}}\relax
\mciteBstWouldAddEndPuncttrue
\mciteSetBstMidEndSepPunct{\mcitedefaultmidpunct}
{\mcitedefaultendpunct}{\mcitedefaultseppunct}\relax
\EndOfBibitem
\bibitem{Decker:1994kw}
R.~Decker and M.~Finkemeier,
  \href{http://dx.doi.org/10.1016/0920-5632(95)00170-E}{Nucl. Phys. Proc.
  Suppl. {\bf 40},  453 (1995)},
  \href{http://arxiv.org/abs/hep-ph/9411316}{{\tt arXiv:hep-ph/9411316
  [hep-ph]}}\relax
\mciteBstWouldAddEndPuncttrue
\mciteSetBstMidEndSepPunct{\mcitedefaultmidpunct}
{\mcitedefaultendpunct}{\mcitedefaultseppunct}\relax
\EndOfBibitem
\bibitem{Antonelli:2010yf}
{FlaviaNet working group on kaon decays}, M.~Antonelli {\em et al.},
  \href{http://dx.doi.org/10.1140/epjc/s10052-010-1406-3}{Eur. Phys. J. {\bf
  C69},  399 (2010)}, \href{http://arxiv.org/abs/1005.2323}{{\tt
  arXiv:1005.2323 [hep-ph]}}\relax
\mciteBstWouldAddEndPuncttrue
\mciteSetBstMidEndSepPunct{\mcitedefaultmidpunct}
{\mcitedefaultendpunct}{\mcitedefaultseppunct}\relax
\EndOfBibitem
\bibitem{Gamiz:2002nu}
E.~Gamiz, M.~Jamin, A.~Pich, J.~Prades, and F.~Schwab,
  \href{http://dx.doi.org/10.1088/1126-6708/2003/01/060}{JHEP {\bf 01},  060
  (2003)}, \href{http://arxiv.org/abs/hep-ph/0212230}{{\tt arXiv:hep-ph/0212230
  [hep-ph]}}\relax
\mciteBstWouldAddEndPuncttrue
\mciteSetBstMidEndSepPunct{\mcitedefaultmidpunct}
{\mcitedefaultendpunct}{\mcitedefaultseppunct}\relax
\EndOfBibitem
\bibitem{Gamiz:2004ar}
E.~Gamiz, M.~Jamin, A.~Pich, J.~Prades, and F.~Schwab,
  \href{http://dx.doi.org/10.1103/PhysRevLett.94.011803}{Phys. Rev. Lett. {\bf
  94},  011803 (2005)}, \href{http://arxiv.org/abs/hep-ph/0408044}{{\tt
  arXiv:hep-ph/0408044 [hep-ph]}}\relax
\mciteBstWouldAddEndPuncttrue
\mciteSetBstMidEndSepPunct{\mcitedefaultmidpunct}
{\mcitedefaultendpunct}{\mcitedefaultseppunct}\relax
\EndOfBibitem
\bibitem{Gamiz:2006xx}
E.~Gamiz, M.~Jamin, A.~Pich, J.~Prades, and F.~Schwab,
  \href{http://dx.doi.org/10.1016/j.nuclphysbps.2007.02.053}{Nucl. Phys. Proc.
  Suppl. {\bf 169},  85 (2007)},
  \href{http://arxiv.org/abs/hep-ph/0612154}{{\tt arXiv:hep-ph/0612154}}\relax
\mciteBstWouldAddEndPuncttrue
\mciteSetBstMidEndSepPunct{\mcitedefaultmidpunct}
{\mcitedefaultendpunct}{\mcitedefaultseppunct}\relax
\EndOfBibitem
\bibitem{Gamiz:2007qs}
E.~Gamiz, M.~Jamin, A.~Pich, J.~Prades, and F.~Schwab, PoS {\bf KAON},  008
  (2008), \href{http://arxiv.org/abs/0709.0282}{{\tt arXiv:0709.0282
  [hep-ph]}}\relax
\mciteBstWouldAddEndPuncttrue
\mciteSetBstMidEndSepPunct{\mcitedefaultmidpunct}
{\mcitedefaultendpunct}{\mcitedefaultseppunct}\relax
\EndOfBibitem
\bibitem{Maltman:2010hb}
K.~Maltman, \href{http://dx.doi.org/10.1016/j.nuclphysbps.2011.06.025}{Nucl.
  Phys. Proc. Suppl. {\bf 218},  146 (2011)},
  \href{http://arxiv.org/abs/1011.6391}{{\tt arXiv:1011.6391 [hep-ph]}}\relax
\mciteBstWouldAddEndPuncttrue
\mciteSetBstMidEndSepPunct{\mcitedefaultmidpunct}
{\mcitedefaultendpunct}{\mcitedefaultseppunct}\relax
\EndOfBibitem
\bibitem{Hardy:2014qxa}
J.~C. Hardy and I.~S. Towner,
  \href{http://dx.doi.org/10.1103/PhysRevC.91.025501}{Phys. Rev. {\bf C91},
  025501 (2015)}, \href{http://arxiv.org/abs/1411.5987}{{\tt arXiv:1411.5987
  [nucl-ex]}}\relax
\mciteBstWouldAddEndPuncttrue
\mciteSetBstMidEndSepPunct{\mcitedefaultmidpunct}
{\mcitedefaultendpunct}{\mcitedefaultseppunct}\relax
\EndOfBibitem
\bibitem{Hardy:2016vhg}
J.~Hardy and I.~S. Towner, \href{http://dx.doi.org/10.22323/1.291.0028}{PoS
  {\bf CKM2016},  028 (2016)}\relax
\mciteBstWouldAddEndPuncttrue
\mciteSetBstMidEndSepPunct{\mcitedefaultmidpunct}
{\mcitedefaultendpunct}{\mcitedefaultseppunct}\relax
\EndOfBibitem
\bibitem{Banerjee:2008hg}
{BaBar} collaboration, S.~Banerjee, \href{http://arxiv.org/abs/0811.1429}{{\tt
  arXiv:0811.1429 [hep-ex]}} (2008),
  \url{http://www-public.slac.stanford.edu/sciDoc/docMeta.aspx?slacPubNumber=SLAC-PUB-14714}\relax
\mciteBstWouldAddEndPuncttrue
\mciteSetBstMidEndSepPunct{\mcitedefaultmidpunct}
{\mcitedefaultendpunct}{\mcitedefaultseppunct}\relax
\EndOfBibitem
\bibitem{Dowdall:2013rya}
R.~J. Dowdall, C.~T.~H. Davies, G.~P. Lepage, and C.~McNeile,
  \href{http://dx.doi.org/10.1103/PhysRevD.88.074504}{Phys. Rev. {\bf D88},
  074504 (2013)}, \href{http://arxiv.org/abs/1303.1670}{{\tt arXiv:1303.1670
  [hep-lat]}}\relax
\mciteBstWouldAddEndPuncttrue
\mciteSetBstMidEndSepPunct{\mcitedefaultmidpunct}
{\mcitedefaultendpunct}{\mcitedefaultseppunct}\relax
\EndOfBibitem
\bibitem{Carrasco:2014poa}
N.~Carrasco {\em et al.},
  \href{http://dx.doi.org/10.1103/PhysRevD.91.054507}{Phys. Rev. {\bf D91},
  054507 (2015)}, \href{http://arxiv.org/abs/1411.7908}{{\tt arXiv:1411.7908
  [hep-lat]}}\relax
\mciteBstWouldAddEndPuncttrue
\mciteSetBstMidEndSepPunct{\mcitedefaultmidpunct}
{\mcitedefaultendpunct}{\mcitedefaultseppunct}\relax
\EndOfBibitem
\bibitem{Bazavov:2017lyh}
A.~Bazavov {\em et al.},
  \href{http://dx.doi.org/10.1103/PhysRevD.98.074512}{Phys. Rev. {\bf D98},
  074512 (2018)}, \href{http://arxiv.org/abs/1712.09262}{{\tt arXiv:1712.09262
  [hep-lat]}}\relax
\mciteBstWouldAddEndPuncttrue
\mciteSetBstMidEndSepPunct{\mcitedefaultmidpunct}
{\mcitedefaultendpunct}{\mcitedefaultseppunct}\relax
\EndOfBibitem
\bibitem{Cirigliano:2011tm}
V.~Cirigliano and H.~Neufeld,
  \href{http://dx.doi.org/10.1016/j.physletb.2011.04.038}{Phys. Lett. {\bf
  B700},  7 (2011)}, \href{http://arxiv.org/abs/1102.0563}{{\tt arXiv:1102.0563
  [hep-ph]}}\relax
\mciteBstWouldAddEndPuncttrue
\mciteSetBstMidEndSepPunct{\mcitedefaultmidpunct}
{\mcitedefaultendpunct}{\mcitedefaultseppunct}\relax
\EndOfBibitem
\bibitem{Marciano:2004uf}
W.~J. Marciano, \href{http://dx.doi.org/10.1103/PhysRevLett.93.231803}{Phys.
  Rev. Lett. {\bf 93},  231803 (2004)},
  \href{http://arxiv.org/abs/hep-ph/0402299}{{\tt arXiv:hep-ph/0402299}}\relax
\mciteBstWouldAddEndPuncttrue
\mciteSetBstMidEndSepPunct{\mcitedefaultmidpunct}
{\mcitedefaultendpunct}{\mcitedefaultseppunct}\relax
\EndOfBibitem
\bibitem{Bazavov:2014wgs}
{Fermilab Lattice, MILC} collaboration, A.~Bazavov {\em et al.},
  \href{http://dx.doi.org/10.1103/PhysRevD.90.074509}{Phys. Rev. {\bf D90},
  074509 (2014)}, \href{http://arxiv.org/abs/1407.3772}{{\tt arXiv:1407.3772
  [hep-lat]}}\relax
\mciteBstWouldAddEndPuncttrue
\mciteSetBstMidEndSepPunct{\mcitedefaultmidpunct}
{\mcitedefaultendpunct}{\mcitedefaultseppunct}\relax
\EndOfBibitem
\bibitem{Rosner:2015wva}
J.~L. Rosner, S.~Stone, and R.~S. Van~de Water,
  \href{http://arxiv.org/abs/1509.02220}{{\tt arXiv:1509.02220 [hep-ph]}}
  (2015)\relax
\mciteBstWouldAddEndPuncttrue
\mciteSetBstMidEndSepPunct{\mcitedefaultmidpunct}
{\mcitedefaultendpunct}{\mcitedefaultseppunct}\relax
\EndOfBibitem
\bibitem{Mohr:2015ccw}
P.~J. Mohr, D.~B. Newell, and B.~N. Taylor,
  \href{http://dx.doi.org/10.1103/RevModPhys.88.035009}{Rev. Mod. Phys. {\bf
  88},  035009 (2016)}, \href{http://arxiv.org/abs/1507.07956}{{\tt
  arXiv:1507.07956}}\relax
\mciteBstWouldAddEndPuncttrue
\mciteSetBstMidEndSepPunct{\mcitedefaultmidpunct}
{\mcitedefaultendpunct}{\mcitedefaultseppunct}\relax
\EndOfBibitem
\bibitem{Jamin:2008qg}
M.~Jamin, A.~Pich, and J.~Portoles,
  \href{http://dx.doi.org/10.1016/j.physletb.2008.04.049}{Phys. Lett. {\bf
  B664},  78 (2008)}, \href{http://arxiv.org/abs/0803.1786}{{\tt
  arXiv:0803.1786 [hep-ph]}}\relax
\mciteBstWouldAddEndPuncttrue
\mciteSetBstMidEndSepPunct{\mcitedefaultmidpunct}
{\mcitedefaultendpunct}{\mcitedefaultseppunct}\relax
\EndOfBibitem
\bibitem{Antonelli:2013usa}
M.~Antonelli, V.~Cirigliano, A.~Lusiani, and E.~Passemar,
  \href{http://dx.doi.org/10.1007/JHEP10(2013)070}{JHEP {\bf 10},  070 (2013)},
  \href{http://arxiv.org/abs/1304.8134}{{\tt arXiv:1304.8134 [hep-ph]}}\relax
\mciteBstWouldAddEndPuncttrue
\mciteSetBstMidEndSepPunct{\mcitedefaultmidpunct}
{\mcitedefaultendpunct}{\mcitedefaultseppunct}\relax
\EndOfBibitem
\bibitem{Hudspith:2017vew}
R.~J. Hudspith, R.~Lewis, K.~Maltman, and J.~Zanotti,
  \href{http://dx.doi.org/10.1016/j.physletb.2018.03.074}{Phys. Lett. {\bf
  B781},  206--212 (2018)}, \href{http://arxiv.org/abs/1702.01767}{{\tt
  arXiv:1702.01767 [hep-ph]}}\relax
\mciteBstWouldAddEndPuncttrue
\mciteSetBstMidEndSepPunct{\mcitedefaultmidpunct}
{\mcitedefaultendpunct}{\mcitedefaultseppunct}\relax
\EndOfBibitem
\bibitem{Boyle:2018dwv}
{RBC, UKQCD} collaboration, P.~Boyle, R.~J. Hudspith, T.~Izubuchi, A.~Jüttner,
  C.~Lehner, R.~Lewis, K.~Maltman, H.~Ohki, A.~Portelli, and M.~Spraggs,
  \href{http://dx.doi.org/10.1103/PhysRevLett.121.202003}{Phys. Rev. Lett. {\bf
  121},  202003 (2018)}, \href{http://arxiv.org/abs/1803.07228}{{\tt
  arXiv:1803.07228 [hep-lat]}}\relax
\mciteBstWouldAddEndPuncttrue
\mciteSetBstMidEndSepPunct{\mcitedefaultmidpunct}
{\mcitedefaultendpunct}{\mcitedefaultseppunct}\relax
\EndOfBibitem
\bibitem{Read:2002hq}
A.~L. Read, \href{http://dx.doi.org/10.1088/0954-3899/28/10/313}{J. Phys. {\bf
  G28},  2693 (2002)}\relax
\mciteBstWouldAddEndPuncttrue
\mciteSetBstMidEndSepPunct{\mcitedefaultmidpunct}
{\mcitedefaultendpunct}{\mcitedefaultseppunct}\relax
\EndOfBibitem
\bibitem{Banerjee:2007is}
S.~Banerjee, B.~Pietrzyk, J.~M. Roney, and Z.~Was,
  \href{http://dx.doi.org/10.1103/PhysRevD.77.054012}{Phys. Rev. {\bf D77},
  054012 (2008)}, \href{http://arxiv.org/abs/0706.3235}{{\tt arXiv:0706.3235
  [hep-ph]}}\relax
\mciteBstWouldAddEndPuncttrue
\mciteSetBstMidEndSepPunct{\mcitedefaultmidpunct}
{\mcitedefaultendpunct}{\mcitedefaultseppunct}\relax
\EndOfBibitem
\bibitem{junk:2007:cdfnote}
{CDF} collaboration, CDF note 8128, 2007, {\small
  \url{{http://www-cdf.fnal.gov/physics/statistics/notes/cdf8128_mclimit_csm_v2.pdf}}}\relax
\mciteBstWouldAddEndPuncttrue
\mciteSetBstMidEndSepPunct{\mcitedefaultmidpunct}
{\mcitedefaultendpunct}{\mcitedefaultseppunct}\relax
\EndOfBibitem
\bibitem{Aubert:2009ag}
{\babar} collaboration, B.~Aubert {\em et al.},
  \href{http://dx.doi.org/10.1103/PhysRevLett.104.021802}{Phys. Rev. Lett. {\bf
  104},  021802 (2010)}, \href{http://arxiv.org/abs/0908.2381}{{\tt
  arXiv:0908.2381 [hep-ex]}}\relax
\mciteBstWouldAddEndPuncttrue
\mciteSetBstMidEndSepPunct{\mcitedefaultmidpunct}
{\mcitedefaultendpunct}{\mcitedefaultseppunct}\relax
\EndOfBibitem
\bibitem{Hayasaka:2007vc}
{Belle} collaboration, K.~Hayasaka {\em et al.},
  \href{http://dx.doi.org/10.1016/j.physletb.2008.06.056}{Phys. Lett. {\bf
  B666},  16 (2008)}, \href{http://arxiv.org/abs/0705.0650}{{\tt
  arXiv:0705.0650 [hep-ex]}}\relax
\mciteBstWouldAddEndPuncttrue
\mciteSetBstMidEndSepPunct{\mcitedefaultmidpunct}
{\mcitedefaultendpunct}{\mcitedefaultseppunct}\relax
\EndOfBibitem
\bibitem{Aubert:2006cz}
{\babar} collaboration, B.~Aubert {\em et al.},
  \href{http://dx.doi.org/10.1103/PhysRevLett.98.061803}{Phys. Rev. Lett. {\bf
  98},  061803 (2007)}, \href{http://arxiv.org/abs/hep-ex/0610067}{{\tt
  arXiv:hep-ex/0610067 [hep-ex]}}\relax
\mciteBstWouldAddEndPuncttrue
\mciteSetBstMidEndSepPunct{\mcitedefaultmidpunct}
{\mcitedefaultendpunct}{\mcitedefaultseppunct}\relax
\EndOfBibitem
\bibitem{Miyazaki:2007jp}
{Belle} collaboration, Y.~Miyazaki {\em et al.},
  \href{http://dx.doi.org/10.1016/j.physletb.2007.03.027}{Phys. Lett. {\bf
  B648},  341 (2007)}, \href{http://arxiv.org/abs/hep-ex/0703009}{{\tt
  arXiv:hep-ex/0703009 [hep-ex]}}\relax
\mciteBstWouldAddEndPuncttrue
\mciteSetBstMidEndSepPunct{\mcitedefaultmidpunct}
{\mcitedefaultendpunct}{\mcitedefaultseppunct}\relax
\EndOfBibitem
\bibitem{Aubert:2009ys}
{\babar} collaboration, B.~Aubert {\em et al.},
  \href{http://dx.doi.org/10.1103/PhysRevD.79.012004}{Phys. Rev. {\bf D79},
  012004 (2009)}, \href{http://arxiv.org/abs/0812.3804}{{\tt arXiv:0812.3804
  [hep-ex]}}\relax
\mciteBstWouldAddEndPuncttrue
\mciteSetBstMidEndSepPunct{\mcitedefaultmidpunct}
{\mcitedefaultendpunct}{\mcitedefaultseppunct}\relax
\EndOfBibitem
\bibitem{Miyazaki:2010qb}
{Belle} collaboration, Y.~Miyazaki {\em et al.},
  \href{http://dx.doi.org/10.1016/j.physletb.2010.07.012}{Phys. Lett. {\bf
  B692},  4 (2010)}, \href{http://arxiv.org/abs/1003.1183}{{\tt arXiv:1003.1183
  [hep-ex]}}\relax
\mciteBstWouldAddEndPuncttrue
\mciteSetBstMidEndSepPunct{\mcitedefaultmidpunct}
{\mcitedefaultendpunct}{\mcitedefaultseppunct}\relax
\EndOfBibitem
\bibitem{Aubert:2009ap}
{\babar} collaboration, B.~Aubert {\em et al.},
  \href{http://dx.doi.org/10.1103/PhysRevLett.103.021801}{Phys. Rev. Lett. {\bf
  103},  021801 (2009)}, \href{http://arxiv.org/abs/0904.0339}{{\tt
  arXiv:0904.0339 [hep-ex]}}\relax
\mciteBstWouldAddEndPuncttrue
\mciteSetBstMidEndSepPunct{\mcitedefaultmidpunct}
{\mcitedefaultendpunct}{\mcitedefaultseppunct}\relax
\EndOfBibitem
\bibitem{Miyazaki:2011xe}
{Belle} collaboration, Y.~Miyazaki,
  \href{http://dx.doi.org/10.1016/j.physletb.2011.04.011}{Phys. Lett. {\bf
  B699},  251 (2011)}, \href{http://arxiv.org/abs/1101.0755}{{\tt
  arXiv:1101.0755 [hep-ex]}}\relax
\mciteBstWouldAddEndPuncttrue
\mciteSetBstMidEndSepPunct{\mcitedefaultmidpunct}
{\mcitedefaultendpunct}{\mcitedefaultseppunct}\relax
\EndOfBibitem
\bibitem{Aubert:2007kx}
{\babar} collaboration, B.~Aubert {\em et al.},
  \href{http://dx.doi.org/10.1103/PhysRevLett.100.071802}{Phys. Rev. Lett. {\bf
  100},  071802 (2008)}, \href{http://arxiv.org/abs/0711.0980}{{\tt
  arXiv:0711.0980 [hep-ex]}}\relax
\mciteBstWouldAddEndPuncttrue
\mciteSetBstMidEndSepPunct{\mcitedefaultmidpunct}
{\mcitedefaultendpunct}{\mcitedefaultseppunct}\relax
\EndOfBibitem
\bibitem{Miyazaki:2008mw}
{Belle} collaboration, Y.~Miyazaki {\em et al.},
  \href{http://dx.doi.org/10.1016/j.physletb.2009.01.058}{Phys. Lett. {\bf
  B672},  317 (2009)}, \href{http://arxiv.org/abs/0810.3519}{{\tt
  arXiv:0810.3519 [hep-ex]}}\relax
\mciteBstWouldAddEndPuncttrue
\mciteSetBstMidEndSepPunct{\mcitedefaultmidpunct}
{\mcitedefaultendpunct}{\mcitedefaultseppunct}\relax
\EndOfBibitem
\bibitem{Lees:2010ez}
{\babar} collaboration, J.~P. Lees {\em et al.},
  \href{http://dx.doi.org/10.1103/PhysRevD.81.111101}{Phys. Rev. {\bf D81},
  111101 (2010)}, \href{http://arxiv.org/abs/1002.4550}{{\tt arXiv:1002.4550
  [hep-ex]}}\relax
\mciteBstWouldAddEndPuncttrue
\mciteSetBstMidEndSepPunct{\mcitedefaultmidpunct}
{\mcitedefaultendpunct}{\mcitedefaultseppunct}\relax
\EndOfBibitem
\bibitem{Hayasaka:2010np}
{Belle} collaboration, K.~Hayasaka {\em et al.},
  \href{http://dx.doi.org/10.1016/j.physletb.2010.03.037}{Phys. Lett. {\bf
  B687},  139 (2010)}, \href{http://arxiv.org/abs/1001.3221}{{\tt
  arXiv:1001.3221 [hep-ex]}}\relax
\mciteBstWouldAddEndPuncttrue
\mciteSetBstMidEndSepPunct{\mcitedefaultmidpunct}
{\mcitedefaultendpunct}{\mcitedefaultseppunct}\relax
\EndOfBibitem
\bibitem{Aad:2016wce}
{ATLAS} collaboration, G.~Aad {\em et al.},
  \href{http://dx.doi.org/10.1140/epjc/s10052-016-4041-9}{Eur. Phys. J. {\bf
  C76},  232 (2016)}, \href{http://arxiv.org/abs/1601.03567}{{\tt
  arXiv:1601.03567 [hep-ex]}}\relax
\mciteBstWouldAddEndPuncttrue
\mciteSetBstMidEndSepPunct{\mcitedefaultmidpunct}
{\mcitedefaultendpunct}{\mcitedefaultseppunct}\relax
\EndOfBibitem
\bibitem{Aaij:2014azz}
{LHCb} collaboration, R.~Aaij {\em et al.},
  \href{http://dx.doi.org/10.1007/JHEP02(2015)121}{JHEP {\bf 02},  121 (2015)},
  \href{http://arxiv.org/abs/1409.8548}{{\tt arXiv:1409.8548 [hep-ex]}}\relax
\mciteBstWouldAddEndPuncttrue
\mciteSetBstMidEndSepPunct{\mcitedefaultmidpunct}
{\mcitedefaultendpunct}{\mcitedefaultseppunct}\relax
\EndOfBibitem
\bibitem{Aubert:2005tp}
{\babar} collaboration, B.~Aubert {\em et al.},
  \href{http://dx.doi.org/10.1103/PhysRevLett.95.191801}{Phys. Rev. Lett. {\bf
  95},  191801 (2005)}, \href{http://arxiv.org/abs/hep-ex/0506066}{{\tt
  arXiv:hep-ex/0506066 [hep-ex]}}\relax
\mciteBstWouldAddEndPuncttrue
\mciteSetBstMidEndSepPunct{\mcitedefaultmidpunct}
{\mcitedefaultendpunct}{\mcitedefaultseppunct}\relax
\EndOfBibitem
\bibitem{Miyazaki:2012mx}
{Belle} collaboration, Y.~Miyazaki {\em et al.},
  \href{http://dx.doi.org/10.1016/j.physletb.2013.01.032}{Phys. Lett. {\bf
  B719},  346 (2013)}, \href{http://arxiv.org/abs/1206.5595}{{\tt
  arXiv:1206.5595 [hep-ex]}}\relax
\mciteBstWouldAddEndPuncttrue
\mciteSetBstMidEndSepPunct{\mcitedefaultmidpunct}
{\mcitedefaultendpunct}{\mcitedefaultseppunct}\relax
\EndOfBibitem
\bibitem{Miyazaki:2005ng}
{Belle} collaboration, Y.~Miyazaki {\em et al.},
  \href{http://dx.doi.org/10.1016/j.physletb.2005.10.024}{Phys. Lett. {\bf
  B632},  51 (2006)}, \href{http://arxiv.org/abs/hep-ex/0508044}{{\tt
  arXiv:hep-ex/0508044 [hep-ex]}}\relax
\mciteBstWouldAddEndPuncttrue
\mciteSetBstMidEndSepPunct{\mcitedefaultmidpunct}
{\mcitedefaultendpunct}{\mcitedefaultseppunct}\relax
\EndOfBibitem
\bibitem{Aaij:2013fia}
{LHCb} collaboration, R.~Aaij {\em et al.},
  \href{http://dx.doi.org/10.1016/j.physletb.2013.05.063}{Phys. Lett. {\bf
  B724},  36 (2013)}, \href{http://arxiv.org/abs/1304.4518}{{\tt
  arXiv:1304.4518 [hep-ex]}}\relax
\mciteBstWouldAddEndPuncttrue
\mciteSetBstMidEndSepPunct{\mcitedefaultmidpunct}
{\mcitedefaultendpunct}{\mcitedefaultseppunct}\relax
\EndOfBibitem
\end{mcitethebibliography}
\end{small}

\end{document}